\documentclass[12pt]{book}

\usepackage{makeidx}

\usepackage{amsfonts}
\usepackage{amsmath}
\usepackage{fullpage}

\usepackage{amssymb}
\usepackage{graphicx}
\usepackage{makeidx}
\usepackage{hyperref}
\usepackage{fancyhdr}
\usepackage[calcwidth]{titlesec}
\usepackage[font={footnotesize},labelfont={bf}]{caption}
\usepackage{setspace}

\usepackage{extarrows}

\usepackage[most]{tcolorbox}
\newtcolorbox{myt}[2][]{%
  attach boxed title to top center
               = {yshift=-4pt},
  colback      = blue!5!white,
  colframe     = blue!75!black,
  halign       = flush left,
  fonttitle    = \bfseries\sffamily,
  colbacktitle = blue!65!black,
  title        = #2,#1,
  enhanced,
}
\newtcolorbox{myd}[2][]{%
  attach boxed title to top center
               = {yshift=-4pt},
  colback      = violet!5!white,
  colframe     = violet!75!black,
  halign       = flush left,
  fonttitle    = \bfseries\sffamily,
  colbacktitle = violet!65!black,
  title        = #2,#1,
  enhanced,
}
\newtcolorbox{mye}[2][]{%
  attach boxed title to top center
               = {yshift=-4pt},
  colback      = purple!5!white,
  colframe     = purple!75!black,
  halign       = flush left,
  fonttitle    = \bfseries\sffamily,
  colbacktitle = purple!65!black,
  title        = #2,#1,
  enhanced,
}

\newtcolorbox{myg}[2][]{%
  attach boxed title to top center
               = {yshift=-4pt},
  colback      = green!5!white,
  colframe     = green!50!black,
  halign       = flush left,
  fonttitle    = \bfseries\sffamily,
  colbacktitle = green!65!black,
  title        = #2,#1,
  enhanced,
}

\providecommand{\U}[1]{\protect\rule{.1in}{.1in}}

\usepackage{graphics}

\usepackage{amsthm}
\usepackage{color}
\usepackage{dsfont}

\usepackage{mathtools}

\DeclarePairedDelimiter\ceil{\lceil}{\rceil}
\DeclarePairedDelimiter\floor{\lfloor}{\rfloor}

\def\cmo{{\rm CMO}}
\def\stab{{\rm Stab}}
\def\ppt{{\rm PPT}}
\def\wit{{\rm WIT}}

\def\spec{{\rm spec}}
\def\reg{{\rm reg}}
\def\rel{{\rm rel}}
\def\pure{{\rm Pure}}
\def\pos{{\rm Pos}}
\def\cp{{\rm CP}}
\def\cptp{{\rm CPTP}}

\def\im{{\rm Im}}
\def\ker{{\rm Ker}}
\def\irr{{\rm Irr}}
\def\sym{{\rm Sym}}
\def\asy{{\rm Asy}}

\def\sign{{\rm sign}}
\def\tho{{\rm TO}}
\def\gp{{\rm GPO}}
\def\gpc{{\rm GPC}}
\def\noisy{{\rm Noisy}}
\def\cto{{\rm CTO}}
\def\distill{{\rm Distill}}
\def\cost{{\rm Cost}}
\def\stoc{{\rm STOC}}
\def\type{{\rm Type}}
\def\st{{\rm st}}

\def\cds{{\rm CDS}}

\def\dual{{\rm dual}}
\def\sep{{\rm SEP}}
\def\pio{{\rm PIO}}
\def\sr{{\rm SR}}
\def\slip{{\rm SLIP}}
\def\stable{{\rm Stable}}
\def\diag{{\rm Diag}}
\def\tot{{\rm tot}}

\def\LR{\mathbf{LR}}
\def\LN{\mathbf{LN}}

\def\N{\mathbf{N}}

\def\BB{\boldsymbol{\{}}
\def\EE{\boldsymbol{\}}}

\def\1{\mathbf{1}}

\def\up{\underline{\mathbf{p}}}

\def\M{\mathbf{M}}
\def\E{\mathbf{E}}
\def\R{\mathbf{R}}
\def\A{\mathbb{A}}
\def\F{\mathbf{F}}

\def\mbD{\mathbf{D}}

\def\>{\rangle}
\def\<{\langle}
\def\id{\mathsf{id}}
\def\mB{\mathcal{B}}
\def\mE{\mathcal{E}}

\def\mF{\mathcal{F}}
\def\mN{\mathcal{N}}
\def\mC{\mathcal{C}}
\def\mL{\mathcal{L}}
\def\mP{\mathcal{P}}
\def\mS{\mathcal{S}}
\def\mD{\mathcal{D}}
\def\mT{\mathcal{T}}
\def\mV{\mathcal{V}}
\def\mG{\mathcal{G}}

\def\mX{\mathcal{X}}
\def\mY{\mathcal{Y}}
\def\mZ{\mathcal{Z}}

\newcommand{\supp}{\operatorname{supp}}

   \newcommand{\rl}{\rangle\langle}
   \newcommand{\lp}{\big\langle\!\!\!\big\langle\!\!\!\big\langle\!\!\!\big\langle}
   \newcommand{\rp}{\big\rangle\!\!\!\big\rangle\!\!\!\big\rangle\!\!\!\big\rangle}

\renewcommand{\qedsymbol}{\nobreak \ifvmode \relax \else
	\ifdim \lastskip<1.5em \hskip-\lastskip \hskip1.5em plus0em
	minus0.5em \fi \nobreak \vrule height0.75em width0.5em
	depth0.25em\fi}

\renewcommand{\le}{\leqslant}
\renewcommand{\geq}{\geqslant}
\renewcommand{\leq}{\leqslant}

\theoremstyle{definition}
\newtheorem{theorem}{Theorem}[chapter]
\newtheorem{corollary}{Corollary}[chapter]
\newtheorem{lemma}{Lemma}[chapter]

\theoremstyle{definition}
\newtheorem{definition}{Definition}[chapter]

\theoremstyle{remark}
\newtheorem*{remark}{Remark}

\theoremstyle{exercise}
\newtheorem{exercise}{Exercise}[chapter]

\newcommand{\bea}{\begin{eqnarray}}
\newcommand{\eea}{\end{eqnarray}}
\newcommand{\be}{\begin{equation}}
\newcommand{\ee}{\end{equation}}
\newcommand{\ba}{\begin{equation}\begin{aligned}}
\newcommand{\ea}{\end{aligned}\end{equation}}
\newcommand{\bs}{\boldsymbol}

\newcommand{\epm}{\end{pmatrix}}
\newcommand{\bpm}{\begin{pmatrix}}

\newcommand{\ebm}{\end{bmatrix}}
\newcommand{\bbm}{\begin{bmatrix}}

\newtheorem*{example}{Example}

\newcommand{\bex}{\begin{exercise}}
\newcommand{\eex}{\end{exercise}}

\newcommand{\ben}{\begin{enumerate}}
\newcommand{\een}{\end{enumerate}}

\def\be{\begin{equation}}
\def\ee{\end{equation}}

\newcommand{\rank}{{\rm Rank}}
\newcommand{\spa}{{\rm span}}
\newcommand{\mU}{\mathcal{U}}
\newcommand{\mA}{\mathcal{A}}
\newcommand{\error}{{\rm error}}
\newcommand{\herm}{{\rm Herm}}
\newcommand{\uD}{\underline{\mathbb{D}}}
\newcommand{\oD}{\overline{\mathbb{D}}}
\newcommand{\inv}{{\rm INV}}
\newcommand{\cov}{{\rm COV}}

\newcommand{\G}{\mathbf{G}}
\newcommand{\bS}{\mathbf{S}}
\newcommand{\bH}{\mathbf{H}}

\newcommand{\mI}{\mathcal{I}}
\newcommand{\mH}{\mathcal{H}}

\newcommand{\ml}{\mathfrak{L}}
\newcommand{\mb}{\mathfrak{B}}
\newcommand{\md}{\mathfrak{D}}
\newcommand{\mf}{\mathfrak{F}}
\newcommand{\ms}{\mathfrak{S}}

\newcommand{\mfu}{\mathfrak{U}}
\newcommand{\mt}{\mathfrak{T}}
\newcommand{\mc}{\mathfrak{C}}
\newcommand{\mk}{\mathfrak{K}}

\newcommand{\mi}{\mathfrak{I}}

\newcommand{\bD}{\overline{D}}

\newcommand{\uF}{\underline{F}}

\newcommand{\mR}{\mathcal{R}}
\newcommand{\mM}{\mathcal{M}}

\newcommand{\mW}{\mathcal{W}}
\newcommand{\mK}{\mathcal{K}}

\newcommand{\D}{\mathbb{D}}
\newcommand{\h}{\mathbb{H}}

\newcommand{\lr}{\rangle\langle}
\newcommand{\la}{\langle}
\newcommand{\ra}{\rangle}
\newcommand{\tr}{{\rm Tr}}

\newcommand{\ua}{\uparrow}
\newcommand{\da}{\downarrow}

\newcommand{\conv}{{\rm Conv}}
\newcommand{\locc}{{\rm LOCC}}

\newcommand{\slocc}{{\rm SLOCC}}

\newcommand{\vect}{{\rm\bf vec}}

\newcommand{\eps}{\varepsilon}
\newcommand{\teps}{\tilde{\varepsilon}}

\newcommand{\mbb}[1]{\mathbb{#1}}

\newcommand{\mr}{\mathfrak{R}}
\newcommand{\muu}{\mathfrak{U}}

  \def\H{\mathbb{H}}

\newcommand{\eqdef}{\coloneqq}

\def\r{\mathbf{r}}
\def\s{\mathbf{s}}
\def\p{\mathbf{p}}
\def\q{\mathbf{q}}
\def\g{\mathbf{g}}
\def\e{\mathbf{e}}
\def\x{\mathbf{x}}
\def\y{\mathbf{y}}
\def\z{\mathbf{z}}
\def\t{\mathbf{t}}
\def\u{\mathbf{u}}
\def\v{\mathbf{v}}
\def\w{\mathbf{w}}
\def\0{\mathbf{0}}
\def\a{\mathbf{a}}
\def\b{\mathbf{b}}
\def\m{\mathbf{m}}
\def\n{\mathbf{n}}

\def\tA{\tilde{A}}
\def\tB{\tilde{B}}

\def\tD{\tilde{D}}
\def\tE{\tilde{E}}
\def\tX{\tilde{X}}

\def\tV{\tilde{V}}
\def\tQ{\tilde{Q}}

\def\tp{\tilde{\mathbf{p}}}
\def\tq{\tilde{\mathbf{q}}}
\def\ta{\tilde{\mathbf{a}}}

\def\tH{\tilde{H}}
\def\tmE{\tilde{\mathcal{E}}}
\def\tmN{\tilde{\mathcal{N}}}
\def\trho{\tilde{\rho}}
\def\tsigma{\tilde{\sigma}}
\def\teta{\tilde{\eta}}

\def\ttau{\tilde{\tau}}
\def\tomega{\tilde{\omega}}
\def\tpsi{{\tilde{\psi}}}
\def\tphi{\tilde{\phi}}

\def\tV{\tilde{V}}
\def\tvarphi{\tilde{\varphi}}
\def\tgamma{\tilde{\gamma}}

\def\tg{\tilde{\mathbf{g}}}

\def\eff{{\rm Eff}}
\def\pr{{\rm Pr}}
\def\rng{{\rm RNG}}
\def\prob{{\rm Prob}}
\def\drng{\rm dRNG}

\newcommand{\GG}[1]{\rm \textcolor{red}{ #1}{\color{red} \to}}

\newcommand{\Gg}[1]{
  \text{\fcolorbox{black}{red!20}{\textcolor{black}{\scriptsize$#1$}}}{\color{red} \xrightarrow{\hspace{0.5cm}\;\hspace{0.5cm}}}
}

\makeindex

\begin{document}

	%\SetWatermarkText{NOT FOR DISTRIBUTION}
	%\SetWatermarkAngle{60}
	%\SetWatermarkScale{0.5}
	
	\title{Resources of the Quantum World\\ \large A modern textbook on quantum resource theories\\ $\;$\\\textbf{Volume 1: Static Resources}}

	\author{Gilad Gour}
	
	\date{\today}

	\maketitle

\pagestyle{fancy}
\fancyhf{}
\fancyhead[RE]{\slshape \leftmark}
\fancyhead[LO]{\slshape \rightmark}
\fancyhead[RO,LE]{\thepage}
\fancyfoot[LO,LE]{{\scriptsize Gilad Gour, Technion - Israel Institute of Technology, University of Calgary}}
\setlength{\headsep}{30pt}
\setlength{\voffset}{-50pt}
\titleformat{\section}[hang]{\bfseries}
{\Large\thesection}{12pt}{\Large}[{\titlerule[0.5pt]}]
\titleformat{\chapter}[display]
{\normalfont\Large\filcenter}
{\titlerule[1pt]
\vspace{1pt}
\titlerule
\vspace{1pc}
\bfseries\LARGE\MakeUppercase{\chaptertitlename} \thechapter}
{1pc}
{\titlerule
\vspace{1pc}
\Huge\bfseries}
\setcounter{tocdepth}{1}
%\makeindex

%\frontmatter  % Start front matter

\begin{titlepage}
    \centering
    \vspace*{2cm}
   % {\Huge \textbf{\textit{Dedication}}}
    
    \vspace{4cm}
    
    \emph{Dedicated to the memory of my beloved father, \\ [0.2cm]
    Gideon Gour\\ [0.2cm]
    1949 -- 2021}

    \vfill

    \vspace{2cm}
\end{titlepage}

\tableofcontents

\makeatother

\chapter*{Acknowledgements}

I extend my deepest gratitude to the numerous colleagues who have profoundly influenced my understanding of quantum information and quantum resource theories. Through countless discussions, their insights have enriched my perspective and deepened my knowledge. I am particularly indebted to Fernando G.S.L. Brandão, S. Brandsen, Francesco Buscemi, Giulio Chiribella, Eric Chitambar, Nilanjana Datta, Julio De Vicente, Runyao Duan, Kun Fang, Shmuel Friedland, Yu Guo, Aram Harrow, Michał Horodecki, David Jennings, Amir Kalev, Barbara Kraus, Ludovico Lami, Iman Marvian, David A. Meyer, Markus P. Müller, Varun Narasimhachar, Jonathan Oppenheim, Carlo Maria Scandolo, Bartosz Regula, Robert W. Spekkens, Ryuji Takagi, Marco Tomamichel, Nolan Wallach, Xin Wang, Mark M. Wilde, Andreas Winter, and Nicole Yunger Halpern for their invaluable contributions.

My sincere appreciation goes to Julio Inigo De Vicente Majua for his meticulous review of the chapter on multipartite entanglement, offering numerous improvements and corrections. Special thanks to Thomas Theurer and Gershon Wolansky, whose relentless feedback on various drafts has been instrumental in refining this work. Additionally, I am grateful to Mark M. Wilde for his advice on enhancing the notations and clarity of presentation. I am also grateful to Ryuji Takagi for identifying errors and typos in the initial version of the book, as well as for offering a streamlined proof for Theorem~\ref{thmdminpe}.

I owe a debt of gratitude to the many students I've had the privilege of interacting with over the years. Their keen observations and identification of errors and typos in various drafts have been crucial in shaping the final content of this book. I thank John Burniston, Nuiok Decaire, Raz Firanko, Kimberly Golubeva, Michael Grabowecky, Alexander Hickey, Takla Nateeboon, Gaurav Saxena, Kuntal Sengupta, Guy Shemesh, Samuel Steakley, Goni Yoeli, and Elia Zanoni for their contributions.

Lastly, but most importantly, I wish to express my heartfelt appreciation to my family. To my parents, Iris and Gideon Gour, whose unwavering support and belief in my pursuits have been the bedrock of my resilience and determination. Their love and encouragement have been a constant source of strength throughout this journey. To my children, Sophia and Elijah Gour, who have been my greatest source of inspiration. Their curiosity and enthusiasm for life remind me daily of the joy of discovery and the importance of sharing knowledge. And to my life partner, Eve Zhang, whose endless support and understanding have been nothing short of miraculous. Her presence and encouragement have been my guiding light, helping me navigate the challenges of this endeavor and crossing the finish line of completing this book. My journey is as much theirs as it is mine, and I am eternally grateful for their love, patience, and sacrifice.

\chapter*{Notations}

First letters of the English alphabet, such as $A$, $B$, and $C$, are used to denote \emph{both} quantum physical systems and their corresponding Hilbert spaces. The letter $R$ is used to denote a quantum \emph{reference} physical system (and its corresponding Hilbert space), and sometimes the letter $E$ is used to denote the \emph{environment} system. The last letters of the English alphabet, such as $X$, $Y$, and $Z$ are used to denote classical systems or classical registers. The dimension of a Hilbert space is denoted with vertical lines; e.g.\ the dimension of systems $A$, $B$, and $X$, are denoted respectively as $|A|$, $|B|$, and $|X|$. The tilde symbol above a system always represents a replica of the system. For example, $\tA$ represents another copy of system $A$ and in particular $|A|=|\tA|$.

We use $|\Omega^{A\tA}\ra$ to denote the unnormalized maximally entangled state $\sum_{x\in[m]}|xx\ra$,
and
$|\Phi^{A\tA}\ra$ the \emph{normalized} maximally entangled state $\frac1{\sqrt{|A|}}\sum_{x\in[m]}|xx\ra$.
We also use notations $\psi,\phi,\Omega,\Phi$ to denotes respectively the rank one pure states $|\psi\lr\psi|,|\phi\lr\phi|,|\Omega\lr\Omega|,|\Phi\lr\Phi|$.

\chapter*{List of Symbols}
\addcontentsline{toc}{chapter}{List of Symbols}

\begin{tabular}{|c|l|}
\hline
$[n]$ & The set $\{1,\ldots,n\}$\\ \hline
  $|A|$ & Dimension of the Hilbert space $A$\\ \hline
    $\mbb{C}^{m\times n}$ , $\mbb{R}^{m\times n}$ & The set of all $m\times n$ complex and real matrices\\ \hline
    $\0_{m,n}$ & The  $m\times n$ zero matrix\\ \hline
  $\mbb{R}^{m\times n}_+$ & The set of all matrices in $\mbb{R}^{m\times n}$ with non-negative components.\\ \hline
  $[a,b]^n$ & The set of all vectors in $\mbb{R}^n$ whose components are between $a$ and $b$.\\
  \hline
  $\prob(n)$ & The set of all probability vectors in $\mbb{R}^n_+$\\ \hline
  $\prob_{>0}(n)$ & The vectors in $\prob(n)$ whose components are all positive.\\ \hline
  $\prob(m,n)$ & The set of matrices in $\mbb{R}_{+}^{m\times n}$ whose components sum to one\\ \hline
$\prob^\da(n)$ & The set of all vectors in $\prob(n)$ whose components are arranged in\\ & non-increasing order\\ \hline  
$\pr(X=x)$ & The probability that a random variable $X$ equals $x$\\ \hline  
$\stoc(m,n)$ & The set of all $m\times n$ column stochastic matrices\\ \hline 
  $\diag(c_1,\ldots,c_n)$ &An $n\times n$ diagonal matrix with $c_1,\ldots,c_n$ on its diagonal\\ \hline
  $\ml(A,B)$ & The set of all linear operators from the Hilbert space $A$ to $B$\\\hline  
  $\ml(A)$ & The set of all linear operators from the Hilbert space $A$ to itself \\ \hline
  $\pos(A)$ & The set of all positive semidefinite operators in $\ml(A)$ \\ \hline
  $\pos_{>0}(A)$ & The set of all positive definite operators in $\ml(A)$ \\ \hline
  $\md(A)$ & The set of all density matrices (mixed quantum states) in $\pos(A)$\\ \hline
  $\md_{\leq}(A)$ & The set of all the elements in $\pos(A)$ with trace no greater than one\\ \hline
    \end{tabular}

\begin{tabular}{|c|l|}
\hline
$\pure(A)$ & The set of all pure (i.e.\ rank one) density matrices in $\md(A)$\\ \hline
$\herm(A)$ & The set of all Hermitian operators in $\ml(A)$ \\ \hline
$\ml(A\to B)$ & The set of all linear operators form $\ml(A)$ to $\ml(B)$\\ \hline
$\herm(A\to B)$ & The subset $\{\mE\in\ml(A\to B)\;:\;\mE(\rho)\in\herm(B)\quad\forall\;\rho\in\herm(A)\}$\\ \hline
  $\cp(A\to B)$ & The set of all completely positive\index{completely positive} maps in $\ml(A\to B)$\\ \hline
  $\cptp(A\to B)$ & The set of all quantum channels in $\ml(A\to B)$\\ \hline
  $\pos(A\to B)$ & The set of all positive maps in $\ml(A\to B)$\\ \hline
$\id^A$ & The identity element (channel) of $\ml(A\to A)$\\ 
\hline
$\#_f$ & The Kubo-Ando Operator Mean (Definition~\ref{kaom})\\
\hline
$\irr(\pi)$ & The set of all irreps (up to equivalency) appearing in the\\ & decomposition of $\pi$\\
\hline
$I_n$ & $n\times n$ identity matrix\\
\hline
$I^A$ & The identity operator in $\ml(A)$\\
\hline
$\u^A$ & The maximally mixed state in $\ml(A)$\\
\hline
$\u^{(n)}$ & The uniform probability vector in $\prob(n)$\\
\hline
$\mathbf{1}_n$ & The column vector $(1,\ldots 1)^T$ in $\mbb{R}^n$\\
\hline
$|\Omega^{A\tA}\ra$ & The \emph{unnormalized} maximally entangled state $\sum_{x\in[m]}|xx\ra$\\ \hline
$|\Phi^{A\tA}\ra$ & The \emph{normalized} maximally entangled state $\frac1{\sqrt{|A|}}\sum_{x\in[m]}|xx\ra$\\
\hline
$\eff(A)$ & The set of all effects in $\pos(A)$; i.e. $\Lambda\in\eff(A)$ if and only if $\0\leq \Lambda\leq I^A$.\\
\hline
$\im(T)$ & The image of $T\in\ml(A,B)$.\\
\hline
$\ker(T)$ & The kernel of $T\in\ml(A,B)$.\\
\hline
$\supp(T)$ & The support subspace of $T\in\ml(A,B)$.\\
\hline
$\supp(\p)$ & The set $\{x\in[n]\;:\;p_x>0\}$, where $\p=(p_1,\ldots,p_n)^T\in\prob(n)$.\\
\hline
$\rho\ll\sigma$ & Inclusion of supports; $\supp(\rho)\subseteq\supp(\sigma)$ for $\rho,\sigma\in\md(A)$.\\
\hline
$\spec(H)$ & The set of all distinct eigenvalues of an Hermitian operator $H\in\herm(A)$.\\
\hline
\end{tabular}

\begin{tabular}{|c|l|}
\hline
$\|\cdot\|_p$ & The Schatten\index{Schatten}  $p$-norm with $p\in[1,\infty]$ (see~\eqref{1pnorm}).\\
\hline
$\|\cdot\|_{(k)}$ & The Ky Fan norm with $k\in[n]$ (see Definition~\ref{def:kyfan} and~\eqref{kfnorm}).\\
\hline
$\mb_{\eps}(\p)$ & The set $\{\p'\in\prob(n)\;:\;\frac12\|\p-\p'\|_1\leq\eps\}$, where $\p\in\prob(n)$ and $\eps\in[0,1]$.\\
\hline
$\mb_{\eps}(\rho)$ & The set $\{\rho'\in\md(A)\;:\;\frac12\|\rho-\rho'\|_1\leq\eps\}$, where $\rho\in\md(A)$ and $\eps\in[0,1]$.\\
\hline
$\sigma\approx_\eps\rho$ & Short notation for $\sigma\in\mb_\eps(\rho)$.\\
\hline
$\Pi_\rho$ & Projection to the support of $\rho$.\\
\hline
$\mt_{\eps}(X^n)$ & The set of $\eps$-typical sequences.\\
\hline
$\mt_{\eps}(A^n)$ & The $\eps$-typical subspace of $A^n$.\\
\hline
$\mt_{\eps}^\st(X^n)$ & The set of strongly $\eps$-typical sequences.\\
\hline
$\mt_{\eps}^\st(A^n)$ & The strongly $\eps$-typical subspace of $A^n$.\\
\hline
$\ppt(AB)$ & The set of density matrices in $\md(AB)$ with positive partial transpose.\\
\hline
$\rho^\Gamma$ & The partial transpose of a bipartite state $\rho\in\md(AB)$ w.r.t.\ system $B$.\\
\hline
$a \gtrapprox b$ & $a \geq \log\left\lfloor 2^b \right\rfloor$, $a,b\in\mbb{R}_{+}$\\
\hline
\end{tabular}

\chapter{Introductory Material}

\section{Introduction}

A recurring theme in the field of physics is the endeavor to unify a variety of distinct physical phenomena into a comprehensive framework that can offer both descriptions and explanations for each of them. One of the most astounding achievements in this endeavor is the unification of fundamental forces. When physicists realized that the forces of electricity and magnetism could be elegantly described using a single framework, it not only substantially enhanced our comprehension of these forces but also gave birth to the expansive domain of electromagnetism.

The remarkable success in unifying forces serves as a testament to the fact that seemingly unrelated phenomena can often be traced back to a common origin. This approach extends beyond the realm of forces and finds resonance in the burgeoning field of quantum information science. Within this field, a novel discipline has emerged, which seeks to identify shared characteristics among seemingly disparate quantum phenomena. The overarching theme of this approach lies in the recognition that various attributes of physical systems can be defined as ``resources." This recognition not only alters our perspective on these phenomena but also seamlessly integrates them within a comprehensive framework known as ``quantum resource theories." 

For example, take the case of quantum entanglement. In the 1990s, it was transformed from a topic of philosophical debates and discussions into a valuable resource. This transformative shift revolutionized our perception of entanglement; it evolved from being an intriguing and non-intuitive phenomenon into the essential driving force behind numerous quantum information tasks. This new perspective on entanglement opened up a vast array of possibilities and applications, starting with its utilization in quantum teleportation and superdense coding. Today, entanglement stands as a fundamental resource in fields such as quantum communication, quantum cryptography, and quantum computing.

Given the success of entanglement theory, it is only natural to explore other physical phenomena that can also be recognized as valuable resources. Currently, there are several quantum phenomena that have been identified as such. These encompass areas such as quantum and classical communication, athermality (within the realm of quantum thermodynamics), asymmetry, magic (in the context of quantum computation), quantum coherence, Bell non-locality, quantum contextuality, quantum steering, incompatibility of quantum measurements, and many more. The recognition of all these phenomena as resources enables us to unify them under the umbrella of quantum resource theories.

Resource theories serve as a crucial framework for addressing complex questions. They aim to unravel puzzles such as determining which sets of resources can be transformed into one another and the methods by which such conversions can occur. Additionally, they explore how to measure and detect different resources. If a direct transformation between particular resources is not feasible, resource theories examine the possibility of non-deterministic conversions and the computation of their associated probabilities. The introduction of catalysts into the equation further deepens the inquiry.

This investigative approach often yields profound insights into the underlying nature of the physical or information-theoretic phenomena under scrutiny---such as entanglement, asymmetry, athermality, and more. Furthermore, this perspective provides a structured framework for organizing theoretical findings pertaining to these phenomena. As demonstrated by the evolution of entanglement theory, the resource-theoretic perspective possesses the potential to revolutionize our understanding of familiar subjects.

In this context, chemistry exemplifies this framework, elucidating how abundant collections of chemicals can be converted into more valuable products. Similarly, thermodynamics fits this mold by addressing inquiries about the conversion of various types of nonequilibrium states---thermal, mechanical, chemical, and more---into one another, including the extraction of useful work from heat baths at differing temperatures.

\begin{figure}[h]
\centering
    \includegraphics[width=0.8\textwidth]{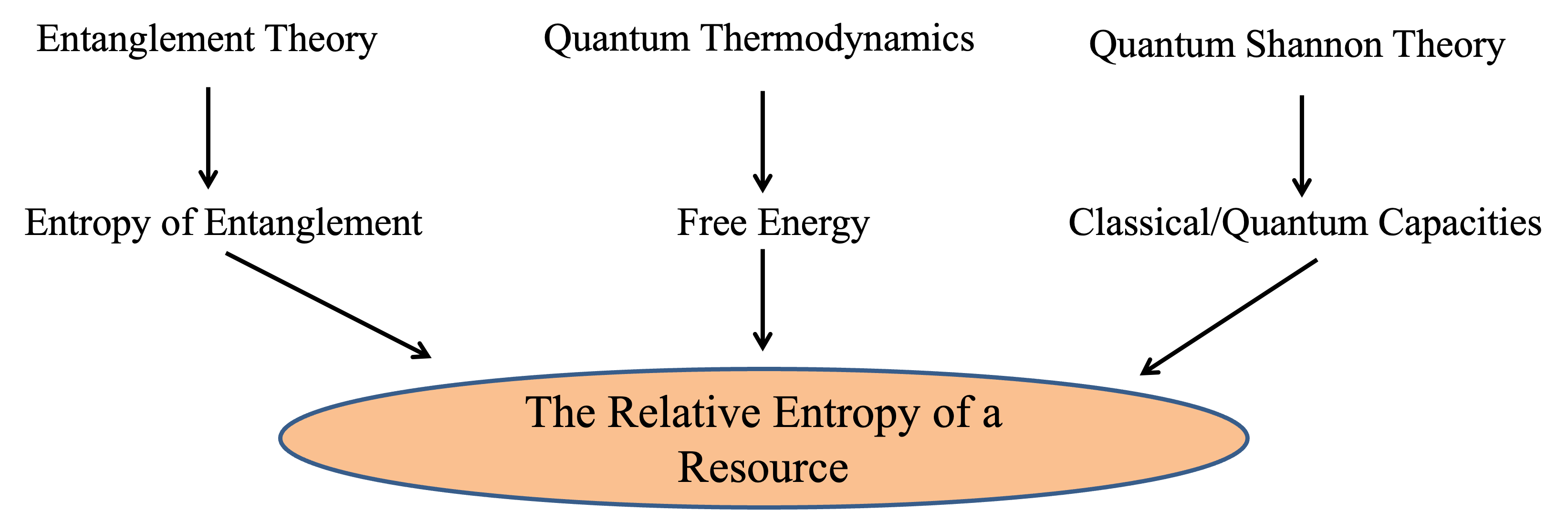}
  \caption{\linespread{1}\selectfont{Unification of Resources}}
  \label{unification}
\end{figure}

Within the realm of quantum resource theories, a fundamental challenge arises in identifying equivalence classes of quantum systems that can be reversibly interconvert (or simulate each other) when considering an abundance of resource copies, and determining the rates at which these interconversions occur. The relative entropy of a resource plays a pivotal role in such reversible transformations, gauging the resourcefulness of a system by quantifying its deviation from the set of free (non-resourceful) systems. Remarkably, this function unifies essential (pseudo) metrics across seemingly disparate scientific domains. For instance, the relative entropy of a resource manifests as free energy in thermodynamics, entanglement entropy in pure state entanglement theory, and the entanglement-assisted capacity of a quantum channel in quantum communication; see Fig.~\ref{unification}.

\section{About This Book}

As mentioned in the introduction, quantum resource theories have recently emerged as a vibrant research area within the quantum information science community. Initially, the emphasis was on understanding the resources used in quantum information processing tasks. However, it has become increasingly evident that quantum resources have broad relevance, extending from quantum computing to quantum thermodynamics and the fundamental principles of quantum physics. This realization has spurred rapid developments in the field, resulting in a proliferation of publications and the development of new tools and mathematical methods that firmly underpin this area of study.

In light of the extensive literature in the field of quantum information science, one might understandably question the need for yet another book on quantum resources. Isn't this territory already covered in existing quantum information textbooks? For instance, quantum Shannon theory can be seen as a theory of interconversions among different types of resources, and Wilde~\cite{Wilde2013} and Watrous~\cite{Watrous2018} have produced outstanding books delving into these topics. Additionally, detailed treatments of subjects covered in this book, such as quantum divergences and Rényi entropies, can be found in Tommamichel's noteworthy work~\cite{Tomamichel2015}.

While it is accurate to say that many of the topics covered in this book are available elsewhere, what distinguishes this book is its unique approach. It explores well-trodden subjects like entropy, uncertainty, divergences, non-locality, entanglement, and energy from a fresh perspective rooted in resource theories. Specifically, the book adopts an axiomatic approach\index{axiomatic approach} to rigorously introduce these concepts, providing illustrative examples. Only then does it transition to operational aspects that involve the examples discussed.

Take, for instance, the topic of conditional entropy, a subject widely covered in numerous textbooks in both classical and quantum information theory. This book, however, offers a distinctive approach by presenting this concept from three distinct perspectives: axiomatic, constructive, and operational. Notably, all three perspectives converge to the same notion of conditional entropy. This approach not only provides the reader with a deeper understanding of the concept but also underscores its robust foundation.

The primary goal of this book is pedagogical in nature, with the hope of providing readers with a contemporary perspective on quantum resource theories. It aspires to equip readers with the necessary physical principles and advanced mathematical techniques required to comprehend recent advancements in this field. Upon completing this book, readers should have the ability to explore open problems and research directions within the field, some of which will be highlighted in the text.

In anticipation of a diverse readership, this book is designed to be inclusive, targeting both graduate students and senior undergraduate students who possess a foundational understanding of linear algebra. It aims to provide them with a comprehensive resource for delving into this fascinating field. Simultaneously, the book serves as a reference, offering fresh insights and innovative approaches that researchers in the early stages of their careers may find valuable. With numerous examples and exercises, it aims to serve as a textbook for courses on the subject, enhancing the learning experience for students.

While the primary audience for this textbook consists of entry-level graduate students interested in pursuing research at the master's or Ph.D. level in quantum resource theories, encompassing quantum information science, it may also prove valuable to researchers in fields influenced by quantum information and resource theories, such as quantum thermodynamics and condensed matter physics. They may find this book to be a useful and accessible reference source.

Although we have endeavored to make the book self-contained, a basic understanding of linear algebra is essential. The goal was to create a resource accessible to graduate students from diverse backgrounds in mathematics, physics, and computer science. As a result, the book includes preliminary chapters and several appendices that fill potential knowledge gaps, given the interdisciplinary nature of the subject matter.

Quantum resource theories constitute a vast research area, with new properties of physical systems continually being recognized as resources. Consequently, the aim of this book is not to exhaustively cover all resource theories but rather to select those that illustrate the techniques used in quantum resource theories effectively.
On the technical front, we have chosen to begin with the modern single-shot approach and employ it to derive asymptotic rates. Historically, asymptotic rates were studied first, but from a pedagogical standpoint, it is more intuitive to start with the single-shot regime.

To the best of our knowledge, there are currently no dedicated books specifically focused on quantum resource theories. With this book, we hope to contribute to the field by providing a comprehensive overview and integrating both new and existing results within a unified framework. While we do not claim this book to be the ultimate authority, we believe it can serve as a valuable reference that consolidates ideas scattered across various journal articles, addressing the need for a centralized resource in the field of quantum resource theories.

\section{The Structure of the Book}

In this book, we delve deep into the comprehensive framework of quantum resource theories, offering a detailed study of their general principles and equipping readers with the necessary tools and methodologies. We extensively cover three illustrative examples of resource theories---Entanglement, Asymmetry, and Thermodynamics---chosen for their pedagogical value in showcasing the diverse facets of quantum resource theories. While we do not have a dedicated chapter solely focused on quantum coherence, this concept is seamlessly woven into our broader discussions. It serves as a recurring illustrative example that enriches our understanding of various aspects of quantum resource theories throughout the book.

\begin{figure}[h]
\centering
    \includegraphics[width=0.6\textwidth]{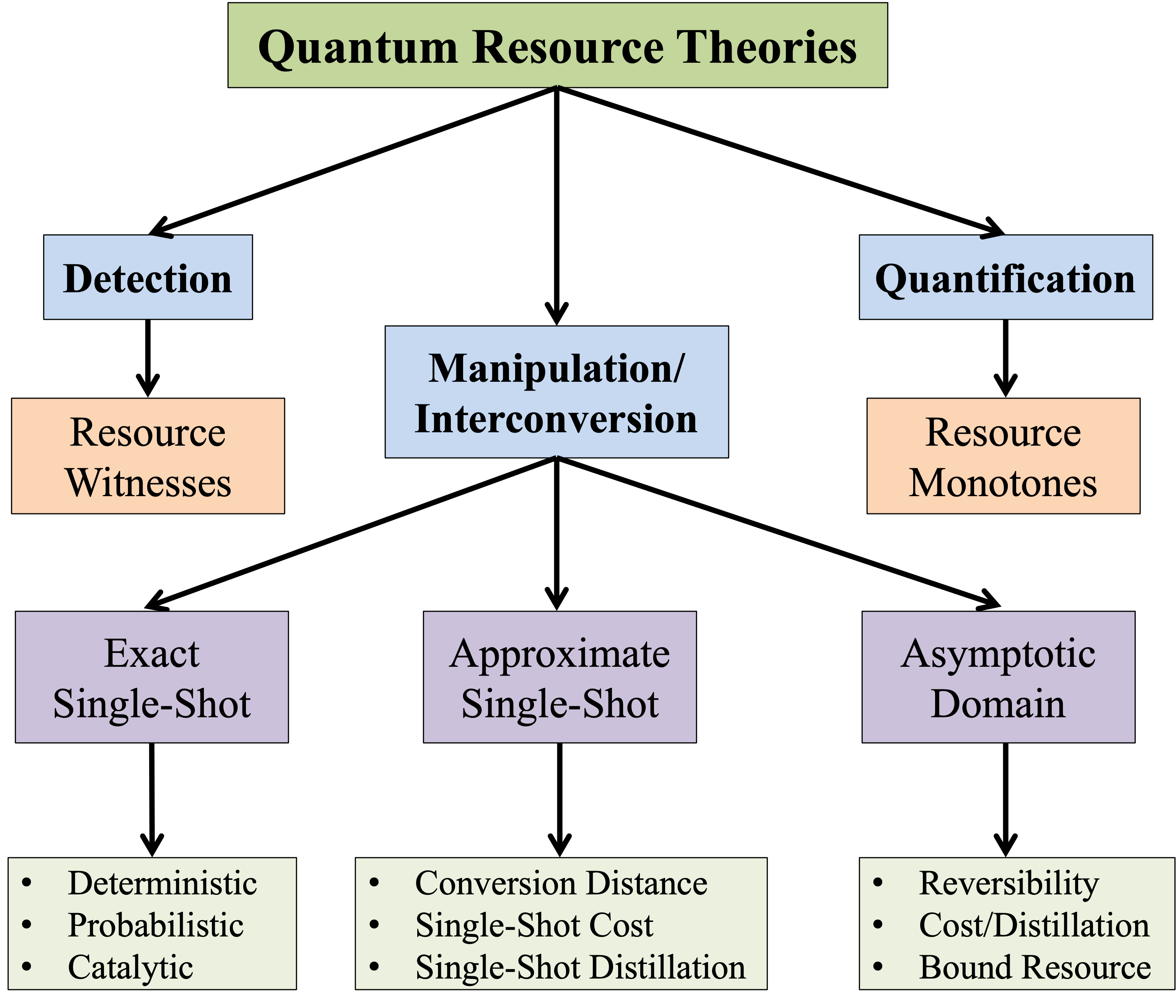}
  \caption{\linespread{1}\selectfont{The Structure of Quantum Resource Theories}}
  \label{qrt}
\end{figure} 

The initial volume of this book is structured into five main parts, with an additional sixth part containing supplementary materials.
\ben
\item[\textbf{Part I:}] The opening section of this book is thoughtfully designed to cater to readers who may not possess prior knowledge of quantum mechanics or quantum information. Within this segment, we embark on a rigorous mathematical journey through quantum theory, emphasizing precise definitions and mathematical proofs of fundamental physical theorems. Key subjects covered in this section encompass quantum states, generalized quantum measurements, quantum channels, POVMs, and more. Moreover, this section extends its reach beyond the boundaries of quantum theory, delving into topics such as Ky Fan norms, the Strømer-Woronowicz theorem, the Pinching Inequality, the Reverse Hölder Inequality, certain hidden variable models, and other subjects that may not commonly cross the paths of graduate students in physics, mathematics, or computer science. Therefore, even those well-versed in these topics may find it beneficial to skim through this chapter briefly, as it has the potential to reveal previously undiscovered insights.
\item[\textbf{Part II:}] The second section of this book delves deep into the methodologies and tools employed within the realm of quantum resource theories and quantum information. While it explores numerous quantum information concepts, it distinguishes itself from conventional quantum information theory textbooks. The introductory chapter of this section provides an all-encompassing mathematical review of majorization theory, encapsulating recent groundbreaking discoveries, such as relative majorization, conditional majorization, and the intersection of probability theory with this field.

Subsequent chapters in this section adopt a distinctive approach to elucidate concepts associated with metrics, divergences, and entropies. These notions are introduced and dissected using techniques and insights drawn from the framework of quantum resource theories. For instance, entropy, conditional entropy, relative entropies, and other divergences are introduced as additive functions that adhere to monotonicity under the set of free operations, a foundational concept in quantum resource theories.

The final chapter in this part of the book is dedicated to the asymptotic regime, focusing on the consequences of the ``law of large numbers" in quantum information and quantum resource theories. This chapter introduces concepts such as weak and strong typicality, the method of types, classical and quantum hypothesis testing, and the symmetric subspace. These tools prove particularly valuable in the asymptotic domain of quantum resource theories when exploring inter-conversion rates among infinitely many resources.

In summary, although the contents of this second section share some commonalities with conventional quantum information theory textbooks, they diverge significantly by presenting concepts and tools in a unique manner. Rather than employing Venn diagrams to define key concepts like entropy, this part of the book aims to provide a comprehensive and rigorous approach to precisely define these concepts by employing axiomatic, constructive, and operational approaches. Leveraging the framework of quantum resource theories, this section offers a fresh and innovative perspective on these familiar topics.
\item[\textbf{Part III:}] In the third section of the book, we delve into the fundamental framework of quantum resource theories. Our journey begins with a meticulous mathematical elucidation of a quantum resource theory. We proceed to examine its foundational principles, including but not limited to the Chernoff\index{Chernoff} of free operations, resource non-generating operations, physically implementable operations, convex and affine resource theories, state-based resource theories, as well as resource witnesses and their associated properties.

Next, we delve into the quantification of quantum resources. In this context, we introduce a plethora of resource measures and resource monotones, delving deep into their properties, which include additivity, sub-additivity, convexity, strong monotonicity, and asymptotic continuity. These concepts form the bedrock of quantum resource theories, and understanding them is pivotal.

Resource monotones and resource measures offer a valuable means of quantifying resources. Our emphasis is on divergence-based resource measures, such as the relative entropy of a resource, given their operational interpretations across various resource theories. We also explore techniques for computing these measures, including semidefinite programming, and delve into a practical approach for ``smoothing" these measures, a technique commonly employed in single-shot quantum information science.

Concluding this section of the book, we introduce a rich array of resource interconversion scenarios. These encompass exact interconversions, stochastic (probabilistic) interconversions, approximate interconversions, and asymptotic interconversions. We delve into essential tools intricately linked to resource interconversions, such as the conversion distance within the single-shot regime, the asymptotic equipartition\index{asymptotic equipartition} property, and the quantum Stein's lemma within the asymptotic domain. Additionally, we explore the uniqueness of the Umegaki relative entropy within the context of quantum resource theories. Our investigation extends to the evaluation of both the cost and distillation of resources, examining these processes within both the single-shot and asymptotic regimes. We have encapsulated the essence of this section of the book in Figure~\ref{qrt}.

\item[\textbf{Part IV:}] The fourth section of this book is dedicated to the quintessential exemplar of quantum resource theories, often referred to as the ``poster child" – entanglement theory. This section comprises three chapters, each focusing on distinct facets of entanglement. The first chapter delves into the realm of pure bipartite entanglement, followed by the second chapter, which explores mixed bipartite entanglement. The third chapter, in turn, delves into the intricacies of multipartite entanglement.

Within these chapters, we leverage the techniques and concepts developed in parts II and III to delve into the theory of entanglement. This enables us to furnish a precise definition of quantum entanglement and undertake a comprehensive examination of its detection, manipulation, and quantification. Notably, the first of these three chapters serves as the cornerstone, offering an in-depth exploration of pure bipartite entanglement, which forms the foundational knowledge upon which the subsequent chapters on mixed and multipartite entanglement build.

\item[\textbf{Part V:}] The fifth part comprises three chapters, with the first two chapters focusing on asymmetry and non-uniformity, laying the groundwork for the third chapter on quantum thermodynamics. In this part of the book, we reveal that athermality, the resource essential for thermodynamic tasks, consists of two components: time-translation asymmetry and non-uniformity.

The first chapter explores the resource theory of asymmetry, introducing an operational framework that arises from practical constraints when multiple parties lack a common shared reference frame. This theory has found numerous applications in quantum information and beyond.

The second chapter delves into the resource theory of non-uniformity. In this theory, maximally mixed states are considered free, while all other states are regarded as valuable resources. This theory can be seen as a unique variant of thermodynamics, involving completely degenerate Hamiltonians. Indeed, we introduce this chapter to serve as a gentle introduction to the world of quantum thermodynamics.

Finally, in the third chapter of this section, we dive into quantum thermodynamics. Throughout the book, whenever we introduce a new quantum resource theory, we adhere to the structured framework outlined in Figure~\ref{qrt}.

\item[\textbf{Part VI:}] The final section of the book serves as a comprehensive resource aimed at ensuring the self-containment of the entire text. It exclusively includes material that directly complements the core content of the book.

In the initial three chapters, we delve into key subjects: convex analysis, operator monotonicity, and representation theory. It's important to note that each of these topics is vast in its own right, with numerous dedicated books solely focused on representation theory or convex analysis, for example. In this section, we have thoughtfully curated and presented the aspects of these topics that are pertinent to our book's core themes. Our approach emphasizes utilizing quantum notations and placing a strong emphasis on furnishing all the essential elements needed to ensure the book's self-contained nature.
\een

\section{Resurrection of Quantum Entanglement: The Birth of a Fundamental Resource}\label{ch:ent0}\index{quantum entanglement}

In this section, we delve into the transformative protocols of quantum teleportation and super-dense coding. These groundbreaking techniques marked a pivotal moment in the history of quantum physics, elevating entanglement from a purely theoretical curiosity to a precious resource with tangible applications. This paradigm shift, akin to the ``resurrection" of quantum entanglement, carried profound implications for the burgeoning field of quantum information. In essence, it played a significant role in catalyzing the emergence of quantum information science. If you're new to the formalism of quantum mechanics, we recommend starting with Chapter~\ref{quantum0} before delving into the following two sections.

Even in the early days of quantum mechanics, entanglement stood out as a distinctive and defining feature of the theory. As articulated by Schrödinger, he remarked, ``I would not call [entanglement] one but rather the characteristic trait of quantum mechanics, the one that enforces its entire departure from classical lines of thought." This statement underscores the profound departure from classical physics that entanglement embodies. Significantly, the intriguing properties of entanglement were recognized well before Bell's seminal paper on the exclusion of local hidden variable models (as discussed in Section~\ref{hvm}).

To illustrate this point, consider a composite system\index{composite system} consisting of two $1/2$-spin particles, such as electrons, in the singlet state: 
\be
|\Psi_{-}\ra=\frac{1}{\sqrt{2}}\left(|\ua_\n\ra|\da_\n\ra-|\da_\n\ra\ua_\n\ra\right)\;.
\ee
Here, $\{|\!\ua_\n\ra,|\!\da_\n\ra\}$ forms an orthonormal basis in the complex vector space $\mathbb{C}^2$, representing the two eigenvectors of the spin observable\index{observable} corresponding to the ``up" and ``down" orientations along a direction $\mathbf{n}\in\mathbb{R}^3$. Notably, a remarkable property of this state is its independence from the specific spin direction $\mathbf{n}$ (see Chapter~\ref{quantum0}).

Now, if Alice performs a measurement in the $\mathbf{n}$ direction, it will instantaneously dictate Bob's post-measurement state to align with the opposite $\mathbf{n}$ direction. This peculiar phenomenon allows Alice to exert immediate influence on Bob's state by simply choosing whether to conduct a Stern-Gerlach measurement (as discussed in Section~\ref{stern}) along the $\mathbf{n}$ or $\mathbf{m}$ direction. This non-intuitive behavior of entangled composite quantum systems prompted Einstein to describe it as a ``spooky action at a distance."

Beyond its profound implications from a fundamental standpoint, entanglement has gained recognition as a valuable and indispensable resource for the realization of specific quantum information processing tasks. This shift in perspective has given rise to a substantial body of research, as entanglement is no longer solely a philosophical curiosity but a powerful tool with remarkable practical applications. These applications encompass protocols like quantum teleportation, superdense coding, and numerous innovations in quantum cryptography and quantum computing.

In this section, we embark on a journey through some of these protocols, known as unit protocols, as they exclusively rely on unit noiseless resources. These protocols serve as a testament to the versatility of entanglement and employ three distinct resource types: a noiseless quantum communication channel, a noiseless classical communication channel, and the entangled bit, abbreviated as ebit.

\subsection{Quantum Teleportation}\label{sec:qtele}

Quantum teleportation \index{quantum teleportation} is a groundbreaking protocol enabling Alice to transmit an unknown quantum state $|\psi\rangle$ to Bob, all without the need for a dedicated quantum communication channel. Instead, it relies on the clever utilization of entanglement and a classical communication channel to achieve this remarkable feat, as illustrated in Figure~\ref{teleportation} below. To elucidate, consider the scenario where Alice and Bob share a composite system\index{composite system} comprising two electrons initially prepared in the singlet state:
\be
|\Psi_{-}^{AB}\ra=\frac{1}{\sqrt{2}}(|01\ra-|10\ra)\;.
\ee

Furthermore, let's consider the scenario where Alice possesses an additional electron in her system, characterized by a quantum state $|\psi^{\tA}\rangle = a|0\rangle + b|1\rangle$. Importantly, both Alice and Bob lack knowledge regarding the spin state of this electron, which means they are unaware of the specific values of $a$ and $b$. According to the principles of quantum mechanics, the collective quantum state of these three electrons---two under Alice's control and one under Bob's---is described by the tensor product: 
\ba
|\psi^{\tA}\rangle\otimes|\Psi_{-}^{AB}\rangle&=\frac{1}{\sqrt{2}}(a|0\ra+b|1\ra)\otimes (|01\ra-|10\ra)\\
\GG{\substack{\text{Openning}\\ \text{Parentheses}}}&=\frac{1}{\sqrt{2}}\big(a|001\ra+b|101\ra-a|010\ra-b|110\ra\big)\;.
\ea

It's noteworthy that in our description above, we represented the state $|\psi^{\tA}\rangle\otimes|\Psi_{-}^{AB}\rangle$ using the computational basis of the vector space $\tA AB$. However, we can achieve a more insightful representation by substituting the computational basis ${|00\rangle,|01\rangle,|10\rangle,|11\rangle}$ of system $\tA A$ with the Bell basis\index{Bell basis} consisting of $|\Phi_{\pm}^{\tA A}\rangle=\frac{1}{\sqrt{2}}(|00\rangle\pm|11\rangle)$ and $|\Psi_{\pm}^{\tA A}\rangle=\frac{1}{\sqrt{2}}(|01\rangle\pm|10\rangle)$. This substitution allows us to express the state as follows:
\ba
|\psi\ra^{\tA}\otimes|\Psi_{-}^{AB}\ra
&=\frac{1}{2}\Big[a\big(|\Phi_{+}^{\tA A}\ra+|\Phi_{-}^{\tA A}\ra\big)|1\ra+b\big(|\Psi_{+}^{\tA A}\ra-|\Psi_{-}^{\tA A}\ra\big)|1\ra\\
&\quad\quad\quad\quad\quad\quad-a\big(|\Psi_{+}^{\tA A}\ra+|\Psi_{-}^{\tA A}\ra\big)|0\ra-b\big(|\Phi_{+}^{\tA A}\ra-|\Phi_{-}^{\tA A}\ra\big)|0\ra\Big]\\
\GG{Collecting\; terms}&=\frac{1}{2}\Big[|\Phi_{+}^{\tA A}\ra(a|1\ra-b|0\ra)+|\Phi_{-}^{\tA A}\ra(a|1\ra+b|0\ra)\\
&\quad\quad\quad\quad\quad\quad\quad+|\Psi_{+}^{\tA A}\ra(b|1\ra-a|0\ra)-|\Psi_{-}^{\tA A}\ra(a|0\ra+b|1\ra)\Big]
\ea 
Therefore, if Alice performs  the Bell measurement on her two qubits $\tA A$, i.e. the basis (projective) measurement \be
\Big\{P_0=|\Psi_{-}^{\tA A}\lr \Psi_{-}^{\tA A}|,\;P_1=|\Phi_{-}^{\tA A}\lr \Phi_{-}^{\tA A}|,\;P_2=|\Phi_{+}^{\tA A}\lr \Phi_{+}^{\tA A}|,\;P_3=|\Psi_{+}^{\tA A}\lr \Psi_{+}^{\tA A}|\Big\}\;,\ee she will get with equal probability four possible outcomes (denoted $x=0,1,2,3$, and global phase is ignored):

\begin{center}
  \begin{tabular}{|c|c|c|}
    \hline
    \textbf{Outcome} & \textbf{Post-Measurement State} & $\substack{\textbf{Simplification}\\\text{(Up to a global phase)}}$ \\
    \hline
    $x=0$ & $|\Psi_{-}^{\tA A}\ra\otimes(a|0\ra+b|1\ra)$ & $|\Psi_{-}^{\tA A}\ra\otimes|\psi\ra$ \\
    \hline
    $x=1$ & $|\Phi_{-}^{\tA A}\ra\otimes(a|1\ra+b|0\ra)$ & $|\Phi_{-}^{\tA A}\ra\otimes\sigma_1|\psi\ra$ \\
    \hline
    $x=2$ & $|\Phi_{+}^{\tA A}\ra\otimes(a|1\ra-b|0\ra)$ & $|\Phi_{+}^{\tA A}\ra\otimes\sigma_2|\psi\ra$ \\
    \hline
    $x=3$ & $|\Psi_{+}^{\tA A}\ra\otimes(b|1\ra-a|0\ra)$ & $|\Psi_{+}^{\tA A}\ra\otimes\sigma_3|\psi\ra$ \\
    \hline
\end{tabular}
\end{center}
where we denoted by $\{\sigma_x\}_{x=0,1,2,3}$, the identity matrix $\sigma_0=I_2$, and the 3 Pauli\index{Pauli} matrices $\sigma_1,\sigma_2,\sigma_3$. Hence, up to a global phase, Bob's state after outcome $x$ occurred is $\sigma_x|\psi\ra$. After Alice sends (via a classical communication channel) the measurement outcome $x$ to Bob, Bob can then perform the unitary operation $U_x=\sigma_x$ to obtain the state
\be
\sigma_x\left(\sigma_x|\psi\ra\right)=\sigma_{x}^{2}|\psi\ra=|\psi\ra\;.
\ee
Therefore, by using shared entanglement, and after transmitting two classical bits (cbits), Alice was able to transfer her unknown qubit state $|\psi\ra$ to Bob's side.

\begin{figure}[h]
\centering
    \includegraphics[width=0.6\textwidth]{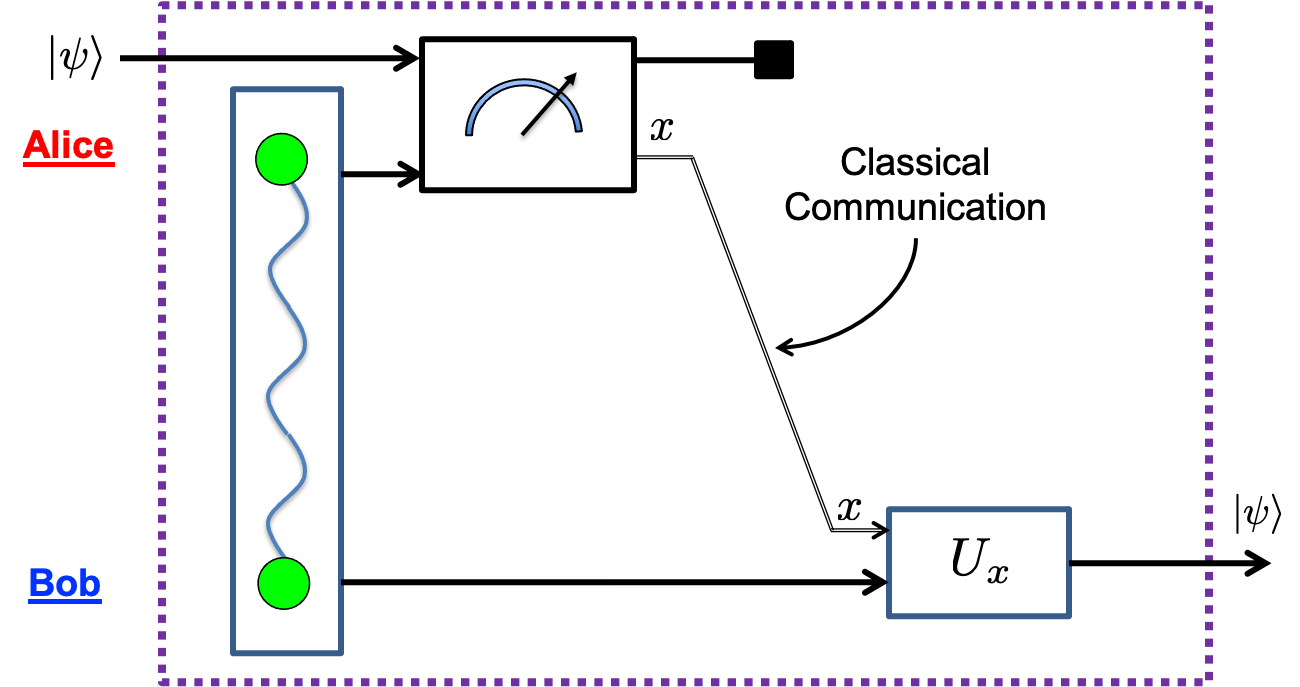}
  \caption{\linespread{1}\selectfont{\small Quantum teleportation. \index{quantum teleportation} Single-line arrows correspond to quantum systems. Double line arrows correspond to classical systems.}}
  \label{teleportation}
\end{figure}

If Bob did not receive the classical message from Alice, then his state is one of the four states $\{\sigma_x|\psi\ra\}_{x=0,1,2,3}$. Since he does not know $x$, from his perspective his state is (see Exercise~\ref{exnch})
\be
\rho=\frac{1}{4}\sum_{x=0}^{3}\sigma_x|\psi\lr\psi|\sigma_x=\frac{1}{2}I\;.
\ee
That is, without the knowledge of $x$, Bob's resulting state is the maximally mixed state, and contains no information about $|\psi\ra$.

\begin{exercise}\label{exnch}
Show that for any density matrix $\rho\in\md(\mbb{C}^2)$, 
\be
\frac{1}{4}\sum_{x=0}^{3}\sigma_x\rho\sigma_x=\frac{1}{2}I\;.
\ee
Hint: Prove first that the left-hand side of the equation above is invariant under a conjugation by $\sigma_x$.
\end{exercise}
\begin{exercise}
Show that if instead of the singlet state $|\Psi_{-}^{AB}\ra$ above, 
Alice and Bob share another maximally entangled state $|\Phi^{AB}\ra$ (i.e. the reduced density matrix of
$|\Phi^{AB}\ra$ is the maximally mixed state), then, by modifying slightly the protocol, they can still teleport an unknown quantum state from Alice to Bob.
\end{exercise}

The protocol above can be generalized in several different ways. First, in Exercise~\ref{teled} you will generalize it to $d$-dimensions. Moreover, in general, if Alice and Bob do not share the singlet state, but instead their particles are prepared in some other non-seperable state (i.e. entangled state, but not maximally entangled state) $\rho^{AB}\in\md({AB})$, then typically perfect/faithful teleportation  will not be possible. Still in this case one can design a protocol achieving quantum teleportation with probability that is less than one (see Exercise~\ref{telep}), and/or in the end of the protocol the state in Bob's lab is not exactly equal to Alice's original state $|\psi\ra^{\tA}$ but only close to it up to some treshold. Thus,
the protocol described above is called \emph{faithful} teleportation, since the protocol teleport perfectly $|\psi\ra$ form Alice to Bob with 100\% success rate.

\begin{exercise}\label{teled}
Let $
|\Phi^{AB}\ra\eqdef\frac{1}{\sqrt{d}}\sum_{z\in[d]}|zz\ra$ be a 2-qudit (normalized) maximally entangled state in ${AB}\cong\mathbb{C}^d\otimes\mathbb{C}^d$.
Consider a family of $d^2$ states in ${AB}$ defined by
\be
|\psi_{xy}^{AB}\rangle=T^x\otimes S^y|\Phi^{AB}\ra\;,\quad\quad x,y\in[d]
\ee
where $T$ and $S$ are the phase and shift operators defined by $T|z\ra=e^{i\frac{2\pi z}{d}}|z\ra$ and $S|z\ra=|z\oplus 1\ra$, where $\oplus$ is the plus modulo $d$, and $z\in[d]$.
\begin{enumerate}
\item Show that $\{|\psi_{xy}^{AB}\ra\}_{x,y\in[d]}$ is an orthonormal basis of ${AB}$.
\item Show that the reduced density matrix of $|\psi_{xy}^{AB}\rangle$ is the maximally mixed state for all $x,y\in[d]$.
\item Find a protocol for faithful teleportation of a qudit from Alice's lab to Bob's lab. Assume that the joint measurement that Alice performs on her two qudits is a basis measurement in the basis $\{|\psi_{xy}^{AB}\ra\}_{x,y\in[d]}$.  What are the unitary operators performed by Bob?
How many classical bits (cbits) Alice transmits to Bob?
\end{enumerate}
\end{exercise}

\begin{exercise}\label{telep}
Suppose Alice and Bob share the state $|\psi^{AB}\ra=\frac{1}{2}|00\ra+\frac{\sqrt{3}}{2}|11\ra$. Show that there exists a 2-outcome (basis) measurement that Alice can perform, such that with some probability greater than zero, the state of Alice and Bob after the measurement becomes the maximally entangled state $|\Phi^{AB}_+\ra=\frac{1}{\sqrt{2}}(|00\ra+|11\ra)$.
\end{exercise}

So far we assumed that the teleported state is a pure state. However, the exact same protocol works even if the unknown state $|\psi\ra$ is replaced with a mixed state $\rho$. This is because we can view any mixed state as some ensemble of pure states $\{p_x,|\psi_x\ra\}$ in which the parameter $x$ is unknown. Irrespective of the value of $x$, the protocol above will teleport $|\psi_x\ra$ from Alice to Bob, and thereby, given that the value of $x$ is unknown, Alice effectively teleported to Bob the mixed state $\rho\eqdef\sum_xp_x|\psi_x\lr\psi_x|$. Alternatively, note that the quantum teleportation protocol in Fig.~\ref{teleportation} can be described as a realization of the identity quantum channel $\id\in\cptp(A\to B)$ (with $|A|=|B|\eqdef d$) given by
\be
\id^{A\to B}(\rho^{A})=\sum_{x\in[d^2]}\tr_{A\tA}\left[\left(P_x^{\tA A}\otimes U_x^B\right)\left(\rho^A\otimes\Phi^{\tA B}\right)\left(P_x^{\tA A}\otimes U_x^B\right)^*\right]
\ee
where $\{P_x^{\tA A}\}_{x\in[d^2]}$ corresponds to the measurement on systems $\tA$ and $A$ in the maximally entangled basis,
$U_x$ is the unitary performed by Bob after he received the value $x$ from Alice, and $\Phi^{AB}$ is the maximally entangled state on system $AB$. The quantum teleportation protocol states that there exists $\{P_x^{\tA A}\}$ and $\{U_x\}$ such that the quantum channel $\id^{A\to B}$ above is indeed the identity channel. Although, in the protocol above we proved it only for pure input states $|\psi\lr\psi|$, from the linearity of the quantum channel $\id^{A\to B}$, it follows that $\id^{A\to B}$ is the identity quantum channel on all mixed states. 

\begin{figure}[h]\centering
    \includegraphics[width=0.5\textwidth]{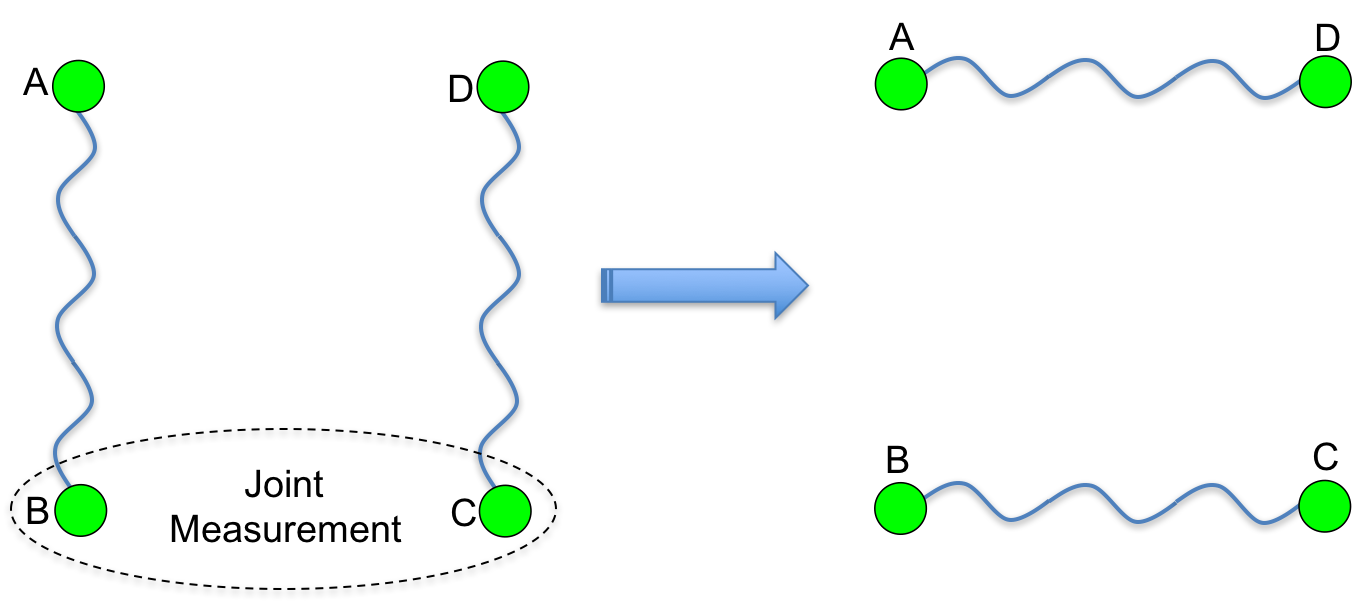}
  \caption{\linespread{1}\selectfont{\small Entanglement Swapping.}}
  \label{swapping}
\end{figure}

\begin{exercise}{\rm \bf [Entanglement Swapping]}\index{entanglement swapping}
Consider four qubit systems $A$, $B$, $C$, and $D$, in the double-singlet state $|\Psi_{-}^{AB}\ra\otimes|\Psi_{-}^{CD}\ra$. 
\begin{enumerate}
\item Show that a joint Bell measurement on system $BC$ generates a maximally entangled state on system $AD$ (along with another maximally entangled state on system $BC$) for all four possible outcomes of the measurement (see Fig.~\ref{swapping}).
\item Show that the singlet state in AD can be generated by quantum teleportation between system $BC$ and system $D$.
\item Generalize the entanglement swapping protocol to four qudit systems each of dimension $d$.
\end{enumerate}
\end{exercise}

\subsection{Superdense Coding}\index{superdense coding}

How much classical information can be transmitted with a single qubit? Suppose Alice wants to transmit a classical message to Bob, but all she has at her disposal is a single qubit (e.g. spin of an electron) and a perfect noiseless quantum communication channel that she can use to transmit the electron to Bob. She can prepare the single electron in the spin that she wants and send the electron to Bob over the noiseless quantum channel. Alice and Bob agree at the beginning that a message $0$ corresponds to spin up in the $z$ direction, and a message $1$ corresponds to spin down in the $z$ direction. In this way, Alice can transmit one cbit with one use of a perfectly noiseless quantum channel. Can they do better?
We will see later on that it is not possible to encode more than one classical bit into a single electron, as long as Alice's electron is not entangled with another electron in Bob's system.

Suppose now that Alice's electron is maximally entangled with another electron in Bob lab, so that Alice and Bob share
the singlet state $|\Psi_{-}^{AB}\ra$. In the first step of the protocol, Alice encode a message $x$ into a her qubit. She is doing it by performing one out of several (unitary) rotations $\{U_x^A\}_{x=0}^{m-1}$ on her qubit (electron). If Alice chose to do the $x$ rotation, the state of the system after the rotation is
\be
|\psi_x^{AB}\ra\eqdef \left(U_x^A\otimes I^B\right)|\Psi_{-}\ra^{AB}
\ee
Taking $U_x=\sigma_x$ to be the four Pauli\index{Pauli} matrices (with $\sigma_0=I_2$) we get that the four states $\{|\psi_{x}^{AB}\ra\}_{x=0}^{3}$ are orthonormal and form a basis of $\mbb{C}^2\otimes\mbb{C}^2$. In fact, this is the Bell basis\index{Bell basis} we encountered in the previous subsection. 
In the next step of the protocol, Alice sends her electron (over a noiseless quantum communication channel) to Bob. Upon receiving Alice's electron, Bob has in his lab two electrons in the state $|\psi_x^{AB}\ra$. Given that the set of states $\{|\psi_x^{AB}\ra\}_{x=0}^3$ form an orthonormal basis, in the last step of the protocol, Bob performs a joint basis measurement on his two electrons, in the basis $\{|\psi_x^{AB}\ra\}_{x=0}^{3}$, and thereby learns the outcome $x$. The outcome $x$ is the message that Alice intended to send Bob.

\begin{figure}[h]
\centering
    \includegraphics[width=0.6\textwidth]{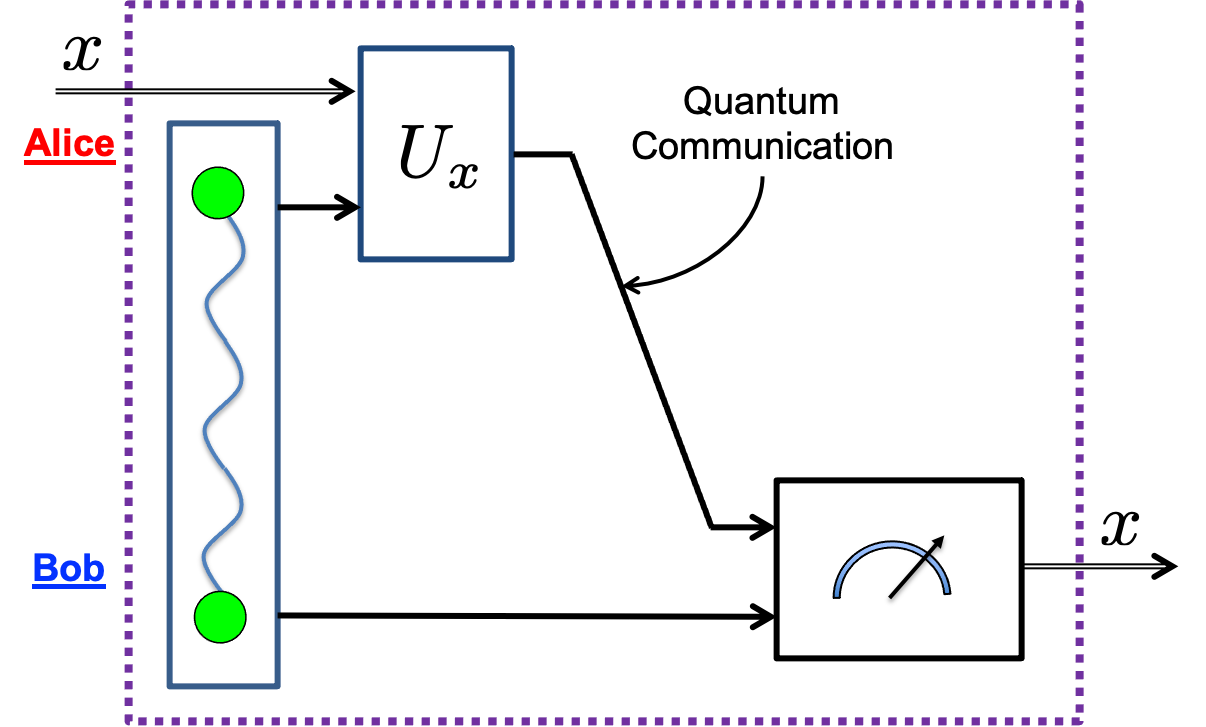}
  \caption{\linespread{1}\selectfont{\small Superdense Coding. Double-lines corresponds to classical systems, and single lines to quantum systems.}}
  \label{superdense}
\end{figure} 

\begin{exercise}
Show that the set of states $\{|\psi_{x}^{AB}\ra\}_{x=0}^{3}$ is an orthonormal basis of $\mbb{C}^2\otimes\mbb{C}^2$. 
\end{exercise}

\begin{exercise}
Let $
|\Phi^{AB}\ra\eqdef\frac{1}{\sqrt{d}}\sum_{z\in[d]}|zz\ra$ be a maximally entangled state in $\mbb{C}^d\otimes\mbb{C}^d$. Show that Alice can use it to transmit to Bob $2\log_2(d)$ cbits.
\end{exercise}

\section{Resource Analysis and Reversibility}\index{reversibility}

The previous two protocols demonstrate that entanglement is a valuable resource with which certain tasks will not be possible without it. We have seen in the protocols above that entanglement can be converted to other types of resources such as quantum or classical communication channels. We will use the following notations to denote these resources: 
\ben
\item  $[qq]$ denotes one ebit; i.e. a unit of a static noiseless resource comprising of two qubits in a maximally entangled state. 
\item $[q\to q]$  denotes one use of an ideal (noiseless) qubit channel
\item $[c\to c]$ denotes one use of a classical bit channel capable of transmitting perfectly one classical bit. 
\item $[cc]$ denotes one bit of shared randomness. 
\een
Observe that the resource units above can be classified into being classical or quantum, and static (such as $[qq]$ and $[cc]$) or dynamic (such as $[q\to q]$ or $[c\to c]$).

With the above notations, the teleportation can be viewed as a process in which one ebit plus two uses of a classical bit channel are consumed to simulate a qubit channel. 
In resource symbols this can be characterized as the following resource inequality
\be\label{r1}
[qq]+2[c\to c]\geq [q\to q]\;.
\ee
Note that the use of the inequality above is justified by the fact that a single use of a quantum channel cannot generate both an entangled state \emph{and} a double use of a classical channel.

For superdense coding, an ebit plus one use of a quantum channel is used to simulate two uses of a classical channel. This can be expressed as the resource inequality
\be\label{r2}
[qq]+[q\to q]\geq 2[c\to c]\;.
\ee
Note that if entanglement is not considered as a resource, that is, the parties are supplied with unlimited singlet states, then we can remove the ebit cost $[qq]$ in~\eqref{r1} and~\eqref{r2} and get that for teleportation
$
2[c\to c]\geq [q\to q]
$
and for superdense coding 
$
[q\to q]\geq 2[c\to c]
$.
This makes teleportation and superdense coding the dual protocols of each other, and in this case we can say that $[q\to q]= 2[c\to c]$.

However, in almost all practical scenarios, entanglement is an expensive resource that can be difficult to generate over long distances and that is also highly sensitive to decoherence and noise. Therefore, specifically pure maximally entanglement is scarce, and must be treated as a resource. The question then becomes if it is possible to change slightly the protocols of teleportation and superdense coding making them more symmetric, in the sense that the two resource inequalities in~\eqref{r1} and~\eqref{r2} merge into a single resource equality. This is indeed possible if we replace $2[c\to c]$ in the right-hand side of~\eqref{r2} with two uses of an isometry channel known as the coherent bit channel.

\subsection{The Coherent Bit Channel}\index{coherent bit channel}

We introduce here another unit resource that is called the \emph{coherent bit channel}, or in short the \emph{cobit} channel, and is denoted by $[q\to qq]$. As the symbol suggest, this unit resource represents one use of a channel. The channel is defined by the isometry,
$V:A\to A\otimes B$, with $|A|=|B|=2$, according to the following action on the basis of $A$:
\be
V|x\ra^A=|x\ra^A|x\ra^B\;
\quad\forall\;x\in\{0,1\}\quad
\text{or equivalently}
\quad
V=\sum_{x=0}^{1}|x\lr x|^A\otimes|x\ra^B\;.
\ee
We will denote by
\be
\mV_Z(\rho)\eqdef V\rho V^*\;,\quad\quad\forall\;\rho\in\ml(A)\;,
\ee
where the subscript $Z$ indicate that the basis $\{|0\ra,|1\ra\}$ is an eigenbasis of the third Pauli operator (i.e., eigenvectors of the spin observable\index{observable} in the $z$-direction).
One can define $V$ with respect to other bases. For example, we will denote by $\mV_X(\cdot)=U(\cdot)U^*$ the coherent bit channel with respect to the basis $\{|+\ra,|-\ra\}$, where $U$ is the isometry defined by $U|\pm\ra^A=|\pm\ra^A|\pm\ra^B$.  

How is this resource related to other resources? First note that with such a resource Alice can transmit a classical bit to Bob. Indeed, 
Alice can encode a cbit $x\in\{0,1\}$ in the state $|x\ra^A$ and send it over the channel $\mV_Z$. Then, Bob receives $|x\ra^B$ on his system and performs a basis measurement to learn $x$. We therefore have
\be\label{greater}
[q\to qq]\geq [c\to c]\;.
\ee
The exercise below shows that we also have 
$
[q\to qq]\geq [qq]\;.
$
Among other things, this also implies that $[c\to c]\not\geq [q\to qq]$ or in other words, $[c\to c]$ is a strictly less resourcefull than $[q\to qq]$.
\begin{exercise}
Show that $\mV_Z\left(|+\lr +|^A\right)=|\Phi_{+}^{AB}\lr\Phi_{+}^{AB}|$.
\end{exercise}

\subsection{Coherent Superdense Coding}\index{coherent superdense coding}

For superdense coding, we saw that an ebit, $[qq]$, plus one use of a quantum channel, $[q\to q]$, can be used to simulate two uses of a classical channel, $2[c\to c]$. We now show that the same resources can also be used to simulate two uses of the coherent map $\mV$. That is, we will show that
\be\label{r2c}
[qq]+[q\to q]\geq 2[q\to qq]\;.
\ee
Note that due to~\eqref{greater}, the above equation also implies the resource inequality~\eqref{r2}. The quantum protocol that achieves this resource conversion is called \emph{coherent} superdense coding.

\begin{figure}[h]
\centering
    \includegraphics[width=0.5\textwidth]{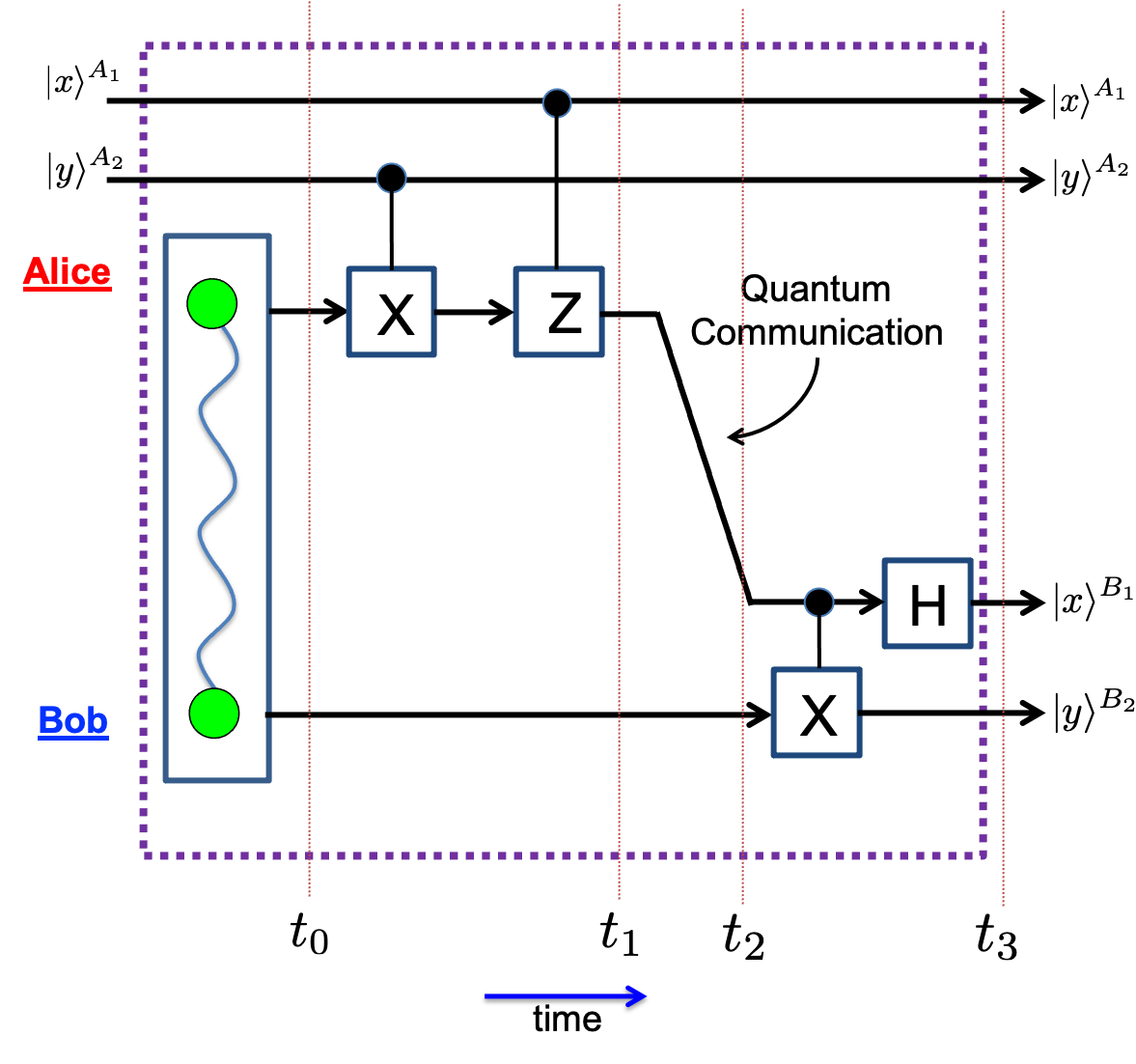}
  \caption{\linespread{1}\selectfont{\small Coherent Superdense Coding. One ebit plus one use of a noiseless qubit channel are implemented to realize two uses of the cobit channel.}}
  \label{csuperdense}
\end{figure} 

Coherent superdense coding protocol (see Fig.~\ref{csuperdense}) consists of several steps. Initially, Alice and Bob share the maximally entangled state $|\Phi_{+}^{AB}\ra$. Alice then prepares an input state $|x\ra^{A_1}|y\ra^{A_2}$ so that Alice and Bob's initial state (time $t_0$ is the figure) is  
\be
|x\ra^{A_1}|y\ra^{A_2}|\Phi_{+}^{AB}\ra\;.
\ee
Alice then performs a sequence of two controlled unitary\index{controlled unitary} gates, controlled $X$ on system $A_2$ and $A$, followed by controlled $Y$ gate on system $A_1$ and $A$. The resulting state at time $t_1$ is
\be
|x\ra^{A_1}|y\ra^{A_2}|\phi_{xy}^{AB}\ra\quad\text{where}\quad |\phi_{xy}^{AB}\ra\eqdef \left(Z^xX^y\otimes I^B\right)|\Phi_{+}^{AB}\ra\quad\text{and}\quad x,y\in\{0,1\}\;,
\ee
where $Z^x$ equals the identity matrix for $x=0$, and the third Pauli matrix for $x=1$ ($X^y$ is defined similarly).
A key observation is that  $\{|\phi_{xy}^{AB}\ra\}_{x,y\in\{0,1\}}$ is precisely the Bell basis, and therefore forms an orthonormal basis for $\mbb{C}^2\otimes\mbb{C}^2$. Note also that this encoding $(x,y)\to|\phi_{xy}^{AB}\ra$ is done by Alice alone, and therefore essentially identical to the superdense coding protocol we encountered earlier. 

In the next step Alice uses a noiseless qubit channel to transmit system $A$ to Bob. Therefore, at time $t_2$ the state of the system is $|x\ra^{A_1}|y\ra^{A_2}|\phi_{xy}^{B_1B_2}\ra$, where as before,
\be
|\phi_{xy}^{B_1B_2}\ra\eqdef \left(Z^xX^y\otimes I^{B_2}\right)|\Phi_{+}^{B_1B_2}\ra\;.
\ee
Since both $\{|\phi_{xy}^{B_1B_2}\ra\}_{x,y\in\{0,1\}}$ and $\{|xy\ra^{B_1B_2}\}_{x,y\in\{0,1\}}$ are orthonormal bases of Bob's two qubit space ${B_1B_2}$,
we conclude that there exists a unitary matrix $U^{B_1B_2}$ such that
\be\label{b1b2}
|xy\ra^{B_1B_2}=U^{B_1B_2}|\phi_{xy}^{B_1B_2}\ra\quad\forall\;x,y\in\{0,1\}\;.
\ee
It turns out that the unitary $U^{AB}$ as defined above can be expressed as a CNOT gate\index{CNOT gate} followed by a Hadamard gate on system $B_1$ (see Bob's side in Fig.~\ref{csuperdense} between time steps $t_2$ and $t_3$). Explicitly,
\ba
U^{B_1B_2}&=\left(H|0\lr 0|^{B_1}\right)\otimes I^{B_2}+\left(H|1\lr 1|^{B_1}\right)\otimes X^{B_2}\\
&=|+\lr 0|^{B_1}\otimes I^{B_2}+|-\lr 1|^{B_1}\otimes X^{B_2}
\ea
Hence, after the application of the unitary $U^{B_1B_2}$ on Bob's system, Alice and Bob state is
$
|x\ra^{A_1}|y\ra^{A_2}|x\ra^{B_1}|y\ra^{B_2}
$.
That is, the quantum circuit in Fig.~\ref{csuperdense} simulates the linear transformation 
\be
|x\ra^{A_1}|y\ra^{A_2}\to|x\ra^{A_1}|x\ra^{B_1}\otimes|y\ra^{A_2}|y\ra^{B_2}
\ee
which is equivalent to two coherent channels. The resources we used to simulate these two coherent channels are precisely the same ones used in superdense coding to simulate two noiseless classical channels.
\begin{exercise}
Show that the unitary matrix $U^{B_1B_2}$ above satisfies~\eqref{b1b2}. 
\end{exercise}

\begin{exercise}
Suppose the initial ebit shared between Alice and Bob was given in the singlet state $|\Psi_{-}^{AB}\ra$ instead of $|\Phi_{+}^{AB}\ra$, and consider the exact same protocol as in Fig.~\ref{csuperdense}, until time step $t_2$. Revise the unitary matrix $U^{B_1B_2}$ after time step $t_2$ so that the protocol still simulates 2 coherent channels.
\end{exercise}

\subsection{Coherent Teleportation}\index{coherent teleportation}

\begin{figure}[h]
\centering
    \includegraphics[width=0.5\textwidth]{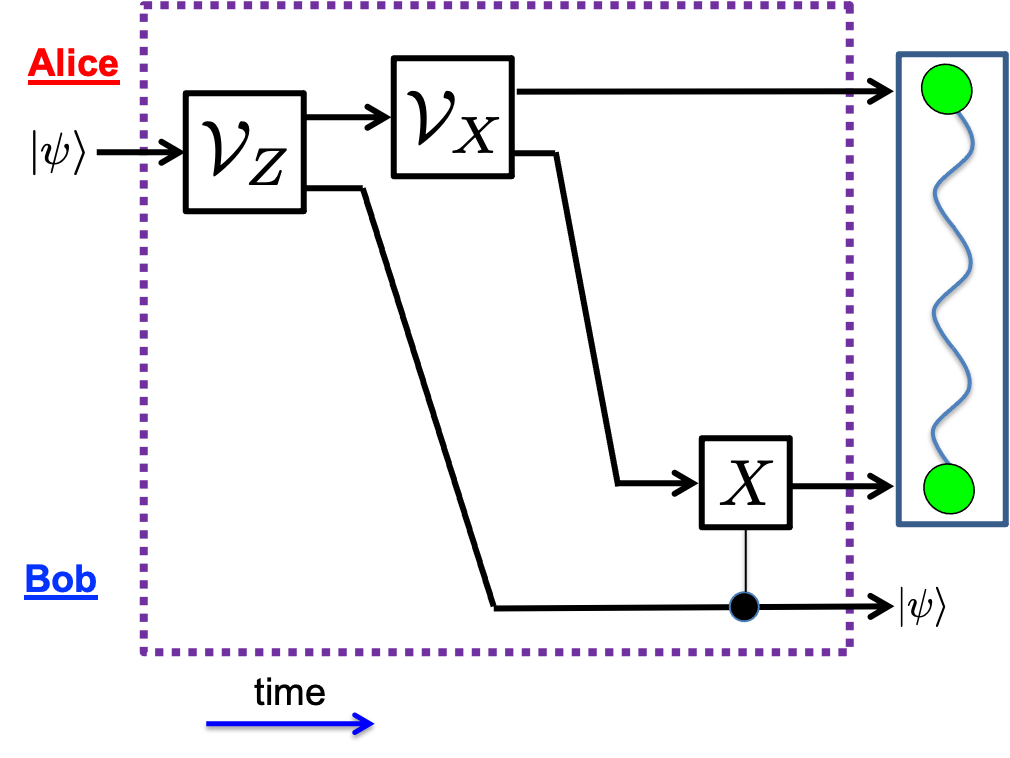}
  \caption{\linespread{1}\selectfont{\small Coherent Quantum Teleportation. Two cobit channels produce one ebit plus one use of a noiseless qubit channel.}}
  \label{cteleportation}
\end{figure}

The coherent teleportation protocol is the resource reversal of the coherent superdense coding protocol. Particularly, it reveals that two uses of a cobit channel are sufficient to generate one ebit and at the same time simulate one use of a qubit channel.
That is, the coherent teleportation protocol demonstrates that
\be\label{r2c}
2[q\to qq]\geq [qq]+[q\to q] \;.
\ee
The protocol achieving this resource inequality is depicted in Fig.~\ref{cteleportation}.
 
In the first step of the protocol Alice sends a qubit $|\psi^A\ra=a|0\ra^A+b|1\ra^A$ into the first cobit channel.
The cobit channel $\mV_Z$ transforms this state into the state
\be
a|0\ra^A|0\ra^{B_1}+b|1\ra^A|1\ra^{B_1}=\frac{1}{\sqrt{2}}\left[|+\ra^{A}\left(a|0\ra^{B_1}+b|1\ra^{B_1}\right)+
|-\ra^{A}\left(a|0\ra^{B_1}-b|1\ra^{B_1}\right)\right]\;.
\ee 
In the next step system $A$ goes through the second cobit channel, $\mV_X$, yielding the state
\be
\frac{1}{\sqrt{2}}\left[|+\ra^{A}|+\ra^{B_2}\left(a|0\ra^{B_1}+b|1\ra^{B_1}\right)+
|-\ra^{A}|-\ra^{B_2}\left(a|0\ra^{B_1}-b|1\ra^{B_1}\right)\right]\;.
\ee
Finally, in the last step Bob sends his systems through a CNOT gate. Note that $X|+\ra=|+\ra$ so that
the CNOT gate\index{CNOT gate} only changes $|-\ra^{B_2}|1\ra^{B_1}$ to $ -|-\ra^{B_2}|1\ra^{B_1}$ while keeping all the other terms intact. Hence, after Bob's CNOT gate, Alice and Bob share the state
\be
\frac{1}{\sqrt{2}}\left[|+\ra^{A}|+\ra^{B_2}\left(a|0\ra^{B_1}+b|1\ra^{B_1}\right)+
|-\ra^{A}|-\ra^{B_2}\left(a|0\ra^{B_1}+b|1\ra^{B_1}\right)\right]=|\Phi_{+}^{AB_2}\ra|\psi^{B_1}\ra\;.
\ee
That is, at the end of the protocol Alice teleported her quantum state $|\psi\ra$ to Bob's system $B_1$, and also share with Bob's system $B_2$ the maximally entangled state $\Phi_{+}^{AB_2}$.

Coherent quantum teleportation and coherent superdense coding demonstrate that two cobit channels have the same resource value as one ebit and one use of a qubit channel. 
\be
[qq]+[q\to q]=2[q\to qq]\;.
\ee
This means that coherent teleportation is the reversal process of coherent superdense coding and vice versa.
 
 \section{Notes and References}
 
 Quantum teleportation was discovered by~\cite{BBC+1993}, and quantum superdense coding by~\cite{BW1992}. These seminal papers paved the way for the development of quantum Shannon theory and entanglement theory, as they demonstrated that entanglement, besides of being interesting from a fundamental point of view, is a resource that can be consumed to achieve certain exotic tasks such as quantum teleportation. Moreover, these protocols are considered as the \emph{unit} protocols (see e.g.~\cite{Wilde2013}), since they form the building blocks with which one studies the capabilities of noisy quantum channels to transmit information in asymptotic settings involving many uses of the channels.
 
The resource analysis using notations such as $[q\to q]$ was first introduced in~\cite{DHW2008} were the rules of this `resource calculus' developed. The coherent bit, coherence teleportation, and coherent superdense coding is due to~\cite{Harrow2004}. More details on coherent communication can be found in the book of Wilde~\cite{Wilde2013}.

\part{Preliminaries}

\chapter{Elements of Quantum Mechanics I: Closed Systems}\label{quantum0}\index{quantum mechanics}

Quantum mechanics, which was discovered during the first quarter of the twentieth century, has profoundly transformed our understanding of the world around us. The theory predicts a plethora of non-intuitive phenomena, including entanglement, quantum non-locality, wave-particle duality, coherence, the uncertainty principle, quantum contextuality, quantum steering, and no-cloning, to name a few. This remarkable departure from classical physics has led to numerous thought-provoking papers and interpretations of quantum mechanics. To date, there is no consensus on which view of quantum mechanics is the most natural one to adopt. Additionally, sub-fields of science, such as quantum logic, have arisen from these phenomena, particularly the inconsistency of classical logic and the uncertainty principle.

The development of quantum mechanics was a gradual process that involved many trials and errors. It began in 1900 with Max Planck's discretization of energy values used to solve the black body radiation. The process continued with Einstein's 1905 paper on the correspondence between energy and frequency, which provided a quantum explanation for the photoelectric effect. The process ultimately ended with the formalism that was developed in the mid-1920s by Erwin Schrödinger, Werner Heisenberg, Max Born, and others.

In this book, we will not review quantum mechanics from a historical or traditional perspective. Instead, we will study it in the context of the modern field of quantum information science, which emerged and developed in the early 1990s. We will discuss the theory's basic postulates, its corresponding mathematical structure, and its many consequences and applications, particularly to information theory, resource theories, and more broadly to physics and science.

The interplay between the concept of `information' and the field of quantum science is a complex and multifaceted one. The fundamental principles of quantum mechanics, such as the superposition of states and entanglement, have led to the development of quantum information theory, which studies the processing and transmission of information using quantum systems. Moreover, the very act of observing a quantum system can alter its state, and this observation is itself a form of information. This has profound implications for our understanding of the nature of reality and the limits of our ability to measure it. Thus, the relationship between `information' and quantum science is a rich and nuanced one, encompassing a broad range of topics from the foundations of quantum mechanics to the practical applications of quantum technologies.

In Shannon's terms, information is defined as ``that which can distinguish one thing from another." For instance, a coin has two sides, ``head" and ``tail," and the ability to differentiate between the two implies that the coin can store information. Typically, when referring to the distinguishability of two elements, we denote the options as $0$ and $1$, and information is then measured in bits, where the number of bits represents the number of distinguishable elements. For example, two bits correspond to four possible elements.

The definition of information is abstract and detached from any specific implementation or labeling. For instance, one bit may correspond to the head or tail of a coin, or the 5-volt versus 0-volt of an electrical circuit. All information processing, such as communication, computation, and manipulation of information, can be performed with either coins or electrical circuits. While storing information in coins is impractical (especially for large numbers of bits), from an information-theoretic perspective, any object can be used to implement classical bits, and we say that information is \emph{fungible}.

\begin{figure}[b]
\centering
    \includegraphics[width=0.2\textwidth]{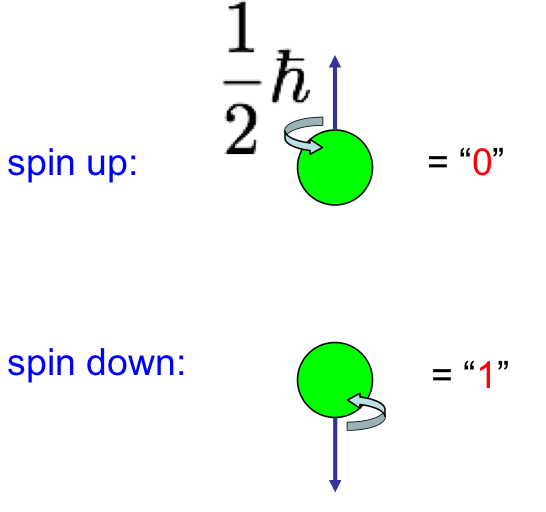}
  \caption{\linespread{1}\selectfont{\small Encoding information with the spin of an electron}}
  \label{fig1}
\end{figure}

However, what happens when we attempt to encode information in the spin of an electron? The electron, being an elementary particle, is uniquely determined by its mass and spin. The magnitude of its spin can only take one value, $\frac{1}{2}\hbar$ (where $\hbar$ is a unit of angular momentum), and its spin can point in any direction. The Stern-Gerlach experiment, discussed below), demonstrates that it is possible to differentiate between ``up" and ``down" along the $z$-direction of the spin of an electron (see Fig.~\ref{fig1}). This implies that information can be encoded in the spin of an electron as well.

Are there any advantages to encoding information in the spins of quantum particles, such as electrons, as opposed to larger classical systems like coins or electrical circuits? If information is fungible, why should encoding information in quantum particles make any difference? Before addressing these questions, we will introduce the Stern-Gerlach experiment and the necessary elements from linear algebra for the study of quantum physics.

\section{The Stern-Gerlach Experiment}\label{stern}\index{Stern-Gerlach experiment}

The Stern-Gerlach (SG) experiment, conducted in Frankfurt, Germany in 1922 by Otto Stern and Walther Gerlach, is a groundbreaking demonstration of the quantization of the spatial orientation of angular momentum in physical systems. The experiment has become a paradigm of quantum measurement, and is considered one of the most significant experiments of the 20th century. It paved the way for numerous subsequent experiments, complementing the double-slit experiment and contributing to our understanding of the behavior of quantum systems.

\begin{figure}[h]\centering
    \includegraphics[width=0.5\textwidth]{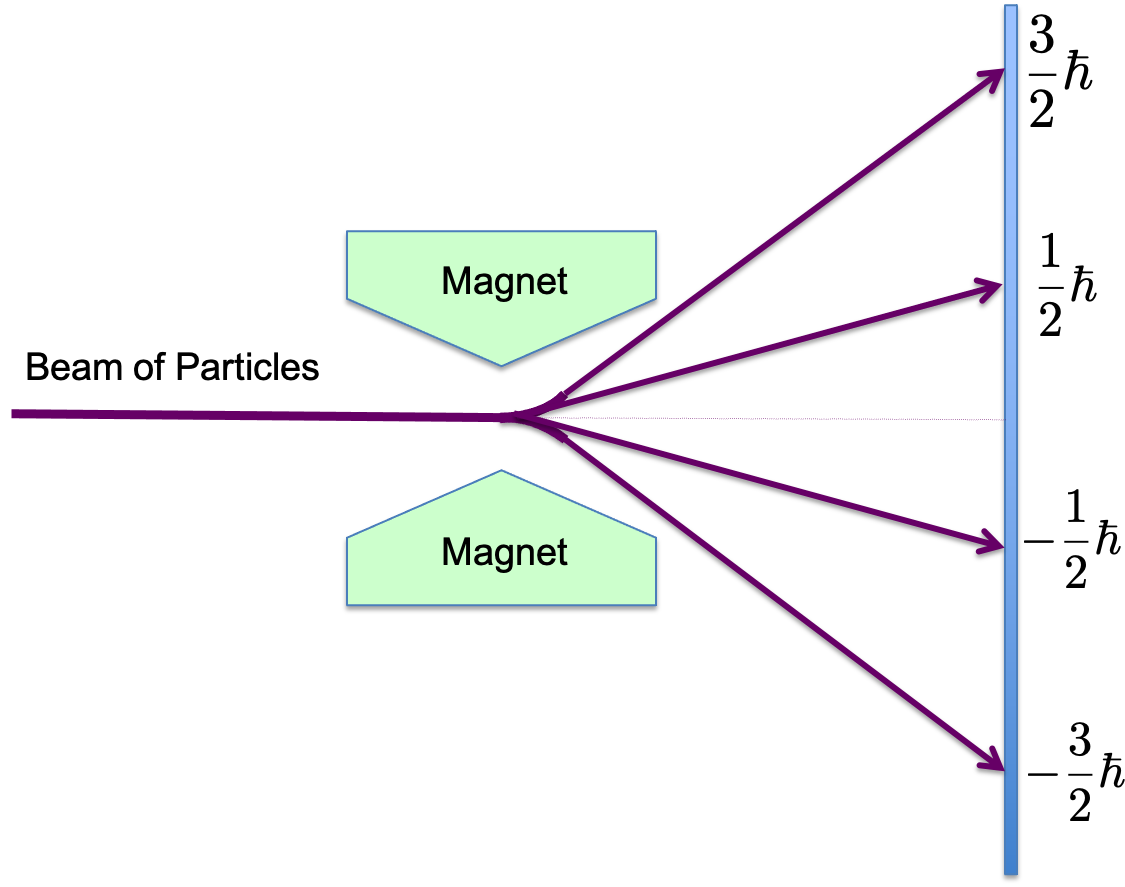}
  \caption{\linespread{1}\selectfont{\small The Stern-Gerlach Experiment.}}
  \label{fig2}
\end{figure}

The SG experiment's relative simplicity and its ability to measure discrete properties of physical systems in finite dimensions make it an excellent example of how information can be extracted from a physical system. In the experiment, a beam of particles is passed through a spatially varying magnetic field, which causes the beam to split into discrete components that are detected on a screen. This behavior indicates that the angular momentum of the particles is quantized; i.e., taking only discrete values. A heuristic illustration of the SG experiment is provided in Fig.~\ref{fig2}. 

The SG experiment involves emitting electrons or atoms from an oven and passing them through a non-homogeneous magnetic field. Their spin orientation determines whether they hit the screen above or below the horizontal line, with the distance from the line reflecting the magnitude of their spin. The experiment can be performed on any physical system, and always yields a discrete spectrum for the angular momentum, given by $\frac{1}{2}n\hbar$ where $n$ is an integer. Regions on the screen in Fig.~\ref{fig2} where no particles hit indicate this quantization, which is true regardless of the type of particle used in the experiment.

The SG experiment has two noteworthy implications. Firstly, electrons always hit the screen at the same distance from the horizontal line, indicating a consistent magnitude of $\frac{1}{2}\hbar$ for their spin. Secondly, all electrons hit the screen in the same two areas, irrespective of their initial spin direction. Even if an electron's initial spin is pointing in the $x$-direction, it should have zero spin in the $z$-direction and therefore should not be deflected by the non-zero $z$-gradient of the magnetic field. However, the fact that electrons still hit the screen in the same two areas shows that measuring the spin in one direction affects its value in another direction.

\begin{figure}[h]\centering
    \includegraphics[width=0.6\textwidth]{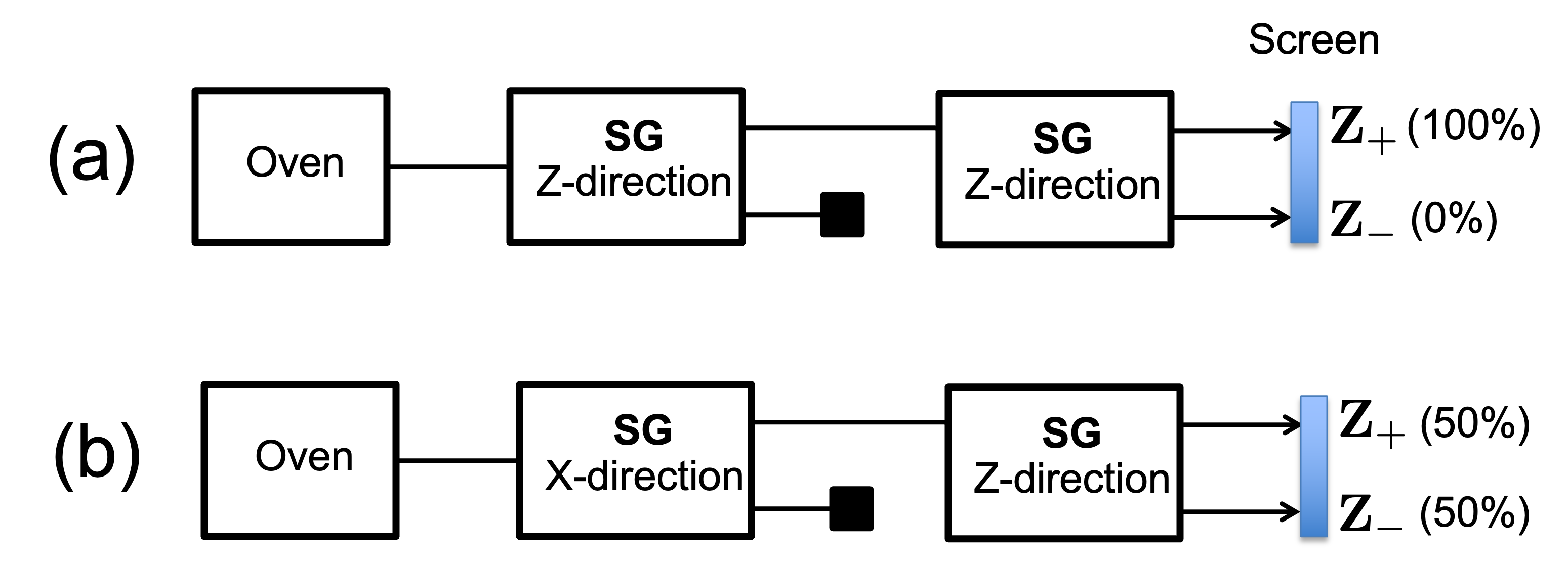}
  \caption{\linespread{1}\selectfont{\small Combination of Stern-Gerlach (SG) experiments}}
  \label{fig3}
\end{figure}

To verify that an electron's spin remains intact after it is deflected by the SG experiment, we can concatenate two SG experiments in a variety of ways. For example, suppose we want to confirm that an electron deflected upwards in the first SG box has a spin pointing in the upward $z$-direction. We can set up the experiment as shown in Fig.~\ref{fig3} (a). After passing through the first SG box, only electrons deflected upwards will continue on to the second SG box, while those deflected downward are blocked. After passing through the second SG box, the electrons hit the screen. We observe that no electrons hit the screen below the horizontal line, indicating that all electrons deflected upwards in the first SG box also have a spin pointing in the upward $z$-direction. This implies that after an electron's spin has been measured, it remains intact by further measurements in the same direction.

Another interesting feature of the SG experiment is that it allows us to measure the spin of particles in different directions. For instance, we can measure the spin in the $x$-direction by using a modified version of the experiment. In this case, the electrons are deflected to the left or to the right, depending on their spin in the $x$-direction. Figure~\ref{fig3} (b) shows a schematic of this experiment, where the first SG box measures the spin in the $x$-direction, and the second SG box measures the spin in the $z$-direction.

The results of this experiment are surprising from a classical point of view. According to classical physics, any physical system with its angular momentum pointing in the $x$-direction should have zero angular momentum pointing in the $z$-direction. However, in the SG experiment, we find that 50\% of the electrons are deflected upward and 50\% are deflected downward after passing through the second SG box. This shows that the spin in different directions of a quantum particle is not fully determined by its spin in one particular direction, and that measurements of spin along different axes can yield nontrivial and unexpected results.

\begin{figure}[h]\centering    \includegraphics[width=0.6\textwidth]{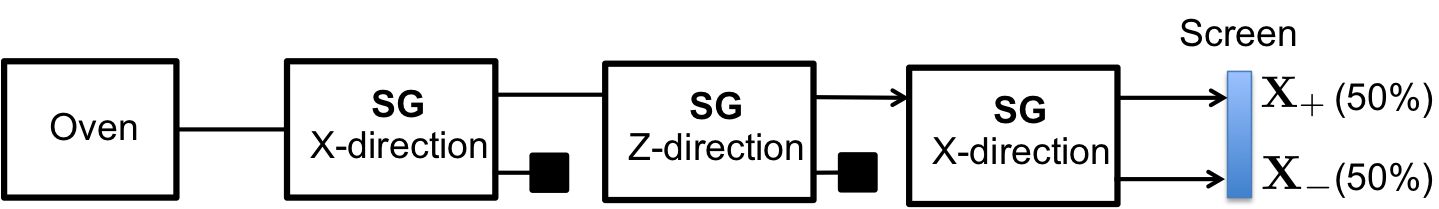}
  \caption{\linespread{1}\selectfont{\small Erasing information with measurement.}}
  \label{fig4}
\end{figure}

What happens to a particle with spin in the $x$-direction when it passes through an SG-experiment in the $z$-direction, followed by an SG-experiment in the $x$-direction? Will its resulting spin in the $x$-direction remain the same? Figure \ref{fig4} illustrates such a scenario with three SG-boxes. The first box filters only particles with spins in the positive (right) $x$-direction. The second box measures their spin in the $z$-direction, and the third box measures their spin in the $x$-direction. Remarkably, there is a 50\% chance that the resulting particle will have spin in the negative $x$-direction. This implies that the measurement in the $z$-direction erases the information about the spin in the $x$-direction completely.

This phenomenon, known as quantum mechanical complementarity, is a fundamental feature of quantum mechanics and is one of the most profound concepts in physics. It implies that it is impossible to simultaneously measure certain pairs of physical quantities with arbitrary precision, such as position and momentum, or spin in different directions. Any measurement of one quantity necessarily disturbs the other, and the uncertainty principle sets a fundamental limit on the precision with which they can be measured simultaneously.

To summarize, the SG experiments discussed above demonstrate that measurements can disturb the state of a system. For instance, an electron with spin initially pointing upward may end up pointing downward after a sequence of SG-experiments. However, this disturbance is not unique to a specific measurement apparatus; the same behavior and statistical outcomes will occur regardless of the device used. In other words, if a particle's spin is known to be pointing in the positive $z$-direction, then a measurement in the $x$ or $y$ directions will erase all knowledge about its spin in the $z$-direction, irrespective of which measurement apparatus is employed. Therefore, it is the \emph{knowledge} of the spin in one direction that prevents its knowledge in another orthogonal direction. This phenomenon is a special instance of the uncertainty principle.

\section{Inner Product Spaces and Hilbert Spaces}\index{inner product}\index{Hilbert space}

In this section and the following one, we discuss essential topics from linear algebra that serve as foundations for quantum theory. While most of the material is covered in standard textbooks on linear algebra, we introduce the Dirac notation for vectors and operators, as well as concepts such as quantum states and maximally entangled states that are used throughout this book. Therefore, even readers familiar with the topics covered here may benefit from reviewing and acquainting themselves with the notations we utilize.

We consider vector spaces over either the real or complex numbers and use the notation $\mbb{F}$ when referring to either field.

\begin{myd}{Inner Product Space}\index{inner product space}
\begin{definition}
An inner product space is a vector space $A$ over a field $\mbb{F}$ equipped with a map 
\be
\la\; |\;\ra \;:A\times A\to\mbb{F}
\ee
that satisfies for all vectors $\psi,\phi,\chi\in A$ and every scalar $c\in\mbb{F}$ the following three axioms: 
\begin{enumerate}
\item Conjugate symmetry: $\la\psi|\phi\ra=\overline{\la\phi|\psi\ra}$.
\item Linearity in the second argument: $\la\psi|c\phi\ra=c\la\psi|\phi\ra$ and $\la\psi|\phi+\chi\ra=\la\psi|\phi\ra+\la\psi|\chi\ra$.
\item Positive-definiteness: $\la\psi|\psi\ra\geq 0$ with equality if and only if $|\psi\ra=\0$.
\end{enumerate}
\end{definition}
\end{myd}
\begin{remark}
We adopt the notation $\la\; |\;\ra$ instead of $\la\; ,\;\ra$, as it conforms with the Dirac notation that we will define shortly. Additionally, while most mathematics textbooks define the linearity property with respect to the first argument, we will use the aforementioned notation to suitably align with the Dirac notation. We use the letters $A$, $B$, and $C$ to denote Hilbert spaces since, in quantum physics, Hilbert spaces correspond to physical systems that are operated on by parties such as Alice, Bob, Charlie, etc.\index{Dirac notation}
\end{remark}

The inner product induces a norm defined by
\be\label{norm}
\|\psi\|_2\eqdef \la\psi|\psi\ra^{1/2}\quad\quad\forall\;\psi\in A\;,
\ee
and a metric 
\be
d(\psi,\phi)\eqdef\|\psi-\phi\|_2\quad\quad\forall\;\psi,\phi\in A\;.
\ee 

\begin{myd}{A Norm}\index{norm}
\begin{definition} A norm is a real-valued function defined on the vector space, $A$, that has the following properties:
\begin{enumerate}
\item For all $\psi\in A$, $\|\psi\|\geq 0$ with equality if and only if $\psi=0$.
\item For all $c\in\mbb{F}$ and $\psi\in A$, $\|c\psi\|=|c|\|\psi\|$.
\item The triangle inequality\index{triangle inequality}: for all $\psi,\phi\in A$, $\|\psi+\phi\|\leq\|\psi\|+\|\phi\|$.
\end{enumerate}
A vector space equipped with a norm is called a normed space.
\end{definition}
\end{myd}
\begin{exercise}
Show that the norm defined by the inner product in equation \eqref{norm} indeed satisfies the three fundamental properties of a norm.
\end{exercise}

\bex\label{nrmp}
Let $A$ be a normed space, and let $\psi,\phi\in A$ with $\|\phi\|=1$. Show that
\be
\left\|\frac1{\|\psi\|}\psi-\phi\right\|\leq 2\|\psi-\phi\|\;.
\ee
Hint: Write $\frac1{\|\psi\|}\psi-\phi=\frac1{\|\psi\|}\psi-\psi+\psi-\phi$ and use the norm properties.
\eex

It is important to note that not all norms are derived from an inner product, and as a result, not every metric necessarily originates from either an inner product or a norm.

\begin{exercise}[The $p$-Norms and The H\"older Inequality] \label{lp}\index{H\"older inequality}
The normed space $\ell^p(\mbb{C}^n)$ (with $p\in[1,\infty]$) is the vector space $\mbb{C}^n$ equipped with the $p$-norm, $\|\cdot\|_p$, defined on all $\psi=(a_1,\ldots,a_n)^T\in\mbb{C}^n$ as:
\be\label{0pnorm}
\|\psi\|_p\eqdef\left(|a_1|^p+\cdots+|a_n|^p\right)^{1/p}\;.
\ee
The H\"older inequality states that for any $p,q\in[1,\infty]$ with $\frac1p+\frac1q=1$ and any $\psi=(a_1,\ldots,a_n)^T\in\mbb{C}^n$ and $\phi=(b_1,\ldots,b_n)^T\in\mbb{C}^n$ we have
\be\label{holder}
\sum_{x\in[n]}|a_xb_x|\leq \|\psi\|_p\|\phi\|_q\;.
\ee
Use this to show that $\|\cdot\|_p$ is a norm, and that for $p\neq 2$ this norm is not induced from an inner product.
\end{exercise}

The $p$-norms above have numerous applications in many fields of science. Of particular importance, are the extreme cases of $p=1$ and $p=\infty$: 
\be
\|\phi\|_1\eqdef |c_1|+\cdots+|c_n|\quad\text{and}\quad\|\phi\|_\infty\eqdef \max_{x\in[n]}|c_x|\;,
\ee
where throughout the book we use the notation $[n]\eqdef\{1,\ldots,n\}$ for every integer $n\in\mbb{N}$.
For these cases, the norms above behave monotonically under stochastic matrices as we show now.

Let $S=(s_{xy})\in\stoc(m,n)$ be an $m\times n$ column stochastic matrix; i.e. a matrix whose components  are non-negative real numbers, and all the columns sums to one. Then, the $1$-norm behaves monotonically under such matrices; i.e.
\be\label{dpisv}\index{trace norm}
\|S\v\|_1\leq\|\v\|_1\quad\forall\v\in\mbb{C}^n\;.
\ee
Indeed, by definition
\ba
\|S\v\|_1=\sum_{x\in[m]}\Big|\sum_{y\in[n]}s_{xy}v_y\Big|\leq\sum_{x\in[m]}\sum_{y\in[n]}s_{xy}\left|v_y\right|=\sum_{y\in[n]}\left|v_y\right|=\|\v\|_1\;.
\ea
where we used the triangle inequality.

\begin{exercise}
Let $R\in\mbb{R}^{m\times n}_+$ be a row stochastic matrix (i.e. a matrix with non-negative components whose rows sum to one). Show that 
\be\index{operator norm}
\|R\v\|_\infty\leq\|\v\|_\infty\quad\quad\forall\;\v\in\mbb{C}^n\;.
\ee
\end{exercise}

\subsection{Hilbert Spaces}\index{Hilbert space}

A \emph{Cauchy sequence} in an inner product space, $A$, is a sequence of vectors $\{\psi_n\}_{n\in\mbb{N}}$ in $A$ with the property that for any $\eps>0$ there exists $N\in\mbb{N}$ such that for any $n,m>N$\index{Cauchy sequence}
\be
\|\psi_n-\psi_m\|<\eps\;.
\ee
An inner product space $A$ is said to be \emph{complete} if any Cauchy sequence in $A$ converges in $A$, with respect to the metric induced by the inner product. That is, $A$ is complete if for any Cauchy sequence, $\{\psi_n\}_{n\in\mbb{N}}\subseteq A$, there exists a state $\psi\in A$ such that $\lim_{n\to\infty}\|\psi_n-\psi\|=0$. Complete inner product spaces are called Hilbert spaces and complete normed spaces are called Banach spaces.

Common examples of Hilbert spaces are $\mbb{R}^n$ and $\mbb{C}^n$ with the standard inner products. In finite dimensions, all Hilbert spaces are isomorphic to these spaces. A common example in quantum mechanics is the Hilbert space of square integrable functions\index{$L^2(\mbb{R})$}\index{Hilbert space}
\be
L^2(\mbb{R})\eqdef\Big\{f:\mbb{R}\to\mbb{C}\;\;\;:\;\;\;\int_{-\infty}^{\infty}|f(x)|^2dx<\infty\Big\}\;,
\ee
where the inner product between two function $f,g\in L^2(\mbb{R})$ is defined as
\be
\la g, f\ra\eqdef\int_{-\infty}^{\infty}f(x)\overline{g(x)}dx\;.
\ee

\begin{exercise}
Show that $L^2(\mbb{R})$ satisfies all the axioms of a Hilbert space.
\end{exercise}

To distinguish more clearly between Hilbert spaces and inner product spaces, consider the space $\mc([a,b])$ of continuous complex-valued functions. This infinite-dimensional vector space is equipped with an inner product given by: \
\be
\la g, f\ra\eqdef\int_{a}^{b}f(t)\overline{g(t)}dt\quad\quad\forall\;f,g\in\mc([a,b])\;.
\ee
However, this space is not a Hilbert space since it is not complete with respect to the metric induced by this inner product. For example, consider the following sequence of continuous functions in $\mc([-1,1])$:
\be
f_k(t)\eqdef \begin{cases}
       0&\quad\text{if} \quad t\in[-1,0) \\
       kt&\quad\text{if} \quad t\in[0,\frac{1}{k}) \\ 
       1 &\quad\text{if}\quad t\in[\frac{1}{k},1]
     \end{cases}\;.
\ee
Note that while the sequence is Cauchy, its limit does not exist in $\mc([-1,1])$ as $f_k$ cannot converge to a continuous function. However, this book mostly considers finite-dimensional Hilbert spaces, in which all inner product spaces are complete and isomorphic to $\mbb{C}^n$ (or $\mbb{R}^n$). Thus, such examples are not relevant for our purposes, and we will use the terms `Hilbert space' and `inner product space' interchangeably. Similarly, in finite dimensions, all normed spaces are Banach spaces (i.e., complete normed spaces).

Another important example of a Hilbert space, 
is the space of $n\times m$ complex matrices, denoted by $\mbb{C}^{n\times m}$. 
In this space, the inner product is defined as follows: for all $M,N\in\mbb{C}^{n\times m}$
\be\label{xy}
\la M|N\ra\eqdef\tr\left[M^*N\right]
\ee
where $M^*$ denotes the adjoint of $M$\index{adjoint operator}. This specific inner product is referred to as the `Hilbert-Schmidt' inner product, and it is occasionally denoted with a subscript as $\la\; |\;\ra_{HS}$.\index{Hilbert-Schmidt inner product}
\begin{exercise} Consider the space $\mbb{C}^{m\times n}$.
\begin{enumerate}
\item Show that the definition in~\eqref{xy} satisfies the three axioms of an inner product.
\item  Find an isometrical isomorphism $\vect:\mbb{C}^{m\times n}\to\mbb{C}^{mn}$ such that for all $M,N\in\mbb{C}^{m\times n}$
\be
\la M|N\ra_{HS}=\la \vect(M)|\vect(N)\ra_2
\ee
where $\la\; |\;\ra_{HS}$ is the Hilbert-Schmidt inner product of $\mbb{C}^{m\times n}$, and $\la\; |\;\ra_2$ is the standard inner product of $\mbb{C}^{mn}$.
\item Express the norm induced by the inner product~\eqref{xy}.
\end{enumerate}
\end{exercise} 

\begin{exercise} Prove the following two properties of a Hilbert space $A$:
\begin{enumerate}
\item The Cauchy-Schwarz inequality: for all $\psi,\phi\in A$
\be
|\la\phi,\psi\ra|\leq \|\phi\|\|\psi\|\;.
\ee 
\item Pythagorean theorem: If $\psi_1,\ldots,\psi_n$ are orthogonal vectors, that is, $\la\psi_x|\psi_y\ra=0$ for distinct indices $x,y\in[n]$, then
\be
\Big\|\sum_{x\in[n]}\psi_x\Big\|^2=\sum_{x\in[n]}\|\psi_x\|^2\;.
\ee
\end{enumerate}
\end{exercise} 

\subsection{The Dirac Notations}\index{Dirac notation}

We will employ Dirac notation along with the ket symbol to represent vectors in a Hilbert space $A$. Consequently, going forward, we will represent the vectors $\psi$, $\phi$, and $\chi$ in $A$ as $|\psi\rangle$, $|\phi\rangle$, and $|\chi\rangle$. The standard basis, also known as the computational basis of $\mathbb{C}^n$, will be denoted as $\{|1\rangle$, $|2\rangle$, $\ldots$, $|n\rangle\}$ or sometimes as $\{|0\rangle$, $|1\rangle$, $\ldots$, $|n-1\rangle\}$. The latter notation is typically used for $n=2$, in which case the standard basis is given by
\be
|0\ra\eqdef\begin{pmatrix}
1 \\
0
\end{pmatrix}
\quad\text{and}\quad
|1\ra\eqdef\begin{pmatrix}
0 \\
1
\end{pmatrix}.
\ee

The \emph{dual space}, $A^*$, of a Hilbert space, $A$, is defined as the set of all linear functionals on $A$. A linear functional is a function $f:A\to\mbb{F}$ with the property that for all $|\psi\ra,|\phi\ra\in A$ and $a,b\in\mbb{F}$ we have $f(a|\psi\ra+b|\phi\ra)=af(|\psi\ra)+bf(|\phi\ra)$. For a fixed vector $|\chi\ra\in A$, the function $f_\chi:A\to\mbb{F}$ defined by $f_\chi(|\psi\ra)\eqdef\la\chi|\psi\ra$ is a linear functional, and every linear functional has this form. It is therefore convenient  
to denote the linear functionals with a `bra' notation. That is, instead of $f_\chi$, we denote this functional simply by $\la\chi|$, so that its action on an element $|\psi\ra$ is given by the inner product $\la\chi|\psi\ra$. Hence, $A^*$ consists of bra vectors. 

As an example, the dual space of $\mbb{C}^2$ is spanned by the standard bra basis
\be
\la0|\eqdef (1,0)
\quad\text{and}\quad
\la1|\eqdef (0,1).
\ee
Note that there is a one-to-one correspondence between $A=\mbb{C}^n$ and its dual $A^*$, via the bijective mapping 
\be
|\psi\ra=\sum_{x\in[n]}c_x|x\ra\quad\mapsto\quad\la\psi|=\sum_{x\in[n]}\bar{c}_{x}\la x|\;,
\ee
where $c_x\in\mbb{C}$ and $\bar{c}_x$ denotes the complex conjugate of $c_x$. Moreover, we will denote $\la\psi|\rho|\phi\ra\eqdef\la\psi|\rho\phi\ra$, where $\rho$ is a linear transformation (see below).

\subsection{Direct Sum of Hilbert Spaces}\index{direct sum}

Given two subspaces $A$ and $B$ of a Hilbert space $C$, we say that $C$ is the \emph{direct sum}
of $A$ and $B$, and write $C=A\oplus B$ if $C=A+B$ and $A\cap B=\{\0\}$. 
Note that in this case $|C|=|A|+|B|$, where we use the notation $|A|$ to denote the dimension of $A$. If $A$ and $B$ are two arbitrary Hilbert spaces we can always construct a third Hilbert space, $C$, such that $C=A\oplus B$. For example, given two Hilbert spaces $A$ and $B$ define (abstractly)
\be
C\eqdef\left\{(\psi,\phi)\;:\;|\psi\ra\in A\quad,\quad|\phi\ra\in B\right\}
\ee
with addition rule $(\psi_1,\phi_1)+(\psi_2,\phi_2)\eqdef (\psi_1+\psi_2,\phi_1+\phi_2)$, scalar multiplication
$c(\psi,\phi)=(c\psi,c\phi)$, and inner product,
\be
\la (\psi_1,\phi_1)|(\psi_2,\phi_2)\ra\eqdef\la\psi_1|\psi_2\ra+\la\phi_1|\phi_2\ra\;.
\ee
Identifying an element $|\psi\ra\in A$ with an element $(\psi,0)\in C$ and an element $|\phi\ra\in B$ with an element $(0,\phi)\in C$, we conclude that $C=A\oplus B$. For example, $\mbb{R}^3$ can be decomposed as $\mbb{R}^2\oplus\mbb{R}$ or as $\mbb{R}\oplus\mbb{R}\oplus\mbb{R}$.  

\subsection{Tensor Product of Hilbert Spaces}\index{tensor product}

Another way to combine two Hilbert spaces is through their \emph{tensor product}. In quantum information, a composite physical system is often described by an element of a tensor product of Hilbert spaces. We will only consider tensor products of finite dimensional Hilbert spaces, although it's worth noting that the tensor product can be defined in a basis-independent manner as a quotient space of a free vector space.

Let $A$ and $B$ be two finite dimensional Hilbert spaces with dimensions $|A|$ and $|B|$, respectively. We define a \emph{bilinear} function $\otimes$ that takes two vectors $|\psi\ra \in A$ and $|\phi\ra \in B$ and returns an element of the form $|\psi\ra\otimes|\phi\ra$. The bilinearity of $\otimes$ means that for all $c\in\mbb{F}$, and all vectors 
$|\psi_1\ra,|\psi_2\ra\in A$, and $|\phi_1\ra,|\phi_2\ra\in B$,
\begin{enumerate}
\item $\left(|\psi_1\ra+|\psi_2\ra\right)\otimes |\phi\ra=|\psi_1\ra\otimes |\phi\ra+|\psi_2\ra\otimes |\phi\ra$
\item $|\psi\ra\otimes\left(|\phi_1\ra+|\phi_2\ra\right)=|\psi\ra\otimes|\phi_1\ra+|\psi\ra\otimes|\phi_2\ra$
\item $c\left(|\psi\ra\otimes|\phi\ra\right)=\left(c|\psi\ra\right)\otimes|\phi\ra=|\psi\ra\otimes\left(c|\phi\ra\right)$
\end{enumerate}
The span of all such objects, $|\psi\ra\otimes|\phi\ra$, with $|\psi\ra\in A$ and $|\phi\ra\in B$ is denoted by $A\otimes B$.
Alternatively,
let $\{|x\ra^{A}\}_{x\in[m]}$ and  $\{|y\ra^B\}_{y\in[n]}$ be two corresponding orthonormal bases of $A$ and $B$, where $m\eqdef|A|$ and $n\eqdef|B|$ denote the dimensions of $A$ and $B$, respectively. Then, $A\otimes B$ can be defined as the collection
\be\label{deften}
A\otimes B\eqdef\Big\{\sum_{x\in[m]}\sum_{y\in[n]}\mu_{xy}|x\ra^{A}\otimes|y\ra^B\;\;:\;\mu_{xy}\in\mbb{F}\Big\}
\ee
From its definition above, it follows that ${A}\otimes B$ is a vector space with an orthonormal basis $\{|x\ra^{A}\otimes|y\ra^B\}$. In particular, note that $|{A}\otimes B|=|AB|\eqdef |{A}||B|$ (we will therefore sometimes use the notation $AB$ to mean $A\otimes B$). The inner product between two elements $|\psi_1\ra\otimes|\phi_1\ra$ and $|\psi_2\ra\otimes|\phi_2\ra$ is simply given by the product of the inner products; i.e. $\la\psi_2|\psi_1\ra\la\phi_2|\phi_1\ra$. More generally, given two states
\be
|\psi^{{A}B}\ra=\sum_{x\in[m]}\sum_{y\in[n]}\mu_{xy}|x\ra^{A}\otimes|y\ra^B\quad\text{and}\quad
|\phi^{{A}B}\ra=\sum_{x\in[m]}\sum_{y\in[n]}\nu_{xy}|x\ra^{A}\otimes|y\ra^B\;,
\ee
their inner product is given by
\be\label{xy2}
\la\psi^{{A}B}|\phi^{{A}B}\ra=\tr\left[M^*N\right]
\ee
where the matrices $M\eqdef(\mu_{xy})$ and $N\eqdef(\nu_{xy})$.
Using these definitions, the set ${A}\otimes B$ forms a Hilbert space. Notably, the inner product defined in Equation~\eqref{xy2} is the same as the one defined in Equation~\eqref{xy}. This is because each element $|\psi\ra\in {A}\otimes B$ can be represented as a matrix $M=(\mu_{xy})$. It can be easily demonstrated that this mapping between bipartite vectors\index{bipartite vector} and matrices is an isometric isomorphism.

\begin{exercise}\label{iso}
Show that $\mbb{C}^{m\times n}\cong\mbb{C}^{m}\otimes\mbb{C}^{n}$.
\end{exercise}

\subsection{The Kronecker Tensor Product}\index{tensor product}

We will use the notation $|\psi\ra|\phi\ra$ to mean $|\psi\ra\otimes|\phi\ra$. For basis elements, we will use interchangeably the notations $|xy\ra^{{A}B}\eqdef|x\ra^{A}|y\ra^B\eqdef|x\ra^{A}\otimes|y\ra^B$.
Since all Hilbert spaces in finite dimensions are isomorphic to $\mbb{C}^n$ (for some $n\in\mbb{N}$), any vector $|\psi^A\ra=\sum_{x\in[m]}a_x|x\ra^{A}\in {A}$ (where $m\eqdef|A|$) corresponds to a vector $\a\eqdef(a_1,\ldots,a_{m})^T\in\mbb{C}^{m}$, and any vector $|\phi^B\ra=\sum_{y\in[n]}b_y|y\ra^B\in B$ (where $n\eqdef|B|$) corresponds to a vector $\b\eqdef(b_1,\ldots,b_{n})^T\in\mbb{C}^{n}$. Hence, the vector $|\psi^A\ra|\phi^B\ra=\sum_{x\in[m]}\sum_{y\in[n]}a_xb_y|xy\ra^{{A}B}$ corresponds to the vector $\a\otimes\b$ given by
\be
\a\otimes\b\eqdef\begin{bmatrix}
a_{1}\b  \\
\vdots \\
a_{m}\b
\end{bmatrix}\in\mbb{C}^{mn}\;.
\ee

The above definition of a tensor product between vectors in $\mbb{C}^{m}$ and $\mbb{C}^{n}$ is the Kronecker product which is also denoted with the symbol $\otimes$. The Kronecker product is defined on arbitrary matrices as follows. Let $M=(\mu_{xy})\in\mbb{C}^{k\times \ell}$ and $N\in\mbb{C}^{p\times q}$. The Kronecker product $M\otimes N$ is a matrix in $\mbb{C}^{k p\times \ell q}$ defined by
\be
M\otimes N=\begin{bmatrix}
\mu_{11}N  & \cdots & \mu_{1\ell} N\\
\vdots & \ddots  & \vdots\\
\mu_{k1}N  & \cdots & \mu_{k \ell} N
\end{bmatrix}\;.
\ee
It is simple to check that the tensor product above is bilinear and associative, however, it is not commutative. 
Also, note that for ${A}=\mbb{C}^{n\times m}$ and $B=\mbb{C}^{p\times q}$ the tensor product given in~\eqref{deften} is equivalent to the Kronecker product. We will therefore use the terms tensor product and Kronecker product interchangeably.

\begin{exercise}
Let $M$ be an $m\times n$ matrix and $N$ be an $n\times k$ matrix. Find \textbf{all} the values of $m,n,k$ for which $M\otimes N=MN$.
\end{exercise}

\begin{exercise}
Show that the Kronecker product is not commutative, and that there always exist permutations matrices $P$ and $Q$ in appropriate dimensions such that $M\otimes N=P\left(N\otimes M\right) Q$.
\end{exercise}

\begin{exercise}
Prove the following properties. For any matrices $K,L,M,N$ in appropriate dimensions (in some cases square matrices):
\begin{enumerate}
\item $(K\otimes L)(M\otimes N)=KM\otimes LN$.
\item $K\otimes L$ is invertible if and only if $K$ and $L$ are invertible and in this case $(K\otimes L)^{-1}=K^{-1}\otimes L^{-1}$.
\item $(K\otimes L)^T=K^T\otimes L^T$ and $(K\otimes L)^{*}=K^{*}\otimes L^{*}$.
\item For $K\in\mbb{C}^{m\times m}$ and $L\in\mbb{C}^{n\times n}$, $\tr\left[K\otimes L\right]=\tr[K]\tr[L]$ 
and $\det(K\otimes L)=(\det(K))^n(\det(L))^m$.
\item Ket $K\in\mbb{C}^{m\times m}$ and $L\in\mbb{C}^{n\times n}$, and let $\lambda_1,\ldots,\lambda_m$ be the eigenvalues of $K$ and $\mu_1,\ldots,\mu_n$ be those of L (listed according to multiplicity). Then the eigenvalues of $K\otimes L$ are $\lambda_x\mu_y$ with $x\in[m]$ and $y\in[n]$.
\item $\rank(K\otimes L)=\rank(K)\rank(L)$.
\end{enumerate}
\end{exercise}

Kronecker also defined a direct sum that is closely related to the definition above. Given two matrices $M\in\mbb{C}^{m\times m}$ and $N\in\mbb{C}^{n\times n}$ their Kronecker sum is defined by
\be
M\oplus N\eqdef M\otimes I_n+I_m\otimes N\;.
\ee 
The sum appears naturally in physics, typically when describing the Hamiltonian of a composite system\index{composite system} consisting of non-interacting subsystems. A celebrated result connecting the Kronecker product and Kronecker sum is given in the following exponential relation:  
\be
e^{M\oplus N}=e^M\otimes e^N\;.
\ee
\begin{exercise}
Prove the above equality. Hint: Use the formula $e^M=\sum_{n=0}^{\infty}\frac{M^n}{n!}$ and the commutativity of $M\otimes I_n$ and $I_m\otimes N$.
\end{exercise}

\section{Linear Operators on Hilbert Spaces}\index{linear operator}

An operator (i.e. a map) $M:{A}\to B$ is said to be \emph{linear} if and only if for all $|\psi\ra,|\phi\ra\in {A}$ and $a,b\in\mbb{F}$
\be
M(a|\psi\ra+b|\phi\ra)=aM|\psi\ra+bM|\phi\ra\;.
\ee
When $|B|=1$ the linear operator $M$ is called a functional.
Given a basis $\{|x\ra^{A}\}_{x\in[m]}$ of ${A}$ and $\{|y\ra^B\}_{y\in[n]}$ of $B$, $M$ can be represented in terms of a matrix
$(\mu_{xy})$ with components $\mu_{yx}\eqdef{}^B\!\la y|M|x\ra^{A}$ which is a notation for the inner product between $M|x\ra^{A}$ and $|y\ra^B$. Specifically, $M$ can be expressed as:
\be\label{rep}
M=\sum_{x\in[m]}\sum_{y\in[n]}\mu_{yx}|y\ra\la x|\;.
\ee
For improved clarity in our exposition, we have omitted the superscripts ${A}$ and $B$ from $|x\ra^{A}$ and $|y\ra^B$. This use of the Dirac notations has the advantage that the action of $M$ on a vector $|\psi\ra\in {A}$ becomes
\be
M|\psi\ra=\sum_{y\in[n]}\sum_{x\in[m]}\mu_{yx}|y\ra\la x|\psi\ra=\sum_{y\in[n]}\Big(\sum_{x\in[m]}\mu_{yx}\la x|\psi\ra\Big)|y\ra
\ee
Note that the numbers $\mu_{yx}$ form a matrix which is known as the matrix representation\index{matrix representation} of $M$. Sometimes we will identify the matrix $(\mu_{yx})$ with the operator $M$ and write $M=(\mu_{yx})$. However, note that different choices of orthonormal bases $\{|x\ra^{A}\}_{x\in[m]}$ and $\{|y\ra^B\}_{x\in[n]}$, correspond to different matrix representations $(\mu_{yx})$ of the \emph{same} linear operator $M$ (see Exercise~\ref{repequiv}).

\begin{exercise}
Show that any linear operator $M:A\to B$ can be expressed as
\be\label{generaldeco}
M=\sum_{z\in[k]}\lambda_z|v_z^B\lr u_z^A|\;,
\ee 
where $\{\lambda_z\}_{z\in[k]}$ are the singular values of $M$, and $\{|u_z^A\ra\}_{z\in[k]}$ and $\{|v_z^B\ra\}_{z\in[k]}$
are orthonormal sets of vectors in $A$ and $B$, respectively. Hint: Use the singular value decomposition of a complex matrix.
\end{exercise}

The adjoint of a linear operator $M:{A}\to B$ is itself a linear operator $M^*:B\to {A}$ defined by the relation\index{adjoint operator}
\be
\la\phi|M\psi\ra=\la M^*\phi|\psi\ra\quad\quad\forall\;|\psi\ra\in {A}\;\text{ and }\;|\phi\ra\in B\;.
\ee
If $M$ has the form~\eqref{rep} then
\be
M^*=\sum_{x\in[m]}\sum_{y\in[n]}\overline{\mu}_{yx}|x\ra\la y|
\ee
where the bar indicates complex conjugation. 

\begin{exercise}
Let $U:A\to A$ be a linear operator on a vector space $A$. Show that $U$ is unitary, i.e. $U^{*}=U^{-1}$ if and only if there exist two orthonormal bases of $A$, $\{|v_x\ra\}_{x\in[m]}$ and $\{|u_x\ra\}_{x\in[m]}$, such that\index{unitary operator} 
\be
U=\sum_{x\in[m]}|v_x\lr u_x|\;.
\ee
\end{exercise}

\begin{exercise}\label{repequiv}
Let $M$ be a linear operator as in~\eqref{rep}, and denote by $\tilde{M}$ the \emph{matrix} whose components are $\mu_{yx}\eqdef\la y|M|x\ra$ (i.e. $M$ is an operator whereas $\tilde{M}$ is a matrix). Let $\{|a_{x}\ra\}$ and $\{|b_{y}\ra\}$ be two orthonormal bases of ${A}$ and $B$, respectively. Show that there exists two unitary matrices $U$ and $V$ (not necessarily of the same size) such that
\be
M=\sum_{x\in[m]}\sum_{y\in[n]}\nu_{yx}|b_{y}\ra\la a_{x}|
\ee
where $\{\nu_{yx}\}$ are the components of the matrix $N\eqdef V\tilde{M}U$.
\end{exercise}

For any linear operator $T:A\to B$ its \emph{kernel}, denoted $\ker(T)$, is the subspace of $A$ consisting of all vectors $|\psi\ra\in A$ such that $T|\psi\ra=\0$. 
The \emph{image} of $T$, denoted by $\im(T)$, is the set of vectors $\{T|\psi\ra\}$ over all vectors $|\psi\ra\in A$. Finally, the \emph{support} of $T$, denoted $\supp(T)$, is also a subspace of $A$ consisting of all the vectors that are orthogonal to all the elements in $\ker(T)$. In particular, for any non-zero vector $|\psi\ra\in\supp(T)$ we have $T|\psi\ra\neq\0$.\index{image of an operator}\index{support of an operator}\index{kernel of an operator}

\begin{exercise}\label{exschur}
Let $A$ and $B$ be two Hilbert spaces, and let $T:A\to B$ be a linear transformation.
\begin{enumerate}
\item Show that if $\ker(T)=\{\0\}$ and $\im(T)=B$ then $A=B$ and $T$ is invertible.
\item Show that if $\ker(T)=A$ then $T=\0$.
\end{enumerate}
\end{exercise}

\subsection{Isometries and Partial Isometries}\index{isometry}\index{partial isometry}

A linear operator $V:A\to B$ is called an isometry if 
\be
\la V\psi|V\psi\ra=\la\psi|\psi\ra\quad\quad\forall\;|\psi\ra\in A\;.
\ee
Using the dual $V^*:B\to A$ we can express the equation above as
\be\label{v115}
\la \psi|V^*V|\psi\ra=\la\psi|\psi\ra\quad\quad\forall\;|\psi\ra\in A\;.
\ee
We therefore conclude (see Exercise~\ref{exab115}) that $V$ is an isometry if and only if $V^*V=I^A$, where $I^A$ is the identity operator on $A$. The condition that $V^*V=I^A$ implies that $|A|\leq |B|$ and if $|A|=|B|$ then $V$ is necessarily a unitary operator. One can view isometries as `embeddings' of the Hilbert space $A$ into $B$. In particular, they preserve the inner product; indeed, for any $|\phi\ra,|\psi\ra\in A$ 
\be
\la V\phi|V\psi\ra=\la \phi|V^*V|\psi\ra=\la\phi|\psi\ra\;.
\ee

We say that $V:A\to B$ is a partial isometry\index{partial isometry} if $V$ is an isometry when restricted to its support. Therefore, an isometry is a partial isometry but the converse is not necessarily true. Let $V:A\to B$ be a partial isometry, and let $\{|a_z\ra\}_{z\in[k]}$ be an orthonormal basis of $\supp(V)$. Then, since $V$ is an isometry on $\supp(V)\subseteq A$, it follows that the vectors $|b_z\ra\eqdef V|a_z\ra$, with $z\in[k]$, form an orthonormal set of vectors in $B$. Therefore, $k$ cannot exceed $|B|$, and $V$ can be expressed as 
\be\label{pisometry}
V=\sum_{z\in[k]}|b_z\lr a_z|
\ee
where $k\eqdef\left|\supp(V)\right|\leq\min\{|A|,|B|\}$, $\{|a_z\ra\}_{z\in[k]}$ is an orthonormal set of vectors in $A$, and $\{|b_z\ra\}_{[k]}$ is an orthonormal set of vectors in $B$. In other words, a linear operator $V:A\to B$ is a partial isometry\index{partial isometry} if and only if there exists two set of orthonormal vectors, $\{|a_z\ra\}_{z\in[k]}\subset A$ and $\{|b_z\ra\}_{z\in[k]}\subset B$ such that~\eqref{pisometry} holds.

\begin{exercise}\label{exab115}
Show that if Eq.~\eqref{v115} holds then $V^*V=I^A$.
\end{exercise}

\begin{exercise}
Use~\eqref{generaldeco} to show~\eqref{pisometry}.
\end{exercise}

\begin{exercise}\label{ppp} 
A linear operator $\Pi:A\to A$ is called an orthogonal projection if and only if $\Pi^2=\Pi=\Pi^*$.
\begin{enumerate}
\item Show that $\Pi:A\to A$ is an orthogonal projection if and only if $\Pi^*\Pi=\Pi$.
\item Show that $V:A\to B$ is a partial isometry if and only if $V^*V$ is an orthogonal projection in A, and $VV^*$ is an orthogonal projection in B. 
\end{enumerate}
\end{exercise}

\begin{exercise}\label{partialiso}
Let $A\subseteq B$ be a subspace of $B$, and let $V:A\to B$ be an isometry satisfying $V^*V=\Pi$, where $\Pi:B\to A$ is the projection onto the subspace $A$. Show that there exists a unitary matrix $U:B\to B$ such that
\be
U\Pi=V\;.
\ee
\end{exercise}

\subsection{Hermitian and Positive Operators}\index{Hermitian operator}\index{positive operator}

A linear operator $H:A\to A$ ($A$ is a Hilbert space) is called Hermitian if $H=H^*$. Any Hermitian operator $H$ has a spectral decomposition
\be
H=\sum_{x\in[m]}\lambda_x|v_x\lr v_x|\;,
\ee
where $\{|v_x\ra\}_{x\in[m]}$ is an orthonormal basis of $A$.
The coefficients $\{\lambda_x\}_{x\in[m]}$ are the eigenvalues of $H$. We denote by $\tr[H]$ the trace of Hermitian operator $H:A\to A$. That is,
\be
\tr[H]\eqdef\sum_{x\in[m]}\la x|H|x\ra\;.
\ee 
Note that the definition above is independent on the choice of the orthonormal basis $\{|x\ra\}$ of $A$.

We say that a linear operator $\rho:A\to A$ is positive semidefinite, and write $\rho\geq 0$, if and only if  
\be
\la\psi|\rho|\psi\ra\geq 0\quad\quad\forall\;|\psi\ra\in A\;.
\ee
If the above inequality is strict for all non-zero $|\psi\ra\in A$ then we say that $\rho$ is positive definite and write $\rho>0$.
 We will also write $\rho\geq \sigma$ to mean $\rho-\sigma\geq 0$, and will use the Greek letters such as $\rho$ and $\sigma$ to denote linear operators that are positive semidefinite. The set of all positive semidefinite operators acting on Hilbert space $A$ will be denoted by $\pos(A)$.
 
 Every positive linear operator $\rho:A\to A$ is necessarily Hermitian. To see why, observe that the positivity\index{positivity}  property above implies
 \be
 \la\psi|\rho|\psi\ra=\overline{\la\psi|\rho|\psi\ra}=\la\psi|\rho^*|\psi\ra\;.
 \ee
 Therefore, for all $|\psi\ra\in A$ we have
 \be
 \la\psi|\rho-\rho^*|\psi\ra=0\;.
 \ee
 Now, observe that the operator $N\eqdef\rho-\rho^*$ satisfies $N^*N=NN^*$. Such operators are called \emph{normal} operators and are known to be diagonalizable. Therefore, taking $|\psi\ra$ above to be an eigenvector of $N$ we conclude that all the eigenvalues of $N$ are zero. Hence, $N=0$ or equivalently $\rho=\rho^*$.

\begin{exercise} Let $\rho:A\to A$ be a linear operator. Show that the following are equivalent:
\begin{enumerate}
\item $\rho\geq 0$ 
\item $\rho$ is Hermitian and all its eigenvalues are non-negative.
\item There exists a linear map $M:A\to A$ such that
$
\rho=M^*M.
$
\end{enumerate}
\end{exercise}

\begin{exercise}\label{posmab}
Let $\rho\in\pos(A)$. Show that for any complex matrix $M:A\to B$ we have $M\rho M^*\in\pos(B)$.
\end{exercise}

\begin{exercise}
Show that for \emph{any} two vectors $|\psi\ra,|\phi\ra\in A$
\be
\tr\left[|\psi\lr\phi|\right]=\la\phi|\psi\ra\;.
\ee
\end{exercise}

Note that the identity operator $I^A:A\to A$ is a positive operator (i.e. an operator with all eigenvalues strictly greater than zero) given by
\be
I^A=\sum_{x\in[m]}|x\lr x|\;.
\ee
Note that for any orthonormal basis $\{|v_x\ra\}_{x\in[m]}$ of $A$ we have
\be
\left(\sum_{x\in[m]}|v_x\lr v_x|\right)|\psi\ra=\sum_{x\in[m]}\la v_x|\psi\ra|v_x\ra=|\psi\ra.
\ee
Therefore, $\sum_{x\in[m]}|v_x\lr v_x|=I^A$ for any orthonormal basis $\{|v_x\ra\}_{x\in[m]}$ of $A$.

\subsubsection{Decomposition of Hermitian Operators}

Let $H:A\to A$ be an Hermitian operator.
Since every Hermitian operator is diagonalizable, we can express $H$ as
\be
H=\sum_{x\in[n]}\lambda_x|\phi_x\lr \phi_x|
\ee
where $\{|\phi_x\ra\}_{x\in[n]}$ is an orthonormal basis, and the eigenvalues $\{\lambda_x\}_{x\in[n]}$ are all real. Therefore, it is possible to decompose $H$ as 
 \be\label{decomherm}
H=H_+-H_-
\ee
where
\be\label{pmherm}
H_+\eqdef\sum_{x:\;\lambda_x\geq 0}\lambda_x|\phi_x\lr \phi_x|\geq 0\quad
\text{and}\quad
H_-\eqdef \sum_{x:\;\lambda_x< 0}|\lambda_x||\phi_x\lr \phi_x|\geq 0\;.
\ee
By definition $H_+,H_-\geq 0$ and $H_+H_-=H_-H_+=0$. Further, denote by $\Pi_-\eqdef \sum_{x:\;\lambda_x< 0}|\phi_x\lr \phi_x|$ the projection to the negative eigenspace  of $H$, and by $\Pi_+=I-\Pi_-$ the projection to the non-negative eigenspace of $H$. Then, $H_\pm=H\Pi_{\pm}=\Pi_{\pm}H$.

\bex
Let $H:A\to A$ be an Hermitian operator as above. Show that
\be
H_\pm=\frac{|H|\pm H}{2}\;,
\ee
where $H_{\pm}$ are the positive and negative parts of $H$ as defined in~\eqref{pmherm}.
\eex

\subsection{Quantum States}\index{quantum states}

A positive semidefinite operator $\rho:A\to A$ (which may be denoted as $\rho^A$ to indicate its underlying Hilbert space $A$) is known as a \emph{quantum state} or a \emph{density matrix} if its trace is equal to one. In this case, we denote its eigenvalues as $\{p_x\}_{x\in[m]}$, as they represent the components of a probability distribution.

A quantum state $\rho$ is called \emph{pure} if its rank is equal to one. Consequently, pure states are projections onto 1-dimensional subspaces of $A$, and the normalized vector they project onto is also referred to as a quantum state. It should be noted that a positive semidefinite matrix $\rho$ is a pure state if and only if $\tr[\rho^2]=\tr[\rho]=1$. We often denote pure states by $\phi\eqdef|\phi\lr \phi|$ or $\psi\eqdef|\psi\lr\psi|$, where $|\psi\ra$ and $|\phi\ra$ are normalized vectors in $A$.

\begin{exercise}
Show that any Hermitian linear operator $\rho:A\to A$ (with $A$ being a finite dimensional Hilbert space)
is a pure quantum state if and only if
\be
\tr[\rho^3]=\tr[\rho^2]=\tr[\rho]=1\;.
\ee
Give an example of Hermitian operator $H:A\to A$ such that $\tr[H^2]=\tr[H]=1$ but $H$ is not a pure quantum state.
\end{exercise}

\bex\label{contraction}
A linear operator $G\in\ml(A)$ is called a contraction if $G^*G\leq I^A$. 
\ben
\item Show that $G\in\ml(A)$ is a contraction if and only if $G^*$ is  a contraction.
\item Show that $G$ is a contraction if and only ig $\big\|G|\psi\ra\big\|_2\leq \big\||\psi\ra\big\|_2$ for all $|\psi\ra\in A$. 
\item Show that if $G$ is a contraction then the real part of $G$, $P\eqdef\frac12(G+G^*)$, satisfies $P\leq I^A$. Hint: Use the triangle inequality to show that $\big\|P|\psi\ra\big\|_2\leq \big\||\psi\ra\big\|_2$.
\een
\eex

Every collection of $m$ normalized vectors $\{|\psi_x\ra\}_{x\in[m]}$ in $A$, along with a probability distribution 
$\{p_x\}_{x\in[m]}$ is called an \emph{ensemble} of states. To every ensemble of states, $\{|\psi_x\ra,\;p_x\}_{x\in[m]}$ there is a corresponding quantum state defined by
\be
\rho=\sum_{x\in[m]}p_x|\psi_x\lr\psi_x|\;.
\ee
Note that the above \emph{pure state decomposition} of $\rho$ is not necessarily the spectral decomposition since the pure states $|\psi_x\ra$ are not necessarily orthogonal. In fact, any quantum state corresponds to infinitely many ensembles of quantum states. For example, consider a quantum state $\rho:\mbb{C}^2\to\mbb{C}^2$ defined by:
\be\label{0e1}
\rho=\frac{1}{4}|0\lr 0|+\frac{3}{4}|1\lr 1|\;.
\ee
Clearly, this is the spectral decomposition of $\rho$. Now, it is simple to check that $\rho$ can also be expressed as
\be\label{0e2}
\rho=\frac{1}{2}|u\lr u|+\frac{1}{2}|v\lr v|\;,
\ee
where
\be
|u\ra\eqdef\sqrt{\frac{1}{4}}|0\ra+\sqrt{\frac{3}{4}}|1\ra\quad,\quad |v\ra\eqdef\sqrt{\frac{1}{4}}|0\ra-\sqrt{\frac{3}{4}}|1\ra\;.
\ee
Note that $|u\ra$ and $|v\ra$ are not orthogonal, and both ensembles in~\eqref{0e1} and~\eqref{0e2} corresponds to the \emph{same} quantum state $\rho$. 

\begin{exercise}\label{ensembles}\index{pure-state decomposition}
Let $\{|\psi_x\ra,\;p_x\}_{x\in[m]}$ and $\{|\phi_y\ra,\;q_y\}_{y\in[n]}$ be two ensembles of quantum states in $A$ with $m\geq n$. Show that they correspond to the same density matrix 
\be
\rho\eqdef\sum_{x\in[m]}p_x|\psi_x\lr\psi_x|=\sum_{y\in[n]}q_y|\phi_y\lr\phi_y|
\ee
if and only if there exists an $m\times n$ isometry matrix $V=(v_{xy})$ (i.e. $V^*V=I_n$) such that
\be
\sqrt{p_x}|\psi_x\ra=\sum_{y\in[n]}v_{xy}\sqrt{q_y}|\phi_y\ra\quad\quad\forall\;x\in[m]\;.
\ee
\end{exercise}

\subsection{The Space of Linear Operators}

We will denote by $\ml(A,B)$ the set of all linear operators from $A$ to $B$, and set $\ml(A)\eqdef\ml(A, A)$. The space $\ml(A, B)$ is itself a Hilbert space. Let $m\eqdef|A|$ and $n\eqdef|B|$, and observe that $\ml(A,B)$ is isomorphic to $\mbb{C}^{n\times m}$ with inner product defined for all $M,N\in\ml(A, B)$ by\index{Hilbert-Schmidt inner product}
\be
\la M|N\ra\eqdef\tr\left[M^{*}N\right]\;.
\ee
Note that $M^{*}N\in\ml(A)$ for which the trace is well defined ($\tr[M]$ is not well defined if $m\neq n$). The standard basis of $\ml(A, B)$ is given by $\{|y\lr x|\}$, with indices running over $x\in[m]$ and $y\in[n]$, where $|x\ra\in A$ and $|y\ra\in B$. Hence, the dimension of  
$\ml(A, B)$ is $mn$.

Most of the objects in quantum mechanics, like quantum states, observables, positive operator\index{positive operator} valued measures (POVM), and so on, are described with Hermitian operators. It is therefore useful to characterize the space of Hermitian operators in $\ml(A)$. We denote this space by $\herm(A)$. That is, $H\in\herm(A)$ if and only if $H\in\ml(A)$ with $H^{*}=H$. Note that $\herm(A)$ cannot be a vector space over the complex numbers. This is because if $c\in\mbb{C}$ is a non-real complex number, and $0\neq H\in\herm(A)$ then 
\be
(cH)^*=\bar{c}H^{*}=\bar{c}H\neq cH\;.
\ee 
Therefore, $cH\not\in\herm(A)$ so that $\herm(A)$ cannot be a vector space over the complex numbers.

However, it is simple to verify that $\herm(A)$ is a vector space over the real numbers. Particularly, if $r_1,r_2$ are real numbers and $H_1,H_2\in\herm(A)$ are two Hermitian operators, then also $r_1H_1+r_2H_2$ is an Hermitian operator. The inner product between two elements of $\herm(A)$ is induced from $\ml(A)$, and is given by
\be
\la H_1|H_2\ra\eqdef\tr\left[H_1H_2\right]\;.
\ee 
To summarize, $\herm(A)$ is a real Hilbert space.

\begin{exercise}
Show that the dimension of $\herm(A)$ is $|A|^2$.
\end{exercise}

\begin{exercise}\label{hermitianex}
Show that $\ml(A)$ has an orthonormal basis consisting of Hermitian matrices.
\end{exercise}

\begin{exercise}\label{propounit2}\index{Hilbert-Schmidt inner product}
Let $\ml_0(A)$ be the space of all matrices in $\ml(A)$ with zero trace.
\begin{enumerate}
\item Show that $\ml_0(A)$ is an inner product space (under the Hilbert-Schmidt inner product) of dimension $d\eqdef |A|^2-1$.
\item Show that if $\{\eta_1,\ldots,\eta_d\}$ is an orthonormal basis of $\ml_0(A)$, then
$\{\frac1{\sqrt{|A|}}I^A,\eta_1,\ldots,\eta_d\}$ is an orthonormal basis of $\ml(A)$.
\item Let $\omega\in\ml(A)$. Show that $\omega$ is proportional to the identity matrix if and only if $\tr\left[\omega \eta\right]=0$ for all traceless matrices $\eta\in\ml_0(A)$.
\end{enumerate}
\end{exercise}

\begin{exercise}\label{pauli}
Let $A$ be a 2-dimensional Hilbert space (e.g. $A=\mbb{C}^2$) with an orthonormal basis $\{|0\ra,|1\ra\}$ and let
\be
\sigma_0\eqdef I^A=|0\lr 0|+|1\lr 1|\;\;,\;\;\sigma_1=|0\lr 1|+|1\lr 0|\;\;,\;\;\sigma_2=-i|0\lr 1|+i|1\lr 0|\;\;,\;\;\sigma_3=|0\lr 0|-|1\lr 1|\;.
\ee
The three operators $\sigma_1,\sigma_2,\sigma_3$ are known as the Pauli\index{Pauli} operators.
\begin{enumerate}
\item Show that all the operators above are Hermitian, unitary, and have a norm equals to $\sqrt{2}$. 
\item Show that the set $\{\sigma_0,\sigma_1,\sigma_2,\sigma_3\}$ form an orthogonal basis of $\herm(A)$. 
\item Show that the commutator $[\sigma_i,\sigma_j]=2i\sum_{k\in[3]}\varepsilon_{ijk}\sigma_k$, where the structure constant $\varepsilon_{ijk}$ is the Levi-Civita symbol, and $i,j,k\in\{1,2,3\}$.
\item Show that the anti-commutator $\{\sigma_i,\sigma_j\}=2\delta_{ij}I$.
\end{enumerate}  
\end{exercise}

Like the inner product, also norms in $\mbb{C}^n$ have a natural extension to the space $\ml(A, B)$. In particular, the $\ell^p$ norms as defined in Exercise~\eqref{lp} on vectors in $\mbb{C}^n$ can be extended to elements of $\ml(A, B)$. 

\begin{myd}{The Schatten Norms}\index{Schatten}
\begin{definition}\label{schatt}
Let $A$ and $B$ be two Hilbert spaces, $M\in\ml(A, B)$, and $p\in[1,\infty]$. The Schatten $p$-norm of $M$ is defined as
\be\label{1pnorm}
\|M\|_p\eqdef\big(\tr|M|^p\big)^{\frac1p}\quad\text{where}\quad|M|\eqdef \sqrt{M^*M}\;.
\ee
\end{definition}
\end{myd}
The case $p=1$ is often called the \emph{trace norm} and we will discuss it in details in Sec.~\ref{sectracenorm}. The case $p=\infty$ is understood in terms of the limit $p\to\infty$. It is often called the \emph{operator norm} and is given by\index{operator norm}
\be
\|M\|_\infty=\lambda_{\max}(|M|)\;,
\ee
where $\lambda_{\max}(|M|)$ is the largest eigenvalue of $|M|$, or equivalently, the largest singular value of $M$.
The Schatten norms appear quite often in quantum Shannon theory due to their relation to the R\'enyi entropies that we will study later on. We leave it as an exercise for the reader to prove some of their key properties. 
\begin{exercise}
Let $A$ and $B$ be two Hilbert spaces, $M,N\in\ml(A, B)$, and $p,q\in[1,\infty]$ such that $\frac1p+\frac1q=1$. Show that the $p$-Schatten norm is indeed a norm satisfying the following properties:
\begin{enumerate}\index{invariance}
\item \textbf{Invariance.} For any two Hilbert spaces $A',B'$ with $|A'|\geq |A|$ and $|B'|\geq |B|$, and any isometries $V\in\ml(B, B')$ and $U\in\ml(A, A')$
\be\label{invprosch}
\|VMU^*\|_p=\|M\|_p\;.
\ee
\item \textbf{H\"older Inequality.} \index{H\"older inequality}
\be\label{holin}
\|MN\|_1\leq\|M\|_p\|N\|_q\;.
\ee
\item \textbf{Sub-Multiplicativity.}
\be
\|MN\|_p\leq\|M\|_p\|N\|_p\;.
\ee
\item \textbf{Monotonicity.} If $p\leq q$
\be
\|M\|_1\geq\|M\|_p\geq\|M\|_q\geq\|M\|_{\infty}\;.
\ee
\item \textbf{Duality.}\index{duality}
\be
\|M\|_p=\sup\Big\{\left|\tr\left[M^*L\right]\right|\;:\;\|L\|_q=1\;,\;\;L\in\ml(A, B)\Big\}
\ee
\end{enumerate}
\end{exercise}

\begin{exercise}[Young’s Inequality]\index{Young's inequality}
Let $A$ and $B$ be two Hilbert spaces, $M,N\in\pos(A)$, and $p,q\in[1,\infty)$ such that $\frac1p+\frac1q=1$. Use the H\"older inequality of the Schatten norm to show that
\be\label{young}
\tr[MN]\leq\frac1p\tr[M^p]+\frac1q\tr[N^q]\;. 
\ee
with equality if and only if $M^p=N^q$. Hint: Take the logarithm on both sides of H\"older inequality and use the concavity property of the logarithm. 
\end{exercise}

\bex\label{optrrel}\index{trace norm}\index{operator norm}
Show that for any $M\in\herm(A)$, the operator norm and the trace norm can be expressed as
\be
\|M\|_1=\max_{\substack{\eta\in\herm(A)\\ \|\eta\|_\infty\leq 1}}\tr[\eta M]\quad\quad{\rm and}\quad\quad\|M\|_\infty=\max_{\substack{\eta\in\herm(A)\\ \|\eta\|_1\leq 1}}\tr[\eta M]\;.
\ee
\eex

\begin{myd}{The Ky Fan Norms}\index{Ky Fan norm}
\begin{definition}\label{def:kyfan}
Let $A$ and $B$ be two Hilbert spaces, $M\in\ml(A, B)$, $n\eqdef\min\{|A|,|B|\}$, and $k\in[n]$. The Ky Fan $k$-norm of $M$ is defined as
\be
\|M\|_{(k)}\eqdef s_1+s_2+\cdots+s_k\;,
\ee
where $s_1\geq s_2\geq\cdots\geq s_k$ are the $k$ largest singular values of $M$.
\end{definition}
\end{myd}

\begin{remark}
When restricting $M$ to be diagonal real matrix we get the following definition of the Ky Fan norm\index{Ky Fan norm} on $\mbb{R}^n$:
For any $\r\in\mbb{R}^n$ and $k\in[n]$ the $k$th-Ky Fan norm of $\r$ is defined as
\be\label{kfnorm}
\|\p\|_{(k)}\eqdef\sum_{x\in[k]}|r_x^\da|\;,
\ee
where $\{r_x^\da\}_{x\in[n]}$ are the components of $\r$ arranged such that $|r_1^\da|\geq|r_2^\da|\geq\cdots\geq |r_n^\da|$.
\end{remark}

The Ky Fan norms plays an important role in the resource theory of entanglement. We leave it as an exercise to prove that the Ky Fan norms are indeed norms.

\begin{exercise}
Show that the Ky Fan norms are indeed norms that have the following invariance property. Using the same notations as in the definition above, show that for any two Hilbert spaces $A',B'$ with $|A'|\geq |A|$ and $|B'|\geq |B|$, and any isometries $V\in\ml(B, B')$ and $U\in\ml(A, A')$
\be
\|VMU^*\|_{(k)}=\|M\|_{(k)}\;.
\ee
\end{exercise}

\bex
Show that the Ky Fan $k$-norm can be expressed as
\be
\|M\|_{(k)}=\sup\tr\big[\Pi|M|\big]\;,
\ee
where the supremum is over all orthogonal projections $\Pi$ with rank no greater than $k$.
\eex

\begin{exercise}[The Ky Fan norms on $\mbb{R}^n$] Consider the variant of the Ky Fan norm as defined in~\eqref{kyfan}.
\ben
\item Show that the Ky Fan norms are indeed norms in $\mbb{R}^n$. 
\item Let $D$ be an $n\times n$ matrix that can be written as a convex combination of permutation matrices (such matrices are known as doubly-stochastic matrices; see Appendix~\ref{sec:bir} for more details). Show that for all $k\in[n]$ and all $\p\in\mbb{R}_+^n$
\be\label{dkyfan}
\|\p\|_{(k)}\geq \|D\p\|_{(k)}\;.
\ee
\item Show that for any two probability vectors $\p,\q\in\prob(n)$ and any $k\in[n]$ we have
\be
\|\p-\q\|_{(k)}\leq \frac12\|\p-\q\|_1
\ee
and conclude that
\be\label{e5p18}
\|\p\|_{(k)}-\|\q\|_{(k)}\leq \frac12\|\p-\q\|_1.
\ee
Hint: For the first inequality use the fact that $\frac12\|\p-\q\|_1=\sum_{x\in[n]}(p_x-q_x)_+$, and for the second inequality use the properties of a norm. For any $r\in\mbb{R}$ the symbol $(r)_+\eqdef r$ if $r\geq 0$ and otherwise $(r)_+\eqdef 0$.
\een
\end{exercise}

\subsection{Linear Operators as Bipartite Vectors}\index{bipartite vector}

As seen in Exercise~\ref{iso}, given two finite dimensional Hilbert spaces $A$ and $B$ of dimensions $m\eqdef|A|$ and $n\eqdef|B|$, the space $A\otimes B$ is (isometrically) isomorphic to the Hilbert space of $m\times n$ matrices $\mbb{C}^{m\times n}$. For convenience, we will denote ${AB}\eqdef A\otimes B$ and call it a bipartite Hilbert space and its elements bipartite vectors. Moreover, we will use the tilde notation to indicate spaces with the same dimension as without the tilde. For example, $A$ and ${\tA}$ both will have the same dimension $m\eqdef|A|$, and consequently ${A\tA}\eqdef A\otimes {\tA}$ will have dimension $m^2$. The isomorphism $A\otimes B\cong\mbb{C}^{m\times n}\cong\ml(B,A)$ indicates that we can think of bipartite vectors\index{bipartite vector} in ${AB}$ as linear operators from $B$ to $A$. This correspondence will be very useful later on, so we discuss it in more details here.

Any bipartite vector\index{bipartite vector} $|\psi\ra\in {AB}$ can be expressed in terms of the orthonormal basis $\{|x\ra^A\otimes|y\ra^B\}$ as
\be
|\psi^{AB}\ra=\sum_{x\in[m]}\sum_{y\in[n]}\mu_{xy}|x\ra^A\otimes|y\ra^B=\sum_{y\in[n]}\Big(\sum_{x\in[m]}\mu_{xy}|x\ra^A\Big)\otimes|y\ra^B\;,
\ee
where $\mu_{xy}\in\mbb{C}$.
Let $M_\psi$ be a linear map from $\tB$ to $A$ defined below by its action on the basis elements $\{|y\ra^{\tB}\}$ of ${\tB}$: 
\be
M_\psi|y\ra^{\tB}\eqdef \sum_{x\in[m]}\mu_{xy}|x\ra^{A}\;.
\ee 
With this definition we get that
\be
|\psi^{AB}\ra=\sum_{y\in[n]}\left(M_\psi|y\ra^{\tB}\right)\otimes|y\ra^{B}=\left(M_\psi\otimes I^{B}\right)\sum_{y\in[n]}|yy\ra^{\tB B}
\ee
Denoting by 
\be
|\Omega^{\tB B}\ra\eqdef \sum_{y\in[n]}|yy\ra^{\tB B}
\ee
we conclude that
\be\label{psiab}
|\psi^{AB}\ra=M_\psi\otimes I^B|\Omega^{\tB B}\ra\;.
\ee
Therefore, for any bipartite vector\index{bipartite vector} $|\psi\ra^{AB}$ there is a corresponding linear map $M_\psi:{\tB}\to A$ and vice versa. In other words, the mapping
\be
|\psi^{AB}\ra\mapsto M_\psi
\ee
is an (isometrically) isomorphism map from the space $AB$ and the space $\mbb{C}^{m\times n}$.
The vector $|\Omega^{\tB B}\ra$ has many interesting properties, and later on we will see that, physically, its normalized version corresponds to a composite system\index{composite system} of two maximally entangled subsystems.

\begin{exercise}\label{bipartite}
Prove the following properties of $|\Omega^{A\tA}\ra$:
\begin{enumerate}
\item For any matrix $N\in\ml(\tA)$
\be
\la\Omega^{A\tA}|I\otimes N|\Omega^{A\tA}\ra=\tr\left[N\right]\;.
\ee
\item Let $M:{B}\to A$ be a linear map and denote its transpose map by $M^{T}:{A}\to {B}$. Show that
\be\label{ptrans}
I^A\otimes M^T|\Omega^{A\tA}\ra=M\otimes I^B|\Omega^{\tB B}\ra
\ee
\item Show that for any invertible matrices $M,N\in\ml(A)$
\be
M\otimes N|\Omega^{A\tA}\ra=|\Omega^{A\tA}\ra\quad\iff\quad M=\left(N^{-1}\right)^T\;.
\ee
\item Show that if $|\psi\ra=L\otimes R|\varphi\ra$ for some matrices $L$ and $R$, then
\be
M_{\psi}=LM_{\varphi}R^T\;.
\ee
\item \textbf{Schmidt Decomposition:}\index{Schmidt decomposition} Show that for any normalized vector $|\psi^{AB}\ra\in {AB}$  there exists orthonormal sets of $k\leq\min\{|A|,|B|\}$ vectors, $\{|u_{z}^{A}\ra\}_{z\in[k]}$ and $\{|v_{z}^{B}\ra\}_{z\in[k]}$, in $A$ and $B$, respectively, such that
\be
|\psi^{AB}\ra=\sum_{z\in[k]}\sqrt{p_z}|u_{z}^{A}\ra\otimes|v_{z}^{B}\ra\;.
\ee
where $p_z> 0$ for all $z\in[k]$, and $\sum_{z\in[k]}p_z=1$. \emph{Hint:} Use the singular value decomposition of the matrix $M_\psi=UDV$, where $U,V$ are unitary matrices and $D$ is a $|B|\times |A|$ diagonal matrix with diagonal consisting of the singular values of $M_\psi$.
\end{enumerate}
\end{exercise}

\subsubsection{The Reduced Density Matrix}\index{reduced density matrix}

For any bipartite state $|\psi^{AB}\ra$ as in~\eqref{psiab}, the density matrix $\rho^{A}_{\psi}\eqdef M_{\psi}M_\psi^*$ is called the \emph{reduced density matrix} on system $A$ of the bipartite state $|\psi^{AB}\ra$, and the density matrix
$\rho^{B}_\psi\eqdef \left(M_\psi^* M_{\psi}\right)^T$ is called the \emph{reduced density matrix} on system $B$ of the bipartite state $|\psi^{AB}\ra$. Note that $\rho^A_\psi$ is an $|A|\times |A|$ matrix while $\rho^B_\psi$ is an $|B|\times |B|$ matrix. 

\begin{exercise}
Show that the two reduced density matrices above, $\rho^{A}_{\psi}$ and $\rho^{B}_\psi$, have the same non-zero eigenvalues.
\end{exercise}
\begin{mye}{The Partial Trace}\index{partial trace}
\begin{exercise}\label{partialtrace} Let   $\tr_B:\ml(AB)\to\ml(A)$ be a linear map defined by its action
on the basis elements of $\ml(AB)$ as:
\be
\tr_B\left[|x\lr y|^A\otimes|z\lr w|^B\right]\eqdef |x\lr y|^A\;\tr\left[|z\lr w|^B\right]=\delta_{zw}|x\lr y|^A\;.
\ee
Show that the reduced density matrix $\rho^{A}_{\psi}$ of a bipartite pure state $|\psi\ra\in AB$ can be expressed as
\be
\rho^{A}_{\psi}=\tr_B\left[|\psi^{AB}\lr\psi^{AB}|\right]\;.
\ee
\end{exercise}
\end{mye}
It is well known that the trace remains invariant under cyclic permutation of product of matrices. The following exercise states that this remain true also for the partial trace.
\begin{exercise}\label{cycpt}
Show that for any  $\rho\in\ml(A B)$ and any two matrices $\eta,\zeta\in\ml(B)$ 
\be
\tr_B\left[(I^A\otimes \eta^B)\rho^{AB}(I^A\otimes \zeta^B)\right]=\tr_B\left[(I^A\otimes \zeta^B\eta^B)\rho^{AB}\right]\;.
\ee
\end{exercise}

For any pure bipartite state as in~\eqref{psiab} there is a unique reduced density matrix $\rho^{A}_\psi$. On the other hand, for any density matrix $\rho\in\md(A)$ there are many bipartite pure states $|\psi^{AB}\ra$ with the same reduced density matrix $\rho$ (see the exercise below).

\begin{exercise}\label{propounit}
Let $\rho\in\ml(AB)$. 
Show that if for all $\eta\in\ml(B)$ the matrix
\be
\tr_{B}\left[\left(I^{A}\otimes \eta^B\right)\rho^{AB}\right]
\ee 
is proportional to the identity matrix,  then $\rho^{AB}=\u^A\otimes\rho^B$, where $\u^A\eqdef\frac1{|A|}I^A$ is the uniform density matrix also known as the maximally mixed state. Hint: Use Part~3 of Exercise~\ref{propounit2}.
\end{exercise}

\bex\label{abapbp}
Let $A,B,A',B'$ be four Hilbert spaces and let $\Lambda\in\ml(AB,A'B')$. Show that if
\be
\tr_{B}\left[\left(I^{A'}\otimes T\right)\Lambda\right]=0\quad\quad\forall\;T\in\ml(B',B)
\ee
then $\Lambda=\0$. Observe that the operator $\left(I^{A'}\otimes T\right)\Lambda$ belongs to $\ml(AB, A'B)$, so the partial trace above over $B$ is well defined. Hint: Let $\{N_x\}$ be an orthonormal basis (w.r.t.\ the Hilbert-Schmidt inner product) of $\ml(B,B')$ and write $\Lambda=\sum_xM_x\otimes N_x$, where $\{M_x\}$ are some matrices in $\ml(A,A')$. Then show that by taking $T$ above to
be $N_y$ you get $M_y=0$. \index{Hilbert-Schmidt inner product}
\eex

\begin{myd}{}
\begin{definition}
Let $\rho\in\md(A)$ be a density operator. A normalized bipartite pure quantum state $|\psi^{AB}\ra\in A\otimes B$ is called a \emph{purification} of $\rho^A$, if $\rho^A$ is the reduced density matrix of $|\psi^{AB}\ra$.
\end{definition}
\end{myd}

\begin{exercise}\label{purification}
Let $\rho\in\md({A})$ be a density matrix.
\begin{enumerate}
\item Show that $\sqrt{\rho}\otimes I^{\tA}|\Omega^{A\tA}\ra$ is a purification of $\rho$.
\item Show that $|\psi^{AB}\ra\in AB$ and $|\phi^{AC}\ra\in AC$ (assuming $|B|\leq |C|$) are two purifications of $\rho\in\md({A})$ 
if and only if there exists an isometry matrix $V:{B}\to C$ such that
\be
|\phi^{AC}\ra=I^A\otimes V^{B\to C}|\psi^{AB}\ra\;.
\ee
\item Use Part 2 to provide alternative (simpler!) proof of the claim in Exercise~\ref{ensembles}.
\end{enumerate}
\end{exercise}

\bex\label{osd}
\textbf{Operator Schmidt Decomposition:} \index{operator Schmidt decomposition}Let $A$ and $B$ be two Hilbert spaces of dimensions $m\eqdef|A|$ and $n\eqdef|B|$ and denote by $k\eqdef\min\{m^2,n^2\}$. Show that for every $\rho\in\herm(AB)$ there exists $k$ non-negative real numbers $\{\lambda_z\}_{z\in[k]}$, and two orthonormal sets of Hermitian matrices (w.r.t.\ the Hilbert-Schmidt inner product) $\{\eta_z\}_{z\in[k]}\subset\herm(A)$ and $\{\zeta_z\}_{z\in[k]}\subset\herm(B)$ such that
\be
\rho^{AB}=\sum_{z\in[k]}\lambda_z\eta_z^A\otimes\zeta_z^B\;.
\ee
Hint: Use similar lines as in part five of Exercise~\ref{bipartite}.
\eex

\section{Encoding Information in Quantum States}

The first postulate of quantum mechanics states that to any  physical system there is a (separable) Hilbert space $A$ that is associated with it, and that the information about the system is completely described by quantum states; that is, unit-trace positive semidefinite operators in $\ml(A)$. Furthermore, for isolated physical systems, the information is described by a pure state $|\psi\lr\psi|\in\ml(A)$.
Such pure states can be described by \emph{rays} 
of the form $\{e^{i\theta}|\psi\ra\;:\;\theta\in[0,2\pi]\}$ where $|\psi\ra$ is a unit vector in $A$.
Therefore, aside from an irrelevant phase, isolated systems are described with unit vectors in a Hilbert space. 
As an example, we start with the building block of quantum information: the quantum bit.

\subsection{The Quantum Bit}\label{0qubit0}\index{qubit}

The quantum bit, or in short the \emph{qubit}, is the quantum generalization of the classical bit. We will use here the example of a spin of an electron to describe the qubit. The spin of an electron in some fixed direction can take two possible values. We therefore associate with it a two dimensional Hilbert space $A\cong\mbb{C}^2$. If the spin of the electron is pointing in the positive $z$-direction it is described by a pure state, say $|0\lr 0|$. The question now is how to represent the spin of the electron in all other directions if we choose $|0\lr 0|$ to correspond to the positive $z$-direction. To answer this question we will make use of a representation of the rotation group SO$(3)$ on the space $\mbb{C}^2$.

Consider a counterclockwise rotation along an axis of rotation that is described by the unit vector $\n\in\mbb{R}^3$.
In $\mbb{R}^3$, such a rotation by an angle $\theta$ is described by an orthogonal $3\times 3$ matrix, $R^{(\n)}_{\theta}$, rotating a vector $\v\in\mbb{R}^3$ to the vector $R^{(\n)}_{\theta}\v$ (see~\eqref{orthogonalmatrix} for the explicit form of $R^{(\n)}_{\theta}$ in terms of $\n$ and $\theta$). For example, $\x=R^{(\y)}_{\frac{\pi}{2}}\z$, where $\x,\y,\z$ are the unit vectors in the $x$, $y$, and $z$ directions. 
Our goal now is to find a $2\times 2$ complex matrix $T^{(\n)}_{\theta}$ such that if $|0\ra\in\mbb{C}^2$ corresponds to the spin in the $z$-direction, then $T^{(\n)}_{\theta}|0\ra$ corresponds to the spin in the direction $R^{(\n)}_{\theta}\z$. More generally, we look for a matrix $T^{(\n)}_{\theta}$ that has the following property: if the qubit state $|\psi\ra\in\mbb{C}^2$ corresponds to a spin in a direction $\m$, the state 
$T^{(\n)}_{\theta}|\psi\ra$ corresponds to the spin in the direction $R^{(\n)}_{\theta}\m$. Note that in this definition we made an assumption that it is the same matrix $T_{\theta}^{(\n)}$ that is applied to describe a rotation by an angle $\theta$ along the $\n$-axis irrespective if the initial state of the system was in the $z$-direction or any other $\m$-direction. This is justified physically since the physical process that causes the spin of the particle to rotate is \emph{external} to the particle and therefore is independent on the direction that the initial spin of the particle is pointing at. This assumption is also related to the fourth (evolution) axiom\index{axiom} of quantum systems that we will discuss later on.

The matrices $T_\theta^{(\n)}$ are not unique since for any phase $\alpha\in[0,2\pi]$ the matrices $e^{i\alpha}T_\theta^{(\n)}$ also transform $\psi\eqdef|\psi\lr\psi|$ to $T_\theta^{(\n)}\psi (T_\theta^{(\n)})^*$. Hence, $T_\theta^{(\n)}$ is determined uniquely up to a phase. Moreover, since $\|T^{(\n)}_{\theta}|\psi\ra\|=1$ for all normalized vectors $|\psi\ra\in\mbb{C}^2$,  $T^{(\n)}_{\theta}$ must be a unitary matrix. This is a special case of Wigner theorem that states that physical symmetries act on the Hilbert space of quantum states unitarily (or antiunitarily). 
Moreover, repeating the previous arguments, we conclude that the state $T^{(\n_2)}_{\theta_2}T^{(\n_1)}_{\theta_1}|\psi\ra$ is the quantum state that corresponds to thespin in the direction $R^{(\n_2)}_{\theta_2}R^{(\n_1)}_{\theta_1}\m$. Combining everything we conclude that the mapping
\be
R_{\theta}^{(\n)}\mapsto \mT_{\theta}^{(\n)}\quad\text{where}\quad \mT_{\theta}^{(\n)}(\rho)\eqdef T_{\theta}^{(\n)}\rho(T_{\theta}^{(\n)})^*\quad\forall\rho\in\ml(A)\;,
\ee
is a group representation of $SO(3)$ on the Hilbert space $\ml(A\to A)$ (i.e.\ the space of linear operators from $\ml(A)$ to $\ml(A)$).
 
It will be more convinient to work with a unitary representation on the space of $\ml(A)$ itself rather than the Hilbert space $\ml(A\to A)$. For this purpose we need to eliminate the freedom in the choice of the phase so that $R_{\theta}^{(\n)}$ is mapped to a unique $T_{\theta}^{(\n)}$. We therefore assume without loss of generality that $\det\left(T^{(\n)}_{\theta}\right)=1$ so that $T^{(\n)}_{\theta}\in\text{SU}(2)$. This almost eliminates completely the ambiguity in the phase, although note that if $T^{(\n)}_{\theta}\in\text{SU}(2)$ then also $-T^{(\n)}_{\theta}\in\text{SU}(2)$. This would mean that both $\pm T_{\theta}^{(\n)}$ correspond to the same $R_{\theta}^{(\n)}$. To summarize, up to a sign factor, the collection of matrices $\{T^{(\n)}_{\theta}\}_{\n,\theta}$ form a group representation of SO$(3)$. Such 
a $2:1$ and onto homomorphism $h:\text{SU}(2)\to \text{SO}(3)$ with the property that $h(T)=h(-T)$ for any $T\in \text{SU}(2)$ was found by Cornwell in 1984 (see the Exercise~\ref{2to1}). We now discuss the explicit form of $T_{\theta}^{(\n)}$.

In Appendix~\ref{sec:rep} we show that the most general unitary matrix in SU$(2)$ has the form $e^{-i\frac{\theta}{2}(\n\cdot\bs{\sigma})}$ (see~\eqref{c15}), where the factor $1/2$ implies that under a $2\pi$ addition to $\theta$ we get $e^{-i\frac{\theta+2\pi}{2}(\n\cdot\bs{\sigma})}=-e^{-i\frac{\theta}{2}(\n\cdot\bs{\sigma})}$. This property will be consistent with the identification $T^{(\n)}_{\theta}=e^{-i\frac{1}{2}\theta(\n\cdot\bs{\sigma})}$ which we motivate below (recall that $T^{(\n)}_{\theta}\in\text{SU}(2)$) since a rotation by $\theta$ or by $\theta+2\pi$ along any axis $\n$ should have the same effect on any qubit state;
that is,
\be
e^{-i\frac{\theta+2\pi}{2}(\n\cdot\bs{\sigma})}|\psi\lr\psi| e^{i\frac{\theta+2\pi}{2}(\n\cdot\bs{\sigma})}=e^{-i\frac{\theta}{2}(\n\cdot\bs{\sigma})}|\psi\lr\psi|e^{i\frac{\theta}{2}(\n\cdot\bs{\sigma})}\;.
\ee
On the other hand, if we didn't include the factor $1/2$, then the identification 
$T^{(\n)}_{\theta}=e^{-i\theta(\n\cdot\bs{\sigma})}$ would imply an undesired property that a rotation by $\theta$ or by $\theta+\pi$ (along any fixed axis $\n$) would have the same effect on a qubit $|\psi\lr\psi|$.

To justify the identification $T^{(\n)}_{\theta}=e^{-i\frac{1}{2}\theta(\n\cdot\bs{\sigma})}$, 
recall that any rotation around the $z$-axis should not change the state $|0\lr 0|$, as it represents spin in the $z$-direction. Taking $\n=\z$ we get
\be
e^{-i\frac{\theta}{2}(\z\cdot\bs{\sigma})}|0\ra
=\big(\cos\left(\theta/2\right)I-i\sin\left(\theta/2\right)\sigma_3\big)|0\ra
=e^{-i\frac{\theta}{2}}|0\ra\;,
\ee
where we used~\eqref{nsig}. Recall that the vector $e^{i\frac{\theta}{2}}|0\ra$ corresponds to the same quantum state $|0\lr 0|$. Therefore, although there are many possible representations for $SO(3)$ in $\mbb{C}^2$ (such as 
$e^{i0.7\theta(\n\cdot\bs{\sigma})}$ for example), the representation 
$T^{(\n)}_{\theta}=e^{-i\frac{1}{2}\theta(\n\cdot\bs{\sigma})}$ is the only one that have the following essential properties:
\begin{enumerate}
\item The mapping $T_{\theta}^{(\n)}\mapsto R_{\theta}^{(\n)}$ is an onto homomorphism between $SU(2)$ and $SO(3)$.
\item For any $|\psi\ra\in\mbb{C}^2$ we have $T_{\theta+2\pi}^{(\n)}|\psi\lr\psi|\left(T_{\theta+2\pi}^{(\n)}\right)^*=T_{\theta}^{(\n)}|\psi\lr\psi|\left(T_{\theta}^{(\n)}\right)^*$.
\item $T_{\theta}^{(\z)}|0\lr 0|\left(T_{\theta}^{(\z)}\right)^*=|0\lr 0|$ for all $\theta\in\mbb{R}$.
\end{enumerate}

With this representation at hand, we are ready to identify spins in different directions. We start with a few examples. 
The spin in the negative $z$-direction can be obtained by rotating $|0\ra$ by $180^\circ$ along the $x$ (or $y$) axis.
It is therefore given by
\be
T_{\pi}^{(\x)}|0\ra=\left(\cos\left(\frac{\pi}{2}\right)I-i\sin\left(\frac{\pi}{2}\right)\x\cdot\bs{\sigma}\right)|0\ra=-i|1\ra
\ee
Therefore, the quantum state $|1\lr 1|$ corresponds to the negative $z$-direction. Recall from the previous section  that using the SG experiment one can determine with certainty if an electron was prepared in the positive $z$-direction or negative $z$-direction. This ability to distinguish between the two possible spins of the electron is reflected mathematically by the orthogonality of the vectors $|0\ra$ and $|1\ra$. This is a general property of quantum mechanics that any two distinguishable states of a physical system are described mathematically by orthogonal vectors. Other examples include:
\begin{itemize}
\item Spin in the positive $x$-direction;
\be
T_{\frac{\pi}{2}}^{(\y)}|0\ra=\left(\cos\left(\frac{\pi}{4}\right)I-i\sin\left(\frac{\pi}{4}\right)\sigma_2\right)|0\ra=\frac{1}{2}\left(|0\ra+|1\ra\right)\eqdef|+\ra\;.
\ee
\item Spin in the negative $x$-direction;
\be
T_{-\frac{\pi}{2}}^{(\y)}|0\ra=\left(\cos\left(\frac{\pi}{4}\right)I+i\sin\left(\frac{\pi}{4}\right)\sigma_2\right)|0\ra=\frac{1}{2}\left(|0\ra-|1\ra\right)\eqdef|-\ra\;.
\ee
\item Spin in the positive $y$-direction;
\be
T_{-\frac{\pi}{2}}^{(\x)}|0\ra=\left(\cos\left(\frac{\pi}{4}\right)I+i\sin\left(\frac{\pi}{4}\right)\sigma_1\right)|0\ra=\frac{1}{2}\left(|0\ra+i|1\ra\right)\eqdef|+i\ra\;.
\ee
\item Spin in the negative $y$-direction;
\be
T_{\frac{\pi}{2}}^{(\x)}|0\ra=\left(\cos\left(\frac{\pi}{4}\right)I-i\sin\left(\frac{\pi}{4}\right)\sigma_1\right)|0\ra=\frac{1}{2}\left(|0\ra-i|1\ra\right)\eqdef|-i\ra\;.
\ee
\end{itemize}

In general, rotations along the $\n$-axis do not change a spin that points in the positive or negative $\n$-direction. We can use this physical property to compute the qubit representing a spin in the $\n$-direction. Specifically, a quantum state $|\psi\ra$ represents an electron with spin in the positive or negative $\n$-direction if and only if
$T_{\theta}^{(\n)}|\psi\ra=e^{i\alpha}|\psi\ra$ for some phase $e^{i\alpha}$. Now, since $T_{\theta}^{(\n)}=\cos(\theta/2)I-i\sin(\theta/2)\n\cdot\bs{\sigma}$ we get that $|\psi\ra$ is an eigenvector of the matrix $T_{\theta}^{(\n)}$ if and only if it is an eigenvector of the \emph{spin matrix} $S_{\n}\eqdef\frac{1}{2}\n\cdot\bs{\sigma}$.\index{spin matrix}

\begin{exercise}
Let $S_\n$ be the spin matrix in direction $\n=(\sin(\alpha)\cos(\beta),\sin(\alpha)\sin(\beta),\cos(\alpha))^T$, with $\alpha$ and $\beta$ being its spherical coordinates.
\begin{enumerate}
\item Show that $S_{\n}^2=\frac{1}{4}I$.
\item Show that the eigenvalues of $S_\n$ are $\pm\frac{1}{2}$.
\item Show that if $S_{\n}|\psi\ra=\frac{1}{2}|\psi\ra$ then up to global phase
\be\label{qubit}
|\psi\ra=\cos\left(\frac{\alpha}{2}\right)|0\ra+e^{i\beta}\sin\left(\frac{\alpha}{2}\right)|1\ra\;.
\ee
\end{enumerate}
\end{exercise}

\begin{figure}[h]\centering    \includegraphics[width=0.3\textwidth]{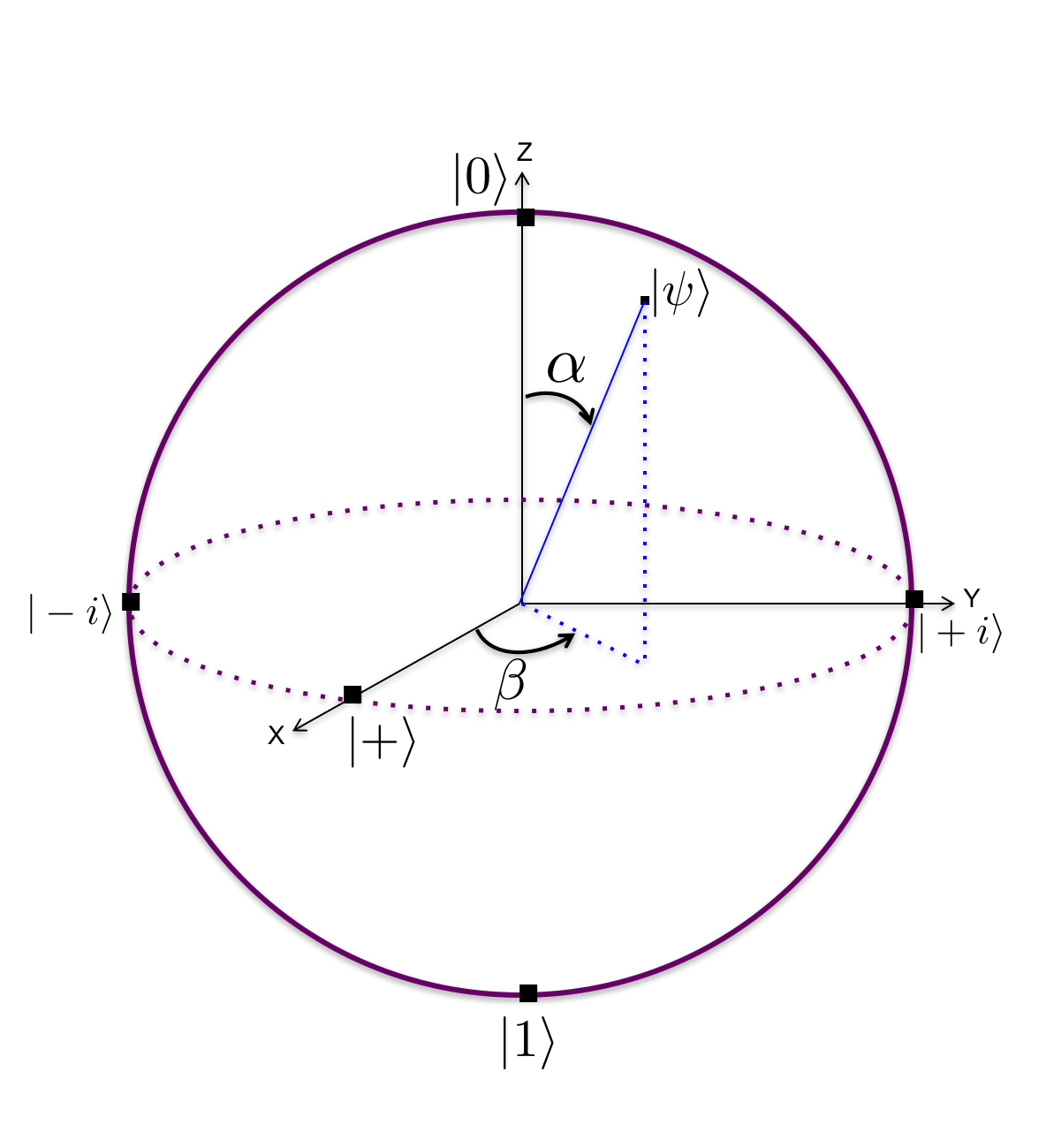}
  \caption{\linespread{1}\selectfont{\small The Bloch Sphere}}
  \label{fig5}
\end{figure}

From the exercise above it follows that any qubit is characterized as in~\eqref{qubit} and corresponds to spin in the positive direction of  $\n=(\sin(\alpha)\cos(\beta),\sin(\alpha)\sin(\beta),\cos(\alpha))^T$. This correspondence between the point on the sphere and a qubit is known in the community as the Bloch representation\index{Bloch representation} of a qubit. In Fig.~\ref{fig5} we show some of the popular qubit states and their location on the Bloch sphere.\index{Bloch sphere}

Note that although we focused here on the spin of an electron, the qubit corresponds to any two level quantum system. For example, one can implement a qubit with a photon, using, say, the $|0\ra$ to correspond to positive (or left) circular polarization, and the $|1\ra$ to corresponds to the negative (or right) circular polarization.
Any linear combination of $|0\ra$ and $|1\ra$ will then correspond to different types of polarizations. Other examples are atoms, molecules, and nucleuses, with two energy levels (excited state vs ground state). All these examples demonstrate that the qubit can be implemented in many different ways and in this sense, we can claim that quantum information is fungible!

\subsection{The Quantum Dit and Observables}\index{qudit}\index{observable}

The quantum dit (qudit) represents any $d$-level quantum system with $d>2$. To any such physical system we associate a $d$-dimensional Hilbert space $A\cong\mbb{C}^d$. How should we interpret the quantum states in $A$? Recall that in the qubit case we associated to any orthonormal basis of $\mbb{C}^2$ a spin in some direction $\n$, with the two possibilities of ``spin up" and ``spin down" corresponding to the two elements of the basis. Moreover, we were able to construct the ``spin matrix" $S_\n$ whose eigenvectors are the basis elements, and its eigenvalues give the spins (i.e. $\pm1/2$) associated with the two basis elements. 
 
In a similar way, any orthonormal basis of $A\cong\mbb{C}^d$ corresponds to $d$ possible outcomes that can be, at least in principle, observed in some experiment. Moreover, the second postulate of quantum mechanics states that  any observable\index{observable} (a dynamic variable that can be measured, like position, momentum, spin, energy, etc)  is represented with an Hermitian operator whose eigenvalues correspond to the values of the observable. 
Recall that for any Hermitian operator, $H$, there exists an orthonormal basis of $A$ consisting only of eigenvectors of
$H$. This basis corresponds to the possible outcomes in the measurement of the observable $H$.

For example, in the qubit case, the spin matrix $S_\n$ is an observable corresponding to the measurement of spin with the SG experiment. In physical systems of $d$ energy levels, the Hamiltonian $H=\sum_{x\in[d]}E_x|x\lr x|$, is an observable corresponding to the measurement of energy. This particular observable\index{observable} states that the values for the energy of the system (i.e. $E_x$) are discrete, and that the eigenvectors $\{|x\ra\}_{x\in[d]}$ correspond to these energy levels.

\subsection{Composite Systems}\index{composite system}

The third postulate of quantum mechanics states that the Hilbert space associated with a composite system\index{composite system}  is the Hilbert space formed by the tensor product of the state spaces associated with the component subsystems. For example, the Hilbert space associated with three electrons is given by $ABC=\mbb{C}^2\otimes\mbb{C}^2\otimes\mbb{C}^2$.
Note that although $\mbb{C}^2\otimes\mbb{C}^2\otimes\mbb{C}^2\cong\mbb{C}^8$, the tensor product structure has a physical significance as the component subsystems correspond to individual particles. For example, the state
\be
|0\ra^A|-\ra^B|+i\ra^C
\ee
corresponds to three spin particles (e.g. electrons) with spin A pointing in the positive $z$-direction, spin B pointing in the negative $x$-direction, and spin C pointing in the positive $y$-direction. Of course, not all states have the same tensor product form as the state above. For example, the Greenberger-Horne-Zeilinger (GHZ) state of three qubits \index{GHZ state}
\be
|GHZ\ra\eqdef\frac{1}{\sqrt{2}}\left(|0\ra^A|0\ra^B|0\ra^C+|1\ra^A|1\ra^B|1\ra^C\right)
\ee
cannot be written as a tensor product of three vectors. States like this will be called entangled.

\begin{exercise}
Show that the GHZ state above cannot be written as a tensor product of three vectors; i.e.
\be
|GHZ\ra\neq |\psi^A\ra|\phi^B\ra|\chi^C\ra
\ee
for any three qubit states $|\psi^A\ra$, $|\phi^B\ra$, and $|\chi^C\ra$. 
\end{exercise}

\begin{exercise}\label{singlet}
Show that for any unit vector $\n\in\mbb{R}^3$ the singlet state\\ $|\Psi_{-}^{AB}\ra\eqdef\left(|01\ra-|10\ra\right)/\sqrt{2}$
can be expressed as
\be
|\Psi_{-}^{AB}\ra=\frac{1}{\sqrt{2}}\left(|\ua_\n\ra|\da_\n\ra-|\da_\n\ra|\ua_\n\ra\right)
\ee
where $|\ua_\n\ra$ and $|\da_\n\ra$ are the eigenvalues of the spin matrix $S_\n$. In other words, for any $2\times 2$ unitary matrix $U$ we have $U\otimes U|\Psi_{-}\lr\Psi_{-}|U^*\otimes U^*=|\Psi_{-}\lr\Psi_{-}|$.
\end{exercise}

\begin{exercise}
Consider the Hilbert space of two electrons $AB\cong\mbb{C}^2\otimes\mbb{C}^2$. For any unit vector $\n\in\mbb{R}^3$ denote by $J_\n\eqdef S_\n\otimes I+I\otimes S_\n$, and by $J^2\eqdef J_{\x}^{2}+J_{\y}^{2}+J_{\z}^{2}$.
\begin{enumerate}
\item Show that the eigenvalues of $J_\n$ are $1$, $0$, and $-1$.
\item Find the eigenvalues of $J^2$.
\item Show that $[J^2,J_z]=0$.
\item Show that each of the following four 2-qubit states are eigenvectors of both $J^2$ and $J_z$:
\be
|00\ra\;\;,\quad|11\ra\;\;,\quad\text{and}\quad
|\Psi_{\pm}\ra\eqdef\frac{1}{\sqrt{2}}\left(|01\ra\pm|10\ra\right)
\ee
\end{enumerate}
\end{exercise}

\begin{exercise}\label{exchsh}
Let $S_\n=\frac{1}{2}\n\cdot\bs{\sigma}$ and 
$S_{\m}=\frac{1}{2}\m\cdot\bs{\sigma}$ be the spin matrices (observables)
in the directions of the unit vectors $\n$ and $\m$, respectively. 
\begin{enumerate}
\item Show that the commutator $[S_\n,S_\m]=iS_{\r}$, where $\r\in\mbb{R}^3$ is a unit vector. What is the direction of $\r$?
\item Calculate
\be
\langle\Psi_{-}|S_{\n}\otimes S_{\m}|\Psi_{-}\rangle\quad\text{where}\quad |\Psi_{-}\rangle=\frac{1}{\sqrt{2}}\left(|0\rangle\otimes|1\rangle-|1\rangle\otimes|0\rangle\right).
\ee
\item Let $\n'$ and $\m'$ be two more unit vectors, and let
\be
B\eqdef
S_{\n}\otimes S_{\m}+
S_{\n}\otimes S_{\m'}+
S_{\n'}\otimes S_{\m}-
S_{\n'}\otimes S_{\m'}\;.
\ee
Show that
\be
B^{2}=\frac{1}{4}I-[S_{\n},S_{\n'}]\otimes[S_{\m},S_{\m'}]\;,
\ee
and use it to prove that
\be
\left|\langle\psi|B|\psi\rangle\right|\leq \frac{1}{\sqrt{2}}\;,
\ee 
for any state in $|\psi\rangle\in\mathbb{C}^2\otimes\mathbb{C}^2$.
\end{enumerate}
\end{exercise}

\section{Quantum Measurements}\index{quantum measurements}

Since quantum mechanics aim to study the behaviour of subatomic particles, the process of measurement is essential to the theory and requires a rigorous treatment.  The SG experiment 
demonstrates that physical systems are not separated from the apparatuses that are measuring them.
From a philosophical standpoint, ``the observer is not separated from that which is being observed". 
The effect of observation on physical systems is not unique to quantum mechanics.  
The following story from Jostein Gaarder's novel, ``Sophie's World", shows, among many other things, that even the behaviour of a centipede is effected by measurements and observations. 
\begin{quote}
 ``Once upon a time there was a centipede that was amazingly good at 
dancing with all hundred legs. All the creatures of the forest gathered to watch 
every time the centipede danced, and they were all duly impressed by the 
exquisite dance. But there was one creature that didn't like watching the 
centipede dance - that was a tortoise. 

How can I get the centipede to stop dancing? thought the tortoise. He 
couldn't just say he didn't like the dance. Neither could he say he danced 
better himself, that would obviously be untrue. So he devised a fiendish plan.

He sat down and wrote a letter to the centipede. `O incomparable 
centipede,' he wrote, `I am a devoted admirer of your exquisite dancing. I 
must know how you go about it when you dance. Is it that you lift your left leg 
number 28 and then your right leg number 39? Or do you begin by lifting your 
right leg number 17 before you lift your left leg number 44? I await your 
answer in breathless anticipation. Yours truly, Tortoise.' "
\end{quote}

You can easily guess now what happened next!

\subsection{Born's Rule and von-Neumann Projective Measurements}\index{Born's rule}\index{von-Neumann measurement}

Consider the SG experiment, which involves measuring the spin of an electron along an arbitrary direction denoted as $\mathbf{n}$. It is well-established that the act of making this measurement can impact the state of the electron. This change occurs when the electron's initial spin is not aligned with either the positive or negative directions of $\mathbf{n}$.

Consider the initial quantum state $|\psi\ra = a|0\ra + b|1\ra$, where $a$ and $b$ are complex numbers, subject to the normalization condition $|a|^2 + |b|^2 = 1$. Additionally, let $|\ua_\n\ra$ and $|\da_\n\ra$ denote the eigenvectors of the spin operator $S_\n$, corresponding to the electron's spin aligned with the upward and downward directions along $\mathbf{n}$, respectively. Under an SG-experiment in the direction $\n$, the transformation of the state $|\psi\ra$ follows this simple rule:
\ben
\item If the SG experiment yields an outcome in the upward direction along $\mathbf{n}$, the state evolves to $|\ua_\n\ra$.
\item If the SG experiment yields an outcome in the downward direction along $\mathbf{n}$, the state evolves to $|\da_\n\ra$.
\een
It's crucial to emphasize that this transformation is independent of the specific form of $|\psi\ra$. What does vary with $|\psi\ra$ is the probability associated with each possible outcome. For instance, consider the case where $\mathbf{n}=\mathbf{z}$ and $|\psi\ra$ is initially prepared as $|\ua_\x\ra$. In this case, both the upward and downward outcomes are equally probable, each with a 50\% chance.

The general rule governing the probability of obtaining a particular outcome in the measurement is known as Born's rule. According to Born's rule, the probability, denoted as $\pr(\psi,\n)$, of observing the outcome $\ua_\n$ (i.e., the electron's spin aligned with the positive direction of $\mathbf{n}$) in an SG experiment along the $\mathbf{n}$ direction, when the electron is initially prepared in the state $|\psi\ra$, is given by: 
\be\label{brul}
 \pr(\psi,\n)=\left|\la\psi|\!\ua_\n\ra\right|^2\;.
\ee
This fundamental principle provides a mathematical framework for determining the likelihood of various outcomes in quantum measurements, and it plays a central role in quantum mechanics.
For example, suppose an electron in the state $|\psi\ra=a|0\ra+b|1\ra$ is sent through a SG-experiment in the $z$-direction. Then, using the Born's rule~\eqref{brul} we get that $\left|\la\psi|\!\ua_\z\ra\right|^2=|a|^2$ is the probability to obtain spin up (in the $z$-direction), and $\left|\la\psi|\!\da_\n\ra\right|^2=|b|^2$ is the probability to obtain spin down. 

Similarly, we can extend the Born's rule\index{Born's rule} for any qudit $|\psi\ra\in A\cong\mbb{C}^m$, and any quantum measurement that corresponds to an orthonormal basis $\{|\phi_x\ra\}_{x\in[m]}$ of $A$. The probability to obtain an outcome $x$ is given by 
\be
\pr(\psi,\phi_x)=\left|\la\psi|\phi_x\ra\right|^2\;.
\ee
Note that the above assignment of probability to each $\phi_x$ is indeed a probability; that is,
\ba
\sum_{x\in[m]}\pr(\psi,\phi_x)=\sum_{x\in[m]}\left|\la\psi|\phi_x\ra\right|^2&=\sum_{x\in[m]}\la\psi|\phi_x\lr \phi_x|\psi\ra\\
&=\la\psi|\Big(\sum_{x\in[m]}|\phi_x\lr \phi_x|\Big)|\psi\ra=\la\psi|\psi\ra=1\;.
\ea
We call every such a measurement that corresponds to an orthonormal basis a \emph{basis measurement}.

To establish a connection between a basis measurement and a physical observable, let's consider the energy of a physical system. Energy is a fundamental observable in quantum mechanics and therefore can be measured. 
As previously discussed, any observable in quantum mechanics is represented by an Hermitian operator acting on the Hilbert space $A$. We denote the energy operator, often referred to as the Hamiltonian, as:
\be
H=\sum_{x\in[m]}a_x|\phi_x\lr \phi_x|\;,
\ee
where $\{|\phi_x\ra\}_{x\in[m]}$ is an orthonormal basis of $A$.
Therefore, in order to measure the energy, one has to perform a basis measurement corresponding to the orthonormal basis $\{|\phi_x\ra\}_{x\in[m]}$, since the energy $a_x$ is determined by the value of $x$.
However, this system of $m$-energy levels, can be degenerate as it happens quite often in many physical systems. 
In this case, not all of the energy values $\{a_x\}_{x\in[m]}$ are distinct. Suppose for example that $a_1=a_2<a_3<\cdots<a_{m}$; i.e. the state with minimum energy (the ground state) is degenerate. In this case, both outcomes $1$ and $2$ correspond to the same ground state, so that the probability that the energy is equal $a_1=a_2\eqdef b_1$ is given by
\be\label{two}
\pr(\psi,\phi_1)+\pr(\psi,\phi_2)=\la\psi|\big(|\phi_1\lr \phi_1|+|\phi_2\lr \phi_2|\big)|\psi\ra\eqdef\la\psi|\Pi|\psi\ra\;,
\ee
where $\Pi\eqdef |\phi_1\lr \phi_1|+|\phi_2\lr \phi_2|$. More generally, if we have degeneracy in other energy levels, we can always express the observable\index{observable} $H$ as
\be
H=\sum_{y\in[r]}b_y\Pi_{y}\;,
\ee
where $b_1<b_2<\cdots<b_r$, and each $\Pi_y$ is a sum of rank one projections from $\{|\phi_x\lr \phi_x|\}$ that correspond to the same energy level $b_y$. With this at hand, the probability to measure an energy of value $b_y$ is given by
\be\label{2point4}
\pr\left(\psi,\Pi_y\right)=\la\psi|\Pi_y|\psi\ra\;.
\ee
Therefore, the basis-measurement that we considered so far, can be extended to \emph{projective von-Neumann measurement} which is defined as follows.
\begin{myd}{}
\begin{definition}
A  von-Neumann projective measurement (or, in short, projective measurement) on a Hilbert space $A$, is a collection of mutually orthogonal projections $\{\Pi_x\}_{x\in[r]}$
satisfying $\sum_{x\in[r]}\Pi_x=I^A$ and for all $x,y\in[r]$
\be
\Pi_x\Pi_y=\delta_{xy}\Pi_x\;.
\ee 
\end{definition} 
\end{myd}

Historically, the Born's rule\index{Born's rule} above (see~\eqref{2point4}) was determined essentially from consistency with experiments. That is, one can perform many experiments, like the SG-experiment for example, collect the data, and find a rule that is consistent with the data. Later on, however, Gleason came up with a theorem showing how to calculate probabilities in quantum mechanics, and loosely speaking \emph{derived} the Born's rule\index{Born's rule} above from a few fundamental principles involving measures of a Hilbert space. Gleason's theorem is applicable for general (separable) Hilbert spaces in any dimension, but for us, only the finite dimensional case, i.e. the qudit, will be relevant.
We postpone the discussion on Gleason's theorem for the next chapter, after we discuss other types of quantum measurements, in order to prove a slightly more generalized version of Gleason's theorem, that will be applicable to all types of measurements (not only to projective von-Neumann measurements).

\begin{exercise}
Let $\Pi$ be a projection on a Hilbert space $A$. Show that $\{\Pi,I-\Pi\}$ is a two-outcome von-Neumann projective measurement.
\end{exercise}

\begin{exercise}
Let $\{\Pi_x\}_{x\in[r]}$ be a projective von-Neumann measurement on a finite dimensional Hilbert space $A$. Show that the collection of all the linearly independent normalized eigenvectors, of all the projections $\{\Pi_x\}_{x\in[r]}$, form an orthonormal basis of $A$.
\end{exercise}

\subsection{The Post-Measurement State}\index{post-measurement state}

Recall that the basis measurements we discussed earlier lead to a change in the system's state according to the rule $|\psi\ra\to|\phi_x\ra$ if outcome $x$ occurs. However, when it comes to projective measurements, it's possible to encounter a scenario where a degenerate energy value, such as $b_1=a_1=a_2$, occurs, as illustrated in~\eqref{two}. In this case, after the measurement yields outcome $b_1$, all we can ascertain is that the system's state belongs to the subspace $B$, defined as $B\eqdef\spa\{|\phi_1\ra,|\phi_2\ra\}$, since any state within this subspace is an eigenvector of the Hamiltonian $H$ associated with the same energy eigenvalue $b_1$. However, it raises the question of which specific state within this subspace will become the post-measurement state.

Notably, $B$ is a subspace of $A$, and $\Pi_1=|\phi_1\lr \phi_1|+|\phi_2\lr \phi_2|$ projects states from $A$ to $B$. Furthermore, if the pre-measurement state $|\psi\ra\in B$, then $|\psi\ra$ is already an eigenvector of $H$ and should remain unaffected by the measurement of $H$. Consequently, if we denote by $\Lambda$  the transformation that converts the pre-measurement state $|\psi\ra$ into the post-measurement state (which may not be normalized), it follows that $\Lambda|\psi\ra=|\psi\ra$ for $|\psi\ra\in B$.

On the other hand, if $|\psi\ra\in B^{\perp}$ (the orthogonal complement of $B$ in $A$), the energy $b_1$ has zero probability of occurring, and therefore, we assume that its corresponding post-measurement state is $\Lambda|\psi\ra=0$. In this context, $\Lambda$ must be equivalent to $\Pi_1$. However, unless $|\psi\ra$ is within $B$, the state $\Pi_1|\psi\ra$ is not normalized. Thus, the rule stipulates that the pre-measurement state $|\psi\ra$ undergoes transformation to the normalized post-measurement state $\Pi_{1}\|\psi\ra/|\Pi_{1}|\psi\ra\|$.

To summarize, for any physical system that is prepared in a state $|\psi\ra\in A$, and for any von-Neumann projective measurement, $\{\Pi_x\}_{x\in[r]}$, of the observable $H=\sum_{x\in[r]}a_x\Pi_x$, the rules of quantum mechanics state that the probability to obtain a value $a_x$ is given by the Born's rule
\be
\pr\left(\psi,\Pi_x\right)=\la\psi|\Pi_x|\psi\ra\;,
\ee 
and the quantum state of the system after the outcome $x$ occurred is 
\be
\frac{1}{\|P_x|\psi\ra\|_2}P_x|\psi\ra\;.
\ee

\begin{exercise}
Consider the space $A\cong\mbb{C}^3$.
\begin{enumerate}
\item Find the projection $\Pi_0$ to the 2-dimensional subspace $B\eqdef\spa\{|0\ra+|1\ra,|0\ra+|2\ra\}$.
\item Use the projection $\Pi_0$ that you found in part 1 to construct a two-outcome projective measurement $\{\Pi_0,\Pi_1\}$, with $\Pi_1=I-\Pi_0$. If a physical system was prepared initially in the state $|\psi\ra=\frac{1}{\sqrt{3}}\left(|0\ra+|1\ra+|2\ra\right)$, what is the probability that this projective measurement yields an outcome 0 (i.e. corresponds to $\Pi_0$)? What will be the post-measurement state in this case?
\end{enumerate}
\end{exercise} 
 
\begin{exercise}\label{distinguish}
Let $A$ be a $d$-dimensional Hilbert space, and let $|\psi\ra,|\phi\ra\in A$ be two quantum states.
\begin{enumerate}
\item Show that if $|\psi\ra$ and $|\phi\ra$ are orthogonal, then there exists a projective measurement that distinguishes them. That is, there exists a two-outcome projective measurement $\{\Pi_0,\Pi_1\}$ such that
\be
\pr(\psi,\Pi_0)=1\quad \text{and}\quad \pr(\phi,\Pi_1)=1.
\ee 
\item Show that if $|\psi\ra$ and $|\phi\ra$ are \emph{not} orthogonal, then there is no projective measurement that distinguishes them.
\end{enumerate}
\end{exercise}

\section{Hidden Variable Models}\label{hvm}\index{hidden variables}

The axioms of quantum mechanics that we considered so far have profound consequences. For example, suppose an electron has been prepared with a spin in the positive $z$-direction. 
The rules of quantum mechanics tells us that if we where to measure its spin in other directions (including $x$ or $y$ directions) there is a non-zero probability to find out the spin pointing in those directions. This raises the question of whether it is possible to model the quantum system (i.e. the electron in this case) with a (possibly uncountable) collection of classical random variables, sometimes called \emph{hidden} variables, each contains information about the spin of the electron in some direction. Remarkably, such attempts to model physical systems with local random variables instead of quantum states lead to inconsistencies with experiments. 
That is, the axioms of quantum mechanics are inconsistent with local hidden variable models of reality.

\subsection{The CHSH Inequality and Local Realism}\index{CHSH inequality}\index{local realism}

Let $0$ and $1$ represent positive and negative directions of a spin of an electron, respectively, and let $\n$ be a unit vector in $\mbb{R}^3$. Denote by $A_{\n}\in\{0,1\}$ the value of the spin in the $\n$ direction. That is, we replace the Hilbert space $A$ of the electron, with a collection of classical variables $\{A_\n\}_\n$, where $\n$ is running over all unit vectors in $\mbb{R}^3$. Suppose the electron is prepared at the positive $\z$-direction. This means that $A_{\z}=0$, while from the SG-experiment above we already saw that $A_{\x}=0$ with probability $50\%$ and $A_{\x}=1$ also with  $50\%$ probability. That is, $A_{\n}$ is a random bit, and the probability that $A_{\n}=a$ (with $a\in\{0,1\}$) is denoted by $p(a|\n)$. The conditional probabilities $p(a|\n)$ provide all the information about the spin of an electron that was prepared initially in the $z$-direction.

Aside from being somewhat artificial (i.e. the model does not provide a mechanism to derive $p(a|\n)$
from a set of axioms), a-priori, it seems to provide a valid description of the information about the spin. One can view $A_{\n}$ as a \emph{hidden variable} that the observer does not know (unless $\n=\pm\z$) until she/he performs an SG-experiment (or any other experiment for that matter). Note that after every measurement the observer will need to update the conditional probability $p(a|\n)$. 

For any such a \emph{hidden variable model}, there is an inherent assumption that the
values of the hidden variables are fixed, predetermined, and corresponds to an \emph{element of reality}. 
It is just the observer's lack of knowledge about this element of reality that leads to statistical behaviours.  
Historically, hidden variable theories were promoted by some physicists who argued that the formulation of quantum mechanics (as we will discuss in the rest of this book), does not provide a complete description for the system. Along with Albert Einstein, they argued that quantum mechanics is ultimately incomplete, and that a complete theory would  avoid any indeterminism.
Indeed, hidden variable models as the one described above for the spin of one electron cannot be ruled out, although, as we discuss now,
\emph{local} hidden variable models can!

Consider two friends, Alice and Bob, that are located far from each other, and each one of them posses an electron in their lab. How can we describe the spins of the two electrons? Following the same line of thoughts as above, we denote by $A_{\n}$ the random variable associated with the spin of Alice's electron in the $\n$-direction, and by $B_{\m}$ the random variable associated with the spin of Bob's electron in the $\m$-direction. We denote by $p(ab|\n\m)$, with $a,b\in\{0,1\}$, the joint probability that the two SG experiments in Alice's lab and Bob's lab will yield respectively $A_{\n}=a$ and $B_{\m}=b$.

Since it is possible that the spins are correlated in some way, we are \emph{not} assuming that $p(ab|\n\m)$ has the form $p^A(a|\n)p^B(b|\m)$, where $p^A(a|\n)$ is the probability that Alice will get the value $a$ in a SG experiment in the $\n$-direction (and $p^B(b|\m)$ is defined similarly). Instead, since  in general $A_{\n}$ and $B_{\m}$ can be correlated, there exists a
parameter $\lambda$ ($\lambda$ can describe a collection of variables) and a probability distribution $q_\lambda$ over it, such that 
\be\label{local}
p(ab|\n\m)=\int d\lambda \;q_\lambda \;p_{\lambda}^{A}(a|\n)\;p_{\lambda}^{B}(b|\m)\;,
\ee 
where $p_{\lambda}^{A}(a|\n)$ and $p_{\lambda}^{B}(b|\m)$ are probability distributions that depend on the correlating parameter $\lambda$. The parameter $\lambda$ can be either continuous or discrete, and for the latter the integral above is replaced with a sum. Note that the distribution above is more general than the form $p^A(a|\n)p^B(b|\m)$ as it allows for correlations between Alice's and Bob's spins. Yet, it is a \emph{local} probability distribution depending only on the local variables $A_\n$ and $B_\m$. 
We now discuss a crucial consequence of this local hidden variable model for the spin of two electrons. 

\begin{exercise}
Let $\n$, $\n'$, $\m$, and $\m'$, be four unit vectors in $\mbb{R}^3$, and use the tilde symbol over a random variable $X$ to mean $\tilde{X}=2X-1$ (i.e. $\tX$ takes values $\pm1$ while $X$ takes values $0,1$). Show that
\be
\left|\tA_{\n}\tB_{\m}+\tA_{\n'}\tB_{\m}+\tA_{\n}\tB_{\m'}-\tA_{\n'}\tB_{\m'}\right|\leq 2\;.
\ee
\end{exercise}

\begin{exercise}
Denote by
\be
\la AB\ra_{\n\m}\eqdef \la A_{\n}B_{\m}\ra\eqdef\sum_{a,b\in\{0,1\}}ab\;p(ab|\n\m)\;\;.
\ee
\item Show that any local probability distribution $p(ab|\n\m)$ as in~\eqref{local} satisfies
\be\label{chsh}
\left|\la \tA\tB\ra_{\n\m}+\la \tA\tB\ra_{\n\m'}+\la \tA\tB\ra_{\n'\m}-\la \tA\tB\ra_{\n'\m'}\right|\leq 2\;.
\ee
\end{exercise}
The inequality in the exercise above is called the CHSH inequality\index{CHSH inequality} after Clauser, Horne, Shimony and Holt, and it generalizes a similar inequality that was proved in a seminal paper by John Bell from 1964. 
As we will see in the exercise below, not all probability distributions $p(ab|\n\m)$ satisfy this inequality.
One obvious property of the local distribution~\eqref{local} is that if we sum over $a$ the dependance on $\n$ disappears, and similarly if we sum over $b$ the dependance in $\m$ disappear. This property is called ``no-signalling" since by choosing different directions of $\n$, Alice cannot signal Bob, since the marginal distribution on his side remains intact. The no-signalling property can be stated as follows:
\begin{align}\label{nonsig}
& \sum_{a}p(ab|\n\m)=\sum_ap(ab|\n'\m)\eqdef p^B(b|\m) \quad\forall\;b,\n,\n',\m\nonumber\\
& \sum_{b}p(ab|\n\m)=\sum_bp(ab|\n\m')\eqdef p^A(a|\n)\quad\forall\;a,\n,\m,\m'\;.
\end{align}
The following exercise shows that there exists a probability distribution that on one hand, is non-signalling, and on the other hand, is violating the CHSH inequality~\eqref{chsh}.

\begin{exercise}\label{PR}
Denote the two directions in Alice's side by $\n_0\eqdef\n$ and $\n_1\eqdef\n'$, and the two direction vectors in Bob's side by $\m_0\eqdef\m$ and $\m_1\eqdef\m'$. Denote also by $p(ab|xy)=p(ab|\n_x\m_y)$ with $x,y\in\{0,1\}$.
Consider the probability distribution given by
\be\label{pr}
p(ab|xy)= \begin{cases}
       \frac{1}{2} &\quad\text{if} \quad a\oplus b=xy \\
       0&\quad\text{otherwise} 
     \end{cases}\;,
\ee
where the $\oplus$ denotes addition modulo 2.
\begin{enumerate}
\item Show that $p(ab|xy)$ above is non-signalling; i.e. satisfies~\eqref{nonsig}.
\item Show that $p(ab|xy)$ is non-local by showing that it violates the CHSH inequality~\eqref{chsh}.
\item Show that no other probability distribution (even a signalling distribution) can provide a higher violation than the one achieved by the distribution~\eqref{pr}.
\end{enumerate}
\end{exercise}

To summarize, any local hidden variable model have two main assumptions. The first one is called the \emph{realism} assumption, corresponding to our assumption that the spins of the electrons in all directions have definite values which exist independently of observation.
The second assumption is called the \emph{locality} assumption corresponding to our implicit assumption that if, say   Alice, is performing a measurement on her electron, it does not influence the result of Bob's measurement (on the spin of the electron in his lab). The following violation of the CHSH inequality\index{CHSH inequality}  demonstrates that \emph{local realism} does not hold! \index{local realism}
 
\subsection{Quantum Violation of the CHSH Inequality}

\begin{figure}[h]\centering
    \includegraphics[width=0.5\textwidth]{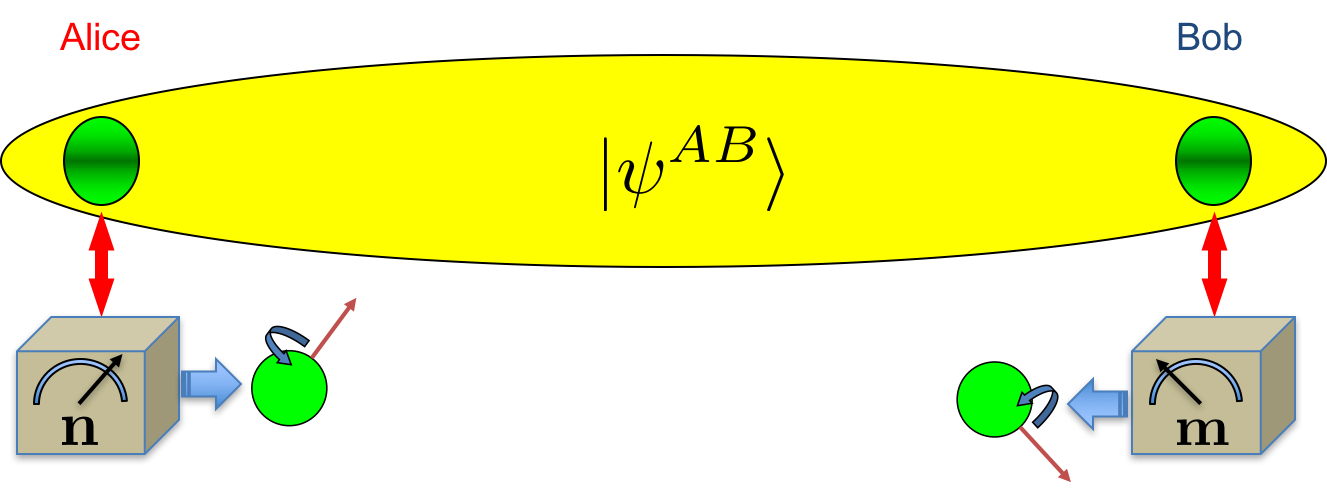}
  \caption{\linespread{1}\selectfont{\small Measurements of the spins of two electrons}}
  \label{fig6}
\end{figure}

The violation of the Bell and CHSH inequalities is one of the most profound results of the 20th century. It states that the formalism of quantum mechanics allows for a violation of the inequality in~\eqref{chsh}. This means that a local hidden variable model cannot account for quantum correlations. To see the violation, consider two electrons one located in Alice's lab and the other in Bob's lab, that are prepared in some state $|\psi^{AB}\ra\in {AB}\cong\mbb{C}^2\otimes\mbb{C}^2$. Both Alice and Bob perform a SG experiment with Alice in the direction $\n$ and Bob in the direction $\m$  (see Fig.~\ref{fig6}). As in the previous section, denote by $p(ab|\n\m)$ the probability that Alice obtains an outcome $a$ (with $a=0$ for positive $\n$-direction and $a=1$ for negative $\n$-direction) and Bob obtains an outcome $b$ (again with the same correspondence of positive and negative directions for $b=0$ and $b=1$, respectively). According to Born's rule, this probability is given by
\be\label{probqm}
p_\psi(ab|\n\m)=\left|\la\psi^{AB}|\phi_{a}^{A}\otimes\varphi_{b}^{B}\ra\right|^2
\ee
where $|\phi_{a}^{A}\ra\in\mbb{C}^2$ (with $a=0,1$) are the eigenvectors of the spin matrix $S_\n$, and $|\varphi_{b}^{B}\ra\in\mbb{C}^2$ (with $b=0,1$) are the eigenvectors of the spin matrix $S_\m$. The corresponding eigenvalues are given by, $\frac{1}{2}-a$, for $|\phi_{a}^{A}\ra$ and by, $\frac{1}{2}-b$, for $|\varphi_{b}^{B}\ra$. Note the relation with the previous notations; for example, $|\phi_{0}^{A}\ra=|\ua_\n\ra$ and $|\phi_{1}^{A}\ra=|\da_\n\ra$.

\begin{exercise}
Show that the product spin average is given by
\be
\sum_{a,b\in\{0,1\}}\left(\frac{1}{2}-a\right)\left(\frac{1}{2}-b\right)p_\psi(ab|\n\m)=\la\psi^{AB}|S_\n\otimes S_\m|\psi^{AB}\ra\;.
\ee 
\end{exercise}
Recall that $\tA_{\n}$ and $\tB_{\m}$ in~\eqref{chsh} are random variables taking the values $\pm1$, whereas the eigenvalues of $S_\n$ and $S_\m$ are $\pm\frac{1}{2}$. Keeping this in mind,   
from the exercise above we conclude that the probability distribution $p_\psi(ab|\n\m)$ as given in~\eqref{probqm} violates the CHSH inequality~\eqref{chsh} if 
\be
\left|\la\psi|B|\psi\ra\right| > \frac{1}{2}\;,
\ee
where $B$ is the Bell/CHSH operator from Exercise~\ref{exchsh}. From Exercise~\ref{exchsh} it follows
that for any state $|\psi\ra\in\mbb{C}^2\otimes\mbb{C}^2$, we have $\left|\la\psi|B|\psi\ra\right| \leq 1/\sqrt{2}$.
From the next exercise it follows that there exists directions $\n,\m,\n',\m'$ such that this bound is saturated, thereby violating the CHSH inequality\index{CHSH inequality} since $\frac{1}{\sqrt{2}}>\frac{1}{2}$.
This bound is called the Tsirelson bound. 

\begin{exercise}
Prove that the Tsirelson bound can be achieved by taking $|\psi^{AB}\ra$ to be the singlet state 
$|\Psi_{-}^{AB}\ra=\left(|01\ra-|10\ra\right)/\sqrt{2}$. That is, find four directions $\n,\n',\m,\m'$ such that
\be
\left|\la\Psi_{-}|B|\Psi_{-}\ra\right| = \frac{1}{\sqrt{2}}\;.
\ee
\end{exercise}

Note that the violation of the CHSH inequality\index{CHSH inequality} implies that the quantum probability distribution $p_\psi(ab|\n\m)$ is in general not local; i.e. not of the form~\eqref{local}. Such non-local probability distributions have other non-intuitive consequences as we discuss below.

\subsection{John Preskill's Example: Quantum Coins}\index{quantum coins}

Consider two electrons, one in Alice's lab and the other in Bob's lab, prepared in the singlet state $|\Psi_{-}^{AB}\ra=\left(|01\ra-|10\ra\right)/\sqrt{2}$.
Suppose Alice wants to measure the spin of her electron in two directions $\n_1$ and $\n_2$. 
She knows that if she performs a measurement in the $\n_1$-direction that will affect the state of her electron, and she will not be able to determine what would have happened if she did the $\n_2$-measurement instead.
Therefore, she asks Bob to perform the $\n_2$-measurement on his electron, while she performs the $\n_1$-measurement on her electron. From Exercise~\ref{singlet}, the 
singlet state can be written as $\left(|\ua_{\n_2}\ra|\da_{\n_2}\ra-|\da_{\n_2}\ra|\ua_{\n_2}\ra\right)/\sqrt{2}$. This means that if Bob's measurement output is $\ua_{\n_2}$ then Alice would have measured $\da_{\n_2}$ had she chose to do the $\n_2$ measurement instead. Similarly, if Bob's measurement output is $\da_{\n_2}$ then Alice would have measured $\ua_{\n_2}$. This gives a way for Alice to determine what would be the output of the measurement if she chose to perform the $\n_2$-measurement, and thereby know simultaneously the values of the spins in directions $\n_1$ and $\n_2$ of her electron. This idea of determining the outputs of several possibly \emph{counterfactual} measurements by the use of entangled states (like the singlet) is the key technique used in many experiments including the famous \emph{delayed choice quantum eraser experiment}.

\begin{exercise}
Using the method described above, show that the probability $p_{\rm same}(\n_1,\n_2)$, that Alice will obtain both spins of her (single) electron pointing in the same direction (i.e. both positive $\ua_{\n_{1}}^{A}\ua_{\n_{2}}^{A}$ or both negative $\da_{\n_{1}}^{A}\da_{\n_{2}}^{A}$)  is:
\be
p_{\rm same}(\n_1,\n_2)\eqdef\left|\la\Psi_{-}^{AB}|\ua_{\n_1}^{A}\da_{\n_2}^B\ra\right|^2+\left|\la\Psi_{-}^{AB}|\da_{\n_1}^A\ua_{\n_2}^B\ra\right|^2=\frac{1}{2}(1+\cos(\theta))
\ee
where $\theta$ is the angle between the unit vectors $\n_1$ and $\n_2$.
\end{exercise}

\begin{figure}[h]\centering
    \includegraphics[width=0.3\textwidth]{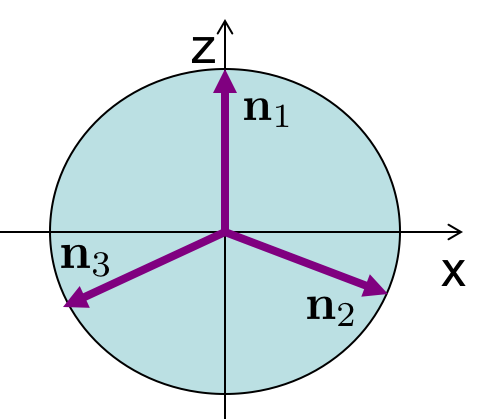}
  \caption{\linespread{1}\selectfont{\small Three directions with an angle of $120^\circ$ between any two.}}
  \label{fig7}
\end{figure}

Consider now three unit vectors $\n_1,\n_2,\n_3\in\mbb{R}^3$ with an angle of $120^\circ$ between any two; see Fig.~\ref{fig7}. From the exercise above we get that $p_{\rm same}(\n_1,\n_2)=p_{\rm same}(\n_1,\n_3)=p_{\rm same}(\n_2,\n_3)=\frac{1}{4}$ since $\cos(120)=-1/2$. Therefore, 
\be\label{3coins}
p_{\rm same}(\n_1,\n_2)+p_{\rm same}(\n_1,\n_3)+p_{\rm same}(\n_2,\n_3)=\frac{3}{4}<1\;.
\ee
On the other hand, suppose it was possible to describe Alice's electron $\n_1$-spin, $\n_2$-spin, and $\n_3$-spin, with three random variables $X_1$, $X_2$, and $X_3$ (with some underlying probability distribution over the three variables). Each of the three random variables can take the values $\pm\frac{1}{2}$ determining if the spin is pointing in the positive or negative direction. Then, irrespective of the underlying probability distribution, the probabilities $\text{Pr}(X_j=X_k)$ (with $j\neq k$ and $j,k\in\{1,2,3\}$) must satisfy
\be
\text{Pr}(X_1=X_2)+\text{Pr}(X_1=X_3)+\text{Pr}(X_2=X_3)\geq 1\;.
\ee
This problem is analogous to the problem of flipping 3 coins and asking what is the probability that at least two of them are the same (either two heads or two tails). Clearly, flipping three coins will always yield two that show the same symbol (either head or tail). Eq.~\eqref{3coins} shows that this is not the case for quantum coins (i.e. spins of an electron).

So far we have seen a contradiction between quantum mechanics and local realism\index{local realism} through the violation of the CHSH inequality~\eqref{chsh}, and the inequality in~\eqref{3coins}. The next two paradoxes show that this inconsistency between quantum mechanics and local realism can be expressed without inequalities. In the literature they are referred to as ``Bell non-locality without inequalities".

\subsection{Hardy's Paradox}\index{Hardy's paradox}

Using the same notations as in Exercise~\eqref{PR}, we consider as before two electrons, one on Alice's side and the other on Bob's side. We denote  two directions in Alice's side by $\n_0$ and $\n_1$, and the two directions in Bob's side by $\m_0$ and $\m_1$. Further, we denote by $p(ab|xy)=p(ab|\n_x\m_y)$ (with $x,y\in\{0,1\}$) the probability that two SG experiments one on Alice's side and the other on Bob's side, yield an outcome $a$ for Alice's spin-measurement in the direction $\n_x$, and $b$ for Bob's spin-measurement in the direction $\m_y$. 

Recall that the probability distribution $p(ab|xy)$ is said to be local if there exists conditional probabilities $p^{A}_{\lambda}(a|x)$ and $p^{B}_{\lambda}(b|y)$ such that 
\be
p(ab|xy)=\int d\lambda \;q_\lambda \;p_{\lambda}^{A}(a|x)\;p_{\lambda}^{B}(b|y).
\ee
Suppose now that the probability distribution $p(ab|xy)$ satisfies
\be\label{zeros0}
p(00|00)=p(01|10)=p(10|01)=0\;.
\ee

\begin{exercise}
Show that if $p(ab|xy)$ is local and satisfies~\eqref{zeros0} then
$
p(00|11)=0\;.
$
\end{exercise}

We now show that the logical implication of the exercise above does not hold for quantum mechanics.
Unlike the use of the singlet in the previous examples, here we consider a bipartite state $|\psi_\theta^{AB}\ra$
that has the form:
\be
|\psi^{AB}_{\theta}\ra=\frac{\tan(\theta)}{\sqrt{1+2\tan^2(\theta)}}(|01\ra+|10\ra)-\frac{1}{\sqrt{1+2\tan^2(\theta)}}|11\ra\;,
\ee
with $\theta\in[0,2\pi]$ being some angle. Note that the state above is normalized for all $\theta$. Suppose that Alice and Bob perform the same measurements, and in particular $\n_0=\m_0=\z$ corresponds to a measurement in the computational basis,
while $\n_1=\m_1$ corresponds to a measurement in the orthonormal basis $|u_0\ra\eqdef\cos(\theta)|0\ra+\sin(\theta)|1\ra$ and $|u_1\ra\eqdef\sin(\theta)|0\ra-\cos(\theta)|1\ra$.
 
\begin{exercise}
Verify that the above choices satisfy:
\ba
&p_\psi(00|00)=|\la\psi^{AB}|0\ra|0\ra|^2=0\\
&p_\psi(01|10)=|\la\psi^{AB}|u_0\ra|1\ra|^2=0\\
&p_\psi(10|01)=|\la\psi^{AB}|1\ra|u_0\ra|^2=0
\ea
while
\be
p_{{}_{\rm Hardy}}(\theta)\eqdef p_\psi(00|11)=|\la\psi^{AB}|u_0\ra|u_0\ra|^2=\frac{\sin^4(\theta)}{1+2\tan^2(\theta)}\;.
\ee
\end{exercise}
We therefore see that for this example $p_{{}_{\rm Hardy}}(\theta)>0$ for all $0<\theta<\frac{\pi}{2}$. Interestingly, $p_{{}_{\rm Hard}}>0$ for all non-product states in $\{|\psi^{AB}_\theta\ra\}_\theta$ except for the maximally entangled state 
$|\psi_{\theta=\pi/2}^{AB}\ra=(|01\ra+|10\ra)/\sqrt{2}$ for which $p_{{}_{\rm Hard}}(\pi/2)=0$. The maximum value of 
the function $p_{{}_{\rm Hard}}(\theta)$ can easily be computed to give
\be
\max_{\theta\in[0,2\pi]}p_{{}_{\rm Hardy}}(\theta)=\frac{1}{2}(5\sqrt{5}-11)\approx0.09
\ee

\subsection{The GHZ Paradox}\index{GHZ paradox}

The Hardy paradox shows that the inconsistency of local realism\index{local realism} with quantum mechanics can be demonstrated without  inequalities as the CHSH inequality, but it is still probabilistic; i.e. provide constraints on $p_\psi(ab|xy)$.
Our final example of this inconsistency is due to Greenberger, Horne, and Zeilinger. Perhaps this is the example for which the contradiction between local realism and quantum mechanics is the sharpest.

Consider three electrons shared between Alice's, Bob's, and Charlie's labs, and prepared in the state
\be\label{ghz}
|{\rm GHZ}\ra\eqdef\frac{1}{\sqrt{2}}\left(|000\ra+|111\ra\right)
\ee
The state $|{\rm GHZ}\ra$ in written above in the $zzz$ basis; that is, $|0\ra\eqdef|\!\ua_{\z}\ra$ and $|1\ra\eqdef |\!\da_\z\ra$ are the eigenvectors of $S_\z$. We can also rewrite this vector in many other bases such as the $yyx$-basis or the $xxx$-basis.

\begin{exercise}
Show that the GHZ state as defined in~\eqref{ghz} can be expressed in the $yyx$-basis as
\be\label{yyx}
|{\rm GHZ}\ra=\frac{1}{2}\left[\left(|\ua_{\y}^{A}\ua_{\y}^{B}\ra+|\da_{\y}^{A}\da_{\y}^{B}\ra\right)\otimes|\da_{\x}^{C}\ra
+\left(|\ua_{\y}^{A}\da_{\y}^{B}\ra+|\da_{\y}^{A}\ua_{\y}^{B}\ra\right)\otimes|\ua_{\x}^{C}\ra\right]
\ee
and in the $xxx$-basis as
\be\label{xxx}
|{\rm GHZ}\ra=\frac{1}{2}\left[\left(|\ua_{\x}^{A}\ua_{\x}^{B}\ra+|\da_{\x}^{A}\da_{\x}^{B}\ra\right)\otimes|\ua_{\x}^{C}\ra
+\left(|\ua_{\x}^{A}\da_{\x}^{B}\ra+|\da_{\x}^{A}\ua_{\x}^{B}\ra\right)\otimes|\da_{\x}^{C}\ra\right]\;.
\ee
\end{exercise}

Denote by $A_\x$ (and similarly $A_\y$) the random variables that take the value $+1$ if the spin of the first electron in the $\x$-direction is positive, and take the value $-1$ if it is in the negative $\x$-direction.
The random variables $B_\x$, $B_\y$, $C_\x$, and $C_\y$, are defined similarly.

Now, according to~\eqref{xxx}, if Alice, Bob, and Charlie perform the $xxx$-measurement, the results of their measurements, given by $A_\x$, $B_\x$, and $C_\x$, must satisfy
\be
A_\x B_\x C_\x=1\;.
\ee
On the other-hand, if they chose to do the $yyx$-measurement, according to~\eqref{yyx} they would get
\be
A_\y B_\y C_\x=-1\;.
\ee
Moreover, since the GHZ state~\eqref{ghz} is invariant under any permutation of the three subsystems, we conclude also that both
\be
A_\y B_\x C_\y=-1\quad\text{and}\quad A_\x B_\y C_\y=-1\;.
\ee
With all this at hand, we get the following contradiction:
\ba
-1=(-1)(-1)(-1)&=(A_\y B_\y C_\x)(A_\y B_\x C_\y)(A_\x B_\y C_\y)\\
&=(A_\x B_\x C_\x)A_{\y}^{2}B_{\y}^{2}C_{\y}^{2}\\
&=A_\x B_\x C_\x=1\;,
\ea
where we used  $A_{\x}^{2}=B_{\y}^{2}=C_{\y}^{2}=1$ since these variable can only take the two values $\pm1$.
To summarize, according to quantum mechanics, an $xxx$-measurement can only yield one of the four possible outcomes:
\be
|\ua_{\x}^{A}\ua_{\x}^{B}\ua_{\x}^{C}\ra\;\;,\;\;|\da_{\x}^{A}\da_{\x}^{B}\ua_{\x}^{C}\ra\;\;,\;\;
|\ua_{\x}^{A}\da_{\x}^{B}\da_{\x}^{C}\ra\;\;,\;\;|\da_{\x}^{A}\ua_{\x}^{B}\da_{\x}^{C}\ra\;.
\ee
On the other hand, local realism\index{local realism} predicts that an $xxx$-measurement yields  the four possible outcomes:
\be
|\ua_{\x}^{A}\ua_{\x}^{B}\da_{\x}^{C}\ra\;\;,\;\;|\da_{\x}^{A}\da_{\x}^{B}\da_{\x}^{C}\ra\;\;,\;\;
|\ua_{\x}^{A}\da_{\x}^{B}\ua_{\x}^{C}\ra\;\;,\;\;|\da_{\x}^{A}\ua_{\x}^{B}\ua_{\x}^{C}\ra\;,
\ee
in maximal contrast with quantum mechanics. One may argue that we used quantum mechanics to express the GHZ state in the form~\eqref{yyx}, but this does not affect the conclusion that \index{local realism} cannot co-exist with the quantum mechanical formalism.

\subsection{The CHSH Game}\index{CHSH game}

Consider the following  game known as the CHSH game played by two players, Alice and Bob, along with a referee. 
The referee chooses at random (sampled from a uniform distribution) two bits $x$ and $y$, and sends $x$ to Alice and $y$ to Bob. After receiving the bits from the referee, Alice sends back to the referee the number $a$, and Bob sends back the number $b$ (see Fig.~8). 
\begin{figure}[h]\centering
    \includegraphics[width=0.3\textwidth]{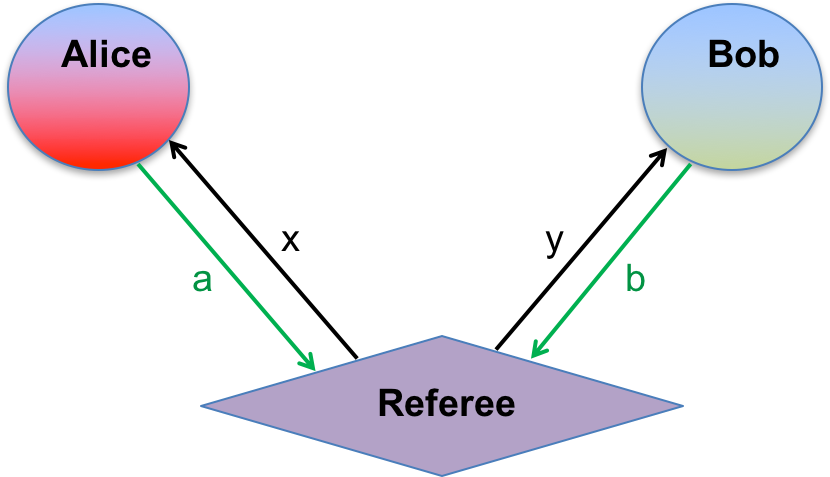}
  \caption{\linespread{1}\selectfont{\small The CHSH game.}}
  \label{fig8}
\end{figure}
The rule of the game is that Alice and Bob win the game if $a\oplus b=xy$, where $\oplus$ is addition modulus 2. The following table summarizes the desired value for $a\oplus b$ for each of the values of $x$ and $y$:
\begin{center}
  \begin{tabular}{ | l | c | r |}
    \hline
    $x$ & $y$ & $a\oplus b$\\ \hline
    0 & 0  &  0$\;\;\;$ \\ 
    0 & 1  &  0$\;\;\;$ \\ 
    1 & 0  &  0$\;\;\;$  \\ 
    1 & 1   &  1$\;\;\;$ \\     \hline
  \end{tabular}
\end{center}
Clearly, from the table above it is obvious that if Alice and Bob always choose $a=b=0$ (no matter what the values of $x$ and $y$) then they will win the game 3/4 of the times. Can they do better?

\begin{exercise}
Show that Alice and Bob cannot win more than 3/4 of the times even if they use some randomness (i.e. they share some correlated random variable).
\end{exercise}

Suppose now that Alice and Bob share quantum correlations; in particular, suppose they each posses an electron in their lab, and that the two electrons are prepared in the some bipartite state $|\psi^{AB}\ra$. With this state at hand, they use the following strategy. Based on the bits $x$ and $y$ that they receive from the referee, they choose to perform spin-measurements in the direction $\n_x$ for Alice, and in the direction $\m_y$ for Bob. They then send to the referee the outcomes of their corresponding measurements. The probability that Alice and Bob win this CHSH game\index{CHSH game} is given by
\be
p_{\rm win}\eqdef\frac{1}{4}\sum_{x,y,a,b}p_{\psi}(ab|xy)\;\delta_{xy,a\oplus b}\;,
\ee
where the factor $1/4$ represent the (uniform) probability that the referee sends $x$ to Alice and $y$ to Bob.
From the following exercise it follows that for appropriate choices of $\n_x$, $\m_y$, and $\psi^{AB}$, Alice and Bob can win the game with a probability greater than $3/4$. 

\begin{exercise}
Consider the quantum strategy described above, and denote  by $p_{\rm lose}=1-p_{\rm win}$ the probability that Alice and Bob lose the game. Recall the Bell operator $B$ as defined in Exercise~\ref{exchsh} with $\n\eqdef\n_0$, $\n'\eqdef \n_1$, $\m\eqdef\m_0$, and $\m'\eqdef\m_1$.
\begin{enumerate}
\item Show that 
\be
\la\psi|B|\psi\ra=p_{\rm win}-p_{\rm lose}\;.
\ee
\item Use Part~1 together with the Tsirelson bound to show that there exists a quantum strategy (i.e. directions $\n_x,\m_y$ and a quantum state $|\psi^{AB}\ra$) such that
\be
p_{\text{win}}=\frac{1}{2}+\frac{1}{2\sqrt{2}}>\frac{3}{4}
\ee
\end{enumerate}
\end{exercise}

\subsection{All Bell inequalities\index{Bell inequalities}}

The CHSH inequality\index{CHSH inequality} is one out of many similar inequalities known as Bell inequalities\index{Bell inequalities}. 
Any Bell inequality can be expressed in the following form
\be\label{bellnl}
\s\cdot\p\eqdef\sum_{a,b,x,y}s_{abxy}p(ab|xy)\leq c\;,
\ee
where $0<c\in\mbb{R}$ and $\p$ is the vector whose components are the conditional probabilities $p(ab|xy)$ and $\s$ is any real vector with the same dimension as $\p$. Note that for any real vector $\s$ one can take $c$ in~\eqref{bellnl} to be
\be\label{c}
c=\max_{\p\in\mL(n)}\s\cdot\p
\ee
where $\mL(n)\subset{R}^n$ is the set of all vectors $\p=\{p(ab|xy)\}$ whose components have the form (cf.~\eqref{local})
\be\label{0local0}
p(ab|xy)=\int d\lambda \;q_\lambda \;p_{\lambda}^{A}(a|x)\;p_{\lambda}^{B}(b|y)\;.
\ee 
We can therefore identify any Bell inequality with a single real vector $\s$ (since the constant $c$ is determined from above). In the general case, $x=1,\ldots,|X|$, $y=1,\ldots,|Y|$, $a=1,\ldots,|A|$, and $b=1,\ldots,|B|$, can take more than two values. We also denoted by $n\eqdef|X|\cdot|Y|\cdot|A|\cdot|B|$ the dimension of the vectors $\p$ and $\s$.
This corresponds to higher dimensional systems, and in the quantum case $a$ corresponds to the outcome of a projective von-Neumann measurement that is labeled by $x$ on Alice's subsystem, and similarly $b$ corresponds to the outcome of a projective measurement that is labeled by $y$ on Bob's subsystem. Note that the definition of local distribution as in~\eqref{0local0} remains unchanged in higher dimensions.
Therefore, there are many Bell inequalities\index{Bell inequalities}, and in recent years much effort has been made to characterize and understand the structure of all them. 

The Bell inequalities\index{Bell inequalities} that we consider here are those that can be used to test if a given distribution vector $\p$ is local (i.e. has the form~\eqref{0local0}). If a given distribution vector $\p$ violates a Bell inequality $\s$ (i.e. a Bell inequality of the form~\eqref{bellnl}) then we learn from it that $\p$ is non-local. However, if a probability distribution does not violate a particular Bell inequality, $\s$, this alone does not mean that the distribution is local. 

Given a probability vector $\p$, how can we decide if it is local (i.e. can be written in the form~\eqref{0local0})?
To answer this question, we first discuss the convexity property of local distributions.

\begin{exercise}
Denote by $\mP(n)\subset\mbb{R}^{n}$ the space of all real vectors in dimension $n=|ABXY|$ whose components are given in terms of conditional probabilities $\{p(ab|xy)\}$, and let $\mL(n)\subset\mP(n)$ be the set of all local vectors as in~\eqref{0local0}. Show that $\mL(n)$ is a convex set.
\end{exercise}

\begin{exercise}
Show that $\mP(n)$ is a polytope in $\mbb{R}^n$. Hint: Recall the definition of a polytope in Sec.~\ref{secpoly}.
\end{exercise}

Consider now a vector $\p\in\mP(n)$, and define the set $\{\p\}$ consisting of exactly one vector. As such, it is (trivially) a convex set in $\mbb{R}^{n}$. Suppose now that $\p\not\in\mL(n)$. This means that $\{\p\}\cap\mL(n)=\emptyset$, or in other words, $\{\p\}$ and $\mL(n)$ are two disjoint convex sets. Therefore, from the hyperplane separation theorem (see Theorem~\ref{hyper}) it follows that there exists a vector $\s\in\mbb{R}^{n}$ and a real number $r$ such that
\be
\s\cdot \q\leq r< \s\cdot\p\quad\quad\forall\;\q\in\mL(n)\;.
\ee
The above equation can be interpreted as follows. If $\p\not\in\mL(n)$ then there exists a Bell inequality $\s$ that it violates.
We summarize it in the following theorem.
\begin{myt}{}
\begin{theorem}
Let $\mL(n)\subset\mP(n)$ be the set of all local probability vectors as in~\eqref{0local0} with fixed cardinalities $|X|$, $|Y|$, $|A|$, $|B|$, and $n=|ABXY|$. Then, $\p\not\in\mL(n)$ if and only if it violates at least one Bell inequality.
\end{theorem}
\end{myt}

The theorem above implies that in order to determine if a probability distribution is local, one has to check that it doesn't violate all Bell inequalities\index{Bell inequalities} as in~\eqref{bellnl}. This, at first, may seem as an impossible task as one will have to check infinite number of inequalities corresponding to each vector $\s\in\mbb{R}^n$. However, as we know from Sec.~\ref{secconvex} (convex analysis) this is not necessary. First, note that there are many cases that are redundant (e.g, if we checked the Bell inequality for $\s$ there is no need to check it for $2\s$). More importantly, it follows that $\mL(n)$ in addition to being convex, is in fact a \emph{polytope}.

\begin{exercise}
Show that $\mL(n)$ is a polytope in $\mbb{R}^n$ (i.e. a convex hull of a finite number of points). Hint: Show first that the set of vectors $\p^A$, whose components are any conditional probabilities $\{p(a|x)\}$, is itself a polytope (i.e. find its extreme points and show that there are a finite number of them).
\end{exercise}

From Theorem~\ref{mink},  the polytope $\mL(n)$ can be represented as an intersection of finitely
many half-spaces. Denoting by $\s^{(j)}$ (with $j=1,\ldots,m$) the normal vectors to these half spaces, we therefore conclude that $\p\in\mL(n)$ if and only if
\be
\s^{(j)}\cdot\p\leq c_j\quad\forall j=1,\ldots,m
\ee
where $c_j\eqdef\max_{\q\in\mL(n)}\s^{(j)}\cdot \q$. In other words, there exists finitely many Bell inequalities\index{Bell inequalities} that can determine if a vector $\p\in\mL(n)$.

This analysis may give the impression that deciding if $\p$ is local is easy. Therefore, it is important to note first  that the computation of $\s^{(j)}$ may be hard, and that the number $m$ may grow exponentially with the cardinalities $|X|$,$|Y|$, $|A|$, and $|B|$. In particular, already for the case that $|A|=|B|=2$ with arbitrary large $|X|=|Y|$, it was shown that the decision problem of whether $\p$ is local is NP-complete~\cite{AIIS2004}. On the other hand, the simplest case in which $|A|=|B|=|X|=|Y|=2$ was fully characterized in~\cite{Froissart1981}, and independently by~\cite{Fine1982}, and, in particular, it was shown that the only non-trivial Bell inequality is the CHSH inequality. That is, for bits $x,y,a,b\in\{0,1\}$, the 16-dimensional vector $\p=(p(ab|xy))$ is local if and only if it does not violate any of the CHSH inequalities.

\section{Unitary Evolution and the Schr\"{o}dinger Equation}\index{unitary evolution}\index{Schr\"odinger equation}

The last postulate of quantum mechanics  is about time evolution of physical systems. It states that the state of a  closed physical system evolves unitarily in time. That is, if $|\psi(t)\ra$ is the state describing the system at time $t$, then there exists a parameterized family of unitary matrices $U(t)$ with parameter $t\in\mbb{R}$ such that
\be\label{0unitary0}
|\psi(t)\ra=U(t)|\psi(0)\ra\;.
\ee
We emphasize here that $U(t)$ (with $t>0$) does not depend on the initial state (i.e. on the preparation of the system at time $t=0$).
It is important also to note that the formalism of quantum mechanics does not propose which unitary family $U(t)$ one should choose to describe a particular evolution of a quantum system. It just states the evolution (whatever the specific causes for it) is described with a unitary matrix. We saw earlier something similar about quantum states of the spin of an electron. The first postulate of quantum mechanics did not tell us which states to assign to a specific system. It only stated that all the information about the system is encoded in a quantum state. We then used, as in the example of the spin of an electron, further symmetry properties, to assign the physical interpretation of any qubit state like $|0\ra,\;|1\ra,\;|+\ra\;|-i\ra$.

One can view the unitary evolution postulate as a principle of \emph{distinguishability preserving}. Recall from Exercise~\ref{distinguish} that if two quantum states are orthogonal then they can be perfectly distinguished by a suitable projective measurement.  The principle of distinguishability preserving asserts that if a closed system is prepared in one out of two or more distinguishable states, then the ability to distinguish between them remains intact throughout the evolution, unless some type of external noise is pumped into the system. Therefore, one can view a unitary evolution as a distinguishability preserving map. Alternatively, since information quantifies the ability to distinguish between one thing from another, the unitary evolution postulate of quantum mechanics, loosely speaking, is the statement that closed systems don't loose information (i.e. the ability to distinguish) if they don't interact with the external world.

We now discuss the form of the parametrized family of the unitaries $U(t)$ given in~\eqref{0unitary0}.
We will assume here that the function $t\mapsto U(t)$ is continuous, and even differentiable. 
Moreover, $U(0)=I$ is the identity matrix so that we can express for a small $t=\eps>0$
\be
U(\eps)=I-iH\eps+O(\eps^2)
\ee
where $H$ is some Hermitian matrix. Note that $H$ must be Hermitian since otherwise $U(\eps)$ will not be a unitary matrix; i.e.
\be
U^*(\eps)U(\eps)=(I+iH\eps)(I-iH\eps)+O(\eps^2)=I+O(\eps^2)
\ee
where we assumed that $H$ is Hermitian. Therefore, taking the derivative on both sides of~\eqref{0unitary0} and setting $t=0$ gives:
\be
\frac{d}{dt}|\psi(t)\ra\Big|_{t=0}=-iH|\psi(0)\ra\;.
\ee
Now, since the system is isolated, the state $|\psi(t)\ra$ must evolve according to the same rule as the state $|\psi(0)\ra$. Hence, this homogeneity assumption implies that for all $t>0$
\be\label{Sch}
\frac{d}{dt}|\psi(t)\ra=-iH|\psi(t)\ra\;.
\ee

\begin{exercise}
Show that from the equation above it follows that:
 \be
U(t)=e^{-iHt}\;.
\ee
\end{exercise}

As we discussed before, in quantum mechanics, any Hermitian operator corresponds to an observable. The observable $H$ above is known as the Hamiltonian of the system and it corresponds to the energy of the system.
There are many books in physics from which you can learn how to construct the Hamiltonian $H$ for specific physical systems, but generally speaking, quantum mechanics itself does not provide the prescription on how to construct the Hamiltonian of a specific physical system. Hamiltonians are also constructed in classical physics.

The Hamiltonian has the units of energy.  Therefore, when incorporating the physical dimensions, Eq.~\eqref{Sch} takes the form of the celebrated Schr\"{o}dinger equation,
\begin{mye}{The Schr\"odinger Equation}\index{Schr\"odinger equation}
\be
i\hbar\frac{d}{dt}|\psi(t)\ra=H|\psi(t)\ra\;,
\ee
\end{mye}
\noindent where the Plank's constant $\hbar=h/2\pi$ has the units of energy$\times$time so that both sides of the equation have the same dimensions.
Since the Hamiltonian $H$ is an Hermitian matrix it can be diagonalized as $H=\sum_{x}E_x|\varphi_x\lr \varphi_x|$, where $\{E_x\}$ are the energy levels of the system, and $\{|\varphi_x\ra\}$ are the corresponding eigenstates. The eigenstate $|\varphi_x\ra$ that corresponds to the lowest energy level is called the \emph{ground state} of the system.  
 
 Finally, we assumed above that the system is closed, i.e. does not interact with the environment in any way. 
 This led us to assume a continuous uniform evolution. However, many physical systems are not closed, and even us, the experimenters, can change the Hamiltonian by changing parameters in the lab at different times. We leave this discussion to the next chapter that covers evolution of open systems.

\subsection{The Measurement Problem}\index{measurement problem}

Quantum mechanics allows for two types of evolutions for isolated systems. One is a probabilistic evolution, in which quantum measurements such as the projective von-Neumann measurement, transform a state $|\psi\ra\in A$
to another post-measurement state $|\psi_x\ra$ with some probability $p_x$. The other is a deterministic evolution, in which a quantum state $|\psi\ra$ evolves unitarily and deterministically to another state $U|\psi\ra$, where the unitary matrix $U$ is determined from the Hamiltonian of the system. It is therefore natural to ask if both evolutions can co-exist or if they lead to inconsistencies within quantum theory.

We already learned that a physical system is not really isolated if it is being measured, since the measurement apparatus can be viewed as external system that  interacts with the system. In fact, we saw that the measurement can change the state of the system. 
Therefore, at first glance it seems that there is no contradiction  between quantum measurements and the assertion that closed systems evolve unitarily. However, we can consider both the system and its measuring device as a single composite system.

Any measuring device (including the device itself and us, the experimenters) consists of numerous atoms and molecules. This means that practically it is impossible to write down its Hamiltonian as one will need to include the contributions from all the $~10^{23}$ (even more) particles constituting the device. Yet, according to the rules of quantum mechanics, there exists a Hamiltonian, $H^{AE}$, associated with the measuring device (environment system E) + the quantum system (system A) that is being measured.
Then, according  to Schr\"{o}dinger equation, since the system  + environment form a closed composite system, they must evolve unitarily according to the joint unitary\index{joint unitary} matrix $U^{AE}=e^{-iH^{AE}t}$.
On the other hand, according to Born's rule\index{Born's rule} they must evolve probabilistically. Which evolution will occur, the deterministic one or the probabilistic one?

Let's consider the SG experiment for measuring the spin in the $z$-direction of a single electron. In this case, the quantum system $A$ is described with a unit vector $|\psi^A\ra=a|0\ra+b|1\ra$ in $A\cong \mbb{C}^2$, where $|a|^2+|b|^2=1$. Denote by $E$ the Hilbert space associated with the measuring device (plus the experimenters and the rest of the universe for that matter). 
If the measurement apparatus is treated externally, according to Born's rule\index{Born's rule} one obtains the outcome $|0\ra$ with probability $|a|^2$ and the outcome $|1\ra$ with probability $|b|^2$.

If the measurement apparatus is treated internally, then we can assume that prior to the experiment, the measuring device was given in some ``ready" state. That is, according to the first postulate of quantum mechanics there exists a vector $|\text{ready}\ra\in E$ containing all the information about the measuring device prior to the measurement. Now, suppose first that the state of the system was $|0\ra^A$. Then, the initial state of the system+device is $|0\ra^A|\text{ready}\ra^E$. After the measurement, the joint system evolves to
\be\label{con1}
|0\ra^A|\text{ready}\ra^E\to U^{AE}|0\ra^A|\text{ready}\ra^E=|0\ra^A|\text{output ``0"}\ra^E
\ee
where the equality follows from the fact that the measurement is performed in the $z$-direction, so the system state $|0\ra^A$ must remain intact, while the vector $|\text{ready}\ra^E$ of the measuring device is transformed to another vector  $|\text{output ``0"}\ra^E$ in $E$, containing the information that the output was $0$. 
Similarly, if the initial state of the system was $|1\ra^A$, then the initial state of the system+device is $|1\ra^A|\text{ready}\ra^E$, and after the measurement, the joint system would evolve to
\be\label{con2}
|1\ra^A|\text{ready}\ra^E\to U^{AE}|1\ra^A|\text{ready}\ra^E=|1\ra^A|\text{output ``1"}\ra^E\;.
\ee

Now, lets consider the case in which the initial state of the system is $|\psi\ra^A=a|0\ra+b|1\ra$. In this case, 
as before, the initial state of the system+device is $|\psi\ra^A|\text{ready}\ra^E$. However, after the measurement,
the system evolves unitarily to the state
\ba
|\psi\ra^A|\text{ready}\ra^E\to U^{AE}\left[|\psi\ra^A|\text{ready}\ra^E\right]
&=U^{AE}\left[(a|0\ra^A+b|1\ra^A)|\text{ready}\ra^E\right]\\
&=aU^{AE}\left[|0\ra^A|\text{ready}\ra^E\right]+bU^{AE}\left[|1\ra^A|\text{ready}\ra^E\right]\\
&=a|0\ra^A|\text{output ``0"}\ra^E+b|1\ra^A|\text{output ``1"}\ra^E\;,
\ea
where in the last equality we used~\eqref{con1} and~\eqref{con2}. The above state is an entangled state between the system and the measuring device representing quantum correlations between the two. 

We therefore see a sharp contrast between the two types of evolution of a quantum state. 
Although this problem haunt quantum mechanics right from its early formulation at the beginning of the 20th century, there is a controversy on how to resolve this problem. This is also related to the different interpretations of quantum mechanics. For example, the Everett ``many worlds" interpretation adopts the unitary evolution whereas others adopt the probabilistic nature of it. This is a fascinating topic, but it goes far beyond the scope of this book.

\subsection{The No-Cloning Theorem}\index{no-cloning}

\begin{figure}[h]\centering
    \includegraphics[width=0.6\textwidth]{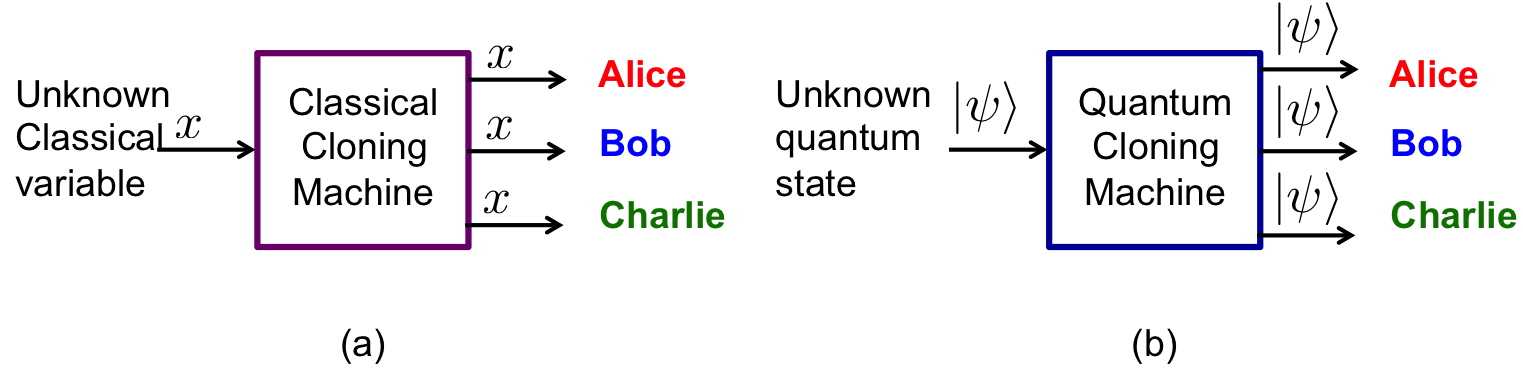}
  \caption{\linespread{1}\selectfont{\small Classical and quantum cloning machines.}}
  \label{fig11}
\end{figure}

One of the very useful properties of classical information is that it can be cloned; that is, it can be copied and for example broadcast to several other parties (see Fig.~\ref{fig11}a). We now explore if quantum information also has this property. Quantum information is encoded in quantum states, so cloning of quantum information corresponds to copying of an unknown quantum state $|\psi\ra\in A$ (see Fig.~\ref{fig11}b).

Suppose there exists a quantum machine that takes an unknown state $|\psi\ra\in A$ and output two copies of it 
$|\psi\ra|\psi\ra$. Since it maps any normalized vector to a normalized vector, it can be modelled by an isometry
$V:A\to A\otimes A$ with the property that $V^{*}V=I^A$. Consider now two arbitrary states $|\psi\ra,|\phi\ra\in A$. Then, from our assumption
\be
V|\psi\ra=|\psi\ra|\psi\ra\quad\text{and}\quad V|\phi\ra=|\phi\ra|\phi\ra\;.
\ee
Therefore,  
\be
\la\psi|\phi\ra=\la\psi|V^{*}V|\phi\ra=\big(\la\psi|\phi\ra\big)^2
\ee
But any complex number that satisfies $c=c^2$ must be equal to 0 or 1. Hence, $\la\psi|\phi\ra\in\{0,1\}$ which means that either $|\psi\ra=|\phi\ra$ or that $|\psi\ra$ is orthogonal to $|\phi\ra$. Therefore, there is no quantum machine that is capable of generating two copies of an arbitrary unknown quantum state. This result is known by the term \emph{the no-cloning theorem}.

\subsection{The Controlled Unitary Operation}\index{controlled unitary}\index{CNOT gate}

The control unitary operation is a particular quantum evolution that is a key building block in quantum circuits.
It is used quite often in quantum computing, and in particular, the controlled not (CNOT) gate, which is a key component in the construction of a quantum computer, is a special type of a controlled unitary map.
In its most general form, a unitary map $U:{AB}\to {AB}$ (recall that ${AB}=A\otimes B$) is a \emph{controlled} unitary map (or gate) if $U^{AB}$ can be written as
\be
U^{AB}=\sum_{x\in[m]}|x\lr x|^A\otimes U_{x}^{B}
\ee
where $\{|x\ra^A\}_{x\in[m]}$ is some orthonormal basis of $A$, and $\{U_x^B\}$ is a collection of $|A|$ unitary matrices in $B$. Note that $U^{AB}\left(|x\ra^A|\psi^B\ra\right)=|x\ra^A\otimes \left(U_x^B|\psi^B\ra\right)$, so that by choosing the input $|x\ra^A$, Alice controls the unitary that is acted on $|\psi\ra^B$.
\begin{exercise}
Verify that $U^{AB}$ in the equation above is indeed a unitary matrix.
\end{exercise}

\begin{figure}[h]\centering
    \includegraphics[width=0.4\textwidth]{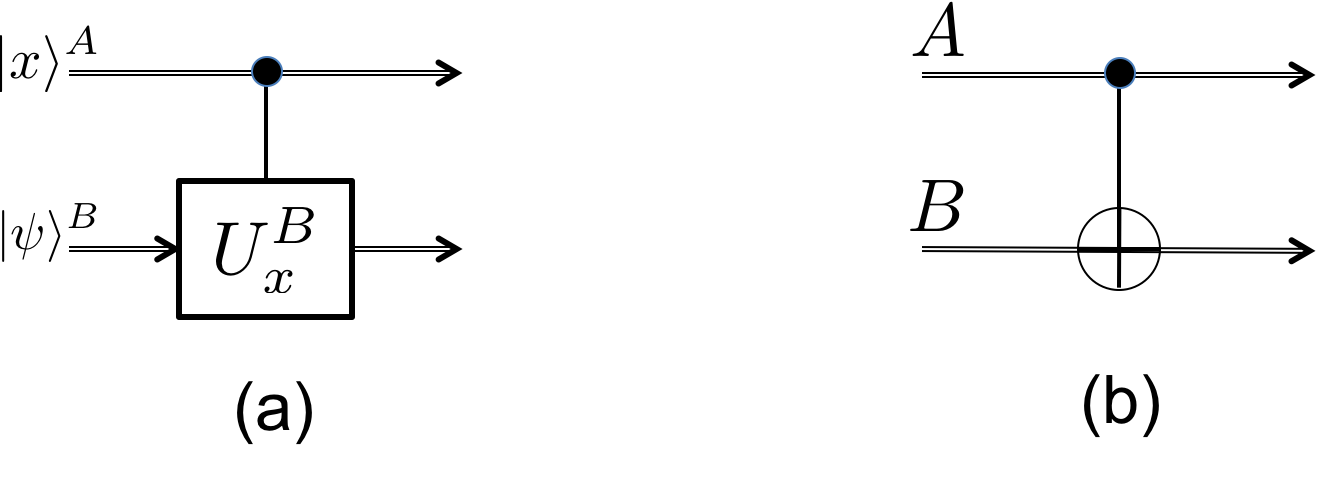}
  \caption{\linespread{1}\selectfont{\small (a) Controlled unitary gate. (b) Controlled NOT (CNOT) gate.}}
  \label{cunitary}
\end{figure}

In quantum circuits, the controlled unitary\index{controlled unitary} is depicted as in Fig.~\ref{cunitary}a. The CNOT gate\index{CNOT gate} is the controlled unitary
map
\be
U^{AB}=|0\lr 0|\otimes I+|1\lr 1|\otimes \sigma_1
\ee
where $\sigma_1$ is the first Pauli (unitary) matrix. The CNOT gate\index{CNOT gate} is depicted in Fig.~\ref{cunitary}b.
\begin{exercise}
Show that the CNOT gate can generate the maximally entangled state $(|00\ra+|11\ra)/\sqrt{2}$ from a tensor product of two vectors (i.e. from a product state of the form $|\psi\ra^A|\phi\ra^B$).
\end{exercise}
\begin{figure}[h]\centering    \includegraphics[width=0.4\textwidth]{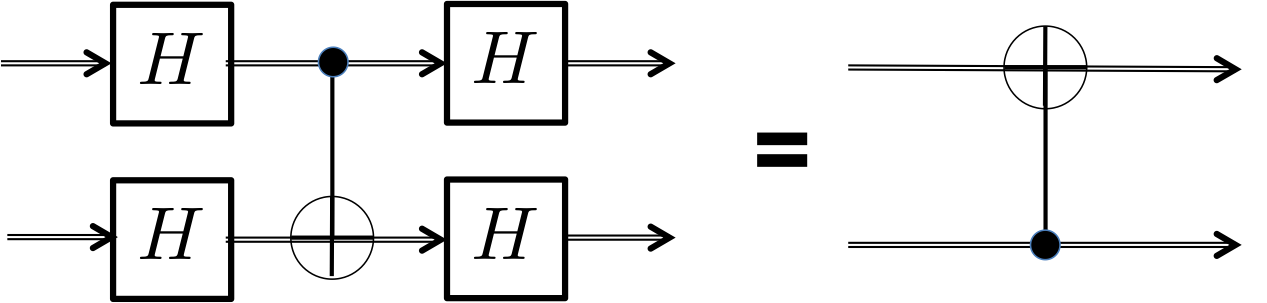}
  \caption{\linespread{1}\selectfont{\small CNOT gate\index{CNOT gate} in Hadamard basis}}
  \label{cunitary2}
\end{figure}

\begin{exercise}
Show the equivalence of the two circuits in Fig.~\ref{cunitary2}, where 
\be
H\eqdef\frac{1}{\sqrt{2}}\begin{bmatrix}
1 & 1\\
1 & -1
\end{bmatrix}
\ee
is the Hadamard unitary matrix.
\end{exercise}

\section{Notes and References}

Many books on quantum mechanics contains much of the material presented here.
More details on the Stern-Gerlach experiment can be found for example in traditional books on quantum mechanics such as~\cite{Messiah1967} and~\cite{Sakurai1994}. 

For topics on inner product spaces in finite dimensions that include many of the concepts used in this book we refer to~\cite{Bhatia1997,HJ2012}.
The treatment of linear algebra with Dirac notations can be found in many text books on quantum physics and quantum information including, for example,~\cite{NC2000,Wilde2013,Watrous2018}.
Each of these books also provide a review on quantum mechanics.

The example of the three quantum coins was taken from~\cite{Preskill2015}, and the Hardy paradox can be found in~\cite{Hardy1992}. More details on Bell nonlocality and many related references can be found in the review article~\cite{BCP+2014}. Readers interested to learn more about the measurement problem and the different interpretations of quantum mechanics may find the review article~\cite{Schlosshauer2005} as useful starting point. 

Much more details on the no-cloning theorem can be found in~\cite{SIGA2005}.

\chapter{Elements of Quantum Mechanics II: Open Systems}\index{open systems}

 Open physical systems are systems that have interactions with other external systems.
 These external systems, which we will refer to as `the environment', can either be correlated with the system, 
 and/or exchange information, energy, or matter, with it. Consequently, the description and evolution of such systems can be very different than those we discussed for isolated systems. 
 Yet, there is no need to introduce new postulates in order to develop the theory of open quantum systems. 
 Instead, we will see that \emph{all} the postulates of quantum mechanics on isolated systems are sufficient to determine the evolution, the measurements, and the description of open systems.

\section{Generalized Measurements}\label{sec:gm}\index{generalized measurement}

In the previous chapter we saw that isolated physical systems can undergo two types of evolutions: 
the unitary (Schr\"{o}dinger) evolution and the probabilistic measurement evolution. The combination of these two evolutions yield another type of evolution.
Explicitly, let $|\psi\ra\in A$ be a pure state of an isolated physical system, $U$ be a unitary operator, and $\{P_x\}_{x\in[m]}$ be a projective measurement. Applying the projective measurement after the state $|\psi\ra$ evolved to the state $U|\psi\ra$ yields the state
$
|\psi_x\ra\eqdef\frac{1}{\sqrt{p_x}}P_xU|\psi\ra
$
with probability $p_x\eqdef\la\psi| U^*P_xU|\psi\ra$. Denoting by $M_x\eqdef P_xU$ we get 
\be\label{stat}
|\psi_x\ra=\frac{1}{\sqrt{p_x}}M_x|\psi\ra\quad\text{and}\quad p_x=\la\psi| M_{x}^*M_x|\psi\ra\;.
\ee
Therefore, the combination of a unitary evolution followed by a projective measurement can be modelled by a collection of complex matrices $\{M_x=P_xU\}_{x\in[m]}$. Note that $M_x$ are not projections, although they have a very special form given by $P_xU$.
If additional ancillary systems are available (i.e. the system is not closed), then the combination of a unitary evolution on both the system and ancilla, followed by a projective (or basis) measurement  yields an even more general type of evolution known by the name \emph{generalized measurement} as it generalizes the von-Neumann projective measurement. 

\begin{figure}[h]\centering
    \includegraphics[width=0.6\textwidth]{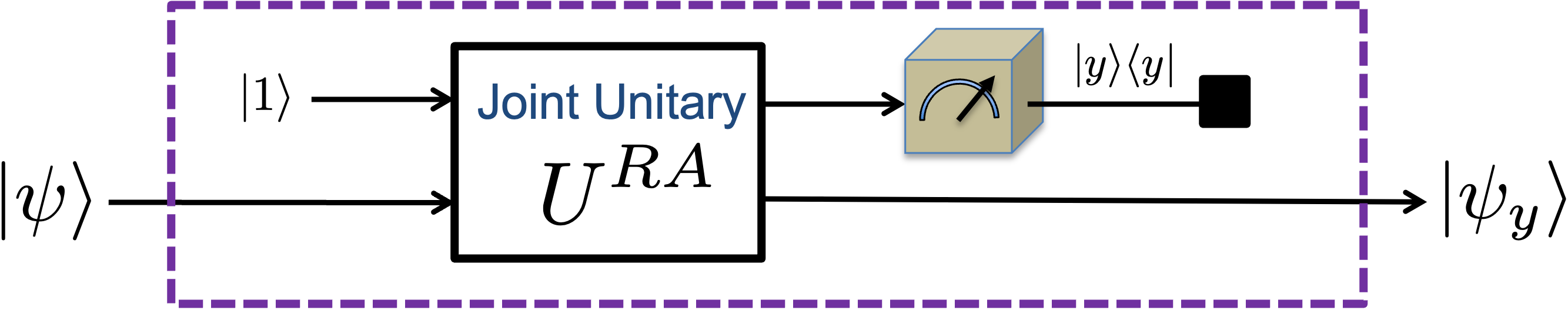}
  \caption{\linespread{1}\selectfont{\small Realization of a generalized measurement}}
  \label{fig9}
\end{figure}

In Fig.~\ref{fig9} we describe the following evolution of a quantum state $|\psi\ra\in A$. In the first step of the evolution, an ancillary system is introduced which is prepared in some state $|1\ra\in R$. Consequently, the state of the joint system is $|1\ra^R|\psi^A\ra$. Next, a joint unitary\index{joint unitary} evolution, $U^{RA}$ is applied to the joint state $|1\ra|\psi\ra$ yielding the bipartite state $ U^{RA}|1\ra|\psi\ra$. Finally, a basis measurement $\{|y\lr y|^R\}_{y\in[n]}$ is applied on the reference system $R$.\index{joint unitary}

We discuss now how the output state $|\psi_y\ra$ is related to the input state $|\psi\ra$, and what is the probability to obtain an outcome $y$. Denote by $m\eqdef |A|$ and $n\eqdef|R|$. As an operator in the vector space $\ml(R\otimes A)$, the unitary matrix $U^{RA}$ can be express as 
\be\label{3232}
U^{RA}=\sum_{y,y'\in[n]}|y\lr y'|\otimes\Lambda_{yy'}\quad\text{where}\quad \Lambda_{yy'}\in\ml(A)\;.
\ee
Note that any operator in $\ml(R\otimes A)$ has the above form, but
since $U^{RA}$ is unitary we have the equivalence
\be\label{unitar}
U^*U=I^{RA}\quad\iff\quad\sum_{y\in[n]}\Lambda_{yz}^{*}\Lambda_{yx}=\delta_{xz}I^A\quad\quad\forall\;x,z\in[n]\;.
\ee
Now, observe that
\be
\left(|y\lr y|^R\otimes I^A\right)U^{RA}|1\ra^R|\psi^A\ra= |y\ra^R\big(\Lambda_{y1}|\psi^A\ra\big)\;.
\ee
Therefore, denoting by $M_x\eqdef\Lambda_{x1}$, we get
\be
|\psi_y\ra=\frac{1}{\sqrt{p_y}}M_{y}|\psi\ra\quad\text{with}\quad p_y\eqdef \la\psi|M_{y}^{*}M_{y}|\psi\ra\;.
\ee
Moreover, from~\eqref{unitar} it follows that $\sum_{y\in[n]}M^{*}_{y}M_{y}=I^A$, so that $\sum_{y\in[n]}p_y=1$, where $p_y$ is the probability to obtain an outcome $y$. Note that the post-measurement state $|\psi_y\ra$ with its associated probability $p_y$, has a very similar form to the form of $|\psi_x\ra$ and $p_x$ in~\eqref{stat}. However,  unlike the form $P_xU$ of $M_x$ in~\eqref{stat},
the only condition on $\{M_{y}\eqdef\Lambda_{y1}\}$ is that they 
can be extended to a family of matrices $\{\Lambda_{yy'}\}$
that satisfies~\eqref{unitar}. This will ensure that $U^{AB}$ is unitary.
We now show that any set of complex matrices $\{M_{y}\}_{y\in[n]}$ with the property 
$\sum_{y\in[n]}M^{*}_{y}M_{y}=I^A$ can be completed to a full family of matrices $\{\Lambda_{yy'}\}$ that satisfies~\eqref{unitar}. 
To see this, observe that the matrix $U^{RA}$ can be expressed in the following block form
\be\label{uab}
U^{RA}=\begin{bmatrix}
\Lambda_{11} &\Lambda_{12} &\cdots &\Lambda_{1n}\\
\Lambda_{21} &\Lambda_{22} &\cdots &\Lambda_{2n}\\
\vdots &\vdots &\ddots &\vdots\\
\Lambda_{n1} &\Lambda_{n2} &\cdots &\Lambda_{nn}
\end{bmatrix}
\ee
with the matrices $\{M_y\eqdef\Lambda_{y1}\}_{y\in[n]}$ appearing in the first block column. Moreover, this first column satisfies
\be
\begin{bmatrix}
M_{1}^{*} & M_{2}^{*} & \cdots & M_{n}^*
\end{bmatrix}
\begin{bmatrix}
M_{1} \\
M_{2}\\
\vdots \\
M_{n}
\end{bmatrix}
=\sum_{y\in[n]}M^{*}_{y}M_{y}=I^A\;.
\ee
Therefore, the first column block in~\eqref{uab} consists of $m\eqdef|A|$ orthonormal vectors. Any such set of $m$ orthonormal vectors in $\mbb{C}^{mn}$ can be completed to a full orthonormal basis of  $\mbb{C}^{mn}$(for example, by the Gram-Schmidt process). Therefore, it is always possible to construct a unitary matrix $U^{RA}$ as above, from a set of matrices $\{\Lambda_{y1}\}_{y\in[n]}$ that satisfy $\sum_{y\in[n]}\Lambda^{*}_{y1}\Lambda_{y1}=I^A$.

\begin{myd}{Generalized Measurement}\index{generalized measurement}
\begin{definition}
A generalized measurement is a collection of $m\in\mbb{N}$ complex matrices $\{M_{x}\}_{x\in[m]}\subset\ml(A)$ with the property that
\be\label{nor}
\sum_{x\in[m]}M_{x}^{*}M_x=I^A\;.
\ee 
\end{definition}
\end{myd}

When a generalized measurement is applied to a physical system, it transforms the state of the system, $|\psi\ra$, to the post-measurement state 
\be\label{stat}
|\psi_x\ra\eqdef \frac{1}{\sqrt{p_x}}M_x|\psi\ra\quad\text{with probability}\quad p_x\eqdef\la\psi| M_{x}^*M_x|\psi\ra\;.
\ee
We showed above that a generalized measurement can always be realized as in Fig.~\ref{fig9}. Further, both projective measurements, and  the measurements described in~\eqref{stat} with $M_x=P_xU$, are special types of generalized measurements. 

How do we know that the generalized measurement described above is indeed the most general one? Can we construct another circuit like the one in Fig.~\ref{fig9} which would yield perhaps a more general measurement? Note that the generalized measurement described in Fig.~\ref{fig9} make use of the two types of evolutions in quantum mechanics: a unitary evolution followed by a projective measurement. Since these are the \emph{only} two types of evolution in quantum mechanics, any evolution can be decomposed into  a sequence of these two types of processes. Since both unitary evolution and projective measurements are themselves generalized measurement, we conclude that the most general measurement on a quantum system can be described as a sequence of generalized measurements. In the following exercise it is argued that any such sequence of generalized measurements can be simulated by a single generalized measurement. Hence, the generalized measurement described above 
is indeed general enough to describe the most general measurement in quantum mechanics.

\begin{exercise}
Show that if $\{M_x\}_{x\in[m]}$ and $\{N_y\}_{y\in[n]}$ are two generalized measurements then $\{M_xN_y\}$ is also a generalized measurement. Use this to show that a sequence of generalized measurements can be simulated by a single generalized measurement.
\end{exercise}

\begin{exercise}
Show that the matrices (operators) $M_x$ do not have to be square. That is, show that any collection of $m$ operators $\{M_{x}\}_{x\in[m]}\subset\ml(A,B)$ that satisfy~\eqref{nor}  can also be realized as a generalized measurement as depicted in Fig.~\ref{fig9}. Hint: Consider a unitary operator $U:RA\to R'B$ where the reference systems $R$ and $R'$ are such that $|RA|=|R'B|$.
\end{exercise}

\begin{exercise}
Let $A=\mbb{C}^2$, and let $M_0=a|+\lr 0|$ and $M_1=b|0\lr +|$ be two operators in $\ml(A)$ with $a,b\in\mbb{C}$. Find the precise conditions on $a$ and $b$ for the existence of a third operator $M_2\in \ml(A)$ such that $\{M_0,M_1,M_2\}$ form a generalized measurement.
\end{exercise}

\begin{exercise}
Consider $d$ (rank-one) operators $\{M_x=|\psi_x\lr\phi_x|\}_{x\in[d]}$ in $\ml(\mbb{C}^d)$, where $\{|\psi_x\ra\}_{x\in[d]}$ and $\{|\phi_x\ra\}_{x\in[d]}$ are some normalized states in $\mbb{C}^{d}$. Show that $\{M_x\}_{x\in[d]}$ is a generalized measurement if and only if $\{|\phi_x\ra\}_{x\in[d]}$ is an orthonormal basis of $\mbb{C}^d$.
\end{exercise}

\section{The Mixed Quantum State}\index{mixed state}

The first postulate of quantum mechanics states that the information about closed physical systems is encoded in pure quantum states. Here we show how this postulate implies that for open quantum systems, mixed quantum states encode all the information that can be extracted from the system. We derive this conclusion in two different ways,  one by considering a system that is correlated \emph{classically} to another ancillary system, and the other by considering  \emph{quantum} correlations with the ancillary system. 

\subsection{The Emergence of Density Operators from Classical Correlations}\index{density matrix}

So far we considered isolated systems that we described with a pure state $|\psi\ra\in A$, or more precisely, with the rank one matrix $|\psi\lr \psi|\in\md(A)$. Suppose now, that in addition to thequantum system, Alice also has access to classical systems, like coins or dice, that can generate random numbers. In this case, Alice can roll a dice with $m$ possible outcomes, and based on the outcome, prepare one of the $m$ states $\{|\psi_x\lr\psi_x|\}_{x\in[m]}\subset\md(A)$. This way, Alice can prepare the state $|\psi_x\ra$ with some probability $p_x$ (the classical systems, i.e. the coins or dice, do not have to be unbiased). Now, suppose that Alice forgot the value of $x$. 
Then, Alice knows that her state is one out of the $m$ states in the \emph{ensemble} of states $\{|\psi_x\lr\psi_x|,\;p_x\}_{x\in[m]}$. How should we characterize the ensemble $\{|\psi_x\lr\psi_x|,\;p_x\}_{x\in[m]}$? We will see that there exists many other ensemble of states that contains the exact same information as the ensemble $\{|\psi_x\lr\psi_x|,\;p_x\}_{x\in[m]}$. Therefore, instead of characterizing the information with a particular ensemble (such as $\{|\psi_x\lr\psi_x|,\;p_x\}_{x\in[m]}$), we will characterize it with a mathematical object that remains invariant under exchanges of such equivalent ensembles.

To gather information about her system, Alice can execute a generalized measurement, denoted as $\{M_y\}_{y\in[n]}$, on her system, characterized by the ensemble $\{|\psi_x\rangle\langle\psi_x|,\;p_x\}_{x\in[m]}$. This measurement results in an outcome $y$ with a corresponding probability denoted as $q_y$. Furthermore, following the occurrence of outcome $y$, there emerges a post-measurement ensemble that describes the state of Alice's system. We will now delve into these details to demonstrate that the dependencies of these quantities rely solely on a density matrix associated with the ensemble $\{|\psi_x\rangle\langle\psi_x|,\;p_x\}_{x\in[m]}$.

If the pre-measurement state was $|\psi_x\ra$ then the post-measurement state after outcome $y$ occurred is
\be\label{defpxy}
|\phi_{xy}\ra\eqdef\frac{1}{\sqrt{p_{y|x}}}M_y|\psi_x\ra
\ee
with probability
\be 
p_{y|x}\eqdef\la\psi_x| M_{y}^*M_y|\psi_x\ra=\tr\left[M_{y}^*M_y|\psi_x\lr\psi_x|\right]\;.
\ee 
However, since Alice does not know the value of $x$, if she performs a measurement  
$\{M_y\}_{y\in[n]}$ on her system, she will get the outcome $y$ with probability
\be
q_y\eqdef\sum_{x\in[m]}p_{y|x}p_x=\sum_{x\in[m]}p_x\tr\left[M_{y}^*M_y|\psi_x\lr\psi_x|\right]
=\tr\left[M_{y}^*M_y\rho\right]
\ee
where 
\be
\rho\eqdef\sum_{x\in[m]}p_x|\psi_x\lr\psi_x|\;,
\ee 
is the density matrix associated with the ensemble $\{|\psi_x\lr\psi_x|,\;p_x\}_{x\in[m]}$.

Note that $p_{y|x}p_x$ is the probability that both the pre-measurement state is $|\psi_x\ra$ \emph{and} that the outcome of its measurement is $y$. Therefore, using the Bayesian rule of probabilities, the probability that the pre-measurement state is $|\psi_x\ra$ given that the measurement outcome is $y$, can be expressed as
$
q_{x|y}\eqdef p_{y|x}p_x/q_y
$.
Consequently, after outcome $y$ occurred the ensemble $\{p_x,|\psi_x\lr\psi_x|\}$ changes to
\be
\{q_{x|y},|\phi_{xy}\rl\phi_{xy}|\}_{x\in[m]}\;.
\ee 
Note that the density operator, $\sigma_y$, that is associated with the above ensemble is given by
\ba
\sigma_y&\eqdef\sum_{x\in[m]}q_{x|y}|\phi_{xy}\lr\phi_{xy}|\\
\Gg{q_{x|y}\eqdef p_{y|x}p_x/q_y}&=\frac{1}{q_y}\sum_{x\in[m]}p_xp_{y|x}|\phi_{xy}\lr\phi_{xy}|\\
\GG{\eqref{defpxy}}&=\frac{1}{q_y}\sum_{x\in[m]}p_xM_{y}|\psi_{x}\lr\psi_{x}|M_{y}^{*}\\
&=\frac{1}{q_y}M_{y}\rho M_{y}^{*}\;.
\ea

To summarize, the outcome $y$, of any generalized measurement $\{M_y\}_{y\in[n]}$, occurs with probability $q_y=\tr\left[M_{y}^*M_y\rho\right]$, when applied to an ensemble 
$\{p_x,|\psi_x\lr\psi_x|\}_{x\in[m]}$.  Recall from Exercise~\ref{ensembles} that aside from the ensemble $\{p_x,|\psi_x\lr\psi_x|\}_{x\in[m]}$, there are infinitely many other ensembles that also correspond to the same density operator $\rho$. Therefore, the dependance of $q_y$ only on $\rho$ demonstrates that the statistics of any measurement outcome depends only on the density operator, and not on the particular ensemble that realizes it. To clarify, suppose $\{p_x,|\psi_x\lr\psi_x|\}_{x\in[m]}$ and $\{r_z,|\varphi_z\lr\varphi_z|\}_{z\in[k]}$ are two ensembles that correspond to the same density operator $\rho$. Then, the probability to obtain an outcome $y$ is the same for both ensembles, and therefore there is no way to distinguish between the two ensembles. One may argue that maybe there is a way to distinguish between the post-measurement ensembles, however, as can be seen in the above equation, any post-measurement ensemble is also associated with a unique density operator $\frac{1}{q_y}M_{y}\rho M_{y}^{*}$ that depends only on $\rho$ and not on the particular ensemble $\{p_x,|\psi_x\lr\psi_x|\}_{x\in[m]}$ or $\{r_z,|\varphi_z\lr\varphi_z|\}_{z\in[k]}$ that realizes $\rho$. 

Given that the formalism of quantum mechanics lacks the means to differentiate between ensembles of states corresponding to the same density operator, all the information about the physical system accessible to us as observers is encapsulated within the density operator. Consequently, instead of characterizing physical systems using ensembles of states, we shall henceforth employ density operators for their descriptions.

\begin{mye}{The Rules of Quantum Measurements on Density Operators} \index{density matrix}\index{generalized measurement}
To any physical system (open or closed) there is a Hilbert space $A$ that is associated with it. The information about a physical system is encoded in a density operator $\rho\in\md(A)$.  Information about the system can be extracted with an $m$-output generalized measurement, $\{M_x\}_{x\in[m]}$. The probability to obtain an outcome
$x$ is given by
\be\label{m1}
p_x=\tr\left[M^{*}_xM_x\rho\right]\;,
\ee
and the post-measurement state of the system, $\sigma_x$, after output $x$ occurred is 
\be\label{m2}
\sigma_x=\frac{1}{p_x}M_{x}\rho M_{x}^{*}\;.
\ee
\end{mye}

As an example, consider a qubit state $\rho\in\md(\mbb{C}^2)$. That is, $\rho\geq 0$ and $\tr[\rho]=1$. \index{Pauli}
Any such qubit state can be expressed as a linear combination of the Pauli basis of $\herm(\mbb{C}^2)$
\be
\rho=r_0\sigma_0+r_1\sigma_1+r_2\sigma_2+r_3\sigma_3\;,
\ee
where $\sigma_0\eqdef I_2$.
Now, since the Pauli matrices $\sigma_1$, $\sigma_2$, and $\sigma_3$, are traceless, the condition $\tr[\rho]=1$ gives $r_0=1/2$. What are the conditions on $\r\eqdef (r_1,r_2,r_3)^T\in\mbb{R}^3$ that ensure that $\rho\geq 0$ ? Since $\rho$ has two eigenvalues, say $\lambda$ and $1-\lambda$, it follows that $\rho\geq 0$ if and only if $0\leq \lambda\leq 1$. This condition is equivalent to
$\tr[\rho^2]=\lambda^2+(1-\lambda)^2\leq 1$. Therefore, $\rho\geq 0$ if and only if
\be
1\geq\tr[\rho^2]=\frac{1}{2}+\sum_{j,k}r_jr_k\tr[\sigma_j\sigma_k]=\frac{1}{2}+2\|\r\|^2_2\;.
\ee
That is, $\|\r\|_2\leq1/2$. Therefore, after the renaming $\r\to\frac{1}{2}\r$ we conclude that all qubit quantum states has the form:
\be\label{bloch}
\rho=\frac{1}{2}\left(I_2+\r\cdot\bs{\sigma}\right)
\ee
with $\|\r\|_2\leq 1$. Moreover, since $\tr[\rho^2]=1$ if and only if $\|r\|_2=1$ we get that $\rho$ above is pure if and only if $\|r\|_2=1$. Hence, a qubit can be represented by the \emph{Bloch Sphere}\index{Bloch sphere} (see Fig.~\ref{fig5}) with the pure states represented on the boundary of the sphere and mixed states in the interior of the sphere. Note that the center of the sphere, i.e. $\r=0$, corresponds to the state $\rho=\frac{1}{2}I$, which is called the \emph{maximally mixed state}.

\begin{exercise}
Show that for $\r=(\sin(\alpha)\cos(\beta),\sin(\alpha)\sin(\beta),\cos(\alpha))^T$, $\rho$ in~\eqref{bloch} is given by 
the state $\rho=|\psi\lr\psi|$ with $|\psi\ra$ as in~\eqref{qubit}. 
\end{exercise}

\begin{exercise}
Consider a density operator for a qutrit; that is, $\rho\in\md(\mathbb{C}^{3})$, $\rho\geq 0$, and $\tr[\rho]=1$. Let $\bs{\lambda}=(\lambda_1,\lambda_2,\ldots,\lambda_8)$ be a vector of matrices with $\{\lambda_j\}_{j\in[8]}$
being some \emph{Hermitian traceless} $3\times 3$  matrices satisfying the condition $\text{Tr}(\lambda_i\lambda_j)=2\delta_{ij}$
(note that also the Pauli\index{Pauli} matrices satisfy this orthogonality condition).
\begin{enumerate}
\item Show that $\rho$ can be written as:
\be
\rho=\frac{1}{3}I_3+\t\cdot\bs{\lambda}\;,
\ee
where $\t\in\mathbb{R}^8$ and $I_3$ is the $3\times 3$ identity matrix.
\item Show that $\|\t\|_2\leq\frac{1}{\sqrt{3}}$.
\item Show that if $\rho$ is a pure state then $\|\t\|_2=\frac{1}{\sqrt{3}}$.\\
\item Is it true that for every $\t$ with $\|\t\|_2\leq\frac{1}{\sqrt{3}}$, $\rho$ above corresponds to a density matrix?
If yes prove it, otherwise give a counter example.
\end{enumerate}
\end{exercise}

\subsection{The Emergence of Density Operators from Quantum Correlations}\index{density matrix}

The density operator can be interpreted from a different perspective, distinct from the ensemble-based interpretation discussed earlier.
Let's delve into this interpretation within the context of a composite system\index{composite system} comprising two particles distributed between two separate entities, namely Alice and Bob.
If the system is prepared in a pure state $|\psi^{AB}\ra\in {A B}$, how should Alice represent the marginal state that corresponds to the electron in her lab? We assume here that the labs are very far from each other (perhaps on two different galaxies!) and Alice and Bob cannot even communicate. 

Without loss of generality, we will assume that $|A|\leq |B|$ and
write $|\psi^{AB}\ra$ as (see Exercise~\ref{purification})
\be\label{sqrtrhot}
|\psi^{AB}\ra=\sqrt{\rho}\otimes V|\Omega^{A\tA}\ra\;,
\ee
where $V:\tA\to B$ is an isometry and $\rho\eqdef\tr_B\left[\psi^{AB}\right]$ is the reduced density matrix of $\psi^{AB}=|\psi^{AB}\lr\psi^{AB}|$.
Suppose now that Alice performs a generalized measurement, $\{N_x\}_{x\in[m]}\subset\ml(A)$, on her subsystem.
Then, the probability that outcome $x$ occurs is given by
\ba
p_x&\eqdef\la\psi^{AB}|N_{x}^{*}N_x\otimes I^B|\psi^{AB}\ra\\
\GG{\eqref{sqrtrhot}}&=\big\la\Omega^{A\tA}\big|\sqrt{\rho}N_{x}^{*}N_x\sqrt{\rho}\otimes V^*V\big|\Omega^{A\tA}\big\ra\\
\Gg{V^*V=I^{\tA}}&=\big\la\Omega^{A\tA}\big|\sqrt{\rho}N_{x}^{*}N_x\sqrt{\rho}\otimes I^{\tA}\big|\Omega^{A\tA}\big\ra\\
\GG{Part~1\; of\; Exercise~\ref{bipartite}}&=\tr\left[N_{x}^{*}N_x\rho^A\right]\;.
\ea
Thus, the outcome probability $p_x$ depends only on the reduced density matrix  
$\rho^{A}$ and not (directly) on the bipartite state $|\psi^{AB}\ra$. Moreover, the post measurement state after outcome $x$ occurred is given by
\ba\label{expsixfot}
|\psi_{x}^{AB}\ra&=\frac{1}{\sqrt{p_x}}N_x\otimes I^B|\psi^{AB}\ra\\
\GG{\eqref{sqrtrhot}}&=\frac{1}{\sqrt{p_x}}N_x\sqrt{\rho}\otimes V\big|\Omega^{A\tA}\big\ra\;.
\ea
Therefore, the reduced density matrix $\sigma_x^A$ of $\psi^{AB}_x=|\psi^{AB}_x\lr\psi^{AB}_x|$ is given by
\be
\sigma_{x}^{A}\eqdef\tr_B\left[\psi_x^{AB}\right]
=\frac{1}{p_x}N_x\sqrt{\rho}\;\tr_B\left[\left(I^A\otimes V\right)\Omega^{A\tA}\left(I^A\otimes V\right)^*\right]\sqrt{\rho}N_x^*\;,
\ee
where we substitute the expression in~\eqref{expsixfot} for $\psi^{AB}_x$. Now, from the cyclic property of the partial trace (see Exercise~\ref{cycpt}) we have that
\be
\tr_B\left[\left(I^A\otimes V\right)\Omega^{A\tA}\left(I^A\otimes V\right)^*\right]=\tr_{\tA}\left[\left(I^A\otimes V^*V\right)\Omega^{A\tA}\right]=\tr_{\tA}\left[\Omega^{A\tA}\right]=I^A\;.
\ee
Combining this with the previous equation we conclude that
\be
\sigma_{x}^{A}=\frac{1}{p_x}N_{x}\rho^{A}N^{*}_{x}\;.
\ee
This demonstrates that the reduced density matrix $\rho^{A}$ along with the measurement operators $\{N_x\}_{x\in[m]}$
determine the post-measurement reduced density matrices in the exact same way as we saw in the previous section. For the same reasons as before, we conclude
that all the information that can be extracted from Alice's subsystem (via quantum generalized measurements) is encoded in the marginal state $\rho^{A}$. Therefore, if Alice has no accesses to Bob's subsystem, then from her perspective, the state of her subsystem can be characterized by the marginal density operator $\rho^{A}$, and the fact that her subsystem is entangled with Bob's can be ignored.

\subsection{The Classical-Quantum State}\index{cq-state}

Any ensemble of quantum states $\{p_x, |\psi_x\lr\psi_x|\}_{x\in[m]}$ in $\md(A)$ can be viewed from two distinct perspectives, depending on the treatment of the variable $x$.This duality emerges when considering whether $x$ remains unknown and unrecorded, or if it is explicitly stored in a classical system.

When $x$ remains both unknown and unrecorded, as discussed previously, the complete characterization of the system is encapsulated by the density operator $\rho^A= \sum_{x\in[m]} p_x|\psi_x\rangle\langle\psi_x|$.
Conversely, when $x$ is recorded within the classical system $X$ using the mapping $x\mapsto |x\rangle\langle x|^X,$ the description of the system adopts a classical-quantum state, abbreviated as a `cq-state', represented by:
\be
\rho^{XA}\eqdef\sum_{x\in[m]}p_x|x\lr x|^X\otimes |\psi_x\lr\psi_x|^A\;.
\ee

To establish the equivalence between $\rho^{XA}$ and the ensemble $\{p_x, |\psi_x\rangle\langle\psi_x|\}_\{x\in[m]\}$, we demonstrate that it is possible to transform $\rho^{XA}$ into $\{p_x, |\psi_x\rangle\langle\psi_x|\}_{x\in[m]}$ and vice versa.
Firstly, consider performing a measurement in the $|x\rangle$ basis on system $X$ of a composite system\index{composite system} $XA$ in the cq-state\index{cq-state} $\rho^{XA}$. This measurement yields the state $|\psi_{x}\rangle$ with a probability of $p_x$. Consequently, this process reconstructs the ensemble $\{p_x, |\psi_x\rangle\langle\psi_x|\}_{x\in[m]}$ from $\rho^{XA}$.

Conversely, imagine that we have a state $|\psi_x\rangle$ randomly selected from the ensemble ${p_x, |\psi_x\rangle\langle\psi_x|}_{x\in[m]}$. If Alice possesses knowledge of which state was selected (i.e., she knows the value of $x$), she can encode this information by introducing $|x\rangle\langle x|^X$, resulting in her state transitioning to $|x\rangle\langle x|^X \otimes |\psi_x\rangle\langle\psi_x|^A$. When Alice opts to forget the specific value of $x$, her quantum state becomes identical to $\rho^{XA}$.
Furthermore, it is worth noting that when we only have access to the marginal state $\rho^A = \tr_X\left[\rho^{XA}\right] = \sum_{x\in[m]}p_x|\psi_x\rangle\langle\psi_x|$, it is generally impossible to perfectly recover the value of $x$.

Cq-states play a pivotal role in quantum information science, particularly when describing the outcomes of quantum measurements. Let's consider a physical system characterized by the density operator $\rho\in\md(A)$ and a generalized quantum measurement $\{M_x\}_{x\in[m]}$.
As previously discussed, the application of this generalized measurement to the state $\rho^A$ results in the state $\sigma_{x}^{A}$, as outlined in~\eqref{m2}, with the associated probability $p_x$ as defined in~\eqref{m1}. Because we know the outcome $x$, we have the option to record it within a classical system denoted as $X$. In this context, we can perceive the measurement's effect as a transformation, mapping the state $\rho^A$ to a cq-state, represented by $\sigma^{XA},$ defined as follows:
\be
\sigma^{XA}\eqdef\sum_{x\in[m]}p_x|x\lr x|^X\otimes\sigma_{x}^{A}
=\sum_{x\in[m]}|x\lr x|^X\otimes M_x\rho^{A}M_{x}^{*}\;,
\ee
where we substitute $p_x\sigma^{A}_{x}=M_x\rho^{A}M_{x}^{*}$. In essence, by employing a cq-state, we can describe the impact of a generalized measurement $\{M_x\}_{x\in[m]}$ on $\rho^A$ as a ``deterministic" process, transforming $\rho^A$ into $\sigma^{XA}$. It's important to note that we use the term ``deterministic" to describe the transformation $\rho^A\to\sigma^{XA}$, not to characterize the mapping $\rho^A\to\sigma^A_x$, which is inherently indeterministic in nature.

If the information about the measurement outcome $x$ is lost, or if Alice has no access to the value of $x$, then the post-measurement state is given by
\be
\sigma^A=\tr_X\left[\sigma^{XA}\right]=\sum_{x\in[m]}M_x\rho^{A}M_{x}^{*}\;.
\ee
We will see later on that in this case, the measurement acts as a quantum channel, converting one density operator, $\rho^{A}$, to another, $\sigma^{A}$ (see Fig.~\ref{fig10}c below).

\subsection{Separable Density Operators}\index{separable states}

\begin{figure}[h]\centering    \includegraphics[width=0.4\textwidth]{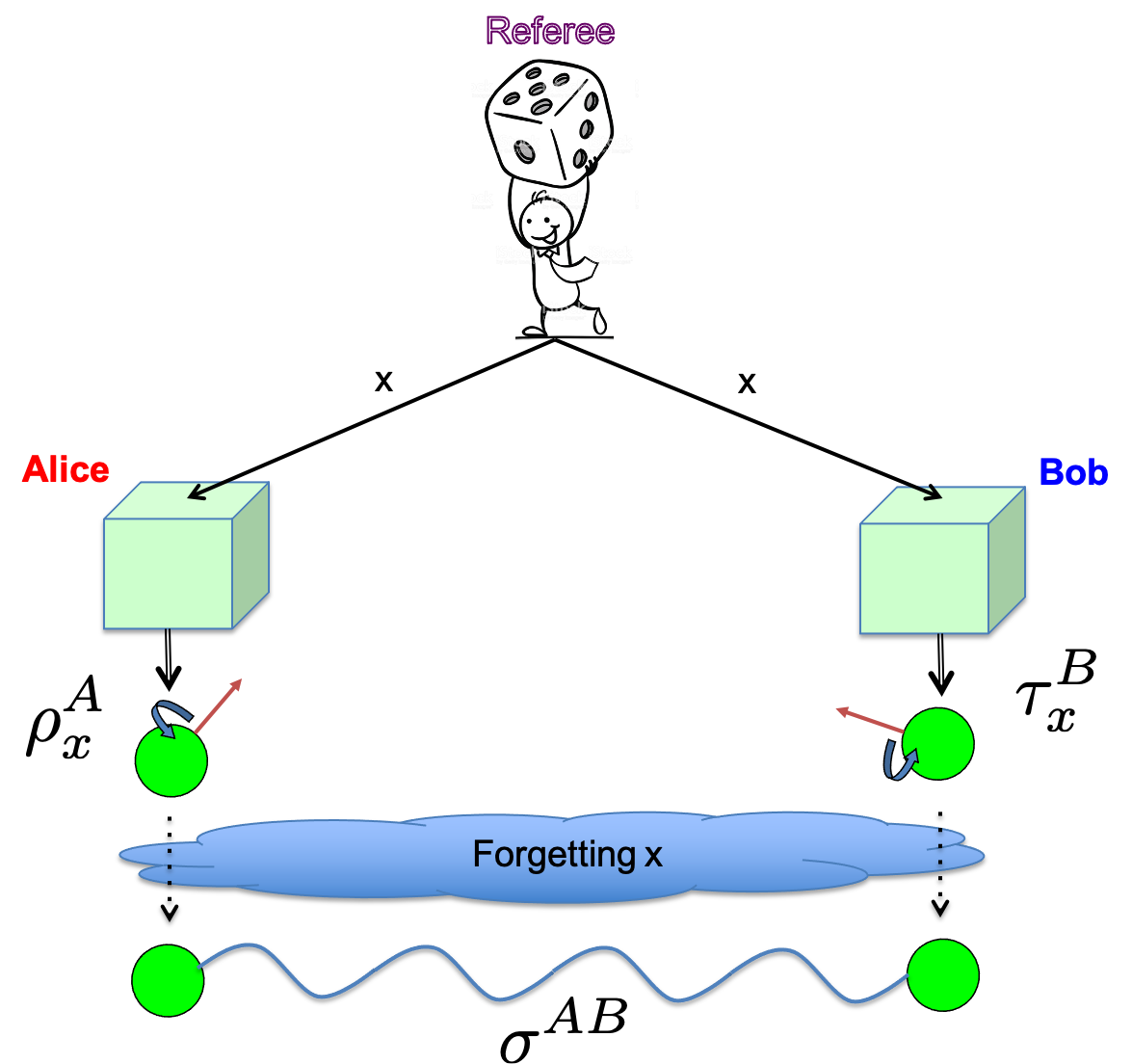}
  \caption{\linespread{1}\selectfont{\small Preparation of a separable state with shared randomness.}}
  \label{figsep}
\end{figure}

Density operators that are acting on a bipartite Hilbert space ${A}\otimes B$ can be divided into two types:
separable and entangled. A separable density matrix can be prepared in the following way (see Fig.~\eqref{figsep}).
A referee samples a number $x$ with a probability distribution $p_x$ (e.g. roll a possiblly biased dice, or flip a coin) and sends the number $x$ to Alice and Bob who are spatially separated. Based on this value, Alice prepares the state $\rho_x^A$, and Bob prepares the state $\tau_x^B$. Then,  if Alice and Bob forget the value of $x$, but still know the distribution $p_x$ from which $x$ was sampled, the state of their composite system\index{composite system} becomes
\be\label{sep}
\sigma^{AB}=\sum_{x\in[m]}p_x\rho_x^A\otimes\tau_x^B\in\md(A\otimes B)\;.
\ee

Note that the role of the referee above is to provide Alice and Bob with a shared randomness. Therefore, any separable state as in~\eqref{sep} can be prepared by local operations assisted with shared randomness. Bipartite density matrices that do not have this form are called entangled and we will discuss them in details in the following chapters on entanglement theory.

\begin{exercise}
Show that the maximally entangled state $\Phi^{AB}\eqdef |\Phi^{AB}\lr\Phi^{AB}|\in\md(A\otimes B)$ is not separable.
\end{exercise}
\begin{exercise}\label{ex:sepstate}
Show that if $\sigma\in\md(A\otimes B)$ is separable then there exists an integer $k\in\mbb{N}$, a probability distribution $\{q_z\}_{z\in[k]}$, a set of $k$ pure states 
$\{\psi_z\}_{z\in[k]}$ in Alice's Hilbert space (i.e. each $\psi_z\in\pure(A)$), and a set of $k$ pure states $\{\phi_z\}_{z\in[k]}$ on Bob's Hilbert space $B$, such that
\be
\sigma^{AB}=\sum_{z\in[k]}q_z\psi_z^A\otimes\phi_z^B\;.
\ee
\end{exercise}

\section{Positive Operator Valued Measure (POVM)}\index{POVM}

\begin{figure}[h]\centering    \includegraphics[width=0.7\textwidth]{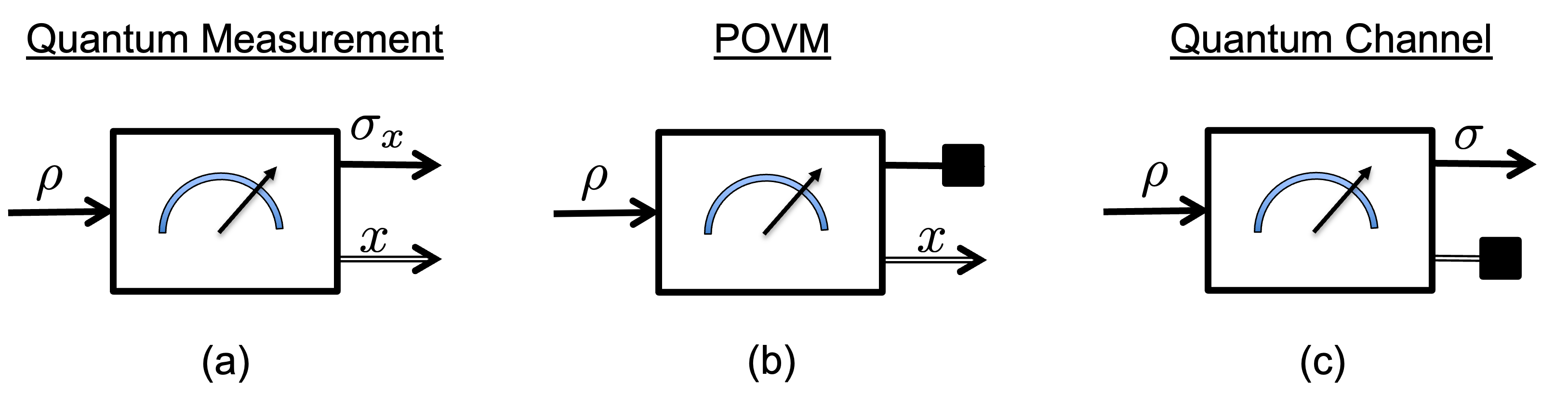}
  \caption{\linespread{1}\selectfont{\small Three types of generalized quantum measurements}}
  \label{fig10}
\end{figure}

Every quantum measurement can be viewed as a box (see Fig.~\ref{fig10}a) that takes as its input a quantum state $\rho$, and outputs a classical variable $x$ and a post-measurement state $\sigma_x$. 
In the SG-experiment that we discussed earlier, the electrons get absorbed by the screen, and all there is left
after the measurements are the spots on the screen. Therefore, the SG-experiment can be viewed as a special type of measurement in which the quantum output is ``traced out" (see Fig.~\ref{fig10}b). Such quantum measurements with only classical output are called positive operator\index{positive operator} valued measures, or in short POVM.

Recall that the Born's rule\index{Born's rule} (adapted to density matrices and generalized measurements) states that the probability to obtain an outcome $x$, when a measurement $\{M_x\}_{x\in[m]}$ is performed on a system described by a density operator $\rho$, is given by $p_x=\tr\left[M_{x}^{*}M_x\rho\right]$. Therefore, to describe a POVM we only need to consider the operators, $\{\Lambda_x\eqdef M_{x}^{*}M_x\}_{x\in[m]}$, since we are only interested in the statistics of the measurement and not the post-measurement state. The POVM operators $\Lambda_x$, are called \emph{effects}, and have the following two properties:
\be
\Lambda_x\geq 0\quad\text{and}\quad\sum_{x\in[m]}\Lambda_x=I^A\;.
\ee
To every generalized measurement there exists a \emph{unique} POVM that corresponds to it via the relation $\Lambda_x= M_{x}^{*}M_x$. However, for every POVM there are many quantum measurements corresponding to it.

\begin{exercise}\label{polar}{\bf \rm [Polar Decomposition]}\index{polar decomposition}
\begin{enumerate}
\item Show that for any $n\times n$ complex matrix $A$ there exists an $n\times n$ unitary matrix $U$ such that
\be
A=U|A|\quad\text{where}\quad |A|\eqdef\sqrt{A^*A}\;.
\ee 
\item Show that for $A$ and $U$ as above,
\be
\max_{U}\tr\left[AU\right]=\tr[|A|]\;,
\ee
where the maximum is over all unitary matrices $U$.
\end{enumerate}
\end{exercise}

\begin{exercise}
Let $\{\Lambda_x\}_{x\in[m]}$ be a POVM in $\pos(A)$. Show that a generalized measurement $\{M_x\}_{x\in[m]}\subset\ml(A)$ corresponds to the POVM $\{\Lambda_x\}_{x\in[m]}$ if and only if there exists $m$ unitary matrices, $\{U_x\}_{x\in[m]}$, in $\ml(A)$ such that
\be
M_x=U_x\sqrt{\Lambda_x}\;.
\ee
Hint: Use the polar decomposition of a complex matrix.
\end{exercise}

\begin{exercise}
Consider the following POVM in $\pos(\mbb{C}^2)$
\be
\Lambda_1=a|1\lr 1|\quad,\quad \Lambda_2=b|-\lr -|\quad,\quad \Lambda_3=I-\Lambda_1-\Lambda_2\quad\quad a,b\in\mbb{R}.
\ee
\begin{enumerate}
\item Find all the possible values of $a$ and $b$ for which the set $\{\Lambda_1,\Lambda_2,\Lambda_3\}$ is a POVM. 
\item Which values of $a$ and $b$ that you found in part 1 correspond to a rank 1 POVM (i.e. all the POVM elements have rank 1)? 
\end{enumerate}  
\end{exercise}

\begin{exercise}
Suppose Alice and Bob share a composite quantum system in the state $\rho^{AB}$. Alice performs a measurement on her system described by a POVM $\{\Lambda_x\}_{x\in[m]}$, and record the outcome $x$ in a classical system $X$. Show that the post-measurement state can be expressed as a cq-state\index{cq-state} of the form 
\be
\sigma^{XB}=\sum_{x\in[m]}p_x|x\lr x|^X\otimes\sigma_{x}^{B}\;.
\ee
Express the probabilities $p_x$, and density matrices $\sigma_{x}^{B}$, in terms of $\Lambda_x$ and $\rho^{AB}$.
\end{exercise}

\subsection{Informationally Complete POVMs and Quantum Tomography}\index{informationally complete}

Consider a scenario where you have access to a machine that consistently produces an unknown quantum state $\rho\in\md(A)$. Your objective is to learn the identity of this state $\rho$ by employing quantum measurements. Since we assume that this machine can be used repeatedly, generating an abundance of copies of $\rho$, it is sufficient to focus on POVMs. This is due to the fact that post-measurement states cannot reveal more information about $\rho$ than $\rho$ itself. 

One effective strategy for learning $\rho$ is to carry out basis measurements $\{|x\rangle\langle x|\}_{x\in[d]}$ (where $d\equiv|A|$). By conducting these measurements multiple times, you can approximate the probabilities associated with each outcome $x$. Given that $p_x=\tr[|x\lr x|\rho]=\la x|\rho|x\ra$, this approach allows you to estimate the diagonal elements of $\rho$ within the basis $\{|x\rangle\}_{x\in[d]}$. Repeating this procedure using various bases enables you to ultimately learn the complete structure of $\rho$ by capturing its diagonal elements with respect to multiple bases. As we illustrate now, opting for POVMs over projective measurements in specific bases enables the construction of a single POVM that can be employed to fully identify the state $\rho$.

\begin{myd}{}
\begin{definition}
A POVM $\{\Lambda_x\}_{x\in[m]}$ in $\herm(A)$ is said to be \emph{informationally complete} if 
\be
\spa_{\mbb{R}}\left\{\Lambda_1,\Lambda_2,\ldots,\Lambda_{m}\right\}=\herm(A)\;.
\ee 
\end{definition}
\end{myd}
The span above is with respect to the real numbers since $\herm(A)$ is a real vector space. By definition, if a POVM $\{\Lambda_x\}_{x\in[m]}$ is informationally complete then $m\geq d^2$, where
 $d\eqdef|A|$. Moreover, if $m=d^2$ then the informationally complete POVM form a basis of $\herm(A)$.
 Clearly, the basis is \emph{not} orthonormal since $\Lambda_x\geq 0$ for all $x\in[d^2]$. A theorem from linear algebra states that for any basis of a vectors space there exists a \emph{dual basis}. That is, if $\left\{\Lambda_1,\Lambda_2,\ldots,\Lambda_{d^2}\right\}$ is a basis of $\herm(A)$ then there exists another basis $\left\{\Gamma_1,\Gamma_2,\ldots,\Gamma_{d^2}\right\}$ of $\herm(A)$, 
 such that the Hilbert Schmidt inner products
\be\label{dual}
\tr\left[\Lambda_x\Gamma_y\right]=\delta_{xy}\quad\forall x,y\in[d^2]\;.
\ee 
\begin{exercise}
Consider the following four elements of $\pos(\mbb{C}^{2})$
\be
\Lambda_0\eqdef|0\lr 0|\quad,\quad \Lambda_1\eqdef |1\lr 1|\quad,\quad \Lambda_2\eqdef|+\lr +|\quad,\quad \Lambda_3\eqdef|+i\lr +i|\;.
\ee
\begin{enumerate}
\item Show that $\{\Lambda_0,\Lambda_1,\Lambda_2,\Lambda_3\}$ is a basis of $\herm(\mbb{C}^2)$.
\item Find its dual basis $\{\Gamma_0,\Gamma_1,\Gamma_2,\Gamma_3\}$.
\item Set $\Lambda\eqdef\sum_{x=0}^{3}\Lambda_x$ and show that it is invertible, and that the operators $\{\tilde{\Lambda}_0,\tilde{\Lambda}_1,\tilde{\Lambda}_2,\tilde{\Lambda}_3\}$, with $\tilde{\Lambda}_x\eqdef \Lambda^{-1/2}\Lambda_x \Lambda^{-1/2}$, form a rank 1 informationally complete POVM.  
\end{enumerate}
\end{exercise}

\begin{exercise}\label{ex:povmicb}$\;$
\ben
\item Construct a rank 1 informationally complete POVM in $\herm(A)$. Hint: Try to generalize the qubit example in the previous exercise to any (finite) dimension.
\item Let $\Gamma\in\herm(AB)$. Show that if
\be
\tr\left[\Gamma^{AB}\left(\psi^A\otimes\phi^B\right)\right]=0\quad\quad\forall\;\psi\in\pure(A)\;\;\;\;\forall\;\phi\in\pure(B)\;,
\ee
then $\Gamma^{AB}=0$.
\een
\end{exercise}

Informationally complete POVM can be used to learn an unknown quantum state. Let $\rho\in\md(A)$ be an unknown quantum state, and let $\{\Lambda_x\}_{x\in[m]}$ be an informationally complete POVM in $\herm(A)$. 
If $m>d^2$ than $\{\Lambda_x\}_{x\in[m]}$ is an over-complete basis, also referred to, in linear algebra, as a \emph{frame}.
Every frame, like a basis, has a unique dual frame, $\{\Gamma_y\}_{y\in[m]}$, 
defined by the condition
\be
\sum_{x\in[m]}\tr[\Gamma_xM]\Lambda_x=\sum_{x\in[m]}\tr[\Lambda_xM]\Gamma_x\quad\quad\forall\;M\in\herm(A)\;.
\ee
However, unlike a basis, in general a frame does not satisfy the relation~\eqref{dual} with its dual. 

Consider now the case that $m=d^2$ so that $\{\Lambda_x\}_{x\in[m]}$ is a basis of $\herm(A)$. Since its dual  $\{\Gamma_y\}_{y\in[m]}$ also span $\herm(A)$ it follows that the density matrix $\rho$ can be written in terms of the linear combination
\be\label{rho3}
\rho=\sum_{y\in[m]}p_y \Gamma_y
\ee
with some \emph{real} coefficients $p_y$. A key observation is that
\ba
\tr\left[\Lambda_x\rho\right]&=\sum_{y\in[m]}p_y \tr\left[\Lambda_x\Gamma_y\right]\\
\GG{\eqref{dual}}&=\sum_{y\in[m]}p_y\delta_{xy}=p_x\;.
\ea
That is, the coefficients $p_y$ in~\eqref{rho3} are given by $p_y=\tr[\Lambda_y\rho]\geq 0$, and can be interpreted as the probability  to obtain an outcome $y$. The significance of~\eqref{rho3} with $p_y=\tr[\Lambda_y\rho]\geq 0$, is that by repeating the POVM $\{\Lambda_x\}_{x\in[m]}$
on many copies of $\rho$, one can estimate from the measurement outcomes the values of the $p_y$s, and thereby learn $\rho$ due to the relation~\eqref{rho3}.

\begin{exercise}
Let $\{\Lambda_x\}_{x\in[m]}$ be an informationally complete POVM in $\herm(A)$, and let its dual frame be
$\{\Gamma_y\}_{y\in[m]}$.
\begin{enumerate}
\item Show that at least one of the matrices in the dual frame is not positive semidefinite. That is, there exists $y\in[m]$ such that $\Gamma_y\not\geq 0$.
\item Show that $\tr[\Gamma_y]=1$ for all $y\in[m]$.
\end{enumerate}
\end{exercise}

\begin{exercise} {\rm \bf [Symmetric Informationally Complete (SIC) POVM]}\index{SIC POVM}\\
Let $d\eqdef|A|$, $m\eqdef d^2$, and $\{\Lambda_x\}_{x\in[m]}$ be an informationally complete POVM in $\herm(A)$, with the following properties: 
\ben
\item Each $\Lambda_x$ is rank one.
\item  $\tr\left[\Lambda_x\right]=\tr\left[\Lambda_y\right]$ for all $x,y\in[m]$. 
\item $\tr\left[\Lambda_x\Lambda_y\right]=\tr\left[\Lambda_{x'}\Lambda_{y'}\right]$ for all $x,x',y,y'\in[m]$ with $x\neq y$ and $x'\neq y'$.
\een
Such a POVM is called \emph{symmetric} informationally complete POVM or in short SIC-POVM.
\ben
\item Show that 
\be
\tr\left[\Lambda_x\right]=\frac{1}{d}\quad\text{and}\quad\tr\left[\Lambda_x\Lambda_y\right]=\frac{d\delta_{xy}+1}{d^2(d+1)}\quad\quad\forall\; x,y\in[m]\;.
\ee
Hint: Denote $a_x\eqdef \tr[\Lambda_x]$ and $b\eqdef\tr[\Lambda_x\Lambda_y]$ for any $x\neq y$. Then, $a_{x}^{2}=\tr[\Lambda_{x}^2]$ (why?) so that $a_x=\tr[\Lambda_xI]=\sum_{y\in[m]}\tr[\Lambda_x\Lambda_y]=a_{x}^{2}+(d^2-1)b$ and conclude that $a_x=a_y$ for all $x,y\in[m]$.
\item Find the dual frame $\{\Gamma_x\}_{x\in[m]}$. Hint: Express each $\Gamma_x$ as a linear combination of $\Lambda_x$ and the identity.
\item Show that any density operator $\rho\in\herm(A)$ can be expressed as
\be
\rho=\sum_{x\in[m]}\big(d(d+1)p_x-1\big)\Lambda_x\quad\text{with}\quad p_x\eqdef\tr[\Lambda_x\rho]\;.
\ee  
\item Show that
\begin{align}
& \Lambda_{1}=\frac{1}{12\sqrt{3}}\left(\begin{array}{cc}3\sqrt{3}+1&-5+i\\-5-i&3\sqrt{3}-1\end{array}\right)\;\;,\;\;
\Lambda_{2}=\frac{1}{12\sqrt{3}}\left(\begin{array}{cc}3\sqrt{3}+1&1-5i\\1+5i&3\sqrt{3}-1\end{array}\right)\nonumber\\
& \Lambda_{3}=\frac{1}{12\sqrt{3}}\left(\begin{array}{cc}3\sqrt{3}-5&1+i\\1-i&3\sqrt{3}+5\end{array}\right)\;\;,\;\;
\Lambda_{4}=\frac{1}{4\sqrt{3}}\left(\begin{array}{cc}\sqrt{3}+1&1+i\\1-i&\sqrt{3}-1\end{array}\right).
\end{align}
form a SIC POVM in $\herm(\mbb{C}^2)$. Moreover, show that the four pure state $\{2\Lambda_x\}_{x\in[4]}$ are the vertices
of a tetrahedron in the Bloch sphere.\index{Bloch sphere} 
\end{enumerate}
\end{exercise}

 \subsection{Gleason's Theorem}\index{Gleason theorem}
 
One of the enigmas of quantum mechanics concerns with the emergence of probabilities at a very fundamental level.
One may wonder if the Born's rule, which specifies how to assign probabilities to different measurement outcomes, could have been different. Since we are only interested in the statistics of the measurement we will consider here a POVM $\{\Lambda_x\}_{x\in[m]}$. According to Born's rule, the probability to obtain an outcome $x$ is given by the formula 
$
p_x=\tr[\Lambda_x\rho]
$.
Is this formula unique? 

Specifically, consider the convex set of all effects 
\be
\eff(A)\eqdef\left\{\Lambda\in\pos(A)\;:\;\Lambda\leq I^A\right\}\;.
\ee
A measure on the set $\eff(A)$ is a function $\mu: \eff(A)\to\mbb{R}$ with the following three properties:
\begin{enumerate}
\item $0\leq \mu(\Lambda)\leq 1$ for all $\Lambda\in\eff(A)$.
\item $\mu(I^A)=1$.
\item For any possibly incomplete POVM $\{\Lambda_x\}_{x\in[m]}$ with $\sum_{x\in[m]}\Lambda_x\leq I^A$ 
\be\label{measure}
\mu\Big(\sum_{x\in[m]}\Lambda_x\Big)=\sum_{x\in[m]}\mu(\Lambda_x)\;.
\ee
\end{enumerate}
We now show that any function $\mu$ with these properties \emph{must} have the form $\mu(\Lambda)=\tr[\rho \Lambda]$ for some fixed density operator $\rho\in\md(A)$. This means that there is no way to assign probabilities to effects other than the Born's rule. Originally, this remarkable result was proved by Gleason for the case of projective measurements, i.e. the set $\eff(A)$ was replaced with the set of all projections on $A$, and consequently the requirement on $\mu$ was much weaker, assuming only that~\eqref{measure}  holds for orthogonal projections. 
Nonetheless, Gleason was able to derive the probability formula for systems of dimension $d\geq 3$, and in the qubit case he showed that there are counter examples. Gleason also considered in his proof the infinite dimensional case, and derived similar results. Gleason's proof for projective measurements goes beyond the scope of this book, and we will follow here a much simpler proof of the above generalization of Gleason theorem (see the section on Notes and References for more details).  
%which is due to Busch (2003).
\begin{myt}{}
\begin{theorem}
Let $A$ be a finite dimensional Hilbert space, and $\mu$ a measure on the set $\eff(A)$. Then, there exists $\rho\in\md(A)$ such that 
\be
\mu(\Lambda)=\tr[\rho \Lambda]\quad\quad\forall\;\Lambda\in\eff(A)\;.
\ee
\end{theorem}
\end{myt}

The idea of the proof is as follows. First we will show that for any $r\in[0,1]$ $\mu(r\Lambda)=r\mu(\Lambda)$, and use it to show that $\mu$ can be extended to a linear functional on the space of Hermitian operators $\herm(A)$. Then, as a linear functional, it can be expressed as $\mu(\Lambda)=\tr[\Lambda\rho]$, and we end by showing that $\rho$ must be a density operator.

\begin{proof}
Note that for any effect $\Lambda\in\eff(A)$ and any integer $n$, we have $\Lambda=\frac{1}{n}\Lambda+\cdots\frac{1}{n}\Lambda\leq I^A$, where the sum contains $n$ terms. Therefore,  from~\eqref{measure}, $\mu(\Lambda)=n\mu(\frac{1}{n}\Lambda)$. Multiplying this equation by an integer $m\leq n$ and dividing by $n$ gives $\frac{m}{n}\mu(\Lambda)=m\mu(\frac{1}{n}\Lambda)=\mu\left(\frac{m}{n}\Lambda\right)$, where we used the third property of a measure $\mu$ as defined above. So far we showed that for any rational number $p\in[0,1]$, we must have $\mu(p\Lambda)=p\mu(\Lambda)$. Let $r\in[0,1]$ be a real number and let $\{p_j\}$ and $\{q_k\}$ be two sequences of \emph{rational} numbers in $[0,1]$ that converge to $r$ and have the property that $p_j\leq r\leq q_k$ for all $j$ and $k$. We therefore have
\be
p_j\mu(\Lambda)=\mu(p_j\Lambda)\leq\mu(r\Lambda)\leq \mu(q_k\Lambda)=q_k\mu(\Lambda)\;,
\ee
where the inequalities above follows from the fact that if two effects satisfy $\Lambda\geq \Gamma$ (i.e. $\Lambda-\Gamma\geq 0$) then
\be
\mu(\Lambda)=\mu\big(\Gamma+(\Lambda-\Gamma)\big)=\mu(\Gamma)+\mu(\Lambda-\Gamma)\geq\mu(\Gamma)\;.
\ee
Taking the limit $j,k\to\infty$ gives $\mu(r\Lambda)=r\mu(\Lambda)$ for all $r\in[0,1]$.

We now extend the definition of $\mu$ to any element in $\herm(A)$. First, for any positive semidefinite
matrix $P\geq 0$ that is \emph{not} in $\eff(A)$ there always exists $r>1$ such that $ \frac{1}{r}P\in\eff(A)$. 
Define $\mu(P)\eqdef r\mu(\frac{1}{r}P)$. To show that $\mu(P)$ is well define, let $r'>1$ be another number such that 
$\frac{1}{r'}P\in\eff(A)$ and assume without loss of generality that $r'>r$ so that $\frac{r}{r'}<1$. Then,
\be
r\mu\left(\frac{1}{r}P\right)=r'\frac{r}{r'}\mu\left(\frac{1}{r}P\right)=r'\mu\left(\frac{1}{r'}P\right)\;.
\ee

Note that this extension of the domain of $\mu$ to any element of $\pos(A)$ preserves the linearity of $\mu$; i.e.
for any two matrices $M,N\in\pos(A)$ and large enough $r$ such that$\frac{1}{r}(M+N)\in\eff(A)$,
\be\label{proof1}
\mu(M+N)=r\mu\left(\frac{1}{r}(M+N)\right)=r\mu\left(\frac{1}{r}M\right)+r\mu\left(\frac{1}{r}N\right)=\mu(M)+\mu(N)\;.
\ee
Finally, we extend the definition of $\mu$ to include in its domain any matrix $L\in\herm(A)$. Any such matrix can be expressed as $L=M-N$, with $M,N\in\pos(A)$. We therefore define:
\be
\mu(L)\eqdef\mu(M)-\mu(N)\;.
\ee
To show that $\mu(L)$ is well defined, we need to show that for any other decomposition of $L=M'-N'$  with $M',N'\in\pos(A)$, we have
\be\label{linfun}
\mu(M)-\mu(N)=\mu(M')-\mu(N')\;.
\ee
This is indeed the case since the equality $L=M-N=M'-N'$ implies that $M+N'=M'+N$ and from the additivity\index{additivity} property in~\eqref{proof1} we conclude that
\be
\mu(M)+\mu(N')=\mu(M')+\mu(N)
\ee
which is equivalent to~\eqref{linfun}. 

To summarize, we where able to extend $\mu$ to a linear functional $\mu:\herm(A)\to\mbb{R}$. Since the dual space of $\herm(A)$ is itself, any linear functional on $\herm(A)$ can be expressed as
$
\mu(\Lambda)=\tr\left[\rho \Lambda\right]
$,
where $\rho\in\herm(A)$ is a fixed matrix. Now, from the fact that $\mu(\Lambda)\geq 0$ for all effects $\Lambda\in\eff(A)$ we conclude that $\tr[\rho \Lambda]\geq 0$ for all $\Lambda\geq 0$ which implies that $\rho\geq 0$ (see Exercise below).
Finally, to show that $\rho$ is a density operator, note that the condition $\mu(I)=1$ gives $\tr[\rho]=1$. This completes the proof.
\end{proof}

\begin{exercise}
Show that $\rho\geq 0$ if and only if $\tr[\rho \Lambda]\geq 0$ for all $\Lambda\geq 0$.
\end{exercise}

\subsection{Naimark's Dilation Theorem}\index{Naimark}

Naimark's dilation theorem reveals that any POVM can be implemented with a von-Neumann projective measurement on a larger Hilbert space. This is not much of a surprise to us as we already saw that generalized measurements can be implemented by a bipartite unitary map followed by a projective measurement. However, Naimark's theorem deals directly with POVMs and makes the connection between POVMs and projective measurements more transparent.
Moreover, Naimark's dilation theorem is applicable to infinite dimensional systems, although we will prove it here only in the finite dimensional case.

\begin{myt}{\color{yellow}Naimark's Theorem}\index{Naimark}
\begin{theorem}\label{naimark}
Let $A$ be a Hilbert space and $\{\Lambda_x\}_{x\in[m]}\subset \eff(A)$ be a POVM. Then, there exists an extended Hilbert space $B$ (i.e. $|B|\geq |A|$), an isometry $V:A\to B$, and a von-Neumann projective measurement $\{P_x\}_{x\in[m]}\subset\eff(B)$, such that $\Lambda_x=V^*P_xV$ for all $x\in[m]$.
\end{theorem}
\end{myt}

\begin{proof}
Every POVM element $\Lambda_x$ can be expressed as $\Lambda_x=M_x^*M_x$ where $\{M_x\}$ is a generalized measurement. In Sec.~\ref{sec:gm} we saw that for every generalized measurement there exists an ancillary system $R$, and unitary matrix $U^{RA}$, such that (cf.~\eqref{3232})
\be\label{fomr}
M_x^A={}^R\big\la x\big|U^{RA}\big|1\big\ra^R
\ee
where $|1\ra^R$ is some fixed state in $R$. Define $B\eqdef RA$ and define the operator $V:A\to B$ by $V\eqdef U^{RA}\big|1\big\ra^R$. That is, for any $|\psi\ra\in A$
\be
V|\psi\ra^A\eqdef U^{RA}\big|1\big\ra^R\big|\psi\big\ra^A\in B\;.
\ee
Clearly, $V^*V=I^A$ so that $V$ is an isometry.
With this definition we get that 
\ba
\Lambda_x&=M_x^*M_x\\
\GG{\eqref{fomr}}&={}^R\big\la 1\big|U^{*RA}\left( |x\ra\la x|^R\otimes I^A\right)U^{RA}\big|1\big\ra^R\\
&=V^*\left( |x\ra\la x|^R\otimes I^A\right)V
\ea
Denoting by $P_x^B\eqdef |x\ra\la x|^R\otimes I^A$ we conclude that $\Lambda_x=V^*P_x^BV$. This completes the proof.
\end{proof}

We point out that any  POVM $\{\Lambda_x\}_{x\in[m]}\subset\eff(A)$ can be implemented with a rank one POVM in the following simple way. Since each $\Lambda_x$ is positive semidefinite, 
it can be expressed in terms of its (unnormalized) eigenvectors $\{|\phi_{xy}\ra\}_{y\in[n]}$ as 
\be
\Lambda_x=\sum_{y\in[n]}\phi_{xy}\;.
\ee 
Moreover, the set of rank one matrices $\{\phi_{xy}\}$ also form a POVM (i.e. $\sum_{x,y}\phi_{xy}=I^A$). Therefore, one can implement the POVM $\{\Lambda_x\}_{x\in[m]}$ by first implementing the rank one POVM $\{\phi_{xy}\}$, with corresponding $(x,y)$ outcomes, and then forgetting/ignoring the outcome $y$. 

\begin{exercise}
Show that if $\{\Lambda_x\}_{x\in[m]}$ is a rank 1 POVM in $\eff(A)$, where $m\eqdef |A|$, then $\{\Lambda_x\}_{x\in[m]}$ is a basis measurement (i.e. rank one von-Neumann projective measurement).
\end{exercise}

\begin{exercise}
Consider therank one matrices
\be
\Gamma_x=\frac{2}{3}|\uparrow_{\n_x}\rangle\langle\uparrow_{\n_x}|\;\;,\;\;\;\;
x=1,2,3\;\;\;;\;\;|\uparrow_{\n_x}\rangle\in\mathbb{C}^2
\ee
were the unit vectors $\n_x\in\mathbb{R}^3$ satisfies $\n_1+\n_2+\n_3=0$.\\
\begin{enumerate}
\item Show that the set $\{\Gamma_x\}_{x\in[3]}$ is a POVM.
\item Find an orthonormal basis $\{|\varphi_x\rangle\}_{x\in[3]}$ in $\mathbb{C}^3$ such that $\{\Gamma_x\}_{x\in[3]}$ is realized by $\{\varphi_x\}_{x\in[3]}$.
\end{enumerate}
\end{exercise}

\section{Evolution of Open Systems}\index{evolution}

What is the most general evolution that a (possibly open) quantum system can undergo? We will first tackle this problem axiomatically, using minimal physical assumptions, and then provide several ways to demonstrate how to physically realize such an evolution. We will see that any evolution of a physical system can be described with a quantum channel. Quantum channels lie at the heart of quantum information theory, and we will devote much of this section to describe their different representations.

\subsection{The axiomatic approach\index{axiomatic approach}}\index{axiomatic approach\index{axiomatic approach}}

For closed systems, a pure state $|\psi\ra$ can evolve either deterministically (described by a
unitary matrix $U$) to $U|\psi\ra$,  or probabilistically to $|\phi_x\ra$ with some probability $p_x$.
The latter can be described as an evolution from the density matrix $|\psi\lr\psi|$ to the classical quantum state
$
\sigma^{XA}=\sum_x p_x|x\lr x|\otimes |\phi_x\lr\phi_x|\;.
$
Since any open system is described with a density operator, any evolution of a quantum system can be described with a transformation $\mE$ that takes density operators in $\md(A)$ to density operators in $\md(B)$.
Note that since the systems are open, the transformation $\mE$ can change the dimension (e.g. particles added or discarded) so that the input dimension $|A|$ can be different than the output dimension $|B|$.

Recall that we use the notation $\ml(A,B)$ to indicate linear transformations from $A$ to $B$. However, our current focus lies in the set $\ml(A\to B)$, which denotes transformations from $\ml(A)$ to $\ml(B)$. We will typically employ calligraphic letters like $\mE$, $\mF$, $\mN$, and $\mM$ to denote the elements of $\ml(A\to B)$ and sometimes include a superscript, such as $\mE^{A\to B}$, to emphasize the underlying input and output Hilbert spaces. The identity element of $\ml(A\to A)$ will be denoted as $\id^A$ or $\id^{A\to A}$.
 
 \begin{exercise}
 Let $A$ and $B$ be two Hilbert spaces, $m\eqdef |A|$, and  $\{\eta_j\}_{j\in[m^2]}$ be an orthonormal basis of $\ml(A)$ (in the Hilbert-Schmidt inner product). For any two elements $\mE,\mF\in\ml(A\to B)$ define 
 \be\label{innerpc}
 \la\mE,\mF\ra\eqdef\sum_{j\in[m^2]}\left\la\mE(\eta_j),\mF(\eta_j)\right\ra_{HS}\;,
 \ee 
 where $\la\;,\;\ra_{HS}$ denote the Hilbert-Schmidt inner product between matrices.
 \begin{enumerate}
 \item Show that the function  above is well defined in the sense that it is independent on the choice of the orthonormal basis $\{\eta_j\}_{j\in[m^2]}$ of $\ml(A)$.
 \item Show that the function above is an inner product in the vector space $\ml(A\to B)$. Hence, $\ml(A\to B)$ is a Hilbert space.
 \end{enumerate}
 \end{exercise}

We are now ready to introduce the axiomatic approach\index{axiomatic approach} describing a physical evolution from system $A$ into $B$. Below each axiom\index{axiom} we provide the physical justification.\\

\noindent\textbf{Axiom\index{axiom} 1:}\index{axiom}
\emph{A physical evolution can be described with a linear transformation.}\\

Consider the scenario where Alice rolls a dice to obtain a classical variable $x\in[m]$, with associated probabilities $\{p_x\}_{x\in[m]}$. If she obtains the value $x$, she prepares her system in the state $\rho_x$. After a quantum evolution takes place, if the initial state was $\rho_x$, it will evolve to $\mE(\rho_x)$.
Now, if Alice forgets which state she prepared (i.e., she forgets the outcome of the dice roll), from her new perspective, the input state is given by $\sum_{x\in[m]} p_x\rho_x$. In the same vein, the post-evolution state of the system becomes $\sum_{x\in[m]}p_x\mE(\rho_x)$. Consequently, we must have:
\be
\mE\Big(\sum_{x\in[m]} p_x\rho_x\Big)=\sum_{x\in[m]}p_x\mE(\rho_x)\;.
\ee
The above equation signifies that $\mE$ is convex-linear (i.e., linear under convex combinations) on the set of density matrices and can be extended linearly to act on the entire space $\ml(A)$ (not limited to $\md(A)$ and not limited to convex combinations). Therefore, we infer that every map describing a physical evolution is an element of $\ml(A\to B)$.\\

\noindent\textbf{Axiom\index{axiom} 2:}\index{trace preserving}
\emph{A physical evolution  is trace preserving.}\\

Since a physical evolution $\mE\in\ml(A\to B)$ takes density matrices to density matrices it must satisfy for all $\rho\in\md(A)$,
$\tr[\mE(\rho)]=1$, since $\mE(\rho)$ is a density matrix. Now, let $\eta\in\herm(A)$ be an arbitrary Hermitian matrix. Then, $\eta$ can always be written as $\eta=t\rho-s\sigma$, where $t,s\geq 0$ and $\rho,\sigma\in\md(A)$. Therefore, from the linearity of $\mE$ we get that \be
\tr\left[\mE(\eta)\right]=t\tr\left[\mE(\rho)\right]-s\tr\left[\mE(\sigma)\right]=t-s=\tr[\eta]\;,
\ee
so that $\mE$ preserve the trace of hermitian matrices.
Moreover, if $M\in\ml(A)$ is not Hermitian it can still be expressed as $M= \eta_0+i\eta_1$, where both $\eta_0\eqdef (M+M^{*})/2$ and $\eta_1\eqdef (M+M^{*})/2$ are Hermitian matrices, so that
\be
\tr\left[\mE(M)\right]=\tr\left[\mE(\eta_0)\right]+i\tr\left[\mE(\eta_1)\right]=\tr\left[\eta_0\right]+i\tr\left[\eta_1\right]=\tr[M]\;.
\ee
We therefore conclude that any physical evolution $\mE$ is a \emph{trace preserving} (TP) linear map.\\

\noindent\textbf{Axiom\index{axiom} 3:}\index{completely positive}
\emph{A physical evolution is completely positive.}\\

Since a physical evolution $\mE$ takes density matrices to density matrices it also preserves positivity\index{positivity} . That is, if $\rho\in\pos(A)$ is positive semidefinite matrix then also $\mE(\rho)$ is a positive semidefinite matrix in $\pos(B)$.
We call such linear maps \emph{positive} maps. There is yet one more  property that $\mE$ has to satisfy 
if it describes an evolution a physical system. 

Consider a composite system\index{composite system} consisting of two subsystems $A$ and $B$. Such a system is described by a bipartite density operator $\rho^{AB}\in\md(A\otimes B)$. If the subsystem $B$ undergoes a physical evolution described by a linear map $\mE\in\ml(B\to B')$, while system $A$ does not evolve and remain intact, then the state $\rho^{AB}$ will evolve to the state 
\be
\sigma^{AB'}\eqdef\id^A\otimes\mE^{B\to B'} \left(\rho^{AB}\right)\;.
\ee
Therefore, if $\mE$ represents a physical evolution, then both $\mE$ and $\id^A\otimes\mE$ must takes density matrices to density matrices.
In particular, the linear map $\id\otimes\mE\in\ml(AB\to AB')$ must also be a positive map for any system $A$. It turns out that there are linear maps $\mE$ that are positive while $\id^A\otimes\mE$ is not positive. One such example is the transposition map.

Consider the linear map $\mT\in\ml(A\to A)$ defined by\index{partial transpose}
\be
\mT(\rho)\eqdef \rho^T\quad\quad\forall\;\rho\in\ml({A})\;.
\ee
The transpose map preserves the eigenvalues and therefore is a trace-preserving positive map. Now, take $A=\mbb{C}^2$ and consider the matrix $\Omega^{A\tA}\eqdef|\Omega^{A\tA}\lr \Omega^{A\tA}|\in\ml(\mbb{C}^2\otimes\mbb{C}^2)$. The matrix $\Omega^{A\tA}$ is a rank one, positive semidefinite matrix that can be expressed as
\be
\Omega^{A\tA}=|00\lr 00|+|00\lr 11|+|11\lr 00|+|11\lr 11|=\begin{bmatrix} 1 & 0 & 0 & 1\\
0 & 0 & 0 & 0\\
0 & 0 & 0 & 0\\
1 & 0 & 0 & 1
\end{bmatrix}\;.
\ee
On the other hand, its \emph{partial transpose\index{partial transpose}} on system $\tA$ is given by
\be
\mT^{\tA\to \tA}\big(\Omega^{A\tA}\big)=|00\lr 00|+|01\lr 10|+|10\lr 01|+|11\lr 11|=\begin{bmatrix} 1 & 0 & 0 & 0\\
0 & 0 & 1 & 0\\
0 & 1 & 0 & 0\\
0 & 0 & 0 & 1
\end{bmatrix}\;.
\ee
It is relatively easy to check (see the exercise below) that $\mT^{\tA\to \tA}\big(\Omega^{A\tA}\big)$ has three eigenvalues equals to $1$, and one eigenvalue
equals $-1$. Hence, $\mT^{\tA\to \tA}\big(\Omega^{A\tA}\big)$ is not positive semidefinite!

\begin{exercise}\label{petnpt}
Let $\psi^{A\tA}\eqdef|\psi\lr \psi|$ with $|\psi^{A\tA}\ra=\sum_{x\in[d]}\sqrt{p_x}|x\ra|x\ra\in A\otimes {\tA}$. Find the eigenvalues and eigenvectors of $\mT^{\tA\to \tA}\big(\psi^{A\tA}\big)$.
For which values of $p_x$, the matrix $\mT^{\tA\to \tA}\big(\psi^{A\tA}\big)$ is positive semidefinite?  
\end{exercise}

\begin{myd}{Complete Positivity}\index{completely positive}
\begin{definition}\label{def:cp}
A linear map $\mE\in\ml(A\to B)$ is called $k$-positive\index{$k$-positive} if $\id_k\otimes\mE$ is positive, where 
$\id_k\in\ml(\mbb{C}^k\to \mbb{C}^k)$ is the identity map. Furthermore, $\mE$ is called completely positive\index{completely positive} (CP) if it is $k$-positive\index{$k$-positive} for all $k\in\mbb{N}$.
\end{definition}
\end{myd}

By definition, every map that is $k$-positive\index{$k$-positive} is also $k'$-positive if $k'\leq k$. On the other hand, there are maps that are $k$-positive\index{$k$-positive} but not $(k+1)$-positive. The canonical example of such a map is the map $\mE^{(k)}\in\ml(A\to A)$ (with $d\eqdef|A|$) defined by
\be\label{kpositive}
\mE^{(k)}(\eta)=k\tr[\eta]I_d-\eta\quad\quad\forall\;\eta\in\ml(A)\;.
\ee
From Exercise~\ref{exkpos} in the next few subsections it follows that $\mE^{(k)}$ is $k$-positive\index{$k$-positive}, and yet, if $k<d$ then $\mE^{(k)}$  is \emph{not} $(k+1)$-positive.
If $k\geq d$ then this map is completely positive. More generally, we will see later that a map is $d$-positive if and only if it is completely positive. 

In conclusion, any evolution of a physical system has to be (1) linear, (2) trace-preserving (TP), and (3) completely positive\index{completely positive} (CP). Such a linear CPTP map is called a quantum channel. The set of all quantum channels in $\ml(A\to B)$ will be denoted by $\cptp(A\to B)$. In the next subsections we discuss several representations of quantum channels, and along the way show that \emph{any} quantum channel has a physical realization; that is, it can be implemented by physical processes. Therefore, the axiomatic approach\index{axiomatic approach} above led us to the precise conditions on the evolution of a physical system that are both necessary and sufficient for the existence of its physical realization. 

\bex\label{flcp}
Let $\Lambda\in\ml(A)$ and define a map $\mF_\Lambda:\ml(A)\to\ml(A)$ via 
\be
\mF_\Lambda(\omega)\eqdef \Lambda\omega\Lambda^*\quad\quad\forall\;\omega\in\ml(A)\;.
\ee
Show that $\mF_\Lambda$ is a completely positive\index{completely positive} \emph{linear} map. Hint: Prove it first that $\mF_\Lambda^{A\to A}(\psi^{RA})\geq 0$ for a pure state $|\psi^{RA}\ra=M\otimes I^{A}|\Omega^{\tA A}\ra$.
\eex

\subsubsection{The Dual of a Linear Map in $\ml(A\to B)$}\index{dual map}

We saw earlier that any vector in a Hilbert space has a dual vector. In particular, we saw that any matrix $M\in\ml(A,B)$ has a dual or adjoint matrix $M^*\in\ml(B,A)$\index{adjoint operator}. Since the space $\ml(A\to B)$ is also a Hilbert space with respect to the inner product given in~\eqref{innerpc}, it follows that any map in $\ml(A\to B)$ has a dual map\index{dual map} in $\ml(B\to A)$.

\begin{myd}{}
\begin{definition}
Let $\mE\in\ml(A\to B)$ be a linear map. Its dual or adjoint map\index{adjoint map} is the map $\mE^{*}\in\ml(B\to A)$
that satisfies
\be\label{dualxz}
\tr\left[\sigma^*\mE(\rho)\right]=\tr\left[\left(\mE^{*}(\sigma)\right)^*\rho\right]\quad\quad\forall\;\rho\in\ml(A)\quad\text{and}\quad\forall\;\sigma\in\ml(B)\;.
\ee
Furthermore, we say that $\mE$ is self-adjoint\index{self-adjoint} if $|A|=|B|$ and $\mE=\mE^{*}$. 
\end{definition}
\end{myd}

The definition above of the dual map\index{dual map} is analogous to the definition of a dual map in $\ml(A,B)$. Recall that the dual of a matrix $M\in\ml(A,B)$ is defined via the relation
\be
\la\phi|M\psi\ra=\la M^*\phi|\psi\ra\quad\quad\forall\;|\psi\ra\in A,\;|\phi\ra\in B\;.
\ee
Similarly, the definition of $\mE^*$ in~\eqref{dualxz} can be expressed as
\be
\la \sigma,\mE(\rho)\ra_{HS}=\la \mE^*(\sigma),\rho\ra_{HS}\quad\quad\forall\;\rho\in\ml(A),\;\sigma\in\ml(B)\;,
\ee
where $\la\cdot,\cdot\ra_{HS}$ is the Hilbert-Schmidt inner product.\index{Hilbert-Schmidt inner product}

\begin{exercise}\label{showdual}
Let $\mE\in\ml(A\to B)$ be a linear map.
\begin{enumerate}
\item Show that $\mE$ is trace preserving if and only if its dual $\mE^*$ is unital; i.e. $\mE^*(I^B)=I^A$.
\item Show that $\mE$ is trace non-increasing (i.e., $\tr[\mE(\eta)]\leq\tr[\eta]$ for all $\eta\in\pos(A)$) if and only if its dual $\mE^*$ is sub-unital; i.e. $\mE^*(I^B)\leq I^A$.
\item Show that $\mE$ is positive if and only if $\mE^*$ is positive.
\item Show that $\mE$ is completely positive\index{completely positive} if and only if $\mE^*$ is completely positive.
\end{enumerate}
\end{exercise}

\begin{exercise}
Show that a unitary evolution is a CPTP map. Specifically, show that a unitary map $\mU\in\ml(A\to A)$ defined by 
\be
\mU(\rho)\eqdef U\rho U^*\quad\quad\forall\;\rho\in\ml(A)
\ee 
where $U\in\muu(A)$ is a unitary operator, is a quantum channel.
\end{exercise}

\begin{exercise}
The replacement map is a map $\mE\in\ml(A\to B)$ defined by
\be
\mE(\rho)\eqdef \tr[\rho]\;\sigma \quad\quad\forall\;\rho\in\ml(A)
\ee
where $\sigma\in\md(B)$ is some fixed density matrix. 
\begin{enumerate}
\item Show that $\mE$ is a quantum channel.
\item Show that for any two quantum states $\rho\in\md(A)$ and $\sigma\in\md(B)$ there exists a quantum channel $\mE$ such that $\mE(\rho)=\sigma$.
\end{enumerate}
\end{exercise}

\begin{exercise}\label{exposcone}
Let $A$ and $B$ be two finite dimensional Hilbert spaces, and denote the set of positive maps in $\ml(A\to B)$ by
\be
\pos(A\to B)\eqdef\Big\{\mE\in\ml(A\to B)\;:\;\mE(\rho)\in\pos(B)\quad\forall\;\rho\in\pos(A)\Big\}
\ee
Denote also by $\mR\in\cptp(A\to B)$ the replacement channel $\mR(\rho^A)\eqdef\tr[\rho^A]\frac1{|B|}I^B$ for all $\rho\in\ml(A)$. 
\begin{enumerate}
\item Show that $\pos(A\to B)$ is a convex cone in the Hilbert space $\ml(A\to B)$.
\item Prove the equivalence of the following properties of a map $\mE\in\pos(A\to B)$:
\begin{enumerate}
\item $\mE$ belongs to the interior of the cone $\pos(A\to B)$.
\item $\mE=(1-t)\mF+t\mR$ for some $t\in(0,1]$ and some $\mF\in\pos(A\to A)$.
\item $\mE^*$ belongs to the interior of the cone $\pos(A\to B)$.
\item For any non-zero $\rho\in\pos(A)$ we have $\mE(\rho)>0$.
\end{enumerate} \end{enumerate}
\end{exercise}

\subsection{The Matrix Representation}\index{matrix representation}

We begin our exploration of linear maps with the most familiar representation: the matrix representation. In the context of linear transformations between two Hilbert spaces, every element in the set $\ml(A\to B)$ possesses a corresponding matrix representation.
While this matrix representation\index{matrix representation} can prove valuable in certain applications, especially when dealing with low-dimensional cases, we will discover later on that it is, in fact, the least intuitive representation and finds limited use in the realm of quantum information. The reason for this lies in the fact that the concept of complete positivity\index{positivity} , a crucial property of quantum channels, does not translate naturally and tends to become cumbersome within this matrix-based framework. 

Let $\mE\in\ml(A\to B)$ be a linear map, $\rho\in\ml(A)$, and $\sigma\eqdef\mE(\rho)\in\ml(B)$. The relationship $\sigma=\mE(\rho)$ can be expressed in matrix form as $\s_\sigma=M_\mE\r_\rho$, where $\r_\rho$ and $\s_\sigma$ denote column vectors representing $\rho$ and $\sigma$, respectively, and $M_\mE$ is a complex matrix representing the linear map $\mE$.  To illustrate it explicitly,
let $\{\Lambda_x\}_{x\in[m^2]}$ with $m\eqdef|A|$ be a fixed orthonormal basis of
$\ml(A)$, and $\{\Gamma_y\}_{y\in[n^2]}$ with $n\eqdef|B|$ be a fixed orthonormal basis of $\ml(B)$. 
Any operator $\rho\in\ml(A)$ can be expressed as a linear combination of the basis elements
\be
\rho=\sum_{x\in[m^2]}r_x\Lambda_x\;.
\ee
with $\r_\rho\eqdef (r_1,\ldots,r_{m^2})^T\in\mbb{C}^{m^2}$. Observe that the mapping $\rho\mapsto\r_\rho$ defines an isomorphism between $\ml(A)$ and $\mbb{C}^{m^2}$. Similarly, for any
$\sigma=\sum_{x\in[n^2]}s_y\Gamma_y\in\ml(B)$ we define $\s_\sigma\eqdef (s_1,\ldots,s_{n^2})^T$.
Finally, for any linear map $\mE\in\ml(A\to B)$ we define the $n^2\times m^2$ matrix $M_\mE$ whose components are given by
\be
\left(M_\mE\right)_{yx}\eqdef\la\Gamma_{y}|\mE(\Lambda_x)\ra_{HS}\eqdef\tr\left[\Gamma_{y}^{*}\mE(\Lambda_x)\right]\;.
\ee
With these notations we have
\be
\sigma=\mE(\rho)\iff\s_\sigma=M_\mE\r_\rho\;.
\ee
Therefore, $M_\mE$ is the matrix representation\index{matrix representation} of the linear map $\mE$.

Suppose now that the map $\mE$ is a quantum channel, and that the orthonormal bases $\{\Lambda_x\}_{x\in[m^2]}$  and $\{\Gamma_y\}_{y\in[n^2]}$ consist of Hermitian matrices (see Exercise~\ref{hermitianex}). Then, $\rho$ is Hermitian if and only if $\r_{\rho}$ is a real vector, and similarly, $\sigma$ is Hermitian if and only if $\s_\sigma$ is a real vector. Therefore, $\mE$ is Hermitian preserving if and only if $M_\mE$ is a real matrix. Moreover, since $\mE$ is trace preserving, it will be convenient to choose the orthonormal bases  $\{\Lambda_x\}_{x\in[m^2]}$  and $\{\Gamma_y\}_{y\in[n^2]}$ with the first element proportional to the identity. That is, we choose
\be
\Lambda_1\eqdef\frac{1}{\sqrt{m}}I^A\quad\text{and}\quad \Gamma_1\eqdef\frac{1}{\sqrt{n}}I^B\;,
\ee
so that the remaining elements of the orthonormal bases, $\{\Lambda_x\}_{x=2}^{m^2}$  and $\{\Gamma_y\}_{y=2}^{n^2}$, are all traceless (since they have to be orthogonal in the Hilbert-Schmidt inner product to the first element). With these choices,  an operator $\rho\in\ml(A)$ has trace one if and only if $\r_\rho$ has the form $(\frac{1}{\sqrt{m}},r_2,\ldots,r_{m^2})^T$. We therefore conclude that the linear map $\mE$ is both trace preserving and Hermitian preserving if and only if its matrix representation\index{matrix representation} has the form
\be\label{t}
M_\mE =\begin{bmatrix}\sqrt{\frac{m}{n}} & \0\\
\t  & N_{\mE} 
\end{bmatrix}\quad\text{where}\quad\t \in\mbb{R}^{n^2-1}\quad\text{and}\quad N_{\mE}\in\mbb{R}^{(n^2-1)\times(m^2-1)}\;.
\ee
What are the conditions on the matrix $N_{\mE} $ and the vector $\t$ that correspond to the condition that $\mE$ is completely positive? These conditions can be very complicated since, in general, even the set of all vectors $\r_\rho$ for which $\rho\geq 0$ doesn't have a simple characterization. 

\begin{exercise}\label{showdual2}
Let $\mE\in\ml(A\to B)$ be a linear map.
\begin{enumerate}
\item Show that
\be
M_{\mE^{*}}=M_{\mE}^{*}\;.
\ee
\item Show that $\mE$ is a self-adjoint map\index{adjoint map} (i.e. $\mE=\mE^*$) that is both trace-preserving and Hermitian preserving if and only if the matrix $M_\mE$ in~\eqref{t} has $\t=0$ and $N_{\mE}$ is a real symmetric matrix (i.e. $N_{\mE}^T=N_{\mE}$).
\end{enumerate}
\end{exercise}

\subsubsection{The Qubit Case}\index{qubit}

In the qubit case, where $|A|=|B|=2$, the Bloch representation\index{Bloch representation} simplifies the situation. In particular, $\rho\in\ml(\mbb{C}^2)$ is a density matrix if and only if $\r_\rho=\frac{1}{\sqrt{2}}(1,\r)^T$, where $\r$ corresponds to the Bloch vector belonging to $\mbb{R}^3$ and its length $\|\r\|\leq 1$. Therefore, in this case, $\sigma=\mE(\rho)$ if and only if the Bloch vector of $\sigma$, denoted by $\s$, is related to $\r$ via
\be
\s=\t+\big(N_{\mE}\big)\r\;.
\ee
From the relation above, $\mE$ is a positive map if the matrices $\t$ and $N_{\mE}$ are such that whenever $|\r|\leq 1$, $|\s|\leq 1$ also holds. It's important to note, however, that this criterion pertains solely to the positivity\index{positivity}  of $\mE$ and does not address its complete positivity. In the specific scenario of qubits, it is feasible to articulate the conditions governing $N_{\mE}$ and $\t$ for $\mE$ to be completely positive. Nevertheless, these conditions tend to be rather intricate, and we direct the interested reader to the pertinent literature found in the Notes and References section at the end of this chapter. In the exercises below, you will demonstrate that these conditions become more straightforward in the case of doubly-stochastic maps.

Doubly stochastic maps encompass mappings that possess two key properties: trace-preservation and unitality, meaning they preserve the identity operator. One of the simplest examples of such maps is the unitary map $\mU(\rho)\eqdef U\rho U^*$, where $U$ is a unitary matrix. Another example includes convex combinations of unitary maps in the form $\sum_{x\in[m]}p_x\mU_x$, where $\{p_x\}_{x\in[m]}$ forms a probability distribution, and each $\mU_x$ represents a unitary quantum channel. These instances exemplify completely positive\index{completely positive} maps that also qualify as doubly stochastic.

Conversely, the transpose map $\mT(\rho)=\rho^T$ or its combination with a unitary map, denoted as $\mU\circ\mT$, serve as examples of doubly stochastic maps that are positive but not completely positive. In the subsequent set of problems, we will see that for the qubit case,  all positive doubly stochastic maps can be expressed as convex combinations of such maps.

\begin{exercise}\label{ex:ds}
Let $\mE\in\pos\left(\mbb{C}^2\to\mbb{C}^2\right)$ be a positive linear map. 
\begin{enumerate}
\item Show that $\mE$ is both trace-preserving and unital (i.e. doubly-stochastic) if and only if
\be\label{1ds1}
M_\mE =\begin{bmatrix}1 & \0\\
\0  & N_{\mE} 
\end{bmatrix}
\ee
where $N_{\mE}\in\mbb{R}^{3\times 3}$ has the property that $\|N_{\mE}\r\|_2\leq 1$ for all $\r\in\mbb{R}^3$ with $\|\r\|_2= 1$ (in particular, the absolute value of the eigenvalues of $N_\mE$ cannot exceed one).
\item Suppose $\mE=\mU$ is the doubly stochastic unitary map given by $\mU(\rho)=U\rho U^{*}$, where $U\in SU(2)$ can be expressed as 
$
U=wI_2+i(x\sigma_1+y\sigma_2+z\sigma_3)
$
 with $w,x,y,z\in\mbb{R}$ and $w^2+x^2+y^2+z^2=1$ (cf.~\eqref{unitar2}). Show that the matrix $N_{\mU}$ of~\eqref{1ds1}
 is an orthogonal matrix in $SO(3)$ given by
 \be\label{orthogonalmatrix2}
N_{\mU}=\begin{pmatrix}
1-2y^2-2z^2 & 2xy+2zw & 2xz-2yw\\
2xy-2zw & 1-2x^2-2z^2 & 2yz+2xw\\
2xz+2yw & 2yz-2xw & 1-2x^2-2y^2
\end{pmatrix}
\ee
(cf.~\eqref{orthogonalmatrix}). Hint: Calculate directly the components $\frac{1}{2}\tr\left[\sigma_iU\sigma_jU^*\right]$ for $i,j\in\{1,2,3\}$.
\item Use the previous parts to show that for any map of the form $\mU(\rho)=U\rho U^{*}$, we have that $U\in U(2)$ if and only if the matrix $N_\mU\in SO(3)$. Hint: Every unitary $U\in U(2)$ can be written as $U=\exp(i\theta)\tilde{U}$, where $\tilde{U}\in SU(2)$ and $\theta\in[0,2\pi)$.  
\end{enumerate}
\end{exercise}

\begin{exercise}
Let $\mT\in\ml\left(\mbb{C}^2\to\mbb{C}^2\right)$ be the transpose map defined by $\mT(\rho)=\rho^T$ for all $\rho\in\ml(\mbb{C}^2)$.
\begin{enumerate} 
\item Show that the matrix representation of the transpose map with respect to the Pauli\index{Pauli} basis of $\ml(\mbb{C}^2)$ is given by
\be
M_\mT=\begin{bmatrix}
1 & 0 & 0 & 0\\
0 & 1 & 0 & 0\\
0 & 0 & -1 & 0\\
0 & 0 & 0 & 1
\end{bmatrix}
\ee
\item Show that if $\mE\in\ml(\mbb{C}^2\to\mbb{C}^2)$ is a positive doubly stochastic linear map with $N_\mE\in O(3)$ and with $\det(N_\mE)=-1$ then $\mE=\mT\circ\mU$ for some unitary map $\mU$. Hint: Use the fact that any $3\times 3$ orthogonal matrix can be expressed as a matrix product of an element in $SO(3)$ with $N_\mT$.
\end{enumerate}
\end{exercise}

\begin{exercise}\label{exds2}
Use the exercises above in conjunction with Exercise~\ref{ex:2by2} to conclude that any doubly stochastic positive map $\mE\in\pos(\mbb{C}^2\to\mbb{C}^2)$ can be expressed as 
\be
\mE=t\mN_1+(1-t)\mT\circ\mN_2\;,
\ee where $t\in[0,1]$ and both $\mN_1$ and $\mN_2$ are mixtures of unitary maps; i.e. maps of the form $\sum_{j\in[m]} p_j\;\mU_j$ with each $\mU_j$ being a unitary map and $\{p_j\}_{j\in[m]}$ is a probability distribution. Hint: Use Exercise~\ref{ex:2by2} to show that $N_\mE$ can be expressed as a finite convex combination of orthogonal matrices.
\end{exercise}

\subsubsection{Positivity vs Complete Positivity in Low Dimensions}
\index{positivity}

We have seen before that the transpose map $\mT$ is positive, but not 2-positive (and consequently, the transpose map is not completely positive). The following theorem that was proved originally by Str\o mer and Woronowicz shows essentially that in low dimensions combinations of the transpose map with completely positive\index{completely positive} maps are the only maps that can be positive but not completely positive. 

\begin{myt}{\color{yellow} St\o rmer-Woronowicz Theorem}\index{St\o rmer-Woronowicz theorem}
\begin{theorem}\label{thm:sw}
Let $\mE\in\pos(A\to B)$ be a positive linear map. If $|A|=2$ and $|B|\leq 3$ then there exists two CP maps $\mN_1,\mN_2\in\cp(A\to B)$ such that
\be\label{formenn}
\mE=\mN_1+\mT\circ\mN_2
\ee
where $\mT\in\pos(B\to B)$ is the transpose map.
\end{theorem}
\end{myt}
\begin{remark}
The case $|B|=2$ was proven by St\o rmer, and the case $|B|=3$ was proven by Woronowicz. Here, we will only prove St\o rmer theorem (i.e. $|A|=|B|=2$) and refer the reader to the section `Notes and References' (at the end of this chapter) for more details. 
\end{remark}

\begin{proof} 
 We prove the theorem for the case that $\mE$ is in the interior of $\pos(A\to A)$ (the more general case will then follow from a continuity argument; see Exercise~\ref{contargq}). From Exercise~\ref{exposcone} it follows that also $\mE^*$ is in the interior of $\pos(A\to A)$, and furthermore, $\mE(\rho)>0$ for any non-zero $\rho\in\pos(A)$. 
 
 The key idea of the proof is to find two positive definite operators $\Lambda,\Gamma>0$ with the property that the channel
 \be\label{dfn0}
\mD\eqdef\mF_\Lambda\circ\mE\circ\mF_\Gamma
\ee
is doubly stochastic, where
\be
\mF_\Lambda(\omega)\eqdef \Lambda\omega \Lambda\quad\text{and}\quad\mF_\Gamma(\omega)\eqdef \Gamma\omega \Gamma\quad\quad \forall\;\omega\in\ml(A)\;.
\ee
From Exercise~\ref{flcp} (see also the section on operator sum representation below) it follows that the above maps are completely positive.
Apriori it is not clear if such positive definite matrices $\Lambda$ and $\Gamma$ exists, but if they do then from~\eqref{dfn0} we have $\mE=\mF_{\Lambda^{-1}}\circ\mD\circ\mF_{\Gamma^{-1}}$, and since all doubly stochastic positive maps have the form~\eqref{formenn} (see Exercise~\ref{exds2}) it follows that also $\mE$ has the form~\eqref{formenn}. It is therefore left to show that such $\Lambda$ and $\Gamma$ do exist.

By definition, the channel $\mD$ is doubly stochastic if and only if both $\mD$ and its dual $\mD^*$ are unital channels. Since the dual of $\mD$ is given by
$\mD^*= \mF_\Gamma\circ\mE^{*}\circ\mF_\Lambda$ (see Exercise~\ref{ex:mds}) we conclude that $\mD$ is doubly stochastic if and only if the matrices $\Lambda$ and $\Gamma$ satisfies
\be
I=\mD(I)=\Lambda\mE(\Gamma^2)\Lambda\quad\text{and}\quad I=\mD^*(I)=\Gamma\mE^*(\Lambda^2)\Gamma\;.
\ee
By conjugating with the inverses of $\Lambda$ and $\Gamma$, the two equations above can be expressed as
\be
\Lambda^{-2}=\mE(\Gamma^2)\quad\text{and}\quad\Gamma^{-2}=\mE^*(\Lambda^2)\;.
\ee
It is therefore left to show that there exists $\rho\eqdef\Lambda^{-2}>0$ and $\sigma\eqdef\Gamma^2>0$ 
such that
\be\label{3p91}
\rho=\mE(\sigma)\quad\text{and}\quad\sigma^{-1}=\mE^*\left(\rho^{-1}\right)\;.
\ee
Observe that if $\rho$ and $\sigma$ satisfy the two equations above then for any $s>0$ also $s\rho$ and $s\sigma$ satisfy the two equations. Hence, without loss of generality we can assume that if there exists $\rho$ and $\sigma$ that satisfy the equation above then $\sigma$ is normalized and since $\mE$ is trace preserving this implies that also $\rho\eqdef\mE(\sigma)$ is normalized.

The equation $\rho=\mE(\sigma)$ can be taken to be the definition of $\rho$. Substituting this $\rho$ into the second equality of~\eqref{3p91} implies that 
\be\label{rmin1}
\sigma^{-1}=\mE^*\left(\big(\mE(\sigma)\big)^{-1}\right)\;.
\ee
To show that such a $\sigma$ exists, define the function $f:\md(A)\to\pos(A)$ via
\be
f(\omega)\eqdef\left(\mE^*\left(\big(\mE(\omega)\big)^{-1}\right)\right)^{-1}\quad\forall\;\omega\in\md(A)\;,
\ee
and observe that~\eqref{rmin1} is equivalent to $f(\sigma)=\sigma$.
We also  define the normalized version of $f$, the function $g:\md(A)\to\md(A)$, as
\be
g(\omega)\eqdef\frac{f(\omega)}{\tr\left[f(\omega)\right]}\quad\quad\forall\;\omega\in\md(A)\;.
\ee
Then, from Brouwer's fixed-point theorem (see Theorem~\ref{fixedpoint}) there exists a density matrix $\sigma\in\md(A)$ such that $g(\sigma)=\sigma$. Denoting by $t\eqdef\tr[f(\sigma)]>0$ this is equivalent to 
\be\label{tisig}
f(\sigma)=t\sigma\;.
\ee 
It is therefore left to show that $t=1$. For this purpose, observe first that with the definition $\rho\eqdef\mE(\sigma)$ we can express the above equation as
\be
(t\sigma)^{-1}=\mE^*\left(\rho^{-1}\right)\;,
\ee
so that
\be\label{d89i0}
t^{-1}I=\sigma^{\frac12}\mE^*\left(\rho^{-1}\right)\sigma^{\frac12}=\mD^*(I)\;,
\ee
where $\mD$ is defined in~\eqref{dfn0} with $\Lambda=\rho^{-1/2}$ and $\Gamma=\sigma^{1/2}$. On the other hand, the relation $\rho=\mE(\sigma)$ can be written as
\be
I=\rho^{-\frac12}\mE\left(\sigma\right)\rho^{-\frac12}=\mD(I)
\ee
which implies that the linear map $\mD$ is unital. Since the dual of a unital map is trace preserving (see the first part of Exercise~\ref{showdual}) we conclude that $\mD^*$ is trace preserving. Hence, by taking the trace on both sides of~\eqref{d89i0} and using the fact that $\mD^*$ is trace preserving we conclude that $t=1$. This completes the proof.
\end{proof}

\begin{exercise}\label{ex:mds}
Show that if $\mD\eqdef \mF_\Gamma\circ\mE\circ\mF_\Lambda$ as in the proof above, then
\be
\mD^*\eqdef \mF_\Lambda\circ\mE^{*}\circ\mF_\Gamma\;.
\ee
\end{exercise}

\begin{exercise}\label{contargq}
Use a continuity\index{continuity} argument to prove that if all maps in the interior of $\pos(A\to A)$ have the form~\eqref{formenn} then all the maps in $\pos(A\to A)$ has this form.
\end{exercise}

\subsection{The Choi Representation}\index{Choi-Jamiolkowski isomorphism}

The Choi representation, also known as the Choi-Jamiolkowski isomorphism, is another method to characterize linear maps using matrices, offering a simpler way to identify complete positivity\index{positivity} . In this representation, quantum channels are associated with positive semidefinite matrices. This characteristic makes the Choi representation particularly useful in various applications within quantum information science, especially because it allows for the translation of certain optimization problems into semidefinite programs.

In subsequent discussions and throughout the rest of the book, we will adopt the shorthand notation $\mE^{B\to B'}(\rho^{AB})$ to denote $\left(\id^A\otimes\mE^{B\to B'}\right)(\rho^{AB})$. This notation simplifies expressions and discussions involving the application of a quantum channel $\mE$ from system $B$ to system $B'$ on part of a bipartite state $\rho^{AB}$. 

Given a linear map $\mE\in\ml(A\to B)$, the Choi matrix\index{Choi matrix} is defined by the action of $\mE$ on one subsystem of a maximally entangled state. Setting $m\eqdef|A|$ and denoting by 
\be
\Omega^{A\tA}=|\Omega^{A\tA}\lr\Omega^{A\tA}|=\sum_{x,y\in[m]}|x\lr y|^{A}\otimes |x\lr y|^{\tA}\;,
\ee
the Choi matrix of $\mE$ is defined by
\be\label{choi1}
J_{\mE}^{AB}\eqdef \mE^{\tA\to B}\big(\Omega^{A\tA}\big)=\sum_{x,y\in[m]}|x\lr y|\otimes \mE\big(|x\lr y|\big)\;.
\ee
One of the key properties of the Choi matrix\index{Choi matrix} is that it satisfies the relation
\be\label{choi2}
\mE(\rho)=\tr_A\left[J_{\mE}^{AB}\left(\rho^T\otimes I\right)\right]\quad\quad\forall\;\rho\in\ml(A)\;.
\ee
To see this, let $\rho=\sum_{x,y\in[m]}r_{xy}|x\lr y|$ be a linear operator in $\ml(A)$ with components $r_{xy}\in\mbb{C}$ and $m\eqdef|A|$. From the definition of the Choi matrix in~\eqref{choi1} we get
\ba
\tr_A\left[J_{\mE}^{AB}\left(\rho^T\otimes I\right)\right]&=\sum_{x,y\in[m]}\tr\left[\rho^T|x\lr y|\right] \mE\left(|x\lr y|\right)\\
&=\sum_{x,y\in[m]}r_{xy} \mE\left(|x\lr y|\right)\\
&=\mE\Big(\sum_{x,y\in[m]}r_{xy}|x\lr y|\Big)=\mE(\rho)\;.
\ea
The two relations~(\ref{choi1},\ref{choi2}) demonstrate that the mapping $\mE\mapsto J_\mE$ is a linear bijection\index{bijection} (i.e. an isomorphism). This isomorphism is between the vector space $\ml(A\to B)$ of linear operators from $\ml(A)$ to $\ml(B)$, and the space of bipartite matrices/operators $\ml(AB)$. In the following exercise you show that the mapping $\mE\mapsto J_\mE$ is in fact isometrically isomorphism between these two spaces.

\begin{exercise}
Let $A$ and $B$ be two finite dimensional Hilbert spaces and consider the Hilbert space $\ml(A\to B)$ equiped with the inner product defined in~\eqref{innerpc}. Show that this inner product can be expressed as follows. For all $\mE,\mF\in\ml(A\to B)$ we have
\be
\la\mE,\mF\ra=\la J_{\mE}^{AB}, J_{\mF}^{AB}\ra_{HS}
\ee
where on the right-hand side we have the Hilbert-Schmidt inner product between the two Choi matrices of $\mE$ and $\mF$.\index{Hilbert-Schmidt inner product}
\end{exercise}

\begin{myt}{}
\begin{theorem}
A linear map $\mE\in\ml(A\to B)$ is completely positive\index{completely positive} if and only if $J_\mE^{AB}\geq 0$.
\end{theorem}
\end{myt}

\begin{proof}
If $\mE$ is completely positive\index{completely positive} then by definition $J_\mE^{AB}\eqdef \mE^{\tA\to B}(\Omega^{A\tA})\geq 0$. Suppose now that $J_\mE^{AB}\geq 0$. Let
 $k\in\mbb{N}$, and $|\psi^{RA}\ra\in\mbb{C}^k\otimes\mbb{C}^d$, where $R$ is a $k$-dimensional (reference) system. Recall that any bipartite vector\index{bipartite vector} $|\psi\ra^{RA}$ can be expressed as
 \be
 |\psi\ra^{RA}=M\otimes I^A|\Omega^{\tA A}\ra\;,
 \ee
 where $M:{\tA}\to R$ is some linear operator. We therefore have
\ba
 (\id_k\otimes\mE)\left(|\psi^{RA}\lr\psi^{RA}|\right)&=(\id_k\otimes\mE)\left((M\otimes I^{A})\Omega^{A \tA}(M^{*}\otimes I^{A})\right)\\
 &=M\otimes I^{B}\left(\mE^{\tA\to B}\big(\Omega^{A\tA}\big)\right)M^{*}\otimes I^{B} \\
 &=\left(M\otimes I^{B}\right)J_\mE^{AB}\left(M^{*}\otimes I^{B}\right)\\
\GG{Exercise~\ref{posmab}}& \geq 0\;.
\ea 
Finally, any operator $\rho^{RA}\geq 0$ can be diagonalized as $\rho^{RA}=\sum_{x\in[m]}|\psi_x^{RA}\lr \psi_x^{RA}|$, where  $|\psi^{RA}_x\ra\in\mbb{C}^k\otimes\mbb{C}^d$ are some (possibly unnormalized) pure states. Since $|\psi^{RA}\ra$ above was arbitrary, we conclude that
\be
(\id_k\otimes\mE)\left(\rho^{RA}\right)=\sum_{x\in[m]}(\id_k\otimes\mE)\left(|\psi^{RA}_x\lr\psi^{RA}_x|\right)\geq 0\;,
\ee
since each term in the sum is positive semidefinite. This completes the proof.
\end{proof}

\begin{myg}{}
\begin{corollary}
A linear map $\mE\in\ml(A\to B)$ is $|A|$-positive if and only if it is completely positive.
\end{corollary}
\end{myg}
\begin{proof}
By definition if $\mE$ is a CP map then it is $|A|$-positive. Conversely, if $\mE$ is $|A|$-positive, then its Choi matrix\index{Choi matrix} $\mE^{\tA\to B}(\Omega^{A\tA})$ is positive semidefinite. Hence, from the theorem above $\mE$ is a CP map.
\end{proof}

\begin{myt}{}
\begin{theorem}
A linear map $\mE\in\ml(A\to B)$ is trace preserving if and only if the marginal state $J_\mE^{A}\eqdef\tr_B\left[J_\mE^{AB}\right]=I^A$.
\end{theorem}
\end{myt}

\begin{proof}
Suppose $\mE$ is trace preserving and set $m\eqdef|A|$. Then, from~\eqref{choi1}
\ba
\tr_B\left[J_\mE^{AB}\right]&=\sum_{x,y\in[m]}|x\lr y| \tr\left[\mE\left(|x\lr y|\right)\right]\\
\GG{\mE\;is\;trace\text{-}preserving}&=\sum_{x,y\in[m]}|x\lr y| \tr\left[|x\lr y|\right]\\
&=\sum_{x,y\in[m]}|x\lr y|\delta_{xy}=I^A\;.
\ea
Conversely, suppose $J_\mE^{A}=I^A$, then from~\eqref{choi2} for every $\rho\in\ml(A)$
\be
\tr\left[\mE(\rho)\right]=\tr\left[J_{\mE}^{AB}\left(\rho^T\otimes I\right)\right]=\tr\left[J_{\mE}^{A}\rho^T\right]=\tr[\rho^T]=\tr[\rho]\;.
\ee
This completes the proof.
\end{proof}

We therefore conclude that a linear map $\mE$ is a quantum channel if and only if its Choi matrix $J_\mE^{AB}\geq 0$, and its marginal $J_\mE^{A}=I^A$. In particular, the Choi matrix\index{Choi matrix} has trace $|A|$ so that $\frac{1}{|A|}J_\mE^{AB}\in\md(A\otimes B)$. Hence, the Choi representation reveals that quantum channels can be represented with bipartite quantum states. This equivalence between quantum channels and bipartite quantum states is used very often in quantum information science.

\begin{exercise}\label{exkpos}
Show that the linear map, $\mE^{(k)}$, defined in~\eqref{kpositive}, with $k<d$, is $k$-positive\index{$k$-positive} but not $(k+1)$-positive. 
\end{exercise}

\begin{exercise}
Let $\mE\in\ml(A\to A)$ be a quantum channel with the property that $\mE(U\rho U^*)=U\mE(\rho)U^*$ for any unitary matrix $U\in\muu(A)$. Show that its Choi matrix\index{Choi matrix} $J^{A\tA}_\mE$ must satisfy
\be
\left(\bar{U}\otimes U\right)J^{A\tA}_\mE\left(\bar{U}\otimes U\right)^*=J^{A\tA}_\mE
\ee
for all unitary matrices $U$.
\end{exercise}

\begin{exercise}
Show that any density matrix $\rho\in\md(AB)$ can be expressed as
\be
\rho^{AB}=\mE^{\tA\to B}\left(\psi^{A\tA}\right)
\ee
where $\psi\in\pure(A\tA)$ is pure bipartite state that has a marginal $\psi^A=\rho^A$, and $\mE\in\cptp(A\to B)$ is a quantum channel. Hint: look at $(\rho^A)^{-\frac12}\rho^{AB}(\rho^A)^{-\frac12}$.
\end{exercise}

\begin{exercise}
Show that if instead of the standard Choi representation above, one defines
\be
J^{AB}_{\mE}\eqdef\mE^{\tA\to B}\big(\psi^{A\tA}\big)
\quad
\text{where}
\quad
|\psi^{A\tA}\rangle\eqdef\sum_{x\in[m]}\sqrt{p_x} |x\rangle^{A}|x\rangle^{\tA}\ee
with $\{p_x\}$ a probability distribution, then
\be 
\mE\left(\rho\right)=\tr_{A}\left[J^{AB}_{\mE}\left(\sigma^{-1/2}U\rho^TU^*\sigma^{-1/2}\otimes I^B\right)\right]\;\;,
\ee
where $U$ is some fixed diagonal unitary on system $A$ and $\sigma\eqdef\tr_{\tA}\big[\psi^{A\tA}\big]$.
\end{exercise}

\subsection{The Operator Sum Representation}\label{osr}\index{operator sum}

In Fig.~\ref{fig10}c we considered a generalized measurement $\{M_x\}_{x\in[m]}$ on a physical system described by the state $\rho\in\md(A)$.
If outcome $x$ occurred then the state of the system $\rho$ changes to
$
\left(M_x\rho M_{x}^{*}\right)/p_x
$
, where $p_x$ is the probability that outcome $x$ occurred. However, if the value of $x$ is erased after the measurement, then the post-measurement state is given by the average over all the possible outcomes; that is,
\be
\sum_{x\in[m]}p_x\left(\frac{M_x\rho M_{x}^{*}}{p_x}\right)=\sum_{x\in[m]}M_x\rho M_{x}^{*}\;.
\ee
We now show that the mapping $\rho\mapsto \sum_{x\in[m]}M_x\rho M_{x}^{*}$ is a quantum channel and that every quantum channel can be realized in this way. This representation of a quantum channel is called the \emph{operator sum representation}, and the elements $\{M_x\}_{x\in[m]}$ (with $\sum_{x\in[m]}M_{x}^{*}M_x=I^A$) are called the \emph{Kraus operators}.
\begin{myt}{}
\begin{theorem}
A linear map $\mE\in\ml(A\to B)$ is a quantum channel if and only if it has an operator sum representation. That is, $\mE$ is a quantum channel if and only if there exists a set of Kraus operators $\{M_x\}_{x\in[m]}\subset\ml(A,B)$, such that 
\be\label{kraus}
\mE(\rho)=\sum_{x\in[m]}M_x\rho M_{x}^{*}\;.
\ee 
\end{theorem} 
\end{myt}

\begin{proof}
Suppose $\mE$ is a quantum channel. Since the Choi matrix\index{Choi matrix} of a quantum channel is positive semidefinite we can always express it as
\be\label{ini}
J_{\mE}^{AB}=\sum_{x\in[m]}|\psi_x^{AB}\lr\psi^{AB}_{x}|
\ee
for some integer $m$ and some (possibly unnormalized) vectors $|\psi_{x}^{AB}\ra\in A\otimes B$. Recall that any bipartite state $|\psi_{x}^{AB}\ra$ can be expressed as
\be\label{krauso}
|\psi_{x}^{AB}\ra=I^A\otimes M_x|\Omega^{A\tA}\ra=M_{x}^{T}\otimes I^B|\Omega^{\tB B}\ra\;,
\ee
where $M_x\in\ml(A,B)$ is a linear operator. Moreover, since the marginal Choi matrix\index{Choi matrix} $J_{\mE}^{A}=I^A$ we get
\ba
I^A
=\tr_B\big[J_\mE^{AB}\big]&=\tr_B\left[\sum_{x\in[m]}\left(M_{x}^{T}\otimes I^B\right)\Omega^{\tB B}\left((M_{x}^{*})^T\otimes I^B\right)\right]\\
&=\sum_{x\in[m]}M_{x}^{T}\;\tr_B\left[\Omega^{\tB B}\right](M_{x}^{*})^T=\sum_{x\in[m]}M_{x}^{T}(M_{x}^{*})^T
\ea
By taking the transpose on both sides of the equation above we get
\be
I^A=\sum_{x\in[m]}M_{x}^{*}M_x\;.
\ee
Moreover, substituting~\eqref{ini} into~\eqref{choi2} gives
\ba
\mE(\rho)&=\sum_{x\in[m]}\tr_A\left[|\psi_x^{AB}\lr\psi^{AB}_{x}|\left(\rho^T\otimes I\right)\right]\\
&=\sum_{x\in[m]}\tr_A\left[\left(I^A\otimes M_x\right)|\Omega^{A\tA}\lr \Omega^{A\tA}|\left(I^A\otimes M_{x}^{*}\right)\left(\rho^T\otimes I\right)\right]\\
&=\sum_{x\in[m]}M_x\left(\tr_A\left[|\Omega^{A\tA}\lr \Omega^{A\tA}|\left(\rho^T\otimes I\right)\right]\right)M_{x}^{*}
\ea
To simplify this last term, note that
\ba
\tr_A\left[|\Omega^{A\tA}\lr \Omega^{A\tA}|\left(\rho^T\otimes I\right)\right]&=\tr_A\left[|\Omega^{A\tA}\lr \Omega^{A\tA}|\left(I\otimes \rho\right)\right]\\
&=\tr_A\left[|\Omega^{A\tA}\lr \Omega^{A\tA}|\right]\rho=I^{\tA}\rho=\rho\;.
\ea
We therefore conclude that $\mE(\rho)=\sum_{x\in[m]}M_x\rho M_{x}^{*}$. 

To prove the converse, suppose that $\mE$ has the form~\eqref{kraus}. Then, clearly the Choi matrix\index{Choi matrix} 
$J_{\mE}^{AB}\eqdef\mE^{\tA\to B}(\Omega^{A\tA})$ has the form~\eqref{ini} with $|\psi_{x}^{AB}\ra\eqdef I^A\otimes M_x|\Omega^{A\tA}\ra$. Hence, $J_{\mE}^{AB}\geq 0$ and $J_{\mE}^{A}=I^A$ since $\sum_{x\in[m]}M_x^{*} M_{x}=I^A$. This completes the proof.
\end{proof}

\begin{exercise}
Let $\mE\in\ml(A\to B)$ be a linear map. 
\begin{enumerate}
\item Show that there exists two sets of matrices $\{M_x\}_{x\in[m]}$ and $\{N_x\}_{x\in[m]}$ such that
\be
\mE(\rho)=\sum_{x\in[m]}M_x\rho N_{x}^{*}\;.
\ee
Hint: Start by showing that it is possible to express the complex Choi matrix\index{Choi matrix} as $J_{\mE}^{AB}=\sum_{x\in[m]}|\psi_x\lr\phi_x|$, and then follow similar lines as in the proof above.
\item Show that the dual (adjoint) map $\mE^{*}:\ml(B\to A)$ is given by:\index{adjoint map}
\be
\mE^*(\rho)=\sum_{x\in[m]}M_{x}^{*}\rho N_x\;.
\ee
\item Show that $\mE\in\ml(A\to B)$ is completely positive\index{completely positive} if and only if its dual map $\mE^*\in\ml(B\to A)$ is completely positive.
\end{enumerate}
\end{exercise}

We next prove a uniqueness theorem of operator-sum representations.\index{uniqueness}

\begin{myt}{}
\begin{theorem}\label{uniqkraus}
Let $m,n\in\mbb{N}$ with $m\leq n$. Consider a generalized measurement $\{M_x\}_{x\in[m]}\subset\ml(A,B)$,  be a generalized measurement, and a set of matrices $\{N_y\}_{y\in[n]}\subset\ml(A,B)$. The following statements are equivalent:
\ben
\item The sets $\{M_x\}_{x\in[m]}$ and $\{N_y\}_{y\in[n]}$ constitute two operator sum representations of the same quantum channel.
\item There exists an $n\times m$ isometry $V=(v_{yx})$ such that for all $y\in[n]$:
\be\label{nxmxv}
N_y\eqdef\sum_{x\in[m]}v_{yx}M_x\;.
\ee
\een
\end{theorem}
\end{myt}

\begin{proof}
We first prove the implication $2\Rightarrow 1$. From~\eqref{nxmxv} it follows that
\ba
\sum_{y\in[n]}N_{y}^{*}N_y&=\sum_{x,x'\in[m]}\sum_{y\in[n]}\bar{v}_{yx}v_{yx'}M_{x}^{*}M_{x'}\\
\Gg{V^*V=I_m}&=\sum_{x,x'\in[m]}\delta_{xx'}M_{x}^{*}M_{x'}=\sum_{x\in[m]}M_x^{*} M_{x}=I^A\;,
\ea
so that $\{N_y\}_{y\in[n]}$ is a generalized measurement.
Similarly, for any $\rho\in\ml(A)$ we have
\ba
\sum_{y\in[n]}N_{y}\rho N_{y}^{*}&=\sum_{x,x'\in[m]}\sum_{y\in[n]}v_{yx}\bar{v}_{yx'}M_x\rho M_{x'}^{*}\\
\Gg{V^*V=I_m}&=\sum_{x,x'\in[m]}\delta_{xx'}M_x\rho M_{x'}^{*}=\sum_{x\in[m]}M_x\rho M_{x}^{*}\;.
\ea
Hence, the sets $\{M_x\}_{x\in[m]}$ and $\{N_y\}_{y\in[n]}$ are two operator sum\index{operator sum} representations of the same quantum channel. 

Next, we proof the implication $1\Rightarrow 2$. From the assumption we have in particular that for every $\psi\in\pure(A)$ we have
\be
\rho\eqdef\sum_{x\in[m]}M_x|\psi\lr\psi| M_{x}^{*}=\sum_{y\in[n]}N_y|\psi\lr\psi| N_{y}^{*}\;.
\ee
Since both $\{M_x|\psi\ra\}_{x\in[m]}$ and $\{N_y|\psi\ra\}_{y\in[n]}$ form an unnormalized pure-state decomposition of the same density matrix $\rho$, we get from Exercise~\ref{ensembles} that there exists an $n\times m$ isometry matrix $V=(v_{yx})$ such that for all $y\in[n]$
\be
N_y|\psi\ra=\sum_{x\in[m]}v_{yx}M_x|\psi\ra\;.
\ee
Since the above equality holds for all $\psi\in\pure(A)$ the relation~\eqref{nxmxv} must hold.
This completes the proof.
\end{proof}

\bex
Show that for every quantum channel $\mE\in\cptp(A\to B)$ there exists an operator-sum representation with no more than $|AB|$ elements.
\eex

\subsubsection{The Canonical Operator Sum Representation} \index{canonical representation}

Recall from Exercise~\ref{ensembles} that $J_{\mE}^{AB}$ has pure state decompositions that are all related via some isometry in the same manner as the two sets of the Kraus operators above are related.
Therefore, each operator sum representations of $\mE$ corresponds to a particular pure state decomposition of $J_{\mE}^{AB}$ as in~\eqref{ini}.  The \emph{canonical} operator sum representation of the quantum channel $\mE$ is the one corresponding to the diagonalization of $J_{\mE}^{AB}$. That is, in the canonical\index{canonical representation} representation we take the vectors $\{|\psi_x^{AB}\ra\}_{x\in[m]}$ in~\eqref{ini} to be orthogonal. This means that they are linearly independent and consequently we must have $m\leq |AB|$ since the rank of $J_{\mE}^{AB}$ cannot exceed $|A\otimes B|=|AB|$. Moreover, the orthogonality of the vectors $\{|\psi_x^{AB}\ra\}_{x\in[m]}$ implies that for $x\neq x'$
\be
0=\la\psi_{x}^{AB}|\psi_{x'}^{AB}\ra=\la\Omega^{A\tA}|M_{x}^{*}M_{x'}\otimes I^{\tA}|\Omega^{A\tA}\ra=\tr\left[M_{x}^{*}M_{x'}\right]\;.
\ee
That is, the Kraus operators are also orthogonal in the Hilbert-Schmidt inner product. We therefore arrived at the following corollary.\index{Hilbert-Schmidt inner product}
\begin{myg}{}
\begin{corollary}\label{canonical}
Let $\mE\in\ml(A\to B)$ be a quantum channel. Then $\mE$ has a \emph{canonical} operator sum\index{operator sum} representation $\{M_{x}\}_{x\in[m]}$, with $m\leq |AB|$, $\sum_{x\in[m]}M_x^{*} M_{x}=I^A$, and for each
$x\neq x'$
\be
\tr\left[M_{x}^{*}M_{x'}\right]=0\;.
\ee
\end{corollary}
\end{myg}
\noindent In particular, there are always operator sum representations with no more than $|AB|$ Kraus operators.

\begin{exercise}\label{canfree}
Let $\{M_x\}_{x\in[m]}$ be a canonical\index{canonical representation} Kraus decomposition of $\mE\in\cptp(A\to B)$. Show that for any $m\times m$ unitary matrix $U=(u_{yx})$, also $\{N_y\}_{y\in[n]}$ with 
\be
N_y\eqdef\sum_{x\in[m]}u_{yx}M_x\;,
\ee
is a canonical Kraus decomposition of $\mE$.
\end{exercise}

\subsection{The Unitary Representation}\label{sec:stine}\index{unitary representation}

\begin{figure}[h]\centering
    \includegraphics[width=0.4\textwidth]{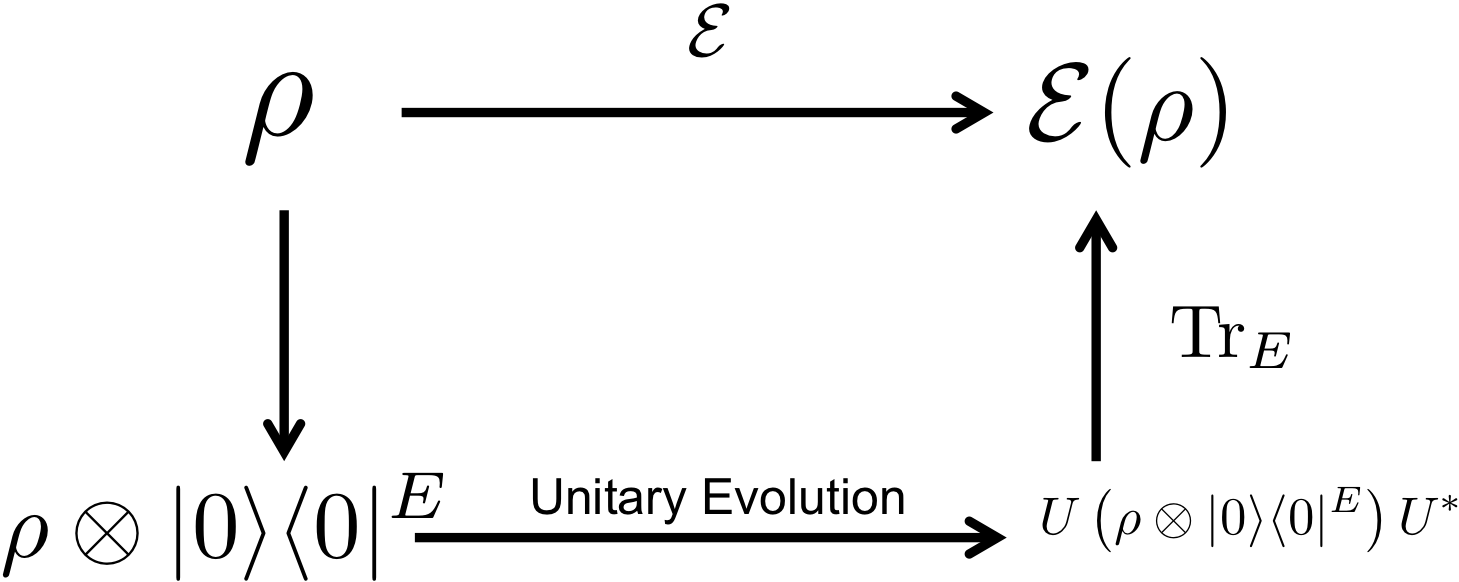}
  \caption{\linespread{1}\selectfont{\small Stinespring Dilation.}}
  \label{fig12}
\end{figure}

In the previous section we saw that quantum measurement can be used to realize any CPTP map. Here we show that a deterministic unitary evolution can be used to realize a quantum channel. Specifically, in Fig.~\ref{fig12} below a system $A$ is assumed to be initially uncorrelated with the environment. The initial state of the system is denoted by $\rho^A$ and the initial state of the environment by $|0\lr 0|^E$. Then, the system+environment undergoes a unitary evolution  which converts the initial state $\rho^A\otimes |0\lr 0|^E$ to the state
\be
U\left(\rho^A\otimes |0\lr 0|^E\right)U^*\quad;\quad U^*U=I^{AE}\;.
\ee
Finally, the environment system is traced out yielding the final state
\be\label{unitary}
\mE(\rho^{A})\eqdef\tr_E\left[U\left(\rho^A\otimes |0\lr 0|^E\right)U^*\right]\;.
\ee

We show now that every quantum channel can be realized in this way, giving a new interpretation for quantum channels as joint unitary\index{joint unitary} evolutions on the system plus environment. Recall that typically, the degrees of freedom of the environment are not accessible and therefore they are traced out at the end of the process.

\begin{myt}{\color{yellow} Stinespring Dilation Theorem}\index{Stinespring dilation}
\begin{theorem}\label{thm:stine}
A linear map $\mE\in\ml(A\to B)$ is a quantum channel if and only if there exist an environment (ancillary) system (and corresponding Hilbert space) $E$ of dimension $|E|\leq |AB|$, and an isometry $V:A\to B\otimes E$ with $V^*V=I^{A}$ such that
\be\label{isometry0}
\mE(\rho^{A})\eqdef\tr_{E}\left[V\rho^AV^*\right]\quad\quad\forall\;\rho\in\ml(A)\;.
\ee
\end{theorem}
\end{myt}

\begin{remark}
The theorem above is an adaptation of Stinespring Dilation Theorem to the finite dimensional case.
\end{remark}

\begin{proof}
Suppose $\mE$ has the form~\eqref{isometry0}. 
To show that $\mE$ is a quantum channel we denote by $\{|\phi^E_z\ra\}_{z\in[k]}$ an orthonormal basis of $E$, where $k\eqdef|E|$, and by $M_z\eqdef \la \phi^E_z|V$. By definition, for every $z\in[k]$, $M_z:A\to B$, and from~\eqref{isometry0} we get
\be\label{lilatov}
\mE(\rho^{A})=\sum_{z\in[k]}\la \phi^E_z|V\rho^AV^*|\phi^E_z\ra=\sum_{z\in[k]}M_z\rho M_z^{*}\;.
\ee
Moreover, since $V^*V=I^A$ we obtain
\be
\sum_{z\in[k]}M_{z}^{*}M_z=\sum_{z\in[k]}V^*|\phi^E_z\lr \phi^E_z|V=V^{*}I^EV=V^*V=I^A\;.
\ee

Conversely, suppose $\mE$ is a quantum channel. Then, from the previous section it has an operator sum representation $\mE(\rho)=\sum_{z\in[k]}M_z\rho M_{z}^{*}$, with $k\leq |A||B|$ and $\sum_{z\in[k]}M_{z}^{*}M_z=I^A$.
Set $E\eqdef\mbb{C}^k$ and define the map $V:A\to BE$ via
\be\label{v123form}
V=\sum_{z\in[k]}M_z\otimes|\phi^E_z\ra\;.
\ee
From its definition,
\be
V^*V=\sum_{z',z\in[k]}M_{z'}^*M_z\la \phi_{z'}^E|\phi^E_z\ra=\sum_{z\in[k]}M_{z}^*M_z=I^{A}\;,
\ee
so that $V$ is an isometry. Moreover, from the definition of $V$ we get $M_z= \la \phi^E_z|V$, so that
\ba\label{3143}
\tr_{E}\left[V\rho^AV^*\right]&=\sum_{z\in[k]}\la \phi^E_z|V\rho^AV^*|\phi^E_z\ra\\
&=\sum_{z\in[k]}M_z\rho M_{z}^{*}=\mE(\rho)\;.
\ea
Hence, there exists an isometry $V:A\to BE$ such that~\eqref{isometry0} holds. This completes the proof.
\end{proof}

\bex
Consider the isometry $V:A\to BE$ as expressed in~\eqref{v123form}, where each $M_z:A\to B$. Let $\{|x\ra^A\}_{x\in[m]}$ be an orthonormal basis of $A$, and $\{|\psi_{x}^{BE}\ra\}_{x\in[m]}$ be an orthonormal set of vectors in $B E$, such that
\be
V=\sum_{x\in[m]}|\psi_{x}^{BE}\ra{}^A\!\la x|\;.
\ee
Finally, express each $|\psi_{x}^{BE}\ra$  as
\be
|\psi_{x}^{BE}\ra=\sum_{z\in[k]}|\phi_{z|x}^B\ra|\phi^E_z\ra
\ee
with some vectors $\{|\phi_{z|x}^B\ra\}_{z\in[k]}\subset B$. Show that with these notations
\be
M_z=\sum_{x\in[m]}|\phi_{z|x}^B\ra{}^A\!\la x|\;.
\ee
\eex

\begin{exercise}
Show that a linear map $\mE\in\ml(A\to A)$ is a quantum channel if and only if there exists an environment system $E$ and a unitary matrix $U:AE\to AE$ such that $\mE$ has the form~\eqref{unitary}. Hint: Complete the isometry $V$ in the above theorem into a unitary operator.
\end{exercise}

Note that in the proof above we defined the isometry $V$ in~\eqref{v123form} using the Kraus operators. Therefore,
Eq.~\eqref{v123form} provides a direct relationship between the Stinespring representation and the operator sum\index{operator sum} representation. Moreover, the operator sum representation is directly related to the Choi representation via the relationship in Eqs.(\ref{ini},\ref{krauso}). Therefore, together with~\eqref{v123form} we can establish a direct relationship among all three representations. We will use these relationships quite often in the next sections.

\subsubsection{How unique is Stinespring dilation?}\index{uniqueness} 

In the following theorem we address the uniqueness of the isometry $V$ that appears in Stinespring dilation theorem.

\begin{myt}{}
\begin{theorem}
Let $\mE\in\cptp(A\to B)$, $V:A\to BE$ be an isometry satisfying~\eqref{isometry0}, and let $W:A\to BE$ be a linear operator. The following statements are equivalent:
\ben
\item There exists a unitary matrix $U^E:E\to E$ such that
\be
W=\left(I^B\otimes U^E\right)V\;.
\ee 
\item For all $\rho\in\ml(A)$
\be\label{isometry00}
\mE(\rho)=\tr_E\left[W\rho W^*\right]\;.
\ee
\een
\end{theorem}
\end{myt}
\begin{proof}
The proof of the implication $1\Rightarrow 2$ is straightforward and is left as an exercise. We will now focus on proving the implication $2\Rightarrow 1$.
Following the methodology used in Stinesprings' dilation theorem, let us denote by $\{|\phi^E_z\ra\}_{z\in[k]}$ an orthonormal basis of $E$, where $k\eqdef|E|$. For each $z\in[k]$, we define $M_z\eqdef \la \phi^E_z|V$ and $N_z\eqdef \la \phi^E_z|W$.
Note that for all $z\in[k]$, $M_z$ and $N_z$ are linear operators from $A$ to $B$. In the proof of Stinesprings' theorem, particularly in~\eqref{3143}, it was demonstrated that $\{M_z\}_{z\in[k]}$ constitutes an operator sum\index{operator sum} representation of $\mE$. Similarly, from~\eqref{isometry00}, we get
\be\label{lilatov}
\mE(\rho^{A})=\sum_{z\in[k]}\la \phi^E_z|W\rho^AW^*|\phi^E_z\ra=\sum_{z\in[k]}N_z\rho N_z^{}\;.
\ee
Therefore, $\{N_z\}_{z\in[k]}$ also forms an operator sum representation of $\mE$. Observe in particular that the property that $\mE$ is trace preserving implies that
\ba
I^A&=\sum_{z\in[k]} N_z^{*}N_z\\
\Gg{N_z\eqdef \la \phi^E_z|W}&=\sum_{z\in[k]}W^*|\phi_z^E\lr \phi_z^E|W\\
\Gg{\sum_{z\in[k]}\phi_z^E=I^E}&=W^*W\;.
\ea
Thus, $W$ is an isometry.
Given that both $\{M_z\}_{z\in[k]}$ and $\{N_z\}_{z\in[k]}$ are operator-sum representations of $\mE$, Theorem~\ref{uniqkraus} implies the existence of a $k\times k$ unitary matrix $V=(v_{zw})$, such that for every $z\in[k]$
\ba
N_z=\sum_{w\in[k]}u_{zw}M_w\;.
\ea
This leads us to (cf.~\eqref{v123form})
\ba
W&=\sum_{z\in[k]}N_{z}\otimes|\phi^E_z\ra\\
&=\sum_{w,z\in[k]}v_{zw}M_{w}\otimes|\phi^E_z\ra\\
&=\sum_{w\in[k]}M_{w}\otimes\sum_{z\in[k]}v_{zw}|\phi^E_z\ra\;.
\ea
Defining the matrix $U^E\eqdef V^T$, we find for all $w\in[k]$
\be
U^E|\phi_w^E\ra=\sum_{z\in[k]}v_{zw}|\phi^E_z\ra\;.
\ee
Thus, we conclude
\ba
W&=\left(I^B\otimes U^E\right)\sum_{w\in[k]}M_w\otimes |\phi^E_w\ra\\
\GG{\eqref{v123form}}&=\left(I^B\otimes U^E\right)V\;.
\ea
This concludes the proof.
\end{proof}

\section{Examples of Quantum Channels}

In this section we list and discuss briefly several examples of quantum channels. These channels have many interesting properties that make the interplay among the different representations of a quantum channel more apparent. These channels appear quite often in the field of quantum information, and we will also encounter some of them later on in the book.

\subsection{Qubit Channels}\index{qubit channels}

We discuss a few common examples of qubit channels; i.e. CPTP maps from $\ml(\mbb{C}^2)$ to itself.
Unlike quantum channels in higher dimensions, qubit channels can be characterized by their effect on the Bloch vector.
This is a convenient property that is useful for some applications. We start with a communication channel that represents the most basic error in information theory; namely, the bit flip.

\subsubsection{The Classical Bit Flip}\index{bit-flip}

\begin{figure}[h]\centering
    \includegraphics[width=0.13\textwidth]{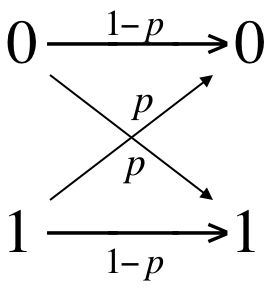}
  \caption{\linespread{1}\selectfont{\small The symmetric classical bit-flip channel.}}
  \label{bitflip}
\end{figure}

The classical bit flip channel (see Fig.~\ref{bitflip}) is a process that flips a classical bit according to some probability distribution. Specifically, the zero is flipped to one with some probability $p$ and remain unchanged with probability $1-p$. The bit-flip channel is symmetric if the probability to flip one is given by the same probability $p$.

\subsubsection{The Quantum Bit Flip}\index{bit-flip}

The quantum bit-flip channel act in a similar way. The bit flip operator is the Pauli first matrix 
$X=\begin{pmatrix} 0 & 1\\ 1 & 0\end{pmatrix}$, since $X|0\lr 0|X=|1\lr 1|$ and $X|1\lr 1|X=|0\lr 0|$.
Denote by 
\be
M_0=\sqrt{1-p}I\quad\text{and}\quad M_1=\sqrt{p} X\;,
\ee
and note that $M^{*}_0M_0+M_1^*M_1=I$. The qubit channel $\mE\in\ml(A\to B)$ (with $|A|=|B|=2$) 
defined by $\mE(\rho)=M_0\rho M_0^*+M_1\rho M_1^*$ is called the quantum bit flip channel. Its Choi representation is given by
\be\label{3p146}
J^{AB}_\mE=\mE^{\tA\to B}(\Omega^{A\tA})=2(1-p)\Phi^{AB}_++2p\;\Psi_{+}^{AB}\;,
\ee
where $|\Phi_{\pm}^{AB}\ra=\frac1{\sqrt{2}}(|00\ra\pm|11\ra)$ and $|\Psi_{\pm}^{AB}\ra=\frac1{\sqrt{2}}(|01\ra\pm|10\ra)$ form the \emph{Bell basis} of the two qubit system.\index{Bell basis}

\subsubsection{The Depolarizing Channel}\index{depolarizing channel}

Another very common example of a qubit channel is the depolarizing channel. It is defined by
\be\label{depo}
\mE(\rho)=\frac{p}{2}\tr[\rho]I+(1-p)\rho\quad\quad\forall\;\rho\in\ml(\mbb{C}^2).
\ee
Note that this channel has the unique property that for any unitary matrix  $U\in\ml(\mbb{C}^2)$
\be
\mE(U\rho U^*)=U\mE(\rho)U^*\;.
\ee
\begin{exercise}
Show that the depolarizing channel is indeed a quantum channel, by showing that it has an operator sum representation that is given in terms of the following four Kraus operators:
\be
M_0=\sqrt{1-\frac{3p}{4}}I\quad\text{and for }j\in[3]\;,\;\;M_j=\frac{\sqrt{p}}{2} \sigma_j\;,
\ee
where $\sigma_1$, $\sigma_2$, and $\sigma_3$, are the three Pauli\index{Pauli} matrices.
\end{exercise}
From the exercise above it also follows that the normalized version of the Choi matrix of the depolarizing channel has the form
\be
\frac12J^{AB}_\mE=\mE^{\tA\to B}\left(\Phi_{+}^{A\tA}\right)
=\left(1-\frac{3p}{4}\right)\Phi_{+}^{AB}+\frac{p}{4}\left(\Phi_{-}^{AB}+\Psi_{+}^{AB}+\Psi_{-}^{AB}\right)\;.
\ee
The state above is known (up to local unitary) as the 2-qubit isotropic state\index{isotropic state} and is used quite often in quantum information as it has several interesting properties. We will discuss it in more details later on.

The Stinespring isometry of the depolarizing channel can also be computed from~\eqref{v123form} and the exercise above; it is given by
\be\label{depolarz}
V=\sum_{j=0}^{3}M_j\otimes|j\ra^E=\sqrt{1-\frac{3p}{4}}I\otimes|0\ra+\frac{\sqrt{p}}2\sum_{j\in[3]}\sigma_j\otimes |j\ra
\ee
where $\sigma_1$, $\sigma_2$, and $\sigma_3$, are the three Pauli\index{Pauli} matrices.
Interestingly, the following exercise shows that the Bloch representation\index{Bloch representation} is in some sense the simplest representation of the depolarizing channel.

\begin{exercise}
Let $\rho=\frac{1}{2}\left(I+\r\cdot\bs{\sigma}\right)$ and $\rho'=\frac{1}{2}\left(I+\r'\cdot\bs{\sigma}\right)$ be two Bloch representations of two quantum states, and let $\mE$ be the depolarizing channel~\eqref{depo}. Show that if $\rho'=\mE(\rho)$ then $\r'=(1-p)\r$.
\end{exercise}

\subsection{The Completely Dephasing Channel}\label{cdcha}\index{dephasing channel}

The completely dephasing map, sometimes also referred to as the completely decohering map, is the channel
$
\Delta\in\cptp(A\to A)\;,
$
that removes the off-diagonal terms from any matrix $\rho\in\ml(A)$, with respect to some fixed orthonormal basis $\{|x\ra\}_{x\in[m]}$, where $m\eqdef |A|$.
Specifically, its Kraus representation is given by $\{M_x\eqdef |x\lr x|\}_{x\in[m]}$. Denoting by $\{r_{xy}\}_{x,y\in[m]}$ the matrix elements of $\rho$ with respect to the basis $\{|x\ra\}_{x\in[m]}$, we get from the operator sum representation of $\Delta$ that
\be
\Delta(\rho)=\sum_{x\in[m]}|x\lr x|\rho|x\lr x|=\sum_{x\in[m]}r_{xx}|x\lr x|\;.
\ee
From the above representation of $\Delta$, it is clear that the completely dephasing map is \emph{idempotent}, that is, it satisfies \be\Delta^2\eqdef\Delta\circ\Delta=\Delta\;.\ee
Its Choi matrix is
\be
J_{\Delta}=\Delta^{\tA\to \tA}(\Omega^{A\tA})=\sum_{x\in[m]}|x\lr x|\otimes|x\lr x|\;.
\ee
Its Stinespring isometry $V_\Delta:A\to A\otimes E$ is given by:
\be
V_\Delta|x\ra^A=|x\ra^A|x\ra^E
\ee
where $\{|x\ra^E\}$ is an orthonormal basis of $E$.

\begin{exercise}
A generalized dephasing channel $\mN\in\cptp(A\to A)$ is a channel that transmit some preferred basis  
$\{|x\ra\}_{x\in[m]}$ of $A$ without error. That is,
\be
\mN(|x\lr x|)=|x\lr x|\;.
\ee
\begin{enumerate}
\item Show that the completely dephasing channel $\Delta$ is a generalized dephasing channel.
\item Show that $\mN$ above has a Stinespring isometry 
\be
V_\mN|x\ra^A=|x\ra^A|\phi_x^E\ra\quad\quad\forall\; x\in[m]\;,
\ee 
where $\{|\phi_x^E\ra\}_{x\in[m]}$ are some normalized vectors in $E$ (note that if $\{|\phi_x^E\ra\}$ is an orthonormal set then $\mN=\Delta$).
\item Show that 
\be
\Delta\circ\mN=\mN\circ\Delta=\Delta\;.
\ee
\end{enumerate}
\end{exercise}

\subsection{Classical Channels}\index{classical channel}

A classical channel is a stochastic process that converts an input (classical) variable $X=x$ into an output variable $Y=y$. 
The classical ``noise" is modelled by some conditional probability distribution $\pr(Y=y|X=x)$ that determines the probability that the output variable $Y$ is equal to $y$ given the input variable $X$ equals $x$. Denote by $p_x\eqdef\pr(X=x)$ with $x\in[m]$, and $q_y=\pr(Y=y)$ with $y\in[n]$, the (marginal) distributions associated with the random variables $X$ and $Y$. 
Then the probability vectors  $\p=(p_1,\ldots,p_{m})^T$ and $\q=(p_1,\ldots,p_{n})^T$ are related by
\be\label{classical}
\q=T\p
\ee
where the $n\times m$ evolution matrix $T=(t_{y|x})$, also known as the \emph{transition matrix}, has components $t_{y|x}=\pr(Y=y|X=x)$. This is the Shannon's model for a classical channel. \index{transition matrix}

A classical channel\index{classical channel} can be viewed as a very special type of a quantum channel. In particular, we say that
a quantum channel $\mE\in\cptp(A\to B)$ is a classical channel if 
\be
\Delta^B\circ\mE\circ \Delta^A=\mE\;,
\ee
where $\Delta^A\in\cptp(A\to A)$ and $\Delta^B\in\cptp(B\to B)$ are the completely dephasing channels with respect to some fixed classical bases $\{|x\ra\}_{x\in[m]}$ and $\{|y\ra\}_{y\in[n]}$ of $A$ and $B$, respectively. From its definition it is clear that $\mE(\rho)=\mE(\Delta^A(\rho))$.
Therefore, it is sufficient to consider only diagonal input states. Moreover, if $\sigma=\mE(\rho)$ then $\Delta^{B}(\sigma)=\sigma$; i.e. the output state is always diagonal in the basis 
$\{|y\lr y|\}_{y\in[n]}$. Hence, taking $\rho=\sum_{x\in[m]}p_x|x\lr x|^A$ we get
\be
\sigma\eqdef\mE(\rho)=\Delta^B\left((\mE(\rho)\right)=\sum_{y\in[n]}\la y\left|\mE(\rho)\right|y\ra |y\lr y|^B\;.
\ee
Denote by $t_{y|x}\eqdef \la y\left|\mE(|x\lr x|)\right|y\ra \geq 0$, and note that $\sum_{y\in[n]}t_{y|x}=\tr[\mE(|x\lr x|)]=1$ since $\mE$ is trace preserving. We therefore have
\be
q_y\eqdef\la y|\sigma|y\ra=\la y|\mE(\rho)|y\ra=\sum_{x\in[m]}p_x \la y|\mE(|x\lr x|)|y\ra =\sum_{x\in[m]}t_{y|x}p_x\;.
\ee
In other words, we get that $\q=T\p$ as in~\eqref{classical}, with $\p$ and $\q$ being the vectors whose components are the diagonal elements of $\rho$ and $\sigma$, respectively, and $T=(t_{y|x})$ being the column stochastic matrix whose components are $\la y|\mE(|x\lr x|)|y\ra$.

\begin{exercise}
Let $\{\rho_z\}_{z\in[k]}$ and $\{\sigma_z\}_{z\in[k]}$
be two sets of $k$ \textbf{diagonal} density operators with respect to two fixed bases $\{|x\ra\}_{x\in[m]}$ and $\{|y\ra\}_{y\in[n]}$ of $A$ and $B$, respectively. Show that there exist a quantum channel $\mE\in\cptp(A\to B)$ such that $\sigma_z=\mE(\rho_z)$ for all $z\in[k]$ if and only if there exists a classical channel\index{classical channel} with the same property.
\end{exercise}

\subsection{POVM Channels}\label{sec:povmc}\index{POVM channel}

A quantum channel $\mE\in\cptp(A\to B)$ is called a \emph{quantum to classical} channel, or POVM channel, if
\be
\Delta^B\circ\mE^{A\to B}=\mE^{A\to B}\;,
\ee
where $\Delta^B$ is the completely dephasing map on the output space with respect to some fixed basis $\{|y\ra\}_{y\in[n]}$ of $B$, where $n\eqdef|B|$. Such channels have the property that for any $\rho\in\md(A)$
\be
\mE(\rho)=\Delta\left(\mE(\rho)\right)=\sum_{y\in[n]}\la y|\mE(\rho)|y\ra|y\lr y|\;.
\ee
Note that
\be
\la y|\mE(\rho)|y\ra=\tr\left[|y\lr y|\mE(\rho)\right]=\tr\left[\mE^{*}(|y\lr y|)\rho\right]\eqdef\tr\left[\Lambda_y\rho\right]
\ee
where $\{\Lambda_y\eqdef\mE^{*}(|y\lr y|)\}_{y\in[n]}$ are  positive semidefinite matrices in $\pos(A)$ satisfying
\be
\sum_{y\in[n]}\Lambda_y=\sum_{y\in[n]}\mE^{*}(|y\lr y|)=\mE^{*}(I^B)=I^A\;,
\ee
since the dual map\index{dual map} of any CPTP map is unital (see Exercise~\ref{showdual}).
We therefore get that a POVM channel has the form
\be\label{chpovm}
\mE^{A\to B}\left(\rho^A\right)=\sum_{y\in[n]}\tr\left[\Lambda_y^A\rho^A\right]|y\lr y|^B\;.
\ee

\begin{exercise}
Let $\mE\in\cptp(A\to B)$ be a POVM channel as in~\eqref{chpovm}. 
\begin{enumerate}
\item Show that its Choi matrix is given by
\be
J^{AB}_{\mE}=\sum_{y\in[n]}\Lambda_{y}^{T}\otimes |y\lr y|\;.
\ee
\item Show that the operators $M_{xy}:{A}\to B$ with 
\be
M_{xy}\eqdef |y^B\ra\la x^A|\sqrt{\Lambda_y}
\ee 
can be used to form an operator sum representation of $\mE$.
\item Show that the map $V:A\to B\otimes \tA\tB$  given by
\be
V=\sum_{y\in[n]}|y\ra^B\otimes\sqrt{\Lambda_y^{A\to\tA}}\otimes|y\ra^{\tB}
\ee
is a Stinespring isometry (the environment system is $\tA\tB$) satisfying
\be
\tr_{\tA\tB}\left[V\rho V^{*}\right]=\sum_{y\in[n]}\tr\left[\Lambda_y\rho\right]|y\lr y|^B=\mE(\rho)\quad\text{and}\quad V^{*}V=I^A\;.
\ee
Hint: Use~\eqref{v123form} and show that $\sum_x{}^A\la x|\sqrt{\Lambda_y^{A\to A}}\otimes |x\ra^{\tA}=\sqrt{\Lambda_y^{A\to \tA}}$.
\end{enumerate}
\end{exercise}

\subsection{Preparation Channels}\index{preparation channel}

A preparation channel, commonly referred to as a classical-quantum (cq) channel, is a linear map that transforms classical inputs into quantum states. To elaborate, a channel $\mE\in\cptp(A\to B)$ is classified as a cq-channel if it fulfills the following criterion:
\be
\mE^{A\to B}\circ \Delta^{A\to A}=\mE^{A\to B}\;.
\ee
Hence, for every $\rho\in\md(A)$ we get
\be\label{prech}
\mE(\rho)=\sum_{x\in[m]}p_x\mE(|x\lr x|)=\sum_{x\in[m]}p_x\sigma_x
\ee
where $m\eqdef|A|$, $p_x\eqdef \la x|\rho|x\ra$ for all $x\in[m]$, 
and each $\sigma_x\eqdef\mE(|x\lr x|)$ is a fixed density matrix in $\md(B)$. We can therefore view $\mE$ 
as the mapping $x\mapsto\sigma_x$. The Choi matrix of a cq-channel has the form
\ba
J^{AB}_{\mE}=\mE^{\tA\to B}\big(\Omega^{A\tA}\big)&=\mE^{\tA\to B}\circ\Delta^{\tA\to\tA}\big(\Omega^{A\tA}\big)\\
&=
\mE^{\tA\to B}\Big(\sum_{x\in[m]}|xx\lr xx|^{A\tA}\Big)\\
&=\sum_{x\in[m]}|x\lr x|^A\otimes\sigma_x^{B}\;.
\ea
Therefore, $\mE\in\cptp(A\to B)$ is a cq-channel if and only if its Choi matrix  $J^{AB}_{\mE}$ is 
a cq-state.

\begin{exercise}
Let $\mE\in\cptp(A\to B)$ be a cq-channel as above, and set $m\eqdef|A|$ and $n\eqdef|B|$. 
\begin{enumerate}
\item Show that the set of Kraus operators $\{M_{xy}\}$ with $M_{xy}:A\to B$
\be
M_{xy}\eqdef\sqrt{\sigma_x}|y^B\lr x^A|\quad x\in[m]\;\;,\;\;y\in[n]\;,
\ee 
forms an operator sum representation of $\mE$.
\item Show that the map $V:A\to B\otimes {\tA\tB}$  given by
\be
V=\sum_{x\in[m]}\left(\sqrt{\sigma_x^{B}}\otimes I^{\tB}\right)|\Omega^{B\tB}\ra\otimes |x^{\tA}\lr x^A|
\ee
is a Stinespring isometry (the environment system is $\tA\tB$) satisfying
\be
\tr_{\tA\tB}\left[V\rho V^{*}\right]=\sum_{x\in[m]}p_x\sigma_x=\mE(\rho)\quad\text{and}\quad V^{*}V=I^A\;.
\ee
\end{enumerate}
\end{exercise}

\subsection{Measurement-Prepare Channels and Entanglement Breaking Channels}\index{measurement-prepare channel}\index{entanglement breaking}

A measurement-prepare channel represents a process where a quantum system is first measured, and then, depending on the outcome of this measurement, a specific quantum state is prepared. More specifically, a measurement-prepare channel, $\mE\in\cptp(A\to B)$, is a type of quantum channel characterized by the following form:
\be\label{eb}
\mE(\rho^A)=\sum_{z\in[k]}\tr\left[\Lambda_z^A\rho^A\right]\sigma^{B}_{z}
\ee
where $\{\Lambda_z\}_{z\in[k]}\subset\pos(A)$ is a POVM with $m$ outcomes, and $\{\sigma^{B}_{z}\}_{z\in[k]}$ are $k$ quantum states in $\md(B)$. Clearly, the map $\mE$ above is linear and trace preserving. Moreover, setting $m\eqdef|A|$ we get that its Choi matrix is given by
\ba
J^{AB}_{\mE}=\mE^{\tA\to B}(\Omega^{A\tA})&=\sum_{x,x'\in[m]}|x\lr x'|\otimes \mE(|x\lr x'|)\\
&=\sum_{z\in[m]}\sum_{x,x'\in[m]}\tr\left[\Lambda_z|x\lr x'|\right]|x\lr x'|\otimes\sigma^{B}_{z}\\
&=\sum_{z\in[k]}\Lambda_z^T\otimes\sigma^{B}_{z}\geq 0\;.
\ea
Therefore, $\mE$, in~\eqref{eb} is a quantum channel. Observe that the Choi matrix above is separable.

Measurement-prepare channels can be viewed as a combination of a POVM channel (i.e. qc-channel) followed by a preparation Channel\index{preparation channel} (i.e. cq-channel). Specifically, let $\mM\in\cptp(A\to Z)$ be a POVM Channel\index{POVM channel}  given as in~\eqref{chpovm} by
\be
\mM^{A\to Z}(\rho^A)\eqdef\sum_{z\in[k]}\tr\left[\Lambda_z^A\rho^A\right]|z\lr z|^Z\;,
\ee
for some POVM $\{\Lambda_z\}_{z\in[k]}$ in $\pos(A)$, and a classical system $Z$ of dimension $|Z|=k$. Let $\mP\in\cptp(Z\to B)$ be a preparation Channel\index{preparation channel} given as in~\eqref{prech} via
\be
\mP^{Z\to B}(|z\lr z|^Z)=\sigma_z^B\quad\quad\forall\;z\in[k]\;,
\ee
where $\sigma_z\in\md(B)$ are some density matrices. Then, the measurement-prepare channel $\mE$ given in~\eqref{eb} can be expressed as
\be\label{mpc}
\mE^{A\to B}=\mP^{Z\to B}\circ\mM^{A\to Z}\;.
\ee

\begin{exercise}
Let $A$, $B$ and $R$, be three Hilbert spaces and let $\Delta\in\cptp(R\to R)$ be the completely dephasing map with respect to some fixed basis of $R$. Show that for any two quantum channels $\mE\in\cptp(A\to R)$ and $\mF\in\cptp(R\to B)$ the channel
\be
\mF^{R\to B}\circ\Delta^{R}\circ\mE^{A\to R}
\ee
is a measurement prepare channel in $\cptp(A\to B)$.
\end{exercise}

\begin{exercise}
Let $\mE\in\cptp(B\to B')$ be a measurement-prepare channel as above, and let $\rho^{AB}\in\md(A\otimes {B})$ be a bipartite density operator. Show that the density operator
\be
\sigma^{AB'}\eqdef \mE^{B\to B'}(\rho^{AB})
\ee
is separable. Hint: Use~\eqref{mpc} and consider the cq-state $\tau^{ZB}\eqdef\mM^{A\to Z}(\rho^{AB})$.
\end{exercise}

The above exercise demonstrates that a measurement-prepare channel breaks the entanglement when applied to a subsystem of a composite bipartite system. Channels with this property are called \emph{entanglement breaking channels}. Therefore, measurement-prepare channels are entanglement breaking. It turns out that the converse is also true! That is, any entanglement breaking\index{entanglement breaking} channel can be represented as a measurement-prepare channel. For this reason, we will use, depending on the context, both terms interchangeably. 

\begin{exercise}
Show that any entanglement breaking channel is a measurement-prepare channel. Hint: start by observing that the Choi matrix of any entanglement breaking channel must be separable.
\end{exercise} 

\bex
Let $\mE^{A\to B}$ be the quantum channel defined in~\eqref{eb}. Show that its dual is given by
\be
\mE^{*B\to A}\left(\eta^B\right)=\sum_{z\in[m]}\tr\left[\eta^B\sigma_z^B\right]\Lambda_z^A\;.
\ee 
\eex

\subsection{The Partial Trace}\index{partial trace}

The partial trace as defined in Exercise~\ref{partialtrace} is a linear map $\tr_B\in\ml(AB\to A)$. What is its Choi matrix? Let 
\be\label{3p197}
\Omega^{(AB)(\tA\tB)}=\Omega^{A\tA}\otimes\Omega^{B\tB}
\ee 
be the unnormalized maximally entangled state between system $AB$ and system $\tA\tB$. Then, by definition, its Choi matrix is
\ba
J^{(AB)\tA}_{\mE}&=\id^{AB}\otimes\tr_{\tB}\left(\Omega^{(AB)(\tA\tB)}\right)\\
\GG{\eqref{3p197}}&=\Omega^{A\tA}\otimes I^{B}\geq 0.
\ea
Moreover, note that the marginal $\tr_{\tA}\left[J^{(AB)\tA}_{\mE}\right]=I^{AB}$. We therefore conclude that the partial trace is a quantum channel. Physically, it represents the process of discarding a subsystem from a composite system.

\begin{exercise}
Let $A$ and $B$ be two Hilbert spaces, and let $\{|y\ra^B\}_{y\in[n]}$ be an orthonormal basis of $B$ (with $n\eqdef|B|$). Show that the set of operators $\{M_{y}\}_{y\in[n]}\subset\ml(AB,A)$ given by
\be
M_{y}=I^{A}\otimes\la y|^{B}\;,
\ee
form an operator sum representation of the partial trace. 
\end{exercise}

\subsection{Isometry Channels}\label{isometry}\index{isometry channel}

Let $V:A\to B$ be an isometry matrix; i.e. $V^*V=I^A$ and $|A|\leq |B|$.
We define an isometry channel $\mV\in\cptp(A\to B)$ as
\be
\mV\left(\rho^A\right)=V\rho^A V^*\quad\quad\forall\;\rho\in\ml(A)\;.
\ee
Such an isometry channel can be viewed as an \emph{embedding} of system $A$ into $B$.
Note that like unitary channels, isometry channels have an operator sum representation with a single Kraus operator.

Inrestingly, isometry channels have inverses. Specifically, for any $\tau\in\md(A)$ define
\be\label{isocha}
\mV^{-1}_\tau\left(\sigma^B\right)\eqdef V^*\sigma^B V+\tr\left[\left(I^B-VV^*\right)\sigma^B\right]\tau^A\quad\quad\forall\;\sigma\in\ml(B)\;.
\ee
The linear map above is a quantum channel in $\cptp(B\to A)$ and it is an inverse of the isometry channel $\mV$ above (see the following Exercise).

\begin{exercise}
Show that for all $\tau\in\md(A)$ the linear map $\mV_\tau^{-1}$ as defined above is a channel in $\cptp(B\to A)$ that satisfies
\be
\mV^{-1}_\tau\circ\mV=\id^{A}\;.
\ee 
\end{exercise}

\subsection{Unital Channels}\label{unital}\index{unital channel}

A map $\mE\in\ml(A\to A)$ is termed a unital quantum channel, or more specifically, a \emph{doubly stochastic} channel, if it satisfies two conditions: it is a CPTP map, and it is a unital map (meaning it preserves the identity, as in $\mE(I) = I$). In Exercise~\ref{showdual}, you demonstrated that a linear map is unital if, and only if, its dual is trace-preserving. Consequently, $\mE$ is a doubly stochastic channel if and only if both $\mE$ and its dual $\mE^*$ are quantum channels. 

If $\mE\in\cptp(X\to X)$ is a \emph{classical} doubly stochastic channel, then $\mE$ can be represented by the $m\times m$ evolution column-stochastic matrix $E$, where $m\eqdef|X|$, and the unital condition $\mE(I^X)=I^X$ translates to
$
E\1_m=\1_m
$,
with $\1_m$ is the $m$-dimensional vector $\1_m\eqdef(1,\ldots,1)^T$. This means that the sum of each row of $E$ is equal to one.
Given that $E$ is inherently a column-stochastic matrix, it follows that classical unital channels correspond to doubly-stochastic matrices. A doubly-stochastic matrix is characterized as a square, real matrix with non-negative entries, where the sum of the elements in each row and each column is one. Consequently, in a classical framework, the input distribution $\p$ transitions to the output distribution $\q$ through the doubly stochastic matrix\index{doubly stochastic matrix} relationship $\q = D\p$. Here, the matrix $D$ is used to denote the evolution matrix, emphasizing its nature as doubly stochastic.

In the quantum case there is also a similar relation between the input state $\rho$ and the output state $\sigma=\mE(\rho)$ of a doubly stochastic quantum channel $\mE\in\cptp(A\to A)$. To see this relation, set $m\eqdef|A|$, and denote by $\{|\psi_x\ra\}_{x\in[m]}$ and $\{|\phi_y\ra\}_{y\in[m]}$ the eigenvectors of $\rho$ and $\sigma$, respectively, and by $\{p_x\}_{x\in[m]}$ and $\{q_y\}_{y\in[m]}$, their corresponding eigenvalues. With these notations we have
\be
\rho=\sum_{x\in[m]}p_x|\psi_x\lr \psi_x|\quad\text{and}\quad\sigma=\sum_{y\in[m]}q_y|\phi_y\lr \phi_y|\;,
\ee
The relation $\sigma=\mE(\rho)$ is equivalent to
\be\label{rel}
q_y=\la \phi_y|\sigma|\phi_y\ra=\la \phi_y|\mE(\rho)|\phi_y\ra=\sum_{x\in[m]}p_x\la \phi_y|\mE(|\psi_x\lr \psi_x|)|\phi_y\ra
\ee
Let $D=(d_{xy})$ be the $m\times m$ matrix whose components are 
$
d_{yx}\eqdef \la \phi_y|\mE(|\psi_x\lr \psi_x|)|\phi_y\ra\;,
$
and note that $d_{yx}\geq 0$ for all $x$ and $y$. Moreover, 
\ba
\sum_{x\in[m]}d_{yx}&=\la \phi_y|\mE(I)|\phi_y\ra=\la \phi_y|I|\phi_y\ra=1\quad\text{and}\\
\sum_{y\in[m]}d_{yx}&=\tr\left[\mE(|\psi_x\lr \psi_x|)\right]=\tr\left[|\psi_x\lr \psi_x|\right]=1\;.
\ea
Hence, $D$ is a doubly-stochastic matrix and~\eqref{rel} becomes $\q=D\p$, where $\p$ and $\q$ are the probability vectors consisting of the eigenvalues of $\rho$ and $\sigma$, respectively. 

\begin{exercise}
Let $\mE\in\cptp(A\to B)$ be a quantum channel. Show that if $\mE(I^A)=I^B$ then we must have $|A|=|B|$.
\end{exercise}

\subsubsection{No Quantum Analogue to Birkhoff Theorem}\index{Birkhoff theorem}

Birkhoff theorem (see Theorem~\ref{birkhoff}) states that any doubly-stochastic matrix can be expressed as a convex combination of permutation matrices. For example,
\be
\begin{bmatrix}
\frac{5}{6} & \frac{1}{6} & 0\\
0 & \frac{1}{2} & \frac{1}{2}\\
\frac{1}{6} & \frac{1}{3} & \frac{1}{2}
\end{bmatrix}
=\frac{1}{3}
\begin{bmatrix}
1 & 0 & 0\\
0 & 1 & 0\\
0 & 0 & 1
\end{bmatrix}
+\frac{1}{2}
\begin{bmatrix}
1 & 0 & 0\\
0 & 0 & 1\\
0 & 1 & 0
\end{bmatrix}
+\frac{1}{6}
\begin{bmatrix}
0 & 1 & 0\\
0 & 0 & 1\\
1 & 0 & 0
\end{bmatrix}\;.
\ee 
In general, from Birkhoff theorem\index{Birkhoff theorem} (Theorem~\ref{birkhoff}), any doubly stochastic matrix\index{doubly stochastic matrix} can be expressed as $D=\sum_{w\in[k]}t_w\Pi_w$, where $\{t_w\}_{w\in[k]}$ is a probability distribution and $\{\Pi_w\}_{w\in[k]}$ are permutation matrices. Therefore, the relation $\q=D\p=\sum_{w\in[k]}t_w\Pi_w\p$ implies that $\q$ is a convex combination of permuted versions of $\p$. We now discuss whether there exists a quantum analogue to this property. 

A fundamental example of a unital quantum channel is the unitary evolution. This is analogous to the permutation evolution matrix of a classical channel. 
Indeed, if $\sigma=U\rho U^{*}$ then the vector $\q$, consisting of eigenvalues of $\sigma$, is related by a permutation matrix to the vector $\p$, consisting of the eigenvalues of $\rho$.
Clearly, the unitary channel $\mE(\rho)=U\rho U^*$ satisfies $\mE(I)=UIU^{*}=UU^*=I$, where $U$ is a unitary matrix. One can extend this definition to include  mixture of unitaries. Such channels are called \emph{mixed-unitary channels} (or random-unitary channels). That is, a mixed-unitary channel is a quantum channel 
$\mE\in\cptp(A\to A)$ that has the form
\be\label{mu}
\mE(\rho)=\sum_{w\in[k]}t_w U_w\rho U^{*}_w
\ee
where $\{U_w\}_{w\in[k]}$ is a set of $k$ unitary matrices, and $\{t_w\}_{w\in[k]}$ is a probability distribution. This is the quantum version of a convex combination of permutation matrices.
One can implement such a quantum channel, for example, by rolling a dice and based on the outcome $w$ of the dice apply the evolution $\rho\mapsto U_w\rho U_{w}^*$. After forgetting the value of $w$, such a process can be described by the equation above.

\begin{exercise}
Find the operator sum representation of the mixed-unitary channel~\eqref{mu}. 
\end{exercise}

One may wonder whether all unital channels can be expressed as mixed-unitary channels. To answer this question, consider the following example given by Peter Shor (2010). Let $A=\mbb{C}^3$ and $\mE\in\cptp(A\to A)$ 
be the quantum channel 
\be\label{m1234}
\mE(\omega)= M_1\omega M_{1}^{*}+M_2\omega M_2^*+M_3\omega M_3^*\quad\quad\forall\;\omega\in\md(A)\;,
\ee
where the Kraus operators
\be\label{m123}
M_z\eqdef\frac{|z\lr z+1|+|z+1\lr z|}{\sqrt{2}}\quad\quad\forall\;z\in[3]\;,
\ee
with $|4\ra\eqdef |1\ra$ (i.e.\ the summation in $|z+1\ra$ is modulo 3).
\begin{exercise}
Show that the quantum channel $\mE\in\cptp(A\to A)$ as defined in~\eqref{m1234} and~\eqref{m123} is unital.
\end{exercise}

The unital channel $\mE$ as defined in~\eqref{m1234} and~\eqref{m1234} is \emph{not} a mixed-unitary channel.
To see this, suppose by contradiction that $\mE$ can be expressed as in~\eqref{mu}. Then, $\{M_z\}_{z\in[3]}$ and $\{\sqrt{t_w}U_w\}_{w\in[k]}$ are operator sum\index{operator sum} representations of the \emph{same} channel $\mE$. Therefore, there exists an $m\times 3$ isometry channel $V=(v_{wz})$ such that
\be\label{3p199}
\sqrt{t_w}U_w=\sum_{z\in[3]}v_{wz}M_z\;.
\ee
Multiplying each side of the equation above by its conjugate we get
\be
t_wI^A=\sum_{z,z'\in[3]}\bar{v}_{wz'}v_{wz}M_{z'}^*M_z\;.
\ee
Now, taking all summations to be modulo 3, we have by definition that $M_{z}M_{z+1}=|z\lr z+2|$ and $M_{z}M_{z+2}=|z+1\lr z+2|$ (by definition $M_z=M_z^*$). Hence, the equation above implies that 
\be
\bar{v}_{wz}v_{w(z+1)}=\bar{v}_{z}v_{w(z+2)}=0\quad\quad\;\forall\;z\in[3]\;.
\ee
In other words, we must have $v_{wz'}v_{wz}=0$ for all $z\neq z'\in[3]$. This, in turn, implies that $v_{wz}=0$ for at least two values of $z\in[3]$. Hence, the relation~\eqref{3p199} implies that $U_z$ is proportional to one of the three matrices $\{M_z\}_{z\in[3]}$, in contradiction with the fact that all $\{M_z\}_{z\in[3]}$ have rank two, whereas $U_w$ has a full rank. Therefore, the channel $\mE$ is \emph{not} a mixed-unitary channel.

From the exmple above it follows that there is no quantum analogue to Birkhoff theorem\index{Birkhoff theorem}, as there are unital channels that are not mix-unitary channels. What is the distinction between unital channels and mix-unitary channels  in the Choi representation? Consider a unital quantum channel $\mE\in\cptp(A\to A)$, and let 
\be
J^{A\tA}_{\mE}=\mE^{\tA\to \tA}(\Omega^{A\tA})
\ee 
be its Choi matrix. Then, since it is trace preserving, $J^{A}_{\mE}=I^A$. On the other hand, since it is unital,
\be
J^{\tA}_{\mE}=\tr_{A}\left[J^{A\tA}_{\mE}\right]=\mE(I^{\tA})=I^{\tA}\;.
\ee
Therefore, $\mE$ is unital if and only if both marginals of its Choi matrix are equal to the identity matrix.

If $\mE\in\cptp(A\to A)$ is a mixture of unitaries, then its Choi matrix is proportional to a convex combination of maximally entangled states. A maximally entangled state $\phi^{A\tA}\in\pure(A\tA)$ is a normalized vector with the property that its reduced density matrix is the maximally mixed state; that is,
$\phi^A=\u^A$.
From Exercise~\ref{purification} it follows that all maximally entangled states in $\pure(A\tA)$ must have the form
\be
\frac{1}{\sqrt{m}}\left(I^A\otimes U^{\tA}\right)|\Omega^{A\tA}\ra\;,
\ee
where $m\eqdef|A|$, and $U$ is some unitary matrix. Now, observe that the Choi matrix of the mix-unitary channel~\eqref{mu} is given by
\be
J^{A\tA}_{\mE}=\mE^{\tA\to \tA}(\Omega^{A\tA})=\sum_{w\in[k]}t_w(I^A\otimes U_w)\Omega^{A\tA}(I^A\otimes U_{w}^{*})=m\sum_{w\in[k]}t_w |\phi_{w}^{A\tA}\lr\phi_{w}^{A\tA}|\;,
\ee 
where each
\be
|\phi_{w}^{A\tA}\ra\eqdef\frac{1}{\sqrt{m}}\left(I^A\otimes U^{\tA}_{w}\right)|\Omega^{A\tA}\ra\;,
\ee
is a maximally entangled states.

\begin{exercise}
Let $\mE\in\cptp(A\to A)$ be a quantum channel of the form
\be\label{no}
\mE^{A\to A}(\rho^A)\eqdef\tr_B\left[U^{AB}\left(\rho^A\otimes\u^B\right)U^{*AB}\right]\;,
\ee
where $U\in\ml(AB)$ is a unitary operator. 
\begin{enumerate}
\item Show that the above map is a unital quantum channel.
\item Show that every mixed-unitary channel, such as in~\eqref{mu}, with {\bf rational} probabilities $\{t_w\}_{w\in[k]}$, can be expressed as in~\eqref{no}. That is, there exists a system $B$ and a joint unitary matrix $U^{AB}$ such that the expression for $\mE(\rho)$ in~\eqref{no} becomes~\eqref{mu}.
\item Determine if the Shor example above can be expressed in the form~\eqref{no}.
\end{enumerate}
\end{exercise}

\subsection{Quantum Instruments}\label{qinst}\index{quantum instrument}

A quantum instrument is a quantum channel that takes a quantum state as its input and outputs a cq-state.
Consequently, it can be viewed as a mathematical abstraction of a quantum measurement with the classical output as the recorded measurement outcome, and with the quantum outcome as the post measurement state.
Mathematically, let $A$ be the input system, ${X}$ the classical register of the output system,  and $B$ be the quantum output. Then a quantum channel $\mN\in\cptp(A\to{X}B)$ is a quantum instrument if
\be
\Delta^{X\to X}\circ\mN^{A\to XB}=\mN^{A\to XB}\;,
\ee
where $\Delta\in\cptp(X\to X)$ is the completely dephasing channel with respect to the classical basis of $X$ (see Fig.~\ref{instrument}).

\begin{figure}[h]\centering    \includegraphics[width=0.5\textwidth]{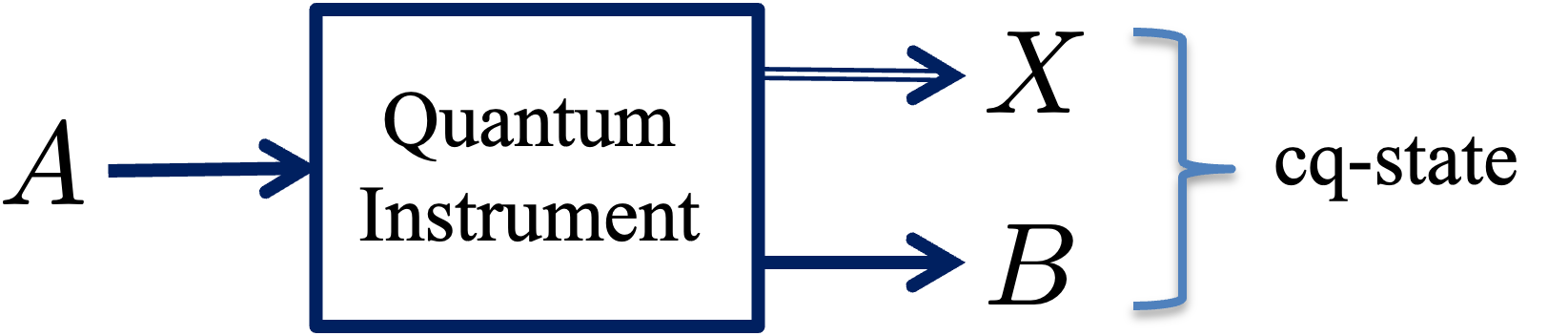}
  \caption{\linespread{1}\selectfont{Quantum instrument.}}
  \label{instrument}
\end{figure}

Set $m\eqdef|X|$ and observe that for all $\omega\in\md(A)$\ba\label{instrument}
\mN^{A\to XB}\left(\omega^A\right)&=\Delta^{X\to X}\left(\mN^{A\to XB}\left(\omega^A\right)\right)\\
&=\sum_{x\in[m]}|x\lr x|^X\otimes \mN_x^{A\to B}\left(\omega^A\right)\;,
\ea
where for each $x\in[m]$ we define the linear map $\mN_x\in\ml(A\to B)$ via 
\be
\mN_x^{A\to B}\left(\omega^A\right)\eqdef \tr_X\left[\left(|x\lr x|^X\otimes I^B\right)\mN^{A\to XB}\left(\omega^A\right)\right]\quad\quad \forall\;\omega\in\ml(A)\;.
\ee
By definition, the marginal channel
\be
\mN^{A\to B}\eqdef\tr_X\circ\mN^{A\to XB}=\sum_{x\in[m]}\mN_x^{A\to B}\;.
\ee
Observe further that $\mN^{A\to B}$ is a quantum channel since it can be expressed as a combination of the two quantum channels $\tr_X$ and $\mN^{A\to BX}$. Moreover, each $\mN_x^{A\to B}$ is a CP map. Indeed, let $J^{AXB}_\mN=\mN^{\tA\to BX}\big(\Omega^{A\tA}\big)$ be the Choi matrix of the quantum instrument $\mN$, then the Choi matrix of $\mN_x$ is given by
\be
J_{\mN_x}^{AB}= \tr_X\left[\left(I^A\otimes|x\lr x|^X\otimes I^B\right)J^{AXB}_\mN\right]\;,
\ee
which is positive semidefinite since $J^{AXB}_\mN\geq 0$.
We therefore conclude that $\{\mN_x^{A\to B}\}_{x\in[m]}$ are trace non-increasing CP maps that sums up to a CPTP map. 

\begin{exercise}
Let $\mE\in\cp(A\to B)$ be a trace non-increasing map. 
\begin{enumerate}
\item Show that any operator sum representation $\{M_x\}_{x\in[m]}$ of $\mE$ satisfies
$
\sum_{x\in[m]}M_x^*M_x\leq I\;.
$
\item Show that the marginal of the Choi matrix $J^{AB}_\mE$ satisfies
$
J^{A}_\mE\eqdef\tr_B\left[J^{AB}_\mE\right]\leq I^A\;.
$
\end{enumerate} 
\end{exercise}

\begin{exercise}
Find an operator sum representation of the quantum instrument $\mN^{A\to XB}$ discussed above.
\end{exercise}

\subsection{Complementary Channel}\index{complementary channel}

\begin{figure}[h]\centering    \includegraphics[width=0.4\textwidth]{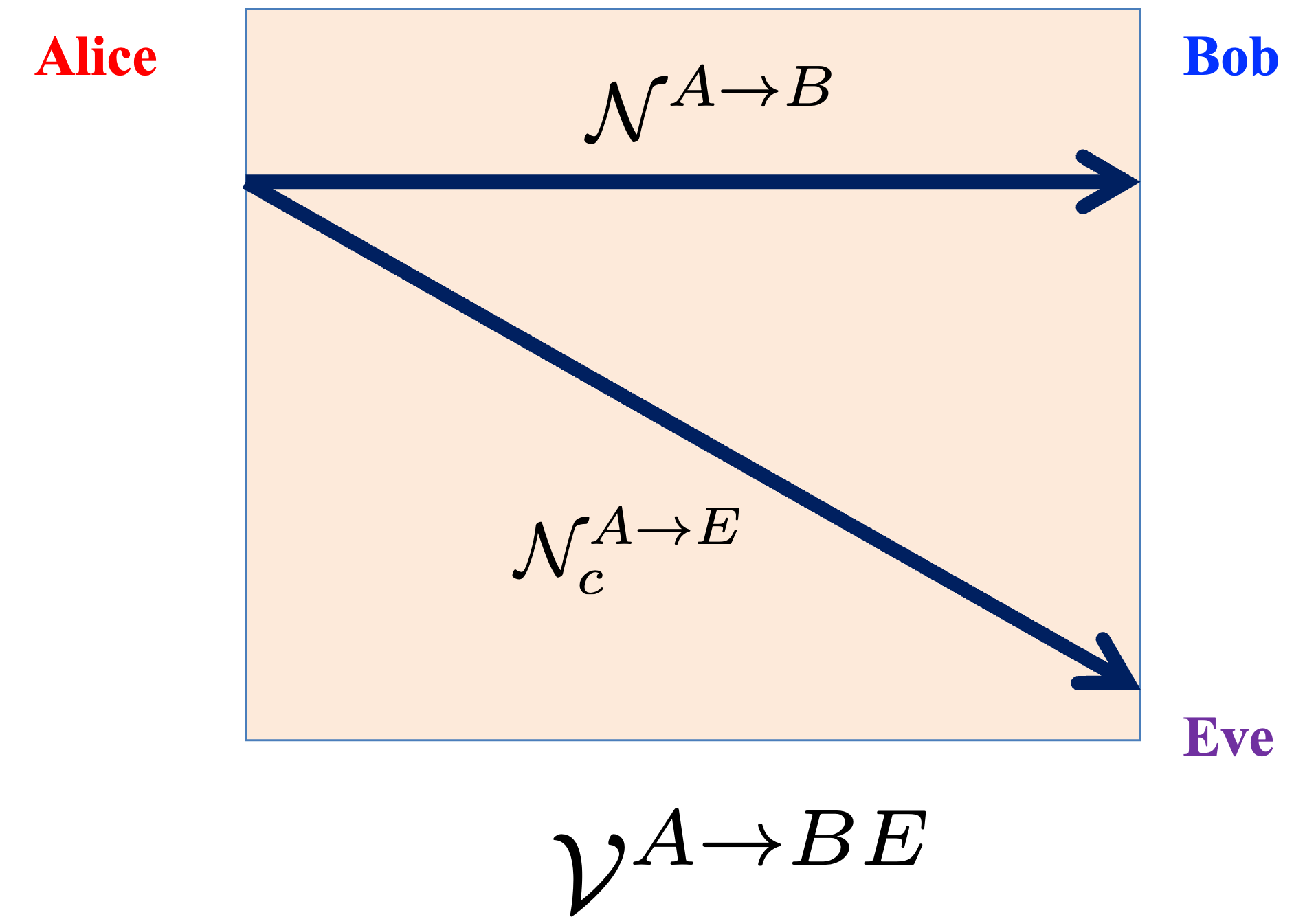}
  \caption{\linespread{1}\selectfont{The complementary channel of $\mN^{A\to B}$.}}
  \label{complementary}
\end{figure}

Due to the Stinespring dilation theorem, we can associate to any quantum channel $\mN\in\cptp(A\to B)$ an isometry channel $\mV\in\cptp(A\to BE)$ such that
\be
\mN^{A\to B}=\tr_E\circ\mV^{A\to BE}\;.
\ee 
The complementary channel of $\mN$, denoted by $\mN_c$, is a channel in $\cptp(A\to E)$ obtained by tracing system $B$ from $\mV^{A\to BE}$ (see Fig.~\ref{complementary}); i.e.
\be
\mN_c^{A\to E}\eqdef\tr_B\circ\mV^{A\to BE}\;.
\ee

As an example, let $\mN^{A\to B}$ be the qubit depolarizing channel with $|A|=|B|=2$.  From it's Stinespring isometry in~\eqref{depolarz} we can get its complementary channel.  For this purpose, we denote by $t_0\eqdef \sqrt{1-\frac{3p}{4}}$ and $t_1=t_2=t_3=\frac{\sqrt{p}}2$, and by $\{\sigma_j\}_{j=0}^3$ the four Pauli\index{Pauli} matrices (including $\sigma_0\eqdef I$). We therefore get
\ba\label{comdep}
\mN_c^{A\to E}(\rho^A)&=\tr_B\left[\mV\rho^A V^*\right]=\sum_{j,k=0}^3t_jt_k\tr\left[\sigma_j\sigma_k\rho\right]|j\lr k|^E\quad\quad\forall\;\rho\in\ml(A)\;.
\ea

\bex
Use the Bloch representation\index{Bloch representation} $\rho=\frac{1}{2}\left(I+\r\cdot\bs{\sigma}\right)$ to simplify the expression in~\eqref{comdep} for the complementary channel of the depolarizing channel.
\eex

\subsection{The Pinching Channel}\label{sec:pinching}\index{pinching channel}

The pinching channel is a generalization of the completely dephasing channel. For any observable\index{observable} $H\in\herm(A)$ with $m$ distinct eigenvalues $\spec(H)=\{\lambda_1,\ldots,\lambda_m\}$ we define the pinching channel $\mP_H\in\cptp(A\to A)$ associated with $H$ as
\be\label{pinchingdef}
\mP_H(\rho)\eqdef\sum_{x\in[m]}P_x\rho P_x\quad\forall\;\rho\in\ml(A)\;,
\ee
where for each $x\in[m]$, $P_x$ is the projector the eigenspace of $\lambda_x$. Note that $\mP_H$ is indeed a quantum channel since $\{P_x\}_{x\in[m]}$ form an orthogonal set of projectors that sum to the identity. Moreover, if $m=|A|$ (i.e. $H$ has  $|A|$ distinct eigenvalues) then $\mP_H=\Delta$, where $\Delta$ is the completely dephasing quantum channel in the basis comprising of the eigenvectors of $H$.

\bex\label{compinch}
Let $H\in\herm(A)$ and $\rho\in\md(A)$. Show that $\rho=\mP_H(\rho)$ if and only if $[\rho,H]=0$.
\eex

\bex\label{srcommute}
Let $A$ be a quantum system and $\rho,\sigma\in\md(A)$ be two density matrices, with
$\sigma=\sum_{y\in[n]}\lambda_y\Pi_y$, where $\{\Pi_y\}_{y\in[n]}$ form an orthogonal projective von-Neumann measurement on system $A$, and $\{\lambda_y\}_{y\in[n]}$ is the set of \emph{distinct} eigenvalues of $\sigma$. For each $y\in[n]$, let $m_y\eqdef\tr[\Pi_y]$ be the multiplicity of the eigenvalue $\lambda_y$.
\ben
\item Show that $\sigma$ commutes with $\mP_\sigma(\rho)$; i.e.,
\be
\left[\mP_\sigma(\rho),\sigma\right]=0\;.
\ee
\item Show that there exists an orthonormal basis of $A$, denoted by $\{|\phi_{xy}\ra\}_{x,y}$, with the property that
\be
\Pi_y=\sum_{x\in[m_y]}\phi_{xy}\quad\text{and}\quad\mP_\sigma(\rho)=\sum_{y\in[n]}\sum_{x\in[m_y]}r_{xy}\phi_{xy}
\ee
for some $r_{xy}\geq 0$ with $\sum_{y\in[n]}\sum_{x\in[m_y]}r_{xy}=1$.
\item Let $y\in[n]$ and $x,x'\in[m_y]$. Show that if $x\neq x'$ then
\be
\la\phi_{x'y}|\rho|\phi_{xy}\ra=0\;.
\ee
\item Let $\Delta\in\cptp(A\to A)$ be the completely dephasing channel in the basis $\{|\phi_{xy}\ra\}_{x,y}$. Show that
\be\label{deltars}
\Delta(\rho)=\mP_\sigma(\rho)\quad\text{and}\quad\Delta(\sigma)=\sigma\;.
\ee
\een
\eex

\begin{exercise}\label{ex:pinch3}
Let $A$ be a quantum system and $\rho,\sigma\in\md(A)$ be two density matrices satisfying $\rho\not\ll\sigma$ (i.e. $\supp(\rho)\not\subseteq\supp(\sigma)$). Show that also 
\be
\mP_\sigma(\rho)\not\ll \sigma\;.
\ee
\end{exercise}

The following exercise demonstrate the pinching channel is a special type of mixture of unitaries.

\begin{exercise}\label{pinchex}
Let $H\in\herm(A)$ be an observable with  $\spec(H)=\{\lambda_1,\ldots,\lambda_k\}$ as above, and let $\mP_H$ be its associated pinching channel as given in~\eqref{pinchingdef}. Show that for any $\rho\in\ml(A)$
\be\label{pinex}
\mP_H(\rho)=\frac1m\sum_{y\in[m]}U_y\rho U_y^{*}\quad\text{where}\quad U_y\eqdef\sum_{x\in[m]}e^{i\frac{2\pi xy}m}P_x\;.
\ee
\end{exercise}

The above exercise demonstrate the pinching channel is a mixed unitary channel. 
Observe also that for $y=m$ we have $U_m=I_m$ so that 
\be
\mP_H(\rho)=\frac1m\rho+\frac1m\sum_{y\in[m-1]}U_y\rho U_y^{*}\geq \frac1m\rho\;.
\ee
Since $m\eqdef|\spec(H)|$ we get  the following inequality known as the \emph{pinching inequality}. 
\begin{mye}{The Pinching Inequality}
For any $H\in\herm(A)$ and $\rho\in\md(A)$
\be\label{pinch}
\mP_H(\rho)\geq \frac1{|\spec(H)|}\rho\;.
\ee
\end{mye}
Since we got the inequality above by removing $m-1$ terms, it may give the impression that this inequality is not very useful since it is  never saturated and not tight. However, we will see that in applications, this inequality can be used to provide good enough approximation to $\mP_H(\rho)$ when we consider the asymptotic case in which $H=\sigma^{\otimes n}$ where $\sigma$ is some quantum state and $n$ is a very large integer. In particular, we will see in Chapter~\ref{ch:qs} (particularly Sec.~\ref{sec:pinch0}) that in this case $m=|\spec(\sigma^{\otimes n})|$ grows polynomially with $n$. The fact that it is not an exponential growth with $n$ is one of the key reasons why the pinching inequality is quite useful.

The pinching map can also be used to prove the reverse H\"older inequality\index{H\"older inequality} with $p\in(0,1)$.  In this case, we still use the notation $\|M\|_p\eqdef\left(\tr\left[|M|^p\right]\right)^{\frac1p}$ for any $M\in\ml(A)$. However, one has to be careful since for $p\in(0,1)$, $\|\cdot\|_p$ is not a norm.
\begin{myg}{Reverse H\"older Inequality}\index{reverse H\"older inequality}\index{reverse H\"older inequality}
\begin{lemma}
Let $\tau,\omega\in\pos(A)$ and $p\in(0,1)$. If $\tau\ll\omega$ then
\be\label{reverse}
\tr[\omega\tau]\geq\frac{\|\tau\|_p}{\left\|\omega^{-1}\right\|_{\frac p{1-p}}}
\ee
where the inverse is a generalized inverse.
\end{lemma}
\end{myg}

\begin{proof}
We first prove the theorem for the case that $\tau$ and $\omega$ commutes. In this case,
\ba
\|\tau\|_p^p=\tr[\tau^p]=\tr\left[\tau^p\omega^p\omega^{-p}\right]=&\left\|\tau^p\omega^p\omega^{-p}\right\|_1\\
{\color{red} \text{H\"older inequality\index{H\"older inequality} }\eqref{holin}\rightarrow}&\leq \|\tau^p\omega^p\|_{\frac1p}
\left\|\omega^{-p}\right\|_{\frac1{1-p}}\\
&=\left(\tr[\tau\omega]\right)^p\left\|\omega^{-p}\right\|_{\frac1{1-p}}\;.
\ea
This prove the theorem for the case that $\omega$ and $\tau$ commutes.
On the other hand, from Exericse~\ref{srcommute} we get that  $\mP_\omega(\tau)$ and $\omega$ commutes so that
\ba\label{3235}
\tr\left[\tau\omega\right]=\tr\left[\mP_\omega(\tau)\omega\right]\geq \frac{\|\mP_\omega(\tau)\|_p}{\left\|\omega^{-1}\right\|_{\frac p{1-p}}}
\ea
Finally, since $t\mapsto t^p$ is operator concave for $p\in(0,1)$ (see Tabel~\ref{table:1}) we conclude that 
\be
\|\mP_\omega(\tau)\|_p^p=\tr\left[(\mP_\omega(\tau))^p\right]\geq \tr\left[\mP_\omega(\tau^p)\right]=\tr[\tau^p]\;.
\ee
That is, $\|\mP_\omega(\tau)\|_p\geq \|\tau\|_p$. Substituting this into~\eqref{3235} completes the proof.
\end{proof}

\begin{exercise}[Reverse Young’s Inequality]
Let $A$ and $B$ be two Hilbert spaces, $M,N\in\pos(A)$, $p\in(0,1)$, and $q$ defined via $\frac1p+\frac1q=1$ (hence $q<0$). Use the reverse H\"older inequality of the Schatten\index{Schatten}  norm to show that
\be\label{young2}
\tr[MN]\geq\frac1p\tr[M^p]+\frac1q\tr[N^q]\;. 
\ee
with equality if and only if $M^p=N^q$. Hint: Take the logarithm on both sides of the reverse H\"older inequality and use the concavity property of the logarithm. 
\end{exercise}

\subsection{Twirling Channels}\index{twirling}

In Chapter~\ref{Ch:Asymmetry}, particularly in Sec.\ref{secgt}, we will explore a channel associated with any compact Lie group through its Haar measure (refer to Appendix~\ref{sec:rep} for more definitions and details on Haar measures and compact Lie groups). This channel is known as the $\G$-twirling operation. A simple example of such a channel, defined for all $\rho\in\ml(A)$, is:
\be\label{8249b}
\mG(\rho)\eqdef\int_{\muu(A)}dU\;U\rho U^*\;,
\ee
where $dU$ is the Haar measure of the unitary group $\muu(A)$.
In other words, $\rho$ is ``twirled" over all unitary matrices $U\in\muu(A)$.

\bex
Consider $\rho\in\ml(A)$ and $\sigma\eqdef\mG(\rho)$, where $\mG$ is the channel defined in~\eqref{8249b}. Demonstrate that the commutator $[U,\sigma]=0$. Hint: Utilize the properties of the Haar measure described in Sec.~\ref{sec:inv} of the appendix.
\eex

The exercise above illustrates that the channel defined in~\eqref{8249b} is not unfamiliar to us. In fact, it can be represented as the replacement channel
\be\label{8249}
\mG(\rho)\eqdef\int_{\muu(A)}dU\;U\rho U^*=\tr[\rho]\u^A\;,
\ee
where $\u^A=I^A/|A|$ is the maximally mixed state. To understand why, remember from the previous exercise that $\sigma\eqdef\mG(\rho)$ commutes with all unitary matrices in $\muu(A)$ and, thus, must be proportional to the identity matrix. We now consider a more complex example with numerous applications in quantum information science.

Let $B$ be a replica of $A$, and set $m\eqdef|A|=|B|$. Consider the twirling\index{twirling} map $\mG\in\cptp(AB\to AB)$, defined for all $\rho\in\ml(AB)$ as:
\be\label{gtwi}
\mG(\rho^{AB})\eqdef\int_{\muu(m)}dU\;(U\otimes {U})\rho^{AB}(U\otimes U)^*\;.
\ee
As in the previous example, for every $\rho\in\ml(AB)$, the matrix $\sigma^{AB}\eqdef\mG(\rho^{AB})$ commutes with $U\otimes U$ for all $U\in \muu(A)$. The ensuing question is: which matrices $\sigma^{AB}$ commute with all matrices of the form $U\otimes U$, where $U$ is a unitary matrix? Clearly, any matrix proportional to the identity matrix satisfies this criterion. However, there exists another type of operator that fulfills this property, known as the \emph{swap} (or flip) operator:
\be
F^{AB}\eqdef\sum_{x,y\in[m]}|x\lr y|^A\otimes |y\lr x|^B\;.
\ee
\bex
Prove that the swap operator $F^{AB}$ satisfies
\be
\left[U\otimes U, F^{AB}\right]=0\quad\quad\forall\;U\in\muu(m)\;.
\ee
\eex
Our analysis in Sec.~\ref{subsec:bip} indicates that any operator commuting with all matrices in the set $\{U\otimes U\}_{U\in\muu(m)}$ can be expressed as a linear combination of the identity and swap operators. Consequently, $\mG(\rho^{AB})=aI^{AB}+bF^{AB}$ for some $a,b\in\mbb{R}$, determinable from the requirement that $\mG$ is CPTP (see the following exercise). Furthermore, in Sec.~\ref{subsec:bip} of Appendix~\ref{sec:rep}, we demonstrate that the representation $U\mapsto U\otimes U$ decomposes into two irreps, corresponding to the symmetric and antisymmetric subspaces, each with a multiplicity of one. Hence, from Theorem~\ref{deop} in Appendix~\ref{sec:rep}, it follows that for all $\rho\in\ml(AB)$,\be\label{prewerner}
\mG(\rho)=\tr\left[\rho\Pi_\sym\right]\frac{\Pi_\sym}{\tr\left[\Pi_\sym\right]}+\tr\left[\rho\Pi_\asy\right]\frac{\Pi_\asy}{\tr\left[\Pi_\asy\right]}
\ee
where $\Pi_\sym\eqdef\frac12(I^{AB}+F^{AB})$ and $\Pi_\asy\eqdef\frac12(I^{AB}-F^{AB})$ are the projections onto the symmetric and antisymmetric subspaces of $AB$ (see\eqref{sw1} and~\eqref{sw2}). When $\rho\in\md(AB)$ is a density matrix, the right-hand side above represents a density matrix known as the \emph{Werner quantum state}.

\bex
Let $F^{AB}$ be the swap operator, with $m\eqdef|A|=|B|$.
\ben
\item Starting with $\mG(\rho^{AB})=aI^{AB}+bF^{AB}$ for some $a,b\in\mbb{R}$,
prove that $\mG$ is CPTP if and only if
for all $\rho\in\ml(AB)$,
\be\label{twirling}
\mG\left(\rho^{AB}\right)=\frac{m\tr[\rho^{AB}]-\tr\left[\rho^{AB} F^{AB}\right]}{m(m^2-1)}I^{AB}+\frac{m\tr\left[\rho^{AB}F^{AB}\right]-\tr[\rho^{AB}]}{m(m^2-1)}F^{AB}\;.
\ee
\item Derive~\eqref{twirling} from~\eqref{prewerner} by expressing the projections onto the symmetric and antisymmetric subspaces in terms of the swap operator.
\een
\eex

\bex\label{m2fab}
Show that for all $M\in\ml(A)$ and $B\cong A$ we have
\be\label{0m2fab}
\tr\left[M^2\right]=\tr\left[\left(M\otimes M\right)F^{AB}\right]\;.
\ee
\eex

\bex
Let $B\eqdef \tA$, $\mG\in\cptp(AB\to AB
)$ be as in~\eqref{gtwi}, and $\mE\in\cptp(AB\to AB)$.
\ben
\item Show that $\mE=\mE\circ\mG$ if and only if there exists $\omega_1,\omega_2\in\eff(AB)$ such that for all $\rho\in\ml(AB)$
\be
\mE\left(\rho^{AB}\right)=\omega_1^{AB}\tr\left[\rho\Pi_\sym^{AB}\right]+\omega_2^{AB}\tr\left[\rho\Pi_\asy^{AB}\right]\;.
\ee
\item Show that $\mE=\mG\circ\mE$ if and only if there exists $\Lambda\in\eff(AB)$ such that for all $\rho\in\ml(AB)$
\be
\mE\left(\rho^{AB}\right)=\frac{1}{\tr\left[\Pi_\sym^{AB}\right]}\Pi_\sym^{AB}\tr\left[\rho^{AB}\Lambda^{AB}\right]+\frac{1}{\tr\left[\Pi_\asy^{AB}\right]}\Pi_\asy^{AB}\tr\left[\rho^{AB}\Lambda^{AB}\right]\;.
\ee
\een
\eex

\bex
Let $\mG:\cptp(A^2\to A^2)$ be the twirling\index{twirling} channel
\be\label{gtwi2}
\mG(\rho)\eqdef\int_{U(m)}dU\;(U\otimes \overline{U})\rho(U\otimes \overline{U})^*\quad\quad\forall\;\rho\in\ml(A^2)\;,
\ee
where $\overline{U}\eqdef (U^*)^T$. 
\ben
\item Show that
\be
\mG(\Phi^{A\tA})=\Phi^{A\tA}\;,
\ee
where $\Phi\in\md(A\tA)$ is the maximally entangled state.
\item Show that for $m\eqdef|A|=|B|$ and $\rho\in\md(AB)$
\be\label{6148}
\mG(\rho^{AB})=m^2\frac{1-\tr\left[\Phi^{AB}\rho^{AB}\right]}{m^2-1}\u^{AB}+\frac{m^2\tr\left[\Phi^{AB}\rho^{AB}\right]-1}{m^2-1}\Phi^{AB}\;.
\ee
 Hint: Use~\eqref{twirling} and observe that $\Phi^{AB}=\mT^{B\to B}(F^{AB})$ where $\mT^{B\to B}$ is the transpose map.
 \een
\eex

\bex
Let $m\eqdef|A|=|B|$, $\mG\in\cptp(AB\to AB
)$ be as in~\eqref{gtwi2}, and $\mE\in\cptp(AB\to AB)$. Define also the state
\be
\tau^{AB}\eqdef\frac{I^{AB}-\Phi^{AB}}{m^2-1}\;.
\ee
\ben
\item Show that for all $\rho\in\md(AB)$
\be\label{newform2}
\mG(\rho^{AB})=\left(1-\tr\left[\Phi^{AB}\rho^{AB}\right]\right)\tau^{AB}+\tr\left[\Phi^{AB}\rho^{AB}\right]\Phi^{AB}\;.
\ee
\item Show that $\mE=\mE\circ\mG$ if and only if there exists $\omega_1,\omega_2\in\eff(AB)$ such that for all $\rho\in\md(AB)$
\be\label{om1om2}
\mE\left(\rho^{AB}\right)=\left(1-\tr\left[\Phi^{AB}\rho^{AB}\right]\right)\omega_1^{AB}+\tr\left[\Phi^{AB}\rho^{AB}\right]\omega_2^{AB}\;.
\ee
\item Show that $\mE=\mG\circ\mE$ if and only if there exists $\Lambda\in\eff(AB)$ such that for all $\rho\in\md(AB)$
\be\label{8248}
\mE\left(\rho^{AB}\right)=\left(1-\tr\left[\Lambda^{AB}\rho^{AB}\right]\right)\tau^{AB}+\tr\left[\Lambda^{AB}\rho^{AB}\right]\Phi^{AB}\;.
\ee
\een
\eex

\begin{exercise}\label{comrand}
Let $A$ be a Hilbert space of dimension $m\eqdef|A|$, and consider the channel given~\eqref{8249b}
Show that for all $\rho\in\ml(A)$
\be
\mE(\rho)=\frac1{m^2}\sum_{p,q\in[m]}W_{p,q}\rho W_{p,q}^*\;,
\ee
where $W_{p,q}$ are the are the Hiesenberg-Weyl operators defined in~\eqref{wpq}.
\end{exercise}

\section{Notes and References}

Many books on quantum information contains much of the material presented here. Informationally complete POVMs were first introduced in~\cite{Prugovecki1977,Busch1991} and since then they became an integral part of quantum tomography. In particular, SIC POVMs have been studied intensively and more details and references about them can be found in the review article by~\cite{Rastegin2014}. 

Gleason's theory was proved originally for projective von-Neumann measurements in~\cite{Gleason1957}. The proof in this case holds only in dimension greater than two, and there are counter examples in the qubit case. The version of Gleason's theorem that we considered here is due to~\cite{Busch2003}. In this case the proof is much simpler (and holds in all dimensions) since   \emph{effects} replace orthogonal projections (and effects are simpler to work with).  Gleason's theorem is of particular importance in the foundations of quantum physics as well as the field of quantum logic in its effort to minimize the number of axioms needed to formulate quantum mechanics. It also closes the bridge between some of the axioms of quantum mechanics and Born's rule. More details, references, and History of Gleason's theorem can be found in a Wikipedia article entitled ``Gleason's theorem".

Naimark's and Stinspring's dilation theorems are results from operator theory and are valid also in infinite dimensional Hilbert spaces. Here we only studied their adaptation to the finite dimensional case. More details on their infinite dimensional version can be found in many books on operator theory; e.g. the book by~\cite{Paulsen2003}.

The St\o rmer-Woronowicz theorem was first proved for the qubit-to-qubit case by~\cite{Stormer1963} and later on for the qubit-to-qutrit case by~\cite{Woronowicz1976}. Counter examples exists in higher dimensions so these dimensions are optimal.  Both proofs involve somewhat complicated calculations, and the simplified proof of the St\o rmer case presented here is due to~\cite{AS2015}. It is an open problem to find a simpler proof of Woronowicz theorem.

More information on the pinching channel and its properties can be found in the book by~\cite{Tomamichel2015}.

\part{Tools and Methods}

\chapter{Majorization}\label{ch:majorization}\index{majorization}

A pre-order is a binary relation between objects that is \emph{reflexive} and \emph{transitive}. For example, consider  the inclusion relation $\supseteq$ between subsets of $[n]\eqdef\{1,\ldots,n\}$ for some $n\in\mbb{N}$. Then, $\supseteq$ is reflexive since for any subset $\mA$ of $[n]$ we have $\mA\supseteq\mA$. The relation $\supseteq$ is transitive since for any three subsets $\mA,\mB,\mC\subseteq[n]$ with $\mA\supseteq\mB$ and $\mB\supseteq\mC$ it follows that $\mA\supseteq\mC$. Furthermore, the relation $\supseteq$ has yet another property known as \emph{symmetry}. That is, if $\mA,\mB\subseteq[n]$ satisfies both $\mA\supseteq\mB$ and $\mB\supseteq\mA$ then necessarily $\mA=\mB$. A pre-order that satisfies this additional symmetry property is called a \emph{partial order}. 

Partial orders plays a fundamental role in quantum resource theories. They typically stem from a set of restrictions imposed on quantum operations. For example, we saw in quantum teleportation that Alice and Bob are restricted to act locally and cannot communicate quantum particles. We will see in Chapter~\ref{entanglement} that this restriction imposes a partial order between two entangled states, determining if one entangled state can be converted to another under operations that are restricted to be local.
It turns out that there is one partial order with variants that appear in many resource theories.
This partial order, known as majorization, has been studied extensively in quantum information and other fields, particularly in the field of matrix analysis, and there are several books on the topic (see section `Notes and References' at the end of this chapter).

\section{Majorization Between Probability Vectors}\label{majobetween}\label{gambling game}

Consider a gambling game in which a host rolls a biased dice, and a player has to guess the outcome. Denote by $\p=(p_1,\ldots,p_n)^T\in\prob(n)$ the probability vector corresponding to the $n$ possible outcomes. 
The player wins the game if he or she guesses correctly the outcome. Clearly, the player will guess the value $x\in[n]$ that satisfies 
$p_x=\max\{p_1,\ldots,p_n\}$.  In other games, the player is allowed to provide more than one guess, and wins the game if the outcome belongs to the set of numbers (guesses) he or she provided to the host.  For example, if the player is allowed to provide a set with two numbers as his/her guesses, then in order to have the highest odds to win, he or she 
will choose the two numbers out of the $n$ numbers that have the highest probability to occur. Hence, the player will
win the game with probability $p_1^{\downarrow}+p_2^{\downarrow}$, where we denote by $\p^{\downarrow}=(p_1^{\downarrow},\ldots,p_n^{\downarrow})^T$ the vector obtained from $\p$ by rearranging its components in non-increasing order. 
We call a game in which the player is allowed to provide a set with $k$-numbers as guesses, a \emph{$k$-gambling game}. Note that the highest probability to win a $k$-gambling game is given by $\sum_{x\in[k]}p_x^{\downarrow}$.\index{gambling game}

Suppose now that at the beginning of each game, the player is allowed to choose between two dice with corresponding probabilities $\p$ and $\q$. Clearly, the player will choose the dice that has better odds to win the game.
For a $k$-game the player will choose the $\p$-dice if \label{dice}
\be\label{a11}
\sum_{x\in[k]}p_x^{\downarrow}\geq\sum_{x\in[k]}q_x^{\downarrow}\;.
\ee 
If the relation above holds for all $k\in[n]$, then the player will choose the $\p$-dice for any $k$-gambling game. In this case we say that $\p$ majorizes $\q$ and write $\p\succ\q$.

\begin{myd}{Majorization}\index{majorization}
\begin{definition}
Let $\p,\q\in\mbb{R}^n$. We say that $\p$ majorizes $\q$ and write $\p\succ\q$ if~\eqref{a11} holds for all $k\in[n]$ with equality for $k=n$.
\end{definition}
\end{myd}

\begin{remark}
Note that in the definition above we did not assume that $\p$ and $\q$ are probability vectors, however, in the applications we consider in this book, $\p$ and $\q$ will always be probability vectors. 
\end{remark}

Majorization is a pre-order. That is, given three real vectors $\p,\q,\r\in\mbb{R}^n$ we have
$\p\succ\p$ (reflexivity), and if $\p\succ\q$ and $\q\succ \r$ then $\p\succ\r$ (transitivity). Moreover, if both $\p\succ\q$ and $\q\succ\p$ then $\p$ and $\q$ are related by a permutation matrix. Therefore, using the notation $\prob^\da(n)$ to denote the subset of $\prob(n)$ consisting of all vectors $\p\in\prob(n)$ with the property that $\p=\p^\da$, we get that the majorization relation $\succ$ is a partial order on the set $\prob^\da(n)$. 

\begin{exercise}
Show that for any $n$-dimensional probability vector $\p$ we have
\be\label{spread}
(1,0,\ldots,0)^T\succ\p\succ(1/n,\ldots,1/n)^T\;.
\ee
\end{exercise}

\begin{exercise}\label{mixdec}
Let $\p\in\prob(n)$, $\u^{(n)}\eqdef(1/n,\ldots,1/n)^T\in\prob(n)$ be the uniform probability vector, and $t\in[0,1]$. Show that
\be
\p\succ t\p+(1-t)\u^{(n)}\;.
\ee
\end{exercise}

\bex
Find an example of two vectors $\p,\q\in\prob(3)$ such that $\p$ does not majorize $\q$, and $\q$ does not majorize $\p$.
Vectors with such a property that $\p\not\succ\q$ and $\q\not\succ\p$ are said to be incomparable.
\eex

\bex\label{pum}
Let $n\in\mbb{N}$ and $\p\in\prob(n)$. Show that for sufficiently large $m\in\mbb{N}$  we have
\be
\p\succ\left(\u^{(2)}\right)^{\otimes m}\;,
\ee
where $\u^{(2)}\eqdef (1,1)^T$ is the 2-dimensional uniform distribution.
\eex

\bex\label{ex:mml}
Let $m\in\mbb{N}$, $\p,\q\in\prob(m)$, and $L$ be the $m\times m$  lower triangular matrix
\be
L\eqdef
\begin{pmatrix}
1 & 0 & 0 & \cdots & 0\\
1 & 1 & 0 & \cdots & 0\\
1 & 1 & 1 & \cdots & 0\\
\vdots & \vdots & \; &\ddots & \vdots\\
1 & 1 & 1 & \cdots & 1
\end{pmatrix}\;.
\ee
\ben
\item Show that $\p\succ\q$ if and only if $L\p^\da\geq L\q^\da$, where the inequality is entrywise.
\item Show that $L$ is invertible and find $L^{-1}$.
\item Show that $L^{-1}\p\geq 0$ (entrywise) if and only if $\p=\p^\ua$.
\item Show that $(L^T)^{-1}\p\geq 0$ (entrywise) if and only if $\p=\p^\da$.
\een
\eex

\bex\label{convexmajo}
Let $\p_1,\ldots,\p_k\in\prob^\da(m)$ and $\q_1,\ldots,\q_k\in\prob(m)$ be such that $\p_x\succ\q_x$ for all $x\in[k]$. Show that for any $(r_1,\ldots,r_k)^T\in\prob(k)$ we have
\be
\sum_{x\in[k]}r_x\p_x\succ\sum_{x\in[k]}r_x\q_x\;.
\ee
\eex

\begin{exercise}\label{exindices}
Let $\p,\q\in\prob^\da(n)$ and suppose $\p\neq\q$ and $\p\succ\q$. Show that the largest integer $z\in[n]$ for which $p_z\neq q_z$ must satisfy $q_z>p_z$.
\end{exercise}

\bex\label{ex:majorest}
Let $\p,\q\in\prob^\da(n)$ and $k,m\in[n-1]$ be such that $k\leq m$. Suppose that  $\q$ has the form 
\be
\q=(\underbrace{a,a,\ldots,a}_{k\text{-times}},\;
q_{k+1},\ldots,q_m,\underbrace{b,b,\ldots,b}_{(n-m)\text{-times}})^T\;.
\ee
Show that $\p\succ\q$ if and only if 
\be
\|\p\|_{(\ell)}\geq\|\q\|_{(\ell)}\quad\quad\forall\;\ell\in\{k,k+1,\ldots,m\}\;.
\ee
(i.e.\, there is no need to consider the cases $\ell<k$ or  $\ell>m$). Hint: Use the fact that $\p=\p^\da$ and $\q=\q^\da$.
\eex

\subsection{Characterization of Majorization}\label{sec:rrl}\index{majorization}

Eq.~\eqref{spread} corresponds to the fact that majorization can be used to determine if one probability distribution is more spread out than another one. This intuition can be deepened with further analysis of the dice games. Specifically, consider the $\p$-dice and $\q$-dice of the discussion above with $\p,\q\in\prob(n)$.  
We saw earlier that if $\p\succ\q$ then a player has better odds to win any $k$-gambling game with the $\p$-dice  than with the $\q$-dice. One can therefore conclude that the outcomes obtained by rolling the $\q$-dice are more uncertain than those obtained by rolling the $\p$-dice. 

To make this notion of \emph{uncertainty} precise, consider a game of chance in which the player is allowed to permute the symbols on the dice (for example, permute the stickers on the dice). Clearly, such a permutation cannot change the odds in any $k$-gambling game. This \emph{relabeling} of the outcomes, is equivalently described by a permutation matrix $P$ that is acting on $\p$; i.e. after the relabeling (permutation) of the symbols on the $\p$-dice, the new probability vector is given by $P\p$. 

Consider now a (somewhat unrealistic) scenario in which the player chooses to perform the relabeling at random. For example, the player can flip an unbiased coin and if the outcome is ``head" the player does nothing to the $\p$-dice whereas if the outcome is a ``tail" the player performs relabeling described by the permutation matrix $P$. Moreover, suppose also that the player forget the outcome of the coin flipping. Hence, the player changed his odds in winning the game since with probability 1/2 he did nothing to the $\p$-dice and with probability 1/2 he changed the order of the stickers on the dice. This means that now, effectively, the player holds a $\q$-dice with
\be
\q\eqdef\frac12\p+\frac12P\p\;.
\ee
Since by ``forgetting" the outcome of the coin flip the player cannot decrease the uncertainty associated with the $\p$-dice, we must conclude that the new $\q$-dice is more uncertain  than the initial $\p$-dice.

The relation above between the $\p$ and $\q$ can be expressed as
\be
\q=D\p\;,
\ee
where $D\eqdef\frac12I_n+\frac12P$. More generally, if instead of an unbiased coin the player uses a random device that produces the outcome $j\in[m]$ with probability $t_j$ and a relabeling corresponding to a permutation matrix $P_j$, then the matrix $D$ can be expressed as
\be\label{0ds}
D=\sum_{j\in[m]}t_jP_j\;.
\ee
We therefore conclude that for any such matrix and any probability vector $\p$, the vector $\q\eqdef D\p$ corresponds to more uncertainty than $\p$. The matrix $D$ above has the property that all its components are non-negative and each row and column sums to one. Such matrices are called \emph{doubly stochastic} (see Appendix~\ref{sec:bir}).\index{doubly stochastic matrix}

\begin{exercise}
Show that the matrix $D$ in~\eqref{0ds} is doubly stochastic.
\end{exercise}

The converse of the statement of the exercise above is equally valid. Owing to Birkhoff's theorem (see Theorem~\ref{birkhoff}), we know that any $n \times n$ doubly stochastic matrix\index{doubly stochastic matrix} can be expressed as a convex combination of no more than $m \leq (n-1)^2 + 1$ permutation matrices (see Exercise~\ref{ex:mun}). By integrating this theorem with our prior analysis, we deduce that $\q$ is more uncertain than $\p$ if and only if $\q = D\p$. Shortly, we will demonstrate that the relationship $\q = D\p$ corresponds to majorization. However, before we present this, it's essential to introduce the concept of a $T$-transform.

\subsubsection{$T$-Transform}\index{$T$-transform}

A $T$-transform \index{$T$-transform}is a special kind of a linear transformation from $\mbb{R}^n$ to itself. The matrix representation\index{matrix representation} of a $T$-transform \index{$T$-transform}is an $n\times n$ matrix
\be\label{ttransform2}
T=tI_n+(1-t)P\;,
\ee 
where $t\in[0,1]$ and $P$ is a permutation matrix that just exchanges two components of the vector it acts upon. Therefore, for $T$ as above there exist $x,y\in[n]$ with $x<y$ such that for every $\p=(p_1,\ldots,p_n)^T\in\prob(n)$ and every $z\in[n]$ the $z$-component of the vector $\r\eqdef T\p$ is given by
\be\label{4p14}
r_z=\begin{cases}p_z & \text{for }z\not\in\{x,y\}\\
tp_x+(1-t)p_y & \text{for }z=x\\
tp_{y}+(1-t)p_x & \text{for }z=y
\end{cases}\;.
\ee 

\bex\label{exitec}
Let $\p$ and $\r\eqdef T\p$ be as above. Show that $\p\succ\r$. 
\eex

\begin{myg}{}
\begin{lemma}\label{ttransform}
Let $\p,\q\in\mbb{R}^n$ be such that $\p\succ\q$. Then, there exists a finite $m\in\mbb{N}$ and $n\times n$ $T$-transforms $T_1,\ldots,T_m$ such that
\be
\q=T_1\cdots T_m\p\;.
\ee
\end{lemma}
\end{myg}

\begin{proof}
If $\p$ and $\q$ are related by a permutation matrix then the lemma follows from the fact that any permutation matrix on $n$ elements is a product of transposition matrices (i.e. matrices  that only exchange two elements and keep the rest unchanged). We therefore assume now that $\p$ is not a permutation of $\q$, and without loss of generality assume that $\p=\p^\da$ and $\q=\q^\da$.

The main idea of the proof is to construct a $T$-transform \index{$T$-transform}of the form given in~\eqref{ttransform2} such that the vector $\r\eqdef T\p$ as given in~\eqref{4p14} has the following three properties: 
\ben
\item $\p\succ\r\succ\q$.
\item $r_x\neq p_x$ and $r_y\neq p_y$.
\item $r_x=q_x$ or $r_y=q_y$.
\een
From the third property at least one of the components of $\r$ is equal to one of the components of $\q$. Therefore, if such a $T$-transform \index{$T$-transform}exists, by a repetition of the above process, $\q$ can be obtained from $\p$ by a finite number of such $T$-transforms. 
It is therefore left to show that a $T$-transform \index{$T$-transform}with the above three properties exists. 

Given that both $\p$ and $\q$ are probability vectors satisfying $\p\neq\q$, it is impossible for their components to satisfy $p_x \geq q_x$ for all $x\in[n]$. If this were the case, it would lead to a contradiction, as the sum of the components of both $\p$ and $\q$ must equal one, implying $p_x = q_x$ for all $x\in[n]$. Consequently, we define $x \in [n]$ as the largest integer for which $p_x > q_x$.

Similar arguments to those discussed above imply that the reverse scenario, where $p_x \leq q_x$ for all $x\in[n]$, is also not feasible.
Moreover, since $\p\neq\q$ and $\p\succ\q$, it follows from Exercise~\ref{exindices} that the largest integer $z\in[n]$ for which $p_z\neq q_z$ must satisfy $q_z>p_z$. Therefore, there exists an integer $y\in[n]$ with the property that $y>x$ and $q_y>p_y$. We take $y$ to be the \emph{smallest} integer that satisfies these two criteria.

Since $x \in [n]$ is the largest integer for which $p_x > q_x$ we get that $p_w\leq q_w$ for all $w>x$. Similarly, since $y$ is the \emph{smallest} integer that satisfy $y>x$ and $q_y>p_y$ we get that for all $x<w<y$ we have $p_w\leq q_w$. Combining these two observations we conclude that
\be\label{4p16}
p_w=q_w\quad\quad\forall\;w\in[n]\;\text{ such that }\;x<w<y\;.
\ee
Moreover,
from the definitions of $x$ and $y$, along with the fact that $x < y$ and $\p = \p^\da$ and $\q = \q^\da$, we deduce the following inequality:
\be\label{ineqxy}
p_x > q_x \geq q_y > p_y\;.
\ee

Utilizing Equation~\eqref{4p14}, we find that $r_x = tp_y + (1-t)p_x$ and $r_y = tp_x + (1-t)p_y$. By choosing $t \in (0,1)$, which means $t$ is strictly between zero and one, we ensure that the second condition, $r_x \neq p_x$ and $r_y \neq p_y$, is satisfied, given that $p_x \neq p_y$. It is important to note that $\p \succ \r$ for any $T$-transform \index{$T$-transform}(as per Exercise~\ref{exitec}). Consequently, our remaining task is to demonstrate the existence of a $t \in (0,1)$ such that $\r \succ \q$, and either $r_x = q_x$ or $r_y = q_y$ is true.

Set $\eps\eqdef\min\{p_x-q_x,q_y-p_y\}>0$, and define 
$
t\eqdef 1-\frac\eps{p_x-p_y}
$.
By definition, $0<t<1$ (due to~\eqref{ineqxy}), and if $\eps=p_x-q_x$ then $t=\frac{q_x-p_y}{p_x-p_y}$ so that
\ba
r_x=tp_x+(1-t)p_y&=p_y+t(p_x-p_y)\\
&=p_y+q_x-p_y=q_x\;.
\ea
Similarly, if $\eps=q_y-p_y$ then $t=\frac{p_x-q_y}{p_x-p_y}$ so that
\ba
r_y=tp_y+(1-t)p_x&=p_x-t(p_x-p_y)\\
&=p_x-p_x+q_y=q_y\;.
\ea
We therefore conclude that for this choice of $t\in(0,1)$ we get that $r_x=q_x$ or $r_y=q_y$. 

Before showing that $\r\succ\q$, we argue that $r_x$ can never be strictly smaller that $q_x$. Indeed, we saw above that if $\eps=p_x-q_x$ then $r_x=q_x$. Moreover, for the second option that $\eps=q_y-p_y$ we have $t=\frac{p_x-q_y}{p_x-p_y}$ so that
\ba
r_x=tp_x+(1-t)p_y&=p_y+t(p_x-p_y)\\
&=p_y+p_x-q_y\\
\Gg{\eps=q_y-p_y}&=p_x-\eps\\
\Gg{\eps\leq p_x-q_x}&\geq p_x-(p_x-q_x)=q_x\;.
\ea
We therefore conclude that for both options $r_x\geq q_x$.

Finally, to show that $\r\succ\q$ we show that $\|\r\|_{(k)}\geq\|\q\|_{(k)}$ for all $k\in[n]$. We show it in three cases:
\ben
\item For $1\leq k<x$ we have $\|\r\|_{(k)}=\|\p\|_{(k)}\geq \|\q\|_{(k)}$ since $\p\succ\q$ and $r_w=p_w$ for $w\in[k]$. \item For $x \leq k < y$, we have the following relation:
\be\label{eq4p2}
\|\r\|_{(k)} \geq \sum_{w \in [k]} r_w = \|\p\|_{(x-1)} + r_x + \sum_{w=x+1}^k p_w\;.
\ee
The first term on the right-hand side, $\|\p\|_{(x-1)}$, satisfies $\|\p\|_{(x-1)} \geq \|\q\|_{(x-1)}$ since $\p \succ \q$. For the second term, we have already established that $r_x \geq q_x$, and for the third term we get from~\eqref{4p16} that $\sum_{w=x+1}^k p_w= \sum_{w=x+1}^k q_w$. Incorporating these three relations into~\eqref{eq4p2} yields $\|\r\|_{(k)} \geq \|\q\|_{(k)}$.
\item For $y\leq k\leq n$ we use the fact that $r_x+r_y=p_x+p_y$ so that
\be
\|\r\|_{(k)}\geq \sum_{w\in[k]}r_w\\
=\sum_{w\in[k]}p_w=\|\p\|_{(k)}\\
\geq\|\q\|_{(k)}\;,
\ee
where the last inequality follows from the fact that $\p\succ\q$.
\een
Hence, $\r\succ\q$. This completes the proof.
\end{proof}

\bex
Prove the converse of the statement presented in Lemma~\ref{ttransform}. Specifically, demonstrate that for any vector $\p \in \mathbb{R}^n$ and a sequence of $m$ $n \times n$ $T$-transforms, denoted as $T_1, \ldots, T_m$, the resulting vector $\q \eqdef T_1\cdots T_m\p$ fulfills the condition $\p \succ \q$. Hint: Refer to Exercise~\ref{exitec} for guidance.
\eex

\subsubsection{The Fundamental Theorem of Majorization}\index{majorization}

We are now prepared to introduce the fundamental theorem of majorization, which delineates the relationship between doubly-stochastic matrices and majorization. For clarity, we adopt the notation $\mathbf{1}_n \eqdef (1,\ldots,1)^T \in \mathbb{R}^n$. This allows for a succinct expression of the sum of the components of any vector $\p \in \mathbb{R}^n$ as the dot product $\mathbf{1}_n \cdot \p$.

\begin{myt}{\color{yellow} Characterization}
\begin{theorem}\label{chmaj}
Let $\p,\q\in\mbb{R}^n$. The following are equivalent:
\begin{enumerate}
\item $\p\succ\q$.
\item There exists an $n\times n$ doubly stochastic matrix\index{doubly stochastic matrix} $D$ such that $\q=D\p$.
\item $\1_n\cdot\p=\1_n\cdot\q$ and for every $t\in\mbb{R}$
\be
\left\|\p-t\mathbf{u}^{(n)}\right\|_1\geq \left\|\q-t\mathbf{u}^{(n)}\right\|_1\;.
\ee
\end{enumerate}
\end{theorem}
\end{myt}

\begin{remark}
In some books the last condition is expressed as
\be\label{sholds}
\sum_{x\in[n]}\left(p_x-t\right)_+\geq \sum_{x\in[n]}\left(q_x-t\right)_+\;,
\ee
where for all $r\in\mbb{R}$ the notation $(r)_+=r$ if $r\geq 0$ and otherwise $(r)_+=0$. 
To see the equivalence note that $(r)_+=\frac12(|r|+r)$ and ``absorb" the factor $1/n$ into $t$.
\end{remark}

\begin{proof} We divide the proof into three parts:

{\it The implication $1\Rightarrow2$:}
From Lemma~\ref{ttransform} we get that if $\p\succ\q$ then  $\q=T_1\cdots T_m\p$ for some $T$-transforms $T_1,\ldots,T_m$. Since the product of the $T$-transforms $D\eqdef T_1\cdots T_m$ is doubly stochastic (see Exercise~\ref{pofdub}) it follows that $\q=D\p$ and $D$ is doubly stochastic. 

{\it The implication $2\Rightarrow3$:} Suppose $\q=D\p$ for some doubly stochastic matrix\index{doubly stochastic matrix} $D$. Then, since $D\u^{(n)}=\u^{(n)}$ we get for all $t\in\mbb{R}$
\ba
\left\|\q-t\u^{(n)}\right\|_1=\left\|D\p-t\u^{(n)}\right\|_1&=\left\|D\p-tD\u^{(n)}\right\|_1\\
\GG{\eqref{dpisv}}&\leq \left\|\p-t\u^{(n)}\right\|_1\;,
\ea
where in the last inequality we used the property~\eqref{dpisv} of the norm $\|\cdot\|_1$, in conjunction with the fact that the doubly-stochastic matrix $D$ is particularly column stochastic.

{\it The implication $3\Rightarrow1$:} Suppose~\eqref{sholds} holds for all $t\in\mbb{R}$. Without loss of generality suppose that $\p=\p^\da$ and $\q=\q^\da$. Fix $k\in[n-1]$ and observe that for $t=p_{k+1}$ the left-hand side of~\eqref{sholds} can be expressed as:
\ba
\sum_{x\in[n]}\left(p_x-t\right)_+&=\sum_{x\in[n]}\left(p_x-p_{k+1}\right)_+\\
\Gg{\p=\p^\da}&=\|\p\|_{(k)}-kp_{k+1}\;.
\ea
Furthermore, the right-hand side of~\eqref{sholds} satisfies 
\ba
\sum_{x\in[n]}\left(q_x-p_{k+1}\right)_+&\geq \sum_{x\in[k]}\left(q_x-p_{k+1}\right)_+\\
\Gg{\left(q_x-p_{k+1}\right)_+\geq q_x-p_{k+1}}&\geq \sum_{x\in[k]}\left(q_x-p_{k+1}\right)\\
\Gg{\q=\q^\da}&=\|\q\|_{(k)}-kp_{k+1}\;.
\ea
Hence, the combination of the two equations above with our assumption that~\eqref{sholds} holds for all $t\in\mbb{R}$, and in particular for $t=p_{k+1}$, gives $\|\p\|_{(k)}\geq\|\q\|_{(k)}$. 
Since $k\in[n-1]$ was arbitrary we conclude that $\p\succ\q$. This completes the proof.
\end{proof}

\bex\label{pofdub}
Show that the product of two $n\times n$ doubly stochastic matrices is itself a doubly stochastic matrix.
\eex

\subsection{Three Equivalent Approaches}\label{3approaches}

We have seen that majorization is a pre-order relationship between vectors in $\prob(m)$. Given that a probability vector can represent a classical system $X$, we can understand majorization in terms of \emph{mixing operations} applied to system $X$. To approach this, let's set aside our initial definition of majorization and attempt to redefine it as follows: We state that a vector $\p$ majorizes another vector $\q$ (denoted as $\p\succ\q$) if there exists a mixing operation $M\in\stoc(m,m)$ that fulfills the condition:
\be
\q=M\p\;.
\ee
The pivotal question then becomes how to define these mixing operations. Conceptually, mixing operations are processes that increase the uncertainty of system $X$. In this context, we propose that the mixing operation $M$ can be conceptualized in three distinct manners:

\ben

\item \emph{The axiomatic approach\index{axiomatic approach}:} Since mixing operations can only increase the uncertainty of system $X$, the uniform distribution remains invariant under mixing operations, as its uncertainty cannot be increased further. Hence, $M\in\stoc(n,n)$ can be defined as a mixing operation if \index{axiomatic approach\index{axiomatic approach}}
\be
M\u^X=\u^X\;.
\ee
That is, $M=D$ is doubly stochastic.
\item \emph{The Constructive Approach:}\index{constructive approach} In this approach the mixing operations are defined intuitively as a convex combination of permutation matrices. Indeed, mixing a pack of cards literally corresponds to the action of a random permutation. Therefore, in this approach $M\in\stoc(n,n)$ corresponds to a mixing operation if there exists $k\in\mbb{N}$ $m\times m$ permutation  matrices $\{P_j\}_{j\in[k]}$ such that
\be
M=\sum_{j\in[k]}s_jP_j\;,
\ee 
for some $\s\in\prob(k)$.
\item \emph{Operational Approach:}\index{operational approach} In this approach, a mixing operation is defined as a stochastic process $M\in\stoc(m,m)$ that cannot increase the chances to win a game of chance. Specifically, $M\in\stoc(m,m)$ is a mixing operation if and only if for all $\p\in\prob(m)$ and for all $k\in[m]$ the probability, $\pr_k$, to win a $k$-gambling game satisfies
\be
\pr_k(\p)\geq\pr_k(M\p)\;.
\ee
\een
As we proved in the preceding subsections, all the three approaches above are equivalent, leading to the same pre-order given in~\eqref{a11}. Furthermore, the established equivalence of these approaches solidifies the conceptual foundation of uncertainty. This, in turn, validates functions that exhibit monotonic behavior under majorization as reliable quantifiers of uncertainty. Such measures of uncertainty are known as Schur concave functions.

\subsection{Schur Convexity}\label{sec:schurconvex}\index{Schur's convexity}

A function $f:\mbb{R}^n\to \mbb{R}$ that preserves the majorization order is said to be \emph{Schur convex}. That is, $f$ is Schur convex if and only if for any two vector $\p,\q\in\mbb{R}^n$ such that $\p\succ\q$ we have
\be
f(\p)\geq f(\q)\;.
\ee
We also say that $f$ is Schur concave if $-f$ is Schur convex.
Note that a Schur convex function $f$ is symmetric with respect to permutations, since if $\q=P\p$ for some permutation matrix $P$ then we have $\p\succ\q$ and $\q\succ\p$ so that $f(\p)=f(\q)$. Moreover, from the fundamental theorem of majorization if $\p\succ\q$ then there exists a set of permutation matrices $\{P_j\}_{j\in[m]}$ and a probability distribution $\{t_j\}_{j\in[m]}$ such that
\be
\q=\sum_{j\in[m]}t_jP_j\p\;.
\ee
Therefore, if a function $f:\mbb{R}^n\to \mbb{R}$ is symmetric (under permutations) and convex then it is necessarily Schur convex. To see this, observe that
\ba
f(\q)&=f\Big(\sum_{j\in[m]}t_jP_j\p\Big)\\
\GG{{\it f}\;is\;convex}&\leq\sum_{j\in[m]}t_jf\left(P_j\p\right)\\
\GG{{\it f}\;is\;symmetric}&=\sum_{j\in[m]}t_jf\left(\p\right)=f(\p)\;.
\ea
As an example, consider the Shannon entropy\index{Shannon entropy}, defined for any probability vector $\p\in\prob(n)$ as
\be
H(\p)\eqdef-\sum_{x\in[n]}p_x\log_2(p_x)\;.
\ee 
This function is clearly symmetric under any permutation of the components of $\p$, and it is also concave. Therefore, the Shannon entropy is an example of a Schur concave function.

\begin{exercise}
Show that the geometric mean function\index{geometric mean}
\be
G(\p)\eqdef \big(p_1p_2\cdots p_n\big)^{\frac1n}\quad\quad\forall\;\p\in\prob(n)\;,
\ee 
is Schur concave. Hint: Show first that $\log G(\p)$ is Schur concave by showing that it is both symmetric and concave (what is its Hessian matrix?).
\end{exercise}

From the following exercise it follows that not all Schur convex functions are symmetric and convex. Therefore, in this sense, the notion of Schur convexity is weaker than (standard) convexity.

\begin{exercise}
Consider the function
\be
f(\p)=\log \max\{p_1,\ldots,p_n\}\quad\quad\forall\;\p\in\prob(n)\;.
\ee
\begin{enumerate}
\item Show that $f$ is symmetric.
\item Show that $f$ is \emph{not} convex.
\item Show that $f$ is Schur convex.
\end{enumerate}
\end{exercise} 

One way to test if a given multivariable function is convex is to check if its Hessian matrix is positive semidefinite. Since Schur convex functions are not necessarily convex, this test cannot always be used to determine if a function is Schur convex. Instead, we can use the theorem below to test if a given symmetric function is Schur convex.

\begin{myt}{\color{yellow} Schur's Test}\index{Schur's test}
\begin{theorem}\label{scht}
Let $f:\prob(n)\to\mbb{R}$ be a continuous function that is also continuously differentiable on the interior of $\prob(n)$. Then, $f$ is Schur convex if and only if the following two conditions hold:
\begin{enumerate}
\item $f$ is symmetric in $\prob(n)$; i.e. for every $n\times n$ permutation matrix $P$
\be
f(\p)=f(P\p)\quad\quad\forall\;\p\in\prob(n)\;.
\ee
\item For all $0<\p\in\prob(n)$ 
\be\label{crite}
(p_1-p_2)\left(\frac{\partial f(\p)}{\partial p_1}-\frac{\partial f(\p)}{\partial p_2}\right)\geq 0\;.
\ee
\end{enumerate}
\end{theorem}
\end{myt}
\begin{remark}
Since the function $f$ is symmetric, the condition in~\eqref{crite} is equivalent the following condition. For all $0<\p\in\prob(n)$ and all $x\neq y\in[n]$
\be\label{crite2}
(p_x-p_y)\left(\frac{\partial f(\p)}{\partial p_x}-\frac{\partial f(\p)}{\partial p_y}\right)\geq 0\;.
\ee
\end{remark}
\begin{proof}

Suppose $f$ is Schur convex. We need to show that~\eqref{crite} holds. Let $\p>0$ and observe that if $p_1=p_2$ the condition clearly holds. Therefore, without loss of generality we assume that $p_1>p_2$ (and recall that $p_2>0$ since $\p>0$). Let $0<\eps< p_1-p_2$ and define
\be\label{ptil}
\tp_\eps\eqdef(p_1-\eps,p_2+\eps,p_3,\ldots,p_n)^T\;.
\ee
Let $a\eqdef \frac\eps{p_1-p_2}<1$ and observe that $\tp_\eps=D\p$, where
$D=D_2\oplus I_{n-2}$  with 
\be
D_2\eqdef \begin{pmatrix}
1-a &a\\
a & 1-a
\end{pmatrix}\;.
\ee
Since $D$ is doubly stochastic, we get  from Theorem~\ref{chmaj} that $\p\succ\tp_\eps$. Therefore, from the assumption that $f$ is Schur convex we conclude that for all $0<\eps< p_1-p_2$
\ba
0&\leq \frac{f(\p)-f(\tp_\eps)}{\eps}\\
&=\frac{f(\p)-f(p_1-\eps,p_2,\ldots,p_n)}\eps+\frac{f(p_1-\eps,p_2,\ldots,p_n)-f(\tp_\eps)}\eps\\
&\xrightarrow{\eps\to 0^+}\frac{\partial f(\p)}{\partial p_1}-\frac{\partial f(\p)}{\partial p_2}\;.
\ea
Hence,  we get that $\frac{\partial f(\p)}{\partial p_1}\geq\frac{\partial f(\p)}{\partial p_2}$ which is equivalent to~\eqref{crite} (since $p_1>p_2$).

For the converse, let $0<\p,\q\in\prob(n)$ and suppose $\p\succ\q$. We need to show that $f(\p)\geq f(\q)$.  From Lemma~\ref{ttransform} we know that $\q$ can be obtained from $\p$ by a sequence of $T$-transforms. Therefore, it is sufficient to show that $f$ is non-increasing under a single action of $T$-transform. Explicitly, it is sufficient to show that $f(T\p)\leq f(\p)$ for any $T=tI_n+(1-t)P$ where $t\in[0,1]$ and $P$ is a permutation matrix exchanging only two components. Moreover, we can assume that $t\in[\frac12,1]$ since $f$ is symmetric under permutations and therefore $f(T\p)=f(PT\p)$, where $PT$ is the same $T$ transform but with $1-t$ replacing $t$ (since $P^2=I_n$).
For simplicity of the exposition, we also assume that $P$ is the matrix exchanging the first and second components, so that 
\be
T\p=(tp_1+(1-t)p_2,tp_2+(1-t)p_1,p_3,\ldots,p_n)^T\;.
\ee
Note that if $p_1=p_2$ then the transformation does not effect $\p$. Therefore, without loss of generality suppose that $p_1>p_2$ (recall that $f$ is symmetric, so we can exchange between $p_1$ and $p_2$ if necessary). Now, from~\eqref{crite} we get that
\be
\frac d{d\eps}f(p_1-\eps,p_2+\eps,p_3,\ldots,p_n)\leq 0\;,
\ee
for any $0\leq \eps\leq \frac12(p_1-p_2)$ (note that in this domain $p_1-\eps\geq p_2+\eps$). We therefore conclude that the function \be
g(\eps)\eqdef f(p_1-\eps,p_2+\eps,p_3,\ldots,p_n)
\ee
is non-increasing in the domain
$0\leq \eps\leq \frac12(p_1-p_2)$. Taking $\eps\eqdef (1-t)(p_1-p_2)$ we conclude that
\ba
f(\p)=g(0)\geq g(\eps)&=f(p_1-\eps,p_2+\eps,p_3,\ldots,p_n)\\
\GG{By\;definition\;of\;\eps}&=f(tp_1+(1-t)p_2,tp_2+(1-t)p_1,p_3,\ldots,p_n)\\
&=f(T\p)
\ea
This completes the proof.
\end{proof}

As an example, consider the family of R\'enyi entropies defined for any $\alpha\in[0,\infty]$ and all $\p\in\prob(n)$ as
\be
H_\alpha(\p)\eqdef\frac1{1-\alpha}\log\sum_{x\in[n]}p_x^\alpha\;,
\ee 
where the cases $\alpha=0,1,\infty$ are defined in terms of their limits.
In the next chapter we will study these functions in more details. Here we show that for all $\alpha\in(0,\infty)$ the R\'enyi entropies are Schur concave. Due to the monotonicity of the log function, it is enough to show that $f_\alpha(\p)\eqdef\sum_{x\in[n]}p_x^\alpha$ is Schur convex for $\alpha\in(0,1)$ and Schur concave for $\alpha\in(1,\infty)$. Indeed,
\be
(p_1-p_2)\left(\frac{\partial f(\p)}{\partial p_1}-\frac{\partial f(\p)}{\partial p_2}\right)
=\alpha(p_1-p_2)\left(p_1^{\alpha-1}-p_2^{\alpha-1}\right)
\ee
which is always non-negative for $\alpha>1$ and non-positive for $\alpha\in(0,1)$. Hence, from the Schur's test it follows that $H_\alpha(\p)$ is Schur concave for all $\alpha\in[0,\infty]$ (the cases $\alpha=0,1,\infty$ follow from the continuity of $H_\alpha$ in $\alpha$).

As another example, consider the elementary symmetric functions defined for each $k\in[n]$ by
\be\label{531}
f_k(\p)\eqdef\sum_{\substack{x_1<\cdots< x_k\\ x_1,\ldots,x_k\in[n]}}p_{x_1}\cdots p_{x_k}\quad\quad\forall\;\p\in\prob(n)\;.
\ee
For example, for $k=2$ they take the form
\be
f_2(\p)=\sum_{\substack{x<y\\ x,y\in[n]}}p_xp_y\;,
\ee
and for $k=n$ we have
$
f_n(\p)=p_1\cdots p_n
$.
From  Schur's test it follows that the elementary symmetric functions are Schur concave.
\begin{exercise}
Use  Schur's test to verify that the elementary symmetric functions are Schur concave.
\end{exercise}

\section{Approximate Majorization}\label{sec:am}\index{approximate majorization}

We saw in the previous section that given two probability vectors $\p,\q\in\prob(n)$ it is possible to have both $\p\not\succ\q$ and $\q\not\succ\p$ (i.e. $\p$ and $\q$ are incomparable). For some applications in resource theories, it is often useful to know how much one has to perturb $\p$ so that $\p\succ\q$. To make this idea rigour,  
we first introduce several notions of ``greatest" and ``least" elements in a subset of probability vectors. 

\begin{myd}{Maximal and Minimal Elements}
\begin{definition}
Let $\mc\subseteq\prob(n)$ be a subset of $n$-dimensional probability vectors.
\begin{enumerate}
\item[$\bullet$] A vector $\p\in\mc$ is said to be a maximal element of $\mc$ if for all $\q\in\mc$ we have $\p\succ\q$.
\item[$\bullet$] A vector $\p\in\mc$ is said to be a minimal element of $\mc$ if for all $\q\in\mc$ we have $\q\succ\p$.
\end{enumerate}
\end{definition}
\end{myd}
In general, partial orders don't always have maximal and minimal elements. For example, the set 
\be
\mc\eqdef\left\{\begin{bmatrix}1/2\\ 1/4 \\ 1/4\end{bmatrix},
\begin{bmatrix}2/5\\ 2/5 \\ 1/5\end{bmatrix}\right\}
\ee
has no maximal nor minimal elements since none of the two vectors majorize the other. On the other hand, one can define upper and lower \emph{bounds} on a set of probability vectors. Specifically, given a subset $\mc\subseteq\prob(n)$
\begin{enumerate}
\item[$\bullet$] A vector $\p\in\prob(n)$ is said to be an upper bound of $\mc$ if for all $\q\in\mc$ we have $\p\succ\q$.
\item[$\bullet$] A vector $\p\in\prob(n)$ is said to be a lower bound of $\mc$ if for all $\q\in\mc$ we have $\q\succ\p$.
\end{enumerate}
Note that lower and upper bounds always exist since the vector $\e_1=(1,0,\ldots,0)^T$ is always an upper bound and the vector $\u^{(n)}$ is always a lower bound.  Less trivial bounds are those that are optimal: Given a subset $\mc\subseteq\prob(n)$,
\begin{enumerate}
\item[$\bullet$] An upper bound $\p\in\prob(n)$ of $\mc$ is said to be \emph{optimal} if for any other upper bound $\p'\in\prob(n)$ of $\mc$ we have $\p'\succ\p$.
\item[$\bullet$] A lower bound $\p\in\prob(n)$ of $\mc$ is said to be \emph{optimal} if for any other lower bound $\p'\in\prob(n)$ of $\mc$ we have $\p\succ\p'$.
\end{enumerate}

\begin{exercise}
 Let 
 \be
\mc\eqdef\big\{\p_1,\ldots,\p_m\big\}\subset\prob(n)\;.
\ee
be a set consisting of $m$ probability vectors, and for each $z\in[n]$ denote by
 \be
s_z\eqdef \max_{y\in[m]}\|\p_y\|_{(z)}
\ee
where $\|\cdot\|_{(z)}$ is the Ky Fan norm (see~\eqref{kfnorm}). Finally, denote by 
\be\label{qdz5}
q_z\eqdef s_z-s_{z-1}\quad\quad\forall\;z\in[n]\;,
\ee
with $s_0\eqdef 0$. Show that the vector $\q=(q_1,\ldots,q_n)^T$ is a probability vector in $\prob(n)$ that is also an upper bound  of $\mc$. Is it optimal?
\end{exercise}

The study of the above notions of extrema under majorization of a set $\mc$ goes beyond the scope of this book. Here, we are only interested in a particular set of vectors, namely, a ball of a small radius around some probability vector.
Let $\p\in\prob(n)$ be a probability vector and for any $\eps\in[0,1]$ define a ``ball" of radius $\eps$ around it as
\be
\mb_{\eps}(\p)\eqdef\left\{\p'\in\prob(n)\;:\;\frac12\|\p-\p'\|_1\leq\eps\right\}\;.
\ee
Remarkably, the above subset of $\prob(n)$ has both minimal and maximal elements, known as the \emph{flattest} and \emph{steepest} $\eps$-approximations of $\p$.

\subsection{The Steepest $\eps$-Approximation}\index{steepest approximation}
In this subsection we find the maximal element (under majorization) of $\mb_{\eps}(\p)$, where $\p\in\prob(n)$. 
Recall that the vector $\e_1\eqdef(1,0,\ldots,0)^T\in\prob(n)$ satifies $\e_1\succ\q$ for all $\q\in\prob(n)$. Therefore,
if $\frac12\|\p-\e_1\|_1\leq\eps$, i.e., $\e_1\in\mb_{\eps}(\p)$, then the maximal element of $\mb_{\eps}(\p)$ is unique up to permutation of the components,  and is given by  $\e_1$. We will therefore assume now that $\frac12\|\p-\e_1\|_1>\eps$ and that the components of $\p$ are arranged in non-increasing order; i.e. $\p=\p^\da$.

\begin{exercise}
Let $\{\e_z\}_{z=1}^n$ be the elementary basis of $\mbb{R}^n$. Show that if $\p=\p^\da$ and $\e_1\not\in\mb_{\eps}(\p)$ then for all $z\in[n]$ we have $\e_z\not\in\mb_{\eps}(\p)$.
Hint: Show first that  $\frac12\|\p-\e_z\|_1=1-p_z$.
\end{exercise}

Fix $\eps\in(0,1)$ and let $k\in[n]$ be the integer satisfying 
\be\label{secinein}
\|\p\|_{(k)}\leq 1-\eps< \|\p\|_{(k+1)}\;.
\ee
With this index $k$ we define the steepest $\eps$-approximation of $\p$, denoted by $\overline{\p}^{(\eps)}$, whose components are
\be\label{a556}
\overline{p}_x^{(\eps)}\eqdef 
\begin{cases}
p_1+\eps &\text{if }x=1\\
p_x &\text{if }x\in\{2,\ldots,k\}\\
1-\eps-\|\p\|_{(k)} &\text{if }x=k+1\\
0 &\text{otherwise}
\end{cases}\;.
\ee
Note that from its definition above, $\overline{\p}^{(\eps)}$ is indeed a probability vector whose components are arranged in non-increasing order. 

\begin{exercise}
Utilize the definition of $k$ as provided in~\eqref{secinein} and the definition of $\overline{\p}^{(\eps)}$ as outlined in~\eqref{a556} to demonstrate the following two properties:
\ben
\item The components of $\overline{\p}^{(\eps)}$ are arranged in non-increasing order. In particular, $\overline{p}_k^{(\eps)}>\overline{p}_{k+1}^{(\eps)}$.
\item The components of $\overline{\p}^{(\eps)}$ satisfy
\be\label{inequcom}
\overline{p}_x^{(\eps)}\leq p_x\quad\quad\forall\;x\in\{2,\ldots,n\}\;.
\ee
In particular, $\overline{p}_{k+1}^{(\eps)}< p_{k+1}$.
\een
\end{exercise}

The intuition behind the definition above is that we want to alter $\p$ in a way that it becomes more similar to $\e_1$. However, since $\overline{\p}^{(\eps)}$ must be close to $\p$ we cannot increase $p_1$ by too much. Indeed, the vector $\overline{\p}^{(\eps)}$ as defined above is $\eps$-close to $\p$. To see this, observe that from its definition
\ba
\frac12\left\|\overline{\p}^{(\eps)}-\p\right\|_1&=\sum_{x\in[n]}\left(\overline{p}_x^{(\eps)}-p_x\right)_+\\
\GG{\eqref{inequcom}}&=p_1+\eps-p_1=\eps\;.
\ea
Therefore, the vector $\overline{\p}^{(\eps)}$ is indeed in $\mb_{\eps}(\p)$. 

\begin{myt}{}
\begin{theorem}\label{steepest}
Let $\p\in\prob^\da(n)$ be such that 
$\frac12\|\p-\e_1\|_1>\eps$. Then, the vector $\overline{\p}^{(\eps)}$ as defined in~\eqref{a556} is the maximal element (under majorization) of $\mb_{\eps}(\p)$.
\end{theorem}
\end{myt}

\begin{proof}
Since we already showed that $\overline{\p}^{(\eps)}\in\mb_{\eps}(\p)$ it is left to show that for any $\q\in\mb_{\eps}(\p)$  we have $\overline{\p}^{(\eps)}\succ\q$. Indeed, since $\q\in\mb_{\eps}(\p)$ it follows from~\eqref{e5p18} that for every $\ell\in[n]$
\ba
\|\q\|_{(\ell)}-\|\p\|_{(\ell)}\leq \frac12\|\q-\p\|\leq\eps\;.
\ea
Therefore, for $\ell\in[k]$ we get 
\ba
\|\q\|_{(\ell)}&\leq \|\p\|_{(\ell)}+\eps\\
\GG{\eqref{a556}}&=\|\overline{\p}^{(\eps)}\|_{(\ell)}\;.
\ea
Combining this with the fact that for $k+1\leq\ell\leq n$, $\|\overline{\p}^{(\eps)}\|_{(\ell)}=1$, we conclude that $\overline{\p}^{(\eps)}\succ\q$. This completes the proof.
\end{proof}

\bex\label{ex:mubz}
Let $m,n\in\mbb{N}$ be such that $m<n$, and let $\p\in\prob^\da(n)$ be such that $\u^{(m)}\not\succ\p$.
Show that a minimal element of the set
 \be
 \mc_{\p,m}\eqdef\big\{\q'\in\prob(m)\;:\;\q'\succ\p\big\}
\ee
is given by a probability vector $\q\in\prob^\da(m)$ of the form
\be
\q=\big(p_1,\ldots,p_\ell,p_\ell,\ldots p_\ell,1-(m-\ell-1)p_\ell\big)^T\;,
\ee
where $\ell\in[m]$ is the largest integer satisfying
\be
(m-\ell)p_\ell\geq 1-\|\p\|_{(\ell)}\;.
\ee
\eex

One can use the standard\index{steepest approximation} $\eps$-approximation to compute the distance of a vector $\p\in\prob(n)$ to the set of all vectors $\r\in\prob(n)$ that majorizes $\q$. Specifically, let
\be
{\rm Majo}(\q)\eqdef\{\r\in\prob(n)\;:\;\r\succ\q\}\;,
\ee
denotes the set of all vectors in $\prob(n)$ that majorize $\q$, and define the distance between $\p\in\prob(n)$ and the set ${\rm Majo}(\q)$ as:
\be
T\big(\p,{\rm Majo}(\q)\big)\eqdef\min_{\r\in{\rm Majo}(\q)}\frac12\|\p-\r\|_1\;.
\ee
\begin{myt}{}
\begin{theorem}\label{thm:dlocc0}
Using the same notations as above, for all $\p,\q\in\prob(n)$
\be\label{1224t0}
T\big(\p,{\rm Majo}(\q)\big)=\max_{\ell\in[n]}\big\{\|\q\|_{(\ell)}-\|\p\|_{(\ell)}\big\}\;.
\ee
\end{theorem}
\end{myt}

\begin{proof}
Without loss of generality we will assume that $\p,\q\in\prob^\da(n)$. For any $\eps\in(0,1)$, let $\overline{\p}^{(\eps)}$ be the steepest $\eps$-approximation of $\p$; see~\eqref{a556}. Observe that by definition, for any $m\in[d]$ we have
$
\left\|\overline{\p}^{(\eps)}\right\|_{(m)}\leq \|\p\|_{(m)}+\eps
$
with equality if $m\in[k]$.
In Theorem~\ref{steepest} we showed that  $\overline{\p}^{(\eps)}$ is the maximal element of $\mb_{\eps}(\p)$ as long as $\eps<\frac12\|\p-\e_1\|_1$ (otherwise, $\e_1$ is the maximal element).
Hence,
\ba\label{1228}
T\big(\p,{\rm Majo}(\q)\big)&\eqdef\min\Big\{ \frac12\|\p-\r\|_1\;:\;\r\succ\q\;,\quad\r\in\prob(n)\Big\}\\
&=\min\Big\{ \eps\in[0,1]\;:\;\r\succ\q\;,\quad\r\in\mb_\eps(\p)\Big\}\\
\Gg{\overline{\p}^{(\eps)}\succ\r\;\;\quad\forall\;\r\in\mb_\eps(\p)}&=\min\Big\{\eps\in[0,1]\;:\;\overline{\p}^{(\eps)}\succ\q\Big\}\;.
\ea
That is, it is left to compute the smallest $\eps$ that satisfy $\overline{\p}^{(\eps)}\succ\q$. We will show that this smallest $\eps$ equals 
\be\label{defdel}
\delta\eqdef\max_{\ell\in[n]}\big\{\|\q\|_{(\ell)}-\|\p\|_{(\ell)}\big\}\;.
\ee 
We first show that $\overline{\p}^{(\delta)}\succ\q$. Let $k$ be the integer satisfying~\eqref{secinein} but with $\delta$ replacing $\eps$. Then, from the definition of $\overline{\p}^{(\delta)}$ it follows that for $m>k$ we have 
\be
\left\|\overline{\p}^{(\delta)}\right\|_{(m)}=1\geq \|\q\|_{(m)}\;.
\ee
Moreover, for $m\in[k]$ the definition in~\eqref{defdel} gives $\delta\geq \|\q\|_{(m)}-\|\p\|_{(m)}$ so that 
\ba
\left\|\overline{\p}^{(\delta)}\right\|_{(m)}&=\|\p\|_{(m)}+\delta\geq
\|\q\|_{(m)}\;.
\ea
Hence, $\overline{\p}^{(\delta)}\succ\q$. To prove the optimality of $\delta$, we show that for any $0<\delta'<\delta$ we must have $\overline{\p}^{(\delta')}\not\succ\q$. Indeed, since $\delta'<\delta$ there exists $m\in[d]$ such that
\be\label{dep3}
\delta'<\|\q\|_{(m)}-\|\p\|_{(m)}\;.
\ee
Combining this with the observation that $\big\|\overline{\p}^{(\delta')}\big\|_{(\ell)}\leq \|\p\|_{(\ell)}+\delta'$ for all $\ell\in[n]$, we get for this $m$
\ba
\big\|\overline{\p}^{(\delta')}\big\|_{(m)}&\leq \|\p\|_{(m)}+\delta'\\
\GG{\eqref{dep3}}&<\|\q\|_{(m)}\;.
\ea
Hence, $\overline{\p}^{(\delta')}\not\succ\q$. This concludes the proof.
\end{proof}

\subsection{The Flattest $\eps$-Approximation}\label{sec:flattest}\index{flattest approximation}

In this subsection, we aim to identify the minimal element within the set $\mb_{\eps}(\p)$, given that $\p\in\prob(n)$. Specifically, our objective is to locate the vector in $\mb_{\eps}(\p)$ that exhibits the most uniform (or ``flattest") distribution.
It's evident that if the uniform distribution vector $\u^{(n)}$ is a member of $\mb_\eps(\p)$, then it is the minimal element of $\mb_{\eps}(\p)$. Therefore, our analysis will focus on the scenario where $\u^{(n)}\not\in\mb_\eps(\p)$. In this context, the parameter $\eps$ satisfies the following condition:
\be\label{5p38}
0<\eps<\frac12\left\|\p-\u^{(n)}\right\|_1\;.
\ee
Additionally, we will assume that the components of the vector $\p$ are sorted in a non-increasing order; i.e., $\p=\p^\da$.

\bex\label{exl}
Let $\eps\in(0,1)$, $\p\in\prob^\da(n)$, and $\ell\in[n]$ be the integer satisfying $p_\ell\geq\frac1n>p_{\ell+1}$. Show that the inequality in~\eqref{5p38} holds if and only if 
\be\label{oeps}
\eps<\|\p\|_{(\ell)}-\frac\ell n\;.
\ee
Hint: Start by expressing $\frac12\left\|\p-\u^{(n)}\right\|_1$ as $\sum_{x\in[n]}\big(p_x-1/n\big)_+$.
\eex

The minimal element of $\mb_{\eps}(\p)$ can be found by ``flattening" the tip of $\p$ (i.e. first few components of $\p$) and tail of $\p$ (i.e. the last few components of $\p$). The intuition behind this idea is to alter  the vector $\p$ so that it becomes more similar to the uniform distribution $\u^{(n)}$. This process involves replacing the first $k$ components of $\p$ with a constant $a$, and substituting the last $n-m$ components with another constant $b$. We denote by $\up^{(\eps)}\in\prob(n)$ the resulting vector. Its components are given by
\be\label{599a}
\underline{p}_x^{(\eps)}\eqdef 
\begin{cases}
a &\text{if }x\in[k]\\
p_x &\text{if }k<x\leq m\\
b &\text{if }x\in\{m+1,\ldots,n\}
\end{cases}\;.
\ee
The objective is to select suitable values for $a$, $b$, $k$, and $m$, ensuring that $\up^{(\eps)}$ forms the flattest $\eps$-approximation of $\p$. 

To find the coefficients $a,b,k,m$ we outline the properties that $\up^{(\eps)}$ has to satisfy:
\ben
\item The vector $\up^{(\eps)}$ is a probability vector in $\prob(n)$. Since all of its components are non-negative, we just need to require that they sum to one. Using the relation $\sum_{x=k+1}^mp_x=\|\p\|_{(m)}-\|\p\|_{(k)}$ we get that the coefficients $a,b,k,m$ must satisfy
\be\label{kanmb}
1=\sum_{x\in[n]}\underline{p}_x^{(\eps)}=ka+\|\p\|_{(m)}-\|\p\|_{(k)}+(n-m)b\;.
\ee
\item The vector $\up^{(\eps)}\in\prob^\da(n)$; i.e.\ its components are arranged in non-decreasing order. Since $\p=\p^\da$ it is sufficient to require that  $a>p_{k+1}$ and $b<p_m$ (these inequalities are strict since we want $k$ and $m$ to mark the indices in~\eqref{599a} in which that the ``flattening" process ends and begins, respectively). We therefore conclude that
\be\label{545}
a\in(p_{k+1},p_k]\quad\text{and}\quad b\in[p_{m+1},p_m)\;.
\ee
\item The vector $\up^{(\eps)}\in\mb^{(\eps)}(\p)$. Moreover, since $\up^{(\eps)}$ is an optimal vector, we would expect it to be $\eps$-close (and not $\delta$-close with $\delta<\eps$) to $\p$. Therefore, we require that
\ba\label{546}
\eps=\frac12\left\|\p-\up^{(\eps)}\right\|_1&=\sum_{x\in[n]}\left(p_x-\underline{p}_x^{(\eps)}\right)_+\\
\Gg{\substack{a\in(p_{k+1},p_k]\\b\in [p_{m+1},p_m)}}&=\sum_{x\in[k]}\left(p_x-a\right)\\
&=\|\p\|_{(k)}-ka\;.
\ea
\een 
In addition to the three conditions above, we need to require that $\up^{(\eps)}$ is the minimal element of $\mb_{\eps}(\p)$. However, we first show that the three conditions above already determine uniquely the coefficients $a,b,k,m$. Indeed, from~\eqref{kanmb} it follows that
\be
\|\p\|_{(k)}-ka=(n-m)b+\|\p\|_{(m)}-1\;.
\ee 
Comparing this equality with~\eqref{546} implies that $\eps=\|\p\|_{(k)}-ka$ and $\eps=(n-m)b+\|\p\|_{(m)}-1$. We therefore conclude that
\be\label{ab5p45}
a=\frac{\|\p\|_{(k)}-\eps}{k}\quad\text{and}\quad b=\frac{1+\eps-\|\p\|_{(m)}}{n-m}\;.
\ee
That is, the equation above can be viewed as the definitions of $a$ and $b$, and it is left to determine $k$ and $m$. 

Substituting the above definitions of $a$ and $b$ into~\eqref{545} and isolating $\eps$ gives that
\be\label{5pp52}
\eps\in[r_k,r_{k+1})\quad\text{and}\quad\eps\in[s_{m+1},s_m)\;,
\ee
where for all $z\in[n]$
\be\label{defrz}
r_z\eqdef\|\p\|_{(z)}-z p_{z}\quad\text{and}\quad
s_z\eqdef(n-z)p_z+\|\p\|_{(z)}-1\;.
\ee
Moreover, in the exercise below you show that the components of the vectors $\r\eqdef(r_1,\ldots,r_n)^T$ and $\s\in(s_1,\ldots,s_n)^T$ are non-negative and satisfy $\r=\r^\ua$ and $\s=\s^\da$. Thus, the relations in~\eqref{5pp52} uniquely specify $k$ and $m$. However, it is left to show that $k\leq m$ since otherwise $\up^{(\eps)}$ would not be well defined.

For this purpose, let $\ell\in[n]$ be the integer defined in Exercise~\ref{exl}. We will show that $k\leq\ell\leq m$. To prove $k\leq \ell$, suppose by contradiction that $k\geq\ell+1$. Since $\eps\in[r_k,r_{k+1})$ we have $\eps\geq r_k\geq r_{\ell+1}$, where the second inequality follows from the fact that $\r=\r^\ua$ and our assumption that $k\geq\ell+1$. Combining this with the definition of $r_{\ell+1}$ in~\eqref{defrz}, we get
\ba
\eps&\geq \|\p\|_{(\ell+1)}-(\ell+1)p_{\ell+1}\\
&=\|\p\|_{(\ell)}-\ell p_{\ell+1}\\
\Gg{p_{\ell+1}<\frac1n}&>\|\p\|_{(\ell)}-\frac{\ell}{n}\;, 
\ea
which is in contradiction with~\eqref{oeps}. Therefore, the assumption that $k\geq\ell+1$ cannot hold and we conclude that $k\leq\ell$.

Similarly, to prove that $m\geq \ell$, suppose by contradiction that $m\leq\ell-1$. Since $\eps\in[s_{m+1},s_{m})$ we have $\eps\geq s_{m+1}\geq s_{\ell}$, where the second inequality follows from the fact that $\s=\s^\da$ and our assumption that $m+1\geq\ell$. Combining this with the definition of $s_\ell$ in~\eqref{defrz}, we get
\ba
\eps&\geq (n-\ell)p_\ell+\|\p\|_{(\ell)}-1\\
\Gg{p_{\ell}\geq\frac1n}&\geq\|\p\|_{(\ell)}-\frac{\ell}{n}\;, 
\ea
which is again in contradiction with~\eqref{oeps}. Therefore, the assumption that $m\leq\ell-1$ cannot hold and we conclude that $m\geq\ell$. Combining this with our earlier result that $k\leq \ell$ we conclude that $k\leq m$.

\bex
Show that the vectors  $\r$ and $\s$, whose components are given in~\eqref{defrz} satisfy:
\be
0=r_1\leq r_2\leq \cdots \leq r_n=1-np_n\quad{\rm and}\quad
np_1-1= s_1\geq s_2\geq\cdots\geq s_{n}=0\;.
\ee
\eex

It's important to note that the index $k$ is characterized by its role as the maximizer of the function $\ell\mapsto t_{\ell}\eqdef\frac{\|\p\|_{(\ell)}-\eps}{\ell}$. To put it another way,
$
t_k=\max_{\ell\in[n]}\{t_\ell\}
$.
This implies that the coefficient $a$ can be straightforwardly defined as:
\be\label{altera}
a\eqdef\max_{\ell\in[n]}\left\{\frac{\|\p\|_{(\ell)}-\eps}{\ell}\right\}\;.
\ee
To understand this, let $\ell$ be the largest integer that satisfies $t_\ell=\max_{\ell'\in[n]}\{t_{\ell'}\}$. The inequality $t_{\ell}> t_{\ell+1}$ leads to (see Exercise~\ref{veria}):
\ba
0< t_{\ell}-t_{\ell+1}=\frac{r_{\ell+1}-\eps}{\ell(\ell+1)}\;,
\ea
where $r_\ell\eqdef\|\p\|_{(\ell)}-\ell p_\ell$ as previously defined.
This implies that $r_{\ell+1}>\eps$. Conversely, by following a similar reasoning, the condition $t_{\ell}\geq t_{\ell-1}$ yields $r_{\ell}\leq\eps$. Therefore, we conclude that $\ell$ is the integer for which $\eps$ falls in the interval $[r_{\ell},r_{\ell+1})$, which leads us to deduce that $\ell=k$.

\bex\label{veria}
Verify the equality $t_{\ell}-t_{\ell+1}=\frac{r_{\ell+1}-\eps}{\ell(\ell+1)}$.
\eex

\bex
Using the same notations as above, show that for every $\eps\in(0,1)$ and $\p\in\prob(n)$ the coefficient $b$ can be expressed as:
\be
b=\min_{\ell\in[n-1]}\frac{1+\eps-\|\p\|_{(\ell)}}{n-\ell}\;.
\ee
\eex

\begin{myt}{}
\begin{theorem}\label{flattest}
Let $\eps\in(0,1)$ and $\p\in\prob^\da(n)$ be a probability vector such that~\eqref{5p38} holds.  Let $k,m\in[n-1]$ be the integers satisfying~\eqref{5pp52}, and $a$ and $b$ be the numbers defined in~\eqref{ab5p45}.
Then, for these choices of $k$, $m$, $a$, and $b$, the vector $\up^{(\eps)}$ as defined in~\eqref{599a} is the minimal element (under majorization) of $\mb_{\eps}(\p)$. 
\end{theorem}
\end{myt}
\begin{proof}
We already showed that $\up^{(\eps)}\in\mb_\eps(\p)$. It is therefore left show that if $\q\in\mb_\eps(\p)$  then $\q\succ\up^{(\eps)}$. To establish that $\|\q\|_{(\ell)}\geq\left\|\up^{(\eps)}\right\|_{(\ell)}$ for every $\ell\in[n]$, we partition the proof into three distinct cases:
\ben
\item The case $\ell\in[k]$. In this scenario, since $\left\|\up^{(\eps)}\right\|_{(\ell)}=\ell a$, the condition $\|\q\|_{(\ell)}\geq\left\|\up^{(\eps)}\right\|_{(\ell)}$ simplifies to $\frac1\ell\|\q\|_{(\ell)}\geq a$. Since the components of $\q^\da$ are arranged in non-increasing order, for every integer $\ell\leq k$ the average
$
\frac1\ell\|\q\|_{(\ell)}
$
is no smaller than the average 
$
\frac1k\|\q\|_{(k)}
$. Therefore, it is sufficient to show that $\frac1k\|\q\|_{(k)}\geq a$. 
Indeed, since $\q\in\mb_\eps(\p)$ we get from~\eqref{e5p18} that 
\be\label{iwi}
\|\p\|_{(k)}-\|\q\|_{(k)}\leq\frac12\|\p-\q\|\leq\eps\;.
\ee
Isolating $\|\q\|_{(k)}$ gives
\ba
\|\q\|_{(k)}&\geq\|\p\|_{(k)}-\eps\\
\GG{\eqref{ab5p45}}&=ka\;,
\ea
so that  $\frac1k\|\q\|_{(k)}\geq a$.
\item The case $k<\ell\leq m$. We use again~\eqref{e5p18} (with $\ell$ replacing $k$) to get
\ba
\|\q\|_{(\ell)}&\geq\|\p\|_{(\ell)}-\eps\\
&=\|\p\|_{(k)}-\eps+\sum_{x=k+1}^\ell p_x\\
\GG{\eqref{ab5p45}}&=ka+\sum_{x=k+1}^\ell p_x\\
\GG{\eqref{599a}}&=\left\|\underline{\p}^{(\eps)}\right\|_{(\ell)}\;.
\ea
\item The case $m<\ell\leq n$. We use once more~\eqref{e5p18} (with $m$ replacing $k$) to get
$
\|\q\|_{(m)}\geq\|\p\|_{(m)}-\eps
$.
Moreover, observe that in this case we have for all $m<\ell\leq n$
\be\label{5p66}
\|\q\|_{(\ell)}=1-\sum_{x=\ell+1}^nq_x^\da\quad\text{and}\quad\left\|\up^{(\eps)}\right\|_{(\ell)}=1-(n-\ell)b\;.
\ee
Therefore, in order to prove that $\|\q\|_{(\ell)}\geq\left\|\up^{(\eps)}\right\|_{(\ell)}$ it is sufficient to show that
\be\label{needtosh}
\frac1{n-\ell}\sum_{x=\ell+1}^nq_x^\da\leq b.
\ee
The inequality in~\eqref{needtosh} is equivalent to the statement that the average of the last $n-\ell$ components $\q^\da$ is no greater than $b$. Since the components of $\q^\da$ are arranged in non-increasing order, this average is no greater than the average of the last $n-m$ components of $\q^\da$ (recall that $n-m>n-\ell$). Hence,
\ba\label{needtosh2}
\frac1{n-\ell}\sum_{x=\ell+1}^nq_x^\da&\leq \frac1{n-m}\sum_{x=m+1}^nq_x^\da\\
&=\frac{1-\|\q\|_{(m)}}{n-m}\\
\Gg{\|\q\|_{(m)}\geq\|\p\|_{(m)}-\eps}&=\frac{1+\eps-\|\p\|_{(m)}}{n-m}\\
\GG{\eqref{ab5p45}}&=b\;.
\ea
\een
We therefore concludes  that $\q\succ\up^{(\eps)}$.
\end{proof}

One can use the flatest $\eps$-approximation to compute the distance of a vector $\p\in\prob(n)$ to the set of all vectors $\r\in\prob(n)$ that are majorized by $\q$. Specifically, let
\be\label{defmaj6}
{\rm majo}(\q)\eqdef\{\r\in\prob(n)\;:\;\q\succ\r\}\;,
\ee
denotes the set of all vectors in $\prob(n)$ that are majorized by $\q$, and define the distance between $\p\in\prob(n)$ and the set ${\rm majo}(\q)$ as:
\be
T\big(\p,{\rm majo}(\q)\big)\eqdef\min_{\r\in{\rm majo}(\q)}\frac12\|\p-\r\|_1\;.
\ee
\begin{myt}{}
\begin{theorem}\label{thm:majog}
Using the same notations as above, for all $\p,\q\in\prob(n)$
\be\label{1224t02}
T\big(\p,{\rm majo}(\q)\big)=\max_{\ell\in[n]}\big\{\|\p\|_{(\ell)}-\|\q\|_{(\ell)}\big\}\;.
\ee
\end{theorem}
\end{myt}

\begin{proof}
Without loss of generality we will assume that $\p,\q\in\prob^\da(n)$ and $\q\not\succ\p$. For any $\eps\in(0,1)$, let $\underline{\p}^{(\eps)}$ be the flattest $\eps$-approximation of $\p$; see~\eqref{599a}.
By definition,
\ba\label{1228new}
T\big(\p,{\rm majo}(\q)\big)&\eqdef\min\Big\{ \frac12\|\p-\r\|_1\;:\;\q\succ\r\;,\quad\r\in\prob(n)\Big\}\\
&=\min\Big\{ \eps\in[0,1]\;:\;\q\succ\r\;,\quad\r\in\mb_\eps(\p)\Big\}\\
\Gg{\r\succ\underline{\p}^{(\eps)}\;\;\quad\forall\;\r\in\mb_\eps(\p)}&=\min\Big\{\eps\in[0,1]\;:\;\q\succ\underline{\p}^{(\eps)}\Big\}\;.
\ea 
That is, it is left to compute the smallest $\eps$ that satisfy $\q\succ\underline{\p}^{(\eps)}$. We will show that this smallest $\eps$ equals 
\be\label{defdel}
\delta\eqdef\max_{\ell\in[n]}\big\{\|\p\|_{(\ell)}-\|\q\|_{(\ell)}\big\}\;.
\ee 
We first show that $\q\succ\underline{\p}^{(\delta)}$. Let $k,m\in[n-1]$ be the integers satisfying~\eqref{5pp52}, and $a$ and $b$ be the numbers defined in~\eqref{ab5p45}, but with $\delta$ replacing $\eps$. 
From Exercise~\ref{ex:majorest}  we have $\q\succ\underline{\p}^{(\delta)}$ if and only if
\be\label{pell}
\|\q\|_{(\ell)}\geq \left\|\up^{(\delta)}\right\|_{(\ell)}\quad\quad\forall\;\ell\in\{k,k+1,\ldots,m\}\;.
\ee 
Now, for $k\leq\ell\leq m$ 
\ba
\|\q\|_{(\ell)}- \left\|\up^{(\delta)}\right\|_{(\ell)}&=\|\q\|_{(k)}-ka+\sum_{x=k+1}^{\ell}(q_x-p_x)\\
\Gg{ka=\|\p\|_{(k)}-\delta}&=\delta+\|\q\|_{(k)}-\|\p\|_{(k)}+\sum_{x=k+1}^{\ell}(q_x-p_x)\\
&=\delta+\|\q\|_{(\ell)}-\|\p\|_{(\ell)}\;.
\ea
Hence, $\q\succ\up^{(\delta)}$ if and only if for all $\ell\in\{k,\ldots,m\}$ we have $\delta\geq \|\p\|_{(\ell)}-\|\q\|_{(\ell)}$. 
From its definition, $\delta\geq\|\p\|_{(\ell)}-\|\q\|_{(\ell)}$ for all $\ell\in[n]$. Hence, $\q\succ\up^{(\delta)}$.

To prove the optimality of $\delta$, we use the fact that $\up^{(\delta)}$ is $\delta$-close to $\p$ so that from~\eqref{e5p18} we get for any $\ell\in[n]$ 
\ba
\delta&\geq\|\p\|_{(\ell)}-\left\|\up^{(\delta)}\right\|_{(\ell)}\\
\GG{\q\succ\up^{(\delta)}}&\geq \|\p\|_{(\ell)}-\left\|\q\right\|_{(\ell)}\;.
\ea
Since the above inequality holds for all $\ell\in[n]$ we conclude that $\delta$ is optimal. 
This concludes the proof.
\end{proof}

\subsubsection{Two Key Functions}

\begin{figure}[h]\centering    \includegraphics[width=1\textwidth]{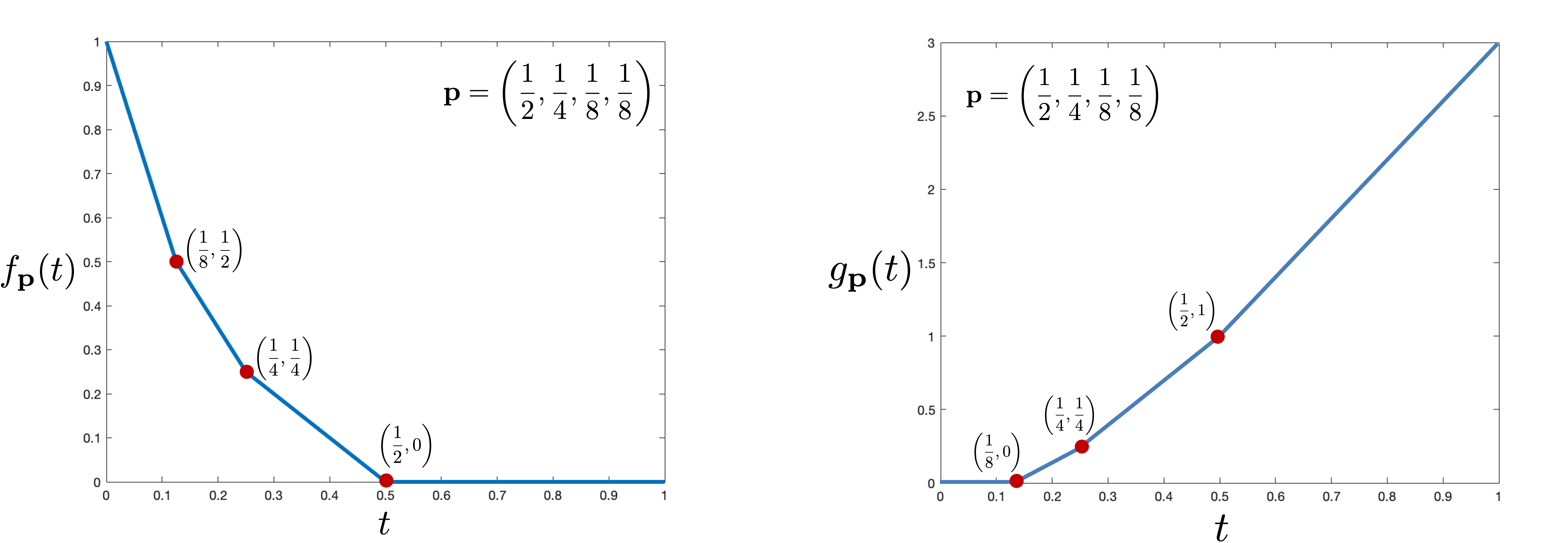}
  \caption{\linespread{1}\selectfont{\small The functions $f_\p(t)$ and $g_\p(t)$. The dots indicate the points at which the slop of the functions changes.}}
  \label{fp}
\end{figure}

We end the section by introducing two functions frequently used in majorization theory, enabling the study of approximate majorization\index{approximate majorization} more effectively.
For a given $\p\in\prob(n)$ we define $f_\p:[0,1]\to[0,1]$ and $g_\p:[0,1]\to[0,n-1]$ via
\be
f_{\p}(t)\eqdef\sum_{x\in[n]}(p_x-t)_+\quad\text{and}\quad g_\p(t)\eqdef\sum_{x\in[n]}(t-p_x)_+\;.
\ee
Recall that $(s-t)_+=\frac12(|s-t|+s-t)$, and since the absolute value is a continuous function, these functions are continuous (although not differentiable). Observe that  $f_\p(t)=0$ for $t\geq p_1$, whereas $g_\p(t)=0$ for $t\in[0,p_n]$ and $g_\p(t)=nt-1$ for $t\geq p_1$.
The function $f_\p(t)$ is non-increasing in $t$ while $g_\p(t)$ is non-decreasing in $t$. See Fig.~\ref{fp} for examples of $f_\p(t)$ and $g_\p(t)$.

\bex\label{ex522}
In this exercise we use the same notations used in this subsection with a fix $\p\in\prob^\da(n)$. 
\ben
\item Show that for any $z\in[n]$
\be
f_\p(p_z)=r_z\quad\text{and}\quad
g_\p(p_z)=s_z\;\;.
\ee
\item Use Part 1 to provide an alternative proof that $\r=\r^\ua$ and $\s=\s^\da$\;.
\item Show that $f_\p(a)=g_\p(b)=\eps$.
\item Show that
$
f_\p\left(1/n\right)=g_\p\left(1/n\right)=\|\p\|_{(\ell)}-\frac {\ell} n
$,
where $\ell\in[n]$ is the largest integer satisfying $p_{\ell}\geq\frac1n$.
\item Show that
\be
\|\p\|_{(\ell)}-\frac {\ell} n\leq 1-np_n\quad\text{and}\quad\|\p\|_{(\ell)}-\frac {\ell} n\leq np_1-1\;.
\ee
\een
\eex

The functions $f_\p(t)$ and $g_\p(t)$  are one-to-one when restricted to the domains $[0,p_1]$ and $[p_n,1]$, respectively (and therefore in these domains they are monotonically decreasing and monotonically increasing, respectively).  Therefore, the functions $f_\p:[0,p_1]\to[0,1]$ and $g_\p:[p_n,1]\to[0,n-1]$  have inverse functions.
Using the same notations as above, the inverse function $f_\p^{-1}:[0,1]\to[0,p_1]:r\mapsto f_\p^{-1}(r)$ is given by (see Exercise~\ref{inversefunction})
\be\label{1440}
f_\p^{-1}(r)=\begin{cases}\frac{\|\p\|_{(k)}-r}{k}&\text{if }r>0\\
1 &\text{if }r=0
\end{cases}\;,
\ee
where $k\in[n]$ is the integer satisfying $r\in(r_{k},r_{k+1}]$.
The inverse function $g_\p^{-1}:[0,n-1]\to[p_n,1]:s\mapsto g_\p^{-1}(s)$ is given by (see Exercise~\ref{inversefunction})
\be\label{1441}
g_\p^{-1}(s)=\begin{cases}\frac{1+s-\|\p\|_{(m)}}{n-m} &\text{if }s>0\\
0 &\text{if }s=0
\end{cases}\;,
\ee
where $m\in\{0,1,\ldots,n-1\}$ is the integer satisfying $s\in(s_{m+1},s_{m}]$.

\bex\label{inversefunction}
Consider the functions $f_\p:[0,p_1]\to[0,1]$ and $g_\p:[p_n,1]\to[0,n-1]$ as defined above and let $f_\p^{-1}$ and $g_\p^{-1}$ be as defined in~\eqref{1440} and~\eqref{1441}. 
\ben
\item Show that for any $t\in[0,p_1]$ and $r\in[0,1]$
\be
f_\p^{-1}\left(f_\p(t)\right)=t\quad\text{and}\quad f_\p\left(f_\p^{-1}(r)\right)=r\;.
\ee
\item Show that for any $t\in[p_n,1]$ and $s\in[0,n-1]$
\be
g_\p^{-1}\left(g_\p(t)\right)=t\quad\text{and}\quad g_\p\left(g_\p^{-1}(s)\right)=s\;.
\ee
\een
\eex

\section{Relative Majorization}\label{secrm}\index{relative majorization}

In the first section of this chapter we compared between a $\p$-dice and a $\q$-dice via the degree of uncertainty they each posses. In this section we study the degree of distinguishability between the two dice. Unlike arbitrary vectors, objects like probability vectors (as well as quantum states, quantum channels, etc) contain information about physical systems and therefore their distinguishability is typically quantified with functions that are sensitive to this information. For example,
suppose a player receives  a biased dice whose probability distribution is either $\p\eqdef\{p_x\}_{x\in[n]}$ or $\q\eqdef\{q_x\}_{x\in[n]}$ . The player can estimate which of the two distributions corresponds to the dice by rolling the dice many times. The intuition is that if $\p$ and $\q$ are very distinguishable it would be easier (i.e. quicker) to determine which one of them corresponds to the dice. 

One of the key observations in any distinguishability task as above, is that by sending the information source (i.e. the outcomes of the dice) through a communication channel, the player cannot increase his or her ability to distinguish between the two distributions $\p$ and $\q$. This means that if $E$ is the column stochastic matrix that corresponds to a classical communication channel, then the resulting distributions $E\p$ and $E\q$ are less distinguishable than $\p$ and $\q$. 

\begin{myd}{Relative Majorization}\index{relative majorization}
\begin{definition}
Let $\p,\q\in\prob(n)$ and $\p',\q'\in\prob(m)$ be two pairs of probability distributions. We say that $(\p,\q)$ relatively majorize $(\p',\q')$ and write
\be
(\p,\q)\succ(\p',\q')\quad\iff\quad\p'=E\p\;\text{ and }\;\q'=E\q\;,
\ee
where $E$ is an $m\times n$ column stochastic matrix. Further, if $(\p,\q)\succ(\p',\q')\succ(\p,\q)$ we then write
\be\label{533}
(\p,\q)\sim(\p',\q')\;.
\ee
\end{definition}
\end{myd}

Relative majorization is a pre-order. The property $(\p,\q)\succ (\p,\q)$ (i.e. reflexivity) follows by taking $E$ in the definition above to be the identity matrix. The transitivity of relative majorization follows from the fact that the product of two column stochastic matrices is also a column stochastic matrix.

\begin{exercise} Consider the equivalence relation $\sim$ as defined in~\eqref{533}.
\begin{enumerate}
\item Let $\p,\q\in\prob(n)$ be two $n$-dimensional probability vectors and let $P$ be an $n\times n$ permutation matrix. Show that\index{direct sum}
\be\label{secorel}
(P\p,P\q)\sim(\p,\q)\quad\text{and}\quad(\p\oplus 0,\q\oplus 0)\sim(\p,\q).
\ee
\item Show that for any $m,n\in\mbb{N}$ and any probability vectors, $\p\in\prob(n)$ and $\q\in\prob(m)$,
\be
(\p,\p)\sim(\q,\q)\;.
\ee
\item Let $\p_1,\p_2\in\prob(n)$ and $\q_1,\q_2\in\prob(m)$. Show that if $\p_1\cdot\p_2=\q_1\cdot\q_2=0$ then
\be
(\p_1,\p_2)\sim(\q_1,\q_2)\;.
\ee
\end{enumerate}
\end{exercise}

From the second relation in~\eqref{secorel} it follows that without loss of generality we can always assume that there is no $x\in[n]$ such that $p_x=q_x=0$ (since any $x$-component with $p_x=q_x=0$ can be removed from the vectors $\p$ and $\q$ without changing the equivalency). 

\begin{myd}{Standard Form}\index{standard form}
\begin{definition}\label{def:sf}
A pair $(\p,\q)$ of probability vectors in $\prob(n)$ is said to be given in a \emph{standard form} if there is no $x\in[n]$ such that $p_x=q_x=0$, and the components of the vectors $\p$ and $\q$ are arranged such that
\be\label{order}
\frac{p_1}{q_1}\geq\frac{p_2}{q_2}\geq\cdots\geq\frac{p_n}{q_n}\;,
\ee
where we used the convention $p_x/q_x=\infty$ for $x\in[n]$ with $p_x> 0$ and $q_x=0$.
\end{definition}
\end{myd}

Observe that since $(P\p,P\q)\sim(\p,\q)$ for every permutation matrix $P$, any pair of vectors is equivalent (under relative majorization) to its standard form. The choice of the order given in~\eqref{order} will be clear later on when we characterize relative majorization with testing regions.

\bex
Let $\{\e_1,\e_2\}$ be the standard basis of $\mbb{R}^2$. Express the pair $(\e_1,\e_2)$ in the standard form.
\eex

\begin{exercise}\label{tgv}
Show that if $\p,\q\in\prob(n)$ and $\q$ has the form $\q=(q_1,\ldots,q_r,0,\ldots,0)$ for some $r<n$ then
\be
(\p,\q)\sim(\p',\q)
\ee
for any $\p'\in\prob(n)$ whose first $r$ components equal the first $r$ components of $\p$.
\end{exercise}

\subsection{Lower and Upper Bounds}

We saw earlier that the maximal and minimal elements of $\prob(n)$ under majorization are $\e_1$ and $\u^{(n)}$, respectively. In the following exercise you find the maximal and minimal elements of $\prob(n)\times\prob(n)$ under relative majorization.

\begin{exercise}
Let $\p,\q\in\prob(n)$ be two $n$-dimensional probability vectors, and let $\e_1,\e_2\in\prob(m)$ be two $m$-dimensional probability vectors with orthogonal support; i.e. $\e_1\cdot\e_2=0$. Show that for any $k\in\mbb{N}$ and any $\r\in\prob(k)$
\be
(\e_1,\e_2)\succ(\p,\q)\succ (\r,\r)\;.
\ee
\end{exercise}

In the following theorem we bound any pair of probability vectors by pairs of two dimensional vectors. For any $n\in\mbb{N}$ and $\p,\q\in\prob(n)$, we denote by
\be\label{4p99}
\lambda_{\min}\eqdef\sum_{x\in\supp(\p)}q_x\quad\text{and}\quad\lambda_{\max}\eqdef\min_{x\in\supp(\p)}\frac{q_x}{p_x}\;,
\ee
where $\supp(\p)\eqdef\{x\in[n]\;:\;p_x\neq 0\}$.
Later in the book we will see that $\lambda_{\max}$ and $\lambda_{\min}$ are related to the min and max relative entropies. In the following theorem we use the notations $\e_1\eqdef(1 , 0)^T$ and $\e_2\eqdef(0 , 1)^T$, and in addition denote by
\be
\v_{\max}\eqdef\lambda_{\max}\e_1+(1-\lambda_{\max})\e_2\quad\text{and}\quad\v_{\min}\eqdef\lambda_{\min}\e_1+(1-\lambda_{\min})\e_2\;.
\ee
\begin{myt}{}
\begin{theorem}\label{bminmax}
Using the same notations as above we have for any $\p,\q\in\prob(n)$
\be\label{lmaxmin}
\left(\e_1,\v_{\max}\right)\succ(\p,\q)\succ\left(\e_1,\v_{\min}\right)
\ee
\end{theorem}
\end{myt}
\begin{remark}
Note that the bounds are not symmetric, meaning that if we swap $\p$ with $\q$, the vector $\e_1\eqdef(1,0)^T$ will appear second in the bounding pairs, and $\lambda_{\min}$ and $\lambda_{\max}$ will also change. 
Moreover, if $\p>0$ (i.e.\ all the components of $\p$ are strictly positive)  than $\lambda_{\min}=1$ and consequently the lower bound becomes trivial.
\end{remark}

\begin{proof}
We first prove the upper bound. For this purpose, we need to find an $n\times 2$ column stochastic channel $E\in\stoc(n,2)$ with the property that 
\be\label{sidra}
E\e_1=\p\quad\text{and}\quad E\v_{\max}=\q\;,
\ee
where $\e_1\eqdef(1 , 0)^T$ and $\e_2\eqdef(0 , 1)^T$. Observe that the first condition above implies that the first column of $E$ must be equal to $\p$. Combining this with the definition of $\v_{\max}$ and with the second condition above we get that 
\be
E\v_{\max}=\lambda_{\max}\p+(1-\lambda_{\max})E\e_2=\q\quad\quad\Rightarrow\quad\quad
E\e_2=\frac{\q-\lambda_{\max}\p}{1-\lambda_{\max}}\;.
\ee
By definition, $\lambda_{\max}\in[0,1]$ and it has the property that $\q\geq\lambda_{\max}\p$ (i.e.\  $q_x\geq\lambda_{\max}p_x$ for each $x\in[n]$). Therefore, the right-hand side of the equation above is a probability vector. To summarize, the $n\times 2$ column stochastic matrix $E$, whose first column is $\p$, and its second column is $\frac{\q-\lambda_{\max}\p}{1-\lambda_{\max}}$ satisfies~\eqref{sidra} so that by definition the upper bound in~\eqref{lmaxmin} holds.

We now prove the lower bound. By definition, it is sufficient to show that there exists a channel $E\in\stoc(2,n)$ such that
\be
E\p=\e_1\quad\text{and}\quad E\q=\v_{\min}=\lambda_{\min}\e_1+(1-\lambda_{\min})\e_2\;.
\ee
Since $E$ must be a column stochastic matrix with two rows, it follows that if its first row is $\t^T$ then its second row is $(\textbf{1}_n-\t)^T$, where $\textbf{1}^T_n=(1,\ldots,1)$. Hence, $E$ satisfies the above conditions if and only if
\be\label{tttt}
\t\cdot\p=1\quad\text{and}\quad\t\cdot\q=\lambda_{\min}\;.
\ee
Note also that we must have $0\leq\t\leq\mathbf{1}_n$ (element-wise) since $E$ is column stochastic. We therefore choose $\t=(t_1,\ldots,t_n)^T$ with
\be
t_x=\begin{cases}
1 &\text{if }p_x>0\\
0 &\text{if }p_x=0
\end{cases}\quad\quad\forall\;x\in[n]\;.
\ee
It is simple to check that this $\t$ satisfies~\eqref{tttt}. This completes the proof.  
\end{proof}

Note that when $\supp(\p)=\supp(\q)$, the value of $\lambda_{\max}$, as defined in~\eqref{4p99}, is constrained to the range $0<\lambda_{\max}<1$, ensuring that $\v_{\max}>0$. In this case we can improve the upper bound $(\e_1,\v_{\max})$.
Indeed, let $0<s<t\leq1$ be such that
\be\label{tcin}
\frac{1-t}{1-s}\p\leq\q\leq\frac{t}{s}\p\;.
\ee 
Note that such $s$ and $t$ exists since we assume that $\p$ and $\q$ have the same support, and we can take $s$ close enough to zero and $t$ close enough to one. Now, define a stochastic evolution matrix $E=[\v_1\;\v_2]\in\stoc(m,2)$, with the two columns $\v_1,\v_2\in\prob(m)$ given by
\ba
\v_1 \eqdef\frac{(1-s)\q-(1-t)\p}{t-s}\quad\text{and}\quad\v_2\eqdef\frac{t\p-s\q}{t-s}\;.
\ea
Note that the conditions in~\eqref{tcin} implies that $E$ is indeed a column stochastic matrix since $t\p-s\q\geq 0$ (entrywise)
and $(1-s)\q-(1-t)\p\geq 0$. Moreover, denoting by $\s\eqdef(s,1-s)^T$ and $\t\eqdef(t,1-t)^T$ we have by direct calculation (Exercise~\ref{verifyexst})
\be\label{alreadyver}
E\s=\p\quad\text{and}\quad E\t=\q\;.
\ee
We therefore conclude that for every $\p,\q\in\prob(n)$ with equal support, i.e., $\supp(\p)=\supp(\q)$, there exist $\s,\t\in\prob_{>0}(2)$ satisfying the relation:
\be\label{stpq}
(\s,\t)\succ(\p,\q)\;.
\ee
Observe that the relation above hold as long as $s,t\in[0,1]$ satisfies
$s<t$ and
\be
(1-s)\q\geq(1-t)\p\quad{and}\quad t\p\geq s\q\;.
\ee
\bex
\label{verifyexst}
Verify by direct calculation the relations in~\eqref{alreadyver}.
\eex

\bex
Show that by taking $t=1-\lambda_{\max}$ and $s=0$ the relation $(\s,\t)\succ(\p,\q)$ is equivalent to the upper bound in~\eqref{lmaxmin}.
\eex

\subsection{Majorization vs Relative Majorization}\index{majorization}

Majorization and relative majorization are interrelated concepts, rather than independent ones. Specifically, they become equivalent when one of the probability vectors in each pair has a uniform distribution. In this section, we will explore the deep connection between these two concepts. This exploration will allow us, in the following section, to understand provide a geometrical characterization of relative majorization. Moreover, the insights gained will be used later in the book to prove and explore other related findings.

Relative majorization generalizes majorization between vectors. Specifically, for any two probability vectors $\p,\q\in\prob(n)$
\be\label{maju}
(\p,\u)\succ(\q,\u)\quad\iff\quad\p\succ\q\;,
\ee
where $\u\eqdef(\frac1n,\ldots,\frac1n)^T$ is the uniform distribution. In the exercise below you will prove this assertion using Theorem~\ref{chmaj}.

\begin{exercise}
Use the equivalence between the first two conditions in Theorem~\ref{chmaj} to prove~\eqref{maju}.
\end{exercise}

\subsubsection{The Special Case of Vectors with Rational Components}\index{rational components}

If one of the vectors has positive rational components, the relationship between majorization and relative majorization becomes even closer than what we have seen so far. Consider a pair of vectors $(\p,\q)$, where $\p\in\prob(n)$ and 
\be\label{rational}
\q\eqdef\left(\frac{k_1}{k},\ldots,\frac{k_n}{k}\right)^T\quad\quad\;k_1,\ldots,k_n\in\mbb{N}\;,
\ee 
and $k\eqdef k_1+\cdots+k_n$. Define the vector $\r\in\prob(k)$ via
\be\label{directsum}
\r\eqdef\bigoplus_{x\in[n]}p_x\u^{(k_x)}=\Big(\underbrace{\frac{p_1}{k_1},\ldots,\frac{p_1}{k_1}}_{k_1\text{-times}},\underbrace{\frac{p_2}{k_2},\ldots,\frac{p_2}{k_2}}_{k_2\text{-times}},\ldots,
\underbrace{\frac{p_n}{k_n},\ldots,\frac{p_n}{k_n}}_{k_n\text{-times}}\Big)^T
\ee
where $\u^{(k_x)}$ is the uniform probability vector in $\prob(k_x)$. We then have the following theorem.

\begin{myt}{}
\begin{theorem}\label{onlyr}
Let $\p,\q\in\prob(n)$ and $\r\in\prob(k)$ be as above with $\q$ having positive rational components. Then,
\be\label{ruk}
(\p,\q)\sim(\r,\u^{(k)})\;.
\ee
\end{theorem}
\end{myt}
\begin{remark}
Observe that any vector $0<\q\in\prob(n)\cap\mbb{Q}^n$ can be expressed as in~\eqref{rational} for sufficiently large $k$.
This $k$ is a common denominator for all the components of $\q$.
\end{remark}

\begin{proof}
We first show that $(\p,\q)\succ(\r,\u^{(k)})$.
For any $x\in[n]$, let $E^{(x)}$ be the $k_x\times n$ matrix whose $x$-column is $\u^{(k_x)}$ and all the remaining $n-1$ columns are zero. Moreover, let $E$ be the $k\times n$ matrix given by
\be\label{cf149}
E\eqdef \begin{bmatrix}E^{(1)}\\ E^{(2)}\\\vdots\\ E^{(n)}\end{bmatrix}\;.
\ee
By definition, $E^{(x)}\p=p_x\u^{(k_x)}$ so that $E\p=\r$. Similarly, $E^{(x)}\q=\frac{k_x}{k}\u^{(k_x)}=\frac1k\mathbf{1}_{k_x}$ so that $E\q=\u^{(k)}$. Therefore, since $E$ is column stochastic we get that $(\p,\q)\succ(\r,\u^{(k)})$.

For the converse, let $F^{(x)}$ be the $n\times k_x$ matrix whose $x$-row is $[1,\ldots,1]$ and all the remaining $n-1$ rows are zero. Moreover, denote by $F$ the $n\times k$ column stochastic matrix given by
\be
F\eqdef \begin{bmatrix}F^{(1)} & F^{(2)} & \cdots & F^{(n)}\end{bmatrix}\;.
\ee
Observe that $FE=I_n$. Therefore, $F\r=FE\p=\p$ and similarly $F\u^{(k)}=FE\q=\q$. In other words, $(\r,\u^{(k)})\succ (\p,\q)$. But since we already proved that $(\p,\q)\succ(\r,\u^{(k)})$ we conclude that  $(\p,\q)\sim(\r,\u^{(k)})$.
\end{proof}

\begin{exercise}
Verify all steps in the proof above; i.e. show that $E$ and $F$ are indeed column stochastic and $FE=I_n$.
\end{exercise}

We have seen that relative majorization reduces to majorization when one of the vectors is the uniform vector (see~\eqref{maju}). 
In the following excerise, you show that the remarkable equivalence between $(\p,\q)$ and $(\r,\u^{(k)})$ implies that relative majorization reduces to majorization if one of the vectors has positive rational components.

\begin{exercise}
Let $\p\in\prob(n)$, $0<\q\in\prob(n)\cap\mbb{Q}^n$, $\p'\in\prob(m)$ and $0<\q'\in\prob(m)\cap\mbb{Q}^m$. Show that there exists $k\in\mbb{N}$ and vectors $\r,\r'\in\prob(k)$ such that
\be\label{4p106}
(\p,\q)\succ(\p',\q')\iff \r\succ\r'\;.
\ee 
\end{exercise}

\bex
Let $\p\in\prob(n)$ and $0<\q\in\prob(n)\cap\mbb{Q}^n$. Show that the pair $(\p,\q)$ is given in the standard form (i.e., satisfies~\eqref{order}) if and only if the vector $\r$ as defined in~\eqref{directsum} satisfies $\r=\r^\da$.
\eex

\subsection{Testing Regions}\index{testing region}

\begin{figure}[h]\centering
    \includegraphics[width=0.5\textwidth]{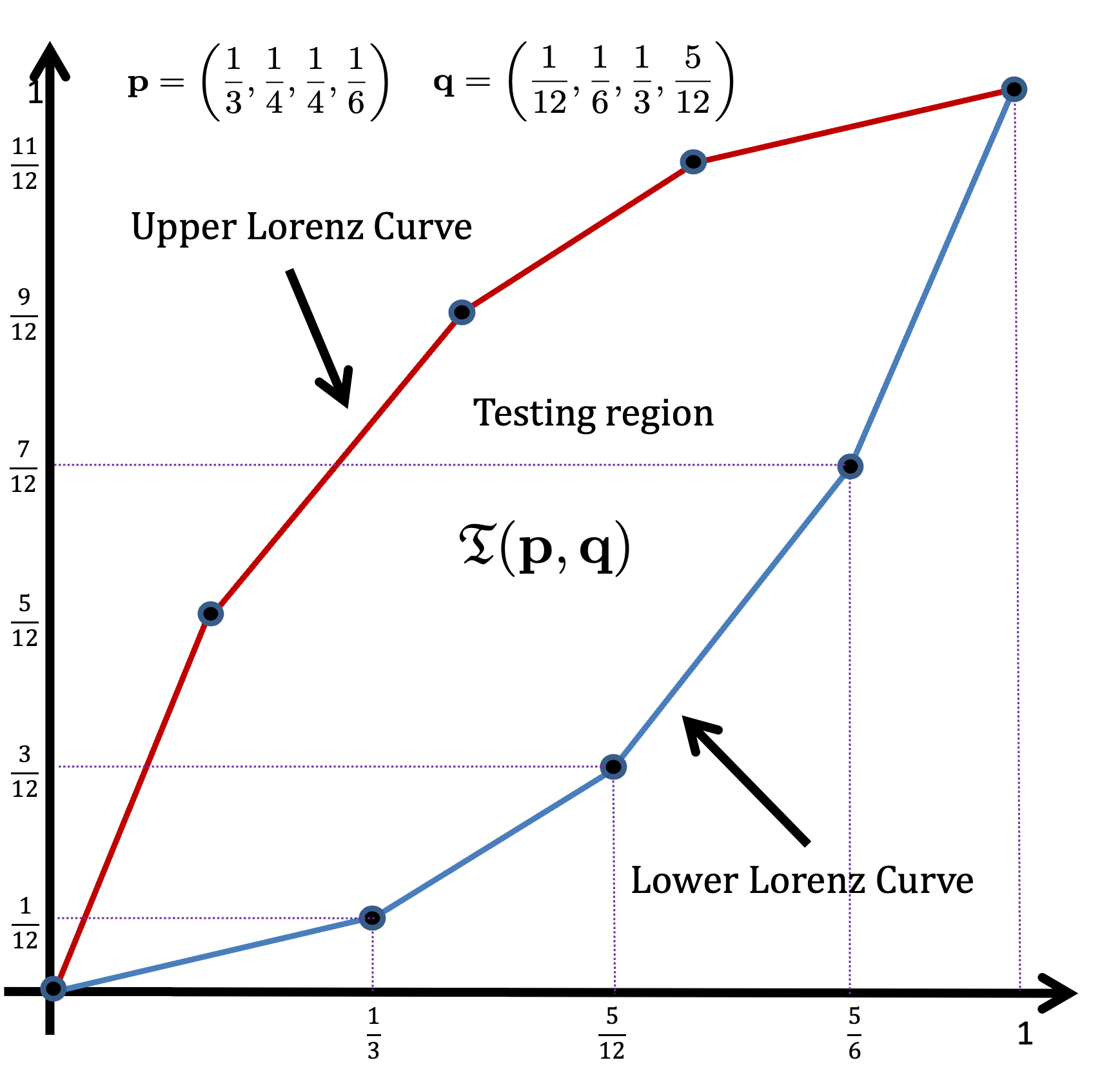}
  \caption{\linespread{1}\selectfont{\small Testing Region.}}
  \label{tregion}
\end{figure}

Relative majorization can be characterized geometrically in terms of testing regions (also known as zonotopes).\index{zonotope} Testing regions are regions in $\mbb{R}^2$ that have several applications in statistics, particularly the area of hypothesis testing. The testing region associated with a pair of probability vectors $\p,\q\in\prob(n)$ is a region in $\mbb{R}^2$ defined by
\be\label{testingr}
\mt(\p,\q)\eqdef\Big\{(\p\cdot\t,\q\cdot\t)\;:\;\t\in[0,1]^n\Big\}\;,
\ee
where $\t$ (also known as a probabilistic hypothesis test) is an $n$-dimensional vector with entries between 0 and 1. Note that for any pair of probability vectors the points $(0,0)$ and $(1,1)$ belong to its testing region. Explicitly, $(0,0)$ is obtained by taking $\t$ to be the zero vector, and $(1,1)$ is obtained by taking $\t=(1,\ldots,1)^T$. An example of a testing region is plotted in Fig.~\ref{tregion}. \index{Lorenz curve}

\begin{exercise}
Show that the testing region is convex, and it has the symmetry that if $(x,y)\in\mt(\p,\q)$ then also $(1-x,1-y)\in\mt(\p,\q)$.
Hint: For the latter property, consider the vector $\t'=(1,\ldots.,1)-\t$.
\end{exercise}
 
The testing region is bounded by two curves known as lower and upper Lorenz curves.  Due to the symmetry that $(1-x,1-y)\in\mt(\p,\q)$ for any $(x,y)\in\mt(\p,\q)$, the upper Lorenz curve\index{Lorenz curve} can be
obtained from the lower Lorenz curve\index{Lorenz curve} through a 180-degree rotation centered at the midpoint $(1/2,1/2)$.  Consequently, either the lower or the upper Lorenz curve\index{Lorenz curve} is sufficient to uniquely define the entire testing region.

Since the testing region is convex it can be characterized by its extreme points. It is tempting to draw a parallel with the convex set $[0,1]^n$, which possesses $2^n$ extreme points encapsulated within the set ${0,1}^n$. However, this analogy can be misleading in the context of our testing region. In reality, only $2n$ points are necessary to fully characterize the testing region.  We will focus on the extreme points characterizing 
the lower Lorenz curve. Note that the lower Lorenz curve\index{Lorenz curve} is a convex curve (while the upper Lorenz curve is concave).

\begin{exercise}
Let $\p,\q\in\prob(n)$ and let $P$ be an $n\times n$ permutation matrix. 
\ben
\item Show that
\be
\mt(P\p,P\q)=\mt(\p,\q)\;.
\ee
\item Show that
\be
\mt(\p\oplus 0,\q\oplus 0)=\mt(\p,\q)\;.
\ee
\een
\end{exercise}

Since for any permutation matrix $P$, $(\p,\q)$ and $(P\p,P\q)$ have the same testing region, we can assume without loss of generality that $(\p,\q)$ are always given in the standard form.  This is also justified by the fact that under relative majorization we have $(\p,\q)\sim(P\p,P\q)$ for any permutation matrix $P$. 
 
\begin{myt}{}
\begin{theorem}\label{vertices}
Given $\p,\q\in\prob(n)$ in standard form, the extreme points on the lower boundary of the testing region $\mt(\p,\q)$ (specifically, on its lower Lorenz curve) are the $n+1$ vertices: 
\be\label{akbk}
(a_k,b_k)\eqdef\Big(\sum_{x\in[k]}p_x\;,\;\sum_{x\in[k]}q_x\Big)\quad\quad k=0,1,\ldots,n\;,
\ee
where $a_0\eqdef 0$ and $b_0\eqdef 0$.
\end{theorem}
\end{myt}
\begin{remark}
In general, the sum $\sum_{x\in[k]}p_x$ does not equal to $\|\p\|_{(k)}$ since the components of $\p$ are not necessarily arranged in a non-increasing order. Instead, the components of $\p$ and $\q$ are arranged such that the order in~\eqref{order} holds.
\end{remark}
\begin{proof}
Let $f:[0,1]\to [0,1]$ be the function whose graph is the lower Lorenz curve\index{Lorenz curve} of $(\p,\q)$. Then, by definition, for every $a\in[0,1]$, $(a,f(a))$ is the lowest point in $\mt(\p,\q)$ whose $x$-coordinate is $a$. Therefore, for any $r\in[0,1]$ we can express $f(r)$ as
\be\label{fx}
f(r)=\min\big\{\q\cdot\t\;:\;\t\in[0,1]^n\;\;,\;\; \p\cdot\t=r\big\}\;.
\ee
Our objective is to demonstrate that the function $f(r)$ defines the segment connecting two adjacent vertices. Specifically, consider a fixed $k\in{0,1,\ldots,n}$. Our aim is to establish that for any $r$ in the interval $[a_k,a_{k+1})$, the function $f(r)$ corresponds to the line segment joining the points $(a_k,b_k)$ and $(a_{k+1},b_{k+1})$. Mathematically, this means that for all $r\in[a_k,a_{k+1})$, we have:
\be\label{closedform}
f(r)=s_{k+1}(r-a_{k})+b_{k}\;,
\ee
where $s_x\eqdef q_x/p_x$ for all $x\in[n]$, adhering to the convention that $s_x\eqdef\infty$ if $p_x=0$ and $q_x>0$ (recall that we assume that there is no $x\in[n]$ such that both $p_x$ and $q_x$ are equal to zero). Successfully proving this relationship implies that the set of points $\{(a_k,b_k)\}_{k=0}^n$ are indeed the extreme points on the lower Lorenz curve\index{Lorenz curve} of the testing region $\mt(\p,\q)$.

To prove~\eqref{closedform}, observe that the optimization problem in~\eqref{fx} is a linear program. In Exercise~\ref{lowlor} you will apply methods discussed in Sec.~\ref{app:sdp}, specifically the dual problem\index{dual problem} framework, to express $f(r)$ as:
\be\label{552}
f(r)=\max\Big\{rs-\v\cdot\textbf{1}_n\;:\;\v\in\mbb{R}^n_{+}\;\;,\;\; s\p-\q\leq\v\;\;,\;\;s\in\mbb{R}_+\Big\}
\ee 
where $\textbf{1}_n\eqdef(1,\ldots,1)^T$ and the inequality is entry-wise. The maximization in~\eqref{552} can be simplified since the vector $\v$ with the smallest non-negative components that satisfies the constraint $\v\geq s\p-\q$ is given by $\v=(s\p-\q)_+$ (the components of $(s\p-\q)_+$ are $\{(sp_x-q_x)_+\}_{x\in[n]})$. Hence,
\be\label{5101}
f(r)=\max_{s\geq 0}\big\{sr-(s\p-\q)_+\cdot\textbf{1}_n\big\}\;.
\ee
To simplify further, note that
\be
(s\p-\q)_+\cdot\textbf{1}_n=\sum_{x\in[n]}(sp_x-q_x)_+=\sum_{x\in[n]}p_x\left(s-s_x\right)_+\;.
\ee
Now, from~\eqref{order} it follows that
$
s_1\leq s_2\leq\cdots\leq s_n
$.
Therefore, for any $s\geq 0$ there exists $\ell\in \{0,1,\ldots,n\}$ with the property that $s_\ell\leq s<s_{\ell+1}$, where we added the definitions $s_0\eqdef 0$ and $s_{n+1}\eqdef\infty$. With this definition of $\ell$ we get
\be
(s\p-\q)_+\cdot\textbf{1}_n=\sum_{x\in[\ell]}p_x\left(s-s_x\right)=sa_{\ell}-b_{\ell}\;.
\ee
Therefore, by splitting the maximization in~\eqref{5101} into maximization over all $\ell\in\{0,1,\ldots,n\}$ and all $s\in[s_{\ell},s_{\ell+1})$ we get
\be
f(r)=\max_{\ell\in\{0,1\ldots,n\}}\sup_{s\in[s_{\ell},s_{\ell+1})}\big\{s(r-a_{\ell})+b_{\ell}\big\}\;.
\ee
If the optimal $\ell$ above satisfies $\ell\leq k$ then $r-a_\ell\geq 0$ so that
\be \label{5p105}
\sup_{s\in[s_{\ell},s_{\ell+1})}\big\{s(r-a_{\ell})+b_{\ell}\big\}=s_{\ell+1}(r-a_{\ell})+b_{\ell}\;.
\ee
Moreover, among all $\ell\in\{0,1,\ldots,k\}$ the choice $\ell=k$ yields the greatest value since the right-hand side above is increasing in $\ell$ as long as $\ell\leq k$ (see Exercise~\ref{ex537}); hence,
\be\label{5p106}
\max_{\ell\in\{0,1,\ldots,k\}}\sup_{s\in[s_{\ell},s_{\ell+1})}\big\{s(r-a_{\ell})+b_{\ell}\big\}=s_{k+1}(r-a_{k})+b_{k}\;.
\ee
On the other hand, if $\ell>k$ then $r-a_\ell<0$ so that
\be\label{5p107}
\sup_{s\in[s_{\ell},s_{\ell+1})}\big\{s(r-a_{\ell})+b_{\ell}\big\}=s_{\ell}(r-a_{\ell})+b_{\ell}
\ee
Furthermore, among all $\ell\in\{k+1,\ldots,n\}$ the choice $\ell=k+1$ yields the greatest value since the right-hand side above is decreasing in $\ell$ as long as $\ell\geq k+1$ (see Exercise~\ref{ex537}); that is,
\be\label{5p108}
\max_{\ell\in\{k+1,\ldots,n\}}\sup_{s\in[s_{\ell},s_{\ell+1})}\big\{s(r-a_{\ell})+b_{\ell}\big\}=s_{k+1}(r-a_{k+1})+b_{k+1}\;.
\ee
The right-hand side of~\eqref{5p106} is in fact equal to the right-hand side of~\eqref{5p108} (see Exercise~\ref{ex537}). We therefore conclude that for any $k\in\{0,1,\ldots,n\}$ and any $r\in[a_k,a_{k+1})$ we have $f(r)=s_{k+1}(r-a_{k})+b_{k}$.
This completes the proof. 
\end{proof}

\begin{exercise}\label{lowlor}
Consider the function $f(r)$ as defined in~\eqref{fx}.
\ben
\item Show that the condition $\p\cdot\t=r$ in~\eqref{fx} can be replaced with $\p\cdot\t\geq r$. Hint:  Observe that any $\t$ satisfying $\p\cdot\t>r$ can be rescaled to give $\p\cdot\t=r$ (and this rescaling can only decrease $\q\cdot\t$).
\item Prove the equality in~\eqref{552}. Hint: First express the minimization in~\eqref{fx}
(after replacing $\p\cdot\t=r$ with $\p\cdot\t\geq r$) as a conic linear programming\index{linear programming} of the form~\eqref{primal} (with vectors in $\mbb{R}^n$ replacing Hermitian matrices, and the dot product replacing the Hilbert-Schmidt inner product). Then use~\eqref{dual123} and the strong duality\index{duality} to get~\eqref{552}. 
\een
\end{exercise}

\bex\label{ex537}
Show that the right-hand side of~\eqref{5p105} is increasing in $\ell\in[k]$, and the right-hand side of~\eqref{5p107} is decreasing in $\ell\in\{k+1,\ldots,n\}$. Moreover, show that the two expressions are the same for $\ell=k$ and $\ell=k+1$ (i.e., show that the right-hand side of~\eqref{5p106} is equal to the right-hand side of~\eqref{5p108}).
\eex

\bex
Let $\p,\q\in\prob(n)$ and $t\in\mbb{R}$. Prove the following equalities:
\ba\label{exinw}
{\it 1.}\quad& (\p-t\q)_+\cdot\1_n=\frac12\big(\|\p-t\q\|_1+1-t\big)\;.\\
{\it 2.}\quad& (t\p-\q)_+\cdot\1_n=\frac12\big(\|t\p-\q\|_1+t-1\big)\;.
\ea
Hint: Use the relation $(a-b)_+=\frac12|a-b|+\frac12(a-b)$.
\eex

\begin{exercise}
Compute the vertices of the lower Lorenz curve of the example given in Fig.~\ref{tregion}. If necessary, rearrange the components of $\p$ and $\q$ so that~\eqref{order} holds. 
\end{exercise}

\bex
For a given $\p,\q,\in\prob(n)$, find the vertices of $\mt(\p,\q)$ that are located on the upper Lorenz curve of $(\p,\q)$.
\eex

\begin{exercise}\label{vertices2}
Let $\p\in\prob(n)$. Show that the vertices of the lower Lorenz curve of the pair $(\p,\u^{(n)})$ are given by  
\be\label{vert4}
\left(\|\p\|_{(k)}\;,\;\frac{k}{n}\right)\quad\quad k=0,1,\ldots,n\;,
\ee
with the convention that for $k=0$, $\|\p\|_{(0)}\eqdef 0$.
\end{exercise}

Theorem~\ref{vertices} has the following interesting corollary.

\begin{myg}{}
\begin{corollary}\label{corptq}
Let $\p,\q\in\prob(n)$ and $\p',\q'\in\prob(n')$.
If for all $t\geq 1$ we have $\|\p-t\q\|_1\geq\|\p'-t\q'\|_1$ then $\mt(\p,\q)\supseteq\mt(\p',\q')$.
\end{corollary}
\end{myg}
\begin{remark}
We will see shortly that the converse to the statement in the corollary above is also true.
\end{remark}
\begin{proof}
The proof follows immediately from the expression for the lower Lorenz curve\index{Lorenz curve} in~\eqref{5101}. Explicitly, by using the variable $t\eqdef\frac1s$ in~\eqref{5101}, and using the notation $f_{\p,\q}(r)$ for $f(r)$ we get that for all $r\in[0,1]$
\ba
f_{\p,\q}(r)&=\max_{t\geq 1}\left\{\frac{r-(\p-t\q)_+\cdot\textbf{1}_n}{t}\right\}\\
\GG{\eqref{exinw}}&=\max_{t\geq 1}\left\{\frac{2r-\|\p-t\q\|_1+t-1}{2t}\right\}\;.
\ea
Therefore, if $\|\p-t\q\|_1\geq\|\p'-t\q'\|_1$ for all $t\geq 1$ then $f_{\p,\q}(r)\leq f_{\p',\q'}(r)$ for all $r\in[0,1]$; i.e., the lower Lorenz curve\index{Lorenz curve} of the pair $(\p,\q)$ is nowhere above the lower Lorenz curve\index{Lorenz curve} of $(\p',\q')$ so that $\mt(\p,\q)\supseteq\mt(\p',\q')$.
\end{proof}

\bex\label{exdpi1norm}
Let $\p,\q\in\prob(n)$ and $\p',\q'\in\prob(n')$. Show that if $(\p,\q)\succ(\p',\q')$ then for all $t\in\mbb{R}$ we have $\|\p-t\q\|_1\geq\|\p'-t\q'\|_1$. Hint: Use the property~\ref{dpisv} of the 1-norm $\|\cdot\|_1$.
\eex

\bex\label{tbelongtp}
Let $\p,\q\in\prob(n)$ and $\p',\q'\in\prob(n')$. 
\ben 
\item Show that if $(\p,\q)\succ(\p',\q')$ then  $\mt(\p',\q')\subseteq\mt(\p,\q)$. 
\item Show that if $(\p,\q)\sim(\p',\q')$ then $\mt(\p',\q')=\mt(\p,\q)$. 
\een
Hint: For the first part, let $E\in\stoc(n',n)$ be such that $\p'=E\p$ and $\q'=E\q$, and show first that for any $\t'\in[0,1]^n$ the vector $\t\eqdef E^T\t'$ belongs to $[0,1]^n$ and satisfies $(\t'\cdot\p',\t'\cdot\q')=(\t\cdot\p,\t\cdot\q)$.
\eex

\subsection{Characterization of Relative Majorization}

Relative majorization possesses several valuable characterizations, all of which are succinctly summarized in the theorem that follows. In Appendix~\ref{AlternativeBP}, we present a more extensive and traditional proof of this theorem, employing concepts from convex analysis such as support functions and sublinear functionals. In contrast to this comprehensive approach, we also provide a considerably shorter proof here. This brief proof avoids reliance on the aforementioned concepts from convex analysis, and instead leverages the intricate relationship that we previously examined between majorization and relative majorization.

\begin{figure}[h]\centering
    \includegraphics[width=0.4\textwidth]{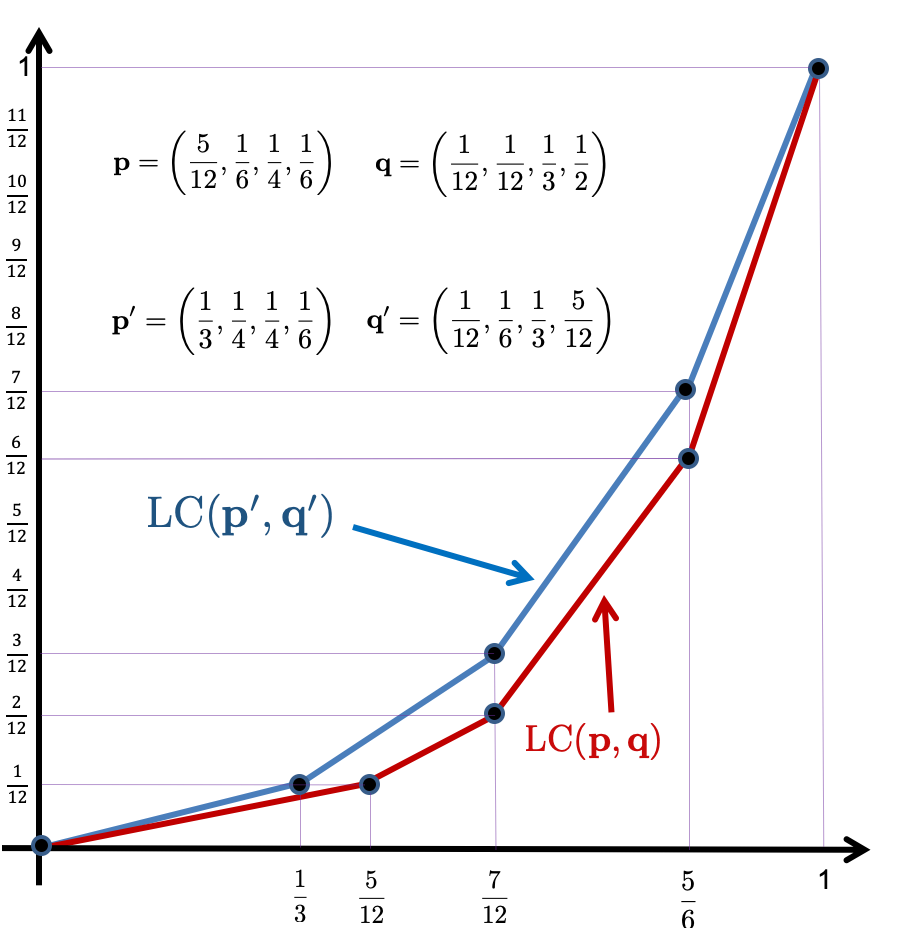}
  \caption{\linespread{1}\selectfont{\small Lower Lorenz Curves. The red lower Lorenz curve $\text{LC}(\p,\q)$ is nowhere above the blue lower Lorenz curve $\text{LC}(\p',\q')$. This means that the pair $(\p,\q)$ relatively majorizes the pair $(\p',\q')$. Note that aside from the vertices $(0,0)$ and $(1,1)$, the vertices of $\text{LC}(\p,\q)$ are $(\frac{5}{12},\frac1{12})$, $(\frac7{12},\frac1{6})$, $(\frac56,\frac12)$, and the vertices of $\text{LC}(\p',\q')$ are $(\frac{1}{3},\frac1{12})$, $(\frac{7}{12},\frac1{4})$, $(\frac56,\frac7{12})$.}}
  \label{lower}
\end{figure}

\begin{myt}{\color{yellow} Characterization}
\begin{theorem}\label{chararm}
Let $n,n'\in\mbb{N}$, $\p,\q\in\prob(n)$, and $\p',\q'\in\prob(n')$.  Then, the following are equivalent:
\begin{enumerate}
\item $(\p,\q)\succ(\p',\q')$.
\item For all $t\in\mbb{R}$ we have $\|\p-t\q\|_1\geq\|\p'-t\q'\|_1$\;.
\item $\mt(\p,\q)\supseteq\mt(\p',\q')$. 
\end{enumerate}
\end{theorem}
\end{myt}
\begin{remark}
The equivalence between 1 and 3 in  Theorem~\ref{chararm} provides a very simple geometrical characterization of relative majorization. Denoting by $\text{LC}(\p,\q)$ and $\text{LC}(\p',\q')$ the two lower Lorenz curves associated with the two testing regions, we have that $(\p,\q)\succ(\p',\q')$ if and only if $\text{LC}(\p,\q)$ is nowhere above $\text{LC}(\p',\q')$. An example illustrating this property is depicted in Fig.~\ref{lower}.\index{Lorenz curve}
\end{remark} 

The implication $1 \Rightarrow 2$ can be easily deduced from the monotonicity property of the norm $\|\cdot\|_1$, as discussed in~\eqref{dpisv}. This part of the proof is straightforward and hence, is suggested as an exercise for the reader (refer to Exercise~\ref{exdpi1norm}). Having previously established the implication $2 \Rightarrow 3$ in Corollary~\ref{corptq}, our remaining task is to demonstrate that $3 \Rightarrow 1$. We begin this proof by focusing on the case where both $\q$ and $\q'$ consist of positive rational components.

\begin{myg}{}
\begin{lemma}\label{olemo}
Let $\q,\p\in\prob(n)$ and $\p',\q'\in\prob(n')$, and suppose that $\q$ and $\q'$ have positive rational components.
If $\mt(\p,\q)\supseteq\mt(\p',\q')$ then $(\p,\q)\succ(\p',\q')$.
\end{lemma}
\end{myg}

\begin{proof}
From Theorem~\ref{onlyr} we get that there exist vectors $\r,\r'\in\prob(k)$ such that $(\p,\q)\sim(\r,\u^{(k)})$ and $(\p',\q')\sim(\r',\u^{(k)})$, where $k$ is a common denominator of all the components of $\q$ and $\q'$. Therefore, from the second part of Exercise~\ref{tbelongtp} we get that $\mt(\p,\q)=\mt(\r,\u^{(k)})$ and $\mt(\p',\q')=\mt(\r',\u^{(k)})$. Moreover, since we assume that $\mt(\p,\q)\supseteq\mt(\p',\q')$ we get that $\mt(\r,\u^{(k)})\supseteq\mt(\r',\u^{(k)})$.
Hence, the Lorenz curve\index{Lorenz curve} ${\rm LC}(\r,\u^{(k)})$ is nowhere above the Lorenz curve\index{Lorenz curve} ${\rm LC}(\r',\u^{(k)})$. In addition, the non-zero vertices of 
${\rm LC}(\r,\u^{(k)})$ and ${\rm LC}(\r',\u^{(k)})$ are given respectively by (cf.~\eqref{vert4}):
\be
\left\{\left(\|\r\|_{(\ell)}\;,\;\frac\ell k\right)\right\}_{\ell\in[k]}\quad\text{and}\quad\left\{\left(\|\r'\|_{(\ell)}\;,\;\frac\ell k\right)\right\}_{\ell\in[k]}\;.
\ee
Therefore, since the vertex $(\|\r\|_{(\ell)},\ell/k)$ has the same $y$-coordinate as the vertex $(\|\r\|_{(\ell)},\ell/ k)$, and since the convex curve ${\rm LC}(\r,\u^{(k)})$ is nowhere above the convex curve ${\rm LC}(\r',\u^{(k)})$, we get that 
$\|\r\|_{(\ell)}\geq \|\r'\|_{(\ell)}$ for all $\ell\in[k]$. That is, $\r\succ\r'$ and from~\eqref{4p106} this is equivalent to $(\p,\q)\succ(\p',\q')$. This completes the proof.
\end{proof}

In order to completes the proof of Theorem~\ref{chararm} we will need a continuity\index{continuity} argument that extends the lemma above to the general case of arbitrary $\q$ and $\q'$.

\begin{myg}{}
\begin{lemma}\label{continuity101}
Let $\q,\p\in\prob(n)$ and $\p',\q'\in\prob(n')$ and suppose $\mt(\p,\q)\supseteq\mt(\p',\q')$. Then, for every $\eps\in(0,1)$ there exist two vectors $\q^{(\eps)}\in\prob(n)$ and $\q'^{(\eps)}\in\prob(n')$ with positive rational components\index{rational components} such that $\q^{(\eps)}\approx_\eps\q$, $\q^{\prime(\eps)}\approx_\eps\q'$, and
\be
\mt\left(\p,\q^{(\eps)}\right)\supseteq\mt\left(\p',\q'^{(\eps)}\right)\;.
\ee
\end{lemma}
\end{myg}
\begin{proof}
We provide  a geometrical proof using Fig.~\ref{proof}. \index{Lorenz curve}
\begin{figure}[h]\centering
    \includegraphics[width=0.8\textwidth]{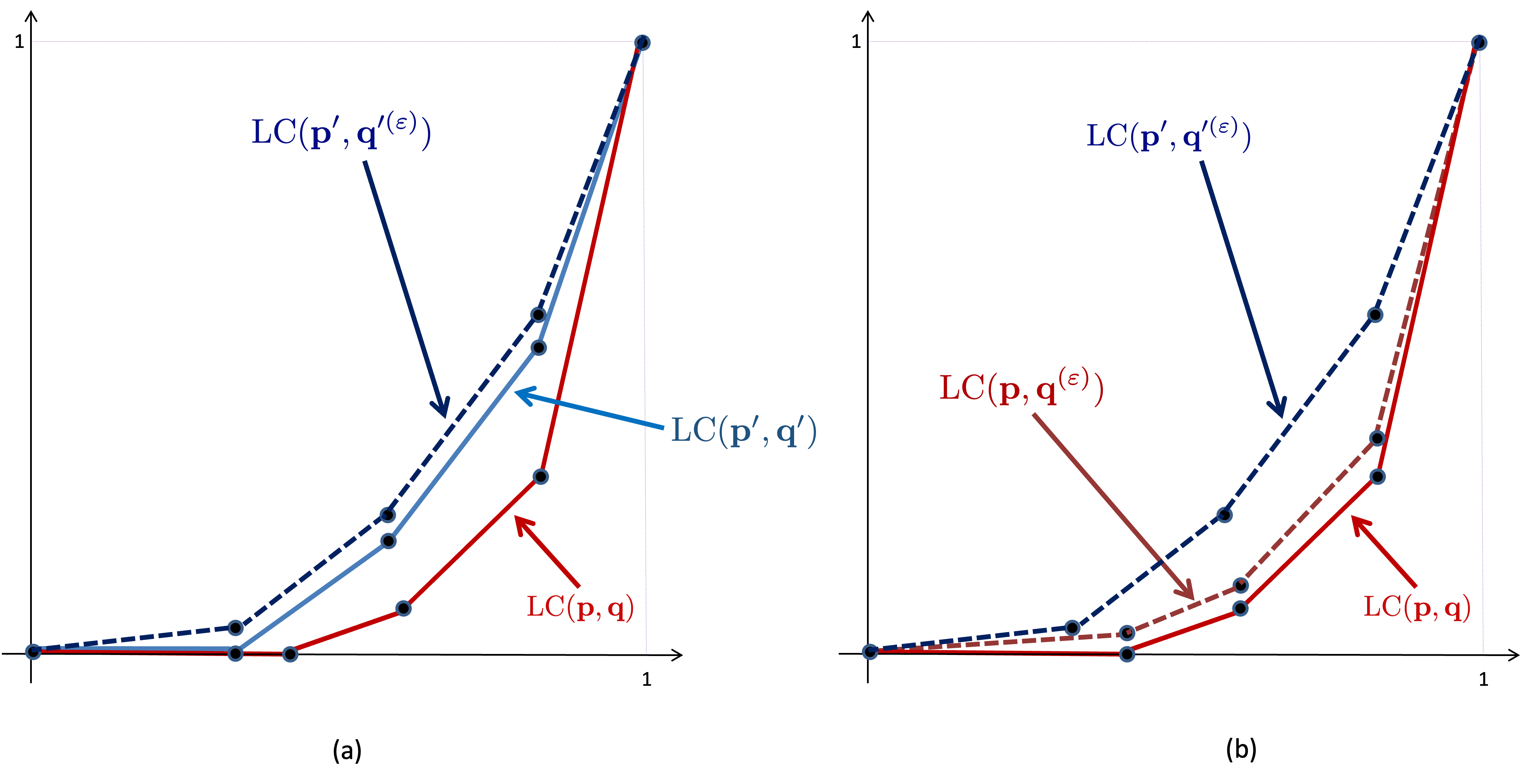}
  \caption{\linespread{1}\selectfont{\small Geometrical Proof of Lemma~\ref{continuity101}.}}
  \label{proof}
\end{figure}
By keeping $\p'$ unchanged, we can raise slightly and vertically the vertices of $\text{LC}(\p',\q')$ to get the lower Lorenz curve\index{Lorenz curve} of $\text{LC}(\p',\q^{\prime (\eps)})$, where $\q^{\prime (\eps)}$ has positive rational components\index{rational components} and is $\eps$-close to $\q'$; see Fig.~\ref{proof}(a) for an illustration. Explicitly, let $\eps_1,\ldots,\eps_{n-1}$ be small enough positive numbers such that for all $x\in[n-1]$, ${q'}_x^{(\eps)}\eqdef q_x'+\eps_x$ is a rational number. Furthermore, we can always choose $\eps_1,\ldots,\eps_{n-1}$ to be small enough such that their sum $\delta\eqdef\sum_{x\in[n-1]}\eps_x<\eps$ and satisfies
${q'}_n^{(\eps)}\eqdef q'_n-\delta>0$ (due to the standard form of $(\p',\q')$ we have $q_n'>0$). For these choices, the $\q'^{(\eps)}\eqdef(q_1'^{(\eps)},\ldots,q_n'^{(\eps)})^T$ has positive rational numbers, and $\q'^{(\eps)}$ is also $\eps$-close to $\q$. Furthermore, by construction, $\text{LC}(\p',\q^{\prime (\eps)})$ is everywhere above $\text{LC}(\p',\q')$.

Once we established $\text{LC}(\p',\q^{\prime (\eps)})$ we construct $\q^{(\eps)}$ in a similar way; see Fig.~\ref{proof}(b). Explicitly, let $\nu_1,\ldots,\nu_{n-1}$ be small enough positive numbers such that for all $x\in[n-1]$, ${q}_x^{(\eps)}\eqdef q_x+\nu_x$ is a rational number. Furthermore, we can always choose $\nu_1,\ldots,\nu_{n-1}$ to be small enough such that their sum $\nu\eqdef\sum_{x\in[n-1]}\nu_x<\eps$ and satisfies
${q}_n^{(\eps)}\eqdef q_n-\nu>0$. For these choices, the vector $\q^{(\eps)}\eqdef(q_1^{(\eps)},\ldots,q_n^{(\eps)})^T$ has positive rational numbers, is $\eps$-close to $\q$, and as long as $\nu_1,\ldots,\nu_{n-1}$ are sufficiently small, $\text{LC}(\p,\q^{(\eps)})$ is everywhere below $\text{LC}(\p',\q^{\prime (\eps)})$. This completes the proof.
\end{proof}

With these lemmas at hand, we are now ready to prove the theorem.

\subsubsection{Proof of Theorem~\ref{chararm}}

\begin{proof}
Recall that it is left to prove that $3\Rightarrow 1$.
Since we assume that $\mt(\p,\q)\supseteq\mt(\p',\q')$, Lemma~\ref{continuity101}  implies that for any $\eps\in(0,1)$ there exists two vectors $\q^{(\eps)}\in\prob(n)$ and $\q'^{(\eps)}\in\prob(n')$ with positive rational components\index{rational components} such that
$
\mt(\p,\q^{(\eps)})\supseteq\mt(\p',\q'^{(\eps)})
$.
Thus, applying Lemma~\ref{olemo} with the two pairs  $(\p,\q^{(\eps)})$ and $(\p',\q'^{(\eps)})$ replacing $(\p,\q)$ and $(\p',\q')$ gives
\be\label{qepe}
(\p,\q^{(\eps)})\succ(\p',\q'^{(\eps)})\;.
\ee
Now, let $\{\eps_{\ell}\}_{\ell\in\mbb{N}}$ be a sequence of numbers in $(0,1)$ with zero limit, and denote by $\q_\ell\eqdef{\q}^{(\eps_\ell)}$ and $\q_\ell'\eqdef{\q'}^{(\eps_\ell)}$. With these notations for each $\ell\in\mbb{N}$ we have $(\p,\q_{\ell})\succ(\p',\q'_{\ell})$ and $\lim_{\ell\to\infty}\q_\ell=\q$ and $\lim_{\ell\to\infty}\q'_\ell=\q'$. The condition $(\p,\q_{\ell})\succ(\p',\q'_{\ell})$ implies that for each $\ell\in\mbb{N}$ there exists a matrix $E^{(\ell)}\in\stoc(n',n)$ such that 
\be\label{bothofthe}
\p'=E^{(\ell)}\p\quad\text{and}\quad\q'_\ell=E^{(\ell)}\q_\ell\;.
\ee
Since $\stoc(n',n)$ is a compact set, the sequence $\{E^{(\ell)}\}_{\ell\in\mbb{N}}$ has a converging subsequence. For simplicity of the exposition we assume that $\{E^{(\ell)}\}_{\ell\in\mbb{N}}$ is itself a convergent sequence (otherwise, just replace everywhere $\{\ell\}_{\ell\in\mbb{N}}$ with a subsequence of elements $\{\ell_j\}_{j\in\mbb{N}}$). Therefore, there exists $E\in\stoc(n',n)$ such that $E=\lim_{\ell\to\infty}E^{(\ell)}$. By taking the limit $\ell\to\infty$ on both of the equations in~\eqref{bothofthe} we get $\p'=E\p$ and $\q'=E\q$. That is, $(\p,\q)\succ(\p',\q')$.
This completes the proof.
\end{proof}

\bex 
Consider the second statement of Theorem~\ref{chararm}.
\ben
\item Show that $t\in\mbb{R}$  can be restricted to $t\geq 1$. Hint: Recall Corollary~\ref{corptq}.
\item Show that $t\in\mbb{R}$  can be restricted to $t\in[0,1]$. Hint: Recall that $(\p,\q)\succ(\p',\q')$ if and only if $(\q,\p)\succ(\q',\p')$.
\een
\eex

\begin{exercise}\label{ex:68}
Let $\p=(p,1-p)^T$ and $\q=(q,1-q)^T$ be two probability vectors satisfying $p>q$. 
Let $\p'=(p',1-p')^T$ and $\q'=(q',1-q')^T$ be another pair of probability vectors with $p'\geq q$ and $,q'\leq p$. 
Show that
\ba\label{68111}
(\p',\q)\prec(\p,\q)\quad\iff\quad  p'\leq p\\
(\p,\q')\prec(\p,\q)\quad\iff\quad q\leq q'
\ea
\end{exercise}

\begin{exercise}\label{themaj}
Let $0<\q\in\prob(n)$ and suppose $\q=\q^\da$. Show that
\be
(\e_n,\q)\succ(\p,\q)\quad\quad\forall\;\p\in\prob(n)\;,
\ee
where $\e_n=(0,\ldots,0,1)^T$.
\end{exercise}

\subsection{Continuity of Relative Majorization}\label{TGC}\index{continuity}

In this subsection we investigate several continuity properties of relative majorization.
We start by noting that
in the case that $\q$ doesn't have positive rational components, the equivalence of the form~\eqref{ruk} does not hold in general.
However, it is still possible to approximate this relation as the following exercise demonstrates.

\bex\label{1lem1}
Let $\eps>0$, and $\p,\q\in\prob(n)$. Show that:
\ben 
\item If $\p\neq \q$ then there exists $0<\q'\in \prob(n) \cap \mbb{Q}^n$ such that $\q'$ is $\eps$-close to $\q$ and $(\p,\q)\succ(\p,\q')$.
\item There exists a vector $\q'\in \prob(n) \cap \mbb{Q}^n$ such that
$\q'$ is $\eps$-close to $\q$, $\supp(\q')=\supp(\q)$, and 
$(\p,\q')\succ (\p,\q)$.
\een
Hint: Take $\q'$ to be $\q^{(\eps)}$ or~$\q^{\prime(\eps)}$ as defined in Lemma~\ref{continuity101}.
\eex

In the exercise above we approximated $(\p,\q)$ with a pair of vectors $(\p,\q')$, where $\q'$ has some desired properties (particularly, rational components), and $\p$ is fixed. In the next two lemmas we remove some of the assumptions on $\q$ by allowing $\p$ to vary. We will only assume that $\q'$ is close to $\q$.

Specifically, consider two probability vectors $\p,\q\in\prob(n)$ and let $\eps\in(0,1)$ be a sufficiently small number to be determined later. Our objective is to show that for every $\p'\in\mb_\eps(\p)$ there exists $\delta\in(0,1)$ and $\q'\in\mb_\delta(\q)$ such that $(\p',\q')\succ(\p,\q)$. 
We will be able to show that such a $\q'$ exists if we assume that $\q>0$ and that $\supp(\p')\subseteq\supp(\p)$. In the following lemma we will use the notation
$\eps_0\eqdef\frac12p_{\min}q_{\min}$,
where $p_{\min}$ and $q_{\min}$ are the smallest non-zero components of $\p$ and $\q$, respectively.

\begin{myg}{}
\begin{lemma}\label{3lem3}
Let $\p,\q\in\prob(n)$, $\q>0$, $\eps\in(0,\eps_0)$, $\delta\eqdef\eps/{q_{\min}}$, and $\p'\in\mb_{\eps}(\p)$ with $\supp(\p')\subseteq\supp(\p)$.
Then, there exists $\q'\in\mb_\delta(\q)$ such that $(\p',\q')\succ (\p,\q)$.
\end{lemma}
\end{myg}

\begin{proof}
We need to define $\q'\in\mb_{\delta}(\q)$ and a channel $E\in\stoc(n,n)$ such that $E\p'=\p$ and $E\q'=\q$. The key idea is to look for a matrix $E$ of the form
\be\label{eslash}
E\t\eqdef\p+s\left(\t-\p'\right)\quad\quad\forall\;\t\in\prob(n)\;,
\ee
where $s\in\mbb{R}_+$ is some coefficient. Observe that we define the matrix $E$ by its action on probability vectors. Clearly, by construction, $E\p'=\p$. However, $E$ above is not necessarily
a stochastic matrix since for arbitrary $s\in\mbb{R}$ the vector $\p+s\left(\t-\p'\right)$ could have negative components. We therefore choose $s\geq 0$ to be such that $\p\geq s\p'$ (entry-wise) so that $E$, as defined above, is indeed a column stochastic matrix. We therefore take $s$ to be  the optimal $s$ that satisfies $\p\geq s\p'$; that is, we define 
\be
s\eqdef\min_{x\in\supp(\p')}\frac{p_x}{p_x'}\;.
\ee
Observe that $s\leq 1$ (since we cannot have $p_x>p_x'$ for all $x\in\supp(\p')$), and since $\supp(\p')\subseteq\supp(\p)$ we have $s>0$. In fact, since $\p'$ is $\eps$-close to $\p$ we must have $p_x'\leq p_x+\eps$ for all $x\in[n]$ so that
\be\label{5122}
s\geq \min_{x\in\supp(\p')}\frac{p_x}{p_x+\eps}\geq\frac{p_{\min}}{p_{\min}+\eps}\;.
\ee

Our next goal is to define $\q'$ such that $E\q'=\q$. Observe that according to~\eqref{eslash} we have $E\q'=\p+s\left(\q'-\p'\right)$ so that $E\q'=\q$ if $\p+s\left(\q'-\p'\right)=\q$. Isolating $\q'$ we get that $\q'$ should have the form
\be\label{5699}
\q'\eqdef\p'+\frac1s(\q-\p)
\ee
(indeed, check that with this $\q'$ we have $E\q'=\q$). However, it is not obvious that $\q'$ has non-negative components. Therefore, we next show that $\eps$ is small enough so that $\q'\geq 0$ (i.e. $\q'$ is a probability vector). 

Observe that $\q'\geq 0$ if and only if for all $x\in[n]$
we have $q_x\geq p_x-sp_x'$. Now, since $\p$ is $\eps$-close to $\p'$ we get that
\ba
p_x-sp_x'\leq p_x-s(p_x-\eps)&=(1-s)p_x+s\eps\\
\Gg{p_x,s\leq 1}&\leq 1-s+\eps\\
\GG{\eqref{5122}}&\leq\frac{\eps}{p_{\min}+\eps}+\eps\leq \frac{2\eps}{p_{\min}}\leq q_{\min}\;.
\ea
Hence, $\p'$ is a probability vector. By definition,
\ba
\frac12\left\|\q'-\q\right\|_1&=\frac1{2s}\left\|(1-s)\q-(\p-s\p')\right\|_1\\
\GG{Triangle\; inequality}&\leq\frac1{2s}\left(1-s+\left\|\p-s\p'\right\|_1\right)\\
\Gg{\p\geq s\p'}&=\frac{1-s}s=\frac1s-1\\
\GG{\eqref{5122}}&\leq\frac\eps{p_{\min}}=\delta\;.
\ea
This completes the proof.
\end{proof}
\bex
Using the same notations as in Lemma~\ref{3lem3}, let
\be
\eps_1\eqdef\min_{x\in\supp(\p)}\frac{p_{\min}q_x}{p_{\min}+(p_x-q_x)_+}\;.
\ee
\ben
\item Show that $\eps_0<\eps_1$.
\item Show that the Lemma~\ref{3lem3} still holds if we replace $\eps_0$ with $\eps_1$.
\een
\eex

\begin{myg}{}
\begin{lemma}\label{4lem40}
Let $\p,\q,\q'\in\prob(n)$ and denote by
$
\delta\eqdef1-\min_{x\in[n]}\frac{q_x'}{q_x}
$.
Then, there exists $\p'\in\mb_\delta(\p)$ such that $(\p,\q)\succ (\p',\q')$.
\end{lemma}
\end{myg}

\begin{remark}
Note that if $\frac12\|\q-\q'\|_1\leq\eps$ for some $\eps>0$ then 
\be
\delta=\max_{x\in[n]}\frac{q_x-q_x'}{q_x}\leq\frac\eps{q_{\ell}}
\ee
where $\ell$ is the integer satisfying $\delta=1-q_\ell'/q_\ell$. Therefore, if $\q$ and $\q'$ are very close to each other so are $\p$ and $\p'$.
\end{remark}

\begin{proof}
Let $s\eqdef 1-\delta=\min_{x\in[n]}\frac{q_x'}{q_x}$.
From the definition of $s$ we have  $\q'\geq s\q$ (entry-wise), so that the mapping
\be\label{eteq}
E\t\eqdef\q'+s\left(\t-\q\right)\quad\quad\forall\;\t\in\prob(n)
\ee
is a channel. By definition,  $E\q=\q'$. Define
\ba
\p'&\eqdef E\p\\
\GG{\eqref{eteq}}&=\q'+s\left(\p-\q\right)\;,
\ea
so that $(\p,\q)\succ(\p',\q')$.
Then,
\ba
\frac12\left\|\p'-\p\right\|_1&=\frac{1}{2}\left\|\q'-s\q-(1-s)\p\right\|_1\\
\GG{Triangle\;inequality}&=\frac{1}{2}\left\|\q'-s\q\right\|_1+\frac12(1-s)\\
\Gg{\q'\geq s\q}&=1-s\;.
\ea
This completes the proof.
\end{proof}

\bex\label{ex:4lem4}
Prove the following theorem: Let $\p,\q,\p'\in\prob(n)$ and denote by
$
\delta\eqdef1-\min_{x\in[n]}\frac{p_x'}{p_x}
$.
Then, there exists $\q'\in\mb_\delta(\q)$ such that $(\p,\q)\succ (\p',\q')$.
\eex

\section{The Trumping Relation}\label{sec:tr}\index{trumping relation}
Another pre-order that plays an important role in several resource theories is the following variant of majorization.
\begin{myd}{The Trumping Relation}\index{trumping relation}
\begin{definition}
For any $\p,\q\in\prob(n)$ we say that $\p$ \emph{trumps} $\q$ and write 
\be
\p\succ_{*}\q\;,
\ee
if there exists an integer $m\in\mbb{N}$ and a vector $\r\in\prob(m)$, known as the catalyst vector, such that 
\be\label{condr}
\p\otimes\r\succ\q\otimes\r\;.
\ee
\end{definition}
\end{myd}

\begin{remark}
Observe that while we require that the dimension $m<\infty$, it is still unbounded. We will see later that this implies that the trumping relation is very sensitive to perturbations, and also leads to a phenomenon know as `embezzlement' of entanglement.
\end{remark}

By definition, if $\p\succ\q$ then necessarily $\p\succ_*\q$. The example below demonstrates that the opposite direction does not hold in general. In this sense, the trumping relation impose a weaker constraint than majorization.

\begin{example}\nonumber
Consider the probability vectors
\be\nonumber
\p\eqdef\begin{bmatrix}
2/5\\ 2/5 \\ 1/10\\ 1/10
\end{bmatrix}\quad\text{and}\quad\q=\begin{bmatrix}
1/2\\ 1/4 \\ 1/4\\ 0
\end{bmatrix}
\ee
It is simple to check (see exercise below) that $\p\not\succ\q$ and $\q\not\succ\p$. Yet, $\p\succ_{*}\q$ since the vector $\r=(3/5,2/5)^T$ satisfies~\eqref{condr}.
\end{example}

\begin{exercise}\label{ex551}
Let $\p,\q,\r$ be as in the example above. 
\begin{enumerate}
\item Verify that $\p\not\prec\q$ and $\q\not\prec\p$.
\item Verify that~\eqref{condr} holds.
\end{enumerate}
\end{exercise}

\begin{exercise}
Show that if $\p,\q\in\prob(3)$ and $\p\succ_{*}\q$ then $\p\succ\q$. 
\end{exercise}

\begin{exercise}
Show that the uniform probability vector $\u$ cannot act as a catalyst verctor; that is, show that if $\p\in\prob(n)$ and $\q\in\prob(m)$ are such that $\p\not\succ\q$ then for any $k\in\mbb{N}$, $\p\otimes\u^{(k)}\not\succ\q\otimes\u^{(k)}$ . 
\end{exercise}

\begin{exercise}
Let $f:\prob(n)\to\mbb{R}$ be a Schur convex function that is additive under tensor product; i.e.  for all $\p\in\prob(n)$ and $\q\in\prob(m)$
\be
f(\p\otimes\q)=f(\p)+f(\q)\;.
\ee
Show that
\be
\p\succ_{*}\q\quad\Rightarrow\quad f(\p)\geq f(\q)\;.
\ee
\end{exercise}

The tramping relation can also be extended to pairs of probability vectors.

\begin{myd}{Relative Trumping}\index{relative trumping}
\begin{definition}
Let $\p,\q\in\prob(n)$ and $\p',\q'\in\prob(m)$. We say that the pair $(\p,\q)$ relatively trumps the pair $(\p',\q')$, and write
\be
(\p,\q)\succ_*(\p',\q')\;,
\ee
if there exists $k\in\mbb{N}$ and vectors $\r,\s\in\prob(k)$ with $\s>0$ such that
\be
(\p\otimes\r,\q\otimes\s)\succ(\p'\otimes\r,\q'\otimes\s)\;.
\ee
\end{definition}
\end{myd}

\begin{exercise}
Show that if we did not impose $\s>0$ (or alternatively that $\r>0$) in the definition above then there would always exists a catalyst. Hint: Take $\r$ and $\s$ to be orthogonal.
\end{exercise}

\begin{remark}
In the above definition, an alternative approach could have been to require that the vectors $\r$ and $\s$ are not orthogonal instead of enforcing $\s>0$. Nevertheless, opting for the stricter criterion of $\s>0$ brings two distinct benefits. Firstly, within the rational field, where vectors like $\q, \q', \r$ and others comprise rational components, Theorem~\ref{onlyr} implies that $(\r, \s)\sim(\t, \u)$, for some vector $\t$ of higher dimensionality. This equivalence effectively simplifies relative trumping to standard trumping in this scenario. Secondly, when applying the concept of relative trumping to thermodynamic contexts, the vector $\s$ typically represents a Gibbs state, which is inherently positive by nature. This correspondence ensures that the mathematical model is in harmony with the underlying physical principles of Gibbs states in thermodynamics.
\end{remark}

In the next chapter we will study functions that behaves monotonically under relative trumping.
A well known family of such functions are R\'enyi divergences.\index{R\'enyi divergence} R\'enyi divergences are defined for any $\alpha\in[0,\infty]$ and any $\p,\q\in\prob(n)$ as
\be
D_{\alpha}(\p\|\q)\eqdef\begin{cases}
\frac{1}{\alpha-1}\log\sum_{x\in[m]}p_x^\alpha q_x^{1-\alpha} &\text{if }\supp(\p)\subseteq\supp(\q)\;.\\
\infty &\text{otherwise}
\end{cases}
\ee
The cases $\alpha=0,1,\infty$ are defined by taking the appropriate limits
(more details will be given in the next chapter).
Both the trumping and relative trumping relations can be characterized with the above family of R\'enyi divergences. 
\begin{myt}{\color{yellow} Characterization of the Trumping Relation}\index{trumping relation}
\begin{theorem}\label{ktt}
Let $\p,\q\in\prob(n)$ with either $\p>0$ or $\q>0$ and $\p\neq\q$. Then,
\be
\p\succ_*\q
\ee
if and only if for all $\alpha\geq\frac12$
\be\label{stri}
D_\alpha(\p\|\u)>D_{\alpha}(\q\|\u)\quad\text{ and }\quad D_\alpha(\u\|\p)>D_{\alpha}(\u\|\q)\;.
\ee
\end{theorem}
\end{myt}
The proof of the theorem above is rather complicated and goes beyond the scope of this book. In the section `Notes and References' at the end of this chapter, we discuss its history and provide relevant references for further reading. In the corollary below we show that this theorem can be extended to relative tramping for the case that one of the vectors in each pair has positive rational components. We use the notation
\be
(\p,\q)\otimes (\p'\otimes\q')\eqdef(\p\otimes\p',\q\otimes\q')\quad\quad\forall\;\p,\q\in\prob(n)\;,\;\;\forall\;\p',\q'\in\prob(m)\;.
\ee

\begin{myg}{}
\begin{corollary}\label{coltrump}
Let $m,n\in\mbb{N}$, $\p,\q\in\prob(n)$, $\p',\q'\in\prob(m)$, and suppose that both $\q$ and $\q'$ have positive rational components.
Then, the following are equivalent:
\begin{enumerate}
\item There exists $k\in\mbb{N}$ and a vector $\s\in\prob(k)$ such that 
\be\label{trucon}
(\p,\q)\otimes\left(\s,\u^{(k)}\right)\succ\left(\p',\q'\right)\otimes\left(\s,\u^{(k)}\right)
\ee
\item For all $\alpha\geq\frac{1}{2}$
\be\label{dal}
D_{\alpha}(\p\|\q)> D_{\alpha}(\p'\|\q')\quad\text{and}\quad D_{\alpha}(\q\|\p)> D_{\alpha}(\q'\|\p')\;.
\ee
\end{enumerate}
\end{corollary}
\end{myg}
Note that the condition 1 above implies in particular that $(\p,\q)\succ_{*}(\p',\q')$. We leave the proof of the corollary as an exercise.

\begin{exercise}
Prove Corollary~\ref{coltrump} using the combination of Theorem~\ref{ktt} and Theorem~\ref{onlyr}.
\end{exercise}

\section{Catalytic Majorization}\index{catalytic majorization}

In this section we study a variant of the trumping relation that we call catalytic majorization.
We will see that catalytic majorization is robust under small perturbations and therefore is more useful in some applications,
particularly in thermodynamics.

\subsection{Robustness Under Small Perturbations}

Both majorization and relative majorization are robust under small perturbations. To be precise, let $\{\p_k\}_{k\in\mbb{N}}$ and $\{\q_k\}_{k\in\mbb{N}}$ be sequences in $\prob(n)$ with limits $\p$ and $\q$, respectively, and let $\{\p_k'\}_{k\in\mbb{N}}$ and $\{\q_k'\}_{k\in\mbb{N}}$ be sequences in $\prob(m)$ with limits $\p'$ and $\q'$. Suppose now that $(\p_k,\q_k)\succ(\p_k',\q_k')$ for all $k\in\mbb{N}$ and recall from Theorem~\ref{chararm} that this means that $\mt(\p_k,\q_k)\supseteq\mt(\p_k',\q_k')$. Since this inclusion of testing regions is robust under taking the limit, we conclude that $\mt(\p,\q)\supseteq\mt(\p',\q')$
so that necessarily $(\p,\q)\succ(\p',\q')$. 

The above argument cannot be applied to the trumping and relative trumping relations. To see why, consider two sequences $\{\p_k\}_{k\in\mbb{N}}\subseteq\prob(n)$ and $\{\q_k\}_{k\in\mbb{N}}\subseteq\prob(m)$ with limits $\p$ and $\q$, and suppose that  $\p_k\succ_{*}\q_k$ for all $k\in\mbb{N}$. This means that for each $k\in\mbb{N}$ there exists a catalyst vector $\r_k\in\prob(\ell_k)$ such that $\p_k\otimes\r_k\succ\q_k\otimes\r_k$, where $\ell_k\in\mbb{N}$ is the dimension of $\r_k$ that can depend on $k$.
Without invoking additional arguments, one cannot rule out the possibility that the dimension $\ell_k$ goes to infinity as $k$ goes to infinity.  This means that we cannot conclude that $\p\succ_{*}\q$. However, at the time of writing this book, it is left open to find an example with convergent sequences satisfying $\p_k\succ_{*}\q_k$ for all $k\in\mbb{N}$, whereas their limits satisfy $\p\not\succ_{*}\q$.

\begin{exercise}
Show that if there exists an example as above, then there is also a similar example for relative trumping. That is, there exists sequences $\{\p_k\}_{k\in\mbb{N}}$, $\{\q_k\}_{k\in\mbb{N}}$, $\{\p_k'\}_{k\in\mbb{N}}$, and $\{\q_k'\}_{k\in\mbb{N}}$, in $\prob(n)$, with limits $\p$, $\q$, $\p'$, and $\q'$, respectively, such that $(\p_k,\q_k)\succ_*(\p_k',\q_k')$ for all $k\in\mbb{N}$ and $(\p,\q)\not\succ_*(\p',\q')$.
\end{exercise}

\subsection{Robust Version of Relative Trumping}\index{trumping relation}

In the definition below we define a robust version of relative trumping. Similarly, one can define a robust version for trumping itself, however, we do not do it here since the robust version of relative trumping is the concept that we will use later on in applications to resource theories. 

\begin{myd}{Catalytic Majorization}\index{catalytic majorization}
\begin{definition}\label{cmajo}
Let $m,n\in\mbb{N}$ and $\p,\q\in\prob(n)$ and $\p',\q'\in\prob(m)$. We say that $(\p,\q)$ catalytically majorizes $(\p',\q')$ and write
\be
(\p,\q)\succ_c(\p',\q')\;,
\ee
if for any $\eps>0$ there exists four vector $\p_\eps\in\mb_{\eps}(\p)$, $\q_\eps\in\mb_{\eps}(\q)$, $\p'_\eps\in\mb_{\eps}(\p')$, and $\q_\eps'\in\mb_{\eps}(\q')$ such that
$(\p_\eps,\q_\eps)\succ_*(\p_\eps',\q_\eps')$.
\end{definition}
\end{myd}

\begin{exercise}
Show that $\succ_c$ is indeed a pre-order, and if $(\p,\q)\succ_*(\p',\q')$ then necessarily $(\p,\q)\succ_c(\p',\q')$.
\end{exercise}

The relation $\succ_c$ is robust under perturbation essentially by definition.

\begin{exercise}[Robustness of Catalytic Majorization]\label{rcm}
Show that if for any $\eps>0$ there exists four vector $\p_\eps\in\mb_{\eps}(\p)$, $\q_\eps\in\mb_{\eps}(\q)$, $\p'_\eps\in\mb_{\eps}(\p')$, and $\q_\eps'\in\mb_{\eps}(\q')$ such that
$(\p_\eps,\q_\eps)\succ_c(\p_\eps',\q_\eps')$ then $(\p,\q)\succ_c(\p',\q')$.
\end{exercise}

In the definition of catalytic majorization, we considered four balls of radius $\eps$ around $\p$, $\q$, $\p'$, and $\q'$. We now show that under certain support conditions, it is sufficient to consider only two such balls.
\begin{myg}{}
\begin{lemma}\label{oneside}
Let $m, n \in \mbb{N}$, $\p, \q \in \prob(n)$ and $\p', \q' \in \prob(m)$. Suppose further that both $\p'>0$ and $\q>0$. Then, the following are equivalent:
\begin{enumerate}
\item $(\p, \q) \succ_c (\p', \q')$.
\item For any $\eps>0$ there exist $\q_\eps\in\mb_{\eps}(\q)$ and  $\q_\eps'\in\mb_{\eps}(\q')$ such that
$(\p, \q_\eps) \succ_* (\p', \q'_\eps)$.
\end{enumerate}
\end{lemma}
\end{myg}

\begin{proof}
Clearly, if for all $\eps>0$ the two vectors in the second statement exist, then $(\p, \q) \succ_c (\p', \q')$ since we can define $\p_\eps\eqdef\p$ and $\p'_\eps\eqdef\p'$, so that the four vectors $\p_\eps$, $\q_\eps$, $\p_\eps'$, and $\q_\eps'$ satisfy the conditions in Definition~\ref{cmajo}. It thus remains to show the converse implication.

Suppose $(\p, \q) \succ_c (\p', \q')$ and let $\p_\eps$, $\q_\eps$, $\p'_\eps$, and $\q_\eps'$ be as in Definition~\ref{cmajo}. In particular, $(\p_\eps,\q_\eps)\succ_*(\p_\eps',\q_\eps')$ and we choose $\eps>0$ to be sufficiently small so that $\q_\eps>0$ (recall that $\q>0$ and $\q_\eps$ is $\eps$-close to $\q$) and $\p'_\eps>0$. From Lemma~\ref{3lem3} it follows that there exists $\r\in\mb^{\delta}(\q_\eps)$ with $\delta\eqdef\frac\eps{q_{\eps,\min}}$ ($q_{\eps,\min}$ being the smallest component of $\q_\eps$) such that $(\p,\r)\succ(\p_\eps,\q_\eps)$. Similarly, from the version of Lemma~\eqref{4lem40} that is given in Exercise~\ref{ex:4lem4} it follows that there exists a vector $\r'\in\mb^{\delta'}(\q_\eps')$ with $
\delta'\eqdef1-\min_{x\in[n]}\frac{p_x'}{(\p_\eps')_x}
$, such that $(\p'_\eps, \q'_\eps)\succ(\p', \r')$.
Observe that $\r$ and $\r'$ satisfy
\begin{align}\label{4succ}
(\p, \r)\succ(\p_\eps, \q_\eps)\succ_*(\p'_\eps, \q'_\eps)\succ(\p', \r')\;.
\end{align}
Hence, in particular $(\p, \r)\succ_*(\p', \r')$. Now, recall that $\r$ is $\delta$-close to $\q_{\eps}$ and therefore $(\delta+\eps)$-close to $\q$. Similarly, $\r'$ is $(\delta'+\eps)$-close to $\q'$.
The proof is therefore concluded by the observation that both $\delta$ and $\delta'$ go to zero in the limit $\eps\to 0$ so that $\r$ and $\r'$ can be made arbitrarily close to $\q$ and $\q'$, respectively.
\end{proof}

\begin{exercise}\label{suppoex}
Show that under the assumption that $\q>0$ and $\p'>0$ both $\delta$ and $\delta'$ in the lemma above goes to zero as $\eps$ goes to zero.
\end{exercise}

\subsection{Characterization of Catalytic Majorization}\index{catalytic majorization}

This subsection concludes with an insightful characterization of catalytic majorization. This characterization is articulated through R\'enyi divergences, a topic we will delve into in greater detail in subsequent chapters. We have previously encountered R\'enyi divergences in Theorem~\ref{ktt} and Corollary~\ref{coltrump}. Both Theorem~\ref{ktt} and Corollary~\ref{coltrump} incorporate strict inequalities among R\'enyi divergences. These strict inequalities highlight the fact that relative trumping may not be resilient to noise, though a counterexample has not yet been identified. The upcoming theorem expands upon Theorem~\ref{ktt}, replacing the concept of the trumping relation with the robust relative trumping relation, what we define as catalytic majorization.

\begin{myt}{\color{yellow} Characterization of Catalytic Majorization}
\begin{theorem}\label{thm:char2}
  	Let $m, n \in \mbb{N}$, $\p, \q \in \prob(n)$, and $\p', \q' \in \prob(m)$ such that either $\p>0$ or $\q>0$. Then the following statements are equivalent:
\begin{enumerate}
		\item $(\p, \q) \succ_c (\p', \q')$
		\item For all $\alpha\geq\frac12$ we have
		$D_{\alpha}(\p\|\q) \geq D_{\alpha}(\p'\|\q')$ and $D_{\alpha}(\q\|\p) \geq D_{\alpha}(\q'\|\p')$.
		\end{enumerate}
\end{theorem}
\end{myt}

\begin{proof}\index{catalytic majorization}\index{R\'enyi divergence}
The proof of the theorem for the special case that either $\p=\q$ or $\p'=\q'$ is very simple and is left as an exercise. Therefore, we assume now that both $\p\neq\q$ and $\p'\neq\q'$. Due to the symmetry in the roles of $\p$ and $\q$, and since one of them has full support, we assume without loss of generality (without loss of generality) that it is $\q$; that is, we assume $\q>0$. The proof of the monotonicity property of $D_\alpha$ under catalytic majorization will be detailed in Chapter~\ref{ch:relent}, where we extensively study the properties of the Rényi divergences. Consequently, this section will only cover the proof of the implication $2\Rightarrow1$.

From Exercise~\ref{1lem1} it follows that for any $\eps>0$ there exist $\q_\eps\in\mb_{\eps}(\q)$ and $\q_\eps'\in\mb_{\eps}(\q')$ with $0<\q_\eps\in \prob(n) \cap \mbb{Q}^n$ and $0<\q_\eps'\in \prob(m) \cap \mbb{Q}^m$ such that  
\begin{align}
(\p,\q_\eps)\succ(\p,\q)\quad\text{and}\quad (\p',\q')\succ(\p',\q_\eps')\,.
\end{align}
In Chapter~\ref{ch:relent} we will see that $D_\alpha$ behaves monotonically under both relative majorization and catalytic majorization\index{catalytic majorization}. Therefore, the relations above combined with our assumption that $(\p, \q) \succ_c (\p', \q')$, lead to the conclusion that 
\begin{align}
D_{\alpha}(\p\|\q_\eps)\geq D_{\alpha}(\p\|\q)\geq D_{\alpha}(\p'\|\q')\geq D_{\alpha}(\p'\|\q_\eps')\;,
\end{align}
and similarly
\begin{align}
D_{\alpha}(\q_\eps\|\p)\geq D_{\alpha}(\q\|\p)\geq D_{\alpha}(\q'\|\p')\geq D_{\alpha}(\q_\eps'\|\p')\;.
\end{align}
Now, since both $\q_\eps$ and $\q_\eps'$ have positive rational components, there exists two finite dimensional probability vectors $\r_\eps,\r_\eps'\in\prob(m_\eps)$, with $m_\eps\in\mbb{N}$,  such that (see Theorem~\ref{onlyr})
\begin{align}
\big(\r_\eps,\u^{(m_\eps)}\big)\sim(\p,\q_\eps)\quad\text{and}\quad\big(\r_\eps',\u^{(m_\eps)}\big)\sim(\p',\q_\eps')\;.\label{conpq}
\end{align}
Hence, for all $\alpha\geq \frac12$
\begin{align}\label{580}
D_{\alpha}\big(\r_\eps\big\|\u^{(m_\eps)}\big) \geq D_{\alpha}\big(\r_\eps' \big\|\u^{(m_\eps)}\big)\quad\text{and}\quad D_{\alpha}\big(\u^{(m_\eps)}\big\|\r_\eps\big) \geq D_{\alpha}\big(\u^{(m_\eps)} \big\|\r_\eps'\big)\;.
\end{align}
Our strategy is to use Theorem~\ref{ktt} in conjunction with the inequalities above in order to obtain a majorization relation between $\r_\eps'$ and $\r_\eps$. However, since the inequalities above are not strict, we cannot use Theorem~\ref{ktt} and will need to tweak a bit the vector $\r_\eps'$.

We first rule out the possibility that $\r_\eps' = \u^{(m_\eps)}$. Consulting the construction in Theorem~\ref{onlyr}, we see that $\r_\eps' = \u^{(m_\eps)}$ implies $\p' = \q_\eps'$. However, this cannot occur for sufficiently small $\eps>0$ since $\q_\eps' \xrightarrow{\eps\to 0^+} \q' \neq \p'$ by our assumption.
Hence, we can assume $\r_\eps' \neq \u^{(m_\eps)}$ for sufficiently small $\eps>0$. Moreover, observe that for any $\eps\in(0,1)$, we have (see Exercise~\ref{mixdec}) 
\begin{align}\label{rpk}
\r_\eps'\succ \s_\eps\eqdef\left(1-\eps\right)\r_\eps'+\eps\u^{(m_\eps)}\;.
\end{align}
A combination of the relation $\r_\eps'\succ \s_\eps$ (note that $\s_\eps>0$) with Theorem~\ref{ktt} and Eq.~\eqref{580},  gives the following strict inequalities for all $\alpha\geq \frac12$
\begin{align}
D_{\alpha}\big(\r_\eps\big\|\u^{(m_\eps)}\big) > D_{\alpha}\big(\s_\eps \big\|\u^{(m_\eps)}\big)\quad\text{and}\quad
 D_{\alpha}\big(\u^{(m_\eps)}\big\|\r_\eps\big)> D_{\alpha}\big(\u^{(m_\eps)} \big\|\s_\eps\big)\;.
\end{align}
Since the condition above is equivalent to the condition given in Theorem~\ref{ktt} it follows that $\r_\eps\succ_* \s_\eps$. 
Hence, 
\begin{align}\label{99}
(\p,\q_\eps)\sim(\r_\eps,\u^{(m_\eps)})\succ_* (\s_\eps,\u^{(m_\eps)})\sim(\p'_\eps,\q_\eps')\;,
\end{align}
where 
\begin{align}
\p_\eps'\eqdef\left(1-\eps\right)\p'+\eps\q_\eps'\;.
\end{align}
The equivalence $(\s_\eps,\u^{(m_\eps)})\sim(\p'_\eps,\q_\eps')$ can be verified from the construction in Theorem~\ref{onlyr}.
Since catalytic majorization\index{catalytic majorization} is robust to small perturbations (see Exercise~\ref{rcm}), the relation $(\p,\q_\eps)\succ_c(\p'_\eps,\q_\eps')$ that follows from~\eqref{99} (and holds for any sufficiently small $\eps>0$) gives $(\p,\q)\succ_c(\p',\q')$. This concludes the proof.
\end{proof}

\begin{exercise}
Prove the theorem above for the special case that $\p=\q$. Similarly, prove it also for the case that $\p'=\q'$.
\end{exercise}

\begin{exercise}
Prove the equivalence $(\s_\eps,\u^{(m_\eps)})\sim(\p'_\eps,\q_\eps')$. Hint: Use the construction in Theorem~\ref{onlyr}.
\end{exercise}

\section{Conditional Majorization}\label{sec:cm}\index{conditional majorization}

Conditional majorization is a concept that helps us understand the uncertainty inherent in a physical system when we have access to another correlated system. This idea is formalized as a pre-order relationship within the set of probability distributions denoted as $\prob(mn)$, where $m$ represents the dimension of a classical system denoted as $X$ (held by Alice), and $n$ is the dimension of the ``conditioning" system referred to as $Y$ (held by Bob). In essence, this pre-order quantifies the uncertainty associated with system $X$ when we possess information about system $Y$. As we delve further into this topic in the book, we'll find that conditional majorization serves as the foundation for defining conditional entropy.

We will use the notation $\p^{XY}$ to denote a probability vector in $\prob(mn)$, which is $mn$-dimensional. It's important to note that we view $\p^{XY}$ as a probability distribution associated with the \emph{joint} system $XY$. To clarify this perspective, we denote the components of $\p^{XY}$ as $\{p_{xy}\}$ and express $\p^{XY}$ as follows:
\be\label{pcapxy}
\p^{XY}=\sum_{x\in[m]}\sum_{y\in[n]}p_{xy}\e_x^X\otimes\e_y^Y\;,
\ee
where $\{\e_x^X\}_{x\in[m]}$ is the standard basis of $\mbb{R}^m$, and  $\{\e_y^Y\}_{y\in[n]}$ is the standard basis of $\mbb{R}^n$. While mathematically, $\p^{XY}$ is a vector in $\prob(mn)$, conceptually, we treat it as a joint probability distribution.

To introduce the concept of conditional majorization, we build upon the foundation of conditionally mixing\index{conditionally mixing}  operations. Our objective is to characterize a set of evolution matrices that possess a specific property: they increase the conditional uncertainty associated with system $X$ when provided access to system $Y$.
To embark on this journey, consider three classical systems: $X$, $Y$, and $Y'$, each with dimensions $m$, $n$, and $n'$, respectively. Additionally, consider two probability vectors  $\p^{XY}\in\prob(mn)$ and $\q^{XY'}\in\prob(mn')$. We say that $\p^{XY}$ conditionally majorizes $\q^{XY'}$ and denote it as $\p^{XY}\succ_X\q^{XY'}$ when there exists a conditionally mixing\index{conditionally mixing} operation (to be define shortly), denoted as $M\in\stoc(mn',mn)$, such that:
\be
\q^{XY'}=M\p^{XY}\;.
\ee
The challenge lies in crafting a meaningful definition for $M$ that aligns with the concept of ``conditionally mixing." In this context, we demonstrate that there exist three distinct approaches to defining $M$, mirroring the three methodologies introduced in Section~\ref{3approaches}. Remarkably, all of these approaches converge to the same definition of conditionally mixing\index{conditionally mixing} and conditional majorization, thereby establishing a solid foundation for the notion of conditional majorization. 

This section is structured as follows: Initially, we introduce both the axiomatic and constructive approaches\index{constructive approach}, showing that they both lead to the identical definition of a conditionally mixing operation. We then use this definition to establish conditional majorization and to examine some of its key properties. Following that, we explore a useful characterization of conditional majorization, paying special attention to cases in smaller dimensions. Lastly, in the final subsection on this topic, we investigate the operational approach\index{operational approach}  to conditional majorization and demonstrate its consistency with the axiomatic and constructive approaches.

\subsection{The axiomatic approach\index{axiomatic approach}}\label{sec:axiomatica}

In this approach, we build our foundation upon two minimalistic axioms, which we regard as fundamental for any reasonable definition of a conditionally mixing\index{conditionally mixing} operation. Much like in the previous context of mixing operations, our framework operates under the assumption that the conditionally mixing\index{conditionally mixing} operation can be represented by a stochastic matrix denoted as $M$, which belongs to the set $\stoc(mn',mn)$. This matrix $M$ serves as a representation of a classical channel\index{classical channel} that transforms the joint system $XY$ into $XY'$. In a conceptual sense, we can envision Alice as the possessor of system $X$, while Bob holds systems $Y$ and $Y'$.

\subsubsection{Axiom\index{axiom} 1: No-Signaling from Alice to Bob.}\index{no-signaling}
A key condition that the conditionally mixing\index{conditionally mixing} operation $M\in\stoc(mn, mn')$ must adhere to in order to avoid decreasing conditional uncertainty relates to the prevention of information leakage from subsystem $X$ to subsystem $Y$. Such leakage could potentially reduce uncertainty about system $X$. Conditional uncertainty specifically pertains to the notion of uncertainty about system $X$ when one has access to system $Y$. To address this, we introduce a minimalistic causality assumption that accounts for the property that system $X$ has no causal effect on system $Y'$. In mathematical terms, this assumption implies that the components of the stochastic matrix $M=(\mu_{x'y'|xy})$ satisfy the following equation for all $x\in[m]$, $y\in[n]$, and $y'\in[n']$:
\be\label{0w2}
\sum_{x'\in[m]}\mu_{x'y'|xy}=r_{y'|y}\;,
\ee
where $\{r_{y'|y}\}$ (with $y\in[n]$, and $y'\in[n']$) is some conditional probability distribution independent on $x$. We refer to this condition as non-signalling from $X$ to $Y'$ or in short $X\not\to Y'$-signalling (see~\eqref{nonsig} for a similar definition). 

Matrices that satisfies the above non-signalling condition has a relatively simple form. Specifically, for every $y\in[n]$ and $y'\in[n']$ let $T^{(y,y')}\in\mbb{R}^{m\times m}_+$ be the matrix whose components are
\be
t^{(y,y')}_{x'|x}\eqdef\frac{\mu_{x'y'|xy}}{r_{y'|y}}\quad\quad\forall x,x'\in[m]\;.
\ee 
From~\eqref{0w2} it then follows that $T^{(y,y')}$  is column stochastic; i.e., for every $y\in[n]$ and $y'\in[n']$ we have $T^{(y,y')}\in\stoc(m,m)$. With these notations we can express $M$ as (summations runs over all $x,x'\in[m]$ and all $y\in[n]$ and $y'\in[n']$)
\ba
M&=\sum_{x,x',y,y'}\mu_{x'y'|xy}|x'\lr x|\otimes|y'\lr y|\\
&= \sum_{y,y'}r_{y'|y}\sum_{x,x'}t^{(y,y')}_{x'|x}|x'\lr x|\otimes |y'\lr y|\;,
\ea
where we employed quantum notations by denoting $|x'\lr x|$ (and similarly $|y'\lr y|$) as the $m\times m$ rank-one matrix $\e_{x'}^T\e_x$, which is a matrix with a one at the $(x',x)$-position and zeros elsewhere.  We therefore conclude that $M\in\stoc(mn, mn')$ is $X\not\to Y'$-signalling if and only if there exists $nn'$ stochastic matrices $T^{(y,y')}\in\stoc(m,m)$ (with $y\in[n]$ and $y'\in[n']$), and another stochastic matrix $R=(r_{y'|y})\in\stoc(n',n)$ such that
\be\label{onewy}
M=\sum_{y,y'}r_{y'|y}T^{(y,y')}\otimes |y'\lr y|\;.
\ee
In the following exercise you show that the above form of $M$ represents a bipartite channel that can be realized with one-way communication from Bob to Alice.

\begin{figure}[h]
\centering
    \includegraphics[width=0.4\textwidth]{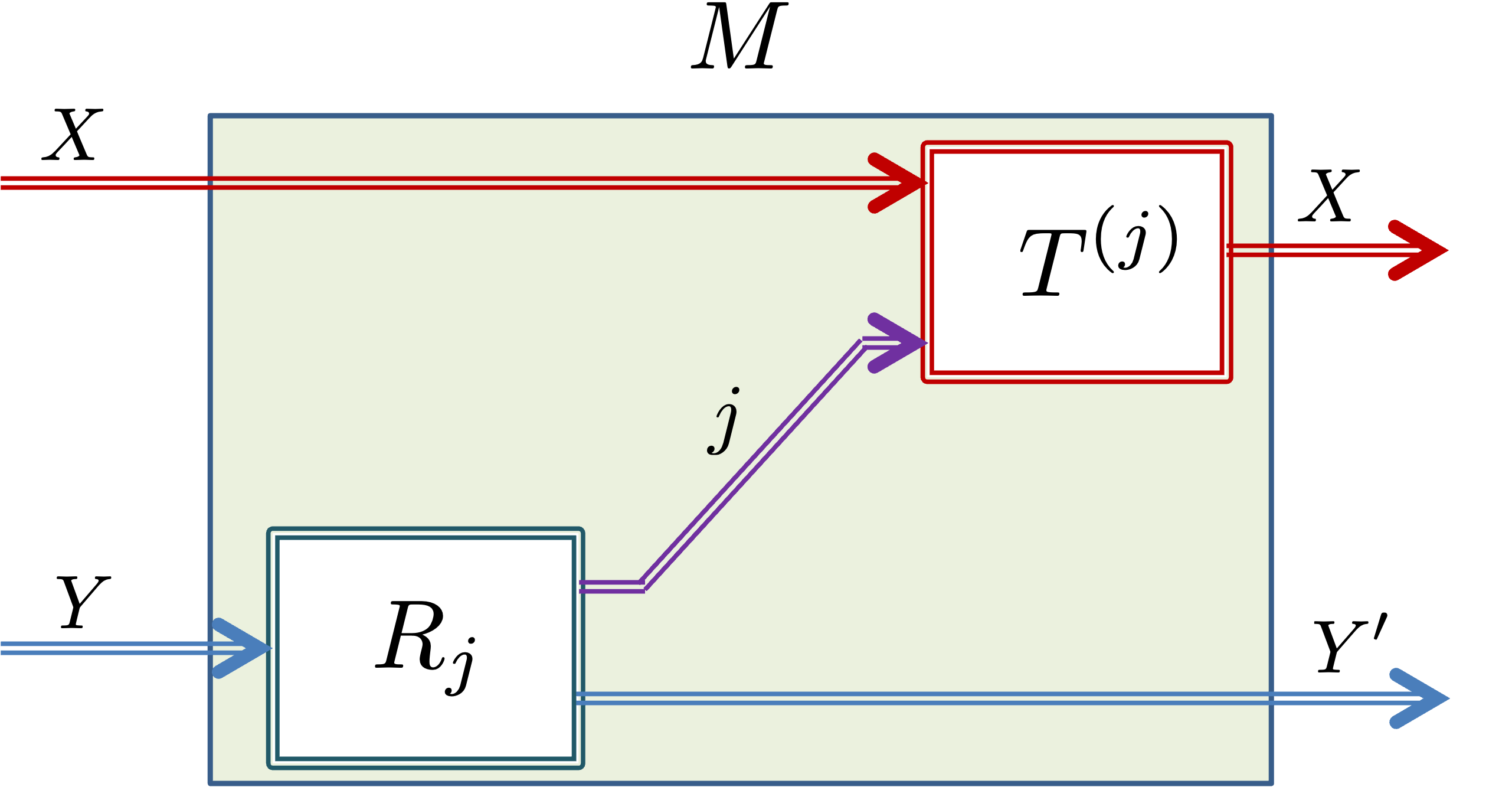}
  \caption{\linespread{1}\selectfont{\small An $X\not\to Y'$-signalling bipartite channel $M\in\stoc(mn',mn)$.}}
  \label{semi-causal0}
\end{figure}

\bex
Let $M\in\stoc(mn',mn)$. Show that the following two statements are equivalent:
\ben
\item $M$ is $X\not\to Y'$-signalling.
\item $M$ can be realized with one-way communication from Bob to Alice. That is, $M$ can be expressed as (refer to Fig.~\ref{semi-causal0}):
\be\label{onewayform}
M=\sum_{j\in[k]}T^{(j)}\otimes R_j
\ee
where $k\in\mbb{N}$, and for each $j\in[k]$, $T^{(j)}\in\stoc(m,m)$, $R_j\in\mbb{R}_+^{n'\times n}$, and $R\eqdef\sum_{j\in[k]}R_j\in\stoc(n',n)$.
\een 
\eex

The term $X\not\to Y'$-signalling is sometimes referred to as $X\not\to Y'$ semi-causal. The following exercise introduces the notion of $X\not\to Y'$ semi-causal.

\begin{figure}[h]
\centering
    \includegraphics[width=0.6\textwidth]{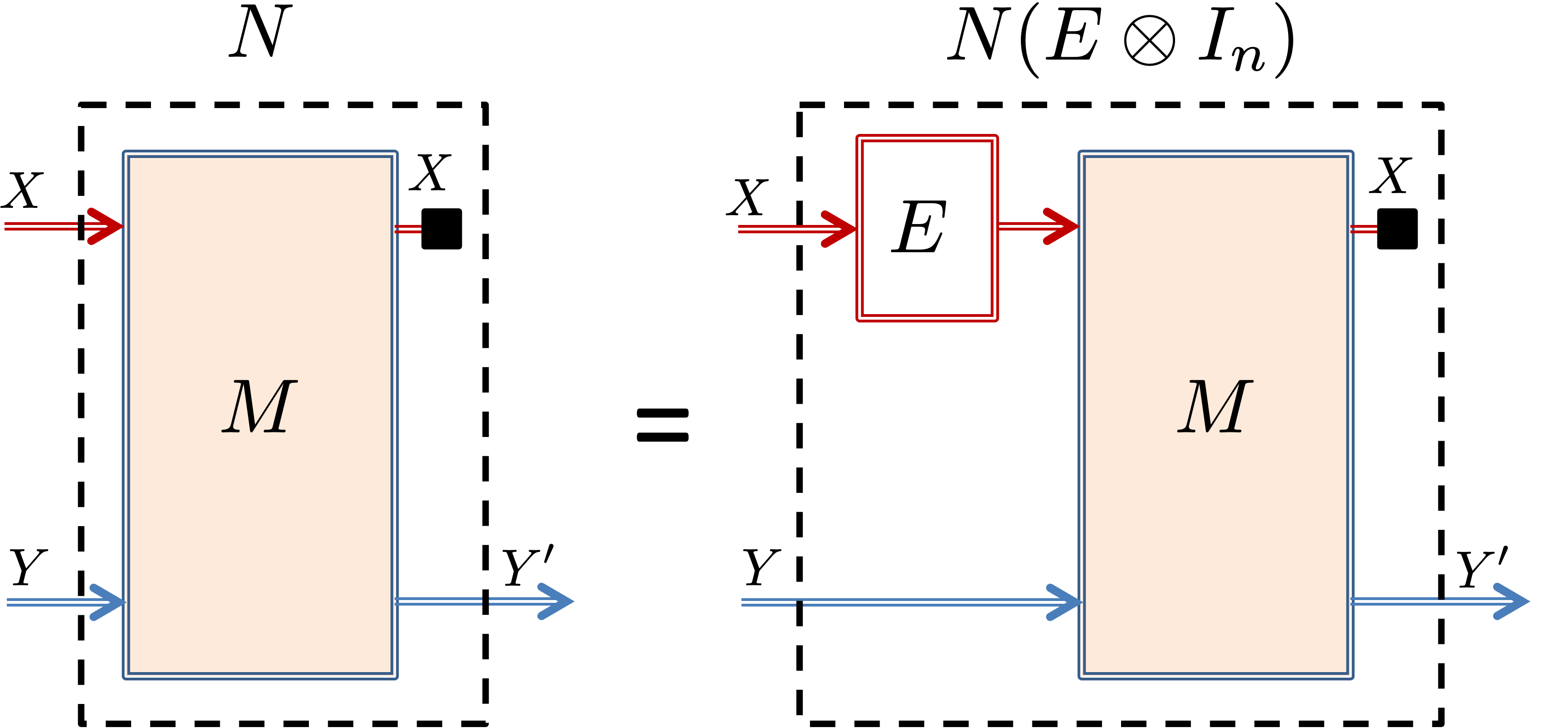}
  \caption{\linespread{1}\selectfont{\small An illustration of a semi-causal classical bipartite channel $M$. The marginal channel $N$ equals $N(E\otimes I_n)$ for any choice of $E\in\stoc(m,m)$.}}
  \label{semicfc}
\end{figure}

\bex
Let $M=(\mu_{x'y'|xy})\in\stoc(mn', mn)$, and $N=(\nu_{y'|xy})\in\stoc(n',mn)$, where $\nu_{y'|xy}\eqdef\sum_{x'\in[m]}\mu_{x'y'|xy}$.
\ben
\item Show that if $M$ satisfies~\eqref{0w2} then for every evolution matrix $E\in\stoc(m,m)$ the marginal channel $N$ satisfies
\be\label{semic2}
N(E\otimes I_n)=N\;.
\ee 
This condition ensure that any operation $E$ that Alice (system $X$) may chose to apply to her system cannot be detected by Bob (system $Y$). Such a condition is also called $X\not\to Y'$ semi-causal. See Fig.~\ref{semicfc} for an illustration of a semi-causal channel.
\item Show that if $N$ satisfies~\eqref{semic2} for all $E\in\stoc(m,m)$ then $M$ satisfies~\eqref{0w2}.
\een
\eex 

\subsubsection{Axiom\index{axiom} 2: Preservation of Maximal Uncertainty}

Consider a joint probability vector in the form of $\u^X\otimes\p^Y$, where $\u^X\in\prob(m)$ represents the uniform probability vector, and $\p^Y$ is some probability vector in $\prob(n)$. Since this joint probability distribution is uncorrelated, having access to $Y$ cannot aid in reducing the uncertainty associated with $X$. Therefore, we can assume that such a joint probability vector exhibits the highest degree of conditional uncertainty concerning $X|Y$ (i.e., $X$ given access to $Y$).
Therefore, any stochastic (evolution) matrix $M\in\stoc(mn, mn')$ that does not decrease conditional uncertainty must map states with maximal conditional uncertainty to states that also possess maximal conditional uncertainty. In explicit terms, such a stochastic map $M\in\stoc(mn, mn')$ must satisfy, for all $\p^Y\in\prob(n)$:
\be\label{0w0}
M\left(\u^X\otimes\p^Y\right)=\u^{X}\otimes\q^{Y'}
\ee
where $\q^{Y'}$ is a probability vector in $\prob(n')$. It is important to note that this assumption is exceedingly minimalistic, as it merely posits that if $X$ is initially maximally uncertain, then applying a conditionally mixing\index{conditionally mixing} operation should not diminish this maximal uncertainty.

The vector $\q^{Y'}$ in the equation above is not independent of $M$ and $\p^Y$. In fact, by left-multiplying both sides of the equation with $\1_m^T\otimes I^{Y'}$, we can express the vector $\q^{Y'}$ as follows:
\be\label{formqyp}
\q^{Y'}=\left(\1_m^T\otimes I^{Y'}\right)M\left(\u^X\otimes\p^Y\right)\;.
\ee
Note that the multiplication by row vector $\1_m^T$ in the expression above effectively functions as the ``tracing out" of system $X$.

\subsubsection{Definition of Conditionally Mixing Operations (CMO)}\index{conditionally mixing}

By combining the two axioms presented above, we arrive at the following definition for conditionally mixing operations. In this definition, we consider three classical systems: $X$, $Y$, and $Y'$, with dimensions $m$, $n$, and $n'$, respectively.
 
\begin{myd}{}
\begin{definition}\label{def:ccmo}
A stochastic matrix $M\in\stoc(mn',mn)$ is called a conditionally mixing operation if it satisfies the two conditions given in~\eqref{0w0} and~\eqref{0w2}. The set of all conditionally mixing\index{conditionally mixing} operations in $\stoc(mn',mn)$ is denoted by $\cmo(mn',mn)$.
\end{definition}
\end{myd}

Before we delve into characterizing the maps within $\cmo(mn',mn)$, let's explore an alternative approach, which we refer to as the constructive approach. We will demonstrate that this approach ultimately yields the same set of conditionally mixing operations.

\subsection{The Constructive Approach}\index{constructive approach}

\begin{figure}[h]
\centering
    \includegraphics[width=0.4\textwidth]{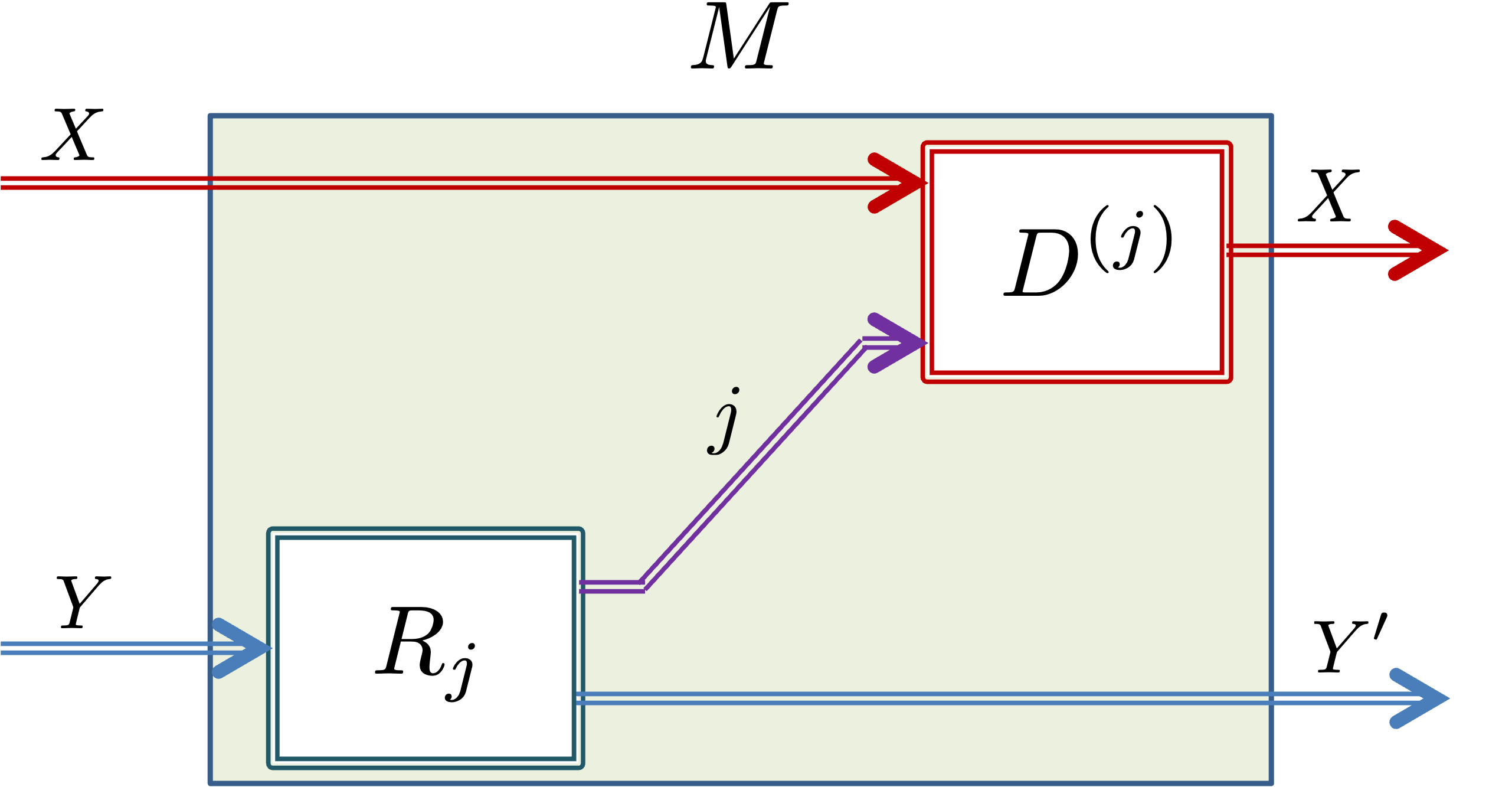}
  \caption{\linespread{1}\selectfont{\small A conditionally doubly stochastic (CDS) channel $M\in\stoc(mn',mn)$.}}
  \label{cds}
\end{figure}

In the constructive approach we propose to construct conditionally mixing operation as it is intuitively suggests.
More precisely, a conditionally mixing\index{conditionally mixing} operation $M$ is a stochastic map obtained by Alice applying a mixing operation to her system (i.e., doubly stochastic map) conditioned on information received from Bob (see Fig.~\ref{cds}). Mathematically, 
\be\label{onewayform2}
M=\sum_{j\in[k]}D^{(j)}\otimes R_j
\ee
where $j\in[k]$ is the information Bob sends to Alice after he processes his input $y$ via $R_j=(r_{y'j|y})$. Upon receiving $j$ Alice applies a mixing operation to her input $x$ described by the $m\times m$ doubly-stochastic matrix $D^{(j)}$. 
\begin{myd}{Conditionally Doubly Stochastic (CDS)}\index{conditionally doubly stochastic}
\begin{definition}\label{def:cds}
A stochastic matrix $M\in\stoc(mn',mn)$ is said to be conditionally doubly stochastic (CDS) if it has the form given in~\eqref{onewayform2} with $k\in\mbb{N}$, and for each $j\in[k]$, $D^{(j)}$ is an $m\times m$ doubly-stochastic matrix, and $R_j\in\mbb{R}_+^{n'\times n}$ is a sub-stochastic matrix such that $\sum_{j\in[k]}R_j\in\stoc(n',n)$. The set of all conditionally doubly stochastic matrices in $\stoc(mn',mn)$ is denoted by $\cds(mn',mn)$.
\end{definition}
\end{myd}

It is important to note that expression in~\eqref{onewayform2} is very similar to the one given in~\eqref{onewayform} except that for any fixed $j\in[k]$ the stochastic matrix $T^{(j)}$ is replaced with the doubly stochastic matrix\index{doubly stochastic matrix} $D^{(j)}$. 
Therefore, CDS channels are necessarily $X\not\to Y'$ semi-causal. Moreover, since each $D^{(j)}$ is doubly stochastic we get that for any $\p^Y\in\prob(n)$ 
\ba
M\left(\u^X\otimes\p^Y\right)&=\sum_{j\in[k]}D^{(j)}\u^X\otimes R_j\p^Y\\
\Gg{D^{(j)}\u^X=\u^X}&=\u^X\otimes R\p^Y\;, 
\ea
where $R=\sum_{j\in[k]}R_j$. That is, $M$ satisfies  the condition given in~\eqref{0w0}. We therefore conclude that every CDS channel $M$ is necessarily a CMO.
In the next theorem we prove that the converse also holds. Therefore, we get that both the axiomatic and the constructive approaches\index{constructive approach} leading to the same set of conditionally mixing operations.

\bex
Let $M$ be as in~\eqref{onewayform2}. Show that without loss of generality we can assume that the matrices $D^{(j)}$ are permutation matrices. Hint: Recall that every doubly-stochastic matrix is a convex combination of permutation matrices.
\eex

\bex
Show that for any doubly stochastic matrix $D\in\stoc(m,m)$ and any stochastic matrix $R\in\stoc(n',n)$ we have that $R\otimes D$ is CDS.
\eex

\begin{myt}{}
\begin{theorem}\label{thmcds}
Let $m,n,n'\in\mbb{N}$. Then,
\be
\cds(mn',mn)=\cmo(mn',mn)\;.
\ee
\end{theorem}
\end{myt}
\begin{proof}\index{conditionally mixing}\index{conditionally doubly stochastic}
We already proved the inclusion $\cds(mn',mn)\subseteq\cmo(mn',mn)$ (see the discussion below Definition~\ref{def:cds}). To prove the opposite inclusion, suppose $M\in\cmo(mn',mn)$. We want to show that $M\in\cds(mn',mn)$. For this purpose we will use the form given in~\eqref{onewy} for $X\not\to Y'$ semi-causal, and show that each $T^{(y,y')}$ is an $m\times m$ doubly stochastic matrix. Observe that if $r_{y'|y}=0$ for some $y'\in[n']$ and $y\in[n]$, then replacing $T^{(y,y')}$ with the identity matrix will not affect $M$, since $r_{y'|y}=0$. Consequently, it suffices to demonstrate that $T^{(y,y')}$ is doubly stochastic for those indices $y$ and $y'$ where $r_{y'|y}\neq 0$.

Let $\{\e_y^Y\}_{y\in[n]}$ be the standard basis of $\mbb{R}^n$ and consider the condition given in~\eqref{0w0}. Fix $y\in[n]$ and observe that from~\eqref{onewy} we get
\be\label{ontoh}
M\left(\u^X\otimes \e_y^{Y}\right)=\sum_{y'\in[n']}r_{y'|y}T^{(y,y')}\u^X\otimes \e_{y'}^{Y'}\;.
\ee
On the other hand, by taking $\p^Y=\e_y^Y$ in~\eqref{0w0} we get that
\ba\label{ontoh2}
M\left(\u^X\otimes \e_y^{Y}\right)&=\u^X\otimes \q_{y}^{Y'}\\
&=\sum_{y'\in[n']}q_{y'|y}\u^X\otimes \e_{y'}^{Y'}\;,
\ea
for some vector $\q_y^{Y'}\eqdef\sum_{y'\in[n']}q_{y'|y} \e_{y'}^{Y'}$ in $\prob(n')$. Since $\{\e_y^{Y'}\}_{y'\in[n']}$ is an orthonarmal basis of $\mbb{R}^{n'}$, a comparison of~\eqref{ontoh} and~\eqref{ontoh2} reveals that for all $y\in[n]$ and $y'\in[n']$ that
\be
r_{y'|y}T^{(y,y')}\u^X=q_{y'|y}\u^X\;.
\ee
Observe that since $T^{(y,y')}$ is column stochastic, $T^{(y,y')}\u^X$ is a probability vector so its dot product with the vector $\1_m$ equals one. Thus,
by taking the dot product on both sides of the equation above with the vector $\1_m$ we get $r_{y'|y}=q_{y'|y}$ for all $y\in[n]$ and $y'\in[n']$. We therefore conclude that $T^{(y,y')}\u^X=\u^X$ (recall we assume $r_{y'|y}\neq0$). Hence, $T^{(y,y')}$ is doubly stochastic.
This completes the proof.
\end{proof}

In the proof above we showed that the matrices $T^{(y,y')}$ as appear in~\eqref{onewy} are doubly stochastic. We therefore conclude that $M\in\stoc(mn',mn)$ is a conditionally mixing operation  if and only if there exists a column stochastic matrix $R\in\stoc(n',n)$ and $nn'$ doubly stochastic matrices $D^{(y,y')}\in\stoc(m,m)$, with $y\in[n]$ and $y'\in[n']$, such that
\be\label{maincds}
M=\sum_{y\in[n]}\sum_{y'\in[n']}r_{y'|y}D^{(y,y')}\otimes |y'\lr y|\;.
\ee
With the above conclusion we arrive at the definition of conditional majorization.

\begin{myd}{Conditional Majorization}\index{conditional majorization}
\begin{definition}\label{def:cmc}
Let $X$, $Y$, and $Y'$, be three classical systems of dimensions $m$, $n$, and $n'$, respectively. Further, let $\p^{XY}\in\prob(mn)$ and $\q^{XY'}\in\prob(mn')$. We say that $\p^{XY}$ conditionally majorizes $\q^{XY'}$ with respect to $X$, and write
\be
\p^{XY}\succ_X\q^{XY'}\;,
\ee
if there exists $M\in\cmo(mn',mn)$ such that
$
\q^{XY'}=M\p^{XY}
$.
We further write $\q^{XY'}\sim_X\p^{XY}$ if both $\p^{XY}\succ_X\q^{XY'}$ and $\q^{XY'}\succ_X\p^{XY}$.
\end{definition}
\end{myd}

\subsection{Basic Properties}

In this subsection, we study a few basic properties of conditional majorization. For this purpose, we will express $\p^{XY}$ that appear in~\eqref{pcapxy} in a more concise form as:
\be\label{oi1}
\p^{XY}=\sum_{y\in[n]}\p_{y}^X\otimes\e_y^Y,\quad\text{where}\quad\p_{y}^X\eqdef\sum_{x\in[m]}p_{xy}\e_x^X\;.
\ee
Likewise, if we denote the components of $\q^{XY'}$ as $\{q_{xy'}\}$, we can represent $\q^{XY'}$ as:
\be\label{oi2}
\q^{XY'}=\sum_{y'\in[n']}\q_{y'}^X\otimes\e_{y'}^{Y'},\quad\text{where}\quad\q_{y'}^X\eqdef\sum_{x\in[m]}q_{xy'}\e_x^X\;.
\ee
Using these notations, the pre-order $\p^{XY}\succ_X\q^{XY'}$ can be expressed as a relationship between the two sets of vectors $\{\p_y^X\}_{y\in[n]}$ and $\{\q_{y'}^X\}_{y'\in[n']}$. Observe further that all these vectors has non-negative components and their sums are given by the marginal probability vectors:
\be
\p^X\eqdef\sum_{y\in[n]}\p_y^X\in\prob(m)\quad\text{and}\quad\q^X\eqdef\sum_{y'\in[n']}\q_{y'}^X\in\prob(m)\;.
\ee

By definition, if $\p^{XY}\succ_X\q^{XY'}$ then there exists a matrix $M$ of the form~\eqref{maincds} such that $\q^{XY'}=M\p^{XY}$. Using the notations above this relation can be expressed as (see Exercise~\ref{ex:formofm})
\be\label{mte1}
\q_{y'}=\sum_{y\in[n]}r_{y'|y}D^{(y,y')}\p_y\quad\quad\forall\;y'\in[n']\;,
\ee
where $R=(r_{y'|y})\in\stoc(n',n)$ and each $D^{(y,y')}$ is an $m\times m$ doubly stochastic matrix.

\bex\label{ex:formofm}
Prove the relation~\eqref{mte1} using the above form of $\p^{XY}$ and $\q^{XY'}$, and the form~\eqref{maincds} of $M$.
\eex

The relation~\eqref{mte1} implies that if $\tp^{XY}=\sum_{y\in[n]}\tp^X_y\otimes\e_y^{Y}$ is a probability vector in $\prob(mn)$ obtained from $\p^{XY}$ by permutation of the components of the vectors $\{\p^{X}_y\}_{y\in[n]}$, i.e., for each $y\in[n]$, $\tp^X_y=\Pi^{(y)}\p_y^X$ for some $m\times m$ permutation matrix $\Pi^{(y)}$, then 
\be\label{relsimx}
\tp^{XY}\sim_X\p^{XY}\;.
\ee
\bex
Prove the relation~\eqref{relsimx}. Hint: take in~\eqref{mte1}, $Y'=Y$, and for each $y',y\in[n]$ take $r_{y'y}=\delta_{y'y}$ and $D^{(y',y')}=\Pi^{(y')}$.
\eex

The relation~\eqref{relsimx} implies that without loss of generality we can assume that the components of the vectors $\{\p^X_y\}_{y\in[n]}$ and $\{\q^X_{y'}\}_{y'\in[n']}$ are arranged in non-increasing order. We will therefore assume this order in the rest of this section. 

\begin{myt}{}
\begin{theorem}\label{lemcarc}
Let $\p^{XY}\in\prob(mn)$ and $\q^{XY'}\in\prob(mn')$ with $m\eqdef|X|$, $n\eqdef|Y|$, and $n'\eqdef|Y'|$. Then, $\p^{XY}\succ_X\q^{XY'}$ if and only if there exists $R=(r_{y'|y})\in\stoc(n',n)$ such that
\be\label{lqlpt}
\sum_{y\in[n]}r_{y'|y}\p_y^X\succ\q_{y'}^X\quad\quad\quad\forall\;y'\in[n']\;,
\ee
where $\{\p^X_y\}_{y\in[n]}$ and $\{\q^X_{y'}\}_{y'\in[n']}$ are defined in~\eqref{oi1} and~\eqref{oi2}, respectively. 
\end{theorem}
\end{myt}

\noindent{\it Remarks:}
\ben
\item We  assume in the theorem above that for each $y\in[n]$ we have $\left(\p_y^X\right)^\da=\p_y^X$. 
\item The vectors $\p_{y}^X$ and $\q_{y'}^X$ are not necessarily probability vector since the sum of their components $p_{y}\eqdef\1_m\cdot\p_y^X$ and $q_{y'}\eqdef\1_m\cdot\q_{y'}^X$ are in general smaller than one. Therefore, the majorization relation in~\eqref{lqlpt} implies in particular that
\be\label{sumall}
\sum_{y\in[n]}r_{y'|y}p_y=q_{y'}\quad\quad\quad\forall\;y'\in[n']\;.
\ee
We obtained this equality by summing the components on both sides of~\eqref{lqlpt}.
\item Observe that the relation~\eqref{lqlpt} can be expressed also as
\be\label{also}
\sum_{y\in[n]}r_{y'|y}L\p_y^X\geq L\q_{y'}^X\quad\quad\quad\forall\;y'\in[n']\;,
\ee 
where the inequality is entry-wise, and $L$ is the $m\times m$ matrix defined in Exercise~\ref{ex:mml}. We will later demonstrate that the inequality~\eqref{also} is instrumental in characterizing conditional majorization as a semidefinite program.
\een

\begin{proof}
Suppose $\p^{XY}\succ_X\q^{XY'}$ so that the relation~\eqref{mte1} holds. Since each $D^{(y,y')}$ is doubly stochastic we have $\p_y^X\succ D^{(y,y')}\p_y^X$. Multiplying both sides of this relation by $r_{y'|y}$ and summing over $y\in[n]$ gives (see Exercise~\ref{convexmajo}) 
\ba
\sum_{y\in[n]}r_{y'|y}\p_y&\succ\sum_{y\in[n]}r_{y'|y}D^{(y,y')}\p_y\\
\GG{\eqref{mte1}}&=\q_{y'}^X\;.
\ea

Conversely, suppose~\eqref{lqlpt} holds. Then, from Theorem~\ref{chmaj} for every $y'\in[n']$ there exists a doubly stochastic matrix\index{doubly stochastic matrix} $D^{(y')}\in\stoc(m,m)$ such that
\be
\q_{y'}^X=D^{(y')}\sum_{y\in[n]}r_{y'|y}\p_y^X=\sum_{y\in[n]}r_{y'|y}D^{(y')}\p_y^X\;.
\ee
Defining $D^{(y,y')}\eqdef D^{(y')}$ we conclude that the above relation is a special case of the relation~\eqref{mte1}. Hence, $\p^{XY}\succ\q^{XY'}$.
This completes the proof.
\end{proof}

To get a better intuition about conditional majorization, we first consider the cases in which one of the systems $X$, $Y$, and $Y'$, is trivial: 
\ben
\item The Case $|X|=1$. This is a trivial case in which there is no uncertainty about system $X$. We therefore expect the pre-order to be trivial as well. Indeed,
in this case $\p^{XY}=\p^Y\in\prob(n)$ and $\q^{XY'}=\q^{Y'}\in\prob(n')$, so the relation~\eqref{lqlpt} becomes $\q^{Y'}=R\p^{Y}$. Since for any $\p^Y\in\prob(n)$ and any $\q^{Y'}\in\prob(n')$ there exists a row stochastic matrix $R$ that satisfies $\q^{Y'}=R\p^{Y}$, we conclude that $\p^{XY}\sim_X\q^{XY'}$ for any probability vectors $\p^{XY}$ and $\q^{XY'}$ with $|X|=1$.

\item The Case $|Y|=1$.
In this case $\p^{XY}=\p^X\in\prob(m)$, and the stochastic matrix $R$ that appear in Theorem~\ref{lemcarc} is a vector $R=\r\eqdef(r_1,\ldots,r_{n'})^T\in\prob(n')$. Therefore,
the relation~\eqref{lqlpt} becomes $r_{y'}\p^{X}\succ\q_{y'}^X$ for all $y'\in[n']$. Moreover,
denoting by $q_{y'}\eqdef\mathbf{1}_{n'}\cdot\q_{y'}$ the sum of the components of $\q_{y'}$, we get from~\eqref{sumall} that $r_{y'}=q_{y'}$ for all $y'\in[n']$.
We therefore conclude that for $|Y|=1$,
\be
\p^{X}\succ_X\q^{XY'}\quad\quad\iff\quad\quad\p^X\succ\q_{|y'}^X\quad\quad\forall\;y'\in[n']\;,
\ee
where
$\q_{|y'}^X\eqdef\frac{1}{q_{y'}}\q_{y'}^X$ is the vector whose components are $\{q_{x|y'}\}_{x\in[m]}$. 
\item The Case $|Y'|=1$.
In this case, $\q^{XY'}=\q^{X}\in\prob(m)$, and since $R$ in Theorem~\ref{lemcarc} has to be an $1\times n$ column stochastic matrix it must be equal to the row vector $(1,\ldots,1)$. Therefore, denoting by $\p^X\eqdef\sum_{y\in[n]}\p^X_y$ we get from Theorem~\ref{lemcarc} that
\be
\p^{XY}\succ_{X}\q^X\quad\quad\iff\quad\quad\p^X\succ\q^X\;.
\ee
\een

\bex\label{exrxsy}
Let $\p^{XY}\in\prob(mn)$ and $\q^{XY'}\in\prob(mn')$. Show that if $\p^{XY}\succ_X\q^{XY'}$ then $\p^X\succ\q^X$.
\eex

\subsection{Standard Form}\index{standard form}

Consider the majorization relation between probability vectors in $\prob(n)$. This relation is not a partial order since if two vectors $\p,\q\in\prob(n)$ satisfy both $\p\succ\q$ and $\q\succ\p$ it is still possible that $\p\neq\q$. However, the majorization relation becomes a partial order if we assume that the components of the vectors are given in some \emph{standard form}. For example, we can define the standard form of a vector $\p\in\prob(n)$ to be $\p^\da$. Then, the majorization is a partial order when restricted to this standard form; i.e. to vectors in $\prob^\da(n)$.

Similarly, we would like to define a standard form for conditional majorization. For this purpose, we will consider two probability vectors $\p^{XY}\in\prob(mn)$ and $\q^{XY'}\in\prob(mn')$ and characterize the relation $\p^{XY}\sim_X\q^{XY'}$, meaning $\p^{XY}\succ_X\q^{XY'}$ and $\q^{XY'}\succ_X\p^{XY}$.
We start by discussing three examples of vectors in $\prob(mn)$ that are equivalent under conditional majorization to the vector $\p^{XY}$ as defined in~\eqref{oi1} :
\ben
\item In~\eqref{relsimx} we saw that if $\tp^{XY}$ is obtained from $\p^{XY}$ by applying an arbitrary permutation matrix to each of the vectors $\{\p_y^X\}_{y\in[n]}$, we get that $\tp^{XY}\sim_X\p^{XY}$. 
\item Similarly, for any permutation/bijection\index{bijection} $\pi:[n]\to [n]$ we get that (see Exercise~\ref{ex:sbl})
\be\label{equivpxyr}
\p^{XY}\;\sim_X\;\sum_{y\in[n]}\p^X_{\pi(y)}\otimes\e_y^{Y}\;.
\ee
\item
Suppose that there exists $\lambda\in\mbb{R}_+$ such that $\p_{1}^X=\lambda\p_{2}^X$. Then, (see Exercise~\ref{ex:reduc})
\be\label{reduc2}
\p^{XY}\;\sim_X\;(\p_1^X+\p_2^X)\otimes\e_1^{Y'}+\sum_{y=3}^n\p_{y}^X\otimes\e_{y-1}^{Y'}
\ee
where $Y'$ is a system of dimension $|Y'|=n-1$.
\een

\bex\label{ex:sbl}
Prove~\eqref{equivpxyr}. Hint: Take $r_{y'|y}=\delta_{y'\pi(y)}$ and $D^{(y,y')}=I_m$.
\eex

\begin{exercise}\label{ex:reduc}
Suppose that there exists $\lambda\in\mbb{R}_+$ such that $\p_{1}^X=\lambda\p_{2}^X$. Prove the equivalence relation given in~\eqref{reduc2}. Hint: To show $LHS\succ_X RHS$, find $R\in\stoc(n-1,n)$ that satisfies for $j\in\{1,2\}$, $R\e_j^Y=\e_1^{Y'}$, and for $j\in\{3,\ldots,n\}$, $R\e_{j}^Y=\e_{j-1}^{Y'}$. To show $RHS\succ_X LHS$, find $R\in\stoc(n,n-1)$ that satisfies
$R\e_1^{Y'}=\frac1{1+\lambda}\e_1^{Y}+\frac\lambda{1+\lambda}\e_2^{Y}$ and for $j\in\{2,\ldots,n-1\}$, $R\e_j^{Y'}=\e_{j+1}^{Y}$.
\end{exercise}

Recall that the vectors $\{\p_y^X\}_{y\in[n]}$ associated with $\p^{XY}$, are not probability vectors.  Hence, it will be convenient to denote for every $y\in[n]$
$
p_y\eqdef\mathbf{1}_m\cdot\p_y$ and $p_{x|y}\eqdef p_{xy}/p_y$.
Since the re-ordering of the vectors $\{\p_y^X\}_{y\in[n]}$ is a reversible CMO operation, we can choose a particular order to be the order of the ``standard form". The choice of this order is somewhat arbitrary, but it will be useful (in particular for the spacial case in which $|X|=2$ as we will see later) to order the vectors $\{\p_y^X\}_{y\in[n]}$ such that 
\be\label{porder9}
p_{1|1}\geq p_{1|2}\geq\cdots\geq p_{1|n}\;.
\ee
Moreover, if there exists $y\in[n-1]$ such that $p_{1|y}=p_{1|y+1}$ then, if necessary, we will exchange the vectors $\p_y^X$ and $\p_{y+1}^X$ so that $p_{2|y}\geq p_{2|y+1}$. If the latter inequality is also an equality we continue by induction until we get a $k\in[m-1]$ such that
$p_{x|y}= p_{x|y+1}$ for all $x\in[k]$ but $p_{k+1|y}> p_{k+1|y+1}$.
Combining  these observations with the exercise above  we are ready to define the standard form.
\begin{myd}{Standard Form}\index{standard form}
Let $\p^{XY}\in\prob(mn)$ be as defined in~\eqref{oi1}. We say that $\p^{XY}$ is given in the standard form if the vectors $\{\p_y^X\}_{y\in[n]}$ satisfy the following three conditions:
\ben
\item For all $y\in[n]$, $\p_{y}^X=\left(\p_{y}^X\right)^\da$.
\item The vectors $\{\p_y^X\}_{y\in[n]}$ are arranged as discussed around Eq.~\eqref{porder9}.
\item There is no $\lambda\in\mbb{R}$ such that $\p_y^X=\lambda\p_w^X$ for some $y,w\in[n]$ with $y\neq w$.
\een
\end{myd}

\bex\label{ex:qpl}
Let $\p^{XY}\in\prob(mn)$, $\q^{XY'}\in\prob(mn')$ be two probability vectors given in their standard form, and $L$ be the $m\times m$ matrix defined in Exercise~\ref{ex:mml}. Use Theorem~\ref{lemcarc} to show that $\p^{XY}\succ_X\q^{XY'}$ if and only if there exists $R\in\stoc(n',n)$ such that
\be\label{eqvq}
(L\otimes R)\p^{XY}\geq\left(L\otimes I_{n'}\right)\q^{XY'}\;,
\ee
where the inequality is entrywise.
\eex

Based on the preceding discussion, particularly the three examples provided, we can deduce that any $\p^{XY}$ is, under conditional majorization, equivalent to its standard form. Consequently, without loss of generality, we may always assume that $\p^{XY}\in\prob(mn)$ is presented in its standard form. We will now demonstrate that conditional majorization between vectors in standard form indeed constitutes a partial order.

\begin{myt}{}
\begin{theorem}\label{partialord}
Let $\p^{XY}\in\prob(mn)$ and $\q^{XY'}\in\prob(mn')$ be two probability vectors given in their standard form. Suppose further that $\p^{XY}\sim_X\q^{XY'}$. Then, $\p^{XY}=\q^{XY'}$ (in particular, $Y=Y'$ and $n=n'$).
\end{theorem}
\end{myt}

\begin{proof}
From Exercise~\ref{ex:qpl}, the relation $\p^{XY}\sim_X\q^{XY'}$ implies that there exists $R\in\stoc(n',n)$ and $R'\in\stoc(n,n')$, such that 
\be\label{twoconr}
(L\otimes R)\p^{XY}\geq\left(L\otimes I_{n'}\right)\q^{XY'}\quad\text{and}\quad (L\otimes R')\q^{XY'}\geq\left(L\otimes I_{n}\right)\p^{XY}\;.
\ee
Denote by $S=R'R$ and $S'\eqdef RR'$, and observe that the two equations above implies that (see Exercise~\ref{twoexe})
\be\label{2rel5}
(L\otimes S)\p^{XY}\geq\left(L\otimes I_{n}\right)\p^{XY}\quad\text{and}\quad(L\otimes S')\q^{XY'}\geq\left(L\otimes I_{n}\right)\q^{XY'}\;.
\ee
Denoting by $s_{y|w}$ the $(w,y)$-component of the matrix $S$ we get that the equation above is equivalent to
\be
\sum_{w\in[n]}s_{y|w}L\p_w\geq L\p_y\quad\quad\forall\;y\in[n]\;.
\ee 
On the other hand, observe that by taking the sum over $y\in[n]$ on both sides of the equation above we get an equality between the two sides. Therefore, all the $n$ inequalities above must be equalities! Multiplying both sides by the inverse of $L$ gives
\be\label{5237}
\p_y=\sum_{w\in[n]}s_{y|w}\p_w\quad\quad\forall\;y\in[n]\;.
\ee 
Observe that these equalities can be expressed as
\be
(1-s_{y|y})\p_y=\sum_{\substack{w\in[n]\\w\neq y}}s_{y|w}\p_w\quad\quad\forall\;y\in[n]\;.
\ee
We now argue that $s_{y|y}=1$ for all $y\in[n]$ (which is equivalent to $s_{y|w}=\delta_{yw}$ and $S=I_n$). Otherwise, suppose by contradiction that there exists $y\in[n]$ such that $s_{y|y}<1$. Without loss of generality suppose that this $y$ is $n$. We then get that 
\be\label{5p239}
\p_n=\sum_{w\in[n-1]}\frac{s_{n|w}}{1-s_{n|n}}\p_w\;.
\ee
Substituting this into~\eqref{5237} gives for all $y\in[n-1]$
\be
\p_y=\sum_{w\in[n-1]}s_{y|w}\p_w+s_{y|n}\p_n=\sum_{w\in[n-1]}s_{y|w}\p_w+s_{y|n}\sum_{w\in[n-1]}\frac{s_{n|w}}{1-s_{n|n}}\p_w\;.
\ee
Denoting by
\be
t_{y|w}\eqdef s_{y|w}+\frac{s_{y|n}s_{n|w}}{1-s_{n|n}}\;.
\ee
we conclude that
\be\label{redty}
\p_y=\sum_{w\in[n-1]}t_{y|w}\p_w\quad\quad\forall\;y\in[n-1]\;.
\ee
Observe that $\sum_{y\in[n-1]}t_{y|w}=1$. Next, we rule out the case $t_{y|w}=\delta_{yw}$ for all $y,w\in[n-1]$. Otherwise, this relation implies in particular that for all $y,w\in[n-1]$ with $y\neq w$ we have $s_{y|w}=0$ and $s_{y|n}s_{n|w}=0$. Now, recall that we assumed that $s_{n|n}<1$ so there exists $y\in[n-1]$ such that $s_{y|n}>0$. For this choice of $y\in[n-1]$ the relation $s_{y|n}s_{n|w}=0$ gives $s_{n|w}=0$ for all $w\neq y$. Substituting this into~\eqref{5p239} gives 
\be
\p_n=\frac{s_{n|y}}{1-s_{n|n}}\p_y
\ee
in contradiction with the third property of the standard form of the vector $\p^{XY}$. We therefore conclude that there must exists $y\in[n-1]$ such that $t_{y|y}<1$. 
Observe that we started with the relation~\eqref{5237} with the condition that there exists $s_{y|y}<1$ for some $y\in[n]$, and we reduced it to the relation~\eqref{redty} with the condition that there exists $t_{y|y}<1$ for some $y\in[n-1]$. Continuing by induction until we have only one term in the sum on the right-hand side of~\eqref{5237} (or of~\eqref{redty}) we conclude that one of the vectors of $\{\p_y^X\}_{y\in[n]}$ is proportional to another vector in the same set, in contradiction with the standard form of $\p^{XY}$. Therefore, the assumption that there exists $y\in[n]$ such that $s_{y|y}<1$ is in correct, and we conclude that $S=I_n$ or equivalently $R'R=I_n$. 

Moreover, following the same arguments as above we conclude that also $S'\eqdef RR'=I_{n'}$. Combining this with $R'R=I_n$ we must have $n'=n$ and $R'=R^{-1}$. However, the only stochastic matrix whose inverse is also stochastic is a permutation matrix (that is, doubly stochastic and orthogonal). We therefore conclude that the sets $\{\p_y^X\}_{y\in[n]}$ and $\{\q_y^X\}_{y\in[n]}$ can only differ up to a permutation; i.e., for all $y\in[n]$, $\p_y^X=\q_{\pi(y)}^Y$ for some permutation $\pi:[n]\to [n]$. However, since the $\{\p_y^X\}_{y\in[n]}$ and $\{\q_y^X\}_{y\in[n]}$ are ordered in a specific way given in the second property of the standard form, we conclude that $\pi(y)=y$ for all $y\in[n]$. This completes the proof.
\end{proof}

\bex\label{twoexe}
Prove the relations in~\eqref{2rel5}. Hint: Multiply both sides of the first inequality in~\eqref{twoconr} by $R'$ and the second inequality by $R$.
\eex

\subsection{Conditional Schur Convex Functions}\label{sec:csc}\index{Schur's convexity}

Conditionally Schur convex functions are functions from $\prob(mn)$ to the real line that behave monotonically under conditional majorization. Such functions generalize Schur convex functions, and can be used to quantify the amount of
conditional uncertainty contained in a correlated source; that is, they are measures of conditional uncertainty. In Chapter~\ref{sec:ce} we will study a class of conditionally Schur concave functions known as conditional entropies.

\begin{myd}{}
\begin{definition}
A function 
$
f:\bigcup_{n,m\in\mbb{N}}\prob(mn)\to\mbb{R}
$
is said to be conditionally Schur-convex if for every $\p^{XY}\in\prob(mn)$ and $\q^{XY'}\in\prob(mn')$  
\be
\p^{XY}\succ_X\q^{XY'}\quad\Rightarrow\quad f\left(\p^{XY}\right)\geq f\big(\q^{XY'}\big)\;.
\ee
\end{definition} 
\end{myd}
\begin{remark}
In addition to the definition above, $f$ is said to be conditionally Schur-concave if $-f$ is conditionally Schur-convex.
\end{remark}

Observe that the conditionally Schur convex functions reduce to Schur convex functions when restricted to $\prob(m)$ (i.e. $n=1$). Conversely, in the theorem below we show that every convex symmetric function on the set of probability vectors can be extended to a conditionally Schur convex function (see theorem below). Remember that in Subsection~\ref{sec:schurconvex}, we established that such symmetric convex functions are in particular Schur convex.

In the theorem below, for every convex symmetric function $f:\prob(m)\to\mbb{R}$ we define its extension, $H_f$, to any vector $\p^{XY}\in\prob(mn)$ via
\be\label{condschur}
H_f(\p^{XY})\eqdef \sum_{y\in[n]}p_yf\left(\p_{|y}^X\right)\;,
\ee
where $\p_{|y}\eqdef\frac1{p_y}\p_y$ is the probability vector whose components are $\{p_{x|y}\}_{x\in[m]}$. 

\begin{myt}{}
\begin{theorem}\label{thmschurcon}\index{Schur's convexity}
Let $f:\bigcup_{m\in\mbb{N}}\prob(m)\to\mbb{R}$ be a symmetric convex function. Then, the function $H_f$, as defined in~\eqref{condschur}, is conditionally Schur concave.
\end{theorem}
\end{myt}

\begin{proof}
Recall that $\p^{XY}\succ_X\q^{XY}$ if and only if there exists a stochastic matrix $R\in\stoc(n',n)$ such that~\eqref{lqlpt} holds. By rewriting~\eqref{lqlpt} with $\p_y^X=p_y\p_{|y}^X$ and $\q_{y'}^X=q_{y'}\q_{|y'}^X$ we get that
\be\label{0x7}
\sum_{y\in[n]}\frac{r_{y'|y}p_{y}}{q_{y'}}\p_{|y}^X\succ\q_{|y'}^X\quad\quad\forall\;y'\in[n']\;.
\ee
Therefore, if $\p^{XY}\succ_X\q^{XY}$ then
\ba\label{tewq}
H_f\left(\q^{XY'}\right)&=\sum_{y'\in[n']}q_{y'}f\left(\q_{|y'}^X\right)\\
\Gg{f\;\text{is Schur convex}}&\leq\sum_{y'=1}^{n'}q_{y'}f\Big(\sum_{y\in[n']}\frac{r_{y'|y}p_{y}}{q_{y'}}\p_{|y}^X\Big)\;,
\ea
where we used~\eqref{0x7}.
Moreover, from~\eqref{sumall} we have
\be
\sum_{y\in[n]}\frac{r_{y'|y}p_{y}}{q_{y'}}=1\;.
\ee
Thus, since $f$ is convex we get from~\eqref{tewq}
\ba
H_f\left(\q^{XY'}\right)&\leq \sum_{y'\in[n']}\sum_{y\in[n]}r_{y'|y}p_{y}
f\left(\p_{|y}^X\right)\\
\Gg{R\;\text{is stochastic}}&=\sum_{y\in[n]}p_{y}f\left(\p_{|y}^X\right)=H_f(\p^{XY})\;.
\ea
This completes the proof.
\end{proof}

\subsection{Two Dimensional Cases}\label{twodcases}

We study here the relatively simpler cases in which the dimension of the systems $X$ or $Y$ is two. In these cases, it is possible to get the exact analytical expressions that determine if one given state is conditionally majorized by another. We will see in the next section that in the more general case, conditional majorization can  be characterized with a linear program.

\subsubsection{The Case $|X|=2$}

In this case, the vectors $\{\p_{y}^X\}_{y\in[n]}$ and $\{\q_{w}^X\}_{w\in[n']}$ are all two dimensional. It will be convenient to denote their components as follows:
\be
\p_{y}^X\eqdef \bpm a_y\\ p_y-a_y\epm\quad\text{and}\quad\q_{y'}^X\eqdef \bpm b_{y'}\\ q_{y'}-b_{y'}\epm\;,
\ee
where $y\in[n]$, $y'\in[n']$, $a_y\eqdef p_{1y}$, $b_{y'}\eqdef q_{1y'}$, and $p_y\eqdef p_{1y}+p_{2y}$ and $q_{y'}\eqdef q_{1y'}+q_{2y'}$ are the sums of the components of $\p_y^X$ and $\q_{y'}^X$, respectively. 
With these notations we get for all $y\in[n]$ and $y'\in[n']$
\be
L\p_y^X=\bpm a_{y} \\ p_y \epm
\quad\text{and}\quad L\q_{y'}^X=\bpm b_{y'} \\ q_{y'} \epm\;,
\ee
where $L\eqdef\bpm 1 & 0\\ 1 & 1\epm$. Moreover, since we assume that $\p^{XY}$ and $\q^{XY'}$ are given in their standard form, we have (see~\eqref{porder9})
\be\label{ap1c}
\frac{a_1}{p_1}\geq\cdots\geq\frac{a_n}{p_n}\quad\text{and}\quad\frac{b_1}{q_1}\geq\cdots\geq\frac{b_{n'}}{q_{n'}}\;.
\ee

From Theorem~\ref{lemcarc} we have that $\p^{XY}\succ_X\q^{XY'}$ if and only if there exists a row stochastic matrix $R\in\stoc(n',n)$ such that for all $y'\in[n']$ we have
\be
b_{y'}\leq\sum_{y\in[n]}r_{y'|y}a_y\quad\text{and}\quad q_{y'}\leq\sum_{y\in[n]}r_{y'|y}p_y\;.
\ee
Observe that since $\sum_{y'\in[n']}q_{y'}=\sum_{y\in[n]}p_y=1$, the second inequality above must hold with equality (in fact, we know it already from~\eqref{sumall}) . 

Let $\a,\b,\p,\q$ be the vectors whose components are respectively $\{a_y\}_{y\in[n]}$, $\{b_{y'}\}_{y'\in[n']}$, $\{p_y\}_{y\in[n]}$, and $\{q_{y'}\}_{y'\in[n']}$. To streamline our analysis, we omitted the superscripts $Y$ and $Y'$ when referring to the vectors $\a,\b,\p,\q$. It is important for the reader to bear in mind that the vectors $\a\eqdef\a^Y$ and $\p\eqdef\p^Y$ correspond to a system of dimension $n$ (referred to as system $Y$), while $\b\eqdef\b^{Y'}$ and $\q\eqdef\q^{Y'}$ pertain to a system of dimension $n'$ (referred to as system $Y'$).
With these notations we get that $\p^{XY}\succ_X\q^{XY'}$ if and only if there exists $R\in\stoc(n',n)$ such that\index{relative submajorization}
\be\label{stoch4}
R\a\geq\b\quad\text{and}\quad R\p=\q\;.
\ee
This relation is closely related to relative majorization, however, note that the vectors $\a$ and $\b$ are not probability vectors since their components in general don't sum to one. We therefore say in this case that the pair $(\a,\p)$ \emph{relatively submajorize} the pair $(\b,\q)$. Observe also that if $a\eqdef\|\a\|_1$ equals $b\eqdef\|\b\|_1$ then the inequality $R\a\geq\b$ can be replaced with $R\a=\b$ so that~\eqref{stoch4} becomes equivalent to relative majorization; i.e.
\be
\left(\frac1a\a,\p\right)\succ\left(\frac1b\b,\q\right)\;.
\ee
We therefore assume now that $a>b$ (the case $a<b$ is not possible since $R\a\geq\b$, and $R$ is column stochastic).

\begin{figure}[h]
\centering
    \includegraphics[width=0.5\textwidth]{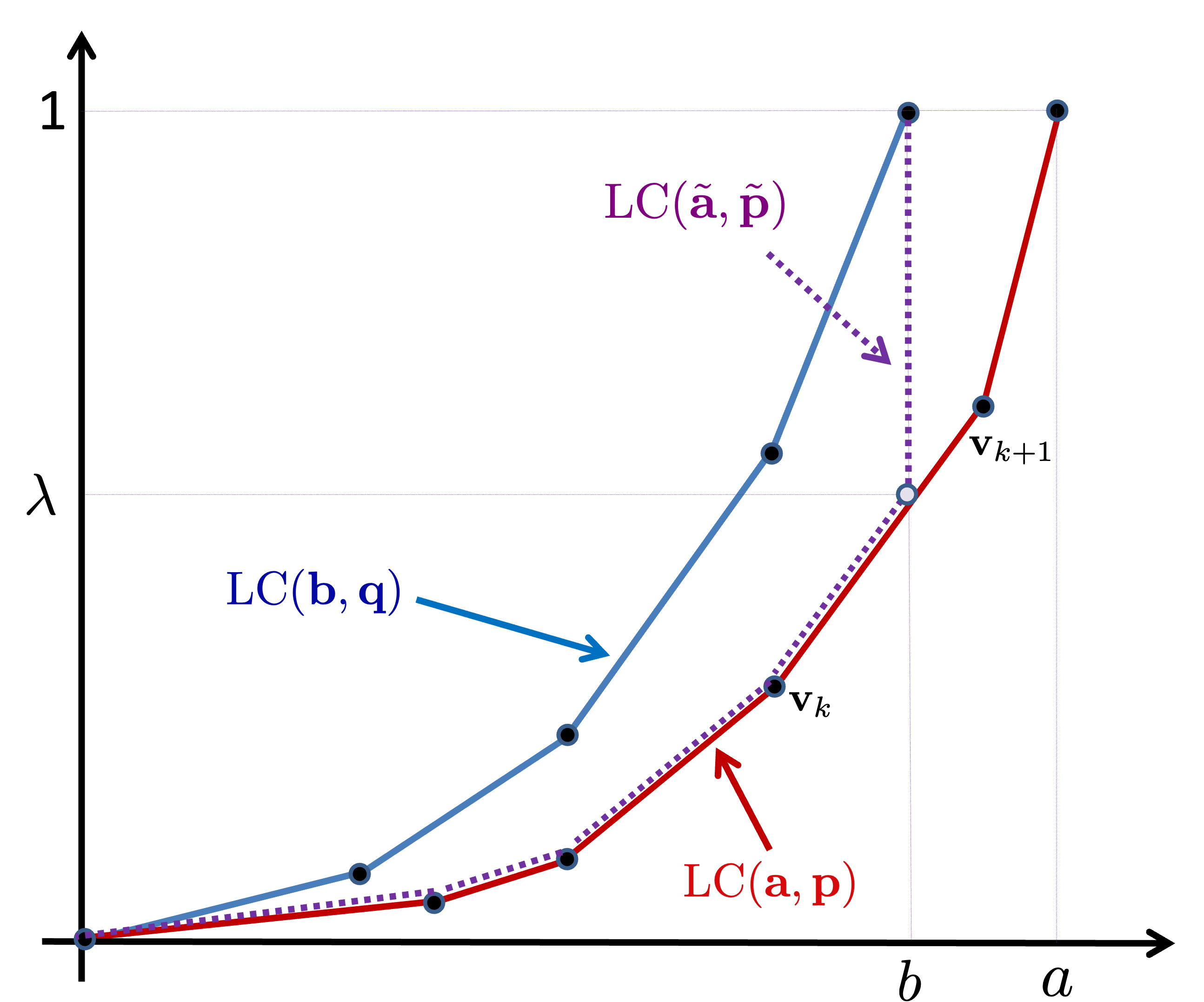}\index{Lorenz curve}
  \caption{\linespread{1}\selectfont{\small Submajorization. Given the Lorenz curve\index{Lorenz curve} LC$(\a,\p)$, we can always construct another Lorenz curve, LC$(\tilde{\a},\tilde{\p})$, with $\|\ta\|_1=b$ such that \emph{any} Lorenz curve LC$(\b,\q)$ that is no where below LC$(\a,\p)$ is also nowhere below LC$(\tilde{\a},\tilde{\p})$. In this example $n=5$ and $k=3$.} }
\label{submajo}
\end{figure}

Even though the components of $\a$  do not sum to one (in general), we can still define its testing region as (see~\eqref{testingr})\index{testing region}
\be\label{testingr2}
\mt(\a,\p)\eqdef\Big\{(\a\cdot\t,\p\cdot\t)\;:\;\t\in[0,1]^n\Big\}\;.
\ee
By taking $\t=\mathbf{1}_n$ we get the point $(a,1)\in\mt(\a,\p)$ as oppose to the point $(1,1)$ that one would get if $\a$ was a probability vector. In fact, the testing region of the pair of probability vectors $(\frac1{a}\a,\p)$ is almost identical to that of $(\a,\p)$ except for a rescaling of the $x$-axis by a factor of $a$; that is, $(r,s)\in\mt(\frac1{a}\a,\p)$ if and only if $(ar,s)\in\mt(\a,\p)$. Therefore, if $(r,s)$ is an extreme point of $\mt(\frac1{a}\a,\p)$ then $(ar,s)$ is an extreme point of $\mt(\a,\p)$. That is, there are $n+1$ extreme points on the lower Lorenz curve of $\mt(\a,\p)$ given by $(0,0)$ and the $n$ points $\{(\mu_\ell,\nu_\ell)\}_{\ell\in[n]}$, where\index{relative submajorization}
\be
\mu_\ell\eqdef\sum_{x\in[\ell]} a_x\quad\text{and}\quad\nu_\ell\eqdef\sum_{x\in[\ell]} p_x\;.
\ee
Recall that since we assume that $\p^{XY}$ is given in its standard form, the components of $\a$ and $\p$ satisfy~\eqref{ap1c}.
See the red line in Fig.~\ref{submajo} for an example of the lower Lorenz curve\index{Lorenz curve} of the pair $(\a,\p)$.

In Fig.~\ref{submajo} we also depicted another Lorenz
curve, by taking it to be identical to the Lorenz curve\index{Lorenz curve} of $(\a,\p)$ if the $x$-coordinate is no greater than $b$, and a vertical line if the $x$-coordinate equals $b$. This curve is a Lorenz curve\index{Lorenz curve} of some pair of vectors $(\ta,\tp)$ for which $\|\ta\|_1=b$ and $\tp$ is a probability vector. Moreover, the Lorenz curve of $(\tilde{\a},\tilde{\p})$ has the property that \emph{any} other Lorenz curve LC$(\b,\q)$ (see the top curve in Fig.~\ref{submajo}) that is no where below the Lorenz curve of $(\a,\p)$ is also nowhere below the Lorenz curve of $(\tilde{\a},\tilde{\p})$. We will see shortly that this implies that the relation $\p^{XY}\succ_X\q^{XY'}$ is equivalent to $(\tilde{\a},\tilde{\p})\succ(\b,\p)$.

The vectors $\ta$ and $\tp$ that corresponds to the purple Lorenz curve\index{Lorenz curve} of Fig.~\ref{submajo} can be expressed as follows. Let $k\in[n-1]$ be the integer satisfying $\mu_k\leq b<\mu_{k+1}$,  or equivalently
\be\label{bakm}
b-a_{k+1}<\mu_k\leq b\;.
\ee
Such an index $k$ exists since we assume that $a>b$.
The line connecting the vertices $\v_k\eqdef(\mu_k,\nu_k)$ with $\v_{k+1}\eqdef(\mu_{k+1},\nu_{k+1})$ contains the point $(b,\lambda)$ (see Fig.~\ref{submajo}), where 
\be
\lambda\eqdef\frac{p_{k+1}}{a_{k+1}}(b-\mu_k)+\nu_k\;.
\ee
The point $(1,\lambda)$ is therefore a vertex of the curve LC$(\tilde{\a},\tilde{\p})$ in Fig.~\ref{submajo}. With this notations, $\ta$ and $\tp$ are given by
\ba
&\ta\eqdef( a_1, \ldots , a_k ,b-\mu_k , 0 )^T\in\mbb{R}_+^{k+2}\\
&\tp\eqdef(p_1,\ldots,p_k,\lambda-\nu_k,1-\lambda)^T\in\prob(k+2)\;.
\ea
 With these notations we have the following characterization of conditional majorization.

\begin{myt}{}
\begin{theorem}
Using the same notations as above, for $|X|=2$ the following statements are equivalent:\index{relative submajorization}
\ben
\item $\p^{XY}\succ_X\q^{XY'}$ 
\item The curve LC$(\a,\p)$ is nowhere above the curve LC$(\b,\p)$.
\item $(\ta,\tp)\succ(\b,\q)$\;.
\een
\end{theorem}
\end{myt}

\begin{proof}
The case $a=b$ is relatively simple and is left as an exercise. We therefore prove the theorem for the case $a>b$.
Suppose $\p^{XY}\succ_X\q^{XY'}$. Then, there exists $R\in\stoc(n',n)$ such that~\eqref{stoch4} holds. Let $(\b\cdot\t',\q\cdot\t')\in{\rm LC}(\b,\q)$ be a point on the lower Lorenz curve\index{Lorenz curve} of the testing region of $(\b,\q)$, where $\t'$ is some vector in $[0,1]^{n'}$. Then, the vector $\t\eqdef R^T\t'\in[0,1]^{n}$ has the property that
\be
\a\cdot\t=\a^TR^T\t'\geq\b\cdot\t'\quad\text{and}\quad\p\cdot\t=\p^TR^T\t'=\q\cdot\t'\;,
\ee
where we used the relations in~\eqref{stoch4}.
The above relation implies that for any point $(\b\cdot\t',\q\cdot\t')$ in ${\rm LC}(\b,\q)$, there exists a point $(\a\cdot\t,\p\cdot\t)$ in the testing region of $(\a,\p)$ that is located to its right (i.e. a point with the same $y$-coordinate and no smaller $x$-coordinate).
Since the lower Lorenz curve\index{Lorenz curve} is convex, this means that ${\rm LC}(\b,\q)$ is a nowhere below ${\rm LC}(\a,\p)$. That is, the second statement of the theorem holds. 

To prove that the second statement implies the third statement of the theorem, observe that by construction of ${\rm LC}(\ta,\tp)$ (see Fig.~\ref{submajo}), the curve ${\rm LC}(\b,\q)$ is nowhere below the curve ${\rm LC}(\ta,\tp)$. Since $\|\ta\|_1=b$ we get from Theorem~\ref{chararm} that $(\ta,\tp)\succ(\b,\q)$. 

It is therefore left to prove that the third statement in the theorem implies the first one. Since we assume that $(\ta,\tp)\succ(\b,\q)$ there exists a matrix $S\in\stoc(n',k+2)$ such that $S\ta=\b$ and $S\tp=\q$. On the other hand, the matrix
\be\label{5p259}
T\eqdef
\bpm
I_{k} & \0_{k,n-k}\\
\0_{2,k} & M
\epm\in\stoc(k+2,n)
\quad\text{where}\quad M\eqdef
\bpm 
\frac{\lambda-\nu_k}{p_{k+1}} & 0 & \cdots & 0\\
\frac{\nu_{k+1}-\lambda}{p_{k+1}} & 1 & \cdots & 1
\epm
\ee
satisfies $T\a\geq\ta$ and $T\p=\tp$ (see Exercise~\ref{tatpp}). Therefore,
\be
ST\a\geq S\ta=\b\quad\text{and}\quad ST\p=S\tp=\q\;,
\ee
so that the matrix $R\eqdef ST\in\stoc(n',n)$ satisfies $R\a\geq \b$ and $R\p=\q$. That is, the condition in~\eqref{stoch4} holds  and therefore $\p^{XY}\succ_X\q^{XY'}$. This completes the proof.
\end{proof}

\bex\label{tatpp}Using the same notations as the proof above:
\ben
\item Prove the theorem above for the case $a=b$.
\item Show that~\eqref{bakm} implies that $\lambda<\nu_{k+1}$.
\item Verify that the matrix $T$ in~\eqref{5p259} is column stochastic and satisfies $T\a\geq\ta$ and $T\p=\tp$.
\een
\eex

\subsubsection{The Case $|Y|=2$}

In this case, $\p^{XY}$ has the form 
\be
\p^{XY}=\p_1\otimes\e_1+\p_2\otimes\e_2\;,
\ee
where $\p_1,\p_2\in\prob(m)$ and $\e_1,\e_2\in\mbb{R}^2$ form the standard basis of $\mbb{R}^2$ (for simplicity, we omitted the superscript $X$ from $\p_1$ and $\p_2$). The matrix $R$ that appears in Theorem~\ref{lemcarc} has the form
$R=[ \a \; \b]$,
where $\a\eqdef(a_1,\ldots,a_{n'})^T$ and $\b\eqdef(b_1,\ldots,b_{n'})^T$ are probability vectors. Therefore, the relation in~\eqref{lqlpt} can be expressed as
\be\label{in5248}
a_{w}\p_{1}+b_{w}\p_2\succ\q_{w} \quad\quad\forall\;w\in[n']\;.
\ee
Moreover, from~\eqref{sumall} we get that
\be
q_{w}= a_{w}p_1+b_{w}p_2\quad\quad\forall\;w\in[n']\;,
\ee
where $p_y\eqdef\mathbf{1}_{m}\cdot\p_{y}$ for all $y\in[n]$, and $q_{w}\eqdef\mathbf{1}_{m}\cdot\q_{w}$ for all $w\in[n']$. 

Let $\p_{|y}\eqdef\frac1{p_{y}}\p_{y}\in\prob(m)$ for $y=1,2$, and $\q_{|w}\eqdef\frac1{q_{w}}\q_{w}\in\prob(m)$ for $w\in[n']$. With these notations, for all $w\in[n']$, the two equations above gives 
\be\label{ineq09}
t_{w}\p_{|1}+(1-t_{w})\p_{|2}\succ\q_{|w},
\quad
\text{where}
\quad
t_{w}\eqdef\frac{a_{w}p_1}{a_{w}p_1+b_{w}p_2}=\frac{a_wp_1}{q_w}\;.
\ee
Observe that if we find a vector $\t\in[0,1]^{n'}$, whose components satisfy $t_{w}\p_{|1}+(1-t_{w})\p_{|2}\succ\q_{|w}$ for all $w\in[n']$, it could still be the case that there are no $\a,\b\in\prob(n')$ that relates to $\t$ as above. 
To find the condition on $\t$ that ensures the existence of such $\a,\b\in\prob(n')$, we rearrange the expression above for $t_w$ to get for all $w\in[n']$
\be\label{hah}
q_wt_w=a_wp_1\;.
\ee
Using the fact that $\a\in\prob(n')$, by summing over $w\in[n']$ both sides of the equation above we get that $\t$ must satisfy
\be\label{tsatp}
\sum_{w\in[n']}q_wt_w=p_1\;.
\ee
We therefore conclude that $\p^{XY}\succ_X\q^{XY'}$ if and only if there exists $\t\in[0,1]^{n'}$ that satisfies~\eqref{tsatp} and for all $w\in[w']$, $t_{w}\p_{|1}+(1-t_{w})\p_{|2}\succ\q_{|w}$.

In order to determine when such $\t\in[0,1]^{n'}$ exists, we assume that $\p^{XY}$ is given in its standard form so that both $\p_{|1}=\p_{|1}^\da$ and $\p_{|2}=\p_{|2}^\da$. With this property, the majorization relation given in~\eqref{ineq09} is equivalent to
\be\label{eabo}
t_{w}\|\p_{|1}\|_{(k)}+(1-t_{w})\|\p_{|2}\|_{(k)}\geq\|\q_{|w}\|_{(k)}\quad\quad\forall\;w\in[n']\;,\;\;\forall\;k\in[m]\;.
\ee
Our next goal is
to characterize the constraints that the equation above impose on $t_w$. For this purpose, we denote by $\mi_+$, $\mi_0$, and $\mi_-$ the set of all indices $k\in[m]$ for which $\|\p_{|1}\|_{(k)}-\|\p_{|2}\|_{(k)}$ is positive, zero, and negative, respectively. With these notations, if $k\in\mi_0$ then~\eqref{eabo} takes the form  $\|\p_{|2}\|_{(k)}\geq\|\q_{|w}\|_{(k)}$. On the other hand, if $k\in\mi_+$ we can isolate $t_w$ to get
\be
t_w\geq \frac{\|\q_{|w}\|_{(k)}-\|\p_{|2}\|_{(k)}}{\|\p_{|1}\|_{(k)}-\|\p_{|2}\|_{(k)}}\;.
\ee
Therefore, since this inequality holds for all $k\in\mi_+$ and since $t_w\geq 0$ we conclude that $t_w\geq\mu_w$ for all $w\in[n']$, where
\be
\mu_w\eqdef\max\left\{0,\max_{k\in\mi_+}\frac{\|\q_{|w}\|_{(k)}-\|\p_{|2}\|_{(k)}}{\|\p_{|1}\|_{(k)}-\|\p_{|2}\|_{(k)}}\right\}\;.
\ee
Simililarly, by isolating $t_w$ in~\eqref{eabo} for the cases that $k\in\mi_-$ we get $t_w\leq \nu_w$ for all $w\in[n']$, where
\be
\nu_{w}\eqdef\min\left\{1,\min_{k\in\mi_-}\frac{\|\p_{|2}\|_{(k)}-\|\q_{|w}\|_{(k)}}{\|\p_{|2}\|_{(k)}-\|\p_{|1}\|_{(k)}}\right\}\;.
\ee
We therefore arrive at the following theorem.
\begin{myt}{}
\begin{theorem}\label{thmy2}
Using the same notations as above, for the case $|Y|=2$ we have $\p^{XY}\succ_{X}\q^{XY'}$ if and only if the following conditions hold:
\ben
\item For all $w\in[n']$ we have $\nu_{w}\geq\mu_{w}$.
\item For all $k\in\mi_0$ we have $\|\p_{|2}\|_{(k)}\geq\max_{w\in[n']}\|\q_{|w}\|_{(k)}$.
\item 
$\sum_{w\in[n']}q_{w}\mu_{w}\leq p_1\leq\sum_{w\in[n']}q_{w}\nu_{w}$.
\een
\end{theorem}
\end{myt}

\bex
Use the arguments above to prove Theorem~\ref{thmy2}.
\eex

\bex
Simplify the conditions in Theorem~\ref{thmy2} for the case that $\mi_+=[m]$.
\eex

\bex
Consider the case $|X|=|Y|=|Y'|=2$ and let $\p^{XY},\q^{XY}\in\prob(4)$ be such that $\p^{Y}=\q^Y=\u^{(2)}$. Simplify the necessary and sufficient conditions given in the theorem above for this case.
\eex

\subsection{Conditional Majorization with Linear Programming}\index{linear programming}\index{conditional majorization}

In Theorem~\ref{lemcarc} we presented a useful characterization of conditional majorization. This characterization posits that $\p^{XY}\succ\q^{XY'}$ if and only if a stochastic matrix $R\in\stoc(n',n)$ exists, satisfying the condition specified in equation~\eqref{lqlpt}.  Furthermore, Theorem~\ref{lemcarc} suggests that determining whether $\p^{XY}\succ_X\q^{XY}$ can be accomplished through linear programming.

To see it explicitly, suppose that $\p^{XY}$ and $\q^{XY'}$ are given in their standard form, and recall that $\p^{XY}\succ_X\q^{XY'}$ if and only if there exists $R\in\stoc(n',n)$ such that
\be\label{7p115}
\sum_{y\in[n]}r_{y'|y}L\p_y^X\geq L\q_{y'}^X\quad\quad\forall\;y'\in[n']\;,
\ee
where $L$ is the $m\times m$ matrix defined in Exercise~\ref{ex:mml}. 
Denote the rows of $R$ by $\r_1,\ldots,\r_{n'}\in\mbb{R}^n_+$; i.e., $\r_{y'}\eqdef(r_{y'|1},\ldots,r_{y'|n})^T$, 
and denote by $\r\in\prob(nn')$ the probability vector 
\be
\r\eqdef\begin{bmatrix}\r_1\\ \vdots \\ \r_{n'}
\end{bmatrix}\;.
\ee
Note that in this vector form of $R$, the condition that $R$ is stochastic is equivalent to $\r\geq 0$ and $\sum_{y'\in[n']}\r_{y'}=\1_n$. Let $P\eqdef\bbm\p_1^X & \cdots & \p_{n}^X\ebm\in\mbb{R}^{m\times n}$, and denote by
\be\label{defM}
M\eqdef\begin{bmatrix}
-LP & \0 & \cdots & \0\\
\0 & -LP & \cdots & \0\\
\vdots & \vdots & \ddots & \vdots\\
\0 & \0 & \cdots & -LP\\
I_{n} & I_{n} & \cdots & I_{n}\\
\end{bmatrix}\in\mbb{R}^{(mn'+n)\times nn'}\quad\text{and}\quad
\b\eqdef\begin{bmatrix}
-L\q_1^X\\
\vdots\\
-L\q_{n'}^X\\
\mathbf{1}_n
\end{bmatrix}\in\mbb{R}^{mn'+n}.
\ee
It is then straight forward to check that the inequalities given in~\eqref{7p115} can be expressed compactly as $M\r\leq\b$. The only other constraint is that $\r\in\mbb{R}_{+}^{nn'}$. The problem of determining if such a vector $\r$ exists is known as a linear programming\index{linear programming} feasibility problem, and there are several algorithms that can be used to solve it. 

\begin{exercise}[Farkas Lemma]\label{farkas}
Show that there exists $\r\in\mbb{R}_{+}^{nn'}$ satisfying $M\r\leq\b$ if and only if
for every $\v\in\mbb{R}_{+}^{mn'+n}$ that satisfies $\v^TM\geq 0$ (entrywise) we have $\v\cdot\b\geq 0$. Hint: For the harder direction, use the hyperplane separation theorem (Theorem~\ref{hyper}).
\end{exercise}

\subsubsection{Dual Characterization}\index{duality}

In the discussion above we saw that the condition $\p^{XY}\succ_X\q^{XY'}$ is equivalent to the existence of a vector $\r\in\mbb{R}_{+}^{nn'}$ such that $M\r\leq\b$. Moreover, from the exercise above it follows that such an $\r$ exists if and only if for every $\v\in\mbb{R}_{+}^{mn'+n}$ that satisfies $\v^TM\geq 0$ we have $\v\cdot\b\geq 0$. We now express this later condition in terms of sub-linear functionals.

For this purpose, we express $\v$ as
\be
\v\eqdef\begin{bmatrix}\v_1\\ \vdots \\ \v_{n'}\\ \t
\end{bmatrix}\;,
\ee
where $\v_1,\ldots,\v_{n'}\in\mbb{R}^{m}_+$ and $\t\in\mbb{R}^n_+$. From the definition of $M$ in~\eqref{defM} we get that the condition $\v^TM\geq 0$ can be expressed as
\be
\t-\v_{w}^TLP\geq 0\quad\quad\forall\;w\in[n']\;,
\ee
which in terms of the components $\{t_y\}_{y\in[n]}$ of $\t$ can be expressed as
\be\label{ryg10}
t_y\geq\max_{w\in[n']}\v_w^TL\p_y^X\quad\quad\forall\;y\in[n]\;.
\ee
Similarly, from the definition of $\b$ in~\eqref{defM} we get that the condition $\v\cdot\b\geq 0$ is equivalent to
\be\label{sy110}
\sum_{y\in[n]}t_y\geq\sum_{y'\in[n']}\v_{y'}^TL\q_{y'}^X\;.
\ee
Hence, the condition that $\v^TM\geq 0$ implies $\v\cdot\b\geq 0$ is equivalent to
\be
\sum_{y\in[n]}\max_{w\in[n']}\v_w^TL\p_y^X\geq\sum_{w\in[n']}\v_{w}^TL\q_{w}^X\;,
\ee
where we took $t_y$ in~\eqref{sy110} to be equal to its smallest possible value as given in~\eqref{ryg10}. Finally, for each $w\in[n']$ let $\s_w\eqdef L^T\v_w$ and observe that the inequality above can be written as
\be
\sum_{y\in[n]}\max_{w\in[n']}\s_w\cdot\p_y^X\geq\sum_{w\in[n']}\s_w\cdot\q_{y'}^X\;.
\ee
Note that since each $\v_w\in\mbb{R}_+^m$ we get that $\s_w\in\mbb{R}_+^m$ and $\s_w=\s_w^\da$ (see Exercise~\ref{ex:mml}).  Finally, by dividing both sides of the inequality above by a sufficiently large number and absorbing it into each $\s_w$ we can assume without loss of generality that the matrix $S\eqdef[\s_1  \cdots \s_{n'}]\in\stoc_{\leq}(m,n')$ is sub-stochastic (i.e., the components of each column sums to a number smaller or equal to one). We therefore arrived at the following theorem.

\begin{myt}{}
\begin{theorem}\label{thm:pcmq}
Let $\p^{XY}\in\prob(mn)$ and $\q^{XY'}\in\prob(mn')$ be given in their standard form. Then, $\p^{XY}\succ_X\q^{XY'}$ if and only if for every sub-stochastic matrix $S\eqdef\left[\s_1  \cdots \s_{n'}\right]\in\stoc_{\leq}(m,n')$, whose columns satisfy $\s_w=\s_w^{\da}$ for all $w\in[n']$, we have
\be\label{dualgames}
\sum_{y\in[n]}\max_{w\in[n']}\s_w\cdot\p_y^X\geq\sum_{w\in[n']}\s_w\cdot\q_{y'}^X\;.
\ee
\end{theorem}
\end{myt}

\bex
Consider the theorem above without the assumption that $\s_w=\s_w^{\da}$ for all $w\in[n']$ and without the assumption that $\p^{XY}$ and $\q^{XY'}$ are given in their standard form. Show that $\p^{XY}\succ_X\q^{XY'}$ if and only if for every sub-stochastic matrix $S\eqdef[\s_1  \cdots \s_{n'}]\in\stoc_{\leq}(m,n')$
\be
\sum_{y\in[n]}\max_{w\in[n']}\s_w^\da\cdot\p_y^\da\geq\sum_{w\in[n']}\s_w^\da\cdot\q_{y'}^\da\;,
\ee
where for simplicity, we removed the superscript $X$ from $\p_y^X$ and $\q_w^{X}$.
\eex

\subsection{The Operational Approach: Games of Chance with a Correlated Source}\index{operational approach}\index{gambling game}

In this subsection we delve into an alternative approach that leads to an equivalent definition of conditional majorization. This particular approach is grounded in operational principles, drawing parallels with games of chance. As a result, it presents a particularly persuasive rationale for the definition of conditional majorization as introduced in the previous subsections.

In the beginning of this chapter we introduced the concept of majorization using games of chance. We saw that two probability vectors, $\p,\q\in\prob(n)$ satisfy $\p\succ\q$ if and only if in all games of chance, a player has better odds to win the game with the $\p$-dice rather than with the $\q$-dice. Similarly, we will see that our definition of conditional majorization as given in Definition~\ref{def:cmc} can be characterized with games of chance that involve a correlated source.

We can think about a correlated source $XY$ as two dice that are connected with a gum. Rolling the two dice results in an outcome $x$ for system $X$ and a correlated outcome $y$ for system $Y$. As before, we denote by $\p^{XY}\in\prob(mn)$ the probability matrix whose $(x,y)$-entry, $p_{xy}$, represents the probability that $X=x$ and $Y=y$. It will be convenient to denote by $p_{x|y}=\frac{p_{xy}}{p_y}$, the conditional probability that $X=x$ given that $Y=y$, where $p_y\eqdef\sum_{x\in[m]}p_{xy}$ for any $y\in[n]$.

Consider now a gambling game with such two correlated dice, in which a player, say Alice, has to provide $k\leq m$ numbers as her guesses for the value of $X$. If Alice has access to
the value $y$ of $Y$, then she will choose the $k$ numbers that has the largest probability to occur relative to the conditional probability $\{p_{x|y}\}_{x\in[m]}$. Therefore, the maximum probability to win such a $k$-gambling game is given by
\be
\sum_{y\in[n]}p_y\sum_{x\in[k]}p_{x|y}^\da\;.
\ee
That is, Alice chooses the $k$ numbers that has the largest probability to occur \emph{after} she learns the value of $Y=y$, which occur with probability $p_y$.

\begin{figure}[h]\centering    \includegraphics[width=0.6\textwidth]{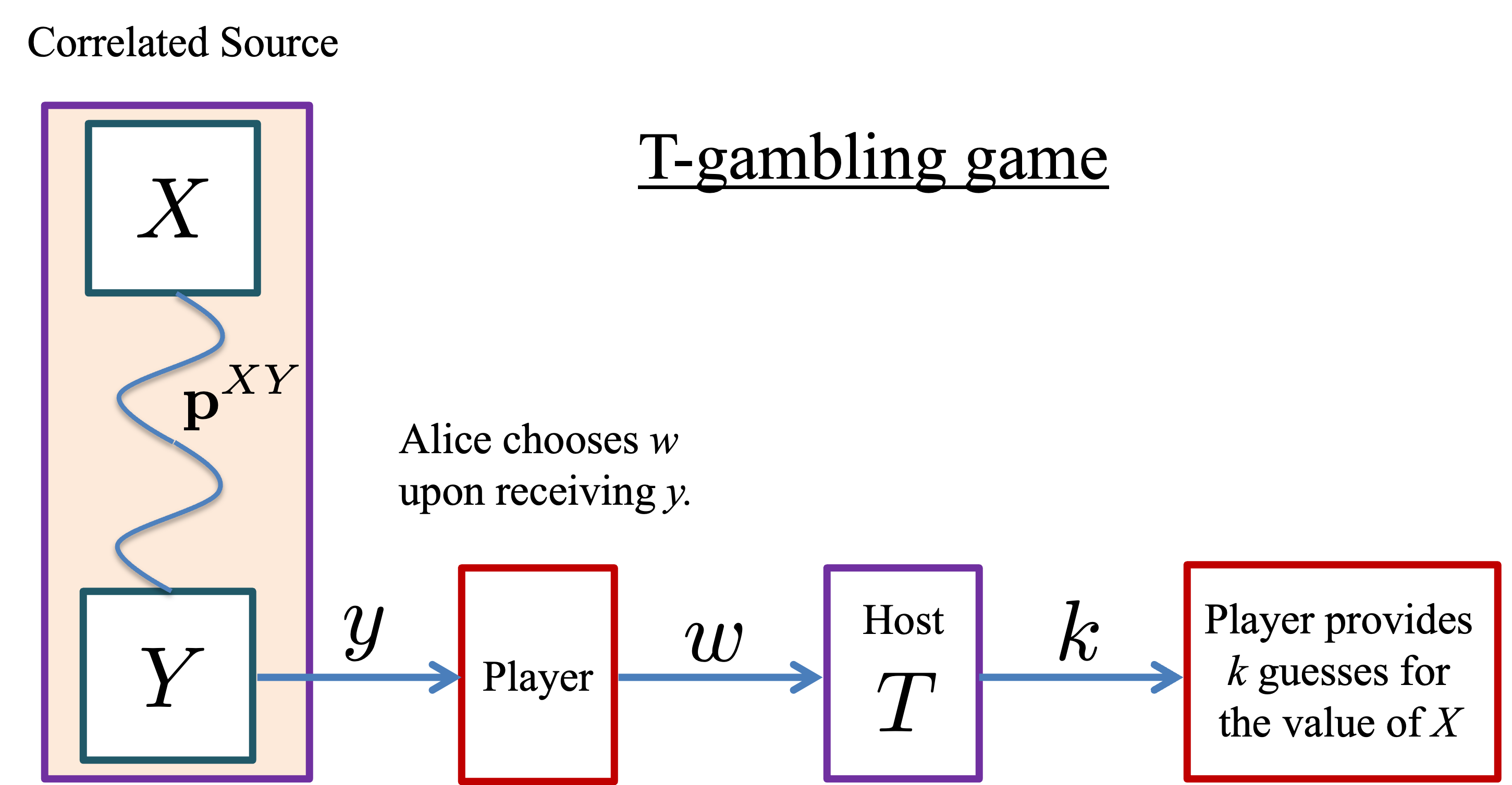}
  \caption{\linespread{1}\selectfont{\small A $T$-gambling game with correlated source $\p^{XY}$. Upon learning the value of $Y$, the player provides the host a number $w$. Then, the host chooses $k$ (at random) according to the distribution $\{t_{k|w}\}_{k\in[m]}$. After that, the player provides her $k$ guesses with the highest probability to occur.}}
  \label{tgambling}\index{gambling game}
\end{figure}

The example provided earlier is not the only kind of gambling game that Alice can engage in with a correlated source, like the two-dice system. More expansively, we can envisage a game where the host randomly determines the value of $k$ according to a certain distribution. In line with our aim to explore the widest range of scenarios in a gambling game with a correlated source, we allow the player a degree of control in choosing which $k$-gambling game will be played. This control is exercised through the player selecting a number $w\in[\ell]$ and communicating it to the game host. Subsequently, the host decides the value of $k$ based on a distribution $T\eqdef(t_{k|w})\in\mbb{R}^{m\times \ell}_+$, a detail known to the player. This distribution adheres to the conditions that $t_{k|w}\geq 0$ and $\sum_{k\in[m]}t_{k|w}\leq 1$ for all $w\in[\ell]$. Notably, we also accommodate the scenario where the set $\{t_{k|w}\}_{k\in[m]}$ does not sum to one, reflecting the possibility of no $k$ value occurring, resulting in the player losing the game from the onset. The procedural steps of such a $T$-gambling game are illustrated in Fig.~\ref{tgambling}.\index{gambling game}

Note that the set encompassing all $T$-gambling games includes all $k$-gambling games as well. This is evident when we consider $T=(t_{k|w})$ with $t_{k|w}=\delta_{kk_0}$, where $k_0$ is a specific integer within $[m]$. In this scenario, the game essentially becomes a $k_0$-gambling game, meaning the host selects $k=k_0$ regardless of $w$. In another example, where $t_{k|w}=\frac1m$, the host picks $k$ from a uniform distribution, also independent of $w$.

Generally, for a given $Y=y$ and a chosen $w\in[\ell]$, the optimal chance Alice has to win a $T$-gambling game can be calculated by the probability:
\be
\sum_{k\in[m]} t_{k|w}\sum_{x\in[k]}p_{x|y}^\da\;.
\ee
Consequently, for each $Y=y$, Alice will select the number $w$ that maximizes this probability. Therefore, the maximum likelihood of winning a $T$-gambling game, as outlined above, is given by:
\be
\pr_{T}\left(\p^{XY}\right)=\sum_{y\in[n]} p_y\max_{w\in[\ell]}\sum_{k\in[m]} t_{k|w}\sum_{x\in[k]}p_{x|y}^\da\;.
\ee

The formula for calculating the winning probability, sometimes referred to as the reward function in game theory, can be simplified. Consider the following transformation:
\be
\sum_{k\in[m]} \sum_{x\in[k]}t_{k|w}p_{x|y}^\da=\sum_{x\in[m]} \sum_{k=x}^mt_{k|w}p_{x|y}^\da\;.
\ee
Let's introduce the matrix $S=(s_{xw})\in\mbb{R}^{m\times \ell}$, whose coefficients are defined by
\be\label{sxw}
s_{xw}\eqdef\sum_{k=x}^mt_{k|w}\;.
\ee
It's important to note that the columns of $S$ are in non-decreasing order; that is, for every $w\in[\ell]$,
\be\label{order1geq}
1\geq s_{1w}\geq s_{2w}\geq\cdots\geq s_{mw}\;.
\ee
With this notation, the probability of winning can be expressed as
\be
\pr_{T}\left(\p^{XY}\right)=\sum_{y\in[n]} p_y\max_{w\in[\ell]} \sum_{x\in[m]}s_{xw}p^\da_{x|y}\;.
\ee
Finally, denoting the columns of $S$ by $\s_1,\ldots,\s_{\ell}$ we conclude that
\be\label{expprt}
\pr_{T}\left(\p^{XY}\right)=\sum_{y\in[n]} \max_{w\in[\ell]}\; \s_w\cdot\p_{y}^\da\;,
\ee
where $\p_{y}\in\mbb{R}^m_+$ is the vector with components $\{p_{xy}\}_{x\in[m]}$. Observe that the formula above for calculating the winning probability coinside with the left-hand side of~\eqref{dualgames}. This observation provides our initial clue about the connection between games of chance and conditional majorization. Another key insight is that conditionally mixing\index{conditionally mixing} operations cannot increase the maximum probability of winning the game, as the following lemma demonstrates.\index{gambling game}

\begin{myg}{}
\begin{lemma}\label{3onedir3}
Let $T\in\stoc_{\leq}(m,\ell)$, $\p^{XY}\in\prob(mn)$, and $\q^{XY'}\in\prob(mn')$. If $\p^{XY}\succ\q^{XY'}$ then the maximal probability to win a $T$-gambling game satisfies 
\be
\pr_{T}\left(\p^{XY}\right)\geq \pr_{T}\left(\q^{XY'}\right)\;.
\ee
\end{lemma}
\end{myg}

\begin{proof}
Let $S$ be the $m\times \ell$ matrix whose components as defined in~\eqref{sxw}. Due to~\eqref{order1geq} the columns of $S=[\s_1  \cdots  \s_{\ell}]$ satisfy $\s_{w}^\da=\s_{w}$ for all $w\in[\ell]$. According to~\eqref{expprt}, the function $\pr_T(\p^{XY})$ has the form~\eqref{condschur} with $f:\prob(m)\to\mbb{R}$ being the sublinear functional 
\be
f(\p)\eqdef\max_{w\in[n']}\s_w\cdot\p^\da\quad\quad\forall\;\p\in\prob(m)\;.
\ee
Since the function above is convex and symmetric (under permutations) we get from Theorem~\ref{thmschurcon} that  $\pr_T(\p^{XY})$ is conditionally Schur concave and in particular $\pr_T(\p^{XY})\geq \pr_T(\q^{XY'})$. 
\end{proof}

The subsequent exercise establishes a one-to-one correspondence (bijection) between the set of all $m\times \ell$ $T$-matrices and all $m\times \ell$ $S$-matrices.

\bex\label{ex:tas}
Use~\eqref{sxw} to find an $m\times m$ matrix $U$ such that $S=UT$. Show that $U$ is invertible by computing its inverse, and use that to show that for any matrix $S\in\mbb{R}^{m\times \ell}_+$ whose components satisfy~\eqref{order1geq}, the matrix $U^{-1}S$ has non-negative entries.
\eex

\subsubsection{Operational Characterization of Conditional Majorization}\index{conditional majorization}

 The reward function above can serve as the basis for defining conditional majorization. Recall that majorization can be formally established through the concept of $k$-gambling games: $\p\succ\q$ if, for every $k\in[m]$, a player has a higher probability of winning a $k$-gambling game using the $\p$-dice rather than the $\q$-dice. In a similar vein, we can operationally define conditional majorization as follows: $\p^{XY}\succ_X\q^{XY}$ if, for all sub-stochastic matrices $T$ (over all dimensions $\ell\in\mbb{N}$), a player enjoys superior odds of winning a $T$-gambling game when using the $\p^{XY}$ dice-pair as opposed to the $\q^{XY'}$ dice-pair. We show now that this operational definition of conditional majorization coincides with Definition~\ref{def:cmc} that follows from the axiomatic and constructive approaches\index{constructive approach}.

\begin{myt}{\color{yellow} Conditional Majorization and Games of Chance}
\begin{theorem}
Let $\p\in\prob(mn)$ and $\q\in\prob(mn')$. We have $\p^{XY}\succ_X\q^{XY'}$ if and only if
\be\label{thm259}
\pr_{T}\left(\p^{XY}\right)\geq \pr_{T}\left(\q^{XY'}\right)\quad\quad\forall\;T\in\stoc_{\leq}(m,n')\;,
\ee
where $\pr_T$ is the maximal probability to win a $T$-gambling game.
\end{theorem}
\end{myt}
\begin{remark}
We emphasize that the theorem above states that $\p^{XY}\succ_X\q^{XY'}$ if and only if with the $\p^{XY}$-dice pair, Alice has better odds to win all $T$-gambling games than with the $\q^{XY'}$-dice pair. Moreover, observe that instead of considering $T$-gambling games with $T\in\stoc_{\leq}(m,\ell)$ over all $\ell\in\mbb{N}$, it is sufficient to consider $\ell=n'$. That is, the dimensions of $T$ are completely determined by $X$ and $Y'$.
\end{remark}

\begin{proof}
Due to Lemma~\ref{3onedir3}, it is sufficient to prove the~\eqref{thm259} implies $\p^{XY}\succ_X\q^{XY'}$. 
Let $S\eqdef[\s_1  \cdots \s_{n'}]\in\stoc_{\leq}(m,n')$, whose columns satisfy $\s_w=\s_w^{\da}$ for all $w\in[n']$. From Exercise~\ref{ex:tas} it follows that there exists a sub-stochastic matrix $T\in\stoc_{\leq}(m,n')$ that satisfies the relation~\eqref{sxw}. Therefore,
\ba
\sum_{y\in[n]}\max_{w\in[n']}\s_w\cdot\p_y&=\pr_{T}\left(\p^{XY}\right)\\
\GG{\eqref{thm259}}&\geq \pr_{T}\left(\q^{XY'}\right)\\
&=\sum_{y'\in[n']}\max_{w\in[n']}\s_{w}\cdot\q_{y'}\\
&\geq\sum_{y'\in[n']}\s_{y'}\cdot\q_{y'}\;.
\ea
Since the above inequality holds for all $S\eqdef[\s_1  \cdots \s_{n'}]\in\stoc_{\leq}(m,n')$, whose columns satisfy $\s_w=\s_w^{\da}$ for all $w\in[n']$, we conclude from Theorem~\ref{thm:pcmq} that $\p^{XY}\succ_X\q^{XY'}$. This completes the proof.
\end{proof}

\section{Notes and References}

The book by~\cite{MOA2011} is dedicated solely to the theory of majorization. Also the book by~\cite{Bhatia1997} is a good source, particularly, the second chapter covers majorization. Lemma~\ref{ttransform} goes back to~\cite{Muirhead1902} and~\cite{HLP1929}, and the Schur's test (Theorem~\ref{scht}) is due to~\cite{Schur1923} and~\cite{Ostrowski1952}.

Approximate majorization\index{approximate majorization} was introduced in~\cite{HOS2018}, and we will see later on that the concept of the flattest $\eps$-approximation plays a useful role in several resource theories. 

Relative majorization is the backbone of the resource theoretic approach to quantum thermodynamics. It was studied under different names such as $d$-majorization in~\cite{Veinott1971}, matrix majorization in~\cite{Dahl1999}, and thermo-majorization in~\cite{HO2013}. The ideas of the proof of main characterization theorem of relative majorization (Theorem~\ref{chararm}) goes back to~\cite{Blackwell1953}. More details were given by~\cite{RSS1978} and~\cite{Joe1990}. Independent proof was also given more recently by~\cite{Dahl1999} by employing techniques from convex analysis. To the author knowledge, the proof we presented here did not appear elsewhere. 

In the proof of Theorem~\ref{vertices} we followed~\cite{Renes2016}.
Theorems~\ref{bminmax} and~\ref{onlyr} as well as Exercise~\ref{1lem1} are due to~\cite{GT2021}.
The vector $\r$ that appears in Theorem~\ref{onlyr} was first introduced in~\cite{BHN+2015} in the context of thermodynamics. 

The characterization of the trumping relation (Theorem~\ref{ktt}) was proved by~\cite{Turgut2007} and independently by~\cite{Klimesh2007}. Both proofs are very complicated and it is an open problem to find a simpler/shorter proof of the theorem. The symmetric rewrite of this theorem in terms of R\'enyi divergences, and the characterization of catalytic majorization\index{catalytic majorization} (Theorem~\ref{thm:char2}), are due to~\cite{GT2021}.

Conditional majorization was first introduced by~\cite{GGH+2018} in the context of the quantum uncertainty principle. More recently, its relation to games of chance was introduced by~\cite{BGG2021}, and its quantum version was given by~\cite{GWB+2022}. In Sec.~\ref{twodcases} we saw that for the case $|X|=2$ conditional majorization becomes equivalent to relative submajorization. Relative submajorization has some applications in thermodynamics, and it was first introduced and studied by~\cite{Renes2016}.

\chapter{Divergences and Distance Measures}\label{chadiv}\index{divergence}\index{metric}

This chapter explores methods to quantify the distinguishability between entities such as probability distributions and quantum states. Unlike generic vectors, mathematical objects like probability vectors and quantum states embody information about physical systems. Consequently, their distinguishability is typically measured using functions attuned to this inherent information.
Consider this example: Alice possesses a system in her laboratory that is either in state $\rho$ (for instance, an electron with its spin oriented in the $z$-direction) or in state $\sigma$ (such as the same electron with spin in the $x$-direction). Alice can attempt to discern the state of her system (whether it is $\rho$ or $\sigma$) by performing a quantum measurement on it. The underlying principle is that the greater the distinguishability between $\rho$ and $\sigma$, the easier (or more likely) it is for Alice to accurately identify which of the two states her system is in.

\begin{figure}[h]\centering
    \includegraphics[width=0.5\textwidth]{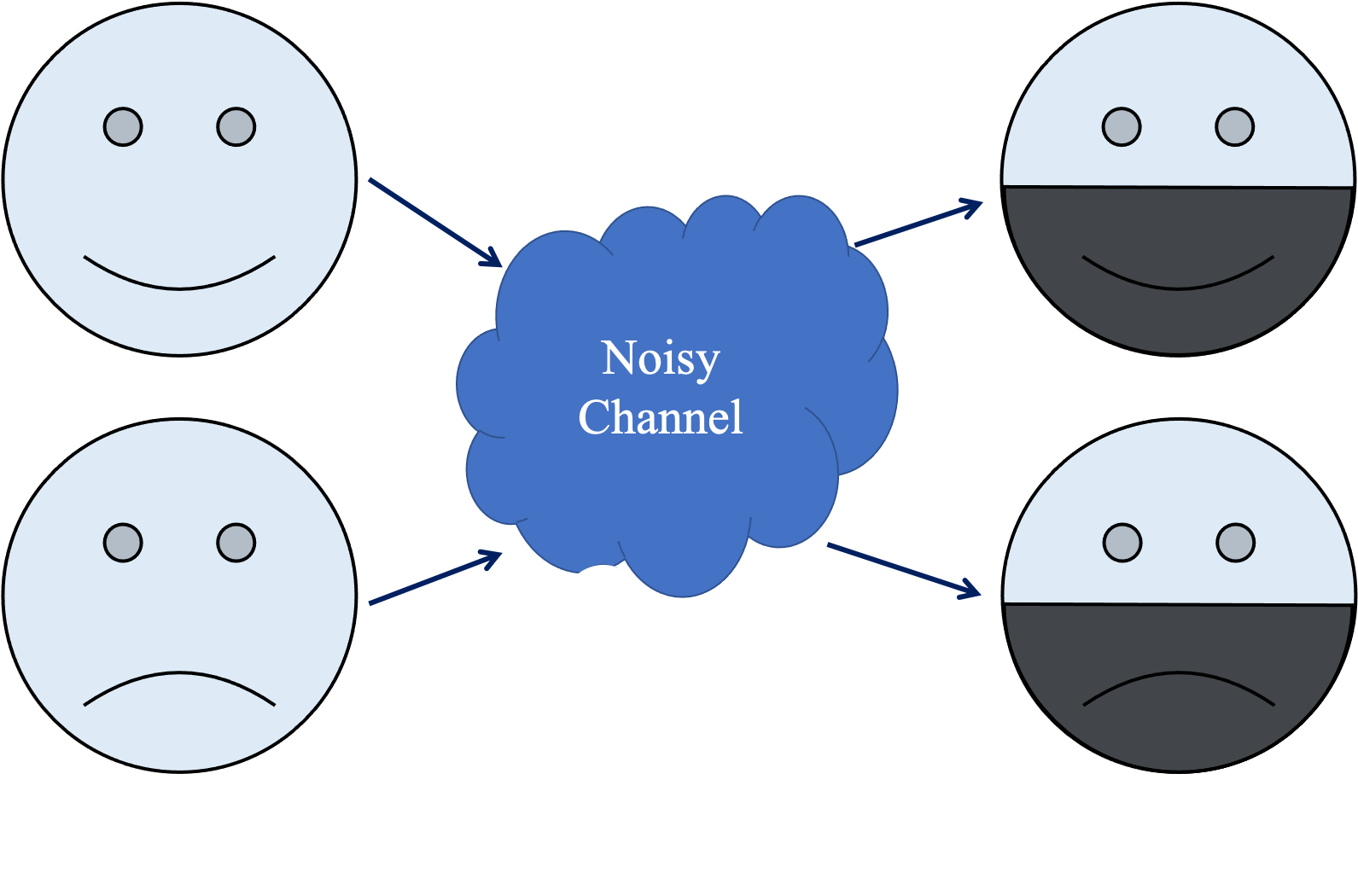}
  \caption{A noisy process makes objects look more similar.}
  \label{smilely}
\end{figure}

In any task involving distinguishability, such as the one mentioned earlier, a key observation is that sending a system (like the electron in Alice's lab) through a quantum communication channel does not enhance Alice's ability to differentiate between two states, $\rho$ and $\sigma$. This implies that if $\mE\in\cptp(A\to B)$ represents a quantum channel, the states $\mE(\rho)$ and $\mE(\sigma)$ that result from this channel are less distinguishable than the original states $\rho$ and $\sigma$ (this concept is visually illustrated in Fig.~\ref{smilely}). In essence, any measure that quantifies the distinguishability between two quantum states $\rho$ and $\sigma$ must decrease (or at most stay the same) under any quantum process that transforms the pair $(\rho,\sigma)$ into $(\mE(\rho),\mE(\sigma))$. Functions that adhere to this principle are known as quantum divergences. Their characteristic of reducing in value under such transformations is often referred to as the \emph{data processing inequality} (DPI).\index{data processing inequality} 

Quantum divergence extends the concept of divergences from classical to quantum realms. In a classical context, divergences are functions that behave monotonically under transformations that map a pair of probability vectors $(\p,\q)$ to $(E\p,E\q)$, with $E$ being a column stochastic matrix. As a result, many metrics in $\mbb{R}^n$, like the Euclidean distance, do not serve well for quantifying distinguishability between two probability vectors. It's also noteworthy that divergences are functions that behave monotonically under relative majorization.  Therefore, the tools developed in Chapter~\ref{ch:majorization} will be very useful in this context as well.

\section{Classical Divergences}

We begin by presenting the formal definition of a classical divergence. Let $\D$ represent a function defined as
\be\label{dbdc}
\D:\bigcup_{n\in\mbb{N}}\Big\{\prob(n)\times\prob(n)\Big\}\to\mbb{R}\cup\{\infty\}\;,
\ee 
which operates on pairs of probability vectors across all finite dimensions.

\begin{myd}{Classical Divergence}\index{classical divergence}
\begin{definition}\label{defd} 
The function $\D$, as defined in~\eqref{dbdc}, is termed a \emph{divergence} provided it fulfills these two conditions:
\begin{enumerate}
\item Data Processing Inequality (DPI)\index{data processing inequality} 
\be\label{cdpi}
\D\big(E\p\big\|E\q\big)\leq \D(\p\|\q)
\ee
for all $n,m\in\mbb{N}$, all $\p,\q\in\prob(n)$, and all $E\in\stoc(m,n)$.
\item Normalization, $\D(1\|1)=0$.
\end{enumerate}
\end{definition}
\end{myd}

Note that for the trivial dimension $n=1$, $\prob(n)$ contains only the number one. In this dimension, we require the divergence to be zero. Functions as above that satisfy the DPI but with $\D(1\|1)\neq 0$ will be called \emph{unnormalized divergences}. Moreover, observe that the DPI property of a divergence $\D$ can also be viewed as monotonicity under relative majorization. That is, we can state the first property above as follows: For all $\p,\q\in\prob(n)$ and $\p',\q'\in\prob(m)$ such that $(\p,\q)\succ (\p',\q')$ we have
\be
\D(\p\|\q)\geq \D(\p'\|\q')\;.
\ee
We now discuss a few basic properties of classical divergences.

\subsection{Basic Properties}

Divergences are non-negative since the one-row matrix $[1,\ldots,1]$ is column stochastic matrix in $\stoc(1,n)$, so that for any pair of probability vectors $\p,\q\in\prob(n)$
\be
\D(\p\|\q)\geq \D\big([1,\ldots,1]\p\big\|[1,\ldots,1]\q\big)=\D(1\|1)=0\;,
\ee
where the inequality follows from the DPI. Moreover, any probability vector $\p\in\prob(n)$ with dimension $n>1$ can be viewed as a preparation Channel\index{preparation channel} (i.e. one-column stochastic matrix) $\p\in\stoc(n,1)$, so that
\be
\D(\p\|\p)=\D\big(\p\cdot1\big\|\p\cdot 1\big)\leq \D(1\|1)=0\;,
\ee
where again we used the DPI for divergences. Combining this with the non-negativity property of divergences, we conclude that for any state $\p\in\prob(n)$ in any dimension $n\in\mbb{N}$
\be
\D(\p\|\p)=0\;.
\ee
This is consistent with the intuition that divergences quantify the distinguishability between two states.
An interesting question remaining is whether the converse of the above property also holds. That is, the question is whether $\D(\p\|\q)=0$ necessarily implies that $\p=\q$. This property is called \emph{faithfulness} and we will see later on that there exists divergences that are not faithful.

Another interesting property that divergences  satisfy is the following lower and upper bounds for any $\p,\q\in\prob(n)$:
\be
\D\left(\e_1\big\|\lambda_{\min}\e_1+(1-\lambda_{min})\e_2\right)\leq\D(\p\|\q)\leq \D\left(\e_1\big\|\lambda_{\max}\e_1+(1-\lambda_{max})\e_2\right)
\ee
where $\lambda_{\max}$ and $\lambda_{\min}$ are defined in~\eqref{4p99}, and $\{\e_1,\e_2\}$ is the computational basis of $\mbb{R}^2$. This property follows from a combination of the DPI with Theorem~\ref{bminmax}.

We now move to discuss some examples of divergences.
In the previous section, particularly Theorem~\ref{chararm}, we found several characterization for relative majorization. Some of the characterizations are given in terms of divergences (although not explicitly). For example, fix $t\geq 1$ and define for all $n\in\mbb{N}$ and all $\p,\q\in\prob(n)$ the function 
\be\label{tfdiv}
D_t(\p\|\q)\eqdef\sum_{x\in[n]}(p_x-tq_x)_+\;.
\ee
Since we assume $t\geq 1$ we have $D_t(\p\|\p)=0$ for all $\p\in\prob(n)$. To show that $D_t$ satisfies the DPI we can use the relation $(r)_+=(|r|+r)/2$ for all $r\in\mbb{R}$ to express $D_t$ as
\be
D_t(\p\|\q)=\frac12\left(\|\p-t\q\|_1+1-t\right)\;.
\ee
Consequently, the property outlined in~\eqref{dpisv} implies that $\{D_t\}_{t\geq 1}$ is a family of divergences. Moreover, 
from Theorem~\ref{chararm} we learn that this family of classical divergences can be used to characterize relative majorization; i.e., for all $\p,\q\in\prob(n)$ and $\p',\q'\in\prob(n')$ we have (see Exercise~\ref{ppqq2})
\be\label{ppqq}
(\p,\q)\succ(\p',\q')\quad\quad\iff\quad\quad D_t(\p\|\q)\geq D_t(\p'\|\q')\quad\quad\forall\;t\geq 1\;.
\ee

\bex\label{ppqq2}
Prove~\eqref{ppqq}. Hint: Use Theorem~\ref{chararm} in conjunction with Corollary~\ref{corptq}.
\eex

\bex
Let $\D$ be a divergence, and $P$ be an $n\times n$ permutation matrix. Show that
\be\label{invup}
\D(\p\|\q)=\D(P\p\|P\q)\quad\quad\forall\;\p,\q\in\prob(n)\;.
\ee
\eex

There is another family of divergences, consisting of numerous divergences (that appear in applications), and also includes the above example as a special case. This family is called the $f$-Divergence\index{$f$-divergence}.

\subsection{The $f$-Divergence}\index{$f$-divergence}

\begin{myd}{}
\begin{definition}\label{cfdiv}
Let $f:(0,\infty)\to\mbb{R}$ be a convex function that satisfy $f(1)=0$. Then, the \emph{$f$-Divergence\index{$f$-divergence}} is defined for any $\p,\q\in\prob(n)$ by
\be\label{ffdiv}
D_f(\p\|\q)\eqdef\sum_{x\in[n]}q_xf\left(\frac{p_x}{q_x}\right)
\ee
with the following conventions:
\be\nonumber
f(0)\eqdef\lim_{r\to 0^+}f(r)\quad,\quad0f\left(\frac00\right)\eqdef 0\quad,\quad0f\left(\frac{a}0\right)\eqdef\lim_{r\to 0^+} rf\left(\frac ar\right)=a\lim_{s\to 0+}sf\left(\frac1s\right)\;.
\ee
\end{definition}
\end{myd}

\begin{remark}
We do not assume that $f$ is positive nor that \be\label{tildef} \tilde{f}(0)\eqdef\lim_{s\to 0+}sf\left(\frac1s\right)\ee is finite. Therefore, for some convex functions as above, we can have $D_f(\p\|\q)=\infty$ for some choices of $\p,\q\in\prob(n)$. From the theorem below it will follow that $D_f$ is a divergence and therefore is always non-negative (even if $f(x)$ is negative for some $x\in(0,\infty)$).
Furthermore, observe that the $f$-Divergence\index{$f$-divergence} can be expressed for any $\p,\q\in\prob(n)$ as
\be\label{ffdiv2}
D_f(\p\|\q)=\tilde{f}(0)\sum_{x\not\in\supp(\q)}p_x+\sum_{x\in\supp(\q)}q_xf\left(\frac{p_x}{q_x}\right)\;,
\ee
where we split the sum in~\eqref{ffdiv} into a sum over all $x\in[n]$ with $q_x=0$ and over all $x\in[n]$ with $q_x\neq 0$. 
\end{remark}

\bex\label{ex:512}
Show that for every $t\geq 1$ the function $D_t$ as defined in~\eqref{tfdiv} is an $f$-Divergence\index{$f$-divergence}. 
\eex

\begin{exercise}\label{cont611}
Show that the definition above for the $f$-Divergence\index{$f$-divergence} is equivalent to the following definition. Let $f:(0,\infty)\to\mbb{R}$ be a convex function and define $f(0)\eqdef\lim_{\eps\to 0^+}f(\eps)$. Then, the $f$-Divergence\index{$f$-divergence} is defined as in~\eqref{ffdiv} for $\p,\q\in\prob(n)$ with $\q>0$, and for $\q\not>0$,
\be\label{0615}
D_{f}(\p\|\q)\eqdef\lim_{\eps\to 0^+}D_{f}\left(\p\big\|(1-\eps)\q+\eps\u\right)\;,
\ee
where $\u$ is the uniform distribution in $\prob(n)$.
\end{exercise}

\begin{exercise}
Let $f:(0,\infty)\to\mbb{R}$ be a convex function that satisfy $f(1)=0$, and let $\tilde{f}(r)\eqdef rf\left(\frac1r\right)$.
Show that $\tilde{f}$ is also convex with $\tilde{f}(1)=0$ and prove that
\be
D_{f}(\p\|\q)=D_{\tilde{f}}(\q\|\p)\;.
\ee
\end{exercise}

\begin{myt}{}
\begin{theorem}\label{fdivt}
Let $f:(0,\infty)\to\mbb{R}$ be a convex function that satisfy $f(1)=0$. 
Then, $D_f$ as defined in Definition~\ref{cfdiv} is a classical divergence.
\end{theorem}
\end{myt}

\begin{proof}
The normalization condition $D_f(1|1)=0$ is directly derived from the requirement that $f(1)=0$. To illustrate the data processing inequality, consider $m, n \in \mbb{N}$, a stochastic matrix $E \in \stoc(m, n)$, and probability vectors $\p, \q \in \prob(n)$. Define $\r \eqdef E\p$ and $\s \eqdef E\q$. For each $x \in [m]$ and $y \in [n]$, let $e_{x|y}$ represent the $(x, y)$-component of $E$.

With these definitions, the $x$-components of $\r$ and $\s$ are respectively $r_x = \sum_{y \in [n]} e_{x|y} p_y$ and $s_x = \sum_{y \in [n]} e_{x|y} q_y$. Assuming initially that $\q > 0$, we find that if $s_x = 0$, then $e_{x|y} = 0$ for all $y \in [n]$. Consequently, if $s_x = 0$, it follows that $r_x$ must also be 0. This leads to the conclusion that
\be\label{twosums}
D_f(E\p\|E\q)=D_f(\r\|\s)=\sum_{ x\in\supp(\s)}s_xf\left(\frac{r_x}{s_x}\right)\;,
\ee
where the summation is limited to all $x \in [m]$ for which $s_x \neq 0$. If $s_x = 0$, then $r_x$ is also 0, which contributes $0f(\frac{0}{0}) = 0$ to the sum in~\eqref{ffdiv}, with $(\r, \s)$ replacing $(\p, \q)$.

The strategy of the proof involves representing $r_x/s_x$, as seen on the right-hand side of~\eqref{twosums}, as a convex combination of the ratios ${p_y/q_y}{y\in[n]}$. This is achieved by defining, for each $x\in\supp(\s)$,
\be\label{tdyix}
t_{y|x}\eqdef\frac{e_{x|y}q_y}{\sum_{y'}e_{x|y'}q_{y'}}=\frac{e_{x|y}q_y}{s_x}\;.
\ee
It is important to note that for every $x\in\supp(\s)$, the set $\{t_{y|x}\}_{y\in[n]}$ forms a probability vector, and for all $x\in\supp(\s)$, it holds that
\be
\frac{r_x}{s_x}=\sum_{y\in[n]}t_{y|x}\frac{p_y}{q_y}\;.
\ee
Inserting this expression into~\eqref{twosums}, we obtain
\ba
D_f(E\p\|E\q)&=\sum_{x\in\supp(\r)}s_xf\Big(\sum_{y\in[n]}t_{y|x}\frac{p_{y}}{q_y}\Big)\\
\GG{\text{by convexity}}&\leq \sum_{x\in\supp(\s)}\sum_{y\in[n]}t_{y|x}s_xf\left(\frac{p_{y}}{q_y}\right)\\
\GG{\eqref{tdyix}}&=\sum_{y\in[n]}\sum_{x\in\supp(\s)}\!\!\!\!\!e_{x|y}q_yf\left(\frac{p_{y}}{q_y}\right)\\
\Gg{\sum_{x\in\supp(\s)}\!\!\!\!\!\!e_{x|y}=1}&=\sum_{y\in[n]}q_yf\left(\frac{p_{y}}{q_y}\right)=D_f(\p\|\q)\;,
\ea
where the equality $\sum_{x\in\supp(\s)} e_{x|y} = 1$ is valid because $e_{x|y} = 0$ if $s_x = 0$.

For the case $\q\geq 0$ define $\q_\eps\eqdef(1-\eps)\q+\eps\u$. We then get that $\q_\eps>0$ so that for any $\eps>0$
\be
D_f(E\p\|E\q_\eps)\leq D_f(\p\|\q_\eps)\;.
\ee
Taking the limit $\eps\to 0^+$ on both sides of the equation above and using continuity of $D_f(\p\|\q)$ in $\q$ (see Exercise~\ref{excontiq}) completes the proof.
\end{proof}

\begin{exercise}[Continuity of $D_f(\p\|\q)$ in $\q$]\label{excontiq}
In the last part of the proof above we used the fact that the $f$-Divergence\index{$f$-divergence} is continuous in $\q$. Show that in general, if $\{\q_k\}_{k\in\mbb{N}}$ is a sequence of probability vectors in $\prob(n)$ that satisfies $\q_k\to\q$ as $k\to\infty$ then
\be
\lim_{k\to\infty}D_f(\p\|\q_k)=D_f(\p\|\q)\;.
\ee
Hint: Observe that $f$ is continuous in $(0,\infty)$ since it is convex.
\end{exercise}

\begin{myg}{}
\begin{corollary}
Let $n,m\in\mbb{N}$, $\p,\q\in\prob(n)$, and $\p',\q'\in\prob(m)$. The condition $(\p,\q)\succ(\p',\q')$ holds if and only if
for every convex function $f(0,\infty)\to\mbb{R}$ with $f(1)=0$
\be
D_f(\p\|\q)\geq D_f(\p'\|\q')\;.
\ee
\end{corollary}
\end{myg}

\begin{exercise}
Prove the corollary above. Hint: Use~\eqref{ppqq}, Exercise~\ref{ex:512}, and Theorem~\ref{fdivt} to prove the above corollary.
\end{exercise}

\subsection{Examples}

In this subsection we give several examples of $f$-Divergence\index{$f$-divergence}s that play important role in applications.

\subsubsection{Kullback–Leibler divergence}\index{KL-divergence}

The Kullback–Leibler divergence (also known as the KL-divergence or the relative entropy) is perhaps the most  well known divergence which appears in numerous applications in statistics, information theory, and as we will see in resource theories. For this reason, it is the only divergence that we will denote simply by $D$ without any subscript. It is the $f$-Divergence\index{$f$-divergence} that corresponds to the function 
$f(r)=r\log r$. For this choice, we get,
\be\label{kldiv}
D(\p\|\q)=
\begin{cases}
\sum_{x\in[n]}p_x\big(\log p_x-\log q_x\big) &\text{if }\p\ll\q\\
\infty &\text{otherwise}
\end{cases}
\ee
where $\p\ll\q$ denotes $\supp(\p)\subseteq\supp(\q)$, and we use the convention $0\log0=0$. In the next chapters we will study the many properties of this divergence.

\subsubsection{The Trace Distance}\index{trace distance}

The trace distance, also known as the total variation distance (sometimes also called statistical distance), is an $f$-Divergence\index{$f$-divergence} with $f(r) = \frac{1}{2}|r-1|$. For this convex function we get
\be\label{fdclass}
D_f(\p\|\q)=\sum_{x\in[n]}q_x\frac12\left|\frac{p_x}{q_x}-1\right|=\frac12\sum_{x\in[n]}\left|p_x-q_x\right|=\frac12\|\p-\q\|_1\;.
\ee
This $f$-Divergence\index{$f$-divergence}, which also functions as a metric, will be examined in detail in the subsequent sections.

\subsubsection{The Hellinger Distance}\index{Hellinger distance}

This is another distance measure that is closely related an $f$-Divergence\index{$f$-divergence} with $f(r)=\frac12(\sqrt{r}-1)^2$. For this $f$ we get
\be
D_f(\p\|\q)=\sum_{x\in[n]}q_x\frac12\left(\sqrt{\frac{p_x}{q_x}}-1\right)^2=\frac12\sum_{x\in[n]}\left(\sqrt{p_x}-\sqrt{q_x}\right)^2
\ee
The Hellinger distance is defined as the square root of the above expression
\be
H(\p,\q)\eqdef\sqrt{\frac12\sum_{x\in[n]}\left(\sqrt{p_x}-\sqrt{q_x}\right)^2}
\ee
We will see later on that the above divergence is also a metric\index{metric} that is closely related to a quantity known as the fidelity.

\subsubsection{The $\alpha$-Divergence}\index{$\alpha$-divergence}

The $\alpha$-Divergence is an $f$-Divergence\index{$f$-divergence} corresponding to $f_\alpha(r)=\frac{r^\alpha-r}{\alpha(\alpha-1)}$, where $\alpha\in[0,\infty)$,  where the case $\alpha=1$ is defined by the limit 
$\lim_{\alpha\to 1}\frac{r^\alpha-r}{\alpha(\alpha-1)}=r\ln(r)$ (which yields the KL-divergence), and similarly the case $\alpha=0$ is given by $-\ln(r)$. For this choice of $f$ we get
\be\label{aldiv}
D_{f_\alpha}(\p\|\q)=\sum_{x\in[n]}q_x\frac1{\alpha(\alpha-1)}\left(\left(\frac{p_x}{q_x}\right)^\alpha-\frac{p_x}{q_x}\right)
=\frac1{\alpha(\alpha-1)}\left(\sum_{x\in[n]}p_x^\alpha q_x^{1-\alpha}-1\right)
\ee
The $\alpha$-Divergence\index{$\alpha$-divergence} can be expressed as a function of the R\'enyi divergences that we will study in the Chapter~\ref{ch:relent}.

\begin{exercise}
Show that all the functions $f$ above are convex and satisfy $f(1)=0$.
\end{exercise}

\begin{exercise}[The Jensen–Shannon Divergence]
Let $f:(0,\infty)\to\mbb{R}$ given by
\be
f(r)=(r+1)\log\left(\frac{2}{r+1}\right)+r\log r\quad\quad\forall\;r\in\mbb{R}\;.
\ee
Show that $f$ is convex with $f(1)=0$ and compute its $f$-Divergence\index{$f$-divergence}.
\end{exercise}

\subsection{Continuity of Divergences}\index{continuity}

Divergences are generally not continuous over $\prob(n)\times\prob(n)$, even when they exhibit continuity in their first and/or second arguments. For example, consider the two sequences of probability vectors 
\be
\p_k\eqdef\left(\frac1k,1-\frac1k\right)^T\quad\text{and}\quad\q_k\eqdef\left(\frac1{k^2},1-\frac1{k^2}\right)^T\quad\quad\forall\;k\in\mbb{N}\;.
\ee
Clearly, we have
\be\label{644}
\p\eqdef\lim_{k\to\infty}\p_k=\begin{bmatrix}0 \\ 1\end{bmatrix}\quad\text{and}
\quad\q\eqdef\lim_{k\to\infty}\q_k=\begin{bmatrix}0 \\ 1\end{bmatrix}\;.
\ee
Consider now the $\alpha$-Divergence\index{$\alpha$-divergence} with $\alpha=2$ as defined in~\eqref{aldiv}. Denote this divergence by 
\be
D'(\p\|\q)=\frac1{2}\left(\sum_{x\in[n]}p_x^2 q_x^{-1}-1\right)
\ee
It is simple to check (see the exercise below) that
\be\label{645}
\lim_{k\to\infty}D'(\p_k\|\q_k)=\frac12\neq 0=D'(\p\|\q)\;.
\ee
For the $\alpha$ divergences, this discontinuity only happens when the limits $\p,\q$ are on the boundary. In fact, we will see in Chapter~\ref{ch:relent} that many divergences are continuous in the interior of $\prob(n)\times\prob(n)$.

\begin{exercise}
Let $D'$ be the $\alpha$-Divergence\index{$\alpha$-divergence} for $\alpha=2$, and let $\p_k$ and $\q_k$ as in~\eqref{644}. Prove Eq.~\eqref{645}.
\end{exercise}

We show now that by utilizing the data processing inequality, if a divergence is continuous in one of its arguments it is necessarily continuous in the second argument as well. 

\begin{myt}{}
\begin{theorem}\label{thmcontii}
Let $\D$ be a divergence. The following statements are equivalent:
\ben
\item For every fixed $\q\in\prob(n)$ the function $\p\mapsto\D(\p\|\q)$ is continuous in $\prob(n)$.
\item For every fixed $\p\in\prob(n)$ the function $\q\mapsto\D(\p\|\q)$ is continuous in $\prob(n)$.
\een
\end{theorem}
\end{myt}
 
\begin{proof}
We will demonstrate the implication of $1\Rightarrow 2$. The converse, $2\Rightarrow 1$, will be established using a similar approach.

Let $\q,\q'\in\prob(n)$, and define the channel $E\in\stoc(n,n)$  by its action on every $\v\in\prob(n)$ as
\be
	E\v \eqdef (1-\eps) (\v -\q) +\q' \quad\text{where}\quad\eps\eqdef1-\min_{x\in[n]}\frac{q_x'}{q_x}\;.  
\ee
By definition, $E\q=\q'$, 
and $\eps\in(0,1)$ can be reduced to an arbitrarily small value by taking $\q$ and $\q'$ to have sufficiently small trace distance $\frac12\|\q-\q'\|_1$. Note also that from the definition of $\eps$ it follows that $\q'-(1-\eps)\q\geq 0$. Therefore, for every $\v\in\prob(n)$
\be
E\v=(1-\eps)\v+\big(\q'-(1-\eps)\q\big)\geq 0\;.
\ee
The inequality above implies that  $E$ is indeed column stochastic. Moreover, by definition
\be
	\| \p - E\p  \|_1 = \big\| \eps( \p - \q ) + \q' - \q \big\|_1 \leq \eps + \| \q - \q' \|_1\;.	
\ee
This inequality demonstrates that as $\q$ converges towards $\q'$, the vector $E\p$  approaches $\p$ (recall that $\eps$ goes to zero as $\q$ approaches $\q'$). Applying the DPI with $E$ we can thus bound
\ba\label{o1o}
	\D(\p\|\q) - \D(\p\|\q') &\geq \D(E\p  \|E\q) - \D(\p\|\q')\\
	&=\D(E\p  \|\q') - \D(\p\|\q') \;.
\ea
Since $\D(\p\|\q)$ is continuous in $\p$, the expression on the right-hand side of the equation above vanish when $\q \to \q'$.

Subsequently, we establish a comparable upper bound for $\D(\p|\q) - \D(\p|\q')$ by introducing $\tE\in\stoc(n,n)$ and $\teps\in(0,1)$. These are defined identically to how $E$ and $\eps$ were defined, but with the roles of $\q$ and $\q'$ reversed. Specifically, $\tE$ is defined by its action on every $\v\in\prob(n)$ as
\be
	\tE\v \eqdef (1-\teps) (\v -\q') +\q \quad\text{where}\quad\teps\eqdef1-\min_{x\in[n]}\frac{q_x}{q_x'}\;.  
\ee
By definition, $\tE\q'=\q$ and $\teps\in(0,1)$. Further,
following similar steps as above, it can be verified that $\tE$ is indeed column stochastic, and $\tE\p$ approaches $\p$ as $\q$ approaches $\q'$. Utilizing the DPI we also have
\ba	
	\D(\p\|\q) - \D(\p\|\q') &\leq \D(\p \|\q) - \D(\tE\p \| \tE\q')\\
	&=\D(\p \|\q) - \D(\tE\p\| \q) \;.
\ea
Therefore, as before, due to the continuity of $\D(\p\|\q)$ in $\p$, the expression on the right-hand side of the equation above vanish when $\q' \to \q$. Combining this with the lower bound in~\eqref{o1o}, we conclude that $\D(\p\|\q)$ is  continuous in $\q$.
\end{proof}

\subsection{Divergences from Measures of Nonuniformity}\index{nonuniformity}

In this subsection, we demonstrate a one-to-one correspondence between divergences and Schur convex functions that fulfill specific criteria. This relationship allows for the construction of divergences from Schur convex functions, thereby expanding the class of $f$-Divergence\index{$f$-divergence}s. This correspondence proves particularly valuable in Chapter~\ref{ch:relent}, where we utilize it to establish a bijection\index{bijection} between entropies and relative entropies.

\begin{myd}{A Measure of Nonuniformity}
\begin{definition}\label{mesnon}
A function 
\be
g: \bigcup_{n \in \mbb{N}} \prob(n) \to \mbb{R} \cup \{\infty\}
\ee 
is called a measure of nonuniformity\index{nonuniformity} if it satisfies the following three properties:
\begin{enumerate}
	\item For every $n \in \mbb{N}$, the restriction of $g$ to $\prob(n)$ is Schur convex and continuous on $\prob(n)$.
	\item For $n = 1$ it is normalized to $g(1) = 0$.
	\item For all $k, n \in \mbb{N}$ and $\r \in \prob(n)$,
	$
		g\big( \r \otimes \u^{(k)} \big) = g(\r) \,.
	$
\end{enumerate}
\end{definition}
\end{myd}

In Chapter~\ref{ch:nonuniformity} we will study the resource theory of nonuniformity in which the functions above quantify the resource of this theory. Specifically, these functions quantify how different a probability vector $\p$ is from the uniform distribution $\u$. As an indication of this,
note that if $\D$ is a classical divergence\index{classical divergence} that is continuous in its first argument, then the function
\be
g_{{}_\D}(\p)\eqdef\D\left(\p\big\|\u^{(n)}\right)\quad\quad\forall\;\p\in\prob(n)\;,
\ee
is a measure of non-uniformity.

\begin{exercise}
Verify that $g_{\D}$ indeed satisfies all the three properties above. Hint: For the third property show that 
\begin{align}
	\D(\p \| \u^{(n)}) = \D\big(\p \otimes \u^{(k)} \big\| \u^{(nk)} \big)
\end{align}
as a consequence of the DPI applied twice for channels introducing and removing an independent distribution $\u^{(k)}$.
\end{exercise}

\begin{exercise}
Let $f:(0,\infty)\to\mbb{R}$ be convex with $f(1)=0$. Show that for the $f$-Divergence\index{$f$-divergence}, $D_f$, we have
\be
g_{{}_{D_{\!f}}}(\p) = \frac{1}{n} \sum_{x\in[n]} f(n \p_x)\quad\quad\forall\;\p\in\prob(n)\;.
\ee
Verify by direct calculation that this expression satisfy the three properties of $g$.
\end{exercise}

In Theorem~\ref{onlyr}, we established that for every $n \in \mbb{N}$, $\p \in \prob(n)$, and $\q \in \prob_{>0}(n) \cap \mbb{Q}^n$, there is a vector $\r\in\prob(k)$ with the property that $(\p,\q)\sim(\r,\u^{(k)})$. To elaborate, let $\q = ( \frac{k_1}{k},\ldots, \frac{k_n}{k})^T$, where each $k_x\in \mbb{N}$ and $k\eqdef k_1+\cdots+k_n$. The vector $\r$ is then expressed as:
\be\label{thervec}
\r\eqdef \bigoplus_{x\in[n]} p_x \u^{(k_x)}\;.
\ee
Building upon this equivalency, the following theorem demonstrates a bijective relationship between divergences and measures of non-uniformity. However, prior to this, it's essential to explore the uniqueness of the vector $\r$ above. 

The vector $\r$ is not unique. To see why, let $\{m_x\}_{x\in[n]}$ be a set of $n$ integers satisfying
\be
q_x=\frac{k_x}{k}=\frac{m_x}m\quad\text{where}\quad m\eqdef\sum_{x\in[n]}m_x\;.
\ee
Given any such set, we can define the probability vector $\s\eqdef\bigoplus_{x\in[n]}p_x\u^{(m_x)}$ so that $(\p,\q)\sim(\s,\u^{(m)})$. This demonstrates that $\r$ is not unique. However, since $km_x=mk_x$, we observe that:
\be\label{5p4500}
\u^{(m)}\otimes\r=\bigoplus_{x\in[n]} p_x \u^{(mk_x)}=\bigoplus_{x\in[n]} p_x \u^{(km_x)}=\u^{(k)}\otimes \s\;.
\ee
The above relation highlights that for any measure of non-uniformity $g$, following the third property of Definition~\ref{mesnon}, we obtain:
\ba
g(\r)&=g\left(\u^{(m)}\otimes\r\right)\\
\GG{\eqref{5p4500}}&=g\left(\u^{(k)}\otimes\s\right)\\
&=g(\s)\;,
\ea
where the last equality again utilizes the third property of Definition~\ref{mesnon}. Thus, in this context, $\r$ and $\s$ have the same non-uniformity.
 
\begin{myt}{\color{yellow} Bijection\index{bijection} between Divergences and Measures of Nonuniformity}\index{nonuniformity}
\begin{theorem}\label{biji}
	Let $g$ be a measure of non-uniformity. For any $n \in \mbb{N}$, $\p \in \prob(n)$ and $\q \in \prob_{>0}(n) \cap \mbb{Q}^n$, define
	\be
		D_g(\p\|\q) \eqdef g (\r) ,  \label{eq:Dg}
	\ee
	where $\r\in\prob(k)$ is the vector defined in~\eqref{thervec}. For general $\q \in \prob(n)$, $D_g$ is defined via a continuous extension. Then, $D_g$ is a divergence and continuous in $\q$ for any fixed $\p \in \prob(n)$.
\end{theorem}
\end{myt}

\begin{remark}
Observe that $D_g$ in~\eqref{eq:Dg} is well defined since $g(\r)=g(\s)$ for any other vector $\s$ as defined above. We will also show in the proof below that the continuous extension for general $\q\in\prob(n)$ is well defined.
\end{remark}

\begin{proof}
We first show that $D_g$ is a divergence on the restricted space in which $\q \in \prob_{>0}(n) \cap \mbb{Q}^n$. The normalization of $D_g$ holds since $D_g(1\|1) = g(1) = 0$. To show the DPI, let $\p \in \prob(n)$, $\q \in \prob_{>0}(n) \cap \mbb{Q}^n$, and $E \in \stoc(m,n)\cap\mbb{Q}^{m\times n}$ be a stochastic matrix (channel) with rational components. Let further $k\in\mbb{N}$ be large enough such that we can express
\begin{align}
	\q= \left(\frac{k_1}k, \ldots, \frac{k_n}k\right)^T, \quad \q' \eqdef E\q  =\left(\frac{k_1'}k, \ldots ,\frac{k_m'}k\right)^T, \label{eq:qq}
\end{align}
where $k_x,k_x'\in\mbb{N}$ with $\sum_{x\in[n]}k_x=\sum_{x\in[m]}k_x'=k$.
Due to Theorem~\ref{onlyr} there exists $\r,\s\in \prob(k)$ such that $(\p,\q) \sim (\r,\u^{(k)})$ and $(E\p , E\q ) \sim (\s,\u^{(k)})$. 
By definition,
$(\p,\q) \succ (E\p , E\q )$ so that $(\r,\u^{(k)}) \succ (\s,\u^{(k)})$, or equivalently, $\r \succ \s$. Hence,
\begin{align}
D_g(E\p  \|E \q ) = g(\s) \leq g(\r) = D_{g}(\p\|\q)\;,
\end{align}
where the inequality follows from the Schur convexity of $g$.

Next, we must demonstrate continuity\index{continuity} in the second argument within $\prob_{>0}(n) \cap \mbb{Q}^n$ to ensure the continuous extension is well defined. As inferred from~\eqref{eq:Dg}, for any fixed $\q\in\prob_{>0}(n)\cap\mbb{Q}^n$, the function $D_g(\p\|\q)$ is continuous in $\p\in\prob(n)$ due to the continuity of $g$. Combining this with Theorem~\ref{thmcontii} we conclude that indeed the function $D_g(\p\|\q)$ is also continuous in the second argument $\q\in\prob_{>0}(n) \cap \mbb{Q}^n$.
Finally, observe that the DPI remains valid when we define the quantity for irrational $\q$ via continuous extension. This completes the proof.
\end{proof}

\begin{exercise}
Describe explicitly the bijection\index{bijection} between divergences that are continuous in their second argument and measures of non-uniformity. That is, for any $g$ express the corresponding $\D$ and vice versa.
\end{exercise}

\section{Quantum Divergences}\index{quantum divergences}

In this section we study the quantum version of a classical divergence. We start with their formal definition and some of their basics properties, and then move to study a systematic approach to extend classical divergences to the quantum domain. We then give examples focusing on the quantum extension of the $f$-Divergence\index{$f$-divergence}.
Similarly to the definition of a classical divergence, a quantum divergence is also defined in terms of the DPI.

In the following definition we consider a function that is acting on pairs of quantum states in all finite dimensions:
\be\label{dbd}\D:\bigcup_{A}\Big\{\md(A)\times\md(A)\Big\}\to\mbb{R}\cup\{\infty\}
\ee 

\begin{myd}{}\index{data processing inequality}
\begin{definition}\label{defd}
The function $\D$, as presented in~\eqref{dbd}, is termed a \emph{quantum divergence} if it fulfills the following two criteria:
\begin{enumerate}
\item Data Processing Inequality (DPI): $\D\big(\mE(\rho)\big\|\mE(\sigma)\big)\leq \D(\rho\|\sigma)$ for all $\mE\in\cptp(A\to B)$ and all $\rho,\sigma\in\md(A)$.
\item Normalization: $\D(1\|1)=0$.
\end{enumerate}
\end{definition}
\end{myd}
\begin{remark}
Note that a classical divergence\index{classical divergence} can be viewed as a quantum divergence whose domain is restricted to classical systems. 
The union in~\eqref{dbd} is over all systems $A$ and particularly over all finite dimensions $|A|$. Therefore, the domain of $\D$ consists of pairs of density matrices $(\rho,\sigma)$ in any dimension $|A|\in\mbb{N}$. For the case of a trivial system $A$ with  $|A|=1$ the only density matrix in $\md(A)$ is the number one. In this case, divergences satisfy $\D(1\|1)=0$.
\end{remark}

Like classical divergences, quantum divergences are non-negative since for any pair of states $\rho,\sigma\in\md(A)$ we have the trace $\tr\in\cptp(A\to 1)$, so that
\be
\D(\rho\|\sigma)\geq \D\big(\tr[\rho]\big\|\tr[\sigma]\big)=\D(1\|1)=0\;,
\ee
where the inequality follows from the DPI. Moreover, recall that any state $\rho\in\md(A)$ with dimension $|A|>1$ can be viewed as a preparation channel\index{preparation channel} $\rho^{1\to A}\in\cptp(1\to A)$. Hence,
\be
\D(\rho^A\|\rho^A)=\D\big(\rho^{1\to A}(1)\big\|\rho^{1\to A}(1)\big)\leq \D(1\|1)=0\;,
\ee
where again we used the DPI for divergences. Combining this with the non-negativity of divergences we conclude that for any state $\rho\in\md(A)$ in any dimension $|A|\in\mbb{N}$
\be
\D(\rho\|\rho)=0\;.
\ee
This is consistent with the intuition that divergences quantify the distinguishability between two states.

An interesting question remaining is whether the converse of the above property also holds. A quantum divergence, $\D$, is said to be \emph{faithful} if for any $\rho,\sigma\in\md(A)$, the condition $\D(\rho\|\sigma)=0$ implies $\rho=\sigma$. We will see later on that not all quantum divergences are faithful. However, in the following lemma we show that a quantum divergence is faithful if and only if its reduction to classical systems is faithful.

\begin{myg}{}
\begin{lemma}\index{faithfulness}
Let $\D$ be a quantum divergence. Then, $\D$ is faithful if and only if its reduction to classical (diagonal) states is faithful.
\end{lemma}
\end{myg}

\begin{proof}
Clearly, if $\D$ is faithful on quantum states it is also faithful on classical states as the latter is a subset of the former. Suppose now that $\D$ is faithful on classical states, and suppose by contradiction that there exists $\rho\neq\sigma\in\md(A)$ such that $\D(\rho\|\sigma)=0$. Then, there exists a basis of $A$ such that the diagonal of $\rho$ in this basis does not equal to the diagonal of $\sigma$ (see exercise below). Let $\Delta\in\cptp(A\to A)$ be the completely dephasing channel in this basis. Then, $\Delta(\rho)\neq\Delta(\sigma)$ and we get
\be
\D\big(\Delta(\rho)\big\|\Delta(\sigma)\big)\leq\D(\rho\|\sigma)=0\;.
\ee
But since $\D$ is faithful on diagonal states we get the contradiction that $\Delta(\rho)=\Delta(\sigma)$.
Hence, $\D$ is faithful also on quantum states.
\end{proof} 

\begin{exercise}
Let $\rho,\sigma\in\md(A)$. Show that if $\rho\neq\sigma$ then there exists a basis of $A$ such that the diagonal of $\rho$ in this basis does not equal to the diagonal of $\sigma$ in the same basis. 
\end{exercise}

The data processing inequality\index{data processing inequality}  implies that quantum divergences are invariant under isometries. 
That is, for any isometry channel $\mV\in\cptp(A\to B)$ we have
\be\label{inviso}
\mbb{D}\big(\mV(\rho)\big\|\mV(\sigma)\big)=\mbb{D}(\rho\|\sigma)\quad\quad\forall\;\rho,\sigma\in\md(A)\;.
\ee
To see why, recall that every isometry channel has a left inverse channel $\mR\in\cptp(B\to A)$ that satisfies $\mR^{B\to A}\circ\mV^{A\to B}=\id^A$ (see Section~\ref{isometry}). Hence, by definition of $\mR$
\ba
D(\rho\|\sigma)&=\mbb{D}\big(\mR\circ\mV(\rho)\big\|\mR\circ\mV(\sigma)\big)\\
\GG{DPI}&\leq\mbb{D}\big(\mV(\rho)\big\|\mV(\sigma)\big)\\
\GG{DPI}&\leq \mbb{D}(\rho\|\sigma)\;.
\ea 
That is, all the inequalities above must be equalities so that~\eqref{inviso} holds. 

\begin{exercise}\label{emb}
Use the invariance property of quantum divergences under isometries, to show that any classical divergence, $\D$, satisfies
\be
\D(\p\oplus 0\|\q\oplus 0)=\D(\p\|\q)\;.
\ee
\end{exercise}

\begin{exercise}
Let $\D$ be a quantum divergence, $\rho,\sigma\in\md(A)$, and $\mE\in\cptp(A\to B)$. 
\begin{enumerate}
\item Show that if there exists a channel $\mF\in\cptp(B\to A)$ such that
$\mF\circ\mE(\rho)=\rho$ and $\mF\circ\mE(\sigma)=\sigma$ then
\be
\D(\rho\|\sigma)=\D\big(\mE(\rho)\big\|\mE(\sigma)\big)\;.
\ee
\item Show that for any $\omega\in\md(B)$
\be\label{tenom}
\D(\rho\otimes\omega\|\sigma\otimes\omega)=\D(\rho\|\sigma)\;.
\ee
\end{enumerate}
\end{exercise}

\begin{exercise}
Let $\rho,\sigma\in\md(A)$ and denote by 
\be
\sup(\rho/\sigma)\eqdef\sup_{0\leq \Lambda\leq I^A}\frac{\tr[\rho \Lambda]}{\tr[\sigma \Lambda]}\;.
\ee
\begin{enumerate}
\item Show that the function above satisfies the DPI; i.e.
\be
\sup\big(\mE(\rho)/\mE(\sigma)\big)\leq \sup(\rho/\sigma)\quad\quad\forall\;\mE\in\cptp(A\to B)\;.
\ee
\item Show that
\be
\sup(\rho/\sigma)=\inf\big\{\lambda\in\mbb{R}\;:\;\lambda\sigma-\rho\geq 0\big\}\;.
\ee
\item Show that the function $f(\rho\|\sigma)\eqdef\sup(\rho/\sigma)-1$ is a quantum divergence.
\end{enumerate}
\end{exercise}

\subsubsection{Joint Convexity}\index{joint convexity}

We say that a quantum divergence $\D$ is jointly convex if for any quantum system $A$, $m\in\mbb{N}$, $\p\in\prob(m)$, and two sets, $\{\rho_x\}_{x\in[m]}$ and $\{\sigma_x\}_{x\in[m]}$ of $m$ density matrices in $\md(A)$ we have
\be
\D\Big(\sum_{x\in[n]}p_x\rho_x\Big\|\sum_{x\in[n]}p_x\sigma_x\Big)\leq \sum_{x\in[n]}p_x\D(\rho_x\|\sigma_x)\;.
\ee
Although not every quantum divergence exhibits joint convexity, the combination of joint convexity with both the property described in~\eqref{tenom}, and the invariance\index{invariance} under isometries, results in a condition that is more stringent than DPI.

\begin{myg}{}
\begin{lemma}\label{lemaa}
Let $\D$ be a function with the same domain and range as a quantum divergence that is invariant under isometries. Suppose further that $\D$ is jointly convex and satisfies~\eqref{tenom} for any quantum systems $A$ and $B$, and quantum states $\rho,\sigma\in\md(A)$ and $\omega\in\md(B)$. Then $\D$ satisfies the DPI.
\end{lemma} 
\end{myg}

\begin{proof}
Due to Stinespring dilation theorem, the invariance under isometries implies that it is sufficient to prove that for any two bipartite states $\rho,\sigma\in\md(AB)$
\be
\D\left(\rho^{AB}\big\|\sigma^{AB}\right)\geq\D\left(\rho^{A}\big\|\sigma^{A}\right)\;.
\ee
Let $k\eqdef|B|^2$ and $\mR\in\cptp(B\to B)$ be the completely randomizing (or equivalently depolarizing) channel. Such a channel takes all states to the maximally mixed state, and in Chapter~\ref{ch:qs} (particularly, Exercise~\ref{comrand}) we will see that it can be expressed as the following uniform mixture of unitary channels,
$
\mR=\frac1{n}\sum_{x\in[n]}\mU_x
$,   
where $n\eqdef|B|^2$ and each $\mU_x\in\cptp(B\to B)$ is a unitary channel.
We then get from~\eqref{tenom} that
\ba
\D\left(\rho^{A}\big\|\sigma^{A}\right)&=\D\left(\rho^{A}\otimes\u^B\big\|\sigma^{A}\otimes\u^B\right)\\
&=\D\left(\mR^{B\to B}\left(\rho^{AB}\right)\Big\|\mR^{B\to B}\left(\sigma^{AB}\right)\right)\\
{\color{red} \text{Joint Convexity }\rightarrow}&\leq \frac1{n}\sum_{x\in[n]}\D\left(\mU^{B\to B}_x\left(\rho^{AB}\right)\Big\|\mU^{B\to B}_x\left(\sigma^{AB}\right)\right)\\
{\color{red} \text{Invariance under unitaries }\rightarrow}&=\D\left(\rho^{AB}\big\|\sigma^{AB}\right)
\ea
This completes the proof.
\end{proof}

\subsection{The Quantum $f$-Divergence}\index{$f$-divergence}

In previous subsections, we observed that a wide range of classical divergences can be represented as an $f$-Divergence\index{$f$-divergence}. This subsection delves into their extension within the quantum realm. We will explore how this quantum extension gives rise to a variety of divergences that are significant in practical applications. However, it is important to note that extending classical divergences to the quantum domain generally does not yield unique results. In particular, the quantum $f$-Divergence\index{$f$-divergence} is not the only possible quantum extension. We will explore other quantum extensions of this concept in the following subsection.

To motivate the formal definition of the quantum $f$-Divergence\index{$f$-divergence}, we first discuss a useful correspondence between pairs of density matrices and pairs of probability distributions. Consider two quantum states $\rho,\sigma\in\md(A)$ with spectral decomposition (here $m\eqdef|A|$)
\be\label{sdc}
\rho=\sum_{x\in[m]}p_x|a_x\lr a_x|\quad\text{and}\quad\sigma=\sum_{y\in[m]}q_y|b_y\lr b_y|
\ee
where $\{|a_x\ra\}_{x\in[m]}$ and $\{|b_y\ra\}_{y\in[m]}$ are orthonormal bases consisting of the eigenvectors of $\rho$ and $\sigma$, respectively.
Define the probability vectors $\tilde{\p},\tilde{\q}\in\prob(m^2)$ whose components are given by
\be\label{corre}
\tilde{p}_{xy}\eqdef p_x|\la a_x|b_y\ra|^2\quad\text{and}\quad\tilde{q}_{xy}=q_y|\la a_x|b_y\ra|^2\quad\quad\forall\;x,y\in[m]\;.
\ee
Now, if $\D$ is a classical divergence\index{classical divergence} then we can extend it to $\rho,\sigma\in\md(A)$ by
\be\label{dq}
\D^q(\rho\|\sigma)\eqdef\D\left(\tilde{\p}\big\|\tilde{\q}\right)\;.
\ee
Clearly, $\D^q(\rho\|\sigma)$ is zero if $\rho=\sigma$, and in the exercise below you show that if  $\rho$ and $\sigma$ are diagonal then $\D^q(\rho\|\sigma)\eqdef\D\left({\p}\big\|{\q}\right)$, where $\p$ and $\q$ are the diagonals of $\rho$ and $\sigma$.
In the following lemma we show that $\D^q$ is  invariant under isometries.
\begin{myg}{}
\begin{lemma}\label{isoqdiv}
Let $\D$ be a classical divergence\index{classical divergence} and define $\D^q$ as in~\eqref{dq}. Then, for any isometry channel $\mV\in\cptp(A\to B)$ and any $\rho,\sigma\in\md(A)$ 
\be
\D^q\big(\mV(\rho)\big\|\mV(\sigma)\big)=\D^q(\rho\|\sigma)\;.
\ee
\end{lemma}
\end{myg}

\begin{proof}
The non-zero components of $\tilde{\p}$ and $\tilde{\q}$ remain unchanged if $\rho$ and $\sigma$ are replaced with $\mV(\rho)$ and $\mV(\sigma)$, for any isometry  $\mV\in\cptp(A\to B)$. Moreover, note that with $n\eqdef|A|$
\be
\mK\eqdef\spa\{V|a_x\ra\;:\;x\in[m]\}=\spa\{V|b_y\ra\;:\;y\in[m]\}\;,
\ee
since both $\{|a_x\ra\}_{x\in[m]}$ and $\{|b_y\ra\}_{y\in[m]}$ are bases of $A$. Hence, denoting by $n\eqdef|B|$, the additional $n-m$ zero eigenvalues of $\mV(\rho)$ (and similarly of $\mV(\sigma)$), corresponds to eigenvectors that are in the orthogonal complement of $\mK$. Hence, if $\tilde{\p},\tilde{\q}$ corresponds to $\rho$ and $\sigma$ as in~\eqref{corre}
then $\tp\oplus\0_{k}$ and $\tq\oplus\0_{k}$ corresponds to $\mV(\rho)$ and $\mV(\sigma)$, respectively, where $\0_k$ is the zero vector in dimension $k\eqdef n^2-m^2$. Hence,
\be
\D^q\big(\mV(\rho)\big\|\mV(\sigma)\big)=\D\big(\tp\oplus\0_{k}\big\|\tq\oplus\0_{k}\big)=\D\left(\tp\big\|\tq\right)=\D^{q}(\rho\|\sigma)\;,
\ee
where the second equality follows from the fact that classical divergences are invariant under embedding (see Exercise~\ref{emb}).
\end{proof}

\begin{exercise}
Show that if  $\rho$ and $\sigma$ are diagonal then $\D^q(\rho\|\sigma)\eqdef\D\left({\p}\big\|{\q}\right)$, where $\p$ and $\q$ are the diagonals of $\rho$ and $\sigma$.
\end{exercise}

Due to Lemma~\ref{isoqdiv}, the Stinespring delation implies that $\D^q(\rho\|\sigma)$ as defined above is a quantum divergence if and only if it is non-increasing under the partial trace. This later property does not hold in general, however, it does hold when $\D$ belongs to a large class of $f$-Divergence\index{$f$-divergence}s.

\begin{myd}{Quantum $f$-Divergence\index{$f$-divergence}}
\begin{definition}\label{qfdiv} Let $f:(0,\infty)\to\mbb{R}$ be an operator convex function satisfying $f(1)=0$.
Let $D_f$ be its corresponding classical $f$-Divergence\index{$f$-divergence} as defined in Definition~\ref{cfdiv}. The quantum $f$-Divergence\index{$f$-divergence}, $D_f^q$, is defined on any $\rho,\sigma\in\md(A)$ as
\be
D_f^q(\rho\|\sigma)\eqdef D_f(\tilde{\p}\|\tilde{\q})\;,
\ee
where $\tilde{\p}$ and $\tilde{\q}$ are the probability vectors in $\prob(m^2)$ as defined in~\eqref{corre}.
\end{definition}
\end{myd}

\begin{remark} 
We will see below that the requirement that $f$ is operator convex (vs just convex) ensures that $D^q_f$ is indeed a quantum divergence. Moreover, from~\eqref{0615},
for any $\rho,\sigma\in\md(A)$ with spectral decomposition as in~\eqref{sdc} we have (see exercise below)
\be\label{formulafdiv}
D_f(\rho\|\sigma)=\lim_{\eps\to 0^+}\sum_{x,y}(q_y+\eps)f\left(\frac{p_x}{q_y+\eps}\right)|\la a_x| b_y\ra|^2\;.
\ee
\end{remark}

\begin{exercise}
Prove~\eqref{formulafdiv} and use it to show that for any $\rho,\sigma\in\md(A)$
\be
D_f(\rho\|\sigma)=\sum_{\substack{ x\in\supp(\p)\\y\in\supp(\q)}}q_yf\left(\frac{p_x}{q_y}\right)|\la a_x| b_y\ra|^2+f(0)\tr\left[(I-\rho^0)\sigma\right]
+\tilde{f}(0)\tr\left[(I-\sigma^0)\rho\right]\;,
\ee
where $\tilde{f}(0)\eqdef\lim_{r\to \infty}\frac{f(r)}{r}$, and $\rho^0$ and $\sigma^0$ are the projections to the supports of $\rho$ and $\sigma$.
Hint: For the first part, note that the components of $(1-\eps)\q+\eps\u$ (see~\eqref{0615}) can be written as $q_y+\eps(\frac1n-q_y)$. Now $\eps(\frac1n-q_y)$ is positive if $q_y=0$, and if $q_y>\frac1n$ then since $f$ is continuous we can still replace $\eps(\frac1n-q_y)$ with $\eps$ in the limit $\eps\to0^+$. 
\end{exercise}

The quantum $f$-Divergence\index{$f$-divergence} has the following quantum formula.
\begin{myt}{\color{yellow} Quantum Formula}
\begin{theorem}\label{quantumformula}
Let $f:(0,\infty)\to\mbb{R}$ be an operator convex function satisfying $f(1)=0$.
For any $\rho,\sigma\in\md(A)$ with $\sigma>0$ the quantum $f$-Divergence\index{$f$-divergence} can be expressed as
\be\label{qfdf}
D_f(\rho\|\sigma)=\tr\left[\phi_{\sigma}^{A\tA}f\left(\sigma^{-1}\otimes\rho^T\right)\right]\;,
\ee
where $|\phi_{\sigma}^{A\tA}\ra\eqdef\sigma^{1/2}\otimes I^{\tA}|\Omega^{A\tA}\ra$ is a purification of $\sigma$.
For $\sigma\geq 0$ the $f$-Divergence\index{$f$-divergence} satisfies (with $\u\in\md(A)$ is the maximally mixed state)
\be
D_f(\rho\|\sigma)=\lim_{\eps\to 0^+}D_f\big(\rho\big\|(1-\eps)\sigma+\eps \u\big)\;.
\ee
\end{theorem}
\end{myt}
\begin{proof}
Suppose first that $\sigma>0$. Then, by definition
\ba
D_f(\rho\|\sigma)\eqdef D_f\left(\tilde{\p}\big\|\tilde{\q}\right)&=\sum_{x,y\in[m]}\tilde{q}_{xy}f\left(\frac{\tilde{p}_{xy}}{\tilde{q}_{xy}}\right)\\
&=\sum_{x,y\in[m]}q_y|\la a_x|b_y\ra|^2f\left(\frac{p_{x}}{q_y}\right)\;.\label{gxg}
\ea
Now, for every $x,y\in[m]$ we can express $q_y|\la a_x|b_y\ra|^2$ as follows:
\ba
q_y|\la a_x|b_y\ra|^2&=\big\la\Omega^{A\tA}\big|q_y|b_y\lr b_y|\otimes|a_x\lr a_x|^T\Big|\Omega^{A\tA}\Big\ra\\
\Gg{q_y|b_y\lr b_y|=\sigma^{\frac12}|b_y\lr b_y|\sigma^{\frac12}}&=\Big\la\phi^{A\tA}_\sigma\Big||b_y\lr b_y|\otimes|a_x\lr a_x|^T\Big|\phi^{A\tA}_\sigma\Big\ra\;.
\ea
Substituting this expression into~\eqref{gxg} we obtain
\ba
D_f(\rho\|\sigma)
&=\Big\la\phi^{A\tA}_{\sigma}\Big|\sum_{x,y\in[m]} f\left(\frac{p_{x}}{q_y}\right)|b_y\lr b_y|\otimes |a_x\lr a_x|^T\Big|\phi^{A\tA}_{\sigma}\Big\ra\\
\Gg{\substack{\left\{|b_y\lr b_y|\otimes |a_x\lr a_x|^T\right\}_{x,y\in[m]}\\ \text{is orthonormal}}}&=\Big\la\phi^{A\tA}_{\sigma}\Big|f\left(\sum_{x,y\in[m]} \frac{p_{x}}{q_y}|b_y\lr b_y|\otimes |a_x\lr a_x|^T\right)\Big|\phi^{A\tA}_{\sigma}\Big\ra\\
&=\Big\la\phi_{\sigma}^{A\tA}\left|f\left(\sigma^{-1}\otimes\rho^T\right)\Big|\phi_{\sigma}^{A\tA}\right\ra\;,
\ea
The case $\sigma\geq 0$ follows directly from Exercise~\ref{cont611} and is left as an exercise.
\end{proof}

We now demonstrate that the expression for the $f$-Divergence\index{$f$-divergence}, as outlined in the preceding theorem, satisfies the data processing inequality when $f$ is operator convex.

\begin{myt}{}
\begin{theorem}\label{dpifdiv}
Let $f:(0,\infty)\to\mbb{R}$ be an operator convex function satisfying $f(1)=0$. Then, $D_f^q$ as defined in Definition~\ref{qfdiv} is a quantum divergence.
\end{theorem}
\end{myt}
\begin{proof}
Due to Stinespring dilation theorem, any $\mE\in\cptp(A\to B)$ can be expressed as an isometry followed by a partial trace. From its definition and Lemma~\ref{isoqdiv} it follows that the quantum $f$-Divergence\index{$f$-divergence} is invariant under isometries. It is therefore left to show that it is monotonic under partial trace.  For this purpose, let $\rho,\sigma\in\md(AB)$ and without loss of generality we assume that $\sigma^{AB}>0$ since the case $\sigma^{AB}\geq 0$ follows from the continuity of $D_f$ in the limit $\eps\to 0^+$. 

We need to show that
\be\label{dfdpi}
D_f(\rho^A\|\sigma^A)\leq D_f(\rho^{AB}\|\sigma^{AB})\;.
\ee
From the quantum formula in~\eqref{qfdf} we see that the left hand side of~\eqref{dfdpi} depends on $(\sigma^{A})^{-1}\otimes (\rho^{\tA})^T$ whereas the right hand side depends on $(\sigma^{AB})^{-1}\otimes (\rho^{\tA\tB})^T$.
In the exercise below you will show that there exists an isometry that relates between these two expressions. Explicitly, you will show that there exists an isometry $V:A\tA\to A\tA B\tB$ (with $V^*V=I^{A\tA}$) such that
\be\label{baq1}
V^*\left(\left(\sigma^{AB}\right)^{-1}\otimes \left(\rho^{\tA\tB}\right)^T\right)V=\left(\sigma^{A}\right)^{-1}\otimes \left(\rho^{\tA}\right)^T\;.
\ee
Combining this with the operator Jensen's inequality~\eqref{jensen} for operator convex functions we get
\ba
f\left(\left(\sigma^{A}\right)^{-1}\otimes \left(\rho^{\tA}\right)^T\right)
&=f\left(V^*\left(\sigma^{AB}\right)^{-1}\otimes \left(\rho^{\tA\tB}\right)^TV\right)\\
\GG{\eqref{jensen}}&\leq V^*f\left(\left(\sigma^{AB}\right)^{-1}\otimes \left(\rho^{\tA\tB}\right)^T\right)V\;.
\ea
Finally, multiplying both sides by $\phi_{\sigma}^{A\tA}$ and taking the trace gives
\ba
D_f\left(\rho^A\big\|\sigma^A\right)&\leq \tr\left[V\phi_{\sigma}^{A\tA}V^*f\left(\left(\sigma^{AB}\right)^{-1}\otimes \left(\rho^{\tA\tB}\right)^T\right)\right]\\
\GG{\eqref{vsaab}}&=\tr\left[\phi_{\sigma}^{AB\tA\tB}f\left(\left(\sigma^{AB}\right)^{-1}\otimes \left(\rho^{\tA\tB}\right)^T\right)\right]\\
&=D_f\left(\rho^{AB}\big\|\sigma^{AB}\right)\;.
\ea
This completes the proof.
\end{proof}

\begin{exercise}
Let $V:\;A\tA\to AB\tA \tB$ be the matrix
\be
V=\sum_{y\in[m]}\left(\sigma^{AB}\right)^{\frac{1}{2}}\left(\left(\sigma^A\right)^{-\frac{1}{2}}\otimes |y\ra^B\right)\otimes I^{\tA}\otimes |y\ra^{\tB}\;.
\ee
\ben
\item Show that $V$ is an isometry, i.e. $V^*V=I^{A\tA}$.
\item Show that $V$ satisfies~\eqref{baq1}.
\item Show that
\be\label{vsaab}
V\left(\left(\sigma^{A}\right)^{\frac{1}{2}}\otimes I^{\tA}\right)\big|\Omega^{A\tA}\big\ra=\left(\left(\sigma^{AB}\right)^{\frac{1}{2}}\otimes I^{\tA\tB}\right)\big|\Omega^{AB\tA\tB}\big\ra\;,
\ee
\een
\end{exercise}

\subsubsection{Examples:}
\begin{enumerate}
\item The Umegaki Divergence. For the function $f(r)=r\log r$
\ba\label{680}
D_f(\rho\|\sigma)&=\lim_{\eps\to 0^+}\sum_{x,y}(q_y+\eps)f\left(\frac{p_x}{q_y+\eps}\right)|\la a_x| b_y\ra|^2\\
&=\lim_{\eps\to 0^+}\sum_{x,y}{p_x}\log\left(\frac{p_x}{q_y+\eps}\right)|\la a_x| b_y\ra|^2\\
&=\sum_{x}p_x\log p_x-\sum_{x,y}p_x|\la a_x| b_y\ra|^2\log q_y\\
&=\tr[\rho\log\rho]-\tr[\rho\log\sigma]\;,
\ea
where in the last line we used the relation $\la b_y|\rho|b_y\ra=\sum_xp_x|\la a_x| b_y\ra|^2$. Since $f(r)=r\log r$ is operator convex, the above expression is a quantum divergence. It is known as the Umegaki divergence or sometimes referred to as \emph{the} relative entropy. We will discuss many of its properties in the following chapters.
\item The Trace Distance? The function $f(r)=\frac12|r-1|$ is convex but it is not operator convex (on any domain that includes $1$). Therefore, we cannot conclude that for this choice $D_f^q$ is a quantum divergence. Moreover, note that for this case
\ba
D_f(\rho\|\sigma)
&=\lim_{\eps\to 0^+}\sum_{x,y}(q_y+\eps)\frac12\left|\frac{p_x}{q_y+\eps}-1\right||\la a_x| b_y\ra|^2\\
&=\frac12\sum_{x,y}\Big|{p_x}|\la a_x| b_y\ra|^2-{q_x}|\la a_x| b_y\ra|^2\Big|\;,
\ea
which cannot be expressed as a simple function of $\rho$ and $\sigma$.
\item The quantum $\alpha$-Divergence\index{$\alpha$-divergence}. The function $f(r)=\frac{r^\alpha-r}{\alpha(\alpha-1)}$ is known to be operator convex for $\alpha\in[0,2]$ (cf. Table~\ref{table:1}). For any $\rho,\sigma\in\md(A)$ let $\tilde{\p},\tilde{\q}\in\prob(m^2)$ (with $m\eqdef|A|$) be the probability vectors defined in~\eqref{corre}. Then, for this $f$ we have
\ba\label{682}
D_f^q(\rho\|\sigma)&=D_f(\tilde{\p}\|\tilde{\q})=\frac1{\alpha(\alpha-1)}\left(\sum_{x,y}\tilde{p}_{xy}^\alpha \tilde{q}_{xy}^{1-\alpha}-1\right)\\
&=\frac1{\alpha(\alpha-1)}\left(\sum_{x,y}\left(p_x|\la a_x| b_y\ra|^2\right)^\alpha 
\left(q_y|\la a_x| b_y\ra|^2\right)^{1-\alpha}-1\right)\\
&=\frac1{\alpha(\alpha-1)}\left(\sum_{x,y}|\la a_x| b_y\ra|^2p_x^\alpha q_y^{1-\alpha}-1\right)\\
&=\frac1{\alpha(\alpha-1)}\left(\tr\left[\rho^{\alpha}\sigma^{1-\alpha}\right]-1\right)\;.
\ea
\end{enumerate}

\begin{exercise}
Use the quantum formula given in Theorem~\ref{quantumformula} to compute the Umegaki divergence and the quantum $\alpha$-Divergence\index{$\alpha$-divergence}.
\end{exercise}

\section{Optimal Quantum Extensions of Divergences}\label{optim}\index{optimal extension}\index{divergence}

In addition to the method for extension introduced above, sometimes it is possible to guess a quantum extension of a classical divergence\index{classical divergence} simply by replacing $\p$ with $\rho$ and $\q$ with $\sigma$. For example, the trace distance divergence $\frac12\|\p-\q\|_1$ can be replaced with its quantum version $\frac12\|\rho-\sigma\|_1$. However, this method is not always simple and, in general, it is not even unique. Moreover, one still needs to verify that the proposed quantum divergence indeed satisfies the DPI. We therefore investigate in this subsection a more systematic approach for quantum extensions.

Given a \emph{classical} divergence $\D$ we denote by $\uD$ its minimal quantum extension, and by $\oD$ its maximal quantum extension. That is, $\uD$ and $\oD$ are quantum divergences that reduce to $\D$ on classical systems, and any other quantum divergence, $D'$, that reduces to $\D$ on classical states, necessarily satisfies
\be\label{bboun}
\uD(\rho\|\sigma)\leq D'(\rho\|\sigma)\leq\oD(\rho\|\sigma)\quad\quad\forall\;\rho,\sigma\in\md(A)\;.
\ee
One may wonder whether such optimal quantum extensions exists. In the next theorem we prove that they do using the following construction.

Let $\D$ be a classical divergence, and for any $\rho,\sigma\in\md(A)$ define
\begin{align}
&\uD(\rho^A\|\sigma^A) \eqdef \sup \D\left(\mE^{A\to X}\left(\rho^A\right)\big\|\mE^{A\to X}\left(\sigma^A\right)\right)\,, \label{ud}\\ 
&\oD(\rho\|\sigma)\eqdef\inf \left\{\D\left(\p^X\big\|\q^X\right)\;:\;\rho^A=\mF^{X\to A}(\p^X),\;\sigma^A=\mF^{X\to A}(\q^X) \right\}\label{od}\,, 
\end{align}
where the optimizations are over the classical system $X$, the channels $\mE \in \cptp(A \to X)$ and $\mF \in \cptp(X \to A)$, as well as the probability distributions (diagonal density matrices) $\p,\q\in\md(X)$. Note that $\mE$ is a POVM Channel\index{POVM channel}  and therefore $\D\big(\mE(\rho)\big\|\mE(\sigma)\big)$ is well defined since $\mE(\rho)$ and $\mE(\sigma)$ are classical states; i.e. they can be viewed as probability vectors or diagonal density matrices. Similarly, $\p^X$ and $\q^X$ can be viewed either as diagonal density matrices or as probability vectors.
Moreover, the supremum and infimum are taken over all dimensions $|X|\in\mbb{N}$.

\begin{myt}{\color{yellow} Optimal Extensions}
\begin{theorem}\label{optim2}
Let $\D$ be a classical divergence, and let $\uD$ and $\oD$ be as in~\eqref{ud} and~\eqref{od}, respectively.
Then, both $\uD$ and $\oD$ are quantum divergences that reduces to $\D$ on classical states. In addition, any other quantum divergence $D'$ that reduces to $\D$ on classical states satisfies~\eqref{bboun}.
\end{theorem}
\end{myt}

\begin{proof}
We first prove the reduction property. Let $\rho,\sigma\in\md(A)$ be classical states. Then, for $\uD$ we can take $X$ in~\eqref{ud} to be a classical system with $|X|=|A|$ and $\mE$ to be  the identity channel. Since this identity channel is not necessarily the optimal channel, we get that
\be\label{q1q}
\uD(\rho\|\sigma)\geq \D(\rho\|\sigma)\;.
\ee
Conversely, since $\rho$ and $\sigma$ are classical, any $\mE$ in~\eqref{ud} can be assumed to be classical since
\be
\mE(\rho)=\mE\circ\Delta(\rho)\quad\text{and}\quad\mE(\sigma)=\mE\circ\Delta(\sigma)
\ee
where $\Delta$ is the completely dephasing channel. Therefore, if $\mE$ is not classical we can replace it with $\mE\circ\Delta$ which is classical (recall that the output of $\mE$ is classical). Now, by the DPI property of the classical divergence\index{classical divergence} $\D$ we have for all such classical $\mE$, $\D\big(\mE(\rho)\big\|\mE(\sigma)\big)\leq \D(\rho\|\sigma)$. Hence, we must have
\be\label{q2q}
\uD(\rho\|\sigma)\leq \D(\rho\|\sigma)\;.
\ee
Combining~\eqref{q1q} with~\eqref{q2q} we conclude that $\uD(\rho\|\sigma)= \D(\rho\|\sigma)$. Similarly, for $\oD$ we can assume that $\mF$ in~\eqref{od} is a classical channel\index{classical channel} since $\rho$ and $\sigma$ are classical. Hence, by the DPI of $\D$ we get the lower bound $\oD(\rho\|\sigma)\geq \D(\rho\|\sigma)$, and this bound can be saturated since we can take $\mF$ in~\eqref{od} to be the identity channel.

We next prove that $\uD$ and $\oD$ both satisfy the DPI. Let $\mN\in\cptp(A\to B)$. Then,
\ba
\uD\big(\mN(\rho)\big\|\mN(\sigma)\big) &= \sup_X \big\{\D\big(\mE\circ\mN(\rho)\big\|\mE\circ\mN(\sigma)\big)\;:\;\mE\in\cptp(B\to X)\big\}\\
\Gg{\mE'\text{ replaces }\mE\circ\mN\;\rightarrow}&\leq \sup_X \big\{\D\big(\mE'(\rho)\big\|\mE'(\sigma)\big)\;:\;\mE'\in\cptp(A\to X)\big\}\\
&=\uD(\rho\|\sigma)\;.
\ea
For $\oD$ we have
\ba
\oD(\rho\|\sigma)&\eqdef\inf_X \big\{\D(\p\|\q)\;:\;\rho=\mF(\p),\;\sigma=\mF(\q),\;\mF\in\cptp(X\to A) \big\}\\
&\geq\inf_X \big\{\D(\p\|\q)\;:\;\mN(\rho)=\mN\circ\mF(\p),\;\mN(\sigma)=\mN\circ\mF(\q),\;\mF\in\cptp(X\to A) \big\}\\
&\geq \inf_X \big\{\D(\p\|\q)\;:\;\mN(\rho)=\mF'(\p),\;\mN(\sigma)=\mF'(\q),\;\mF'\in\cptp(X\to B) \big\}\\
&=\uD\big(\mN(\rho)\big\|\mN(\sigma)\big)\;,
\ea
where the first inequality follows from the fact that if $\rho=\mF(\p)$ then necessarily $\mN(\rho)=\mN\circ\mF(\p)$ (but the converse is not necessarily true), and in the second inequality we replaced $\mN\circ\mF$ with $\mF'$.

Finally, we prove the optimality of $\uD$ and $\oD$. First observe that from the DPI of $D'$ we have for any $\rho,\sigma\in\md(A)$ and any $\mE\in\cptp(A\to X)$ 
\be
D'(\rho\|\sigma)\geq D'\big(\mE(\rho)\big\|\mE(\sigma)\big)=\D\big(\mE(\rho)\big\|\mE(\sigma)\big)\,,
\ee
where the last equality follows from the fact that $D'$ reduces to $\D$ on classical states. 
Since the above inequality holds for all $\mE\in\cptp(A\to X)$ it also holds for the supremum over such $\mE$. We therefore conclude that $D'(\rho\|\sigma)\geq \uD(\rho\|\sigma)$. For the second inequality, let $\rho,\sigma\in\md(A)$ and $\p,\q\in\md(X)$, and suppose there exists $\mF\in\cptp(X\to A)$ such that $\rho=\mF(\p)$ and $\sigma=\mF(\q)$. Then, from the DPI of $D'$ we get
\be
D'(\rho\|\sigma)=D'\big(\mF(\p)\big\|\mF(\q)\big)\leq D'(\p\|\q)=\D(\p\|\q)\;,
\ee
where the last equality follows from the fact that $D'$ reduces to $\D$ on classical states. Since the above inequality holds for all such $\p,\q$ for which there exists an $\mF$ that takes them to $\rho$ and $\sigma$, it must also hold for the infimum over all such $\p,\q$. Hence, $D'(\rho\|\sigma)\leq \oD(\rho\|\sigma)$.
\end{proof}

Since the maximal and minimal extension provides upper and lower bounds on all extensions, it can be useful to have a closed formula for them.  Remarkably, a closed formula for the maximal extension exists if one of the input states is pure, or for the $f$-Divergence\index{$f$-divergence}s if $f$ is operator convex.
On the other hand,  at the time of writing this book, a closed formula for the minimal extension of the $f$-Divergence\index{$f$-divergence} is not known. However, for specific examples such as the trace distance and fidelity, the minimal extension can be computed (see the next section), and as we will see in Chapter~\ref{ch:relent} the regularized minimal extension can also be computed for all known relative entropies. 

\begin{exercise}
Let $f:[0,\infty)\to[0,\infty)$ be a operator convex function, and for all $\rho,\sigma\in\md(A)$ with $\sigma>0$ define 
\be 
D'_f(\rho\|\sigma)\eqdef\tr\left[\rho\#_f\sigma\right]=\tr\left[\sigma f\left(\sigma^{-\frac12}\rho\sigma^{-\frac12}\right)\right]\;,
\ee
where $\#_f$ is the Kubo-Ando operator mean  (see Definition~\ref{kaom}). Finally, let
$\bD_f$ be the maximal $f$-Divergence\index{$f$-divergence}. 
\begin{enumerate}
\item Show that $D'_f$ reduces to the classical $f$-Divergence\index{$f$-divergence} when $\rho$ and $\sigma$ are classical (i.e.\ diagonal in the same basis).
\item Show that $D'_f$ satisfies the DPI in the domain $\md(A)\times\md_{>0}(A)$.
Hint: Show that $D_f'$ satisfies all the conditions of Lemma~\ref{lemaa}.
\item Show that for any $\rho,\sigma\in\md(A)$ with $\sigma>0$  
\be\label{dff}
\bD_f(\rho\|\sigma)\geq D_f'(\rho\|\sigma)\;.
\ee
Hint: Use Theorem~\ref{optim2}
\end{enumerate}
\end{exercise}

\subsection{The Maximal Quantum Extension}\label{qetd}
\index{optimal extension}

The maximal extension can be expressed as
\be\label{maxdpq}
\oD(\rho\|\sigma)=\inf\D(\p\|\q)
\ee
subject to the conditions
\be\label{rsrs}
\rho=\sum_{x\in[n]}p_x\omega_x\quad\text{and}\quad\sigma=\sum_{x\in[n]}q_x\omega_x\;,
\ee
where $n\in\mbb{N}$, $\p,\q\in\prob(n)$, and for each $x\in[n]$, $\omega_x\in\md(A)$.
Note that we replaced $\mF(|x\lr x|)$ with $\omega_x$. 
The infimum above can include vectors $\p$ and $\q$ with zero components. We now show that the number of zeros in each of these vectors can be restricted to be at most one.
\begin{myg}{}
\begin{lemma}\label{zeros}
The infimum in~\eqref{maxdpq} can be restricted to vectors $\p,\q\in\prob(n)$ that has at most one zero component.
\end{lemma}
\end{myg}

\begin{proof}
We first show that $\q$ can have this property. Since divergences are invariant under (joint) permutation of the components of $\p$ and $\q$, without loss of generality we can assume that 
\be
q_1\geq\cdots \geq q_r>q_{r+1}=\cdots=q_n=0\;.
\ee
where $r$ is the number of non-zero components of $\q$. With this order of $\q$ we have (see Exercise~\ref{tgv}) $(\p,\q)\sim(\p',\q)$, 
where 
\be
\p'=\left(p_1,\ldots,p_r,p'_{r+1},0,\ldots,0\right)^T\quad\text{where}\quad p'_{r+1}\eqdef\sum_{x=r+1}^{n}p_x\;.
\ee
Note that the relations in~\eqref{rsrs} can be expressed as
\be
\rho=\sum_{x\in[r]}p_x\omega_x+p'_{r+1}\tau\quad\text{and}\quad\sigma=\sum_{x\in[r]}q_x\omega_x\;,
\ee
where $\tau\eqdef\frac1{p'_{r+1}}\sum_{x=r+1}^{n}p_x\omega_x$. Therefore, the vectors
\be\label{tptq8}
\tp=\left(p_1,\ldots,p_r,p'_{r+1}\right)^T\quad\text{and}\quad\tq=(q_1,\ldots,q_r,0)
\ee
satisfy both~\eqref{rsrs} with $n$ replaced by $r+1$, and $\D(\p\|\q)=\D(\tp\|\tq)$. Repeating the same arguments for $\tp$ completes the proof.
\end{proof}

It's important to note that the lemma mentioned above aids in simplifying the optimization problem described in~\eqref{maxdpq}. This simplification is achieved by assuming, without any loss of generality, that $\p$ and $\q$ have forms similar to $\tp$ and $\tq$ as specified in~\eqref{tptq8}. Consequently, we can redefine the infimum in~\eqref{maxdpq} as an infimum over all $1<n\in\mbb{N}$, $\p\in\prob(n)$, and $0<\q\in\prob(n-1)$, provided there are $n-1$ density matrices $\{\omega_x\}_{x\in[n-1]}\subset\md(A)$ meeting the following criteria:
\be\label{grgr0}
\rho\geq\sum_{x\in[n-1]}p_x\omega_x\quad\text{and}\quad\sigma=\sum_{x\in[n-1]}q_x\omega_x\;,
\ee
where it is understood that the inequality in the first relation is satisfied if and only if there exists a density matrix $\omega_n\in\md(A)$ such that
\be
\rho=\sum_{x\in[n-1]}p_x\omega_x+p_n\omega_n\;.
\ee
As we will see, in certain cases, working with the expression in~\eqref{grgr0} becomes more manageable because $\q>0$. In other situations, it might be preferable to work with $\p>0$. It is worth noting that by applying the same reasoning as above but substituting $\p$ for $\q$ and vice versa, we can also express the infimum in~\eqref{maxdpq} as an infimum over all $1<n\in\mbb{N}$, $0<\p\in\prob(n-1)$ and $\q\in\prob(n)$, with the requirement of having $n-1$ density matrices $\{\omega_x\}_{x\in[n-1]}\subset\md(A)$ that satisfy:
\be\label{grgr2}
\rho=\sum_{x\in[n-1]}p_x\omega_x\quad\text{and}\quad\sigma\geq\sum_{x\in[n-1]}q_x\omega_x\;.
\ee
In the next theorem
we employ this property to calculate the maximal divergence when one of the input states is pure.

\begin{myt}{}
\begin{theorem}\label{thmod}
Let $\D$ be a classical divergence, $\psi\in\pure(A)$, and $\sigma\in\md(A)$. Then, 
\be\label{modtim}
\oD(\psi\|\sigma)=\D\left((1,0)^T\big\|(\lambda_{\max},1-\lambda_{\max})^T\right)
\ee
where 
\be\label{modtim2}
\lambda_{\max}\eqdef\max\Big\{\lambda\in\mbb{R}\;:\;\lambda \psi\leq\sigma\Big\}\;.
\ee
\end{theorem}
\end{myt}

\begin{proof}
Consider the relation~\eqref{grgr2} with the pure state $\psi$ replacing $\rho$. Since $\rho\eqdef\psi$ is a pure state, the first relation in~\eqref{grgr2}
can hold if and only if for any $x\in[n-1]$  we have $\omega_x=\psi$. Substituting this into the second relation in~\eqref{grgr2} we obtain
\be
\sigma\geq\sum_{x\in[n-1]}q_x\psi=(1-q_n)\psi\;.
\ee
Observe that this condition is equivalent to $(1-q_n)\leq \lambda_{\max}$. Consequently, we have reached the following expression:
\be
\oD(\psi\|\sigma)=\inf_{1<n\in\mbb{N}}\Big\{\D(\p\oplus 0\|\q)\;:\;0<\p\in\prob(n-1),\;\q\in\prob(n),\;q_n\geq 1-\lambda_{\max}\Big\}.
\ee
Finally, we simplify the expression above by demonstrating that we can confine the value of $n$ in the optimization above to be equal to two. To achieve this, let $E$ be the $2\times n$ column stochastic matrix 
\be
E\eqdef\begin{bmatrix} 1 & \cdots & 1 & 0\\
0 &  \cdots & 0 & 1
\end{bmatrix}\;.
\ee
Observe that $E\p=(1,0)^T$ and $E\q=(1-q_n,q_n)^T$ so that 
\ba
(\p\oplus 0,\q)\succ(E(\p\oplus 0),E\q)
&=\left((1,0)^T,(1-q_n,q_n)^T\right)\\
\Gg{1-q_n\leq \lambda_{\max}}&\succ
\left((1,0)^T,(\lambda_{\max},1-\lambda_{\max})^T\right)\;.
\ea
Therefore, the minimum is obtained with $n=2$ and with the pair $(\p,\q)$ being equal to the pair on the right hand side of the equation above.
\end{proof}

\subsection{The Maximal $f$-Divergence\index{$f$-divergence}}

In this section, we calculate the maximal $f$-Divergence\index{$f$-divergence} $\bD_f(\rho|\sigma)$ for $\rho,\sigma\in\md(A)$ with the condition that $\sigma>0$. The special case where $\sigma$ possesses zero eigenvalues will be deferred to the appendix.

When $\D$ equals the $f$-Divergence\index{$f$-divergence}, the expression given in Eq.~\eqref{ffdiv2}, along with~\eqref{grgr0}, indicates that the infimum in~\eqref{maxdpq} can be represented as follows:
\be\label{odeq0}
\bD_f(\rho\|\sigma)\eqdef\inf_{1<n\in\mbb{N}}\Big\{ \sum_{x\in[n-1]}q_xf\left(\frac{p_x}{q_x}\right)+\tilde{f}(0)p_{n}\Big\}\;,
\ee
where $\tilde{f}(0)$ is defined in~\eqref{tildef}, and the infimum above is over all $1<n\in\mbb{N}$, $\p\in\prob(n)$ and $0<\q\in\prob(n-1)$, 
such that there exists $n-1$ density matrices $\{\omega_x\}_{x\in[n-1]}\subset\md(A)$ satisfying~\eqref{grgr0}.
Denoting by
$\Lambda_x\eqdef q_x\sigma^{-\frac12}\omega_x\sigma^{-\frac12}$, and applying the conjugation $\sigma^{-\frac12}(\cdot)\sigma^{-\frac12}$ to both sides of~\eqref{grgr0}  gives the relations:
\be\label{lfg}
 \sigma^{-\frac12}\rho\sigma^{-\frac12}\geq\sum_{x\in[n-1]}\frac{p_x}{q_x}\Lambda_x\quad\text{and}\quad\sum_{x\in[n-1]}\Lambda_x=I^A\;.
\ee
With these new notations, the infimum in~\eqref{odeq0} is taken over all $1<n\in\prob(n)$, all $\p\in\prob(n)$, and all POVMs $\{\Lambda_{x}\}_{x\in[n-1]}$ for which  the inequality~\eqref{lfg} holds with $q_x\eqdef\tr\left[\Lambda_x\sigma\right]>0$.

One natural choice/guess for the optimal $n$, $\p$ and $\{\Lambda_x\}_{x\in[n-1]}$, is to choose them such that the inequality in~\eqref{lfg} becomes an equality. This is possible for example by taking $n=|A|+1$,  and for any $x\in[n-1]$ to take $\Lambda_x=\psi_x\in\pure(A)$ with $|\psi_x\ra$ being the 
$x$-eigenvector of $\sigma^{-\frac12}\rho\sigma^{-\frac12}$ corresponding to the eigenvalue $p_x/q_x$ (i.e. $\p$ is chosen such that $p_x/\tr[\sigma \Lambda_x]$ is the $x$-eigenvalue of $\sigma^{-\frac12}\rho\sigma^{-\frac12}$). For this choice we have 
\be\label{6116g}
\sigma^{-\frac12}\rho\sigma^{-\frac12}=\sum_{x\in[n-1]}\frac{p_x}{q_x}|\psi_x\lr\psi_x|
\ee
which forces $p_n$ to be
\ba
p_n&=1-\sum_{x\in[n-1]}p_x=1-\sum_{x\in[n-1]}\frac{p_x}{q_x}\la\psi_x|\sigma|\psi_x\ra=1-\tr[\rho]=0\;,
\ea
where the last equality follows by multiplying both sides of~\eqref{6116g} by $\sigma$ and taking the trace.
Moreover, for these choices of $n$, $\p$ and $\{\Lambda_x\}$, we have
\ba
\sum_{x\in[n-1]}q_xf\left(\frac{p_x}{q_x}\right)&=\sum_{x\in[n-1]}\tr[\sigma|\psi_x\lr\psi_x|] f\left(\frac{p_x}{q_x}\right)\\
\Gg{\forall t\geq 0\;\;f(t|\psi_x\lr\psi_x|)=f(t)|\psi_x\lr\psi_x|}&=\sum_{x\in[n-1]}\tr\left[\sigma f\left(\frac{p_x}{q_x}|\psi_x\lr\psi_x|\right)\right]\\
\Gg{\{|\psi_x\ra\}_{x\in[n-1]}\text{ is orthonormal }}&=\tr\Big[\sigma f\Big(\sum_{x\in[n-1]}\frac{p_x}{q_x}|\psi_x\lr\psi_x|\Big)\Big]\\
\GG{\eqref{6116g}}&=\tr\left[\sigma f\left(\sigma^{-\frac12}\rho\sigma^{-\frac12}\right)\right]\;.
\ea 
Note that we obtained the formula above for a particular choice of $n$, $\p$ and $\{\Lambda_x\}_{x\in[n-1]}$. Therefore, since this is not necessarily the optimal choice (recall $\bD_f$ is defined in terms of an infimum), we must have 
\be\label{showedg}
\bD_f(\rho\|\sigma)\leq \tr\left[\sigma f\left(\sigma^{-\frac12}\rho\sigma^{-\frac12}\right)\right]=\tr\left[\rho\#_f\sigma\right]\;.
\ee
where $\#_f$ is the Kubo-Ando operator mean  (see Definition~\ref{kaom}).
Interestingly, to get this upper bound we did not even assume that $f$ is convex, but if $f$ is operator convex we get an equality above. 

\begin{myt}{\color{yellow} Closed Formula of The Maximal $f$-Divergence\index{$f$-divergence}}\index{optimal extension}
\begin{theorem}\label{t633g}
Let $\rho,\sigma\in\md(A)$ with $\sigma>0$, and let $f\eqdef(0,\infty)\to\mbb{R}$ be operator convex, with $f(0)\eqdef\lim_{\eps\to 0^+}f(\eps)$ and $f(1)=0$.
Then,
\be\label{mainfd0}
\bD_f(\rho\|\sigma)= \tr\left[\rho\#_f\sigma\right]=\tr\left[\sigma f\left(\sigma^{-\frac12}\rho\sigma^{-\frac12}\right)\right]\;.
\ee
\end{theorem}
\end{myt}
\begin{remark}$\;$
\ben
\item In the theorem above we restricted the domain of $\bD_f$ to $\md(A)\times\md_{>0}(A)$. To incorporate the singular case it is necessary to employ additional more delicate arguments (see e.g.~\ref{formoff} in the appendix). Explicitly, for the case that $\sigma$ is singular, we can write $\sigma=\begin{pmatrix}\tsigma & \0\\
\0 & \0\end{pmatrix}$ in a block matrix form with $\tsigma>0$, and the formula for $\bD_f$ becomes
(see Appendix~\ref{singularfd})
\be\label{7124}
\bD_f(\rho\|\sigma)= \tr\left[\trho\#_f\tsigma\right]+(1-\tr[\trho])\tilde{f}(0)
\ee
where $\tilde{f}(0)\eqdef\lim_{\eps\to 0^+}\eps f(\frac1\eps)$, and $\trho\eqdef\rho/\rho_{22}\eqdef\rho_{11}-\zeta\rho_{22}^{-1}\zeta^*$ is the Schur complement (see~\eqref{schurcom2}) of the block $\rho_{22}$ of $\rho=\begin{pmatrix}\rho_{11} & \zeta\\
\zeta^* & \rho_{22}\end{pmatrix}$.  
\item In Theorem~\ref{equivjoint} we proved that for a continuous function $f:[0,\infty)\to[0,\infty)$  the Kubo-Ando operator mean $\#_f$ is operator convex if and only if it is jointly convex. Therefore, at least in the domain $\md(A)\times\md_{>0}(A)$ the maximal $f$ divergence is jointly convex for any operator convex $f$.
\een
\end{remark}

\begin{proof}
The proof of the theorem follows immediately from the inequality~\eqref{showedg} combined with the opposite inequality~\eqref{dff}.
\end{proof}

\subsubsection{Examples:}

\begin{enumerate}
\item The Belavkin–Staszewski divergence. Consider the function $f(r)=r\log r$. In this case we have 
$\tilde{f}(0)=\lim_{\eps\to 0^+}\eps f(1/\eps)=\lim_{\eps\to 0^+}\log(1/\eps)=\infty$. According to the closed form in~\eqref{mainfd}, this means that unless $\supp(\rho)\subseteq\supp(\sigma)$ we have $\bD_f(\rho\|\sigma)=\infty$. For the case $\supp(\rho)\subseteq\supp(\sigma)$ we have
\be
\bD_f(\rho\|\sigma)=\tr\left[\sigma \left(\sigma^{-\frac12}\rho\sigma^{-\frac12}\right)\log\left(\sigma^{-\frac12}\rho\sigma^{-\frac12}\right)\right]
\ee
The expression above can be simplified by using the relation $Mf(M^*M)=f(MM^*)M$ from Exercise~\ref{relfmmst}. Denoting $M\eqdef \rho^{\frac12}\sigma^{-\frac12}$ we get
\ba
\bD_f(\rho\|\sigma)&=\tr\left[\sigma M^*M\log\left(M^*M\right)\right]\\
\GG{\eqref{b1ex}}&=\tr\left[\sigma M^*\log\left(MM^*\right)M\right]\\
\Gg{M\eqdef \rho^{\frac12}\sigma^{-\frac12}}&=\tr\left[\rho\log\left(\rho^{\frac12}\sigma^{-1}\rho^{\frac12}\right)\right]\;.
\ea
This divergence is known as the Belavkin–Staszewski divergence.
\item The maximal $\alpha$-Divergences\index{$\alpha$-divergence}. Consider the function $f_\alpha(r)=\frac{r^\alpha-r}{\alpha(\alpha-1)}$ which is known to be operator convex for $\alpha\in(0,2]$. Therefore, for $\alpha\in(0,2]$
we have (recall we assume $\sigma>0$)
\ba\label{6130}
\bD_{f_\alpha}(\rho\|\sigma)&=\frac{1}{\alpha(\alpha-1)}\tr\left[\sigma\left( \left(\sigma^{-\frac12}\rho\sigma^{-\frac12}\right)^\alpha-\sigma^{-\frac12}\rho\sigma^{-\frac12}\right)\right]\\
&=\frac{1}{\alpha(\alpha-1)}\left(\tr\left[\sigma\left(\sigma^{-\frac12}\rho\sigma^{-\frac12}\right)^\alpha\right]-1\right)\;.
\ea
\een

\section{Divergences that are Metrics}\index{metric}\index{divergence}

We saw that every norm induces a corresponding metric. Metrics are used to measure how close two vectors are, but since there are many norms (e.g. the family of $p$-norms) there are also numerous metrics
that can be used to measure distance.  Mathematically, in finite dimensions, all metrics are topologically equivalent. Yet, physically, only very few have a known operational meaning. In this subsection we study metrics that are also divergences.

We first consider norms that induce metrics that are also divergences.
We saw in Chapter 1 that the $1$-norm satisfies the monotonicity condition
\be\label{monotonicity}
\|E\v\|_1\leq\|\v\|_1\quad\forall\;\v\in\mbb{C}^m\quad\forall\;E\in\stoc(n,m)\;.
\ee
This monotonicity property ensures that the metric\index{metric} $D(\p,\q)\eqdef\frac12\|\p-\q\|_1$ is a divergence.
Remarkably, for vectors in $\mbb{R}^n_+$, up to a multiplication by a constant, the $1$-norm is the only norm with this monotonicity property! 

To see this, let $\|\cdot\|$ be a norm satisfying the same monotonicity property~\ref{monotonicity}, and suppose $\left\|(1,0)^T\right\|=1$. From~\eqref{monotonicity} with $\|\cdot\|$ replacing $\|\cdot\|_1$, it follows that $\|\v\|$ is invariant under permutation of its components, and further more it also implies that $\|\v\|=\|\v\oplus 0\|$. Hence, if $\{\e_1,\ldots,\e_m\}$ is the standard basis of $\mbb{C}^m$ then $\|\e_j\|=1$ for all $j=1,\ldots,m$. Now, let $E$ be the column stochastic matrix whose first row is $[1,1,\ldots,1]$ and all other rows are zero. Then,
\be
|v_1+\cdots+v_m|=\|E\v\|\leq\|\v\|=\Big\|\sum_{j\in[m]}v_j\e_j\Big\|\leq \sum_{j\in[m]}|v_j|=\|\v\|_1
\ee
Note that if all the components of $\v$ are non-negative real numbers then all the inequalities above must be equalities and we get in particular that $\|\v\|=\|\v\|_1$. 

\subsection{The Trace Norm}\label{sectracenorm}\index{trace norm}

The trace norm is the Schatten\index{Schatten}  1-norm as introduced in Definition~\ref{schatt}. Specifically, for any operator $M:A\to B$
it is defined by:
\be\label{tn}
\|M\|_1\eqdef\tr\sqrt{M^*M}\;.
\ee
Therefore, the trace norm is the sum of the singular values of $M$. Recall that all Schatten\index{Schatten}  norms satisfy the invariance\index{invariance} property under isometries (see~\eqref{invprosch}). Particularly, that for any two isometries $U:B\to B'$ and $V:A\to A'$ we have
\be\label{iso3}
\big\|UMV^*\big\|_1=\|M\|_1\;.
\ee

\begin{exercise}$\;$
\begin{enumerate}
\item Show that the trace norm is indeed a norm.
\item Show that the trace norm is always bigger than the norm induced by the inner product~\eqref{xy}.
\end{enumerate}
\end{exercise}

\begin{exercise}\label{tn0}
Show that for any 3 Hermitian operators $M,N,\sigma\in\herm(A)$, with $\sigma> 0$, the following holds:
\begin{enumerate}
\item $\tr(MN)\leq\|M\|_2\|N\|_2$
\item $\|\sqrt{\sigma} M\sqrt{\sigma}\|_1\leq \|M\|_2\|\sigma\|_2$
\item $\|M\|_1\leq \sqrt{\tr[\sigma]}\big\|\sigma^{-1/4}M\sigma^{-1/4}\big\|_2$.\\ Hint: Use part $(b)$ with $M$ replaced by $\sigma^{-1/4}M\sigma^{-1/4}$ and $\sigma$ replaced by $\sqrt{\sigma}$.
\end{enumerate}
where $\|\cdot\|_2$ is the norm induced by the Hilbert-Schmidt inner product.
\end{exercise}

The subsequent two lemmas establish that the trace norm can be formulated as optimization problems. These formulations are instrumental in proving various properties of the trace norm. We begin with an expression that is particularly useful for Hermitian matrices.

\begin{myg}{}
\begin{lemma}\index{trace norm}
Let $M:A\to A$ be an Hermitian operator. The trace norm of $M$ can be expressed as:
\be
\|M\|_1=\max\left\{\tr\left[M\Pi\right]\;:\;-I^A\leq\Pi\leq I^A\;,\;\;\Pi\in\herm(A)\right\}\;.
\ee
\end{lemma}
\end{myg}

\begin{proof}
Let $M_+$ and $M_-$ be the positive and negative parts of $M$ (see~\eqref{decomherm}), and let $\Pi_-$ and $\Pi_+=I-\Pi_-$ the projections to the negative, and non-negative eigenspaces of $M$. With these notations we have $|M|=M_++M_-$, so that the trace norm of $M$ can be expressed as
\ba\label{tn2}
\left\|M\right\|_1&=\tr[M_+]+\tr[M_-]\\
&=\tr\left[M\left(\Pi_+-\Pi_-\right)\right]\\
\GG{Exercise~\ref{exleqw}}&=\max_{-I\leq \Pi\leq I}\tr\left[M\Pi\right]\;,
\ea
where the maximum is over all matrices $\Pi\in\herm(A)$ with eigenvalues between $-1$ and $1$.
\end{proof}

\begin{exercise}\label{exleqw}
Prove the last equality in Eq.~\eqref{tn2}.
\end{exercise}

\begin{exercise}\label{puretd}
Show that for any two (normalized) pure states $|\psi\ra,|\phi\ra\in A$ we have
\be\label{tdex}
T(\psi,\phi)\eqdef\frac12\big\||\psi\lr\psi|-|\phi\lr\phi|\big\|_1=\sqrt{1-|\la\psi|\phi\ra|^2}\;.
\ee
\noindent Hint: Denote $|0\ra\eqdef|\psi\ra$ and express $|\phi\ra\eqdef a|0\ra+b|1\ra$ where $|1\ra$ is some (normalized) orthogonal vector to $|0\ra$.
\end{exercise}

\bex\label{innerprel}
Let $A$ be a Hilbert space and let $|\psi\ra,|\phi\ra\in A$ be two (normalized) states in $A$. Denote $\psi\eqdef|\psi\lr \psi|$ and $\phi\eqdef|\phi\lr\phi|$. Show that 
\be
\frac12\|\psi-\phi\|_1\leq\big\||\psi\ra-|\phi\ra\big\|
\ee
where the norm on the right-hand side is the induced inner-product norm $\||\chi\ra\|\eqdef\la\chi|\chi\ra^{1/2}$.
Hint: Use the previous exercise.
\eex

The trace norm can also be expressed as an optimization over partial isometries.
\begin{myg}{}
\begin{lemma}\label{lem:piso}\index{trace norm}
Let $A$ and $B$ be two finite dimensional Hilbert spaces, and let $M:A\to B$ be a linear operator. Then, the trace norm of $M$ can be expressed as
\be\label{maxm}
\|M\|_1=\max_{V:B\to A}\left|\tr\left[VM\right]\right|\;,
\ee
where the maximum is over all partial isometries $V:B\to A$.
\end{lemma}
\end{myg}

\begin{proof}
Express 
$M=\sum_{x\in[n]}\lambda_x|\phi_x^B\lr \psi_x^A|$, where $\{\lambda_x\}_{x\in[n]}$ are the singular values of $M$, and $\{|\psi_x^A\ra\}_{x\in[n]}$ and $\{|\phi_x^B\ra\}_{x\in[n]}$
are orthonormal sets of vectors in $A$ and $B$, respectively. Let $U:B\to A$ be the partial isometry\index{partial isometry} $U=\sum_{x\in[n]}|\psi_x^A\lr \phi_x^B|$.  We then have
\be
\|M\|_1=\big|\tr\left[UM\right]\big|\leq \max_{V:B\to A}\big|\tr\left[VM\right]\big|
\ee
where the maximum is over all partial isometries $V:B\to A$.
Substituting into the right hand side $M=\sum_{x\in[n]}\lambda_x|\phi_x^B\lr \psi_x^A|$  gives
\ba \label{mu1}
\|M\|_1&\leq\max_{V:B\to A}\Big|\sum_{x\in[n]}\lambda_x\tr\left[V|\phi_x^B\lr \psi_x^A|\right]\Big|\\
&\leq \max_{V:B\to A}\sum_{x\in[n]}\lambda_x\left|\la \psi_x^A|V|\phi_x^B\ra\right|\\
\GG{see~\eqref{seeeq}\; below}&\leq \sum_{x\in[n]}\lambda_x=\|M\|_1\;,
\ea
where we used the Cauchy-Schwarz inequality to get
\ba\label{seeeq}
\left|\la \psi_x^A|V|\phi_x^B\ra\right|&\leq \sqrt{\la \psi_x|\psi_x\ra\la \phi_x|V^*V|\phi_x\ra}\\
&= \sqrt{\la \phi_x|V^*V|\phi_x\ra}\\
\Gg{V^*V\text{ is a projection}}&\leq 1
\ea
Hence, all the inequalities in~\eqref{mu1} must be equalities. This completes the proof.
\end{proof}

\begin{exercise}
Show that if $|A|\geq |B|$ in Lemma~\eqref{lem:piso} 
then the maximization over partial isometries in~\eqref{maxm} can be replaced with maximization over isometries $V:B\to A$. Similarly, show that if $|A|\leq |B|$ then
\be
\|M\|_1=\max_{U:A\to B}\tr\left[U^*M\right]\;,
\ee
where the maximum is over all  isometries $U:A\to B$
\end{exercise}

\subsubsection{Strong Monotonicity Property}

We show here that the trace norm behaves monotonically under a positive linear map (not necessarily CPTP). 
This monotonicity property is stronger than what we discussed so far.

\begin{myt}{\color{yellow} Monotonicity of the Trace Norm}\index{trace norm}\index{monotonicity}
\begin{theorem}
Let $\mE\in\ml(A\to B)$ be a trace non-increasing positive linear map, and let $M\in\ml(A)$. Then, 
\be
\left\|\mE(M)\right\|_1\leq\|M\|_1\;.
\ee
\end{theorem}
\end{myt}
\begin{proof} 
Express $M=\sum_{x\in[n]}\lambda_x|\phi_x^A\lr \psi_x^A|$, where $\{\lambda_x\}_{x\in[n]}$ are the singular values of $M$, and $\{|\psi_x^A\ra\}_{x\in[n]}$ and $\{|\phi_x^A\ra\}_{x\in[n]}$
are orthonormal sets of vectors in $A$. Then, from the triangle inequality of the trace norm we have
\be
\left\|\mE(M)\right\|_1=\Big\|\sum_{x\in[n]}\lambda_x\mE\left(|\phi_x^A\lr \psi_x^A|\right)\Big\|_1\leq\sum_{x\in[n]}\lambda_x\left\|\mE\left(|\phi_x^A\lr \psi_x^A|\right)\right\|_1\;.
\ee
Hence, it will be sufficient to prove that $\left\|\mE\left(|\phi_x^A\lr \psi_x^A|\right)\right\|_1\leq 1$ for all $x$ since this would imply that $\left\|\mE(M)\right\|_1\leq\sum_{x\in[n]}\lambda_x=\|M\|_1$.
For simplicity of the exposition we remove the sub-index $x$ from the rest of the proof, since nothing will depend on it. 

Now, the square matrix $\mE(|\phi\lr \psi|)$ has a polar decomposition\index{polar decomposition}  
\be
\mE(|\phi\lr \psi|)=U\left|\mE(|\phi\lr \psi|)\right|\;.
\ee
Since $U$ is a unitary matrix in $\ml(B)$ it is diagonalizable and can be expressed as
\be\label{eq514}
U=\sum_{y\in[n]}e^{i\theta_y}|\varphi_y\lr \varphi_y|\;,
\ee
where $\{|\varphi_y\ra\}$ is an orthonormal basis of $B$, and $\{\theta_y\}$ are some phases. Hence,
\ba
\left\|\mE(|\phi\lr \psi|)\right\|_1=\tr\left|\mE(M)\right|
&=\tr\left[U^*\mE(|\phi\lr \psi|)\right]\\
\GG{\eqref{eq514}}&=\sum_{y\in[n]}e^{-i\theta_y}\tr\left[|\varphi_y\lr \varphi_y|\mE\left(|\phi\lr \psi|\right)\right]\\
&=\sum_{y\in[n]}e^{-i\theta_y}\tr\left[\mE^*\left(|\varphi_y\lr \varphi_y|\right)|\phi\lr \psi|\right]\;.
\ea
From the Exercise~\ref{showdual}, it follows that $\mE^*$ is positive and sub-unital (i.e. $\mE^*(I)\leq I$). Therefore, the matrices $\Lambda_y\eqdef \mE^*\left(|\varphi_y\lr \varphi_y|\right)\geq 0$ form an incomplete POVM since 
\be
\sum_{y\in[n]}\Lambda_y=\mE^*(I^B)\leq I^A\;.
\ee
With this notation,
\ba
\left\|\mE(|\phi\lr \psi|)\right\|_1=\sum_{y\in[n]}e^{-i\theta_y}\tr\left[\Lambda_y|\phi\lr \psi|\right]
&=\sum_{y\in[n]}e^{-i\theta_y}\la \psi|\Lambda_y|\phi\ra\\
&\leq\sum_{y\in[n]}\left|\la \psi|\Lambda_y|\phi\ra\right|=\sum_{y\in[n]}\left|\la \psi|\sqrt{\Lambda_y}\sqrt{\Lambda_y}|\phi\ra\right|\\
\GG{\text{Cauchy-Schwarz inequality}}&\leq \sum_{y\in[n]}\sqrt{\la \psi|\Lambda_y|\psi\ra\la \phi|\Lambda_y|\phi\ra}\\
\GG{\text{geometric-arithmetic inequality}}&\leq \frac{1}{2}\sum_{y\in[n]}\left(\la \psi|\Lambda_y|\psi\ra+\la \phi|\Lambda_y|\phi\ra\right)\leq 1\;.
\ea
This completes the proof.
\end{proof}

\begin{exercise}
Provide an alternative (simpler) proof of the theorem above for the case that $M$ is Hermitian. Hint: Use the previous lemma and prove first that if $-I^B\leq \Pi\leq I^B$ then $-I^A\leq \mE^*(\Pi)\leq I^A$.
\end{exercise}

\subsection{The Trace Distance}\index{trace distance}

The trace distance between two states $\rho,\sigma\in\md(A)$ is defined by
\begin{myd}{}
\be
T(\rho,\sigma)\eqdef\frac{1}{2}\|\rho-\sigma\|_1
\ee
\end{myd}
The inclusion of the one-half factor is for normalization purposes, specifically to ensure that the distance reaches its maximum value of $1$ when the two states, $\rho$ and $\sigma$, are orthogonal (refer to Exercise~\ref{orthog} for more details).

Consider the case in which $\rho$ and $\sigma$ are classical, or equivalently commute, and therefore diagonal in the same basis. In this case, denoting $\rho=\sum_{x\in[n]}p_{x}|x\lr x|$ and $\sigma=\sum_{x\in[n]}q_x|x\lr x|$ we get
\ba
T(\rho,\sigma)\eqdef\frac{1}{2}\Big\|\sum_{x\in[n]}(p_x-q_x)|x\lr x|\Big\|_1=\frac{1}{2}\sum_{x\in[n]}|p_x-q_x|=: T(\p,\q)\;,
\ea
where $\p\eqdef(p_1,\ldots,p_n)^T$, $\q=(q_1,\ldots,q_n)^T$, and $T(\p,\q)$ as defined above denotes the trace distance between the classical probability vectors $\p$ and $\q$. 

In the general case, since both $\rho$ and $\sigma$ have the same trace 
\be
0=\tr[\rho-\sigma]=\tr[(\rho-\sigma)_+]-\tr[(\rho-\sigma)_-]\;,
\ee
where $(\rho-\sigma)_{\pm}$ are the positive and negative parts of $\rho-\sigma$.
Therefore, denoting by $\Pi_+$ the projection to the positive eigenspace of $\rho-\sigma$, we conclude that 
\be
T(\rho,\sigma)\eqdef\frac{1}{2}\left(\tr[(\rho-\sigma)_+]+\tr[(\rho-\sigma)_-]\right)=\tr[(\rho-\sigma)_+]=\tr[(\rho-\sigma)\Pi_+]
\ee
That is, the trace distance can be written as
\be\label{td}
T(\rho,\sigma)=\max_{0\leq \Pi\leq I}\tr[(\rho-\sigma)\Pi]
\ee
where the maximization is over any matrix $\Pi\in\pos(A)$ (not necessarily a projection) with eigenvalues between 0 and 1. 
The expression above for the trace distance will be useful in some of applications we discuss later on.

The monotonicity of the trace norm under quantum channels (in fact positive maps) implies the monotonicity of the trace distance as well. We summarize it in the following theorem.
\begin{myt}{\color{yellow} Monotonicity of the Trace Distance}\index{monotonicity}
\begin{theorem}\label{monotd}
The trace distance is a quantum divergence. In particular,
for any $\rho,\sigma\in\md(A)$ and $\mE\in\cptp(A\to B)$
\be
T(\mE(\rho),\mE(\sigma))\leq T(\rho,\sigma)\;.
\ee
\end{theorem}
\end{myt}

\begin{exercise}\label{orthog}
Let $\rho,\sigma\in\md(A)$. Show that 
\be
T(\rho,\sigma)=1\quad\iff\quad\rho\sigma=\sigma\rho=0.
\ee
\end{exercise}

\bex\label{ex:m2}
Let $\u\in\md(A)$ be the maximally mixed state and $\psi\in\pure(A)$ be a pure state. Show that the trace distance between these two states is given by
\be
T\left(\u^{A},\psi^{A}\right)=\frac{m-1}{m}\;,
\ee
where $m\eqdef|A|$.
\eex

\subsubsection{A Relationship Between the Trace Distance and the Ky Fan Norm}\index{Ky Fan norm}

There is an interesting relationship between the trace distance and the Ky Fan norm (see Definition~\ref{def:kyfan}) that we will use in our study of entanglement theory.
Consider a quantum state $\rho\in\md(A)$ and let $m<|A|$ be an integer. Further, we denote by $\rho^{(m)}$ the  $m$-pruned version of $\rho$:
\be\label{cfrm}
\rho^{(m)}\eqdef\frac{\Pi_m\rho\Pi_m}{\tr\left[\rho\Pi_m\right]}
\ee
where $\Pi_m\in\pos(A)$ is the projection to the subspace spanned by the $m$ eigenvectors of the $m$ largest eigenvalues of $\rho$. By definition, $\tr\left[\rho\Pi_m\right]=\|\rho\|_{(m)}$ and $\rho$ commutes with $\rho^{(m)}$. In the following exercise you use these properties to show that the trace distance between $\rho$ and $\rho^{(m)}$ is related to the Ky Fan norm\index{Ky Fan norm}.

\bex\label{prune}
Using the same notations as above, show that 
\be\label{tdprun}
T\left(\rho,\rho^{(m)}\right)=1-\|\rho\|_{(m)}\;,
\ee
where $\|\cdot\|_{(m)}$ is the Ky Fan norm. Hint: Use the relations $T\left(\rho,\rho^{(m)}\right)=\tr\left(\rho^{(m)}-\rho\right)_+$ and $\tr\left[\rho\Pi_m\right]=\|\rho\|_{(m)}$, and the fact that $\rho$ commutes with $\rho^{(m)}$.
\eex

In the following theorem we use the notation $\md_m(A)$ to denote the set of all density matrices in $\md(A)$, whose rank is not greater than $m$.

\begin{myt}{}
\begin{theorem}\label{0distdm0}
Using the same notations as above, the trace-distance of $\rho$ to the set $\md_m(A)$ is given by
\be
T\big(\rho,\md_{m}(A)\big)\eqdef\min_{\sigma\in\md_{m}(A)}T(\rho,\sigma)=T\left(\rho,\rho^{(m)}\right)=1-\|\rho\|_{(m)}\;.
\ee
\end{theorem}
\end{myt}

\begin{proof}
Let $\sigma\in\md_m(A)$ and observe that since $\rank(\sigma)\leq m$ we have
\ba\label{5p164}
\|\rho\|_{(m)}&\geq \tr\left[\Pi_\sigma\rho\right]\\
&=\tr\left[\Pi_\sigma\sigma\right]+\tr\left[\Pi_\sigma(\rho-\sigma)\right]\\
&=1+\tr\left[\Pi_\sigma\Big((\rho-\sigma)_+-(\rho-\sigma)_-\Big)\right]\\
\Gg{\tr\left[\Pi_\sigma(\rho-\sigma)_+\right]\geq 0}&\geq 1-\tr\left[\Pi_\sigma(\rho-\sigma)_-\right]\\
\Gg{\Pi_\sigma\leq I^A}&\geq 1-\tr(\rho-\sigma)_-\\
&=1-T(\rho,\sigma)\;.
\ea
Therefore, for every $\sigma\in\md_m(A)$ we get that
\ba
T(\rho,\sigma)&\geq 1-\|\rho\|_{(m)}\\
\GG{Exercise~\ref{prune}}&=T\left(\rho,\rho^{(m)}\right)\;.
\ea
On the other hand, since $\rho^{(m)}\in\md_m(A)$ by taking above $\sigma=\rho^{(m)}$ we can achieve an equality. Hence,
\be
T\big(\rho,\md_{m}(A)\big)= T\left(\rho,\rho^{(m)}\right)=1-\|\rho\|_{(m)}\;.
\ee
This completes the proof.
\end{proof}

\subsubsection{Optimality of the Trace Distance}\index{trace distance}\index{optimal extension}

In~\eqref{fdclass} we saw that for $f(r)=\frac12|r-1|$ we get that the classical $f$-Divergence\index{$f$-divergence} is equal to the classical trace distance.  We show now that the minimal quantum extension of the classical trace distance equals to the quantum trace distance. Explicitly, let $T_{c}$ be the classical trace distance, and let $\underline{T}_c$ be its minimal quantum extension. That is, for any $\rho,\sigma\in\md(A)$
(cf.~\eqref{ud})
\be
\underline{T}_c(\rho,\sigma)\eqdef\sup_{\mE\in\cptp(A\to X)}T_c\big(\mE(\rho),\mE(\sigma)\big)\;,
\ee
where the supremum is over all classical systems X, POVM channels $\mE\in\cptp(A\to X)$, and the diagonal matrices $\mE(\rho)$ and $\mE(\sigma)$ are viewed as probability vectors.

\begin{myt}{}
\begin{theorem}\label{tdpovm}
Using the same notations as above, for all $\rho,\sigma\in\md(A)$
\be
\underline{T}_c(\rho,\sigma)=T(\rho,\sigma)\eqdef\frac12\|\rho-\sigma\|_1\ee
\end{theorem}
\end{myt}
\begin{remark}
The theorem above demonstrates that the quantum trace distance is the smallest divergence that reduces to the classical trace distance on classical states.
\end{remark}
\begin{proof}
From  Sec.~\eqref{optim}, particularly Theorem~\ref{optim2}, it follows that for any $\rho,\sigma\in\md(A)$ we have  $T(\rho,\sigma)\geq \underline{T}_c(\rho,\sigma)$ since $\underline{T}_c$ is the minimal quantum divergence that reduces to the classical trace distance when the input restricted to classical states. To prove the converse inequality $T(\rho,\sigma)\leq \underline{T}_c(\rho,\sigma)$ , let $\Pi_{\pm}$ be the two projections to the positive and negative eigenspaces of $\rho-\sigma$, 
and let $\mE\in\cptp(A\to X)$ with $|X|=2$ be its corresponding POVM Channel\index{POVM channel} ; i.e. 
$\mE(\omega)=\tr[\omega \Pi_+]|0\lr 0|+\tr[\omega \Pi_{-}]|1\lr 1|$ for all $\omega\in\md(A)$.
Then, by definition,
\be
\underline{T}_c\big(\rho,\sigma\big)\geq T_c\big(\mE(\rho),\mE(\sigma)\big)=\frac{1}{2}\left(\tr[(\rho-\sigma)_+]+\tr[(\rho-\sigma)_-]\right)=T(\rho,\sigma)\;.
\ee
This completes the proof.
\end{proof}

\subsubsection{Joint Convexity of the Trace Distance}\index{joint convexity}

The monotonicity of the trace distance under quantum channels implies that the trace distance is jointly convex.
\begin{myt}{}
\begin{theorem}
Let $\p=(p_1,\ldots,p_{m})^T$ and $\q=(q_1,\ldots,q_{m})^T$ be two probability vectors, and let $\{\rho_x\}_{x\in[m]}$ and $\{\sigma_x\}_{x\in[m]}$ be two sets of $m$ density matrices in $\md(A)$. Then,
\be
T\Big(\sum_{x\in[m]}p_x\rho_x,\sum_{x\in[m]}q_x\sigma_x\Big) \leq T(\p,\q)+\sum_{x\in[m]}p_xT(\rho_x,\sigma_x)\;,
\ee
where $T(\p,\q)\eqdef\frac{1}{2}\sum_{x\in[m]}|p_x-q_x|$ is the classical trace distance between two probability vectors.
\end{theorem}
\end{myt}
We provide two proofs for this theorem to illustrate different techniques.
\begin{proof}
Let $X$ be a classical system (register) and define the classical-quantum states
\be\label{cqstatesmon}
\rho^{XA}\eqdef\sum_{x\in[m]}p_x|x\lr x|^X\otimes\rho_x^A\quad\text{and}\quad\sigma^{XA}\eqdef\sum_{x\in[m]}q_x|x\lr x|^X\otimes\sigma_x^A\;.
\ee
Then,
\ba
T\left(\rho^{XA},\sigma^{XA}\right)&=\frac{1}{2}\Big\|\sum_{x\in[m]}|x\lr x|\otimes\left(p_x\rho_x-q_x\sigma_x\right)\Big\|_1\\
&=\frac{1}{2}\sum_{x\in[m]}\|p_x\rho_x-q_x\sigma_x\|_1\\
&=\frac{1}{2}\sum_{x\in[m]}\|p_x\rho_x-p_x\sigma_x+p_x\sigma_x-q_x\sigma_x\|_1\\
\GG{Triangle\;inequality}&\leq \frac{1}{2}\sum_{x\in[m]}\big(\|p_x\rho_x-p_x\sigma_x\|_1+\|p_x\sigma_x-q_x\sigma_x\|_1\big)\\
&=\sum_{x\in[m]}p_xT(\rho_x,\sigma_x)+\frac{1}{2}\sum_{x\in[m]}|p_x-q_x|\;.
\ea
Hence,
\ba
T\Big(\sum_{x\in[m]}p_x\rho_x^A,\sum_{x\in[m]}q_x\sigma_x^A\Big)&=T\left(\tr_X[\rho^{XA}],\tr_X[\sigma^{XA}]\right)\\
\GG{\text{monotonicity (Theorem~\ref{monotd})}}&\leq T\left(\rho^{XA},\sigma^{XA}\right)\\
&\leq \sum_{x\in[m]}p_xT(\rho_x^A,\sigma_x^A)+T(\p,\q)\;.\nonumber
\ea
This completes the proof.
\end{proof}

\begin{proof}[Alternative Proof]
Let $\Pi$ be the optimal projection such that
\be
T\Big(\sum_{x\in[m]}p_x\rho_x,\sum_{x\in[m]}q_x\sigma_x\Big)=\tr\Big[\Pi\sum_{x\in[m]}\left(p_x\rho_x-q_x\sigma_x\right)\Big]
\ee
Therefore,
\ba
T\Big(\sum_{x\in[m]}p_x\rho_x,\sum_{x\in[m]}q_x\sigma_x\Big)&=\sum_{x\in[m]}\big(p_x\tr\left[\Pi(\rho_x-\sigma_x)\right]+(p_x-q_x)\tr[\Pi\sigma_x] \big)\\
&\leq \sum_{x\in[m]}p_xT(\rho_x,\sigma_x)+\sum_{x\in[m]}(p_x-q_x)_{+}\\
& =\sum_{x\in[m]}p_xT(\rho_x,\sigma_x)+T(\p,\q)\;,
\ea
where we used the fact that $\tr[\Pi\sigma_x]\leq 1$ and the fact that $p_x-q_x\leq (p_x-q_x)_+$. This completes the proof.
\end{proof}

We conclude this subsection by discussing a nuanced yet crucial property of the trace distance. This property is highly relevant to certain applications in quantum information, though it is often overlooked. Consider $\rho,\sigma\in\md(A)$ and let us define $\eps\eqdef T(\rho,\sigma)$. If $\eps$ is very small, it implies that $\rho$ and $\sigma$ are nearly identical states. This concept can be articulated as follows: Decompose $\rho-\sigma$ into positive and negative parts, written as $\rho-\sigma=(\rho-\sigma)_+-(\rho-\sigma)_-$. Then, define two states $\omega_\pm\eqdef\frac{1}{\eps}(\rho-\sigma)_{\pm}$. Given that $\eps=T(\rho,\sigma)=\tr(\rho-\sigma)_+=\tr(\rho-\sigma)_-$, it follows that $\omega_{\pm}$ are valid density matrices in $\md(A)$. Furthermore, we can express:
\be\label{decomom}
\rho-\sigma=\eps(\omega_+-\omega_-);.
\ee
The importance of this equation lies in the fact that the matrix $H\eqdef \omega_+-\omega_-$ is bounded, satisfying $-I\leq H\leq I$. Additionally, the equation $\rho=\sigma+\eps H$ does not depend explicitly on the underlying dimension $|A|$.

To further elucidate this point, consider the following straightforward example involving the Schatten\index{Schatten}  2-norm (the norm induced by the Hilbert-Schmidt inner product). Let $\rho_n=\frac{1}{n}I_n$ denote the $n\times n$ maximally mixed state. Observe that its 2-norm is calculated as follows:\index{Hilbert-Schmidt inner product}
\be
\|\rho_n\|_2\eqdef\sqrt{\tr[\rho_n^2]}=\frac{1}{\sqrt{n}}\;.
\ee
Consequently, as $n$ approaches infinity, $\|\rho_n\|_2$ tends towards zero, while the trace norm $\|\rho_n\|_1=1$  for all $n\in\mbb{N}$.

\begin{exercise}
Using the same notations as above,
show that if a set of Hermitian matrices $\{H_n\}$, with each $H_n\in\herm(\mbb{C}^n)$, satisfies $\lim_{n\to \infty}\|H_n\|_1= 0$, then there exists a sequence of positive numbers $\{\eps_n\}$ with a limit $\lim_{n\to\infty}\eps_n=0$ and a set of bounded matrices $\{M_n\}$ with $-I_n\leq M_n\leq I_n$ such that $H_n=\eps_nM_n$.
\end{exercise}

\subsection{The Fidelity}\index{fidelity}

The fidelity is another distance like measure between two quantum states, however, unlike the trace distance or any other divergence, it reaches its maximal value when the two states are the same, and its minimal value when they are orthogonal. 
For any $\rho,\sigma\in\md(A)$ it is defined by
\begin{myd}{}
\be
F(\rho,\sigma)\eqdef\|\sqrt{\rho}\sqrt{\sigma}\|_1=\tr\left[|\sqrt{\rho}\sqrt{\sigma}|\right]=\tr\sqrt{\sqrt{\sigma}\rho\sqrt{\sigma}}\;.
\ee
\end{myd}
Consider first the simpler case that $\rho$ and $\sigma$ commutes. In this case, there exists a basis $\{|x\ra\}_{x\in[n]}$ of $A$ such that 
$\rho=\sum_{x\in[n]}p_{x}|x\lr x|$ and $\sigma=\sum_{x\in[n]}q_x|x\lr x|$. We then have
\ba
F(\rho,\sigma)=\|\sqrt{\rho}\sqrt{\sigma}\|_1=\Big\|\sum_{x\in[n]}\sqrt{p_{x}q_x}|x\lr x|\Big\|_1
=\sum_{x\in[n]}\sqrt{p_{x}q_x}\eqdef F(\p,\q)
\ea
where $\p\eqdef(p_1,\ldots,p_n)^T$, $\q=(q_1,\ldots,q_n)^T$, and $F(\p,\q)$ as defined above denotes the fidelity between the classical probability vectors $\p$ and $\q$. 
 If $\rho=\sigma$ we get that $F(\rho,\rho)=\|\sqrt{\rho}\sqrt{\rho}\|_1=\|\rho\|_1=\tr[\rho]=1$. 
Moreover, for any $\rho,\sigma\in\md(A)$, the fidelity $F(\rho,\sigma)$, cannot be greater than one. This will follow trivially from Uhlmann's theorem below, and can also be seen from the following argument:
\ba
|\sqrt{\rho}\sqrt{\sigma}|^2&=\sqrt{\sigma}\rho\sqrt{\sigma}\\
\Gg{\rho= I-(I-\rho)}&=\sigma-\sqrt{\sigma}(I-\rho)\sqrt{\sigma}\\
\Gg{\sqrt{\sigma}(I-\rho)\sqrt{\sigma}\geq 0}&\leq\sigma\leq I^A\;.
\ea
Therefore, $|\sqrt{\rho}\sqrt{\sigma}|\leq I^A$.

\begin{exercise}
Let $\rho,\sigma\in\md(A)$.
\begin{enumerate}
\item Show that the fidelity is symmetric: $F(\rho,\sigma)=F(\sigma,\rho)$. Hint: Use the fact that for any complex matrix $M$, the matrix $M^*M$ has the same non-zero eigenvalues as $MM^*$.
\item Show that if $\sigma=|\psi\lr\psi|$ is pure then $F(\rho,\sigma)=\sqrt{\la\psi|\rho|\psi\ra}$ 
\end{enumerate}
\end{exercise}

\bex\label{eigfid}
Let $\rho,\sigma\in\md(A)$. Show that if $\lambda$ is an eigenvalue of the matrix $|\sqrt{\rho}\sqrt{\sigma}|$ then $\lambda^2$ is an eigenvalue of the non-Hermitian matrix $\rho\sigma$.
Hint: Let $M=\sqrt{\sigma}\rho\sqrt{\sigma}$ and $N=\rho\sigma$ and find a matrix $\eta\geq 0$ such that $M=\eta^{-1}N\eta$, where $\eta^{-1}$ is the generalized inverse of $\eta$.
\eex

\subsubsection{Uhlmann's Theorem}\index{Uhlmann's theorem}

The last part in the exercise above also implies that if both $\rho=|\psi\lr \psi|$ and $\sigma=|\phi\lr \phi|$ are pure, then the fidelity becomes the absolute value of the inner product between the two states; i.e. $F(\rho,\sigma)=|\la\psi|\phi\ra|$. The following theorem by Uhlmann's shows that this can be extended to mixed states by considering all the possible purifications of $\rho$ and $\sigma$. 

\begin{myt}{\color{yellow} Uhlmann's Theorem}
\begin{theorem}\label{Uhlm} Let $\rho,\sigma\in\md(A)$ be two density matrices, and let $|\psi^{AB}\ra$ and $|\phi^{AC}\ra$ be two purifications of $\rho^A$ and $\sigma^A$, respectively.  Then,
\be
F(\rho^A,\sigma^A)=\max_{V^{B\to C}}|\la\psi^{AB}|V^{*}|\phi^{AC}\ra|
\ee
where the maximum is over all partial isometries $V:B\to C$.  
\end{theorem}
\end{myt}
\begin{remark}
We emphasis that the purifying systems $B$ and $C$ are not necessarily isomorphic. That is, we can have $|B|\neq |C|$.
\end{remark}
\begin{proof}
From Exercise~\eqref{purification} it follows that the purifications $|\psi^{AB}\ra$ and $|\phi^{AC}\ra$ must have the form:
\be
|\psi^{AB}\ra=\sqrt{\rho}\otimes U^{\tA\to B} |\Omega^{A\tA}\ra\quad\text{and}\quad |\phi^{AB}\ra=\sqrt{\sigma}\otimes W^{\tA\to C}|\Omega^{A\tA}\ra\;,
\ee
where $U,W$ are two isometries. Thus,
\ba\label{mpp}
\max_{V^{B\to C}}\left|\la\psi^{AB}|V^{*}|\phi^{AC}\ra\right|&=\max_{V^{B\to C}}\left|\la\Omega^{A\tA}|\sqrt{\rho}\sqrt{\sigma}\otimes U^{*}V^*W|\Omega^{A\tA}\ra\right|\\
\GG{\text{part 2 of Exercise~\ref{bipartite}}}&=\max_{V^{B\to C}}\left|\tr\left[\sqrt{\rho}\sqrt{\sigma}\left(U^{*}V^*W\right)^T\right]\right|\\
&=\max_{V^{B\to C}}\left|\tr\left[\bar{U}\sqrt{\rho}\sqrt{\sigma}W^T\bar{V}\right]\right| \\
\GG{\text{Lemma~\ref{lem:piso}}}&=\left\|\bar{U}\sqrt{\rho}\sqrt{\sigma}W^T\right\|_1&\\
\GG{\eqref{iso3}}&=\left\|\sqrt{\rho}\sqrt{\sigma}\right\|_1&\\
&=F(\rho^A,\sigma^A)\;.
\ea
This completes the proof.
\end{proof}

Uhlmann's theorem has numerous applications in quantum information, and we will use it quite often later on in the book.  The following corollary is an immediate consequence of Uhlmann's theorem. We leave its proof as an exercise.

\begin{corollary}
Let $\rho,\sigma\in\md(A)$. Then, $F(\rho,\sigma)\leq 1$ with equality if and only if $\rho=\sigma$.
\end{corollary}

The next consequence of Uhlmann's theorem is the monotonicity of the fidelity under quantum channels.

\begin{myg}{Monotonicity of the Fidelity}\index{monotonicity}
\begin{corollary}\label{monfid}
Let $\rho,\sigma\in\md(A)$, and let $\mE\in\cptp(A\to B)$ be a quantum channel. Then, 
\be\label{fmono}
F(\rho,\sigma)\leq F\left(\mE(\rho),\mE(\sigma)\right)\;.
\ee
\end{corollary}
\end{myg}

\begin{proof}
Let $|\psi^{AC}\ra$ and $|\phi^{AC}\ra$ be optimal purifications of $\rho$ and $\sigma$ such that the fidelity $F(\rho,\sigma)=|\la\psi^{AC}|\phi^{AC}\ra|$ (i.e. we are using Uhlmann's Theorem). Now, from Stinespring dilation theorem there exists an isometry $V:A\to {BE}$ such that
\ba
&\mE(\rho)=\tr_E\left[V\rho V^*\right]=\tr_{EC}\left[\left(V\otimes I^C\right)|\psi^{AC}\lr \psi^{AC}|\left(V^*\otimes I^C\right)\right]\\
&\mE(\sigma)=\tr_E\left[V\sigma V^*\right]=\tr_{EC}\left[\left(V\otimes I^C\right)|\phi^{AC}\lr \phi^{AC}|\left(V^*\otimes I^C\right)\right]\;.
\ea
Denote by $|\tilde{\psi}^{BEC}\ra\eqdef \left(V^{A\to BE}\otimes I^C\right)|\psi^{AC}\ra$ and by $|\tilde{\phi}^{BEC}\ra\eqdef \left(V^{A\to BE}\otimes I^C\right)|\phi^{AC}\ra$. Therefore the above equation implies that $|\tilde{\psi}^{BEC}\ra$ and $|\tilde{\phi}^{BEC}\ra$ are purifications of $\mE(\rho)$ and $\mE(\sigma)$. We therefore get from Uhlmann's Theorem that 
\be
F\left(\mE(\rho),\mE(\sigma)\right)\geq \left|\la\tilde{\psi}^{BEC}|\tilde{\phi}^{BEC}\ra\right|=\left|\la\psi^{AC}|\phi^{AC}\ra\right|=F(\rho,\sigma)\;.
\ee
This completes the proof.
\end{proof}

Note that since the partial trace is a quantum channel it follows that for any two bipartite states $\rho,\sigma\in\md(AB)$ we have
\be
F(\rho^{A},\sigma^{A})\geq F(\rho^{AB},\sigma^{AB})\;.
\ee
Another corollary of Uhlmann's theorem is the joint concavity of the fidelity.
\begin{myg}{Joint Concavity of the Fidelity}\index{joint concavity}
\begin{corollary}
Let $\{p_x\}_{x\in[m]}$ and $\{q_x\}_{x\in[m]}$ be two probability distributions, and let $\{\rho_{x}\}_{x\in[m]}$
and $\{\sigma_{x}\}_{x\in[m]}$ be two sets of $m$ density matrices in $\md(A)$. Then,
\be\label{jc}
F\Big(\sum_{x\in[m]}p_x\rho_x,\sum_{x\in[m]}q_x\sigma_x\Big)\geq\sum_{x\in[m]}\sqrt{p_xq_x}F(\rho_x,\sigma_x)\;.
\ee
\end{corollary}
\end{myg}

\begin{remark}
Note that from the corollary above it follows in particular that
\be\label{jc}
F\Big(\sum_{x\in[m]}p_x\rho_x,\sum_{x\in[m]}p_x\sigma_x\Big)\geq\sum_{x\in[m]}p_xF(\rho_x,\sigma_x)\;.
\ee
Hence, the fidelity is jointly concave.
\end{remark}

\begin{proof}
Let $\{|\psi_x^{AB}\ra\}_{x\in[m]}$ and $\{|\phi_x^{AB}\ra\}_{x\in[m]}$ be optimal purifications, respectively, of $\{\rho_x^A\}_{x\in[m]}$ and $\{\sigma_x^A\}_{x\in[m]}$ such that
\be
F(\rho_x,\sigma_x)=\la\psi_x^{AB}|\phi_{x}^{AB}\ra\quad\quad\forall\;x\in[m]\;.
\ee
Define also the pure states
\be\label{befa}
|\tilde{\psi}^{ABC}\ra\eqdef\sum_{x\in[m]}\sqrt{p_x}|\psi_x^{AB}\ra|x\ra^C\quad\text{and}\quad
|\tilde{\phi}^{ABC}\ra\eqdef\sum_{x\in[m]}\sqrt{q_x}|\phi_x^{AB}\ra|x\ra^C
\ee
where $C$ is some $m$ dimensional system. Note that the two states above are purifications of $\sum_{x\in[m]}p_x\rho_x^A$ and $\sum_{x\in[m]}q_x\sigma_x^A$, respectively. Therefore, we must have
\ba
F\Big(\sum_{x\in[m]}p_x\rho_x^A,\sum_{x\in[m]}q_x\sigma_x^A\Big)&\geq \left|\la\tilde{\psi}^{ABC}|\tilde{\phi}^{ABC}\ra\right|\\
\GG{\eqref{befa}}&=\sum_{x\in[m]}\sqrt{p_xq_x}\la\psi_x^{AB}|\phi_{x}^{AB}\ra \\
&=\sum_{x\in[m]}\sqrt{p_xq_x}F(\rho_x,\sigma_x)\;.
\ea
This completes the proof.
\end{proof}

\begin{exercise}
Prove the joint concavity of the fidelity by defining classical quantum states as in~\eqref{cqstatesmon} and using the monotonicity of the Fidelity under the partial trace.
\end{exercise}

\bex
The square fidelity on $\prob(n)\times\prob(n)$ is defined for all $\p,\q\in\prob(n)$ as
\be
F(\p,\q)^2=\Big(\sum_{x\in[n]}\sqrt{p_xq_x}\Big)^2=\p\cdot\q+\sum_{\substack{x,y\in[n]\\x\neq y}}\sqrt{p_xq_x}\sqrt{p_yq_y}\;.
\ee
Show that the square of the fidelity is concave on each of its arguments; that is, show that for any $k\in\mbb{N}$, $\{\q_z\}_{z\in[k]}\subset\prob(n)$, and $\t\in\prob(k)$ we have
\be\label{concc}
\sum_{z\in[k]}t_zF(\p,\q_z)^2\leq F\Big(\p,\sum_{z\in[k]}t_z\q_z\Big)^2\;.
\ee
Similarly, show that the square fidelity is concave with respect to the first argument.
\eex

\begin{exercise}
Let $\rho,\sigma\in\md(A)$ and let $\tau,\omega\in\md(B)$. Show that
\be
F\left(\rho^A\otimes\tau^{B},\sigma^A\otimes\omega^{B}\right)=F\left(\rho^A,\sigma^A\right)F\left(\tau^{B},\omega^{B}\right)\;.
\ee
\end{exercise}

\subsubsection{Optimal Extensions of the Classical Fidelity}\index{optimal extension}

The corollaries above also implies that the function
$1-F(\rho,\sigma)$
is a faithful quantum divergence. We can therefore study the optimality of the fidelity by exploring the minimal and maximal quantum extensions of the classical divergence\index{classical divergence} $1-F(\p,\q)$. In particular, the maximal quantum extension of the classical fidelity $F(\p,\q)$, which corresponds to the minimal quantum extension of $1-F(\p,\q)$, is given by (cf.~\eqref{ud})
\be
\overline{F}(\rho^A,\sigma^A)\eqdef\inf_{\mE\in\cptp(A\to X)}F\big(\mE^{A\to X}(\rho^A),\mE^{A\to X}(\sigma^A)\big)\quad\quad\forall\;\rho,\sigma\in\md(A)\;,
\ee
where the infimum is over all classical systems $X$ and all POVM channels $\mE\in\cptp(A\to X)$. By applying Theorem~\ref{optim2} to the classical divergence\index{classical divergence} $1-F(\p,\q)$ we get that any function $f$ that satisfies the same monotonicity property~\eqref{fmono} as the fidelity and that reduces to the fidelity on classical states must satisfy
$f(\rho,\sigma)\leq \overline{F}(\rho,\sigma)$ for all $\rho,\sigma\in\md(A)$. 
Remarkably, the Uhlmann's theorem implies that the fidelity in fact equals to this maximal quantum extension.

\begin{myg}{Optimality}
\begin{corollary}\label{fipovm}
For any $\rho,\sigma\in\md(A)$ 
\be
\overline{F}(\rho,\sigma)={F}(\rho,\sigma)
\ee
\end{corollary}
\end{myg} 

\begin{proof}
As discussed above, the inequality $\overline{F}(\rho,\sigma)\geq {F}(\rho,\sigma)$
follows by applying Theorem~\ref{optim2} to the classical divergence\index{classical divergence} $1-F(\p,\q)$.
To prove the converse, let $\{|x\lr x|\}_{x\in[m]}$ be the orthonormal eigenbasis of the Hermitian matrix
\be
\Lambda=\sigma^{-1/2}\left(\sigma^{1/2}\rho\sigma^{1/2}\right)^{1/2}\sigma^{-1/2}
\ee
such that $\Lambda|x\ra=\lambda_x|x\ra$, with $\{\lambda_x\}_{x\in[m]}$ being the eigenvalues of $\Lambda$.
The key reason for this choice of basis is that the matrix $\Lambda$ satisfies 
\be\label{6163}
\Lambda\sigma \Lambda=\rho\;.
\ee
Taking $\mE\in\cptp(A\to X)$ to be the completely dephasing channel in the basis $\{|x\ra\}_{x\in[m]}$, we obtain
\ba
F\big(\mE(\rho),\mE(\sigma)\big)=\sum_x\sqrt{\la x|\rho|x\ra\la x|\sigma|x\ra}&=\sum_x\sqrt{\la x|\Lambda\sigma \Lambda|x\ra\la x|\sigma|x\ra}\\
&=\sum_x\sqrt{\lambda_x^2\la x|\sigma |x\ra\la x|\sigma|x\ra}=\sum_x\lambda_x\la x|\sigma |x\ra\\
&=\sum_x\lambda_x\tr\left[\sigma|x\lr x|\right]
=\tr[\Lambda\sigma]\\
&=\tr\left[\left(\sigma^{1/2}\rho\sigma^{1/2}\right)^{1/2}\right]=F(\rho,\sigma)\;.
\ea
This completes the proof.
\end{proof}

\begin{exercise}
Prove~\eqref{6163}.
\end{exercise}

Since the function $1-F(\p,\q)$ is a classical divergence, we can also define its maximal quantum extension. This maximal extension corresponds to the minimal quantum extension of the fidelity. The minimal quantum extension of the classical fidelity is given by
\be
\underline{F}(\rho,\sigma)=\sup F(\p,\q)
\ee
where the supremum is over all classical systems $X$, and over all $\p,\q\in\md(X)$ for which there exists a channel $\mE\in\cptp(X\to A)$ such that $\rho=\mE(\p)$ and $\sigma=\mE(\q)$ (depending on the context, we are using the notation $\p,\q$ to indicate either diagonal density matrices in $\md(A)$ or probability vectors in $\prob(n)$ ). 

\begin{myt}{\color{yellow} The Minimal Fidelity}
\begin{theorem}
The minimal quantum extension of the classical fidelity is given by
\be
\underline{F}(\rho,\sigma)=F_M(\rho,\sigma)\eqdef\tr\left[\tsigma\left(\tsigma^{-\frac12}\trho\tsigma^{-\frac12}\right)^{\frac12}\right]\;.
\ee
where $\tsigma=\sigma_{11}$ and $\trho\eqdef\rho_{11}-\zeta\rho_{22}^{-1}\zeta^*$ are as given in~\eqref{7124}.
\end{theorem}
\end{myt}

\begin{proof}
Define
\be
D_{\star}(\p\|\q)\eqdef1-F(\p,\q)=\sum_{x\in[n]}\big(q_x-\sqrt{p_{x}q_x}\big)=\sum_{x\in\supp(\q)}q_x\left(1-\sqrt{\frac{p_{x}}{q_x}}\right)
\ee
Therefore, $D_{\star}$ equals the $f$-Divergence\index{$f$-divergence}, $D_f$, with $f(r)\eqdef 1-\sqrt{r}$. Note that the function $f:[0,\infty)\to\mbb{R}$ is  continuous and operator convex, so we can apply the formula~\eqref{7124} to get a closed form for the maximal quantum extension of $D_f$. Explicitly, from~\eqref{7124} and the fact that $\tilde{f}(0)\eqdef\lim_{\eps\to 0^+}\eps f(1/\eps)=\lim_{\eps\to 0^+}(\eps+\sqrt{\eps})=0$ we get
\ba
\bD_f(\rho\|\sigma)&= \tr\left[\tsigma f\left(\tsigma^{-\frac12}\trho\tsigma^{-\frac12}\right)\right]\\
&= \tr\left[\tsigma \left(1-\left(\tsigma^{-\frac12}\trho\tsigma^{-\frac12}\right)^{\frac12}\right)\right]\\
&=1-\tr\left[\tsigma \left(\tsigma^{-\frac12}\trho\tsigma^{-\frac12}\right)^{\frac12}\right]\;.
\ea
Hence, 
\be
\underline{F}(\rho,\sigma)=1-\bD_f(\rho\|\sigma)=\tr\left[\tsigma\left(\tsigma^{-\frac12}\trho\tsigma^{-\frac12}\right)^{\frac12}\right]\;.
\ee
This completes the proof.
\end{proof}
Note that if $\sigma>0$ then $\tsigma=\sigma$ and $\trho=\rho$, and  
if $\sigma\not>0$ then one can express the minimal fidelity in terms of the limit (cf. Exercise~\ref{a01})
\be
F_M(\rho,\sigma)=\lim_{\eps\to 0^+}F_M(\rho,\sigma+\eps I)
\ee 

\begin{exercise}
Prove the following properties of the minimal fidelity:
\begin{enumerate}
\item Ranges from zero to one; i.e. $0\leq F_M(\rho,\sigma)\leq 1$ for all $\rho,\sigma\in\md(A)$.
\item Symmetry; $F_M(\rho,\sigma)=F_M(\sigma,\rho)$ for all $\rho,\sigma\in\md(A)$ (Hint, use the symmetry of the classical divergence).
\item Attains zero for orthogonal states.
\item Joint Concavity; satisfies~\eqref{jc} with $F$ replaced by $F_M$.
\item Multiplicativity over tensor products; 
\be
F_M(\rho_1\otimes\rho_2,\sigma_1\otimes\sigma_2)=F_M(\rho_1,\sigma_1)F_M(\rho_2,\sigma_2)
\ee
for all $\rho_1,\sigma_1\in\md(A)$ and all $\rho_2,\sigma_2\in\md(B)$.
\end{enumerate}
\end{exercise}

\subsection{The Relation Between the Trace Distance and the Fidelity}\index{trace distance}\index{fidelity}
The trace distance and the fidelity satisfy the following inequalities.

\begin{myt}{}
\begin{theorem}
Let $\rho,\sigma\in\md(A)$, $F$ be the fidelity, and $T$ the trace distance. Then,
\be\label{fitr}
1-F\left(\rho,\sigma\right)\leq T\left(\rho,\sigma\right)\leq\sqrt{1-F\left(\rho,\sigma\right)^2}\;.
\ee
\end{theorem}
\end{myt}
This relation reveals that if the fidelity is close to one then the trace distance is close to zero, and if the fidelity is close to zero then the trace distance is close to one. 

\begin{proof} We first prove the upper bound. Let $\psi^{AB}$ and $\phi^{AB}$ be purifications or $\rho^A$ and $\sigma^A$, such that $F(\rho^A,\sigma^A)=|\la\psi^{AB}|\phi^{AB}\ra|$. Such purifications exists due to Uhlmann's theorem.
We then have from monotonicity of the trace distance under partial trace
\ba
T\left(\rho^A,\sigma^A\right)&\leq T\left(\psi^{AB},\phi^{AB}\right)&\text{}\\
\GG{\text{Exercise~\ref{puretd}}}&=\sqrt{1-|\la\psi^{AB}|\phi^{AB}\ra|^2}\\
&=\sqrt{1-F\left(\rho^A,\sigma^A\right)^2}\;.
\ea
where the last equality follows from the definition of $\psi^{AB}$ and $\phi^{AB}$.

To get the lower bound of~\eqref{fitr}, we start by observing that from Corollary~\ref{fipovm} that there exists a POVM $\{\Lambda_x\}_{x\in[n]}$ such that 
\be
F(\rho,\sigma)=\sum_{x\in[n]}\sqrt{p_xq_x}\quad\text{where}\quad p_x\eqdef\tr[\Lambda_x\rho]\;\;,\;\;q_x\eqdef\tr[\Lambda_x\sigma]\;.
\ee
Hence, from the equality $2\sqrt{p_xq_x}=p_x+q_x-(\sqrt{p_x}-\sqrt{q_x})^2$ it follows
\be\label{b01}
F(\rho,\sigma)=\frac{1}{2}\sum_{x\in[n]}\Big(p_x+q_x-(\sqrt{p_x}-\sqrt{q_x})^2\Big)=1- \frac{1}{2}\sum_{x}(\sqrt{p_x}-\sqrt{q_x})^2\;.
\ee
To bound the last term, observe that
\ba\label{b02}
\frac{1}{2}\sum_{x\in[n]}(\sqrt{p_x}-\sqrt{q_x})^2&=\frac{1}{2}\sum_{x}|\sqrt{p_x}-\sqrt{q_x}||\sqrt{p_x}-\sqrt{q_x}|\\
\Gg{|\sqrt{p_x}-\sqrt{q_x}|\leq\sqrt{p_x}+\sqrt{q_x}}&\leq\frac12\sum_{x\in[n]}|\sqrt{p_x}-\sqrt{q_x}|(\sqrt{p_x}+\sqrt{q_x})\\
&=\frac12\sum_{x\in[n]}|p_x-q_x|=T(\p,\q)\\
\GG{Theorem~\ref{tdpovm}}&\leq T(\rho,\sigma)
\ea
Combining Eqs.~(\ref{b01},\ref{b02}) gives
\be
F(\rho,\sigma)\geq 1-T(\rho,\sigma)\;,
\ee
which is equivalent to the lower bound of~\eqref{fitr}.
\end{proof}

We say that two states $\rho,\sigma\in\md(A)$ are $\eps$-close in trace distance if $T(\rho,\sigma)\leq\eps$ for some $\eps>0$. Similarly, $\rho$ and $\sigma$ are $\eps$-close in fidelity if $F(\rho,\sigma)\geq 1-\eps$.
In the exercise below you will show that the relation between the trace distance and the fidelity can be used to show that the two notions of ``$\eps$-close" are essentially the same.

\begin{exercise}$\;$
\begin{enumerate}
\item Show that if two states $\rho,\sigma\in\md(A)$ are $\eps$-close in trace distance then they are $\eps$-close in fidelity. 
\item Show that if two states $\rho,\sigma\in\md(A)$ are $\eps$-close in fidelity, then they are $\sqrt{2\eps}$-close in fidelity. 
\end{enumerate}
\end{exercise}

The relation between the trace distance and the fidelity can also be used to derive some additional bounds on the trace distance. For example, consider a \emph{pure} state $\rho^{AB}$ whose marginal (mixed) state $\rho^A$ is $\eps$-close to a pure state $\psi^A$. Since $\psi^A$ is pure, this means that the marginal state $\rho^A$ is itself close to being pure, and this in turn means that the pure state $\rho^{AB}$ should be close to a produce state $\psi^{A}\otimes\rho^B$. We make this intuition rigorous in the following lemma.
\begin{myg}{}
\begin{lemma}\label{4rte}
Let $\rho\in\pure(AB)$ be a \emph{pure} state and let $\psi\in\pure(A)$ be another pure state.
If the marginal of $\rho^{AB}$ satisfies $T\left(\rho^A,\psi^A\right)\leq \eps$ then
\be
T\left(\rho^{AB},\psi^A\otimes\rho^B\right)\leq 2\sqrt{2\eps}\;.
\ee
\end{lemma}
\end{myg}
\begin{proof}
From the relation~\eqref{fitr} 
\be\label{6190}
F(\rho^A,\psi^A)\geq 1-T\left(\rho^A,\psi^A\right)\geq 1-\eps\;.
\ee
From Uhlmann's theorem there exists a pure state $|\phi\ra\in B$ such that
\be
F(\rho^A,\psi^A)=F(\rho^{AB},\psi^{A}\otimes\phi^B)
\ee
Hence, applying again~\eqref{fitr} gives
\ba
T\left(\rho^{AB},\psi^A\otimes\phi^B\right)&\leq \sqrt{1-F\left(\rho^{AB},\psi^{A}\otimes\phi^B\right)^2}\\
&= \sqrt{1-F(\rho^A,\psi^A)^2}\\
\GG{\eqref{6190}}&\leq \sqrt{1-\left(1-\eps\right)^2}\leq \sqrt{2\eps}\;.
\ea
From the monotonicity of the trace distance we also have
\be
T\left(\rho^{B},\phi^B\right)\leq T\left(\rho^{AB},\psi^A\otimes\phi^B\right)\leq \sqrt{2\eps}\;.
\ee
Hence, from the triangle inequality of the trace distance we get
\ba
T\left(\rho^{AB},\psi^A\otimes\rho^B\right)&\leq T\left(\rho^{AB},\psi^A\otimes\phi^B\right)+T\left(\psi^A\otimes\phi^B,\psi^A\otimes\rho^B\right)\\
&=T\left(\rho^{AB},\psi^A\otimes\phi^B\right)+T\left(\phi^B,\rho^B\right)\\
&\leq \sqrt{2\eps}+\sqrt{2\eps}=2\sqrt{2\eps}\;.
\ea
This completes the proof.
\end{proof}

\begin{exercise}
Using the same notations as in the theorem above, suppose $F(\rho^A,\psi^A)\geq1-\eps$. What is the best lower bound that you can find for $F(\rho^{AB},\psi^A\otimes \rho^B)$?
\end{exercise}

\subsubsection{The Gentle Measurement Lemma}\index{gentle measurement}

We now use the relationship between the trace distance and fidelity to prove an intuitive phenomenon related to the disturbance of states under quantum measurements. Specifically, let $\rho\in\md(A)$, and let $\Lambda\in\eff(A)$ be one element of a POVM (i.e. $\Lambda$ is an effect). From Born's rule, the quantity $\tr[\rho \Lambda]$ can be interpreted as the probability that the outcome associated with $\Lambda$ will occur. The gentle measurement lemma asserts that if this probability is high then the post measurement state will not change much and remain very close to $\rho$.  This fundamental property lies at the heart of several applications of quantum information and quantum resource theories.

\begin{myg}{}  
\begin{lemma}\label{gentle1}
Let $\eps\in(0,1)$, $\rho\in\md(A)$, and $\Lambda\in\eff(A)$. Suppose $\tr[\rho \Lambda]\geq 1-\eps$. Then, the post measurement state is $\sqrt{\eps}$-close to $\rho$; i.e. 
\be
\frac12\|\rho-\trho\|_1\leq\sqrt{\eps}\quad\text{where}\quad\trho\eqdef\frac{\sqrt{\Lambda}\rho\sqrt{\Lambda}}{\tr[\Lambda\rho]}\;.
\ee
\end{lemma}
\end{myg}
\begin{proof}
Let $|\psi^{AR}\ra$ be a purification of $\rho^A$, and observe that (see Exercise~\ref{ex:puri789})
\be\label{puri789}
|\tpsi^{AR}\ra\eqdef\frac{\sqrt{\Lambda}\otimes I^R|\psi^{AR}\ra}{\sqrt{\tr[\rho \Lambda]}}
\ee
is a purification of $\trho^A$. From Uhlmann's theorem, and the fact that $\Lambda\leq \sqrt{\Lambda}$ for any effect $\0\leq \Lambda\leq I^A$ we have
\ba
F(\rho,\trho)\geq \left|\la\psi^{AR}|\tpsi^{AR}\ra\right|=\frac{\la\psi^{AR}|\sqrt{\Lambda}\otimes I^R|\psi^{AR}\ra}{\sqrt{\tr[\rho \Lambda]}}
&\geq \frac{\la\psi^{AR}|\Lambda\otimes I^R|\psi^{AR}\ra}{\sqrt{\tr[\rho \Lambda]}}\\
&=\frac{\tr[\rho \Lambda]}{\sqrt{\tr[\rho \Lambda]}}=\sqrt{\tr[\rho \Lambda]}\geq\sqrt{1-\eps}\;.
\ea
Therefore, from the relation~\eqref{fitr} between the trace distance and the fidelity we conclude that
\be
\frac12\|\rho-\trho\|_1\leq\sqrt{1-F(\rho,\trho)^2}\leq\sqrt{1-(\sqrt{1-\eps})^2}=\sqrt{\eps}\;.
\ee
\end{proof}
\begin{exercise}\label{ex:puri789}
Prove that the state defined in~\eqref{puri789} is indeed a purification of $\trho^A$.
\end{exercise}

\section{Distance Between Sub-normalized States}\index{subnormalized states}

Sub-normalized states are positive semidefinite matrices with a trace less than or equal to one. These states, acting on a Hilbert space $A$, are denoted as:
\be
\md_{\leq}(A)\eqdef\Big\{\rho\in\pos(A)\;:\;\tr[\rho]\leq 1\Big\}\;.
\ee
Note that $\md(A)\subset\md_{\leq}(A)$.

Although subnormalized states do not represent physical systems, they arise in quantum measurements. For instance, consider a quantum state $\rho\in\md(A)$ and a quantum instrument $\mE=\sum_{x\in[m]}\mE_x^{A\to B}\otimes |x\lr x|^X\in\cptp(A\to BX)$, as discussed in Sec.~\ref{qinst}. The quantum state $\mE(\rho)$ is given by:
\be
\mE^{A\to BX}(\rho^A)=\sum_{x\in[m]}\mE_x^{A\to B}(\rho^A)\otimes |x\lr x|^X\;,
\ee
where $\{\mE_x\}_{x\in[m]}$ are trace non-increasing CP maps, and $\{\mE_x(\rho)\}_{x\in[m]}$ are sub-normalized states. These states provide both the information about the probability $p_x\eqdef\tr[\mE_x(\rho)]$ that an outcome $x\in[m]$ occurs during the quantum measurement, and the post-measurement state $\frac{1}{p_x}\mE_x(\rho)$.

We previously saw that distance measures for normalized states are monotonic under quantum channels and satisfy the DPI, a crucial aspect in applications. Quantum channels map normalized states to normalized states, while trace non-increasing (TNI) CP maps, including CPTP maps, take subnormalized states to subnormalized states. Therefore, it's beneficial to define a distance measure for subnormalized states that is monotonic under TNI-CP maps. We denote by $\cp_\leq(A\to B)$ the set of all TNI maps in $\cp(A\to B)$.

In section~\ref{optim}, we explored extending divergences from the classical to the quantum domain. We now apply a similar approach to extend divergences from normalized to sub-normalized states. However, unlike classical-to-quantum extensions, we will see that there is no analogous `minimal' extension from normalized to sub-normalized states. Thus, we begin by introducing the maximal extension of a quantum divergence to the sub-normalized domain.

\begin{myd}{The Maximal Extension}\index{optimal extension}
\begin{definition} \label{subnp}
Let $\D$ be a quantum divergence. The maximal extension of $\D$ to subnormalized states\index{subnormalized states}, $\oD$, is defined for any $\rho,\sigma\in\md_{\leq}(A)$ by
\be\label{dextd}
\oD(\rho\|\sigma)\eqdef\inf \D(\trho\|\tsigma)
\ee
where the infimum is over all systems $R$, and all density matrices $\trho,\tsigma\in\md(R)$ for which there exists
$\mE\in\cp_{\leq}(R\to A)$ such that 
\be\label{bconb}
\rho=\mE(\trho)\quad\text{and}\quad\sigma=\mE(\tsigma)\;.
\ee
\end{definition}
\end{myd}
\begin{remark}
Note that earlier we used the same notation $\oD$ to denote the maximal extension of a classical divergence\index{classical divergence} to a quantum one.
The bar symbol over $\D$ in our notations will always indicate maximal extensions from one domain to a larger one, whereas the domain of a given extension should be clear from the context.
\end{remark}

\noindent The maximal extension $\oD$ have the following three properties:
\begin{enumerate}
\item {\it Reduction.} For any $\rho,\sigma\in\md(A)$ we have
\be
\oD(\rho\|\sigma)=\D(\rho\|\sigma)\;.
\ee
\item {\it  Monotonicity\index{monotonicity}.} For any $\mE\in\cp_\leq(A\to B)$ and any subnormalized states $\rho,\sigma\in\md_{\leq}(A)$ \be\label{0mono0}
\oD\big(\mE(\rho)\|\mE(\sigma)\big)\leq \oD(\rho\|\sigma)\;.
\ee
\item {\it Optimality.} If $f:\md_{\leq}(A)\times\md_{\leq}(A)\to\mbb{R}_{+}$ is a function that reduces to $\D$ when restricted to normalized states, and behaves monotonically under TNI-CP maps (as in~\eqref{0mono0}) then 
\be\label{0opt0}
f(\rho\|\sigma)\leq \oD(\rho\|\sigma)\quad\quad\forall\;\rho,\sigma\in\md_{\leq}(A)\;.
\ee 
\end{enumerate}
The last property justify the name for $\oD$ as the maximal extension of $\D$ to subnormalized states. 

\begin{exercise}
Prove the three properties above using the same techniques that were used to prove Theorem~\ref{optim2}.
Hint: As you follow the same lines used in the proof of Theorem~\ref{optim2}, replace `classical states' with `normalized quantum states' and `quantum states' with `subnormalized quantum states'.
\end{exercise}

Remarkably, the maximal extension has the following closed formula.

\begin{myt}{\color{yellow} Closed Formula}
\begin{theorem}\label{sns}
Let $\D$ be a quantum divergence and $\oD$ be its maximal extension to sub-normalized states as defined in~\eqref{dextd}. For any pair of sub-normalized states $\rho,\sigma\in\md_{\leq}(A)$
\be\label{hsh}
\oD(\rho\|\sigma)=\D\big(\rho\oplus(1-\tr[\rho])\big\|\sigma\oplus(1-\tr[\sigma])\big)\;.
\ee
\end{theorem}
\end{myt}
\begin{proof}
Let $\trho,\tsigma\in\md(R)$ and $\mE\in\cp_\leq(R\to A)$ be a TNI-CP map such that $\rho=\mE(\trho)$ and $\sigma=\mE(\tsigma)$. Moreover, define $\mN\in\cptp(R\to A\oplus\mbb{C})$ as
\be
\mN(\omega)\eqdef\mE(\omega)\oplus\big(\tr[\omega]-\tr[\mE(\omega)]\big)\quad\quad\forall\;\omega\in\ml(A)\;.
\ee
Then, since $\mN$ is a CPTP map,
\ba
\D(\trho\|\tsigma)&\geq \D\big(\mN(\trho)\|\mN(\tsigma)\big)\\
&=\D\big(\mE(\trho)\oplus(1-\tr[\mE(\trho)])\big\|\mE(\tsigma)\oplus(1-\tr[\mE(\tsigma)])\big)\\
&=\D\big(\rho\oplus(1-\tr[\rho])\big\|\sigma\oplus(1-\tr[\sigma])\big)\;.
\ea
Since the above inequality holds for all such $\trho,\tsigma,\mE$ we must have that $\oD(\trho\|\tsigma)$ is no smaller than the right-hand side on~\eqref{hsh}.
To prove the converse inequality, take $R=A\oplus\mbb{C}$, $\trho=\rho\oplus(1-\tr[\rho])$, $\tsigma=\sigma\oplus(1-\tr[\sigma])$, and $\mE(\cdot)\eqdef P(\cdot)P^\dag$, where $P$ is the projection to the subspace $A$ in $R$. Then, $\rho=\mE(\trho)$ and $\sigma=\mE(\tsigma)$
so that by definition (see~\eqref{dextd}) we must have $\oD(\rho\|\sigma)\leq \D(\trho\|\tsigma)$. Together with the previous inequality, this completes the proof of the equality in~\eqref{hsh}.
\end{proof}

If $\D$ is a quantum divergence, its minimal extension $\uD$ can be defined in analogy with~\eqref{ud}  as
\be
\uD(\rho\|\sigma)\eqdef\sup\D(\mE(\rho)\|\mE(\sigma))\quad\quad\forall\;\rho,\sigma\in\md_{\leq}(A)\;,
\ee
where the supremum is over all systems $R$ and all $\mE\in\cp_{\leq}(A\to R)$ such that and
$\mE(\rho)$ and $\mE(\sigma)$ are normalized states. However, such $\mE$ does not exists if either $\rho$ or $\sigma$ has trace strictly smaller than one. Hence, the minimal extension of $\D$ must satisfy
\be
\uD(\rho\|\sigma)=0
\ee
for all subnormalized states $\rho,\sigma\in\md_{\leq}(A)$ with either $\tr[\rho]<1$ or $\tr[\sigma]<1$. Therefore, this extension is rather pathological and not useful in applications.
The following corollary applies specifically to the case where $\D$ functions as both a divergence and a metric.
\begin{myg}{}
\begin{corollary}
Let $\D$ be a quantum divergence that is also a metric. Then, its maximal extension to sub-normalized states, $\oD$, is also a metric.
\end{corollary}
\end{myg}
\begin{proof}
We need to show that $\oD$ is symmetric and satisfies the trumping\index{trumping relation}. To see it, let $\rho,\sigma,\omega\in\md_{\leq}(A)$. The symmetry of $\oD$ follows from the symmetry of $\D$:
\ba
\oD(\rho\|\sigma)&=\D\big(\rho\oplus(1-\tr[\rho])\;\big\|\;\sigma\oplus\big(1-\tr[\sigma]\big)\big)\\
\GG{\D\;is\; symmetric}&=\D\big(\sigma\oplus(1-\tr[\sigma])\big\|\rho\oplus(1-\tr[\rho])\big)\\
&=\oD(\sigma\|\rho)\;.
\ea
Similarly, the triangle inequality of $\oD$ follows from the triangle inequality of $\D$:
\ba
\oD(\rho\|\sigma)&=\D\big(\rho\oplus(1-\tr[\rho])\big\|\sigma\oplus(1-\tr[\sigma])\big)\\
&\leq \D\big(\rho\oplus\big(1-\tr[\rho])\big\|\omega\oplus(1-\tr[\omega])\big)+\D\big(\omega\oplus(1-\tr[\omega])\big\|\sigma\oplus(1-\tr[\sigma])\big)\\
&=\oD(\rho\|\omega)+\oD(\omega\|\sigma)\;.
\ea
\end{proof}

\subsection{Examples}
\subsubsection{The Generalized Trace Distance}\index{trace distance}

The maximal extension of the trace distance to sub-normalized states is known as the \emph{genaralized} trace distance. From Theorem~\ref{sns} we get that the generalized trace distance has the following simple form.
\begin{myg}{}
\begin{corollary}\label{gtdform}
The generalized trace distance can be expressed for any $\rho,\sigma\in\md_{\leq}(A)$ as
\be\label{bdrh1}
\overline{T}(\rho,\sigma)=\frac{1}{2}\|\rho-\sigma\|_1+\frac{1}{2}\big|\tr[\rho-\sigma]\big|\;.
\ee
\end{corollary}
\end{myg}

\begin{exercise}
Prove the corollary above using the formula given in Theorem~\ref{sns} when $\D$ is replaced by the trace distance.
\end{exercise}

\begin{remark}
The generalized trace distance can also be expressed as
\be\label{bdrh2}
\overline{T}(\rho,\sigma)=\max\Big\{\tr(\rho-\sigma)_+,\tr(\rho-\sigma)_-\Big\}\;.
\ee
To see why, set $a\eqdef\tr(\rho-\sigma)_+$ and
$b\eqdef\tr(\rho-\sigma)_-$, and use the relation $\max\{a,b\}=\frac{1}{2}\big(a+b+|a-b|\big)$.
The formula above is consistent with the fact that the trace distance is the largest extension of the trace distance to sub-normalized states that satisfies the monotonicity property under TNI-CP maps.
\end{remark}

\begin{exercise}
Show that for any $\rho,\sigma\in\md_{\leq}(A)$ the function $f(\rho,\sigma)=\frac{1}{2}\|\rho-\sigma\|_1$ is also an extension of $D$ to sub-normalized states that satisfies the exact same properties satisfied by $\overline{T}$ except for the optimality. Give an example showing that $f(\rho,\sigma)$ can be strictly smaller than $\overline{T}(\rho,\sigma)$.
\end{exercise}

\begin{exercise}\label{puretd2}
Show that for two sub-normalized \emph{pure} states $\psi,\phi\in\md_{\leq}(A)$ the generalized trace distance can be expressed as
\be
\overline{T}(\psi,\phi)=\sqrt{\frac{1}{4}\left(\tr[\psi+\phi]\right)^2-|\la\psi|\phi\ra|^2}+\frac{1}{2}\big|\tr[\psi-\phi]\big|\;.
\ee 
Hint: Use similar techniques as in Exercise~\ref{puretd}.
\end{exercise}

\subsubsection{The Gentle Operator Lemma}\index{gentle operator}

The gentle operator lemma is a variant of the gentle measurement lemma (Lemma~\ref{gentle1}) in which the post-measurement state is taken to be unnormalized.
Specifically, in the theorem below we let $\eps\in(0,1)$, $\rho\in\md(A)$, $\Lambda\in\eff(A)$, and consider the subnormalized state $\trho\eqdef\sqrt{\Lambda}\rho\sqrt{\Lambda}$.

\begin{myg}{}  
\begin{lemma}\label{gentle2}
Using the notations above, if $\tr[\rho \Lambda]\geq 1-\eps$ then
$
\frac12\left\|\rho-\trho\right\|_1\leq \sqrt{\eps}
$.
\end{lemma}
\end{myg}
\begin{remark}
The gentle operator lemma's extension to cases where $\trho=G\rho G^*$, with $G\in\ml(A,B)$ being an arbitrary element of a generalized measurement and $\Lambda\equiv G^*G\in\eff(A)$, may seem promising. However, without imposing further constraints on $G$, such an extension could result in non-informative bounds. Consider, for instance, the scenario where $G$ is a unitary matrix, making $\Lambda=G^*G=I^A$. Here, $\tr[\Lambda\rho]=1\geq 1-\eps$ for any $\eps\geq 0$. But, if we choose $\rho=|0\rangle\langle 0|$ and a unitary $G$ such that $G|0\rangle=|1\rangle$, it follows that $\frac12\|\rho-\trho\|_1=1$. This example illustrates that extending the gentle operator lemma to encompass arbitrary elements of generalized measurements is impractical without specific additional constraints on $G$.
\end{remark}
\begin{proof}
Let $|\psi^{A\tA}\ra=\sqrt{\rho}\otimes I^{\tA}|\Omega^{A{\tA}}\ra$ and $|\tpsi^{A{\tA}}\ra\eqdef\sqrt{\Lambda}\otimes I^{\tA}|\psi^{A{\tA}}\ra$ be purifications $\rho^A$ and $\trho^A$, respectively. Denote by $t\eqdef\tr[\rho \Lambda]\geq 1-\eps$ and observe that 
\ba\label{6219}
\la\psi^{A{\tA}}|\tpsi^{A{\tA}}\ra&=\la\psi^{A{\tA}}|\sqrt{\Lambda}\otimes I^{\tA}|\psi^{A{\tA}}\ra\\
\Gg{\sqrt{\Lambda}\geq \Lambda}&\geq \la\psi^{A{\tA}}|\Lambda\otimes I^{\tA}|\psi^{A{\tA}}\ra\\
\Gg{|\psi^{A{\tA}}\ra=\sqrt{\rho}\otimes I^{\tA}|\Omega^{A{\tA}}\ra}&=\tr[\rho \Lambda]=t\;.
\ea
Due to the DPI of the generalized trace distance it follows that \be\overline{T}(\rho^A,\trho^A)\leq\overline{T}(\psi^{A{\tA}},\tilde{\psi}^{A{\tA}})\;.\ee
From~\eqref{bdrh1} we get that both sides on the equation above contain the same term $\frac{1}{2}\big|\tr[\rho-\tilde{\rho}]\big|=\frac{1}{2}\big|\tr[\psi-\tilde{\psi}]\big|$, so we can cancel it. Combining this with Exercise~\ref{puretd2} we conclude that
\ba
\frac12\left\|\rho-\trho\right\|_1&\leq \sqrt{\frac{1}{4}\left(\tr[\psi+\tilde{\psi}]\right)^2-|\la\psi|\tilde{\psi}\ra|^2}\\
\GG{\eqref{6219}}&\leq\sqrt{\frac{1}{4}\left(1+t\right)^2-t^2}\\
\GG{Exercise~\ref{ftbb}}&\leq\sqrt{1-t}\\
\Gg{t\geq 1-\eps}&\leq\sqrt{\eps}\;.
\ea
This completes the proof.
\end{proof}

\begin{exercise}\label{ftbb}
Show that the function $f(t)\eqdef\sqrt{\frac{1}{4}\left(1+t\right)^2-t^2}$ is smaller than $\sqrt{1-t}$ for all $t\in[0,1]$.
\end{exercise}

\bex
Show that one can use the gentle measurement lemma (Lemma~\ref{gentle1}) to prove a slightly weaker version of the gentle operator lemma (Lemma~\ref{gentle2}). Use only Lemma~\ref{gentle1} and the triangle inequality of the trace norm to show that
\be
\frac12\left\|\rho-\trho\right\|_1\leq\sqrt{\eps}+\frac12\eps\;.
\ee
Hint: Set $\rho'\eqdef\frac{\sqrt{\Lambda}\rho\sqrt{\Lambda}}{\tr[\Lambda\rho]}$ and write $\rho-\trho=\rho-\rho'+\rho'-\trho$.
\eex

\subsubsection{The Generalized Fidelity}\index{fidelity}

We can use the techniques developed above to extend the fidelity to sub-normalized states.
However, since the fidelity achieves its maximum for identical states, the infimum of~\eqref{dextd} will be replaced with a supremum. 
\begin{myd}{} 
\begin{definition}
Let $\rho,\sigma\in\md_{\leq}(A)$ be two sub-normalized states. We define the \emph{generalized fidelity} $\uF:\md_{\leq}(A)\times\md_{\leq}(A)\to \mbb{R}_{+}$ to be
\be\label{dmbd}
\uF(\rho,\sigma)\eqdef\sup F(\trho,\tsigma)\;,
\ee
where the supremum is over all systems $R$, and all density matrices $\trho,\tsigma\in\md(R)$ for which there exists
$\mE\in\cp_\leq(R\to A)$ with the property that 
$
\rho=\mE(\trho)$
and $\sigma=\mE(\tsigma)
$.
\end{definition}
\end{myd}

Since $1-F(\rho,\sigma)$ is a quantum divergence we can use Theorem~\ref{sns} to get a closed formula for the generalized fidelity.
\begin{myg}{}
\begin{corollary}\label{gfform}
The generalized fidelity can be expressed as
\be\label{gf3}
\uF(\rho,\sigma)=\|\sqrt{\rho}\sqrt{\sigma}\|_1+\sqrt{(1-\tr[\rho])(1-\tr[\sigma])}\quad\quad\forall\;\rho,\sigma\in\md_{\leq}(A)\;.
\ee
\end{corollary}
\end{myg}

\begin{remark}
The formula~\eqref{gf3} for the generalized fidelity reveals that we have $\uF(\rho,\sigma)=\|\sqrt{\rho}\sqrt{\sigma}\|_1$ even if only one of the states is normalized. Note that for the trace distance $\overline{T}(\rho,\sigma)=T(\rho,\sigma)$ only if both states are normalized. 
\end{remark}

The generalized fidelity has the following properties:
\begin{enumerate}
\item {\it Reduction.} For any $\rho,\sigma\in\md(A)$ we have
$
\uF(\rho,\sigma)=F(\rho,\sigma)\;.
$
\item {\it  Monotonicity\index{monotonicity}.} For every $\mE\in\cp_\leq(A\to B)$ and every sub-normalized states $\rho,\sigma\in\md_{\leq}(A)$ we have
$
\uF\big(\mE(\rho),\mE(\sigma)\big)\leq \uF(\rho,\sigma)\;.
$
\item {\it Faithfulness.} For every $\rho,\sigma\in\md_{\leq}(A)$,
$
\uF(\rho,\sigma)=1\iff\rho=\sigma.
$
\item {\it Symmetry.} For any $\rho,\sigma\in\md_{\leq}(A)$, 
$
\uF(\rho,\sigma)=\uF(\sigma,\rho).
$
\item {\it Optimality.} If $f:\md_{\leq}(A)\times\md_{\leq}(A)\to\mbb{R}_{+}$ is a function that reduces to $F$ when restricted to normalized states, and behaves monotonically under TNI-CP maps (as above) then 
$
f(\rho,\sigma)\geq \uF(\rho,\sigma)
$
for all $\rho,\sigma\in\md_{\leq}(A)$.
\end{enumerate}
The last property above indicates that the generalized fidelity is the minimal extension of the fidelity to sub-normalized states.
\begin{exercise}
Prove the above 5 properties of the generalized fidelity.
\end{exercise}

\begin{exercise}
Show that for two sub-normalized pure states $\psi\in\md_{\leq}(A)$ and $\phi\in\md_{\leq}(A)$ the generalized fidelity is give by
\be
\uF(\psi,\phi)=|\la\psi|\phi\ra|+\sqrt{(1-\la\psi|\psi\ra)(1-\la\phi|\phi\ra)}\;.
\ee
\end{exercise}
\begin{myt}{\color{yellow} Generalization of Uhlmann's theorem}\index{Uhlmann's theorem}
\begin{theorem}\label{gut}
Let $\rho,\sigma\in\md_{\leq}(A)$ be two subnormalized states\index{subnormalized states}, and let $|\psi^{AB}\ra$ and $|\phi^{AC}\ra$ be two purifications of $\rho^A$ and $\sigma^A$, respectively. 
Suppose also that $|B|\leq |C|$.  Then,
\be
\uF\left(\rho^A,\sigma^A\right)=\max_{\mV^{B\to C}}\uF\left(\mV^{B\to C}\left(\psi^{AB}\right),\phi^{AC}\right)
\ee
where the maximum is over all CPTP maps $\mV(\cdot)\eqdef V(\cdot)V^*$, where $V:B\to C$ is an isometry.  
\end{theorem}
\end{myt}
\begin{proof}
Note first that 
\be
\tr\left[\mV^{B\to C}\left(\psi^{AB}\right)\right]=\tr\left[\psi^{AB}\right]=\tr[\rho^A]\;,
\ee
and  similarly $\tr\left[\mV^{B\to C}\left(\phi^{AB}\right)\right]=\tr[\sigma^A]$.
Therefore, it is sufficient to show that
\be
\left\|\sqrt{\rho}\sqrt{\sigma}\right\|_1=\max_{\mV^{B\to C}}\left\|\sqrt{\mV^{B\to C}\left(\psi^{AB}\right)}\sqrt{\phi^{AC}}\right\|_1.
\ee
Since $\mV^{B\to C}\left(\psi^{AB}\right)$ and $\phi^{AC}$ are rank one sub-normalized states we have (Exercise~\eqref{exqo})
\be\label{3eq3}
\left\|\sqrt{\mV^{B\to C}\left(\psi^{AB}\right)}\sqrt{\phi^{AC}}\right\|_1=\left|\la\psi^{AB}|I^A\otimes V^*|\phi^{AC}\ra\right|\;.
\ee
Hence, the rest of the proof follows the exact same lines as in the proof of Uhlmann's theorem (Theorem~\ref{Uhlm}). In particular, note that all the steps in~\eqref{mpp} holds even if $\rho$ and $\sigma$ are sub-normalized.
\end{proof}

\begin{exercise}\label{exqo}
Prove the equality in Eq.~\eqref{3eq3}.
\end{exercise}

\section{The Purified Distance}\index{purified distance}

The trace distance and fidelity are both highly valuable in numerous applications. The trace distance benefits from being a metric, especially due to its compliance with the triangle inequality. Conversely, the fidelity's advantage lies in its compatibility with Uhlmann's theorem, allowing the simplification of expressions involving fidelity through the use of quantum state purifications. Given the desirability of both these properties in quantum information applications (which will be further explored in this book), it naturally leads to the question: is there a distance measure that embodies both qualities? As we will discuss now, such a measure does indeed exist.

For any two pure state $\psi,\phi\in\pure(A)$ the trace distance is given by (see Exercise~\ref{puretd})
\be
T(\psi,\phi)=\sqrt{1-|\la\psi|\phi\ra|^2}=\sqrt{1-F(\psi,\phi)^2}\;.
\ee
Considering this close relationship between trace distance and fidelity when applied to pure states, we will explore all possible extensions of the trace distance from pure states to mixed states. In this context, we define the purified distance as the maximal extension among all such extensions.\index{trace distance}
\begin{myd}{}
\begin{definition}
Let $T$ be the trace distance. The purified distance\index{purified distance} is defined for all $\rho,\sigma\in\md(A)$ as
\be\label{defpure}
P(\rho,\sigma)\eqdef\inf_{\psi,\phi\in\pure(R)} \Big\{T(\psi,\phi)\;:\;\rho=\mE(\psi)\;,\;\sigma=\mE(\phi)\;,\;\mE\in\cptp(R\to A)\Big\}
\ee
where the infimum is also over all systems $R$.
\end{definition}
\end{myd}

\noindent{\it Remarks:}
\begin{enumerate}
\item Observe that the extension of the trace distance in the definition above is reminiscent to the maximal quantum extension of the trace distance that was discussed in the previous sections. Later on we will develop a framework to extend certain functions (specifically resource monotones) from one domain to a larger one. This framework is very general and all extensions discussed in this chapter (including the above extension of the trace distance, i.e. the purified distance) are just specific applications of the framework.
\item We will see below that the purified distance has a closed formula. Historically, this closed formula has been used as its definition. However, the definition above emphasize its operational meaning as the largest mixed-state extension of the trace distance (see Theorem~\ref{thm:pd} below).
\item The justification for the name ``purified distance" will become clear from the properties discussed below.
\end{enumerate}

We start by showing that the purified distance is an optimal divergence.
\begin{myt}{}
\begin{theorem}\label{thm:pd}
The purified distance is a quantum divergence that reduces to the trace distance on pure states. Moreover, if $\D$ is another quantum divergence that reduces to the trace distance on pure states then for any $\rho,\sigma\in\md(A)$ we have
\be
\D(\rho\|\sigma)\leq P(\rho,\sigma)\;.
\ee 
\end{theorem}
\end{myt}

The proof follows very similar lines as in the proof of Theorem~\ref{optim2} and is left as an exercise.

\bex
Prove Theorem~\eqref{thm:pd}. Hint: Adopt the methodology used in Theorem~\ref{optim2} related to $\oD$. In this process, substitute each occurrence of a classical state on system $X$ with a pure state on system $R$.
\eex

The upcoming lemma demonstrates that the purified distance\index{purified distance} is derived from a purification process, which justifies its name. We will utilize this lemma to derive a closed formula for the purified distance.
\begin{myg}{}
\begin{lemma}\label{lempure}
Let $P$ be the purified distance and $T$ the trace distance. Then, for all $\rho,\sigma\in\md(A)$
\be\label{ppros}
P(\rho^A,\sigma^A)=\inf_{\psi,\phi}T(\psi^{AB},\phi^{AB})\;,
\ee
where the infimum is over all purifications of $\rho^A$ and $\sigma^A$.
\end{lemma}
\end{myg}

\begin{proof}
Let $\psi^{AB}$ and $\phi^{AB}$ be purifications of $\rho^A$ and $\sigma^A$, respectively, and denote by $\mE^{AB\to A}\eqdef\tr_B$. By definition, $\mE^{AB\to A}(\psi^{AB})=\rho^A$ and $\mE^{AB\to A}(\phi^{AB})=\sigma^A$ so that $\psi^{AB}$ and $\phi^{AB}$ satisfies the conditions in~\eqref{defpure} with $R\eqdef AB$. Therefore, $P(\rho^A,\sigma^A)$
cannot be greater than the right-hand side of~\eqref{ppros}. To get the other direction, recall the definition~\eqref{defpure} and let $\rho^A=\mE(\psi^R)$ and $\sigma^A=\mE(\phi^R)$ for some $\psi,\phi\in\pure(R)$ and $\mE\in\cptp(R\to A)$. Let $\mV\in\cptp(R\to AB)$ be the isometry purifying $\mE^{R\to A}$.
Therefore, $\rho^A=\tr_B\left[\mV^{R\to AB}\left(\psi^R\right)\right]$ and $\sigma^A=\tr_B\left[\mV^{R\to AB}\left(\phi^R\right)\right]$. Finally,
since the trace distance is invariant under isometries, denoting by $\chi^{AB}\eqdef \mV^{R\to AB}\left(\psi^R\right)$ and $\varphi^{AB}\eqdef \mV^{R\to AB}\left(\phi^R\right)$ we get
\be
T(\psi^R,\phi^R)=T(\chi^{AB},\varphi^{AB})\geq \inf_{\psi',\phi'} T(\psi'^{AB},\phi'^{AB})
\ee
where the infimum is over all purifications $\psi'^{AB}$ and $\phi'^{AB}$ of $\rho^A$ and $\sigma^A$. Hence, since $\psi^R$ and $\phi^R$ were arbitrary pure states that satisfy the conditions in~\eqref{defpure}, we conclude that $P(\rho^A,\sigma^A)$ is no smaller than the right-hand side of~\eqref{ppros}. This completes the proof.
\end{proof}
\begin{myt}{\color{yellow} Closed Formula}
\begin{theorem}
Let $P$ be the purified distance. Then, for all $\rho,\sigma\in\md(A)$
\be\label{cf6255}
P(\rho,\sigma)=\sqrt{1-F(\rho,\sigma)^2}\;.
\ee
\end{theorem}
\end{myt}\index{fidelity}
\begin{proof}
From Lemma~\ref{lempure} we get by direct computation
\ba
P(\rho,\sigma)&=\min_{\psi,\phi}T(\psi,\phi)\\
\GG{\text{Exercise~\ref{puretd}}}&=\min_{\psi,\phi}\sqrt{1-|\la\psi|\phi\ra|^2}\\
&=\sqrt{1-\max_{\psi,\phi}|\la\psi|\phi\ra|^2}\\
\GG{\text{Uhlamnn's}\; Theorem}&=\sqrt{1-F(\rho,\sigma)^2}\;.
\ea
This completes the proof.
\end{proof}

The maximal extension of purified distance\index{purified distance} from density matrices to sub normalized states follows trivially from Theorem~\ref{sns}. We therefore extend the definition of the purified distance to subnormalized states in the following way.
\begin{myd}{The Purified Distance}
\begin{definition}
Let $\rho,\sigma\in\md_{\leq}(A)$ be two sub-normalized states. The purified distance is defined as
\be\label{dpdis}
P(\rho,\sigma)\eqdef\sqrt{1-\uF(\rho,\sigma)^2}
\ee
where $\uF$ is the generalized fidelity as given in~\eqref{gf3}.
\end{definition}
\end{myd}
\begin{remark}
The purified distance on normalized states has been defined earlier as the maximal extension of the trace distance from pure states to mixed states. Therefore, the purified distance on subnormalized states\index{subnormalized states} can be viewed as the maximal extension of the trace distance from pure states to mixed subnormalized states. Moreover, observe that the purified distance can also be expressed as:
\be
P(\rho,\sigma)\eqdef\sqrt{1-F(\trho,\tsigma)^2}
\ee
where $\trho\eqdef\rho\oplus(1-\tr[\rho])$ and $\tsigma\eqdef\sigma\oplus(1-\tr[\sigma])$.
\end{remark}

Finally, we show that the purified distance is a metric.
\begin{myt}{}
\begin{theorem}
The purified distance\index{purified distance} is a metric\index{metric} on the set of subnormalized states.
\end{theorem}
\end{myt}

\begin{proof}
Since $\uF(\rho,\sigma)\leq 1$ the purified distance is non-negative. Since $\uF(\rho,\sigma)=1$ if and only if $\rho=\sigma$ the purified distance $P(\rho,\sigma)=0$ if and only if $\rho=\sigma$. Since $\uF$ is symmetric also $P$ is symmetric. It is therefore left to show that the purified distance satisfies the triangle inequality. 

Let $\rho,\sigma,\omega\in\md_{\leq}(A)$ and set $\trho\eqdef\rho\oplus(1-\tr[\rho])$, $\tsigma\eqdef\sigma\oplus(1-\tr[\sigma])$, and $\tomega\eqdef\omega\oplus(1-\tr[\omega])$. Moreover,
let
$\psi,\phi,\varphi\in\md(B\tB)$ be the purifications of $\trho$, $\tsigma$, and $\tomega$ such that $F(\trho,\tomega)=F(\psi,\varphi)$ and $F(\tomega,\tsigma)=F(\varphi,\phi)$. Such purifications exist due to Uhlmann's theorem. Moreover, note that from the Uhlmann's theorem we also have $F(\trho,\tsigma)\geq F(\psi,\phi)$. Hence,
\ba
P(\rho,\omega)+P(\omega,\sigma)&=\sqrt{1-F(\trho,\tomega)^2}+\sqrt{1-F(\tomega,\tsigma)^2}\\
&=\sqrt{1-F(\psi,\varphi)^2}+\sqrt{1-F(\varphi,\phi)^2}\\
\GG{\eqref{tdex}}&= T(\psi,\varphi)+T(\varphi,\phi)\\
\GG{\text{Triangle inequality of {\it T}}}&\geq T(\psi,\phi)\\
\GG{\eqref{tdex}}&=\sqrt{1-F(\psi,\phi)^2}\\
\Gg{F(\trho,\tsigma)\geq F(\psi,\phi)} &\geq\sqrt{1-F(\trho,\tsigma)^2}\\
&=P(\rho,\sigma)
\ea
This completes the proof.
\end{proof}
 
Note that the purified distance is monotonic under TNI-CP maps. That is, for every map $\mE\in\cp_\leq(A\to B)$, and any two subnormalized states $\rho,\sigma\in\md_{\leq}(A)$ we have
\be
P\big(\mE(\rho),\mE(\sigma)\big)\leq P(\rho,\sigma)\;.
\ee
This follows trivially from Theorem~\ref{sns} and the monotonicity property in~\eqref{0mono0} (or equivalently from the monotonicity of the generalized fidelity). Moreover, note that from Theorem~\ref{gut} it follows that for any $\rho,\sigma\in\md_{\leq}(A)$ and any purification $\psi^{AB}$ of $\rho^A$, there exists a purification $\phi^{AB}$ of $\sigma^A$ such that
\be
P(\rho^A,\sigma^A)=P(\psi^{AB},\phi^{AB})\;.
\ee
We end this subsection by showing that the purified distance is bounded by the generalized trace distance, $\overline{T}$.

\begin{myt}{}
\begin{theorem}\label{pdthm}
Let $\rho,\sigma\in\md_{\leq}(A)$ be two subnormalized states.
The generalized trace distance and the purified distance\index{purified distance} satisfy the following inequalities:
\be\label{6263}
\overline{T}(\rho,\sigma)\leq P(\rho,\sigma)\leq\sqrt{2\overline{T}(\rho,\sigma)}\;.
\ee
\end{theorem}
\end{myt}

\begin{proof}
Set $\trho=\rho\oplus(1-\tr\rho)$ and $\tsigma=\sigma\oplus(1-\tr\sigma)$ and observe
\ba
P(\rho,\sigma)&=\sqrt{1-F(\trho,\tsigma)^2}\\
\GG{\text{\eqref{fitr}}}&\geq T(\trho,\tsigma)=\overline{T}(\rho,\sigma)\;.
\ea
To get the other inequality observe that
\ba
P(\rho,\sigma)^2&=1-F(\trho,\tsigma)^2\\
\GG{\text{\eqref{fitr}}}&\leq 1-\big(1-T(\trho,\tsigma)\big)^2\\
&\leq 2T(\trho,\tsigma)\\
&=2\overline{T}(\rho,\sigma)\;.
\ea
This completes the proof.
\end{proof}

\bex
Let $\rho,\sigma\in\md_{\leq}(A)$ be two subnormalized states. Let $\tau\in\md_{\leq}(AB)$ be an extension of $\rho^A$. That is, $\tr_B[\tau^{AB}]=\rho^A$. Show that there exists an extension $\omega\in\md_{\leq}(AB)$ of $\sigma^A$ such that  
\be
P\left(\rho^A,\sigma^A\right)=P\left(\tau^{AB},\omega^{AB}\right)\;.
\ee
\eex

\section{Notes and References}

We followed the definition and basic properties of classical and quantum divergences as given by~\cite{GT2021} and~\cite{GT2020}. The classical $f$-Divergence\index{$f$-divergence}s goes back to the work of~\cite{Renyi1961} followed by the independent works of~\cite{Csiszar1963}, \cite{Morimoto1963}, and~\cite{AS1966}.
The quantum version of the $f$-Divergence\index{$f$-divergence}s are a special case of Petz' quasi-entropies defined by~\cite{Petz1985}. Extensive details on their properties, applications in quantum information, and additional references can be found in the papers by~\cite{HMPB2011} and~\cite{HM2017} (see also the appendix in~\cite{TCR2009} for a similar derivation of the closed formula in Theorem~\ref{quantumformula}). The maximal extension of the classical $f$-Divergence\index{$f$-divergence} (Theorem~\ref{t633}) is due to~\cite{Matsumoto2018}. The trace distance and  the fidelity of subnormalized states were introduced in~\cite{Tomamichel2015} and later on developed further in~\cite{GT2020}.

\chapter{Entropies and Relative Entropies}\label{ch:relent}

In Chapter~\ref{chadiv}, we introduced the concept of a divergence as a measure quantifying the distinguishability between probability vectors or quantum states, emphasizing that any measure of distinguishability should satisfy the data processing inequality. This chapter builds upon that foundation by integrating the principle of additivity. The additivity\index{additivity} of certain functions under the tensor product of states is a recurring theme in physics. For instance, entropy's additivity under tensor products is intimately linked to several characteristics of thermal systems, including the second law of thermodynamics.

This chapter approaches the definition of entropies and relative entropies axiomatically. This method is particularly beneficial, as it reveals a multitude of properties common to all entropies and relative entropies. Furthermore, it distinguishes unique properties of certain relative entropies from those that are universally applicable. For instance, we will discover that the KL-divergence, introduced in Chapter~\ref{chadiv}, is the sole relative entropy characterized by asymptotic continuity. This distinction underpins the significant role of KL divergence in information theory.

\section{Entropy}\index{entropy}

Entropy is pivotal in numerous fields, including statistical mechanics, thermodynamics, information theory, black hole physics, cosmology, chemistry, and even economics. This wide range of applications has led to diverse interpretations of entropy. In thermodynamics, it's seen as a measure of energy dispersal at a specific temperature. In contrast, information theory views it as a rate of compression. Other perspectives, explored extensively in literature, link entropy to disorder, chaos, system randomness, and the concept of time's arrow. These varying attributes and contexts give rise to different measures of entropy, such as Gibbs and Boltzmann entropy, Tsallis entropies, R\'enyi entropies, and von-Neumann and Shannon entropies, along with other entropy functions like molar entropy, entropy of mixing, and loop entropy.

The multifaceted nature of entropy calls for a systematic and unifying approach, where entropy is defined rigorously and context-independently. This requires identifying common characteristics across all forms of entropy. One such universal trait is \emph{uncertainty}, whether it's about the state of a physical system or the output of a compression scheme. In various contexts, this uncertainty also encompasses concepts like disorder and randomness. For instance, uncertainty about a system's state correlates with its disorder level.

In Chapter~\ref{ch:majorization}, especially in Sec.~\ref{majobetween}, we delved into the role of majorization in defining uncertainty. We employed three different methodologies -- axiomatic, constructive, and operational -- to determine that every measure of uncertainty should inherently be a Schur concave function. Consequently, it is reasonable to anticipate that entropy functions will exhibit monotonic behavior under majorization.

Besides uncertainty, entropy embodies other attributes. A second key feature, related to the second law of thermodynamics -- especially the Clausius and Kelvin-Planck statements -- involves cyclic processes where a system undergoes a thermodynamic transition while all other systems, including the environment and heat baths, return to their original state. Recent developments in quantum information's approach to small-scale thermodynamics, as referenced in~\cite{BHN+2015}, categorize these as catalytic processes. Consider a thermodynamical evolution where a physical system $A$ in state $\rho^A$ transitions into system $B$ in state $\sigma^{B}$. The encompassing thermal machine, including heat baths, environment, etc., can be represented as an additional system $C$ in state $\tau^C$. Thus, for cyclic processes, the thermodynamic transition can be described as:
\be\label{transition}
\rho^{A}\otimes\tau^{C}\to\sigma^{B}\otimes\tau^{C}\;.
\ee

In this framework, the second law asserts not just that system $A$'s entropy is no greater than that of system $B$, but also that this holds true only if the entropy of the combined state $\rho^{A}\otimes\tau^C$ increases or remains unchanged in such a thermodynamic cyclic process where $\tau^{C}$ is preserved. If entropy is measured with an additive function (under tensor products), then the entropy of $\rho^A$ being no greater than that of $\sigma^{B}$ implies the same relationship between $\rho^{A}\otimes\tau^{C}$ and $\sigma^{B}\otimes\tau^{C}$. Thus, we define entropies as \emph{additive} measures of uncertainty.

\subsection{Classical Entropy}\index{entropy}

In this section we study entropy in the classical domain. We first introduce the formal definition of an entropy in terms of two axioms, and then provide further justification for these axioms. We will consider a function
\be\label{thefn}
\h:\bigcup_{n\in\mbb{N}}\prob(n)\to\mbb{R}
\ee
that maps probability vectors in all finite dimensions to the real numbers.

\begin{myd}{Entropy}\index{entropy}
\begin{definition}\label{entropy}
The function $\h$ as given in~\eqref{thefn} is called an entropy if it is not equal to the constant zero function and it satisfies the following two axioms:
\begin{enumerate}
\item\textbf{Monotonicity under mixing}. For every $n,m\in\mbb{N}$, and every $\p\in\prob(n)$ and $\q\in\prob(m)$ that satisfy $\p\succ\q$ we have
$\h(\p)\leq\h(\q)$.
\item \textbf{Additivity}. For any $n,m\in\mbb{N}$, and any $\p\in\prob(n)$ and $\q\in\prob(m)$,
\be
\h(\p\otimes\q)=\h(\p)+\h(\q)\;.
\ee
\end{enumerate}
\end{definition}
\end{myd}

The first axiom\index{axiom} ensures that an entropy quantifies uncertainty. In Sec.~\ref{majobetween} we arrived at the definition of majorization from a game of chance, indicating that $\p\succ\q$ if $\q$ is more uncertain than $\p$. Note however that we extend here the definition of majorization to vectors that are not necessarily of the same dimension. This can be done by adding zeros to the vector with the smaller dimension to make the vectors in the same dimension. This means in particular that for any $\p\in\prob(n)$, any entropy $\h$ satisfies $\h(\p\oplus 0)=\h(\p)$. Note also that this axiom\index{axiom} also implies that $\h$ is Schur concave.

The additivity\index{additivity} axiom\index{axiom} distinguishes entropy functions from arbitrary measures of uncertainty. For example, in Sec.~\ref{sec:schurconvex} we encounter several Schur concave functions, such as the symmetric elementary functions (see~\eqref{531}) that are in general not additive. Therefore, such functions cannot be entropies. The additivity property is consistent with the extensivity property of entropy in thermodynamics, and particularly, the monotonicity of entropy under cyclic thermodynamical processes. As mentioned above, in such cycles, all degrees of freedom other than the degrees of freedom of the system remains intact at the end of the cycle. Therefore, suppose the system at the beginning and end of the cycle is characterized with some probability vectors $\p$ and $\q$, respectively. If the initial state of the system was described by $\p\otimes\r$, where $\r$ corresponds to the remaining degrees of freedom, then at the end of the cycle the system+environment are described by $\q\otimes\r$ (i.e. with the same $\r$). Since entropy should be monotonic under such cycle in which $\p\otimes\r\succ\q\otimes \r$, it motivates the additivity property of an entropy function so that it is monotonic under the trumping relation. That is, the monotonicity under mixing can be strengthened using the additivity\index{additivity} property such that
\be
\p\succ_*\q\;\;\;\Rightarrow\;\;\;\h(\p)\leq\h(\q)\;.
\ee
There are other arguments to motivate the additivity axiom\index{axiom} that comes from information theory and we will discuss them as we go along. 

In the definition above we allow for the case that $n=1$. In this trivial case, $\prob(n)=\prob(1)$ contains only the 1-dimensional vector (i.e. number) one. Observe that 
for any $\p\in\prob(n)$ we get from the additivity axiom\index{axiom} that $\h(\p)=\h(\p\otimes 1)=\h(\p)+\h(1)$, so that $\h(1)=0$. From the fact that for all $x\in[n]$ 
$1\succ\e_x\succ 1$ (i.e. $1\sim\e_x$), where $\{\e_x\}_{x\in[n]}$ is the standard (elementary) basis of $\mbb{R}^n$, we conclude that also $\h(\e_x)=0$ for all $x\in[n]$. Moreover, since for every $n\in\mbb{N}$ and every $\p\in\prob(n)$ we have $\e_x\succ\p$ we get from the monotonicity axiom\index{axiom} that $\h(\p)\geq\h(\e_x)=0$. That is, entropy functions cannot be negative.

In the definition of entropy above we assumed that the entropy $\h$ is not the zero function. This means that there exists $n\in\mbb{N}$ and $\p\in\prob(n)$ such that $\h(\p)\neq 0$. Since entropy cannot be negative, this means that $\h(\p)>0$. On the other hand, for sufficiently large $m\in\mbb{N}$ we have $\p\succ\left(\u^{(2)}\right)^{\otimes m}$ (see Exercise~\ref{pum}) so that 
\ba
\h(\u^{(2)})&=\frac1m\h\left(\left(\u^{(2)}\right)^{\otimes m}\right)\\
\Gg{\p\succ\left(\u^{(2)}\right)^{\otimes m}}&\geq\frac1m \h(\p)>0\;.
\ea
Therefore, all entropy functions take strictly positive values on $\u^{(2)}\eqdef\frac12(1,1)^T\in\prob(2)$. It will be convenient to normalize all entropy functions such that
\be\label{normasin}
\h\left(\u^{(2)}\right)=1\;.
\ee
Throughout the remainder of the book, we will focus exclusively on entropy functions that are normalized as above.

\begin{myg}{}
\begin{lemma}
Let $\h$ be an entropy normalized as in~\eqref{normasin}, $n\in\mbb{N}$, and $\u^{(n)}\eqdef\frac1n(1,\ldots,1)^T\in\prob(n)$. Then, for all  $\p\in\prob(n)$ we have
\be
\h(\p)\leq\h(\u^{(n)})=\log n\;.
\ee 
\end{lemma}
\end{myg}
\begin{proof}
The inequality follow from the Schur concavity of $\h$ and the fact that 
$
\p\succ\u^{(n)}
$.
To prove the equality, define $f:\mbb{N}\to\mbb{R}$, via $f(n)\eqdef\h(\u^{(n)})$. 
From the normalization~\eqref{normasin} we have $f(2)=1$ and from the additivity $f(2^{k})=k$ for all $k\in\mbb{N}$. More generally, for any $m,n\in\mbb{N}$ the additivity gives
\be
f(n^m)=\h\left(\u^{(n^m)}\right)=\h\left(\left(\u^{(n)}\right)^{\otimes m}\right)=m\h\left(\u^{(n)}\right)=mf(n)\;.
\ee
Moreover, from the monotonicity property of $\h$ and the fact that $\u^{(n)}\succ\u^{(n+1)}$ we get that $f$ is monotonically non-decreasing. Using these properties of $f$ we get for all $k,m\in\mbb{N}$
\ba
f(n)=\frac1mf(n^m)&=\frac1mf\left(2^{m\log(n)}\right)\\
\Gg{f\text{ is non-decreasing}}&\leq\frac1mf\left(2^{\lceil m\log(n)\rceil}\right)\\
&=\frac1m{\lceil m\log(n)\rceil}\;.
\ea
Similarly, taking the floor instead of the ceiling above gives $f(n)\geq \frac1m{\lfloor m\log(n)\rfloor}$. In the limit $m\to\infty$ both of these bounds converge to $\log n$. This concludes the proof.
\end{proof}

\begin{exercise}
Show that any convex combination of entropies is itself an entropy. That is, if $\{\h_x\}_{x=1}^k$ is a set of entropies and $\s\in\prob(k)$ then $\sum_{x\in[k]}s_x\h_x$ is itself an entropy.
\end{exercise}

\subsubsection{The R\'enyi Entropies}\index{R\'enyi entropy}

An important class of entropies is the class of R\'enyi entropies. The R\'enyi entropies are defined for any $n\in\mbb{N}$, $\p\in\prob(n)$, and $\alpha\in[0,\infty]$, as
\be\label{clrenyi}
H_{\alpha}(\p)\eqdef\frac1{1-\alpha}\log\sum_{x\in[n]}p_x^\alpha\;,
\ee
where the cases $\alpha=0,1,\infty$ are defined by the appropriate limits. That is,
for $\alpha=0$ the R\'enyi entropy is also known by the name \emph{the max-entropy\index{max-entropy}} and is given by
\be\label{maxen}
H_{\max}(\p)\eqdef H_0(\p)\eqdef\lim_{\alpha\to 0}H_\alpha(\p)=\log|\supp(\p)| 
\ee
where $|\supp(\p)|$ is the number of non-zero components in $\p$.  For $\alpha=1$ the R\'enyi entropy reduces to the Shannon entropy\index{Shannon entropy}
\be
H(\p)=\lim_{\alpha\to 1}H_{\alpha}(\p)=-\sum_{x\in[n]}p_x\log p_x\;.
\ee
Finally, for the case $\alpha=\infty$ the R\'enyi entropy\index{R\'enyi entropy} is also known by the name the min-entropy\index{min-entropy} and is given by
\be\label{minen}
H_{\min}(\p)\eqdef\lim_{\alpha\to\infty}H_\alpha(\p)=-\log\max_{x\in[n]}\{p_x\}\;.
\ee
\begin{exercise}
Prove the three limits above.
\end{exercise}

The R\'enyi entropies are indeed entropies since they satisfy the three axioms of Definition~\ref{entropy}. To show the monotonicity under mixing (i.e. Schur concavity) observe that for $x\neq y\in[n]$ and $\alpha\in(0,\infty)$
\be
(p_x-p_y)\left(\frac{\partial H_\alpha(\p)}{\partial p_x}-\frac{\partial H_\alpha(\p)}{\partial p_y}\right)=\frac\alpha{1-\alpha}\frac1{\|\p\|_\alpha^\alpha}(p_x-p_y)\left(p_x^{\alpha-1}-p_y^{\alpha-1}\right)< 0\;.
\ee
Hence, from the Schur's test in~\eqref{crite2} this implies that $H_\alpha$ are strictly  Schur concave for $\alpha\in(0,\infty)$. The Schur concavity of $H_\alpha$ for the cases $\alpha=0$ and $\alpha=\infty$ follows by taking the limits $\alpha\to 0^+$ and $\alpha\to\infty$, respectively.

\begin{exercise}
Prove directly, without taking the limits on $\alpha$, that $H_{\min}$ and $H_{\max}$ are Schur concave.
\end{exercise}

Note that the R\'enyi entropy\index{R\'enyi entropy} can be expressed as
\be
H_{\alpha}(\p)\eqdef\frac\alpha{1-\alpha}\log\|\p\|_\alpha
\ee
where $\|\p\|_\alpha$ is the $p$-norm with $p=\alpha$. Hence, for $\alpha> 1$ the function $\|\p\|_\alpha$ is convex and also symmetric, so that $\|\p\|_\alpha$ is in particular Schur convex. Combining this with the fact that the log is monotonically increasing function and recalling our assumption that $1-\alpha< 0$ this provides an alternative proof that $H_\alpha$ is Schur concave for $\alpha>1$. 

\begin{exercise}$\;$
\begin{enumerate}
\item Show that the R\'enyi entropy satisfies the additivity axiom\index{axiom} of an entropy.
\item Show by direct calculation that $H_\alpha(\u^{(n)})=\log n$ for all $n\in\mbb{N}$.
\item Show that for any fixed $\p\in\prob(n)$, $H_\alpha(\p)$ is monotonically decreasing in $\alpha$.
\end{enumerate}
\end{exercise}

The following is a very interesting result proved in~\cite{MPS+2021}. It essentially states that all the entropy functions are R\'enyi entropies. We refer the reader to~\cite{MPS+2021} for the proof as it goes beyond the scope of this book.
\begin{myt}{}
\begin{theorem}\label{thm:MPS+2021}
Let $\H$ be an entropy as defined in Definition~\ref{entropy}. Then, $\H$ can be expressed as a convex combination of the R\'enyi entropies.
\end{theorem}
\end{myt}

\subsection{Quantum Entropies}\index{entropy}

In the definition below we make use of the notation $\rho\succ\sigma$ to indicate that the eigenvalues of $\rho$ form a vector that majorizes the vector consisting of the eigenvalues of $\sigma$. With this notation the extension of Definition~\ref{entropy} to the quantum domain is straightforward. Specifically, we will consider a function 
\be\label{asincup}
\H:\bigcup_{A}\md(A)\to\mbb{R}
\ee 
that maps density matrices in all finite dimensions to the real numbers.

\begin{myd}{Quantum Entropy}\index{monotonicity}
\begin{definition}\label{qentropy}
Let $\H$ be as in~\eqref{asincup} and suppose it is not equal to the constant zero function. Then, $\h$ is called an entropy if it satisfies the following two axioms:
\begin{enumerate}
\item\textbf{Monotonicity}. For every $\rho\in\md(A)$ and $\sigma\in\md(B)$ that satisfy $\rho^A\succ\sigma^B$ we have
$\h(\rho^A)\leq\h(\sigma^B)$.
\item \textbf{Additivity}. For every $\rho\in\md(A)$ and $\sigma\in\md(B)$, 
$
\H(\rho^A\otimes\sigma^B)=\H(\rho^A)+\H(\sigma^B)\;.
$
\end{enumerate}
\end{definition}
\end{myd}

There is a one-to-one correspondence between  quantum entropies and classical entropies. Explicitly,  note that the monotonicity property of entropies implies that they are invariant under unitaries; that it, let $\H$ be an entropy function and let $U\in\muu(A)$ be a unitary matrix. Then,
\be
\H(\rho)=\H\left(U\rho U^*\right)\quad\quad\forall\;\rho\in\md(A)\;.
\ee
The above invariance property implies that $\H(\rho)$ depends only on the eigenvalues of $\rho$ and therefore the quantum entropy of $\rho$ can be viewed as the classical entropy of the probability vector consisting of the eigenvalues of $\rho$. Therefore, any classical entropy $\H_{\rm classical}$ can be extended to the quantum domain via ($m\eqdef|A|$)
\be
\H_{\rm quantum}(\rho)\eqdef\H_{\rm classical}(\lambda_1,\ldots,\lambda_{m})\quad\quad\forall\rho\;\in\md(A)\;,
\ee
where $\{\lambda_x\}_{x\in[m]}$ are the eigenvalues of $\rho$. It is left as a simple exercise to show that $\H_{\rm quantum}$ is indeed a quantum entropy that satisfies the two axioms of the definition above.

As an example, consider the classical R\'enyi entropies as defined in~\eqref{clrenyi}. By replacing the components $\{p_x\}_{x\in[n]}$ with the eigenvalues $\{\lambda_x\}_{x\in[n]}$ of $\rho$, we get the quantum version of the R\'enyi entropies. For any $\alpha\in[0,\infty]$ they are given by
\be\label{qrenyi}
H_{\alpha}(\rho)\eqdef\frac1{1-\alpha}\log\sum_{x\in[n]}\lambda_x^\alpha=\frac1{1-\alpha}\log\tr[\rho^\alpha]\;.
\ee
Similarly, from the classical case we get that the limits $\alpha=0,1,\infty$ are given for all $\rho\in\md(A)$ by:
\ben
\item The max-entropy ($\alpha=0$),\index{max-entropy}
\be\label{qhmax}
H_{\max}(\rho)=\log\tr\left[\Pi_\rho\right]
\ee
where $\Pi_{\rho}$ is the projector to the support of $\rho$.

\item The von-Neumann\index{von-Neumann} entropy ($\alpha=1$), 
\be
H(\rho)=-\tr[\rho\log\rho]\;.
\ee
\item The min-entropy ($\alpha=\infty$), \index{min-entropy}
\be\label{qminen}
H_{\min}(\rho)=-\log\|\rho\|_{\infty}\;.
\ee
\een

\begin{exercise}
Let $\H$ be an entropy function, and let $\mE\in\cptp(A\to B)$ be a random isometry channel; i.e.
\be
\mE^{A\to B}=\sum_{x\in[n]}p_x\mV^{A\to B}_x
\ee
where $\{p_x\}_{x\in[n]}$ is a probability distribution and each $\mV_x\in\cptp(A\to B)$ is an isometry. Show that
\be
\H(\rho)\leq\H\big(\mE(\rho)\big)\quad\quad\forall\;\rho\in\md(A)\;.
\ee
\end{exercise}

\bex\label{ex:unital}
Let $\H$ be a quantum entropy, and let $\mU\in\cptp(A\to A)$ be a unital channel.
Show that
\be
\H(\rho)\leq \H\big(\mU(\rho)\big)\quad\quad\forall\;\rho\in\md(A)\;.
\ee
\eex

\section{Classical Relative Entropies}\index{relative entropy}

Let 
\be\label{sdrele}
\D:\bigcup_{n\in\mbb{N}}\Big\{\prob(n)\times\prob(n)\Big\}\to\mbb{R}\cup\{\infty\}
\ee 
be a function acting on pairs of probability vectors in all finite dimensions. 

\begin{myd}{Relative Entropy}
\begin{definition}\label{cre}
The function $\D$ in~\eqref{sdrele} is called a \emph{relative entropy} if it satisfies the following three conditions:
\begin{enumerate}
\item {\it Data Processing Inequality.} See~\eqref{cdpi}.
\item {\it Additivity.} For any $n,n'\in\mbb{N}$, $\p,\q\in\prob(n)$, and $\p',\q'\in\prob(n')$,
\be
\D(\p\otimes\p'\|\q\otimes\q')=\D(\p\|\q)+\D(\p'\|\q')\;.
\ee
\item {\it Normalization.} $\D(\e_1\|\u^{(2)})=1$, where $\e_1=(1,0)^T$ and $\u^{(2)}\eqdef(\frac12,\frac12)^T$.
\end{enumerate}
\end{definition}
\end{myd}

In the definition above we did not include the normalization condition $\D(1\|1)=0$ (as satisfied by all normalized divergences) since it follows from the additivity\index{additivity} property. Indeed, let $\p,\q\in\prob(n)$ and observe that
\be
\D(\p\|\q)=\D(\p\otimes 1\|\q\otimes 1)=\D(\p\|\q)+\D(1\|1)\;.
\ee 
Therefore, we must have $\D(1\|1)=0$. Hence, relative entropies are divergences.

The normalization condition $\D(\e_1|\u^{(2)})=1$ eliminates the possibility of scaling a relative entropy by a fixed constant. Notably, this normalization condition is asymmetric with respect to its two inputs. Indeed, any alternative normalization would disrupt this symmetry, as we cannot select a symmetric input (bearing in mind that all divergences satisfy $\D(\p|\p)=0$ for every $\p\in\prob(n)$).  

\begin{exercise}
Show that any relative entropy $\D$ must satisfy
\be
\D(\e_1\|\e_2)=\infty\;,
\ee
where $\e_1=(1,0)^T$ and $\e_2=(0,1)^T$.
Hint: Show first that $\D(\e_1\|\e_2)\geq 1$ and then use the additivity property together with the DPI to show that for any $n\in\mbb{N}$, $\D(\e_1\|\e_2)\geq n$.
\end{exercise} 

\subsection{The R\'enyi Relative Entropies}\index{R\'enyi divergence}\index{relative entropy}

In his seminal paper, R\'enyi introduced a one parameter family of relative entropies. We already encountered them in Sec.~\ref{sec:tr}, and here we will study some of their properties. 

\begin{myd}{The R\'enyi Relative Entropies}\index{R\'enyi divergence}

\begin{definition}\label{renyi}
The R\'enyi relative entropy of order $\alpha\in[0,\infty]$ is defined for all $\p,\q\in\prob(n)$ as
\be\nonumber
D_\alpha(\p\|\q)\eqdef\begin{cases}\frac{1}{\alpha-1}\log\sum_{x\in[n]}p_x^{\alpha}q_x^{1-\alpha} &\text{if }\supp(\p)\subseteq\supp(\q)\text{ or }\alpha\in[0,1)\text{ and }\p\cdot\q\neq 0.\\
\infty&\text{otherwise.}
\end{cases}
\ee
The cases $\alpha=0,1,\infty$ are defined in terms of the appropriate limits. 
\end{definition}
\end{myd}
\begin{remark}
We use the convention that if $q_x=p_x=0$ then $p_x^\alpha q_x^{1-\alpha}=0$ even for $\alpha>1$. With this convention, the conditions that $\supp(\p)\subseteq\supp(\q)$ or $\alpha\in[0,1)$ and $\p\cdot\q\neq 0$ are precisely the conditions that the expression $\frac{1}{\alpha-1}\log\sum_{x\in[n]}p_x^{\alpha}q_x^{1-\alpha}$ is well defined. Otherwise, if it is not well defined the R\'enyi relative entropy is set to be infinity.
\end{remark}

For $\alpha=0$ the relative R\'enyi entropy is called the min-relative entropy. It is given by 
\be\label{minrel}
D_{\min}(\p\|\q)\eqdef\lim_{\alpha\to 0^+}D_\alpha(\p\|\q)=-\log\sum_{x\in\supp(\p)}q_x\;.
\ee
Observe that if $D_{\min}(\p\|\q)\neq0$ then $\p$ must have zero components.
For $\alpha=\infty$  the relative R\'enyi entropy is called the max-relative entropy. It is given by 
\be\label{maxrel}
D_{\max}(\p\|\q)\eqdef\lim_{\alpha\to \infty}D_\alpha(\p\|\q)=\log\max_{x\in[n]}\left\{\frac{p_x}{q_x}\right\}\;.
\ee
Finally, for $\alpha=1$ the R\'enyi relative entropy is called the Kullback–Leibler divergence, or in short the KL-divergence. It is given by
\be
D(\p\|\q)\eqdef\lim_{\alpha\to 1}D_\alpha(\p\|\q)=\sum_{x\in[n]}p_x(\log p_x-\log q_x)\;,
\ee
with the convention $0\log 0=0$.

\begin{exercise}
Prove the three limits above.
\end{exercise}

\begin{exercise}\label{ex:23}
Show that the R\'enyi entropies are related to the R\'enyi relative entropies via
\be
H_\alpha(\p)=\log n-D_\alpha(\p\|\u^{(n)})\quad\forall\;\p\in\prob(n)\;.
\ee 
\end{exercise}

\begin{exercise}\label{condmax}
Let $\p,\q\in\prob(n)$ and $r\in\mbb{R}_+$. 
\begin{enumerate}
\item Show that $r\q-\p\geq 0$ (entry-wise) if and only if $r\geq 2^{D_{\max}(\p\|\q)}$.
\item Show that $\p-r\q\geq 0$ if and only if $r\leq 2^{-D_{\max}(\q\|\p)}$.
\item Why in the first inequality above $r$ must be greater than one, whereas in the second it must be smaller than one?
\end{enumerate}
\end{exercise}

\begin{exercise}\label{twoax}
Show that all the R\'enyi entropies satisfy the additivity and normalization properties of a relative entropy as given in Definition~\ref{cre}.
\end{exercise}

\begin{exercise}
A relative entropy $\D$ is said to be pathological if $\D(\u^{(2)}\|\e_1^{(2)})=0$, where $\u^{(2)}\eqdef(\frac12,\frac12)^T$ is the uniform distribution in $\prob(2)$, and $\e_1^{(2)}\eqdef(1,0)^T$.
Show that $D_{\min}$ is pathological and use it to show that $D_{\rm path}$ which is defined for any $n\in\mbb{N}$ and $\p,\q\in\prob(n)$ as
\be
D_{\rm path}(\p\|\q)\eqdef D_{\min}(\p\|\q)+D_{\min}(\q\|\p)\;,
\ee
is a  relative entropy.
\end{exercise}

We now show that in addition to the additivity and normalization, the R\'enyi relative entropies also satisfies the DPI.

\begin{myt}{}
\begin{theorem}
The R\'enyi relative entropy of any order $\alpha\in[0,\infty]$ is a relative entropy; i.e. it satisfies the axioms of DPI, additivity, and normalization, as given in Definition~\ref{cre}. 
\end{theorem}
\end{myt}

\begin{proof}
The additivity and normalization you proved in Exercise~\ref{twoax}. To show DPI recall the $\alpha$ divergences given in~\eqref{aldiv} by
\be
D_{f_\alpha}(\p\|\q)=\frac1{\alpha(\alpha-1)}\left(\sum_{x\in[n]}p_x^\alpha q_x^{1-\alpha}-1\right)
\ee
Since the above expression has been derived from the convex function $f_\alpha(r)=\frac{r^\alpha-r}{\alpha(\alpha-1)}$ it is an $f$-Divergence\index{$f$-divergence} and in particular satisfies the DPI. For $\alpha=1$ the above expression coincide with the R\'enyi relative entropy of that order (i.e. the KL-divergence), so in this case the DPI property follows. For $\alpha\neq 1$ we denote by
\be
Q_\alpha(\p\|\q)\eqdef\sum_{x\in[n]}p_x^\alpha q_x^{1-\alpha}\;.
\ee
Observe that from the DPI of $D_{f_\alpha}$ we get that for $\alpha>1$ the function $Q_\alpha(\p\|\q)$ is monotonically non-decreasing under maps $(\p,\q)\mapsto(E\p,E\q)$ with $E\in\stoc(m,n)$, and for $\alpha<1$ it is monotonically non-increasing under such maps.
Since the R\'enyi relative entropy can be expressed as $D_\alpha(\p\|\q)\eqdef\frac1{\alpha-1}\log Q_\alpha(\p\|\q)$ and the log is monotonically increasing function, we conclude that $D_\alpha(\p\|\q)$ satisfies the DPI.
\end{proof}

\begin{exercise}\label{ex:flip}
Show that for any $\alpha\in(0,1)$ and $\p,\q\in\prob(n)$
\be
D_\alpha(\p\|\q)=\frac\alpha{1-\alpha}D_{1-\alpha}(\q\|\p)\;.
\ee
\end{exercise}

\begin{exercise}\label{ex:clqu}
Show that if $\p,\q\in\prob(n)$ and $\rho,\sigma\in\md(X)$ are two diagonal density matrices with diagonals $\p$ and $\q$, respectively, then
\be
D_\alpha(\p\|\q)=\frac1{\alpha-1}\log\tr[\rho^\alpha\sigma^{1-\alpha}]\;.
\ee 
\end{exercise}

\subsection{Properties of Relative Entropies}\index{relative entropy}

The three axioms of a relative entropy provides it with enough structure that yields many interesting properties.
In this subsection we explore some of these key properties that holds for \emph{all} relative entropies.

\begin{myt}{}
\begin{theorem}\label{822}
Let $\D$ be a relative entropy, and let $\{\e_x\}_{x\in[n]}$ be the standard (elementary) basis of $\mbb{R}^n$.
Then, for any $\p\in\prob(n)$ and $x\in[n]$ we have
\be
\D(\e_x\|\p)=-\log p_x\;,
\ee
where $p_x$ is the $x$-component of $\p$.
\end{theorem}
\end{myt}

The proof of the theorem above is based on the following lemma by Erd\"os.
\begin{myg}{Erd\"os Theorem}\index{Erd\"os theorem}
\begin{lemma}\label{lem:erd}
Let $g:\mbb{N}\to\mbb{R}$ be a function from the set of natural numbers to the real line. Suppose $g$ is non-decreasing and is additive; i.e. $g(mn)=g(n)+g(m)$ for all $n,m\in\mbb{N}$. Then, there exists a constant $c\in\mbb{R}$ such that $g(n)=c\log(n)$ for all $n\in\mbb{N}$.
\end{lemma}
\end{myg}
\begin{proof}
Suppose by contradiction that $\frac{g(n)}{\log n}$ is not a constant. Therefore, there exists $m,n\in\mbb{N}$ such that
\be
\frac{g(m)}{\log m}>\frac{g(n)}{\log n}\;.
\ee
Denote by $a\eqdef \frac{g(m)}{\log m}$ and $b\eqdef \frac{g(n)}{\log n}$ and observe that $a>b$ or equivalently $\frac ba<1$. Multiplying both sides of the inequality $\frac ba<1$ by the positive number $\frac{\log n}{\log m}k$, where $k$ is any integer, gives
\be
\frac ba\frac{\log n}{\log m}k<\frac{\log n}{\log m}k\;.
\ee
Therefore, for sufficiently large $k\in\mbb{N}$ there must exists an integer between the above two numbers; i.e. there exists $\ell\in\mbb{N}$ such that
\be
\frac ba\frac{\log n}{\log m}k<\ell<\frac{\log n}{\log m}k\;.
\ee
The above two inequalities can be expressed as
\be
k\log n> \ell\log m\quad\text{and}\quad kb\log n < \ell a\log m\;.
\ee 
The first equation above implies that the integers $n^k$ and $m^\ell$ satisfies $n^k>m^{\ell}$, and the second equation implies that $kg(n)<\ell g(m)\;.$ From the additivity of $g$ we therefore conclude that $g(n^k)< g(m^\ell)$. To summarize, we got that \be n^k>m^{\ell}\quad\text{and}\quad g(n^k)< g(m^\ell)\;.\ee These two inequalities are in contradiction with the assumption that $g$ is non-decreasing. This completes the proof. 
\end{proof}

\begin{proof}[Proof of Theorem~\ref{822}]
Since divergences (and therefore relative entropies) are invariant under permutations (see~\eqref{invup}), it is sufficient to show that $\D(\e_1\|\p)=-\log p_1$. We first show that for any vector $\r=(r_1,\ldots,r_n)^T\in\prob(n)$ with $r_1=0$ we have
\be\label{j1}
(\e_1,\p)\sim\big(\e_1,p_1\e_1+(1-p_1)\r\big)
\ee
where the symbol $\sim$ corresponds to the equivalence relation under relative majorization. Define $E\eqdef[\e_1,\r,\ldots,\r]\in\stoc(n,n)$ to be the column stochastic matrix whose first column is $\e_1$ and the remaining $n-1$ columns equal $\r$. We then have
\be\label{j2}
(\e_1,\p)\succ(E\e_1,E\p)=\big(\e_1,p_1\e_1+(1-p_1)\r\big)\;.
\ee
Conversely, define $\tp\eqdef\frac{1}{1-p_1}(0,p_2,\ldots,p_n)^T\in\prob(n)$ and $\tilde{E}\eqdef[\e_1,\tp,\ldots,\tp]\in\stoc(n,n)$. Then,
\be\label{j3}
\big(\e_1,p_1\e_1+(1-p_1)\r\big)\succ\big(\tilde{E}\e_1,p_1\tilde{E}\e_1+(1-p_1)\tilde{E}\r\big)=(\e_1,\p)\;.
\ee
Combining~\eqref{j2} and~\eqref{j3} gives~\eqref{j1}.

The relation in~\eqref{j1} implies that
\be
\D(\e_1\|\p)=\D\big(\e_1\big\|p_1\e_1+(1-p_1)\r\big)
\ee
so that the function $f(p_1)\eqdef\D(\e_1\|\p)$ is independent on $p_2,\ldots,p_n$. Moreover, the function $f:[0,1]\to\mbb{R}_+\cup\{\infty\}$ has the following two properties:
\begin{enumerate}
\item $f$ is monotonically non-increasing.
\item $f$ is additive; i.e. $f(st)=f(s)+f(t)$ for all $s,t\in[0,1]$.
\end{enumerate}
The first property of $f$ follows from the fact that for any $E\in\stoc(n,n)$ with $E\e_1=\e_1$ we get $E\p=p_1\e_1+(1-p_1)\s$ for some $\s\in\prob(n)$. This means that the first component of $\q\eqdef E\p$ satisfies $q_1\geq p_1$. From the DPI we have
\be
f(p_1)\eqdef \D(\e_1\|\p)\geq \D(E\e_1\|E\p)=\D(\e_1\|\q)=f(q_1)\;,
\ee
so that $f$ is monotonically non-increasing. The additivity of $f$ follows trivially from the additivity of $\D$ under tensor products (see exercise below).

Define now the function $g:\mbb{N}\to\mbb{R}_+\cup\{\infty\}$ via the relation $g(m)\eqdef f\left(\frac{1}{m}\right)$.
This function is non-decreasing and additive. Therefore, from Erd\"os theorem there exists a constant $c\in\mbb{R}$ such that  $g(m)=c\log m$ for all $m\in\mbb{N}$. The condition $g(2)=f(1/2)=\D(\e_1\|\u^{(2)})=1$ gives $c=1$. Therefore, for any $m\in\mbb{N}$ we have $f(1/m)=\log m$. Furthermore, observe that for any $k\leq m$ the additivity of $f$ gives 
\ba
\log k+f\left(\frac km\right)=f\left(\frac 1k\right)+f\left(\frac km\right)=f\left(\frac 1m\right)=\log m\;.
\ea
Hence, $f\left(\frac km\right)=\log m-\log k=-\log(k/m)$. Hence, $f(r)=-\log r$ for all rationals in $[0,1]$. To prove that this relation holds for any $r\in[0,1]$ (possibly irrational), let $\{s_k\}$ and $\{t_k\}$ be two sequences of rational numbers in $[0,1]$ both with limit $r$ and with $s_k\leq r\leq t_k$ for all $k\in\mbb{N}$. Then, the monotonicity property of $f$ gives for any $k\in\mbb{N}$
\be
-\log s_k=f(s_k)\geq f(r)\geq f(t_k)=-\log t_k\;.
\ee
Taking the limit $k\to\infty$ on both sides and using the continuity of the log function gives $f(r)=-\log r$. This completes the proof.
\end{proof}

The theorem above has the following interesting corollary that justify the terminology of the max and min relative entropies.

\begin{myg}{}
\begin{corollary}
Let $\D$ be a relative entropy. Then for any $n\in\mbb{N}$ and any $\p,\q\in\prob(n)$,
\be
D_{\min}(\p\|\q)\leq\D(\p\|\q)\leq D_{\max}(\p\|\q)\;.
\ee
\end{corollary}
\end{myg}

\begin{proof}
In Theorem~\ref{bminmax} we proved that for any $\p,\q\in\prob(n)$
\be
\left((1,0)^T,(\lambda_{\max},1-\lambda_{\max})^T\right)\succ(\p,\q)\succ\left((1,0)^T,(\lambda_{\min},1-\lambda_{\min})^T\right)
\ee
where $\lambda_{\max}\eqdef\min_{x\in[n]}q_x/p_x=2^{-D_{\max}(\p\|\q)}$ and $\lambda_{\min}\eqdef\sum_{x\in\supp(\p)}q_x=2^{-D_{\min}(\p\|\q)}$. Hence, the monotonicity of $\D$ under relative majorization gives
\ba
\D(\p\|\q)&\leq \D\left(\e_1\big\|\left(2^{-D_{\max}(\p\|\q)},1-2^{-D_{\max}(\p\|\q)}\right)^T\right)\\
\GG{\text{Theorem~\ref{822}} }&=D_{\max}(\p\|\q)\;,
\ea
and
\ba
\D(\p\|\q)&\geq \D\left(\e_1\big\|\left(2^{-D_{\min}(\p\|\q)},1-2^{-D_{\min}(\p\|\q)}\right)^T\right)\\
\GG{\text{Theorem~\ref{822}} }&=D_{\min}(\p\|\q)\;.
\ea
This completes the proof.
\end{proof}

\begin{exercise}
Use the corollary above to show that any relative entropy $\D$ satisfies for all $\p,\q\in\prob(n)$:
\begin{itemize}
\item $\D(\p\|\q)<\infty$ if $\supp(\p)\subseteq\supp(\q)$, and
\item $\D(\p\|\q)=\infty$ if $\p\cdot\q=0$.
\end{itemize}
\end{exercise}

Relative entropies are not metrics but we show now that they satisfy the following variant of the triangle inequality.

\begin{myt}{\color{yellow} Triangle Inequality}\index{triangle inequality}
\begin{theorem}\label{tin}
Let $\D$ be a relative entropy. Then, for any $\p,\q,\r\in\prob(n)$
\be\label{656}
\D(\p\|\q)\leq\D(\p\|\r)+D_{\max}(\r\|\q)\;.
\ee
\end{theorem}
\end{myt}
\begin{proof}
The key idea of the proof is to denote by $\eps\eqdef2^{-D_{\max}(\r\|\q)}$ and observe that the right hand side of~\eqref{656} can be expressed as
\ba
\D(\p\|\r)+D_{\max}(\r\|\q)&=\D(\p\|\r)-\log\eps\\
\GG{Theorem~\ref{822}}&=\D(\p\|\r)+\D\left(\e_1\big\|(\eps,1-\eps)^T\right)\\
\GG{Additivity}&=\D\left(\p\otimes\e_1\big\|\r\otimes (\eps,1-\eps)^T\right)\;.
\ea
Therefore, to prove the inequality~\eqref{656} it is sufficient to show that
\be
\left(\p\otimes\e_1\big\|\r\otimes (\eps,1-\eps)^T\right)\succ (\p\|\q)\;.
\ee
To prove the above relation, we define a channel $E\in\stoc(n,2n)$ that acts on $\prob(n)\otimes\prob(2)$ as an identity upon detecting $\e_1\eqdef(1,0)^T$
 in the second register, and produces a constant output $\t\in\prob(n)$ (to be determined shortly) upon detecting $\e_2\eqdef(0, 1)^T$ in the second register; explicitly, for any $\s\in\prob(n)$
\be\label{840}
E\left(\s\otimes\e_1\right)=\s\quad,\quad E\left(\s\otimes\e_2\right)=\t\;.
\ee
By definition, $E\left(\p\otimes\e_1\right)=\p$. Our objective is to choose $\t$ such that
$E\left(\r\otimes (\eps,1-\eps)^T\right)=\q$. From the definition of $E$ we have
\ba
E\left(\r\otimes (\eps,1-\eps)^T\right)&=\eps E\left(\r\otimes \e_1 \right)+(1-\eps)E\left(\r\otimes \e_2\right)\\
&=\eps\r+(1-\eps)\t\;.
\ea
Therefore, it is left to show that there exists $\t\in\prob(n)$ such that  $\eps\r+(1-\eps)\t=\q$. This means that we need to show that the vector
\be
\t\eqdef\frac{\q-\eps\r}{1-\eps}\;,
\ee
has non-negative components. Indeed, from the definition of $\eps$ we have $\q-\eps\r\geq 0$. This completes the proof.
\end{proof}

\begin{exercise}
Use the theorem above to show that the function
\be
D_T(\p\|\q)\eqdef\max\Big\{D_{\max}(\p\|\q),D_{\max}(\q\|\p)\Big\}
\ee
is a divergence that is also a metric. Show that it satisfies $D_T(\p\|\q)<\infty$ if and only if $\supp(\p)=\supp(\q)$.
This metric\index{metric} is known as the Thompson's metric. 
\end{exercise}

The theorem above can be expressed in terms of the Thompson's metric. 

\begin{myg}{}
\begin{corollary}
Any relative entropy $\D$ satisfies for all $\p,\q,\q'\in\prob(n)$
\be\label{tom}
\big|\D(\p\|\q)-\D(\p\|\q')\big|\leq D_{T}(\q\|\q')\;.
\ee
\end{corollary}
\end{myg}

\begin{exercise}
Let $\p,\q,\r\in\prob(n)$ be 3 probability vectors. Show that
\be
D(\r\|\p)+D(\r\|\q)\geq D_{1/2}(\p\|\q)
\ee
where $D$ is the KL-divergence and $D_{1/2}$ is the R\'enyi relative entropy of order $\alpha=1/2$. Moreover, show that for any choice of $\p$ and $\q$ there exists $\r$ that achieves the equality.
\end{exercise}

\subsubsection{Continuity of Relative Entropies}\index{continuity}\index{relative entropy}

The corollary above demonstrates that any relative entropy is continuous in the second argument when $\q,\q'>0$. 
We will now use it to explore the continuity of a relative entropy in $\prob(n)\times\prob(n)$.
We say that $\D$ is upper semi-continuous at $(\p,\q)\in\prob(n)\times\prob(n)$ if for any sequence $\{(\p_k,\q_k)\}_{k\in\mbb{N}}\subset\prob(n)\times\prob(n)$ that converges to $(\p,\q)$ we have
\be
\limsup_{k\to\infty}\D(\p_k\|\q_k)\leq \D(\p\|\q)\;.
\ee
We say that $\D$ is lower semi-continuous at $(\p,\q)\in\prob(n)\times\prob(n)$ if 
\be
\liminf_{k\to\infty}\D(\p_k\|\q_k)\geq \D(\p\|\q)\;.
\ee
Note that $\D$ is both lower and upper semi-continuous at $(\p,\q)$ if and only if it is continuous at $(\p,\q)$.

\begin{exercise}$\;$\label{ex:829}
\begin{enumerate}
\item Show that the max relative entropy, $D_{\max}(\p\|\q)$, is not  upper semi-continuous when $\q$ does not have full support.
{\it Hint: Consider the sequences $\{\p_k\}_{k\in\mbb{N}}$ and $\{\q_k\}_{k\in\mbb{N}}$ with $\p_k\eqdef \left(\frac1k,1-\frac1k\right)^T$ and $\q_k\eqdef\left(\frac1{k^2},1-\frac1{k^2}\right)^T$.}
\item Show that $D_{\rm path}(\p\|\q)\eqdef D_{\min}(\p\|\q)+D_{\min}(\q\|\p)$ is not lower semi-continuous at the boundary of $\prob(n)\times\prob(n)$.
\end{enumerate}
\end{exercise}

From the exercise above it is clear that we cannot expect relative entropies to be continuous everywhere in $\prob(n)\times\prob(n)$.
However, if we remove some of the points in the boundary, we get the following continuity property.

\begin{myt}{\color{yellow} Continuity of Relative Entropies}
\begin{theorem}\label{continuityrel}
Let $\D$ be a relative entropy. Then, $\D$ is upper semi-continuous at any point in $\prob(n)\times\prob_{>0}(n)$, and is continuous at any point in $\prob_{>0}(n)\times\prob_{>0}(n)$.
\end{theorem}
\end{myt}

\begin{proof}
Let ${(\p_k,\q_k)}_{k\in\mbb{N}}$ be a sequence in $\prob(n)\times\prob(n)$ that converges to $(\p,\q)$.
For any $k\in\mbb{N}$, define a column stochastic matrix $E_k\in\stoc(n,n)$ by its action on every $\s\in\prob(n)$ as
\be\label{846}
E_k\s\eqdef\p_k+2^{-D_{\max}(\p\|\p_k)}(\s-\p)\;.
\ee
Since $\lim_{k\to\infty}\p_k=\p$, for sufficiently large $k$ we have $2^{-D_{\max}(\p\|\p_k)}>0$ (see the exercise below). Moreover, from the definition of $D_{\max}$ we get that
$\p_k-2^{-D_{\max}(\p\|\p_k)}\p\geq 0$ so that $E_k$ is indeed a column stochastic matrix.
Using these notations, we derive the following from the DPI:
\ba
\D(\p\|\q)&\geq\D(E_k\p\| E_k\q)\\
{\color{red}\text{\eqref{846}}\rightarrow}&=\D(\p_k\| E_k\q)\\
{\color{red}\text{Theorem~\ref{tin}}\rightarrow}&\geq \D(\p_k\| \q_k)-D_{\max}(E_k\q\|\q_k)
\ea
Moving the term involving $D_{\max}$ to the other side and taking the supremum limit on both sides gives
\be
\limsup_{k\to\infty}\D(\p_k\| \q_k)\leq\D(\p\|\q)+\limsup_{k\to\infty}D_{\max}(E_k\q\|\q_k)\;.
\ee
The second term on the right-hand side above vanishes since the vector 
\be
\tq_k\eqdef E_k\q=2^{-D_{\max}(\p\|\p_k)}\q+\left(\p_k-2^{-D_{\max}(\p\|\p_k)}\p\right)
\ee
has a limit $\lim_{k\to\infty}\tq_k=\q$ so that  
\be\label{limit56}
\limsup_{k\to\infty}D_{\max}(\tq_k\|\q_k)=0\;.
\ee
Note that we used indirectly the fact that $\q>0$, since for sufficiently large $k$ we must have $\q_k>0$ so the limit above is indeed zero. This completes the proof that $\D$ is upper semi-continuous on $\prob(n)\times\prob_{>0}(n)$. 

We now prove the lower semi-continuity on $\prob_{>0}(n)\times\prob_{>0}(n)$. Note that since we already proved upper semi continuity in this domain, this will imply that $\D$ is continuous on  $\prob_{>0}(n)\times\prob_{>0}(n)$.
For any $k\in\mbb{N}$, we define $E_k$ as before but with the role of $\p_k$ and $\p$ interchanged; i.e. $E_k\in\stoc(n,n)$ is defined by its action on any $\s\in\prob(n)$ as
\be\label{8460}
E_k\s\eqdef\p+2^{-D_{\max}(\p_k\|\p)}(\s-\p_k)\;.
\ee
Note that for all $k$, $2^{-D_{\max}(\p_k\|\p)}>0$ since we assume $\p>0$. Moreover, from the definition of $D_{\max}$ we have
$\p-2^{-D_{\max}(\p_k\|\p)}\p_k\geq 0$, so that $E_k$ is indeed a column stochastic matrix.
With the above notations we get from the DPI
\ba
\D(\p_k\|\q_k)&\geq\D(E_k\p_k\| E_k\q_k)\\
{\color{red}\text{\eqref{8460}}\rightarrow}&=\D(\p\| E_k\q_k)\\
{\color{red}\text{Theorem~\ref{tin}}\rightarrow}&\geq \D(\p\| \q)-D_{\max}(E_k\q_k\|\q)
\ea
Taking the infimum limit on both sides gives
\be
\liminf_{k\to\infty}\D(\p_k\| \q_k)\geq\D(\p\|\q)\;,
\ee
where we used the fact that
\be
E_k\q_k=2^{-D_{\max}(\p_k\|\p)}\q_k+\left(\p-2^{-D_{\max}(\p_k\|\p)}\p_k\right)
\ee
has a limit $\lim_{k\to\infty}E_k\q_k=\q$ so that  
\be\label{limit560}
\limsup_{k\to\infty}D_{\max}(E_k\q_k\|\q)=0\;.
\ee
This completes the proof. 
\end{proof}

\begin{exercise}$\;$
\begin{enumerate}
\item Show that if $\{\p_k\}_{k\in\mbb{N}}$ is a sequences in $\prob(n)$ that converges to $\p\in\prob(n)$ then for sufficiently large $k$ we have $D_{\max}(\p\|\p_k)<\infty$. Hint: Show that for sufficiently large $k$, $\supp(\p)\subseteq\supp(\p_k)$.
\item Prove the limits in~\eqref{limit56} and~\eqref{limit560}.
\end{enumerate}
\end{exercise}

\subsubsection{Faithfulness of Relative Entropies}

We have seen before that any divergence $\D$ has the property that $\D(\p\|\q)=0$ if $\p=\q$. Faithfulness\index{faithfulness} of a divergence refers to the property that this equality holds if and only if $\p=\q$. The minimal relative entropy, $D_{\min}$, provides an example of a relative entropy that is not faithful. Particularly, $D_{\min}(\p\|\q)=0$ for any $\p$ with $\supp(\q)\subseteq\supp(\p)$. However, from the theorem below it follows that $D_{\min}$ is a very unique relative entropy and almost all relative entropies are faithful.

\begin{myt}{\color{yellow} Faithfulness}\index{faithfulness}
\begin{theorem}\label{faithful}
Let $\D$ be a relative entropy. The following statements are equivalent:
\begin{enumerate}
\item $\D$ is not faithful.
\item $\D(\p\|\q)= 0$ for all $m\in\mbb{N}$ and all $\p,\q\in\prob(m)$ with $\supp(\p)=\supp(\q)$. 
\end{enumerate}
\end{theorem}
\end{myt}

\begin{proof}
The direction $2\Rightarrow 1$ is trivial. We therefore prove that $1\Rightarrow2$. Since $\D$ is not faithful there exists $\p,\q\in\prob(m)$ such that $\p\neq\q$ and $\D(\p\|\q)=0$. For any $n\in\mbb{N}$ it follows from the additivity\index{additivity} property of $\D$ that also $\D\left({\p}^{\otimes n}\big\|{\q}^{\otimes n}\right)=0$. Now, in Corollary~\ref{asymajo} of the next chapter we will see that for any $\s,\t\in\prob_{>0}(2)$ and large enough $n$ we have
\be
(\p^{\otimes n},\q^{\otimes n})\succ(\s,\t)\;.
\ee
Therefore,
\be
0=\D\left({\p}^{\otimes n}\big\|{\q}^{\otimes n}\right)\geq\D(\s\|\t)\;.
\ee
We therefore conclude that $\D(\s\|\t)=0$ for all $\s,\t\in\prob_{>0}(2)$. It is left to show that this also holds in dimensions higher than two. 

Indeed, let $\p,\q\in\prob(m)$ with $\supp(\p)=\supp(\q)$, and recall from~\eqref{stpq} that there exists $\s,\t\in\prob_{>0}(2)$ such that $(\s,\t)\succ(\p,\q)$. We therefore get that 
\be
\D(\p\|\q)\leq \D\left(\s\|\t\right)\;.
\ee
Since we already proved that $\D(\s\|\t)=0$ for all $\s,\t\in\prob_{>0}(2)$ we conclude that $\D(\p\|\q)=0$ for all $\p$ and $\q$ with the same support. This completes the proof.
\end{proof}
\begin{myg}{Continuity Implies Faithfulness}\index{continuity}
\begin{corollary}
Let $\D$ be a relative entropy, $2\leq m\in\mbb{N}$, and $\q\in\prob(m)$ be a probability vector whose first component satisfies $q_1\in(0,1)$. If $\D$ is not faithful then the function $f_\q(\p)\eqdef\D(\p\|\q)$  is not lower semi-continuous at $\p=\e_1^{(m)}$.
\end{corollary}
\end{myg}
\begin{remark}
Note that the corollary above in particular implies that relative entropies that are continuous in the first argument must be faithful.
\end{remark}
\begin{proof}
Let $\{\p_k\}_{k\in\mbb{N}}$ be a sequence in $\prob(m)$ such that $\supp(\q)=\supp(\p_k)$ and $\p_k\to\e_1$ and $k\to\infty$. Such a sequence exists since $q_1\in(0,1)$.
From the theorem above it follows that $\D(\p_k\|\q)=0$ so that
\be
\lim_{k\to\infty}\D(\p_k\|\q)=0<-\log(q_1)=\D(\e_1\|\q)\;.
\ee
Therefore, $f_\q(\p)$ cannot be lower-semi-continuous at $\p=\e_1$.
\end{proof}

\subsection{Bijection\index{bijection} Between Entropies and Relative Entropies}\label{sec:bijection}\index{relative entropy}

The R\'enyi entropies are related to the R\'enyi relative entropies via (see Exercise~\eqref{ex:23})
\be
H_\alpha(\p)=\log n-D_\alpha(\p\|\u^{(n)})\quad\forall\;\p\in\prob(n)\;.
\ee 
More generally, every relative entropy $\D$ can be used to define an entropy $\h$ via 
\be\label{hfromd}
\h(\p)\eqdef\log n-\D(\p\|\u^{(n)})\quad\forall\;\p\in\prob(n)\;.
\ee

\begin{exercise}
Show that if $\D$ is a relative entropy then $\h$ as defined in~\eqref{hfromd} satisfies the normalization and additivity axioms of an entropy.
\end{exercise}
 
To show that $\h$ as defined in~\eqref{hfromd} is indeed an entropy, we need to prove the monotonicity property (in addition to the properties you proved in the exercise above).
Recall that if $\p,\q\in\prob(n)$ and $\p\succ\q$ then there exists a doubly stochastic matrix\index{doubly stochastic matrix} $D\in\stoc(n,n)$ such that
$\q=D\p$. Therefore, in this case we get that
\ba
\h(\q)=\h(D\p)&=\log n-\D(D\p\|\u^{(n)})\\
{\color{red} D\text{ is doubly-stochastic}\rightarrow}&=\log n-\D(D\p\|D\u^{(n)})\\
{\color{red} \text{DPI}\rightarrow}&\geq \log n-\D(\p\|\u^{(n)})\\
&=\h(\p)\;.
\ea
That is, $\h$ satisfies the monotonicity property of an entropy if the two vectors have the same dimension. If $\p\in\prob(n)$ and $\q\in\prob(m)$ have different dimensions (i.e. $n\neq m$) then the relation $\p\succ\q$ is equivalent to a majorization relation between two vectors with the same dimension $\max\{n,m\}$ in which one of the vectors is padded with zeros to make the dimensions equal. Therefore, to show that $\h$ above satisfies the monotonicity property of an entropy, it is left to show that it is invariant under embedding; i.e.\, $\h(\p\oplus 0)=\h(\p)$ for all $\p\in\prob(n)$. For this purpose, note that
\be
\h(\e_1^{(n)})=\log n-\D(\e_1^{(n)}\|\u^{(n)})=\log n-\log n=0\;,
\ee
where we used Theorem~\ref{822}. Therefore, from the additivity property that you proved in the exercise above it follows that
for any $\p\in\prob(n)$
\be
\h(\p)=\h\left(\p\otimes\e_1^{(n+1)}\right)=\h\left((\p\oplus 0)\otimes \e_1^{(n)}\right)=\h\left(\p\oplus 0\right)\;.
\ee
Hence, $\h$ satisfies the monotonicity property of an entropy, and when combined with the exercise above we conclude that
Eq.~\eqref{hfromd} demonstrates that for any relative entropy there is a corresponding entropy. Remarkably, the next theorem shows that the converse is also true.

\begin{myt}{\color{yellow} One-To-One Correspondence}
\begin{theorem}
There exists a bijection\index{bijection} $\mathfrak{f}$ with inverse $\mathfrak{f}^{-1}$ mapping between relative entropies that are continuous in the second argument and  entropies. 
\end{theorem}
\end{myt}

\begin{proof}
For any relative entropy $\D$ we define
\be
\mathfrak{f}(\D)\eqdef\h
\ee
where $\h$ is defined as in~\eqref{hfromd}. Since we already established that $\h$ is an entropy it follows that $\mathfrak{f}$ is indeed a mapping between relative entropies and entropies. We therefore need to show that when the domain of $\mathfrak{f}$ is restricted to relative entropies that are continuous in the second arguments then it has an inverse.

Let $\h$ be an entropy and define the function 
\be
D_\h:\bigcup_{n\in\mbb{N}}\prob(n)\times (\prob_{>0}(n) \cap \mbb{Q}^n)\to\mbb{R}\ee via
\be\label{rti}
		D_\h(\p\|\q) \eqdef \log n-\h \left( \bigoplus_{x=1}^n p_x \u^{(k_x)} \right) , 
	\ee
for all $n \in \mbb{N}$, $\p \in \prob(n)$, and $\q \in \prob_{>0}(n) \cap \mbb{Q}^n$ with
$\q = ( \frac{k_1}{k},\ldots, \frac{k_n}{k})^T$ for $k_x\in \mbb{N}$ and $k=k_1+\cdots+k_n$. 
Note that this construction is equivalent to the one given in Theorem~\ref{biji} with $g(\p)\eqdef\log n-\h(\p)$, although we do not assume here that $\h$ is continuous, and therefore also $g$ is not assumed to be continuous. Still, the same arguments given in the proof of Theorem~\ref{biji} imply that $D_\h$ is a divergence in the restricted domain in which the second argument has positive rational components. Moreover, since $\h$ is additive also $D_\h$ as defined above is additive under tensor products; i.e., $D_\h$ is a relative entropy with a restricted domain (see Exercise~\ref{ex:8215} below).
This restricted domain will not change the arguments leading to~\eqref{tom} and we therefore conclude that for any fixed $n\in\mbb{N}$ and $\p\in\prob(n)$, $D_\h(\p\|\q)$ is continuous in $\q \in \prob_{>0}(n) \cap \mbb{Q}^n$. Therefore, the continuous extension of $D_\h$ to $\prob(n)\times\prob(n)$ is well defined. We therefore define $\mathfrak{f}^{-1}(\h)\eqdef D_\h$, where $D_\h$ is the continuous extension of the expression in~\eqref{rti} to the full domain $\prob(n)\times\prob(n)$. Note that data-processing inequality and additivity\index{additivity} are preserved under continuous extensions and thus the resulting quantity $D_\h$ is indeed a relative entropy, concluding the proof.
\end{proof}

\begin{exercise}\label{ex:8215}
Show that $D_\h$ as defined in~\eqref{rti} is a relative entropy on the restricted domain
\be\bigcup_{n\in\mbb{N}}\prob(n)\times (\prob_{>0}(n) \cap \mbb{Q}^n)\;.\ee
Explicitly, show that:
\begin{enumerate}
\item Normalization: $D_\h(\e_1^{(2)}\|\u^{(2)})=1$.
\item DPI: for all $\p\in\prob(n)$, $\q\in\prob_{>0}(n)\cap\mbb{Q}^n$, and $E \in \stoc(m,n)\cap\mbb{Q}^{m\times n}_{>0}$,
\be
D_\h(E\p\|E\q)\leq D_\h(\p\|\q)\;.
\ee
Hint: look at the proof of Theorem~\ref{biji}.
\item Additivity: for all $\p_1,\p_2\in\prob(n)$ and $\q_1,\q_2\in\prob_{>0}(n)\cap\mbb{Q}^n$
\be
D_\h(\p_1\otimes\p_2\|\q_1\otimes\q_2)= D_\h(\p_1\|\q_1)+D_\h(\p_2\|\q_2)\;.
\ee
\end{enumerate}
\end{exercise}

In Exercise~\ref{ex:829} you showed that $D_{\rm path}(\p\|\q)\eqdef D_{\min}(\p\|\q)+D_{\min}(\q\|\p)$, provides a counterexample to lower semi-continuity. Note that $\mathfrak{f}\left(D_{\rm path}\right)=H_{\max}$, and $H_{\max}$ is in turn mapped to $D_{\min}(\p\|\q)$ by its inverse $\mathfrak{f}^{-1}$; i.e.\ $\mathfrak{f}^{-1}(H_{\max})=D_{\min}$ so that the contribution $D_{\min}(\q\|\p)$ that is discontinuous in $\q$ is lost in the process. This underscores why the continuity of relative entropies in the second argument is essential for the existence of the  bijection\index{bijection} $\mathfrak{f}$.
Finally, observe that the correspondence between relative entropies and entropies allows to import certain results from relative entropies to entropies.

\begin{myg}{}
\begin{corollary}\label{entmaxmin}
Let $\h$ be an entropy and $n\in\mbb{N}$. Then, $\h$ is continuous on $\prob_{>0}(n)$ and lower semi-continuous everywhere. Moreover, for all $\p\in\prob(n)$, we have
\be
H_{\min} (\p) \leq H(\p) \leq H_{\max} (\p),
\ee
where the min and max relative entropies have been defined in~\eqref{minen} and~\eqref{maxen}, respectively.
\end{corollary}
\end{myg}

\begin{exercise}
Prove Corollary~\ref{entmaxmin}.
\end{exercise}

We end this section by recalling Theorem~\ref{thm:MPS+2021} proved by~\cite{MPS+2021}. This theorem states that any entropy function can be expressed as a convex combination of R\'enyi entropies. Combining this with the one-to-one correspondence between entropies and relative entropies we get the following uniqueness result.
\begin{myg}{Uniqueness of R\'enyi Divergences}\index{uniqueness}\index{R\'enyi divergence}
\begin{corollary}
Let $\D$ be a relative entropy as defined in Definition~\ref{cre} and that is continuous in its second argument. Then, $\D$ can be expressed as a convex combination of the R\'enyi divergences.
\end{corollary}
\end{myg}

Observe the crucial need for continuity in the second argument. This is highlighted by the fact that $D_{\rm path}(\p|\q)\eqdef D_{\min}(\p|\q) + D_{\min}(\q|\p)$ lacks continuity in its second argument and is not a convex combination of R\'enyi divergences.

\section{Quantum Relative Entropies}\index{quantum relative entropy}

Quantum relative entropies are defined analogously to their classical counterpart. 
Replacing probability vectors with quantum states we will define a quantum divergence as a function 
\be\label{asgivenin}
\D:\bigcup_{A}\Big\{\md(A)\times\md(A)\Big\}\to\mbb{R}\cup\{\infty\}\;.
\ee 
that is acting on pairs of quantum states in all finite dimensions $|A|<\infty$. 

\begin{myd}{Quantum Relative Entropy}
\begin{definition}\label{qcre}
The function $\D$ as given in~\eqref{asgivenin} is called a \emph{quantum relative entropy} if it satisfies the following three conditions:
\begin{enumerate}
\item DPI: For all $\rho,\sigma\in\md(A)$ and all $\mE\in\cptp(A\to B)$
$
\D\big(\mE(\rho)\|\mE(\sigma)\big)\leq \D(\rho\|\sigma)
$.
\item Additivity: For any $\rho,\sigma\in\md(A)$ and $\rho',\sigma'\in\md(B)$
\be\label{additivityd}
\mbb{D}\big(\rho\otimes\rho'\|\sigma\otimes\sigma'\big)= \mbb{D}(\rho\|\sigma)+\mbb{D}(\rho'\|\sigma')\;.
\ee
\item Normalization: For $\D\left(|0\lr 0|\big\|\u^{(2)}\right)= 1$, where $|0\lr 0|,\u^{(2)}\in\md(\mbb{C}^2)$.
\end{enumerate}
\end{definition}
\end{myd}

\begin{remark}
Quantum relative entropies can be viewed as generalizations of classical relative entropies. Particularly, a quantum relative entropy reduces to a classical relative entropy when the domain is restricted to diagonal states in a fixed basis, and the diagonals identified with  probability vectors.
\end{remark}

From the remark above it follows that some of the properties of a quantum relative entropies follow trivially from their classical counterpart.

\begin{exercise}
Let $\D$ be a quantum relative entropy.
\begin{enumerate}
\item Show that $\D(\rho\|\rho)=0$. 
\item Show that if $|0\ra\in A$ is an eigenvector of $\rho$ then
\be
\D\left(|0\lr 0|\big\|\rho\right)=-\log\la 0|\rho|0\ra\;.
\ee 
Hint; use Theorem~\ref{822}.
\end{enumerate}
\end{exercise}

Theorem~\ref{822} has an interesting consequence in the quantum domain.
\begin{myt}{}
\begin{theorem}
Let $\D$ be a quantum relative entropy, $n\in\mbb{N}$, $\p\in\prob(n)$, $\rho\in\md(A)$, and $\sigma^{AX}=\sum_{x\in[n]}p_{x}\sigma^A_{x}\otimes |x\lr x|^X$ a cq-state.
Then, for any $x\in[n]$
\be
\D\left(\rho^A\otimes|x\lr x|^X\big\|\sigma^{AX}\right)=\D\left(\rho^A\big\|\sigma^A_x\right)-\log p_x\;.
\ee
\end{theorem}
\end{myt}
\begin{proof}
Fix $x\in[n]$ and let $\mE\in\cptp(AX\to AX)$ be a quantum channel that acts as the identity channel if the input on the classical system $X$ is $|x\lr x|^X$,  and otherwise acting as a replacement channel on system $A$ with output $\sigma^A_x$. Explicitly, for all $\tau\in\md(A)$ and $w\in[n]$ 
\be\label{def884}
\mE^{AX\to AX}\left(\tau^A\otimes |w\lr w|^X\right)\eqdef \begin{cases} \tau^A\otimes |x\lr x|^X &\text{if }w=x\\
\sigma^A_x\otimes |w\lr w|^X &\text{otherwise}
\end{cases}
\ee 
Then, denoting by $\p^X\eqdef\sum_{w\in[n]}p_{w}|w\lr w|^X$, we get from the DPI of $\D$
\ba\label{880}
\D\left(\rho^A\otimes|x\lr x|^X\big\|\sigma^{AX}\right)&\geq 
\D\left(\mE^{AX\to AX}\left(\rho^A\otimes|x\lr x|^X\right)\Big\|\mE^{AX\to AX}\left(\sigma^{AX}\right)\right)\\
{\color{red} \text{\eqref{def884}}\rightarrow}&=\D\left(\rho^A\otimes|x\lr x|^X\big\|\sigma^{A}_x\otimes\p^X\right)\\
{\color{red}\text{Additivity}\rightarrow}&=\D\left(\rho^A\big\|\sigma^{A}_x\right)+\D\left(|x\lr x|^X\big\|\p^X\right)\\
{\color{red}\text{Theorem~\ref{822}}\rightarrow}&=\D\left(\rho^A\big\|\sigma^{A}_x\right)-\log p_x\;.
\ea
Conversely, for all $w\in[n]$, define $\mF\in\cptp(AX\to AX)$ as
\be\label{def887}
\mF^{AX\to AX}\left(\tau^A\otimes |w\lr w|^X\right)\eqdef \begin{cases} \tau^A\otimes |x\lr x|^X &\text{if }w=x\\
\sigma^A_{w}\otimes |w\lr w|^X &\text{otherwise}
\end{cases}
\ee 
We then get
\ba
\D\left(\rho^A\otimes|x\lr x|^X\big\|\sigma^{A}_x\otimes\p^X\right)&\geq \D\left(\mF^{AX\to AX}\left(\rho^A\otimes|x\lr x|^X\right)\Big\|\mF^{AX\to AX}\left(\sigma^{A}_x\otimes\p^X\right)\right)\\
{\color{red} \text{\eqref{def887}}\rightarrow}&=\D\left(\rho^A\otimes|x\lr x|^X\big\|\sigma^{AX}\right)\;.
\ea
The combination of the above equation with~\eqref{880} concludes the proof.
\end{proof}

\begin{exercise}\label{extsig}
Let $\D$ be a quantum relative entropy, $\rho,\sigma\in\md(A)$, $\omega\in\md(B)$, and $t\in[0,1]$. In addition, let $Z$ be a $|A|\times |B|$ complex matrix such that $\begin{bmatrix}t\sigma & Z\\
Z^* & (1-t)\omega\end{bmatrix}$ is a density matrix in $\md(A\oplus B)$. Show that
\begin{equation}
\D\left(\begin{bmatrix}\rho & \0\\
\0 & \0\end{bmatrix}\Big\|\begin{bmatrix}t\sigma & Z\\
Z^* & (1-t)\omega\end{bmatrix}\right)\geq\D(\rho\|\sigma)-\log t\;,
\end{equation}
with equality if $Z=0$.
\end{exercise}

\bex\label{ex:vn}
Let $\D$ be a quantum relative entropy and $\u\in\md(A)$ be the maximally mixed state. 
\ben
\item Show that 
\be
\H(\rho^A)\eqdef\log|A|-\D(\rho^A\|\u^A)\quad\quad\forall\;\rho\in\md(A)\;,
\ee
is a quantum entropy.
\item Show that if $\D$ is jointly convex then $\H$, as defined above, is concave.
\een
\eex

Before we discuss additional properties of quantum relative entropies, we first consider an example of a family of relative entropies that generalizes the R\'enyi relative entropies.

\subsection{The Petz Quantum R\'enyi Divergence}

The first generalization of the R\'enyi divergences that we consider here is perhaps the most straightforward 
one, in the sense that the expression $\sum_{x}p_x^\alpha q_x^{1-\alpha}$ is simply replaced by $\tr\left[\rho^\alpha\sigma^{1-\alpha}\right]$ which looks most reminiscent to its classical counterpart. 

\begin{myd}{The Petz Quantum R\'enyi Divergence}\index{Petz}\index{R\'enyi divergence}
\begin{definition}\label{def:petz}
For any $\alpha\in[0,2]$ and $\rho,\sigma\in\md(A)$ the Petz quantum Renyi divergence is defined as 
\be\nonumber
D_\alpha(\rho\|\sigma)\eqdef\begin{cases}\frac{1}{\alpha-1}\log\tr\left[\rho^\alpha\sigma^{1-\alpha}\right] &
\text{if }\supp(\rho)\subseteq\supp(\sigma)\text{ or }\alpha<1\text{ and }\rho\sigma\neq 0\\
\infty &\text{otherwise}
\end{cases}
\ee
The cases $\alpha=0,1$ are defined by appropriate limits.
\end{definition}
\end{myd}

\begin{remark}
If $\supp(\rho)\subseteq\supp(\sigma)$ the trace in the definition above is strictly positive for all $\alpha\in[0,\infty]$.
Also, if $\alpha<1$ and $\rho\sigma\neq 0$ (i.e. $\rho$ and $\sigma$ are not orthogonal) then also $\rho^\alpha\sigma^{1-\alpha}\neq 0$ and we have in this case $\tr\left[\rho^\alpha\sigma^{1-\alpha}\right]>0$. In all other cases, the trace in the definition above is either zero or not well defined. One can also extend the definition to $\alpha>2$ however we will see below that the DPI only holds for $\alpha\in[0,2]$.
\end{remark}
\begin{exercise}
Prove all the statements in the remark above (except for the very last one about $\alpha>2$).
\end{exercise}

\bex\label{exarsa}
Let $\rho,\sigma\in\md(A)$ and consider their spectral decomposition as given in~\eqref{sdc}. Set $m\eqdef|A|$, and let $\p^{XY},\q^{XY}\in\prob(m^2)$ be the probability vectors whose components are $\{p_x|\la a_x|b_y\ra|^2\}_{x,y\in[m]}$ and $\{q_y|\la a_x|b_y\ra|^2\}_{x,y\in[m]}$, respectively (cf.~\eqref{corre}). Show that
\be\label{bdrsd}
D_{\alpha}(\rho\|\sigma)=D_\alpha\left(\p^{XY}\big\|\q^{XY}\right)\;.
\ee
where the right-hand side is the classical R\'enyi divergence\index{R\'enyi divergence} between $\p^{XY}$ and $\q^{XY}$.
\eex

The Petz-R\'enyi divergence satisfies all the properties of a relative entropy. The normalization and additivity properties you will prove in the exercise below, and we now prove the data processing inequality. 

\begin{myt}{}
\begin{theorem}
The Petz\index{Petz} quantum $\alpha$-R\'enyi divergence\index{R\'enyi divergence} is a relative entropy for any $\alpha\in[0,2]$.
\end{theorem}
\end{myt}

\begin{proof}
In~\eqref{682} we proved that the quantum $\alpha$-Divergence\index{$\alpha$-divergence},
\be
D_{f_\alpha}^q(\rho\|\sigma)=\frac1{\alpha(\alpha-1)}\left(\tr\left[\rho^{\alpha}\sigma^{1-\alpha}\right]-1\right)
\ee 
is a divergence for $\alpha\in[0,2]$. Therefore, for $\alpha\in[0,1)$ 
the expression $\tr\left[\rho^{\alpha}\sigma^{1-\alpha}\right]$ is monotonically increasing under mappings $(\rho,\sigma)\mapsto(\mE(\rho),\mE(\sigma))$ with $\mE\in\cptp(A\to B)$. Similarly, for $\alpha\in(1,2]$ the expression $\tr\left[\rho^{\alpha}\sigma^{1-\alpha}\right]$ is monotonically decreasing under such mappings. Therefore, for any $1\neq\alpha\in[0,2]$ the Petz\index{Petz} quantum R\'enyi $\alpha$-Divergence\index{$\alpha$-divergence} satisfies the DPI. The DPI for the case $\alpha=1$ has been proven in~\eqref{680}.
\end{proof}

\begin{exercise}
Show that for any $\alpha\geq 0$, the Petz quantum R\'enyi entropy $D_\alpha$ satisfies the normalization and additivity properties of a relative entropy.
\end{exercise}

The calculation of the limits $\alpha\to 0,1$ of the Petz quantum R\'enyi divergence\index{R\'enyi divergence} is a bit more subtle than the classical case. For this purpose, we will use the expression~\eqref{bdrsd} in Exercise~\ref{exarsa}.
For the limit $\alpha\to 0$ observe that
\begin{align}
\lim_{\alpha\to 0^+}D_{\alpha}(\rho\|\sigma)&=\lim_{\alpha\to 0^+}D_\alpha\left(\p^{XY}\big\|\q^{XY}\right)\\
&=-\log\sum_{\substack{x,y\in[m]\\p_x|\la u_x|v_y\ra|^2> 0}}q_y|\la u_x|v_y\ra|^2\\
&=-\log\sum_{\substack{x,y\in[m]\\p_x> 0}}q_y|\la u_x|v_y\ra|^2\\
&=-\log\tr[\Pi_\rho\sigma]\;,
\end{align}
where $\Pi_\rho$ is the projection to the support of $\rho$. The quantity above is also known as the min quantum relative entropy and is denoted by
\begin{myd}{The Min Quantum Relative Entropy}\index{min relative entropy}
\be
D_{\min}(\rho\|\sigma)\eqdef\begin{cases} -\log\tr[\Pi_\rho\sigma] &\text{if }\rho\sigma\neq 0\\
\infty &\text{otherwise.}
\end{cases}
\ee
\end{myd}
Note that the quantum min relative entropy reduces to the classical min relative entropy when the states are classical (i.e. diagonal).

For the limit $\alpha\to 1$ we use again~\eqref{bdrsd} to get
\ba
\lim_{\alpha\to 1}D_{\alpha}(\rho\|\sigma)
=\lim_{\alpha\to 1}D_\alpha\left(\p^{XY}\big\|\q^{XY}\right)
&=D\left(\p^{XY}\big\|\q^{XY}\right)\\
&=\sum_{x,y}p_x|\la u_x|v_y\ra|^2\log\left(\frac{p_x}{q_y}\right)\\
&=\tr[\rho\log\rho]-\tr[\rho\log\sigma]
=D(\rho\|\sigma)\;,
\ea
where $D(\rho\|\sigma)$ is  the Umegaki relative entropy. 

\bex
Prove the last two lines in the equation above; particularly, show that $\sum_{x,y}p_x|\la u_x|v_y\ra|^2\log\left(\frac{p_x}{q_y}\right)
=\tr[\rho\log\rho]-\tr[\rho\log\sigma]$.
\eex

\bex[Quasi-Convexity]
Show that for any $\alpha\in[0,2]$, $\rho,\omega_0,\omega_1\in\md(A)$, and $t\in[0,1]$ we have
\be\label{quascon}
D_\alpha\left(\rho\big\|t\omega_0+(1-t)\omega_1\right)\leq\max\Big\{D_\alpha(\rho\|\omega_0),D_\alpha(\rho\|\omega_1)\Big\}\;.
\ee
\eex

Similar to the definition of the min quantum relative entropy, we can extend the max relative entropy to the quantum domain.

\begin{myd}{The Max Quantum Relative Entropy}\index{max relative entropy}
The max quantum relative entropy is defined for all $\rho,\sigma\in\md(A)$ as 
\be
D_{\max}(\rho\|\sigma)\eqdef\log\min\big\{t\in\mbb{R}\;:\;t\sigma\geq\rho\big\}
\ee
for the case that $\supp(\rho)\subseteq\supp(\sigma)$ and otherwise it is set to $\infty$.
\end{myd}

\begin{exercise}Show that:
\begin{enumerate}
\item The max quantum relative entropy is indeed a relative entropy. 
\item $D_{max}(\rho\|\sigma)$ reduces to the classical max relative entropy when $\rho$ and $\sigma$ commutes.
\item For the case that $\supp(\rho)\subseteq\supp(\sigma)$
$
D_{\max}(\rho\|\sigma)=\log\|\sigma^{-\frac12}\rho\sigma^{-\frac12}\|_{\infty}
$. Hint; conjugate both sides of $t\sigma\geq\rho$ by $\sigma^{-\frac12}(\cdot)\sigma^{-\frac12}$.
\item $D_{max}(\rho\|\sigma)=\lim_{\alpha\to\infty}D_{\alpha}(\rho\|\sigma)$ if $\rho$ and $\sigma$ commutes, and give an example for which $D_{max}(\rho\|\sigma)\neq\lim_{\alpha\to\infty}D_{\alpha}(\rho\|\sigma)$. Here $D_\alpha$ refers to the same formula as the Petz quantum R\'enyi divergence\index{R\'enyi divergence} but with $\alpha>2$. 
\end{enumerate}
\end{exercise}

\subsection{Basic Properties}

In this subsection we will see that several of the properties of classical relative entropies carry over to the quantum domain. However, the proofs of these properties have to be adjusted to incorporate the larger domain.
\begin{myt}{}
\begin{theorem}
Let $\D$ be a relative entropy. Then for any quantum system $A$ and any $\rho,\sigma,\omega\in\md(A)$:
\begin{enumerate}
\item Bounds: 
\be\label{b886}
D_{\min}(\rho\|\sigma)\leq\D(\rho\|\sigma)\leq D_{\max}(\rho\|\sigma)\;.
\ee
\item Triangle Inequality:\index{triangle inequality}
\be\label{tri0}
\D(\rho\|\sigma)\leq\D(\rho\|\omega)+D_{\max}(\omega\|\sigma)\;.
\ee
\end{enumerate}
\end{theorem}
\end{myt}
\begin{proof}
Let  $\Pi_{\rho}$ denotes the projector to the support of $\rho$. Define the POVM Channel\index{POVM channel}  $\mE\in\cptp(A\to X)$ with $|X|=2$ as
\be\label{836}
\mE(\sigma)\eqdef\tr\big[\sigma\Pi_\rho\big]|0\lr 0|^X+\tr\big[\sigma\left(I-\Pi_\rho\right)\big]|1\lr 1|^X\;.
\ee
Then,
\ba\label{substi}
\D(\rho\|\sigma)&\geq \D\big(\mE(\rho)\|\mE(\sigma)\big)\\
{\color{red} \text{\eqref{836}}\rightarrow}&=\D\left(|0\lr 0|\Big\|\tr\big[\sigma\Pi_\rho\big]|0\lr 0|+\tr\big[\sigma\left(I-\Pi_\rho\right)\big]|1\lr 1|\right)\\
{\color{red} \text{Theorem~\ref{822}}\rightarrow}&=-\log\tr\big[\sigma\Pi_\rho\big]\\
&=D_{\min}(\rho\|\sigma)\;.
\ea
For the second inequality, denote by $t=2^{D_{\max}(\rho\|\sigma)}$, and note that in particular, $t\sigma\geq\rho$ (i.e. $t\sigma-\rho\geq 0$). Define a channel $\mE\in\cptp(X\to A)$ with $|X|=2$ by 
\be\label{898}
\mE(|0\lr 0|)\eqdef\rho\quad\text{and}\quad\mE(|1\lr 1|)\eqdef\frac{t\sigma-\rho}{t-1}\;.
\ee
Furthermore, denote 
\be
\q^X\eqdef\frac{1}{t}|0\lr 0|^X+\frac{t-1}{t}|1\lr 1|^X\;,
\ee
 and observe that $\mE(\q^X)=\sigma$.
Hence,
\ba
\D(\rho\|\sigma)&=\D\Big(\mE(|0\lr 0|^X)\big\|\mE(\q^X)\Big)\\
{\color{red} \text{DPI}\rightarrow}&\leq \D\left(|0\lr 0|^X\big\|\q^X \right)\\
{\color{red} \text{Theorem~\ref{822}}\rightarrow}&=-\log \frac1t=D_{\max}(\rho\|\sigma)\;.
\ea
This completes the proof of~\eqref{b886}.

To prove the triangle inequality~\eqref{tri0}, note first that for $|A|=1$ the statement is trivial so we can assume $|A| \geq 2$. Let $\eps\eqdef 2^{-D_{\max}
	(\omega\|\sigma)}\in(0,1)$, and observe that $\sigma \geq \eps \omega$ so that the matrix $\tau \eqdef  (\sigma - \eps\omega)/(1-\eps)$ is a density matrix satisfying
	\be
		\sigma = \eps \omega + (1-\eps) \tau\;.
	\ee
	From  the definition of $\eps$ we have
	\ba
		\D(\rho\|\omega) +D_{\max}(\omega\|\sigma)&=\D(\rho\|\omega) - \log \eps\\
		\GG{Theorem~\ref{822}}&=\D(\rho\|\omega) + \D \big( |0\lr 0| \big\| \eps|0\lr 0| + (1-\eps)|1\lr 1|  \big)\\
{\color{red} \text{Additivity}\rightarrow}&= \D \Big( \rho \otimes |0\lr 0| \Big\| \omega \otimes \big(\eps|0\lr 0|+ (1-\eps)|1\lr 1|\big)\Big)\\
{\color{red} \text{DPI}\rightarrow}		& \geq \D(\rho\|\sigma)\;,
	\ea
	where in the last inequality we used the DPI property of $\D$ with a quantum channel that acts as an identity upon measuring $|0\lr 0|$ in the second register, and produces a constant output $\tau$ upon measuring $|1\lr 1|$ in the second register.
\end{proof}

\begin{exercise}
The quantum Thompson's metric\index{metric} is defined for any $\rho,\sigma\in\prob(n)$ by
\be
D_T(\rho\|\sigma)\eqdef\max\Big\{D_{\max}(\rho\|\sigma),D_{\max}(\sigma\|\rho)\Big\}\;.
\ee
\begin{enumerate}
\item Prove that the quantum Thompson's metric is both a quantum divergence and a metric in $\md(A)\times\md(A)$. 
\item Prove that any quantum relative entropy $\D$ satisfies for all $\rho,\sigma,\sigma'\in\md(A)$
\be\label{tom}
\big|\D(\rho\|\sigma)-\D(\rho\|\sigma')\big|\leq D_{T}(\sigma\|\sigma')\;.
\ee
\end{enumerate}
\end{exercise}

The exercise above demonstrates that quantum relative entropies are continuous in their second argument. One can also get a continuity property in the first argument.
 
 \begin{myg}{}\index{continuity}
 \begin{lemma}\label{drsr}
	Let $\rho, \rho', \sigma \in \md(A)$ be quantum states. Then, we have
	\begin{align}\label{al8112}
		\D(\rho\|\sigma) - \D(\rho'\|\sigma) &\leq \min_{0\leq s\leq 2^{-D_{\max}(\rho'\|\rho)}} D_{\max}\left(\rho+s(\sigma-\rho')\big\|\sigma\right)\\
		&\leq \log \left( 1 + \frac{  \| \rho - \rho' \|_{\infty} } {\lambda_{\min}(\rho') \lambda_{\min}(\sigma) }  \right)
		\end{align}
		where the second inequality holds if $\sigma>0$ and $\lambda_{\min}(\rho')>\|\rho-\rho'\|_\infty$.
		\end{lemma}
		\end{myg}

\begin{proof}
	In somewhat of a variation of the previous theorem, 
	fix $0\leq s\leq 2^{-D_{\max}(\rho'\|\rho)}$ and denote by $\eps\eqdef 2^{-D_{\max}\left(\rho+s(\sigma-\rho')\|\sigma\right)}$. Then, 
	\ba\label{aligh}
		\D(\rho'\|\sigma)+D_{\max}\left(\rho+s(\sigma-\rho')\big\|\sigma\right)&=\D(\rho'\|\sigma) - \log \eps\\
		\GG{Theorem~\ref{822}}& =\D(\rho'\|\sigma)+\D \Big(|0\lr 0| \Big\| \eps|0\lr 0|+ (1-\eps)|1\lr 1| \Big)\\
		\GG{Additivity}&= \D \Big( \rho' \otimes|0\lr 0| \Big\| \sigma \otimes \big(\eps|0\lr 0|+ (1-\eps)|1\lr 1|\big) \Big)\\	
\GG{DPI}& \geq \D\left(\mN(\rho')\big\|\eps\mN(\sigma)+(1-\eps)\omega\right)
	\ea
	where in the last inequality we used the DPI with a channel that acts as some channel $\mathcal{N}\in\cptp(A\to A)$ when measuring $|0\lr 0|$ in the second register and outputs some state $\omega\in\md(A)$ when measuring $|1\lr 1|$. In other words, the inequality above holds for all $\mN\in\cptp(A\to A)$ and all $\omega\in\md(A)$. It is therefore left to show that there exists such $\mN$ and $\omega$ that satisfy $\mN(\rho')=\rho$ and $\eps\mN(\sigma)+(1-\eps)\omega=\sigma$. The latter implies that we can define $\omega$ to be
	\be
	\omega\eqdef\frac{\sigma-\eps\mN(\sigma)}{1-\eps}\;.
	\ee
	Note that we need to choose $\mN$ such that $\sigma-\eps\mN(\sigma)\geq 0$ so that $\omega\in\md(A)$.
	We take $\mN\in\cptp(A\to A)$ to be a measurement-prepare channel\index{measurement-prepare channel} of the form
	\begin{align}
		\mathcal{N}(\eta)\eqdef s \eta + (1-s) \tau\quad\quad\forall\eta\in\ml(A)\;,	
		\end{align}
where we want to choose $\tau$ such that both $\mN(\rho')=\rho$ and $\sigma-\eps\mN(\sigma)\geq 0$.
The condition $\mN(\rho')=\rho$ can be expressed as $\rho=s \rho' + (1-s) \tau$. Isolating $\tau$ we get that
\be\label{8137}
\tau=\frac{\rho-s\rho'}{1-s}\;.
\ee
The above matrix is positive semidefinite if and only if $\rho\geq s\rho'$ which hold since $s\leq 2^{-D_{\max}(\rho'\|\rho)}$. We therefore choose $\tau$ as above so that $\mN(\rho')=\rho$. It is left to check that $\sigma-\eps\mN(\sigma)\geq 0$. Indeed, since $\mathcal{N}(\sigma)\eqdef s \sigma + (1-s) \tau$ we have
\ba
\sigma-\eps\mN(\sigma)&=(1-\eps s)\sigma-\eps(1-s) \tau\\
\GG{\eqref{8137}}&=(1-\eps s)\sigma-\eps(\rho-s\rho')\\
&=\sigma-\eps\big(\rho+s(\sigma-\rho')\big)\\
\GG{\text{By definition of }\eps}&\geq 0\;.
\ea
To summarize, we showed that for any $0\leq s\leq 2^{-D_{\max}(\rho'\|\rho)}$ we have
\be
\D(\rho'\|\sigma)+D_{\max}\left(\rho+s(\sigma-\rho')\big\|\sigma\right)\geq D(\rho\|\sigma)\;.
\ee
Since the above equation holds for all $s\leq2^{-D_{\max}
	(\rho'\|\rho)}$ we conclude that the inequality~\eqref{al8112} holds. 
	
	To prove the second inequality, observe first that the inequality~\eqref{al8112}
	can be expressed as
	\be
	\D(\rho\|\sigma)-\D(\rho'\|\sigma)\leq\log\min\Big\{r\geq 0\;:\;(r-s)\sigma\geq\rho-s\rho'\geq 0\;,\;s\geq 0\Big\}	
	\ee
	Since we assume now that $\mu\eqdef\lambda_{\min}(\sigma)>0$ we can take $r=1+\frac{1-s}{\mu}$. Note that for this choice of $r$ we have
	\be
	(r-s)\sigma=(1-s)(1+\mu)\frac{\sigma}{\mu}\geq (1-s)(1+\mu)I^A\geq \rho-s\rho'
	\ee
	since $\rho-s\rho'$ is a subnormalized state with trace $1-s$. Moreover, if $\lambda_{\min}(\rho')\geq\|\rho-\rho'\|_{\infty}$ then we can take $s=1-\frac{\|\rho-\rho'\|_{\infty}}{\lambda_{\min}(\rho')}$ since in this case $s\leq 2^{-D_{\max}(\rho'\|\rho)}$ (or equivalently $\rho\geq s\rho'$, see Exercise~\ref{confex}). 	We therefore get for these choices of $r$ and $s$
	\begin{align}
	\D(\rho\|\sigma)-\D(\rho'\|\sigma)\leq \log r=\log \left( 1 + \frac{  \| \rho - \rho' \|_{\infty} } {\lambda_{\min}(\rho') \lambda_{\min}(\sigma) }  \right)	\;.\end{align}
	This completes the proof.
	\end{proof}
	
	\bex\label{confex}
	Show that if  $\lambda_{\min}(\rho')\geq\|\rho-\rho'\|_{\infty}>0$ then $\rho\geq s\rho'$ where $s=1-\frac{\|\rho-\rho'\|_{\infty}}{\lambda_{\min}(\rho')}$.	
	\eex
	
	\begin{exercise}
	Show that if $\rho,\sigma\in\md(A)$ and $\lambda_{\min}(\rho)> \| \sigma - \rho \|_{\infty}$ then
	\begin{equation}
	D_{\max}(\rho\|\sigma)\leq - \log \left( 1 - \frac{ \| \sigma - \rho \|_{\infty} }{ \lambda_{\min}(\rho) } \right)
	\end{equation}
Use this to get a bound on $D_{T}(\sigma\|\sigma')$ in~\eqref{tom}.
	\end{exercise}
	
	\begin{myt}{\color{yellow} Continuity of Quantum Relative Entropies}\index{continuity}\index{quantum relative entropy}
\begin{theorem}\label{continuityrel}
Let $\D$ be a quantum relative entropy. Then, $\D$ is upper semi-continuous at any point in $\md(A)\times\md_{>0}(A)$, and is continuous at any point in $\md_{>0}(A)\times\md_{>0}(A)$.
\end{theorem}
\end{myt}

\begin{proof}
Let ${(\rho_k,\sigma_k)}_{k\in\mbb{N}}$ be a sequence in $\md(A)\times\md(A)$ that converges to $(\rho,\sigma)$.
For any $k\in\mbb{N}$, define a quantum channel $\mE_k\in\cptp(A\to A)$ by its action on any $\omega\in\md(A)$ as
\be\label{8462}
\mE_k(\omega)\eqdef\rho_k+2^{-D_{\max}(\rho\|\rho_k)}(\omega-\rho)\;.
\ee
Note that for sufficiently large $k$, $2^{-D_{\max}(\rho\|\rho_k)}>0$ (see the exercise below). Moreover, observe that
$\rho_k-2^{-D_{\max}(\rho\|\rho_k)}\rho\geq 0$ so that $\mE_k$ is indeed a quantum channel.
With the above notations we get from the DPI
\ba
\D(\rho\|\sigma)&\geq\D\big(\mE_k(\rho)\big\| \mE_k(\sigma)\big)\\
{\color{red}\text{\eqref{8462}}\rightarrow}&=\D\big(\rho_k\big\| \mE_k(\sigma)\big)\\
{\color{red}\text{\eqref{tri0}}\rightarrow}&\geq \D(\rho_k\| \sigma_k)-D_{\max}\big(\mE_k(\sigma)\big\|\sigma_k\big)
\ea
Moving the term involving $D_{\max}$ to the other side and taking the supremum limit on both sides gives
\be
\limsup_{k\to\infty}\D(\rho_k\| \sigma_k)\leq\D(\rho\|\sigma)+\limsup_{k\to\infty}D_{\max}\big(\mE_k(\sigma)\big\|\sigma_k\big)\;.
\ee
The second term on the right-hand side above vanishes since the density matrix 
\be
\tsigma_k\eqdef \mE_k(\sigma)=2^{-D_{\max}(\rho\|\rho_k)}\sigma+\left(\rho_k-2^{-D_{\max}(\rho\|\rho_k)}\rho\right)
\ee
has a limit $\lim_{k\to\infty}\tsigma_k=\sigma$ so that  
\be\label{limit562}
\limsup_{k\to\infty}D_{\max}(\tsigma_k\|\sigma_k)=0\;.
\ee
Note that we used indirectly the fact that $\sigma>0$, since for sufficiently large $k$ we must have $\sigma_k>0$ so the limit above is indeed zero. This completes the proof that $\D$ is upper semi-continuous on $\md(A)\times\md_{>0}(A)$. 

We now prove the lower semi-continuity on $\md_{>0}(A)\times\md_{>0}(A)$. Note that since we already proved upper semi continuity in this domain, this will imply that $\D$ is continuous on  $\md_{>0}(A)\times\md_{>0}(A)$.
For any $k\in\mbb{N}$, we define $\mE_k$ as before but with the role of $\rho_k$ and $\rho$ interchanged; i.e. $\mE_k\in\cptp(A\to A)$ is defined by its action on any $\omega\in\md(A)$ as
\be\label{84602}
\mE_k(\omega)\eqdef\rho+2^{-D_{\max}(\rho_k\|\rho)}(\omega-\rho_k)\;.
\ee
Since we assume that $\rho>0$ we get that $2^{-D_{\max}(\rho_k\|\rho)}>0$ for all $k$. Moreover, observe that
$\rho-2^{-D_{\max}(\rho_k\|\rho)}\rho_k\geq 0$ so that $\mE_k$ is indeed a quantum channel.
With the above notations we get from the DPI
\ba
\D(\rho_k\|\sigma_k)&\geq\D\big(\mE_k(\rho_k)\big\| \mE_k(\sigma_k)\big)\\
{\color{red}\text{\eqref{84602}}\rightarrow}&=\D\big(\rho\big\| \mE_k(\sigma_k)\big)\\
{\color{red}\text{\eqref{tri0}}\rightarrow}&\geq \D(\rho\| \sigma)-D_{\max}\big(\mE_k(\sigma_k)\big\|\sigma\big)
\ea
Taking the infimum limit on both sides gives
\be
\liminf_{k\to\infty}\D(\rho_k\| \sigma_k)\geq\D(\rho\|\sigma)\;,
\ee
where we used the fact that
\be
\mE_k(\sigma_k)=2^{-D_{\max}(\rho_k\|\rho)}\sigma_k+\left(\rho-2^{-D_{\max}(\rho_k\|\rho)}\rho_k\right)
\ee
has a limit $\lim_{k\to\infty}\mE_k(\sigma_k)=\sigma$ so that  
\be\label{limit5602}
\limsup_{k\to\infty}D_{\max}(\mE_k(\sigma_k)\|\sigma)=0\;.
\ee
This completes the proof. 
\end{proof}

\begin{exercise}$\;$
\begin{enumerate}
\item Show that if $\{\rho_k\}_{k\in\mbb{N}}$ is a sequences in $\md(A)$ that converges to $\rho\in\md(A)$ then for sufficiently large $k$ we have $D_{\max}(\rho\|\rho_k)<\infty$. Hint: Show that for sufficiently large $k$, $\supp(\rho)\subseteq\supp(\rho_k)$.
\item Prove the limits in~\eqref{limit562} and~\eqref{limit5602}.
\end{enumerate}
\end{exercise}

\section{Optimal Quantum Extensions of Relative Entropies}\label{optim2}\index{optimal extension}\index{relative entropy}

The minimal and maximal quantum extensions, $\uD$ and $\oD$, of a classical divergence\index{classical divergence} $\D$ are the smallest and largest quantum divergences that reduce to $\D$ on classical states. We encountered them in Sec.~\ref{optim} particularly through Eqs.~(\ref{ud},\ref{od}). However, the expressions given in~(\ref{ud},\ref{od}) for $\uD$ and $\oD$ are, in general, not additive under tensor products even if the classical divergence\index{classical divergence} $\D$ is additive. Therefore, in order to get the optimal extensions of relative entropies, we will use regularization\index{regularization} to make the quantum extensions at least partially additive.

Suppose $\D$ is a classical relative entropy and define $\uD$ and $\oD$ as in~\eqref{ud}; i.e.
\begin{align}
&\uD(\rho\|\sigma) \eqdef \sup \D\big(\mE(\rho)\big\|\mE(\sigma)\big)\,,\\ 
&\oD(\rho\|\sigma)\eqdef\inf \big\{\D(\p\|\q)\;:\;\rho=\mF(\p),\;\sigma=\mF(\q) \big\}\,, 
\end{align}
where the optimizations are over the classical system $X$, the channels $\mE \in \cptp(A \to X)$ and $\mF \in \cptp(X \to A)$ as well as the diagonal density matrices $\p,\q\in\md(X)$. The functions $\uD$ and $\oD$ are in general not additive even if the $\D$ is a classical relative entropy (and therefore additive). However, in the following lemma we show that in this case $\uD$ is super-additive while $\oD$ is sub-additive.
\begin{myg}{}
\begin{lemma}
Let $\D$ be a classical relative entropy, and let $\oD$ and $\uD$ be its maximal and minimal quantum extensions as defined in~\eqref{ud}. Then, for all $\rho_1,\sigma_1\in\md(A_1)$ and $\rho_2,\sigma_2\in\md(A_2)$ we have:
\ben
\item Super-Aditivity: $\uD\left(\rho_1\otimes\rho_2\big\|\sigma_1\otimes\sigma_2\right)\geq\uD(\rho_1\|\sigma_1)+\uD(\rho_2\|\sigma_2)$.
\item Sub-Additivity: $\oD\left(\rho_1\otimes\rho_2\big\|\sigma_1\otimes\sigma_2\right)\leq\oD(\rho_1\|\sigma_1)+\oD(\rho_2\|\sigma_2)$.
\een
\end{lemma}
\end{myg}

\begin{proof}
We will prove the sub-additivity property and leave it as an exercise to prove the super-additivity using similar lines. By definition we have
\ba
\uD\left(\rho_1\otimes\rho_2\big\|\sigma_1\otimes\sigma_2\right)&=\sup_{\mE\in\cptp(A_1A_2\to X)}\D\left(\mE(\rho_1\otimes\rho_2)\big\|\mE(\sigma_1\otimes\sigma_2)\right)\\
\Gg{\text{Restricting }\mE=\mE_1\otimes\mE_2\to} &\geq\sup_{\substack{\mE_1\in\cptp(A_1\to X_1)\\ \mE_2\in\cptp(A_2\to X_2)}}\D\left(\mE_1(\rho_1)\otimes\mE_2(\rho_2)\big\|\mE_1(\sigma_1)\otimes\mE_2(\sigma_2)\right)\\
\GG{Additivity\;of\;\D} &=\sup_{\mE_1}\D\left(\mE_1(\rho_1)\big\|\mE_1(\sigma_1)\right)+\sup_{\mE_2}\D\left(\mE_2(\rho_2)\big\|\mE_2(\sigma_2)\right)\\
&=\uD(\rho_1\|\sigma_1)+\uD(\rho_2\|\sigma_2)\;.
\ea
This completes the proof.
\end{proof}

\begin{exercise}
Prove the sub-additivity of $\oD$.
\end{exercise}

Since $\uD$ and $\oD$ are not necessarily additive, we define their regularization as\index{regularization}
\be
\uD^\reg(\rho\|\sigma)\eqdef\lim_{n\to\infty}\frac1n\uD\left(\rho^{\otimes n}\big\|\sigma^{\otimes n}\right)\quad\text{and}\quad\oD^\reg(\rho\|\sigma)\eqdef\lim_{n\to\infty}\frac1n\oD\left(\rho^{\otimes n}\big\|\sigma^{\otimes n}\right)\;.
\ee
In the Exercise~\ref{ex842} below you will show that the limits above exist and that in general $\uD^\reg(\rho\|\sigma)\geq \uD(\rho\|\sigma)$ and $\oD^\reg(\rho\|\sigma)\leq \oD(\rho\|\sigma)$. Moreover, note that by definition, $\uD^\reg$ and $\oD^\reg$ are at least partially additive in the sense that for any $n\in\mbb{N}$ and any $\rho,\sigma\in\md(A)$
\be
\uD^\reg\left(\rho^{\otimes n}\big\|\sigma^{\otimes n}\right)=n\uD^\reg\left(\rho\big\|\sigma\right)\quad\text{and}\quad
\oD^\reg\left(\rho^{\otimes n}\big\|\sigma^{\otimes n}\right)=n\oD^\reg\left(\rho\big\|\sigma\right)\;.
\ee
It is an open problem to determine if $\uD^\reg$ and $\oD^\reg$ are fully additive. 
We will see below that in many examples, $\uD^\reg$ and $\oD^\reg$ turns out to be fully additive so that they are in fact relative entropies. The following theorem shows that these functions remains optimal.

\begin{myt}{}
\begin{theorem}\label{optim3}
Let $\D$ be a classical relative entropy, and let $\uD^\reg$ and $\oD^\reg$ be as above.
Then, both $\uD^\reg$ and $\oD^\reg$ are partially additive quantum divergences that reduces to $\D$ on classical states. In addition, any other quantum relative entropy $D'$ that reduces to $\D$ on classical states, satisfies for all $\rho,\sigma\in\md(A)$
\be\label{regineq}
\uD^\reg(\rho\|\sigma)\leq D'(\rho\|\sigma)\leq\oD^\reg(\rho\|\sigma)\;.
\ee
\end{theorem}
\end{myt}
\begin{remark}
Observe that since in general $\uD^\reg(\rho\|\sigma)\geq \uD(\rho\|\sigma)$ and $\oD^\reg(\rho\|\sigma)\leq \oD(\rho\|\sigma)$, the bounds on $D'$ above are tighter than the bounds given in~\eqref{bboun}. We are able to get tighter bounds since $\D$ is additive.
\end{remark}
\begin{proof}
We already saw that $\uD^\reg$ and $\oD^\reg$ are partially additive quantum divergences that reduces to $\D$ on classical states. It is therefore left to prove the inequality~\eqref{regineq}. From~\eqref{bboun} we have for all $n\in\mbb{N}$
\be
\uD\left(\rho^{\otimes n}\big\|\sigma^{\otimes n}\right)\leq D'\left(\rho^{\otimes n}\big\|\sigma^{\otimes n}\right)\leq\oD\left(\rho^{\otimes n}\big\|\sigma^{\otimes n}\right)
\ee
Since $D'$ is additive under tensor product we get after dividing the equation above by $n$
\be
\frac1n\uD\left(\rho^{\otimes n}\big\|\sigma^{\otimes n}\right)\leq D'\left(\rho\|\sigma\right)\leq\frac1n\oD\left(\rho^{\otimes n}\big\|\sigma^{\otimes n}\right)\;.
\ee
The proof is concluded by taking the limit $n\to\infty$ in the equation above.
\end{proof}

\begin{exercise}\label{ex842}
Let $\rho,\sigma\in\md(A)$ and let $\D$ be a classical relative entropy with maximal and minimal quantum extensions $\oD$ and $\uD$. Denote by 
\be
a_n\eqdef\oD\left(\rho^{\otimes n}\big\|\sigma^{\otimes n}\right)\quad\text{and}\quad
b_n\eqdef\uD\left(\rho^{\otimes n}\big\|\sigma^{\otimes n}\right)\;.
\ee 
\begin{enumerate}
\item Show that the sequences $\{a_n\}$ and $\{b_n\}$ satisfies for all $n,m\in\mbb{N}$ 
\be
a_{n+m}\leq a_n+a_m\quad\text{and}\quad b_{n+m}\geq b_n+b_m\;.
\ee
\item Use the inequalities above to show that
\ba
&\lim_{n\to\infty}\frac{a_n}n=\alpha\eqdef\inf\left\{\frac{a_n}n\;:\;n\in\mbb{N}\right\}\\
&\lim_{n\to\infty}\frac{b_n}n=\beta\eqdef\sup\left\{\frac{b_n}n\;:\;n\in\mbb{N}\right\}\;.
\ea
Hint: Let $\eps>0$, choose $k$ such that $\alpha+\eps>\frac{a_k}k$, and observe that for any integers $n,m\in\mbb{N}$ that satisfies $nk\leq m<(n+1)k$ we have $a_{m}\leq a_{nk}+a_{m-nk}\leq na_k+c$, where $c\eqdef\max\{a_j\}_{j\in[k]}$. Use this to bound $\limsup_{m\to\infty}\frac{a_m}m$.
\end{enumerate}
\end{exercise}

\subsection{The Minimal Quantum Extension}\index{optimal extension}

This section illuminates the remarkable aspect of R\'enyi relative entropies, specifically the existence of a closed formula for the minimal quantum extension of the R\'enyi divergence. It is noteworthy that if $D_\alpha$ represents the \emph{classical} R\'enyi relative entropy of order $\alpha$, then the quantum extension, denoted as $D^\reg_\alpha(\rho\|\sigma)$, can be expressed as follows:
\be\label{mixedopt}
D^\reg_\alpha(\rho\|\sigma)\eqdef\lim_{n\to\infty}\frac1n\sup_{\mE_n\in\cptp(A^n\to X)}D_\alpha\left(\mE_n(\rho^{\otimes n})\big\|\mE_n(\sigma^{\otimes n})\right)\;,
\ee
where the supremum encompasses all dimensions of the classical system $X$.

To derive a single-letter closed formula for the expression above, a two-step approach is required:
\ben
\item First, identify a function $\mE_n$ that approaches optimality as $n \to \infty$.
\item Then, use this selected $\mE_n$ to compute the limit as $n \to \infty$, which will lead to the desired closed formula.
\een
This approach enables the development of a precise and concise formula representing the minimal quantum extension for the R\'enyi divergence.

A natural guess for optimal POVM channels $\mE_n$ are the pinching channels discussed in Sec.~\ref{sec:pinching}. Recall that for any $\rho,\sigma\in\md(A)$, and a pinching channel $\mP_\sigma\in\cptp(A\to A)$, we have that $\mP_{\sigma}\left(\rho\right)$ and $\sigma$ commutes. Therefore, $\mP_{\sigma}\left(\rho\right)$ and $\sigma$ have a common eigenbasis $\{|x\ra\}_{x\in[m]}$ (with $m\eqdef|X|=|A|$) that spans $A$. Let $\Delta\in\cptp(A\to X)$ be the completely dephasing channel in this basis. Then, the channel $\Delta\in\cptp(A\to X)$ is a POVM Channel\index{POVM channel}  that we can take to be $\mE_1$.  From Exercise~\ref{srcommute} it follows that $\Delta(\sigma)=\sigma$ and $\Delta(\rho)=\mP_\sigma(\rho)$ (see~\eqref{deltars}).

In general, for any $n\in\mbb{N}$, we can choose $\mE_n=\Delta_n$,  where $\Delta_n\in\cptp(A^n\to X^n)$ is the completely dephasing channel in the common eigenbasis of $\mP_{\sigma^{\otimes n}}\left(\rho^{\otimes n}\right)$ and $\sigma^{\otimes n}$. We will see shortly that this choice is indeed optimal in the limit $n\to\infty$.

Before we continue with the derivation of the closed formula, we first give a snapshot of what one can expect the formula to be. With $\{|x\ra\}_{x\in[m]}$ being the common eigenbasis of $\mP_{\sigma}\left(\rho\right)$ and $\sigma$ we get (cf.~\eqref{deltars}) that 
\be
D_\alpha\left(\mP_{\sigma}\left(\rho\right)\big\|\sigma\right)=D_\alpha\left(\Delta\left(\rho\right)\big\|\sigma\right)=\frac1{\alpha-1}\log\sum_{x\in[m]}\la x|\rho| x\ra^\alpha\la x|\sigma| x\ra^{1-\alpha}\;.
\ee
where we used the fact that each $|x\ra$ is a common eigenvector of both $\sigma$ and $\mP_\sigma(\rho)$. In particular,
for any $\lambda\in\mbb{R}$ we have
$\la x|\sigma^{\lambda}| x\ra=\la x|\sigma| x\ra^{\lambda}$. 
Therefore, the term inside the sum above can be expressed as
\ba
\la x|\rho| x\ra^\alpha\la x|\sigma| x\ra^{1-\alpha}&=\left(\la x|\sigma^{\frac{1-\alpha}{2\alpha}}| x\ra\la x|\rho| x\ra\la x|\sigma^{\frac{1-\alpha}{2\alpha}}| x\ra\right)^\alpha\\
\Gg{| x\ra\la x|\sigma^{\frac{1-\alpha}{2\alpha}}| x\ra=\sigma^{\frac{1-\alpha}{2\alpha}}| x\ra}&=\left(\la x|\sigma^{\frac{1-\alpha}{2\alpha}}\rho\sigma^{\frac{1-\alpha}{2\alpha}}| x\ra\right)^\alpha\;.
\ea
We therefore conclude that
\be
D_\alpha\left(\mP_{\sigma}\left(\rho\right)\big\|\sigma\right)
=\frac1{\alpha-1}\log\sum_{x\in[m]}\left(\la x|\sigma^{\frac{1-\alpha}{2\alpha}}\rho\sigma^{\frac{1-\alpha}{2\alpha}}| x\ra\right)^\alpha\;.
\ee
Since the function $x\mapsto x^\alpha$ is concave for $\alpha\in(0,1)$ and convex for $\alpha\geq 1$ it follows from the Jensen’s inequality~\eqref{jensen2} that
\ba\label{insp}
D_\alpha\left(\mP_{\sigma}\left(\rho\right)\big\|\sigma\right)&\leq \frac1{\alpha-1}\log\sum_{x\in[m]}\la x|\left(\sigma^{\frac{1-\alpha}{2\alpha}}\rho\sigma^{\frac{1-\alpha}{2\alpha}}\right)^\alpha| x\ra\\
&=\frac1{\alpha-1}\log\tr\left(\sigma^{\frac{1-\alpha}{2\alpha}}\rho\sigma^{\frac{1-\alpha}{2\alpha}}\right)^\alpha\;.
\ea
The expression on the right-hand side is known as the sandwiched R\'enyi relative entropy\index{sandwiched relative entropy}. Remarkably, we will see below that the regularization\index{regularization} of the left-hand side equals the right-hand side in the equation above. For this purpose, it will be convenient to denote the trace in the equation above as
\begin{mye}{}
\be
\tQ_\alpha(\rho\|\sigma)\eqdef\tr\left(\sigma^{\frac{1-\alpha}{2\alpha}}\rho\sigma^{\frac{1-\alpha}{2\alpha}}\right)^\alpha\;.
\ee
\end{mye}

\begin{exercise}\label{1ex}
Show that for any isometry channel $\mV\in\cptp(A\to B)$, any $\rho,\sigma\in\md(A)$, and any $\omega\in\md(C)$,
\be
Q_\alpha\left(\mV(\rho)\big\|\mV(\sigma)\right)=Q_\alpha(\rho\|\sigma)\quad\text{and}\quad
Q_\alpha(\rho\otimes\omega\|\sigma\otimes\omega)=Q_\alpha(\rho\|\sigma)\;.
\ee
\end{exercise}

\begin{myd}{The Sandwiched R\'enyi Relative Entropy}\index{sandwiched relative entropy}
\begin{definition}\label{def:sandwich}
The \emph{sandwiched R\'enyi relative entropy} of order $\alpha\in[0,\infty]$, is defined on any quantum system $A$ and  $\rho,\sigma\in\md(A)$ as
\be\nonumber
\tD_{\alpha}(\rho\|\sigma)=
	\begin{cases}
	\frac1{\alpha-1}\log \tQ_\alpha(\rho\|\sigma)&\text{if }\big(\frac12\leq\alpha<1\text{ and }\rho\not\perp\sigma\big)\text{ or }\rho\ll\sigma\\
	\frac1{\alpha-1}\log \tQ_{1-\alpha}(\sigma\|\rho)&\text{if }0\leq \alpha<\frac12\text{ and }\rho\not\perp\sigma\\
	\infty&\text{otherwise}
	\end{cases}
\ee
The cases $\alpha=0,1,\infty$ are understood in terms of limits.
\end{definition}
\end{myd}

We first show that $\tD_{\alpha}$ is indeed a relative entropy. It's additivity\index{additivity} and normalization properties are relatively easy to show and are left as an exercise.
\begin{exercise}\label{ex:additive}
Show that the sandwiched R\'enyi relative entropy  of order $\alpha\in[0,\infty]$ satisfies the additivity and normalization properties of a quantum relative entropy.
\end{exercise}

\begin{exercise}
Show that for any $\rho,\sigma\in\md(A)$
\be
\tr\left(\sigma^{\frac{1-\alpha}{2\alpha}}\rho\sigma^{\frac{1-\alpha}{2\alpha}}\right)^\alpha=\tr\left(\rho^{\frac12}\sigma^{\frac{1-\alpha}{\alpha}}\rho^{\frac12}\right)^\alpha
\ee
Hint: Recall that for any complex matrix $M$, the matrices $MM^*$ and $M^*M$ have the same non-zero eigenvalues. 
\end{exercise}

\begin{myt}{}
\begin{theorem}
The sandwiched R\'enyi relative entropy of any order $\alpha\in[0,\infty]$ is a quantum relative entropy; i.e., it satisfies the three relative entropy axioms of DPI, additivity, and normalization.
\end{theorem}
\end{myt}

\begin{proof}
 Since $\tD_\alpha(\rho\|\sigma)$ fulfills both additivity and normalization properties (as shown in Exercise~\ref{ex:additive}), our task is to demonstrate its compliance with the DPI. For $\alpha>1$, the DPI of $\tD_\alpha$ is derived from that of $\tQ_\alpha$. For $\alpha\in[\frac12,1)$, it follows from the DPI of $-\tQ_\alpha$. Based on Exercise~\ref{1ex} and Lemma~\ref{lemaa}, we know that if $\tQ_\alpha$ is jointly convex for $\alpha>1$, then it satisfies the DPI. Similarly, for $\alpha\in[\frac12,1)$, if $\tQ_\alpha$ is jointly concave, then $-\tQ_\alpha$ satisfies the DPI. Our objective is therefore to show that for $\alpha>1$, $\tQ_\alpha$ is jointly convex, and for $\alpha\in[\frac12,1)$, $\tQ_\alpha$ is jointly concave.
The case $\alpha\in(0,\frac12]$ is effectively covered by the case $\alpha\in[\frac12,1)$ when we swap $\rho$ with $\sigma$, thus it need not be considered separately. 

Firstly, consider $\alpha>1$ and define $\beta\eqdef \frac{\alpha-1}{2\alpha}$. The proof's central strategy is to decompose the trace $\tr\left[(\sigma^{-\beta}\rho\sigma^{-\beta}\right)^\alpha]$ into two terms, one dependent only on $\rho$ and the other solely on $\sigma$. This decomposition allows us to separately assess the convexity in $\rho$ and $\sigma$. To obtain this decomposition, we utilize Young's inequality~\eqref{young}, choosing $M=\sigma^{-\beta}\rho\sigma^{-\beta}$, $N=\sigma^\beta \eta\sigma^\beta$, $p=\alpha$, and $q=\frac\alpha{\alpha-1}=\frac1{2\beta}$, where $\eta$ is an arbitrary positive semidefinite matrix in $\pos(A)$. With these choices, $\tr[MN]=\tr[\rho \eta]$, leading to the inequality (cf.~\eqref{young})
\be
\tr\left[\rho \eta\right]\leq \frac1\alpha\tr\left(\sigma^{-\beta}\rho\sigma^{-\beta}\right)^\alpha+\frac{\alpha-1}\alpha\tr\left(\sigma^\beta \eta\sigma^\beta\right)^{\frac1{2\beta}}
\ee
Rearranging terms and recalling $\tQ_\alpha(\rho\|\sigma)=\tr\left(\sigma^{-\beta}\rho\sigma^{-\beta}\right)^\alpha$, we obtain
\be\label{qal}
\tQ_\alpha(\rho\|\sigma)\geq \alpha\tr[\rho \eta]-(\alpha-1)\tr\left(\sigma^\beta \eta\sigma^\beta\right)^{\frac1{2\beta}}\;.
\ee
This inequality holds for all $\eta\in\pos(A)$, with equality if $M^p=N^q$, which translates to (Exercise~\eqref{verieta0})
\be\label{verieta}
\eta=\sigma^{-\beta}\left(\sigma^{-\beta}\rho\sigma^{-\beta}\right)^{\alpha-1}\sigma^{-\beta}\;.
\ee
Therefore, $\tQ_\alpha(\rho\|\sigma)$ can be expressed as
\be\label{qaldec}
\tQ_\alpha(\rho\|\sigma)=\sup_{\eta\geq 0}\Big\{ \alpha\tr[\rho \eta]-(\alpha-1)\tr\left(\sigma^\beta \eta\sigma^\beta\right)^{\frac1{2\beta}}\Big\}\;.
\ee
With this expression, we can now analyze the convexity of each term independently.

A consequence of Lieb's concavity theorem, given in Corollary~\ref{yl01}, establishes the concavity of the function
\be\label{liebcon}
\sigma\mapsto\tr\left(\sigma^\beta \eta\sigma^\beta\right)^{\frac1{2\beta}}=\tr\left(\eta^{\frac12}\sigma^{2\beta} \eta^{\frac12}\right)^{\frac1{2\beta}}\;,
\ee
where we used the fact that $LL^*$ and $L^*L$ have the same non-zero eigenvalues, where $L\eqdef\sigma^\beta \eta^{\frac12}$.
Therefore, the term $-(\alpha-1)\tr\left(\sigma^\beta \eta\sigma^\beta\right)^{\frac1{2\beta}}$ is convex in $\sigma$. Furthermore, the linearity of $\alpha\tr[\rho \eta]$ in $\rho$ ensures its convexity in $\rho$.
As a result, for any $\p\in\prob(n)$ and two sets of $n$ density matrices in $\md(A)$, $\{\rho_x\}_{x\in[n]}$ and $\{\sigma_x\}_{x\in[n]}$, it follows that
\ba
\tQ_\alpha\Big(\sum_{x\in[n]} p_x\rho_x\Big\|\sum_{x\in[n]} p_x\sigma_x\Big)&\leq \sup_{\eta\geq 0}\Big\{ \alpha\sum_{x\in[n]}p_x\tr[\rho_x \eta]-(\alpha-1)\sum_{x\in[n]}p_x\tr\left(\eta^{\frac12}\sigma^{2\beta}_x \eta^{\frac12}\right)^{\frac1{2\beta}}\Big\}\\
&\leq\sum_{x\in[n]}p_x\sup_{\eta\geq 0}\Big\{ \alpha\tr[\rho_x \eta]-(\alpha-1)\tr\left(\eta^{\frac12}\sigma^{2\beta}_x \eta^{\frac12}\right)^{\frac1{2\beta}}\Big\}\\
\GG{\eqref{qaldec}}&=\sum_{x\in[n]}p_x\tQ_\alpha( \rho_x\| \sigma_x)\;.
\ea
This proves the case for $\alpha>1$. For $\alpha\in\big[\frac12,1\big)$, we apply similar reasoning using the reverse Young's inequality~\eqref{young2}. Using the same substitutions for $M$ and $N$, we obtain~\eqref{qal} but with the inequality reversed. Consequently, we have
\be
\tQ_\alpha(\rho\|\sigma)=\inf_{\eta\geq 0}\Big\{ \alpha\tr[\rho \eta]+(1-\alpha)\tr\left(\sigma^\beta \eta\sigma^\beta\right)^{\frac1{2\beta}}\Big\}\;.
\ee
Observe that $\beta<0$ since $\alpha<1$.
As with the previous case, the joint concavity\index{joint concavity} of $\tQ_\alpha$ follows from the concavity of the function in~\eqref{liebcon}, completing the proof.
\end{proof}

\bex\label{verieta0}
Using the same notations as in the proof above, show that $M^p=N^q$ if and only if $\eta$ have the form given in~\eqref{verieta}.
\eex

\begin{exercise}
Show that for any $\rho,\sigma\in\md(A)$, the function $\alpha\mapsto\tD_{\alpha}(\rho\|\sigma)$ is continuous for all $\alpha\in[0,\infty]$. 
\end{exercise}

We are now ready to prove the closed formula for the minimal quantum R\'enyi relative entropy.

\begin{myt}{\color{yellow} Single Letter Formula}\index{sandwiched relative entropy}
\begin{theorem}\label{cfor}
For any $\alpha\in[0,\infty]$, the regularized minimal quantum extension of the R\'enyi relative entropy, $D_\alpha$, is given by
the sandwiched R\'enyi relative entropy  of order $\alpha$. That is, for all $\alpha\in[0,\infty]$, quantum system $A$, and  $\rho,\sigma\in\md(A)$, we have
\be\label{thmeq}
\underline{D}^\reg_{\alpha}(\rho\|\sigma)=\tD_{\alpha}(\rho\|\sigma)\;.
\ee
\end{theorem}
\end{myt}
\begin{remark}
Recall that a priori,  $\underline{D}^\reg_{\alpha}$ is only known to be partially additive, however, the theorem above  implies that it is fully additive.
\end{remark}

\begin{proof}
It is sufficient to prove the theorem for all $\alpha\geq \frac12$, since if the theorem holds for this case then the case $\alpha\in(0,\frac12)$ simply follows from Exercise~\ref{ex:flip} via the relation
\ba
\underline{D}^\reg_{\alpha}(\rho\|\sigma)&=\frac\alpha{1-\alpha}\underline{D}^\reg_{1-\alpha}(\rho\|\sigma)\\
{\color{red}\text{\eqref{thmeq}}\rightarrow}& =\frac\alpha{1-\alpha}\tD_{1-\alpha}(\rho\|\sigma)\\
&=\tD_{\alpha}(\rho\|\sigma)\;.
\ea 
We will therefore assume in the rest of the proof that $\alpha\geq\frac12$.
Since $\tD_\alpha$ is a relative entropy that reduces to the R\'enyi relative entropy in the classical domain, it follows from Theorem~\ref{optim3} that 
\be
\underline{D}^\reg_{\alpha}(\rho\|\sigma)\leq \tD_{\alpha}(\rho\|\sigma)\;.
\ee
For the reversed inequality, we first show that
\be\label{underregal2}
\underline{D}^\reg_{\alpha}(\rho\|\sigma)\geq\lim_{n\to\infty}\frac1nD_\alpha\left(\mP_{\sigma^{\otimes n}}\left(\rho^{\otimes n}\right)\big\|\sigma^{\otimes n}\right)\;.
\ee
Indeed, since $\mP_{\sigma^{\otimes n}}\left(\rho^{\otimes n}\right)$ commutes with $\sigma^{\otimes n}$ they have a common eigenbasis that spans $A^n$. Let $\Delta_n\in\cptp(A^n\to A^n)$ be the completely dephasing channel in this basis. From Exercise~\ref{srcommute} we have $\mP_{\sigma^{\otimes n}}\left(\rho^{\otimes n}\right)=\Delta_n\left(\rho^{\otimes n}\right)$. Therefore,
\ba
\sup_{\mE\in\cptp(A^n\to X)}D_{\alpha}\left(\mE\left(\rho^{\otimes n}\right)\big\|\mE\left(\sigma^{\otimes n}\right)\right)&\geq D_{\alpha}\left(\Delta_n\left(\rho^{\otimes n}\right)\big\|\Delta_n\left(\sigma^{\otimes n}\right)\right)
\\
&=D_{\alpha}\left(\mP_{\sigma^{\otimes n}}\left(\rho^{\otimes n}\right)\big\|\sigma^{\otimes n}\right)\;.
\ea
Dividing both sides of the equation above by $n$ and taking the limit $n\to\infty$ proves~\eqref{underregal2}. 

It is left to show that the right-hand side of~\eqref{underregal2} is no smaller than $\tD_\alpha(\rho\|\sigma)$. We will divide this part of the proof into several cases:
\begin{enumerate}
\item The case $\alpha>1$ and $\rho\not\ll\sigma$. Recall that for every $n\in\mbb{N}$ the states $\mP_{\sigma^{\otimes n}}\left(\rho^{\otimes n}\right)$ and $\sigma^{\otimes n}$ have a common eigenbasis. Therefore, both of these states are diagonal in this eigenbasis and the condition that $\rho\not\ll\sigma$ implies that these diagonal states also satisfy $\mP_{\sigma^{\otimes n}}\left(\rho^{\otimes n}\right)\not\ll \sigma^{\otimes n}$ (see Exercise~\ref{ex:pinch3}). Hence, we must have that $D_{\alpha}\left(\mP_{\sigma^{\otimes n}}\left(\rho^{\otimes n}\right)\big\|\sigma^{\otimes n}\right)=\infty$ for all $n\in\mbb{N}$. 
\item The case $\alpha>1$ and $\rho\ll\sigma$. Observe first that
since $\mP_\sigma(\rho)$ and $\sigma$ commutes, we have (cf. Exercise~\ref{ex:clqu})
\be\label{8149}
D_\alpha\left(\mP_{\sigma}\left(\rho\right)\big\|\sigma\right)=\frac1{\alpha-1}\log\tr\left[\left(\sigma^{\frac{1-\alpha}{2\alpha}}\mP_\sigma(\rho)\sigma^{\frac{1-\alpha}{2\alpha}}\right)^\alpha\right]\;.
\ee
Now, from the pinching inequality~\eqref{pinch} we have $\rho\leq|\spec(\sigma)|\mP_\sigma(\rho)$, so that (cf. Exercise~\ref{tracefunc})
\ba\label{8150}
D_\alpha\left(\mP_{\sigma}\left(\rho\right)\big\|\sigma\right)&\geq\frac1{\alpha-1}\log\tr\left[\left(\sigma^{\frac{1-\alpha}{2\alpha}}\frac\rho{|\spec(\rho)|}\sigma^{\frac{1-\alpha}{2\alpha}}\right)^\alpha\right]\\
&=\frac1{\alpha-1}\log\tr\left[\left(\sigma^{\frac{1-\alpha}{2\alpha}}\rho\sigma^{\frac{1-\alpha}{2\alpha}}\right)^\alpha\right]-\frac\alpha{\alpha-1}\log|\spec(\sigma)|\;.
\ea
Hence, replacing $\rho$ and $\sigma$ above with $\rho^{\otimes n}$ and $\sigma^{\otimes n}$, and recalling from~\eqref{specsig} that $|\spec(\sigma^{\otimes n})|\leq (n+1)^{|A|}$ we get in the limit $n\to\infty$
\be\label{happy}
\lim_{n\to\infty}\frac1nD_\alpha\left(\mP_{\sigma^{\otimes n}}\left(\rho^{\otimes n}\right)\big\|\sigma^{\otimes n}\right)\geq\frac1{\alpha-1}\log\tr\left[\left(\sigma^{\frac{1-\alpha}{2\alpha}}\rho\sigma^{\frac{1-\alpha}{2\alpha}}\right)^\alpha\right]\;.
\ee
\item The case $\alpha\in[\frac12,1)$ and $\rho\perp\sigma$. In this case $D_\alpha\left(\mP_{\sigma^{\otimes n}}\left(\rho^{\otimes n}\right)\big\|\sigma^{\otimes n}\right)=\infty$ since $\mP_{\sigma^{\otimes n}}\left(\rho^{\otimes n}\right)=\rho^{\otimes n}$ (and note also that $\rho$ and $\sigma$ commute; i.e., classical).
\item The case $\alpha\in[\frac12,1)$ and $\rho\not\perp\sigma$. In this case, the first inequality in~\eqref{8150} holds in the opposite direction since the factor $\frac1{\alpha-1}$ is negative. We therefore need another argument or trick. First, observe that
\be\label{8153}
\left(\sigma^{\frac{1-\alpha}{2\alpha}} \rho\sigma^{\frac{1-\alpha}{2\alpha}}\right)^{\alpha}=\left(\sigma^{\frac{1-\alpha}{2\alpha}} \rho\sigma^{\frac{1-\alpha}{2\alpha}}\right)^{\alpha-1}\sigma^{\frac{1-\alpha}{2\alpha}} \rho\sigma^{\frac{1-\alpha}{2\alpha}}\;.
\ee
Then, using the pinching inequality $\rho\leq|\spec(\sigma)|\mP_\sigma(\rho)$ we get
\be
\sigma^{\frac{1-\alpha}{2\alpha}}\rho\sigma^{\frac{1-\alpha}{2\alpha}}\leq |\spec(\sigma)|\sigma^{\frac{1-\alpha}{2\alpha}}\mP_\sigma(\rho)\sigma^{\frac{1-\alpha}{2\alpha}}
\ee
Combining this with the fact that the function $t\mapsto t^{\alpha-1}$ is anti-operator monotone for $\alpha\in[\frac12,1)$ (see Table~\ref{table:1}) we get that
\be
\left(\sigma^{\frac{1-\alpha}{2\alpha}}\rho\sigma^{\frac{1-\alpha}{2\alpha}}\right)^{\alpha-1}\geq |\spec(\sigma)|^{\alpha-1}\left(\sigma^{\frac{1-\alpha}{2\alpha}}\mP_\sigma(\rho)\sigma^{\frac{1-\alpha}{2\alpha}}\right)^{\alpha-1}
\ee
Combining the above inequality with~\eqref{8153} gives
\ba
\tr\left(\sigma^{\frac{1-\alpha}{2\alpha}}\rho\sigma^{\frac{1-\alpha}{2\alpha}}\right)^\alpha
&\geq|\spec(\sigma)|^{\alpha-1}\tr\left[\left(\sigma^{\frac{1-\alpha}{2\alpha}}\mP_\sigma(\rho)\sigma^{\frac{1-\alpha}{2\alpha}}\right)^{\alpha-1}\sigma^{\frac{1-\alpha}{2\alpha}}\rho\sigma^{\frac{1-\alpha}{2\alpha}}\right]\\
{\color{red} \text{Exercise~\ref{thepinchex} }\rightarrow}&=|\spec(\sigma)|^{\alpha-1}\tr\left[\left(\sigma^{\frac{1-\alpha}{2\alpha}}\mP_\sigma(\rho)\sigma^{\frac{1-\alpha}{2\alpha}}\right)^{\alpha}\right]\;.
\ea
Using the above inequality in~\eqref{8149} gives
\be
D_\alpha\left(\mP_{\sigma}\left(\rho\right)\big\|\sigma\right)\geq\frac1{\alpha-1}\log\tr\left[\left(\sigma^{\frac{1-\alpha}{2\alpha}}\rho\sigma^{\frac{1-\alpha}{2\alpha}}\right)^\alpha\right]-\log|\spec(\sigma)|\;.
\ee
Finally, replacing $\rho$ and $\sigma$ above with $\rho^{\otimes n}$ and $\sigma^{\otimes n}$, and recalling that $|\spec(\sigma^{\otimes n})|\leq (n+1)^{|A|}$ we get~\eqref{happy} in the limit $n\to\infty$.
\end{enumerate}
This completes the proof.
\end{proof}

\begin{exercise}\label{thepinchex}
Show that for any $a,b\in\mbb{R}$ and any $\rho,\sigma\in\md(A)$
\be
\tr\left[\left(\sigma^a\mP_\sigma(\rho)\sigma^a\right)^b\sigma^a\rho\sigma^a\right]=\tr\left[\left(\sigma^a\mP_\sigma(\rho)\sigma^a\right)^{b+1}\right]
\ee
Hint: Recall that $\mP_\sigma(\rho)$ commutes with $\sigma$ and that all the pinching projectors commutes with all the operators above except for the single $\rho$. 
\end{exercise}

\begin{myg}{}
\begin{corollary}\index{regularization}
Let $A$ be a quantum system and  $\rho,\sigma\in\md(A)$.
For $\alpha\geq 1/2$
\be\label{8139}
\underline{D}^\reg_{\alpha}(\rho\|\sigma)=\lim_{n\to\infty}\frac1nD_\alpha\left(\mP_{\sigma^{\otimes n}}\left(\rho^{\otimes n}\right)\big\|\sigma^{\otimes n}\right)
\ee
\end{corollary}
\end{myg}

\begin{proof}
Follows trivially from a combination of the theorem above with Eqs.~(\ref{underregal2},\ref{happy}).
\end{proof}

Note that the corollary above demonstrates that an optimizer for~\eqref{mixedopt} is $\mE_n=\Delta_n$.

\begin{exercise}
Show that for all $\alpha\in[0,\infty]$
\be
\underline{D}^\reg_{\alpha}(\rho\|\sigma)\geq\limsup_{n\to\infty}\frac1nD_\alpha\left(\rho^{\otimes n}\big\|\mP_{\rho^{\otimes n}}\left(\sigma^{\otimes n}\right)\right)\;.
\ee
Further, show that the equality above holds for all $\alpha\in(0,\frac12)$.
\end{exercise}

\subsection{The Maximal Quantum Extension}\index{optimal extension}\index{max relative entropy}

In this subsection we apply the results of Sec.~\ref{qetd} to relative entropies. In particular, we will see that the maximal quantum extension of the R\'enyi relative entropy has a closed formula for R\'enyi order parameter $\alpha\in[0,2]$. We start with the following corollary of Theorem~\ref{thmod}.

\begin{myg}{}
\begin{corollary}
Let $\D$ be a classical relative entropy, $\rho,\sigma\in\md(A)$, and suppose $\rho=\psi\eqdef|\psi\lr\psi|$. Then, 
\be
\oD(\psi\|\sigma)=D_{\max}(\psi\|\sigma)
\ee
where $D_{\max}$ is the quantum max relative entropy.
\end{corollary}
\end{myg}
\begin{proof}
Let $\{\e_1,\e_2\}$ be the standard basis of $\mbb{R}^2$, and observe that $\lambda_{\max}$ in~\eqref{modtim2} is precisely $2^{-D_{\max}(\psi\|\sigma)}$. Therefore, Theorem~\ref{thmod} gives
\ba
\oD(\psi\|\sigma)&=\D\left(\e_1\|
\big\|2^{-D_{\max}(\psi\|\sigma)}\e_1+\left(1-2^{-D_{\max}(\psi\|\sigma)}\right)\e_2\right)\\
{\color{red} \text{Theorem~\ref{822}}\rightarrow}&=D_{\max}(\psi\|\sigma)
\ea
where the last equality holds since $\D$ is a relative entropy. 
\end{proof}

The corollary above demonstrates that the maximal quantum extension is closely related to $D_{\max}$. The corollary is universal in the sense that it holds for any classical relative entropy $\D$, however, it is quite limited as it holds only for pure $\rho$. In the corollary below we will see that for some of the R\'enyi relative entropies there exists a closed formula for the maximal quantum extension without any restriction on $\rho$ and $\sigma$. This closed formula is given in terms of the family of \emph{geometric relative entropies}.

\begin{myd}{The Geometric Relative Entropy}\index{geometric relative entropy}
\begin{definition}\label{def:geometric}
The geometric relative entropy of order $\alpha\in[0,2]$ is defined for any $\rho\in\md(A)$ and $0<\sigma\in\pos(A)$ as
\be
\widehat{D}_{\alpha}(\rho\|\sigma)\eqdef\frac{1}{\alpha-1}\log\tr\left[\sigma\left(\sigma^{-\frac12}\rho\sigma^{-\frac12}\right)^\alpha\right] 
\ee
and for singular $\sigma\in\pos(A)$ is defined by
\be
\widehat{D}_{\alpha}(\rho\|\sigma)\eqdef\lim_{\eps\to 0^+}\widehat{D}_{\alpha}\left(\rho\|\sigma+\eps I\right)\;.
\ee
\end{definition}
\end{myd}
{\it Remarks:}
\begin{enumerate}
\item Alternatively, one can define the geometric relative entropy  for any $\rho,\sigma\in\md(A)$ using the decomposition~\eqref{decomrs} with $\trho\eqdef\rho_{11}-\zeta\rho_{22}^{-1}\zeta^*$ and $\tsigma\eqdef\sigma_{11}$.
Then, the geometric relative entropy of order $\alpha\in[0,2]$ is given by
\be\label{8170}
\widehat{D}_{\alpha}(\rho\|\sigma)=\begin{cases}\frac{1}{\alpha-1}\log\tr\left[\tsigma\left(\tsigma^{-\frac12}\trho\tsigma^{-\frac12}\right)^\alpha\right] &\text{if }\;\alpha\in[0,1)\;\text{ or }\;\rho\ll\sigma\\
\infty & \text{otherwise}
\end{cases}
\ee
\item The geometric relative entropy can be written differently  using the relation $Mf(M^*M)=f(MM^*)M$ given in Exercise~\ref{relfmmst}. Denoting $M\eqdef \rho^{\frac12}\sigma^{-\frac12}$ we get
\ba
\widehat{D}_\alpha(\rho\|\sigma)&=\frac{1}{\alpha-1}\log\tr\left[\sigma M^*M\left(M^*M\right)^{\alpha-1}\right]\\
&=\frac{1}{\alpha-1}\log\tr\left[\sigma M^*\left(M^*M\right)^{\alpha-1}M\right]\\
&=\frac1{\alpha-1}\log\tr\left[\rho\left(\rho^{\frac12}\sigma^{-1}\rho^{\frac12}\right)^{\alpha-1}\right]\;.
\ea
\item In the limit $\alpha\to 1$ we get for $\rho\ll\sigma$
\be\label{8194}
\widehat{D}(\rho\|\sigma)\eqdef\lim_{\alpha\to 1}\widehat{D}_{\alpha}(\rho\|\sigma)=
\tr\left[\rho\log\left(\rho^{\frac12}\sigma^{-1}\rho^{\frac12}\right)\right]\;,
\ee
which is known as the Belavkin–Staszewski relative entropy.
\item Observe that for $\alpha=2$ the definition of the geometric relative entropy coincides with the Petz\index{Petz} quantum R\'enyi divergence of the same order.
\end{enumerate}

\begin{exercise}
Show that the two definitions above for the geometric relative entropy are equivalent (for any two density matrices $\rho,\sigma\in\md(A)$); i.e., prove~\eqref{8170}.
\end{exercise}

\begin{exercise}
Show that the geometric relative entropy satisfies the properties (axioms) of additivity and normalization of a quantum relative entropy.
\end{exercise}

\begin{exercise}
Show that the geometric relative entropy reduces to the R\'enyi relative entropy in the classical domain.
\end{exercise}

Instead of proving directly that the geometric relative entropy satisfies the DPI, we will show that it is equal to the maximal quantum extension of the R\'enyi relative entropy.
Since the latter satisfies the DPI, this will imply that geometric relative entropy also satisfies the DPI.

\begin{myg}{}
\begin{corollary}\index{geometric relative entropy}
 The regularized maximal quantum extension of the R\'enyi divergence, $D_\alpha$, with $\alpha\in[0,2]$, is given by the geometric relative entropy; specifically, for any $\rho,\sigma\in\md(A)$ and $\alpha\in[0,2]$
\be
\overline{D}^\reg_{\alpha}(\rho\|\sigma)=\widehat{D}_{\alpha}(\rho\|\sigma)\;.
\ee
\end{corollary}
\end{myg}

\begin{proof}
The proof follows directly from Theorem~\ref{t633g}  for the case $\sigma>0$ and from Theorem~\ref{t633} for the general case. We leave the details as an exercise.
\end{proof}

\begin{exercise}
Provide the full details of the proof of the corollary above.
\end{exercise}

\section{Notes and References}

Axiomatic derivations of entropies and relative entropies have a plentiful literature, starting with the seminal work by~\cite{Shannon1948} followed and refined by~\cite{Faddeev56}, \cite{Diderrich75}, and~\cite{AFN1974}, amongst others. These early papers  focussed on the derivation of the Shannon entropy until the scope was extended by~\cite{Renyi1961}. Detailed reviews on the various axiomatic derivations can be found in books by~\cite{AD75} and~\cite{ESS1998}, and for a more recent guide on the topic see~\cite{Csiszar08}. The axiomatic approach\index{axiomatic approach} presented here was formally introduced by~\cite{GT2020} and~\cite{GT2021}. We point out that other functions that were studied in literature, like the Tsallis entropies, are not entropies according to the definition adopted here, as they are not additive in general. Moreover, historically, the terminology of `relative entropy' has been reserved only to the KL-divergence or to the Umegaki relative entropy in the quantum case, whereas we used this terminology to include all additive, monotonic functions, as given in Definitions~\ref{cre} and~\ref{qcre}.

There are several proofs that can be found in the literature on Erd\"os theorem (Lemma~\ref{lem:erd}). The elementary proof we adopted here is due to~\cite{Howe1986}.

There is a rich literature on both classical and quantum R\'enyi divergences. A recent guide on classical R\'enyi divergences with a thorough details of their properties can be found in the review article by~\cite{Erven2014}.
In the quantum domain, the books by~\cite{Tomamichel2015}, and more recently by~\cite{KW2021}, devotes significant portion to the study of the quantum R\'enyi divergences, and we refer the reader to these books for more details on the history and developments of quantum R\'enyi divergences. 

In this chapter we focused on three types of extensions of the R\'enyi relative entropy to the quantum domain: the Petz quantum R\'enyi divergence (introduced by~\cite{Petz1985}), the minimal (sandwiched) quantum R\'enyi divergence (introduced by~\cite{MDS+2013} and independently by~\cite{WWY2014}), and the maximal quantum R\'enyi divergence (introduced by~\cite{Matsumoto2018}). These three extensions are by no means the only quantum extensions of the classical R\'enyi relative entropy. Examples include the two parameter family studied in~\cite{JOPP2012} and~\cite{AD2015}, and the more recent divergence that was introduced by~\cite{FF2021}. 

The min and max quantum relative entropy were first introduced by~\cite{Datta2009}. Given that all quantum relative entropies are bounded by these two divergences, it is not a surprise that the min and max relative entropies have be used extensively in quantum information.

%%%%%%%%%%%%%%%%%%%%%%%%%%

%%%%%%%%%%%%%%%%%%%%%%%%%%

\chapter{Conditional Entropy}\label{sec:ce}\index{conditional entropy}

In this chapter, we delve into a variant of the entropy function, widely prevalent in information theory and quantum resource theories, especially in the realm of dynamical resources, which is the focus of the second volume of this book. This variant is known as conditional entropy, which pertains to the entropy associated with a physical system 
$A$ that shares a correlation with another system $B$. When an observer, say Bob, has access to system $B$, he can reduce his uncertainty about system $A$ by performing a quantum measurement on his subsystem. In essence, conditional entropy quantifies the residual uncertainty of system $A$ when such access to system $B$ is available.

Traditionally, the conditional entropy of a bipartite state $\rho^{AB}$ is defined in terms of the von Neumann entropy associated with system $AB$ minus the von Neumann entropy associated with system $B$. This is given by:
\begin{equation}
H(A|B)_\rho \eqdef H\left(\rho^{AB}\right) - H\left(\rho^B\right).
\end{equation}
See Fig.~\ref{Venn} for a heuristic description of this definition in terms of a Venn diagram.
However, in this chapter, we take a different approach. Here, conditional entropy is defined axiomatically, similar to how we defined entropy and relative entropy. This approach provides a more rigorous definition of conditional entropy, placing the intuitive Venn diagram interpretation on a more solid theoretical foundation.

\begin{figure}[h]
\centering
    \includegraphics[width=0.4\textwidth]{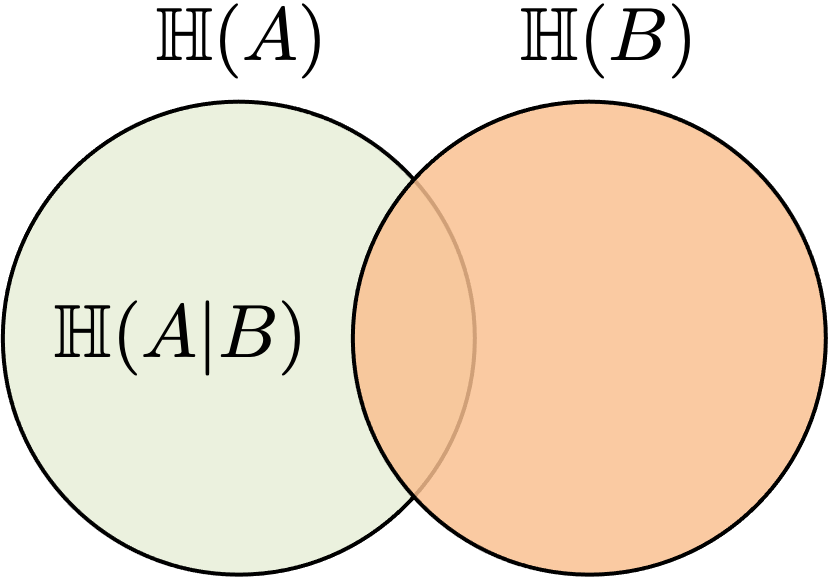}
  \caption{\linespread{1}\selectfont{\small A Venn-diagram for the conditional entropy.}}
  \label{Venn}
\end{figure}

\section{Quantum Conditional Majorization}\label{comajo}\index{conditional majorization}

We start this chapter with a section in which we extend the definition of conditional majorization to the quantum domain. Recall that conditional majorization, as defined in Sec.~\ref{sec:cm}, characterizes the uncertainty associated with a physical system given access to another system that is correlated with it. Therefore, as we will see shortly, conditional majorization provides the foundation for the definition of conditional entropy.

The majorization relation between two probability vectors can be generalized to the quantum domain in a straightforward manner. That is, for any two density matrices $\rho\in\md(A)$ and $\sigma\in\md(A')$ we say that $\rho$ majorizes $\sigma$ and write $\rho^A\succ\sigma^{A'}$ if the probability vector consisting of the eigenvalues of $\rho$, majorizes the probability vector consisting of the eigenvalues of $\sigma$. 
For conditional majorization, such a straightforward extension from the classical to the quantum domain is more complex as it involves two systems that can be correlated quantumly (i.e. entangled). For this reason we employ the axiomatic approach\index{axiomatic approach} to introduce quantum conditional majorization, and then discuss some of its key properties. 

As discussed above, intuitively, conditional majorization is a pre-order on the set $\md(AB)$ that characterizes the uncertainty of system $A$ given access to system $B$. To make this intuition more precise, we employ the axiomatic approach\index{axiomatic approach} determining which set of operations in $\cptp(AB\to AB')$ can only increase the uncertainty of system $A$ (even if one has access to system $B$).
We will now examine two highly intuitive axioms that these channels must adhere to. These two axioms extend the principles explored in Sec.~\ref{sec:axiomatica} to the quantum domain.

\subsection{Conditionally Unital Channels}\index{conditionally unital}

What types of channels are expected to increase the conditional uncertainty associated with a system $A$ given access to $B$? To address this question, let's consider a bipartite quantum state in the form $\u^A \otimes \rho^B$, where $\u^A$ represents the maximally mixed state and $\rho^B$ is a certain density matrix. Given that such a product state is uncorrelated, access to $B$ does not aid in reducing the uncertainty about $A$. Consequently, we can infer that this state exhibits the highest degree of conditional uncertainty on $A|B$ (i.e., $A$ given access to $B$). Now, any channel $\mN \in \cptp(AB \to AB')$ that preserves or increases this conditional uncertainty should transform states with maximal conditional uncertainty into states retaining this property. Specifically, for all $\rho \in \md(B)$, such a channel $\mN \in \cptp(AB \to AB')$ must satisfy:
\be\label{cunital}
\mN^{AB\to AB'}\left(\u^A\otimes\rho^B\right)=\u^{A}\otimes\sigma^{B'}
\ee
where $\sigma^{B'}$ is some density matrix in $\md(B')$. By tracing out system $A$, we can express $\sigma^{B'}$ as
\be\label{cunital00}
\sigma^{B'}= \mN^{AB\to B'}\left(\u^A\otimes\rho^B\right)\;,
\ee
where $\mN^{AB \to B'} \eqdef \tr_{A} \circ \mN^{AB \to AB'}$. Such a channel $\mN \in \cptp(AB \to AB')$ is referred to as \emph{conditionally unital}. It is important to note that both the input and output systems on Alice's side remain the same, whereas on Bob's side, the systems $B$ and $B'$ can be different.

\begin{myg}{}
\begin{lemma}
Let $\mN\in\cptp(AB\to \tA B')$ be a bipartite quantum channel and let $J_{\mN}^{AB\tA B'}$ be its Choi matrix. Then, $\mN^{AB\to\tA B}$ is conditionally unital\index{conditionally unital}   if and only if its Choi matrix satisfies
\be\label{cunital2}
J_{\mN}^{B\tA B'}=J_{\mN}^{B B'}\otimes\u^{\tA}\;.
\ee
\end{lemma}
\end{myg}

\begin{proof}
We begin by proving that the channel $\mathcal{N}^{AB \rightarrow \tA B'}$ is conditionally unital if its Choi matrix has the form in \eqref{cunital2}. To this end, let $\rho\in \mathfrak{D}(B)$ and  consider that
\ba
    \mathcal{N}\left(\mathbf{u} ^{A} \otimes \rho^{B}\right) &= \operatorname{Tr}_{AB}\!\left[ J_{\mathcal{N}}^ {AB\tilde{A}B'} \left( \mathbf{u}^{A} \otimes \left(\rho^{B}\right)^{T} \otimes I^{\tilde{A}B'} \right)\right] \\
    &= \frac{1}{|A|} \operatorname{Tr}_{B}\!\left[ J_{\mathcal{N}}^{B\tilde{A}B'} \left( \left(\rho^{B}\right)^{T} \otimes I^{\tilde{A}B'} \right)\right] \\
    \GG{\eqref{cunital2}}& = \frac{1}{|A|} \operatorname{Tr}_{B}\!\left[\mathbf{u}^{\tilde{A}} \otimes  J_{\mathcal{N}}^{BB'} \left( \left(\rho^{B}\right)^{T} \otimes I^{\tilde{A}B'} \right)\right] \\
    & = \mathbf{u}^{\tilde{A}} \otimes \sigma^B\;,
\ea
where $\sigma^B\eqdef\frac{1}{|A|} \operatorname{Tr}_{B}\!\left[  J_{\mathcal{N}}^{BB'} \left( \left(\rho^{B}\right)^{T} \otimes I^{B'} \right)\right]$.

We next prove that the Choi matrix of $\mathcal{N}^{AB\rightarrow\tilde
{A}B^{\prime}}$ has the form in \eqref{cunital2} if
$\mathcal{N}^{AB\rightarrow\tilde{A}B^{\prime}}$ is conditionally unital.
Recall that the defining property of a conditionally unital\index{conditionally unital}  channel is that
for every state $\rho\in\md(B)$, there exists a state $\sigma\in\md(B^{\prime})$ such
that
\begin{equation}\label{eq14}
\mathcal{N}^{AB\to\tA B'}(\mathbf{u}^{A}\otimes\rho^{B})=\mathbf{u}^{\tilde{A}}\otimes
\sigma^{B^{\prime}}.
\end{equation}
By taking the trace over $\tA$ on both sides of the equation above we get that
\ba
\sigma^{B^{\prime}}=\mathcal{N}^{AB\to B'}(\mathbf{u}^{A}\otimes\rho^{B})
&=\tr_{AB}\!\left[  J^{ABB^{\prime}}_{\mathcal{N}
}\left(  \mathbf{u}^{A}\otimes\left(  \rho^{B}\right)  ^{T}\otimes
I^{B^{\prime}}\right)  \right]\\
&=\frac1{|A|}\tr_{B}\!\left[  J^{BB^{\prime}}_{\mathcal{N}%
}\left( \left(  \rho^{B}\right)  ^{T}\otimes
I^{B^{\prime}}\right)  \right]\;.
\ea
On the other hand, observe that
\ba
  \mathcal{N}^{AB\to\tA B'}(\mathbf{u}^{A}\otimes\rho^{B})
&  =\operatorname{Tr}_{AB}\!\left[  J^{AB\tilde{A}B^{\prime}}_{\mathcal{N}
}\left(  \mathbf{u}^{A}\otimes\left(  \rho^{B}\right)  ^{T}\otimes
I^{\tilde{A}B^{\prime}}\right)  \right]  \\
&  =\frac{1}{|A|}\operatorname{Tr}_{B}\!\left[  J^{B\tilde{A}B^{\prime}
}_{\mathcal{N}}\left(  \left(  \rho^{B}\right)  ^{T}\otimes I^{\tilde
{A}B^{\prime}}\right)  \right]  \;.
\ea
Therefore, from the two expressions above for $\sigma^B$ and $\mathcal{N}(\mathbf{u}^{A}\otimes\rho^{B})$ we conclude that~\eqref{eq14} can be expressed as
\be\label{eq20}
\operatorname{Tr}_{B}\!\left[  J^{B\tilde{A}B^{\prime}
}_{\mathcal{N}}\left(  \left(  \rho^{B}\right)  ^{T}\otimes I^{\tilde
{A}B^{\prime}}\right)  \right]
= \tr_{B}\!\left[ (\u^{\tA}\otimes J^{BB^{\prime}}_{\mathcal{N}
})\left( \left(  \rho^{B}\right)  ^{T}\otimes
I^{\tA B^{\prime}}\right)  \right]\;.
\ee
Denote by $\eta^{B\tA B'}\eqdef J^{B\tilde{A}B^{\prime}
}_{\mathcal{N}}-\u^{\tA}\otimes J^{BB^{\prime}}_{\mathcal{N}}$ and observe that the equation above can be written as
\begin{equation}\label{eq21}
\tr\left[\eta^{B\tA B'}\left(\left(  \rho^{B}\right)  ^{T}\otimes
I^{\tA B^{\prime}}\right)\right]=0\;\;\quad\forall\;\rho\in\mathfrak{D}(B).
\end{equation}
Due to the existence of bases of density operators that span the space of
linear operators acting on ${B}$,
we conclude from \eqref{eq21} that for every 
operator $\zeta\in\mathfrak{L}(B)$ we have
\begin{equation}
\operatorname{Tr}_{B}\!\left[  \eta^{B\tilde{A}B^{\prime}%
}\left(  \zeta^{B}\otimes I^{\tilde{A}B^{\prime}}\right)  \right]
=0\;.
\end{equation}
Note that by multiplying both sides of the equation above by any element $\xi\in\mathfrak{L}(\tA B')$ and taking the trace we get that
\begin{equation}
\tr\big[  \eta^{B\tilde{A}B^{\prime}%
}\left(  \zeta^{B}\otimes \xi^{\tilde{A}B^{\prime}}\right)  \big]
=0\;.
\end{equation}
Since the equation above holds for all $\zeta\in\mathfrak{L}(B)$ and all $\xi\in\mathfrak{L}(\tA B')$ it also holds for any linear combinations of matrices of the form $\zeta^{B}\otimes \xi^{\tilde{A}B^{\prime}}$. Since matrices of the form $ \zeta^{B}\otimes \xi^{\tilde{A}B^{\prime}}$ span the whole space $\mathfrak{L}(B\tA B')$ we conclude that $\eta^{B\tilde{A}B^{\prime}}$ is orthogonal (in the Hilbert-Schmidt inner product) to all the elements of $\ml(B\tA B')$.
Therefore, we must have $\eta^{B\tilde{A}B'}=0$ which is equivalent to~\eqref{cunital2}. This completes the proof. \index{Hilbert-Schmidt inner product}
\end{proof}

\subsection{Semi-Causal Channels}\label{sec:semic}\index{semi-causal}

An additional requirement for the channel $\mN\in\cptp(AB\to AB')$ to not diminish conditional uncertainty is related to preventing information leakage from Alice's subsystem to Bob's. This is crucial since any leaked information could reduce the uncertainty about system $A$. Conditional uncertainty specifically pertains to the uncertainty about system $A$ when only Bob's system is accessible. To support this, we introduce a causality assumption ensuring that system $A$ does not causally influence system $B'$. This is mathematically represented as follows: for all $\mM\in\cptp(A\to A)$,
\be\label{semic}
\mN^{AB\to B'}\circ\mM^{A\to A}=\mN^{AB\to B'}\;,
\ee
where $\mN^{AB\to B'}\eqdef\tr_{A}\circ\mN^{AB\to AB'}$. This condition guarantees that any operation $\mM^{A\to A}$ applied by Alice to her system remains undetected by Bob. We refer to such a condition as $A\not\to B'$ semi-causal. For a visual representation, see Fig.~\ref{semicf} depicting a semi-causal channel.

\begin{figure}[h]
\centering
    \includegraphics[width=0.6\textwidth]{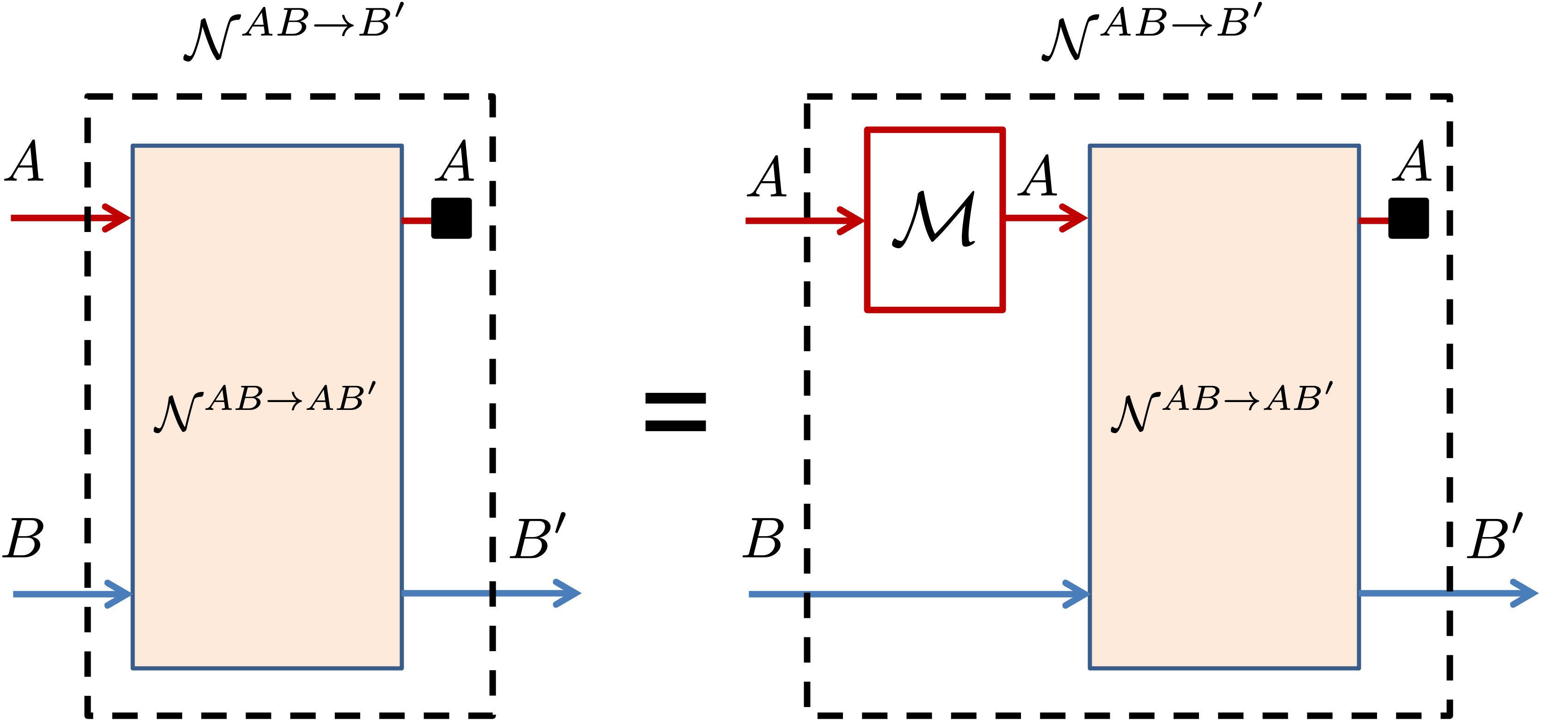}
  \caption{\linespread{1}\selectfont{\small An illustration of an $A\not\to B'$ semi-causal bipartite channel $\mN^{AB\to AB'}$. The marginal channel $\mN^{AB\to B'}$ equals $\mN^{AB\to B'}\circ\mM^{A\to A}$ for any choice of $\mM\in\cptp(A\to A)$.}}
  \label{semicf}
\end{figure}

\begin{exercise}
Let $\mN\in\cptp(AB\to\tA B')$.  Show that $\mN^{AB\to\tA B'}$ is $A\not\to B'$ semi-causal if and only if the marginals of its Choi matrix satisfy
\be\label{semic2}
J_{\mN}^{ABB'}=\u^A\otimes J_{\mN}^{BB'}\;.
\ee
Hint: In one direction, take $\mM\in\cptp(A)$ in~\eqref{semic} to be the replacer channel that always output the maximally mixed state irrespective of the input state (i.e. take $\mM$ to be the completely depolarizing channel, also know as the completely randomizing channel), and compute the Choi matrix for the channels on both sides of~\eqref{semic}.
\end{exercise}

The following theorem establishes the equivalence between $A\not\to B'$ semi-causal channels and $A\not\to B'$ signalling channels. Specifically, a bipartite quantum channel $\mN\in\cptp(AB\to AB')$ is classified as $A\not\to B'$ signalling if it meets the following criteria: there exists a reference system $R$, alongside a quantum channel $\mE\in\cptp(AR\to A)$ and an isometry $\mF\in\cptp(B\to RB')$, such that
\be\label{mainsemic}
\mN^{AB\to AB'}=\mE^{RA\to A}\circ\mF^{B\to RB'};.
\ee
In essence, channels that are $A\not\to B'$ signalling are those that can be implemented via one-way communication from Bob to Alice. An illustration of this concept can be found in Fig.~\ref{semi-causal}.

\begin{myt}{}
\begin{theorem}\label{thm:semic}
Let $\mN\in\cptp(AB\to AB')$ be a bipartite quantum channel. The following two statements are equivalent:\index{signalling}\index{semi-causal}
\ben
\item The channel $\mN^{AB\to AB'}$ is an $A\not\to B'$ signalling, as defined in~\eqref{mainsemic}.
\item The channel $\mN^{AB\to AB'}$ is an $A\not\to B$ semi-causal, as defined in~\eqref{semic}. 
\een
\end{theorem}
\end{myt}
\begin{remark}
The theorem above demonstrate the intuitive assertion that semi-causal bipartite channels are channels that can be realized with one-way communication from Bob to Alice. With such channels, Alice cannot influence Bob's system. We also point out that  the relation in~\eqref{mainsemic} has been written in a compact form; that is, we removed identity channels so that
\be
\mE^{RA\to A}\circ\mF^{B\to RB'}\eqdef \left(\mE^{RA\to A}\otimes\id^{B'\to B'}\right)\circ\left(\id^{A\to A}\otimes\mF^{B\to RB'}\right)\;.
\ee
\end{remark}

\begin{figure}[h]
\centering
    \includegraphics[width=0.4\textwidth]{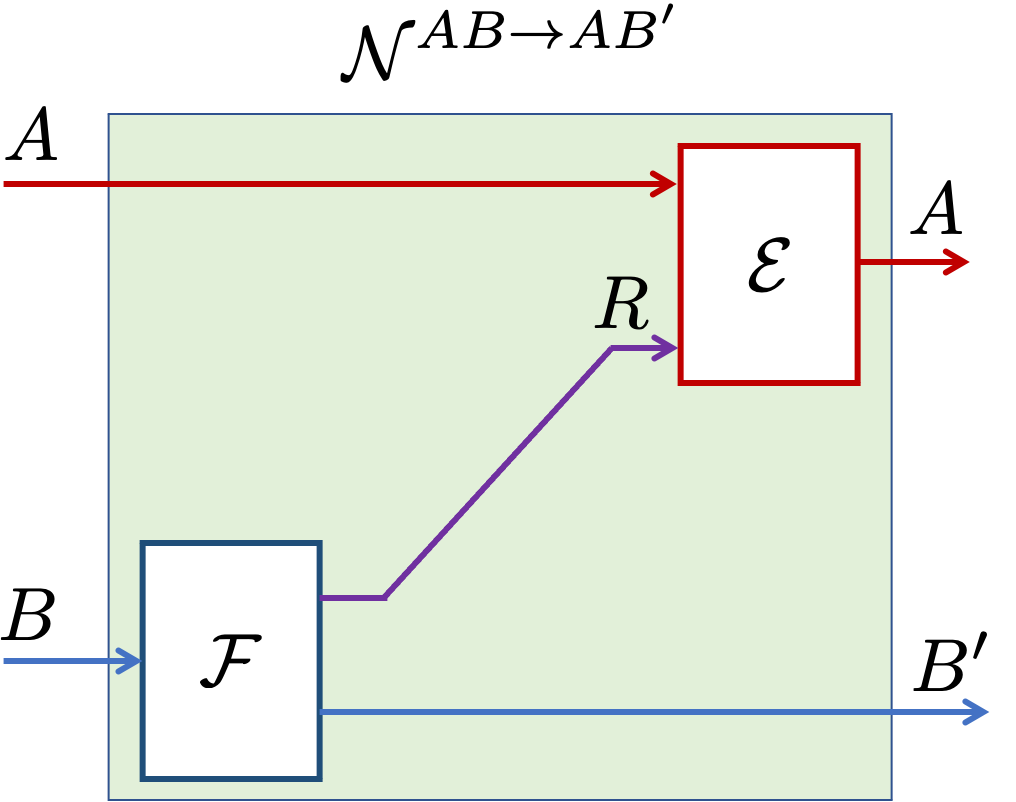}
  \caption{\linespread{1}\selectfont{\small A bipartite channel that is $A\not\to B'$ signalling.}}
  \label{semi-causal}
\end{figure} 

\begin{proof}
We begin by proving the implication $1\Rightarrow 2$. Consider the following marginal of the channel $\mN^{AB\to AB'}$:
\ba
\mN^{AB\to B'}&\eqdef\tr_A\circ\mN^{AB\to AB'}\\
\GG{\eqref{mainsemic}}&=\tr_A\circ\mE^{RA\to A}\circ\mF^{B\to RB'}\\
&=\tr_{RA}\circ\mF^{B\to RB'}\\
&=\tr_A\circ\mF^{B\to B'}\;,
\ea
where we have utilized the trace-preserving property of quantum channels and denoted $\mF^{B\to B'}\eqdef \tr_{R}\circ\mF^{B\to RB'}$. It is evident that this channel satisfies~\eqref{semic} with any trace preserving map $\mM^{A\to A}$, establishing that $\mN^{AB\to AB'}$ is $A\not\to B'$ semi-causal.

Moving on to the implication $2\Rightarrow 1$, we examine two distinct purifications of the marginal Choi matrix $J_{\mN}^{ABB'}$:
\ben
\item Consider a physical system $C$ and a pure (unnormalized) state $\psi^{ABAB'C}$, which acts as a purification of both $J_{\mN}^{ABAB'}$ and its marginal state $J_{\mN}^{ABB'}$. 
\item Denoting by $\varphi^{BB'R}$, an (unnormalized) purification of the operator $\frac{1}{|A|}J_{\mN}^{BB'}$, we get from~\eqref{semic2}, that $\Omega^{A\tA}\otimes\varphi^{BB'R}$ is another purification $J_{\mN}^{ABB'}$.
\een

Since the marginal of $J^{AB}_\mN$ equals $I^{AB}$, it implies that $\varphi^B=\frac{1}{|A|}J_{\mN}^{B}=I^B$. This property implies the existence of an isometry $\mF\in\cptp(B\to B'R)$ that satisfies 
\be\label{oyo}
\varphi^{BB'R}=\mF^{\tB\to B'R}(\Omega^{B\tB})\;.
\ee
Moreover, given that two purifications of the same positive semi-definite matrix are connected by an isometry (as per Exercise~\ref{purification}), there must be an isometry $\mV\in\cptp(R\tA\to AC)$ satisfying
\be
\psi^{ABAB'C}=\mV^{R\tA\to AC}\left(\Omega^{A\tA}\otimes\phi^{BB'R}\right)\;.
\ee 
Finally, tracing out system $C$ and denoting by $\mE^{R\tA\to A}\eqdef\tr_C\circ \mV^{R\tA\to AC}$ gives
\ba
J_{\mN}^{ABAB'}&=\mE^{R\tA\to A}\left(\Omega^{A\tA}\otimes\phi^{BB'R}\right)\\
\GG{\eqref{oyo}}&=\mE^{R\tA\to A}\left(\Omega^{A\tA}\otimes\mF^{\tB\to B'R}\big(\Omega^{B\tB}\big)\right)\\
&=\mE^{R\tA\to A}\circ\mF^{\tB\to B'R}\left(\Omega^{(AB)(\tA\tB)}\right)\;.
\ea
The equation above implies that~\eqref{mainsemic} holds. This completes the proof.
\end{proof}

\begin{exercise}\label{showsemi}
Show that if $|B'|=1$ then any channel in $\cptp(AB\to AB')$ is $A\not\to B'$ semi-causal.
\end{exercise}

\subsection{Quantum Conditionally Mixing Operations}\label{lbsec}\index{conditionally mixing}

We  do not expect conditional uncertainty to decrease under operations that are both conditionally unital\index{conditionally unital}  and $A\not\to B'$ semi-causal. Such operations provides a quantum extension to the operations introduced in Definition~\ref{def:ccmo}.

\begin{myd}{}
\begin{definition}
A quantum channel $\mN\in\cptp(AB\to AB')$ is called a  \emph{conditionally mixing operation} (CMO) if it is both conditionally unital and $A\not\to B'$ semi-causal.  The set of all such conditionally mixing operations in $\cptp(AB\to AB')$ is denoted by $\cmo(AB\to AB')$. 
\end{definition}
\end{myd}

The Choi matrix, $J_\mN^{AB\tA B'}$, of a channel $\mN\in\cmo(AB\to \tA B')$ that is both conditionally unital\index{conditionally unital}  and $A\not\to B'$ semi-causal must satisfies:
\ben
\item $J_{\mN}^{B\tA B'}=J_{\mN}^{B B'}\otimes\u^{\tA}$ (i.e.\ $\mN$ is conditionally unital; see~\eqref{cunital2}).
\item $J_{\mN}^{ABB'}=\u^A\otimes J_{\mN}^{BB'}$ (i.e.\ $\mN$ is semi-causal; see~\eqref{semic2}).
\item $J^{AB}_{\mN}=I^{AB}$ (i.e.\ $\mN$ is trace preserving).
\item $J_\mN^{AB\tA B'}\geq 0$ (i.e.\ $\mN$ is completely positive).
\een
Observe the symmetry of the first two conditions above under exchange of the local input system $A$ and the local output system $\tA$.

As a straightforward example of a CMO, consider that the reference system $R$ in Theorem~\ref{thm:semic} is classical. In such a scenario, we define $X\eqdef R$ and the channel $\mN^{AB\to AB'}$ can be expressed as:
\ba\label{8210}
\mN^{AB\to AB'}&=\mE^{XA\to A}\circ\mF^{B\to XB'}\
&=\sum_{x\in[m]}\mE_{(x)}^{A\to A}\otimes\mF^{B\to B'}_x
\ea
Here, $\{\mF_x^{B\to B'}\}_{x\in[m]}$ constitutes a quantum instrument, and for each $x\in[m]$, the map $\mE_{(x)}^{A\to A}$ is a quantum channel in $\cptp(A\to A)$. We will explore later that this channel typifies one-way LOCC (Local Operations and Classical Communication). Notably, if each $\mE_{(x)}$ is a unital channel, this one-way LOCC is also conditionally unital\index{conditionally unital} .
This channel essentially represents Bob performing a quantum measurement on his system and conveying the result to Alice, who then applies a unital channel to her system.

\bex
Let $\omega\in\md(B)$ and $\mN\in\cmo(AB\to \tA B')$.
Show that the channel
$\mE\in\cptp(A\to \tA B')$ defined for any $\rho\in\ml(A)$ by
\be
\mE^{A\to\tA B'}(\rho^A)\eqdef\mN^{AB\to\tA B'}(\rho^A\otimes\omega^B)
\ee
is also {\rm CMO}; i.e. show that $\mE\in\cmo(A\to \tA B')$.
\eex

\bex\label{upsiex}
Let $\Upsilon\in\ml(AB\tA B'\to AB\tA B')$ be a linear map defined for all $\omega\in\ml(AB\tA B')$ as
\be
\Upsilon\left(\omega^{AB\tA B'}\right)\eqdef\u^A\otimes\left(\omega^{B\tA B'}-\omega^{B B'}\otimes\u^{\tA}\right)+\u^{\tA}\otimes\left(\omega^{AB B'}-\u^{A}\otimes \omega^{B B'}\right)\;.
\ee
\ben
\item Show that a channel $\mN\in\cptp(AB\to \tA B')$ is {\rm CMO} if and only if $\Upsilon(J_\mN^{AB\tA B'})=0$. 
\item Show that $\Upsilon$ is self-adjoint; i.e.\ show that $\Upsilon=\Upsilon^*$.
\item Show that $\Upsilon$ is idempotent; i.e.\ $\Upsilon\circ\Upsilon=\Upsilon$.
\een
\eex

\subsection{Definition of Conditional Majorization}\index{conditional majorization}

We are now prepared to define quantum conditional majorization, grounded on the two aforementioned axioms, for channels that preserve or increase conditional uncertainty.

\begin{myd}{Quantum Conditional Majorization}
\begin{definition}\label{qcomajo}
Let $A,B,B'$ be three quantum systems, and let $\rho\in\md(AB)$ and $\sigma\in\md(AB')$. We say that $\rho^{AB}$ conditionally majorizes $\sigma^{AB'}$, and write $\rho^{AB}\succ_A\sigma^{AB'}$, if there exists a channel $\mN\in\cmo(AB\to AB')$ such that
\be
\sigma^{AB'}=\mN^{AB\to AB'}\left(\rho^{AB}\right)\;.
\ee
\end{definition}
\end{myd}

\bex
Show that quantum conditional majorization as defined above is a pre-order.
\eex

This definition effectively extends the concept of majorization. Specifically, when $|B|=|B'|=1$, the set $\cmo(A\to A)$ coincides with the set of unital channels. Consequently, the relation $\rho^A\succ_A\sigma^A$ as defined above (under the condition $|B|=|B'|=1$) transforms into the well-known majorization relation $\rho^A\succ\sigma^A$. Expanding on this concept, quantum conditional majorization exhibits the following notable property.

\begin{myg}{}
\begin{lemma}
Let $\rho\in\md(AB)$ and $\sigma\in\md(AB')$ be two product states; i.e. $\rho^{AB}=\rho^A\otimes\rho^B$ and $\sigma^{AB'}=\sigma^{A}\otimes\sigma^{B'}$. Then,
\be
\rho^{AB}\succ_A\sigma^{AB'}\iff\rho^A\succ\sigma^A\;.
\ee
\end{lemma}
\end{myg}
\begin{proof}
If $\rho^A\succ\sigma^A$ then there exists a unital channel such that $\sigma^A=\mU^{A\to A}(\rho^A)$. Let $\mE\in\cptp(B\to B')$ be a replacement channel that always outputs $\sigma^{B'}$. Then, the channel 
\be
\mN^{AB\to AB'}\eqdef\mU^{A\to A}\otimes\mE^{B\to B'}
\ee
is {\rm CMO} and satisfies $\sigma^{A}\otimes\sigma^{B'}=\mN^{AB\to AB'}\left(\rho^A\otimes\rho^B\right)$. Hence, $\rho^{AB}\succ_A\sigma^{AB'}$. Conversely, if $\rho^{AB}\succ_A\sigma^{AB'}$ then there exists a semi-causal quantum channel $\mN^{AB\to AB'}=\mE^{RA\to A}\circ\mF^{B\to RB'}$ (that is also conditionally unital\index{conditionally unital} ) that satisfies
\be
\sigma^{A}\otimes\sigma^{B'}=\mE^{RA\to A}\circ\mF^{B\to RB'}\left(\rho^A\otimes\rho^B\right)\;.
\ee
Tracing system $B'$ on both sides, and denoting $\tau^R\eqdef\tr_{B'}\circ\mF^{B\to RB'}\left(\rho^B\right)$ gives
\be
\sigma^{A}=\mE^{RA\to A}\left(\rho^A\otimes\tau^R\right)\;.
\ee
Finally, note that the channel
\be
\mU^{A\to A}(\omega^A)\eqdef\mE^{RA\to A}\left(\omega^A\otimes\tau^R\right)\quad\quad\forall\;\omega\in\md(A)\;,
\ee
is a unital channel since $\mN^{AB\to AB'}$ is conditionally unital\index{conditionally unital} . Since by definition, $\sigma^A=\mU^{A\to A}\left(\rho^A\right)$, we conclude that $\rho^A\succ\sigma^A$. This completes the proof.
\end{proof}

Determining whether $\rho^{AB}\succ_A\sigma^{AB'}$ for two given bipartite quantum states $\rho^{AB}$ and $\sigma^{AB'}$ can be challenging. This task is essentially about verifying the existence of a Choi matrix $J^{AB\tA B'}\mN$ that satisfies both the four initial conditions outlined at the start of this subsection (characteristic of a {\rm CMO}) and the additional criterion:
\be
\tr_{AB}\left[J^{AB\tA B'}_\mN\left(\left(\rho^{AB}\right)^T\otimes I^{\tA B'}\right)\right]=\sigma^{\tA B'}\;.
\ee
These five conditions imposed on $J^{AB\tA B'}\mN$ constitute an SDP (Semidefinite Programming) feasibility problem that can be solved efficiently and algorithmically on a computer.

\subsection{Specific Conditional Majorization Relations}\index{conditional majorization}

Generally, as mentioned above, determining if one state conditionally majorizes another can be resolved through an SDP. However, there are key examples that are useful for later discussions and can be resolved without an SDP. For instance, for any state $\rho\in\md(ABB')$, it holds that
\be
\rho^{ABB'}\succ_A\rho^{AB}
\ee
since partial tracing over system $B'$ qualifies as a {\rm CMO}. Another straightforward case is for any $\rho\in\md(AB)$ and $\sigma\in\md(B)$, where
\be
\rho^{AB}\succ_A\rho^A\otimes\sigma^B\;.
\ee
A more nuanced example is the fact that the maximally entangled state conditionally majorizes all states of the same dimensions.

\begin{myt}{}
\begin{theorem}\label{maxi0}
Let $\rho\in\md(AB)$. Then,
\be
\Phi^{AB}\succ_A\rho^{AB}\;.
\ee
\end{theorem}
\end{myt}

\begin{proof}
In Sec.\ref{sec:qtele}, we discussed how the maximally entangled state $\Phi^{AB}$ can be utilized for teleporting an unknown quantum state. Specifically, a teleportation protocol from Bob to Alice (note that in Sec.\ref{sec:qtele}, we examined teleportation from Alice to Bob) involves Bob performing a joint quantum measurement on his part of $\Phi^{AB}$ and the state to be teleported. This is followed by classical communication to Alice, who then applies a unitary operation. Such a protocol conforms to the structure of {\rm CMO} channels described in~\eqref{8210}, where each $\mE_{(x)}^{A\to A}$ is a unitary channel. This implies that for any bipartite state $\rho^{AB}$, there exists a channel $\mN\in\cmo(AB\to AB)$ such that
\be
\rho^{AB}=\mN^{AB\to AB}\left(\Phi^{AB}\right)\;.
\ee
We emphasize that in the realization of the channel $\mN^{AB\to AB}$, Bob locally prepares the state $\rho^{B'B}$ with $B'\cong A$ and then employs the maximally entangled state $\Phi^{AB}$ to teleport the $B'$ subsystem to Alice, resulting in the state $\rho^{AB}$.
The equation above implies that $\Phi^{AB}\succ_A\rho^{AB}$.  
\end{proof}

The next theorem characterizes the states in $\md(AB)$ that are conditionally majorized by a state in $\pure(A)$. Note that all the states in $\pure(A)$ are equivalent under majorization.

\begin{myt}{}
\begin{theorem}\label{strong}
Let $\rho\in\md(AB)$. Then, the following statements are equivalent:
\ben
\item The state $\rho^{AB}$ satisfies
\be
I^A\otimes\rho^B\geq\rho^{AB}\;.
\ee
\item For every pure state $\psi\in\pure(A)$
\be\label{respsia}
\psi^A\succ_A\rho^{AB}\;.
\ee
\een
\end{theorem}
\end{myt}
\begin{proof}
To demonstrate that the first statement implies the second, we search for a channel $\mN\in\cmo(A\to AB)$ satisfying $\rho^{AB}=\mN^{A\to AB}(\psi^A)$. Our strategy involves considering a binary measurement-and-prepare channel defined as:
\be
\mN^{A \to AB}(\omega^{A}) \eqdef\tr[\psi^{A}\omega^{A}]\rho^{AB}
+\tr[(I^{A}-\psi^{A})\omega^{A}]\tau^{AB}\quad\quad\forall;\omega\in\ml(A);,
\ee
where $\tau^{AB}$ is a density matrix to be determined. By definition, we have $\rho^{AB}=\mN^{A\to AB}(\psi^A)$. The remaining task is to identify a $\tau\in\md(AB)$ that ensures $\mN^{A\to AB}$ is a {\rm CMO} (both $A\not\to B$ semi-causal and conditionally unital\index{conditionally unital} ).

For $\mN^{A\to AB}$ to be $A\not\to B$ semi-causal, it must satisfy the condition $\mN^{A\to B}\circ\mM^{A\to A}=\mN^{A\to B}$ for any $\mM\in\cptp(A\to A)$ (cf.~\eqref{semic}). That is, $\mN^{A\to B}$ must be a replacement channel, independent of its input. Observing that for every input state $\omega\in\md(A)$,
\be\label{stammis}
\mN^{A\to B}(\omega^A)=\operatorname{Tr}[\psi^A\omega^{A}]\rho^{B}
+\operatorname{Tr}[\left( I^{A}-\psi^{A}\right) \omega^{A}
]\tau^{B}\;,
\ee
and given that the left-hand side is independent of $\omega^A$, we conclude $\mN^{A\to AB}$ is $A\not\to B$ semi-causal if and only if $\tau^B=\rho^B$. This is the first condition for $\tau^{AB}$.

The second condition arises from $\mN^{A\to AB}$ being conditionally unital. This is satisfied if and only if (cf.~Eqs.~(\ref{cunital},\ref{cunital00}))
\ba
\mathcal{N}^{A\rightarrow AB}(I^{A})=\rho^{AB}+\left( |A|-1\right) \tau^{AB}
\ea
equals
\be
I^{A}\otimes\mathcal{N}^{A\rightarrow B}(\u^{A})=I^A\otimes\rho^B\;,
\ee
using~\eqref{stammis} with $\tau^B=\rho^B$. Equating these two operators dictates that $\tau^{AB}$ must be
\begin{equation}\label{tauabz2}
\tau^{AB}\coloneqq \frac{I^{A}\otimes\rho^{B}-\rho^{AB}}{|A|-1}\;.
\end{equation}
$\tau^{AB}$ is positive semi-definite, as we assume $I^{A}\otimes\rho^{B}\geq\rho^{AB}$, and has unit trace, qualifying as a density matrix. Also, it satisfies $\tau^B=\rho^B$. This concludes the proof that the first statement implies the second.

Conversely, if $\mN\in\cmo(A\to AB)$ is such that $\rho^{AB}=\mN^{A\to AB}(\psi^A)$, then
\be\label{rbh1}
\rho^{B}=\mN^{A\to B}(\psi^A)\;,
\ee
where $\mN^{A\to B}\eqdef\tr_A\circ\mN^{A\to AB}$ is the marginal channel.
Since $\mN^{A\to AB}$ is $A\not\to B$ semi-causal the marginal channel $\mN^{A\to B}$ must satisfy
\ba\label{finito}
\mN^{A\to B}\left(\u^A\right)&=\mN^{A\to B}\left(\psi^A\right)\\
\GG{\eqref{rbh1}}&=\rho^B\;.
\ea
Moreover, since $\mN^{A\to AB}$ is conditionally unital\index{conditionally unital}  we have
\ba
\mN^{A\to AB}\left(I^A\right)&=I^A\otimes\mN^{A\to B}(\u^A)\\
\GG{\eqref{finito}}&=I^A\otimes\rho^B\;.
\ea
Combining everything we get
\ba
I^A\otimes\rho^B-\rho^{AB}&=\mN^{A\to AB}\left(I^A\right)-\mN^{A\to AB}\left(\psi^A\right)\\
&=\mN^{A\to AB}\left(I^A-\psi^A\right)\geq 0\;,
\ea
where the last inequality follows from the fact that $I^A-\psi^A\geq 0$ and $\mN^{A\to AB}$ is a completely positive\index{completely positive} map. This concludes the proof.
\end{proof}

In Theorem~\ref{maxi0} we saw that under conditional majorization $\Phi^{AB}$ is the maximal element of $\md(AB)$. On the other hand, the maximally mixed state $\u^A$ satisfies the opposite inequality that $\rho^{AB}\succ_A\u^A$ for all $\rho\in\md(AB)$. Combining this maximal and minimal elements gives the state
\be
\Phi^{AB}\otimes\u^{\tA}\;.
\ee
Remarkably, the following theorem shows that under conditional majorization, this state is equivalent to any pure state in $\pure(A\tA)$.
\begin{myt}{}
\begin{theorem}\label{paat}
For all $\psi\in\pure(AB)$ with $|B|=|A|$ we have
\be
\psi^{A\tA}\succ_{A\tA}\Phi^{AB}\otimes\u^{\tA}\succ_{A\tA}\psi^{A\tA}\;.
\ee
\end{theorem}
\end{myt}

\begin{proof}
We first prove that $\Phi^{AB}\otimes\u^{\tA}\succ_{A\tA}\psi^{A\tA}$. We will denote by $\tau^{A\tA}$ the density matrix
\be\label{taudeff}
\tau^{A\tA}\eqdef\frac
{I^{A\tA}-\psi^{A\tA}}{m^{2}-1}\;,
\ee
where $m\eqdef|A|$. 
Let $\mN^{A\tA B\to A\tA}$ be a
quantum channel defined by
\begin{equation}
\mN^{A\tA B\to A\tA}  \eqdef\mE^{AB\to A\tA}
\circ\tr_{\tA},
\label{eq:channel3}
\end{equation}
where for all $\omega\in\ml(AB)$
\be
\mE^{AB\to A\tA}(\omega^{AB})   \eqdef\tr[\Phi
^{AB}\omega^{AB}]\psi^{A\tA}
+\tr[(I^{AB}-\Phi^{AB})\omega^{AB}]\tau^{A\tA}\;.
\ee
By definition, the channel above satisfies 
\be
\mN^{A\tA B\to A\tA}\left(\Phi^{AB}\otimes\u^{\tA}\right)=\psi^{A\tA}\;.
\ee
Therefore, it is left to show that the channel $\mN$ is {\rm CMO}. Since the channel $\mN^{A\tA B\to A\tA}$ does not have an output on Bob's side it is trivially $A\not\to B'$ semi-causal ($|B'|=1$). To show that it is conditionally unital\index{conditionally unital}  observe that for any $\sigma\in\md(B)$ we have
\ba
\mN^{A\tA B\to A\tA }\left(I^{A\tA}\otimes\sigma^B\right)&=\mE^{A B\to A\tA }\left(mI^{A}\otimes\sigma^B\right)\\
&=\psi^{A\tA}
+m\tr\left[\left(mI^{B}-\u^{B}\right)\sigma^B\right]\tau^{A\tA}\\
\GG{\eqref{taudeff}}&=I^{A\tA}\;.
\ea
Hence, $\mN\in\cmo(A\tA B\to A\tA)$. This completes the proof that $\Phi^{AB}\otimes\u^{\tA}\succ_{A\tA}\psi^{A\tA}$.

To prove that $\psi^{A\tA}\succ_{A\tA}\Phi^{AB}\otimes\u^{\tA}$, we set $\tau^{AB}\eqdef (I^{AB}-\psi^{AB})/(m^{2}-1)$ and denote by
$\mN^{A\tA \to A\tA B}$  a
quantum channel defined for all
$\omega\in\ml(A\tA)$ as
\be
\mN^{A\tA \to A\tA B}(\omega^{A\tA})   \eqdef\left(\mN^{A\tA \to A B}(\omega^{A\tA})\right)\otimes\u^{\tA}.
\ee
where
\be
\mN^{A\tA \to A B}(\omega^{A\tA})\eqdef\tr[\psi
^{A\tA}\omega^{A\tA}]\Phi^{AB}
+\tr[(I^{A\tA}-\psi^{A\tA})\omega^{A\tA}]\tau^{AB}\;.
\ee
By definition, the channel above satisfies 
\be
\mN^{A\tA \to A\tA B}\left(\psi^{A\tA}\right)=\Phi^{AB}\otimes\u^{\tA}\;.
\ee
Therefore, it is left to show that the channel $\mN$ is {\rm CMO}.  To show that it is conditionally unital\index{conditionally unital}  observe that 
\ba\label{f756}
\mN^{A\tA \to A\tA B}\left(I^{A\tA}\right)&=\left(\Phi^{AB}+(I^{AB}-\Phi^{AB})\right)\otimes\u^{\tA}\\
&=I^{AB}\otimes\u^{\tA}\;.
\ea
To show that it is $A\not\to B$ semi-causal observe that
for all $\omega\in\md(A\tA)$ the marginal channel $\mN^{A\tA\to B}$ satisfies
\ba\label{lasinq}
\mN^{A\tA\to B}\left(\omega^{A\tA}\right)&=\tr[\psi
^{A\tA}\omega^{A\tA}]\u^{B}
+\tr[(I^{A\tA}-\psi^{A\tA})\omega^{A\tA}]\tau^{B}\\
\GG{Exercise~\eqref{exlasinq}}&=\u^B\;,
\ea
which is independent of $\omega^{A\tA}$, so that $\mN^{A\tA\to A\tA B}$ is $A\not\to B$ semi-causal.
Hence, $\mN\in\cmo(A\tA \to A\tA B)$. This completes the proof.
\end{proof}

\bex\label{exlasinq}
Use the definition of $\tau^{AB}$ to verify the first equality in~\eqref{f756} and the second equality in~\eqref{lasinq}.
\eex

\section{Definition of Conditional Entropy}\index{conditional entropy}

In Sec.~\ref{comajo}, we introduced the concept of conditional majorization between bipartite states $\rho\in\md(AB)$ and $\sigma\in\md(AB')$. This relationship is established if a conditionally unital\index{conditionally unital}  and $A\not\to B'$ signalling channel in $\cptp(AB\to AB')$ can transform $\rho^{AB}$ into $\sigma^{AB'}$. Monotonic functions under this pre-order quantify the conditional uncertainty Bob has about Alice's system $A$. Furthermore, if these functions are additive under tensor products, we refer to them as conditional entropies (see the forthcoming definition).

In the classical entropy's definition (see Definition~\ref{entropy}), we require $\H$ to behave monotonically under majorization. Majorization typically involves comparing two probability vectors of equal dimensions, but this can be extended to vectors of differing dimensions by padding the shorter vector with zeros. In the quantum realm, we expand the concept of conditional majorization by introducing additional isometries to account for varying dimensions of Alice's input and output systems.

\begin{myd}{Extension of Conditional Majorization}
Consider $\rho\in\md(AB)$ and $\sigma\in\md(A'B')$, where $|A|$ may differ from $|A'|$. We state that $\rho^{AB}$ conditionally majorizes $\sigma^{A'B'}$, denoted as $\rho^{AB}\succ_A\sigma^{A'B'}$, if for $|A|\geq |A'|$, an isometry $\mV\in\cptp(A'\to A)$ exists such that $\rho^{AB}\succ_{A}\mV^{A'\to A}\left(\sigma^{A'B'}\right)$. Conversely, if $|A'|> |A|$, an isometry $\mU\in\cptp(A\to A')$ exists so that $\mU^{A\to A'}\left(\rho^{AB}\right)\succ_{A'}\sigma^{A'B'}$.
\end{myd}

With this extension, we can now define conditional entropy. Specifically, we consider
\be\label{tff}
\H:\bigcup_{A,B}\md(AB)\to\mathbb{R}\;,
\ee
as a function mapping the set of all bipartite states across finite dimensions to the real line. The function $\H$ assigns to each bipartite density matrix $\rho^{AB}$ a real number, denoted by $\H(A|B)_\rho$. This notation distinguishes conditional entropy from the entropy of the marginal state $\rho^A$.

Our objective is to identify when $\H$ constitutes a conditional entropy. For systems where $|B|=1$, we denote $\H(A|B)_\rho$ as $\H(A)_\rho\eqdef\H(\rho^A)$, aligning with the notation of conditional entropy. This notation proves useful when examining composite systems with multiple subsystems. Since we define entropy functions as non-constant zero functions, we assume (implicitly throughout this book, and in the definition below) the existence of a quantum system $A$ and a state $\rho\in\md(A)$ such that $\H(A)_\rho\neq 0$.

\begin{myd}{Quantum Conditional Entropy}\index{quantum conditional entropy}
\begin{definition}\label{qce}
The function $\H$ in~\eqref{tff} is termed a conditional entropy if for every $\rho\in\md(AB)$ and $\sigma\in\md(A'B')$, it satisfies the following two properties:
\begin{enumerate}
\item  Monotonicity\index{monotonicity}: If $\rho^{AB}\succ_A\sigma^{A'B'}$, then $\H(A|B)_{\rho}\leq\H(A'|B')_{\sigma}$.
\item Additivity: $\H(AA'|BB')_{\rho\otimes\sigma}=\H(A|B)_{\rho}+\H(A'|B')_\sigma$.
\end{enumerate}
\end{definition}
\end{myd}

There are several properties of conditional entropy that follows from the definition above. First, observe that the case that system $B$ is trivial, i.e.\ $|B|=1$, a conditional entropy function reduces to an entropy function. 
Moreover, if $\rho^{AB}=\omega^A\otimes\tau^B$ is a  product state, then it can be converted reversibly to the product state $\omega^A\otimes\u^B$ by a product channel of the form $\id^{A\to A}\otimes\mE^{B\to B}$ which is in $\cmo(AB\to AB)$. Therefore, from the monotonicity property above it follows that 
\be
\H(A|B)_{\omega^A\otimes\tau^B}=\H(A|B)_{\omega^A\otimes\u^B}\;,
\ee 
so that $\H(A|B)_{\rho}$ depends only on $\omega^A$. Moreover, the function $\omega^A\mapsto \H(A|B)_{\omega^A\otimes\u^B}$ satisfies the two axioms of entropy and therefore can be considered itself as an entropy of $\omega^A$. In other words, conditional entropy reduces to entropy on product states as intuitively expected. 

Next, conditional entropy is invariant under the action of local isometric channels. That is, for a bipartite state~$\rho^{AB}$ and isometric channels~$\mathcal{U}^{A\to A'}$ and $\mathcal{V}^{B\to B'}$,
\be\label{secoiu}
    \H(A|B)_{\rho} = \H(A|B')_{\mV(\rho)}=\H(A'|B)_{\mU(\rho)}\;.
\ee
To see the first equality, observe that $\id^A\otimes\mV^{B\to B'}\in\cmo(AB\to AB')$ so that $\H(A|B)_{\rho} \leq  \H(A|B')_{\mV(\rho)}$. For the converse, let $\mV^{-1}\in\cptp(B'\to B)$ be one of the left-inverses of the isometry $\mV^{B\to B'}$ (see for example~\eqref{isocha}). Again, since  $\id^A\otimes\mV^{-1}\in\cmo(AB'\to AB)$ we conclude that
\be
 \H(A|B')_{\mV(\rho)}\leq \H(A|B)_{\mV^{-1}\circ\mV(\rho)}=\H(A|B)_\rho\;.
\ee
Finally, the second equality in~\eqref{secoiu} (i.e., the invariance\index{invariance} under an isometry on system $A$) follows directly from the monotonicity of conditional entropy under the extended version of conditional majorization.

\bex
Prove the second equality in~\eqref{secoiu}. Hint: Use twice the monotonicity of conditional entropy under the extended version of conditional majorization.
\eex

Consider the product state $\u^A \otimes \rho^B$ where $|A|=2$. Similar to what we found previously for unconditional entropy, the following inequality holds
\begin{equation}
\H(A|B)_{\u \otimes \rho}  = \H(A)_{\u}> 0.
\end{equation}
This strict inequality allows us to set a normalization factor for conditional entropy. To be consistent with the normalization convention for unconditional entropy, we set for the case $|A|=2$ that $\H(A|B)_{\u \otimes \rho} = 1$, which in turn implies that for $|A|>2$
\begin{equation}
\H(A|B)_{\u \otimes \rho} = \log_2 |A|\;.
\end{equation}
In the rest of this book we will always assume that $\H$ is normalized in this way.

\begin{exercise}
Let $\H$ be conditional entropy. 
 Show that for every Hilbert spaces $A$ and $B$
\be
\H(A|B)_\rho\leq\log|A| \quad\quad\forall\;\rho\in\md(AB)\;,
\ee
with equality if $\rho^{AB}=\u^A\otimes\tau^B$ for some $\tau\in\md(B)$.
Hint: Find a channel in $\cmo(AB\to AB)$ that takes $\rho^{AB}$ to $\u^A\otimes\rho^B$.
\end{exercise}

\section{Inevitability of Negative Quantum Conditional Entropy} \label{inevitable}\index{quantum conditional entropy} 

In Theorem~\eqref{maxi0} we saw that the maximally entangled state $\Phi^{AB}$ conditionally majorizes all the states in $\md(AB)$. Therefore, from the monotonicity property of the conditional entropy we get that for every $\rho\in\md(AB)$, with $|A|=|B|$, and every conditional entropy function $\H$,
\be
\H(A|B)_\rho\geq\H(A|B)_{\Phi}\;.
\ee
That is, the maximally entangled state has the least amount of conditional entropy. We will see below that $\H(A|B)_{\Phi}$ is negative.

Unlike entropy, quantum conditional entropy can be negative. This unintuitive phenomena puzzled the community for quite some time until an operational interpretation for the quantum conditional entropy was found. This operational interpretation is given in terms of a protocol known as quantum state merging (which we study in volume 2 of this book). In the following theorem we show that certain entangled states \emph{must} have negative conditional entropy, while classical conditional entropy is always non-negative.

The lower bound in the theorem below is given in terms of the conditional min-entropy\index{conditional min-entropy} defined on every bipartite state $\rho\in\md(AB)$ as
\begin{equation}
H_{\min}(A|B)_{\rho}  \coloneqq - \inf_{\lambda \geq 0} \log_2 \{\lambda : \rho_{AB} \leq \lambda I_A \otimes \rho_B\}\;.
\label{eq:cond-min-ent-def}
\end{equation}
In the next subsection we will see that the conditional min-entropy\index{conditional min-entropy} is indeed  a conditional entropy. 
Originally, this quantity was given the name conditional min-entropy because it was known to be the least among all R\'enyi conditional entropies. The theorem below strengthen this observation by proving that all plausible quantum conditional entropies are not smaller than the conditional min-entropy.

\bex\label{condminent}
Consider the conditional min-entropy\index{conditional min-entropy} as defined in~\eqref{eq:cond-min-ent-def}. 
\ben
\item Show that for the maximally entangled state $\Phi^{AB}$ (with $|A|=|B|$) we have
\be
H_{\min}(A|B)_{\Phi}=-\log|A|\;.
\ee
\item Show that if a density matrix $\rho\in\md(AB)$ satisfies $H_{\min}(A|B)_{\rho}=\log|A|$ then $\rho^{AB}=\u^A\otimes\rho^B$.
\item Show that a state $\rho\in\md(AB)$ has non-negative conditional min-entropy\index{conditional min-entropy} if and only if $I^A\otimes\rho^B\geq\rho^{AB}$.
\item Show that if $\rho^{AB}$ is separable then its conditional min-entropy is non-negative.
\een
\eex

\begin{myt}{}
\begin{theorem}
\label{thm:neg-cond-ent}
Let $\H$ be a quantum conditional entropy\index{quantum conditional entropy} . For all $\rho\in\md(AB)$,
\begin{equation}
\H(A|B)_\rho\geq H_{\min}(A|B)_\rho\;,
\label{eq:main-thm}
\end{equation}
with equality if $\rho^{AB}$ is the maximally entangled state $\Phi^{AB}$.
\end{theorem}
\end{myt}

\begin{remark}
The theorem above states that for the maximally entangled state $\Phi^{AB}$ we have $\H(A|B)_\Phi= H_{\min}(A|B)_\Phi$. Combining this with Exercise~\ref{condminent} we conclude that
\be\label{loga}
\H(A|B)_\Phi=-\log |A|\;.
\ee
That is, \emph{all} conditional entropies are negative on the maximally entangled state and equal to $-\log|A|$. Moreover, in conjunction with the third part of Exercise~\ref{condminent}, the theorem above implies that all conditional entropies are non-negative on separable states (and therefore also on classical states).
\end{remark}

\begin{proof}
Let's start by examining the scenario where $H_{\min}(A|B){\rho}\geq0$, and denote $m\eqdef|A|$. The proof strategy in this case revolves around identifying the largest integer $k\in[m]$ such that a classical system $X$, with dimension $|X|=k$, fulfills the condition $\u^X\succ_A\rho^{AB}$. Once we establish this optimal value of $k$, we can infer that every conditional entropy $\H$ must satisfy
\be
\H(A|B)_\rho\geq H(X)_\u=\log k\;.
\ee
We begin by establishing that it is feasible to set $k \eqdef \left\lfloor 2^{H_{\min}(A|B)_{\rho}}\right\rfloor$.

Let $X$ be a classical system with dimension $k\eqdef \left\lfloor 2^{H_{\min}(A|B)_{\rho}}\right\rfloor$. By the assumption that
$H_{\min}(A|B)_{\rho}\geq0$, and given the dimension bound $H_{\min
}(A|B)_{\rho}\leq\log_{2}m$, it follows that $k\in\left[  m\right]  $. Observe that the case $k=m$ implies that $H_{\min}(A|B)_{\rho}=\log|A|$. In this case, according to the second part of Exercise~\ref{condminent} we must have $\rho^{AB}=\u^A\otimes\rho^B$ so that $\H(A|B)_{\rho}=\log|A|=H_{\min}(A|B)_{\rho}$.

We therefore assume now that $k<m$. We look for a  channel $\mN\in\cmo(A\to AB)$ that satisfies
\be\label{wantrab}
\mN^{A\to AB}\left(\frac1k\Pi^A\right)=\rho^{AB}\;,
\ee
where $\Pi^{A}$ is a projection onto a $k$-dimensional subspace of $A$ (i.e., set $\Pi
^{A}\coloneqq \sum_{x\in[k]}|x\rangle\!\langle x|^{A}$, where $\{|x\rangle\}_{x\in[m]}$ is some
orthonormal basis of $A$). The existence of such a channel will prove that $\u^X\succ_A\rho^{AB}$ since $\u^X$ (with $|X|=k$) is equivalent to $\frac1k\Pi^A$ under conditional majorization. We choose $\mN^{A\to AB}$ to be a measure-and-prepare channel of the form
\be
\mathcal{N}^{A\rightarrow AB}(\omega^{A})\coloneqq \operatorname{Tr}[\Pi^A\omega^{A}]\rho^{AB}
+\operatorname{Tr}[\left(  I^{A}-\Pi^{A}\right)  \omega^{A}
]\tau^{AB}\quad\quad\forall\;\omega\in\ml(A)\;,
\label{eq:ch-def-1}
\ee
where $\tau^{AB}$ is some density matrix that is chosen (see below) such that $\mN^{A\to AB}$ is {\rm CMO}. Indeed, the action of this channel is to perform a measurement according to
the POVM\ $\left\{  \Pi^{X},I^{X}-\Pi^{X}\right\}  $ and prepare the state
$\rho^{AB}$ if the first outcome is obtained and the state $\tau^{AB}$
if the second outcome is obtained. By definition, this channel satisfies~\eqref{wantrab}.

The channel $\mN^{A\to AB}$ is $A\not\to B$ signalling if  and only if the marginal channel $\mN^{A\to B}\eqdef\tr_{A}\circ\mN^{A\to AB}$ satisfies $\mN^{A\to B}\circ\mM^{A\to A}=\mN^{A\to B}$ for all $\mM\in\cptp(A\to A)$. In other words, $\mN^{A\to AB}$ is $A\not\to B$ signalling if and only if the marginal channel $\mN^{A\to B}$ is a replacement channel. Now,
for every $\omega\in\md(A)$ we have that 
\be\label{stammis2}
\mN^{A\to B}(\omega^A)=\operatorname{Tr}[\Pi^A\omega^{A}]\rho^{B}
+\operatorname{Tr}[\left(  I^{A}-\Pi^{A}\right)  \omega^{A}
]\tau^{B}\;.
\ee
Therefore, by taking $\tau^{AB}$ to have the property that its marginal $\tau^B=\rho^B$,  we get that the right-hand side does not depend on $\omega^A$, so that $\mN^{A\to AB}$ is $A\not\to B$ semi-causal.

The channel $\mN^{A\to AB}$ is conditionally unital\index{conditionally unital}  if and only if the state
\ba
\mathcal{N}^{A\rightarrow AB}(I^{A})&=\operatorname{Tr}[\Pi^{A}I^{A}
]\rho^{AB}+\operatorname{Tr}[\left(  I^{A}-\Pi^{A}\right)  I^{A}]\tau^{AB} \\ 
&=k\rho^{AB}+\left(  m-k\right)  \tau^{AB}
\ea
is equal to the state
\be
I^{A}\otimes\mathcal{N}^{A\rightarrow B}(\u^{A})=I^A\otimes\rho^B\;,
\ee
where the we used \eqref{stammis2} with $\tau^B=\rho^B$. The equality between the two states above forces $\tau^{AB}$ to be
\begin{equation}\label{tauabz}
\tau^{AB}\coloneqq \frac{I^{A}\otimes\rho^{B}-k\rho^{AB}}{m-k}.
\end{equation}
The operator $\tau^{AB}$ is positive semi-definite because $m-k>0$ and
\ba
 \frac{1}{k}I^{A}\otimes\rho^{B}-\rho^{AB}& \geq2^{-H_{\min}(A|B)_{\rho}}I^{A}\otimes\rho^{B}-\rho^{AB}\\
\GG{\eqref{eq:cond-min-ent-def}}& \geq0.
\ea
Also, $\tau^{AB}$ has trace equal to one (so it is a density matrix) with marginal $\tau^B=\rho^B$. We therefore proved that
\be\label{7p228}
\H(A|B)_\rho\geq \log k=\log\left\lfloor 2^{H_{\min}(A|B)_{\rho}}\right\rfloor\;.
\ee
Finally, since all conditional entropies are additive for tensor-product states, we conclude that
\ba
\H(A|B)_{\rho}&=\lim_{n\to\infty}\frac1n\H(A^n|B^n)_{\rho^{\otimes n}}\\
\GG{\eqref{7p228}}&\geq \lim_{n\to\infty}\frac1n\log\left\lfloor2^{H_{\min}(A^n|B^n)_{\rho^{\otimes n}}}\right\rfloor\\
&=\lim_{n\to\infty}\frac1n\log\left\lfloor2^{nH_{\min}(A|B)_{\rho}}\right\rfloor\\
&=H_{\min}(A|B)_\rho\;.
\label{eq:alt-cond-ent-main-thm-proof-1-last}
\ea

Next, consider the case $H_{\min}(A|B)<0$. The idea of the proof in this case is to find the largest possible $k\in\mbb{N}$ such that the system $X$ (can be taken to be a classical system) with dimension $|X|=k$ satisfies
\be\label{comajoo}
\psi^{XA}\succ_{AX}\rho^{A}\otimes\u^X\;,
\ee
where $\psi\in\pure(XA)$ is some pure state. Due to the monotonicity property of every conditional entropy $\H$ the above relation implies that
\ba
0&=\H(XA)_{\psi}\\
\GG{\eqref{comajoo}}&\leq \H(XA|B)_{\rho\otimes\u}\\
\GG{Additivity}&=\H(X)_\u+\H(A|B)_\rho\;.
\ea
Finally, since $\H(X)_\u=\log k$ we get that $\H(A|B)_\rho\geq -\log(k)$. We first show that~\eqref{comajoo} holds with $k\eqdef\left\lceil 2^{-H_{\min}(A|B)}\right\rceil$. 

To prove the relation~\eqref{comajoo}, consider the measure-and-prepare channel $\mN\in\cptp(AX\to AXB)$ defined on all $\omega\in\ml(XA)$ as
\be
\mN^{XA\to XAB}\left(\omega^{XA}\right)\eqdef\tr\left[\psi^{XA}\omega^{XA}\right]\u^X\otimes\rho^{AB}
+\tr\left[\Pi^{XA}  \omega^{XA}\right]\tau^{XAB}\;,
\ee
where $\Pi^{AX}\eqdef I^{AX}-\psi^{XA}$ and $\tau^{XAB}$ is chosen (see below) such that $\mN$ is {\rm CMO}. Observe that by definition we have
\be
\mN^{XA\to XAB}\left(\psi^{XA}\right)=\u^X\otimes\rho^{AB}\;.
\ee
We next show that there exists $\tau\in\md(XAB)$ such that $\mN$ as defined above is indeed {\rm CMO}. If $\mN$ is $XA\not\to B$ semi-causal then we must have that the marginal channel $\mN^{XA\to B}\eqdef\tr_{XA}\circ\mN^{XA\to XAB}$ is a replacement channel (i.e., a constant channel). Now, for all $\omega\in\md(XA)$
\be\label{stammis3}
\mN^{XA\to B}\left(\omega^{XA}\right)=\tr\left[\psi^{XA}\omega^{XA}\right]\rho^{B}
+\tr\left[\Pi^{XA}  \omega^{XA}\right]\tau^{B}\;.
\ee
Thus, by choosing $\tau^{XAB}$ to have the property $\tau^B=\rho^B$, we get that the right-hand side of the equation above does not depend on $\omega$, so that $\mN^{XA\to XAB}$ is $XA\not\to B$ semi-causal. 

Next, the channel $\mN^{XA\to XAB}$ is conditionally unital if the operator
\be
\mN^{XA\to XAB}\left(I^{XA}\right)=\u^X\otimes\rho^{AB}
+km\tau^{XAB}
\ee
equals the operator
\be
I^{XA}\otimes \mN^{XA\to B}\left(\u^{XA}\right)=I^{XA}\otimes\tau^{B}\;,
\ee
where the last equality follows from~\eqref{stammis3} with  $\tau^B=\rho^B$. We therefore conclude that $\mN\in\cptp(XA\to XAB)$ as defined above is CMO if and only if $\tau^{XAB}$ equals
\be\label{tauxab}
\tau^{XAB}\eqdef\u^X\otimes\frac{kI^{A}\otimes\rho^{B}-\rho^{AB}}{km-1}.
\ee
Observe that this $\tau^{XAB}$ is indeed a density matrix since by definition of $k$ we have 
\ba
kI^{A}\otimes\rho^{B}&\geq 2^{-H_{\min}(A|B)}I^{A}\otimes\rho^{B}\\
\GG{\eqref{eq:cond-min-ent-def}}&\geq\rho^{AB}\;.
\ea
Moreover, from its definition in~\eqref{tauxab} we have $\tau^B=\rho^B$. Hence, with this $\tau^{XAB}$ the channel $\mN^{XA\to XAB}$ is {\rm CMO} that maps the pure states $\psi^{XA}$ to the state $\u^X\otimes\rho^{AB}$. We therefore conclude that
\be\label{7p239}
\H(A|B)_\rho\geq -\log k=-\log\left\lceil 2^{-H_{\min}(A|B)}\right\rceil\;.
\ee
Finally, from the additivity\index{additivity} property of conditional entropies we get
\ba
\H(A|B)_{\rho}&=\lim_{n\to\infty}\frac1n\H(A^n|B^n)_{\rho^{\otimes n}}\\
\GG{\eqref{7p239}}&\geq- \lim_{n\to\infty}\frac1n\log\left\lceil2^{-H_{\min}(A^n|B^n)_{\rho^{\otimes n}}}\right\rceil\\
&=-\lim_{n\to\infty}\frac1n\log\left\lceil2^{-nH_{\min}(A|B)_{\rho}}\right\rceil\\
&=H_{\min}(A|B)_\rho\;.
\ea

It is left to prove the equality on maximally entangled states.
Since conditional entropy is invariant under local isometries we can assume without loss of generality  that $m \eqdef |A| = |B|$. From Theorem~\ref{paat} we know that the state $\Phi^{AB} \otimes \mathbf{u}^{\tA}$ 
is equivalent under conditional majorization to any pure state in $\pure(A\tA)$. Since the entropy of every pure state in $\pure(A\tA)$ is zero, we conclude that
\ba
    0&=\H(A\tA|B)_{\Phi \otimes \mathbf{u}} \\
    \GG{Additivity}&= \H(A|B)_{\Phi} + \H(\tA )_{\mathbf{u}} \\
    &= \H(A|B)_{\Phi} + \log m\;,
    \ea
where we used the fact that the entropy of the uniform state $\mathbf{u}^{\tA}$ is $\log_2 m$. Hence,
\be
 \H(A|B)_{\Phi}=-\log m= H_{\min}(A|B)_{\Phi}
\ee
This completes the proof.
\end{proof}

We saw in the theorem above that the conditional entropy is positive for all separable states. This does no mean that the conditional entropy is positive just for separable states. In fact, some entangled states (i.e. states that are not separable) have positive conditional entropy. The following corollary provides a simple criterion to determine if a bipartite state has a positive conditional entropy.
\begin{myg}{}
\begin{corollary}\label{rccor}
Let $\rho\in\md(AB)$. Then, the following statements are equivalent:
\ben
\item For any choice of conditional entropy $\H$ we have
$
\H(A|B)_\rho\geq 0
$.
\item The state $\rho^{AB}$ satisfies
\be
I^A\otimes\rho^B\geq\rho^{AB}\;.
\ee
\item For every pure state $\psi\in\pure(A)$
\be\label{respsia}
\psi^A\succ_A\rho^{AB}\;.
\ee
\item The state $\rho^{AB}$ can be obtained by {\rm CMO} channel from a classical distribution; i.e.\ there exists $\omega\in\md(XY)$ such that
\be
\omega^{XY}\succ_A\rho^{AB}\;.
\ee
\een
\end{corollary}
\end{myg}

\begin{proof}
The proof follows directly from Theorem~\ref{thm:neg-cond-ent} in conjunction with Theorem~\ref{strong}.
\end{proof}

As an example, consider the Werner state\index{Werner state} (cf.~\eqref{twirling})
\be
\rho^{AB}_t=t\frac{2}{d(d+1)}\Pi_\sym^{AB} +(1-t)\frac{2}{d(d-1)}\Pi_\asy^{AB}\quad\quad\forall\;t\in[0,1]\;,
\ee
where $\Pi_\sym^{AB}$ and $\Pi_\asy^{AB}$ are the projections, respectively, to the symmetric and anti-symmetric subspaces of $AB$. We will see in Chapter~\ref{entanglement} that these states are entangled if and only if $t<\frac12$. Moreover, observe that the Werner states\index{Werner state} have uniform marginals, particularly, $\rho_t^B=\u^B$ for all $t\in[0,1]$. In addition, the Werner states have only two distinct eigenvalues  given by $\frac{2t}{d(d+1)}$ and $\frac{2(1-t)}{d(d-1)}$. Combining all this information we get that 
\be
I^A\otimes\rho^{B}_t\geq\rho^{AB}_t\quad\iff\quad \frac{3-d}2\leq t\leq\frac{d+1}2\;.
\ee
For $d=2$ this condition holds only for $t\in[\frac12,1]$ in which case $\rho^{AB}_t$ is separable. On the other hand,
for $d\geq 3$ this condition holds for all $t\in[0,1]$.  We therefore conclude that for $d\geq 3$ all Werner states (including the entangled ones) have positive conditional entropy.

\bex
Use Exercise~\ref{exrxsy} to show that in the classical domain, every entropy function $\H$ is also a conditional entropy; that is, show that function
\be
H(X|Y)_\rho\eqdef H(X)_\rho\quad\quad\forall\;\rho\in\md(XY)\;,
\ee
is a conditional entropy of classical states. Moreover, give a counter example to the same statement in the quantum domain.
\eex

\subsection{Negativity Precludes Extensions from the Classical Domain}\index{negativity}

In Sections~\ref{optim} and~\ref{optim2}, we examined the expansion of quantum divergences and relative entropies from the classical to the quantum domain. This process involved a systematic approach to their extensions, incorporating both minimal and maximal forms. This subsection focuses on exploring optimal extensions of classical conditional entropy to the quantum domain. However, we will show that such extensions are not feasible, primarily because quantum conditional entropies exhibit negativity when applied to the maximally entangled state.

Consider a classical conditional entropy $\H$. Following the methodologies in Sections~\ref{optim} and~\ref{optim2}, for any $\rho\in\md(AB)$, we can define the minimal and maximal extensions of $\H$ as:
\ba
&\underline{\H}(A|B)_\rho\eqdef\sup\left\{\H(X|Y)_\omega\;:\;\omega^{XY}\succ_X\rho^{AB}\;,\;\;\omega\in\md(XY)\right\}\\
&\overline{\H}(A|B)_\rho\eqdef\inf\left\{\H(X|Y)_\omega\;:\;\rho^{AB}\succ_A\omega^{XY}\;,\;\;\omega\in\md(XY)\right\}\;.
\ea
At first glance, these functions appear quite reasonable. For example, for any two states $\rho,\sigma\in\md(AB)$ with $\rho^{AB}\succ_A\sigma^{AB}$, if $\omega^{XY}\succ_X\rho^{AB}$, then it necessarily follows that $\omega^{XY}\succ_X\sigma^{AB}$. Consequently,
\ba
\underline{\H}(A|B)_\rho&\leq \sup\left\{\H(X|Y)_\omega\;:\;\omega^{XY}\succ_X\sigma^{AB}\;,\;\;\omega\in\md(XY)\right\}\\
&=\underline{\H}(A|B)_\sigma\;.
\ea
This means $\underline{\H}(A|B)_\rho$ exhibits monotonic behavior under conditional majorization, aligning with expectations for a measure of conditional uncertainty.

However, in general, $\underline{\H}(A|B)_\rho$ is not well defined! This is because CMO channels form a subset of one-way LOCC and thus cannot generate entanglement. Since $\omega\in\md(XY)$ is classical and hence separable, any state $\mN(\omega)$ resulting from a one-way LOCC channel $\mN\in\cmo(XY\to AB)$ lies within $\sep(AB)$. Therefore, $\underline{\H}(A|B)_\rho$ is undefined if $\rho^{AB}$ is entangled.

\bex
Show that $\underline{\H}(A|B)_\rho$ is well defined if and only if $I^A\otimes\rho^B\geq\rho^{AB}$.
Hint: Recall Corollary~\ref{rccor}.
\eex

In contrast to $\underline{\H}(A|B)\rho$, the quantity $\overline{\H}(A|B)\rho$ is well-defined for all $\rho\in\md(AB)$. It also exhibits monotonic behavior under conditional majorization. Specifically, consider two states $\rho,\sigma\in\md(AB)$, where $\rho^{AB}\succ_A\sigma^{AB}$. Then, if $\sigma^{AB}\succ_X\omega^{XY}$ for some $\omega\in\md(XY)$ then $\rho^{AB}\succ_X\omega^{XY}$ is necessarily true. Hence,
\ba
\underline{\H}(A|B)_\sigma&\geq \inf\left\{\H(X|Y)_\omega\;:\;\rho^{AB}\succ_X\omega^{XY}\;,\;\;\omega\in\md(XY)\right\}\\
&=\underline{\H}(A|B)_\rho\;.
\ea
Therefore, $\underline{\H}$ qualifies as a measure of conditional uncertainty. However, since it is defined via an optimization problem, it exhibits only sub-additivity. Consequently, we introduce its regularized version:
\be
\underline{\H}^\reg(A|B)_\rho\eqdef\lim_{n\to\infty}\frac1n \underline{\H}\left(A^n\big|B^n\right)_{\rho^{\otimes n}}\;.
\ee
From its definition, $\underline{\H}^\reg$ satisfies
\be
\underline{\H}^\reg\left(A^n\big|B^n\right)_{\rho^{\otimes n}}=n\underline{\H}^\reg(A|B)_{\rho}\;.
\ee
Furthermore, this measure exhibits monotonic behavior under conditional majorization and is inherently non-negative by definition. Does this not contradict Theorem~\ref{}? The answer is no. The key lies in understanding that $\underline{\H}^\reg$ is only weakly additive, rather than strongly additive, in general.

\bex
Calculate $\underline{\H}^\reg(A\tA|B)_\rho$, where $\rho$ is defined as $\Phi^{AB}\otimes\u^{\tA}$. Question: Does this result in a value of zero?
\eex

Interestingly, the function $\underline{\H}^\reg$ serves as an example of a function meeting all the criteria expected of a quantum conditional entropy\index{quantum conditional entropy} , except it is only weakly additive. Its non-negativity teaches us an important lesson: the tendency of quantum conditional entropies to assume negative values on certain entangled states is intrinsically connected to their property of full additivity.

\section{Conditional Entropies from Relative Entropies}\index{condition entropy}\index{relative entropy}

Any quantum relative entropy can be used to define conditional entropy. In fact, for a given quantum relative entropy $\D$, there are two candidates for conditional entropy given by:
\ba\label{coin1}
& \H(A|B)_\rho\eqdef\log|A|-\D\left(\rho^{AB}\big\|\u^A\otimes\rho^B\right)\\
& \H^\ua(A|B)_\rho\eqdef\log|A|-\min_{\sigma\in\md(B)}\D\left(\rho^{AB}\big\|\u^A\otimes\sigma^B\right)
\ea
The up arrow in the notation above indicates the optimization over $\sigma^B$. By definition,  $\H^\ua(A|B)_\rho\geq \H(A|B)_\rho$ for all $\rho\in\md(AB)$.

\begin{myt}{}
\begin{theorem}
Let $\D$ be a quantum relative entropy. Then, the function $
\H$ as defined in~\eqref{coin1}, is a quantum conditional entropy\index{quantum conditional entropy} .
\end{theorem}
\end{myt}

\begin{proof}
First, we demonstrate that $\H$ satisfies the monotonicity property of conditional entropy. Let $\mathcal{N}^{AB \rightarrow A'B'}$ be a {\rm CMO}, and consider the bipartite density matrix $\rho^{AB}$. We begin by considering the case where $A = A'$ so that
\be\label{subintoi}
    \H(A \big| B)_{\mathcal{N}(\rho)} = 
    \log |A| - \D\left(\mathcal{N}(\rho^{AB}) \big\| \u^{A} \otimes \tr_{A}\!\left[\mN\left(\rho^{AB}\right)\right]\right) \;.
\ee
Since $\mN$ is $A\not\to B'$ semi-causal the marginal channel $\mN^{AB\to B'}\eqdef\tr_{A}\circ\mN^{AB\to AB'}$ satisfies 
\be
\mN^{AB\to B'}\left(\rho^{AB}\right)=\mN^{AB\to B'}\left(\u^A\otimes\rho^{B}\right)\;.
\ee
To see this, take $\mM^{A\to A}$ in~\eqref{semic} to be the completely randomizing channel. With this at hand, we get
\begin{align}
    \mathbf{u}^{A} \otimes \tr_{A}\!\left[\mathcal{N}\left(\rho^{AB}\right) \right] &= \mathbf{u}^{A} \otimes \tr_{A}\!\left[\mN\left(\u^{A} \otimes \rho^{B}\right)\right] \\
    \GG{\mN\text{ is conditionally unital}}&= \mN\left(\u^{A} \otimes \rho^{B}\right)\;.
\end{align}
Substituting this into~\eqref{subintoi} we obtain 
\begin{align}
    \H(A \big| B)_{\mN(\rho)} &= \log |A| - \D\left(\mN\left(\rho^{AB}\right) \big\| \mN\left(\u^{A} \otimes \rho^{B}\right) \right) \notag \\
    \GG{DPI}&\geq \log |A| - \D\left(\rho^{AB} \big\| \u^{A} \otimes \rho^{B} \right)\\
    &=\H(A|B)_\rho \;.
\end{align}

We also need to prove that $\H$ is invariant under the action of a local isometric channel acting on system~$A$. 
Let $\mV\in\cptp(A\to A')$ be an isometry channel, $\rho\in\md(AB)$, and $C$ be a Hilbert space of dimension $|C|=|A'|-|A|$. 
Observe that
\be\label{seq0}
\H(A|B)_{\mV(\rho)}=\log|A'|-\D\left(\mV^{A\to A'}\left(\rho^{AB}\right)\Big\|\u^{A'}\otimes \rho^{B}\right)
\ee
Clearly, if $\mV$ is a unitary channel (i.e. $A\cong A'$) we have $\H(A'|B)_{\mV(\rho)}=\H(A|B)_{\rho}$ since in this case $|A|=|A'|$ and $\mV(\u^A)=\u^{A'}$.
We can therefore assume without loss of generality that 
\be\label{seq1}
\mV^{A\to A'}\left(\rho^{AB}\right)=\rho^{AB}\oplus\0^{CB}\eqdef\begin{pmatrix}\rho^{AB} & \0\\
\0 & \0^{CB}\end{pmatrix}
\ee 
since the conditional entropy of $\mV^{A\to A'}\left(\rho^{AB}\right)$ does not change by a unitary channel on $A'$.
Moreover,
denote by $t\eqdef\frac{|A|}{|A'|}$ and observe that $\u^{A'}$ can be expressed as 
\be\label{seq2}
\u^{A'}=t\u^A\oplus(1-t)\u^C\;.
\ee
Hence, substituting~\eqref{seq1} and~\eqref{seq2} into~\eqref{seq0} gives
\ba
\H(A|B)_{\mV(\rho)}&=\log|A'|-\D\left(\rho^{AB}\oplus\0^{CB}\Big\|\left(t\u^A\otimes \rho^{B}\right)\oplus\left((1-t)\u^C\otimes \rho^{B}\right)\right)\\
\GG{Exercise~\ref{extsig}}&=\log|A'|-\D\left(\rho^{AB}\big\|\u^A\otimes \rho^{B}\right)+\log t\\
\Gg{t=\frac{|A|}{|A'|}}&=\H(A|B)_\rho\;.
\ea

To prove the additivity\index{additivity} property, let $\rho\in\md(A_1B_1)$ and $\sigma\in\md(A_2B_2)$ and observe that since $\u^{A_1A_2}=\u^{A_1}\otimes\u^{A_2}$ we have
\ba
\H(A_1A_2|B_1B_2)_{\rho\otimes\sigma}&=\log|A_1A_2|-\D\left(\rho^{A_1B_1}\otimes\sigma^{A_2B_2}\Big\|\u^{A_1}\otimes\rho^{A_1B_1}\otimes\u^{A_2}\otimes\sigma^{A_2B_2}\right)\\
\GG{\D\;is\;additive}&=\H(A_1|B_1)_{\rho}+\H(A_2|B_2)_{\sigma}\;.
\ea
Finally, the normalization property follows from the fact that when $|B|=1$ and $|A|=2$ we have by definition
\be
\H(A|B)_{\u}=\log 2-\D\left(\u^A\big\|\u^A\right)=1\;.
\ee
This completes the proof.
\end{proof}

\bex
Show that if $\D=D_{\max}$ then its corresponding conditional entropy $\H$ is the conditional min-entropy.
\eex

In the exercise below you will show that $H^\ua$  behaves monotonically under conditionally unital\index{conditionally unital}  channels and consequently behaves monotonically under conditional majorization. However, in general, $H^\ua$ does not necessarily satisfy the additivity property (at least a general proof of additivity of $\H^\ua$ is unknown to the author). Still, this expression has been used extensively by the community, particularly since it can be shown that for $\D=D_\alpha$ or $\D=\tD_{\alpha}$ it is additive (here $\tD_\alpha$ is the sandwiched R\'enyi divergence; see Definition~\ref{def:sandwich}). 
This includes the Umegaki relative entropy, however, as we will see shortly, in this case  $H(A|B)_\rho=H^\ua(A|B)_\rho=H(A|B)_\rho$ for all $\rho\in\md(AB)$.

\bex\label{ex:optex}
Consider the function $\H^\ua$ as defined above.
\ben
\item Show that $\H^\ua$ does not increase under conditionally unital\index{conditionally unital}  channels, and use it to conclude that it satisfies the monotonicity property of conditional entropy.
\item Prove that $\H^\ua$ satisfies the invariance and normalization property of conditional entropy.
\een
\eex

\section{Examples of Quantum Conditional Entropies}\index{quantum conditional entropy}

\subsubsection{The von-Neumann Conditional Entropy}\index{von-Neumann}

Let $D$ be the Umegaki relative entropy. The conditional entropy with respect to this divergence is defined by
\be\label{8267}
H(A|B)_\rho\eqdef\log|A|-D\left(\rho^{AB}\big\|\u^A\otimes\rho^B\right)\;.
\ee
To simplify the expression above, observe that
\ba
D\left(\rho^{AB}\big\|\u^A\otimes\rho^B\right)&=\tr\left[\rho^{AB}\log\rho^{AB}\right]-\tr\left[\rho^{AB}\log\left(\u^A\otimes\rho^B\right)\right]\\
&=-H(\rho^{AB})+\log|A|-\tr\left[\rho^{AB}\log\left(I^A\otimes\rho^B\right)\right]\\
\Gg{\log(I^A\otimes\rho^B)=I^A\otimes\log\rho^B}&=-H(\rho^{AB})+\log|A|-\tr\left[\rho^{B}\log\rho^B\right]\\
&=H(\rho^B)-H(\rho^{AB})+\log|A|\;.
\ea
Therefore, the conditional von-Neumann\index{von-Neumann} entropy can be expressed simply as
\be
H(A|B)_\rho=H\left(\rho^{AB}\right)-H\left(\rho^B\right)\;.
\ee
This formula is consistent with the intuition of conditional entropy as depicted in Fig.~\ref{Venn}. 

The finding from the previous section, which establishes that conditional entropy is non-negative for separable states, leads to an intriguing implication regarding the von Neumann entropy.

\begin{myg}{}
\begin{corollary}\label{cor:ubent}
Let $\{p_x,\rho_x\}_{x\in[m]}$ be an ensemble of quantum states in $\md(A)$. The von-Neumann\index{von-Neumann} entropy satisfies
\be\label{convone}
\sum_{x\in[m]}p_xH(\rho_x)\leq H\Big(\sum_{x\in[m]}p_x\rho_x\Big)\leq \sum_{x\in[m]}p_xH(\rho_x)+H(\p)
\ee
\end{corollary}
\end{myg}
\begin{proof}
The lower bound follows from the concavity of $H$ (see Exercise~\ref{ex:vn}).
To get the upper bound, let $\rho^{XA}\eqdef\sum_{x\in[m]}p_x|x\lr x|^X\otimes\rho_x^A$. Since $\rho^{XA}$ is a cq-state, and in particular separable, it follows that
\be
0\leq H(X|A)_\rho=H(\rho^{AX})-H(\rho^A)\;,
\ee
where $\rho^A=\sum_{x\in[m]}p_x\rho_x^A$ is the marginal state. Therefore,
\ba
H\Big(\sum_{x\in[m]}p_x\rho_x\Big)&\leq H(\rho^{AX})\\
\GG{Exercise~\eqref{ex:ubent}}&=\sum_{x\in[m]}p_xH(\rho_x)+H(\p)\;.
\ea
This completes the proof.
\end{proof}

\bex\label{ex:ubent}
Using the same notations as in the proof above, show that
\be
H(\rho^{AX})=\sum_xp_xH(\rho_x)+H(\p)\;.
\ee
\eex

\begin{exercise}\label{ex:subadditive}
The quantum mutual information\index{mutual information} is a quantity defined for any $\rho\in\md(AB)$ as
\be
I(A:B)_\rho\eqdef D\left(\rho^{AB}\big\|\rho^A\otimes\rho^B\right)\;.
\ee
\ben
\item Express the quantum mutual information in terms of $H(A|B)_\rho$ and $H(A)_\rho$.
\item Show that the von-Neumann\index{von-Neumann} entropy is subadditive; i.e. prove that for all $\rho\in\md(AB)$
\be
H(\rho^{AB})\leq H(\rho^A)+H(\rho^B)\;.
\ee 
Hint: Use the fact that the mutual information is non-negative.
\een
\end{exercise}

The von-Neumann\index{von-Neumann} conditional entropy behaves monotonically even under conditionally unital channels that are not $A\not\to B$ semi-causal. To see this, we first prove the following type of triangle equality of the Umegaki relative entropy.
\begin{myg}{Triangle Equality}\index{triangle inequality}
\begin{lemma}\label{lem:te}
Let $D$ be the Umegaki relative entropy. Then for any $\rho\in\md(AB)$, $\sigma,\tau\in\md(B)$ and $\omega\in\md(A)$, we have
\be
D\left(\rho^{AB}\big\|\omega^A\otimes\sigma^B\right)=D\left(\rho^{AB}\big\|\omega^A\otimes\rho^B\right)+D\left(\rho^{B}\big\|\sigma^B\right)\;.
\ee
\end{lemma}
\end{myg}
\begin{remark}
The relation in the lemma above is equivalent to
\be
D\left(\rho^{AB}\big\|\omega^A\otimes\sigma^B\right)=D\left(\rho^{AB}\big\|\omega^A\otimes\rho^B\right)+D\left(\omega^A\otimes \rho^{B}\big\|\omega^A\otimes \sigma^B\right)
\ee
which explains why we view this relation as a type of triangle equality.
\end{remark}
\begin{proof}
By definition
\be
D\left(\rho^{AB}\big\|\omega^A\otimes\sigma^B\right)=-H\left(\rho^{AB}\right)-\tr\left[\rho^{AB}\log\left(\omega^A\otimes\sigma^B\right)\right]\;.
\ee
Using the property that
\be
\log\left(\omega^A\otimes\sigma^B\right)=\log\omega^A\otimes I^B+I^A\otimes\log\sigma^B\;,
\ee
we get by direct calculation
\ba
D\left(\rho^{AB}\big\|\omega^A\otimes\sigma^B\right)&=-H\left(\rho^{AB}\right)-\tr\left[\rho^{A}\log\omega^A\right]-\tr\left[\rho^{B}\log\sigma^B\right]\\
&=-H\left(\rho^{AB}\right)-\tr\left[\rho^{A}\log\omega^A\right]-\tr\left[\rho^{B}\log\rho^B\right]+D\left(\rho^B\big\|\sigma^B\right)\\
&=-H\left(\rho^{AB}\right)-\tr\left[\rho^{AB}\log\left(\omega^A\otimes\rho^B\right)\right]+D\left(\rho^B\big\|\sigma^B\right)\\
&=D\left(\rho^{AB}\big\|\omega^A\otimes\rho^B\right)+D\left(\rho^{B}\big\|\sigma^B\right)\;.
\ea
This completes the proof.
\end{proof}

Note that by taking in the lemma above $\omega^A=\u^A$ we get that 
\be\label{drabz}
D\left(\rho^{AB}\big\|\u^A\otimes\sigma^B\right)=D\left(\rho^{AB}\big\|\u^A\otimes\rho^B\right)+D\left(\rho^{B}\big\|\sigma^B\right)\;.
\ee
Therefore, 
\ba\label{8281}
H^\ua(A|B)_\rho&\eqdef\log|A|-\min_{\sigma\in\md(B)}D\left(\rho^{AB}\big\|\u^A\otimes\sigma^B\right)\\
\GG{\eqref{drabz}}&=\log|A|-D\left(\rho^{AB}\big\|\u^A\otimes\rho^B\right)-\min_{\sigma\in\md(B)}D\left(\rho^{B}\big\|\sigma^B\right)\\
&=\log|A|-D\left(\rho^{AB}\big\|\u^A\otimes\rho^B\right)\\
&=H(A|B)_\rho\;.
\ea
The equality $H(A|B)_\rho=H^\ua(A|B)_\rho$ reveals that the von-Neumann\index{von-Neumann} conditional entropy is monotonic under conditionally unital channels that are not necessarily $A\not\to B$ semi-causal (see part 1 of Exercise~\ref{ex:optex}). 

Another useful property satisfied by the von-Neumann entropy is known as the strong subadditivity\index{subadditivity} property. Recall from Exercise~\ref{ex:subadditive} that the von-Neumann entropy is subadditive, that is,  for any $\rho\in\md(AB)$, 
\be
H(AB)_\rho\leq H(A)_\rho+H(B)_\rho\;,
\ee
where $H(AB)_\rho$ denotes $H(\rho^{AB})$. A stronger version of this inequality, known as the `strong subadditivity\index{strong subadditivity} of the von-Neumann\index{von-Neumann} entropy' states that for any $\rho\in\md(ABC)$ we have
\be\label{strsa}
H(ABC)_\rho+H(B)_\rho\leq H(AB)_\rho+H(BC)_\rho\;.
\ee
Note that this is a stronger version of the previous inequality since for $|B|=1$ it reduces to subadditivity. The above inequality is unique to the von-Neumann entropy and in general is not satisfied by other entropy functions (at least not in this form). 

We can express the strong subadditivity in terms of conditional entropies. Observe  that since $H(A|BC)_\rho=H(ABC)_\rho-H(BC)_\rho$ and $H(A|B)_\rho=H(AB)_\rho-H(B)_\rho$, the strong subadditivity can be expressed as
\be
H(A|BC)_\rho\leq H(A|B)_\rho\;.
\ee
This version of the strong subadditivity is perhaps more intuitive than~\eqref{strsa} since it can be  interpreted as the statement that by removing the access to system $C$, one can only increase the uncertainty about system $A$. Note also that the above form of the strong subadditivity is satisfied by any conditional entropy function. That is, for any conditional entropy $\H$ and $\rho\in\md(ABC)$ we have
\be\label{condileq}
\H(A|BC)_\rho\leq \H(A|B)_\rho\;.
\ee
The above inequality is a simply consequence of the monotonicity property of conditional entropy, since tracing out system $C$ is a map belonging to $\cmo(ABC\to AB)$. In terms of conditional majorization, we can express it as
\be
\rho^{ABC}\succ_A\rho^{AB}\;.
\ee

\bex\label{ex:dualssa}
Let $\rho\in\md(ABC)$. Show that
\be
H(A|B)_\rho+H(A|C)_\rho\geq 0
\ee
with equality if $\rho^{ABC}$ is a pure state. Hint: If $\rho^{ABC}$ is a mixed state, let $\psi^{ABCD}$ be its purification, and express $H(A|C)_\rho$ in terms of systems $A,B,D$ (for example, $H(AC)_\psi=H(BD)_\psi$). Finally, use~\eqref{strsa} with $D$ replacing $C$.
\eex

\subsubsection{The Conditional R\'enyi Entropies}\index{conditional R\'enyi entropy}

In Chapter~\ref{ch:relent} we encountered three types of quantum relative entropies that generalize the classical R\'enyi divergences: 
\ben
\item The Petz\index{Petz} quantum R\'enyi divergence, $D_\alpha$, as defined in Definition~\ref{def:petz} for $\alpha\in[0,2]$.
\item The sandwiched R\'enyi relative entropy, $\tD_\alpha$, as defined in Definition~\ref{def:sandwich} for all $\alpha\in[0,\infty]$. 
\item The geometric relative entropy, $\widehat{D}_{\alpha}$, as defined in Definition~\ref{def:geometric} for all $\alpha\in[0,2]$.
\een
Each types of the quantum R\'enyi relative entropy above give rise to two types of conditional entropies as given in~\eqref{coin1}. We denote the corresponding six conditional entropies by $H_\alpha$, $H_\alpha^\ua$, $\tilde{H}_\alpha$, 
$\tilde{H}_\alpha^\ua$, $\widehat{H}_{\alpha}$, and $\widehat{H}_{\alpha}^\ua$. The functions $H_\alpha$, $\tilde{H}_\alpha$, 
$\widehat{H}_{\alpha}$ are all quantum conditional entropies since they are additive. We next show that also ${H}_\alpha^\ua$ is additive, and later we will see that this also implies that $\tilde{H}_\alpha^\ua$ is additive. Therefore  ${H}_\alpha^\ua$ and $\tilde{H}_\alpha^\ua$ are conditional entropies as well. However, to the author's knowledge, the additivity\index{additivity} of  $\widehat{H}_{\alpha}^\ua$ has not been explored.

In the following theorem we provide a closed form for the function
\be
H^\ua_\alpha\left(A|B\right)_\rho\eqdef \log|A|-\min_{\sigma\in\md(B)}D_\alpha\left(\rho^{AB}\big\|\u^A\otimes\sigma^B\right)\quad\quad\forall\;\rho\in\md(AB)\;.
\ee

\begin{myt}{\color{yellow} Closed Formula}
\begin{theorem}\label{thmcfce}
Let $\rho\in\md(AB)$ and $\alpha\in[0,2]$. Then,
\be
H^\ua_\alpha\left(A|B\right)_\rho=\frac\alpha{1-\alpha}\log\tr\left[\left(\eta^B_\alpha\right)^{1/\alpha}\right]
\quad\text{where}\quad
\eta^B_\alpha\eqdef\tr_A\left[\left(\rho^{AB}\right)^\alpha\right]\;.
\ee
\end{theorem}
\end{myt}

\begin{proof}
First, observe that 
\ba\label{hupi}
H^\ua_\alpha\left(A|B\right)_\rho&=\max_{\sigma\in\md(B)}\frac1{1-\alpha}\log\tr\left[\left(\rho^{AB}\right)^\alpha\left(I^{A}\otimes\sigma^B\right)^{1-\alpha}\right]\\
&=\max_{\sigma\in\md(B)}\frac1{1-\alpha}\log\tr\left[\eta^B_\alpha\left(\sigma^B\right)^{1-\alpha}\right]
\ea
Set $t\eqdef\tr\left[\left(\eta^B_\alpha\right)^{1/\alpha}\right]$ and denote by
$
\tau^B\eqdef\left(\eta^B_\alpha\right)^{1/\alpha}/t
$
so that $\eta^B_\alpha=t^\alpha\left(\tau^B\right)^\alpha$.
Substituting this into the  previous equation we conclude
\ba
H^\ua_\alpha\left(A|B\right)_\rho&=\frac\alpha{1-\alpha}\log t+\max_{\sigma\in\md(B)}\frac1{1-\alpha}\log\tr\left[\left(\tau^B\right)^\alpha\left(\sigma^B\right)^{1-\alpha}\right]\\
&=\frac\alpha{1-\alpha}\log t-\min_{\sigma\in\md(B)}D_{\alpha}\left(\tau^B\|\sigma^B\right)\\
&=\frac\alpha{1-\alpha}\log t\;.
\ea
This completes the proof.
\end{proof}

\bex\label{halpa}
Use the closed formula above to show that $H^\ua_\alpha$ is additive and therefore a conditional entropy.
\eex

\subsubsection{The Optimized Conditional Min-Entropy}\index{conditional min-entropy}

Among the plethora of families of quantum conditional entropies we saw above, there exists one conditional entropy that appears often in applications, known as the optimized conditional min-entropy. 
The optimized conditional min-entropy is defined as the quantum conditional entropy $\tilde{H}_\alpha^\ua(A|B)_\rho$ with $\alpha=\infty$. That is, 
\be\label{defcmine}
H_{\min}^\ua(A|B)_\rho\eqdef\log|A|-\min_{\sigma\in\md(B)}D_{\max}\left(\rho^{AB}\big\|\u^A\otimes\sigma^B\right)\;.
\ee
The conditional min-entropy\index{conditional min-entropy} is defined in terms of an SDP. To see this, first observe that from the above formula and from the definition of $D_{\max}$ we get
\ba\label{8205}
2^{-H_{\min}^\ua(A|B)_\rho}&=\min\Big\{t\;:\;tI^A\otimes\sigma^B\geq\rho^{AB},\;\sigma\in\md(B),\;t\in\mbb{R}\Big\}\\
\Gg{\Lambda^B\eqdef t\sigma^B}&=\min\Big\{\tr\left[\Lambda^B\right]\;:\;I^A\otimes\Lambda^B\geq\rho^{AB},\;\Lambda\in\pos(B)\Big\}\;.
\ea
Then, using the notations $\mk_1\eqdef\pos(B)$, $\mk_2\eqdef\pos(AB)$, $H_1\eqdef I^B$, $H_2\eqdef\rho^{AB}$, and $\mN\in\ml(A\to AB)$ given by
\be
\mN(\omega^B)\eqdef I^A\otimes\omega^B\quad\quad\forall\;\omega\in\ml(B)\;,
\ee
we conclude that
\be
2^{-H_{\min}^\ua(A|B)_\rho}=\min\big\{\tr[\Lambda H_1]\;:\;\Lambda\in\mk_1\;\;,\mN(\Lambda)-H_2\in\mk_2\big\}\;.
\ee
 The above optimization problem has precisely the same form as the conic linear programming\index{linear programming} given in~\eqref{primal}. Since the cones $\mk_1$ and $\mk_2$ are the sets of positive semidefinite matrices, this conic program is an SDP program. 
 
 The above expression has a dual given by~\eqref{dual123}. Therefore,  the conditional min-entropy\index{conditional min-entropy} can be expressed in terms of the following optimization problem
\ba\label{bb2b}
2^{-H_{\min}^\ua(A|B)_\rho}&=\max\big\{\tr[\eta H_2]\;:\;\eta\in\mk_2^*\;\;,H_1-\mN^*(\eta)\in\mk_1^*\big\}\\
\GG{Exercise~\ref{exbb2b}}&=\max\Big\{\tr\left[\eta^{AB}\rho^{AB}\right]\;:\;\eta\in\pos(AB),\;\eta^B=I^B\Big\}\;.
\ea
Any $\eta^{AB}$ as above is a Choi matrix; hence, it can be expressed as $\eta^{AB}=\mE^{*\tA\to B}\left(\Omega^{A\tA}\right)$ for some channel $\mE\in\cptp(B\to A)$. We therefore get that
\ba\label{r288}
2^{-H_{\min}^\ua(A|B)_\rho}&=\max_{\mE\in\cptp(B\to \tA)}\tr\left[\rho^{AB}\mE^{*\tA\to B}\left(\Omega^{A\tA}\right)\right]\\
&=|A|\max_{\mE\in\cptp(B\to \tA)}\left\la\Phi^{A\tA}\left|\mE^{B\to\tA}\left(\rho^{AB}\right)\right|\Phi^{A\tA}\right\ra\\
&=|A|\max_{\mE\in\cptp(B\to \tA)}F^2\left(\mE^{B\to\tA}\left(\rho^{AB}\right),\Phi^{A\tA}\right)\;,
\ea
where $F$ is the fidelity. That is, the conditional min-entropy\index{conditional min-entropy} can be expressed in terms of the maximal overlap of $\mE^{B\to\tA}\left(\rho^{AB}\right)$ with the maximally entangled state. We now use the above expression to prove that the optimized conditional min-entropy is additive under tensor products, and thereby prove that the optimized conditional min-entropy is indeed a quantum conditional entropy as defined in Definition~\ref{qce}. 

\begin{myg}{}
\begin{lemma}
The optimized conditional min-entropy\index{conditional min-entropy} is a quantum conditional entropy satisfying both  properties of Definition~\ref{qce}.
\end{lemma}
\end{myg}

\begin{proof}
Since the optimized conditional min-entropy equals $\tilde{H}^\ua_\alpha$ with $\alpha=\infty$, it is left to prove that it is additive. Let $\rho\in\md(AB)$, $\tau\in\md(A'B')$, and denote by $Q_{\min}(A|B)_\rho\eqdef 2^{-H_{\min}^\ua(A|B)_\rho}$. Therefore, the additivity\index{additivity} of $H_{\min}$ would follow from the multiplicativity of $Q_{\min}$. On the one hand, from the primal problem~\eqref{8205} we have
\begin{align}
&Q_{\min}(AA'|BB')_{\rho\otimes\tau}
=\min\Big\{\tr\big[\Lambda^{BB'}\big]\;:\;I^{AA'}\otimes\Lambda^{BB'}\geq\rho^{AB}\otimes\tau^{A'B'},\;\Lambda\in\pos(BB')\Big\}\nonumber\\
&\leq \min\Big\{\tr\big[\Lambda^{B}_1\big]\tr\big[\Lambda^{B}_2\big]\;:\;I^{AA'}\otimes\Lambda^{B}_1\otimes\Lambda^{B'}_2\geq\rho^{AB}\otimes\tau^{A'B'},\;\Lambda_1\in\pos(B)\;,\;\Lambda_2\in\pos(B')\Big\}\nonumber\\
&\leq Q_{\min}(A|B)_{\rho}Q_{\min}(A'|B')_{\tau}\;,
\end{align}
where in the first inequality we restricted $\Lambda^{BB'}$ to have the form $\Lambda^{B}_1\otimes\Lambda^{B}_2$, and in the last inequality we replaced the condition $I^{AA'}\otimes\Lambda^{B}_1\otimes\Lambda^{B'}_2\geq\rho^{AB}\otimes\tau^{A'B'}$ with the two conditions $I^{A}\otimes\Lambda^{B}_1\geq\rho^{AB}$ and $I^{A'}\otimes\Lambda^{B'}_2\geq\tau^{A'B'}$. 

To get the opposite inequality we use the dual expression of the conditional min-entropy\index{conditional min-entropy} as given in~\eqref{r288}. Specifically,
\ba
Q_{\min}(AA'|BB')_{\rho\otimes\tau}&=|AA'|\max_{\mE\in\cptp(BB'\to \tA\tA')}F\left(\mE^{BB'\to\tA\tA'}\big(\rho^{AB}\otimes\tau^{BB'}\big),\Phi^{A\tA}\otimes\Phi^{A'\tA'}\right)^2\\
\Gg{\mE=\mE_1\otimes\mE_2}&\geq|AA'|\max_{\substack{\mE_1\in\cptp(B\to \tA)\\\mE_2\in\cptp(B'\to \tA')}}F\left(\mE^{B\to\tA}_1\left(\rho^{AB}\right)\otimes\mE^{B'\to\tA'}_2\big(\tau^{BB'}\big),\Phi^{A\tA}\otimes\Phi^{A'\tA'}\right)^2\\
&=Q_{\min}(A|B)_{\rho}Q_{\min}(A'|B')_{\tau}\;.
\ea
Combining the two equations above we conclude that \be Q_{\min}(AA'|BB')_{\rho\otimes\tau}=Q_{\min}(A|B)_{\rho}Q_{\min}(A'|B')_{\tau}\;,\ee so that $H_{\min}^\ua(AA'|BB')_{\rho\otimes\tau}=H_{\min}^\ua(A|B)_{\rho}+H_{\min}^\ua(A'|B')_{\tau}$.
This completes the proof.
\end{proof}

When system $A$ is classical, the right-hand side of~\eqref{r288} has a simple interpretation as a guessing probability\index{guessing probability}.  Indeed, suppose that $A=X$ is a classical system with $m\eqdef|X|$. Then the state $\rho^{AB}$, which we denote as $\rho^{XB}$, takes the form of a classical-quantum (cq) state:
\be
\rho^{XB}=\sum_{x\in[m]}p_x|x\lr x|^X\otimes\rho_x^B
\ee 
where $\{p_x\}_{x\in[m]}$ is a probability distribution, and each $\rho_x\in\md(B)$. Furthermore, it’s important to note that $\cptp(B\to \tA)$, which is the same as $\cptp(B\to \tX)$, comprises of POVM channels that were initially introduced in Sec.~\ref{sec:povmc}. Under these circumstances, the aforementioned equation simplifies to the following (Exercise\ref{ex:reduction}):
\be\label{r289}
2^{-H_{\min}^\ua(X|B)_\rho}=\max_{\{\Lambda_x\}}\sum_{x\in[m]}p_x\tr\left[\Lambda_x^B\rho_x^B\right]
\ee
where the maximum is over all POVMs $\{\Lambda_x^B\}_{x\in[m]}$ on system $B$. The expression above can be interpreted as the maximum probability for Bob to guess correctly the value of $X$. Specifically, given the cq-state\index{cq-state} $\rho^{XB}$, Bob can try to learn the classical value of $X$ by performing a quantum measurement/POVM, $\{\Lambda_x^B\}_{x\in[m]}$, on his system with $m\eqdef|X|$ possible outcomes. The probability that $X=x$ is $p_x$, and the probability that Bob gets the outcome $y$ given that $X=x$ is given by $\tr\left[\Lambda_y^B\rho_x^B\right]$. If Bob's takes $y$ to be his guess for the value of $X$ then $\tr\left[\Lambda_x^B\rho_x^B\right]$ is the probability that Bob guesses correctly the value of $X$. Given that $X=x$ with probability $p_x$, we get that 
\be
\pr_{\rm g}({X|B})_\rho\eqdef\max_{\{\Lambda_x\}}\sum_{x\in[m]}p_x\tr\left[\Lambda_x^B\rho_x^B\right]
\ee
is the maximal overall probability that Bob's guess of $X$ is correct. With this notation, the conditional entropy of $\rho^{XB}$ can be expressed as
\be
H_{\min}^\ua(X|B)_\rho=-\log\pr_{\rm g}(X|B)_\rho\;.
\ee
Note that $H_{\min}(X|B)_\rho\geq 0$ as expected.

\begin{exercise}\label{exbb2b}
Prove the second equality in~\eqref{bb2b}.
\end{exercise}

\bex\label{ex:reduction}
Prove the reduction of~\eqref{r288} to~\eqref{r289} when $A=X$ is classical.
\eex

\subsubsection{The Conditional Max-Entropy}\index{conditional max-entropy}

In Theorem~\ref{thm:neg-cond-ent}, we demonstrated that the conditional min-entropy\index{conditional min-entropy} $H_{\min}$ represents the lowest possible conditional entropy for a bipartite quantum state. This naturally leads to the question: what is the highest possible conditional entropy? To address this, we explore the upper limit of all conditional entropies.

\begin{myd}{}
\begin{definition}\label{cme}
The conditional max-entropy\index{conditional max-entropy} of a quantum state $\rho\in\md(AB)$ is defined as follows:
\be\label{cme0}
H_{\max}(A|B)_\rho\eqdef\log|A|-\min_{\sigma\in\md(B)}D_{\min}\left(\rho^{AB}\big\|\u^A\otimes\sigma^B\right)\;.
\ee
\end{definition}
\end{myd}
According to Theorem~\ref{thmcfce} and Exercise~\ref{halpa}, the function $H^\ua_{\alpha}$ is additive for all $\alpha\in[0,2]$. Hence, the additivity of the conditional max-entropy\index{conditional max-entropy} (i.e., $H^\ua_{\alpha=0}$) is established. Therefore, we can affirm that the conditional max-entropy\index{conditional max-entropy} qualifies as a legitimate conditional entropy measure.
Moreover, given that the min-relative entropy $D_{\min}$ is the smallest relative entropy, the following inequality holds for all conditional entropies $\H^\ua$ derived from a relative entropy $\D$ (as specified in~\eqref{coin1}), for any quantum state $\rho \in \md(AB)$:
\be
\H^\ua(A|B)_\rho\leq H_{\max}(A|B)_\rho\;.
\ee
This inequality signifies that the conditional max-entropy\index{conditional max-entropy} establishes an upper limit for all conditional entropies defined in relation to a relative entropy. Additionally, as will be explored in subsequent discussions, the conditional max-entropy\index{conditional max-entropy} is essentially the counterpart, or the dual, of the conditional min-entropy.

\section{Duality Relations}\index{duality}

We saw in Exercise~\ref{ex:dualssa} that for a pure state $\varphi\in\pure(ABC)$ the conditional von-Neumann entropy satisfies
\be\label{sduals}
H(A|B)_\varphi+H(A|C)_\varphi=0\;.
\ee 
Such a relation is called a duality relation, and motivates us to define a dual for any conditional entropy. 
\begin{myd}{}
\begin{definition}
Let $\H$ be a conditional entropy. For any $\rho\in\md(AB)$ with a purification $\varphi\in\pure(ABC)$ (where $C$ is the purifying system; i.e. $\rho^{AB}=\tr_C\varphi^{ABC}$) we define the dual of $\H$ as
\be
\H^{\dual}(A|B)_{\rho}\eqdef-\H(A|C)_{\varphi}\;.
\ee 
\end{definition}
\end{myd}

\begin{remark}
Since the conditional entropy is invariant under local isometries (specifically, $\H(A|C)_{\varphi}$ remains invariant under isometries on system $C$) the dual to a conditional entropy is well defined as it does not depend on the choice of the purifying system $C$. 
\end{remark}

By definition, the dual to a conditional entropy satisfies the invariance\index{invariance} and additivity\index{additivity} properties of conditional entropy (see Exercise~\ref{ex:addin}). To see that it satisfies also the normalization property of a conditional entropy, let $\rho^{AB}=\u^A$ with $|A|=2$ and $|B|=1$. A purification of $\rho^{AB}$ can be expressed as the maximally entangled state $\Phi^{AC}$ with $C=\tA$. Therefore,
\ba
\H^{\dual}(A)_{\u}&=\H^{\dual}(A|B)_{\rho}\\
\GG{by\; definition}&=-\H(A|C)_{\Phi}\\
\GG{\eqref{loga}}&=\log2=1\;.
\ea
Therefore, the dual to a conditional entropy would be itself a conditional entropy if it satisfies the monotonicity property. We will see shortly that this is indeed the case for all the conditional entropies studied in literature, although a general proof for all conditional entropy functions is unknown to the author.

\begin{exercise}\label{ex:addin}
Show that the dual to a conditional entropy satisfies the invariance and additivity properties of a conditional entropy.
\end{exercise}

The relation~\eqref{sduals} implies that the conditional von-Neumann\index{von-Neumann} entropy is self dual; i.e. $H^{\dual}(A|B)_\rho=H(A|B)_\rho$ for all $\rho\in\md(AB)$. Consider the Petz\index{Petz} conditional R\'enyi entropy\index{conditional R\'enyi entropy}  of order $\alpha\in[0,2]$ given for all $\rho\in\md(AB)$ by
\be
H_\alpha(A|B)_\rho=\frac1{1-\alpha}\log\tr\left[\left(\rho^{AB}\right)^\alpha\left(I^A\otimes\rho^B\right)^{1-\alpha}\right]\;.
\ee
In the lemma below we compute it's dual.
\begin{myg}{}
\begin{lemma}\label{lem:durel}
For any $\alpha\in[0,2]$, the dual of $H_{\alpha}$ is given by
\be
H^{\dual}_\alpha(A|B)_\rho=H_{2-\alpha}(A|B)_\rho\;.
\ee
\end{lemma}
\end{myg}
\begin{proof}
Let 
\be
\rho^{AB}=\sum_{x\in[n]}p_x|\varphi_x\lr\varphi_x|^{AB}
\ee
be the spectral decomposition of $\rho^{AB}$, and let
 $\rho^{ABC}=|\varphi\lr\varphi|^{ABC}$, with $C\cong AB$,
be the purification of $\rho^{AB}$ given by
\be
|\varphi^{ABC}\ra=\sum_{x\in[n]}\sqrt{p_x}|\varphi_x\ra^{AB}|\varphi_x\ra^C=\left(\rho^{AB}\right)^{\frac12}\otimes I^C|\Omega^{(AB)C}\ra\;.
\ee
where $|\Omega^{(AB)C}\ra=\sum_{x\in[n]}|\varphi_j\ra^{AB}|\varphi_j\ra^C$ is the maximally entangled operator between system $AB$ and $C$. Now, observe that
(see Exercise~\ref{swapex})
\be\label{sweq1}
\left(\rho^{AB}\right)^\alpha\otimes I^C|\Omega^{(AB)C}\ra=I^{AB}\otimes\left(\rho^{C}\right)^\alpha|\Omega^{(AB)C}\ra\;,
\ee
where  $\rho^C\eqdef\tr_{AB}\varphi^{ABC}$.
Therefore, from part 1 of Exercise~\ref{bipartite} we get
\ba\label{8300}
\tr\left[\left(\rho^{AB}\right)^\alpha\left(I^A\otimes\rho^B\right)^{1-\alpha}\right]&=\left\la\Omega^{ABC}\left|\left(\rho^{AB}\otimes I^C\right)^\alpha\left(I^A\otimes\rho^B\otimes I^C\right)^{1-\alpha}\right|\Omega^{ABC}\right\ra\\
\GG{\eqref{sweq1}}&=\big\la\Omega^{ABC}\big|I^A\otimes\left(\rho^B\right)^{1-\alpha}\otimes \left(\rho^C\right)^\alpha\big|\Omega^{ABC}\big\ra\\
&=\big\la\varphi^{ABC}\big|I^A\otimes\left(\rho^B\right)^{1-\alpha}\otimes \left(\rho^C\right)^{\alpha-1}\big|\varphi^{ABC}\big\ra\;,
\ea
where in the last equality we used the fact that $|\varphi^{ABC}\ra=I^{AB}\otimes\left(\rho^{C}\right)^{1/2}|\Omega^{(AB)C}\ra$. Next, let
\be
\rho^B=\sum_{y\in[m]}q_y|y\lr y|^B\quad\text{and}\quad\rho^{AC}=\sum_{y\in[m]}q_y|\chi_y\lr \chi_y|^B
\ee
be the spectral decompositions of $\rho^B$ and $\rho^{AC}$, respectively, and 
consider the following Schmidt decomposition between system $B$ and system $AC$ 
\be
|\varphi^{ABC}\ra=\sum_{y\in[m]}\sqrt{q_y}|y\ra^B|\chi_y\ra^{AC}=\left(\rho^B\right)^{1/2}\otimes I^{AC}\big|\Omega^{B(AC)}\big\ra
\ee
where $|\Omega^{B(AC)}\big\ra=\sum_{y\in[m]}|y\ra^B|\chi_y\ra^{AC}$. Substituting the above expression for $|\varphi^{ABC}\ra$ into~\eqref{8300} gives
\be
\tr\left[\left(\rho^{AB}\right)^\alpha\left(I^A\otimes\rho^B\right)^{1-\alpha}\right]=\big\la\Omega^{B(AC)}\big|I^A\otimes\left(\rho^B\right)^{2-\alpha}\otimes \left(\rho^C\right)^{\alpha-1}\big|\Omega^{B(AC)}\big\ra\;.
\ee
Finally,
using the relation
\be\label{sweq2}
\left(\rho^{B}\right)^{2-\alpha}\otimes I^{AC}|\Omega^{B(AC)}\ra=I^{B}\otimes\left(\rho^{AC}\right)^{2-\alpha}|\Omega^{B(AC)}\ra
\ee
we conclude that
\ba
\tr\left[\left(\rho^{AB}\right)^\alpha\left(I^A\otimes\rho^B\right)^{1-\alpha}\right]&=\big\la\Omega^{B(AC)}\big|I^B\otimes\left(\left(\rho^{AC}\right)^{2-\alpha} \left(I^A\otimes \rho^C\right)^{\alpha-1}\right)\big|\Omega^{B(AC)}\big\ra\\
&=\tr\left[\left(\rho^{AC}\right)^{2-\alpha} \left(I^A\otimes \rho^C\right)^{\alpha-1}\right]\;.
\ea
Therefore,
\ba
H_\alpha(A|B)_\rho&=\frac1{1-\alpha}\log\tr\left[\left(\rho^{AC}\right)^{2-\alpha} \left(I^A\otimes \rho^C\right)^{\alpha-1}\right]\\
&=-H_{2-\alpha}(A|C)_\rho\;.
\ea
Note that the above equality is equivalent to 
\be
H^{\dual}_\alpha(A|B)_\rho\eqdef-H_{\alpha}(A|C)_\rho=H_{2-\alpha}(A|B)_\rho\;.
\ee
This completes the proof.
\end{proof}
\begin{exercise}\label{swapex}
Prove the relations~\eqref{sweq1} and~\eqref{sweq2}.
\end{exercise}

\subsubsection{The Dual of the Optimized Conditional Min-Entropy}\index{conditional min-entropy}

\begin{myg}{}
\begin{lemma}\label{lem:hmin1/2}
Let $\rho\in\pure(ABE)$.  Then,
\be
H_{\min}^\ua(A|B)_\rho=-\tH_{1/2}^{\ua}(A|E)_\rho=-\max_{\tau\in\md(E)}\log F^2\left(\rho^{AE},I^{A}\otimes\tau^{E}\right)\;,
\ee
where $F$ is the fidelity.
\end{lemma}
\end{myg}

\begin{remark}
It is noteworthy that the lemma above establishes $\tH_{1/2}^\ua$ as the dual of $H_{\min}^\ua$. Consequently, $\tH_{1/2}^\ua(A|B)\rho$ is sometimes referred to as the conditional max-entropy. However, we choose not to use this terminology here because, generally speaking, $\tH_{1/2}^\ua(A|B)_\rho\neq H_{\max}(A)_\rho$, particularly when $\rho^{AB}=\rho^A\otimes\rho^B$. In fact, as we will show later, the true dual of $H_{\min}$ (as opposed to $H^\ua_{\min}$) aligns with the conditional max-entropy\index{conditional max-entropy} as defined in~\eqref{cme0}. Additionally, when integrating the above lemma with~\eqref{r288}, we derive the following relationship:
\be\label{fidrel}
\max_{\mE\in\cptp(B\to \tA)}F\left(\mE^{B\to\tA}\left(\rho^{AB}\right),\Omega^{A\tA}\right)=\max_{\tau\in\md(E)}F\left(\rho^{AE},I^{A}\otimes\tau^{E}\right)\;.
\ee
\end{remark}

\begin{proof}
We start with the expression for $Q_{\min}(A|B)_\rho=2^{-H_{\min}^\ua(A|B)_\rho}$, given in~\eqref{r288} as
\be
Q_{\min}(A|B)_\rho=\max_{\mE\in\cptp(B\to \tA)}F^2\left(\mE^{B\to\tA}\left(\rho^{AB}\right),\Omega^{A\tA}\right)\;.
\ee
For any $\mE\in\cptp(B\to \tA)$ let $\mV_\mE\in\cptp(B\to \tA R)$ be its Stinespring's isometry.  Observe that $\mV_\mF^{B\to \tA R}\left(\rho^{ABE}\right)$ is a purification of $\mF^{B\to \tA }\left(\rho^{AB}\right)$.
Moreover, since $\Omega^{A\tA}$ is already pure, any purification of $\Omega^{A\tA}$ in $A\tA RE$ must be of the form $\Omega^{A\tA }\otimes\chi^{RE}$, where $\chi\in\pure(RE)$.  Hence, from the Uhlmann's theorem we get that
\ba
F^2\left(\Omega^{A\tA},\mE^{B\to \tA}\left(\rho^{AB}\right)\right)=
\max_{\chi\in\pure(RE)}F^2\left(\Omega^{A\tA }\otimes\chi^{RE},\mV_\mE^{B\to \tA R}\left(\psi^{ABE}\right)\right)\;.
\ea
Now, observe that \emph{any} purification of the state  $\rho^{AE}\eqdef\tr_B\left[\rho^{ABE}\right]$ in $\pure(A\tA ER)$ has the form $\mV_\mE^{B\to \tA R}\left(\rho^{ABE}\right)$ for some $\mE\in\cptp(B\to \tA )$. Therefore, when we add the maximization over all $\mE\in\cptp(B\to \tA)$ to both sides of the equation above we get
\be
Q_{\min}(A|B)_\rho=
\max_{\substack{\psi\in\pure(A\tA RE)\\ \psi^{AE}=\rho^{AE},\;\chi\in\pure(RE)}}F^2\left(\Omega^{A\tA }\otimes\chi^{RE},\psi^{A\tA RE}\right)\;,
\ee
where on the right-hand side we replaced that maximum over all $\mE\in\cptp(B\to \tA )$ with a maximum over all pure states $\psi\in\pure(A\tA RE)$ with marginal $\psi^{AE}=\rho^{AE}$. Finally, applying the Uhlmann's theorem to the expression above we conclude that
\be
Q_{\min}(A|B)_\rho=\max_{\chi\in\md(E)}F^2\left(I^{A}\otimes\chi^{E},\rho^{AE}\right)\;.
\ee
This completes the proof.
\end{proof}
\bex\label{exconvq}
Show that $Q_{\min}$ is a convex function. That is, show that for every set of $n$ bipartite quantum states $\{\rho^{AB}_x\}_{x\in[n]}$ and every $\p\in\prob(n)$ we have 
\be
Q_{\min}(A|B)_{\rho}\leq\sum_{x\in[n]}p_xQ_{\min}(A|B)_{\rho_x}\quad\text{where}\quad\rho^{AB}=\sum_{x\in[n]}p_x\rho^{AB}_x\;.
\ee
\eex

More generally, the duals of $H^\ua_\alpha$ and $\tilde{H}^\ua_\alpha$ can also be computed and they are given by  (see the section `Notes and References' below for more details)
\ba\label{rel231}
&\tilde{H}_\alpha^{\ua\dual}(A|B)_\rho=\tilde{H}_{\beta}^\ua(A|B)_\rho\quad\text{for}\quad\frac1\alpha+\frac1\beta=2\;\;,\;\;\alpha,\beta\in[1/2,\infty]\\
&\tilde{H}_\alpha^{\dual}(A|B)_\rho=H_{\beta}^\ua(A|B)_\rho\quad\text{for}\quad\alpha\beta=1\;\;,\;\;\alpha,\beta\in[0,\infty]\;.
\ea
Observe that from the first equality above, by taking $\alpha=\infty$ (and hence $\beta=1/2$)  we get the statement given in the lemma above that the dual to the optimized conditional min-entropy is $\tH_{1/2}^\ua$. On the other hand, from the second equality we see that the dual of $H_{\min}$ is $H_{0}^\ua=H_{\max}$ (see Definition~\eqref{cme}). That is, for all $\rho\in\md(AB)$
\be\label{minmaxduality}
H_{\min}^\dual(A|B)_\rho=H_{\max}(A|B)_\rho\;.
\ee
We therefore get the following corollary.
\begin{myg}{}
\begin{corollary}
Let $\H$  be a quantum conditional entropy and suppose its dual $\H^\dual$ is also a quantum conditional entropy. Then, for all $\rho\in\md(AB)$
\be
\H(A|B)_\rho\leq H_{\max}(A|B)_\rho
\ee
\end{corollary}
\end{myg}
\begin{remark}
Previously, we established that conditional entropies defined as in~\eqref{coin1} are upper bounded by the conditional max-entropy\index{conditional max-entropy}. However, the corollary above does not require the conditional entropy $\H$ to be defined with respect to a relative entropy. Instead, it assumes that the dual entropy $\H^\dual$ is also a valid conditional entropy.
It is worth noting that it remains an open problem whether this additional assumption can be removed, i.e., whether the upper bound provided by the conditional max-entropy\index{conditional max-entropy} applies to all conditional entropies or just to those whose dual is also a conditional entropy.
\end{remark}

\begin{proof}
Let $\varphi\in\pure(ABC)$ be a purification of $\rho^{AB}$. From the definition of $\H^\dual$ we get
\ba
\H(A|B)_\rho&=-\H^{\dual}(A|C)_{\varphi}\\
\GG{Theorem~\ref{thm:neg-cond-ent}\; applied\; to\; \H^\dual}&\leq -H_{\min}(A|C)_\varphi\\
\GG{\eqref{minmaxduality}}&=H_{\max}(A|B)_\rho\;.
\ea
This completes the proof.
\end{proof}

\bex
Use the lemma above and the relations~\eqref{rel231} to show that for any $\varphi\in\pure(ABC)$ we have
\ba
&H_{\alpha}(A|B)_\varphi+H_{\beta}(A|C)_\varphi=0\quad\text{for}\quad\alpha+\beta=2,\;\;\alpha,\beta\in[0,2]\\
&\tilde{H}_{\alpha}^\ua(A|B)_\varphi+\tilde{H}_{\beta}^\ua(A|C)_\varphi=0
\quad\text{for}\quad\frac1\alpha+\frac1\beta=2,\;\;\alpha,\beta\in[1/2,\infty]\\
&{H}_{\alpha}^\ua(A|B)_\varphi+\tilde{H}_{\beta}(A|C)_\varphi=0\quad\text{for}\quad\alpha\beta=1,\;\;\alpha,\beta\in[0,\infty]\;.
\ea
\eex

\bex
In the following, use the duality\index{duality} relations above.
\ben
\item Show that the dual of ${H}_\alpha$ is itself a quantum conditional entropy for all $\alpha\in[0,2]$ (i.e. you need to show the monotonicity property).
\item Show that $\tilde{H}^\ua_\alpha$ is a quantum conditional entropy for all $\alpha\in[0,\infty]$ (i.e. you need to show the additivity property).
\item Show that the dual of $\tilde{H}^\ua_\alpha$ is itself a quantum conditional entropy for all $\alpha\in[0,\infty]$. 
\item Use part 2 to provide an alternative proof for the additivity of the optimized conditional min-entropy.
\een
\eex

\section{The Decoupling Theorem}\label{sec:decoupling}\index{decoupling theorem}

The decoupling theorem identifies the conditions under which a system, initially correlated with another system (such as the environment), becomes decoupled from that environment following a physical evolution. The conditional entropy, as examined in the previous sections, can be employed to measure the extent of this decoupling. The decoupling theorem serves as a pivotal instrument in both quantum Shannon theory and quantum resource theories. A ``smoothed" version of this theorem will also be discussed in Sec.~\ref{smoothedd}.

In Sec.~\ref{sec:inv}, we introduced the concept of the $\G$-twirling\index{twirling} operation over a compact Lie group. When the group $\G=\mathfrak{U}(A)$ encompasses all unitary matrices, the resulting $\G$-twirling map, denoted as $\mG\in\cptp(A\to A)$, is given by the channel in~\eqref{8249}. Therefore, the channel $\mG^{A\to A}$ transforms a state $\rho\in\md(AE)$ --- representing a system $A$ correlated with an environment $E$ --- into:
\be
\mG^{A\to A}\left(\rho^{AE}\right)\eqdef\int_{\mathfrak{U}(A)}dU^A\;U^A\rho^{AE} U^{*A}=\u^A\otimes\rho^E\;.
\ee
In this context, the $\G$-twirling map acts as a completely randomizing channel, also known as the completely depolarizing channel.

When a quantum channel $\mN\in\cptp(A\to B)$ is applied to both sides of the equation above, we obtain:
\be\label{7p187}
\int_{\mathfrak{U}(A)}dU^A\;\mN^{A\to B}\left(U^A\rho^{AE} U^{*A}\right)=\tau^B\otimes\rho^E
\ee
where $\tau^B\eqdef\mN^{A\to B}\left(\u^A\right)$. The decoupling theorem estimates how closely
$\mN^{A\to B}\left(U^A\rho^{AE} U^{*A}\right)$ (i.e., removing the integral and considering one specific unitary matrix) can approximate the decoupled state $\tau^A\otimes\rho^E$. Our discussion begins with a lemma using the square of the Frobenius norm for this estimation.
In this lemma, we utilize the function:
\be\label{foab}
f\left(\omega^{AB}\right)\eqdef\frac m{\sqrt{m^2-1}}\left(\tr\left(\omega^{AB}\right)^2-\tr\left(\u^A\otimes\omega^{B}\right)^2\right)\quad\quad\forall\;\omega\in\ml(AB)\;,
\ee
where $m\eqdef|A|$. To simplify the notation in this section, we will omit the square brackets in certain expressions. For instance, in the above formula, we used $\tr\left(\omega^{AB}\right)^2$ instead of $\tr\left[\left(\omega^{AB}\right)^2\right]$. It's important to note that with this revised notation, all powers are included within the trace operation.
\bex
Let $\rho\in\pos(AB)$ (we also assume $\rho^{AB}$ is not the zero matrix) and set $m\eqdef|A|$.
\ben
\item Show that 
\be
\frac{1}{m}\leq\frac{\tr(\rho^{AB})^2}{\tr(\rho^A)^2}\leq m
\ee
Hint: Start by showing $\tr(\rho^{B})^2=\tr\left[\left(\rho^{AB}\otimes I^{\tA}\right)\left(I^{A}\otimes \rho^{\tA B}\right)\right]$ and then use the Cauchy-Schwarz inequality.
For the other side, show first that $\rho^{AB}\leq m I^A\otimes\rho^B$\;.
\item Consider the function $f$ as defined above. 
\ben 
\item Show that  $f\left(\rho^{AB}\right)\geq 0$ with equality if and only if $\rho^{AB}=\u^A\otimes\rho^B$.
\item Show that 
\be\label{frana}
f(\rho^{AB})\leq\tr\left(\rho^{AB}\right)^2\;.
\ee
\een
\een
\eex

\begin{myg}{}
\begin{lemma}\label{lemgood}
Let $\rho\in\ml(AE)$, $m\eqdef|A|$, $\mN\in\ml(A\to B)$, and $\tau^{AB}\eqdef\frac{1}{m}J^{AB}_\mN$, where $J_\mN^{AB}$ is the Choi matrix of $\mN^{A\to B}$. 
\be\label{313q}
\int_{\mathfrak{U}(A)}dU^A\tr\left(\mN^{A\to B}\Big(U^A\rho^{AE}\left(U^A\right)^*\Big)-\tau^B\otimes\rho^E\right)^2=f\left(\rho^{AE}\right)f\left(\tau^{AB}\right)\;,
\ee
where $f$ is defined in~\eqref{foab}.
\end{lemma}
\end{myg}

\begin{remark}
Observe that we do not assume that $\rho^{AE}$ is a density matrix (not even Hermitian) nor that $\mN^{A\to B}$ is a quantum channel (just a linear map). However, if $\rho^{AE}\geq 0$ we can use the bound~\eqref{frana} in conjunction with the lemma above to get the relatively simple upper bound
\be\label{tgft}
\int_{\mathfrak{U}(A)}dU^A\;\left\|\mN^{A\to B}_U\left(\rho^{AE}\right)-\tau^B\otimes\rho^E\right\|_2^2\leq\tr\left(\rho^{AE}\right)^2\tr\left(\tau^{AB}\right)^2\;,
\ee
where $\mN^{A\to B}_U\eqdef\mN^{A\to B}\circ\mU^{A\to A}$, with $\mU^{A\to A}(\cdot)\eqdef U^A(\cdot)U^{*A}$, and we used the fact that any Hermitian matrix $\eta\in\herm(BE)$ satisfies $\|\eta\|_2^2=\tr[\eta^2]$. Moreover, taking the square root on both sides of the equation above and using Jensen's inequality (see Sec.~\ref{opjens}) we obtain that
\ba\label{tgftn}
\sqrt{\tr\left(\rho^{AE}\right)^2\tr\left(\tau^{AB}\right)^2}&\geq \sqrt{\int_{\mathfrak{U}(A)}dU^A\;\left\|\mN^{A\to B}_U\left(\rho^{AE}\right)-\tau^B\otimes\rho^E\right\|_2^2}\\
\GG{Jensen's\; Inequality}&\geq \int_{\mathfrak{U}(A)}dU^A\;\left\|\mN^{A\to B}_U\left(\rho^{AE}\right)-\tau^B\otimes\rho^E\right\|_2
\ea
Finally, it's important to recognize that since the average of the integrand in the above equation is less than the expression on the left-hand side, it implies the existence of at least one unitary $U^A$ for which $\left\|\mN^{A\to B}_U\left(\rho^{AE}\right)-\tau^B\otimes\rho^E\right\|_2$ is smaller than $\tr\left[\rho^{AE}\right]\tr\left[\tau^{AB}\right]$.
\end{remark}

\begin{proof}
For simplicity of the exposition we will omit the superscript from $\mN^{A\to B}_U$ and simply write it as $\mN_U$.
With these notations, the integrand of~\eqref{313q} can be decomposed into three terms:
\ba\label{beint2}
&\tr\left(\mN_U\left(\rho^{AE}\right)-\tau^B\otimes\rho^E\right)^2\\
&=\tr\left(\mN_U\left(\rho^{AE}\right)\right)^2-2\tr\left[\left(\tau^B\otimes\rho^E\right)\mN_U\left(\rho^{AE}\right)\right]+\tr\left(\tau^B\otimes\rho^E\right)^2\;.
\ea
From~\eqref{7p187}, the integral of the second term above can be simplified as
\ba
\int_{\mathfrak{U}(A)}dU\;\tr\left[\left(\tau^B\otimes\rho^E\right)\mN_U\left(\rho^{AE}\right)\right]=\tr\left(\tau^B\otimes\rho^E\right)^2\;.
\ea
Therefore, taking the integral over $\mfu(A)$ on both sides of~\eqref{beint2} gives
\ba\label{linf2}
&\int_{\mathfrak{U}(A)}dU\;\tr\left(\mN_U\left(\rho^{AE}\right)-\tau^B\otimes\rho^E\right)^2\\
&= \int_{\mathfrak{U}(A)}dU\;\tr\Big(\mN_U\left(\rho^{AE}\right)\Big)^2-\tr\left(\tau^B\right)^2\tr\left(\rho^E\right)^2\;.
\ea
To compute the remaining integral we use a linearization technique that is based on Exercise~\ref{m2fab}. That is,
we linearize the square in the integrand by using Exercise~\ref{m2fab} with the flip operator $F^{B\tB E\tE}=F^{B\tB}\otimes F^{E\tE}$. Explicitly,
\ba\label{mesha2}
\tr\Big(\mN_U\left(\rho^{AE}\right)\Big)^2&=\tr\left[\left(\mN_U\left(\rho^{AE}\right)\otimes\mN_U\left(\rho^{AE}\right)\right)F^{B\tB E\tE}\right]\\
&=\tr\left[\mN^{\otimes 2}_U\left(\left(\rho^{AE}\right)^{\otimes 2}\right)F^{B\tB E\tE}\right]\\
&=\tr\left[\left(\rho^{AE}\right)^{\otimes 2}\left(\mN^{*\otimes 2}_U\big(F^{B\tB}\big)\otimes F^{E\tE}\right)\right]\;.
\ea
Taking the integral over $\muu(A)$ on both sides and using the fact that $\mN_U^*=\mU^*\circ \mN^*$ we obtain
\be\label{nottwirl}
\int_{\mathfrak{U}(A)}dU\;\tr\Big(\mN_U\left(\rho^{AE}\right)\Big)^2=\tr\left[\left(\rho^{AE}\right)^{\otimes 2}\left(\mG\big(\mN^{*\otimes 2}\big(F^{B\tB}\big)\big)\otimes F^{E\tE}\right)\right]\;,
\ee
where $\mG\in\cptp(A\tA\to A\tA)$ denotes the twirling channel
\be\label{twirlingw}
\mG(\cdot)\eqdef\int_{\mfu(A)}dU\;U^{*}\otimes U^{*}(\cdot) U\otimes U\;.
\ee
Next, we make use of the fact that the twirling channel turns states to symmetric ones (see~\eqref{twirling}). Specifically,  observe that from~\eqref{twirling} we get
\be\label{hsipo2}
\mG\left(\mN^{*\otimes 2}\big(F^{B\tB}\big)\right)=aI^{A\tA}+bF^{A\tA}\;.
\ee
where the coefficients $a,b\in\mbb{R}$ will be computed shortly using~\eqref{twirling}.
Substituting~\eqref{hsipo2} into~\eqref{nottwirl} gives
\ba\label{givesrho}
\int_{\mathfrak{U}(A)}dU\;\tr\Big(\mN_U\left(\rho^{AE}\right)\Big)^2
&=\tr\left[\left(\rho^{AE}\right)^{\otimes 2}\left(\big(aI^{A\tA}+bF^{A\tA}\big)\otimes F^{E\tE}\right)\right]\\
&=a\tr\left[\left(\rho^{E}\right)^{\otimes 2}F^{E\tE}\right]+b\tr\left[\left(\rho^{AE}\right)^{\otimes 2} F^{A\tA E\tE}\right]\\
\GG{\eqref{0m2fab}}&=a\tr\left(\rho^{E}\right)^{2}+b\tr\left(\rho^{AE}\right)^{2}\;.
\ea
Combining this with~\eqref{linf2} we obtain
\be\label{algebra}
\int_{\mathfrak{U}(A)}dU\;\tr\left(\mN_U\left(\rho^{AE}\right)-\tau^B\otimes\rho^E\right)^2=\left(a-\tr\left(\tau^B\right)^2\right)\tr\left(\rho^{E}\right)^{2}+b\tr\left(\rho^{AE}\right)^{2}\;.
\ee

It is therefore left to compute the coefficients $a$ and $b$. From~\eqref{twirling} they can be expressed as:
\be
a\eqdef\frac{m\tr\left[\mN^{*\otimes 2}\big(F^{B\tB}\big)\right]-\tr\left[\mN^{*\otimes 2}\big(F^{B\tB}\big) F^{A\tA}\right]}{m(m^2-1)}
\ee
and
\be
b\eqdef\frac{m\tr\left[\mN^{*\otimes 2}\big(F^{B\tB}\big)F^{A\tA}\right]-\tr\left[\mN^{*\otimes 2}\big(F^{B\tB}\big)\right]}{m(m^2-1)}\;.
\ee
To simplify the expressions above we use the definition of the adjoint map\index{adjoint map} to get
\ba
\tr\left[\mN^{*\otimes 2}\big(F^{B\tB}\big)\right]
&=\tr\left[\left(\mN(I^A)\right)^{\otimes 2}F^{B\tB}\right]\\
\Gg{\mN\left(I^A\right)=J^{B}_\mN=m\tau^{B}}&=m^2\tr\left[\left(\tau^B\right)^{\otimes 2}F^{B\tB}\right]\\
\Gg{F^B\eqdef\tr_{\tB}\left[F^{B\tB}\right]=I^B}&=m^2\tr\left[\left(\tau^B\right)^{2}\right]\;,
\ea
and
\be
\tr\left[F^{A\tA}\mN^{*\otimes 2}\big(F^{B\tB}\big)\right]
=\tr\left[F^{B\tB}\mN^{\otimes 2}\big(F^{A\tA}\big)\right]\;.
\ee
Moreover, since the Choi matrix of $\mN^{\otimes 2}$ is given by $m^2\left(\tau^{AB}\right)^{\otimes 2}$ we get
\ba
\tr\left[F^{B\tB}\mN^{\otimes 2}(F^{A\tA})\right]&=m^2\tr\left[F^{B\tB}\tr_{A\tA}\left[\left(\tau^{AB}\right)^{\otimes 2}\left(F^{A\tA}\otimes I^{B\tB}\right)\right]\right]\\
&=m^2\tr\left[\left(\tau^{AB}\right)^{\otimes 2}\left(F^{A\tA}\otimes F^{B\tB}\right)\right]\\
\GG{\eqref{0m2fab}}&=m^2\tr\left[(\tau^{AB})^{2}\right].
\ea
We therefore conclude that $a$ and $b$ can be expressed as
\ba
a&=\frac{m^2\tr\left(\tau^{B}\right)^2-m\tr\left(\tau^{AB}\right)^2}{m^2-1}\quad\text{and}\\ b&=\frac{m^2\tr\left(\tau^{AB}\right)^2-m\tr\left(\tau^{B}\right)^2}{m^2-1}=\frac{m}{\sqrt{m^2-1}}f\left(\tau^{AB}\right)\;.
\ea
Finally, substituting these expressions into~\eqref{algebra}, and observing that 
\be
a-\tr\left(\tau^B\right)^2
=\frac{\tr\left(\tau^{B}\right)^2-m\tr\left(\tau^{AB}\right)^2}{m^2-1}=-\frac{1}{\sqrt{m^2-1}}f\left(\tau^{AB}\right)\;,
\ee
we get that the right-hand side of~\eqref{algebra} equals the right-hand side of~\eqref{313q}. This completes the proof.
\end{proof}

\bex
Demonstrate clearly that substituting the expressions in the proof above for $a - \tr\left(\tau^B\right)^2$ and $b$ into~\eqref{algebra} results in the equality~\eqref{313q}.
\eex

\bex
Using the same notations as in the lemma above, with $\rho\in\md(AE)$ and $\mN\in\cp(A\to B)$, show that for all $\sigma\in\md(E)$, 
\be
\int_{\mathfrak{U}(A)}dU^A\tr\left(\mN^{A\to B}\Big(U^A\rho^{AE}\left(U^A\right)^*\Big)-\tau^{B}\otimes\sigma^E\right)^2\geq f\left(\rho^{AE}\right)f\left(\tau^{AB}\right)\;,
\ee
with equality if and only if $\sigma^E=\rho^E$.
\eex

In the proof of the lemma above, we used the twirling\index{twirling} operation $\mG$ as defined in~\eqref{twirlingw}. In Sec.~\ref{secgt} we will see that this channel belong to a family of channels that we call the $\G$-twirling operations. One of the properties of all $\G$ twirling operations, particular the channel $\mG$ as defined in~\eqref{twirlingw}, is that they can be expressed as a finite convex combination of unitary channels. In our context here, it means that there exists $k\in\mbb{N}$, $\p\in\prob(k)$, and a set of unitary matrices $\{U_x\}_{x\in[k]}\subset\muu(A)$ such that $\mG\in\cptp(A\tA\to A\tA)$ as defined in~\eqref{twirlingw} can be expressed as
\be\label{7222}
\mG(\omega^{A\tA})=\sum_{x\in[k]}p_x\left(U_x\otimes U_x\right)\omega^{A\tA}\left(U_x\otimes U_x\right)^*\quad\quad\forall\omega\in\ml(A\tA)\;.
\ee 
Therefore, working with this expression for the twirling map, we obtain the following corollary.

\begin{myg}{}
\begin{corollary}\label{corimpi0}
Let $k\in\mbb{N}$, $\p\in\prob(k)$, and  $\{U_x\}_{x\in[k]}\subset\muu(A)$ be as in~\eqref{7222}. Then, using the same notations as in Lemma~\ref{lemgood}, we have
\be\label{sec313q}
\sum_{x\in[k]}p_x\tr\left(\mN^{A\to B}\Big(U^A_x\rho^{AE}\left(U^A_x\right)^*\Big)-\tau^B\otimes\rho^E\right)^2=f\left(\rho^{AE}\right)f\left(\tau^{AB}\right)\;.
\ee
\end{corollary}
\end{myg}

\bex
Prove the corollary above.
\eex

In the following theorem we make use of the conditional entropy $\tH^\ua_2(A|B)_\omega$ which is defined on every $\omega\in\pos(AB)$ terms of the divergence $\tD_2$ as (c.f.~\eqref{coin1}) 
\ba
\tH_2^\ua(A|B)_\omega&=-\min_{\eta\in\md(B)}\tD_2\left(\omega^{AB}\big\|I^A\otimes\eta^B\right)\\
\GG{Definition~\ref{def:sandwich}}&=-\min_{\eta\in\md(B)}\log\tr\Big(\left(I^A\otimes\eta^{-1/4}\right)\omega^{AB}\left(I^A\otimes\eta^{-1/4}\right)\Big)^2\;.
\ea

\begin{myt}{\color{yellow} Decoupling Theorem} \index{decoupling theorem} 
\begin{theorem}\label{thm1141}
Let $\rho\in\md_{\leq}(AE)$, $\mN\in\cp(A\to B)$, and $\tau^{AB}\eqdef\frac{1}{|A|}J^{AB}_\mN$, where $J_\mN^{AB}$ is the Choi matrix of $\mN^{A\to B}$. 
Then,
\be\label{lhsver1}
\int_{\mathfrak{U}(A)}dU^A\;\left\|\mN^{A\to B}\Big(U^A\rho^{AE}\left(U^A\right)^*\Big)-\tau^B\otimes\rho^E\right\|_1\leq 2^{-\frac{1}{2}\big(\tH_2^\ua(A|E)_\rho+\tH_2^\ua(A|B)_\tau\big)}\;.
\ee
\end{theorem}
\end{myt}

\begin{proof}
In the first step of the proof we upper bound the trace norm with the Hilbert Schmidt norm. Working with the Frobenius norm we will be able to use~\eqref{tgftn}. From the third part of Exercise~\ref{tn0} it follows that for any matrix $M\in\herm(A)$ and $\sigma\in\pos(A)$ 
\be
\|M\|_1\leq \sqrt{\tr[\sigma]}\big\|\sigma^{-1/4}M\sigma^{-1/4}\big\|_2\;.
\ee 
Taking $\sigma=\eta^B\otimes\zeta^E\in\md(BE)$ and 
\be
M=\mN_U\left(\rho^{AE}\right)-\tau^B\otimes\rho^E
\ee 
gives
\ba
&\big\|\mN_U\left(\rho^{AE}\right)-\tau^B\otimes\rho^E\big\|_1\\
&\leq \left\|\left(\eta^B\otimes\zeta^E\right)^{-\frac{1}{4}}\left(\mN_U\left(\rho^{AE}\right)-\tau^B\otimes\rho^E\right)\left(\eta^B\otimes\zeta^E\right)^{-\frac{1}{4}}\right\|_2
\ea
The choice of $\eta^B$ and $\zeta^E$ will be made later.
Denoting by $\tilde{\mN}^{A\to B}_U(\cdot)\eqdef(\eta^B)^{-\frac{1}{4}}\mN^{A\to B}_U(\cdot)(\eta^B)^{-\frac{1}{4}}$, $\trho^{AE}\eqdef(\zeta^E)^{-\frac{1}{4}}\rho^{AE}(\zeta^E)^{-\frac{1}{4}}$, and $\ttau^{AB}\eqdef\frac{1}{m}J^{AB}_{\tmN}$, we get
\be
\big\|\mN_U\left(\rho^{AE}\right)-\tau^B\otimes\rho^E\big\|_1\leq \left\|\tmN_U\left(\trho^{AE}\right)-\ttau^B\otimes\trho^E\right\|_2\;.
\ee
Taking the integral over $U\in\muu(m)$ on both sides we obtain
\ba\label{1234}
\int_{\mathfrak{U}(A)}dU\;\big\|\mN_U\left(\rho^{AE}\right)-\tau^B\otimes\rho^E\big\|_1
&\leq \int_{\mathfrak{U}(A)}dU\;\left\|\tmN_U\left(\trho^{AE}\right)-\ttau^B\otimes\trho^E\right\|_2\\
\GG{\eqref{tgftn}}&\leq \sqrt{\tr\left[(\ttau^{AB})^2\right]\tr\left[(\trho^{AE})^2\right]}\;.
\ea

Finally, choosing $\eta^E$ and $\zeta^B$ such that
\ba
&\tr\left[(\trho^{AE})^2\right]=\tr\left[\left((I^A\otimes(\eta^E)^{-\frac{1}{4}})\rho^{AE}(I^A\otimes(\eta^E)^{-\frac{1}{4}})\right)^2\right]=2^{-\tH_2^\ua(A|E)_\rho}\\
&\tr\left[(\ttau^{AB})^2\right]=\tr\left[\left((I^A\otimes(\zeta^B)^{-\frac{1}{4}})\tau^{AB}(I^A\otimes(\zeta^B)^{-\frac{1}{4}})\right)^2\right]=2^{-\tH_2^\ua(A|B)_\tau}
\ea
completes the proof.
\end{proof}

\begin{myg}{}
\begin{corollary}
Let $k\in\mbb{N}$, $\p\in\prob(k)$, and  $\{U_x\}_{x\in[k]}\subset\muu(A)$ be as in~\eqref{7222}. Then, using the same notations as in Theorem~\ref{thm1141}, we have
\be\label{corimpi}
\sum_{x\in[k]}p_x\left\|\mN^{A\to B}\Big(U^A_x\rho^{AE}\left(U^A_x\right)^*\Big)-\tau^B\otimes\rho^E\right\|_1\leq 2^{-\frac{1}{2}\big(\tH_2^\ua(A|E)_\rho+\tH_2^\ua(A|B)_\tau\big)}\;.
\ee
\end{corollary}
\end{myg}

\bex
Use Theorem~\eqref{thm1141} and Corollary~\eqref{corimpi0} to prove the corollary above.
\eex

\bex
Show that if  $\omega^{AB}\eqdef\frac1t J_\mN^{AB}$, where $t\eqdef\tr\left[J_\mN^{AB}\right]$ (i.e.\ $\omega^{AB}=\frac{|A|}{t}\tau^{AB}$ is a density matrix), and similarly, $\sigma^{AB}\eqdef\frac1r\rho^{AB}$ with $r\eqdef\tr\left[\rho^{AB}\right]$ then the decoupling theorem above can be expressed as
\be
\int_{\mathfrak{U}(A)}dU^A\;\left\|\mN^{A\to B}\Big(U^A\rho^{AE}\left(U^A\right)^*\Big)-\tau^B\otimes\rho^E\right\|_1\leq \frac{rt}{|A|}2^{-\frac{1}{2}\big(\tH_2^\ua(A|E)_\sigma+\tH_2^\ua(A|B)_\omega\big)}\;.
\ee
\eex

\section{Notes and References}

Conditional majorization was first introduced in~\cite{GGH+2018} in the context of the quantum uncertainty principle, but its quantum version was fully defined in~\cite{GWB+2022}.

Many text books on quantum information includes a section on quantum conditional entropies and their properties. Specifically, the books  by~\cite{Tomamichel2015}, and~\cite{KW2021}, provide a comprehensive review on the various quantum R\'enyi conditional entropies and their properties.
The axiomatic approach\index{axiomatic approach} for the quantum conditional entropy, as presented in this chapter, is due to~\cite{GWB+2022}. Semi-causal maps were first introduced by~\cite{BGNP2001}, and was conjectured to have the characterization given in Theorem~\ref{thm:semic}. Shortly after this conjecture was proved by~\cite{ESW2002}. The relatively simplified proof provided here for Theorem~\ref{thm:semic} is due to~\cite{PHHH2006}. The duality relation given in Lemma~\ref{lem:durel} was first introduced in~\cite{TCR2009}. The proofs for the two other relations in Eq.~\ref{rel231} were given for the first relation  by~\cite{MDS+2013} and independently by~\cite{Beigi2013}, and for the second relation of~\eqref{rel231}  by~\cite{TBH2014}. Additional work related to the topics presented in this chapter includes~\cite{SCBG2024,VSGC2022}.

%%%%%%%%%%%%%%%%%%%%%%%%%%

%%%%%%%%%%%%%%%%%%%%%%%%%%%

\chapter{The Asymptotic Regime}\label{ch:qs}

As the dimension of a physical system grows, one can employ several tools from probability theory and statistics (e.g. the law of large numbers), to study its behaviour and properties. Specifically, one of the main goals of quantum resource theories is to determine the rate at which many copies of one resource can be converted into many copies of another. 
The methods and tools developed here provide the foundations for several topics in this asymptotic domain. 
We start by reviewing some of these concepts and their generalizations to the quantum world.

\section{Classical Typicality}\index{typicality}

A central theme in information theory is finding efficient methods for transmitting information from one party (Alice) to another (Bob). Consider a scenario where Alice wishes to send Bob a message $x$ from a set of $m$ possibilities. This transmission would require $\log_2(m)$ classical bits, achievable through $\log_2(m)$ uses of a perfectly noiseless classical bit-channel. At first glance, it seems the resource cost for sending a message of size $m$ is:
\be
\log_2(m)[c\to c]\;,
\ee
where $[c\to c]$ denotes one usage of a noiseless cbit-channel.

However, if Alice's message to Bob consists of English alphabet letters (where $m=26$), additional information about the message emerges. For instance, the letter `E' appears with a frequency of about 12\%, while `Z' occurs only about 0.07\% of the time. Thus, it's more probable for Bob to receive `E' than `Z'. Shannon's groundbreaking 1948 paper illustrated how to leverage this additional information to significantly reduce classical communication costs. This approach is based on the crucial concept of typicality, which plays a major role in information science and, by extension, in quantum resource theories.

\subsection{i.i.d. Information Source}\label{subsec:iid}\index{i.i.d. source}

Building on the above, we model the messages that Alice sends to Bob as being drawn from a \emph{classical information source}. This source can be conceptualized as a sequence of random variables $X_1$, $X_2$, ..., each representing an output. For instance, the word 'PEACE' sent by Alice translates into a five-letter sequence: $X_1=$P, $X_2=$E, $X_3=$A, $X_4=$C, $X_5=$E, with each letter having a specific probability of occurrence. To simplify the mathematics and as a first approximation, we make two assumptions about this source.

Firstly, we assume that the source's various uses are \emph{independent}, meaning that each letter is emitted without being influenced by the previous ones. However, it's clear that this assumption does \emph{not} strictly apply to the English alphabet. Take, for instance, the letter `T', which is the second most frequent letter in English, with an occurrence frequency of about 9\%. If Alice sends the letter `M' to Bob, the likelihood of the next letter being `T' is significantly lower than 9\%, given the relative rarity of the `MT' combination in English words. Thus, for many information sources, including English, the assumption of independence should be considered more as a first order approximation than a definitive rule.

Secondly, we assume the source's uses are \emph{identically distributed}, implying each use of the source can be represented by a random variable, sharing the same alphabet and possessing an identical probability distribution. That is, for all $k \in \mathbb{N}$, the probability, $\pr(X_k = x) \eqdef p_x$, that the random variable $X_k$ is equal to some $x$ in the alphabet will be independent of $k$. We will therefore consider here information sources that are both \emph{independent and identically distributed}, or in short, i.i.d. information sources. An i.i.d. source is thus represented by a single random variable $X$, with an alphabet $x\in\mX$ and a corresponding probability distribution $\{p_x\}_{x\in\mX}$. For simplicity, we consider finite alphabets, taking $\mX=\{1,\ldots,m\}$, and denote the distribution as i.i.d.$\sim \p$, emphasizing the probability vector $\p=(p_1,\ldots,p_m)^T$ of $X$.

Consider, for example, $n$ uses of a binary i.i.d. source, producing a sequence of $n$ bits $X^n=(X_1,\ldots,X_n)$, with $p$ being the probability of outcome ``0" and $1-p$ for ``1" (hence $X_k\in\{0,1\}$). For large $n$, it's highly likely that the sequence $X^n$ will contain roughly $np$ zeros and $n(1-p)$ ones. The occurrence probability of such a typical sequence is approximately:
\be
\Pr\left(X^n=x^n\right)=p_{x_1}p_{x_2}\cdots p_{x_n}\approx p^{np}(1-p)^{n(1-p)}\;.
\ee
This probability can be further simplified to:
\be
\Pr\left(X^n=x^n\right)\approx 2^{-nH(X)}\;,
\ee
where $H(X)$ is the binary Shannon entropy:
\be
H(X)\eqdef-p\log_2(p)-(1-p)\log_2(1-p)\;.
\ee
Despite the variety of typical sequences $x^n$, they all share approximately the same probability of occurrence. This effect is known as \emph{the asymptotic equipartition\index{asymptotic equipartition} property}, a direct result of the (weak) \emph{law of large numbers}.

\subsection{The Law of Large Numbers}\index{law of large numbers}

Let $X_1,X_2,\ldots$ be an i.i.d.$\sim \p$ source with the same distribution as of $X$. Suppose that
\be
\mbb{E}(X)\eqdef \sum_{x\in\mX}p_x x<\infty\quad\text{and}\quad\mbb{E}(X^2)\eqdef \sum_{x\in\mX}p_x x^2<\infty\;.
\ee
Since we only consider in this book sets with finite cardinality, these conditions will trivially hold. 

\begin{myt}{\color{yellow} The Law of Large Numbers}
\begin{theorem}\label{lln}
With the notations as above, for all $\eps>0$ 
\be\label{inprobability}
\lim_{n\to\infty}\Pr\Big(\Big|\frac{1}{n}\sum_{j\in[n]}X_j-\mbb{E}(X)\Big|>\eps\Big)=0\;.
\ee
\end{theorem}
\end{myt}
The law above is very intuitive as it shows that for very large $n$, the probability that $\frac{1}{n}\sum_{j\in[n]}X_j$ is  close to $\mbb{E}(X)$ is almost one. In particular, \eqref{inprobability} is equivalent to the statement that
\be
\lim_{n\to\infty}\frac{1}{n}\sum_{j\in[n]}X_j=\mbb{E}(X)\quad\text{in probability}.
\ee
\begin{proof}
We first prove the theorem for the case that the expectation value of $X$ is zero;  i.e. $\mbb{E}(X)=0$. Denote by $S_n\eqdef\frac{1}{n}\sum_{j\in[n]}X_j$. Then,
\be\label{ave}
\mbb{E}(S_{n}^2)=\frac{1}{n^2}\sum_{j,k=1}^{n}\mbb{E}(X_jX_k)
\ee
A key observation is that for $j\neq k$, the two random variables $X_j$ and $X_k$ are independent, and consequently
\be\mbb{E}(X_jX_k)=\sum_{x_j,x_k\in\mX}x_jx_kp_{x_j}p_{x_k}=\mbb{E}(X_j)\mbb{E}(X_k)=0\ee
Therefore, the only contributing terms in~\eqref{ave} are those with $j=k$. Hence.
\be
\mbb{E}(S_{n}^2)=\frac{1}{n^2}\sum_{j\in[n]}\mbb{E}(X_j^2)=\frac{1}{n}\mbb{E}(X^2)\;.
\ee
The above equation already demonstrates that for very large $n$ the \emph{variance} of $S_n$ is very small, indicating that it will reach a single value in the limit $n\to\infty$. 
On the other hand, $\mbb{E}(S_{n}^2)$ can be splitted into two terms, those for which the value of $S_n$ is close to zero 
and those for which it is at least $\eps$-distance away from zero:
\ba
\mbb{E}(S_{n}^2)=\sum_{s_n^2}s_n^2\Pr(S_n^2=s_n^2)&=\sum_{|s_n|\leq\eps}s_n^2\Pr(S_n^2=s_n^2)
+\sum_{|s_n|>\eps}s_n^2\Pr(S_n^2=s_n^2)\nonumber\\
&\geq\sum_{|s_n|>\eps}s_n^2\Pr(S_n^2=s_n^2)\\
&\geq\eps^2\sum_{|s_n|>\eps}\Pr(S_n^2=s_n^2)=\eps^2\Pr(|S_n|>\eps)\;.
\ea
We therefore conclude that
\be\label{ratetozero}
\Pr(|S_n|>\eps)\leq \frac{\mbb{E}(S_{n}^2)}{\eps^2}=\frac{\mbb{E}(X^2)}{n\eps^2}\xrightarrow[n\to\infty]{}0
\ee
This completes the proof for the case $\mbb{E}(X)=0$. If $\mbb{E}(X)\neq 0$ then define a sequence of i.i.d.$\sim \p$ random variables $Y_j\eqdef X_j-\mbb{E}(X)$. With this definition we get the $\mbb{E}(Y_j)=0$ so that 
\be
\Pr\Big(\Big|\frac{1}{n}\sum_{j\in[n]}X_j-\mbb{E}(X)\Big|>\eps\Big)=\Pr\Big(\Big|\frac{1}{n}\sum_{j\in[n]}Y_j\Big|>\eps\Big)\xrightarrow{n\to\infty}0.
\ee
This completes the proof.
\end{proof}

\begin{exercise}[Markov Inequality]
Prove that for any $t>0$ and any nonnegative random variable $X$
\be\label{markin}
\Pr\left(X\geq t\right)\leq\frac{\mbb{E}(X)}{t}\;.
\ee
\end{exercise}

The law of large numbers above does not tell us much how fast the probability in~\eqref{inprobability} goes to zero. In the proof we saw in~\eqref{ratetozero} that it goes to zero at least as $1/n$. If instead of requiring that the expectation value of $X$ and $X^2$ are finite, we require a stronger condition that the alphabet of $X$ themselves are all bounded,  then it is possible to show that in this case the probability in~\eqref{inprobability} goes to zero exponentially fast with $n$.

\begin{myt}{\color{yellow} Hoeffding's Inequality}\index{Hoeffding's inequality}
\begin{theorem}\label{hoein}
Let $X_1,\ldots,X_n$ be $n$ independent random variable satisfying $a_j\leq X_j\leq b_j$ for all $j=1,\ldots,n$. Then, 
\be\label{hoeffmain}
\Pr\Big(\Big|\frac{1}{n}\sum_{j\in[n]}\Big(X_j-\mbb{E}(X_j)\Big)\Big|>\eps\Big)\leq \exp\left(-\frac{2n^2\eps^2}{\sum_{j\in[n]}(b_j-a_j)^2}\right)\;.
\ee
\end{theorem} 
\end{myt}

Note that Hoeffding's inequality above does not assume that the random variables $X_1,\ldots,X_n$ are identically distributed. If we add this assumption (so that $X_1,\ldots,X_n$ are i.i.d.), then we get a simplified version of Hoeffding's inequality given by
\be\label{hoe2}
\Pr\Big(\Big|\frac{1}{n}\sum_{j\in[n]}X_j-\mbb{E}(X)\Big|>\eps\Big)\leq \exp\left(-\frac{2n\eps^2}{(b-a)^2}\right)\;.
\ee
where we assumed that $a\leq X\leq b$. 

To prove Hoeffding's inequality we will need the following lemma.
\begin{myg}{Hoeffding's Lemma}
\begin{lemma}
Let $X$ be a real valued bounded random variable with expected value $\mbb{E}(X)=\mu$ and $a\leq X\leq b$ for some $a,b\in\mbb{R}$ with $b>a$. Then, for all $t\in\mbb{R}$ we have
\be\label{ineq77}
\mbb{E}\left(e^{tX}\right)\leq\exp\left(t\mu+\frac{t^2(b-a)^2}8\right)\;.
\ee
\end{lemma}
\end{myg}
\begin{proof}
Consider first the case $\mu=0$. We therefore must have $a\leq 0\leq b$. Also, it's important to note that if $a=0$, then the condition $\mathbb{E}(X) = 0$ leads to $\mathbb{E}\left(e^{tX}\right) = 1$ (can you see why?). As a result, the inequality~\eqref{ineq77} is valid under these circumstances. Therefore, we will proceed with the assumption that $a < 0$.
The convexity of the function $f(x)\eqdef e^{tx}$ implies that for any $a\leq x\leq b$ we have
\be
e^{t x}\leq \frac{b-x}{b-a}e^{ta}+\frac{x-a}{b-a}e^{tb}
\ee
where we wrote $x$ as the convex combination $x=a\frac{b-x}{b-a}+b\frac{x-a}{b-a}$. The key idea of the inequality above is that the right-hand side depends linearly on $x$. Applying this inequality to the random variable $X$ we get
\ba
\mbb{E}\left(e^{t X}\right)&\leq \frac{b-\mbb{E}(X)}{b-a}e^{ta}+\frac{\mbb{E}(X)-a}{b-a}e^{tb}\\
\Gg{\mu=0}&=\frac{b}{b-a}e^{ta}-\frac{a}{b-a}e^{tb}\\
\Gg{c\eqdef-\frac a{b-a}}&=(1-c+ce^{t(b-a)})e^{ta}\;.
\ea
Note that $c>0$ since $a<0$. Finally, denote by $s\eqdef t(b-a)$ the right-hand side of the equation above becomes equal to
\be
(1-c+ce^{s})e^{-cs}= e^{f(s)}\;,\quad\text{where}\quad f(s)\eqdef -cs+\log\left(1-c+ce^{s}\right)\;,
\ee
and we used the equality $e^{ta}=e^{-cs}$. Consider the Tylor expansion of $f(s)$
up to its second order
\be
f(s)=f(0)+sf'(0)+\frac12 s^2f''(q)\;,
\ee
where $q$ is some real number between zero and $s$. By straightforward calculation, we get that $f(0)=f'(0)=0$ and $f''(q)\leq\frac14$ (see Exercise~\ref{ex:qfpp} for more details). Combining everything we conclude that
\be
\mbb{E}\left(e^{t X}\right)\leq e^{f(s)}=e^{\frac12 s^2f''(q)}\leq e^{\frac18 s^2}=e^{\frac18 t^2(b-a)^2}\;.
\ee
This completes the proof for the case $\mu=0$. The proof for the case $\mu\neq 0$ is obtained immediately by defining $\tilde{X}\eqdef X-\mu$ and applying the theorem for $\tilde{X}$ (see Exercise~\ref{ex:appl}).
\end{proof}

\begin{exercise}\label{ex:qfpp}
Show that for $f''(q)\leq\frac14$. Hint: Calculate the second derivative $f''(q)$ and show that it can be expressed as $p(1-p)$ for some number  $p>0$ (that depends on $q$) and use the fact that $p(1-p)\leq\frac14$.
\end{exercise}

\begin{exercise}\label{ex:appl}
Show that the proof for the case $\mu\neq 0$ in the lemma above follows immediately by defining $\tilde{X}\eqdef X-\mu$ and applying the theorem for $\tilde{X}$.
\end{exercise}

We are now ready to prove the Hoeffding's inequality.

\begin{proof}[Proof of Theorem~\ref{hoein}.]
Denote by $S_n\eqdef X_1+\cdots+X_n$ and observe that for any $r,t>0$ we have
\ba
\pr\left(S_n-\mbb{E}(S_n)\geq t\right)&=\pr\left(e^{r(S_n-\mbb{E}(S_n))}\geq e^{rt}\right)\\
\GG{\text{Markov's Inequality}~\eqref{markin}}&\leq e^{-rt}\mbb{E}\left(e^{r(S_n-\mbb{E}(S_n))}\right)\\
&=e^{-rt}\mbb{E}\Big(\prod_{j=1}^ne^{r(X_j-\mbb{E}(X_j))}\Big)\\
\GG{\{X_j\}\; are\; independent}&=e^{-rt}\prod_{j=1}^n\mbb{E}\left(e^{r(X_j-\mbb{E}(X_j))}\right)\\
\GG{\text{Hoeffding's Lemma}}&\leq e^{-rt}\prod_{j=1}^ne^{\frac18r^2(b_j-a_j)^2}\\
&=e^{g(r)}
\ea
where $g(r)\eqdef -rt+\frac18r^2\sum_{j\in[n]}(b_j-a_j)^2$ is a quadratic function whose minimum  is given by the right-hand side of~\eqref{hoeffmain}. This completes the proof.
\end{proof}

\begin{exercise}
Show that the minimum of the function $g(r)$ above is given by the right-hand side of~\eqref{hoeffmain}.
\end{exercise}

\subsection{Typical Sequences}

 As we saw above, a typical sequence $x^n=(x_1,\ldots,x_n)$ that is drawn from an i.i.d.$\sim\p$ source has a probability to occur that is close to $2^{-nH(\p)}$. In this section we apply the law of large numbers to make the notion of typical sequences rigorous.

\begin{myd}{Typical Sequence}
\begin{definition}
Let $\eps>0$ and let $X$ be a random variable with cardinality $|\mX|=m$, corresponding to an i.i.d.\ source. A sequence of $n$ source outputs $x^n\eqdef(x_1,\ldots,x_n)$ is called $\eps$-typical if 
\be\label{etypical}
2^{-n\left(H(X)+\eps\right)}\leq \Pr(X^n=x^n)\leq 2^{-n\left(H(X)-\eps\right)}\;.
\ee
where $H(X)\eqdef-\sum_{x\in[m]}p_x\log p_x$ is the Shannon entropy\index{Shannon entropy}.
\end{definition}
\end{myd}

By taking the log on all sides of~\eqref{etypical}, the condition in~\eqref{etypical} can be re-expressed as
\be
\label{tpxn2}
\left|\frac{1}{n}\log_2\left(\frac{1}{\Pr(X^n=x^n)}\right)-H(X)\right|\leq\eps\;.
\ee
We will denote the set of all $\eps$-typical sequences by
\be
\mt_{\eps}(X^n)\eqdef\left\{x^n=(x_1,\ldots,x_n)\;:\;x^n\text{ is }\eps\text{-typical}\right\}\;.
\ee
Therefore, the probability that a sequence is $\eps$-typical is given by
\be
\Pr(\mt_{\eps}(X^n))\eqdef \sum_{x^n\in\mt_{\eps}(X^n)}\Pr(X^n=x^n)\;.
\ee
More generally, for any set of sequences $\mk_n\subseteq[m]^n$ we will use the notation $\pr(\mk_n)$ to denote the probability that a sequence belongs to $\mk_n$. That is, 
\be
\pr(\mk_n)\eqdef\sum_{x^n\in\mt_{\eps}(X^n)}\Pr(X^n=x^n)\;.
\ee
In the following theorem, we denote the constant 
\be\label{c2}
c\eqdef\frac{2}{\big(\log(p_{\max}/p_{\min})\big)^2}>0\;,
\ee
where $p_{\min}>0$ and $p_{\max}$ are the smallest and largest positive (i.e., non-zero) components of $\p\eqdef(p_1,\ldots,p_m)^T$.

\begin{myt}{}
\begin{theorem}\label{tts}
Let $\p\in\prob(m)$, $\eps\in(0,1)$, $\delta_n\eqdef e^{-c\eps^2n}$ where $c$ is defined in~\eqref{c2}, $X$ be a random variable associated with an i.i.d.$\sim\p$ source, and for each $n\in\mbb{N}$, let $\mk_n\subseteq[m]^n$ be a set of sequences with cardinality $|\mk_n|\leq 2^{nr}$ for some $r<H(X)$. Then, for all $n\in\mbb{N}$ the following three inequalities hold:
\begin{enumerate}
\item $\Pr(\mt_{\eps}(X^n))>1-\delta_n$. 
\item $(1-\delta_n)2^{n(H(X)-\eps)}\leq\left|\mt_{\eps}(X^n)\right|\leq 2^{n(H(X)+\eps)}$. 
\item  $\Pr(\mk_{n})\leq e^{-c'n}$, for some $c'>0$.
\end{enumerate}
\end{theorem}
\end{myt}

\begin{proof} 
For the first inequality, we assume
without loss of generality that $\p>0$, since any $x$ with $p_x=0$ never occur and can be removed from the alphabet of $X$. Let $Y\eqdef-\log_2\left(X\right)$ be the random variable whose alphabets symbols are given by $\mY\eqdef\left\{-\log_2\Pr(X=x)\right\}_{x\in[m]}$, with corresponding probabilities $p_x\eqdef\Pr(X=x)$. Let $Y_1,Y_2,\ldots$ be an i.i.d. sequences of random variables where each $Y_j$ corresponds to $X_j$ as above. By definition, each $Y_j$ satisfies $-\log p_{\max}\leq Y_j\leq -\log p_{\min}$. Therefore, from Hoeffding's inequality, particularly~\eqref{hoe2}, we get that
\be\label{pryj}
\Pr\Big(\Big|\frac{1}{n}\sum_{j\in[n]}Y_j-\mbb{E}(Y)\Big|>\eps\Big)\leq e^{-c\eps^2n}\;.
\ee
Observe that
\be
\mbb{E}(Y)=-\sum_{x\in[m]}p_x\log_2\left(\Pr(X=x)\right)=-\sum_{x\in[m]}p_x\log_2 p_x=H(X)\;.
\ee
Moreover,
\be
\frac{1}{n}\sum_{j\in[n]}Y_j=-\frac{1}{n}\sum_{j\in[n]}\log_2\Pr(X_j)=-\frac{1}{n}\log_2\Pr(X^n)=\frac{1}{n}\log_2\left(\frac{1}{\Pr(X^n)}\right)\;.
\ee
We therefore conclude that
\be\label{laep}
\Pr\left(\left|\frac{1}{n}\log_2\left(\frac{1}{\Pr(X^n)}\right)-H(X)\right|>\eps\right)\leq e^{-c\eps^2n}\;.
\ee
The equation above states that the probability that the random variable $X^n=(X_1,\ldots,X_n)$ is not an $\eps$-typical sequence, is no greater than $e^{-c\eps^2n}$. This completes the proof of first part of the theorem.

For the second inequality, we get from the definition of $\eps$-typical sequences that
\be
1\geq\sum_{x^n\in\mt_{\eps}(X^n)}\Pr(X^n=x^n)\geq \sum_{x^n\in\mt_{\eps}(X^n)}2^{-n\left(H(X)+\eps\right)}=\left|\mt_{\eps}(X^n)\right|2^{-n\left(H(X)+\eps\right)}\;.
\ee
Therefore, 
\be
\left|\mt_{\eps}(X^n)\right|\leq 2^{n\left(H(X)+\eps\right)}\;.
\ee
On the other hand, observe that from the first part and the definition of $\eps$-typical sequences we get
\be
1-\delta_n\leq \sum_{x^n\in\mt_{\eps}(X^n)}\Pr(X^n=x^n)\leq \sum_{x^n\in\mt_{\eps}(X^n)}2^{-n\left(H(X)-\eps\right)}=\left|\mt_{\eps}(X^n)\right|2^{-n\left(H(X)-\eps\right)}\;.\ee
Hence,
\be
(1-\delta_n)2^{n(H(X)-\eps)}\leq\left|\mt_{\eps}(X^n)\right|\;.
\ee

For the last part of the proof (i.e., third inequality), let $0<\eps'<\frac12(H(X)-r)$. The probability of $\mk_n$ can be expressed as:
\be
\sum_{x^n\in\mk_n}\Pr(X^n=x^n)=\sum_{x^n\in\mk_n\cap\mt_{\eps'}(X^n)}\Pr(X^n=x^n)+\sum_{\substack{x^n\in\mk_n\\ x^n\not\in\mt_{\eps'}(X^n)}}\Pr(X^n=x^n)
\ee
From the first part of the theorem, the last term can not exceed $\delta_n'\eqdef e^{-c{\eps'}^2n}$ so that
\ba
\sum_{x^n\in\mk_n}\Pr(X^n=x^n)&\leq\sum_{x^n\in\mk_n\cap\mt_{\eps'}(X^n)}\Pr(X^n=x^n)+\delta_n'\\
\GG{{\it x^n}\text{ is }\eps'\text{-typical}}&\leq 2^{-n\left(H(X)-\eps'\right)}|\mk_n|+\delta_n'\\
\Gg{|\mk_n|\leq 2^{nr}}&\leq 2^{-n\left(H(X)-\eps'-r\right)}+\delta_n'\\
\Gg{\eps'<\frac12(H(X)-r)}&\leq 2^{-n\frac12\left(H(X)-r\right)}+\delta_n'\;.
\ea
Since both $\delta_n'$ and  $2^{-n\frac12\left(H(X)-r\right)}$ decrease exponentially fast with zero, there exists $c'>0$ sufficiently small such that $\Pr(\mk_n)\leq e^{-c'n}$. This completes the proof.
\end{proof}

It's also pertinent to mention that~\eqref{laep} can be expressed equivalently as:
\be
\frac{1}{n}\log_2\left(\frac{1}{\Pr(X^n)}\right)\xrightarrow[n\to\infty]{}H(X)\quad\text{in probability}.
\ee
This expression is the exact formulation of the asymptotic equipartition\index{asymptotic equipartition} property, which will be examined in greater detail later in the book.

\begin{exercise}
Using the same notations as in the theorem above, show that if instead of Hoeffding's inequality we use the law of large numbers (i.e., Theorem~\ref{lln}) then we can still show that for any $\delta>0$ and sufficiently large $n\in\mbb{N}$ 
\be
\pr\left(\mt_{\eps}(X^n)\right)>1- \delta\;.
\ee
\end{exercise}

\bex[Variant of Part 3 of Theorem~\ref{tts}]\label{vtts3}
Prove the following variant of part 3 of the theorem above:
Let $r<H(X)$ and let $\{\mk_n\}_{n\in\mbb{N}}$ be sets of sequences of size $n$, and suppose for each $a\in\mbb{N}$ there exists $n>a$ such that $|\mk_n|\leq 2^{nr}$. Then, for any $\delta>0$ and every $b\in\mbb{N}$  there exists $n>b$ such that
$
\Pr(\mk_{n})\leq\delta
$.
\eex

\subsection{Application: Data Compression}\index{data compression}

A data compression scheme is a process by which a sender (Alice) transmits to a receiver (Bob) a message of size $2^n$ by communicating less than $n$ cbits of communication (See Fig.~\ref{compression}). This is possible because Alice draws the message from an i.i.d. information source so that non-typical messages are highly unlikely to occur. Specifically, in a compression scheme of \emph{rate} $r$, Alice encodes (compresses) a message $x^n=(x_1,\ldots,x_n)$ (drawn from an i.i.d. source; particularly, $x_j\in\mX$, where $\mX$ is the alphabet set of the source) into a bit string of size $y^m=(y_1,\ldots,y_m)\in\{0,1\}^n$, with $m=\floor{rn}$, and transmits the sequence $y^m$ to Bob. Bob then decompresses $y^m$ into a sequence 
$z^n=(z_1,\ldots,z_n)$ with $z_j\in\mX$. The goal is that $z^n$ will be almost identical to $x^n$; see Fig.~\ref{compression}. We will denote the compression map by $\mC$ and the decompression map by $\mD$ so that $y^m=\mC(x^n)$ and $z^n=\mD(y^m)$.

\begin{figure}[h]\centering    \includegraphics[width=0.5\textwidth]{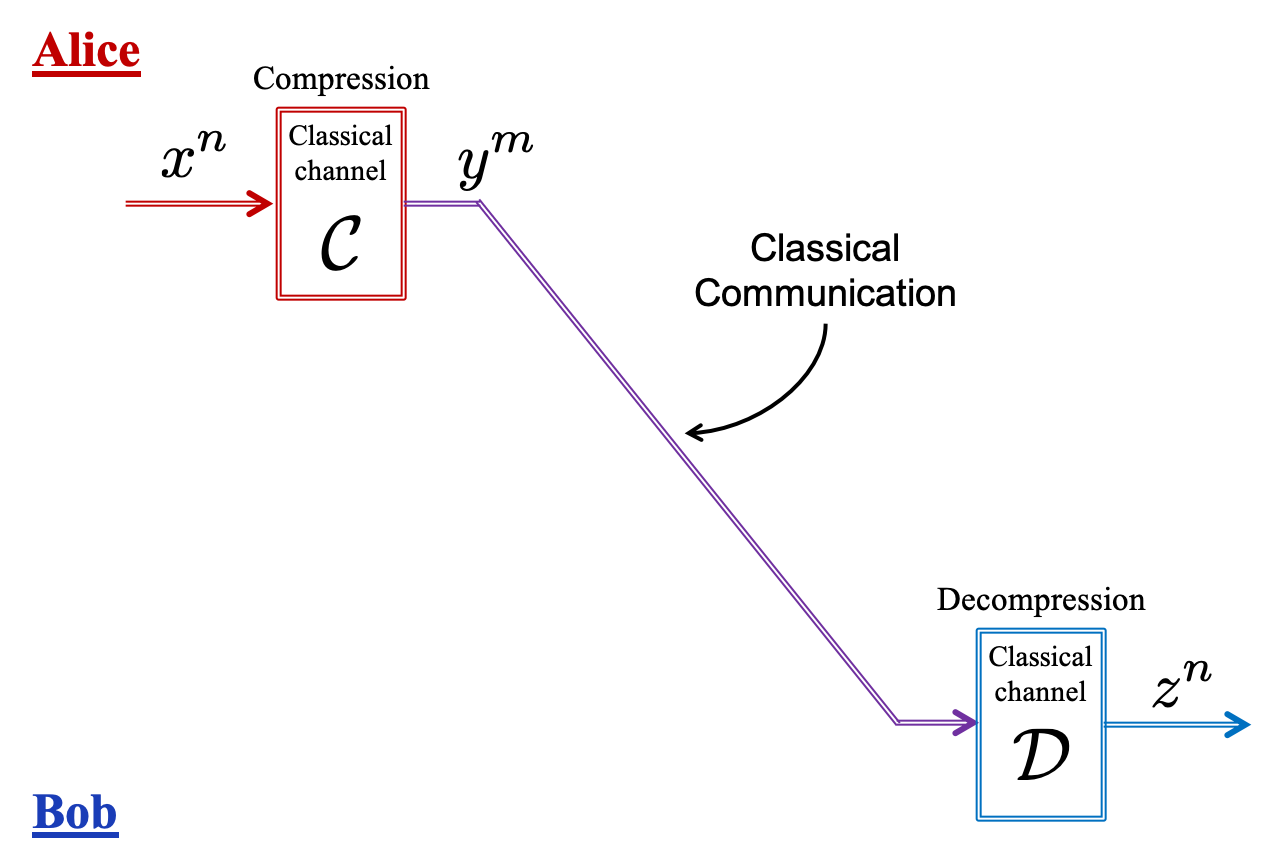}
  \caption{\linespread{1}\selectfont{\small A compression-decompression scheme of rate $r$.}}
  \label{compression}
\end{figure} 

\begin{myd}{}
\begin{definition}
A compression-decompression scheme, $(\mC,\mD)$ of rate $r$ is said to be \emph{reliable} if
\be
\lim_{n\to\infty} \Pr(Z^n=X^n)=\lim_{n\to\infty}\Pr\Big(\mD\left(\mC(X^n)\right)=X^n\Big)=1\;.
\ee
\end{definition}
\end{myd}

\begin{myt}{\color{yellow} Shannon Theorem (1948)}\index{Shannon theorem}
\begin{theorem}
Given an i.i.d. source with entropy $H(X)$, a reliable compression-decompression scheme of rate $r$ exists if and only if $r>H(X)$.
\end{theorem}
\end{myt}

\begin{proof}
Suppose $r>H(X)$. We need to show that there exists a reliable compression scheme of rate $r$. Let $\delta>0$ and let $\eps>0$ be such that $r>H(X)+\eps$. Then, from the first part of Theorem~\ref{tts} for all $n\in\mbb{N}$ we have $\Pr(\mt_{\eps}(X^n))\geq 1-e^{-c\eps^2 n}$, where $c>0$ is a constant defined in~\eqref{c2}. Let $k\in\big\{1,2,\ldots,\left|\mt_{\eps}(X^n)\right|\big\}$ be the index labeling all the $\eps$-typical sequences in $\mt_{\eps}(X^n)$. We assume that  Alice and Bob agreed on the order before hand. Define the compression map $\mC:\mX^n\to\{0,1\}^{m}$, with $m\eqdef\ceil{\log_2\left|\mt_{\eps}(X^n)\right|}$ as follows. If $x^n$ is the $k^{\rm th}$ sequence of $\mt_{\eps}(X^n)$ then $\mC(x^n)$ is the binary representation of $k$. If $x^n\not\in \mt_{\eps}(X^n)$ then $\mC(x^n)=(0,\ldots,0)$; i.e. if Bob receives the zero sequence he knows there is an error.

Now, from the second part of the theorem of typical sequences we know that
\be
\left|\mt_{\eps}(X^n)\right|\leq 2^{n\left(H(X)+\eps\right)}<2^{nr}\;.
\ee
Therefore, for large enough $n$, the sequence $y^m=\mC(x^n)$ is of size $m=\ceil{\log_2\left|\mt_{\eps}(X^n)\right|}\leq nr$.

The decoding scheme $\mD:\{0,1\}^{m}\to\mX^n$ is defined as follows. If $y^m$ is the zero sequence Bob declares an error. Otherwise, if $y^m$ is the binary representation of $k$, then $\mD(y^m)=x^n$ with $x^n$ being the $k^{\rm th}$ sequence of $\mt_{\eps}(X^n)$. It is left to show that the success probability goes to one in the asymptotic limit $n\to\infty$. Indeed, by construction,
\be\label{bcznxn}
\Pr\left(Z^n=x^n\big|X^n=x^n\right)=\begin{cases} 0 & \text{ if }x^n\text{ is not }\eps\text{-typical}\\
1 &\text{ if }x^n\text{ is }\eps\text{-typical}
\end{cases}\;.
\ee
Therefore,
\ba
\Pr(Z^n=X^n)&=\sum_{x^n\in\mX^n}\Pr(X^n=x^n)\Pr\left(Z^n=x^n\big|X^n=x^n\right)\\
\GG{\eqref{bcznxn}}&=\sum_{x^n\in \mt_{\eps}(X^n)}\Pr(X^n=x^n)\\
&=\Pr(\mt_{\eps}(X^n))\geq 1-\delta_n\;.\nonumber
\ea
Since $\lim_{n\to\infty}\delta_n=0$ we conclude that  $\lim_{n\to\infty}\Pr(Z^n=X^n)=1$. Hence, the compression-decompression scheme above is reliable.

Conversely, suppose there exists compression-decompression scheme of rate $r<H(X)$. Then, there are at most $2^{nr}$ outputs for $\mD(y^m)$.
Consequently, the set
\be
\mk_n\eqdef\left\{\ x^n\;:\;\mD\left(\mC(x^n)\right)=x^n\right\}\;,
\ee 
satisfies $\left|\mk_n\right|\leq 2^{nr}$ for all $n$. From the third part of Theorem~\ref{tts} we get that
$\lim_{n\to\infty}\Pr(\mk_n)=0$. Hence, a compression-decompression scheme of rate $r<H(X)$ cannot be reliable. This completes the proof.
\end{proof}

Note that in the proof above we showed that if $r<H(X)$, then not only that the scheme is not reliable, but in fact the probability that $Z^n=X^n$ goes to zero; i.e. $\lim_{n\to\infty}\Pr(\mk_n)=0$. In other words, the error probability goes to one. This type of behaviour is known in classical and quantum Shannon theories as the \emph{strong converse}, whereas the \emph{weak converse} corresponds to a proof in which the error probability is shown to be bounded away from zero as $n$ goes to infinity (but not necessarily goes to one).

\section{Quantum Typicality}\index{typicality}

In this section, we extend the concept of typicality to the quantum realm. We begin by introducing the definition of an i.i.d.\ quantum information source.

\subsection{i.i.d.\ Quantum Source}\index{i.i.d. source}

An i.i.d. quantum source can be considered as an ensemble of states $\left\{q_y,\;\phi_y\right\}_{y\in[k]}$ with each $\phi_y\in\pure(A)$, from which states are independently selected. Consequently, after utilizing the source $n$ times, we obtain a sequence of $n$ quantum states:
\be
|\phi_{y^n}\ra\eqdef |\phi_{y_1}\ra|\phi_{y_2}\ra\cdots|\phi_{y_n}\ra\;.
\ee
In contrast to classical sequences $y^n\eqdef (y_1,\ldots,y_n)$, the quantum sequence above may not be distinguishable, as the states $\{|\phi_{y}\ra\}_{y\in[k]}$ of the source are not orthogonal in general. 

Imagine that Alice wishes to transmit the aforementioned state $|\phi_{y^n}\ra$ to Bob. If Alice is aware of the value of $y^n$, she can employ Shannon's compression coding to send $y^n$ to Bob over a \emph{classical} channel at a rate of $H(Y)$ (meaning the transmission of each source symbol incurs a cost of $H(Y)[c\to c]$). Upon receiving $y^n$, Bob can recreate the state $|\phi_{y^n}\ra$ (assuming Bob knows the quantum source). However, as we will discuss, if Alice and Bob have access to noiseless quantum channels, not only is it unnecessary for Alice to know $y^n$, but she can also transmit the state $|\phi_{y^n}\ra$ to Bob more efficiently! Specifically, each source state transmission costs $H(A)_\rho[q\to q]$, where $H(A)_\rho\eqdef-\tr[\rho^A\log\rho^A]$ denotes the von Neumann entropy of the state
\be\label{rho}
\rho^A\eqdef\sum_{y\in[k]}q_y|\phi_y\lr\phi_y|^A\;.
\ee
Observe that from the upper bound in~\eqref{convone} with $\rho_x$ replaced by the pure state $\phi_y$ and $\p$ replaced by $\q$, we get that $H(A)_\rho\leq H(\q)=H(Y)$.

Without the classical knowledge of $y$, the quantum source generates the state $\rho$ as mentioned above at each usage. Thus, we will denote by i.i.d.$\sim\rho$, an i.i.d.\ quantum source drawn from an ensemble of states whose average state, as described in~\eqref{rho}, is $\rho$. Additionally, after $n$ uses of the source, the produced state is:
\be
\rho\otimes\rho\otimes\cdots\otimes\rho\eqdef\rho^{\otimes n}\;.
\ee

While we assume Alice lacks access to the classical register $y$ of the source, it's plausible that this value is recorded in some register system $R$. If $R$ is classical, each source use generates the cq-state:
\be\label{cr}
\rho^{RA}\eqdef\sum_{y\in[k]}q_y|y\lr y|^R\otimes\phi_y^A\;.
\ee
Alternatively, if the registrar system $R$ is quantum, then each use of the source produces the state
\be\label{entangle}
|\psi^{RA}\ra=\sum_{y\in[k]}\sqrt{q_y}|y\ra^R|\phi_y\ra^A\;.
\ee
Both the classical and quantum register systems record the value of $y$. However, without access to $R$, Alice and Bob cannot distinguish the source $\{q_y,\;\phi_y\}_{y\in[k]}$ from another source $\{r_z,\;\psi_z\}_{z\in[\ell]}$, whose average state is $\sum_{z\in[\ell]}r_z\psi_z=\rho$.

Although we assume that Alice and Bob do not have access to system $R$, it is necessary to think about the quantum source with a recording system. Otherwise, without the knowledge of $y^n$, the states
$|\phi_{y^n}\ra$ become equivalent to $\rho^{\otimes n}$ so that Bob, in principle, can prepare any number of copies of $\rho$ without any communication from Alice. However, if other parties have access to $y^n$, then they can verify that the state $\rho^{\otimes n}$ that Bob prepared is not the original state $|\phi_{y^n}\ra$ that Alice intended to send. 

Note that by applying the completely dephasing map $\Delta^R$ on system $R$, the entangled state $|\psi^{RA}\ra$ in~\eqref{entangle} becomes the cq-state\index{cq-state} in~\eqref{cr}. This demonstrates that taking the registrar to be quantum is more general and we therefore adopt the entangled description $|\psi^{RA}\ra$ of a quantum source.
Note that in this picture, after $n$ uses of the source, Alice shares with the registrar the state
\be
|\psi^{R^nA^n}\ra\eqdef|\psi^{RA}\ra^{\otimes n}\;.
\ee
In the quantum version of the compression scheme discussed above, the task of Alice is to transfer her system $A^n$ to Bob using the smallest possible number of noiseless qubit channels $[q\to q]$. We postpone the full details of this task to volume 2 of the book where we study quantum Shannon theory in more details.

\subsection{Typical Subspaces}\index{typicality}

Consider an i.i.d. quantum source generated from the ensemble $\{q_y,\;\phi_y^A\}_{y\in[k]}$. 
Let 
\be
\rho^A\eqdef\sum_{y\in[k]}q_y|\phi_y\lr\phi_y|^A=\sum_{x\in[m]}p_x|x\lr x|^A\;,
\ee
where $\{p_x\}$ are the eigenvalues of $\rho$ and $\{|x\ra\}$ are the corresponding eigenvectors. We define the classical system (random variable) $X$ to have alphabet symbols $x\in[m]$ corresponding to a probability distribution $\{p_x\}$.
We point out that the alphabet symbols of the system $Y$ that we discussed above corresponds to a different probability distribution $\{q_y\}$.

Now, observe that the state
\ba
\rho^{\otimes n}&=\sum_{x_1\in[m]}\sum_{x_2\in[m]}\cdots\sum_{x_n\in[m]}p_{x_1}p_{x_2}\cdots p_{x_n}|x_1\cdots x_n\lr x_1\cdots x_n|^{A^n}\\
&\eqdef\sum_{x^n\in[m]^n}p_{x^n}|x^n\lr x^n|^{A^n}
\ea
where $|x^n\ra\eqdef |x_1\cdots x_n\ra$ and $p_{x^n}\eqdef p_{x_1}p_{x_2}\cdots p_{x_n}$. We use the notation $\mt_{\eps}(X^n)$ to denote the set of all $\eps$-typical sequences $x^n$ with respect to a \emph{classical} system $X$ corresponding to an i.i.d.$\sim\p$ source. It is important to note that the components of the vector $\p\in\prob(m)$ are the eigenvalues of $\rho$. With this notation,
for every such i.i.d.$\sim\rho^A$ source, we define a corresponding typical subspace 
\be
\mt_{\eps}(A^n)\eqdef \spa\left\{|x_1\cdots x_n\ra\;:\;x^n\in\mt_{\eps}(X^n)\right\}\subseteq A^n\;,
\ee
and a typical projection 
\be
\Pi_{\eps}^{n}\eqdef \sum_{x^n\in\mt_{\eps}(X^n)}|x^n\lr x^n|\;.
\ee

\begin{myt}{}
\begin{theorem}\label{qtypical}
Let $\rho\in\md(A)$, $\eps\in(0,1)$, $\delta_n\eqdef e^{-c\eps^2n}$ where $c$ is defined in~\eqref{c2}, $\mt_{\eps}(A^n)$ and $\Pi_{\eps}^{n}$ be the typical subspaces and projections associated with a quantum i.i.d.$\sim\rho$ source. Further, for each $n\in\mbb{N}$, let $P_n\in\pos(A^n)$ be an orthogonal projection to a subspace with dimension $\tr\left[P_n\right]\leq 2^{nr}$ for some $r<H(A)_\rho$ ($r$ is independent on $n$). Then, for all $n\in\mbb{N}$ the following three inequalities hold:
\begin{align}
1.\quad\quad&\tr\left[\Pi_{\eps}^{n}\rho^{\otimes n}\right]\geq 1-\delta_n\;.\label{643}\\
2.\quad\quad&(1-\delta_n)2^{n(H(A)_\rho-\eps)}\leq\tr\left[\Pi_{\eps}^{n}\right]\leq 2^{n(H(A)_\rho+\eps)}\;.\label{20900}\\
3.\quad\quad &\tr\left[P_{n}\rho^{\otimes n}\right]\leq e^{-c'n}\text{ for some }c'>0\;.
\end{align}
\end{theorem}
\end{myt}
\begin{proof}
The  proofs of the first two parts of the theorem follow from their classical counterparts. Particularly, for the first part
\ba
\tr\left[\Pi_{\eps}^{n}\rho^{\otimes n}\right]&=\sum_{x^n\in\mt_{\eps}(X^n)}p_{x^n}\\
&=\Pr\left(\mt_{\eps}(X^n)\right)\geq 1-\delta_n\;.
\ea
For the second part 
\be
\tr\left[\Pi_{\eps}^{n}\right]=\dim\left( \mt_{\eps}(A^n)\right)=\left|\mt_{\eps}(X^n)\right|\;.
\ee
Therefore, this part follows as well from its classical counterpart. It is therefore left to prove the third part.

We first split the trace into two parts:
\be
\tr\left[P_n\rho^{\otimes n}\right]=\tr\left[P_n\rho^{\otimes n}\Pi_{\eps}^{n}\right]+\tr\left[P_n\rho^{\otimes n}(I-\Pi_{\eps}^{n})\right]\;.
\ee
Our objective is to show that both of these terms are going to zero as $n\to\infty$. For the first one
\ba
\tr\left[P_n\rho^{\otimes n}\Pi_{\eps}^{n}\right]&=\sum_{x^n\in\mt_{\eps}(X^n)}p_{x^n}\left\la x^n\left|P_n\right|x^n\right\ra\\
\Gg{p_{x^n}\leq 2^{-n\left(H(A)-\eps\right)}}&\leq 2^{-n\left(H(A)-\eps\right)}\sum_{x^n\in\mt_{\eps}(X^n)}\left\la x^n\left|P_n\right|x^n\right\ra\\
&\leq 2^{-n\left(H(A)-\eps\right)}\tr\left[P_n\right]\\
\Gg{\tr\left[P_n\right]\leq 2^{nr}}&\leq 2^{-n\left(H(A)-\eps-r\right)}\xrightarrow[n\to\infty]{}0\;,
\ea
where we assumed that $\eps>0$ is small enough so that $H(A)-r>\eps$. For the second term,
\be
\tr\left[P_n\rho^{\otimes n}(I-\Pi_{\eps}^{n})\right]=\sum_{x^n\not\in\mt_{\eps}(X^n)}p_{x^n}\left\la x^n\left|P_n\right|x^n\right\ra\leq 
\sum_{x^n\not\in\mt_{\eps}(X^n)}p_{x^n}\xrightarrow[n\to\infty]{}0\;.
\ee
This completes the proof.
\end{proof}

\section{Relative Typicality}\index{typicality}

As discussed earlier, in the realm of information theory and quantum information, we often encounter scenarios where large sequences or states, denoted as $x^n$ or $\rho^{\otimes n}$ respectively, are drawn from i.i.d.\ sources. We saw that for sufficiently large values of $n$, these sequences or states are highly likely to be typical, adhering to a certain expected pattern. 
Building on this foundation, we now extend this concept of typicality to pairs of sequences and pairs of quantum states. This extension is analogous to how majorization, a mathematical concept used to compare probability vectors based on their spread or dispersion, is extended to relative majorization. In the case of relative majorization, we transitioned from comparing individual probability vectors to comparing pairs of vectors. Similarly, our extension involves analyzing pairs of sequences and quantum states, assessing their typicality not just individually but in relation to each other.

\subsection{The Classical Domain}\label{sec:aep0}

We start with the following definition of a relative $\eps$-typical sequence.
\begin{myd}{}
\begin{definition}\label{rets}
Let $n,m\in\mbb{N}$, $\p,\q\in\prob(m)$, and $\eps\in(0,1)$. A sequence $x^n\in[m]^n$ is said to be relative $\eps$-typical with respect to an i.i.d.$\sim\p$ source, and a probability vector $\q\in\prob(m)$ if 
\be\label{defrelseq}
\left|D(\p\|\q)-\frac{1}{n}\log\frac{p_{x^n}}{q_{x^n}}\right|\leq\eps\;.
\ee
The set of all relative $\eps$-typical sequences is denoted by $\mt_\eps^\rel(X^n)$.
\end{definition}
\end{myd}

\bex
Show that if $x^n\in\mt_\eps^\rel(X^n)$ then
\be\label{ineqrne}
2^{-n\left(D(\p\|\q)+\eps\right)}p_{x^n}\leq q_{x^n}\leq 2^{-n\left(D(\p\|\q)-\eps\right)}p_{x^n}
\ee
\eex

We saw in Sec.~\ref{subsec:iid} that given an i.i.d.$\sim\p$ source, all typical sequences with large size $n$, that are generated by the source, have approximately the same probability to occur given by $\approx 2^{-nH(\p)}$. This phenomenon was dubbed as the asymptotic equipartition\index{asymptotic equipartition} property (AEP). Here we study a variant of this property as described in the following theorem.

\begin{myt}{}
\begin{theorem}\label{ttsrel}
Let $\p,\q\in\prob(m)$ with $\supp(\p)\subseteq\supp(\q)$, $\eps\in(0,1)$, $\delta_n\eqdef e^{-c\eps^2n}$ where $c$ is defined in~\eqref{c2}, $X$ be a random variable associated with an i.i.d.$\sim\p$ source, and for each $n\in\mbb{N}$, let $\mk_n\subseteq[m]^n$ be a set of sequences with cardinality $|\mk_n|\leq 2^{nr}$ for some $r<D(\p\|\q)$. Then, for all $n\in\mbb{N}$ the following three inequalities hold:
\begin{enumerate}
\item $\Pr(\mt_{\eps}^\rel(X^n))>1-\delta_n$. 
\item $(1-\delta_n)2^{n(D(\p\|\q)-\eps)}\leq\left|\mt_{\eps}^\rel(X^n)\right|\leq 2^{n(D(\p\|\q)+\eps)}$. 
\item  $\Pr(\mk_{n})\leq e^{-c'n}$, for some $c'>0$.
\end{enumerate}
\end{theorem}
\end{myt}
\begin{remark}
Roughly speaking, the above theorem indicates that almost all sequences  $x^n\in[m]^n$ have the same ratio $\frac{p_{x^n}}{q_{x^n}}\approx 2^{-nD(\p\|\q)}$.
Observe also that although we consider two probability distributions $\p$ and $\q$, the sequences $X_1,X_2,\ldots$ are drawn from a \emph{single} $\p$-source.
\end{remark}

\begin{proof}
Due to the similarity of this theorem to Theorem~\ref{tts}, we provide here only the proof of the first inequality, leaving the remaining two inequalities as an exercise for the reader. Without loss of generality suppose $\q>0$ and let $Y\eqdef\log\frac{p_{X}}{q_{X}}$ be the random variable with alphabet $\{\log(p_x/q_x)\}_{x\in[m]}$ and corresponding probability $\p=(p_1,\ldots,p_m)^T$.  Consider the sequence $X^n=(X_1,X_2,\ldots,X_n)$ drawn from an i.i.d.$\sim\p$ source, and for each $j\in[n]$ let $Y_j=\log\frac{p_{X_j}}{q_{X_j}}$. Observe that for each $j\in\mbb{N}$
\be
\mbb{E}(Y_j)=\mbb{E}(Y)=\sum_{x\in[m]}p_x\log(p_x/q_x)=D(\p\|\q)\;.
\ee
Therefore, since
\ba
\frac{1}{n}\log\frac{p_{X^n}}{q_{X^n}}&=\frac{1}{n}\log\frac{p_{X_1}\cdots p_{X_n}}{q_{X_1}\cdots q_{X_n}}\\
&=\frac{1}{n}\sum_{j\in[n]}\log\frac{p_{X_j}}{q_{X_j}}
\ea
we conclude that
\ba
\pr\left(\mt_\eps^\rel(X^n)\right)&=\pr\left(\left|\frac{1}{n}\log\frac{p_{X^n}}{q_{X^n}}-D(\p\|\q)\right|<\eps\right)\\
&=\pr\Big(\Big|\frac{1}{n}\sum_{j\in[n]}Y_j-\mbb{E}(Y)\Big|<\eps\Big)\\
\GG{Hoeffding's\; Inequality}&>1-e^{-nc\eps^2}\;.
\ea
This completes the proof of the first inequality.
\end{proof}

\bex
Prove the outstanding inequalities of the aforementioned theorem. Hint: Utilize a method similar to that applied in deriving the analogous inequalities in Theorem~\ref{tts}.
\eex

The theorem above demonstrate that the probability that a sequence is relative $\eps$-typical is very high. Note however that the probability $\pr\left(\mt_\eps^\rel(X^n)\right)$ is computed with respect to an i.i.d.$\sim\p$ source. If on the other hand, we change $\p$ with $\q$ we would get the probability
\be
\pr\left(\mt_\eps^\rel(X^n)\right)_\q\eqdef\sum_{x^n\in\mt_\eps^\rel(X^n)}q_{x^n}\;.
\ee
In the following exercise you show that this probability goes to zero exponentially fast with $n$.

\bex\label{ex:tt} Let $\p,\q\in\prob(m)$ with $\supp(\p)\subseteq\supp(\q)$, and $\eps\in(0,1)$.
Show that for all $n\in\mbb{N}$
\be\label{7196b}
(1-\delta_n)2^{-n\left(D(\p\|\q)+\eps\right)}\leq\pr\left(\mt_\eps^\rel(X^n)\right)_\q\leq 2^{-n\left(D(\p\|\q)-\eps\right)}\;.
\ee
Hint: Take the sum over all $x^n\in\mt_\eps(X^n)$ in all sides of~\eqref{ineqrne}, and use
the inequalities $1\geq \pr\left(\mt_\eps^\rel(X^n)\right)>1-\delta_n$.
\eex

The pair of probability vectors $(\p^{\otimes n},\q^{\otimes n})$ becomes more distinguishable as we increase $n$.  In the following corollary, we use the theorem above to characterize this distinguishability with relative majorization.

\begin{myg}{}
\begin{corollary}\label{asymajo}
Let $m\in\mbb{N}$, $\s,\t\in\prob_{>0}(2)$, and $\p,\q\in\prob(m)$ with $\supp(\p)\subseteq\supp(\q)$ and $\p\neq\q$. Then, for large enough $n\in\mbb{N}$
\be
(\p^{\otimes n},\q^{\otimes n})\succ(\s,\t)\;.
\ee
\end{corollary}
\end{myg}

\begin{proof}
Let $\eps>0$ be a small number and define the stochastic evolution matrix
$E\in\stoc(2,2^n)$ by its action on the standard basis $\{\e_{x^n}\eqdef\e_{x_1}\otimes\cdots\otimes\e_{x_n}\}_{x^n\in[m]^n}$ as
\be
E\e_{x^n}\eqdef\begin{cases}\e_1^{(2)} &\text{if }x^n\in\mt_\eps^\rel\left(X^n\right)\\
\e_2^{(2)} &\text{if }x^n\not\in\mt_\eps^\rel\left(X^n\right)
\end{cases}
\ee
where $\e_1^{(2)}\eqdef(1,0)^T$ and $\e_2^{(2)}=(0,1)^T$ form the standard basis of $\mbb{R}^2$.
We then get
\ba
\left({\p}^{\otimes n},{\q}^{\otimes n}\right)&\succ \left(E{\p}^{\otimes n},E{\q}^{\otimes n}\right)\\
&=\left(\begin{bmatrix} \pr\left(\mt_\eps^\rel(X^n)\right)\\ 1-\pr\left(\mt_\eps^\rel(X^n)\right)\end{bmatrix},\begin{bmatrix} \pr\left(\mt_\eps^\rel(X^n)\right)_{\q}\\ 1-\pr\left(\mt_\eps^\rel(X^n)\right)_{\q}\end{bmatrix}\right)
\ea
Observe that due to the bounds $\pr\left(\mt_\eps^\rel(X^n)\right)\geq 1-\delta_n$ and the bound $\pr\left(\mt_\eps^\rel(X^n)\right)\leq 2^{-n\left(D(\p\|\q)-\eps\right)}$ (see~\eqref{7196b}) the pair of vectors on the right-hand side approaches the pair $\left(\e_1,\e_2\right) $ as $n\to\infty$, where $\{\e_1,\e_2\}$ is the standard basis of $\mbb{R}^2$. Therefore, combining this with
Exercise~\ref{ex:68} we conclude that for any $\s,\t\in\prob_{>0}(2)$, and sufficiently large $n\in\mbb{N}$, 
\be\label{st}
\left({\p}^{\otimes n},{\q}^{\otimes n}\right)\succ\left(\s,\t\right)\;.
\ee
This concludes the proof.
\end{proof}

\subsection{Relative Typical Subspaces}\label{rts}

In this subsection, we aim to expand the concept of relative typical sequences into the quantum domain. However, we soon encounter a significant challenge: unlike classical states, quantum states generally do not commute. This non-commutativity makes a straightforward extension from the classical framework, as shown in~\eqref{defrelseq}, non-trivial. Consequently, the definition of a relative typical subspace in the quantum context will significantly diverge from its classical counterpart.

Let $\rho,\sigma\in\md(A)$ be two density matrices (which we can viewed here as two i.i.d.\ quantum sources).
Suppose also that $\supp(\rho)\subseteq\supp(\sigma)$. Let the spectral decomposition of $\rho$ and $\sigma$ be
\ba
\rho=\sum_{x\in[m]}p_x|\psi_x\lr \psi_x|\quad\text{and}\quad \sigma=\sum_{y\in[m]}q_y|\phi_y\lr \phi_y|\;.
\ea
Similar to the notations in the previous section, for any integer $n$ we will denote by $y^n\eqdef(y_1,\ldots,y_n)$, $q_{y^n}=q_{y_1}\cdots q_{y_n}$, and $|\phi_{y^n}\ra\eqdef |\phi_{y_1}\ra\otimes\cdots\otimes|\phi_{y_n}\ra$.
Then, for any $\eps>0$ and $n\in\mbb{N}$ the \emph{relative typical subspace}, $\mt_{\eps}^{\rel}(A^n)$, is defined as
\be\label{tysub}
\mt^{\rel}_{\eps}(A^n)\eqdef\spa\left\{|\phi_{y^n}\ra\;:\;\left|\tr\left[\rho\log\sigma\right]-\frac{1}{n}\log \left(q_{y^n}\right)\right|\leq\eps\right\}\;.
\ee
We also denote by $\Pi_{\eps}^{\rel,n}$  the projection to the relative typical subspace $\mt_{\eps}^{\rel}(A^n)$.
Note that $\tr\left[\rho\log\sigma\right]$ is well defined since $\supp(\rho)\subseteq\supp(\sigma)$.

The definition provided above clearly does not revert to the classical definition of a relative typical sequence when $\rho$ and $\sigma$ commute. Nevertheless, as we will explore in the theorem below and the subsequent sections, this definition proves to be an effective tool for examining the distinguishability of quantum states. Furthermore, as demonstrated in the upcoming exercise, relative typical subspaces do indeed converge to typical subspaces in the case where $\rho=\sigma$.

\begin{exercise}
Show that if $\rho=\sigma$ then the relative typical subspace reduces to the typical subspace of $\rho$ as defined in the previous section. That is, show that in this case $\mt^{\rel}_{\eps}(A^n)=\mt_{\eps}(A^n)$.
\end{exercise}

\begin{myt}{}
\begin{theorem}\label{tmrelts}
Let $\eps>0$, $\rho,\sigma\in\md(A)$ with $\supp(\rho)\subseteq\supp(\sigma)$, and $c>0$ as defined in~\eqref{c3}. Then, for all $n\in\mbb{N}$
\be\label{647}
\tr\left[\Pi_{\eps}^{\rel,n}\rho^{\otimes n}\right]\geq 1-e^{-nc\eps^2}\;.
\ee  
\end{theorem}
\end{myt}
\begin{proof}
Note that the function $\tr[\rho\log\sigma]$ can be expressed as
\be
\tr[\rho\log\sigma]=\sum_{y\in[m]}\la\phi_y\big|\rho\log\sigma\big|\phi_y\ra=\sum_{y\in[m]}\la\phi_y|\rho|\phi_y\ra\log q_y\;.
\ee
Therefore, denoting the relative distribution $r_y\eqdef \la\phi_y|\rho|\phi_y\ra$, and by $Y$ the random variable whose alphabet is $[m]$, and its corresponding distribution is $\{r_y\}_{y\in[m]}$, we get that
\be
\tr[\rho\log\sigma]=\mbb{E}\left(\log q_{Y}\right)\;.
\ee
Furthermore, we denote by $\mc^{n}_{\eps}$ the classical typical set 
\be
\mc^{n}_{\eps}\eqdef\left\{{y^n}\;:\;\left|\tr\left[\rho\log\sigma\right]-\frac{1}{n}\log \left(q_{y^n}\right)\right|\leq\eps\right\}\;.
\ee
 Then, by definition,
\ba
\tr\left[\Pi_{\eps}^{\rel,n}\rho^{\otimes n}\right]&=\sum_{y^n\in \mc^{n}_{\eps}}\la\phi_{y^n}\big|\rho^{\otimes n}\big|\phi_{y^n}\ra\\
&=\sum_{y^n\in \mc^{n}_{\eps}}r_{y^n}\;,
\ea
where the last term is the probability that a sequence $Y^n$ belongs to $\mc^n_\eps$. Therefore,
\ba
\tr\left[\Pi_{\eps}^{\rel,n}\rho^{\otimes n}\right]&=\pr\left\{\mc^{n}_{\eps}\right\}\\
&=\pr\Big\{\Big|\mbb{E}\left(\log q_{Y}\right)-\frac{1}{n}\sum_{i\in[n]}\log q_{Y_i}\Big|\leq\eps \Big\}\\
\GG{Hoeffding's\; Inequality}&\geq 1-e^{-c n\eps^2}\;.
\ea
This completes the proof.
\end{proof}

\begin{exercise}
Using the same notations as above, show that:
\begin{enumerate}
\item The constant $c$ can be taken to be (cf.~\eqref{c2})
\be\label{c3}
c=\frac{2}{\big(\log(q_{\max}/q_{\min})\big)^2}\;,
\ee
where $q_{\min}$ is the smallest non-zero eigenvalue of $\sigma$, and $q_{\max}$ is the largest eigenvalue of $\sigma$.
\item The relative typical projector $\Pi_{\eps}^{\rel,n}$ satisfies
\be\label{reltyp}
2^{n\left(\tr[\rho\log\sigma]-\eps\right)} \Pi_{\eps}^{\rel,n}\leq \Pi_{\eps}^{\rel,n}\sigma^{\otimes n}\Pi_{\eps}^{\rel,n}\leq 2^{n\left(\tr[\rho\log\sigma]+\eps\right)}\Pi_{\eps}^{\rel,n}\;.
\ee
\end{enumerate}
\end{exercise}

\section{The Method of Types}\index{types}

The method of types is a fundamental concept in information theory that provides a powerful framework for analyzing the statistical properties of sequences of symbols. Originating from the work of Claude Shannon, this method revolves around the idea of categorizing sequences into types based on the frequency of each symbol's occurrence. By treating sequences with similar compositions as a single type, the method simplifies the analysis of large sets of data. This approach is particularly effective in understanding the behavior of random processes, quantifying the efficiency of coding schemes, and in the study of large deviations and typicality. The method of types has become a cornerstone in both classical and quantum information theory, underpinning many key theorems and applications in areas such as data compression, communication theory, and statistical inference.

\begin{myd}{Type of a Sequence}
\begin{definition}
Let $n,m\in\mbb{N}$. For every $x^n\eqdef(x_1,\ldots,x_n)\in[m]^n$ and $z\in[m]$, let $N(z|x^n)$ be the number of elements in the sequence $x^n$ that are equal to $z$. The \emph{type} of the sequence $x^n$ is a probability vector in $\prob(m)$ given by
\be
\t(x^n)\eqdef\big(t_1(x^n),\ldots,t_m(x^n)\big)^T\;,\quad\text{where}\quad t_z(x^n)\eqdef\frac{1}{n}N(z|x^n)\quad\quad\forall\;z\in[m].
\ee
\end{definition}
\end{myd}
For example, for $m=3$, the type of the sequence $x^6=(2,1,1,3,2,2)$ is the probability vector $\t(x^6)=(1/3,1/2,1/6)$.

The significance of types comes into play when considering an i.i.d$\sim \p$ source. In this case, the probability of a sequence $x^n\in[m]^n$ drawn from the source is given by
\ba\label{tpxn}
p_{x^n}\eqdef p_{x_1}\cdots p_{x_n}
&=p_1^{N(1|x^n)}\cdots p_m^{N(m|x^n)}\\
\Gg{\forall\;r>0\;r=2^{\log r}}&=2^{\sum_{z\in[m]}N(z|x^n)\log_2p_z}\\
\Gg{N(z|x^n)=nt_z(x^n)}&=2^{n\sum_{z\in[m]}t_z(x^n)\log_2p_z}\\
&=2^{-n\big(H(\t(x^n))+D\left(\t(x^n)\|\p\right)\big)}\;,
\ea
where $H(\t(x^n))$ is the Shannon entropy\index{Shannon entropy} of the type of the sequence $x^n$, and $D\left(\t(x^n)\|\p\right)$ is the KL-divergence between $\t(x^n)$ and $\p$ (see~\eqref{kldiv}).
The above formula manifest that  the probability distributions of sequences drawn from an i.i.d.\ source only depend on the type of the sequence. As we will see below, this property can lead to a significant simplification in some applications.

We denote  by $\type(n,m)\subseteq\prob(m)$ the set of all types of sequences in $[m]^n$. For example, for sequences of bits (i.e. $m=2$)
\be
\type(n,2)=\left\{(1,0)^T,\left(\frac{n-1}{n},\frac{1}{n}\right)^T,\left(\frac{n-2}{n},\frac{2}{n}\right)^T,\ldots,\left(0,1\right)^T\right\}
\ee
Note that any type $\t\in\type(n,m)$ has $m$ components of the form $\frac{k}{n}$ where $k\in\{0,\ldots,n\}$. Therefore,
the number of types in $\type(n,m)$ cannot exceed $(n+1)^m$, which is polynomial in $n$.
The exact number of types can be computed using the ``stars and bars" method in combinatorics. It is given by
\be\label{typeb}
|\type(n,m)|={n+m-1\choose n}\leq(n+1)^m\;.
\ee
On the other hand, the number all sequences of size $n$ is $m^n$ which is exponential in $n$. 

The set of all sequences $x^n$ of a given type $\t=(t_1,\ldots,t_m)$ will be denoted as $X^n(\t)$. We emphasize that $X^n(\t)$ denotes a \emph{set} of all sequences in $[m]^n$ whose type is $\t$, whereas $\t(x^n)$ denotes a \emph{single} probability vector (i.e. the type of a specific sequence $x^n$). The number of sequences
in the set $X^n(\t)$ is given by the combinatorial formula of arranging  $nt_1,\ldots,nt_m$ objects in a sequence,
\be\label{xnts}
\left|X^n(\t)\right|={n\choose nt_1,\ldots,nt_m}\eqdef\frac{n!}{\prod_{x=1}^m(nt_x)!}\;.
\ee
The above formula is somewhat cumbersome, but by using Stirling's approximation we can find simpler lower and upper bound.
\begin{myg}{}
\begin{lemma}\label{lem721}
Let $\t\in\type(n,m)$. Then
\be\label{738}
\frac{1}{(n+1)^m}2^{nH(\t)}\leq \left|X^n(\t)\right|\leq 2^{nH(\t)}\;.
\ee
\end{lemma}
\end{myg}
\begin{proof}
Let $x^n$ be a sequences of size $n$ drawn from an i.i.d. source according to the distribution $\t$. Then,
\ba
1=\sum_{x^n\in[m]^n}t_{x^n}
&\geq \sum_{x^n\in X^n(\t)}t_{x^n}\\
\Gg{\eqref{tpxn}\;\text{with}\;\p=\t}&=\sum_{x^n\in X^n(\t)}2^{-nH(\t)}\\
&=|X^n(\t)|2^{-nH(\t)}\;.
\ea
This proves that $|X^n(\t)|\leq 2^{nH(\t)}$. 
For the other inequality, we make use the Stirling's bounds
\be
\sqrt{2\pi}n^{n+\frac12}e^{-n}\leq n!\leq e n^{n+\frac12}e^{-n}\;.
\ee
By using the lower bound for $n!$ and the upper bound for each $(nt_x)!$ of~\eqref{xnts} we get that
\ba
\left|X^n(\t)\right|=\frac{n!}{\prod_{x=1}^m(nt_x)!}&\geq\frac{\sqrt{2\pi}n^{n+\frac12}e^{-n}}{\prod_{x=1}^m e (nt_x)^{nt_x+\frac12}e^{-nt_x}}\\
&=\frac{\sqrt{2\pi}n^{\frac12}}{ e^m n^{\frac m2}\prod_{x=1}^m t_x^{nt_x+\frac12}}\\
&=\frac{\sqrt{2\pi}\sqrt{n}}{(e\sqrt{n})^m\sqrt{t_1\cdots t_m}}2^{-nH(\t)}\;.
\ea
It is left as an exercise (see Exercise~\ref{stirling}) to show that 
\be\label{ineqstirling}
\frac{\sqrt{2\pi}\sqrt{n}}{(e\sqrt{n})^m\sqrt{t_1\cdots t_m}}\geq\frac1{(n+1)^m}
\ee
for all $n$, $m$, and $t_1,\ldots,t_m$.
\end{proof}

\begin{exercise}\label{stirling}
Prove the inequality in~\eqref{ineqstirling}. Hint: Use the fact that the product $t_1\cdots t_n$ is Schur concave and achieves its maximum when $t_1=t_2=\cdots =t_m=\frac1m$.
\end{exercise}

\begin{exercise}
Let $\mk\subset\type(n,m)$ be a set of probability distributions (that are types), and define $\mc_{n}\eqdef\{x^n\in[m]^n\;:\;\t(x^n)\in\mk\}$. Fix $\q\in\mk$. Show that
\be
\pr(\mc_n)_\q\eqdef\sum_{x^n\in\mc_n}q_{x^n}
\ee
approaches one in the limit $n\to\infty$. 
Hint: Denote by $\mc_n^c$ the complement of $\mc_n$ in $[m]^n$ and show that
$
\pr(\mc_n^c)_\q
$
approaches zero in the limit $n\to\infty$. Use~\eqref{tpxn}, \eqref{738}, and the fact that $D(\t\|\q)>0$ for any type $\t\neq \q$.
\end{exercise}

\bex
Let $n\in\mbb{N}$. Use the \emph{strong} Stirling's approximation, which states that
\be
\sqrt{2\pi n}\left(\frac ne\right)^n\leq n!\leq\sqrt{2\pi n}\left(\frac ne\right)^ne^{\frac1{12n}}
\ee
to show that:
\ben
\item For any $p\in(0,1)$ such that $np\in\mbb{N}$ we have
\be
{n\choose np}\leq \frac{2^{nh(p)}}{\sqrt{\pi np(1-p)}}
\ee
where $h(p)=-p\log p-(1-p)\log(1-p)$ is the binary Shannon entropy.
\item For any integer $k\leq n/2$
\be\label{nchoosek}
{n\choose k}\leq 2^{nh\left(\frac kn\right)}\;.
\ee
\een
\eex

\subsection{Many Copies of a Quantum State}\label{sec:pinch0}

In this subsection we show that the method of types can be used to simplify the expression of $n$ copies of some quantum state $\sigma\in\md(A)$.
 Let $m$ be the number of distinct eigenvalues of $\sigma$, so that 
\be
\sigma=\sum_{x\in[m]}q_xP_x
\ee
where $\spec(\sigma)=\{q_1,\ldots,q_m\}$, and $\{P_x\}_{x\in[m]}$ are orthogonal projectors. For $n$ copies of sigma
\be\label{originalsum}
\sigma^{\otimes n}=\sum_{x^n\in[m]^n}q_{x^n}P_{x^n}
\ee
where $q_{x^n}\eqdef q_{x_1}\cdots q_{x_n}$ and $P_{x^n}=P_{x_1}\otimes \cdots\otimes P_{x_n}$.
From~\eqref{tpxn} the probability $q_{x^n}=2^{-n\big(H(\t(x^n))+D\left(\t(x^n)\|\q\right)\big)}$ depends only on the type of $x^n$. Therefore, we can express $\sigma^{\otimes n}$ as
\be
\sigma^{\otimes n}=\sum_{\t\in\type(n,m)}2^{-n\big(H(\t)+D\left(\t\|\q\right)\big)}P_{\t}
\ee
where for any $\t\in\type(n,m)$
\be
P_{\t}\eqdef\sum_{x^n\in X^n(\t)}P_{x^n}\;.
\ee
Note that the set $\{P_\t\}_{\t\in\type(n,m)}$ is itself a set of orthogonal projectors. The significance of the formula above is that the number of terms in the sum is given by
\be\label{specsig}
\left|\spec\left(\sigma^{\otimes n}\right)\right|=|\type(n,m)|\leq (n+1)^m\;,
\ee
which is polynomial in $n$. Therefore, the original sum in~\eqref{originalsum} that consists of $m^n$ terms, has been reduced to a sum with a polynomial number of terms. 

This exponential reduction in the number of terms can be applied to the pinching map $\mP_H$ given in Eqs.~(\ref{pinchingdef},\ref{pinex}). For the case that $H=\sigma^{\otimes n}$ for some $\sigma\in\md(A)$ we have $|\spec(H)|=\left|\spec\left(\sigma^{\otimes n}\right)\right|\leq (n+1)^m$. Therefore, when combined with the pinching inequality~\eqref{pinch} we conclude that for all $\rho\in\md(A^n)$
\be\label{pinch2}
\mP_{\sigma^{\otimes n}}\left(\rho^{A^n}\right)\geq\frac1{(n+1)^m}\rho^{A^n}\;.
\ee
We will see later on that this inequality can be very useful as polynomial terms such as $(n+1)^m$ turns out to be ``negligible" in some applications.

\subsection{Sanov's Theorem}\index{Sanov's theorem} 

Let $\mc\subseteq\prob(m)$ be a set of probability vectors. Sanov's theorem provides an estimate on the probability that a given sequence $x^n$, that is drawn from an i.i.d.$\sim\q$ source,  has a type belonging to $\mc$.  Clearly,  this probability can be zero if the set $\mc$ does not contain types\index{types (method)}. It is therefore necessary to assume that $\mc$ is a ``nice" set; particularly, we will assume that the set $\mc$ is non-empty and is such that $\mc$ is the closure of its interior. Thus, the interior of $\mc$ is not empty (otherwise, $\mc$ will also be empty) and consequently $\mc$ contains a ball of non-zero radius. In particular, since the set $\bigcup_{n\in\mbb{N}}\type(n,m)$ is dense in $\prob(m)$ it follows that for sufficiently large $n$ the set $\mc\cap \type(n,m)$ is non-empty.

The probability that a sequence of size $n$ has a type in $\mc$ is given by
\be
\pr_n(\mc)\eqdef\sum_{x^n\in\mk_n}q_{x^n}
\ee
where
\be
\mk_n\eqdef\big\{x^n\in[m]^n\;:\; \t(x^n)\in\mc\cap\type(n,m)\big\}\;.
\ee
When $n$ is very large one can expect the type of $x^n$ to be relatively close to $\q$ (we will make this notion precise in the next subsection when we study strong typicality). Therefore, if $\q\not\in\mc$ one can expect that the probability $\pr_n(\mc)$ decreases with $n$. Indeed,
Sanov's theorem states that for large $n$ we have $\pr_n(\mc)\approx2^{-nD\left(\p^\star\|\q\right)}$ where  the probability vector $\p^\star\in\prob(m)$ is defined as
\be
\p^\star\eqdef\arg\min_{\p\in\mc}D(\p\|\q)\;,
\ee
where $D$ is the KL-divergence. This result has a geometrical interpretation that the exponential decay of $\pr_n(\mc)$ is increasing with the ``distance" (as measured by the KL-divergence) of $\q$ from the set $\mc$ (see Fig.~\ref{sanov}).

\begin{figure}[h]
\centering
    \includegraphics[width=0.4\textwidth]{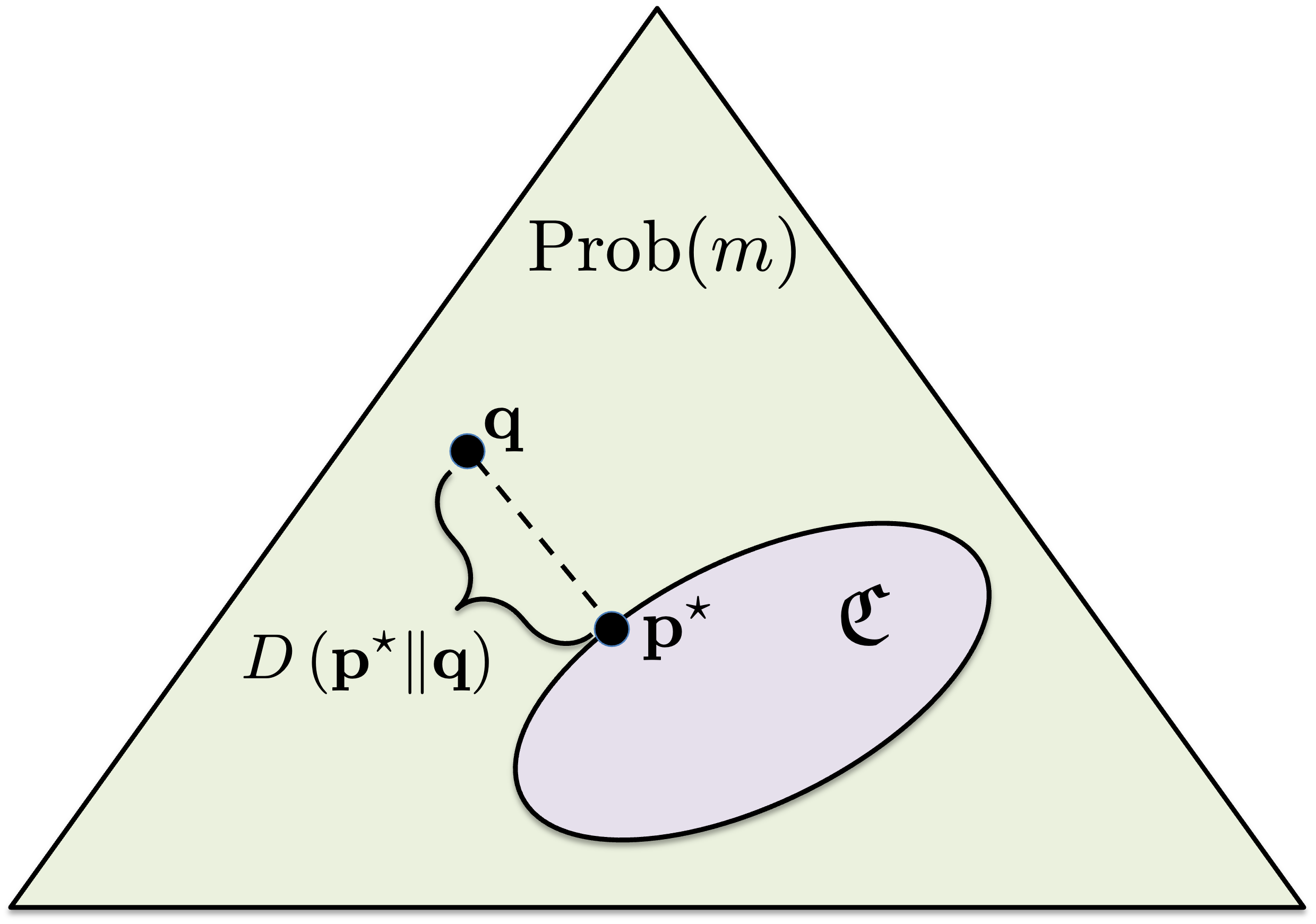}
  \caption{\linespread{1}\selectfont{\small Sanov's Theorem. The exponential decay factor  is determined by the distance of $\q$ from $\mc$ (as measured by the KL-divergence). The triangle represents the probability simplex $\prob(m)$, and the oval shape the set $\mc$.}}
  \label{sanov}
\end{figure}

\begin{myt}{\color{yellow} Sanov's Theorem}
\begin{theorem}\label{sanovt}
Let $\mc\in\prob(m)$ be a non-empty set of probability distributions such that $\mc$ is the closure to its interior, and consider an i.i.d.$\sim\q$ source. Using the same notations as above,
\be
\lim_{n\to\infty}-\frac{1}{n}\log \pr_n(\mc)=\min_{\p\in\mc}D(\p\|\q)\eqdef D(\p^\star\|\q)\;.
\ee
\end{theorem}
\end{myt}
\begin{remark}
Note that while the vector $\p^\star$ is \emph{not} necessarily in $\bigcup_{n\in\mbb{N}}\type(n,m)$, there exists a sequence of types $\{\p_n\}_{n\in\mbb{N}}$, with $\p_n\in\mc\cap\type(n,m)$ for sufficiently large $n$, such that $\p_n\to\p^*$ as $n\to\infty$. 
\end{remark}

\begin{proof}
We will prove the theorem by finding upper and lower bounds for $\pr_n(\mc)$. For the upper bound, using~\eqref{tpxn} we get
\ba\label{756}
\pr_n(\mc)=\sum_{x^n\in\mk_n}2^{-n\big(H(\t(x^n))+D\left(\t(x^n)\|\q\right)\big)}
&=\sum_{\t\in\mc\cap\type(n,m)}|X^n(\t)|2^{-n\big(H(\t)+D\left(\t\|\q\right)\big)}\\
\GG{\eqref{738}}&\leq \sum_{\t\in\mc\cap\type(n,m)}2^{-nD\left(\t\|\q\right)}\\
\GG{By\;definition\;of\;\p^\star}&\leq \sum_{\t\in\mc\cap\type(n,m)}2^{-nD\left(\p^\star\|\q\right)}\\
&\leq|\type(n,m)|2^{-nD\left(\p^\star\|\q\right)}\\
\GG{\eqref{typeb}}&\leq (n+1)^m2^{-nD\left(\p^\star\|\q\right)}\;.
\ea
Note that we got this upper bound without assuming that $\mc$ is the closure of its interior. We only assumed that $\mc$ is non-empty so that $\p^\star$ exists.

For the lower bound, let
$\{\p_n\}_{n\in\mbb{N}}$, with $\p_n\in\mc\cap\type(n,m)$ for sufficiently large $n$, such that $\p_n\to\p^*$ as $n\to\infty$ (see the remark above). Then, for sufficiently large $n$ we have
\ba\label{757}
\pr_n(\mc)&=\sum_{\t\in\mc\cap\type(n,m)}|X^n(\t)|2^{-n\big(H(\t)+D\left(\t\|\q\right)\big)}\\
\GG{\substack{Taking\;only\;the\;term\\ \t=\p_{\it n}\;in\;the\;sum}}&\geq|x^n(\p_n)|2^{-n\big(H(\p_n)+D\left(\p_n\|\q\right)\big)}\\
\GG{\eqref{738}}&\geq \frac{1}{(n+1)^m}2^{-nD\left(\p_n\|\q\right)}\;.
\ea
From the two bounds above we get
\be
D(\p^\star\|\q)-\frac{1}{n}\log (n+1)^m\leq -\frac{1}{n}\log \pr_n(\mc)\leq D\left(\p_n\|\q\right)+\frac{1}{n}\log (n+1)^m\;.
\ee
The proof is concluded by taking the limit $n\to\infty$. 
\end{proof}

\section{Strong Typicality}\index{typicality}

In this section, we employ the method of types\index{types (method)} to define the concepts of strong typical sequences and strong typical subspaces. While weak typicality serves as a valuable tool in various applications due to its simplicity, it often yields less robust results. As we will see in certain examples, there are instances where the application of strong typicality methods becomes necessary to achieve more robust outcomes.

\subsection{Strong Typical Sequences}

Previously, we examined what is often called \emph{weak} typicality. To understand its limitations, consider an i.i.d.$\sim\p$ source sequence that is $\eps$-typical, satisfying~\eqref{tpxn2}. Simplifying~\eqref{tpxn2} using~\eqref{tpxn}, a sequence is (weakly) $\eps$-typical if
\be
\big|H\big(\t\left(x^n\right)\big)+D\left(\t(x^n)\big\|\p\right)-H(\p)\big|\leq \eps\;.
\ee
In other words, $x^n$ is $\eps$-typical if the entropy of its type approximates the entropy of $\p$, with a negligible correction term $D\left(\t(x^n)\|\p\right)$ as $\t(x^n)$ converges to $\p$. To establish a more robust concept of typicality, one might require that the type of $x^n$ is $\eps$-close to $\p$. This is a somewhat more natural requirement and the question is then: which metric to use? The fundamental requirement is that each element of $\t(x^n)$ should be $\eps$-close to its counterpart in $\p$, necessitating
\be
\left|p_z-t_z(x^n)\right|\leq\eps\quad\quad\forall\;z\in[m]\;.
\ee
Note that any sequence $x^n$ that satisfies $\|\p-\t(x^n)\|_p\leq\eps$ (for some $p\geq 1$) will satisfies the above equation, and therefore it impose a slightly stronger condition on the sequence $x^n$. On the other hand, the condition $\|\p-\t(x^n)\|_\infty\leq\eps$ is precisely equivalent to the above condition, and we will therefore use it to measure the distance between $\t(x^n)$ and $\p$.
\begin{myd}{Strongly Typical Sequences}
\begin{definition}
Let $x^n\in[m]^n$ be a sequence of size $n$ drawn from an i.i.d.$\sim\p$ source. The sequence $x^n$ is said to be strongly $\eps$-typical if $N(z|x^n)=0$ for all $z\not\in\supp(\p)$ and
\be
\left\|\p-\t(x^n)\right\|_\infty\leq\eps.
\ee
The set of all strongly typical sequences of size $n$ is denoted by $\mt^{\st}_{\eps}(X^n)$.
\end{definition}
\end{myd}
Note that, in accordance with our previous notation, the condition stipulated in the above definition can be equivalently expressed as:
\be
\left|p_z-\frac{N(z|x^n)}{n}\right|\leq\eps\quad\quad\forall\;z\in[m]\;\text{such that}\;p_z>0\;,
\ee
and $N(z|x^n)=0$ whenever $p_z=0$. This latter condition intuitively implies that a typical sequence should not include any alphabet characters that have a zero probability of occurrence.

In the next theorem we prove several properties of typical sequences. We will use the notation
\be
\Pr(\mt^{\st}_{\eps}(X^n))\eqdef \sum_{x^n\in\mt^{\st}_{\eps}(X^n)}p_{x^n}
\ee
to denote the probability that a sequences is strongly $\eps$-typical (with respect to an i.i.d.$\sim\p$ source). Moreover, we set the constant $a>0$ to be
\be
a\eqdef-\log\prod_{z\in\supp(\p)}p_z\;.
\ee

\begin{myt}{\color{yellow} Properties of Strongly Typical Sequences}
\begin{theorem}\label{stypi}
Let $X$ be the random variable of an i.i.d.$\sim\p$ source, and let $\eps>0$. For all $n\in\mbb{N}$ the following inequalities hold:
\ben
\item  $\Pr(\mt^{\st}_{\eps}(X^n))\geq 1-e^{-2\eps^2n}$.
\item $2^{-n(H(X)+a\eps)}\leq p_{x^n}\leq 2^{-n(H(X)-a\eps)}$, for all $x^n\in\mt^{\st}_{\eps}(X^n)$.
\item $\big(1-e^{-2\eps^2n}\big)2^{n(H(X)-\eps a)}\leq\left|\mt^{\st}_{\eps}(X^n)\right|\leq 2^{n(H(X)+\eps a)}$.
\een
\end{theorem}
\end{myt}
\begin{remark}
Observe that the probability that a sequence is strongly $\eps$-typical approach one exponentially fast with $n$.
The second property highlights the equipartition property, indicating that the probability of every typical sequence approximately $2^{-nH(X)}$. The first and third properties bear resemblance to their counterparts related to weak typical sequences. However, due to the subtle distinctions between the two concepts of typicality, the upcoming proof will incorporate additional tools to address these differences.
\end{remark}
\begin{proof}
For any $z\in[m]$, let $1_z\left(X\right)$ be the \emph{indicator} random variable that equals 1 if $X=z$ and $0$ otherwise.   Fix $z\in[m]$, and let $Z_1,Z_2,\ldots$ be an i.i.d. sequences of random variables where each $Z_j\eqdef 1_z\left(X_j\right)$ is an indicator random variable as above that corresponds to $X_j$. From the law of large numbers, particularly the application of Hoeffding's inequality~\eqref{hoe2}\index{Hoeffding's inequality} to the sequence $Z_1,\ldots,Z_n$ we get
\be
\Pr\Big(\Big|\frac{1}{n}\sum_{j\in[n]}Z_j-\mbb{E}(Z)\Big|>\eps\Big)\leq e^{-2n\eps^2}\;,
\ee
where we used the fact that $0\leq Z_j\leq 1$ so that the constants $a$ and $b$ in~\eqref{hoe2} are given by $a=0$ and $b=1$. By definition,
\be
\mbb{E}(Z)=\sum_{x\in[m]}p_x\delta_{xz}=p_z\;,
\ee
and
\be
\frac{1}{n}\sum_{j\in[n]}Z_j=\frac{1}{n}\sum_{j\in[n]}\delta_{x_jz}=\frac{1}{n}N(z|x^n)=t_z(x^n)\;.
\ee
We therefore conclude that
\be
\Pr\left(\big|t_z(x^n)-p_z\big|>\eps\right)\leq e^{-2n\eps^2}\;.
\ee
The equation above holds for all $z\in[m]$ and all $n\in\mbb{N}$. Hence, it states that the probability that the random variable $X^n=(X_1,\ldots,X_n)$ is not an $\eps$-typical sequence, is no greater than $e^{-2n\eps^2}$. This completes the proof of first part of the theorem.

Suppose now that $x^n\in\mt^{\st}_{\eps}(X^n)$ and observe that
\be
p_{x^n}=p_{x_1}p_{x_2}\cdots p_{x_n}=\prod_{z\in\supp(\p)}p_z^{N(z|x^n)}\;.
\ee
Taking the log on both sides and dividing by $n$ gives
\be
\frac1n\log p_{x^n}=\sum_{z\in\supp(\p)}t_z(x^n)\log p_z
\ee
Now, since $x^n$ is strongly $\eps$-typical sequence, for every $z\in[m]$ 
\be
p_z-\eps \leq t_z(x^n)\leq p_z+\eps\;.
\ee
Combining this with the previous equation gives
\be
\sum_{z\in\supp(\p)}(p_z+\eps)\log p_z\leq\frac1n\log p_{x^n}\leq \sum_{z\in\supp(\p)}(p_z-\eps)\log p_z
\ee
That is,
\be
-\eps a-H(X)\leq\frac1n\log p_{x^n}\leq\eps a-H(X)
\ee
Multiplying both sides by $n$ and raising all sides to the power of $2$ completes the proof of the second part.

To prove the third part, observe that from the lower bound of Part~2 we get
\ba
1\geq\sum_{x^n\in\mt^{\st}_{\eps}(X^n)} p_{x^n}&\geq \sum_{x^n\in\mt^{\st}_{\eps}(X^n)}2^{-n\left(H(X)+\eps a\right)}\\
&=\left|\mt^{\st}_{\eps}(X^n)\right|2^{-n\left(H(X)+\eps a\right)}\;.
\ea
Therefore, 
$
\left|\mt^{\st}_{\eps}(X^n)\right|\leq 2^{n\left(H(X)+\eps a\right)}
$.
On the other hand, from Part~1 we get
\ba
1-e^{-2n\eps^2}&\leq \sum_{x^n\in\mt^{\st}_{\eps}(X^n)} p_{x^n}\\
\GG{Upper\; bound \;of\; Part~2}&\leq \sum_{x^n\in\mt^{\st}_{\eps}(X^n)}2^{-n\left(H(X)-\eps a\right)}\\
&=\left|\mt^{\st}_{\eps}(X^n)\right|2^{-n\left(H(X)-\eps a\right)}\;.
\ea
Hence,
$
\big(1-e^{-2n\eps^2}\big)2^{n(H(X)-\eps a)}\leq\left|\mt^{\st}_{\eps}(X^n)\right|
$. This completes the proof of the third part.
\end{proof}

\bex
Let $x^{n}\in[m]^n$ and $y^k\in[m]^k$ be two sequences of size $n$ and $k$, respectively, drawn from the same i.i.d.$\sim\p$ source. Show that if both $x^n$ and $y^k$ are strongly $\eps$-typical then also the joint sequence $(x^n,y^k)\in[m]^{n+k}$ is strongly $\eps$-typical with respect to the i.i.d.$\sim\p$ source.
\eex

\bex
Let $1<n\in\mbb{N}$, $\eps\in(\frac1n,1)$,  and consider an i.i.d.$\sim\p$ source, where $\p\in\prob(m)$.
\ben
\item Show that if a sequence $x^{n-1}$ is strongly $\eps$-typical then for any $y\in[m]$ the sequence $x^n\eqdef (y,x^{n-1})$ is strongly $(\eps+\frac1n)$-typical.
\item Let $\eps'\eqdef\eps-\frac1n$. Show that $\mt'\subset\mt_\eps^\st(X^n)$, where
\be
\mt'\eqdef\left\{(y,x^{n-1})\;:\;y\in[m]\;,\quad x^{n-1}\in[m]^{n-1}\;,\quad\left\|\t(x^{n-1})-\p\right\|_\infty\leq\eps'\right\}\;.
\ee
\item Show that for any $\q\in\prob(m)$ we have
\be\label{gili9}
\sum_{x^n\in\mt_\eps^\st(X^n)}q_{x_1}p_{x_2}\cdots p_{x_n}\geq 1-e^{-2(n-1)\eps'^2}\;.
\ee
Hint: Use the first part of the theorem above.
\een
\eex

\subsection{Strong Typical Subspace}\label{sec:sts}\index{typicality}
The extension of strong typicality from classical sequences to quantum states follows similar lines to the quantum extension of weak classical typicality. Let $\rho$ be an i.i.d. quantum source with spectral decomposition
\be
\rho^A=\sum_{x\in[m]}p_x|x\lr x|^A\;,
\ee
where $\{p_x\}_{x\in[m]}$ are the eigenvalues of $\rho$ and $\{|x\ra\}_{x\in[m]}$ are the corresponding eigenvectors.
As before we denote
\be
\rho^{\otimes n}=\sum_{x^n\in[m]^n}p_{x^n}|x^n\lr x^n|^{A^n}
\ee
where $|x^n\ra\eqdef |x_1,\ldots,x_n\ra$ and $p_{x^n}\eqdef p_{x_1}p_{x_2}\cdots p_{x_n}$.
For any i.i.d. quantum source $\rho$ we define a corresponding strongly typical subspace 
\be
\mt^{\st}_{\eps}(A^n)\eqdef \spa\left\{|x^n\ra\in A^n\;:\;x^n\in\mt^{\st}_{\eps}(X^n)\right\}
\ee
where $\mt^{\st}_{\eps}(X^n)$ is the set of (classical) sequences of size $n$, drawn from an i.i.d.$\sim\p$ source (with $\p$ being the probability vector whose components are the eigenvalues of $\rho$). The strongly typical projection to this subspace is given by 
\be
\Pi^{n,\st}_{\eps}\eqdef \sum_{x^n\in\mt^{\st}_{\eps}(X^n)}|x^n\lr x^n|\;.
\ee
\begin{myt}{}
\begin{theorem}\label{sqtypical}
Let $\rho\in\md(A)$, $\eps>0$, and for each $n\in\mbb{N}$ let $\mt_{n}^{\eps}(\rho)$ and $\Pi^{n,\st}_{\eps}$ be the strongly typical subspace and projection associated with a quantum i.i.d.$\sim\rho$ source. The following inequalities hold for all $n\in\mbb{N}$:
\ben
\item $\tr\left[\Pi^{n,\st}_{\eps}\rho^{\otimes n}\right]\geq 1-e^{-2\eps^2n}$.
\item $2^{-n(H(A)_\rho+a\eps)}\Pi^{n,\st}_{\eps}\leq\Pi^{n,\st}_{\eps}\rho^{\otimes n}\Pi^{n,\st}_{\eps}\leq 2^{-n(H(A)_{\rho}-a\eps)}\Pi^{n,\st}_{\eps}$.
\item $(1-\delta)2^{n(H(\rho)-\eps a)}\leq\big|\mt^{\st}_{\eps}(A^n)\big|\leq 2^{n(H(\rho)+\eps a)}$.
\een
\end{theorem}
\end{myt}

The  proof follow directly from the classical version of this theorem and is left as an exercise.

\begin{exercise}
Prove the theorem above.
\end{exercise}

\bex
Let $\rho\in\md(A)$, $\eps>0$, integer $m=o(n)$ (e.g. $m=\lfloor n^s\rfloor$ for some $0<s<1$), and $\sigma_m\in\md(A^m)$. Let also $\Pi^{n,\st}_{\eps}$ be the strongly typical projection associated with the quantum i.i.d.$\sim\rho$ source. Show that
\be
\lim_{n\to\infty}\tr\left[\Pi^{n,\st}_{\eps}\left(\rho^{\otimes (n-m)}\otimes\sigma_m\right)\right]=1\;.
\ee
\eex

\section{Classical Hypothesis Testing}\label{cht}\index{hypothesis testing}

Imagine a game where a player, named Alice, is handed one of two biased dice. These dice are characterized by probability vectors, denoted as $\p$ and $\q$. Alice's challenge is to determine if she holds the $\p$-dice or the $\q$-dice. To make an informed decision, she's allowed to roll the dice $n$ times. With a single roll (i.e., $n=1$), the chances of Alice making an incorrect assumption can be high. However, with an increase in the number of rolls, her probability of making a mistake significantly reduces. This prompts a natural question: how rapidly does the error probability decrease as the number of rolls, $n$, gets larger and larger? This situation encapsulates the essence of a classical hypothesis testing problem.

In the realm of hypothesis testing, an observer or player aims to decide between two hypotheses related to two i.i.d.\ sources. These hypotheses are represented as the $\p$-source and the $\q$-source. Upon $n$ independent interactions with this source, the observer receives a sequence denoted as $x^n = (x_1, \ldots, x_n)$ that belongs to the set $[m]^n$. The challenge is to ascertain the correct hypothesis based on this sequence.

The observer's decision-making process can be represented by a function $g_n:[m]^n \to \{0,1\}$. This function divides all potential sequences into two distinct groups:
\ben
\item The set $\{x^n \in [m]^n \;:\; g_n(x^n) = 0\}$ corresponds to the first hypothesis. Here, the observer believes the sequences in this set are from the $\p$-source.
\item The set $\{x^n \in [m]^n \;:\; g_n(x^n) = 1\}$ pertains to the second hypothesis, indicating that the observer surmises the sequences are from the $\q$-source.
\een
Given this decision-making framework, two potential errors can emerge:
\begin{enumerate}
\item \textbf{Type I Error}. The observer incorrectly concludes that the sequence is from the $\q$-source when, in reality, it is from the $\p$-source. The probability of this error occurring is:
\be
\alpha(g_n)\eqdef\sum_{\substack{x^n\in[m]^n\\ g_n(x^n)=1}}p_{x^n}\;.
\ee
\item \textbf{Type II Error}. The observer mistakenly assumes the sequence is from the $\p$-source when it actually originates from the $\q$-source. The likelihood of this error is:
\be\label{msnt}
\beta(g_n)\eqdef \sum_{\substack{x^n\in[m]^n\\ g_n(x^n)=0}}q_{x^n}\;.
\ee
\end{enumerate}

In the two errors above, we considered a deterministic hypothesis test, where the function $g_n:[m]^n\to\{0,1\}$ remains fixed. A more general approach introduces an element of randomness to the problem. Here, the observer randomly selects the function $g_n$
based on a specific probability distribution.
  
To illustrate, consider a set of $\ell$ functions denoted as $\{g_{n,k}\}_{k\in[\ell]}$, where each $g_{n,k}:[m]^n\to\{0,1\}$. Accompanying these functions is a probability vector $\s\in\prob(\ell)$. In this probabilistic framework, when given the sequence $x^n$, the observer first samples a values $k$ according to the distribution $\s$. The observer then attributes the sequence to the $\p$-source if $g_{n,k}(x^n)=0$, and to the $\q$-source if $g_{n,k}(x^n)=1$. It's crucial to note that for each $k\in[\ell]$:
\be
\beta(g_{n,k})=\sum_{\substack{x^n\in[m]^n\\ g_{n,k}(x^n)=0}}q_{x^n}=\q^{\otimes n}\cdot \b_k
\ee
where the $x^n$-component of the bit vector $\b_k\in\{0,1\}^{m^n}$ is one if  $g_{n,k}(x^n)=0$ and zero otherwise.
The vector 
\be\label{phtv}
\t\eqdef\sum_{k\in[\ell]} s_k\b_k
\ee
is termed the \emph{probabilistic hypothesis test}. With the aforementioned notations and considering this broader context, the two types of errors can be described as follows:
\begin{enumerate}
\item \textbf{Type I Error}. This pertains to the likelihood of the observer incorrectly attributing the sequence to the $\q$-source when it originates from the $\p$-source:
\be\label{alnt}
\alpha(\t)\eqdef \sum_{k\in[\ell]}s_k\sum_{\substack{x^n\in[m]^n\\ g_{n,k}(x^n)=1}}p_{x^n}=1-\p^{\otimes n}\cdot\t\;.
\ee
\item \textbf{Type II Error}. This represents the chance of the observer mistakenly deducing the sequence belongs to the $\p$-source when it is from the $\q$-source:
\be\label{bent}
\beta(\t)\eqdef\sum_{k\in[\ell]}s_k\sum_{\substack{x^n\in[m]^n\\ g_{n,k}(x^n)=0}}q_{x^n}=\q^{\otimes n}\cdot\t\;.
\ee
\end{enumerate}
From its definition~\eqref{phtv}, all the components of the probabilistic hypothesis test vector $\t$ are between zero and one (i.e. $\t\in[0,1]^{m^n}$). Conversely, any  vector in $[0,1]^{m^n}$ can be expressed as a convex combination of bit vectors in $\{0,1\}^{m^n}$. Hence, $\t$ uniquely characterizes the probabilistic hypothesis test performed by the observer.

The goal of the observer is therefore to choose a probabilistic test vector $\t$ such that both types of error are very small. There are two common ways to do that, and we discuss both now. The first one is the \emph{asymmetric} method in which the observer minimizes the type II error, $\beta(\t)$, while at the same time keep the type I error, $\alpha(\t)$, below a certain threshold $\eps>0$. The optimal way to do it is characterized by the Stein's lemma. The second method, also known as the \emph{symmetric} way, in which one assumes a prior $\{s_0,s_1\}$ known to the observer in which the $\p$-source occur with probability $s_0$, and the $\q$-source with probability $s_1$. In this case, the goal is to minimize the error probability that is given by $s_0\alpha(\t)+s_1\beta(\t)$. The optimal value of this probability of error is characterized by the Chernoff information. A fundamental instrument in these methods is the divergence used in hypothesis testing.

\subsection{The Classical Hypothesis Testing Divergence}\index{hypothesis testing}

For a single application of the source, specifically when $n=1$ as presented in Eqs.~(\ref{alnt},\ref{bent}), we define the two error types as $\alpha(\t)\eqdef1-\p\cdot\t$ and $\beta(\t)\eqdef\q\cdot\t$. By minimizing $\beta(\t)$ while constraining $\alpha(\t)$ to a specific threshold, we obtain the following divergence.

\begin{myd}{}
\begin{definition}
For any $\p,\q\in\prob(m)$ and $\eps\in[0,1)$,
the \emph{classical} hypothesis testing divergence is defined as
\be\label{tstv}
D_{\min}^\eps(\p\|\q)\eqdef-\log\min\big\{\q\cdot\t\;:\;\p\cdot\t\geq 1-\eps\quad,\quad\t\in[0,1]^m\big\}
\ee
where the minimization is over all probabilistic hypothesis test vectors $\t$ whose components are in the interval $[0,1]$. 
\end{definition}
\end{myd}

The hypothesis testing divergence is always non-negative and equals infinity if $\p\cdot\q=0$.  The reason for the notation $D_{\min}^\eps$ is that for $\eps=0$, it reduces to the min relative entropy:
\ba
D_{\min}^0(\p\|\q)&=-\log\min\big\{\q\cdot\t\;:\;\p\cdot\t\geq 1\quad,\quad\t\in[0,1]^m\big\}\\
\Gg{\substack{\forall\; x\in[m]\\t_x=1\text{ if }p_x\neq 0}}&=-\log\sum_{x\in\supp(\p)}q_x\\
&=D_{\min}(\p\|\q)\;.
\ea
 To see that $D_{\min}^\eps$ in the definition above is indeed an (unnormalized) divergence, let $\p,\q\in\prob(m)$, $E\in\stoc(n,m)$, and observe that
 \ba
 D_{\min}^\eps(E\p\|E\q)&=-\log\min\big\{(E\q)^T\s\;:\;(E\p)^T\s\geq 1-\eps\quad,\quad\s\in[0,1]^n\big\}\\
 \Gg{\substack{\text{Replacing } E^T\s\in[0,1]^m\\ \text{with arbitrary }\t\in[0,1]^m}}&\leq-\log\min\big\{\q\cdot\t\;:\;\p\cdot\t\geq 1-\eps\quad,\quad\t\in[0,1]^m\big\}\\
 &=D_{\min}^\eps(\p\|\q)\;.
 \ea
\bex
Show that the constraint $\p\cdot\t\geq 1-\eps$ in~\eqref{tstv} can be replaced with $\p\cdot\t= 1-\eps$ (i.e. both constraints leads to the same value of $D^\eps_{\min}(\p\|\q)$).
\eex

\bex
Show that for all $\p,\q\in\prob(m)$, $D_{\min}^{\eps}(\p\|\q)$ is non-decreasing in $\eps$, and
\be
D_{\min}^{\eps}(\p\|\q)\geq-\log(1-\eps)\;,
\ee
with equality if $\p=\q$.
\eex

The classical hypothesis testing divergence is closely related to the testing region defined in~\eqref{testingr}. To see the connection, first observe that we can replace the condition $\p\cdot\t\geq 1-\eps$ in~\eqref{tstv} with the equality $\p\cdot\t= 1-\eps$ (since any $\t$ that satisfies $\p\cdot\t> 1-\eps$ is not optimal). With this change, the optimal $\q\cdot\t$ in~\eqref{tstv} can be interpreted as the lowest point of the intersection of the testing region $\mt(\p,\q)$ with the vertical line $x=1-\eps$ (see Fig.~\ref{troptimal}). That is, the optimal $\q\cdot\t$ is the $y$-component of the lower Lorenz curve\index{Lorenz curve} LC$(\p,\q)$ at $x=1-\eps$.

\begin{figure}[h]\centering
    \includegraphics[width=0.6\textwidth]{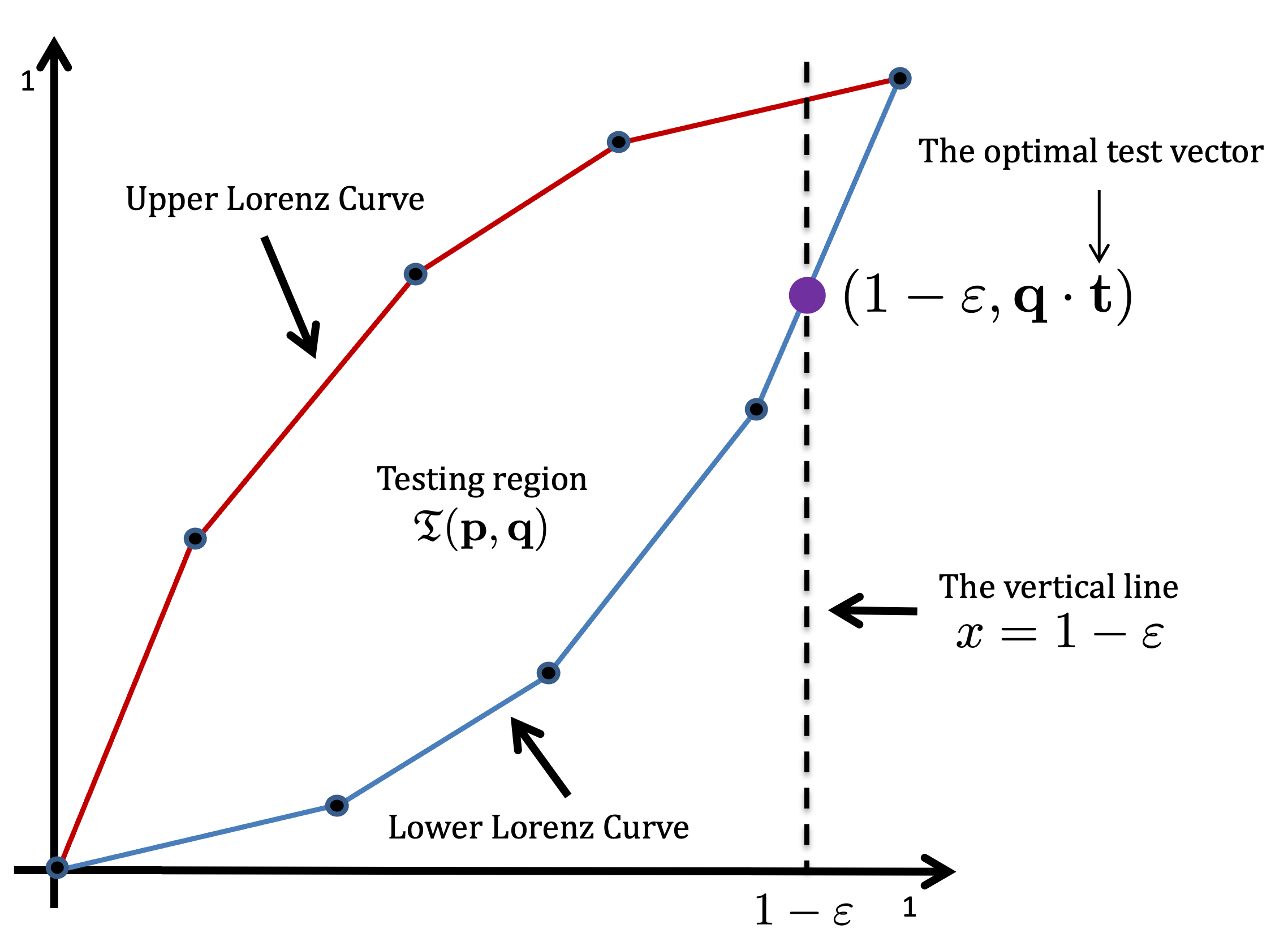}
  \caption{\linespread{1}\selectfont{\small The location of the point in the testing region $\mt(\p,\q)$ with the optimal testing vector $\t$ that minimizes~\eqref{tstv}.}}
  \label{troptimal}
\end{figure}

We can use the above geometrical interpretation of the hypothesis testing divergence to obtain a closed formula for $D_{\min}^\eps$. Without loss of generality, suppose that the components of $\p$ and $\q$ are ordered as in~\eqref{order}. Then, from Theorem~\ref{vertices} we know that the vertices of the lower Lorenz curve\index{Lorenz curve} of $(\p,\q)$ are given by $\{(a_k,b_k)\}_{k=0}^m$ as defined in~\eqref{akbk}. Let $\ell$ be an integer such that $a_\ell< 1-\eps\leq a_{\ell+1}$. Then, the optimal point on the lower Lorenz curve is located between the $\ell$ and the $\ell+1$ vertices. The line between these two vertices has a slop 
\be
\frac{b_{\ell+1}-b_\ell}{a_{\ell+1}-a_\ell}=\frac{q_{\ell+1}}{p_{\ell+1}}\;.
\ee 
Hence, the $y$ component of the optiml point is given by
\be
\q\cdot\t=b_\ell+\frac{q_{\ell+1}}{p_{\ell+1}}(1-\eps-a_{\ell})\;.
\ee
To summarize, we can  express the hypothesis testing divergence as
\be
D^\eps_{\min}\left(\p\big\|\q\right)=-\log\left(b_\ell+\frac{q_{\ell+1}}{p_{\ell+1}}(1-\eps-a_{\ell})\right)
\ee
where $\ell\in\{0,\ldots,m-1\}$ is the integer satisfying $a_\ell< 1-\eps\leq a_{\ell+1}$.
Recall that $D_{\min}^\eps$ is non-decreasing with $\eps$ (see Exercise~\ref{ex:ndeps2}), as it is also evident from the equation above. Therefore, we can bound the hypothesis testing divergences by taking above the two extreme cases $1-\eps=a_\ell$ and $1-\eps=a_{\ell+1}$ to get the simpler bounds
\be\label{blbounds}
-\log b_{\ell+1}\leq D^\eps_{\min}\left(\p\big\|\q\right)\leq-\log b_{\ell}
\ee
where $\ell$, as before, is the integer satisfying $a_\ell< 1-\eps\leq a_{\ell+1}$.

\subsection{The Stein's Lemma}\index{Stein's lemma}

Here we study the minimization of the type II error $\beta_n(\t)$ while at the same time keeping the type I error $\alpha_n(\t)$ below a certain threshold $\eps\geq 0$.

\begin{myt}{\color{yellow} The Stein's Lemma}
\begin{theorem}\label{th:sl}
Using the same notations as above, for any $0<\eps<1$, $m\in\mbb{N}$, and $\p,\q\in\prob(m)$
\be
\lim_{n\to\infty}\frac1nD^\eps_{\min}\left(\p^{\otimes n}\big\|\q^{\otimes n}\right)=D(\p\|\q)\;,
\ee
where $D$ is the KL-divergence.
\end{theorem}
\end{myt}
\begin{remark}
The theorem above states that the type II error can be made as small as $\approx 2^{-nD(\p\|\q)}$ while at the same time keeping the type I error below the threshold $\eps$. The rate of this exponential decay is given by the KL-divergence. We postpone the proof of this theorem to the next section in which we prove the more general theorem known as the quantum Stein's lemma. For the interested reader, we also provide two more direct proofs of this theorem (that applicable only to the classical case) in Appendix~\ref{acht}.
\end{remark}

\subsection{The Chernoff\index{Chernoff} Information}

In the second method of optimizing $\alpha(\t)$ and $\beta(\t)$, also known as the \emph{symmetric} method, there exists a prior $\{s_0,s_1\}$ known to the observer in which the $\p$-source occur with probability $s_0$, and the $\q$-source with probability $s_1$. In this case, for a single use of the source (i.e. $n=1$) the error probability in which the observer guess incorrectly the type of the source is given for any $\p,\q\in\prob(m)$ by
\be
\pr_{\error}(\p,\q,s_0)\eqdef \min\big\{s_0\alpha(\t)+s_1\beta(\t)\;:\;\t\in[0,1]^m\big\}\;.
\ee
Note that this probability of error can be expressed as
\ba\label{sd0p}
\pr_{\error}(\p,\q,s_0)&=\min_{\t\in[0,1]^m}\big\{s_0(1-\p\cdot\t)+s_1\q\cdot\t\big\}\\
&=s_0-\max_{\t\in[0,1]^m}(s_0\p-s_1\q)\cdot\t\\
&=s_0-\sum_{x\in[m]}(s_0p_x-s_1q_x)_+\\
\GG{\lambda\eqdef\frac{s_1}{s_0}}&=s_0\Big(1-\sum_{x\in[m]}(p_x-\lambda q_x)_+\Big)
\ea
where $(s-t)_+\eqdef s-t$ if $s\geq t$ and is zero otherwise.
\bex
Show that the probability of error above can be expressed also as
\be
\pr_{\error}(\p,\q,s_0)=\frac12\big(1-\|s_0\p-s_1\q\|_1\big)\;.
\ee
\eex

\bex
Let $\p,\q\in\prob(m)$ and $\p',\q'\in\prob(m')$. Show that $(\p,\q)\succ(\p',\q')$ if and only if 
\be
\pr_{\error}(\p,\q,s_0)\geq \pr_{\error}(\p',\q',s_0)\quad\quad\forall\;s_0\in[0,1]\;.
\ee
\eex
\bex
Let $\p,\q\in\prob(m)$ and $s_0,\alpha\in(0,1)$.
\ben
\item Show that 
\be\label{7125}
\pr_{\error}(\p,\q,s_0)=\sum_{x\in[m]}\min\{s_0p_x,s_1q_x\}\;.
\ee
Hint: Use the formula $\min\{a,b\}=\frac12(a+b)-\frac12|a-b|$.
\item Use the inequality $\min\{a,b\}\leq a^\alpha b^{1-\alpha}$ to show that for all $\alpha\in[0,1]$
\be
\pr_{\error}(\p,\q,s_0)\leq s_0^\alpha s_1^{1-\alpha}\sum_{x\in[m]}p_x^{\alpha}q_x^{1-\alpha}\;.
\ee
\item Show that 
\be
\liminf_{n\to\infty}-\frac{1}{n}\log\pr_{\error}(\p^{\otimes n},\q^{\otimes n},s_0)\geq -\log\sum_{x\in[m]}p_x^{\alpha}q_x^{1-\alpha}\;.
\ee
\een
\eex

The exercise above demonstrates that in the asymptotic limit the optimal probability of error is bounded by
\be\label{ineqxi}
\liminf_{n\to\infty}-\frac{1}{n}\log\pr_{\error}(\p^{\otimes n},\q^{\otimes n},s_0)\geq \xi(\p,\q)\;,
\ee
where
\ba\label{chernxi}
\xi(\p,\q)&\eqdef-\log\min_{\alpha\in[0,1]}\sum_{x\in[m]}p_x^{\alpha}q_x^{1-\alpha}\\
\GG{Definition~\ref{renyi}}&=\max_{\alpha\in[0,1]}\big\{(1-\alpha)D_\alpha(\p\|\q)\big\}\;.
\ea
In the following theorem we show that the inequality in~\eqref{ineqxi} is in fact an equality. 

\begin{myt}{\color{yellow} The Chernoff\index{Chernoff} Bound}
\begin{theorem}\label{cherthm}
Let $\p,\q\in\prob(m)$ and $s_0\in(0,1)$. Then,
\be
\lim_{n\to\infty}-\frac{1}{n}\log\pr_{\error}(\p^{\otimes n},\q^{\otimes n},s_0)=\xi(\p,\q)\;.
\ee
\end{theorem}
\end{myt}
\begin{remark}
Note that the Chernoff\index{Chernoff} bound $\xi(\p,\q)$ does not depend on $s_0$. Moreover, the theorem above also states that for very large $n$ we have that the probability of error, $\pr_{\error}(\p^{\otimes n},\q^{\otimes n},s_0)\approx 2^{-n\xi(\p,\q)}$, decays exponentially fast with $n$ with an exponential factor given by the Chernoff bound.
\end{remark}
\begin{proof}
We need to prove the opposite inequality of~\eqref{ineqxi}. We will establish it by finding a lower bound on the probability of error $\pr_{\error}(\p^{\otimes n},\q^{\otimes n},s_0)$.
For this purpose,
 set $\lambda\eqdef s_1/s_0$ and $\mk\eqdef\{x\in[m]\;:\;p_x\geq \lambda q_x\}$. From~\eqref{sd0p} we have
\be
\pr_{\error}(\p,\q,s_0)=s_0\sum_{x\in\mk^c}p_x+s_1\sum_{x\in\mk}q_x\geq\min\Big\{\sum_{x\in\mk^c}p_x,\sum_{x\in\mk}q_x\Big\}\;,
\ee
where $\mk^c$ is the complement of $\mk$ in $[m]$. Similarly, for any $n\in\mbb{N}$ we denote  $\mk_n\eqdef \{x^n\in[m]^n\;:\;p_{x^n}\geq \lambda q_{x^n}\}$ so that
\be\label{twolba}
\pr_{\error}(\p^{\otimes n},\q^{\otimes n},s_0)\geq \min\Big\{\sum_{x^n\in\mk^c_n}p_{x^n},\sum_{x^n\in\mk_n}q_{x^n}\Big\}\;.
\ee
Next, we characterize the set $\mk_n$. From~\eqref{tpxn} the inequality $p_{x^n}\geq \lambda q_{x^n}$ holds if and only if
\be
2^{-n\big(H(\t(x^n))+D\left(\t(x^n)\|\p\right)\big)}\geq \lambda2^{-n\big(H(\t(x^n))+D\left(\t(x^n)\|\q\right)\big)}\;,
\ee
which is equivalent to
\be
D\left(\t(x^n)\|\q\right)- D\left(\t(x^n)\|\p\right)\geq\frac{1}{n}\log \lambda\;.
\ee
Therefore, denoting by
\be
\mc_n\eqdef\left\{\t\in\prob(m)\;:\;D\left(\t\|\q\right)- D\left(\t\|\p\right)\geq\frac{1}{n}\log \lambda\right\}
\ee
we get that the error probability can be bounded by
\be
\pr_{\error}(\rho^{\otimes n},\sigma^{\otimes n},s_0)\geq\min\Big\{\sum_{\substack{x^n\in[m]^n\\\t(x^n)\in\mc^c_n}}p_{x^n},\sum_{\substack{x^n\in[m]^n\\ \t(x^n)\in\mc_n}}q_{x^n}\Big\}\;.
\ee
 The first sum corresponds to the probability (with respect to an i.i.d.$\sim\!\p$ source) that a sequence of size $n$ has a type belonging to $\mc^c_n$, whereas the second sum corresponds to the probability (with respect to an i.i.d.$\sim\!\q$ source) that a sequence of size $n$ has a type belonging to $\mc_n$. These probabilities are very similar to the one appearing in Sanov's theorem (see Theorem~\ref{sanovt}) except that here the set $\mc_n$ depends on $n$. To remove this dependancy on $n$, observe that in the limit $n\to\infty$ the set $\mc_n$ approaches the set
\be
\mc\eqdef\left\{\t\in\prob(m)\;:\;D\left(\t\|\q\right)\geq D\left(\t\|\p\right)\right\}\;,
\ee
and similarly, the set $\mc_n^c$ (in $\prob(m)$) approaches the set
\be
\ms\eqdef\left\{\t\in\prob(m)\;:\;D\left(\t\|\q\right)\leq D\left(\t\|\p\right)\right\}\;.
\ee
Let $\p^\star,\q^\star\in\prob(m)$ be the optimizers
\be
D(\p^\star\|\p)\eqdef \min_{\r\in\ms}D(\r\|\p)\quad\text{and}\quad
D(\q^\star\|\q)\eqdef \min_{\r\in\mc}D(\r\|\q)\;.
\ee
From definitions and the continuity of the relative entropy, there exists two sequences of vectors $\{\p_n\}_{n\in\mbb{N}}$ and $\{\q_n\}_{n\in\mbb{N}}$ with limits $\p_n\to\p^*$ and $\q_n\to\q^*$ as $n\to\infty$ such that for each $n\in\mbb{N}$ the vector $\p_n\in\mc_n^c\cap\type(n,m)$ and the vector $\q_n\in\mc_n\cap\type(n,m)$. Therefore, following similar lines as given in Sanov's\index{Sanov's theorem}  theorem we get
for the first sum 
\ba
\sum_{\substack{x^n\in[m]^n\\\t(x^n)\in\mc^c_n}}p_{x^n}&=\sum_{\t\in\mc_n^c\cap\type(n,m)}|X^n(\t)|2^{-n\big(H(\t)+D\left(\t\|\p\right)\big)}\\
\GG{\substack{Taking\;only\;the\;term\\ \t=\p_{\it n}\;in\;the\;sum}}&\geq|X^n(\p_n)|2^{-n\big(H(\p_n)+D\left(\p_n\|\p\right)\big)}\\
\GG{\eqref{738}}&\geq \frac{1}{(n+1)^m}2^{-nD\left(\p_n\|\p\right)}\;.
\ea
Similarly, the second sum is bounded from below by
\be
\sum_{\substack{x^n\in[m]^n\\ \t(x^n)\in\mc_n}}q_{x^n}
\geq \frac{1}{(n+1)^m}2^{-nD\left(\q_n\|\q\right)}\;.
\ee
Substituting the two lower bounds above into~\eqref{twolba} yields
\ba\label{twosums}
\limsup_{n\to\infty}-\frac{1}{n}\log\pr_{\error}(\rho^{\otimes n},\sigma^{\otimes n},s_0)&\leq\limsup_{n\to\infty}-\frac{1}{n}\log\frac{\min\left\{2^{-nD\left(\p_n\|\p\right)},2^{-nD\left(\q_n\|\q\right)}\right\}}{(n+1)^m}\\
&=\min\{D(\p^\star\|\p),D(\q^\star\|\q)\}\;.
\ea
To compute $\p^\star$, observe that
\ba\label{seceqdrp}
\min_{\r\in\mk}D(\r\|\p)&=\min_{\r\in\prob(m)}\big\{D(\r\|\p)\;:\;D\left(\r\|\q\right)\leq D\left(\r\|\p\right)\big\}\\
\GG{Exercise~\ref{exeqq}}&=\min_{\r\in\prob(m)}\big\{D(\r\|\p)\;:\;D\left(\r\|\q\right)= D\left(\r\|\p\right)\big\}\\
&=\min_{\r\in\prob(m)}\Big\{\sum_{x\in[m]}r_x\log\frac{r_x}{p_x}\;:\;\sum_{x\in[m]}r_x\log\frac{q_x}{p_x}=0\Big\}\;.
\ea
Using the method of Lagrange multipliers, we define the Lagrangian
\be
\mL(\r)\eqdef\sum_{x\in[m]}r_x\log\frac{r_x}{p_x}+\mu\sum_{x\in[m]}r_x\log\frac{q_x}{p_x}+\nu\sum_{x\in[m]}r_x\;,
\ee
where $\mu$ and $\nu$ are some coefficients (Lagrange multipliers).
Hence,
\be
\frac{\partial\mL(\r)}{\partial r_x}=\log(e)+\log \frac{r_x}{p_x}+\mu\log\frac{q_x}{p_x}+\nu=0
\ee
which gives after isolating $r_x$
\be
r_x=ap_x\left(\frac{q_x}{p_x}\right)^\mu=ap_{x}^{1-\mu}q^\mu_x
\ee
where $a=e^{-1}2^{-\nu}$ is some constant determined by the condition $\sum_xr_x=1$. Hence,
\be
r_{x}=\frac{p_{x}^{1-\mu}q^\mu_x}{\sum_{x'\in[m]}p_{x'}^{1-\mu}q^\mu_{x'}}
\ee
Hence, denoting by $\r_\mu$ the probability vector whose components are as above we conclude that
\be
\min_{\r\in\mk}D(\r\|\p)=D(\r_\mu\|\p)
\ee
where $\mu$ is the number satisfies
\be
D\left(\r_\mu\|\p\right)=D\left(\r_\mu\|\q\right)\;.
\ee
Moreover, note that for $\mu=0$, $\r_0=\p$, and for $\mu=1$, $\r_1=\q$. Therefore, from the continuity of the  KL-divergence it follows that the equality above is achieved for some $\mu\in[0,1]$. 

From the symmetry of the expression for $\r_\mu$ it is clear that by the repetition of the same computation as above we get $\q^\star=\p^\star=\r_\mu$. We therefore conclude that
\be
\lim_{n\to\infty}-\frac{1}{n}\log\pr_{\error}(\rho^{\otimes n},\sigma^{\otimes n},t_0)
\leq D\left(\r_\mu\|\p\right)
\ee
where $\mu$ is determined by $D\left(\r_\mu\|\q\right)=D\left(\r_\mu\|\p\right)$. In the exercise below you will show that for this $\mu$
\be\label{ls}
D\left(\r_\mu\|\q\right)=-\log\min_{\alpha\in[0,1]}\sum_{x\in[m]}p_x^{1-\alpha}q_x^\alpha\;.
\ee  
This completes the proof.
\end{proof}

\begin{exercise}\label{exeqq}
Prove the second equality of~\eqref{seceqdrp}. Hint: Suppose that the minimum is obtained for some $\r\in\prob(m)$ that satisfies $D\left(\r\|\q\right)< D\left(\r\|\p\right)$ and get a contradiction by showing that the vector $\t=(1-\eps)\r+\eps\p$ (with small $\eps>0$) satisfies $D(\t\|\p)<D(\r\|\p)$.
\end{exercise}

\begin{exercise}
Prove Eq.~\eqref{ls}. Hint: Show that the condition $D(\r_\mu\|\p)=D(\r_\mu\|\q)$ is equivalent to $\sum_{x\in[m]}p_{x}^{1-\mu}q^\mu_x\log\frac{q_x}{p_x}=0$ and compare it with the derivative  of the function $f(s)\eqdef\sum_{x\in[m]}p_x^{1-s}q_x^s$.
\end{exercise}

\section{Quantum Hypothesis Testing}\label{sec:qht}\index{hypothesis testing}

One of the foundational aspects of quantum mechanics is the inability to perfectly distinguish between quantum systems.
To elucidate this concept of distinguishability, let's build upon the ideas presented in the previous section. Consider a scenario where an experimenter, named Alice, possesses a quantum system in her lab (e.g., an electron in a certain spin state) that could be in one of two potential states, denoted as $\rho\in\md(A)$ and $\sigma\in\md(A)$. Alice can carry out a POVM denoted by $\{\Lambda_x\}_{x\in[m]}$ on her system to deduce its state. Depending on the measurement outcome $x$ she may infer the state to be $\rho$ or $\sigma$. In this section, we delve into the best strategy Alice can employ to accurately determine the state of her quantum system.

We note that it's adequate to contemplate POVMs composed of only two elements. We can define $\Lambda\in\eff(A)$ to be the sum of all effects $\{\Lambda_x\}$, where $x$ leads Alice to infer $\rho$. Conversely, $I-\Lambda$ is the sum of the remaining POVM elements, corresponding to $x$ values that result in Alice inferring $\sigma$. Two types of errors might arise:
\ben
\item \textbf{Type I Error:} Alice possesses the state $\rho$ but incorrectly infers it as $\sigma$. The associated probability is:
\be
\alpha(\Lambda)\eqdef\tr\left[\rho (I-\Lambda)\right]
\ee
\item \textbf{Type II Error:} Alice has the state $\sigma$ but mistakenly deduces it as $\rho$. The corresponding probability is:
\be
\beta(\Lambda)\eqdef\tr\left[\sigma \Lambda\right]
\ee
\een
As in the classical scenario, we explore strategies to minimize the error probabilities $\alpha(\Lambda)$ and $\beta(\Lambda)$. With the asymmetric approach, the objective is to minimize the Type II error, $\beta(\Lambda)$, while ensuring that the Type I error, $\alpha(\Lambda)$, stays beneath a specific threshold $\eps>0$. The optimal approach is encapsulated by the quantum Stein's lemma. In the symmetric strategy, the observer is aware of a prior $\{s_0,s_1\}$ where $\rho$ occurs with probability $s_0$, and $\sigma$ with probability $s_1$. The aim here is to reduce the overall error probability represented by $s_0\alpha(\Lambda)+s_1\beta(\Lambda)$.

It's worth noting that any pair of quantum states $\rho,\sigma\in\md(A)$ that aren't identical satisfy $\tr[\rho\sigma]<1$. This implies:
\be
\lim_{n\to\infty}\tr\left[\rho^{\otimes n}\sigma^{\otimes n}\right]=\lim_{n\to\infty}\left(\tr[\rho\sigma]\right)^n=0\;.
\ee
In essence, as $n$ approaches infinity, the states $\rho^{\otimes n}$ and $\sigma^{\otimes n}$ become orthogonal with respect to the Hilbert-Schmidt inner product. Naturally, we might question the rate at which these states turn distinguishable. Both the quantum Stein's lemma and the quantum Chernoff bound address this question, in the asymmetric and symmetric contexts, respectively.\index{Hilbert-Schmidt inner product}

\subsection{The Quantum Hypothesis Testing Divergence}\label{sec:qht}

In asymmetric hypothesis testing, the goal is to minimize the Type II error while simultaneously ensuring that the Type I error remains bounded by a small parameter $\eps > 0$. Under this scenario, the pertinent error probability is defined as:
\be
\beta^*(\eps)\eqdef\min\big\{\beta(\Lambda)\;:\;\alpha(\Lambda)\leq\eps\;\;,\;\;0\leq \Lambda\leq I\big\}\;.
\ee

\begin{exercise}
Show that the optimal probability $\beta^*(\eps)=0$ for all $\eps\geq 0$ if and only if $\rho$ and $\sigma$ satisfies $\rho\sigma=0$.
\end{exercise}

By taking the $-\log$ of the above error probability one obtains a quantity known as the \emph{quantum hypothesis testing divergence}. 
\begin{myd}{}
\begin{definition}
The quantum hypothesis testing divergence is defined for all $\rho,\sigma\in\md(A)$ and $\eps\in[0,1)$ as
\ba\label{dehl}
D^{\eps}_{\min}\left(\rho\|\sigma\right)&\eqdef-\log\min_{\Lambda\in\eff(A)}\big\{\tr\left[\sigma \Lambda\right]\;:\;\tr\left[\rho \Lambda\right]\geq 1-\eps\big\}\;.
\ea
\end{definition}
\end{myd}
The hypothesis testing divergence is invariably non-negative and reaches infinity when $\rho$ and $\sigma$ are orthogonal. Furthermore, Exercise~\ref{exeps0} guides you to demonstrate that when $\rho$ and $\sigma$ are diagonal in the same basis, the quantum hypothesis testing divergence, denoted as $D_{\min}^\eps(\rho\|\sigma)$, simplifies to its classical equivalent, $D_{\min}^\eps(\p\|\q)$. Here, $\p$ and $\q$ represent the diagonal elements of $\rho$ and $\sigma$, respectively. In the classical context, we observed that for $\eps = 0$, the divergence $D_{\min}^\eps$ reduces to the min relative entropy. This observation is consistent in the quantum scenario as well (refer to Exercise~\ref{exeps0}). Such parallels justify the use of the `min' subscript in naming the quantum hypothesis testing divergence. Next, we aim to establish that this function indeed qualifies as an (unnormalized) divergence.

\begin{myt}{\color{yellow} Data Processing Inequality}
\begin{theorem}
Let $\eps\in[0,1)$, $\rho,\sigma\in\md(A)$, and $\mE\in\cptp(A\to B)$.
Then,
\be
D_{\min}^\eps\big(\mE(\rho)\big\|\mE(\sigma)\big)\leq D^{\eps}_{\min}\left(\rho\|\sigma\right)\;.
\ee
\end{theorem}
\end{myt}
\begin{proof}
Let $0\leq \Gamma\leq I^B$ be an optimal effect such that
\be
2^{-D^{\eps}_{\min}\left(\mE(\rho)\|\mE(\sigma)\right)}=\tr\left[\mE(\sigma) \Gamma\right]
\ee
and $\tr\left[\mE(\rho) \Gamma\right]\geq 1-\eps$. Note that $\Lambda\eqdef\mE^*(\Gamma)$ satisfies $0\leq \Lambda\leq I^A$ and $\tr\left[\rho \Lambda\right]=\tr[\mE(\rho) \Gamma]\geq 1-\eps$. Hence,
\ba
2^{-D^{\eps}_{\min}\left(\mE(\rho)\|\mE(\sigma)\right)}&=\tr\left[\mE(\sigma) \Gamma\right]\\
&=\tr\left[\sigma \Lambda\right]\\
&\geq \min\left\{\tr\left[\sigma \Lambda'\right]\;:\;\tr\left[\rho \Lambda'\right]\geq 1-\eps\;\;,\;\;\Lambda'\in\eff(A)\right\}\\
&=2^{-D^{\eps}_{\min}\left(\rho\|\sigma\right)}
\ea
This completes the proof.
\end{proof}

\bex\label{exeps0}
Consider the definition above of the quantum hypothesis testing.
\ben
\item Show that if $\rho,\sigma\in\md(A)$ are diagonal in the same basis of $A$ then $D_{\min}^\eps(\rho\|\sigma)$ reduces to its classical counterpart $D_{\min}^\eps(\p\|\q)$, where $\p$ and $\q$ are the diagonals of $\rho$ and $\sigma$, respectively.
\item Show that for $\eps = 0$, the quantum hypothesis testing divergence simplifies to the quantum min relative entropy. That is, show that 
for all $\rho,\sigma\in\md(A)$
\be
D_{\min}^{\eps=0}(\rho\|\sigma)=D_{\min}(\rho\|\sigma)=-\log\tr\left[\sigma\Pi_\rho\right]\;.
\ee
\een
\eex

\begin{exercise}\label{htcon}
Show that the constraint $\tr\left[\rho \Lambda\right]\geq 1-\eps$ in~\eqref{dehl} can be replaced with $\tr\left[\rho \Lambda\right]= 1-\eps$ (i.e. both constraints leads to the same value of $D^\eps_{\min}(\rho\|\sigma)$).
\end{exercise}

\begin{exercise}\label{ex:smoothed}\label{ex:ndeps2}$\;$
\ben
\item Show that for all $\rho,\sigma\in\md(A)$ we have
\be
D_{\min}^{\eps}(\rho\|\sigma)\geq-\log(1-\eps)\;,
\ee
with equality if $\rho=\sigma$.
\item Show that $D_{\min}^{\eps}(\rho\|\sigma)$ is non-decreasing in $\eps$.
\een
\end{exercise}

\begin{exercise}
Show that the quantum hypothesis testing divergence equals its minimal extension from classical states. That is, show that for all $\rho,\sigma\in\md(A)$
\be\label{clqudmin}
D_{\min}^\eps\left(\rho^A\big\|\sigma^A\right)=\sup_{\mE\in\cptp(A\to X)} D_{\min}^\eps\left(\mE^{A\to X}(\rho^A)\big\|\mE^{A\to X}(\sigma^A)\right)\;,
\ee
where the supremum is over all classical systems $X$ and POVM channels $\mE\in\cptp(A\to X)$ that takes $\rho$ and $\sigma$ to diagonal density matrices (i.e. probability vectors). 
\end{exercise}

\subsubsection{Computation of the Quantum Hypothesis Testing Divergence}\index{hypothesis testing}

The quantum hypothesis testing divergence can be computed using semidefinite programming (SDP) techniques. In particular, we can express it for $\rho,\sigma\in\md(A)$ as in~\eqref{primal} via
\be\label{7137}
2^{-D^{\eps}_{\min}\left(\rho\|\sigma\right)}=\min_{\substack{\mN(\eta)-H_2\geq 0\\ \eta\geq 0}}\tr[\eta H_1]
\ee
with the identifications
$H_1\eqdef\sigma$, $H_2\eqdef (1-\eps)\oplus(-I^A)$, and
$\mN:\herm(A)\to\mbb{R}\oplus\herm(A)$ defined for all $\eta\in\herm(A)$ as
\be
\mN(\eta)\eqdef\tr[\rho \eta]\oplus (-\eta)\;.
\ee
Note that $\mN$ is a linear map and its dual map\index{dual map} $\mN^*:\mbb{R}\oplus\herm(A)\to\herm(A)$ is given by (see the exercise below)
\be\label{dualc}
\mN^*(t\oplus\omega)=t\rho-\omega\;.
\ee
From~\eqref{dual123} it then follows that the dual to the above SDP optimization problem is given by 
\be\label{z0z}
2^{-D^{\eps}_{\min}\left(\rho\|\sigma\right)}=\max_{\substack{H_1-\mN^*(t\oplus\omega)\geq 0\\ t\in\mbb{R}_+,\;\omega\in\pos(A)}}\tr[(t\oplus\omega) H_2]
\ee
Substituting the expressions above for $H_1$, $H_2$, and $\mN^*$, gives
\be\label{z1z}
2^{-D^{\eps}_{\min}\left(\rho\|\sigma\right)}=\max\Big\{t(1-\eps)-\tr[\omega]\;:\;t\rho\leq \sigma+\omega\;\;,\;\; t\in\mbb{R}_+\;\;,\;\;\omega\in\pos(A)\Big\}\;.
\ee
The maximization above is over all $t\in\mbb{R}_+$ and $\omega\in\pos(A)$. For a fixed $t$, we want to minimize $\tr[\omega]$ such that $\omega\geq 0$ and $\omega\geq t\rho-\sigma$. Under these constraints, it follows trom Exercise~\ref{trasdf} that the choice $\omega\eqdef(t\rho-\sigma)_+$ has the minimal trace. We therefore conclude that
\be\label{dminft}
D^{\eps}_{\min}\left(\rho\|\sigma\right)=-\log\max_{t\in\mbb{R}_+}f(t)
\ee
where $f:\mbb{R}_+\to\mbb{R}_+$ is the function
\be\label{fot}
f(t)\eqdef t(1-\eps)-\tr(t\rho-\sigma)_+\;.
\ee
\begin{exercise}$\;$
\ben
\item Verify that indeed~\eqref{7137} is equivalent to~\eqref{dehl}.
\item Prove that the dual map $\mN^*$ is given by~\eqref{dualc}, and use it to derive~\eqref{z1z} from~\eqref{z0z}.
\een
\end{exercise}
\bex\label{trasdf}$\;$
\ben
\item Let $\eta\in\herm(A)$. Show that
\be\label{eta+}
\tr[\eta_+]=\inf\big\{\tr[\zeta]\;:\;\zeta\geq \eta\;\;,\;\;\zeta\in\pos(A)\big\}\;,
\ee
where $\eta_+$ is the positive part of $\eta$ (i.e. $\eta=\eta_+-\eta_-$ with $\eta_+,\eta_-\in\pos(A)$ and $\eta_+\eta_-=\0$).
\item Show that the function in~\eqref{fot} can be expressed as
\be
f(t)=\frac12\big(1+t(1-2\eps)-\|t\rho-\sigma\|_1\big)\;.
\ee
Hint: Recall that $\eta_+=\frac12(|\eta|+\eta)$.
\een
\eex

\subsubsection{The Relationship between the Hypothesis Testing and R\'enyi Divergences}\index{hypothesis testing}

In the subsequent theorem, we establish a connection between the hypothesis testing divergence and the R'enyi relative entropies. For this purpose, we use the notation $D_\alpha$ to represent the Petz\index{Petz} quantum $\alpha$-R'enyi divergence (as defined in Definition~\ref{def:petz}), and $\tD_\alpha$ to denote the quantum sandwiched (minimal) $\alpha$-R'enyi divergence (as per Definition~\ref{def:sandwich}). Additionally, we use $h(\alpha) = -\alpha\log\alpha - (1-\alpha)\log(1-\alpha)$ to denote the binary Shannon entropy.\index{sandwiched relative entropy}

\begin{myt}{}
\begin{theorem}\label{thm:boundsdmin}
Let $\eps\in(0,1)$ and $\rho,\sigma\in\md(A)$.  
\begin{align}
&1.\;\; \text{For all }\alpha>1:\quad\quad\;\;\;
D_{\min}^\eps(\rho\|\sigma)\leq \tD_\alpha(\rho\|\sigma)+\frac\alpha{\alpha-1}\log\left(\frac1{1-\eps}\right)\label{e1}\\
&2.\;\; \text{For all }\alpha\in(0,1):\;\quad
D_{\min}^\eps(\rho\|\sigma)\geq D_\alpha(\rho\|\sigma)+\frac\alpha{1-\alpha}\left(\frac{h(\alpha)}{\alpha}-\log\left(\frac1\eps\right)\right)\label{e2}
\end{align}
\end{theorem}
\end{myt}

\begin{remark}
Given that $\tD_{\alpha}$ represents the minimal quantum extension of the classical $\alpha$-R\'enyi relative entropy, it follows that $\tD_\alpha(\rho\|\sigma) \leq D_\alpha(\rho\|\sigma)$. Consequently, in the upper bound of~\eqref{e1}, we can substitute $\tD_\alpha(\rho\|\sigma)$ with $D_\alpha(\rho\|\sigma)$.
\end{remark}

\begin{proof}
Let $\Lambda\in\eff(A)$ be such that $2^{-D_{\min}^\eps(\rho\|\sigma)}=\tr[\Lambda\sigma]$ and $\tr[\Lambda\rho]=1-\eps$. Set $p\eqdef\tr[\Lambda\sigma]$, and define the binary POVM Channel\index{POVM channel}  $\mE\in\cptp(A\to X)$ via
\be
\mE(\omega)\eqdef\tr[\omega\Lambda]|0\lr 0|+\tr[\omega(I-\Lambda)]|1\lr 1|\quad\quad\forall\;\omega\in\ml(A)\;.
\ee
From the DPI of $\tD_\alpha$ we get that
\ba
\tD_\alpha(\rho\|\sigma)\geq\tD_\alpha\big(\mE(\rho)\big\|\mE(\sigma)\big)
&=D_\alpha\left((1-\eps,\eps)^T\big\|(p,1-p)^T\right)\\
\GG{By\; definition}&=\frac1{\alpha-1}\log\left((1-\eps)^\alpha p^{1-\alpha}+\eps^\alpha(1-p)^{1-\alpha}\right)\\
\Gg{\text{Removing }\eps^\alpha(1-p)^{1-\alpha}}&\geq \frac1{\alpha-1}\log\left((1-\eps)^\alpha p^{1-\alpha}\right)\\
&=\frac\alpha{\alpha-1}\log(1-\eps) -\log p\\
\GG{By\;definition\;of\;{\it p}}&=\frac\alpha{\alpha-1}\log(1-\eps)+D_{\min}^\eps(\rho\|\sigma)\;.
\ea
This concludes the proof of~\eqref{e1}.

To prove~\eqref{e2}, let $\alpha\in(0,1)$. We will use the expression for $D_{\min}^\eps(\rho\|\sigma)$ as given in~\eqref{dminft} and~\eqref{fot}. To bound the expression $\tr(t\rho-\sigma)_+$ in equation \eqref{fot}, we employ the quantum weighted geometric-mean inequality given by \eqref{acm}. This inequality asserts that for any pair of matrices $M, N \in \pos(A)$ and any value of $\alpha$ within the range [0,1]:
\be\label{7167}
\frac{1}{2}\tr\Big[M+N-\big|M-N\big|\Big]\leq \tr\big[M^{\alpha}N^{1-\alpha}\big]\;.
\ee
Since the term $|M-N|$ can be expressed as $|M-N|=2(M-N)_+-(M-N)$, the above inequality is equivalent to
\be
\tr(M-N)_+\geq \tr[M]-\tr\big[M^{\alpha}N^{1-\alpha}\big]
\ee
Taking $M=t\rho$ and $N=\sigma$ we have
\ba\label{sr1t}
\tr(t\rho-\sigma)_+&\geq t-t^{\alpha}\tr\big[\rho^{\alpha}\sigma^{1-\alpha}\big]\\
&=t-t^\alpha2^{(\alpha-1)D_\alpha(\rho\|\sigma)}\;.
\ea
Substituting this into~\eqref{dminft} and~\eqref{fot} we get
\ba\label{optsti}
2^{-D^{\eps}_{\min}\left(\rho\|\sigma\right)}&=\max_{t\in\mbb{R_+}}\Big\{t(1-\eps)-\tr(t\rho-\sigma)_+\Big\}\\
\GG{\eqref{sr1t}}&\leq \max_{t\in\mbb{R_+}}\Big\{-t\eps+t^\alpha2^{(\alpha-1)D_\alpha(\rho\|\sigma)}\Big\}
\ea
It is straightforward to check that for fixed $\alpha,\rho,\sigma,\eps$, the function $t\mapsto -t\eps+t^\alpha2^{(\alpha-1)D_\alpha(\rho\|\sigma)}$ obtains its maximal value at 
\be
t=\left(\frac{\alpha}{\eps}\right)^{\frac1{1-\alpha}}2^{-D_\alpha(\rho\|\sigma)}\;.
\ee
Substituting this value into the optimization in~\eqref{optsti} gives
\be
2^{-D^{\eps}_{\min}\left(\rho\|\sigma\right)}\leq (1-\alpha)\left(\frac{\alpha}\eps\right)^{\frac\alpha{1-\alpha}}2^{-D_\alpha(\rho\|\sigma)}\;.
\ee
By taking $-\log$ on both sides we get~\eqref{e2}. This concludes the proof.
\end{proof}

\subsection{Asymmetric Discrimination of Quantum States}

The subsequent theorem extends Theorem~\ref{th:sl} to encompass the quantum domain. Consequently, the proof we present for the quantum scenario also substantiates the classical Stein's lemma\index{Stein's lemma}, as given in Theorem~\ref{th:sl}.

\begin{myt}{\color{yellow} The Quantum Stein's Lemma}
\begin{theorem}\label{qsl}
Let $A$ be a finite dimensional system, $0<\eps<1$, and $\rho,\sigma\in\md(A)$ with $\supp(\rho)\subseteq\supp(\sigma)$. Then,
\be\label{7171}
\lim_{n\to\infty}\frac{1}{n}D_{\min}^\eps\left(\rho^{\otimes n}\big\|\sigma^{\otimes n}\right)=D(\rho\|\sigma)\;,
\ee
where $D(\rho\|\sigma)\eqdef\tr[\rho\log\rho]-\tr[\rho\log\sigma]$ is known as \emph{the Umegaki relative entropy}.
\end{theorem}
\end{myt}
\begin{remark}
The quantum Stein's lemma indicates that the optimal type II error approximately follows the behavior of $\approx2^{-nD(\rho\|\sigma)}$ with respect to the number of copies, $n$, of $\rho$ and $\sigma$. Specifically, the lemma offers an operational interpretation of the Umegaki divergence, $D(\rho\|\sigma)$, defining it as the maximal rate at which the type II error diminishes to zero in an exponential manner with increasing $n$. Additionally, it's worth noting that the theorem above implies that the limit on the right-hand side of~\eqref{7171} exists and is independent on $\eps$.
\end{remark}
\begin{proof}
The proof follows from the bounds in Theorem~\ref{thm:boundsdmin}. Specifically, from~\eqref{e1} we get for any $\eps\in(0,1)$ and any $\alpha>1$
\ba
\limsup_{n\to\infty}\frac{1}{n}D_{\min}^\eps\left(\rho^{\otimes n}\big\|\sigma^{\otimes n}\right)&\leq \limsup_{n\to\infty}\frac{1}{n}\left(
\tD_\alpha\left(\rho^{\otimes n}\big\|\sigma^{\otimes n}\right)+\frac\alpha{\alpha-1}\log\left(\frac1{1-\eps}\right)\right)\\
&=\tD_\alpha\left(\rho\|\sigma\right)\;,
\ea
where in the last equality we used the additivity\index{additivity} (under tensor products) of $\tD_\alpha$. Since the equation above holds for all $\alpha>1$ we conclude that
\ba\label{secineqq}
\limsup_{n\to\infty}\frac{1}{n}D_{\min}^\eps\left(\rho^{\otimes n}\big\|\sigma^{\otimes n}\right)&\leq \lim_{\alpha\to1^+}\tD_\alpha\left(\rho\|\sigma\right)\\
&=D(\rho\|\sigma)\;,
\ea
where the equality above follows from continuity in $\alpha$ of the function $\alpha\mapsto\tD_\alpha(\rho\|\sigma)$.

For the opposite inequality, we use the bound~\eqref{e2} to get for all $\alpha\in(0,1)$
\ba
\liminf_{n\to\infty}\frac1nD_{\min}^\eps\left(\rho^{\otimes n}\big\|\sigma^{\otimes n}\right)&\geq \liminf_{n\to\infty}\frac{1}{n}\left(D_\alpha\left(\rho^{\otimes n}\big\|\sigma^{\otimes n}\right)+\frac\alpha{1-\alpha}\left(\frac{h(\alpha)}{\alpha}+\log\eps\right)
\right)\\
&=D_\alpha\left(\rho\|\sigma\right)\;,
\ea
where we used the additivity of $D_\alpha$. Since $D_\alpha$ is continuous in $\alpha$, and since the equation above holds for all $\alpha\in(0,1)$, it must also hold for $\alpha=1$; that is,
\be
\liminf_{n\to\infty}\frac{1}{n}D_{\min}^\eps\left(\rho^{\otimes n}\big\|\sigma^{\otimes n}\right)\geq D(\rho\|\sigma)\;.
\ee
Combining this with the inequality~\eqref{secineqq}, we conclude that the limit \be\lim_{n\to\infty}\frac{1}{n}D_{\min}^\eps\left(\rho^{\otimes n}\big\|\sigma^{\otimes n}\right)\ee exists and equals to $D(\rho\|\sigma)$.
\end{proof}

\begin{exercise}\label{ex:umegaki}{\rm [The Umegaki Relative Entropy]}
Let $D$ be the Umegaki relative entropy.
\ben
\item Show that $D$ satisfies the DPI. Hint: Use~\eqref{7171} and the fact that
$D_{\min}^\eps$ satisfies the DPI.
\item Show by direct calculation that for any two cq-states in $\md(AX)$, $\rho^{AX}\eqdef \sum_{x\in[n]}p_x\rho_x^A\otimes|x\lr x|^X$ and $\sigma^{AX}\eqdef\sum_{x\in[n]}q_x\sigma_x^A\otimes|x\lr x|^X$  we have
\be\label{relume}
D\left(\rho^{AX}\big\|\sigma^{AX}\right)=\sum_{x\in[n]}p_xD\left(\rho_x^A\big\|\sigma_x^A\right)+D(\p\|\q)\;,
\ee
where the components of the probability vectors $\p$ and $\q$ are $\{p_x\}_{x\in[n]}$ and $\{q_x\}_{x\in[n]}$, respectively.
\item Use the above two properties to show that for any two ensembles of states
$\{p_x,\rho_x\}_{x\in[n]}$ and $\{q_x,\sigma_x\}_{x\in[n]}$ we have
\be
D\Big(\sum_{x\in[n]}p_x\rho_x\Big\|\sum_{x\in[n]}q_x\sigma_x\Big)\leq\sum_{x\in[n]}p_xD\left(\rho_x\|\sigma_x\right)+D(\p\|\q)\;.
\ee
In particular, show that the Umegaki relative entropy is jointly convex.
\een
\end{exercise}

 \subsection{Symmetric Discrimination of Quantum States}

Consider the following setup, in which an observer (say Alice) is given a quantum state $\rho$ with probability $t_0\eqdef t\in[0,1]$, and a quantum state $\sigma\in\md(A)$ with probability $t_1\eqdef 1-t$. As before, the goal is for Alice  to guess correctly which state she was given. For this purpose Alice perform a binary outcome POVM consisting of two POVM elements  $\Lambda_0\eqdef \Lambda$ and $\Lambda_1\eqdef I-\Lambda$, where $0\leq \Lambda\leq I$. If Alice gets outcome 0 she declares that the state is $\rho$, and if the outcome is 1 she declares that the state is $\sigma$. For any such POVM the probability of error is given by 
\ba\label{8219}
\pr_{\error}(\Lambda,\rho,\sigma,t)&\eqdef t_0\alpha(\Lambda)+t_1\beta(\Lambda)\\
&=t_0\tr[\rho \Lambda_1]+t_1\tr[\sigma \Lambda_0]\\
&=t_0+\tr\left[\big(t_1\sigma-t_0\rho\big)\Lambda\right]
\ea
Minimizing the above expression over all $0\leq \Lambda\leq I$ gives
\ba\label{prerr}
\pr_{\error}(\rho,\sigma,t)&\eqdef\min_{0\leq \Lambda\leq I}\pr_{\error}(\Lambda,\rho,\sigma,t)\\
\GG{\eqref{eta+}}&=t_0-\tr\big(t_1\sigma-t_0\rho\big)_{-}\\
\Gg{\forall\eta\in\herm(A),\;\eta_-=\frac12(|\eta|-\eta)}&=t_0-\frac{1}{2}\left(\big\|t_1\sigma-t_0\rho\big\|_1-1+2t_0\right)\\
&=\frac{1}{2}\left(1-\big\|t_0\rho-t_1\sigma\big\|_1\right)\;.
\ea

\bex
Show that with $t\eqdef t_0$ and $r\eqdef t_1/t_0$ we can express the probability of error as:
\be\label{prerr}
\pr_{\error}(\rho,\sigma,t)=t\big(1-\tr(\rho-r\sigma)_+\big)\;.
\ee
\eex

\bex
Let $\rho,\sigma\in\md(A)$.
\ben
\item Let $\eps\in(0,1)$. Show that
\be
2^{-D_{\min}^\eps(\rho\|\sigma)}=\sup_{t\in(0,1)}\frac{\pr_{\error}(\rho,\sigma,t)-t\eps}{1-t}\;.
\ee
Hint: Use~\eqref{prerr} and~\eqref{fot}.
\item Let $t$ and $r$ be as in~\eqref{prerr}. Show that
\be
\pr_{\error}(\rho,\sigma,t)=t\inf_{\eps\in(0,1)}\left\{\eps+r2^{-D_{\min}^\eps(\rho\|\sigma)}\right\}\;.
\ee
Hint: Recall that $\tr(\rho-r\sigma)_+=\sup_{\Lambda\in\eff(A)}\tr[\Pi(\rho-r\sigma)]$ and split the supremum over all $\eps\in(0,1)$ and all $\Lambda\in\eff(A)$ such that $\tr[\Lambda\rho]=1-\eps$.
\een
\eex

As previously discussed, with increasing $n$ copies of $\rho$ and $\sigma$, the states $\rho^{\otimes n}$ and $\sigma^{\otimes n}$ become more distinguishable. We will demonstrate in the upcoming theorem that the error probability, $\pr_{\error}(\rho^{\otimes n},\sigma^{\otimes n},t)$, diminishes at an exponential rate as $n$ approaches infinity. This rate is characterized by what is known as the quantum Chernoff bound. The classical counterpart of the subsequent theorem, along with its proof, can be found in Section~\ref{cht} (see Theorem~\ref{cherthm}). We will use the notation $\xi_{{\rm Q}}(\rho,\sigma)$ to denote the quantum extension of the classical Chernoff\index{Chernoff} bound $\xi(\p,\q)$ as given in~\eqref{chernxi}. In the quantum domain it is defined as
\ba
\xi_{{\rm Q}}(\rho,\sigma)&\eqdef-\log\min_{0\leq \alpha\leq 1}\tr[\rho^{\alpha}\sigma^{1-\alpha}]\\
\GG{Definition~\ref{def:petz}}&=\max_{\alpha\in[0,1]}\big\{(1-\alpha)D_\alpha(\rho\|\sigma)\big\}\;.
\ea

\begin{myt}{\color{yellow} The Quantum Chernoff\index{Chernoff} Bound}
\begin{theorem}
Let $\rho,\sigma\in\md(A)$. For any probability distribution $\{t_0\eqdef t,t_1\eqdef 1-t\}$ with $0<t<1$,
\be
\lim_{n\to\infty}-\frac{1}{n}\log\pr_{\error}(\rho^{\otimes n},\sigma^{\otimes n},t)=\xi_{{\rm Q}}(\rho,\sigma)\;.
\ee
\end{theorem}
\end{myt}

\begin{proof}
In the proof of Theorem~\ref{thm:boundsdmin} we used~\eqref{sr1t} to bound $\tr(t\rho-\sigma)_+$. Dividing both sides of~\eqref{sr1t} by $t$ and denoting $r\eqdef 1/t$  we get that~\eqref{sr1t} is equivalent to
\be
\tr(\rho-r\sigma)_+\geq 1-r^{1-\alpha}2^{(\alpha-1)D_\alpha(\rho\|\sigma)}\;.
\ee
Combining this with~\eqref{prerr} we get that for all $\alpha\in(0,1)$
\ba
\pr_{\error}(\rho,\sigma,t)&\leq tr^{1-\alpha}2^{(\alpha-1)D_\alpha(\rho\|\sigma)}\\
&=t_0^\alpha t_1^{1-\alpha}\tr\left[\rho^\alpha\sigma^{1-\alpha}\right]\;.
\ea
Hence,
\ba
\pr_{\error}(\rho^{\otimes n},\sigma^{\otimes n},t)\leq t_0^{\alpha}t_1^{1-\alpha}\left(\tr\left[\rho^{\alpha}\sigma^{1-\alpha}\right]\right)^n
\ea
so that 
\be
\liminf_{n\to\infty}-\frac{1}{n}\log\pr_{\error}(\rho^{\otimes n},\sigma^{\otimes n},t)\geq-\log\tr\left[\rho^{\alpha}\sigma^{1-\alpha}\right]\;.
\ee
Since the above equation holds for all $0\leq \alpha\leq 1$ we have
\ba
\lim_{n\to\infty}-\frac{1}{n}\log\pr_{\error}(\rho^{\otimes n},\sigma^{\otimes n},t)&\geq\max_{\alpha\in[0,1]}\big\{-\log\tr\left[\rho^{\alpha}\sigma^{1-\alpha}\right]\big\}\\
&=-\log\min_{\alpha\in[0,1]}\tr\left[\rho^{\alpha}\sigma^{1-\alpha}\right]\;.
\ea

To prove the opposite inequality, let 
\ba
\rho=\sum_{x\in[m]}p_x\psi_x\quad\text{and}\quad \sigma=\sum_{y\in[m]}q_y\phi_y\;,
\ea
be the spectral decomposition of $\rho$ and $\sigma$ (here $m\eqdef|A|$), where $\psi_x,\phi_y\in\pure(A)$ for all $x,y\in[m]$. Then, for any projection $\Pi\in\pos(A)$ (i.e. $\Pi^2=\Pi$) we have
\ba\label{0y0}
\tr\left[\Pi\rho\right]=\sum_{x\in[m]}p_x\la \psi_x|\Pi|\psi_x\ra
&=\sum_{x\in[m]}p_x\la \psi_x|\Pi^2|\psi_x\ra\\
\Gg{\sum_{y\in[m]}\phi_y=I^A}&=\sum_{x\in[m]}p_x\big\la \psi_x\big|\Pi\sum_{y\in[m]}|\phi_y\lr\phi_y|\Pi\big|\psi_x\big\ra\\
&=\sum_{x,y\in[m]}p_x|\la \psi_x|\Pi|\phi_y\ra|^2
\ea
Similarly, since $I-\Pi$ is also a projection we get 
\be\label{0y1}
\tr\left[(I-\Pi)\sigma\right]=\sum_{x,y\in[m]}q_y|\la \psi_x|(I-\Pi)|\phi_y\ra|^2\;.
\ee
Therefore, taking $\Lambda=\Pi$ in~\eqref{8219} gives
\ba
\pr_{\error}(\Pi,\rho,\sigma,t)&=t_0\tr[\rho \Pi]+t_1\tr[\sigma (I-\Pi)]\\
\GG{\eqref{0y0},\eqref{0y1}}&=\sum_{x,y\in[m]}\Big(t_0p_x|\la \psi_x|I-\Pi|\phi_y\ra|^2+t_1q_y|\la \psi_x|\Pi|\phi_y\ra|^2\Big)\\
&\geq\sum_{x,y\in[m]}\min\{t_0p_x,t_1q_y\}\Big(|\la \psi_x|I-\Pi|\phi_y\ra|^2+|\la \psi_x|\Pi|\phi_y\ra|^2\Big)\;.
\ea
Moreover, since for any two complex numbers $c_1$ and $c_2$ satisfies $|c_1|^2+|c_2|^2\geq\frac{1}{2}|c_1+c_2|^2$, we get that
\ba
\pr_{\error}(\Pi,\rho,\sigma,t_0)&\geq\frac{1}{2}\sum_{x,y\in[m]}\min\{t_0p_x,t_1q_y\}\Big|\la \psi_x|I-\Pi|\phi_y\ra+\la \psi_x|\Pi|\phi_y\ra\Big|^2\\
&=\frac{1}{2}\sum_{x,y\in[m]}\min\{t_0p_x,t_1q_y\}\big|\la \psi_x|\phi_y\ra\big|^2\\
&=\frac{1}{2}\sum_{x,y\in[m]}\min\{t_0p_{xy},t_1q_{xy}\}\\
\GG{\eqref{7125}}&=\frac{1}{2}\pr_{\error}(\p,\q,t)\;,
\ea
where $\p=(p_{xy})\in\prob(m^2)$ and $\q=(q_{xy})\in\prob(m^2)$ are probability vectors with components  
\be\label{pqrtp}
p_{xy}\eqdef p_x|\la\psi_x|\phi_y\ra|^2\quad\text{and}\quad q_{xy}\eqdef q_y|\la\psi_x|\phi_y\ra|^2\;.
\ee
Moreover, note that the relation~\eqref{pqrtp} respects tensor products. That is, for $\rho^{\otimes n}$ and $\sigma^{\otimes n}$ the corresponding probability vectors are $\p^{\otimes n}$ and $\q^{\otimes n}$, respectively. Hence,
\ba
 \liminf_{n\to\infty}-\frac{1}{n}\log\pr_{\error}(\rho^{\otimes n},\sigma^{\otimes n},t)&\leq
\liminf_{n\to\infty}-\frac{1}{n}\log\frac{1}{2}\pr_{\error}(\p^{\otimes n},\q^{\otimes n},t)\\
\GG{Theorem~\ref{cherthm}}&=\max_{0\leq \alpha\leq 1}\big\{(1-\alpha)D_{\alpha}(\p\|\q)\big\}\\
\GG{\eqref{bdrsd}}&=\max_{0\leq \alpha\leq 1}\big\{(1-\alpha)D_{\alpha}(\rho\|\sigma)\big\}\\
&=\xi_Q(\rho,\sigma)\;.
\ea
This completes the proof.
\end{proof}

\section{Notes and References}

Many of the classical concepts in this chapter such as typicality of sequences, the method of types, and classical hypothesis testing can be found in standard textbooks on information theory and statistics, e.g.~\cite{Cover2006}. The topic of quantum typicality is covered by many books on quantum information including~\cite{NC2000,Wilde2013,Watrous2018}. The concept of relative typical subspace was first introduced in~\cite{BS2012}.

The expression of the Hypothesis testing divergence in terms of the function in~\eqref{fot} is due to~\cite{BG2017}, whereas other variants can be found in~\cite{DKF+2013}. The direct part of the quantum Stein's lemma was first proved by~\cite{HP1991}, while the strong converse part  was proved almost 10 years later by~\cite{ON2000}. A shorter version for both the direct and strong converse parts was found later in~\cite{BS2012}. However, our extremely shorter version presented in this chapter is based on the moderm approach that involves the inequalities given in~Theorem~\ref{thm:boundsdmin}.

We followed~\cite{NS2009} for the proof of the optimality of the quantum Chenoff bound, and~\cite{ACM+2007} for its achievability.

\part{The General Framework of Resource Theories}

\chapter{Static Quantum Resource Theories}

In this chapter, we present a precise definition of a quantum resource theory (QRT) and explore its general characteristics. As mentioned in the introduction, any set of natural constraints on a physical system results in a QRT. A prime example is the spatial separation between two individuals, Alice and Bob, which naturally leads to the LOCC (Local Operations and Classical Communication) constraint, forming the basis of entanglement theory. In this theory, every physical system is analyzed in the context of spatial separation. This implies that any physical system, for instance, system $A$, is considered a bipartite composite system, denoted as $A=(A_A,A_B)$. Here, $A_A$ represents a subsystem on Alice's side, and $A_B$ is a subsystem on Bob's side. It's important to note that even if $A$ is not inherently a composite system\index{composite system} and is solely located on Alice's side, it can still be regarded in this framework with $A_A\eqdef A$ and $A_B$ being a trivial subsystem (i.e., $|A_B|=1$). For simplicity, in entanglement theory, the notations $A$ for $A_A$ and $B$ for $A_B$ are often used. However, in the context of general resource theories, it is crucial to remember that physical systems, symbolized as $A, B, C$, etc., are interpreted in relation to the constraints applied to them.

\section{The Structure of Quantum Resource Theories}

According to the first axiom\index{axiom} of quantum mechanics, each physical system is uniquely associated with a corresponding Hilbert space. However, the reverse of this statement is not necessarily true. For instance, mathematically, the Hilbert space $\mathbb{C}^4$ is isomorphic to the Hilbert space $\mathbb{C}^2 \otimes \mathbb{C}^2$. Yet, physically, $\mathbb{C}^4$ may represent two entirely distinct physical systems. The space $\mathbb{C}^4$ could describe a single atom with four energy levels, or it might represent a composite system\index{composite system} of two spatially separated electrons (spins). Therefore, it is important to clarify that while we use the notations $A, B, C$, etc., to denote both physical systems and their corresponding Hilbert spaces, in the forthcoming discussion, these notations will primarily refer to specific physical systems.

In the rest of this book, we will use the symbol $1$ to represent the trivial system. In this context, the only element of $\cptp(A \to 1)$ is the trace operation. Additionally, this notation allows us to equate quantum channels in $\cptp(1 \to A)$ with density matrices in $\md(A)$, and conversely. By adopting this identification, we can interpret all entities in quantum mechanics --- such as states, POVMs, quantum instruments, and others --- as specific forms of quantum channels. This integrative perspective aligns with the methodologies utilized in resource theories. We will embrace this approach in our discussions throughout the book.

\begin{myd}{Quantum Resource Theory}
\begin{definition}\label{def:qrt}
Let $\mf$ be a mapping that takes any two physical systems $A$ and $B$ to a set of quantum channels $\mf(A\to B)\subset\cptp(A\to B)$. The mapping $\mf$ is called a \emph{quantum resource theory} if it satisfies the following conditions:
\begin{enumerate}
\item \emph{Doing nothing is free.} For any physical system $A$, the identity channel $\id^A\in\mf(A\to A)$.
\item \emph{Concatenation is free.} For any three physical systems, $A$, $B$, and $C$, if $\mE\in\mf(A\to B)$ and $\mN\in\mf(B\to C)$ then $\mN\circ\mE\in\mf(A\to C)$.
\item \emph{Discarding a system is free.} For any system $A$, the set $\mf(A\to 1)\neq\emptyset$; i.e. $\mf(A\to 1)=\cptp(A\to 1)=\{\tr\}$.
\end{enumerate}
Moreover, the set $\mf(A\to B)$ is called the set of \emph{free operations} from system $A$ to system $B$, and the set $\mf(A)\eqdef\mf(1\to A)$,  is identified as the set of \emph{free states}.
\end{definition}
\end{myd}

The physical interpretation of Definition \ref{def:qrt} is as follows.  Consider a (possibly composite) quantum system held by one agent or distributed to a group of parties.  A QRT models what the parties can physically accomplish given some restrictions or constraints that result from technical or experimental limitations, the rules of some game, or simply the laws of physics.  What operations the agents can still perform given these restrictions is mathematically described by $\mf(A\to B)$, which is typically much smaller than the set of all quantum channels in $\cptp(A\to B)$.  

The first condition in Definition~\ref{def:qrt} simply says that the identity map (i.e., doing nothing) is free, an obvious requirement for any meaningful QRT.  We point out, however, that in some QRTs ``doing nothing" can be considered resourceful, particularly, if the systems involved decohere with the environment and resources degrade in time, so that the preservation or storage of a resource is itself a resource. Nonetheless, in such resource theories, the identity channel $\id^A$ in the definition above corresponds to a channel with zero time delay (i.e. instantaneous)
so that it is indeed free. In general, the time delay of the channels in such resource theories needs to be incorporated into the formalism, and we will discuss it in more details in volume two of this book when we introduce dynamical resource theories. For all the static QRTs that we study in this book, these considerations will not affect the formalism.  

The second property above can be viewed as the defining property of a QRT. It essentially states that free operations cannot generate a resource. A resource in our model is a quantum state that is not free (i.e. $\rho\in\md(A)$ but $\rho\not\in\mf(A)$) or a quantum channel that is not free (i.e. $\mE\in\cptp(A\to B)$ but $\mE\not\in\mf(A\to B)$). In particular, the second property above implies the following rule, known as the ``golden" rule of QRTS.

\begin{mye}{The Golden Rule of QRTs}\index{golden rule}
For any two physical systems $A$ and $B$, if a free channel $\mE\in\mf(A\to B)$ acts on a free state $\rho\in\mf(A)$, the resulting state $\mE(\rho)$ is a free quantum state in $\mf(B)$. 
 \end{mye}

We included in the definition above the property that the trace is a free operation. In all QRTs studied in literature this is indeed the case although one can consider a QRT in which ``waste" or ``trash" is considered a resource. In this book we will not consider such resource theories, and will always consider the trace as a free operation. 
This assumption also leads to the following very useful property of QRTs.

Suppose $\sigma\in\mf(B)$ is a free state, and define the replacement channel
\be
\mN_\sigma^{A\to B}(\rho^A)\eqdef\tr[\rho^A]\sigma^B\quad\quad\forall\;\rho\in\ml(A)\;.
\ee
Then, if we view $\sigma^B$ as a channel, $\sigma^{1\to B}$, from the trivial system $1$ to $B$, the channel $\mN_\sigma^{A\to B}$ can be expressed as a combination of the trace channel and the channel $\sigma^{1\to B}$; specifically, 
\be
\mN_\sigma^{A\to B}=\sigma^{1\to B}\circ\tr\;,
\ee
and since both $\tr$ and $\sigma^{1\to B}$ are free, it follows that $\mN_\sigma^{A\to B}$ is free. Note that this means that  we can convert any state $\rho\in\md(A)$ to any free state $\sigma\in\mf(B)$ by free operations (as intuitively expected).

QRTs emerge from a specific set of limitations or constraints applied to the entire spectrum of quantum operations. The mapping $\mf$ exemplifies this, as the set $\mf(A\to B)$ generally forms a strict subset of all channels in $\cptp(A\to B)$. While every QRT is linked to a unique set of restrictions, these restrictions frequently share common characteristics that contribute to extra structural complexity. These characteristics are so prevalent that some researchers have integrated them into the foundational definition of a QRT.

\subsection{The Axiom of Free Instruments}\index{axiom of free instruments}\label{afi}

In some resource theories, like the QRT of athermality, quantum measurements are not considered free. However, in most QRTs, certain measurements are free. Mathematically, this implies the existence of systems $A, B, X$ --- with $X$ being a classical system --- such that the set $\mf(A\to BX)$ is non-empty. Consider a quantum instrument in $\mf(A\to BX)$, denoted as 
\be
\mE^{A\to BX}=\sum_{x\in[m]}\mE_x^{A\to B}\otimes |x\lr x|^X\;.
\ee
According to the fundamental principle of QRTs, if $\rho\in\mf(A)$ is a free state, then the state $\mE^{A\to BX}(\rho^{A})$ must also be a free state in $\mf(BX)$. This state, expressed as 
\be
\mE^{A\to BX}(\rho^{A})=\sum_{x\in[m]}p_x\sigma_x^B\otimes |x\lr x|^X
\ee
is a classical-quantum (cq) state, where for each $x\in[m]$, $p_x\eqdef\tr[\mE_x(\rho)]$ and $\sigma_x^B\eqdef\frac{1}{p_x}\mE_x^{A\to B}(\rho^A)$. If there existed an $x\in[m]$ for which $p_x\neq 0$ and $\sigma_x\not\in\mf(B)$, then the quantum instrument\index{quantum instrument} $\mE^{A\to BX}$ would create a resource $\sigma_x^B$ from the free state $\rho\in\mf(A)$ with a non-zero probability. To prevent such scenarios in QRTs, in this book we always assumes the axiom\index{axiom} of free instruments.

\begin{mye}{The Axiom\index{axiom} of Free Instruments}
Let $A$ and $B$ be two quantum systems and $X$ be a classical system. If $\mf(A\to BX)$ is non-empty and $\mE\eqdef\sum_{x\in[m]}\mE_x\otimes |x\lr x|\in\mf(A\to BX)$ then for every $\rho\in\mf(A)$ and $x\in[m]$, the state
$
\mE_x^{A\to B}(\rho^A)/\tr[\mE_x^{A\to B}(\rho^A)]
$
belongs to $\mf(B)$.
\end{mye}

Note that the axiom\index{axiom} of free instruments (AFI) reduces to the Chernoff\index{Chernoff} of QRTs when $|X|=1$, thus serving as an extension of this rule to encompass quantum measurements. Additionally, when $|X|>1$, the golden rule of QRTs only ensures that $\mE^{A\to BX}(\rho^{A})$ is a free cq-state. Without further assumptions like the AFI, we cannot infer that each $\frac{\mE_x^{A\to B}(\rho^A)}{\tr[\mE_x(\rho)]}$ is a free state. Since physical QRTs comply with the AFI (as do all QRTs studied in the literature), the rest of this book will proceed under the assumption that QRTs adhere to the AFI, without explicitly stating it each time.
We will use the notation $\mf_{\leq}(A\to B)\subset\cp_{\leq}(A\to B)$ for the set of trace non-increasing CP maps that are part of free quantum instruments. Specifically, $\mE\in\mf_{\leq}(A\to B)$ if there exists a classical system $X$ with dimension $m\in\mathbb{N}$ and maps $\mE_1,\ldots,\mE_m\in\cp_{\leq}(A\to B)$, with the properties that (1) $\mE_x=\mE$ for some $x\in[m]$, and (2) $\sum_{x\in[m]}\mE_x\otimes |x\lr x|\in\mf(A\to BX)$.

\subsection{QRTs with a Tensor Product Structure}

Since the free operations arise from certain physical constraints, it is natural to assume that they can act on a subsystem of a composite system. That is, a free operation $\mE\in\mf(A\to B)$ can act on the state $\rho^{AC}$ as $\mE^{A\to B}(\rho^{AC})\eqdef\mE^{A\to B}\otimes \id^C(\rho^{AC})$. This leads us to the following definition.

\begin{myd}{Tensor Product Structure}
\begin{definition}\label{def:tps}
A QRT, $\mf$, is said to \emph{admit a tensor-product structure} if it fulfills these additional criteria:
\ben
\item[4.] \emph{Completely free operations:} For any three systems $A$, $B$, and $C$, and a channel $\mE\in\mf(A\to B)$, it holds that $\mE^{A\to B}\otimes\id^C\in\mf(AC\to BC)$.
\item[5.] \emph{Freedom of relabeling:} For any integer $n$, a free channel $\mN\in\mf(A^n\to B^n)$, and permutation channels $\mP^{A^n}_{\pi}$ and $\mP^{B^n}_{\pi^{-1}}$ corresponding to a permutation $\pi$ on $n$ elements, the composition $\mP^{B^n}_{\pi}\circ\mN^{A^n\to B^n}\circ\mP^{A^n}_{\pi^{-1}}$ is in $\mf(A^n\to B^n)$.
\een
\end{definition}
\end{myd}

The condition 5 above is very intuitive as it just state that the relabeling of  $A^n=(A_1,\ldots,A_n)$ and $B^n=(B_1,\ldots,B_n)$, respectively, as $(A_{\pi(1)},\ldots,A_{\pi(n)})$ and $(B_{\pi(1)},\ldots,B_{\pi(n)})$, keeps the channel $\mN$ free. Note that we do \emph{not} require that $ \mP_{\pi}\circ\mN\circ\mP_{\pi^{-1}}=\mN$, but only that $\mP_{\pi}\circ\mN\circ\mP_{\pi^{-1}}$ is itself a free channel. See Fig.~\ref{relabeling} for an illustration. 

\begin{figure}[h]
\centering
    \includegraphics[width=0.6\textwidth]{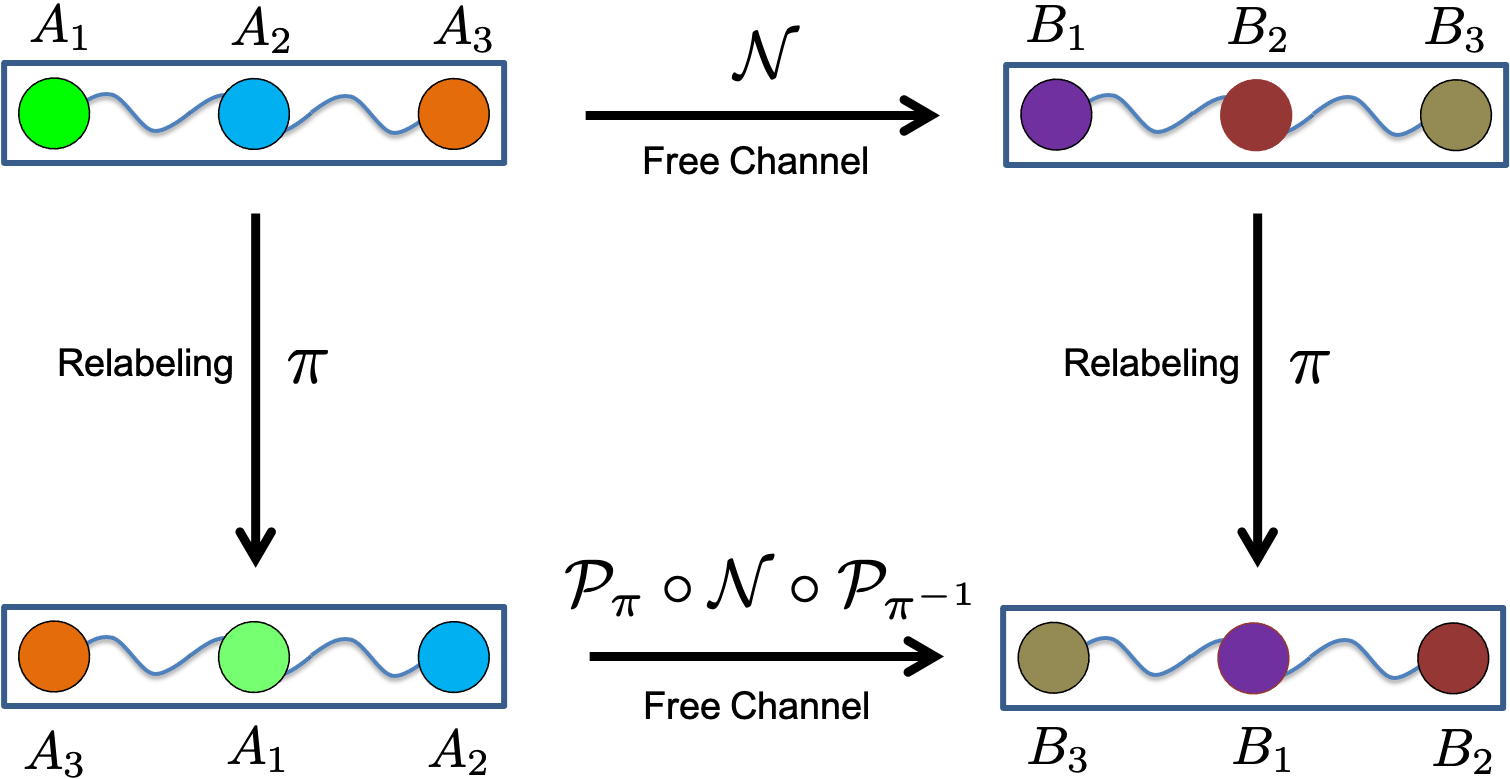}
  \caption{\linespread{1}\selectfont{\small Illustration of the fifth condition. Relabeling maintains the ``freeness" of $\mN$.}}
  \label{relabeling}
\end{figure} 

Conditions 1-5 above have several additional implications. First, note that since the trace is a free channel (property 3), the partial trace is also free (since $\id\otimes\tr$ is free). Second, note that if $\mN\in\mf(A\to B)$ and $\mM\in\mf(A'\to B')$ then
\be
\mN\otimes\mM=\left(\mN\otimes\id\right)\circ\left(\id\otimes\mM\right)\in\mf(AA'\to BB')\;.
\ee
In particular, this means that if two states are free then their tensor product is also free.
Finally, appending a free state is also a free channel. Specifically, let $\sigma\in\mf(B)$ and define
\be\label{appending}
\mN^{A\to AB}_\sigma(\rho^A)\eqdef\rho^A\otimes\sigma^B\quad\quad\forall\;\rho\in\ml(A)\;.
\ee
This channel can be viewed as a tensor product of two free channels, namely $\mN^{A\to AB}=\id^A\otimes\sigma^{1\to B}$, and therefore is free.

\subsection{Two Additional properties: Closedness and Convexity}\label{ccsubs}

Resource theories can vary a lot as the set of free operations is unique to each resource theory. For example LOCC in entanglement theory (Chapter~\ref{entanglement}), seems to be a very different set of operations than the set of thermal operations in thermodynamics (Chapter~\ref{ch:thermal}).
Yet, in addition to the tensor product structure, both of these sets of operations have additional structure that is common to both of them. 
Here we discuss two additional common properties that are satisfied by almost all QRTs studied in literature.

\begin{myd}{}
\begin{enumerate}
\item[6.] For any physical system $A$ the set of free states $\mf(A)$ is closed.
\item[7.] For any physical system $A$ the set of free states $\mf(A)$ is convex.
\end{enumerate}
\end{myd}

Property 6 states that if for a sequence of states $\{\rho_n\}_{n\in\mbb{N}}\subset\mf(A)$  the limit $\rho\eqdef\lim_{n\to\infty}\rho_n$ exists, then that limit is in $\mf(A)$ as well. Equivalently, if $\{\rho_n\}_{n\in\mbb{N}}\subset\mf(A)$ and there exists $\rho\in\md(A)$ such that $\lim_{n\to\infty}T(\rho,\rho_n)=0$, where $T$ is the trace distance (or any other distance measure) then $\rho\in\mf(A)$. Note that if this property does not hold then it would mean that there exists a sequence of free states that approaching a resource $\rho$. However, if $T(\rho,\rho_n)$ is extremely small, say $10^{-100}$, for all practical purposes it is not possible to distinguish between $\rho$ and $\rho_n$. Therefore, the assumption that $\mf(A)$ is closed is very practical and consequently satisfied by all the QRTs studied in literature so far.

Property 7 is not satisfied by all QRTs, e.g. non-Gaussianity in quantum optics, although many resource theories like entanglement do satisfy it and it is quite common. Besides of being a convenient mathematical property, we can develop some intuition for this property. Consider a QRT in which an agent, say Alice, has access to an unbiased coin.
She can flip the unbiased coin and prepares the state $\rho\in\mf(A)$ if she get a head and otherwise prepares the state $\sigma\in\mf(A)$. Since $\rho$ and $\sigma$ are free, she can prepare them at no cost. Suppose now that Alice forgets which state she prepared. We will assume here that this ``forgetting" is itself a free operation. Then, her state of the system is now $\frac{1}{2}\rho+\frac{1}{2}\sigma$. Therefore, we can assume that the convex combination $\frac{1}{2}\rho+\frac{1}{2}\sigma\eqdef\tau$ is also free since Alice prepared it at no cost. Moreover, if $\tau$ is free, Alice can repeat the same process with $\rho$ and $\tau$ to get that $\frac{3}{4}\rho+\frac{1}{4}\sigma$ is also free. Repeating this process, Alice can prepare any combination $\frac{k}{2^n}\rho+(1-\frac{k}{2^n})\sigma$, with $n\in\mbb{N}$ and $k\in[2^n]$. Therefore, such convex combinations must also be free. Finally, since the set $\{\frac{k}{2^n}\}$ is dense in $[0,1]$, Property 6 implies that for any $t\in[0,1]$ the convex combination $t\rho+(1-t)\sigma$ is free.

\begin{exercise} Consider the process described above.
\begin{enumerate}
\item Show that for any $n\in\mbb{N}$ and $k\in[2^n]$ Alice can prepare the state $\frac{k}{2^n}\rho+(1-\frac{k}{2^n})\sigma$. How many times Alice has to flip the coin.
\item Suppose that Alice has access to a \emph{biased} coin, with probability $0<p<1$ to get a head (and $1-p$ to get a tail). Show that Alice can use the coin to prepare any convex combination of free states.
\end{enumerate}
\end{exercise}

The two properties of closedness and convexity can also be applied to quantum channels. That is, we can require that for any two physical systems $A$ and $B$ the set $\mf(A\to B)$ is both closed and convex. However, we postpone the discussion of them to the second volume of this book where we study dynamical QRTs.

\section{State-Based Resource Theories}

Certain types of quantum phenomena can be identified directly on the level of states without involving constraints on quantum processes.  This is particularly true for phenomena such as coherence and certain forms of Bell nonlocality.  Within the framework of QRTs, the challenge becomes identifying sets of free operations that align with a predefined set of free states. It's interesting to note that these free states, represented as $\mf(A)=\mf(1\to A)$, can themselves be considered a unique kind of free operations, specifically as preparation channels. This approach affords substantial flexibility in choosing a consistent set of free operations for any given set of free states, even within QRTs that admits a tensor-product structure.

Consider, for example, the phenomenon of quantum coherence. Quantum coherence epitomizes a key aspect of quantum mechanics, illustrating the principle that particles, such as electrons or photons, can simultaneously exist in multiple states. This phenomenon stems from the principle of superposition, enabling particles to exist in a mixture of states, or in \emph{coherent} superposition, thus allowing them to interfere with one another in predictable manners. However, coherence is a fragile state, easily disturbed by external influences in a process known as decoherence, where quantum systems relinquish their superposition and adopt more classical behaviors. In recent developments, the capability to control and preserve quantum coherence has become crucial for the advancement of cutting-edge quantum technologies, including quantum computing and quantum cryptography, empowering the execution of tasks that surpass the capabilities of classical physics.

Considering the significance of this pivotal phenomenon, extensive efforts have been dedicated to characterizing it within the realm of quantum resource theories. How is this achieved? We start by identifying the set of free states in $\md(A)$. This is accomplished as follows: for any system $A$, a \emph{classical} basis of the system is identified, denoted as $\{|x\ra\}_{x\in[m]}\subset A$. Subsequently, the set of free states, or incoherent states, is defined as all diagonal density matrices in $\md(A)$ with respect to the classical basis. Thus, in the QRT of coherence, the set of free states is clearly defined and is specified for any system $A$ with dimension $m\eqdef|A|$ as:
\be\label{cfree}
\mf(A)=\Big\{\sum_{x\in[m]}p_x|x\lr x|\;:\;\p\in\prob(m)\Big\}\;.
\ee
Consequently, the primary challenge in the resource theory of quantum coherence lies in identifying a set of free operations that aligns consistently with the above set of free states.

\bex\label{ex:coiu}
Let $\mf(A)$ be the set of free states defined in~\eqref{cfree}, and let $\Delta\in\cptp(A\to A)$ be the completely dephasing map defined with respect to the classical basis. Show that for all $\rho\in\md(A)$ we have that $\rho\in\mf(A)$ if and only if $\Delta(\rho)=\rho$.
\eex

Physical factors often play a pivotal role in determining the choice of free operations within the realm of quantum mechanics. Nonetheless, even when these free operations are well-defined and grounded in physical principles, it is advantageous to investigate other classes of free operations that correspond to the same set of free states. This exploration can provide valuable insights and potentially reveal alternative mathematical or theoretical frameworks, as alternate classes might offer simpler or more elegant solutions that are not immediately apparent in operations primarily motivated by physical factors. A pertinent example is found in entanglement theory, where characterizing the class of LOCC is notably complex. To circumvent these complexities, considerable research has focused on entanglement theory within broader and more mathematically accessible sets of operations, such as separable operations and non-entangling operations (refer to Chapter~\ref{entanglement}). A commonality among these resource theories of entanglement is the identification of the set of separable states as the free states. Exploring more advanced operations can result in demonstrating no-go theorems for the less powerful  but physically motivated free operations. In essence, if a quantum information task is unachievable with a more capable class of operations, it will certainly be infeasible with a weaker set. This section delves into various consistent sets of free operations in general QRTs, emphasizing their physical justifications and unique properties. To illustrate these abstract concepts, the QRTs of coherence and entanglement will frequently be used as examples.

Often some physical consideration will motivate a certain choice of free operations. But even in this case, it is valuable to study different classes of free operations for the same set of free states.  This is because different classes may have an easier or more elegant mathematical structure than the physically-motivated class of operations.  This is the case, for example, in entanglement theory where LOCC is a notoriously difficult class of operations to characterize.  To avoid the technical difficulties that arise when using these operations, much work has been devoted to the study of entanglement theory under larger and more analytically-friendly sets of operations such as separable operations, non-entangling operations, and more (see Chapter~\ref{entanglement}).  In all these resource theories of entanglement, a shared characteristic is the designation of separable states as the set of free states. Investigating more advanced operations can facilitate the proof of no-go theorems for less powerful, albeit more intuitive, free operations. This is based on the principle that if a quantum information task is unachievable using a more capable class of operations, it will inevitably be impossible with a weaker set.
In this section, we study different consistent sets of free operations in general QRTs, highlighting their various physical motivations and properties.  As examples to illustrate abstract ideas, we will often use the QRTs of coherence and entanglement to demonstrate them.

\subsection{Resource Non-Generating Operations}\index{resource non-generating}\label{sec:rng}

Let $\mf(A)$ be a set of free density matrices. Any conceivable set of free operations $\mf(A\to B)$ must satisfies the properties given in Definition~\ref{def:qrt}. In particular, any channel $\mN\in\mf(A\to B)$ must satisfy the Chernoff\index{Chernoff} of QRTs. We use this property in the following definition.

\begin{myd}{RNG Operations}
\begin{definition}\label{def:rng}
Let $\mf(A)\subset\md(A)$ be the set of free states on any physical system $A$. The set of \emph{resource non-generating operations} (RNG) between two physical systems $A$ and $B$ is defined as:
\be
\rng(A\to B)\eqdef\Big\{\mN\in\cptp(A\to B)\;:\;\mN\left(\rho\right)\in\mf(B),\;\;\forall\;\rho\in\mf(A)\Big\}
\ee
\end{definition}
\end{myd}

RNG operations form the maximal set of free operations. That is, every other QRT $\mf$ with the same set of free state $\mf(A)$ must satisfy 
\be
\mf(A\to B)\subseteq \rng(A\to B)\;.
\ee
In the QRT of coherence this set of RNG operations is denoted by MIO$(A\to B)$ where the acronym MIO stands for \emph{maximally incoherent operations}. Denoting the $\Delta^A\in\cptp(A\to A)$ and $\Delta^B\in\cptp(B\to B)$ the completely dephasing channels with respect to the classical systems $A$ and $B$, respectively, we get from the definition above in conjunction with Exercise~\ref{excoiu} that
\be\label{mio}
{\rm MIO}(A\to B)=\Big\{\mN\in\cptp(A\to B)\;:\;\Delta^B\circ\mN^{A\to B}\circ\Delta^A=\mN^{A\to B}\circ\Delta^A\Big\}\;.
\ee
\bex
Prove~\eqref{mio}.
\eex

\bex\label{ex:coh}
Consider the QRT of coherence where $\mf(A)$ and $\mf(B)$ are sets of diagonal density matrices with respect to some fixed bases $\left\{|x\ra^A\right\}_{x\in[m]}$ and $\left\{|y\ra^B\right\}_{y\in[n]}$ of $A$ and $B$, respectively. Show that a quantum channel $\mN\in{\rm MIO}(A\to B)$ if and only if there exists conditional probability distribution $\{p_{y|x}\}$ such that for all $x\in[m]$ the state \be\mN^{A\to B}\left(|x\lr x|^A\right)=\sum_{y\in[n]}p_{y|x}|y\lr y|^B\;.\ee
\eex

In entanglement theory the set of RNG operations is called \emph{non-entangling operations}. They consists of all bipartite quantum channels that cannot generate entanglement from separable states. Non-entangling operations form a strict superset of LOCC. For example, the (global) swap operator, that swap all the subsystems of Alice with those of Bob is non-entangling. Such a swap operation is highly non-local and clearly cannot be simulated with LOCC.

In general, RNG operations do not qualify as completely free. This means there are instances where an operation $\mE\in\rng(A\to B)$, when combined with an identity operation on some system $C$, i.e., $\mE^{A\to B}\otimes\id^C$, does not remain a free operation. To illustrate this, consider the context of entanglement theory and a product state $\Phi^{A\tA}\otimes |0\lr 0|^B$ shared between Alice and Bob, where the sizes of $A$ and $B$ are equal. Previously, we noted that a global swap operation is non-entangling. Specifically, applying a global swap to the aforementioned state results in $|0\lr 0|^A\otimes\Phi^{B\tB}$, maintaining its product state nature. However, if a partial swap is applied between subsystems $A$ and $B$ (keeping $\tA$ intact), the resulting state becomes $|0\lr 0|^A\otimes\Phi^{\tA B}$, which is entangled. This demonstrates that while a swap operation does not generate entanglement when applied to the entire system, it can induce entanglement when acting on individual subsystems.

\subsection{Completely Resource Non-Generating Operations}\index{resource non-generating}

The realization that RNG (resource non-generating) operations are generally not completely free (as outlined in condition 4 of Definition~\ref{def:tps}) underscores their somewhat non-physical nature. This leads us to a new definition that not only adheres to the Chernoff\index{Chernoff} of Quantum Resource Theories (QRTs) but also integrates the tensor product structure.

\begin{myd}{Completely RNG Operations}
\begin{definition}\label{def:crng}
Given $k\in\mbb{N}$, a quantum channel $\mN\in\rng(A\to B)$ is defined as $k$-RNG if for any reference system $R$ of dimension $k$, the channel $\id^R\otimes\mN$ belongs to $\rng(RA\to RB)$. Furthermore, the channel $\mN$ is termed \emph{completely} RNG (CRNG) if it is $k$-$\rng$ for all $k\in\mbb{N}$. The set of all CRNG channels in $\cptp(A\to B)$ is denoted by ${\rm C}\rng(A\to B)$. 
\end{definition}
\end{myd}

The definition above generalizes the concepts of $k$-positivity and complete-positivity\index{positivity}  (see Definition~\ref{def:cp}) to QRTs. Specifically, if we take the free set $\mf(A)=\md(A)$ to be the set of \emph{all} density matrices acting on $A$, then maps that are $k$-RNG and completely-RNG are equivalent to maps that are $k$-positive\index{$k$-positive} and completely positive, respectively. Moreover, the set of $k$-RNG maps with $k=1$ is simply the set of RNG maps.

As an example, let $\mf(AB)\eqdef\sep(AB)\subset\md(AB)$ be the set of all separable states, and let $\mN\in{\rm CRNG}(AB\to A'B')$ be a (bipartite) quantum channel that takes separable states to separable states even when acting on subsystems. Specifically, for any composite reference system $R=R_AR_B$ the channel $\id^R\otimes\mN^{AB\to A'B'}$ is non-entangling (i.e. RNG). Recall the discussion at the beginning on this chapter that every system $R$ in entanglement theory is viewed as a bipartite system\index{bipartite system} $R_AR_B$ with $R_A$ on Alice's side and $R_B$ on Bob's side. Taking $R_A\cong A$ and $R_B\cong B$ we get that the state $\Phi^{R_AA}\otimes\Phi^{R_BB}$ is a product state between Alice composite system\index{composite system} $R_AA$  and Bob's composite system\index{composite system} $R_BB$ (see Fig.~\ref{double}a). Therefore, since product states are in particular separable, and since $\id^R\otimes\mN^{AB\to A'B'}$ is non-entangling we get that
\be\label{sepchoi}
\mN^{AB\to A'B'}\left(\Phi^{R_AA}\otimes\Phi^{R_BB}\right)\in\sep(R_AA'R_BB')
\ee
is a separable state between Alice's system $R_AA'$ and Bob's system $R_BB'$.

\begin{figure}[h]
\centering
    \includegraphics[width=0.8\textwidth]{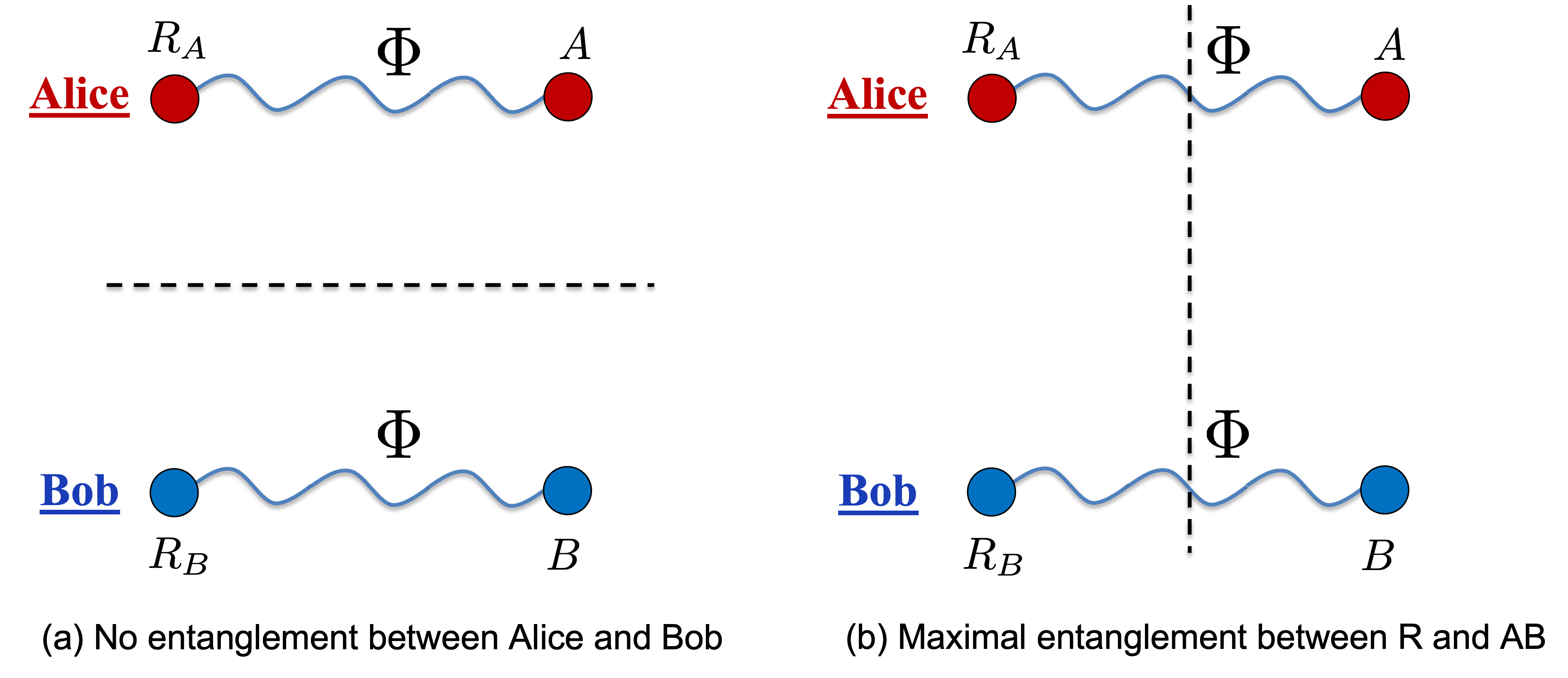}
  \caption{\linespread{1}\selectfont{\small The state $\Phi^{R_AA}\otimes\Phi^{R_BB}$ in the lens of two bipartite cuts.}}
  \label{double}
\end{figure}

A key observation in this example is that the state $\Phi^{R_AA}\otimes\Phi^{R_BB}$ can be viewed as a maximally entangled state between system $R=R_AR_B\cong AB$ and system $AB$ (see Fig.~\ref{double}b); i.e.
\be
\Phi^{(R_AR_B)(AB)}=\Phi^{R_AA}\otimes\Phi^{R_BB}\;.
\ee
Therefore, the state in~\eqref{sepchoi} is proportional to the Choi matrix of $\mN$. Since $R_A\cong A$ and $R_B\cong B$ we conclude that the Choi matrix
\be\label{sq1}
J_\mN^{ABA'B'}\eqdef \mN^{\tA\tB\to A'B'}\left(\Omega^{A\tA}\otimes\Omega^{B\tB}\right)
\ee
is an unnormalized separable state between Alice's system $AA'$ and Bob's system $BB'$. From Exercise~\ref{ex:sepstate} it follows that the Choi matrix can be expressed as
\be\label{sq2}
J_\mN^{ABA'B'}=\sum_{j\in[k]}\psi_j^{AA'}\otimes\phi_j^{BB'}
\ee
where the sets $\{\psi_j^{AA'}\}_{j\in[k]}$ and $\{\phi_j^{BB'}\}_{j\in[k]}$ are sets of (possibly unnormalized) pure states (i.e. rank one operators) in $\pos(AA')$ and $\pos(BB')$, respectively. For each $j\in[k]$, we can write
\be
|\psi_j^{AA'}\ra=\left(I^A\otimes M_j\right)|\Omega^{A\tA}\ra\quad\text{and}\quad|\phi_j^{BB'}\ra=\left(I^B\otimes N_j\right)|\Omega^{B\tB}\ra\;,
\ee
 for some complex matrices $M_j\in\ml(A,A')$ and $N_j\in\ml(B,B')$. Using this notation in~\eqref{sq2} and comparing it with~\eqref{sq1} we conclude that the channel $\mN$ has the following operator sum representation
 \be\label{sepo}
 \mN^{AB\to A'B'}\left(\rho^{AB}\right)=\sum_{j\in[k]}(M_j\otimes N_j)\rho^{AB}(M_j\otimes N_j)^{*}\quad\quad\forall\;\rho\in\ml(AB)\;.
 \ee
 A channel is classified as separable if it possesses at least one operator sum representation where each Kraus operator can be expressed as a tensor product in the form of $M_j\otimes N_j$. The collection of all such separable channels is denoted by $\sep(AB\to A'B')$. This leads to an important conclusion in entanglement theory: the set of completely non-entangling operations, i.e., CRNG, aligns exactly with the set of separable channels. In other words, we have ${\rm CRNG}(AB\to A'B')=\sep(AB\to A'B')$. 

\bex
Let $\rng(AB\to A'B')$ be the set of non-entangling operations (i.e. RNG with respect to the set $\mf(AB)=\sep(AB)$). 
\ben
\item Show that $\mN\in\rng(AB\to A'B')$ is $k$-RNG with $k\geq |A|$ then it is completely RNG. 
\item Show that 
\be
\rng=1\text{-}\rng\supseteq2\text{-}\rng\supseteq\cdots\supseteq |AB|\text{-}\rng={\rm CRNG}=\sep\;.
\ee
\een
\eex

We saw above that for $\mf(AB)=\sep(AB)$ we have ${\rm CRNG}\subset \rng$ where the inclusion is strict since the global swap operator is non-entangling but also not a separable channel. Moreover,
we will see in Chapter~\ref{entanglement} that also the inclusion $\rng\supseteq2\text{-}\rng$ in the exercise above can be strict. However, in other resources theories some of these inclusions can be equalities. For example, in the QRT of coherence, in which $\mf(A)\subset\md(A)$ consists of diagonal states with respect to a fixed basis $\{|x\ra^A\}_{x\in[m]}$, we have that $\rng={\rm CRNG}$. To see this, let $\mN\in{\rm MIO}(A\to B)$ (recall that in the QRT of coherence we denote all RNG operations from system $A$ to $B$ by ${\rm MIO}(A\to B)$). According to~\eqref{mio}, $\Delta^B\circ\mN\circ\Delta^A=\mN\circ\Delta^A$. 
We need to show that for any system $C$, we have $\mN\otimes\id^C\in{\rm MIO}(AC\to BC)$. Let $\Delta^C$ be the completely dephasing channel with respect to the classical basis of system $C$. In the exercise below you show that $\Delta^{AC}=\Delta^{A}\otimes\Delta^C$. Therefore,
\ba
\Delta^{CB}\circ\left(\id^C\otimes\mN\right)\circ\Delta^{CA}&=\left(\Delta^C\otimes\Delta^B\right)\circ\left(\id^C\otimes\mN\right)\circ\left(\Delta^{C}\otimes\Delta^A\right)\\
\Gg{\Delta^C\circ\Delta^C=\Delta^C}&=\left(\id^C\otimes\Delta^B\circ\mN\circ\Delta^A\right)\circ\left(\Delta^{C}\otimes\id^A\right)\\
\Gg{\mN\in{\rm MIO}(A\to B)}&=\left(\id^C\otimes\mN\circ\Delta^A\right)\circ\left(\Delta^{C}\otimes\id^A\right)\\
\Gg{\Delta^{CA}=\Delta^C\otimes\Delta^A}&=\left(\id^C\otimes\mN\right)\circ\Delta^{CA}\;.
\ea
Thus, $\mN\otimes\id^C\in{\rm MIO}(AC\to BC)$.

\bex
Let $\{|x\ra^A\}_{x\in[m]}$ and $\{|y\ra^B\}_{y\in[n]}$, be respectively, two orthonormal bases of $A$ and $B$. Further, let $\Delta^{AB}$ be the completely dephasing channel with respect to the basis $\{|xy\ra^{AB}\}_{x,y}$. Show that $\Delta^{AB}=\Delta^A\otimes\Delta^B$, where $\Delta^A$ and $\Delta^B$ are the completely dephasing channels with respect to the bases $\{|x\ra^A\}_{x\in[m]}$ and $\{|y\ra^B\}_{y\in[n]}$, respectively.
\eex

\subsection{Physically Implementable Operations}\index{physically implementable}

\label{Sect:PhysicallyImplementable}

The use of CPTP maps and generalized measurements in quantum information science is so common that their physical implementations are often taken for granted. Specifically, in Sec.~\ref{sec:stine} we saw that the Stinespring dilation theorem ensures that every quantum channel in $\cptp(A\to A)$ can be implemented with a unitary evolution  on the joint system $AE$ followed by tracing out the environment degrees of freedom (i.e. tracing out system $E$).  Similarly, generalized measurements can be implemented with a joint unitary\index{joint unitary} operation followed by a projective measurement as described in Fig.~\ref{fig9}. In the context of QRTs, such implementations of free channels (or free generalized measurements) may not be free since the joint unitary\index{joint unitary} (or projective measurement) identified in such implementation is itself not necessarily free.  

As an example, consider the QRT of coherence, with $\mf={\rm MIO}$ being the set of free operations. If MIO were physically implementable, one would expect that for any channel $\mN\in{\rm MIO}(A\to A)$ there exists a system $E$, a free unitary channel 
$\mU\in{\rm MIO}(AE\to AE)$, and a diagonal state $\gamma\in\mf(E)$ such that
\be\label{mnaa}
\mN^{A\to A}(\omega^A)=\tr_E\left[U^{AE}\left(\omega^A\otimes\gamma^E\right)U^{*AE}\right]\;.
\ee
Now, from Exercise~\ref{ex:coh} we know that a channel $\mV\in\cptp(A\to A)$ is MIO if and only if $\mV(|x\lr x|^A)$ is a diagonal state in $\md(A)$ for all $x\in[m]$ (here $m\eqdef|A|$). Therefore, if $\mV$ is a unitary channel, then we must have for all $x\in[m]$
\be
\mV^{A\to A}(|x\lr x|^A)\eqdef V|x\lr x|^AV^*=|\pi(x)\lr \pi(x)|^A
\ee
where $\pi$ is some permutation on $m$ elements. This relation implies that the unitary matrix $V$ satisfies $V|x\ra^A=e^{i\theta_x}|\pi(x)\ra^A$. In other words, up to phases, all the free unitary operations in the QRT of coherence are permutations.
Given that permutation matrices form an extremely small set of operations relative to the set of all unitary channels, it is not too hard to show (see the relevant references at the end of this chapter) that there exists channels in ${\rm MIO}(A\to A)$ that do not have the form~\eqref{mnaa} with free (i.e. incoherent)  $U^{AE}$ and free (i.e. diagonal) $\gamma^E$. In other words, it costs coherence (i.e. resources) to implement some free channels in ${\rm MIO}(A\to A)$. 

The above problem does not occur in the QRT of entanglement in which the set of free operations is LOCC.  This is always the case whenever a QRT is defined in terms of a \emph{physical} restriction (e.g. distant labs in entanglement theory) that is imposed on the set of free operations. On the other hand, any QRT such as quantum coherence, in which first the free states are identified, and only then consistent free operations are proposed, may face such an implementation problem. Aside from the QRT of coherence, all the free operations of the QRTs studied in this book will have a physically implementable set of free operations.

\begin{myd}{}
\begin{definition}
Let $\mf$ be a QRT, and $A$ and $B$ two physical systems. We say that $\mf(A\to B)$ is \emph{physically implementable} if any channel in $\mf(A\to B)$ can be generated by a sequence of unitary channels (possibly on composite systems), projective measurements, appending of free states, and processing of the classical outcomes, where each element in the sequence is itself a free action (see Fig.~\ref{PIO} for an illustration). 
\end{definition}
\end{myd}

\begin{remark}
In the definition above we added classical processing as a possible free physically implementable operation. This include for example
classical communication between subsystems, if these were allowed in the QRT (e.g. entanglement theory). Note also that if the free operations in a QRT are not physically implementable (according to the definition above), then the QRT would identify certain maps as being free with no way to physically implement these processes using free operations.  
\end{remark}

\begin{figure}[h]
\centering
    \includegraphics[width=0.7\textwidth]{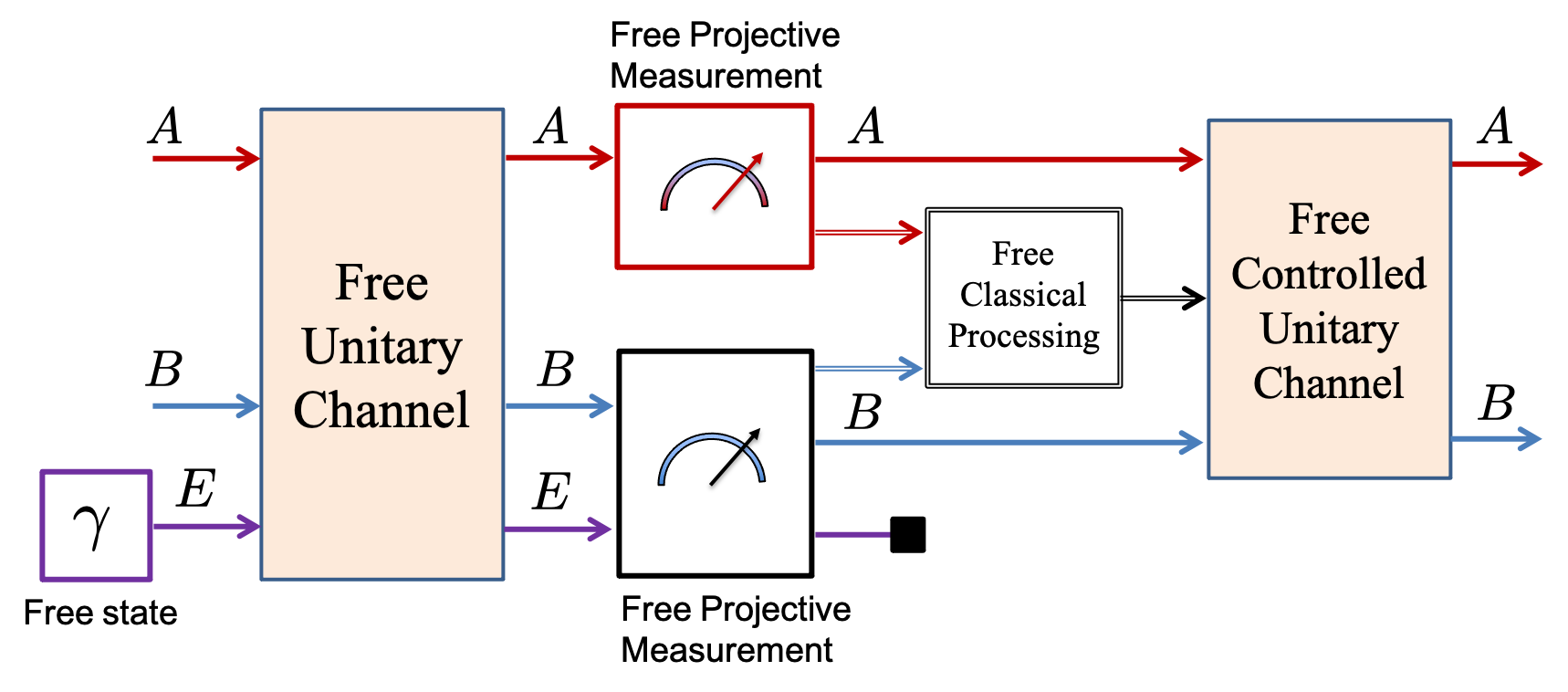}
  \caption{\linespread{1}\selectfont{\small Example of a physically implementable free operation on a composite system\index{composite system} $AB$.}}
  \label{PIO}
\end{figure}

For a given designation of free states $\mf(A)$, it is possible to construct a unique physically implementable QRT that admits a tensor-product structure.  Simply define the free operations to be any composition of (i) appending arbitrary free states, (ii) CRNG unitaries and projective measurements, (iii) discarding subsystems, and (iv) all free classical-processing maps.  
For a given two subsystems $A$ and $B$ we denote this set of \emph{physically implementable operations} (PIO) as $\pio(A\to B)$. By design, $\pio(A\to B)$ is physically implementable and has tensor-product structure.  Most QRTs that were studied in literature have the property that all the isometries in $\rng$ are completely free. In such QRTs, $\pio$ is the minimal set of free operations that is consistent with the set of free states $\mf(A)$.  The class $\pio(A\to B)$ fits into the hierarchy of operations as 
\be
\pio\subseteq {\rm CRNG}\subseteq\rng\;.
\ee

\subsection{Dually RNG Operations}\index{resource non-generating}

Let $\mE\in\mf(A\to B)\subseteq\cptp(A\to B)$ represent a free operation within a specific resource theory $\mf$. As $\mE$ is a completely positive\index{completely positive} map, its dual map\index{dual map} $\mE^*$ is also completely positive. However, it is crucial to note that the dual map $\mE^*$ may not necessarily preserve trace. Despite this, in many resource theories, although $\mE^*$ is not trace-preserving, it satisfies the following condition:
\be\label{drng}
\frac{\mE^*(\sigma)}{\tr\left[\mE^*(\sigma)\right]}\in\mf(A)\quad\quad\forall\;\sigma\in\mf(B)\;.
\ee
This means that $\mE^*$ is an RNG operation up to normalization.

\bex
Consider the resource theory of quantum entanglement.
Show that if $\mE\in\sep(AB\to A'B')$ then
\be
\frac{\mE^*\left(\sigma^{A'B'}\right)}{\tr\left[\mE^*\left(\sigma^{A'B'}\right)\right]}\in\sep(AB)\quad\quad\forall\;\sigma\in\sep(A'B')\;.
\ee
\eex

The need for normalization in~\eqref{drng} can be eliminated by broadening the definition of RNG operations to include cone-preserving operations. Specifically, for each system $A$, let us define $\mk(A)\subseteq\pos(A)$ as the cone represented by:
\be
\mk(A)\eqdef\big\{t\sigma\;:\;\sigma\in\mf(A)\;,\;\;t\in\mbb{R}_+\big\}\;.
\ee 
We then classify a map $\mE\in\cp(A\to B)$ as a $\mk$-preserving operation if, for every $\eta\in\mk(A)$, it holds that $\mE(\eta)\in\mk(B)$. By extending the scope of RNG operations to maps that are not necessarily trace-preserving, the condition in~\eqref{drng} essentially signifies that $\mE^*$ is a $\mk$-preserving operation.

\begin{myd}{}
\begin{definition}
Using the same notations as above, we say that a quantum channel $\mE\in\cptp(A\to B)$ is \emph{dually resource non-generating} if both $\mE$ and its dual map\index{dual map} $\mE^*$ are $\mk$-preserving operations.
\end{definition}
\end{myd}
\begin{remark}
Note that since $\mE$ is a trace-preserving map, the requirement for it to be a $\mk$-preserving operation is essentially equivalent to the condition of it being a RNG operation.
Accordingly, we will use $\drng(A\to B)$ to represent the collection of all dually RNG channels within $\cptp(A\to B)$.
\end{remark}

In a resource theory $\mf$, if every operation $\mE\in\mf(A\to B)$ satisfies~\eqref{drng}, then it follows that:
\be
\mf(A\to B)\subseteq \drng(A\to B)\subseteq\rng(A\to B)\;.
\ee
This implies that dually RNG channels form a subset of the RNG channels, which includes all the free channels. For instance, in the theory of entanglement, the dual of any separable map is clearly separable, making separable maps dually non-entangling (and thus, LOCC channels as well). Protocols involving dually RNG operations are more restricted compared to other operations, such as non-entangling maps. This can provide a more precise approximation to LOCC, potentially leading to improved bounds on their operational power.

Take, for example, the resource theory of coherence.  Let $\mE\in\cptp(A\to B)$ be an RNG operation; that is, $\mE\in {\rm MIO}(A\to B)$. The condition in~\eqref{drng} for this theory is equivalent to:
\be
\Delta^A\circ\mE^*\left(\sigma^B\right)=\mE^*\left(\sigma^B\right)\quad\quad\forall\;\sigma\in\mf(B)\;.
\ee
For any $\rho\in\md(A)$, the above equation leads to:
\be\label{lrlrwq}
\tr\left[\sigma^B\mE\left(\rho^A\right)\right]=\tr\left[\sigma^B\mE\circ\Delta^A\left(\rho^A\right)\right]\;.
\ee
Since $\sigma\in\mf(B)$ if and only if $\sigma=\Delta^B(\tau)$ for some $\tau\in\md(B)$, we can rewrite the left-hand side of the equation as:
\be
\tr\left[\sigma^B\mE\left(\rho^A\right)\right]=\tr\left[\tau^B\Delta^B\circ\mE\left(\rho^A\right)\right]\;,
\ee
utilizing the self-adjoint nature of $\Delta^B$. Similarly, applying the self-adjoint property of $\Delta^B$ to the right-hand side of Equation~\eqref{lrlrwq} results in:
\ba
\tr\left[\sigma^B\mE\circ\Delta^A\left(\rho^A\right)\right]&=\tr\left[\tau^B\Delta^B\circ\mE\circ\Delta^A\left(\rho^A\right)\right]\\
\Gg{\mE\in{\rm MIO}(A\to B)}&=\tr\left[\tau^B\mE\circ\Delta^A\left(\rho^A\right)\right]\;.
\ea
Combining these equations, it is concluded that for all $\rho\in\md(A)$ and $\tau\in\md(B)$, the following holds:
\be
\tr\left[\tau^B\mE\circ\Delta^A\left(\rho^A\right)\right]=\tr\left[\tau^B\Delta^B\circ\mE\left(\rho^A\right)\right]\;.
\ee
Hence, the condition becomes:
\be\label{condcommu2}
\mE^{A\to B}\circ\Delta^A=\Delta^B\circ\mE^{A\to B}\;.
\ee
This means that if $\mE\in\drng(A\to B)$, it must satisfy the above condition. The exercise below demonstrates that the converse is also true, leading to the conclusion that a quantum channel $\mE\in\cptp(A\to B)$ is in $\drng(A\to B)$ if and only if it satisfies the condition in Equation~\eqref{condcommu2}. Notably, for the case where $A=B$, this condition simplifies to the commutation relation $[\mE,\Delta]=0$. 

We note that in the QRT of coherence, quantum channels that satisfy Equation~\eqref{condcommu2} are identified as Dephasing-covariant Incoherent Operations, abbreviated as DIO. Therefore, we have demonstrated that within the resource theory of coherence, the set $\drng(A\to B)$, is equivalent to the set of DIO channels, denoted as ${\rm DIO}(A\to B)$.

\bex
Show that ${\rm DIO}(A\to B)\subseteq \drng(A\to B)$; i.e., let $\mE\in\cptp(A\to B)$ be a quantum channel satisfying~\eqref{condcommu2}, and show that $\mE\in \drng(A\to B)$.
\eex

\bex
Show that the inclusion ${\rm DIO}(A\to B)\subset{\rm MIO}(A\to B)$ is strict.
\eex

\section{Affine  Resource Theories}\label{sec:affine}\index{affine}

Some QRTs satisfy a stronger condition than convexity, and we call them \emph{affine} resource theories (ARTs). We will see later on that ARTs have many desired properties, and in particular, their conversion rates can be computed efficiently and algorithmically with an SDP.  
\begin{myd}{An Affine Set}
\begin{definition}
Let $A$ be a physical system, and $\mf(A)\subseteq\md(A)$ be a set of density matrices. The set $\mf(A)$ is called an \emph{affine} set, 
if every affine combination of $n$ free states
\be
\sigma\eqdef\sum_{x\in[n]}t_x\rho_x\quad\quad\rho_1,\ldots,\rho_n\in\mf(A)\;\;,\;\;t_1,\ldots,t_n\in\mbb{R}\;,
\ee
such that $\sigma\in\md(A)$, satisfies $\sigma\in\mf(A)$.
\end{definition}
\end{myd}

As an example of an affine set, let $\mf$ be the QRT of coherence in which $\mf(A)$ is the set of all diagonal states in $\md(A)$ with respect to a fixed basis of $A$.
The set of diagonal states $\mf(A)$ is affine since for any affine combination of diagonal states is diagonal; that is, if $\sum_{x\in[n]}t_x\rho_x\geq 0$, where each state $\rho_x\in\mf(A)$ (i.e. each $\rho_x$ is diagonal) and $\sum_{x\in[n]}t_x=1$, then also $\sum_{x\in[n]}t_x\rho_x$ is a diagonal state. Therefore, the set of free states in the QRT of coherence is affine.

\begin{exercise}
Let $\mf(A)$ be the set of all density matrices in $\md(A)$ with \emph{real} components with respect to a fixed basis
$\{|x\ra\}_{x\in[m]}$ of $A$. That is, $\rho\in\mf(A)\subset\md(A)$ if and only if 
the number $\la x|\rho|x'\ra$ is real for all $x,x'\in[m]$. Show that $\mf(A)$ is affine.
\end{exercise}
The following exercise provides another characterization of affine sets.
\bex
Let $\mf(A)\subseteq\md(A)$ be a set of density matrices, and let $\mathfrak{K}(A)\eqdef\spa_{\mbb{R}}\left\{\mf(A)\right\}$ be the subspace of $\herm(A)$ consisting of all linear combinations of the elements of $\mf(A)$. 
Show that $\mf(A)$ is affine if and only if
\be
\mf(A)=\mk(A)\cap\md(A)\;.
\ee
\eex

Not all convex sets are affine. For example, the set of separable states in entanglement theory is not affine. To see why, recall that product states of the form $\psi^A\otimes\phi^B$ are free states. Let $m\eqdef|A|$, $n\eqdef|B|$, $\{\psi_x^A\}_{x\in[m^2]}$ be a rank one basis of $\herm(A)$, and  
$\{\phi_y^B\}_{y\in[n^2]}$ be a rank one basis of $\herm(B)$. Then, the $m^2n^2$ states $\{\psi_x^A\otimes \phi_y^B\}_{x,y}$ form a basis of $\herm(AB)$. This means that \emph{any} density matrix $\rho\in\md(AB)$ can be expressed as an affine combination of product states; i.e.
\be\label{txy}
\rho^{AB}=\sum_{x\in[m^2]}\sum_{y\in[n^2]}t_{xy}\psi_x^A\otimes\phi_y^B\;,
\ee
where $\{t_{xy}\}$ is a set of real numbers. Hence, the set of separable states is not affine since even entangled states can be expressed as affine combination of product states. In this sense, the set of separable states is maximally non-affine. More generally, we say that a set $\mf(A)$ is maximally non-affine if 
\be
\herm(A)=\spa_{\mbb{R}}\{\mf(A)\}\;.
\ee

\begin{exercise}
Show that the set $\{t_{xy}\}_{x,y}$ in~\eqref{txy} must satisfy $\sum_{x,y}t_{xy}=1$.
\end{exercise}

\begin{exercise}
Let $|\Phi_{+}^{AB}\ra=\frac{1}{\sqrt{2}}(|00\ra+|11\ra)$ be the maximally entangled states.
\begin{enumerate}
\item Express (explicitly) $\Phi^{AB}$ as an affine combination of product states.
\item Show that  $\Phi^{AB}$ cannot be expressed as a \emph{convex} combination of product states.
\end{enumerate}
\end{exercise}

It is straightforward to extend the definition of an affine set in $\md(A)$ to sets in $\cptp(A\to B)$. 

\begin{myd}{Affine Resource Theories}
\begin{definition}
A QRT, $\mf$, is called an \emph{affine resource theory} (AFT) if for any two systems $A$ and $B$, and any affine combination of $n$ free channels 
\be
\mN^{A\to B}\eqdef\sum_{x\in[n]}t_x\mE^{A\to B}_x\;,\quad\quad\mE_1,\ldots,\mE_n\in\mf(A\to B)\;\;,\;\;t_1,\ldots,t_n\in\mbb{R}\;,
\ee
such that $\mN\in\cptp(A\to B)$, the channel $\mN$ is free (i.e. $\mN\in\mf(A\to B)$).
\end{definition}
\end{myd}

We will see in the next two chapters  that ARTs have several properties that make them much easier to study. Particularly, many problems in ARTs can be solved with semidefinite programming, unlike certain convex QRTs, such as entanglement theory, in which even the determination of whether a state is free or not is very hard (more precisely, belongs to a complexity class known as NP-hard).  In the following theorem we show that if the set of free operations is RNG or CRNG then the QRT is affine if and only if the set of free states is affine.

\begin{myt}{}
\begin{theorem}
Let $\mf(A)\subseteq\md(A)$ be a set of free states on any physical system $A$, and for any two physical systems $A$ and $B$, let $\rng(A\to B)$ and ${\rm CRNG}(A\to B)$ be as defined in Definitions~\ref{def:rng} and~\ref{def:crng} (with respect to the sets $\mf(A)$ and $\mf(B)$). Then, the following statements are equivalent:
\ben
\item For all systems $A$, the set $\mf(A)$ is affine.
\item For all systems $A$ and $B$, the set $\rng(A\to B)$ is affine.
\item For all systems $A$ and $B$, the set ${\rm C}\rng(A\to B)$ is affine.
\een
\end{theorem}
\end{myt}
\begin{proof}
\textbf{The implication $1\Rightarrow 2$:} Let $\mE_1,\ldots,\mE_n\in\rng(A\to B)$ and $t_1,\ldots,t_n\in\mbb{R}$ be such that
\be\label{xix}
\mN^{A\to B}\eqdef\sum_{x\in[n]}t_x\mE^{A\to B}_x\in\cptp(A\to B)\;.
\ee
We need to show that $\mN\in\rng(A\to B)$. Let $\sigma\in\mf(A)$ be a free state. Since $\mN$ is a quantum channel $\mN^{A\to B}(\sigma^A)\in\md(B)$. On the other hand, from the definition of $\mN$, we can write this relation as
\be
\mN^{A\to B}(\sigma^A)=\sum_{x\in[n]}t_x\omega_x^B\in\md(B)\quad\text{where}\quad\omega^B_x\eqdef\mE^{A\to B}_x\left(\sigma^A\right)\;.
\ee
Since $\sigma\in\mf(A)$ and $\mE_x\in\rng(A\to B)$ it follows that each $\omega_x\in\mf(B)$. Finally, using the assumption that $\mf(B)$ is affine we get that $\mN^{A\to B}(\sigma^A)\in\mf(B)$. Since $\sigma$ was arbitrary state in $\mf(A)$ we conclude that $\mN\in\rng(A\to B)$.

\textbf{The implication $2\Rightarrow 3$:} Let $\mE_1,\ldots,\mE_n\in{\rm C}\rng(A\to B)$ be a set of $n$ CRNG channels, and let $t_1,\ldots,t_n\in\mbb{R}$ be such that~\eqref{xix} holds. We need to show that $\mN\in{\rm C}\rng(A\to B)$ or equivalently, that for any reference system $R$, $\id^R\otimes\mN\in\rng(RA\to RB)$. Since each $\id^R\otimes\mE_x\in\rng(RA\to RB)$ and since we assume that $\rng(RA\to RB)$ is affine  it follows that 
\be
\id^R\otimes\mN=\sum_{x\in[n]}t_x\;\id^R\otimes\mE^{A\to B}_x\in\rng(RA\to RB)\;.
\ee
Hence, ${\rm C}\rng(A\to B)$ is affine.

\textbf{The implications $3\Rightarrow 1$:} By taking $A$ to be the trivial system (i.e. $|A|=1$) in  C$\rng(A\to B)$ we get that $\mf(B)$ is affine for all systems $B$. This completes the proof.
\end{proof}

\subsection{Resource Destroying Maps}\label{sec:RDM}\index{resource destroying map}

In specific QRTs, notably those concerning coherence and asymmetry, a distinct transformation exists. This transformation has the ability to map any density matrix into a free state, essentially ``destroying" the resource. At the same time, it functions as the identity channel when applied to free states. This dual capability highlights its unique role in these particular QRTs. We point out that such maps do not necessarily have to be channels, and they might not even be linear. However, in the context of this book, where our exploration is confined to theories within the realm of quantum mechanics, we will consistently assume that these resource-destroying maps are at least linear.

\begin{myd}{Resource Destroying Map}
\begin{definition}\label{def:RDM}
Let $\mf$ be a QRT. A \emph{resource destroying map} (RDM) is a linear map $\Delta\in\ml(A\to A)$ with the following two properties:
\ben
\item For all $\rho\in\md(A)$, $\Delta(\rho)\in\mf(A)$.
\item For all $\sigma\in\mf(A)$, $\Delta(\sigma)=\sigma$.
\een
\end{definition}
\end{myd}
\begin{remark}
It is important to note that the definition of a RDM is relative to the set of free states $\mf(A)$. Consequently, different QRTs with an identical set of free states $\mf(A)$ may share the same RDM. Conversely, there could exist multiple RDMs corresponding to the same set $\mf(A)$.
Furthermore, it is evident that $\Delta\in\pos(A\to A)$. This is because for any positive operator $\Lambda\in\pos(A)$, the transformed state $\Delta(\Lambda/\tr[\Lambda])$ is a free state and therefore positive semidefinite. However, it is crucial to understand that a RDM is not necessarily completely positive.
\end{remark}

As an example of a RDM, consider the QRT of coherence in which $\mf(A)\subset\md(A)$ is the set of diagonal states with respect to the basis $\{|x\ra\}_{x\in[m]}$ (here $m\eqdef|A|$). Relative to this basis, define the completely dephasing channel
\be\label{deltaco}
\Delta(\rho)\eqdef\sum_{x\in[n]}\la x|\rho|x\ra|x\lr x|\;.
\ee
It is simple to check that $\Delta$ as defined above is a RDM with respect to the set of diagonal matrices. In this example $\Delta$ is a quantum channel.

\begin{exercise}
Show that $\Delta$ as defined in~\eqref{deltaco} is a RDM. Moreover, show that it is self-adjoint; i.e. $\Delta^*=\Delta$.
\end{exercise}
\bex
Show that any RDM $\Delta\in\ml(A\to A)$ is idempotent; i.e. $\Delta\circ\Delta=\Delta$.
\eex

In the following lemma we show that not all QRTs have a RDM.
\begin{myg}{}
\begin{lemma}
Let $\mf$ be a QRT and let $A$ be a quantum system. If $\mf(A)$ is not affine then the QRT $\mf$ does not have a RDM.
\end{lemma} 
\end{myg}

\begin{proof}
Suppose by contradiction that $\mf(A)$ is not affine, and yet,  there exists a RDM $\Delta\in\ml(A\to A)$. Then, by definition, there exists $t_1,\ldots,t_n\in\mbb{R}$ and $\sigma_1,\ldots,\sigma_n\in\mf(A)$ such that
\be
\sigma\eqdef\sum_{x\in[n]}t_x\sigma_x\in\md(A)\quad\text{but}\quad\sigma\not\in\mf(A)\;.
\ee 
Moreover, since $\Delta$ is a RDM we must have $\Delta(\sigma)\in\mf(A)$. On the other hand,
\ba
\Delta(\sigma)&=\sum_{x\in[n]}t_x\Delta(\sigma_x)\\
&=\sum_{x\in[n]}t_x\sigma_x\\
&=\sigma\not\in\mf(A)\;,
\ea
in contradiction with the fact that $\Delta(\sigma)\in\mf(A)$. Therefore, if a QRT admits a RDM then it's set of states must be affine. 
\end{proof}

\subsubsection{Characterization of self-adjoint\index{self-adjoint} RDM}

All the RDMs that play a role in well studied QRTs, such as the QRTs of coherence and asymmetry, have the property that they are self-adjoint. 
self-adjoint RDMs have a relatively simple characterization, and they always exists if the set of free states is affine\index{affine}. 
In the definition below we consider an affine set $\mf(A)\subseteq\md(A)$, and denote by $\mathfrak{K}(A)\eqdef\spa_{\mbb{R}}\left\{\mf(A)\right\}$ the subspace of $\herm(A)$ consisting of all linear combinations of the elements of $\mf(A)$. Moreover, we denote by  
$\mk(A)^\perp$ the orthogonal complement of $\mk(A)$ in $\herm(A)$ that satisfies $\herm(A)=\mk(A)\oplus \mk(A)^{\perp}$.

\begin{myd}{self-adjoint RDM}
\begin{definition}\label{def:sardm}
Let $\mf(A)$, $\mk(A)$, and $\mk(A)^\perp$ be as defined above. The linear map $\Delta:\herm(A)\to\herm(A)$
\be\label{sadelta}
\Delta(\eta+\zeta)\eqdef\eta\quad\quad\forall\;\eta\in\mk(A)\;\;\text{and}\;\;\forall\;\zeta\in\mk(A)^{\perp}\;,
\ee
is called the \emph{self-adjoint RDM}.
\end{definition}
\end{myd}

\begin{remark}
Note that in the context above, the map $\Delta$ is referred to as \emph{the} self-adjoint RDM. This designation is justified by demonstrating that for every ART, there exists a unique self-adjoint RDM (see theorem below). This uniqueness within each ART underscores the specific role that such a map plays in the theory.
\end{remark}

\begin{exercise}\label{ex:rdm}
Verify that $\Delta$ in the definition above is indeed a RDM which is self-adjoint (with respect to the Hilbert-Schmidt inner product).
\end{exercise}

\begin{myt}{}
\begin{theorem}
Let $\mf(A)\subseteq\md(A)$ be an affine set. Then, there exists a unique self-adjoint RDM associated with $\mf(A)$ which is given by $\Delta$ as defined in~\eqref{sadelta}.
\end{theorem}
\end{myt}
\begin{proof}
The existence of $\Delta$ follows from Definition~\ref{def:sardm} and Exercise~\ref{ex:rdm}. To prove uniqueness, let $\Delta:\herm(A)\to\herm(A)$ be a self-adjoint RDM. We would like to show that it coincide with the RDM given in Definition~\ref{def:sardm}. Indeed, from the linearity of $\Delta$, and the fact that $\Delta$ is a RDM, we must have $\Delta(\eta)=\eta$ for all $\eta\in\mk(A)$. Moreover, since $\Delta$ is self-adjoint for every $\zeta\in\mk(A)^{\perp}$ and $\eta\in\mk(A)$ we have 
\ba
0=\tr[\eta\zeta]=\tr[\Delta(\eta)\zeta]=\tr[\eta\Delta(\zeta)]\;.
\ea
Since the above equation holds for all $\eta\in\mk(A)$ we conclude that  $\Delta(\zeta)\in\mk(A)^{\perp}$. However, since $\Delta$ is a RDM we must  have $\Delta(\zeta)\in\mk(A)$. Both conditions hold only if $\Delta(\zeta)=0$. To summarize, we got that for all $\eta\in\mk(A)$ and all $\zeta\in\mk(A)^{\perp}$ we have
\ba
\Delta(\eta+\zeta)&=\Delta(\eta)+\Delta(\zeta)\\
\Gg{\Delta(\zeta)=0}&=\Delta(\eta)\\
&=\eta\;.
\ea
Hence, $\Delta$ coincides with the map defined in~\eqref{sadelta}. This completes the proof.
\end{proof}

\bex
Let $\mf(A)\subseteq\md(A)$ be an affine set, and let $\{\eta_x\}_{x\in[m]}$ be an orthonormal basis of $\spa_{\mbb{R}}\{\mf(A)\}$ (w.r.t.\ to Hilbert-Schmidt inner product). Show that the linear map
\be
\Delta(\omega)\eqdef\sum_{x\in[m]}\tr\left[\eta_x\omega\right]\eta_x\quad\quad\forall\omega\in\ml(A)\;,
\ee
is a self-adjoint resource destroying map.
\eex

We have seen that if a $\mf(A)\subseteq\md(A)$ has a RDM then it must be affine. On the other hand, any affine set $\mf(A)$ has a unique self-adjoint RDM. Therefore, if $\mf(A)$ has a RDM then it also has a \emph{self-adjoint} RDM. Note however that the self-adjoint RDM, $\Delta$, as defined in Definition~\ref{def:sardm}, is not necessarily a quantum channel.
To be a quantum channel,  the Choi matrix of $\Delta$, denoted as $J^{A\tA}_\Delta$, must satisfy the two conditions of a quantum channel, namely, $J_\Delta^{A\tA}\geq 0$ and $J^{A}_\Delta=I^A$.
In the QRT of coherence, the completely decohering map as defined in~\eqref{deltaco} is indeed a self-adjoint RDM that is also a quantum channel. 

However, there exists affine sets for which the self-adjoint RDM is not a quantum channel. For example, let $\mf(A)$ be the set of all density matrices in $\md(A)$ whose components are real with respect to a fixed basis $\{|x\ra\}_{x\in[m]}$ (with $m\eqdef|A|$).
A QRT with this set of states have been used in literature to quantify the \emph{imaginarity} of quantum mechanics. Now, in the exercise below you show that the self-adjoint RDM of this resource theory is give by
\be\label{rdmd}
\Delta(\eta)=\frac12\left(\eta+\eta^T\right)\quad\quad\forall\;\eta\in\herm(A)\;,
\ee
where the transpose is taken with respect to the fixed basis $\{|x\ra\}_{x\in[m]}$. Therefore, since the traspose is not completely positive\index{completely positive} it follows that $\Delta$ above is not completely positive\index{completely positive} (see Exercise below).

\bex
Consider the affine set, $\mf(A)$, of all density matrices in $\md(A)$ whose components are real with respect to a fixed basis $\{|x\ra\}_{x\in[m]}$ (with $m\eqdef|A|$).
\ben
\item Show that the self-adjoint RDM associated with $\mf(A)$ is given by~\eqref{rdmd}.
\item Show that this self-adjoint RDM is not completely positive.
\een
\eex

\section{Resource Witnesses}\label{sec:rw}\index{resource witness}

In certain resource theories, the complexity of the set of free states is such that determining whether a given state is part of this set can be challenging. This is particularly evident in the theory of mixed bipartite entanglement. To address this challenge, a valuable tool from convex analysis can be employed to ascertain whether a quantum state qualifies as a free state. This approach provides a practical method for navigating the intricacies of these theories and identifying free states within complex sets.

\begin{myd}{Resource Witness}
\begin{definition}\label{def:wit}
Let $\mf$ be a QRT and let $A$ be a physical system. An operator $W\in\herm(A)$ is called a resource witness if the following two conditions holds:
\begin{enumerate}
\item For any $\sigma\in\mf(A)$
\be\label{defwit}
\tr\left[W\sigma\right]\geq 0\;.
\ee
\item There exists $\rho\in\md(A)$ such that
\be\label{nega}
\tr\left[W\rho\right]<0\;.
\ee
\end{enumerate}
\end{definition}
\end{myd}

Since a resource-witness is an Hermitian matrix it corresponds to an observable\index{observable} and the expressions $\tr[W\rho]$ and $\tr[W\sigma]$ corresponds to \emph{expectation values} of $W$ and in principle can be measured in a laboratory. Therefore, resource witnesses provides a practical method to test if a given quantum system is a resource.

The condition~\eqref{nega} in the definition above is equivalent to the statement that $W$ is not a positive semidefinite matrix.  Clearly, every positive semidefinite matrix satisfies~\eqref{defwit}. Therefore, resource witnesses can be viewed as the elements of $\mf(A)^*$ that are not positive semidefinite. Recall from Sec.~\ref{sec:dualcone} that $\mf(A)^*$ denotes the dual cone of $\mf(A)$ defined by
\be
\mf(A)^*\eqdef\big\{W\in\herm(A)\;:\;\tr\left[W\sigma\right]\geq 0\quad\forall\;\sigma\in\mf(A)\big\}\;.
\ee
Since $\pos(A)\subseteq\mf(A)^*$ we conclude that the set of all resource-witnesses can be viewed as the non-positive semidefinite matrices in $\mf(A)^*$. If $\mf(A)$ is closed and convex then the set of all witnesses completely determines the set of free states.

\begin{myt}{}
\begin{theorem}\label{thm:941}
Let $A$ be a physical system, $\mf(A)\subseteq\md(A)$ be a closed and convex subset of density matrices, and $\sigma\in\md(A)$. Then, $\sigma\in\mf(A)$ if and only if
\be
\tr\left[W\sigma\right]\geq 0
\ee
for \emph{all} resource witnesses $W\in\herm(A)$.
\end{theorem}
\end{myt}
\begin{proof}
This theorem follows from the property that any closed and convex set $\mk\subset\herm(A)$ satisfies $\mk^{**}=\mk$ (see Theorem~\ref{thm:closedconv}). Hence, in particular, $\mf(A)^{**}=\mf(A)$. The latter means that $\sigma\in\mf(A)$ if and only if $\sigma\in\mf(A)^{**}$; i.e. if and only if
\be
\tr[W\sigma]\geq 0\quad\forall\;W\in\mf(A)^*\;.
\ee 
Note that the inequality above holds trivially for all $W\geq 0$. Therefore, it is sufficient to check it for all $0\not\leq W\in\mf(A)^*$; i.e. for all resource witnesses\index{resource witness}.
\end{proof}

We point out that for affine QRTs, determining whether a quantum state is free or not is relatively an easy task. Since any affine set $\mf(A)$ has a self-adjoint resource destroying map (see Definition~\ref{def:sardm}), to determine if a state $\rho\in\md(A)$ is free or not, all we have to do is to check if $\Delta(\rho)=\rho$. Such a simplification does not occur in certain important convex QRTs (e.g. entanglement theory).

\section{Notes and References}

The term ``resource theory" has a bit of a history, starting with the
early recognition that quantum information theory, particularly quantum Shannon theory (which we cover in volume 2 of this book), is a theory of interconversions among different resources; see~\cite{Bennett2004} and~\cite{DHW2008}. 
Originally coined by Schumacher in 2003 (unpublished), the term ``resource theory" first appeared in a paper on the QRT of quantum reference frames by~\cite{GS2008}, although the framework for a QRT of information had already been investigated in a series of earlier papers by~\cite{OHHH2002,HHH+2003,HHH+2005}. The definition given here for a QRT is due to~\cite{BG2015} and~\cite{CG2019}. Other definitions involving symmetric monoidal categories can be found in papers by~\cite{CFS2016,Fritz2017}. However, these definitions involve terms from category theory and goes beyond the scope of this book.

Affine resource theories were introduced by~\cite{Gour2017} and resource destroying maps by~\cite{LHL2017}.

\chapter{Quantification of Quantum Resources}\label{ch:quantification}

One  of  the  most  useful  aspects  of  a  QRT  is  that  it generates precise and operationally meaningful ways to quantify a given physical resource.    Here we study a variety of resource measures that can be  introduced  in  any  QRT.  We start with the definition of a resource measure, and discuss some additional desirable properties that any resource measure should satisfy. After that, we study different families of specific resource measures applicable to any QRT. We put  emphasis on the Umegaki relative entropy of a resource, as it turns out that this resource measure has several operational interpretations, and it plays a major role throughout this book.

\section{Definitions and Properties of Resource Measures}\index{resource measure}

In their definition, the set of free states, $\mf(A)$, and the set of free operations, $\mf(A\to B)$, are defined on \emph{any} Hilbert spaces $A$ and $B$.
Consequently, a resource measure is defined as a function whose domain is the set of all density matrices in every finite dimension.  In the following definition we consider a function
\be\label{m}
\mathbf{M}:\bigcup_A\md(A)\to\mbb{R}\;,
\ee
where the union is over all Hilbert spaces $A$. This union also includes the trivial system $A=\mbb{C}$ (i.e. $|A|=1$) in which case $\md(A)=\{1\}$ consists of only one element, namely the number 1.

\begin{myd}{}
\begin{definition}\label{def:mor}
The function $\M$ in~\eqref{m} is called a \emph{resource measure} if it satisfies:
\begin{enumerate}
\item  Monotonicity\index{monotonicity}: $\M\big(\mE(\rho)\big)\leq \M(\rho)$ for all $\mE\in\mf(A\to B)$ and all $\rho\in\md(A)$.
\item Normalization: $\M(1)=0$.
\end{enumerate}
\end{definition}
\end{myd}

The first property is fundamental in resource theories. It asserts that the value of any resource measure cannot be increased through the use of free operations. This principle, known as monotonicity, is consistent with the ``Chernoff\index{Chernoff}" of QRTs that free operations cannot generate resources. The normalization condition, in conjunction with monotonicity, leads to the positivity of every resource measure $M$, which can be expressed as
\be
\M(\rho)\geq 0\quad\quad\forall\;\rho\in\md(A)\;.
\ee
The positivity follows from the fact that the trace is a free operation in any QRT, and therefore, from the monotonicity of $M$ under free operations we have
\be
\M(\rho)\geq \M(\tr[\rho])=\M(1)= 0\;.
\ee

Similarly, the two conditions of normalization and monotonicity implies that for any finite dimensional system $A$, 
\be\label{zeromono}
\sigma\in\mf(A)\quad\Rightarrow\quad \M\left(\sigma\right)=0\;.
\ee
Indeed, any state $\sigma\in\mf(A)$ can be viewed as a free channel $\sigma^{1\to A}$, where $1$ represent the trivial system corresponding to the Hilbert space $\mbb{C}$. Hence, 
\be
\M\left(\sigma^{A}\right)=\M\left(\sigma^{1\to A}(1)\right)\leq \M(1)=0\;,
\ee
where the inequality follows from the monotonicity property of $\M$. Therefore, since $\M$ is non-negative we must have $\M(\sigma)=0$ for all $\sigma\in\mf(A)$ and all finite dimensional systems $A$. We discuss now several properties that are satisfied by some, but not all, resource measures.

\subsubsection{Faithfulness}\index{faithfulness}

The condition expressed in~\eqref{zeromono} quantitatively defines the notion of ``no resource." Intuitively, one might be inclined to consider that the reverse of~\eqref{zeromono} should also hold true. This concept is referred to as faithfulness. A general resource measure $M$ is deemed \emph{faithful} if $M(\rho)=0$ necessarily implies that $\rho$ is a free state.

However, it's important to recognize that for certain tasks, some resource states may not offer any operational advantage over free states. In such scenarios, these states should be assigned a zero value by any measure that quantifies their utility for performing the specified task. For instance, as we will explore later, the measure of distillable entanglement, which is a significant measure of entanglement, is zero for all bound entangled states. Therefore, although faithfulness\index{faithfulness} is an intuitively attractive property, it is not an essential requirement for a resource measure. This perspective allows for a more nuanced understanding of resource measures and their application in various contexts within quantum resource theories.

\subsubsection{Resource Monotones}\index{resource monotone}

In certain QRTs, quantum measurements do not belong to the set of free operations. One such example is quantum thermodynamics as we will see in Chapter~\ref{ch:thermal}. However, in many other QRTs, like entanglement, quantum measurements can be free, and they represent an important component of the theory. In such QRTs, the set of quantum instruments $\mf(A\to BX)$, where $X$ is a classical `flag' system, is not empty. In particular, any such channel in $\mE\in\mf(A\to BX)$ can be express as $\mE^{A\to BX}=\sum_{x\in[m]}\mE_x^{A\to B}\otimes|x\lr x|^X$, and consequently, any resource measure $\M$ satisfies for all $\rho\in\md(A)$
\be\label{0106}
\M\left(\rho^A\right)\geq \M\left(\mE^{A\to BX}(\rho^A)\right)=\M\Big(\sum_{x\in[m]}p_x\sigma^B_x\otimes |x\lr x|^X\Big)\;,
\ee 
where 
\be\label{pxsx}
p_x\eqdef\tr[\mE_x^{A\to B}(\rho^A)]\quad\text{and}\quad\sigma_x^B\eqdef\frac{1}{p_x}\mE_x^{A\to B}(\rho^A)\;.
\ee 
We say that $\M$ is convex linear on qc-states if for all $\sigma\in\md(BX)$ as above
\be\label{0107}
\M\Big(\sum_{x\in[m]}p_x\sigma^B_x\otimes |x\lr x|^X\Big)=\sum_{x\in[m]}p_x\M(\sigma^B_x\otimes |x\lr x|^X)\;.
\ee
Almost all resource measures studied in literature are convex linear on QC states. One reason for that is that the equality above is satisfied by many functions, like the von-Neuman entropy, R\'enyi entropies, all the Schatten\index{Schatten}  $p$-norms, etc. If a QRT admits a tensor product structure then the partial trace is considered free so that for every $x\in[m]$
\be\label{0108}
\M(\sigma^B_x\otimes |x\lr x|^X)\geq \M(\sigma^B_x)\;.
\ee
Combining this with~\eqref{0106} and~\eqref{0107} we get that in such QRTs 
\be\label{1010sm}
\M\left(\rho^A\right)\geq\sum_{x\in[m]}p_x\M(\sigma^B_x)\;.
\ee
This property is sometimes referred to as \emph{strong monotonicity}. An intuitive justification for requiring strong monotonicity is to prevent $M$ from increasing on average when the experimenter can post-select or ``flag" the multiple outcomes of a quantum measurement.

We point out that in many QRTs, the classical flag states $\{|x\lr x|^X\}_{x\in[m]}$ are themselves considered as free states in $\mf(X)$. As physical QRTs admits a tensor product structure, appending or discarding flags is considered a free operations (recall that the partial trace is free and appending free states as in~\eqref{appending} is a free operation). 
For such QRTs we have in fact equality in~\eqref{0108}.

\begin{myd}{Resource Monotone}\index{resource monotone}
\begin{definition}
A resource measure $\M$ is called a resource monotone if it satisfies:
\begin{enumerate}
\item \emph{Strong monotonicity}. For any resource $\rho\in\md(A)$, and any free quantum instrument\index{quantum instrument} $\mE=\sum_{x\in[m]}\mE_x\otimes|x\lr x|\in\mf(A\to BX)$ 
\be\label{stmon}
\M\left(\rho^A\right)\geq \sum_{x\in[m]}p_x\M\left(\sigma^B_x\right)\;.
\ee
where $\{p_x\}_{x\in[m]}$ and $\{\sigma_x^B\}_{x\in[m]}$ are defined in~\eqref{pxsx}.
\item \emph{Convexity}. For any ensemble of states $\{p_x, \rho_x\}_{x\in[m]}$ in $\md(A)$
\be
\M\Big(\sum_{x\in[m]}p_x\rho_x\Big)\leq \sum_{x\in[m]}p_x\M\left(\rho_x\right)\;.
\ee
\end{enumerate}
\end{definition}
\end{myd}

As we delve further, we will see that the convexity property above is extremely useful from a mathematical perspective in calculating the resource monotone\index{resource monotone} for a specific state. Concurrently, a common physical interpretation of convex measures is that the process of mixing states does not result in an increase in the resource quantity. However, it's important to be cautious in drawing parallels between this mathematical notion of convexity and the physical process of mixing states, as the latter typically involves discarding information. We have previously discussed this important distinction in Section~\ref{ccsubs}. Additionally, it's worth noting that in QRTs where a freely available classical (flag) basis does not exist, and thus strong monotonicity is not a relevant concept, convex resource measures will be referred to as resource monotones\index{resource monotone}.

\subsubsection{Subadditivity}\index{subadditivity}

Some resource measures have additional properties that are mathematically convenient. One of such properties is \emph{subadditivity}.
A resource measure $M$ is said to be subadditive if for any $\rho\in\md(A)$ and $\sigma\in\md(B)$,
\be\label{subaddmes}
\M(\rho\otimes\sigma)\leq \M(\rho)+\M(\sigma)\;.
\ee  
While subadditivity is a natural property to expect from a resource measure, it will not hold for all measures in a general QRT. In particular, we will see examples of that when we discuss superactivation. 

\subsubsection{Additivity}\index{additivity}

An even stronger property of a resource measure is \emph{additivity}. That is, $\M$ is said to be additive when equality holds in~\eqref{subaddmes} for all states. While most resource measures do not satisfy this property, there exists a procedure known as \emph{regularization} that allows for the general construction of measures that are additive on multiple copies of the same state. We have already encountered this procedure implicitly in few places of previous chapters. The regularization\index{regularization} of a resource measure $\M$ is defined for all $\rho\in\md(A)$ as
\be\label{reg1}
\M^\reg(\rho)\eqdef\lim_{n\to\infty}\frac{1}{n}\M\left(\rho^{\otimes n}\right)\;,
\ee
provided the limit exists. In the following exercise you show that the limit above exists if $\M$ satisfies a weaker form of subadditivity\index{subadditivity}.
\begin{exercise}\label{ex:reg0}
Show that the limit in~\eqref{reg1} exists if $M$ satisfies for all $n,m\in\mbb{N}$ and any density matrix $\rho$
\be
\M\left(\rho^{\otimes (m+n)}\right)\leq \M\left(\rho^{\otimes m}\right)+\M\left(\rho^{\otimes n}\right)\;.
\ee
Hint: Use Exercise~\ref{ex842}.
\end{exercise}

\subsubsection{Asymptotic Continuity}\index{asymptotic continuity}

It's a reasonable expectation for any resource measure with physical significance to exhibit continuity. This expectation stems from the idea that if one quantum state is a slight perturbation of another, their resource contents should be very similar. However, it's important to note that a function $f:\bigcup_A\md(A)\to\mbb{R}_+$ satisfying the following condition:
\be\label{0720}
|A|\|\rho-\sigma\|_1\leq \big|f(\rho)-f(\sigma)\big|\leq |A|^2\|\rho-\sigma\|_1\quad\quad\forall\;\rho,\sigma\in\md(A)\;,
\ee
is indeed continuous. Yet, in the context of very large dimensions (i.e., $|A|\gg 1$), this type of continuity may not be practically useful. This is because, for the difference  $|f(\rho)-f(\sigma)|$ to be small, $\rho$ and $\sigma$ need to be so closely aligned that they are virtually identical for all practical purposes.
Therefore, a more robust notion of continuity, known as asymptotic continuity, is often considered. Asymptotic continuity is especially pertinent in the realm of large dimensions. It limits the dependence on dimension to a logarithmic scale, thereby providing a more practical and realistic measure of continuity when dealing with high-dimensional quantum states. This concept is particularly useful in assessing the continuity of resource measures in quantum systems where the dimensionality plays a significant role.

\begin{myd}{Asymptotic Continuity}
\begin{definition}\label{def:ac}
A resource measure $\M$ is said to be asymptotically continuous if for any $\rho,\sigma\in\md(A)$, and $\eps\eqdef\frac{1}{2}\|\rho-\sigma\|_1$,
\be\label{ascn}
\big|\M(\rho)-\M(\sigma)\big|\leq f(\eps)\log|A|\;,
\ee
where  $f:\mbb{R}\to\mbb{R}$ is some continuous function, independent on the dimensions, and satisfies $\lim_{\eps\to 0^+}f(\eps)=0$.
\end{definition}
\end{myd}

Note that the above notion of continuity is stronger than regular notion of continuity in the sense that the right-hand side of~\eqref{ascn} depends on the dimension through a log function.
This in particular implies that the regularization\index{regularization} of $\M$, if exists, is bounded. To see why, note that if $\M$ is an asymptotically continuous resource measure then for any $n\in\mbb{N}$
\ba
\left|\frac{1}{n}\M\left(\rho^{\otimes n}\right)-\frac{1}{n}\M\left(\sigma^{\otimes n}\right)\right|&\leq \frac{1}{n}\log(|A|^n)f\left(\frac12\|\rho^{\otimes n}-\sigma^{\otimes n}\|_1\right)\\
&=f\left(\frac12\|\rho^{\otimes n}-\sigma^{\otimes n}\|_1\right)\log|A|\\
\Gg{\frac12\|\rho^{\otimes n}-\sigma^{\otimes n}\|_1\leq 1}&\leq \max_{0\leq\eps\leq 1}f(\eps)\log|A|\;.&\nonumber
\ea
Moreover, since $f$ is continuous, $ \max_{0\leq\eps\leq 1}f(\eps)\eqdef c<\infty$. Hence, taking $\sigma\in\mf(A)$ to be free we get
from the above equation that for all $n\in\mbb{N}$
\be
\frac{1}{n}\M\left(\rho^{\otimes n}\right)\leq  c\log(|A|)\;.
\ee
Hence, taking the limit $n\to\infty$ we get $\M^{\reg}(\rho)<\infty$. 

\begin{exercise}
Let $f:\cup_A\md(A)\to\mbb{R}_{+}$ be a function that satisfy~\eqref{0720}, and suppose there exists a state $\sigma\in\md(A)$ such that $\lim_{n\to\infty}\frac{1}{n}f(\sigma^{\otimes n})<\infty$. Show that for all other $\rho\in\md(A)$ we must have 
\be
\lim_{n\to\infty}\frac{1}{n}f\left(\rho^{\otimes n}\right)=\infty.
\ee
Hint: Prove first that $\left\|\rho^{\otimes n}-\sigma^{\otimes n}\right\|_1\geq \|\rho-\sigma\|_1$ for all $n\in\mbb{N}$.
\end{exercise}

If the set of free states, $\mf(A)$, contains a full rank state for any system $A$, then one can define a slightly weaker version of asymptotic continuity\index{asymptotic continuity} that will be very useful for our study, since most QRTs have this property. 

\begin{myd}{Asymptotic Continuity (Alternative Definition)}
\begin{definition}\label{def:ac2}
A resource measure $\M$ is said to be asymptotically continuous if for all $\rho,\sigma\in\md(A)$, and $\eps\eqdef\frac{1}{2}\|\rho-\sigma\|_1$,
\be\label{ascn2}
\big|\M(\rho)-\M(\sigma)\big|\leq f(\eps)\log\min_{\eta\in\mf(A)}\left\|\eta^{-1}\right\|_\infty
\ee
where  $f:\mbb{R}\to\mbb{R}$ is some continuous function, independent on the dimensions, and satisfies $\lim_{\eps\to 0^+}f(\eps)=0$.
\end{definition}
\end{myd}

Observe that any density matrix $\eta\in\md(A)$ satisfies $\left\|\eta^{-1}\right\|_\infty\geq|A|$. Therefore, the above notion of asymptotic continuity\index{asymptotic continuity} is a weaker one than the version given in Definition~\ref{def:ac}. On the other hand, if the QRT $\mf$ has the property that there exists a constant $0<c<\infty$, independent of the dimensions, such that
\be\label{miet}
\min_{\eta\in\mf(A)}\|\eta^{-1}\|_\infty\leq c|A|
\ee
for any choice of system $A$ (and $c$ is independent on $|A|$) then the two notions of asymptotic continuity\index{asymptotic continuity} become equivalent. Since all the QRTs studied in this book satisfies the above condition, we will use these two notions of asymptotic continuity\index{asymptotic continuity} interchangeably.

\bex
Let $\mf$ be a QRT in which the maximally mixed state is free. Show that the two notions of asymptotic continuity\index{asymptotic continuity} coincide in this case.
\eex

Asymptotic continuity is a property that is extensively utilized in QRTs, especially in the asymptotic regime. Functions that are asymptotically continuous often incorporate the von Neumann entropy or the Umegaki\index{Umegaki} relative entropy. This reliance is partly because the Umegaki relative entropy is the only asymptotically continuous relative entropy, making it a unique and pivotal tool in QRTs.
The proof of this uniqueness theorem, which establishes the singular nature of the Umegaki relative entropy in terms of asymptotic continuity, is an important aspect of these theories. However, we will delve into the details of this proof later in Section~\ref{sec:unique}. For now, our focus will shift to introducing key examples of resource measures. These examples will provide a practical illustration of how the theoretical concepts discussed above applied in QRTs.

\section{Distance-Based Resource Measures}\label{sec:dbm}

In this section we introduce a general distance-based recipe for constructing resource measures in QRTs. The idea is to quantify the amount of a resource in a quantum state by ``how far" it is from the set of free states. In Chapter~\ref{chadiv} we saw several examples of well-defined measures that satisfy the mathematical requirements of distance between two density matrices. We also saw that from an operational perspective, the property of monotonicity under quantum channels (i.e. data processing inequality) offers a more useful foundation for quantifying distance than standard metric space approaches. We will therefore employ quantum divergences to define resource measures.

\begin{myd}{Divergence-Based Resource Measure}
\begin{definition}
Let $\mf$ be a QRT, and let $\D$ be a quantum divergence.  The $\D$-\emph{divergence of a resource} is the function
\be\label{rmasdf}
\D(\rho\|\mf)\eqdef \inf_{\sigma\in\mf(A)}\D\left(\rho\|\sigma\right)
\ee
Moreover, if $\D(\cdot\|\cdot)$ is a quantum relative entropy then $\D(\cdot\|\mf)$ is called a \emph{relative entropy of a resource}.
\end{definition}
\end{myd}

\begin{remark}
In this book, we consistently regard the set of free states, $\mf(A)$, as a closed and compact set. Consequently, the infimum in~\eqref{rmasdf} can be substituted with a minimum. This means that there exists an optimal state $\sigma^\star\in\mf(A)$ which fulfills the following equation:
\be
\D(\rho\|\mf)= \D\left(\rho\big\|\sigma^\star\right)\;.
\ee 
The state $\sigma^\star$ is called a \emph{closest free state} (CFS); see Fig.~\ref{cfs} for an illustration.
\end{remark}

\begin{figure}[h]
\centering
    \includegraphics[width=0.4\textwidth]{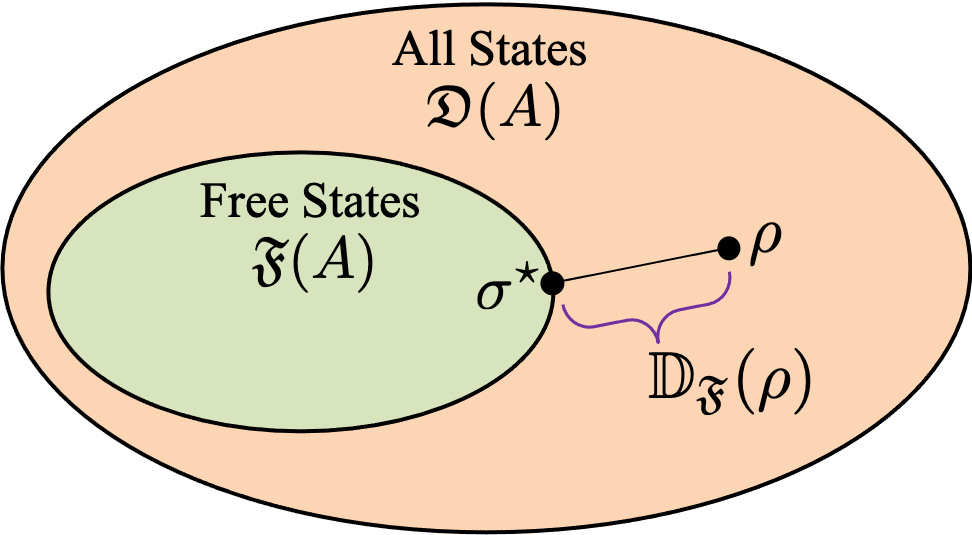}
  \caption{\linespread{1}\selectfont{\small The closest free state (CFS).}}
  \label{cfs}
\end{figure}

Remarkably, the two conditions that a quantum divergence has to satisfy, namely, DPI and normalization, are sufficient to guarantee that $\D(\cdot\|\mf)$ is a resource measure. Indeed, $\D(\cdot\|\mf)$ is non-negative since $\D$ is non-negative, and if $\rho\in\mf(A)$ then $\D(\rho\|\mf)=0$. To see the monotonicity of $\D(\cdot\|\mf)$ under free operations observe that for any $\mE\in\mf(A\to B)$ and any $\rho\in\md(A)$ we have
\ba
\D\left(\mE^{A\to B}(\rho^A)\big\|\mf\right)&\eqdef\inf_{\omega\in\mf(B)}\D\left(\mE^{A\to B}(\rho^A)\big\|\omega^B\right)\\
\Gg{\text{restricting }\omega=\mE(\sigma)}&\leq \inf_{\sigma\in\mf(A)}\D\left(\mE^{A\to B}(\rho^A)\big\|\mE^{A\to B}(\sigma^A)\right)\\
\GG{DPI}&\leq  \inf_{\sigma\in\mf(A)}\D\left(\rho^A\big\|\sigma^A\right)\\
&=\D\left(\rho^A\big\|\mf\right)\;.
\ea
Hence, $\D(\cdot\|\mf)$ is indeed a resource measure.

Several of the properties of the divergence $\D$ carry over to $\D(\cdot\|\mf)$. For example, suppose that $\D$ is faithful, which is the case for almost all quantum relative entropies. Then, also $\D(\cdot\|\mf)$ is faithful; i.e. for any $\rho\in\md(A)$,
\be
\D(\rho\|\mf)=0\iff\rho\in\mf(A)\;.
\ee
Indeed, if $\D(\rho\|\mf)=0$ it means that there exists a $\sigma\in\mf(A)$ such that $\D(\rho\|\sigma)=0$, but this is possible only if $\rho=\sigma$ which means that $\rho\in\mf(A)$. 

Also subadditivity\index{subadditivity} carry over from $\D$ to $\D(\cdot\|\mf)$. To see this, suppose that $\D$ is subadditive so that for any $\rho_1,\sigma_1\in\md(A)$ and $\sigma_1,\sigma_2\in\md(B)$
\be
\D\left(\rho_1\otimes\rho_2\big\|\sigma_1\otimes\sigma_2\right)\leq\D(\rho_1\|\sigma_1)+\D(\rho_2\|\sigma_2)\;.
\ee
Then,
\ba
\D\left(\rho_1^A\otimes\rho_2^B\big\|\mf\right)&=\inf_{\sigma\in\mf(AB)}\D\left(\rho_1^A\otimes\rho_2^A\big\|\sigma^{AB}\right)\\
\Gg{\text{restricting }\sigma=\sigma_1\otimes\sigma_2}&\leq\inf_{\substack{\sigma_1\in\mf(A)\\ \sigma_2\in\mf(B)}}\D\left(\rho_1^A\otimes\rho_2^B\big\|\sigma_1^A\otimes\sigma_2^B\right)\\
\Gg{\text{subadditivity of }\D}&\leq \inf_{\sigma_1\in\mf(A)}\D\left(\rho_1^A\big\|\sigma_1^A\right)+\inf_{\sigma_2\in\mf(B)}\D\left(\rho_2^B\big\|\sigma_2^B\right)\\
&=\D\left(\rho_1^A\big\|\mf\right)+\D\left(\rho_2^B\big\|\mf\right)\;.
\ea

If $\D$ is a relative entropy (in particular, it is additive) then $\D(\cdot\|\mf)$ is not necessarily additive since the restriction $\sigma^{AB}=\sigma_1^A\otimes\sigma^B_2$ in the equation above can lead to a strict inequality  $\D(\rho_1\otimes\rho_2\big\|\mf)<\D(\rho_1\|\mf)+\D(\rho_2\|\mf)$.
Still, subadditive resource measures can be regularized (see Exercise~\ref{ex:reg0}) so that the limit in the function
\be
\D^\reg(\rho\|\mf)\eqdef\lim_{n\to\infty}\frac1n\D\left(\rho^{\otimes n}\big\|\mf\right)
\ee
exists. Note that $\D^\reg(\cdot\|\mf)$ is at least weakly additive in the sense that $\D^\reg\left(\rho^{\otimes m}\big\|\mf\right)=m\D^\reg(\rho\|\mf)$ for any integer $m$.

If the set $\mf(A)$ is convex, and the quantum divergence $\D$ is jointly convex, then the resulting $\D$-divergence of a resource is convex as well. To see this, let $\{p_x,\rho_x\}_{x\in[m]}$ be an ensemble of quantum states in $\md(A)$, and for each $x\in[m]$ let $\sigma_x$ be the corresponding CFS of $\rho_x$. We then have
\ba\label{1027}
\D\Big(\sum_{x\in[m]}p_x\rho_x\Big\|\mf\Big)&=\inf_{\sigma\in\mf(A)}\D\Big(\sum_{x\in[m]}p_x\rho_x\Big\|\sigma\Big)\\
\Gg{\text{Taking }\sigma=\sum_{x\in[m]}p_x\sigma_x}&\leq \D\Big(\sum_{x\in[m]}p_x\rho_x\Big\|\sum_{x\in[m]}p_x\sigma_x\Big)\\
\GG{Joint\;convexity}&\leq \sum_{x\in[m]}p_x\D(\rho_x\|\sigma_x)\\
\Gg{\text{Each }\sigma_x\text{ is CFS}}&=\sum_{x\in[m]}p_x\D(\rho_x\|\mf)\;.
\ea

\subsection{The Umegaki\index{Umegaki} Relative Entropy of a Resource}

The Umegaki relative entropy of a resource (sometime we will call it in short \emph{the} relative entropy of a resource) is a resource measure that plays a major role in QRTs. It demonstrates 
that many seemingly unrelated properties of physical systems can be characterized with the same mechanism of resource theories. 
For instance, the entropy of entanglement that we will encounter in Chapter~\ref{entanglement}, the free energy in quantum thermodynamics (Chapter~\ref{ch:thermal}),
and the various communication capacities of a quantum channel (to appear in volume 2), all can be characterized as a relative entropy of a resource.

The relative entropy of a resource is a $D$-divergence of a resource, where $D$ is the Umegaki relative entropy. It is defined as
\be\label{rer}
D\left(\rho\|\mf\right)\eqdef\inf_{\sigma\in\mf(A)}D\left(\rho\|\sigma\right)\quad\quad\forall\rho\in\md(A)
\ee
where the Umegaki relative entropy $D(\rho\|\sigma)\eqdef\tr[\rho\log\rho]-\tr[\rho\log\sigma]$, and the infimum is taken over all free states $\sigma\in\mf(A)$.
Under the construction of $\D$-divergence of a resource, strong monotonicity~\eqref{stmon} is not guaranteed to be satisfied, but for the relative entropy of a resource this is indeed the case.

\begin{myt}{}
\begin{theorem}
Let $D$ be the Umegaki relative entropy and $\mf$ be a convex QRT.  Then, the relative entropy of a resource, $D(\cdot\|\mf)$, is a resource monotone.\index{resource monotone}
\end{theorem}
\end{myt}

\begin{proof}
Since $\mf$ is convex, and since the Umegaki relative entropy is jointly convex (see Exercise~\ref{ex:umegaki}), it follows from~\eqref{1027} that $D(\cdot\|\mf)$ is convex. It is therefore left to show that $D(\cdot\|\mf)$ satisfies the strong monotonicity property.
Consider
a free quantum instrument $\mE=\sum_{x\in[m]}\mE_x\otimes|x\lr x|^X\in\mf(A\to BX)$, with each $\mE_x\in\cp(A\to B)$, and $\sum_{x\in[m]}\mE_x$ is trace-preserving. For any resource state $\rho\in\md(A)$, denote $\sigma^{BX}\eqdef\sum_{x\in[m]}p_x\sigma_x^B\otimes|x\lr x|^X$, where $p_x\eqdef\tr[\mE_x^{A\to B}(\rho^A)]$ and $\sigma_x^B\eqdef\frac{1}{p_x}\mE_x^{A\to B}(\rho^A)$. Then,
\be
D\left(\rho^A\big\|\mf\right)\geq D\left(\mE^{A\to BX}\left(\rho^A\right)\big\|\mf\right)=D\left(\sigma^{BX}\big\|\mf\right)=\min_{\omega\in\mf(BX)}D\left(\sigma^{BX}\big\|\omega^{BX}\right)\;,
\ee
where we used the monotonicity under free operations of the relative entropy of a resource. Denoting $\omega^{BX}\eqdef\sum_{x\in[m]}q_x\omega_x^B\otimes|x\lr x|^X$ we continue
\ba
D\left(\rho^A\big\|\mf\right)&\geq\min_{\omega\in\mf(BX)}D\Big(\sum_{x\in[m]}p_x\sigma_x^B\otimes|x\lr x|^X\Big\|\sum_{x\in[m]}q_x\omega_x^B\otimes|x\lr x|^X\Big)\\
\GG{Exercise~\ref{ex:umegaki}}&=\min_{\omega\in\mf(BX)}\sum_{x\in[m]}p_xD\left(\sigma_x^B\big\|\omega_x^B\right)+D(\p\|\q)\\
\Gg{D(\p\|\q)\geq 0}&\geq \min_{\{\omega_x\}\subset\mf(B)}\sum_{x\in[m]}p_xD\left(\sigma_x^B\big\|\omega_x^B\right)\\
&= \sum_{x\in[m]}p_x\min_{\omega\in\mf(B)}D\left(\sigma_x^B\big\|\omega^B\right)=\sum_{x\in[m]}p_xD\left(\sigma_x^B\big\|\mf\right)\;.\nonumber
\ea
This completes the proof.
\end{proof}

So far we learned that the relative entropy of a resource is a faithful, subadditive resource monotone assuming $\mf$ is a QRT whose set of free states $\mf(A)$ is convex.  
The final property that we prove is that the Umegaki relative entropy of a resource is also asymptotically continuous. Later on, we will see that this measure is the \emph{only} asymptotically continuous relative entropy of a resource, making the Umegaki\index{Umegaki} relative entropy of a resource very unique. Moreover,  from the following chapters it will follow that asymptotic continuity\index{asymptotic continuity} has several important applications in QRTs.

The relative entropy of a resource is not always bounded. As a very simple example, suppose the set of free states $\mf(A)$ consists  
of only one pure state $|0\lr 0|$. In this case, we get that 
\be
\D(|1\lr 1|\|\mf)=\infty\;.
\ee
Hence, for such a (pathological) example the relative entropy of a resource is not bounded, and in particular, cannot be asymptotically continuous. However, in most QRTs, the set of free states $\mf(A)$ contains a full rank state. If such a state exists, say $\eta$, then since $\eta^{-1}$ exists and satisfies $\eta^{-1}\leq\|\eta^{-1}\|_{\infty}I^A$ it follows that
\ba
D(\rho\|\mf)&\leq D(\rho\|\eta)=\tr[\rho\log\rho]-\tr[\rho\log\eta]\\
\GG{\tr[\rho\log\rho]<0}&\leq-\tr[\rho\log\eta]=\tr\left[\rho\log(\eta^{-1})\right]\\
\GG{log\;is\;an\;operator\;monotone}&\leq \tr\left[\rho\log\left(\|\eta^{-1}\|_{\infty}I^A\right)\right]\\
&=\log\|\eta^{-1}\|_{\infty}\;.\nonumber
\ea
For example, if the set of free states contains the maximally mixed state then $D\left(\rho^A\big\|\mf\right)\leq\log|A|$.

\begin{myt}{\color{yellow} Asymptotic Continuity of the Relative Entropy of a Resource}
\begin{theorem}\label{thm:ac}
Let $\mf(A)$ be a closed and convex set, $\rho,\sigma\in\md(A)$, and set $\eps\eqdef\frac{1}{2}\|\rho-\sigma\|_1$. If
\be\label{1kappa1}
\kappa\eqdef\max_{\omega\in\md(A)}D(\omega\|\mf)<\infty
\ee
then
\be\label{asyfor}
\big|D(\rho\|\mf)-D(\sigma\|\mf)\big|\leq\eps\kappa+(1+\eps)h\left(\frac{\eps}{1+\eps}\right)
\ee
where $h(x)=-x\log x-(1-x)\log(1-x)$ is the binary Shannon entropy.
\end{theorem}
\end{myt}

\begin{proof}
Let $\rho,\sigma\in\md(A)$ and write $\rho-\sigma=\eps(\omega_+-\omega_-)$ as in~\eqref{decomom}
with $\omega_\pm\eqdef\frac{1}{\eps}(\rho-\sigma)_{\pm}$, and with $\eps\eqdef\frac{1}{2}\|\rho-\sigma\|_1$ being the trace distance between $\rho$ and $\sigma$. Set $t\eqdef1/(1+\eps)$ so that
\be\label{gamtr}
\gamma\eqdef t\rho+(1-t)\omega_-=t\sigma+(1-t)\omega_+\;,
\ee
where for the second equality we used~\eqref{decomom}. The key idea of the proof is to find lower and upper bounds for $D(\gamma\|\mf)$ in terms of $D(\gamma\|\mf)$, $D(\sigma\|\mf)$, and $\kappa$. 

Since $\mf(A)$ is convex, we saw above that the relative entropy of a resource is convex, so that
\ba\label{up00}
D(\gamma\|\mf)&\leq tD(\sigma\|\mf)+(1-t)D(\omega_+\|\mf)\\
&\leq tD(\sigma\|\mf)+(1-t)\kappa \;.
\ea 
To get a lower bound, let $\eta$ be such that $D(\gamma\|\mf)=D(\gamma\|\eta)$. Then, from the definition of the Umegaki relative entropy we have
\begin{align}
D(\gamma\|\mf)&=-H(\gamma)-\tr[\gamma\log\eta]\nonumber\\
\GG{\eqref{gamtr}}&=-H\big(t\rho+(1-t)\omega_-\big)-t\tr[\rho\log\eta]-(1-t)\tr[\omega_-\log\eta]\;.
\end{align}
From Corollary~\ref{cor:ubent} we have that the first term on the right-hand side above satisfies
\be
-H\left(t\rho+(1-t)\omega_-\right)\geq -tH(\rho)-(1-t)H\left(\omega_-\right)-h\left(t\right)\;.
\ee
Substituting this into the previous equation gives
\ba
D(\gamma\|\mf)&\geq -t\big(H\left(\rho\right)+\tr[\rho\log\eta]\big)-(1-t)\big(H\left(\omega_-\right)+\tr[\omega_-\log\eta]\big)-h\left(t\right)\\
&= tD(\rho\|\eta)+(1-t)D(\omega_-\|\eta)-h\left(t\right)\\
&\geq tD(\rho\|\mf)-h\left(t\right)\;,
\ea
where in the last line we removed the term $(1-t)D(\omega_-\|\eta)\geq 0$, and replaced  $D(\rho\|\eta)$ with $D(\rho\|\mf)$.
Combining the lower bound above with the upper bound in~\eqref{up00} we conclude that
\ba
D(\rho\|\mf)-D(\sigma\|\mf)&\leq t^{-1} h\left(t\right)+\frac{1-t}{t}\kappa\\
&= (1+\eps) h\left(\frac{\eps}{1+\eps}\right)+\eps\kappa\;.
\ea
The same upper bound holds for $D(\sigma\|\mf)-D(\rho\|\mf)$ by exchanging between $\rho$ and $\sigma$ everywhere above. Hence, this completes the proof.
\end{proof}

\bex\label{ex:bef2}
Let $\mf$ be as in Theorem~\ref{thm:ac} and suppose further that there exists a free state $\eta\in\mf(A)$ that is full rank; i.e. $\eta>0$. 
Show that there exists a continuous function $f:[0,1]\to\mbb{R}_+$, independent on the dimension of $A$, such that $f(0)=0$ and
\be
\big|D(\rho\|\mf)-D(\sigma\|\mf)\big|\leq f(\eps)\log\|\eta^{-1}\|_\infty\;.
\ee
\eex

\subsubsection{Asymptotic Continuity of the Umegaki\index{Umegaki} Relative Entropy}

In Def.~\ref{def:ac} we defined asymptotic continuity\index{asymptotic continuity} on resource measures. This definition can be extended to divergences and relative entropies. Since a relative entropy $\D$ takes two density matrices as its input, we will consider its continuity on the first argument. Moreover,
we saw that relative entropies can be unbounded, so this property needs to be accommodated into the definition.

\begin{myd}{}
\begin{definition}\label{def:ac7}
A relative entropy $\D$ is said to be \emph{asymptotically continuous} if there exists a continuous function $f:[0,1]\to\mbb{R}_+$ such that $f(0)=0$, and for all $\rho,\rho',\sigma\in\md(A)$, with $\supp(\rho)\subseteq\supp(\sigma)$ and $\supp(\rho')\subseteq\supp(\sigma)$
\be\label{asy}
\left|\D(\rho\|\sigma)-\D(\rho'\|\sigma)\right|\leq f(\eps)\log\|\sigma^{-1}\|_\infty
\ee
where $\eps\eqdef\frac{1}{2}\|\rho-\rho'\|_1$. We emphasize that $f$ is independent of $|A|$.
\end{definition}
\end{myd}

We now argue that the Umegaki relative entropy is asymptotically continuous. This is a simple consequence of Theorem~\ref{thm:ac}.

\begin{myg}{}
\begin{corollary}\label{cor:acume}
The Umegaki relative entropy is asymptotically continuous.
\end{corollary}
\end{myg}

\begin{proof}
Let $\rho,\rho',\sigma\in\md(A)$ and set $\eps\eqdef\frac{1}{2}\|\rho-\rho'\|_1$. Since $\supp(\rho)\subseteq\supp(\sigma)$ and $\supp(\rho')\subseteq\supp(\sigma)$ we can assume without loss of generality that $\sigma>0$. Let $\mf(A)\eqdef\{\sigma\}$ be the set consisting of $\sigma$ (i.e., $\mf(A)$ contains only one density matrix). The set $\mf(A)$ is trivially closed and convex. Moreover, note that for this $\mf$ we get $D(\rho\|\mf)=D(\rho\|\sigma)$ and similarly $D(\rho'\|\mf)=D(\rho'\|\sigma)$. Therefore, applying Theorem~\ref{thm:ac} gives
\be
\left|D(\rho\|\sigma)-D(\rho'\|\sigma)\right|\leq \eps\kappa+(1+\eps)h\left(\frac{\eps}{1+\eps}\right)\;,
\ee
with
\ba
\kappa=\max_{\omega\in\md(A)}D(\omega\|\sigma)\leq \max_{\omega\in\md(A)}-\tr[\omega\log\sigma]\;,
\ea
where we droped the term $\tr[\omega\log\omega]=-H(\omega)$ as it is negative.
Moreover, note that $-\tr[\omega\log\sigma]=\tr[\omega\log(\sigma^{-1})]$, and since $\sigma^{-1}\leq \|\sigma^{-1}\|_\infty I^A$ we get that 
\be
\kappa\leq\max_{\omega\in\md(A)}\tr[\omega\log(\sigma^{-1})]\leq\log\|\sigma^{-1}\|_\infty 
\ee
where we used the operator monotonicity of the log function. This completes the proof (see Exercise~\ref{ex:bef0}).
\end{proof}

\bex\label{ex:bef0}
Show that there exists function $f:[0,1]\to\mbb{R}_+$, independent on the dimension of $A$, such that $f(0)=0$ and
\be
\eps\log\|\sigma^{-1}\|_\infty+(1+\eps)h\left(\frac{\eps}{1+\eps}\right)\leq f(\eps)\log\|\sigma^{-1}\|_\infty\;.
\ee
\eex

\subsubsection{Asymptotic Continuity of the von-Neumann\index{von-Neumann} Conditional Entropy}

The conditional entropy (see Chapter~\ref{sec:ce}) can be viewed (at least mathematically) as a resource measure in a QRT in which the free operations are conditionally mixing operations ({\rm CMO}) channels and the free states are given by
\be
\mf(AB)=\Big\{\u^A\otimes\sigma^B\;:\;\sigma\in\md(B)\Big\}\;.
\ee

\bex
Show that any conditional entropy $\H$ (see Definition~\ref{qce}) is a resource measure in the QRT in which $\mf(AB\to AB')=\cmo(AB\to AB')$.
\eex

For the case $\mf\eqdef\cmo$ we have
\ba
D\left(\rho^{AB}\big\|\mf\right)&=\min_{\sigma\in\md(B)}D\left(\rho^{AB}\big\|\u^A\otimes\sigma^B\right)\\
\GG{{\it cf.}~\eqref{8281}}&=D\left(\rho^{AB}\big\|\u^A\otimes\rho^B\right)\\
\GG{\eqref{8267}}&=\log|A|-H(A|B)_\rho\;.
\ea
Moreover, observe that $\kappa$ as defined in~\eqref{1kappa1} satisfies
\ba
\kappa&\eqdef\max_{\omega\in\mf(AB)}D\left(\omega^{AB}\big\|\mf\right)\\
&=\log|A|-\max_{\omega\in\mf(AB)}H(A|B)_\rho\\
\GG{Theorem~\ref{thm:neg-cond-ent}}&\leq 2\log|A|
\ea
Therefore, taking this choice of $\mf$ in Theorem~\ref{thm:ac} yields the following corollary.

\begin{myg}{}
\begin{corollary}
Let $\rho,\sigma\in\md(AB)$ and $\eps\eqdef\frac12\|\rho^{AB}-\sigma^{AB}\|_1$.
Then,
\be\label{condentcon}
\left|H(A|B)_\rho-H(A|B)_\sigma\right|\leq2\eps\log|A|+(1+\eps)h\left(\frac\eps{1+\eps}\right)
\ee
\end{corollary}
\end{myg}

\bex
Use Theorem~\ref{thm:ac} and the expressions above to prove the corollary.
\eex

\bex
Using the same notations as in the corollary above, show that there exists a continuous function $f:[0,1]\to\mbb{R}_+$ satisfying $\lim_{\delta\to 0^+}f(\delta)=0$ and
\be\label{condentcon2}
H(A|B)_\rho\geq H(A|B)_\sigma-f(\eps)\log|A|\;.
\ee
\eex

\bex[Asymptotic Continuity of the von-Neumann Entropy]
Let $\rho,\sigma\in\md(A)$ and $\eps\eqdef\frac12\|\rho-\sigma\|_1$. 
\ben
\item Show that the von-Neumann entropy $H$ satisfies 
\be\label{fennes}
\big|H(\rho)-H(\sigma)\big|\leq\eps\log|A|+(1+\eps)h\left(\frac\eps{1+\eps}\right)\;,
\ee
Hint: Consider the QRT in which $\mf(A\to A)$ is the set of all unital channels.
\item Show that there exists a continuous function $f:[0,1]\to\mbb{R}_+$ satisfying $\lim_{\delta\to 0^+}f(\delta)=0$ and
\be
H(\rho)\geq H(\sigma)-f(\eps)\log|A|\;.
\ee
\een
\eex

\subsection{The Robustness of a Resource}\label{robustness}\index{robustness}

Given a QRT $\mf$, the robustness and global robustness of a resource $\rho\in\md(A)$ are defined as
 \ba\label{robust}
 &\R(\rho)\eqdef\inf\left\{s\geq 0\;:\;\frac{\rho+s\omega}{1+s}\in\mf(A)\;\;,\;\;\omega\in\mf(A)\right\}\\
&\R_g(\rho)\eqdef\inf\left\{s\geq 0\;:\;\frac{\rho+s\omega}{1+s}\in\mf(A)\;\;,\;\;\omega\in\md(A)\right\}\;.
 \ea
 Moreover, if there is no such $s\geq 0$ and $\omega\in\mf(A)$ such that $\frac{\rho+s\omega}{1+s}\in\mf(A)$ then $\R(\rho)\eqdef\infty$.
The robustness and global robustness of a resource  measure of how robust a resource $\rho$ is to mixing with noise. In other words, they quantify the ability of the resource to maintain its usefulness in the presence of disturbances. The term ``global" refers to the fact that the noise can be any density matrix $\omega\in\md(A)$, which represents a wide range of possible disturbances. However, if we limit the density matrix $\omega$ to only represent free states, then the resulting quantity is called the robustness.

By definition, $\R_g(\rho)\leq\R(\rho)$ (see exercise below) and if  $\rho\in\mf(A)$ then $\R(\rho)=\R_g(\rho)=0$ since $s$ above can be taken to be zero. Furthermore, from the following exercise, the converse of this statement is also true; that is, $\R_g$ is faithful.

\bex\label{robfaith}
Consider the robustness and global robustness as defined above.
\ben
\item Show that   $\R_g(\rho)\leq \R(\rho)$ for all $\rho\in\md(A)$.
\item Show that if $\mf$ is affine then $\R(\rho)=\infty$ for all $\rho\in\md(A)$ that is not free.
\item Show that $\R_g(\rho)=0$ if and only if $\rho\in\mf(A)$.
\een
\eex

\bex
Let $\rho\in\md(A)$ and suppose $\R(\rho)<\infty$.
\ben
\item   Show that there exist $\tau,\omega\in\mf(A)$ such that
\be\label{rhoasie}
\rho=\big(1+\R(\rho)\big)\tau-\R(\rho)\omega\;.
\ee
\item Show that there exist $\tau\in\mf(A)$ and $\omega\in\md(A)$ such that
\be\label{rhoasie}
\rho=\big(1+\R_g(\rho)\big)\tau-\R_g(\rho)\omega\;.
\ee
\een
The above decompositions of $\rho$ are sometimes referred to as pseudo-mixtures of states.
\eex

Note that from the exercise above it follows that $\R_g(\rho)$ can also be expressed as
\be
 \R_g(\rho)\eqdef\min\big\{s\geq 0\;:\;\rho=(1+s)\tau-s\omega\;\;,\;\;
\tau\in\mf(A)\;\;,\;\;\omega\in\md(A)\big\}\;,
 \ee
so that we can think of the pseudo-mixture in~\eqref{rhoasie} as the optimal one achieved with $s=\R_g(\rho)$.

\begin{myt}{}
\begin{theorem}\label{thm:robust}
The global robustness of a resource, $\R_g$, is a resource measure satisfying the strong monotonicity property. Moreover, if the set of free states is convex then $\R_g$ is a resource monotone.\index{resource monotone}
\end{theorem}
\end{myt}
\begin{proof}
Since we already saw that $\R_g(\rho)=0$ for all $\rho\in\mf(A)$,
we prove now the strong monotonicity property. 
Let $\mE\eqdef\sum_{x\in[m]}\mE_x\otimes |x\lr x|\in\mf(A\to BX)$ be a free quantum instrument (if $\mf(A\to BX)$ is an empty set then strong monotonicity holds trivially). Let $\rho\in\md(A)$ be as in~\eqref{rhoasie} and for all $x\in[m]$ denote by
\ba\label{extau}
\sigma_x\eqdef\frac1{\tr[\mE_x(\rho)]}\mE_x(\rho)&=\frac1{\tr[\mE_x(\rho)]}\Big(\big(1+\R_g(\rho)\big)\mE_x(\tau)-\R_g(\rho)\mE_x(\omega)\Big)\\
&=(1+s)\frac{\mE_x(\tau)}{\tr[\mE_x(\tau)]}-s\frac{\mE_x(\omega)}{\tr[\mE_x(\omega)]}\;,
\ea
where $s\eqdef\frac{\tr[\mE_x(\omega)]}{\tr[\mE_x(\rho)]}\R_g(\rho)$. On the other hand,
the state $\sigma_x$ can also be expressed in its optimal pseudo-mixture as in~\eqref{rhoasie} via
\be\label{secpsedo}
\sigma_x=\big(1+\R_g(\sigma_x)\big)\tau_x-\R_g(\sigma_x)\omega_x
\ee
for some $\omega_x\in\md(B)$ and $\tau_x\in\mf(B)$. Therefore, from the two expressions above for $\sigma_x$, and the optimality of the pseudo-mixture in~\eqref{secpsedo}, we get that $\R_g(\sigma_x)$ is no greater than $s$; that is,
\be
\R_g(\sigma_x)\leq \frac{\tr[\mE_x(\omega)]}{\tr[\mE_x(\rho)]}\R_g(\rho)\;.
\ee
From the above inequality we conclude that
\be
\sum_{x\in[m]}\tr[\mE_x(\rho)]\R_g(\sigma_x)\leq\sum_{x\in[m]}\tr[\mE_x(\omega)]\R_g(\rho)=\R_g(\rho)\;,
\ee
where in the last equality we used the fact that $\sum_{x\in[m]}\mE_x$ is trace-preserving. This completes the proof of strong monotonicity.

Next, suppose that $\mf(A)$ is convex, and let $\{p_x,\rho_x\}_{x\in[m]}$ be an ensemble of quantum states in $\md(A)$. Express each $\rho_x$ as a pseudo-mixture
\be
\rho_x=\big(1+\R_g(\rho_x)\big)\tau_x-\R_g(\rho_x)\omega_x
\ee
for some $\tau_x\in\mf(A)$ and $\omega_x\in\md(A)$. Denote by $\bar{\rho}\eqdef\sum_{x\in[m]}p_x\rho_x$. Then, from the equation above we have
\be\label{oppsedo}
\bar{\rho}=\sum_{x\in[m]}p_x\Big(\big(1+\R_g(\rho_x)\big)\tau_x-\R_g(\rho_x)\omega_x\Big)=(1+r)\tau-r\omega
\ee
where 
\be
r\eqdef\sum_{x\in[m]}p_x\R_g(\rho_x),\quad\tau\eqdef \frac{1}{1+r}\sum_{x\in[m]}p_x\big(1+\R_g(\rho_x)\big)\tau_x,\quad\omega\eqdef\frac1r\sum_{x\in[m]}p_x\R_g(\rho_x)\omega_x\;.
\ee
Note that $\tau\in\mf(A)$ since each $\tau_x\in\mf(A)$ and $\mf(A)$ is convex. Since the pseudo-mixture in~\eqref{oppsedo} is not necessarily the optimal one we conclude that
\be
\R_g(\bar{\rho})\leq r=\sum_{x\in[m]}p_x\R_g(\rho_x)\;.
\ee
That is, $\R_g$ is a convex function. This completes the proof.
\end{proof}

\bex
Let $\mf$ be a convex QRT, and let $\R$ be the corresponding robustness measure. Prove that $\R$ is a resource monotone. Hint: Follow similar steps as in the proof of Theorem~\ref{thm:robust}.
\eex

\subsubsection{The Logarithmic Global Robustness of a Resource}

The logarithmic global robustness\index{robustness} is another important distance based measure given by replacing $D$ in~\eqref{rer} with the max relative entropy\index{max relative entropy} $D_{\max}$. It is given by 
\be
D_{\max}(\rho\|\mf)\eqdef \min_{\sigma\in\mf(A)}D_{\max}\left(\rho\|\sigma\right)
\ee
The terminology of $D_{\max}(\rho\|\mf)$ is due to the following connection between $D_{\max}(\rho\|\mf)$ and $\R_g$.

\begin{myg}{}
\begin{lemma} Let $\mf$ be a QRT. Then, for any $\rho\in\md(A)$
\be
D_{\max}(\rho\|\mf)=\log\big(1+\R_g(\rho)\big)\;.
\ee
\end{lemma}
\end{myg}
\begin{proof}
By definition of $D_{\max}$ and the logarithmic global robustness\index{robustness} of a resource, we have for all $\rho\in\md(A)$
\ba
D_{\max}(\rho\|\mf)&=\min\Big\{\log t\;:\;t\sigma\geq\rho\;,\;\sigma\in\mf(A)\Big\}\\
&=\min\Big\{\log t\;:\;t\sigma-\rho=(t-1)\omega\;,\;\sigma\in\mf(A),\;\;\omega\in\md(A),\;\;t\geq 1\Big\}\;,
\ea
since $t\sigma-\rho\geq 0$ implies that $t\sigma-\rho=(t-1)\omega$ for some density matrix $\omega$. Denoting by $s=t-1$ we continue
\ba
D_{\max}(\rho\|\mf)&=\min\Big\{\log (1+s)\;:\;(1+s)\sigma-\rho=s\omega\;,\;\sigma\in\mf(A),\;\;\omega\in\md(A),\;\;s\geq 0\Big\}\\
\GG{Isolating\;\sigma}&=\min\Big\{\log (1+s)\;:\;\frac{\rho+s\omega}{1+s}\in\mf(A),\;\;\omega\in\md(A),\;\;s\geq 0\Big\}\\
&=\log\big(1+\R_g(\rho)\big)\;.
\ea
This completes the proof.
\end{proof}

\subsection{The Hypothesis Testing and $\alpha$-Relative Entropy of a Resource}\index{hypothesis testing}
  
 Let $\mf$ be a quantum resource theory such that the set of free states $\mf(A)\subseteq\md(A)$ is closed and convex.
We define the hypothesis testing measure of a resource as
\be\label{1067}
D_{\min}^{\eps}(\rho\|\mf)\eqdef\min_{\omega\in\mf(A)}D_{\min}^{\eps}(\rho\|\omega)\quad\quad\forall\;\rho\in\md(A)\;.
\ee
For any $\alpha\in[0,2]$, we also define the $\alpha$-R\'enyi relative entropy of a resource as
\be\label{1068}
D_{\alpha}(\rho\|\mf)\eqdef\min_{\omega\in\mf(A)}D_{\alpha}(\rho\|\omega)\quad\quad\forall\;\rho\in\md(A)\;.
\ee
Note that the case $\alpha=0$ corresponds to $D_{\min}(\rho\|\mf)$.
The special case of $\alpha=1$ is $D_{\alpha=1}(\rho\|\mf)= D(\rho\|\mf)$ (\emph{the} relative entropy of a resource). 
Since the Petz quantum R\'enyi divergence, $D_\alpha(\cdot\|\cdot)$, is non-decreasing with $\alpha$, also $D_{\alpha}(\cdot\|\mf)$ is not decreasing in $\alpha$. The continuity of $D_\alpha(\cdot\|\cdot)$ in $\alpha$ also carries over to $D_{\alpha}(\cdot\|\mf)$ including the continuity at $\alpha=1$. This result is a simple consequence of  Sion's minimax theorem.

\begin{lemma}[Sion's Minimax Theorem] \label{lem: sion}
Let $X$ be a compact convex subset of a linear topological space, and let $Y$ be a convex subset of a topological space. Let $f:X\times Y\to\mbb{R}\cup\{-\infty,+\infty\}$ be a real valued function satisfying
\begin{enumerate}
\item For every fixed $y\in Y$, the function $x\mapsto f(x,y)$ is lower semicontinuous and quasi-convex on X. 
\item For every fixed $x\in X$, the function $y\mapsto f(x,y)$ is upper semicontinuous and quasi-concave on Y.
\end{enumerate}
Then 
\be
\min_{x\in X}\sup_{y\in Y}f(x,y)=\sup_{y\in Y}\min_{x\in X}f(x,y)\;.
\ee
\end{lemma}

\begin{myg}{}
\begin{lemma}\label{lem:cont}
Let $\rho\in\md(A)$ be a fixed density matrix, and define $g:[0,2]\to\mbb{R}_+$ as $g(\alpha)\eqdef D_{\alpha}(\rho\|\mf)$ for all $\alpha\in[0,2]$. Then, $g(\alpha)$ is a continuous function.
\end{lemma}
\end{myg}

\begin{proof}
Let $\beta\in(0,2)$. Since $D_\alpha$ is monotonically non-decreasing in $\alpha$ we have that
\ba
\lim_{\alpha\to\beta^+}g(\alpha)&=\lim_{\alpha\to\beta^+}\min_{\omega\in\mf(A)}D_{\alpha}(\rho\|\omega)\\
&=\inf_{\alpha\in(\beta,2)}\min_{\omega\in\mf(A)}D_{\alpha}(\rho\|\omega)\\
&=\min_{\omega\in\mf(A)}\inf_{\alpha\in(\beta,2)}D_{\alpha}(\rho\|\omega)\\
&=\min_{\omega\in\mf(A)}D_{\beta}(\rho\|\omega)=g(\beta)\;.
\ea
When approaching $\beta$ from below observe that
\ba\label{19}
\lim_{\alpha\to\beta^-}g(\alpha)&=\lim_{\alpha\to\beta^-}\min_{\omega\in\mf(A)}D_{\alpha}(\rho\|\omega)\\
&=\sup_{\alpha\in(0,\beta)}\min_{\omega\in\mf(A)}D_{\alpha}(\rho\|\omega)
\ea
In order to switch the order between the sup and min above we need to verify that all the conditions in Sion's minimax theorem are satisfied.
Indeed,  the function $f(\omega,\alpha)\eqdef D_{\alpha}(\rho\|\omega)$ has the property that it is continuous in $\omega$ (and therefore lower semi-continuous). Moreover, note that for a fixed $\alpha\in[0,2]$, the function $\omega\mapsto f(\omega,\alpha)$ is a quasi-convex function since for any $t\in[0,1]$ and $\omega_0,\omega_1\in\mf(A)$ we have
\ba
f(t\omega_0+(1-t)\omega_1,\alpha)&=D_{\alpha}\left(\rho\big\|t\omega_0+(1-t)\omega_1\right)\\
\GG{\eqref{quascon}}&\leq\max\big\{D_{\alpha}(\rho\|\omega_0),D_{\alpha}(\rho\|\omega_1)\big\}\\
&=\max\big\{f(\omega_0,\alpha),f(\omega_1,\alpha)\big\}\;.
\ea
On the other hand, for a fixed $\omega\in\mf(A)$ the function $\alpha\mapsto f(\omega,\alpha)$ is a continuous function (and therefore upper semi-continuous) and quasi-concave since for any $t\in[0,1]$ and $\alpha_0,\alpha_1\in[0,2]$ we have
\ba
f(\omega,t\alpha_0+(1-t)\alpha_1)&=D_{t\alpha_0+(1-t)\alpha_1}\left(\rho\|\omega\right)\\
\GG{monotonicity\;of\;\text{$D_\alpha$}\;in\;\alpha}&\geq D_{\min\{\alpha_0,\alpha_1\}}\left(\rho\|\omega\right)\\
&=\min\big\{f(\omega,\alpha_0),f(\omega,\alpha_1)\big\}\;.
\ea
Therefore, $f(\omega,\alpha)$ satisfies all the requirements of Sion's minimax theorem. This means that we can switch the order of the sup and min in~\eqref{19} to get
\ba
\lim_{\alpha\to\beta^-}g(\alpha)&=\min_{\omega\in\mf(A)}\sup_{\alpha\in(0,\beta)}D_{\alpha}(\rho\|\omega)\\
\GG{continuity\;of\;\text{$D_\alpha$}\;in\;\alpha}&=\min_{\omega\in\mf(A)}D_{\beta}(\rho\|\omega)=g(\beta)\;.
\ea
This completes the proof of the lemma.
\end{proof}

Using the bounds~\eqref{e1} and~\eqref{e2} we get that
\begin{myg}{}
\begin{corollary}\label{corht01}
Let $\eps\in(0,1)$ and $\rho,\sigma\in\md(A)$.  
\begin{enumerate}
\item For all $\alpha\in(1,2]$
\be\label{ee1}
D_{\min}^\eps(\rho\|\mf)\leq D_{\alpha}(\rho\|\mf)+\frac\alpha{\alpha-1}\log\left(\frac1{1-\eps}\right)
\ee
\item For all $\alpha\in(0,1)$
\be\label{ee2}
D_{\min}^\eps(\rho\|\mf)\geq D_{\alpha}(\rho\|\mf)+\frac\alpha{1-\alpha}\left(\frac{h(\alpha)}{\alpha}-\log\left(\frac1\eps\right)\right)\ee
\end{enumerate}
\end{corollary}
\end{myg}

The regularized $\alpha$-R\'enyi relative entropy of a resource is defined as
\be
D_{\alpha}^\reg(\rho\|\mf)\eqdef \lim_{n\to\infty}\frac1nD_{\alpha}\left(\rho^{\otimes n}\big\|\mf\right)\quad\quad\forall\;\rho\in\md(A)\;.
\ee
The limit above exists since the $\alpha$-R\'enyi relative entropy of a resource is subadditive (see Exercise~\ref{ex:reg0}).  From Corollary~\ref{corht01} it follows that
\be
\limsup_{n\to\infty}\frac1nD_{\min}^\eps\left(\rho^{\otimes n}\big\|\mf\right)\leq \lim_{\alpha\to 1^+}D_{\alpha}^\reg(\rho\|\mf)\;,
\ee
and similarly
\be\label{llowl}
\liminf_{n\to\infty}\frac1nD_{\min}^\eps\left(\rho^{\otimes n}\big\|\mf\right)\geq \lim_{\alpha\to 1^-}D_{\alpha}^\reg(\rho\|\mf)\;.
\ee
Observe that in general we do not know if $D_{\alpha}^\reg(\rho\|\mf)$ is continuous at $\alpha=1$, but we can show continuity from the right.

\begin{myg}{}
\begin{lemma}\label{imlim}
Let $\rho\in\md(A)$ and let $\mf$ be a quantum resource theory admitting a tensor product structure, and has the property that $\mf(A)\subseteq\md(A)$ is closed and convex. Then,
\be
\lim_{\alpha\to 1^+}D_{\alpha}^\reg(\rho\|\mf)=D^\reg(\rho\|\mf)\;.
\ee
\end{lemma}
\end{myg}

\begin{proof}
Observe that
\ba
\lim_{\alpha\to 1^+}D_{\alpha}^\reg(\rho\|\mf)&=\inf_{\alpha\in(1,\infty)}\inf_{n\in\mbb{N}}\frac1n D_{\alpha}\left(\rho^{\otimes n}\big\|\mf\right)\\
&=\inf_{n\in\mbb{N}}\inf_{\alpha\in(1,\infty)}\frac1n D_{\alpha}\left(\rho^{\otimes n}\big\|\mf\right)\\
\GG{Lemma~\ref{lem:cont}}&=\sup_{n\in\mbb{N}}\frac1n D\left(\rho^{\otimes n}\big\|\mf\right)\\
&=D^\reg(\rho\|\mf)\;.
\ea
\end{proof}

We therefore conclude that 
\be\label{1083}
\limsup_{n\to\infty}\frac1nD_{\min}^\eps\left(\rho^{\otimes n}\big\|\mf\right)\leq D^\reg(\rho\|\mf)\;.
\ee
In several resource theories  the opposite inequality also holds, but in general we do not know
if 
the limit $\lim_{\alpha\to 1^-}D_{\alpha}^\reg(\rho\|\mf)$ equals to $D^\reg(\rho\|\mf)$. At the time of writing this book it is a big open problem in the field to determine under what conditions the inequality in the equation above can be replaced with an equality.

\section{Computation of the Relative Entropy of a Resource}\label{sec:crer}

The computation of the relative entropy of a resource can be hard, depending of course on the complexity of the set $\mf(A)$. As we will see in the next Chapter, in entanglement theory its computation belong to a class of problems knowns as NP hard. If $\mf(A)$ is closed and convex, some techniques from convex analysis can be employed to compute the relative entropy. Particularly, in this case the converse problem\index{converse problem} can be computed efficiently as we discuss now.

\subsection{The Converse Problem}\index{converse problem}

Let $\sigma\in\mf(A)$ be the free state that optimize~\eqref{rer} for a given resource state $\rho\in\md(A)$. We will see shortly that, as it suggests intuitively,  $\sigma$ must be on the \emph{boundary} of the set $\mf(A)$. The boundary of $\mf(A)$ is define as
\be
\partial\mf(A)\eqdef\Big\{\omega\in\md(A)\;:\;\forall\eps>0\;\;\exists\eta,\zeta\in\mb_{\eps}(\omega)\;\;\text{s.t.}\;\;\eta\in\mf(A)\;\text{and}\;\zeta\not\in\mf(A)\Big\}
\ee
where $\mb_{\eps}(\omega)$ is the set of all density matrices that are $\eps$-close (in trace distance) to $\omega$. That is, in any neighbourhood of a state on the boundary of $\mf(A)$ there exists at least one state in $\mf(A)$ and at least one state not in $\mf(A)$.  

The state $\sigma$ can be thought of as the closest free state (CFS) to $\rho$, when we measure the ``distance" with the relative entropy. As we already mentioned, the computation of $\sigma$ can be very hard. However, for a given  state $\omega\in\partial\mf(A)$ we can compute all the resource states in $\md(A)$ for which $\omega$ is the CFS. This converse problem\index{converse problem} has several applications and can be used to produce examples of resource states for which one knows the value of the relative entropy of a resource. 

Note that if $0<\rho\not\in\mf(A)$ and $D(\rho\|\mf)=D(\rho\|\sigma)$ (i.e. $\sigma$ is a CFS) then $\sigma>0$ or otherwise $D(\rho\|\mf)=\infty$. For simplicity of the exposition here, we will always assume that $\sigma$ has full rank, and refer the interested reader to the end of this chapter for more details and references on the singular case. 
We start by showing that if $0<\sigma\in\mf(A)$ is a CFS then $\sigma\in\partial \mf(A)$.

\begin{myt}{}
\begin{theorem}
Let $0<\sigma\in\mf(A)$ be a closest free state of a resource state $\rho\in\md(A)$. Then, $\sigma\in\partial \mf(A)$.
\end{theorem}
\end{myt}

\begin{proof}
Consider the following Taylor expansion of the logarithmic function. This expansion is based on the divided difference approach discussed in Appendix~\ref{A:DD}. For any $t>0$, $0<\sigma\in\md(A)$, and $\eta\in\herm(A)$ we have
\be\label{expansion}
\log(\sigma+t\eta)=\log\sigma+t\mL_\sigma(\eta)+O(t^2)
\ee
where $\mL_\sigma:\herm(A)\to\herm(A)$ is a linear operator defined as follows. Let $\{p_x\}_{x\in[m]}$ (with $m\eqdef|A|$) be the eigenvalues of $\sigma$, and let $\{\eta_{xy}\}_{x,y\in[m]}$ be the matrix components of a matrix $\eta\in\herm(A)$ in the eigenbasis of $\sigma$. Then, the matrix components of $\mL_\sigma(\eta)$ are given by
\be
\big[\mL_\sigma(\eta)\big]_{xy}\eqdef\eta_{xy}\frac{\log p_x-\log p_y}{p_x-p_y}\quad\forall x,y\in[m]\;,
\ee
where the case $p_x=p_y$ is understood in terms of the limit \be\frac{\log p_x-\log p_y}{p_x-p_y}\eqdef\lim_{p_y\to p_x}\frac{\log p_x-\log p_y}{p_x-p_y}=\frac{1}{p_x}\;.\ee  In Exercise~\ref{exsg0} below you show that 
$\mL_\sigma$ is a linear self-adjoint map\index{adjoint map} that satisfies $\mL_\sigma(\sigma)=I$.

Now, suppose by contradiction that $0<\sigma\in\mf(A)$ is a CFS of $\rho\in\md(A)$, and $\sigma\not\in\partial\mf(A)$. This means that $\sigma$ is in the interior of $\mf(A)$, and in particular, there exists $\eps>0$ such that $\mb_{\eps}(\sigma)$ does not contain any resource state (i.e. $\mb_{\eps}(\sigma)\subset\mf(A)$). Moreover, since $\sigma>0$ it follows that for any $\sigma'\in\md(A)$ (i.e. not necessarily free) and small enough $|t|$, where $t\in\mbb{R}$ can be negative, the state $\omega\eqdef (1-t)\sigma+t\sigma'\in\mb_{\eps}(A)\subset\mf(A)$.
Hence, for small enough $|t|$,
\be
D(\rho\|\mf)=D(\rho\|\sigma)\leq D(\rho\|\omega)\;,
\ee
since $\sigma$ is a CFS of $\rho$. The above expression is equivalent to
$
f(t)\leq\tr[\rho\log\sigma]\;,
$
where
\be\label{ftt}
f(t)\eqdef\tr\left[\rho\log\big(\sigma+t(\sigma'-\sigma)\big)\right]\;.
\ee
Since $f(0)=\tr[\rho\log\sigma]$ achieves the maximum value, we must have $f'(0)=0$. Using~\eqref{expansion} we get
\ba\label{fttt}
f'(0)&=\tr\left[\rho\mL_\sigma(\sigma'-\sigma)\right]\\
&=\tr\left[\rho\big(\mL_\sigma(\sigma')-I\big)\right]\\
&=\tr\left[\mL_\sigma(\rho)\sigma'\right]-1
\ea
Therefore, the condition that $f'(0)=0$  implies that
$\tr\left[\mL_\sigma(\rho)\sigma'\right]=1$ for all $\sigma'\in\md(A)$. This means that $\mL_\sigma(\rho)=I$ which is possible only if $\rho=\sigma$. But since we assume that $\rho\not\in\mf(A)$ we get a contradiction. This completes the proof.
\end{proof}

\begin{exercise}\label{exsg0}
Let $0<\sigma\in\md(A)$ and $m\eqdef|A|$.
\begin{enumerate}
\item Show that $\mL_\sigma(\sigma)=I^A$.
\item Show that $\mL_\sigma$ is a linear self-adjoint map\index{adjoint map}. That is, show that for any $\eta,\zeta\in\herm(A)$
\be
\tr\left[\eta\mL_\sigma(\zeta)\right]=\tr\left[\mL_\sigma(\eta)\zeta\right]\;.
\ee
\item Show that $\mL_\sigma$ is invertible, and its inverse, $\mL^{-1}_\sigma$, is also self-adjoint and is given by
\be
\big[\mL_\sigma^{-1}(\zeta)\big]_{xy}\eqdef\zeta_{xy}\frac{p_x-p_y}{\log p_x-\log p_y}\quad\quad\forall\; x,y\in[m]\;\text{,}\quad\forall\;\zeta\in\herm(A)\;.
\ee
\end{enumerate}
\end{exercise}

The next theorem provides a formula for all the resource states that have the same CFS. We will use the notation $\wit_\mf(A)$ to denote the subset of $\herm(A)$ that consists of all the normalized resource witnesses\index{resource witness}
 of the QRT $\mf$. Explicitly,
\be
\wit_\mf(A)\eqdef\Big\{\eta\in\mf(A)^*\;:\;\eta\not\geq 0\;,\;\|\eta\|_1=1\Big\}\;.
\ee
Note that we normalized the resource witnesses to have a unit trace norm since if $\eta\in\herm(A)$ is a resource witness also $a\eta$ with $0<a\in\mbb{R}$ is a resource witness, and for our purposes it will be sufficient to consider only one representative of the set $\{a\eta\}_{a>0}$.

\begin{myt}{\color{yellow} Closed Formula for the Relative Entropy of a Resource}
\begin{theorem}\label{2cfre2}
Let $\mf(A)$ be convex and $0<\sigma\in\partial\mf(A)$. Then, the set $\mathfrak{R}(\sigma)$ of all resource states in $\md(A)$ for which $\sigma$ is the closest free state is given by
\be\label{cformula}
\mathfrak{R}(\sigma)=\Big\{\sigma-a\mL_\sigma^{-1}(\eta)\;:\;\eta\in\text{WIT}_\mf(A)\;\;,\;\;\tr[\sigma\eta]=0\;\;,\;\;0<a\leq a_{\max}\Big\}
\ee
where $a_{\max}$ is the largest positive number that satisfies $a_{\max}\mL_\sigma^{-1}(\eta)\leq\sigma$.
\end{theorem} 
\end{myt}
\begin{remark}
The conditions $\tr[\sigma\eta]=0$ and $a\leq a_{\max}$ ensure that the state $\sigma-a\mL_\sigma^{-1}(\eta)$ is a density matrix. Indeed, the condition $a\leq a_{\max}$ ensures that it is positive semidefinite, and its trace is one since the self-adjointness of $\mL_\sigma^{-1}$ gives
\be
\tr\left[\mL_\sigma^{-1}(\eta)\right]=\tr\left[\mL_\sigma^{-1}(I)\eta\right]=\tr[\sigma\eta]=0\;.
\ee
\end{remark}
\begin{proof}
From the supporting hyperplane theorem, (see Theorem~\ref{supporting}) it follows that for any (fixed) $\sigma\in\partial\mf(A)$ there exists an Hermitian matrix $\eta\in\herm(A)$ such that
\be
\tr[\sigma'\eta]\geq\tr[\sigma\eta]\quad\quad\forall\;\sigma'\in\mf(A).
\ee 
Moreover, since both $\sigma$ and $\sigma'$ are normalized, if $\eta$ satisfies the equation above, also $\eta+aI$ satisfies it for any $a\in\mbb{R}$. We will therefore assume without loss of generality that $\tr[\sigma\eta]=0$ which means that $\tr[\sigma'\eta]\geq 0$ for all $\sigma'\in\mf(A)$; i.e. $\eta$ is a resource witness (observe that the condition $\tr[\sigma\eta]=0$ implies that $\eta\not\geq 0$ since $\sigma>0$). Note also that we can always normalize $\eta$ such that $\|\eta\|_1=1$. Quite often, such a resource witness that satisfies these three conditions (i.e. $\tr[\sigma\eta]=0$, $\tr[\sigma'\eta]\geq 0$ for all $\sigma'\in\mf(A)$, and $\|\eta\|_1=1$) is unique, although for some special boundary points $\sigma\in\partial\mf(A)$, there is a cone of such witnesses of dimension greater than one (see Fig.~\ref{figa3}).

\begin{figure}[h]
\centering
    \includegraphics[width=0.4\textwidth]{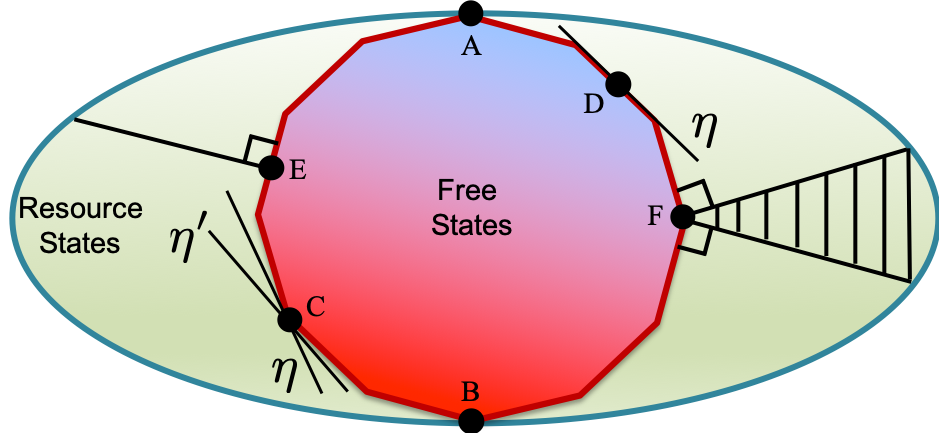}
  \caption{\linespread{1}\selectfont{\small A schematic diagram of free states (dodecagon) and resource states (oval).  Most points on the boundary, like the  points  D  and  E,  have  a  unique  supporting  hyperplane (which is also the tangent plane).  The point E is the closest free state of all  the  points  on  the  vertical  line from it.   Some  of  the  points,  like  the points C and F, have more than one supporting hyperplane.The point F is the closest free state of all the points in the shaded grid area.  Some points on the boundary, like the points A and B, can not be a closest free states; for example, separable states of rank 1 (i.e.product states) are on the boundary of separable states, but can never be the closest separable states of some entangled state.}}
  \label{figa3}
\end{figure} 

Let $\rho$ be a resource state in $\md(A)$ for which $\sigma$ is the closest free state.
The main idea of the proof is the observation that $\eta'\eqdef I^A-\mL_\sigma(\rho)$ is a affine\index{affine}. To see that, first observe that
\ba
\tr[\eta'\sigma]&=1-\tr[\sigma\mL_\sigma(\rho)]\\
\GG{\text{$\mL_\sigma$ is self adjoint}}&=1-\tr[\mL_\sigma(\sigma)\rho]\\
\Gg{\mL_\sigma(\sigma)=I^A}&=1-\tr[\rho]=0\;.
\ea
Moreover, for every $\sigma'\in\mf(A)$, define $f(t)$ as in~\eqref{ftt}, but with non-negative $t\in[0,1]$ (recall that here $\sigma$ is a boundary point not in the interior of $\mf(A)$, so that we can only conclude that $\omega\eqdef(1-t)\sigma+t\sigma'$ is a free state for non-negative $t\in[0,1]$). Since $\sigma$ is the closest free state to $\rho$, we must have that $f'(0)\leq 0$ (we cannot conclude that the derivative is zero since $t$ cannot be negative). From~\eqref{fttt} we get for all $\sigma'\in\mf(A)$
\ba
0\leq -f'(0)&=1-\tr\left[\mL_\sigma(\rho)\sigma'\right]\\
&=\tr\left[\big(I-\mL_\sigma(\rho)\big)\sigma'\right]\\
&=\tr[\eta'\sigma']\;.
\ea
Hence, $\eta'$ is a affine\index{affine}. We can then normalize it $\eta\eqdef\frac{1}{a}\eta'$ with $a>0$ such that $\|\eta\|_1=1$. We then conclude from
the definition $\eta'\eqdef I^A-\mL_\sigma(\rho)$ that 
\ba
\rho&=\mL_\sigma^{-1}\left(I^A-\eta'\right)\\
\Gg{\eta'=a\eta}&=\mL_\sigma^{-1}\left(I^A\right)-a\mL_\sigma^{-1}\left(\eta\right)\\
\Gg{\mL_\sigma^{-1}\left(I\right)=\sigma}&=\sigma-a\mL_\sigma^{-1}\left(\eta\right).
\ea

Conversely, suppose $\rho=\sigma-a\mL_\sigma^{-1}\left(\eta\right)$ for some $\eta\in\wit_\mf(A)$ and $a>0$. We need to show that $D(\rho\|\mf)=D(\rho\|\sigma)$.
For this purpose, let $\sigma'\in\mf(A)$ be any free state, and observe that $D(\rho\|\sigma)\leq D(\rho\|\sigma')$ if and only if $f(0)\geq f(1)$, where $f(t)$ is defined in~\eqref{ftt}.
From the joint convexity\index{joint convexity} of the relative entropy (and particularly its convexity in the second argument) it follows that the function $f(t)$ is concave (see Exercise~\ref{comproofb}). This means that if $f'(0)\leq 0$ then we must have $f(0)\geq f(1)$ (Exercise~\ref{comproofb}). Now, note that from~\eqref{fttt} 
\ba
f'(0)&=\tr\left[\mL_\sigma(\rho)\sigma'\right]-1\\
\Gg{\rho=\sigma-a\mL_\sigma^{-1}\left(\eta\right)}&=\tr\left[\mL_\sigma\big(\sigma-a\mL_\sigma^{-1}\left(\eta\right)\big)\sigma'\right]-1\\
&=-a\tr\left[\eta\sigma'\right]\\
\GG{\eta\text{ is a resource witness}}&\leq 0\;.
\ea
Hence, $f(0)\geq f(1)$ which is equivalent to $D(\rho\|\sigma)\leq D(\rho\|\sigma')$. Since $\sigma'$ was arbitrary state in $\mf(A)$, this completes the proof.
\end{proof}

The significance of the theorem above is that if for a given resource state $\rho$ we have a candidate $\sigma$ that we believe to be a closest free state, then we can check it with the formula in~\eqref{cformula}. Specifically, what needs to be checked is whether the matrix $I-\mL_\sigma(\rho)$ is a affine\index{affine}. We will see how this can be done when we compute the relative entropy of entanglement on pure bipartite states .
We also point out that the techniques used above are not limited to the Umegaki\index{Umegaki} relative entropy, and similar results can be obtained for the $\alpha$-R\'enyi relative entropy of a resource as defined in~\eqref{1068}.

\bex\label{comproofb}
Let $f(t)$ be the function defined in~\eqref{ftt}.
\ben
\item Show that $f(t)$ is concave. Hint: Use the convexity of $D(\rho\|\sigma)$ in $\sigma$ (with fixed $\rho$).
\item Show that if $f'(0)\leq 0$ then $f(0)\geq f(1)$.
\een
\eex

\begin{exercise}
Let $0<\rho\in\md(A)$ be a full rank resource state (i.e. $\rho\not\in\mf(A)$). Show that the closest free state to $\rho$ is unique.
Hint: Let $\sigma\neq\sigma'$ be two closest free states, define $t\sigma+(1-t)\sigma'$, and use the strict concavity of the function $f(\sigma)=\tr[\rho\log\sigma]$.
\end{exercise}

\subsection{Exact Formula for Special Cases}

In~\eqref{1068}, for any $\alpha\in[0,2]$, we  defined the $\alpha$-R\'enyi relative entropy of a resource as
\be
D_{\alpha}(\rho\|\mf)\eqdef\min_{\omega\in\mf(A)}D_{\alpha}(\rho\|\omega)\quad\quad\forall\;\rho\in\md(A)\;.
\ee
This quantity has a simple closed formula if the set of free states is affine\index{affine} (see Sec.~\ref{sec:affine}) and in addition satisfy for any $\alpha\in[0,2]$
\be\label{regular5}
\frac{\sigma^\alpha}{\tr[\sigma^\alpha]}\in\mf(A)\quad\quad\forall\;\sigma\in\mf(A)\;.
\ee

\begin{myt}{}
\begin{theorem}
Let $\alpha\in[0,2]$ and $\mf(A)\subseteq\md(A)$ be an  affine set  satisfying~\eqref{regular5}. Then, for all $\rho\in\md(A)$ we have
\be\label{cfrelative}
D_{\alpha}(\rho\|\mf)=\frac{1}{\alpha-1}\log\left\|\Delta\left(\rho^\alpha\right)\right\|_{1/\alpha}\ee
where $\Delta\in\pos(A\to A)$ is the self-adjoint RDM associated with the affine\index{affine} set $\mf(A)$ (see Definition~\ref{def:sardm}).
\end{theorem}
\end{myt}
\begin{proof}
Observe that for all $\sigma\in\mf(A)$ we have
\ba
\tr\left[\rho^\alpha\sigma^{1-\alpha}\right]&=\tr\left[\rho^\alpha\Delta\left(\sigma^{1-\alpha}\right)\right]\\
&=\tr\left[\Delta\left(\rho^\alpha\right)\sigma^{1-\alpha}\right]\\
&=\left\|\Delta\left(\rho^\alpha\right)\right\|_{1/\alpha}\tr\left[\gamma^\alpha\sigma^{1-\alpha}\right]
\ea
where 
\be
\gamma\eqdef\frac{\big(\Delta\left(\rho^\alpha\right)\big)^{1/\alpha}}{\tr\left[\big(\Delta\left(\rho^\alpha\right)\big)^{1/\alpha}\right]}
\ee
The proof is concluded with the observation that for $\alpha\leq 1$ we have 
$
\tr\left[\gamma^\alpha\sigma^{1-\alpha}\right]\leq 1$ (H\"older's inequality), and for $\alpha>1$ we have $
\tr\left[\gamma^\alpha\sigma^{1-\alpha}\right]\geq 1$ (reverse H\"older's inequality),\index{reverse H\"older inequality} where equality holds in both cases for $\sigma=\gamma$. Therefore, $\sigma=\gamma$ is the optimizer.
\end{proof}

\bex
Let $\alpha\in[0,2]$. Give a closed expression for $D_{\alpha}(\rho\|\mf)$ for the following cases:
\ben
\item $\rho\in\md(A)$ and $\mf(A)$ consists of a set of diagonal density matrices in some fixed basis.
\item $\rho\in\md(AB)$ and $\mf(AB)=\{\sigma^A\otimes\u^B\;:\;\sigma\in\md(A)\}$.
\item $\rho\in\md(A)$ and $\mf(A)$ consists of a set of symmetric density matrices (i.e.\ $\sigma\in\mf(A)$ if and only if $\sigma=\sigma^T$ where the transpose is taken in some fixed basis).
\item $G$ is a unitary group, $\rho\in\md(A)$, and $\mf(A)$ consists of the set of $G$-invariant states (i.e.\ $\sigma\in\mf(A)$ if and only if $\sigma=U\sigma U^*$ for all $U\in G$).
\een
\eex

\bex
Show that the expression given in~\eqref{cfrelative} for the $\alpha$-relative entropy of a resource can be rewritten as
\be\label{cutefor}
D_{\alpha}(\rho\|\mf)=H_{1/\alpha}\big(\Delta(\rho_\alpha)\big)-H_\alpha(\rho)
\ee
where $H_\alpha$ is the R\'enyi entropy\index{R\'enyi entropy} and $\rho_\alpha\eqdef\rho^\alpha/\tr[\rho^\alpha]$. Conclude that for $\alpha=1$
\be
D(\rho\|\mf)=H\big(\Delta(\rho)\big)-H(\rho)\;.
\ee
\eex

\section{Smoothing of Resource Measures}\index{smoothing}

Resource measures, by definition, do not necessarily have to be smooth. An illustrative example is the logarithmic robustness, expressed as:
\be
D_{\max}(\rho\|\mf)=\min_{\sigma\in\mf(A)}D_{\max}(\rho\|\sigma)\;.
\ee
This lack of smoothness in the logarithmic robustness can be traced back to the discontinuity present in the max relative entropy. Contrasting with the Umegaki\index{Umegaki} relative entropy, $D_{\max}$, does not exhibit asymptotic continuity. In fact, more broadly, it is not continuous with respect to its first argument. For example, take 
\be
\sigma=\frac12|0\lr 0|+\frac12|1\lr 1|\quad\text{and}\quad\rho_\eps=\left(\frac12-\eps\right)|0\lr 0|+\frac12|1\lr 1|+\eps|2\lr 2|\;.
\ee
For these choices we have $D(\rho_\eps\|\sigma)=\infty$ for all $\eps\in(0,1/2]$ whereas $D(\rho_{\eps=0}\|\sigma)=0$. On the other hand, in the laboratory, the preparation of a physical system in a state $\rho$ always results in some error
such that the intended state $\rho$ is differ (in trace distance) from the prepared state by some small $\eps>0$. Therefore, discontinuous resource measures are unlikely to have practical physical significance unless some smoothing process has been applied to them.In the following definition we provide a simple method to smooth a resource measure.

\begin{myd}{} 
\begin{definition}
Let $\mf$ be a QRT, $\M$ be a resource measure, and $\eps>0$. The $\eps$-smoothed version of $\M$ is the function
\be\label{mer}
\M^\eps(\rho)\eqdef\min_{\rho'\in\mb_{\eps}(\rho)}\M(\rho')\;.
\ee
\end{definition}
\end{myd}

\begin{remark}
In the  definition we employed the notation
$
\mb_{\eps}(\rho)\eqdef\left\{\rho'\in\md(A)\;:\;\frac12\|\rho'-\rho\|_1\leq\eps\right\}
$
to denote the ``ball" of states $\rho'$ that are $\eps$-close to $\rho$ in trace distance. The rationale for choosing the minimum over the ball $\mb_\eps(\rho)$ stems from the intention to identify the minimum amount of resource present within this ball. This approach ensures that the value $\M^\eps(\rho)$ represents the minimum guaranteed resource level in the system, even when our knowledge is limited to the state of the system being $\eps$-close to $\rho$. Essentially, this method accounts for uncertainty in the system's state by considering the least amount of resource that can be confidently ascribed to states within an $\eps$-radius of $\rho$. This approach is both cautious and practical, as it provides a conservative estimate of the resource quantity in the practical situations where exact state information is not available.
\end{remark}

\begin{myg}{}
\begin{lemma}\label{lemsmot}
Let $\M$ be a resource measure. Then, its smoothed version, $\M^\eps$, is also a resource measure for all $\eps>0$.
\end{lemma}
\end{myg}

\begin{proof}
By definition, $\M^\eps$ is non-negative since $\M$ is non-negative. On the other hand, for any $\rho\in\mf(A)$ we have $\M^\eps(\rho)\leq\M(\rho)=0$, were we took $\rho'=\rho$ in~\eqref{mer}. Thus, we must have $\M^\eps(\rho)=0$ for all $\rho\in\mf(A)$. It is therefore left to prove the monotonicity of $\M^\eps$.
 
Let $\mN\in\mf(A\to B)$ and $\rho\in\md(A)$. Then,
\ba
\M^\eps\big(\mN(\rho)\big)&=\min_{\sigma\in\mb_{\eps}(\mN(\rho))}\M(\sigma)\\
&=\min\Big\{\M(\sigma)\;:\;\frac12\|\sigma-\mN(\rho)\|_1\leq\eps,\;\;\sigma\in\md(B)\Big\}\\
\GG{Restricting\;\sigma=\mN(\rho')}&\leq\min\Big\{\M\big(\mN(\rho')\big)\;:\;\frac12\|\mN(\rho')-\mN(\rho)\|_1\leq\eps,\;\;\rho'\in\md(A)\Big\}\\
\GG{Monotonicity\;of\;\M}&\leq\min\Big\{\M(\rho')\;:\;\frac12\|\mN(\rho')-\mN(\rho)\|_1\leq\eps,\;\;\rho'\in\md(A)\Big\}\\
\GG{DPI\;of\;trace\;distance}&\leq\min\Big\{\M(\rho')\;:\;\frac12\|\rho'-\rho\|_1\leq\eps,\;\;\rho'\in\md(A)\Big\}\\
&=\M^\eps(\rho)\;,
\ea
where in the last inequality we used the DPI of the trace distance, particularly, the property that if $\frac12\|\rho'-\rho\|_1\leq\eps$ then $\frac12\|\mN(\rho')-\mN(\rho)\|_1\leq\eps$. Therefore, the former impose a stronger constraint than the latter which give rise to the last inequality. This completes the proof.
\end{proof}

\subsection{Smoothing of Divergence-Based Resource Measures}\index{smoothing}

Let $\eps>0$, $\D$ be a quantum divergence, and $\D(\cdot\|\mf)$ its associated resource measure as defined in~\eqref{rmasdf}. Then, the $\eps$-smoothed version of $\D(\cdot\|\mf)$ is given for all $\rho\in\md(A)$ by
\ba
\D^\eps(\rho\|\mf)&\eqdef\min_{\rho'\in\mb_{\eps}(\rho)}\D(\rho'\|\mf)\\
&=\min_{\rho'\in\mb_{\eps}(\rho)}\inf_{\sigma\in\mf(A)}\D\left(\rho'\|\sigma\right)\\
&=\inf_{\sigma\in\mf(A)}\D^\eps\left(\rho\|\sigma\right)\;,
\ea
where $\D^\eps$ is defined as
\be\label{divsmoo}
\D^\eps\left(\rho\|\sigma\right)\eqdef\min_{\rho'\in\mb_{\eps}(\rho)}\D\left(\rho'\|\sigma\right)\;.
\ee
The quantity $\D^\eps$ is called the $\eps$-smoothed version of the quantum divergence $\D$.
Smoothed divergences play key roles in QRTs and in the next theorem we prove some useful relationships among some of them.

\bex
Let $\D$ be a quantum divergence and $\eps>0$. Show that $\D^\eps$ is itself a quantum divergence.
\eex

We denote the $\eps$-smoothed version of $D_{\max}$ as $D_{\max}^\eps$. For $D_{\min}$, the notation $D_{\min}^\eps$ already signifies the quantum hypothesis testing divergence (see Sec.\ref{sec:qht}). This aligns with the notion that the quantum hypothesis testing divergence is a smoothed version of $D_{\min}$ (refer to Exercise~\ref{ex:smoothed}). Additionally, when smoothing\index{smoothing} $D_{\min}$ in the form $\min_{\rho'\in\mb_{\eps}(\rho)}D_{\min}\left(\rho'|\sigma\right)$, the result is always zero. This is because for any $\eps>0$ and $\rho\in\md(A)$, there's a $\rho'\in\md(A)$ that's $\eps$-close to $\rho$ with $\rho'>0$, making $D_{\min}(\rho'\|\sigma)=0$. Henceforth, $D_{\min}^\eps$ will exclusively represent the quantum hypothesis testing divergence in this book. 

In the forthcoming theorem, we will establish specific inequalities that involve $D_{\min}^\eps$ and $D_{\max}^\eps$. These inequalities are crucial and will play a significant role in the subsequent discussions and analyses. The relationships between $D_{\min}^\eps$ and $D_{\max}^\eps$ are fundamental in understanding various aspects of quantum resources both in the single-shot regime as well as in the asymptotic domain. Furthermore, later on we will use some of these relationships to provide operational interpretation to both $D_{\min}^\eps$ and $D_{\max}^\eps$.

\begin{myt}{}
\begin{theorem}\label{ubo}
Let $\rho,\sigma\in\md(A)$ and $\eps,\eps_1,\eps_2\in(0,1)$ such that $\eps
_1+\eps_2\leq 1$. Then, the following relations hold:
\begin{align}
& D_{\max}^{\eps_1}(\rho\|\sigma)\geq D_{\min}^{\eps_2}(\rho\|\sigma)+\log\left(1-\eps_1-\eps_2\right)\label{ubo1}\\
& D^\eps_{\min}(\rho\|\sigma)\leq \frac{D(\rho\|\sigma)+h(\eps)}{1-\eps}\label{ubo2}\\
& D_{\min}^\eps(\rho\|\sigma)\geq D_{\max}^{\sqrt{1-\eps^2}}(\rho\|\sigma)+\log\left(\frac\eps{1-\eps}\right)\label{ubo3}
\end{align}
where $h(\eps)\eqdef-\eps\log\eps-(1-\eps)\log(1-\eps)$ is the binary Shannon entropy.
\end{theorem}
\end{myt}
\noindent We prove each of the inequalities separately:
\begin{proof}[Proof of inequality~\eqref{ubo1}]
Let $\trho\in\md(A)$ be such that $D_{\max}^{\eps_1}(\rho\|\sigma)=D_{\max}(\trho\|\sigma)$ and $\frac12\|\trho-\rho\|_1\leq\eps_1$. From~\eqref{decomom} it follows that there exists $\omega_\pm\in\md(A)$ such that
\be
\trho=\rho+\eps_1(\omega_+-\omega_-)
\ee
where without loss of generality we assumed the equality $\frac12\|\trho-\rho\|_1=\eps_1$. Denote $r\eqdef D_{\max}^{\eps_1}(\rho\|\sigma)$ so that $\trho\leq 2^r\sigma$. Combining this with the inequality $\rho\leq\trho+\eps_1\omega_-$ (that follows from the equation above) gives
\be\label{gh01}
\rho \leq2^r\sigma+\eps_1\omega_-\;.
\ee
Finally, let $\Lambda\in\eff(A)$ be the optimal operator satisfying 
\be\label{gh02}
D_{\min}^{\eps_2}(\rho\|\sigma)=-\log\tr\left[\sigma \Lambda\right]\quad\text{and}\quad\tr[\rho \Lambda]\geq 1-\eps_2\;.
\ee
Then, 
\ba
1-\eps_2&\leq\tr[\rho \Lambda]\\
\GG{\eqref{gh01}}&\leq 2^r\tr[\sigma \Lambda]+\eps_1\tr[\omega_-\Lambda]\\
\Gg{\eqref{gh02}\text{ and }\Lambda\leq I}&\leq 2^{r-D_{\min}^{\eps_2}(\rho\|\sigma)}+\eps_1\;.
\ea
Recalling the definition of $r$, we get from the equation above that
\be
D_{\max}^{\eps_1}(\rho\|\sigma)-D_{\min}^{\eps_2}(\rho\|\sigma)\geq\log(1-\eps_1-\eps_2)\;.
\ee
This completes the proof.
\end{proof}

\begin{proof}[Proof of inequality~\eqref{ubo2}]
Let $\Lambda\in\eff(A)$ be the optimal effect such that $D^\eps_{\min}(\rho\|\sigma)=-\log\tr[\Lambda\sigma]$ and $\tr[\Lambda\rho]= 1-\eps$. Define the channel $\mE\in\cptp(A\to X)$, with $|X|=2$, as
\be
\mE(\omega)\eqdef\tr[\Lambda\omega]|0\lr 0|^X+\tr[(I-\Lambda)\omega]|1\lr 1|^X\quad\quad\forall\;\omega\in\ml(A)\;.
\ee
By definition, $\mE(\rho)=\diag(1-\eps,\eps)$ and $\mE(\sigma)=\diag(t,1-t)$, where
$t\eqdef2^{-D^\eps_{\min}(\rho\|\sigma)}$ and $\diag(\cdot,\cdot)$ denotes $2\times 2$ diagonal matrix.
With these definitions we get from the DPI that
\ba
D(\rho\|\sigma)\geq D\left(\mE(\rho)\big\|\mE(\sigma)\right)&=D\left(\diag(1-\eps,\eps)\big\|\diag(t,1-t)\right)\\
&=-h(\eps)-(1-\eps)\log t-\eps\log\left(1-t\right)\\
\Gg{-\eps\log\left(1-t\right)\geq 0,\;\log t=-D_{\min}^\eps(\rho\|\sigma)}&\geq -h(\eps)+(1-\eps)D_{\min}^\eps(\rho\|\sigma)\;.
\ea
This completes the proof.
\end{proof}

\begin{proof}[Proof of inequality~\eqref{ubo3}]
Recall from~\eqref{z1z} that there exists $t\in\mbb{R}_+$ and $\omega\in\pos(A)$ such that $t\rho\leq\sigma+\omega$ and
\be\label{zxa}
2^{-D^{\eps}_{\min}\left(\rho\|\sigma\right)}=(1-\eps)t-\tr[\omega]\;.
\ee
Observe that since $2^{-D^{\eps}_{\min}\left(\rho\|\sigma\right)}\geq 0$ the equation above implies in particular that
\be\label{hela}
\tr[\omega]\leq (1-\eps)t\;.
\ee
Since we would like to relate between $D^{\eps}_{\min}\left(\rho\|\sigma\right)$ and a smoothing\index{smoothing} of $D_{\max}\left(\rho\|\sigma\right)$, we rewrite $t\rho\leq\sigma+\omega$ in a way that involves $D_{\max}(\rho'\|\sigma)$ for some $\rho'\in\md(A)$ and then estimate how close $\rho'$ to $\rho$. For this purpose, we define the operator
$G\eqdef\sigma^{\frac12}(\sigma+\omega)^{-\frac12}$ and observe that 
that
\ba
tG\rho G^*&\leq G(\sigma+\omega)G^*\\&=\sigma\;.
\ea
Moreover, denoting by $\Lambda\eqdef G^*G$ we get that
\ba\label{10114r}
\Lambda&=(\sigma+\omega)^{-\frac12}\sigma(\sigma+\omega)^{-\frac12}\\
\GG{\sigma=\sigma+\omega-\omega\to}
&\leq I^A-(\sigma+\omega)^{-\frac12}\omega(\sigma+\omega)^{-\frac12}\\
&\leq I^A\;.
\ea
Hence, $\Lambda\in\eff(A)$. Finally, denoting by $\rho'\eqdef \frac{G\rho G^*}{\tr[\Lambda\rho]}$ (so that $\rho'\in\md(A)$), the condition $t\rho\leq\sigma+\omega$ implies that $t\tr[\Lambda\rho]\rho'\leq\sigma$, so that
\be
D_{\max}(\rho'\|\sigma)\leq-\log\big(t\tr[\Lambda\rho]\big)\;.
\ee
Next, we estimate $\tr[\Lambda\rho]$:
\ba
\tr[\Lambda\rho]&=1-\tr\left[(I-\Lambda)\rho\right]\\
\Gg{t\rho\leq\sigma+\omega}&\geq 1-\frac1t\tr\left[(I-\Lambda)(\sigma+\omega)\right]\\
\Gg{\tr\left[\Lambda(\sigma+\omega)\right]=1}&=1-\frac{\tr[\omega]}{t}\\
\GG{\eqref{hela}}&\geq \eps\;,
\ea
where we used the expression for $\Lambda$ in~\eqref{10114r} to get that $\tr\left[\Lambda(\sigma+\omega)\right]=1$. Combining the two equations above gives
$D_{\max}(\rho'\|\sigma)\leq-\log(t\eps)$. To estimate $t$, we use the fact that $\tr[\omega]\geq 0$ to get from the relation in~\eqref{zxa} that $2^{-D^{\eps}_{\min}\left(\rho\|\sigma\right)}\leq (1-\eps)t$. Substituting this lower bound on $t$, into the inequality $D_{\max}(\rho'\|\sigma)\leq-\log(t\eps)$ gives
\ba
D_{\max}(\rho'\|\sigma)&\leq-\log\left(\frac\eps{1-\eps}2^{-D^{\eps}_{\min}\left(\rho\|\sigma\right)}\right)\\
&=D^{\eps}_{\min}\left(\rho\|\sigma\right)-\log\left(\frac\eps{1-\eps}\right)\;.
\ea
It is therefore left to show that $\rho'$ is $\sqrt{1-\eps^2}$-close to $\rho$.

Let $|\psi^{A\tA}\ra\eqdef(\sqrt{\rho}\otimes I)|\Omega^{A\tA}\ra$ and $|\tpsi^{A\tA}\ra\eqdef (G\otimes I)|\psi^{A\tA}\ra$. Observe that $\psi^{A\tA}$ and $\tpsi^{A\tA}$ are purifications of $\rho$ and $\trho\eqdef G\rho G^*$, respectively. Moreover, observe that $|{\psi'}^{A\tA}\ra\eqdef \frac1{\sqrt{\tr[\trho]}}|\tpsi^{A\tA}\ra$ is a purification of $\rho'$. 
From Uhlmann's theorem the fidelity between $\rho$ and $\rho'$ satisfies:
\ba\label{10116a}
F(\rho,\rho')&\geq\big|\la{\psi'}^{A\tA}|\psi^{A\tA}\ra\big|\\
\Gg{\tr[\trho]\leq 1}&\geq\big|\la\tpsi^{A\tA}|\psi^{A\tA}\ra\big|\\
\GG{\text{Real Part}}&\geq\frac12\left(\la\tpsi^{A\tA}|\psi^{A\tA}\ra+\la\psi^{A\tA}|\tpsi^{A\tA}\ra\right)\\
\Gg{P\eqdef\frac12(G+G^*)}&=\la\psi^{A\tA}|P\otimes I^{\tA}|\psi^{A\tA}\ra\\
&=\tr[\rho P]\;.
\ea
Combining this with the fact that $P\leq I^A$ (see Exercise~\ref{contraction}) we obtain
\ba\label{10116}
F(\rho,\rho')\geq\tr[\rho P]&=1-\tr[\rho(I-P)]\\
\Gg{t\rho\leq\sigma+\omega}&\geq 1-\frac1t\tr[(\sigma+\omega)(I-P)]\\
\GG{\text{By definition of }P}&=1-\frac1t-\frac{\tr[\omega]}{t}+\frac1{t}\tr\left[\sigma^{\frac12}(\sigma+\omega)^{\frac12}\right]\\
\GG{(\sigma+\omega)^{\frac12}\geq\sigma^{\frac12}}&\geq 1-\frac{\tr[\omega]}{t}\\
\GG{\eqref{hela}}&\geq\eps\;.
\ea
Therefore, from the relation~\eqref{fitr} between the trace distance and the fidelity we get $\frac12\|\rho-\rho'\|_1\leq\sqrt{1-\eps^2}$.
This completes the proof.
\end{proof}

The relation between the smoothed max and min entropies can be used to obtained a generalized version of the AEP property.

\begin{myg}{}
\begin{corollary}\label{coraep}
For any $\rho,\sigma\in\md(A)$ with $\supp(\rho)\subseteq\supp(\sigma)$, and any $\eps\in(0,1)$
\be\label{qaep}
\lim_{n\to\infty}\frac1nD^\eps_{\max}\left(\rho^{\otimes n}\big\|\sigma^{\otimes n}\right)=D(\rho\|\sigma)\;.
\ee
\end{corollary}
\end{myg}

\begin{proof}
From~\eqref{ubo1} we get that for any $\eps_1,\eps_2\in(0,1)$ with $\eps_1+\eps_2<1$ 
\ba
\liminf_{n\to\infty}\frac1nD_{\max}^{\eps_1}\left(\rho^{\otimes n}\|\sigma^{\otimes n}\right)&\geq \liminf_{n\to\infty}\frac1n\left(D_{\min}^{\eps_2}\left(\rho^{\otimes n}\|\sigma^{\otimes n}\right)+\log\left(1-\eps_1-\eps_2\right)\right)\\
&=\liminf_{n\to\infty}\frac1n D_{\min}^{\eps_2}\left(\rho^{\otimes n}\|\sigma^{\otimes n}\right)\\
\GG{\eqref{7171}}&=D(\rho\|\sigma)\;.
\ea
Conversely, from~\eqref{ubo3}, after replacing the roles between $\delta\eqdef\sqrt{1-\eps^2}$ and $\eps$, we get for every $\eps\in(0,1)$
\ba
\limsup_{n\to\infty}\frac1nD_{\max}^{\eps}\left(\rho^{\otimes n}\|\sigma^{\otimes n}\right)&\leq \limsup_{n\to\infty}\frac1n\left(D_{\min}^{\delta}\left(\rho^{\otimes n}\|\sigma^{\otimes n}\right)-\log\left(\frac{\delta}{1-\delta}\right)\right)\\
&=\limsup_{n\to\infty}\frac1n D_{\min}^{\delta}\left(\rho^{\otimes n}\|\sigma^{\otimes n}\right)\\
\GG{\eqref{7171}}&=D(\rho\|\sigma)\;.
\ea
From the two equations above it follows that~\eqref{qaep} must hold. 
\end{proof}

The technique applied in the aforementioned theorem, especially in the proof of \eqref{ubo3}, is also applicable for upper-bounding the smoothed max relative entropy\index{max relative entropy}. This can be achieved by smoothing the second argument of $D_{\max}$. For further details, please refer to Appendix~\ref{SmoothSec}.

\bex
Let $\rho\in\md(AB)$ and $\eps\in(0,1)$. The smoothed version of $H_{\min}$ and $H_{\max}$ (see Definition~\ref{cme}) are defined, respectively, as
\be
H_{\min}^{\eps}(A|B)_{\rho}\eqdef\max_{\rho'\in\mb_\eps(\rho)}H_{\min}(A|B)_{\rho'}\quad\text{and}\quad H_{\max}^{\eps}(A|B)_{\rho}\eqdef\min_{\rho'\in\mb_\eps(\rho)}H_{\max}(A|B)_{\rho'}\;.
\ee
\ben
\item Show that 
\be
\lim_{n\to\infty}\frac1nH_{\min}^{\eps}\left(A^n\big|B^n\right)_{\rho^{\otimes n}}=H(A|B)_\rho\;.
\ee
Hint: Use Corollary~\ref{coraep}.
\item Show that 
\be\label{hmaxaep}
\lim_{n\to\infty}\frac1nH_{\max}^{\eps}\left(A^n\big|B^n\right)_{\rho^{\otimes n}}=H(A|B)_\rho\;.
\ee
Hint: Use the duality relation between $H_{\min}(A|B)$ and $H_{\max}(A|B)$.
\een
\eex

\subsubsection{Closed Formula for the Smoothing of an Entropy Function}\index{smoothing}

Generally, the task of smoothing resource measures or relative entropies tends not to result in functions with closed forms. Yet, this scenario is notably different when it comes to the smoothing of entropy functions. Here, the utilization of the concepts of standard\index{steepest approximation} and flattest approximations of a probability vector, as delineated in Sec.~\ref{sec:am} on approximate majorization\index{approximate majorization}, becomes particularly relevant. Intriguingly, these approximations reveal that a smoothed entropy function can indeed be articulated in a closed form. This insight introduces an element of simplicity and exactness into a domain typically characterized by its complexity.

Let $\H$ be a quantum entropy and $\eps\in[0,1)$. Then, the $\eps$-smoothed version of $\H$ is defined as
\be\label{optimir}
\H^\eps(\rho)\eqdef\max_{\rho'\in\mb_\eps(\rho)}\H\left(\rho'\right)\;.
\ee
Observe that this definition is consistent with the definition of a smoothed relative entropy. Specifically, if $\H$ is related to a quantum divergence $\D$ as $\H(\rho)=\log|A|-\D(\rho\|\u)$ then $\H^\eps$ as defined above is related to the smooth relative entropy $\D^\eps$ as $\H^\eps(\rho)=\log|A|-\D^\eps(\rho\|\u)$.

\begin{myt}{}
\begin{theorem}
Let $\H$ be a quantum entropy and let $\eps\in[0,1)$. Then, the $\eps$-smoothed version of $\H$ is given by
\be\label{fromjjj}
\H^\eps(\rho)=\H\left(\up^{(\eps)}\right)\quad\quad\forall\;\rho\in\md(A)\;,
\ee
where $\up^{(\eps)}$ is the probability vector defined in~\eqref{599a}.
\end{theorem}
\end{myt}

\begin{proof}
Let $\rho'$ be an optimal quantum state in $\mb_\eps(\rho)$ such that $\H^\eps(\rho)=\H(\rho')$. We first argue that without loss of generality we can assume that $\rho'$ commutes with $\rho$. To see this, let $\Delta\in\cptp(A\to A)$ be the completely dephasing channel in the eigenbasis of $\rho$. Then, since $\Delta$ is a doubly stochastic channel we have
$\H\left(\Delta(\rho')\right)\geq\H\left(\rho'\right)$. Moreover, since $\Delta(\rho)=\rho$ we have
\ba
\frac12\left\|\Delta\left(\rho'\right)-\rho\right\|_1&=\frac12\left\|\Delta\left(\rho'\right)-\Delta(\rho)\right\|_1\\
\GG{DPI}&\leq \frac12\left\|\rho'-\rho\right\|_1\leq\eps\;.
\ea
Hence, $\Delta\left(\rho'\right)$ is also $\eps$-close to $\rho$, so that $\Delta(\rho')$ is also an optimizer of~\eqref{optimir}. Hence, without loss of generality we can assume that $\rho'$ is diagonal in the same eigenbasis of $\rho$.

Let $\p=\p^\da$ be the vector consisting of the eigenvalues of $\rho$. From the argument above we can epress the smoothed entropy in~\eqref{optimir} as
\be\label{optimir2}
\H^\eps(\rho)=\max_{\p'\in\mb_\eps(\p)}\H\left(\p'\right)\;.
\ee
Now, since $\mb_\eps(\p)$ has the property that for every $\p'\in\mb_\eps(\p)$ the vector $\up^{(\eps)}$ as defined in~\eqref{599a} satisfies $\p'\succ\up^{(\eps)}$ so that $\H(\p')\leq  \H(\up^{(\eps)})$. Therefore, the choice $\p'=\up^{(\eps)}$ gives the maximum value.
\end{proof}

As a simple example, consider the min-entropy\index{min-entropy} as defined in~\eqref{qminen}; i.e., 
$H_{\min}(\rho)=-\log\|\rho\|_{\infty}$ for all $\rho\in\md(A)$. Note that this entropy is related to the max relative entropy\index{max relative entropy} via $H_{\min}(\rho)=\log|A|-D_{\max}(\rho\|\u)$. From the theorem above, $H_{\min}^\eps(\rho)=-\log\left\|\up^{(\eps)}\right\|_{\infty}$. Using the definition of $\up^{(\eps)}$ in~\eqref{599a} we get from~\eqref{fromjjj} that
\ba
H_{\min}^\eps(\rho)&=-\log(a)\\
\GG{\eqref{ab5p45}}&=\log\left(\frac{k}{\|\p\|_{(k)}-\eps}\right)\;,
\ea
where $k$ is the integer satisfying~\eqref{5pp52} which is equivalent to
\be
p_{k+1}<\frac{\|\p\|_{(k)}-\eps}{k}\leq  p_{k}\;.
\ee
Alternatively, observe that from~\eqref{altera} we can also express $H^{\eps}_{\min}(\rho)$ as
\be\label{fofhmn}
H_{\min}^\eps(\rho)=-\log\max_{\ell\in[n]}\left\{\frac{\|\p\|_{(\ell)}-\eps}{\ell}\right\}\;.
\ee

It is worth noting that for the case that $\H=H_{\max}$ (recall that $H_{\max}(A)_\rho\eqdef\log\rank(\rho)$ for all $\rho\in\md(A)$), the definition in~\eqref{optimir} results with a quantity that always equals $\log|A|$ since for any $\eps\in(0,1)$ and any $\rho\in\md(A)$, there always exists a full rank state $\rho'\in\md(A)$ that is $\eps$-close to $\rho$. Therefore, in this case instead of taking the maximum in~\eqref{optimir} we take the minimum, so that the smoothed version of $H_{\max}$ is defined as
\be\label{maxhmax}
H^\eps_{\max}(A)_\rho\eqdef\min_{\rho'\in\mb_\eps(\rho)}H_{\max}\left(A\right)_{\rho'}\;.
\ee
Also this formula has a closed form.
\begin{myg}{}
\begin{lemma}\label{lemofhmax}
Let $\eps\in[0,1)$ and $\rho\in\md(A)$. Then, the $\eps$-smoothed max-entropy\index{max-entropy} is given by
\be
H^\eps_{\max}(A)_\rho=\log(m)
\ee
where $m$ is the integer satisfying $\|\rho\|_{(m-1)}<1-\eps\leq \|\rho\|_{(m)}$.
\end{lemma}
\end{myg}
\begin{proof}
From its definition we get
\ba
H_{\max}^\eps\left(\rho^A\right)&=\min\left\{\log\rank(\rho')\;:\;T(\rho,\rho')\leq\eps\right\}\\
\Gg{m\eqdef\rank(\rho')}&=\min\left\{\log m\;:\;T\big(\rho,\md_m(A)\big)\leq\eps\right\}\;,
\ea
where we used the notation $\md_m(A)$ to denote the set of all density matrices in $\md(A)$, whose rank is not greater than $m$. In Theorem~\ref{0distdm0}
we showed that $T\big(\rho,\md_{m}(A)\big)=1-\|\rho\|_{(m)}$. Substituting this to the equation above we conclude that
\ba
H_{\max}^\eps\left(A\right)_\rho=\min\left\{\log m\;:\;\|\rho\|_{(m)}\geq1-\eps\right\}\;.
\ea
This completes the proof.
\end{proof}

\bex Using the same notations as above:
\ben
\item Show that if we replace the maximum in~\eqref{optimir2} with a minimum we get
\be
\max_{\p'\in\mb_\eps(\p)}\H\left(\p'\right)=\H\left(\overline{\p}^{(\eps)}\right)\;,
\ee
where $\overline{\p}^{(\eps)}$ is the steepest $\eps$-approximation of $\p$ as defined in~\eqref{a556}. 
\item Show that if $\rho,\sigma\in\md(A)$ are such that $\rho\succ\sigma$ then
\be
H_{\max}^\eps\left(A\right)_\rho\leq H_{\max}^\eps\left(A\right)_\sigma\;.
\ee
\een
\eex

\subsubsection{Another Smoothed Version of $D_{\min}$}\index{hypothesis testing}

As previously mentioned, the smoothed version of $D_{\min}$ given by $\min_{\rho'\in\mb_{\eps}(\rho)}D_{\min}\left(\rho'\|\sigma\right)$  is always zero. Hence, we've recognized the hypothesis testing divergence as its smoothed variant. Yet, one might seek to define the smoothed version of the min relative entropy\index{min relative entropy} using a maximum instead of a minimum. Specifically, for any  $\rho,\sigma\in\md(A)$ and $\eps\in[0,1)$, we introduce another smoothed version of $D_{\min}$ as
\be\label{defdmine}
D_{\min}^{(\eps)}(\rho\|\sigma)\eqdef\max_{\rho'\in\mb_{\eps}(\rho)}D_{\min}\left(\rho'\|\sigma\right)\;.
\ee
It's crucial to emphasize, however, that this quantity is not necessarily a divergence, given the optimization uses a maximum instead of a minimum (thus, Lemma~\ref{lemsmot} isn't applicable). Nevertheless, in this book, we will encounter this function in some applications.

\bex\label{exhmaxm}
Show that for every $\rho\in\md(A)$ we have
\be
H_{\max}^\eps(A)_\rho=\log|A|-D_{\min}^{(\eps)}(\rho\|\u)\;.
\ee
\eex

\begin{myt}{}
\begin{theorem}\label{thmdminpe}
Let $\rho,\sigma\in\md(A)$ and $\eps\in[0,1)$. Then,
\be
D_{\min}^{(\eps)}(\rho\|\sigma)\leq D^\eps_{\min}\left(\rho\|\sigma\right)\;.
\ee
\end{theorem}
\end{myt}

\begin{proof}
Let $\trho\in\md(A)$ be a state that satisfies $D_{\min}^{(\eps)}(\rho\|\sigma)=D_{\min}(\trho\|\sigma)$. By definition, $\trho$ is $\eps$ close to $\rho$ and from the relation~\eqref{td} we get that
\be
\max_{0\leq \Lambda\leq I}\tr[(\trho-\rho)\Lambda]\leq\eps\;.
\ee
In particular, taking $\Lambda=\Pi_{\trho}$ to be the projection to the support of $\trho$ we obtain 
\be
\tr\left[\rho\Pi_{\trho}\right]\geq 1-\eps\;.
\ee
Therefore, taking $\Lambda=\Pi_{\trho}$ in the definition~\eqref{dehl} we conclude
\ba
D_{\min}^\eps(\rho)&\geq-\log\tr\left[\sigma\Pi_{\trho}\right]\\
&=D_{\min}(\trho\|\sigma)=D_{\min}^{(\eps)}(\rho\|\sigma)\;.
\ea
This completes the proof.
\end{proof}

\bex\label{exdminrs}
Let $\rho,\sigma\in\md(A)$ and $\eps\in(0,1)$. Show that
\be
D_{\min}^{(\eps)}\left(\rho\big\|\sigma\right)\geq D_{\min}\left(\frac{\sqrt{\Lambda}\rho\sqrt{\Lambda}}{\tr[\Lambda\rho]}\Big\|\sigma\right)\;,
\ee
for any $\Lambda\in\eff(A)$ that satisfies $\tr[\Lambda\rho]\geq 1-\eps^2$.
Hint: Use the gentle measurement lemma (Lemma~\ref{gentle1}).
\eex

In the quantum Stein's lemma\index{Stein's lemma} (Theorem~\ref{qsl}) we saw that the regularization\index{regularization} of $D_{\min}^\eps$ yields the Umegaki relative entropy. We show now that the same holds also for $D_{\min}^{(\eps)}$. We will use this result later on when we discuss the uniqueness of the Umegaki relative entropy.

\begin{myt}{}
\begin{theorem}\label{lemma1142}
Let $\eps\in(0,1)$, and $\rho,\sigma\in\md(A)$ with $\supp(\rho)\subseteq\supp(\sigma)$. Then,
\be
\lim_{n\to\infty}\frac{1}{n}D_{\min}^{(\eps)}(\rho^{\otimes n}\|\sigma^{\otimes n})= D(\rho\|\sigma)\;.
\ee
\end{theorem}
\end{myt}{}

\begin{proof}
From Theorem~\ref{thmdminpe} we get that
\ba
\limsup_{n\to\infty}\frac{1}{n}D_{\min}^{(\eps)}(\rho^{\otimes n}\|\sigma^{\otimes n})&\leq \limsup_{n\to\infty}\frac{1}{n}D_{\min}^{\eps}(\rho^{\otimes n}\|\sigma^{\otimes n})\\
\GG{Theorem~\ref{qsl}}&=D(\rho\|\sigma)\;.
\ea
In order to prove the opposite inequality, we make use of  the method of relative typical subspace introduced in Sec.~\ref{rts}. Set $\eps,\delta\in(0,1)$ and let $\Pi_{\delta}^{\rel,n}$ be the projection to the relative typical subspace given in~\eqref{tysub}, $P_{\delta}^n$ be the projection to the $\delta$-typical subspace associated with $\rho$, and define 
	\be
	\rho_n\eqdef\frac{\Pi_{\delta}^{\rel,n}P_{\delta}^n\rho^{\otimes n}P_{\delta}^n\Pi_{\delta}^{\rel,n}}{\tr\left[\Pi_{\delta}^{\rel,n}P_{\delta}^n\rho^{\otimes n}\right]}\;.
	\ee
Observe that for any $\delta_1,\delta_2>0$ we have for large enough $n$ 
\ba
\tr\left[\Pi_{\delta}^{\rel,n}P_{\delta}^n\rho^{\otimes n}\right]&=\tr\left[\Pi_{\delta}^{\rel,n}\rho^{\otimes n}\right]-\tr\left[\Pi_{\delta}^{\rel,n}\left(I^{A^n}-P_{\delta}^n\right)\rho^{\otimes n}\right]\\
&\geq\tr\left[\Pi_{\delta}^{\rel,n}\rho^{\otimes n}\right]-\tr\left[\left(I^{A^n}-P_{\delta}^n\right)\rho^{\otimes n}\right]\\
\GG{\eqref{647},\eqref{643}}&\geq1-\delta_1-\delta_2\;,
\ea
where we use the properties of typical and relative-typical projectors. Taking $\delta_1+\delta_2=\eps^2$ and using  the gentle measurement lemma (see Lemma~\ref{gentle1}) we get that	 $\frac12\left\|\rho_n-\rho^{\otimes n}\right\|_1\leq\eps$.
Hence,
\be
D_{\min}^{(\eps)}(\rho^{\otimes n}\|\sigma^{\otimes n})\geq D_{\min}(\rho_ n\|\sigma^{\otimes n})=-\log\tr\left[\Pi_{\rho_n}\sigma^{\otimes n}\right]\;.
\ee
Now, we make two observations about $\Pi_{\rho_n}$: (1) It projects into a subspace of $\mt^{\rel}_{\eps}(A^n)$, and (2) its trace is no greater than 
\be\label{ranrn}
\rank(\rho_n)\leq\tr\left[P_{\delta}^n\right]\leq2^{n(H(A)_\rho+\delta)}\;.
\ee 
Due to the first property, it follows by definition~\eqref{tysub} of the relative-typical subspace that $\tr\left[\Pi_{\rho_n}\sigma^{\otimes n}\right]\leq2^{n\left(\tr[\rho\log\sigma]+\delta\right)}\tr\left[\Pi_{\rho_n}\right]$.
Combining this with the above equation we conclude that for sufficiently large $n$
\ba
D_{\min}^{(\eps)}(\rho^{\otimes n}\|\sigma^{\otimes n})&\geq-\log\left(2^{n\left(\tr[\rho\log\sigma]+\delta\right)}\tr\left[\Pi_{\rho_n}\right]\right)\\
&=-n\left(\tr[\rho\log\sigma]+\delta\right)-\log\tr\left[\Pi_{\rho_n}\right]\\
\GG{\eqref{ranrn}}&\geq -n\left(\tr[\rho\log\sigma]+\delta\right)-n\left(H(A)_\rho+\delta\right)\\
&=n\left(D(\rho\|\sigma)-2\delta\right)
\ea
Dividing both sides by $n$ and taking the limit $n\to\infty$ we conclude that
\be
\liminf_{n\to\infty}\frac1nD_{\min}^{(\eps)}(\rho^{\otimes n}\|\sigma^{\otimes n})\geq D(\rho\|\sigma)-2\delta\;.
\ee
Since the above inequality holds for all $\delta>0$ it must also hold for $\delta=0$. This completes the proof.
\end{proof}

As a simple application of the result above, consider the smoothed max-entropy\index{max-entropy} as defined in~\eqref{maxhmax}. Then, from Exercise~\ref{exhmaxm} and the theorem above we get the following version of the AEP: For all $\eps\in(0,1)$ and all $\rho\in\md(A)$
\be\label{AEPmaxversion}
\lim_{n\to\infty}\frac1nH_{\max}^\eps(A^n)_{\rho^{\otimes n}}=H(A)_\rho\;,
\ee
where $H(A)_\rho$ is the von-Neumann entropy of $\rho$.

\bex
Prove this AEP version, and compare it with~\eqref{hmaxaep} for the case $|B|=1$.
\eex

\subsection{Smoothed Decoupling Theorem}\label{smoothedd}\index{decoupling theorem} 

In this subsection we show that the decoupling theorem as given in Theorem~\ref{thm1141} can be expressed with smoothed quantities. In particular, we will replace the optimized conditional entropies $\tilde{H}_2^\ua(A|E)$ and $\tilde{H}_2^\ua(A|B)$ with the smoothed conditional min-entropies.

\begin{myg}{Decoupling Theorem (Smoothed Version)}
\begin{corollary}\label{thm1141b}
Let $\rho\in\md_{\leq}(AE)$, $\mE\in\cp_{\leq}(A\to B)$, and $\tau^{AB}\eqdef\frac{1}{|A|}J^{AB}_\mE$, where $J_\mE^{AB}$ is the Choi matrix of $\mE^{A\to B}$. 
Then, for any $\eps>0$
\be\label{decoupling}
\int_{\mathfrak{U}(A)}dU^A\;\left\|\mE^{A\to B}\Big(U^A\rho^{AE}\left(U^A\right)^*\Big)-\tau^B\otimes\rho^E\right\|_1\leq 2^{-\frac{1}{2}\big(H_{\min}^\eps(A|E)_\rho+H_{\min}^\eps(A|B)_\tau\big)}+8\eps
\ee
where $\mathfrak{U}(A)$ is the group of all unitary matrices acting on $A$, and $\int_{\mathfrak{U}(A)}dU^A$ denotes the integral over the Haar measure on $\mathfrak{U}(A)$.
\end{corollary}
\end{myg}

\begin{proof}
This corollary concerns with the replacement of the terms involving $H_2$ in the decoupling theorem with the smoothed min-entropy. For this purpose,  let $\trho^{AE}$ and $\ttau^{AB}$ be such that $H_{\min}^\eps(A|E)_\rho=H_{\min}(A|E)_{\trho}$ and $H_{\min}^\eps(A|B)_\tau=H_{\min}(A|B)_{\ttau}$. Note also that by definition $\|\rho^{AE}-\trho^{AE}\|_1\leq2\eps$ and $\|\tau^{AB}-\ttau^{AB}\|_1\leq2\eps$. Denoting by $\tilde{\mE}$ the CP map whose Choi matrix is $\ttau^{AB}$ we get 
\ba
2^{-\frac{1}{2}\big(H_{\min}^\eps(A|E)_\rho+H_{\min}^\eps(A|B)_\tau\big)}&=2^{-\frac{1}{2}\big(H_{\min}(A|E)_{\trho}+H_{\min}(A|B)_{\ttau}\big)}\\
\Gg{H_{\min}\leq H_2}&\geq 2^{-\frac{1}{2}\big(H_{2}(A|E)_{\trho}+H_{2}(A|B)_{\ttau}\big)}\\
\GG{Theorem~\ref{thm1141}}&\geq \int_{\mathfrak{U}(A)}dU^A\;\left\|\tilde{\mE}^{A\to B}\Big(U^A\trho^{AE}\left(U^A\right)^*\Big)-\ttau^B\otimes\trho^E\right\|_1\\
\GG{See~\eqref{seeabelow}\;below}&\geq \int_{\mathfrak{U}(A)}dU^A\;\left\|\tilde{\mE}^{A\to B}\Big(U^A\trho^{AE}\left(U^A\right)^*\Big)-\tau^B\otimes\rho^E\right\|_1-4\eps
\ea
where in the last inequality we used the fact that $\eta\eqdef\tilde{\mE}\left(U\trho U^*\right)\in\pos(BE)$ satisfies
\ba\label{seeabelow}
\|\eta-\tau\otimes\rho\|_1&=\|\eta-(\ttau+\tau-\ttau)\otimes\rho\|_1\\
\GG{Triangle\;inequality}&\leq \|\eta-\ttau\otimes\rho\|_1+\|(\tau-\ttau)\otimes\rho\|_1\\
\Gg{\tau\approx_\eps\ttau}&\leq  \|\eta-\ttau\otimes(\trho+\rho-\trho)\|_1+2\eps\\
\GG{Triangle\;inequality}&\leq \|\eta-\ttau\otimes\trho\|_1+\|\ttau\otimes(\rho-\trho)\|_1+2\eps\\
\Gg{\rho\approx_\eps\trho}&\leq  \|\eta-\ttau\otimes\trho\|_1+4\eps\;.
\ea
Next, observe that 
\be
\tilde{\mE}\Big(U\trho U^*\Big)=\mE\left(U\rho U^*\right)+\mE\Big(U(\trho-\rho)U^*\Big)+(\tmE-\mE)\Big(U\trho U^*\Big)\;.
\ee
so that from the triangle inequality of the trace norm we get
\ba
\big\|\tilde{\mE}\Big(U\trho U^*\Big)&-\tau\otimes\rho\big\|_1\\
&\geq \left\|{\mE}\Big(U\rho U^*\Big)-\tau\otimes\rho\right\|_1-\left\|\mE\Big(U(\trho-\rho) U^*\Big)\right\|_1-\left\|(\tmE-\mE)\Big(U\trho U^*\Big)\right\|_1
\ea
It is therefore left to  bound the average of the last two terms over the group $\mfu(A)$. Denote by $\eta_{\pm}\eqdef(\trho^{AE}-\rho^{AE})_{\pm}$ and $\zeta_{\pm}\eqdef (\ttau^{AB}-\tau^{AB})_{\pm}$. Since $\trho$ and $\ttau$ are $\eps$-close to $\rho$ and $\tau$, respectively, we have $\tr[\eta_{+}+\eta_{-}]\leq 2\eps$ and $\tr[\zeta_{+}+\zeta_{-}]\leq 2\eps$.
Now, denote by $\mN_{\pm}$ the CP maps whose Choi matrices are 
$\zeta_{\pm}$, respectively.
We then have $\tmE-\mE=\mN_+-\mN_-$ and $\trho-\rho=\eta_+-\eta_-$, so that
\ba
 \int_{\mathfrak{U}(A)}dU\;\left\|\mE\Big(U(\trho-\rho) U^*\Big)\right\|_1&=\int_{\mathfrak{U}(A)}dU\;\left\|\mE\Big(U(\eta_+-\eta_-) U^*\Big)\right\|_1\\
 \GG{Triangle\;inequality}&\leq \int_{\mathfrak{U}(A)}dU\;\tr\left[\mE\Big(U\eta_+ U^*\Big)\right]+ \int_{\mathfrak{U}(A)}dU\;\tr\left[\mE\Big(U\eta_- U^*\Big)\right]\\
 \Gg{\int_{\mathfrak{U}(A)}dU\;U^A\eta_\pm^{AE} U^{A*}=\u^A\otimes\eta^E_\pm}&=\tr\left[\mE(\u)\right]\tr\left[\eta_+\right]+\tr\left[\mE(\u)\right]\tr\left[\eta_-\right]\\
 &=\tr[\tau]\big(\tr\left[\eta_+\right]+\tr[\eta_-]\big)\\
&\leq 2\eps\;. 
\ea
Similarly,
\ba
 \int_{\mathfrak{U}(A)}dU\;\left\|(\tmE-\mE)\Big(U\rho U^*\Big)\right\|_1&=\int_{\mathfrak{U}(A)}dU\;\left\|(\mN_+-\mN_-)\Big(U\rho U^*\Big)\right\|_1\\
 &\leq \int_{\mathfrak{U}(A)}dU\;\tr\left[\mN_+\Big(U\rho U^*\Big)\right]+ \int_{\mathfrak{U}(A)}dU\;\tr\left[\mN_-\Big(U\rho U^*\Big)\right]\\
 &=\tr\left[\mN_+(\u^A)\otimes\rho^E\right]+ \tr\left[\mN_-(\u^A)\otimes\rho^E\right]\\
 &=\tr[\zeta_++\zeta_-]\tr[\rho^E]\\
&\leq 2\eps\;. 
\ea
Combining everything we get
\be
2^{-\frac{1}{2}\big(H_{\min}^\eps(A|E)_\rho+H_{\min}^\eps(A|B)_\tau\big)}\geq 
\int_{\mathfrak{U}(A)}dU\;\left\|{\mE}\Big(U\rho U^*\Big)-\tau\otimes\rho\right\|_1-8\eps\;.
\ee
This completes the proof.
\end{proof}

\section{Resource Monotones and Support Functions}\index{support function}\index{resource monotone}\label{sec:geta}

Let $\mf$ be a convex QRT and for any fixed $\eta\in\herm(B)$ define the function $G_\eta:\bigcup_A\md(A)\to\mbb{R}$ as
\be\label{9146}
G_\eta(\rho^A)\eqdef \sup_{\mE\in\mf(A\to B)}\tr\left[\eta^B\mE^{A\to B}\left(\rho^A\right)\right]-\sup_{\omega\in\mf(B)}\tr\left[\eta^B\omega^B\right]\quad\quad\forall\rho\in\md(A)\;.
\ee
\begin{myt}{}
\begin{theorem}
For any $\eta\in\herm(A)$ the function $G_\eta$ is a resource monotone.
\end{theorem}
\end{myt}
\begin{proof}
We prove the following properties:
\ben
\item {\it  Monotonicity\index{monotonicity}.} Let $\mN\in\mf(A\to A')$ and denote $c_\eta\eqdef\sup_{\omega\in\mf(B)}\tr\left[\eta^B\omega^B\right]$. Then,
\ba
G_\eta\left(\mN^{A\to A'}(\rho^A)\right)&=\sup_{\mM\in\mf(A'\to B)}\tr\left[\eta^B\mM^{A'\to B}\left(\mN^{A\to A'}(\rho^A)\right)\right]-c_\eta\\
\Gg{\text{Replacing }\mM\circ\mN\text{ with }\mE}&\leq\sup_{\mE\in\mf(A\to B)}\tr\left[\eta^B\mE^{A\to B}\left(\rho^A\right)\right]-c_\eta\\
&=G_\eta\left(\rho^A\right)\;.
\ea
\item {\it Normalization.} Let $\sigma\in\mf(A)$. Then,
\be
\sup_{\mE\in\mf(A\to B)}\tr\left[\eta^B\mE^{A\to B}\left(\sigma^A\right)\right]=\sup_{\omega\in\mf(B)}\tr\left[\eta^B\omega^B\right]
\ee
where $\omega^B=\mE^{A\to B}\left(\sigma^A\right)\in\mf(B)$ can be taken to be any free state (by choosing $\mE$ to be a replacement channel in $\mf(A\to B)$ that outputs $\omega^B$). Hence, $G_\eta\left(\sigma^A\right)=0$ for all $\sigma\in\mf(A)$.
\item {\it Strong monotonicity.} Let $\mN=\sum_{x\in[m]}\mN_x\otimes|x\lr x|\in\mf(A\to A'X)$ be a free quantum instrument. Then, observe that for any free channel $\mM\in\mf(A'X\to B)$ we have
\be
\mM^{A'X\to B}\circ\mN^{A\to A'X}=\sum_{x\in[m]}\mM^{A'\to B}_x\circ\mN_x^{A\to A'}
\ee
where $\mM_x\in\mf(A'\to B)$ is define by 
\be
\mM_x^{A'\to B}(\omega^{A'})\eqdef\mM^{A'X\to B}\left(\omega^{A'}\otimes|x\lr x|^X\right)\quad\quad\forall\;\omega\in\ml(A')\;.
\ee
We therefore get that
\ba
G_\eta\left(\mN^{A\to A'X}(\rho^A)\right)&=\sup_{\mM\in\mf(A'X\to B)}\tr\left[\eta^B\mM^{A'X\to B}\left(\mN^{A\to A'X}(\rho^A)\right)\right]-c_\eta\\
&=\sum_{x\in[m]}\sup_{\mM_x\in\mf(A'\to B)}\tr\left[\eta^B\mM^{A'\to B}_x\left(\mN^{A\to A'}_x(\rho^A)\right)\right]-c_\eta\\
&=\sum_{x\in[m]}p_xG_\eta\left(\sigma^{A'}_x\right)\;,
\ea
where $\sigma_x^{A'}\eqdef\frac1{p_x}\mN^{A\to A'}_x(\rho^A)$ and $p_x\eqdef\tr\left[\mN^{A\to A'}_x(\rho^A)\right]$. Combining this with the monotonicity property of $G_\eta$ we conclude that
\be
G_\eta\left(\rho^A\right)\geq \sum_{x\in[m]}p_xG_\eta\left(\sigma^{A'}_x\right)\;.
\ee
\item {\it Convexity.} Let $\rho^{AX}\eqdef\sum_{x\in[m]}p_x\rho_x^A\otimes|x\lr x|^X$ be a cq-state in $\md(AX)$. Using a similar argument as above we have
\be
G_\eta\left(\rho^{AX}\right)=\sum_{x\in[m]}p_xG_\eta\left(\rho^{A}_x\right)\;.
\ee
On the other hand, since the partial trace is a free operation we get that
\be
G_\eta\left(\rho^{A}\right)\leq G_\eta\left(\rho^{AX}\right).
\ee
This equation is equivalent to
\be
G_\eta\Big(\sum_{x\in[m]}p_x\rho^{A}_x\Big)\leq \sum_{x\in[m]}p_xG_\eta\left(\rho^{A}_x\right)\;.
\ee
\een
\end{proof}
Recall that the combination of monotonicity and normalization properties ensures that $G_\eta(\rho) \geq 0$ for all density matrices. Additionally, if we define $\mc_\rho \eqdef {\mE(\rho) : \mE \in \mf(A\to B)}$, then the \emph{support function} of $\mc_\rho$ in the space of Hermitian matrices $\herm(B)$ is described by:
\be
f_\rho(\eta)\eqdef\sup_{\omega\in\mc_\rho}\la\omega^B,\eta^B\ra=\sup_{\mE\in\mf(A\to B)}\tr\left[\eta^B\mE^{A\to B}\left(\rho^A\right)\right]\;.
\ee
As we will explore later, this family of resource monotones is \emph{complete}, meaning that it can be utilized to fully determine exact interconversions among resources. Furthermore, these monotones are formulated as conic linear programming\index{linear programming} problems, and in some QRTs, they reduce to semidefinite programming, which are comparatively simpler to compute.
 
 \subsubsection{Example: Measures of Conditional Uncertainty}
 
 Consider a QRT in which the free operations are {\rm CMO} channels (see Sec.~\ref{lbsec}).
 Given that conditional majorization can be determined by an SDP feasibility problem, it follows that also the corresponding support functions can be computed with an SDP. Recall that in this QRT the set of free states is given by
 \be
 \mf(AB)=\left\{\u^A\otimes\sigma^B\;:\;\sigma\in\md(B)\right\}\;.
 \ee
 Therefore, for this QRT, for any $\eta\in\md(AB')$ the coefficient $c_\eta$ is given by 
 \be
 c_\eta=\sup_{\sigma\in\md(B')}\tr\left[\eta^{AB'}\left(\u^A\otimes\sigma^{B'}\right)\right]=\frac1{|A|}\big\|\eta^{B'}\big\|_{\infty}
 \ee
Hence, for every $\eta\in\md(AB')$  the function $G_\eta$ as defined above can be expressed as
\be\label{10155}
G_\eta(A|B)_\rho\eqdef\sup_{\mN\in\cmo(AB\to AB')}\tr\left[\eta^{AB'}\mN^{AB\to AB'}\left(\rho^{AB}\right)\right]-\frac1{|A|}\big\|\eta^{B'}\big\|_{\infty}\;.
\ee 
We denote by $f_\eta$ the first term on the right-hand side above. In terms of the Choi matrix of $\mN$, this function can be expressed as
\be
f_\eta(A|B)_\rho=\sup_{J^{AB\tA B'}}\tr\left[J^{AB\tA B'}\left(\rho^{AB}\otimes\eta^{\tA B'}\right)\right]
\ee
where the maximization is over all Choi matrices of a {\rm CMO}. 

Recall from Exercise~\ref{upsiex} that a Choi matrix $J^{AB\tA B'}$ is a Choi matrix of a {\rm CMO} if and only if $\Upsilon\left(J^{AB\tA B'}\right)=0$. Moreover, observe that the trace-preserving condition $J^{AB}=I^{AB}$ follows from $J^B=|A|I^B$ and $J^{AB B'}=\u^{A}\otimes J^{B B'}$. Hence,
\be
f_\eta(A|B)_\rho=|A|\sup_{J^{AB\tA B'}\geq 0}\left\{\tr\left[J^{AB\tA B'}\left(\rho^{AB}\otimes\eta^{\tA B'}\right)\right]\;:\;\Upsilon\left(J^{AB\tA B'}\right)=0,\;J^B=I^B\right\}\;.
\ee
The above optimization problem can be solved with an SDP. 
\bex
Use the strong duality relation of an SDP to show that the function $f_\eta$ can also be expressed as:
\be
f_\eta(A|B)_\rho=|A|\inf_{\xi^{AB\tA B'}\geq 0}\left\{\tr\left[\xi^{B}\right]\;:\;\u^{A\tA B'}\otimes\xi^B+\Upsilon\left(\xi^{AB\tA B'}\right)\geq \rho^{AB}\otimes\eta^{\tA B'}\right\}\;.
\ee
\eex
\bex
Show that if $|A|=|B|$ then for the maximally entangled state $\rho^{AB}=\Phi^{AB}$ $G_\eta$ as defined in~\eqref{10155} is given by
\be
G_\eta(A|B)_\Phi=\big\|\eta^{AB'}\big\|_{\infty}-\frac1{|A|}\big\|\eta^{B'}\big\|_{\infty}\;.
\ee
Hint: Recall that for all states $\rho\in\md(AB)$ we have $\Phi^{AB}\succ_A\rho^{AB}$.
\eex

\section{Notes and References}

Measures of quantum resources in general QRTs where introduced in~\cite{BG2015} and studied intensively by~\cite{BCV+2018} and~\cite{Regula2018}. 
The asymptotic continuity\index{asymptotic continuity} of the relative entropy of a resource as given in~\eqref{asyfor} is due to~\cite{Winter2016}. The robustness of a resource as defined here is sometimes called the \emph{global} robustness and it was first introduced by~\cite{HN2003} for entanglement theory.

The converse problem introduced here for the relative entropy of a resource was introduced in~\cite{MI2008} for the case of two-qubit entangled states, in~\cite{FG2011} for all finite dimensions in entanglement theory, and finally in~\cite{GGF2014} for arbitrary QRTs. This technique was also used in~\cite{CDGG2020} to compute the $\alpha$-relative entropy of entanglement for pure bipartite states . More additivity properties of the relative entropy of a resource can be found in~\cite{RT2022}.

The inequality~\eqref{ubo1} is due to~\cite{}, the inequality~\eqref{ubo2} is due to~\cite{WR2012}, and the inequality~\eqref{ubo3} is due to~\cite{DKF+2013}. The decoupling theorem presented here is due to~\cite{DBWR2014}. The relations between resource monotones and support functions as presented here is due to~\cite{GS2021}. The streamlined proof of Theorem~\ref{thmdminpe}, as outlined here, was graciously shared with the author by Ryuji Takagi (private communication). An earlier version of the book had featured a substantially more complicated proof.

%%%%%%%%%%%%%%%%%%%%%%%%%%%%%%%%%%%%%%%%%%%%%%%%%%%%%%%%%%%%%%%%%%%%%%%%%%%%%%%%%%%%%%%%%%%%%%%%%%%%%%%%%%%%%%%%%%%%%%%%%%%%

\chapter{Manipulation of Resources}

One of the central goals of QRTs is to understand optimal and efficient ways to convert one resource to another. A resource in this context correspond to a class of equivalent resource-states. We say that two resource-states $\rho,\sigma\in\md(A)$ are equivalent if both $\rho\xrightarrow{\mf}\sigma$ (i.e.\ $\rho$ can be converted to $\sigma$ by free operations) and $\sigma\xrightarrow{\mf}\rho$. In this chapter we study the conversion of resources in two regimes: the single-shot regime and the asymptotic regime. 

\section{Single-Shot Interconversions}\index{single-shot}

The single-shot regime encompasses exact, probabilistic, and approximate interconversions. By ``exact" interconversions, we refer to the transformation of a resource, such as $\rho$, into another target resource, like $\sigma$, with both 100\% success probability and accuracy. However, in practical scenarios, it is frequently unfeasible to achieve such perfect conversion of $\rho$ to $\sigma$, necessitating a tolerance for slight errors. Additionally, we will discover that allowing for a small margin of error not only accommodates practical limitations but also provides theoretical insight. This flexibility facilitates a smoother transition from the single-shot regime to the asymptotic regime, highlighting the interconnectedness and practicality of these concepts in resource theory.

\subsection{Exact Interconversions}

Different QRTs have different sets of free operations. Consequently, one cannot expect to find a set of simple necessary and sufficient conditions that can be used in any QRT to determine the conversion of one resource $\rho$ to another resource $\sigma$. Still, in the theorem below we show that there exists a common dual characterization for the problem of exact state-conversion that is given in terms of the resource monotones discussed in Sec.~\ref{sec:geta}. We will assume that $\mf$ is a closed convex QRT meaning that for every two Hilbert spaces $A$ and $B$ the set $\mf(A\to B)$ is a closed convex set in the real vector space $\herm(A\to B)$. In particular, $\mf(A)$ is a closed convex set in $\herm(A)$.

\begin{myt}{}
\begin{theorem}\label{thm:1111}
Let $\mf$ be a closed convex QRT, $\rho\in\md(A)$, and $\sigma\in\md(B)$. The following are equivalent:
\ben
\item There exists $\mN\in\mf(A\to B)$ such that $\sigma^B=\mN^{A\to B}\left(\rho^A\right)$.
\item For all $\eta\in\md(B)$
\be\label{tri111}
G_\eta\left(\rho^A\right)\geq G_\eta\left(\sigma^B\right)\;,
\ee
where $G_\eta$ are the resource monotones defined in~\eqref{9146}.
\een
\end{theorem}
\end{myt}

\begin{proof}
Let $\mc_\rho\eqdef\left\{\mE^{A\to B}\left(\rho^A\right)\;:\;\mE\in\mf(A\to B)\right\}$. Observe that $\mc_\rho$ is a convex set in $\herm(B)$. From the hyperplane separation theorem (see Theorem~\ref{hypert}), $\sigma\not\in\mc_\rho$ if and only if there exists a hyperplane $\eta\in\herm(B)$ that separates them; that is,
\be
\tr\left[\eta^B\sigma^B\right]>\max_{\omega\in\mc_\rho}\tr\left[\eta^B\omega^B\right]\;.
\ee
Alternatively, $\sigma\in\mc_\rho$ if and only if for \emph{all} $\eta\in\herm(B)$ we have
\ba\label{inholds}
\tr\left[\eta^B\sigma^B\right]&\leq \max_{\omega\in\mc_\rho}\tr\left[\eta^B\omega^B\right]\\
&=\max_{\mE\in\mf(A\to B)}\tr\left[\eta^B\mE^{A\to B}\left(\rho^A\right)\right]\;.
\ea
Note that if the equation above holds for some $\eta\in\herm(B)$ then it also holds if we replace $\eta^B$ with $\eta^B+cI^B$ and vice versa (here $c$ is any real number). Therefore, the equation above holds for all $\eta\in\herm(B)$ if and only if it holds for all $\eta\in\pos(B)$. Similarly, by dividing both sides of the equation above by $\tr[\eta]$ we conclude that  $\sigma\in\mc_\rho$ if and only if~\eqref{inholds} holds for all density matrices $\eta\in\md(B)$. 

Now, observe that $\sigma\in\mc_\rho$ if and only if
for all $\mM\in\mf(B\to B)$ we have that $\mM(\sigma)\in\mc_\rho$. To see this, suppose $\sigma\in\mc_\rho$ so that $\sigma^B=\mE^{A\to B}\left(\rho^A\right)$ for some $\mE\in\mf(A\to B)$. Then, 
$\mM^{B\to B}\left(\sigma^B\right)=\mM^{B\to B}\circ\mE^{A\to B}\left(\rho^A\right)$
and since $\mM\circ\mE\in\mf(A\to B)$ we conclude that $\mM(\sigma)\in\mc_\rho$. Conversely, if $\mM(\sigma)\in\mc_\rho$ for all $\mM\in\mf(B\to B)$, by taking the identity channel $\mM=\id^B\in\mf(B\to B)$ we get immediately that $\sigma\in\mc_\rho$. 

Finally, from~\eqref{inholds} we get that for any $\mM\in\mf(B\to B)$ we have $\mM(\sigma)\in\mc_\rho$ if and only if for all $\eta\in\md(B)$
\be
\tr\left[\eta^B\mM^{B\to B}\left(\sigma^B\right)\right]\leq \max_{\mE\in\mf(A\to B)}\tr\left[\eta^B\mE^{A\to B}\left(\rho^A\right)\right]\;.
\ee
Hence, $\sigma\in\mc_\rho$ if and only if the above equation holds for all $\mM\in\mf(B\to B)$. Taking the maximum over all such $\mM\in\mf(B\to B)$ we conclude that 
$\sigma\in\mc_\rho$ if and only if for all $\eta\in\md(B)$
\be
\max_{\mM\in\mf(B\to B)}\tr\left[\eta^B\mM^{B\to B}\left(\sigma^B\right)\right]\leq \max_{\mE\in\mf(A\to B)}\tr\left[\eta^B\mE^{A\to B}\left(\rho^A\right)\right]\;.
\ee
The proof is concluded by recognizing that the above inequality is equivalent to~\eqref{tri111}.
\end{proof}

In general, the theorem above does not provide an efficient way to determine if one resource can be converted to another by free operations. This is the case even if the resource monotones $G_\eta$ themselves can be computed efficiently, as we need to check the conditions for all $\eta\in\md(B)$. Therefore, instead, we can use~\eqref{inholds} to conclude that $\rho^A$ can be converted to $\sigma^B$ by free operations if and only if
\be\label{abcxz}
\min_{\eta\in\md(B)}\max_{\mE\in\mf(A\to B)}\tr\left[\eta^B\Big(\mE^{A\to B}\left(\rho^A\right)-\sigma^B\Big)\right]\geq 0\;.
\ee
For some QRTs, the optimization problem above is an SDP and therefore can be solved efficiently.

\subsection{Stochastic (Probabilistic) Interconversions}

Probabilistic interconversion of one resource to another can also be considered as an \emph{exact} interconversion except that the conversion does not occur with a 100\% success.  Specifically, let $\rho\in\md(A)$ and $\sigma\in\md(B)$ be two resource states, and let $\pr(\rho\xrightarrow{\mf}\sigma)$ be the maximum probability that $\rho$ can be converted to $\sigma$ by free operations. For some states it may be that $\pr(\rho\xrightarrow{\mf}\sigma)=0$ in which case $\rho$ cannot be converted to sigma even with small probability. In another extreme case, $\pr(\rho\xrightarrow{\mf}\sigma)=1$, meaning that
$\rho$ can be converted to $\sigma$ deterministically. If $0<\pr(\rho\xrightarrow{\mf}\sigma)<1$ we say that $\rho$ can be converted to $\sigma$ \emph{stochastically}.

Any free probabilistic transformation can be characterized with a channel/instrument $\mE\in\mf(A\to BX)$ of the form
\be\label{decompo09}
\mE^{A\to BX}=\sum_{x\in[n]}\mE_x^{A\to B}\otimes |x\lr x|^X\;,
\ee
where each $\mE_x\in\cp(A\to B)$ is trace non-increasing such that $\sum_{x\in[m]}\mE_x$ is trace preserving. From the axiom\index{axiom} of free instruments we require that $\mE_x^{A\to B}(\omega^A)/\tr[\mE_x^{A\to B}(\omega^A)]$ is a free state in $\mf(B)$ whenever $\omega\in\mf(A)$. If there exists such a free quantum instrument with the property that 
\be
\sigma^B=\frac{\mE_x^{A\to B}(\rho^A)}{\tr[\mE_x^{A\to B}(\rho^A)]}
\ee
for some $x\in[n]$, then we say that $\rho$ can be converted to $\sigma$ by free operations with probability $p_x\eqdef \tr[\mE_x^{A\to B}(\rho^A)]$.

Recall that we used the notation $\mf_{\leq}(A\to B)$ to denote the set of trace non-increasing CP maps in $\cp_{\leq}(A\to B)$ that are part of free quantum instruments. With this notation,  $\rho$ can be converted to $\sigma$ by free operations with probability $p$ if and only if there exists $\mE\in\mf_{\leq}(A\to B)$ such that $p=\tr[\mE(\rho)]$ and $\sigma=\mE(\rho)/p$.
Therefore, $\pr(\rho\xrightarrow{\mf}\sigma)$ can be defined as
\be\label{defmaxpro}
\pr(\rho\xrightarrow{\mf}\sigma)\eqdef \max_{\mE\in\mf_{\leq}(A\to B)}\left\{\tr[\mE(\rho)]\;:\;\sigma=\frac{\mE(\rho)}{\tr[\mE(\rho)]}\right\}\;.
\ee
We will now demonstrate that this probability is, in fact, a resource monotone.

\begin{myt}{}
\begin{theorem}\label{prmon}
Let $\sigma\in\md(B)$ be such that $\sigma\not\in\mf(B)$ (i.e. $\sigma$ is a resource state). Then, the function $f_\sigma:\md(A)\to[0,1]$, defined via
\be
f_\sigma(\rho)\eqdef \pr(\rho\xrightarrow{\mf}\sigma)\quad\quad\forall\;\rho\in\md(A)\;,
\ee 
is a resource measure satisfying the strong monotonicity property.
\end{theorem}
\end{myt}

\begin{proof}
First observe that from the axiom\index{axiom} of free instruments and the fact that $\sigma$ is a resource state, we must have $f_\sigma(\rho)=0$ for all $\rho\in\mf(A)$. Next, we show that $f_\sigma$ is a resource measure.
Let $\mN\in\mf(A\to C)$ be a free channel.  Let $\mM\in\mf_{\leq}(C\to B)$ be an optimal free instrument satisfying 
\be
\pr(\mN(\rho)\xrightarrow{\mf} \sigma)=\tr\left[\mM\big(\mN\left(\rho\right)\big)\right]\;. 
\ee
Define $\mE\eqdef\mM\circ\mN$, and observe that $\mE\in\mf_{\leq}(A\to B)$ and $\pr(\mN(\rho)\xrightarrow{\mf} \sigma)=\tr\left[\mE\left(\rho\right)\right]$. Hence, from the definition of $\pr(\rho\xrightarrow{\mf}\sigma)$ in~\eqref{defmaxpro} we get
\be
\pr(\rho\xrightarrow{\mf}\sigma)\geq \pr(\mN(\rho)\xrightarrow{\mf} \sigma)\;.
\ee
By definition, this is equivalent to $f_\sigma(\rho)\geq f_\sigma\big(\mN(\rho)\big)$.
We therefore established that $f_\sigma$ is a resource measure. 

To prove strong monotonicity, let $\mN\in\mf(A\to BY)$ and denote by 
\be
\tau^{BY}\eqdef\mN^{A\to CY}(\rho^A)=\sum_{y\in[n]}t_y\tau_y^C\otimes |y\lr y|^Y
\ee
where each $\tau_y\in\md(C)$ and $\{t_y\}_{y\in[n]}$ is a probability distribution. From the monotonicity of $f_\sigma$ under free channels (in particular, under $\mN$) we get
\ba\label{1113}
f_{\sigma}\left(\rho^A\right)&\geq f_\sigma\left(\tau^{CY}\right)\\
&=\pr\Big(\sum_{y\in[n]}t_y\tau_y^C\otimes |y\lr y|^Y\xrightarrow{\mf}\sigma^B\Big).
\ea
Let $\mE^{(y)}\in\mf_{\leq }(C\to B)$ be an optimal trace non-increasing CP map such that $\pr\left(\tau_y^C\xrightarrow{\mf}\sigma^B\right)=\tr\left[\mE^{(y)}(\tau_y)\right]$ and $\mE^{(y)}(\tau_y)$ is proportional to $\sigma$. We also define $\mM\in\mf_{\leq }(CY\to B)$ as:
\be
\mM^{CY\to B}\left(\omega^C\otimes |y\lr y|^Y\right)\eqdef\mE^{(y)}\left(\omega^C\right)\quad\quad\forall\;\omega\in\ml(C)\quad\forall\;y\in[n]\;.
\ee
In Exercise~\ref{eleof} you show that $\mM^{CY\to B}$ is indeed an element of $\mf_{\leq}(CY\to B)$.
By definition, the state $\mM^{CY\to B}\left(\sum_{y\in[n]}t_y\tau_y^C\otimes |y\lr y|^Y\right)$ is proportional to $\sigma^B$. Therefore,
\ba
\pr\Big(\sum_{y\in[n]}t_y\tau_y^C\otimes |y\lr y|^Y\xrightarrow{\mf}\sigma^B\Big)&\geq\tr\Big[\mM^{CY\to B}\Big(\sum_{y\in[n]}t_y\tau_y^C\otimes |y\lr y|^Y\Big)\Big]\\
&=\sum_{y\in[n]}t_y\tr\left[\mE^{(y)}(\tau_y)\right]\\
&=\sum_{y\in[n]}t_y\pr\Big(\tau_y^C\xrightarrow{\mf}\sigma^B\Big)\\
&=\sum_{y\in[n]}t_yf_\sigma\left(\tau_y^C\right).
\ea
Combining this with~\eqref{1113} we conclude that
\be
f_{\sigma}\left(\rho^A\right)\geq\sum_{y\in[n]}t_yf_\sigma\left(\tau_y^C\right).
\ee
This completes the proof.
\end{proof}

\bex\label{eleof}
Show that $\mM^{CY\to B}$ as defined above belongs to $\mf_{\leq}(CY\to B)$. Hint: Define a free channel $\mF^{CY\to BX}=\sum_{x\in[m]}\mF_{x}^{CY\to B}\otimes |x\lr x|^X$ such that $\mF_{1}^{CY\to B}=\mM^{CY\to B}$.
\eex

\bex
Let $\rho\in\md(A)$ and $\sigma\in\md(B)$. Show that for any $\mM\in\mf(B\to C)$ 
\be
\pr\big(\rho\xrightarrow{\mf}\mM(\sigma)\big)\geq \pr\big(\rho\xrightarrow{\mf}\sigma\big)\;.
\ee
\eex

As discussed above, if $\rho^A$ can be converted by free operations to $\sigma^B$, with some non-zero probability, then there exists a free quantum instrument\index{quantum instrument} $\mE=\sum_{x\in[m]}\mE_x\otimes |x\lr x|\in\mf(A\to BX)$ such that $\mE_1^{A\to B}(\rho^A)$ is proportional to $\sigma^B$. For every $x\in[m]$, let $p_x\eqdef \tr\left[\mE_x^{A\to B}(\rho^A)\right]$,  and for $x\geq 2$ let $\omega_x^B\eqdef\frac1{p_x}\mE_x^{A\to B}(\rho^A)$. With this notations we have
\be
\mE^{A\to BX}(\rho^A)=p_1\sigma^B\otimes |1\lr 1|^X+\sum_{x=2}^{|X|}p_x\omega^B_x\otimes|x\lr x|^X\;.
\ee
Now, let $\M$ be a resource measure that satisfies the strong monotonicity property. Then, by definition $\M$ satisfies
\be
\M(\rho^A)\geq p_1\M(\sigma^B)+\sum_{x=2}^{|X|}p_x\M(\omega_x^B)\geq p_1\M(\sigma^B)\;.
\ee
In other words, the probability $p_1$ to convert $\rho^A$ to $\sigma^B$ cannot exceed the ratio $\M(\rho^A)/\M(\sigma^B)$. Since this is true for all resource measures that satisfies the strong monotonicity property we get that
\be
\pr(\rho^A\xrightarrow{\mf}\sigma^B)\leq\inf_\M\frac{\M(\rho^A)}{\M(\sigma^B)}\;,
\ee
where the infimum is over all resource measures, $\M$, that satisfy the strong monotonicity property. Moreover, from the theorem above, for a fixed $\sigma$, the function $\M_\sigma(\omega^A)\eqdef\pr(\omega^A\xrightarrow{\mf}\sigma^B)$ is itself a resource measure that satisfies the strong monotonicity property. Hence,
\be
\inf_\M\frac{\M(\rho^A)}{\M(\sigma^B)}\leq \frac{\M_\sigma(\rho^A)}{\M_\sigma(\sigma^B)}=\M_\sigma(\rho^A)=\pr(\rho^A\xrightarrow{\mf}\sigma^B)\;.
\ee
We therefore arrive at the following corollary.
\begin{myg}{}
\begin{corollary}\label{cor:mono}
Let $\rho\in\md(A)$ and $\sigma\in\md(B)$. Then,
\be\label{cormono}
\pr(\rho^A\xrightarrow{\mf}\sigma^B)=\inf_\M\frac{\M(\rho^A)}{\M(\sigma^B)}\;,
\ee
where the infimum is over all resource measures, $\M$, that satisfy the strong monotonicity property.
\end{corollary}
\end{myg}

\subsection{Approximate Interconversion}

In this section we provide the precise definitions of cost and distillation of a resource in the single-shot\index{single-shot} regime. We start with the definitions of conversion distance and the ``golden" unit of resource theories.

\subsubsection{The Conversion Distance}\index{conversion distance}

The conversion distance quantifies the proximity to which a resource state $\rho\in\md(A)$ can be transformed into another resource state $\sigma\in\md(B)$ using free operations (refer to Fig.~\ref{conversiondistance}). Mathematically, it is defined as follows:
\be\label{cd}
T\left(\rho\xrightarrow{\mf} \sigma\right)\eqdef\min_{\mE\in\mf(A\to B)}\frac12\left\|\sigma^B-\mE^{A\to B}\left(\rho^A\right)\right\|_1\;.
\ee
It's evident that the conversion distance is zero if $\rho\xrightarrow{\mf} \sigma$ is achievable. However, deterministic conversion from $\rho$ to $\sigma$ is often not feasible, raising the question of how closely $\sigma$ can be approximated by applying free operations to $\rho$. As such, conversion distance not only provides a meaningful way to evaluate the efficiency of these conversions but, as the following lemma demonstrates, also serves as a resource measure in its own right.

\begin{figure}[h]
\centering
    \includegraphics[width=0.5\textwidth]{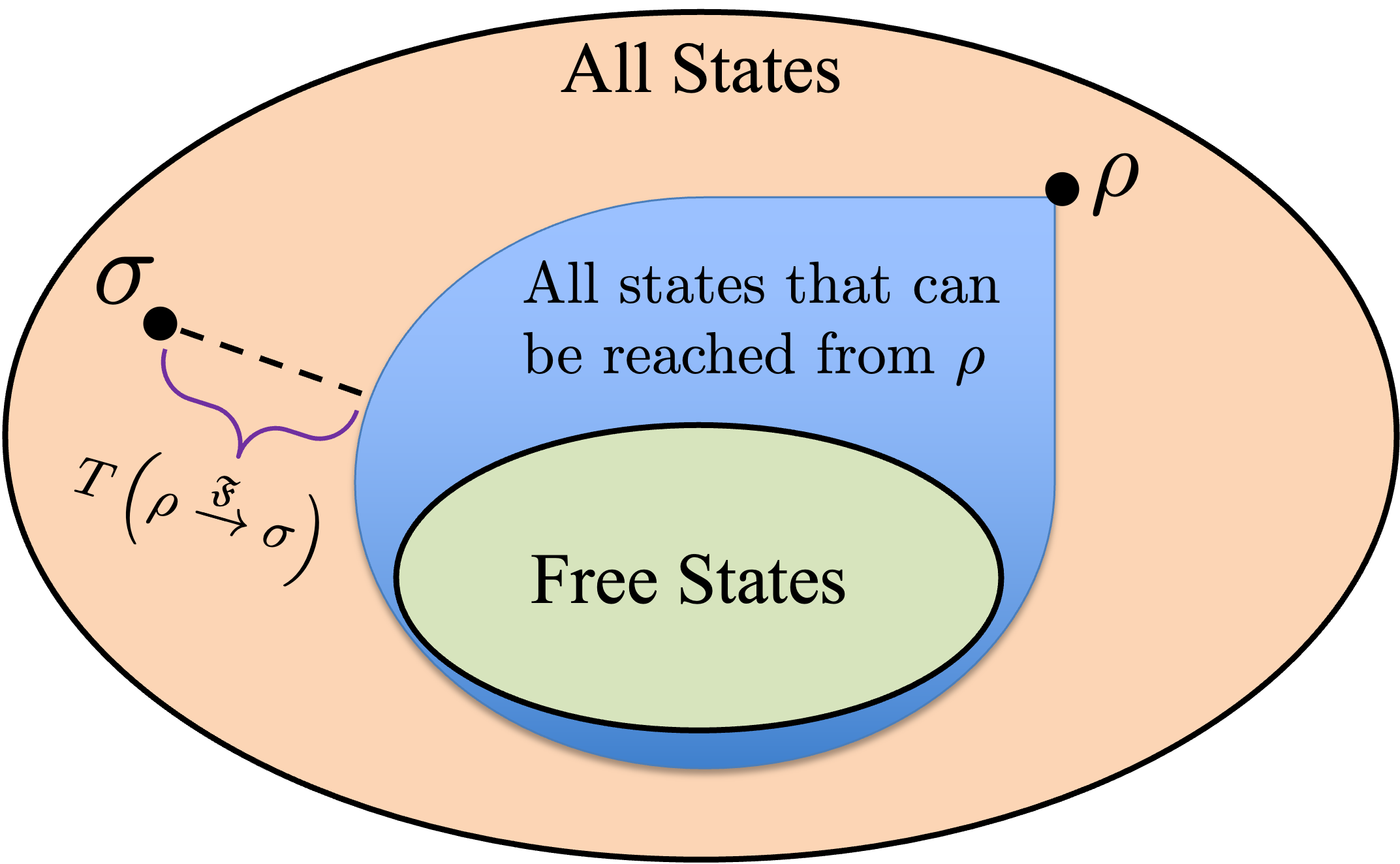}
  \caption{\linespread{1}\selectfont{\small The conversion distance from $\rho$ to $\sigma$.}}
  \label{conversiondistance}
\end{figure} 

\begin{myg}{}
\begin{lemma}\label{lemmonin}
Let $\rho\in\md(A)$, $\sigma\in\md(B)$, $\mM\in\mf(A\to A')$, and $\mN\in\mf(B\to B')$. Then,
\be
T\left(\mM(\rho)\xrightarrow{\mf}\sigma\right)\geq T\left(\rho\xrightarrow{\mf}\sigma\right)\quad\text{and}\quad T\left(\rho\xrightarrow{\mf}\mN(\sigma)\right)\leq T\left(\rho\xrightarrow{\mf}\sigma\right)\;.
\ee
\end{lemma}
\end{myg}

\begin{proof}
For the first inequality, by definition, for any $\mM\in\mf(A\to A')$
\ba
T\left(\mM(\rho)\xrightarrow{\mf}\sigma\right)&=\min_{\mE'\in\mf(A'\to B)}\frac12\left\|\sigma-\mE'\circ\mM\left(\rho\right)\right\|_1\\
\GG{Replacing\;\mE'\circ\mM\;with\;\mE}&\geq \min_{\mE\in\mf(A\to B)}\frac12\left\|\sigma-\mE\left(\rho\right)\right\|_1\\
&=T\left(\rho\xrightarrow{\mf}\sigma\right)\;.
\ea
For the second inequality, we have for any $\mN\in\mf(B\to B')$
\ba
T\left(\rho\xrightarrow{\mf}\mN(\sigma)\right)&=\min_{\mE'\in\mf(A\to B')}\frac12\left\|\mN(\sigma)-\mE'\left(\rho\right)\right\|_1\\
\GG{Taking\;\mE'=\mN\circ\mE}&\leq \min_{\mE\in\mf(A\to B)}\frac12\left\|\mN(\sigma)-\mN\circ\mE\left(\rho\right)\right\|_1\\
\GG{DPI}&\leq \min_{\mE\in\mf(A\to B)}\frac12\left\|\sigma-\mE\left(\rho\right)\right\|_1\\
&=T\left(\rho\xrightarrow{\mf}\sigma\right)\;.
\ea
\end{proof}

\bex\label{ex:tensork}
Let $\mf$ be a QRT, $\rho,\sigma\in\md(A)$, $\eps\in[0,1]$, and $k\in\mbb{N}$. Show that if $T(\rho\xrightarrow{\mf}\sigma)\leq\eps$ then 
\be
T\left(\rho^{\otimes k}\xrightarrow{\mf}\sigma^{\otimes k}\right)\leq k\eps\;.
\ee
\eex

The subsequent lemma highlights an additional property of the conversion distance: small changes in $\rho$ result in only minor variations in the conversion distance. This property underscores the stability of the conversion distance measure against slight perturbations in the resource state.

\begin{myg}{}
\begin{lemma}\label{cdub}
Let $\eps\in(0,1)$, $\rho\in\md(A)$, $\sigma\in\md(B)$, and $\trho\in\mb_\eps(\rho)$. Then,
\be
\Big|T\left(\rho\xrightarrow{\mf}\sigma\right)-T\left(\trho\xrightarrow{\mf}\sigma\right)\Big|\leq\eps\;.
\ee
\end{lemma}
\end{myg}
\begin{proof}
By definition we get
\ba
T\left(\trho\xrightarrow{\mf}\sigma\right)&=\min_{\mE\in\mf(A\to B)}\frac12\left\|\sigma-\mE(\trho)\right\|_1\\
\GG{Triangle\;inequality}&\geq\min_{\mE\in\mf(A\to B)}\left\{\frac12\left\|\sigma-\mE(\rho)\right\|_1-\frac12\left\|\mE(\trho-\rho)\right\|_1\right\}\\
\GG{DPI}&\geq\min_{\mE\in\mf(A\to B)}\frac12\left\|\sigma-\mE(\rho)\right\|_1-\frac12\left\|\trho-\rho\right\|_1\\
&\geq T\left(\rho\xrightarrow{\mf}\sigma\right)-\eps\;.
\ea
The proof is concluded by repeating the same lines as above after exchanging between $\rho$ with $\trho$.
\end{proof}

\subsubsection{The Golden Unit}\index{golden unit}

Most resource theories contains a special type of a resource state, that we call here a ``golden unit". For example, in entanglement theory, the maximally entangled state, $\Phi^{AB}\in\md(AB)$ with $|A|=|B|$ is a golden unit. Maximally entangled states are desirable since they can be used to accomplish quantum information processesing tasks such as quantum teleportation and superdense coding. Maximally entangled states has the property that they are closed under tensor products; that is, a tensor product of two maximally entangled states is itself a maximally entangled state. This motivates us the extend this property to all resource theories.

In the definition below, for any integer $m\in\mathbb{N}$, we denote $\Phi_m\in\md(A)$, where $m$ is defined as $m\equiv|A|$, to represent a resource state, indicating that $\Phi_m$ does not belong to the set of free states $\mf(A)$. Importantly, we do not presuppose its specific form (such as assuming it to be the maximally entangled state, which is common in entanglement theory). Instead, we focus on outlining the necessary properties it must satisfy. Additionally, we use the equivalence notation $\rho\sim\sigma$ to signify that both conversions $\rho\xrightarrow{\mf}\sigma$ and $\sigma\xrightarrow{\mf}\rho$ are possible. This notation helps clarify the relationship between resource states in terms of their interconvertibility within the given resource theory.

\begin{myd}{}
\begin{definition}\label{def:gu}
 The sequence of resource states $\{\Phi_m\}_{m\in\mbb{N}}$ is called \emph{a golden unit\index{golden unit}} if for all $m,n\in\mbb{N}$ the following two conditions hold:
\ben \item $\Phi_n\otimes\Phi_m\sim\Phi_{nm}$. 
\item If $n\geq m$ then $\Phi_n\xrightarrow{\mf}\Phi_m$.
\een
\end{definition}
\end{myd}

A golden unit\index{golden unit}, can be used as a scale to measure the resourcefulness of a given state $\rho\in\md(A)$. There are two distinct ways to do that.

\begin{myd}{}
\begin{definition}\label{def:dceps}
Let $\rho\in\md(A)$, $\{\Phi_m\}_{m\in\mbb{N}}$ be \emph{a golden unit}, and $\eps\in(0,1)$. 
\ben
\item The $\eps$-single-shot resource cost of $\rho$  is defined as
\be
\cost^\eps\left(\rho\right)\eqdef\min\left\{\log m\;:\;T\left(\Phi_m\xrightarrow{\mf}\rho\right)\leq\eps\;,\;\;\;m\in\mbb{N}\right\}\;.
\ee
\item The $\eps$-single-shot distillable resource of $\rho$ is defined as
\be\label{epssins}
\distill^\eps\left(\rho\right)\eqdef\max\left\{\log m\;:\;T\left(\rho\xrightarrow{\mf}\Phi_m\right)\leq\eps\;,\;\;\;m\in\mbb{N}\right\}\;.
\ee
\een
\end{definition}
\end{myd}

\begin{remark}
Both the $\eps$-resource cost and the $\eps$-distillable resource are defined relative to the golden unit\index{golden unit} $\{\Phi_m\}_{m\in\mbb{N}}$. If there is no $m\in\mbb{N}$ such that $T\left(\Phi_m\xrightarrow{\mf}\rho\right)\leq\eps$ then we define $\cost^\eps\left(\rho\right)\eqdef\infty$. On the other hand, for the trivial dimension $m=1$ we must have $\Phi_m=1$ so that in this case $T\left(\rho\xrightarrow{\mf}\Phi_m\right)=0$. Hence, trivially, there is always $m\in\mbb{N}$ such that $T\left(\rho\xrightarrow{\mf}\Phi_m\right)\leq\eps$.
\end{remark}

\bex
Let $\mE\in\mf(A\to B)$ and $\rho\in\md(A)$. Show that
\be
\cost^\eps\left(\mE(\rho)\right)\leq \cost^\eps\left(\rho\right)\quad\text{and}\quad\distill^\eps\left(\mE(\rho)\right)\leq \distill^\eps\left(\rho\right)\;.
\ee
Hint: Use Lemma~\ref{lemmonin}.
\eex

\bex\label{smoothedvers}
Let $\rho\in\md(A)$. Show that
\be\label{smooths}
\cost^\eps\left(\rho\right)=\min_{\rho'\in\mb_\eps(\rho)}\cost^{\eps=0}\left(\rho'\right)\;.
\ee
That is, the $\eps$-single-shot cost can be seen as the smoothed version of its respective zero-error counterpart. Why something similar does not hold for $\distill^\eps(\rho)$?
\eex

In some resource theories there exists a golden unit\index{golden unit} $\{\Phi_m\}_{m\in\mbb{N}}$ with a property that 
\be
\kappa\eqdef\max_{\omega\in\md(A)}D(\omega\|\mf)=D\left(\Phi_m\|\mf\right)=\log(m)\;.
\ee
In such QRTs, one can use the asymptotic continuity\index{asymptotic continuity} of the relative entropy of a resource to obtain an upper bound on the single-shot $\eps$-distillable resource of some resource state $\rho\in\md(A)$. Specifically, let $m\in\mbb{N}$ be such that  $T\left(\rho\xrightarrow{\mf}\Phi_m\right)\leq\eps$. Then, for such $m$ there exists $\sigma\in\md(A')$ with $m\eqdef|A'|$ such that $\rho\xrightarrow{\mf}\sigma$ and $\sigma\approx_\eps\Phi_m$. Now, since the relative entropy of a resource is an entanglement monotone we get
\ba
D(\rho\|\mf)&\geq D(\sigma\|\mf)\\
\GG{\eqref{asyfor}}&\geq D\left(\Phi_m\|\mf\right)-\eps\kappa-(1+\eps)h\left(\frac\eps{1+\eps}\right)\;.
\ea
In many resource theories there exists a golden unit\index{golden unit} $\{\Phi_m\}_{m\in\mbb{N}}$ with a property that $\kappa=D\left(\Phi_m\|\mf\right)=\log(m)$. Therefore, for such resource theories the inequality above take the form
\be
(1-\eps)\log(m)\leq D(\rho\|\mf)+(1+\eps)h\left(\frac\eps{1+\eps}\right)\;.
\ee
Since $m$ was an arbitrary integer satisfying $T\left(\rho\xrightarrow{\mf}\Phi_m\right)\leq\eps$, the inequality above implies that
\be\label{ub2way}
\distill^\eps\left(\rho\right)\leq \frac1{1-\eps}D(\rho\|\mf)+\frac{1+\eps}{1-\eps}h\left(\frac\eps{1+\eps}\right)\;.
\ee
Note that for a small $\eps>0$ the upper bound is close to the relative entropy of a resource.

\section{Generalized Asymptotic Equipartition Property}

In Sec.~\ref{subsec:iid} we saw that given an i.i.d.$\sim\p$ source, all typical sequences with large size $n$, that are generated by the source, have approximately the same probability to occur given by $\approx 2^{-nH(\p)}$. This phenomenon is known as the asymptotic equipartition\index{asymptotic equipartition} property (AEP).
In Sec.~\ref{sec:aep0} we saw a variant of this property that involved the relative entropy. In this subsection we generalize further this property, and express it in terms of the relative entropy of a resource.

For any physical system $A$, let $\mf(A)\subseteq\md(A)$ be a convex closed subset of density matrices.
Recall that the relative entropy of a resource, and the logarithmic robustness\index{robustness}, are defined respectively as
\ba
&D(\rho\|\mf)\eqdef \min_{\sigma\in\mf(A)}D\left(\rho\|\sigma\right)\\
&D_{\max}(\rho\|\mf)\eqdef \min_{\sigma\in\mf(A)}D_{\max}\left(\rho\|\sigma\right)\;.
\ea
Recall also that for any $\eps>0$ the smoothed version of the logarithmic robustness is defined as
\be
D_{\max}^\eps(\rho\|\mf)\eqdef \min_{\rho'\in\mb_{\eps}(\rho)}D_{\max}(\rho'\|\mf)\;,
\ee
and the regularization\index{regularization} of $D\left(\cdot\|\mf\right)$ as
\be
D^{\reg}(\rho\|\mf)\eqdef\lim_{n\to\infty}\frac{1}{n}D\left(\rho^{\otimes n}\big\|\mf\right)\;.
\ee
From the exercise below it follows that $D^{\reg}(\rho\|\mf)$ is well defined since the limit on the right-hand side of the equation above exists.
 \begin{exercise}$\;$
 \begin{enumerate}
 \item Show that for any sequence of real numbers, $\{a_j\}_{n=1}^{\infty}$, that is sub-additive, i.e. $a_{n+m}\leq a_n+a_m$, the limit $\lim_{n\to\infty}a_n$ exists.
 Hint: See the hint given in Exercise~\ref{ex842}.
 \item Show that the limit of the sequence $\{a_n\}$, with $a_n\eqdef \frac{1}{n}D\left(\rho^{\otimes n}\big\|\mf\right)$, exists. 
 \end{enumerate}
 \end{exercise}

\begin{myt}{\color{yellow} The Generalized AEP}
\begin{theorem}\label{AEP}
Let $\mf$ be a QRT whose set of free states has the five properties above. Then, for all $\eps\in(0,1)$ and all $\rho\in\md(A)$ 
\be\label{aep}
\lim_{n\to\infty}\frac{1}{n}D_{\max}^\eps\left(\rho^{\otimes n}\|\mf\right)=D^{\reg}(\rho\|\mf)\;.
\ee
\end{theorem}
\end{myt}

\begin{remark}
At first glance it may not be very clear why the theorem above corresponds to the AEP property. Therefore, after the proof we will will give examples demonstrating  that for different choices of $\mf(A)$, the above theorem reduces to the various variants of the AEP studied in literature. In other words, the above theorem \emph{unifies} all the variants of AEP into a single formula.
\end{remark}

We divide the proof into two lemmas.
\begin{myg}{}
\begin{lemma}\label{lem:lb}
For any $\rho\in\md(A)$ and $\eps\in(0,1)$
\be\label{lowerbound1}
D^{\reg}(\rho\|\mf)\geq\limsup_{n\to\infty}\frac{1}{n}D_{\max}^\eps\left(\rho^{\otimes n}\|\mf\right)\;.
\ee
\end{lemma}
\end{myg}

\begin{proof}
 From~\eqref{ubo3} we have for any $0<\eps<1$
\ba
\limsup_{n\to\infty}\frac1n D_{\max}^\eps\left(\rho^{\otimes n}\|\mf\right)&= 
\limsup_{n\to\infty}\frac1n\min_{\sigma_n\in\mf(A^n)}D_{\max}^{\eps}(\rho^{\otimes n}\|\sigma_n)\\
\GG{\eqref{ubo3}}&\leq\limsup_{n\to\infty}\frac1n\min_{\sigma_n\in\mf(A^n)}D_{\min}^{\sqrt{1-\eps^2}}(\rho^{\otimes n}\|\sigma_n)\\
\GG{\eqref{1067}}&=\limsup_{n\to\infty}\frac1n D_{\min}^{\sqrt{1-\eps^2}}(\rho^{\otimes n}\|\mf)\;.
\ea
Combining this with the inequality~\eqref{ee1}, we then get for all $\alpha\in(1,2)$
\be
\limsup_{n\to\infty}\frac1n D_{\max}^\eps\left(\rho^{\otimes n}\|\mf\right)\leq \limsup_{n\to\infty}\frac1nD_{\alpha}\left(\rho^{\otimes n}\big\|\mf\right)=D_{\alpha}^\reg(\rho\|\mf)\;,
\ee
where $D_{\alpha}(\cdot\|\mf)$ is the $\alpha$-relative entropy of a resource as defined in~\eqref{1068}. Finally, since the above inequality holds for all $\alpha\in(1,2)$ we conclude that for all $\eps\in(0,1)$
\ba\label{strconv}
\limsup_{n\to\infty}\frac1n D_{\max}^\eps\left(\rho^{\otimes n}\|\mf\right)&\leq \lim_{\alpha\to 1^+}D_{\alpha}^\reg(\rho\|\mf)\\
\GG{Lemma~\ref{imlim}}&=D^\reg(\rho\|\mf)\;.
\ea
This completes the proof.
\end{proof}

\begin{myg}{}
\begin{lemma}\label{lem:1062}
For any $\rho\in\md(A)$
\be\label{upperbound2}
D^{\reg}(\rho\|\mf)\leq \liminf_{n\to\infty}\frac{1}{n}D_{\max}^\eps\left(\rho^{\otimes n}\|\mf\right)\;.
\ee
\end{lemma}
\end{myg}

\begin{proof}
Recall the relation~\eqref{ubo1} from Lemma~\ref{ubo}. The relation~\eqref{ubo1} implies that
for every $\eps_1,\eps_2\in(0,1)$ with $\eps_1+\eps_2<1$ we have
\be\label{htep}
D_{\max}^{\eps_1}(\rho\|\mf)\geq D_{\min}^{\eps_2}(\rho\|\mf)+\log\left(1-\eps_1-\eps_2\right)\;.
\ee
This in turn gives
\ba
\liminf_{n\to\infty}\frac{1}{n}D_{\max}^{\eps_1}\left(\rho^{\otimes n}\big\|\mf\right)&\geq\liminf_{n\to\infty}\frac{1}{n}D_{\min}^{\eps_2}\left(\rho^{\otimes n}\big\|\mf\right)\\
\GG{Theorem~\ref{gsl}}&=D^{\reg}(\rho\|\mf)\;,
\ea
where we the generalized quantum Stein's lemma (Theorem~\ref{gsl}) that will be proved in the next subsection.
This completes the proof.
\end{proof}

\subsection{Two Examples of AEP}

In this subsection we give two simple examples of QRTs that satisfy the AEP property. We start with the case that for each $n\in\mbb{N}$ the set $\mf(A^n)$ consists of the single state $\sigma^{\otimes n}$, where $\sigma$ is some fixed state in $\md(A)$. Note that in this case, $D^\reg(\rho\|\mf)=D(\rho\|\sigma)$ and $D_{\max}^\eps(\rho\|\mf)=D_{\max}^\eps(\rho\|\sigma)$. We therefore get the following corollary.

\begin{myg}{}
\begin{corollary}
Let $\rho,\sigma\in\md(A)$. Then, for any $\eps\in(0,1)$
\be\label{seeaep}
D(\rho\|\sigma)=\lim_{n\to\infty}\frac1nD_{\max}^\eps\left(\rho^{\otimes n}\big\|\sigma^{\otimes n}\right)\;.
\ee
\end{corollary}
\end{myg}

To see how this corollary relates the AEP discussed in Sec.~\ref{subsec:iid}, take $\sigma^A=\u^A$ and observe that $D(\rho\|\u)=\log|A|-H(\rho)$ and $D_{\max}^\eps(\rho\|\u)=\log|A|-H_{\min}^\eps(\rho)$, where
\be
H_{\min}^\eps(\rho)\eqdef\max_{\rho'\in\mb_{\eps}(\rho)}H_{\min}(\rho')\;,
\ee
is known as the \emph{smoothed min-entropy} of $\rho$.
Therefore, the corollary above implies in particular that for any $\eps\in(0,1)$
\be\label{vergaep}
\lim_{n\to\infty}\frac1nH_{\min}^\eps\left(\rho^{\otimes n}\right)=H(\rho)\;.
\ee
Recall that $H_{\min}(\rho)=-\log\lambda_{\max}(\rho)$ so that the above equation states that for any $\eps>0$ and sufficiently large $n$, there exists a state $\rho'_n\in\md(A^n)$ that is $\eps$-close to $\rho^{\otimes n}$, and for which $\lambda_{\max}(\rho'_n)\approx 2^{-nH(\rho)}$. In other words, it only requires a small perturbation to make all of the eigenvalues of $\rho^{\otimes n}$ to be bounded from above by $2^{-nH(\rho)}$.

The second example we consider here is a variant of the AEP involving the conditional entropy.
In this variant, we take the set of free states to be 
\be\label{1152}
\mf(AB)\eqdef\big\{\u^A\otimes\rho^B\;:\;\rho\in\md(A)\big\}\;.
\ee
With this set of free states we get
\ba\label{1153}
D\left(\rho^{AB}\big\|\mf\right)&\eqdef \min_{\sigma\in\mf(AB)}D\left(\rho^{AB}\big\|\sigma^{AB}\right)\\
&=\min_{\sigma\in\md(B)}D\left(\rho^{AB}\big\|\u^{A}\otimes\sigma^{B}\right)\\
\GG{\eqref{8281}}&=\log|A|-H(A|B)_\rho\\
\GG{Additivity\;of\;the\;conditional\;entropy}&=D^{\reg}\left(\rho^{AB}\big\|\mf\right)\;.
\ea
Similarly,
\ba
D_{\max}\left(\rho^{AB}\|\mf\right)&\eqdef\min_{\substack{\sigma\in\mf(AB)\\ \trho\in\mb_{\eps}(\rho)}}D_{\max}\left(\trho^{AB}\|\sigma^{AB}\right)\\
&=\min_{\substack{\sigma\in\md(B)\\ \trho\in\mb_{\eps}(\rho)}}D_{\max}\left(\trho^{AB}\|\u^{A}\otimes\sigma^{B}\right)\\
&=\min_{ \trho\in\mb_{\eps}(\rho)}\left\{\log|A|-H_{\min}^\ua\left(A|B\right)_{\trho}\right\}\\
&=\log|A|-H_{\min}^{\ua\eps}(A|B)_\rho\;.
\ea
We can therefore apply Theorem~\ref{AEP} to get the following corollary.

\begin{myg}{}
\begin{corollary}\label{eandr}
Let $\eps\in(0,1)$ and $\rho\in\md(AB)$. Then,
\ba\label{aep2}
H(A|B)_\rho= \lim_{n\to\infty}\frac{1}{n}H_{\min}^{\ua\eps}(A^n|B^n)_{\rho^{\otimes n}}\;.
\ea
\end{corollary}
\end{myg}

\bex
Prove the corollary.
\eex

\section{The Generalized Quantum Stein's Lemma\index{Stein's lemma}}

In this section we consider a generalization of the quantum Stein's lemma given in~\eqref{7171} by optimizing over $\sigma\in\mf(A)$.
Specifically, recall the hypothesis-testing relative entropy of a resource defined in~\eqref{1067} as
\be
D_{\min}^{\eps}(\rho\|\mf)\eqdef\min_{\sigma\in\mf(A)}D_{\min}^\eps(\rho\|\sigma)\;.
\ee
The following theorem applies for a QRT whose set of free states $\mf_n\eqdef\mf(A^n)$ has the following properties for all $n,m\in\mbb{N}$: 
\ben
\item The set $\mf_n$ is closed and convex .
\item The set $\mf_n$ is closed under permutations of the subsystems of $A^n$.
\item If $\sigma\in\mf_n$ and $\omega\in\mf_m$ then $\sigma\otimes\omega\in\mf_{nm}$.
\item There exists $\gamma\in\mf_1$ with $\gamma>0$.
\een
\begin{myt}{\color{yellow} The Generalized Quantum Stein's Lemma}
\begin{theorem}\label{gsl}
Let $\mf$ be a QRT whose sets of free states satisfy the properties above.
Then, for all $\eps\in(0,1)$ and all $\rho\in\md(A)$
\be\label{1158}
\lim_{n\to\infty}\frac{1}{n}D_{\min}^{\eps}\left(\rho^{\otimes n}\big\|\mf\right)=D^{\reg}(\rho\|\mf)\;.
\ee
\end{theorem}
\end{myt}

Recall from Lemma~\ref{ubo} that for any $\eps_1,\eps_2\in(0,1)$ with $\eps_1+\eps_2<1$ we have (cf.~\eqref{ubo1})
\be
D_{\min}^{\eps_1}(\rho\|\sigma)\leq D_{\max}^{\eps_2}(\rho\|\sigma)-\log\left(1-\eps_1-\eps_2\right)\;.
\ee
Then, from the definitions it follows that for any such $\eps_1,\eps_2\in(0,1)$ with $\eps_1+\eps_2<1$ we have
\be\label{1160l}
D_{\min}^{\eps_1}(\rho\|\mf)\leq D^{\eps_2}_{\max}(\rho\|\mf)-\log\left(1-\eps_1-\eps_2\right)\;,
\ee
so that
\ba\label{1161}
\limsup_{n\to\infty}\frac{1}{n}D^{\eps_1}_{\min}\left(\rho^{\otimes n}\big\|\mf\right)&\leq \limsup_{n\to\infty}\frac{1}{n}D^{\eps_2}_{\max}\left(\rho^{\otimes n}\big\|\mf\right)\\
\GG{Lemma~\ref{lowerbound1}}&\leq D^{\reg}(\rho\|\mf)\;.
\ea
This provides a proof for the strong converse of the theorem above (note that we already showed it in~\eqref{1083} using a different approach). We therefore focus now on the opposite inequality; specifically, we would like to prove that
\be\label{inii}
s\eqdef\liminf_{n\to\infty}\frac{1}{n}D^{\eps}_{\min}\left(\rho^{\otimes n}\big\|\mf\right)\geq D^{\reg}(\rho\|\mf)\;.
\ee
To prove the inequality given in~\eqref{inii}, we will develop several tools. We begin with a property of the hypothesis testing divergence. 
For every two operators $\Gamma,\Lambda\in\herm(A)$ we use the notation $\BB \Lambda\geq \Gamma\EE$ to denote the orthogonal projection to the positive part of $\Lambda-\Gamma$; in other words, $\BB \Lambda\geq \Gamma\EE$ denotes the projection to the support of $(\Lambda-\Gamma)_+$. 

\begin{myg}{}
\begin{lemma}\label{lem118}
Let $\eps,\delta\in(0,1)$, and let $\{\rho_n\}_{n\in\mbb{N}}$ and $\{\sigma_n\}_{n\in\mbb{N}}$ be two sets of density matrices, with $\rho_n,\sigma_n\in\md(A_n)$, where $A_n$ is a finite dimensional Hilbert space whose dimension can depend on $n$. If
\ba\label{limitexd}
t\eqdef\liminf_{n\to\infty}\frac1nD_{\min}^\eps(\rho_n\|\sigma_n)\;,
\ea
then 
\be
\limsup_{n\to\infty}\tr\left[\BB2^{n(t+\delta)}\sigma_n\geq\rho_n\EE\rho_n\right]\geq\eps\label{01}\;.
\ee
\end{lemma}
\end{myg}

\begin{remark}
Note that if $\sigma_n$ is chosen as the optimal state such that $D^{\eps}_{\min}\left(\rho^{\otimes n}\big\|\mf\right)=D^{\eps}_{\min}\left(\rho^{\otimes n}\big\|\sigma_n\right)$, then the value of $t$, as defined in~\eqref{limitexd}, is equal to $s$, as defined in~\eqref{inii}. In the proof of the generalized quantum Stein's lemma, we will also consider other sequences $\{\sigma_n\}_{n\in\mbb{N}}$ where, in those cases, $s\neq t$.
\end{remark}

\begin{proof}
Set $\Pi_n\eqdef\BB2^{n(t+\delta)}\sigma_n\geq\rho_n\EE$ and suppose by contradiction that
\be\label{xv}
\limsup_{n\to\infty}\tr\left[\Pi_n\rho_n\right]<\eps\;.
\ee
Let $\{n_k\}_{k\in\mbb{N}}$ be a subsequence such that
\be\label{r87}
t=\lim_{k\to\infty}\frac1{n_k}D_{\min}^\eps(\rho_{n_k}\|\sigma_{n_k})\;.
\ee
From the assumption in~\eqref{xv} it follows that there exists an integer $k_0$ such that for all $k>k_0$
\be
\tr\left[\Pi_{n_k}\rho_{n_k}\right]<\eps\;.
\ee
Denote by $\Lambda_{n_k}\eqdef I-\Pi_{n_k}$, so that $\tr\left[\Lambda_{n_k}\rho_{n_k}\right]> 1-\eps$. On the other hand, from the definition of $\Pi_n$ we get that the projection $\Lambda_{n_k}$ projects to the negative part of $2^{n_k(t+\delta)}\sigma_{n_k}-\rho_{n_k}$.
 Thus,
\be
\tr\left[\Lambda_{n_k}\left(2^{-n(t+\delta)}\rho_{n_k}-\sigma_{n_k}\right)\right]\geq 0\;,
\ee
so that
\ba
\tr\left[\Lambda_{n_k}\sigma_{n_k}\right]&\leq 2^{-n_k(t+\delta)}\tr\left[\Lambda_{n_k}\rho_{n_k}\right]\\
&\leq 2^{-n_k(t+\delta)}
\ea
The above inequality implies that for all $k>k_0$
\ba
t+\delta&\leq -\frac1{n_k}\log \tr\left[\Lambda_{n_k}\sigma_{n_k}\right]\\
&\leq \frac1{n_k}D_{\min}^{\eps}\left(\rho_{n_k}\|\sigma_{n_k}\right)
\ea
where in the last inequality we used the definition of $D_{\min}^\eps$. However, since the above inequality holds for all $k>k_0$ it also hold for the limit $k\to\infty$ which according to~\eqref{r87} exists and equal to $t$. We therefore get the contradiction $t+\delta\leq t$. This completes the proof.
\end{proof}

Next we show that $\mf$ can be replaced with its symmetric counterpart.

\begin{myg}{}
\begin{lemma}
Let $\{\mf_n\}_{n\in\mbb{N}}$ be sets of free states satisfying  the conditions outlined above Theorem~\ref{gsl}, and for every $n\in\mbb{N}$ let $\sym(\mf_n)$ be the set of all states in $\mf_n$ that are symmetric under permutations of the $n$ subsystems of $A^n$. Then, for all $\rho\in\md(A)$, $n\in\mbb{N}$, and a quantum divergence $\D$,  we have
\be
\D\left(\rho^{\otimes n}\big\|\mf_n\right)=\D\left(\rho^{\otimes n}\big\|\sym(\mf_n)\right)\;.
\ee
\end{lemma}
\end{myg}
\begin{proof}
Let $\mG_n\in\cptp(A^n\to A^n)$ be the $\mathbf{S}_n$-twirling operation with respect to $\mathbf{S}_n$, the group of permutations on $n$-elements. That is,
\be
\mG_n\left(\omega^{A^n}\right)\eqdef\frac1{n!}\sum_{\pi\in\mathbf{S}_n}P_\pi^{A^n}\omega^{A^n}P_{\pi^{-1}}^{A^n}\;,
\ee
where $\{P_\pi^{A_n}\}_{\pi\in\mathbf{S}_n}$ for the set of all $n!$ matrices that permutes the $n$ subsystems of $A^n$ (see~\eqref{ppi} for more details).
Since we assume that $\mf_n$ is both convex and closed under permutation, for every $\sigma_n\in\mf_n$ we have necessarily that $\mG_n(\sigma_n)\in\mf_n$. Thus, since $\mG_n(\rho^{\otimes n})=\rho^{\otimes n}$ we get from the DPI that for all $\sigma_n\in\mf_n$ and every quantum divergence $\D$ (in particular for $D_{\min}^\eps$ and $D$) we have
\ba
\D\left(\rho^{\otimes n}\big\|\sigma_n\right)&\geq \D\left(\mG_n\left(\rho^{\otimes n}\right)\big\|\mG_n(\sigma_n)\right)\\
&=\D\left(\rho^{\otimes n}\big\|\mG_n(\sigma_n)\right)\\
&\geq \D\left(\rho^{\otimes n}\big\|\sym(\mf_n)\right)\;.
\ea
This complets the proof since the above inequality holds for all $\sigma_n\in\mf_n$.
\end{proof}

The proof outlined above can be adapted for any compact Lie group $\G$, where the $\G$-twirling channel is well-defined. In our case, this implies that proving Theorem~\ref{gsl} is sufficient if all free states are symmetric under permutations. Consequently, although not explicitly stated, we now assume that every $\sigma_n \in \mf_n$ satisfies $\mG_n(\sigma_n) = \sigma_n$.

Our goal is to find a sequence of free symmetric states, $\{\sigma_n\}_{n\in\mbb{N}}$, with each $\sigma_n\in\sym(\mf_n)$, 
such that
\be\label{2rsn2}
r\eqdef\liminf_{n\to\infty}\frac{1}{n}D\left(\rho^{\otimes n}\big\|\sigma_n\right)\leq s\;.
\ee
 Note that in general, $\rho^{\otimes n}$ does not commute with $\sigma_n$, but in the following lemma we show that we can replace $\rho^{\otimes n}$ with $\mP_n\left(\rho^{\otimes n}\right)$, where $\mP_n$ is the pinching channel (see Sec.~\ref{sec:pinching}) with respect to the state $\sigma_n$.

\begin{myg}{}
\begin{lemma}\label{lempinch}
Let $\{\sigma_n\}_{n\in\mbb{N}}$ be a sequence of free symmetric states with each $\sigma_n\in\sym(\mf_n)$, and for every $n\in\mbb{N}$ let $\mP_n\in\cptp(A^n\to A^n)$ be the pinching channel associated with the state $\sigma_n$. Then,
\be\label{1175m}
\liminf_{n\to\infty}\frac{1}{n}D\left(\rho^{\otimes n}\big\|\sigma_n\right)=\liminf_{n\to\infty}\frac{1}{n}D\left(\mP_n\left(\rho^{\otimes n}\right)\big\|\sigma_n\right)\;.
\ee
\end{lemma}
\end{myg}

\begin{proof}
Observe that
\ba\label{stam0}
D\left(\mP_n\left(\rho^{\otimes n}\right)\big\|\sigma_{n}\right)&=-H\left(\mP_n\left(\rho^{\otimes n}\right)\right)-\tr\left[\mP_n\left(\rho^{\otimes n}\right)\log\sigma_{n}\right]\\
\GG{\mP_{\it n}\;is\;\text{self-adjoint}}&=-H\left(\mP_n\left(\rho^{\otimes n}\right)\right)-\tr\left[\rho^{\otimes n}\mP_n\left(\log\sigma_{n}\right)\right]\\
\GG{\sigma_{{\it n}}\;is\;symmetric}&=-H\left(\mP_n\left(\rho^{\otimes n}\right)\right)-\tr\left[\rho^{\otimes n}\log\sigma_{n}\right]\\
&=D\left(\rho^{\otimes n}\big\|\sigma_{n}\right)+H\left(\rho^{\otimes n}\right)-H\left(\mP_n\left(\rho^{\otimes n}\right)\right)\;.
\ea
Now, from~\eqref{pinex} it follows that the pinching channel can be written as a uniform mixture of unitary channels with $|\spec(\sigma_{n})|$ terms. Since $\sigma_{n}$ is symmetric, its support is a subspace of the symmetric subspace $\sym_n(A)$ (see~Appendix~\ref{sec:symsub} for many properties of the symmetric subspace). Thus,
\ba
|\spec(\sigma_{n})|&\leq\dim\left(\sym_n(A)\right)\\
\GG{\it cf.~\eqref{specsig}}&\leq (n+1)^{|A|}\;.
\ea
Combining this with the upper bound in Corollary~\ref{cor:ubent} we obtain
\be\label{comlll}
H\left(\mP_n\left(\rho^{\otimes n}\right)\right)\leq H\left(\rho^{\otimes n}\right)+|A|\log(n+1)\;.
\ee
Since $\mP_n$ is a doubly stochastic channel we also have $H\left(\mP_n\left(\rho^{\otimes n}\right)\right)\geq H\left(\rho^{\otimes n}\right)$. Combining this with~\eqref{comlll} we obtain
\be
\lim_{n\to\infty}\frac1n\Big(H\left(\mP_n\left(\rho^{\otimes n}\right)\right)- H\left(\rho^{\otimes n}\right)\Big)=0\;.
\ee
Thus, dividing both sides of~\eqref{stam0} by $n$ and taking the $\liminf_{n\to\infty}$ we obtain~\eqref{1175m}.
This completes the proof.
\end{proof}

From Lemma~\ref{lempinch} we get that in order to prove Theorem~\ref{gsl} it is sufficient to find a sequence of free symmetric states, $\{\sigma_n\}_{n\in\mbb{N}}$, with each $\sigma_n\in\sym(\mf_n)$, 
such that
\be\label{3rsn3}
r=\liminf_{n\to\infty}\frac{1}{n}D\left(\mP_n\left(\rho^{\otimes n}\right)\big\|\sigma_n\right)\leq s\;.
\ee
The idea of the proof is to pick an arbitrary sequence of symmetric states $\{\sigma_n\}_{n\in\mbb{N}}$, and look at the difference $r-s$. Clearly if this difference is non-positive then we are done. Thus, we assume that $r-s>0$ and then construct another sequence $\{\sigma_{n,1}\}_{n\in\mbb{N}}$ with the property that
\be\label{3rsn1}
r_1\eqdef\liminf_{n\to\infty}\frac{1}{n}D\left(\mP_n\left(\rho^{\otimes n}\right)\big\|\sigma_{n,1}\right)\;,
\ee
satisfies
\be\label{rat1}
\frac{r_1-s}{r-s}\leq 1-\eps\;.
\ee
Before proving the existence of the sequence $\{\sigma_{n,1}\}_{n\in\mbb{N}}$, let us first demonstrate how the inequality in~\eqref{rat1} can be applied to prove the generalized quantum Stein's lemma.

Observe that unless $s>r_1$ (in which case the proof is done) we get that $r_1$ is closer to $s$ than $r$ is.  Repeating the process by starting from the sequence $\{\sigma_{n,1}\}_{n\in\mbb{N}}$ and constructing the sequence $\{\sigma_{n,2}\}_{n\in\mbb{N}}$ we can obtain that $r_2\eqdef\liminf_{n\to\infty}\frac{1}{n}D\left(\mP_n\left(\rho^{\otimes n}\right)\big\|\sigma_{n,1}\right)$ satisfies
\be\label{rat2}
\frac{r_2-s}{r_1-s}\leq 1-\eps\;.
\ee
The combination of~\eqref{rat1} and~\eqref{rat2} results with
\be
\frac{r_2-s}{r-s}\leq (1-\eps)^2\;.
\ee
Similarly, by repetition, for every $k\in\mbb{N}$ there exists a sequence of symmetric free states $\{\sigma_{n,k}\}_{n\in\mbb{N}}$ such that
\be
r_k-s\leq (1-\eps)^k\left(r- s\right)\quad\text{where}\quad r_k\eqdef\liminf_{n\to\infty}\frac{1}{n}D\left(\mP_n\left(\rho^{\otimes n}\right)\big\|\sigma_{n,k}\right)\;.
\ee
Thus, from the definition of $r_k$ it follows that for every sequence $\{\delta_k\}_{k\in\mbb{N}}\subset(0,1)$ with limit $\lim_{k\to\infty}\delta_k=0$, there exists a sequence of integers $\{n_k\}_{k\in\mbb{N}}$ such that for all $k\in\mbb{N}$
\be\label{11p90p}
\frac{1}{n_k}D\left(\mP_{n_k}\left(\rho^{\otimes n_k}\right)\big\|\sigma_{n_k}\right)-s\leq (1-\eps)^k\left(r- s\right)+\delta_k
\ee
Since the right hand side goes to zero as $k$ goes to infinity we obtain the desired result that
\ba
D^{\reg}\left(\rho\big\|\sym(\mf)\right)-s&\leq \liminf_{k\to\infty}\frac{1}{n_k}D\left(\mP_{n_k}\left(\rho^{\otimes n_k}\right)\big\|\sigma_{n_k}\right)-s\\
\GG{\eqref{11p90p}}&\leq 0\;.
\ea
We therefore conclude that in order to prove the theorem, starting from some sequence of symmetric states $\{\sigma_n\}_{n\in\mbb{N}}$, it is sufficient to construct another sequence of symmetric states $\{\sigma_{n,1}\}_{n\in\mbb{N}}$ that satisfies the inequality in~\eqref{3rsn1}.

\subsubsection{Construction of the Sequence $\{\sigma_{n,1}\}_{n\in\mbb{N}}$:}

If the sequence $\{\sigma_{n}\}_{n\in\mbb{N}}$ satisfies~\eqref{3rsn3} we can take $\sigma_{n,1}=\sigma_n$ for all $n\in\mbb{N}$. We therefore assume that $r>s$. 
Let $\eps_0>0$ and observe that from the definition of $r$ there exists integer $m\in\mbb{N}$ such that
\be\label{rn0}
\frac1m D\left(\rho^{\otimes m}\big\|\sigma_m\right)< r+\eps_0\;.
\ee
For all $n\in\mbb{N}$ denote by $k_n\eqdef\left\lfloor\frac nm\right\rfloor$ and by
\be
\sigma_{n,1}\eqdef\frac13\left(\sigma_n^\star+\mG_n\left(\sigma_m^{\otimes k_n}\otimes\gamma^{\otimes (n-k_nm)}\right)+\gamma^{\otimes n}\right)\;,
\ee
where $\sigma_n^\star\in\sym(\mf_n)$ is the state satisfying
\be
D^{\eps}_{\min}\left(\rho^{\otimes n}\big\|\mf\right)=D^{\eps}_{\min}\left(\rho^{\otimes n}\big\|\sigma^\star_n\right)\;.
\ee
Observe that from its definition, for every $n\in\mbb{N}$ we have that  $\sigma_{n,1}\in\sym(\mf_n)$.

\subsubsection{Properties of $\{\sigma_{n,1}\}_{n\in\mbb{N}}$:}

Let  $\delta>0$ to be sufficiently small such that $r>s+\delta$ (recall that we assume that $r>s$), and denote by
\be \label{operators}
P_n\eqdef \BB2^{n(r+\eps_0)}\sigma_{n,1}\geq \mP_n\left(\rho^{\otimes n}\right)\EE
\quad\text{and}\quad
Q_n\eqdef \BB2^{n(s+\delta)}\sigma_{n,1}\geq\mP_n\left(\rho^{\otimes n}\right)\EE\;,
\ee
\begin{myg}{}
\begin{lemma}
Let $\eps_0,\eps_1,\delta\in(0,1)$ with $\delta$ sufficiently small such that $r>s+\delta$. Then, the operators in~\eqref{operators} satisfy:
\ba
&\limsup_{n\to\infty}\tr\left[Q_n\mP_n\left(\rho^{\otimes n}\right)\right]\geq\eps\;,\\
&\limsup_{n\to\infty}\tr\left[P_n\mP_n\left(\rho^{\otimes n}\right)\right]\geq1-\eps_1\;.
\ea
\end{lemma}
\end{myg}
\begin{proof}
First, observe that from Lemma~\ref{lem118}, it is sufficient to show that
\begin{align}
&\liminf_{n\to\infty}\frac1n D_{\min}^{\eps}\left(\mP_n\left(\rho^{\otimes n}\right)\big\|\sigma_{n,1}\right)\leq s\label{zxz1}\\
&\limsup_{n\to\infty}\frac1n D_{\min}^{1-\eps_1}\left(\mP_n\left(\rho^{\otimes n}\right)\big\|\sigma_{n,1}\right)\leq r+\eps_0\label{zxz2}\;.
\end{align}
To show these two inequalities we use the property that if $\sigma'\geq t\sigma$ then $D_{\min}^\eps(\rho\|\sigma')\leq D_{\min}^\eps(\rho\|\sigma)-\log t$ (see Exercise~\ref{exsiggt}). Indeed, to prove~\eqref{zxz1}, we use the inequality
$\sigma_{n,1}\geq\frac13\sigma_n^\star$ in conjunction with the DPI to get
\ba
D_{\min}^{\eps}\left(\mP_n\left(\rho^{\otimes n}\right)\big\|\sigma_{n,1}\right)&=D_{\min}^{\eps}\left(\mP_n\left(\rho^{\otimes n}\right)\big\|\mP_n\left(\sigma_{n,1}\right)\right)\\
\GG{DPI}&\leq D_{\min}^{\eps}\left(\rho^{\otimes n}\big\|\sigma_{n,1}\right)\\
\Gg{\sigma_{n,1}\geq\frac13\sigma_n^\star}&\leq D_{\min}^{\eps}\left(\rho^{\otimes n}\big\|\sigma_n^\star\right)+\log 3\\
\GG{\text{By definition of }\sigma_{\it n}^\star}&=D_{\min}^{\eps}\left(\rho^{\otimes n}\big\|\mf\right)+\log 3\;,
\ea
so that
\ba
\liminf_{n\to\infty}\frac1n D_{\min}^{\eps}\left(\mP_n\left(\rho^{\otimes n}\right)\big\|\sigma_{n,1}\right)&\leq\liminf_{n\to\infty}\frac1n D_{\min}^{\eps}\left(\rho^{\otimes n}\big\|\mf\right)\\
&=s\;.
\ea
Similarly, to prove~\eqref{zxz2}, we use 
the inequality $\sigma_{n,1}\geq \frac13\mG_n\left(\sigma_m^{\otimes k_n}\otimes\gamma^{\otimes (n-k_nm)}\right)$ in conjunction with the DPI to get
\ba\label{11p90}
D_{\min}^{1-\eps_1}\left(\mP_n\left(\rho^{\otimes n}\right)\big\|\sigma_{n,1}\right)&=D_{\min}^{1-\eps_1}\left(\mP_n\left(\rho^{\otimes n}\right)\big\|\mP_n\left(\sigma_{n,1}\right)\right)\\
\GG{DPI}&\leq D_{\min}^{1-\eps_1}\left(\rho^{\otimes n}\big\|\sigma_{n,1}\right)\\
\Gg{\sigma_{n,1}\geq \frac13\mG_n\left(\sigma_m^{\otimes k_n}\otimes\gamma^{\otimes (n-k_nm)}\right)}&\leq D_{\min}^{1-\eps_1}\left(\rho^{\otimes n}\big\|\mG_n\left(\sigma_m^{\otimes k_n}\otimes\gamma^{\otimes (n-k_nm)}\right)\right)+\log 3\;,
\ea 
Substituting $\rho^{\otimes n}=\mG_n(\rho^{\otimes n})$ into the equation above and applying the DPI gives
\ba
D_{\min}^{1-\eps_1}\left(\mP_n\left(\rho^{\otimes n}\right)\big\|\sigma_{n,1}\right)&\leq D_{\min}^{1-\eps_1}\left(\rho^{\otimes n}\big\|\sigma_m^{\otimes k_n}\otimes\gamma^{\otimes (n-k_nm)}\right)+\log 3\\
\GG{\eqref{e1}}& \leq \tD_\alpha\left(\rho^{\otimes n}\big\|\sigma_m^{\otimes k_n}\otimes\gamma^{\otimes (n-k_nm)}\right)-\frac\alpha{\alpha-1}\log\left(1-\eps\right)+\log 3\\
&=k_n\tD_\alpha\left(\rho^{\otimes m}\big\|\sigma_m\right)+(n-k_nm)\tD_\alpha\left(\rho\big\|\gamma\right)-\frac\alpha{\alpha-1}\log\left(1-\eps\right)+\log 3\;.
\ea
By definition, $k_n/n$ goes to $1/m$ as $n$ goes to infinity. Thus, by dividing both sides of the equation above by $n$ and taking the limit $n\to\infty$ we obtain
\be
\limsup_{n\to\infty}\frac1n D_{\min}^{1-\eps_1}\left(\mP_n\left(\rho^{\otimes n}\right)\big\|\sigma_{n,1}\right)\leq \frac1m\tD_\alpha\left(\rho^{\otimes m}\big\|\sigma_m\right)\;.
\ee
Since the equation above holds for all $\alpha>1$, taking the limit $\alpha\to 1^+$ we get from the continuity of $\tD_\alpha$ at $\alpha=1$ that
\ba
\limsup_{n\to\infty}\frac1n D_{\min}^{1-\eps_1}\left(\mP_n\left(\rho^{\otimes n}\right)\big\|\sigma_{n,1}\right)&\leq \frac1mD\left(\rho^{\otimes m}\big\|\sigma_m\right)\\
\GG{\eqref{rn0}}&< r+\eps_0\;.
\ea
This completes the proof of the lemma.
\end{proof}

\bex\eqref{exsiggt}
Show that if $\sigma'\geq t\sigma$ then $D_{\min}^\eps(\rho\|\sigma')\leq D_{\min}^\eps(\rho\|\sigma)-\log t$.
\eex

We are finally ready to prove the generalized quantum Stein's lemma.

\subsubsection{Proof of Theorem~\ref{gsl}:}

Since
$\mP_n\left(\rho^{\otimes n}\right)$ commutes with $\sigma_{n,1}$, the three projections $\Pi_n^{(1)}\eqdef I-P_n$, $\Pi_n^{(2)}\eqdef P_n-Q_n$, and $\Pi_n^{(3)}\eqdef Q_n$, form a 3-outcome orthogonal von-Neumann projective measurement of system $A^n$. The key idea of the proof is to split the computation of $D\left(\mP_n\left(\rho^{\otimes n}\right)\big\|\sigma_{n,1}\right)$ into three parts based on these projections. Explicitly, we write $D\left(\mP_n\left(\rho^{\otimes n}\right)\big\|\sigma_{n,1}\right)=a_n^{(1)}+a_n^{(2)}+a_n^{(3)}$, where for all $j\in\{1,2,3\}$
\ba
a_n^{(j)}\eqdef\tr\left[\Pi_n^{(j)}\mP_n\left(\rho^{\otimes n}\right)\left(\log\mP_n\left(\rho^{\otimes n}\right)-\log\sigma_{n,1}\right)\right]\;.
\ea
\noindent{\it Upper bound for $a_n^{(1)}$:} Recall  that 
$
\sigma_{n,1}\geq\frac13\gamma^{\otimes n}>0$, and let $\lambda$ be such that $\gamma^{-1}\leq 2^\lambda I^A$. For reasons to be clear later on, we choose $\lambda$ to be sufficiently large so that we also have $\lambda\geq r+\eps_0$. Then,
\ba\label{lamn}
\sigma_{n,1}^{-1}&\leq 3(\gamma^{-1})^{\otimes n}\\
&\leq 2^{n\lambda+\log 3}I^{A^n}\;.
\ea
We therefore bound the first term as:
\ba
a_n^{(1)}&\leq \tr\left[\Pi_n^{(1)}\mP_n\left(\rho^{\otimes n}\right)\log\sigma_{n,1}^{-1}\right]\\
\GG{\eqref{lamn}}&\leq \big(n\lambda+\log3\big)\tr\left[\Pi_1\mP_n\left(\rho^{\otimes n}\right)\right]\;.
\ea
\noindent{\it Upper bound for $a_n^{(2)}$:} When restricted to the support of $\Pi_n^{(2)}\eqdef P_n-Q_n$, we get from the definitions of $P_n$ and $Q_n$ that $\mP_n(\rho^{\otimes n})\leq 2^{n(r+\eps_0)}\sigma_{n,1}$ or equivalently $\mP_n(\rho^{\otimes n})\sigma_{n,1}^{-1}\leq 2^{n(r+\eps_0)}I^{A^n}$ (recall that $\mP_n(\rho^{\otimes n})$ and $\sigma_{n,1}$ commutes). Thus,
\be
a_n^{(2)}\leq n(r+\eps_0)\tr\left[\Pi_2\mP_n\left(\rho^{\otimes n}\right)\right]
\ee
\noindent{\it Upper bound for $a_n^{(3)}$:}  When restricted to the support of $\Pi_n^{(3)}\eqdef Q_n$, we get that $\mP_n(\rho^{\otimes n})\leq 2^{n(s+\delta)}\sigma_{n,1}$ or equivalently $\mP_n(\rho^{\otimes n})\sigma_{n,1}^{-1}\leq 2^{n(s+\delta)}I^{A^n}$. Thus,
\be
a_n^{(3)}\leq n(s+\delta)\tr\left[\Pi_n^{(3)}\mP_n\left(\rho^{\otimes n}\right)\right]\;.
\ee

\noindent{\it Upper bound for $r_1$:}
From the upper bounds for $a_n^{(1)}$, $a_n^{(2)}$, and $a_n^{(3)}$, we get that
\ba
r_1&= \liminf_{n\to\infty}\frac1n(a_n^{(1)}+a_n^{(2)}+a_n^{(3)})\\
&\leq\liminf_{n\to\infty}\Big\{\lambda\tr\left[\Pi_n^{(1)}\mP_n\left(\rho^{\otimes n}\right)\right]+(r+\eps_0)\tr\left[\Pi_n^{(2)}\mP_n\left(\rho^{\otimes n}\right)\right]+(s+\delta)\tr\left[\Pi_n^{(3)}\mP_n\left(\rho^{\otimes n}\right)\right]\Big\}\\
&=\lambda-\limsup_{n\to\infty}\Big\{(\lambda-r-\eps_0)\tr\left[P_n\mP_n\left(\rho^{\otimes n}\right)\right]+(r+\eps_0-s-\delta)\tr\left[Q_n\mP_n\left(\rho^{\otimes n}\right)\right]\Big\}
\ea
Finally, since both $\lambda-r-\eps_0\geq 0$ and $r+\eps_0-s-\delta\geq 0$ we apply Lemma~\ref{lem118} to the equation above to get the upper bound:
\be
r_1\leq\lambda-(\lambda-r-\eps_0)(1-\eps_1)-(r+\eps_0-s-\delta)\eps\;
\ee
Note that the inequality above holds for all $\eps_0,\eps_1,\delta\in(0,1)$ where $\delta$ is sufficiently small such that $s+\delta<r$. Thus, by
 taking the limits $\delta,\eps_0,\eps_1\to 0$ we obtain $r_1\leq r-(r-s)\eps$
which is equivalent to~\eqref{rat1}. This completes the proof of the generalized Stein's lemma.

\bex
Without using Theorem~\ref{gsl},  show that all $\rho\in\md(A)$ we have
\be
\lim_{\eps\to 1^-}\liminf_{n\to\infty}\frac{1}{n}D_{\min}^{\eps}\left(\rho^{\otimes n}\big\|\mf\right)=\lim_{\eps\to 1^-}\limsup_{n\to\infty}\frac{1}{n}D_{\min}^{\eps}\left(\rho^{\otimes n}\big\|\mf\right)=D^{\reg}(\rho\|\mf)\;.
\ee
\eex

\bex
Provide an alternative proof of Theorem~\ref{gsl} for the special case that the set of free states is the one defined in~\eqref{1152}; i.e.\  
\be\label{fgfd9}
\mf(AB)\eqdef\big\{\u^A\otimes\rho^B\;:\;\rho\in\md(A)\big\}\;.
\ee
Follow these steps:
\ben
\item First, show that
\be
D_{\alpha}\left(\rho^{AB}\big\|\mf\right)=\log|A|-H_{\alpha}^{\ua}(A|B)_\rho\;.
\ee
\item Use the additivity of $H_{\alpha}^\ua$ (see Theorem~\ref{thmcfce} and Exercise~\ref{halpa}) to show that $D_{\alpha}(\rho\|\mf)=D_{\alpha}^\reg(\rho\|\mf)$.
\item Use~\eqref{llowl} to conclude that
\be
\liminf_{n\to\infty}\frac1nD_{\min}^\eps\left(\rho^{\otimes n}\big\|\mf\right)\geq D(\rho\|\mf)\;.
\ee
\item Complete the proof of the theorem.
\een
\eex

\section{The Uniqueness of the Umegaki\index{Umegaki} Relative Entropy}\label{sec:unique}

In this section we use the Stein's lemma\index{Stein's lemma} and the AEP property proved in the previous sections, to show the uniqueness of the Umegaki relative entropy.
Recall that the Umegaki relative entropy is defined for any $\rho,\sigma\in\md(A)$ with $\supp(\rho)\subseteq\supp(\sigma)$ as
\be
D(\rho\|\sigma)\eqdef\tr[\rho\log\rho]-\tr[\rho\log\sigma]\;.
\ee
This quantum relative entropy plays a key role in numerous applications in quantum information theory and beyond. We already saw in the quantum Stein's lemma that it can be interpreted as the optimal decay rate of the type-II error exponent. Among all relative entropies, it is the most well known, and
in this section we show that the Umegaki relative entropy can be singled out as the only quantum relative entropy that is asymptotically continuous. We will see later on in the book that this is the key reason of its ``popularity".

Following Definition~\ref{def:ac7}, we say that a relative entropy $\D$ is asymptotically continuous if there exists a continuous function $f:[0,1]\to\mbb{R}_+$ such that $f(0)=0$ and for all $\rho,\rho',\sigma\in\md(A)$, with $\supp(\rho)\subseteq\supp(\sigma)$ and $\supp(\rho')\subseteq\supp(\sigma)$
\be\label{asy}
\left|\D(\rho\|\sigma)-\D(\rho'\|\sigma)\right|\leq f(\eps)\log\|\sigma^{-1}\|_\infty
\ee
where $\eps\eqdef\frac{1}{2}\|\rho-\rho'\|_1$, and $\sigma^{-1}$ is the \emph{generalized} inverse of $\sigma$. We emphasize that $f$ is independent of $|A|$.

\begin{myt}{\color{yellow} Uniqueness of the Umegaki Relative Entropy}
\begin{theorem}\label{uniqueness}
Let $\D$ be a relative entropy that is asymptotically continuous. Then, $\D=D$, where $D$ is the Umegaki relative entropy.
\end{theorem}
\end{myt}

Recall from Corollary~\ref{cor:acume} that the Umegaki relative entropy is shown to be asymptotically continuous. The theorem we are discussing asserts that no other relative entropy possesses this property of asymptotic continuity. To substantiate this claim, we will utilize the following lemma, which introduces a notation for any relative entropy $\D$:
\be
\D^{(\eps)}(\rho\|\sigma) \coloneqq \max_{\rho' \in \mb_\eps(\rho)} \D(\rho'\|\sigma)\;.
\ee
In other words, $\D^{(\eps)}$ represents a form of smoothing, albeit using the maximum rather than the minimum over all states that are $\eps$-close to $\rho$. Consequently, unlike $\D^\eps$, the function $\D^{(\eps)}$ does not qualify as a divergence. It's also worth noting that this specific notation was previously used in the context of the min relative entropy\index{min relative entropy} in Theorem~\ref{lemma1142}. Both the lemma and this notation are crucial for our proof, as they facilitate the examination of how relative entropies respond to minor perturbations in the state $\rho$.

\begin{myg}{}
\begin{lemma}
Let $\rho,\sigma\in\md(A)$ with $\supp(\rho)\subseteq\supp(\sigma)$, and $\D$ be a quantum relative entropy satisfying~\eqref{asy} (i.e. $\D$ is asymptotically continuous).
Then,
\be\label{gg}
\D(\rho\|\sigma)=\lim_{\eps\to 0^+}\liminf_{n\to\infty}\frac{1}{n}\D^\eps\left(\rho^{\otimes n}\big\|\sigma^{\otimes n}\right)=\lim_{\eps\to 0^+}\limsup_{n\to\infty}\frac{1}{n}\D^{(\eps)}\left(\rho^{\otimes n}\big\|\sigma^{\otimes n}\right)\;.
\ee
\end{lemma}
\end{myg}
\begin{proof}
Let $\eps\in(0,1)$ and for each $n\in\mbb{N}$ let
 $\rho_n'\in\md(A^n)$ be such that $\frac{1}{2}\|\rho_{n}'-\rho^{\otimes n}\|_1\leq\eps$.  Then, applying~\eqref{asy} to $n$ copies of $\rho$ and $\sigma$ gives
\be
\left|\D(\rho\|\sigma)-\frac{1}{n}\D(\rho_{n}'\|\sigma^{\otimes n})\right|\leq f(\eps)\log\|\sigma^{-1}\|_{\infty}\;.
\ee
Therefore, by taking the $\liminf_{n\to\infty}$ or $\limsup_{n\to\infty}$ on both sides of the equation above followed by $\lim_{\eps\to 0^+}$ completes the proof.
\end{proof}

We are now ready to prove Theorem~\ref{uniqueness}.
\begin{proof}[Proof of Theorem~\ref{uniqueness}]

Since the lemma above states that~\eqref{asy} implies~\eqref{gg}, it is sufficient to prove that the Umegaki relative entropy is the only relative entropy that satisfies~\eqref{gg}. Let $\D(\rho\|\sigma)$ be a relative entropy satisfying~\eqref{gg}. Therefore, 
\ba
\D(\rho\|\sigma)&=\lim_{\eps\to 0^+}\liminf_{n\to\infty}\frac{1}{n}\D^\eps\left(\rho^{\otimes n}\big\|\sigma^{\otimes n}\right)\\
\GG{\eqref{b886}}&\leq\lim_{\eps\to 0^+}\liminf_{n\to\infty}\frac{1}{n}D_{\max}^{\eps}\left(\rho^{\otimes n}\big\|\sigma^{\otimes n}\right)\\
\GG{AEP; see~\eqref{seeaep}}&=D(\rho\|\sigma)\;.
\ea
Conversely, 
\ba
\D(\rho\|\sigma)&=\lim_{\eps\to 0^+}\limsup_{n\to\infty}\frac{1}{n}\D^{(\eps)}\left(\rho^{\otimes n}\big\|\sigma^{\otimes n}\right)\\
\GG{\eqref{b886}}&\geq\lim_{\eps\to 0^+}\limsup_{n\to\infty}\frac{1}{n}D_{\min}^{(\eps)}\left(\rho^{\otimes n}\big\|\sigma^{\otimes n}\right)\\
\GG{Theorem~\ref{lemma1142}}&= D(\rho\|\sigma)\;.
\ea
This completes the proof.
\end{proof}

\section{Asymptotic Interconversions}

In the first section of this chapter we studied inter-conversions of a single copy of a resource into another resource state by free operations. In this section we study the asymptotic rate at which many copies of a given state can be converted, by free operations, to many copies of another state. We will consider two types of rates. In one type, the goal is to distill (by free operations) as many copies as possible of a desirable state $\sigma$ from $n$ copies of a less desirable resource state $\rho$. That is, this type of rate is defined as the maximization of the ratio $\frac mn$ given that the conversion $\rho^{\otimes n}\xrightarrow{\mf}\sigma^{\otimes m}$ is possible. We will therefore call this rate the distillation rate of converting $\rho$ into $\sigma$.
In the second type of rate, the goal is to find the smallest number $n$ for which the conversion $\rho^{\otimes n}\xrightarrow{\mf}\sigma^{\otimes m}$ is possible.
We therefore call this rate the cost rate of converting $\rho$ into $\sigma$, and define it as the minimization of the ration $\frac nm$ such that $\rho^{\otimes n}\xrightarrow{\mf}\sigma^{\otimes n}$ is possible (see the precise definition below).

As we discussed earlier, in many resource theories, the set of free operations is not so large, so that the exact conversion of one resource state to another is typically not possible. Therefore, instead of considering the exact conversion of $\rho^{\otimes n}$ to $\sigma^{\otimes m}$ we will allow the output state to be $\eps$-close to $\sigma^{\otimes m}$ as long as the error $\eps$ goes to zero in the asymptotic limit $n,m\to\infty$. This idea is made rigorous in the following definition.

\begin{myd}{}
\begin{definition}\label{def:discos}
Let $\rho\in\md(A)$ and $\sigma\in\md(B)$ be two resource states. 
\ben
\item The \emph{asymptotic distillable rate} of $\rho$ into $\sigma$ is defined as
\be\label{dist5}
\distill(\rho\to\sigma)\eqdef\lim_{\eps\to 0^+}\sup_{n,m\in\mbb{N}}\left\{\frac mn\;:\;T\left(\rho^{\otimes n}\xrightarrow{\mf} \sigma^{\otimes m}\right)\leq\eps\right\}
\ee
\item The \emph{asymptotic cost rate} of $\rho$ into $\sigma$ is defined as
\be\label{cost5}
\cost(\rho\to\sigma)\eqdef\lim_{\eps\to 0^+}\inf_{n,m\in\mbb{N}}\left\{\frac nm\;:\;T\left(\rho^{\otimes n}\xrightarrow{\mf} \sigma^{\otimes m}\right)\leq\eps\right\}
\ee
\een
\end{definition}
\end{myd}

\begin{remark}
The two definitions above are not independent of each other. Specifically, observe that
\be\label{dcrel}
\distill(\rho\to\sigma)=\frac1{\cost(\rho\to\sigma)}\;.
\ee
This relationship is consistent with the intuition that if $\rho$ is a free state and $\sigma$ is a resource state, then $\cost(\rho\to\sigma)$ is equal to infinity, while $\distill(\rho\to\sigma)$ equals zero. This is because, in the former case, no matter how many copies of $\rho$ you have, they are insufficient to prepare even a single copy of $\sigma$. In the latter case, it is impossible to distill or extract a resource state $\sigma$ from a free state $\rho$.
\end{remark}

In the above definitions of cost and distillation, we did not impose any constraints on the integers $m$ and $n$. However, as intuition suggests, it is typically the case that $m$ and $n$ are both very large. In fact, for any natural number $a$, we can include the condition $n,m\geq a$ in the aforementioned definitions without altering their value. Specifically, we argue that:
\be\label{11211}
\cost(\rho\to\sigma)=\lim_{\eps\to 0^+}\inf_{\substack{n,m\in\mbb{N}\\ n,m\geq a}}\left\{\frac nm\;:\;T\left(\rho^{\otimes n}\xrightarrow{\mf} \sigma^{\otimes m}\right)\leq\eps\right\}\;,
\ee
(and similarly we can add $n,m\geq a$ to $\distill(\rho\to\sigma)$).
To see why, observe first that the left-hand side of the equation above cannot be greater than the right-hand side since by adding the restriction $n,m\geq a$ one can only increase the infimum. To prove that we must have equality, recall from Exercise~\ref{ex:tensork} that for any such $a\in\mbb{N}$, if $T\left(\rho^{\otimes n}\xrightarrow{\mf} \sigma^{\otimes m}\right)\leq\eps$ then
$T\left(\rho^{\otimes na}\xrightarrow{\mf} \sigma^{\otimes ma}\right)\leq a\eps$. Therefore, 
\ba
\cost(\rho\to\sigma)&\geq \lim_{\eps\to 0^+}\inf_{n,m\in\mbb{N}}\left\{\frac {n}{m}\;:\;T\left(\rho^{\otimes na}\xrightarrow{\mf} \sigma^{\otimes ma}\right)\leq a\eps\right\}\\
\GG{replacing\;{\it na,ma}\; with\; {\it n',m'}}&\geq \lim_{\eps\to 0^+}\inf_{\substack{n',m'\in\mbb{N}\\ m',n'\geq a}}\left\{\frac {n'}{m'}\;:\;T\left(\rho^{\otimes n'}\xrightarrow{\mf} \sigma^{\otimes m'}\right)\leq a\eps\right\}\\
\Gg{\eps'\eqdef a\eps}&= \lim_{\eps'\to 0^+}\inf_{\substack{n',m'\in\mbb{N}\\ m',n'\geq a}}\left\{\frac {n'}{m'}\;:\;T\left(\rho^{\otimes n'}\xrightarrow{\mf} \sigma^{\otimes m'}\right)\leq \eps'\right\}
\ea
Hence, the equality in~\ref{11211}.

In the following exercise you show that the asymptotic cost and distillable rates are themselves resource measures.

\bex
Consider the asymptotic cost and distillable rates defined above.
\ben
\item Show that for a fixed resource state $\sigma\in\md(B)$, the function $f_\sigma(\rho)\eqdef \distill(\rho\to\sigma)$ is a resource measure.
\item Show that for a fixed resource state $\rho\in\md(B)$, the function $g_\rho(\sigma)\eqdef \cost(\rho\to\sigma)$ is a resource measure.
\een
\eex

\bex\label{equivdistcost}
Let $T'$ be another metric that is topologically equivalent to the trace distance $T$ (i.e., there exists $a,b>0$ such that $aT\leq T'\leq b T$). Further, for every $\rho\in\md(A)$ and $\sigma\in\md(B)$ let $\distill'(\rho\to\sigma)$ and $\cost'(\rho\to\sigma)$ be the distillation and cost rates obtained by replacing the trace distance in~\eqref{dist5} and~\eqref{cost5}with the metric $T'$. Show that for all $\rho\in\md(A)$ and all $\sigma\in\md(B)$ 
\be
\distill'(\rho\to\sigma)=\distill(\rho\to\sigma)\quad\text{and}\quad \cost'(\rho\to\sigma)=\cost(\rho\to\sigma)\;.
\ee
\eex

Typically, in the process of converting $n$ copies of $\rho$ into $m$ copies of $\sigma$, some resource is consumed. This is reflected by the fact that the target state $\sigma^{\otimes m}$ (up to a small error) is less resourceful than the source state $\rho^{\otimes n}$. If this loss of a resource is not too high (e.g. sublinear in $n$), then typically one can use the $m$ copies of $\sigma$ to recover the $n$ copies of $\rho$. If the $n$ copies of $\rho$ can always be recovered (up to an error that goes to zero asymptotically) we say that the resource theory is asymptotically reversible.

\begin{myd}{}
\begin{definition}
A QRT $\mf$ is called \emph{asymptotically reversible} if for all $\rho\in\md(A)$ and $\sigma\in\md(B)$
\be
\distill(\rho\to\sigma)=\cost(\sigma\to\rho)\;.
\ee
\end{definition}
\end{myd}

Note that due to~\eqref{dcrel} the condition that a QRT is reversible can also be expressed as
\be
\distill(\rho\to\sigma)\distill(\sigma\to\rho)=1\;,
\ee
or as
\be
\cost(\rho\to\sigma)\cost(\sigma\to\rho)=1\;.
\ee

\begin{myt}{}
\begin{theorem}\label{thm11}
Let $\mf$ be a QRT, $\rho\in\md(A)$ and $\sigma\in\md(B)$. Then,
\be\label{dfrsdf}
\distill(\rho\to\sigma)\leq \frac{D^\reg(\rho\|\mf)}{D^\reg(\sigma\|\mf)}
\ee
and equality holds if the QRT $\mf$ is reversible.
\end{theorem}
\end{myt}
\begin{remark}
The theorem above can also be expressed in terms of the asymptotic cost rate. Specifically, we have the bound
\be\label{11199}
\cost(\rho\to\sigma)\geq \frac{D^\reg(\sigma\|\mf)}{D^\reg(\rho\|\mf)}\;,
\ee
where we used~\eqref{dcrel} in~\eqref{dfrsdf}.
\end{remark}

\begin{proof}
Let $\{\eps_n\}_{n\in\mbb{N}}$ be a sequence of positive numbers with zero limit, let $\{m_n\}_{n\in\mbb{N}}$ be a sequence of integers, and $\{\mE_n\}_{n\in\mbb{N}}$ be a sequence of free channels with $\mE_n\in\mf(A^n\to B^{m_n})$, such that
\be
\left|\distill(\rho\to\sigma)-\frac{m_n}n\right|\leq\eps_n\quad\text{and}\quad\frac12\left\|\sigma^{\otimes m_n}-\mE_n\left(\rho^{\otimes n}\right)\right\|_1\leq\eps_n\;.
\ee
Recall that the relative entropy of a resource is asymptotically continuous. Therefore, for all $n\in\mbb{N}$
\ba
D\left(\rho^{\otimes n}\big\|\mf\right)&\geq D\left(\mE_n\left(\rho^{\otimes n}\right)\big\|\mf\right)\\
\GG{\eqref{asyfor}}&\geq D\left(\sigma^{\otimes m_n}\big\|\mf\right)-\eps_n\kappa_n-(1+\eps_n)h\left(\frac{\eps_n}{1+\eps_n}\right)
\ea
where $\kappa_n\eqdef\max_{\omega\in\md(B^{m_n})}D(\omega\|\mf)$.
Dividing both sides by $n$ and taking the limit $n\to\infty$ yields
\be
D^\reg(\rho\|\mf)\geq \lim_{n\to\infty}\frac{m_n}n\frac1{m_n}D\left(\sigma^{\otimes m_n}\big\|\mf\right)=\distill(\rho\to\sigma)D^\reg\left(\sigma\|\mf\right)
\ee
where we used the assumption that
\be
\limsup_{n\to\infty}\frac{\kappa_n}{n}<\infty\;.
\ee
This completes the proof of the inequality~\eqref{dfrsdf}.
For the equality, observe first that both bounds~\eqref{dfrsdf} and~\eqref{11199} can be written as
\be
\distill(\rho\to\sigma)\leq \frac{D^\reg(\rho\|\mf)}{D^\reg(\sigma\|\mf)}\leq \cost(\sigma\to\rho)\;.
\ee
Hence, if $\mf$ is reversible then both the inequalities above must be equalities. This completes the proof.
\end{proof}

\subsection{Asymptotic Cost and Distillation of a Resource}\label{sec:asymptoticregime}

Some QRTs contains a golden unit\index{golden unit} (see Definition~\ref{def:gu}) like the maximally entangled states in entanglement theory. In such cases, quite often one is interested in computing asymptotic conversion rates when $\rho$ or $\sigma$ are taken to be elements of the golden unit. Specifically, let $\{\Phi_k\}_{k\in\mbb{N}}$ be a golden unit, and take $\sigma=\Phi_2$ be the two dimensional element of the golden unit. For this choice, the rate  $\distill(\rho\to\Phi_2)$ quantifies the number of resource units (i.e. copies of $\Phi_2$) that can be \emph{distilled} from each copy of $\rho$. For this reason the quantity $\distill(\rho\to\Phi_2)$ is called the distillable resource of $\rho$, and denoted by 
\be\label{dist2}
\distill(\rho)\eqdef \distill(\rho\to\Phi_2)\;.
\ee
Conversely, one can use the asymptotic cost rate to quantify the cost (in  resource units $\Phi_2$) of a resource state $\rho$. Specifically, the quantity
\be\label{cost2}
\cost(\rho)\eqdef \cost(\Phi_2\to \rho)
\ee
quantifies the cost in resource units (i.e. copies of $\Phi_2$) that are needed to prepare each copy of $\rho$. The asymptotic cost and distillation of a resource are related to their single-shot versions as follows.

\begin{myg}{}
\begin{lemma}\label{lem:1152}
Let $\mf$ be a QRT and $\rho\in\md(A)$. Then,
\begin{align}
\cost(\rho)&=\lim_{\eps\to 0^+}\liminf_{n\to\infty}\frac1n\cost^\eps\left(\rho^{\otimes n}\right)\label{deseq}\\
\distill(\rho)&=\lim_{\eps\to 0^+}\limsup_{n\to\infty}\frac1n\distill^\eps\left(\rho^{\otimes n}\right)\;.\label{disteq}
\end{align}
where $\distill^\eps$ and $\cost^\eps$ have been defined in Definition~\ref{def:dceps}.
\end{lemma}
\end{myg}

\begin{proof}
We prove the first equality and leave the second one to Exercise~\ref{ex:dist}.
By definition,
\ba
\inf_{n\in\mbb{N}}\frac1n\cost^\eps\left(\rho^{\otimes n}\right)&=\inf\left\{\frac{\log m}n\;:\;T\left(\Phi_m\xrightarrow{\mf}\rho^{\otimes n}\right)\leq\eps\;,\;\;\;n,m\in\mbb{N}\right\}\\
\GG{restricting\;{\it m=2^k}}&\leq\inf\left\{\frac{k}n\;:\;T\left(\Phi_{2^k}\xrightarrow{\mf}\rho^{\otimes n}\right)\leq\eps\;,\;\;\;n,k\in\mbb{N}\right\}\\
\GG{property\; of\; a\; golden\; unit}&=\inf\left\{\frac{k}n\;:\;T\left(\Phi_{2}^{\otimes k}\xrightarrow{\mf}\rho^{\otimes n}\right)\leq\eps\;,\;\;\;n,k\in\mbb{N}\right\}\;.
\ea
Hence,
\ba
\cost(\rho)&\geq\lim_{\eps\to 0^+}\inf_{n\in\mbb{N}}\frac1n\cost^\eps\left(\rho^{\otimes n}\right)\\
\GG{See\;Exercise~\ref{ex:a11211}\; below}&=\lim_{\eps\to 0^+}\inf_{a\leq n\in\mbb{N}}\frac1n\cost^\eps\left(\rho^{\otimes n}\right)\quad\quad\forall\;a\in\mbb{N}\;.
\ea 
Since the above inequality holds for all $a\in\mbb{N}$ we must have
\be\label{cost112}
\cost(\rho)\geq \lim_{\eps\to 0^+}\liminf_{n\to\infty}\frac1n\cost^\eps\left(\rho^{\otimes n}\right)\;.
\ee

Conversely, let $\eps,\delta\in(0,1)$ and let $a\in\mbb{N}$ be large enough such that $\frac1a<\delta$. 
Then,
\ba\label{11211c}
\liminf_{n\to\infty}\frac1n\cost^\eps\left(\rho^{\otimes n}\right)&\geq\inf_{a\leq n\in\mbb{N}}\frac1n\cost^\eps\left(\rho^{\otimes n}\right)\\
\GG{by\; definition}&=\inf_{\substack{n,m\in\mbb{N}\\n\geq a}}\left\{\frac{\log(m)}n\;:\;T\left(\Phi_m\xrightarrow{\mf}\rho^{\otimes n}\right)\leq\eps\right\}\;.
\ea
Combining this with the inequality $\log(m)\geq \lceil\log m\rceil-1$ gives
\ba
\liminf_{n\to\infty}\frac1n\cost^\eps\left(\rho^{\otimes n}\right)&\geq\inf_{\substack{n,m\in\mbb{N}\\n\geq a}}\left\{\frac{\left\lceil\log m\right\rceil-1}n\;:\;T\left(\Phi_m\xrightarrow{\mf}\rho^{\otimes n}\right)\leq\eps\right\}\\
\Gg{k\eqdef\left\lceil\log m\right\rceil}&\geq\inf_{\substack{n,k\in\mbb{N}\\n\geq a}}\left\{\frac{k-1}n\;:\;T\left(\Phi_{2^k}\xrightarrow{\mf}\rho^{\otimes n}\right)\leq\eps\right\}\;,
\ea
where we used the fact that $m\leq 2^k$ so that $\Phi_{2^k}\xrightarrow{\mf}\Phi_m$ and consequently $T\left(\Phi_{2^k}\xrightarrow{\mf}\rho^{\otimes n}\right)\leq T\left(\Phi_{m}\xrightarrow{\mf}\rho^{\otimes n}\right)$.
Now, observe that  for $n\geq a$ we have $(k-1)/n\geq k/n-\delta$ so that
\ba
\lim_{\eps\to 0^+}\liminf_{n\to\infty}\frac1n\cost^\eps\left(\rho^{\otimes n}\right)&\geq\lim_{\eps\to 0^+}\inf_{n,k\in\mbb{N}}\left\{\frac{k}n\;:\;T\left(\Phi_{2^k}\xrightarrow{\mf}\rho^{\otimes n}\right)\leq\eps\right\}-\delta\\
&=\cost(\rho)-\delta\;.
\ea
Since the above inequality holds for all $\delta\in(0,1)$ we conclude that
\be\label{cost1122}
\lim_{\eps\to 0^+}\liminf_{n\to\infty}\frac1n\cost^\eps\left(\rho^{\otimes n}\right)\geq \cost(\rho)\;.
\ee
The two inequalities~\eqref{cost112} and~\eqref{cost1122} then gives the desired equality~\eqref{deseq}.
\end{proof}

\bex\label{ex1293}
Show that for any $m\in\mbb{N}$ and $\rho\in\md(A)$ we have
\be\label{11120}
\frac1m\cost\left(\rho^{\otimes m}\right)\geq\cost(\rho)\quad\text{and}\quad\frac1m\distill\left(\rho^{\otimes m}\right)\leq\distill(\rho)\;.
\ee
\eex

\bex\label{ex:a11211}
Let $\rho\in\md(A)$.
 Show that for any $a\in\mbb{N}$   we have
\be
\lim_{\eps\to 0^+}\inf_{n\in\mbb{N}}\frac1n\cost^\eps\left(\rho^{\otimes n}\right)=\lim_{\eps\to 0^+}\inf_{a\leq n\in\mbb{N}}\frac1n\cost^\eps\left(\rho^{\otimes n}\right)\;.
\ee
Hint: Use similar arguments that were used to prove the equality in~\eqref{11211}.
\eex

\bex\label{ex:dist}
Prove the equality in~\eqref{disteq}.
\eex

In many QRTs it is possible to choose the golden unit\index{golden unit} such that $D^\reg_\mf(\Phi_2)=1$. With this normalization we get from Theorem~\ref{thm11} that
\be\label{dculb}
\distill(\rho)\leq D^\reg(\rho\|\mf)\leq\cost(\rho)\;.
\ee
Particularly, if the QRT $\mf$ is reversible then both the asymptotic cost and the asymptotic distillation of the resource $\rho$ equals $D^\reg(\rho\|\mf)$. Therefore, for reversible QRTs, the regularized relative entropy of a resource is the unique measure of a resource in the asymptotic domain. We make this statement rigorous in the following corollary.

\begin{myg}{}
\begin{corollary}
Let $\mf$ be a reversible QRT with a golden unit $\{\Phi_k\}_{k\in\mbb{N}}$ such that $D(\Phi_2\|\mf)=1$, and let $\M$ be a resource measure that is asymptotically continuous and normalized such that $\M(\Phi_2)=1$. Then,
\be
\M^\reg(\rho)=D^\reg(\rho\|\mf)\quad\quad\forall\;\rho\in\md(A)\;.
\ee
\end{corollary}
\end{myg}

\bex
Prove the corollary above. Hint: Follow all the lines leading to~\eqref{dculb}, but with $\M$ replacing everywhere $D(\cdot\|\mf)$. 
\eex

\subsection{Achieving Reversibility}\index{reversibility}

The reversibility property of a QRT is extremely desirable, as quantum resources are expensive and reversibility ensures resources are not wasted during quantum information processing tasks. Moreover, in the previous subsections we saw that if a QRT is reversible then the relative entropy of a resource characterize uniquely all asymptotic interconversions. However, many QRTs are not asymptotically reversible. This typically happens when the set of free operations is not large enough to enable efficient interconversion of resources. It is therefore natural to ask if a QRT is reversible under the maximal set of free operations; i.e. under RNG operations.

\subsubsection{Asymptotically RNG Operations}

In Sec.~\ref{sec:rng} we discussed RNG operations and argued that they form the largest possible set of free operations. However, in the asymptotic regime, when one consider many copies of resources, one can define a set of operations that are RNG only in the asymptotic limit. That is, the operations become closer to RNG operations when we take the number of copies of the resources involved to infinity.
To make this idea rigorous, we first define a set of operations that are approximately RNG.

\begin{myd}{}
\begin{definition}
Let $\delta\in[0,1]$ and $\mf$ be a QRT. We say that a quantum channel $\mN\in\cptp(A\to B)$ is $\rng_\delta$ if it belong to the set
\be
\rng_\delta(A\to B)\eqdef\Big\{\mE\in\cptp(A\to B)\;:\;\R_g\big(\mE(\sigma)\big)\leq\delta\quad\forall\;\sigma\in\mf(A)\Big\}\;,
\ee
where $\R_g$ is the global robustness of a resource as defined in~\eqref{robust}.
\end{definition} 
\end{myd}

We have used the global robustness in the definition above since it is a resource monotone that is faithful (see Exercise~\ref{robfaith}) so that the inequality $\R_g\big(\mE(\sigma)\big)\leq\delta$ implies that $\mE(\sigma)$ is close to a free state.
Specifically, suppose $\mu\eqdef \R_g\big(\mE(\sigma)\big)\leq\delta$. Then, 
from~\eqref{rhoasie} it follows that
\be
\mE(\sigma)=(1+\mu)\tau-\mu\omega
\ee
for some $\tau\in\mf(B)$ and $\omega\in\md(B)$. Hence, from the above equality we get
\be
\frac12\big\|\mE(\sigma)-\tau\big\|_1=\frac12\big\|\mu(\tau-\omega)\big\|_1\leq\mu\leq\delta\;.
\ee
In other words, if $\R_g\big(\mE(\sigma)\big)\leq\delta$ then $\mE(\sigma)$ is $\delta$-close to a free state. 

\begin{myd}{}
\begin{definition}
Let $\mf$ be a QRT, and for each $n\in\mbb {N}$ let $A_n$ and $B_n$ be two physical systems. A sequence of quantum channel $\{\mE_n\}_{n\in\mbb{N}}$, with $\mE_n\in\cptp(A_n\to B_n)$, is said to be asymptotically RNG   if there exists a sequence of non-negative real numbers $\{\delta_n\}_{n\in\mbb{N}}$ with $\lim_{n\to\infty}\delta_n=0$ such that for each $n\in\mbb{N}$, $\mE_n\in\rng_{\delta_n}(A_n\to B_n)$.
\end{definition}
\end{myd}

Note that in the definition above we do not specify how quickly $\delta_n$ goes to zero. The main result of this section will not be effected even if we require in addition that $\delta_n$ goes to zero exponentially fast with $n$. However, to keep the notion of asymptotically RNG in its most generality we did not include such a condition in the definition above.

%Let $\mf$ be a convex QRT admitting the tensor product structure. For any $\rho\in\md(A)$ and $\sigma\in\md(B)$ define the RNG conversion distance as
%\be
%d_{\rng}\left(\rho\to \sigma\right)\eqdef\min\left\{\frac12\left\|\mE(\rho)-\sigma\right\|_1\;:\;\mE\in\rng(A\to B)\right\}\;.
%\ee

\subsubsection{Cost and Distillation} 

The asymptotic distillable rate of $\rho$ into $\sigma$ under asymptotically RNG operations is defined slightly different than the definitions given in Definition~\ref{def:discos}.  Recall that the distillable rate is defined as the supremum of the ratio $\frac mn$ under the constraints that the conversion distance between $\rho^{\otimes n}$ and $\sigma^{\otimes m}$ is not too high. 
In the context of \emph{asymptotically} RNG, instead of taking the \emph{supremum} of $\frac mn$ we take the \emph{limit}  when $n$ goes to infinity since we are only interested in the optimal \emph{asymptotic} behaviour of this ratio. 

For any $\eps\in(0,1)$ we will use the notation $\mr_\eps(\rho\to\sigma)$ to denote the set of all $r\in\mbb{R}_+$ such that there exists a sequence $\{m_n\}_{n\in\mbb{N}}\subset\mbb{N}$  fulfilling the two criteria:
\ben
\item $\lim_{n\to\infty}\frac{m_n}{n}=r$.
\item There exists another sequence $\{\delta_n\}_{n\in\mbb{N}}\subset\mbb{R}_+$  with a limit of zero, such that for every $n\in\mbb{N}$
\be\label{deltaneps}
T\left(\rho^{\otimes n}\xrightarrow{{\rng_{\delta_n}}} \sigma^{\otimes {m_n}}\right)\leq \eps\;.
\ee
\een
That is, the sequence $\{m_n\}_{n\in\mbb{N}}$ is such that $\rho^{\otimes n}$ can be converted by $\rng_{\delta_n}$ to a state that is $\eps$-close to $\sigma^{\otimes {m_n}}$. Hence, the set $\mr_\eps(\rho\to\sigma)\subset\mbb{R}_+$ consists of all achievable conversion rates under asymptotically RNG that tolerate an $\eps$-error.
To get the optimal distillable rate we will have to take the limit $\eps\to 0^+$.

\bex
Let $\rho\in\md(A)$, $\sigma\in\md(B)$, and
\be\label{rdasheps}
r_\eps\eqdef\sup\big\{r\;:\;r\in\mr_\eps(\rho\to\sigma)\big\}\;.
\ee
\ben
\item Show that  $\mr_\eps(\rho\to\sigma)=[0,r_\eps]$; in particular, show that the supremum in the definition of $r_\eps$ can be replaced with a maximum.
\item Show that $r_\eps$ is non-increasing in $\eps$.
\een
\eex

\begin{myd}{}
\begin{definition}
Let $\mf$ be a QRT, $\rho\in\md(A)$ and $\sigma\in\md(B)$.  Using the notation given in~\eqref{rdasheps} of the exercise above, the asymptotically RNG distillable rate is defined as
\be\label{deflimsup}
\distill(\rho\to\sigma)\eqdef\lim_{\eps\to 0^+}r_\eps\;.
\ee
 \end{definition}
\end{myd}

The condition~\eqref{deltaneps}  implies that there exists $\mE_n\in\rng_{\delta_n}(A^n\to B^{m_n})$ such that $\mE_n(\rho^{\otimes n})\approx_{\eps}\sigma^{\otimes m_n}$. Moreover, the condition that $\lim_{n\to\infty}\delta_n=0$ implies that this sequence of channels $\{\mE_n\}$ is asymptotically RNG. Therefore, for a given $\eps\in(0,1)$, we get that  $\rho^{\otimes n}$ can be converted by $\rng_{\delta_n}$ to $\sigma^{\otimes m_n}$ up to an $\eps$-error. The reason that we require $\lim_{n\to\infty}\frac {m_n}n=r$ instead of just $\sup\{\frac{m_n}n\}=r$ is that the supremum can be achieved with a finite $n$ in which case $\delta_n$ may not be very small. Taking the limit $n\to\infty$ ensures that the conversion $\rho^{\otimes n}\xrightarrow{\rng_{\delta_n}}\sigma^{\otimes m_n}$ (up to an $\eps$-error) is achieved with a very small $\delta_n$. 

\bex
Show that if in the definition of $\mr_\eps(\rho\to\sigma)$ we require $\sup\{\frac{m_n}n\}=r$ instead of $\lim_{n\to\infty}\frac {m_n}n=r$ then we will get that $\distill(\rho\to\sigma)=\infty$. Hint: Let $n_0$ be a large integer and take $\delta_n=n_0$ for $n\leq n_0$ and $\delta_n=0$ if $n>n_0$.
\eex

For simplicity of the notation, we did not include a subscript in $\distill(\rho\to\sigma)$ to indicate that the asymptotic distillable rate  is calculated with respect to asymptotically RNG operations. Similarly, we denote by $\cost(\rho\to\sigma)=1/\distill(\rho\to\sigma)$ the asymptotic cost rate of $\rho$ into $\sigma$ under asymptotic RNG operations.

%\begin{myg}{}
%\begin{lemma}\label{lem1111}
%Let $\mf$ be a QRT, $\rho\in\md(A)$ and $\sigma\in\md(B)$. Then,
%\be
%\distill(\rho\to\sigma)\leq \frac{D^\reg(\rho\|\mf)}{D^\reg(\sigma\|\mf)}\leq \cost(\rho\to\sigma)\;.
%\ee
%\end{lemma}
%\end{myg}

\subsubsection{Towards Reversibility}\index{reversibility}

In this book, we will restrict our attention to QRTs that meet the following condition:
\be
\kappa(A)\eqdef\limsup_{n\to\infty}\frac1n\max_{\omega\in\md(A^n)}D(\omega\|\mf)<\infty\;.
\ee
It's worth noting that this assumption is extremely lenient and is fulfilled by the majority, if not all, of the QRTs discussed in the existing literature. In fact, for many QRTs $\kappa(A)=0$.

\begin{myt}{}
\begin{theorem}\label{disthm}
For any $\rho\in\md(A)$ and $\sigma\in\md(B)$,  the asymptotic distillable rate of $\rho$ into $\sigma$ under asymptotic RNG is bounded by
\be
\distill(\rho\to\sigma)\leq \frac{D^{\reg}(\rho\|\mf)}{D^{\reg}(\sigma\|\mf)}\;.
\ee
\end{theorem}
\end{myt}
\begin{remark}
The theorem above does \emph{not} follow from Theorem~\ref{thm11} since $\distill(\rho\to\sigma)$ is calculated with respect to asymptotically RNG operations. Since these operations allow for the generation of a  resource (although small amount that vanishes asymptotically), the proof of Theorem~\ref{thm11} cannot be applied directly, and a revised version is necessary to accommodate this case.
\end{remark}
%\begin{remark}
%The proof of the theorem above has two parts. The first one is the direct coding part of the proof which states that $\distill(\rho\to\sigma)\geq \frac{D^{\reg}(\rho\|\mf)}{D^{\reg}(\sigma\|\mf)$, and the second is the converse part which states that $\distill(\rho\to\sigma)\leq \frac{D^{\reg}(\rho\|\mf)}{D^{\reg}(\sigma\|\mf)$.
%\end{remark}

\begin{proof}
Suppose by contradiction that 
\be
\distill(\rho\to\sigma)> \frac{D^{\reg}(\rho\|\mf)}{D^{\reg}(\sigma\|\mf)}+2\delta
\ee
for some small positive $\delta$. By definition, this means in particular that for sufficiently small $\eps\in(0,1)$ there exists  $r\in\mr_\eps(\rho\to\sigma)$ such that
\be
r> \frac{D^{\reg}(\rho\|\mf)}{D^{\reg}(\sigma\|\mf)}+\delta\;.
\ee
Since $r\in\mr_\eps(\rho\to\sigma)$  there exists a sequence $\{m_n\}_{n\in\mbb{N}}\subset\mbb{N}$ satisfying both $r=\lim_{n\to\infty}\frac{m_n}n$ and~\eqref{deltaneps}. From~\eqref{deltaneps} it follows that there exists $\mE_n\in\rng_{\delta_n}\left(A^n\to B^{m_n}\right)$ such that
\be
\mE_n\left(\rho^{\otimes n}\right)\approx_\eps\sigma^{\otimes m_n}\;.
\ee
Now, since $D(\cdot\|\mf)$ is asymptotically continuous it follows that (cf.~\eqref{asyfor})
\be
\left|D\left(\mE_n\left(\rho^{\otimes n}\right)\big\|\mf\right)- D\left(\sigma^{\otimes m_n}\big\|\mf\right)\right|\leq c_n\eps+(1+\eps)h\left(\frac{\eps}{1+\eps}\right)
\ee
where 
\be
c_n\eqdef\max_{\omega\in\md(B^{m_n})}D(\omega\|\mf)\;.
\ee
Dividing both sides by $m_n$ and taking the limit $n\to\infty$ gives
\be\label{11130}
D^{\reg}(\sigma\|\mf)\leq \lim_{n\to\infty}\frac1{m_n}D\left(\mE_n\left(\rho^{\otimes n}\right)\big\|\mf\right)+\kappa(B)\eps\;.
\ee
For each $n\in\mbb{N}$, let $\omega_n\in\mf\left(A^{n}\right)$ be an optimizer state satisfying
\be
D\left(\rho^{\otimes n}\big\|\omega_n\right)=D\left(\rho^{\otimes n}\big\|\mf\right)\;.
\ee
Then, for each $n\in\mbb{N}$
\ba\label{11132}
D\left(\mE_n\left(\rho^{\otimes n}\right)\big\|\mf\right)&=\min_{\tau_n\in\mf\left(B^{m_n}\right)}D\left(\mE_n\left(\rho^{\otimes n}\right)\big\|\tau_n\right)\\
\GG{\eqref{tri0}}&\leq D\left(\mE_n\left(\rho^{\otimes n}\right)\big\|\mE_n(\omega_n)\right)+\min_{\tau_n\in\mf\left(B^{m_n}\right)}D_{\max}\left(\mE_n\left(\omega_n\right)\big\|\tau_n\right)\\
\GG{DPI}&\leq D\left(\rho^{\otimes n}\big\|\mf\right)+D_{\max}\left(\mE_n\left(\omega_n\right)\big\|\mf\right)
\ea
Since $\omega_n$ is a free state and since each $\mE_n$ is $\rng_{\delta_n}$, the global robustness\index{robustness} of  $\mE_n\left(\omega_n\right)$  cannot exceed $\delta_n$, and in particular $D_{\max}\left(\mE_n\left(\omega_n\right)\big\|\mf\right)\leq\log(1+\delta_n)$. Therefore,
\be\label{1163}
D\left(\mE_n\left(\rho^{\otimes n}\right)\big\|\mf\right)\leq D\left(\rho^{\otimes n}\big\|\mf\right)+\log(1+\delta_n)\;.
\ee
Substituting this into~\eqref{11130} gives
\ba\label{donotcm}
D^{\reg}(\sigma\|\mf)&\leq\lim_{n\to\infty}\frac1{m_n}\Big(D\left(\rho^{\otimes n}\big\|\mf\right)+\log(1+\delta_n)\Big)+\kappa(B)\eps\\
&=\lim_{n\to\infty}\frac{n}{m_n}\frac1{n}D\left(\rho^{\otimes n}\big\|\mf\right)+\kappa(B)\eps\\  
\Gg{\lim_{n\to\infty}\frac{m_n}n=r}&=\frac1rD^{\reg}(\rho\|\mf)+\kappa(B)\eps\;.
\ea
However, since $r>\frac{D^{\reg}(\rho\|\mf)}{D^{\reg}(\sigma\|\mf)}+\delta$ for sufficiently small $\eps\in(0,1)$ we get the contradiction
\ba
D^{\reg}(\sigma\|\mf)&\leq\frac1rD^{\reg}(\rho\|\mf)+\kappa(B)\eps\\
\GG{Exercise~\ref{contraex}}&<D^{\reg}(\sigma\|\mf)\;.
\ea
This completes the proof.
\end{proof}

\bex\label{contraex}
Set $a\eqdef D^{\reg}(\rho\|\mf)$ and $b\eqdef D^{\reg}(\sigma\|\mf)$. Show that if $\eps\in(0,1)$ is chosen small enough such that
\be
\kappa(B)\eps<\frac{a^2}{b}\delta\;,
\ee
then for $r>\frac ab+\delta$
\be
\frac ar+\kappa(B)\eps<b\;.
\ee
\eex

If the channels $\{\mE_n\}_{n\in\mbb{N}}$ in the proof above where generating a sublinear amount of a resource (instead of being asymptotically RNG) the result would still not change. That is, in the proof above we could replace the condition $\mE_n\in\rng_{\delta_n}(A^{n}\to B^{m_n})$ with the weaker condition that
\be\label{slasi}
\lim_{n\to\infty}\max_{\omega_n\in\mf(A^{n})}\frac{D_{\max}\left(\mE_n(\omega_n)\big\|\mf\right)}{m_n}=0\;.
\ee
To see why, observe that the only change in the proof above would be to replace the term $\frac{\log(1+\eps)}n$ in the first line of~\eqref{donotcm} with the ratio $D_{\max}\left(\mE_n(\omega_n)\big\|\mf\right)/m_n$ which also goes to zero in the limit $n\to\infty$.
Hence, if the logarithmic robustness of $\mE_n(\omega_n)$ grows sublinearly with $m_n$ the bound on the distillable rate would still hold. This observation is consistent with the intuition that a sublinear amount of a resource becomes negligible in the asymptotic limit and therefore cannot increase the distillable rate.

 \begin{myt}{}
\begin{theorem}
For any $\rho\in\md(A)$, $\sigma\in\md(B)$, and $\eps\in(0,1)$, the asymptotic distillable rate of $\rho$ into $\sigma$ under asymptotic RNG is given by
\be
\distill(\rho\to\sigma)=\frac{D^{\reg}(\rho\|\mf)}{D^{\reg}(\sigma\|\mf)}\;.
\ee
\end{theorem}
\end{myt}

\begin{proof}
Due to Theorem~\ref{disthm}, it is sufficient to prove that
\be
\distill(\rho\to\sigma)\geq \frac{D^{\reg}(\rho\|\mf)}{D^{\reg}(\sigma\|\mf)}\;.
\ee 
For this purpose, let $r$ be a positive number satisfying $r< D^{\reg}(\rho\|\mf)/D^{\reg}(\sigma\|\mf)$, fix $\eps\in(0,1)$, and denote by $m_n\eqdef\left\lceil nr\right\rceil$ so that $\lim_{n\to\infty}\frac {m_n}{n}= r$.
We will construct a sequence of channels $\{\mE_n\}_{n\in\mbb{N}}$ with the following two properties:
\ben
\item For sufficiently large $n\in\mbb{N}$, the channel $\mE_n\in\rng_{\delta_n}(A^n\to B^{m_n})$ with $\delta_n\eqdef 2^{-n\delta}$ (for some $\delta>0$). Hence, the sequence $\{\mE_n\}_{n\in\mbb{N}}$ is asymptotically RNG.
\item For sufficiently large $n\in\mbb{N}$, we have $\mE_n(\rho^{\otimes n})\approx_{\eps}\sigma^{\otimes m_n}$.
\een
Note that from the definition of $\distill(\rho\to\sigma)$, if for any choice of $\eps\in(0,1)$ there exists a sequence $\{\mE_n\}_{n\in\mbb{N}}$ that satisfies the above two conditions then we must have $\distill(\rho\to\sigma)\geq r$.

The idea behind the construction of the channels $\{\mE_n\}_{n\in\mbb{N}}$ is to try to achieve the rate $r$ with a (two-outcome) measurement-prepare channel\index{measurement-prepare channel} of the form
\be\label{e-n}
\mE_n\left(\eta\right)\eqdef\tr\left[\Lambda_n\eta\right]\sigma_n+\tr\left[\left(I^{A^n}-\Lambda_n\right)\eta\right]\omega_n\quad\quad\forall\;\eta\in\ml(A^n)\;,
\ee
for some $\sigma_n,\omega_n\in\md(B^{m_n})$ and some $\Lambda_n\in\eff(A^n)$. We therefore need to check if there exists $\sigma_n$, $\omega_n$ and $\Lambda_n$ that satisfy both $\mE_n(\rho^{\otimes n})\approx_{\eps}\sigma^{\otimes m_n}$ and $\mE_n\in\rng_{\delta_n}(A^n\to B^{m_n})$. Note that if we choose $\Lambda_n$ such that $\tr[\Lambda_n\rho^{\otimes n}]$ is close to one then $\mE_n(\rho^{\otimes n})$ will be close to $\sigma_n$. Therefore, if $\sigma_n$ is close to $\sigma^{\otimes m_n}$ we will get in this case that $\mE_n(\rho^{\otimes n})$ is also close to $\sigma^{\otimes m_n}$. 
We take $\omega_n\in\mf\left(B^{m_n}\right)$ to be any free density matrix, and define now $\Lambda_n$ and $\sigma_n$.
\ben
\item \textbf{Definition of $\Lambda_n$}:  Denote by $a\eqdef D^\reg(\rho\|\mf)$ and observe that from the generalized quantum Stein's lemma (see Theorem~\ref{gsl}) for every $\delta>0$ and sufficiently large $n\in\mbb{N}$ we get
\be
\min_{\tau_n\in\mf(A^{n})}D_{\min}^{\eps/2}\left(\rho^{\otimes n}\big\|\tau_n\right)\geq 
n\left(a-\delta\right)\;,
\ee
(the choice $\eps/2$ instead of $\eps$ will be clear shortly).
The left-hand side of the inequality above can be expressed as
\ba
\min_{\tau_n\in\mf(A^n)}D_{\min}^{\eps/2}\left(\rho^{\otimes n}\big\|\tau_n\right)&=-\log\max_{\tau_n\in\mf(A^n)}\min_{\substack{\Lambda_n'\in\eff(A^n)\\
\tr[\rho^{\otimes n}\Lambda_n']\geq 1-{\eps/2}}}\tr\left[\Lambda_n'\tau_n\right]\\
\GG{\substack{\rm Minimax\; Theorem\\ \rm See\; Lemma~\ref{lem: sion}} }&=-\log\min_{\substack{\Lambda_n'\in\eff(A^n)\\
\tr[\rho^{\otimes n}\Lambda_n']\geq 1-{\eps/2}}}\max_{\tau_n\in\mf(A^n)}\tr\left[\Lambda_n'\tau_n\right]\\
\GG{Definition\;of\;{\it \Lambda_n}}&\eqdef-\log\max_{\tau_n\in\mf(A^n)}\tr\left[\Lambda_n\tau_n\right]\;.
\ea
 Combining the two equations above implies that the optimal effect $\Lambda_n$ satisfies (for and $\delta>0$ and sufficiently large $n\in\mbb{N}$)
\be\label{tng}
\max_{\tau_n\in\mf(A^n)}\tr\left[\Lambda_n\tau_n\right]\leq 2^{-n\left(a-\delta\right)}\quad\text{and}\quad\tr[\rho^{\otimes n}\Lambda_n]= 1-\frac{\eps}2\;.
\ee
\item \textbf{Definition of $\sigma_n$}:
First, observe that if we choose $\sigma_n$ to be $\eps/2$-close to $\sigma^{\otimes m_n}$ we get from the triangle inequality
\ba
\frac12\left\|\mE_n\left(\rho^{\otimes n}\right)-\sigma^{\otimes m_n}\right\|_1
&\leq \frac12\left\|\mE_n\left(\rho^{\otimes n}\right)-\sigma_n\right\|_1+\frac12\left\|\sigma_n-\sigma^{\otimes m_n}\right\|_1\\
\GG{\eqref{e-n}}&= \frac12\left\|\frac\eps2(\omega_n-\sigma_n)\right\|_1+\frac12\left\|\sigma_n-\sigma^{\otimes m_n}\right\|_1\\
&\leq \frac\eps2+\frac\eps2=\eps\;.
\ea
That is, $\mE_n\left(\rho^{\otimes n}\right)\approx_{\eps}\sigma^{\otimes m_n}$. Therefore, we would like to define $\sigma_n$ that is $\eps/2$-close to $\sigma^{\otimes m_n}$ such that $\mE_n\in\rng_{\delta_n}(A^n\to B^{m_n})$.

We take $\sigma_n\in\md\left(B^{m_n}\right)$ to be a density matrix that satisfies $D_{\max}(\sigma_n\|\mf)=D_{\max}^{\eps/2}\left(\sigma^{\otimes m_n}\big\|\mf\right)$. The intuition behind this choice is that besides of being $\eps/2$-close to $\sigma^{\otimes m_n}$, the density matrix $\sigma_n$ does not have ``too much" robustness\index{robustness}. To see why, recall first that from Lemma~\ref{lem:lb} it follows that
\ba
D^{\reg}(\sigma\|\mf)&\geq\limsup_{n\to\infty}\frac1{m_n}D_{\max}^{\eps/2}\left(\sigma^{\otimes m_n}\big\|\mf\right)\\
&=\limsup_{n\to\infty}\frac1{m_n}D_{\max}(\sigma_n\|\mf)\\
\Gg{\lim_{n\to\infty}\frac {m_n}n=r}&=\frac1{r}\limsup_{n\to\infty}\frac1{n}D_{\max}(\sigma_n\|\mf)\;.
\ea
Now, since the inequality $r<a/D^{\reg}(\sigma\|\mf)$ is strict, there exists $\delta>0$ sufficiently small such that $r<(a-2\delta)/D^{\reg}(\sigma\|\mf)$, or equivalently
\be\label{3delta} 
rD^{\reg}(\sigma\|\mf)<a-2\delta\;.
\ee
Hence, by combining the two equations above we get that for sufficiently large $n$
\be
D_{\max}(\sigma_n\|\mf)\leq n\left(a-2\delta\right)\;.
\ee
That is, the global robustness\index{robustness} (as defined in~\eqref{robust}) of $\sigma_n$ satisfies
\be\label{rsn}
\R_g(\sigma_n)\leq 2^{n\left(a-2\delta\right)}-1\;.
\ee
\een

To show that for these choices the channel $\mE_n$ is $\rng_{\delta_n}$,
let $\eta\in\mf(A^n)$ be a free state, and denote by $t_n\eqdef\tr\left[\Lambda_n\eta\right]$ and $r_n\eqdef \R_g(\sigma_n)$. Then, from the convexity of the global robustness we get
\ba
\R_g\big(\mE_n\left(\eta\right)\big)&\leq t_n\R_g(\sigma_n)+(1-t_n)\R_g(\omega_n)\\
\Gg{\omega_n\in\mf\left(B^{m_n}\right)}&=t_nr_n\leq t_n(1+r_n)\;.
\ea
Now, from~\eqref{rsn} we have $r_n+1\leq 2^{n\left(a-2\delta\right)}$ and from~\eqref{tng}
\be
t_n\eqdef\tr\left[\Lambda_n\eta\right]\leq \max_{\tau_n\in\mf(A^n)}\tr\left[\Lambda_n\tau_n\right]\leq 2^{-n\left(a-\delta\right)}\;.
\ee
Combining everything we get
\be
\R_g\left(\mE_n\left(\eta\right)\right)\leq 2^{-n\delta}=\delta_n\;.
\ee
That is, $\mE_n\in\rng_{\delta_n}\left(A^n\to B^{m_n}\right)$. This completes the proof.
\end{proof}

\bex
Let $\sigma_n$ and $\sigma^{\otimes m_n}$ be as in the proof above.
\ben
\item Show that the robustness of $\sigma^{\otimes m_n}$ is bounded by
\be\label{boundex}
\R_g\left(\sigma^{\otimes m_n}\right)\leq 2^{m_nD_{\max}(\sigma\|\mf)}-1
\ee 
\item Show that the right-hand side of~\eqref{boundex} is larger than the bound on $R(\sigma_n)$ given in~\eqref{rsn}.
\een
\eex

\subsubsection{Examples}\index{reversibility}

As an example, consider the resource theory of quantum coherence. In this example, the set of free states are diagonal with respect to a fixed basis. In this example, the $\alpha$-R\'enyi relative entropy of a resource is additive. To see this recall that (see~\eqref{cfrelative}) 
\be
D_{\alpha}(\rho\|\mf)=\frac{1}{\alpha-1}\log\left\|\Delta\left(\rho^\alpha\right)\right\|_{1/\alpha}
\ee
where $\Delta\in\cptp(A\to A)$ is the completely dephasing map. Moreover, on $n$ copies of $A$, the completely dephasing map $\Delta_n\in\cptp(A^n\to A^n)$ satisfy $\Delta_n=\Delta^{\otimes n}$, where $\Delta$ is the completely dephasing map on a single copy of $A$. We therefore get that
\ba
D_{\alpha}\left(\rho^{\otimes n}\big\|\mf\right)&=\frac{1}{\alpha-1}\log\left\|\Delta_n\left(\left(\rho^\alpha\right)^{\otimes n}\right)\right\|_{1/\alpha}\\
&=\frac{1}{\alpha-1}\log\left\|\left(\Delta\left(\rho^\alpha\right)\right)^{\otimes n}\right\|_{1/\alpha}\\
&=n\frac{1}{\alpha-1}\log\left\|\Delta\left(\rho^\alpha\right)\right\|_{1/\alpha}=nD_{\alpha}(\rho\|\mf)\;.
\ea
Therefore, in the case,
\be
D_{\alpha}^\reg(\rho\|\mf)=D_{\alpha}(\rho\|\mf)\;.
\ee
Since $D_{\alpha}(\cdot\|\mf)$ is continuous at $\alpha=1$ (see Lemma~\ref{lem:cont}) we conclude that  the QRT of quantum coherence is reversible with
\be
\distill(\rho\to\sigma)=\frac{H\big(\Delta(\rho)\big)-H(\rho)}{H\big(\Delta(\sigma)\big)-H(\sigma)}\;.
\ee
\bex
Prove the equality above.
\eex

As a second example, consider the QRT consisting of conditionally unital\index{conditionally unital}  channels. In this QRT the free states are given by
\be
\mf(AB)=\left\{\u^A\otimes\sigma^B\;:\sigma\in\md(B)\right\}
\ee
Since this QRT is also an affine\index{affine} QRT it has a self-adjoint resource destroying channel given by
\be
\Delta^{AB\to AB}\left(\omega^{AB}\right)\eqdef\u^A\otimes\omega^B\quad\quad\forall\;\omega\in\ml(AB)\;.
\ee
Also in this QRT the $\alpha$-relative entropy of a resource is additive so that we get the distillable rate to be
\be
\distill\left(\rho^{AB}\to\sigma^{AB}\right)=\frac{\log|A|-H(A|B)_\rho}{\log|A|-H(A|B)_\sigma}\;.
\ee
\bex
Prove the equality above.
\eex
\section{Notes and References}

Theorem~\ref{thm:1111} is credited to~\cite{GS2021}. In deriving Corollary~\ref{cor:mono}, we adopted the approach of~\cite{Vidal2000}, which utilized similar methodologies to establish the same corollary within entanglement theory. The generalized Asymptotic Equipartition Property (AEP) presented in Theorem~\ref{AEP} is a novel addition to this book, although a less robust version was initially proven by~\cite{BP2010}. In the same publication, the authors also supported the validity of the generalized Stein's lemma (Theorem~\ref{gsl}); however, subsequent analyses identified a flaw in the proof, as discussed in~\cite{FGW2022} and~\cite{BBG+2022}. Fortunately, the gap was recently addressed by two independent corrections found in~\cite{HY2024} and~\cite{Lami2024}, with this book adopting the proof strategy from~\cite{HY2024}. The proof for the uniqueness of the Umegaki relative entropy, as outlined in~\cite{GT2020}, traces its origins back to the foundational work by~\cite{Matsumoto2010}.

%%%%%%%%%%%%%%%%%%%%%%%%%%%%%%%%%%%%%%%%%%%%%%%%%%%%%%%%%%%%%%%%%%%%%%%%%%%%%%%%%%%%%%%%%%%%%%%%%%%%%%%%%%%%%%%%%%%%%%%%%%%%

\part{Entanglement Theory}

\chapter{Pure-State Entanglement}\label{entanglement}

Entanglement theory is the poster child of quantum resource theories. As we explored in Chapter~\ref{ch:ent0}, entanglement not only piques our curiosity from a fundamental perspective but also represents a valuable resource that can facilitate specific quantum information processing tasks. In the chapters ahead, we embark on a journey to rigorously define entanglement and formulate its corresponding resource theory. Our focus encompasses the classification, detection, quantification, and manipulation of entanglement.

\section{Definition of Quantum Entanglement}\index{quantum entanglement}

As discussed earlier, entanglement can be regarded as a resource with practical utility in specific quantum information processing tasks. Take, for instance, quantum teleportation, a process where entanglement is harnessed to simulate a quantum channel. Naturally, if Alice and Bob already possess a noiseless quantum channel that they can freely employ an unlimited number of times, entanglement holds no value for them since they can generate it without constraints. In this context, entanglement serves as the means by which parties surmount the limitations imposed by their apparatuses, enabling them to perform operations that extend beyond local quantum operations (e.g., quantum measurements) aided by classical communication. Thus, we can operationally define entanglement as follows:

\begin{myd}{Quantum Entanglement}
\begin{definition}\label{ent}
Entanglement is a characteristic of a composite physical system that cannot be created or enhanced through local (quantum) operations and classical communication (LOCC).
\end{definition}
\end{myd}

This definition precisely captures the intuition that entanglement is a quantum property of a composite system\index{composite system} that corresponds to correlations that are not classical. Historically, this intuition led many researchers to associate entanglement with the non-local correlations exhibited by composite physical systems. These correlations find expression in the probability distribution $p(ab|xy)$ observed in a Bell-type scenario. However, as we will explore later, while the above definition of entanglement relates to Bell's non-local correlations, it is not an exact replica of the same property. Consequently, in general, entanglement and Bell non-locality represent subtly different concepts.

To better understand the properties of entanglement, it is essential first to grasp the structure of LOCC. We first encountered LOCC in the context of quantum teleportation. In this process, Alice performs a quantum measurement on her two systems (a local quantum operation) and then transmits the measurement's outcome to Bob (using classical communication). At the end of the protocol, Bob executes a local (unitary) operation on his system. The LOCC in the teleportation protocol is particularly unique because it involves only a one-way classical communication channel from Alice to Bob. We refer to this restricted set of LOCC as $\locc_1$.

Generally, LOCC allows for unlimited rounds of classical communication. LOCC limited to $n$ rounds of classical communication is denoted as $\locc_n$. We also use the notation ${\rm LO}=\locc_0$ for local operations that occur without any communication. An example of an LO map is the local unitary operation $U^{AB}=U^A\otimes U^B$. More broadly, an LO operation can be defined as a quantum channel $\mE\in\locc(AB\to A'B')$ of the form $\mE=\mM\otimes\mN$, where $\mM\in\cptp(A\to A')$ and $\mN\in\cptp(B\to B')$ are channels on Alice's and Bob's sides, respectively.

The most general quantum operation involving Alice performing a local operation (such as a generalized measurement) and then sending classical information to Bob can be characterized by a quantum instrument, $\mE\in\cptp(A\to A'Y)$. Here, $Y$ represents the classical system that Bob receives from Alice. Upon receiving $Y$, Bob can implement a local operation, $\mF\in\cptp(BY\to B')$. Consequently, the set of all $\locc_1$ operations is defined mathematically as follows:
\be
\locc_1(AB\to A'B')\eqdef\left\{\mF^{BY\to B'}\circ\mE^{A\to A'Y}\;:\;\substack{\mE\in\cptp(A\to A'Y)\\ \mF\in\cptp(BY\to B')}\;,\;|Y|<\infty\right\}\;.
\ee
Note that we do not impose any constraint on the classical system $Y$, only that it is finite dimensional. Setting $n\eqdef|Y|$,  an $\locc_1$ channel can be expressed as
\be
\mF^{BY\to B'}\circ\mE^{A\to A'Y}=\sum_{y\in[n]}\mF^{B\to B'}_{(y)}\otimes\mE^{A\to A'}_y\;.
\ee
Here, for every $y\in[n]$, the operation $\mF_{(y)}\in\cptp(B\to B')$, and $\mE_y\in\cp(A\to A')$. Furthermore, the sum $\sum_{y\in[n]}\mE_y$ is trace preserving.

By incorporating an additional round of communication from Bob to Alice, we obtain channels in $\locc_2$. Specifically, a channel $\mN\in\locc_2(AB\to A'B')$ can be expressed as follows:
\be\label{onewayn}
\mN^{AB\to A'B'}=\mE_1^{A_{1}X_1\to A'}\circ\mF^{BY_1\to B_1X_1}\circ\mE_0^{A\to A_1Y_1}\;,
\ee
where $A_1$ represents an additional system on Alice's side, $\mE_0$ and $\mE_1$ are channels on Alice's side, and $\mF$ is a channel on Bob's side. It's important to note that without the second round of communication, which corresponds to the case when $|X_1|=1$, the description reverts to a channel in $\locc_1$.

\bex
Show that if $|X_1|=1$ then the channel in~\eqref{onewayn} belongs to $\locc_1$.
\eex

In the same fashion, one can continue and express the most general protocol in $\locc_n$. Clearly, from the construction above it is obvious that the expression of LOCC protocols can be very complicated particularly if it involves a large number of classical communication rounds (see Fig.~\ref{locc}). Moreover, it is also known that $\locc_n$ is a strict subset of $\locc_{n+1}$ for all $n\in\mbb{N}$. Due to this notorious complexity of LOCC, and despite the enormous body of work in recent years on the study of LOCC, there are still many open problems in entanglement theory.
For this reason, it is sometimes convenient to consider a slightly larger class of operations that contains LOCC and have a simpler characterization. We will consider in the next chapter two such sets of operations known as the \emph{separable set} and the PPT set. However, as we will see in this  chapter, the complexity of LOCC is reduced dramatically when the bipartite system\index{bipartite system} is initially in a \emph{pure} state.

\begin{figure}[h]\centering    \includegraphics[width=0.6\textwidth]{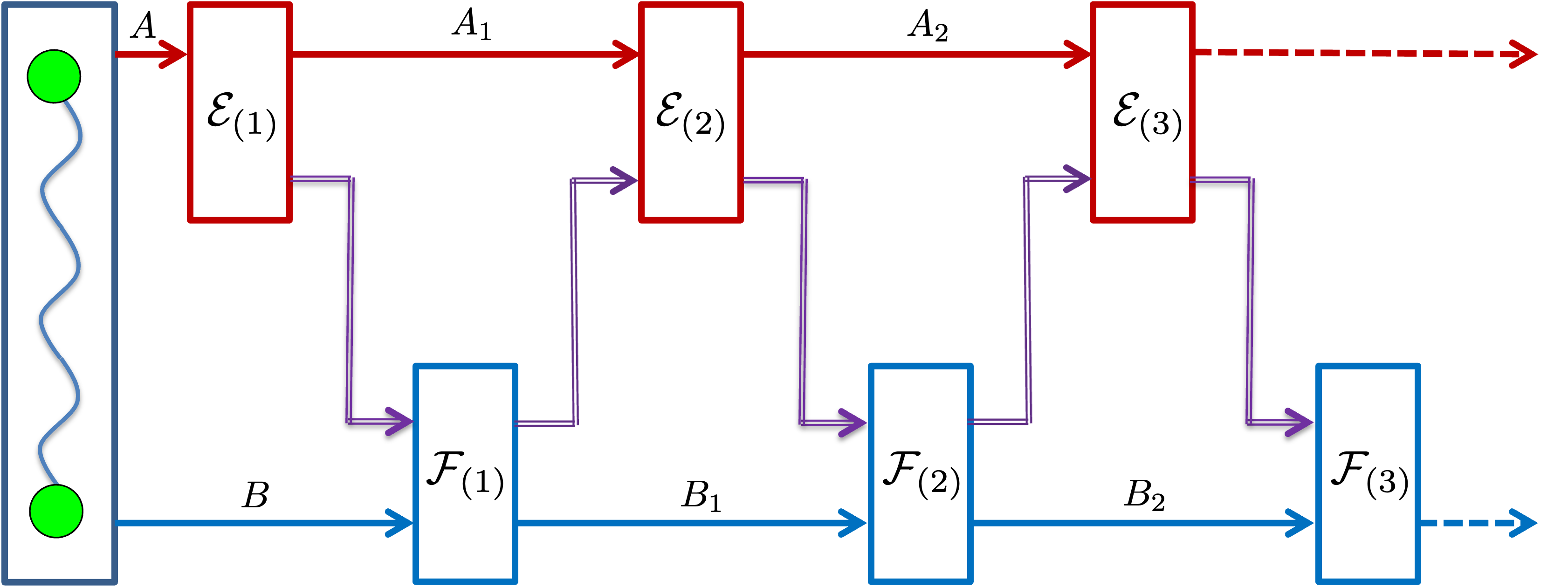}
  \caption{\linespread{1}\selectfont{\small An LOCC operation. The double lines represents classical communication between the parties.}}
  \label{locc}
\end{figure} 

In entanglement theory we consider all systems to be composite systems consisting of at least two subsystems that we will denote by $A$ (operated by Alice) and $B$ (operated by Bob). This means that even a single system on Alice's side that is described by a density matrix $\rho\in\md(A)$ can be treated mathematically as a bipartite system\index{bipartite system} with $\rho$ acting on $AB$, with $B$ being the trivial system (i.e. $|B|=1$) in this case. The set of free operations of this theory is $\mf=\locc$, and we will typically use the notation  $\locc(AB\to A'B')$ to indicated bipartite channels in $\cptp(AB\to A'B')$ that take bipartite states in $\md(AB)$ to bipartite states in $\md(A'B')$ (i.e. $A$ and $A'$ represents Alice's subsystems, and $B$ and $B'$ Bob's subsystems).

\bex[Separable Operations]
Let $\sep(AB\to A'B')\subset\cptp(AB\to A'B')$ be the set of all quantum channels that has a separable operator sum representation; i.e. $\mE\in\sep(AB\to A'B')$ if and only if there exist sets of operators $M_x:A\to A'$ and $N_x:B\to B'$ such that
\be
\mE^{AB\to A'B'}\left(\rho^{AB}\right)\eqdef\sum_x\left(M_x\otimes N_x\right)\rho^{AB}\left(M_x\otimes N_x\right)^*\quad\quad\forall\;\rho\in\ml(AB)\;.
\ee
\ben
\item Show that 
\be\label{ex12p6}
\locc(AB\to A'B')\subseteq\sep(AB\to A'B')\;.
\ee
\item Show that if $\mE\in\sep(AB\to A'B')$ then its normalized Choi matrix is a separable density matrix in $\md(AA'BB')$ (the seperability is between Alice system $AA'$ and Bob's system $BB'$).
\een
\eex

\section{Exact Manipulations of Entanglement}
 
Since entanglement is a resource, it is typically not available in its purest form as described for example by the pure (bipartite) singlet state. Instead, entangled systems are quite often only partially entangled, and given in a form that is mixed with noise. Consequently, such systems are not ideal resources for certain quantum information processing tasks (e.g. quantum teleportation). To identify which entangled states are more resourceful than others, we study here manipulations of entanglement; i.e. the conversion of one (partially) entangled state to another by LOCC. 

The most general LOCC protocol can be characterized by a quantum instrument $\{\mE_x\}$, with each $\mE_{x}\in\cp(AB\to A'B')$ being a trace non-increasing map. The parameter $x$ corresponds to all the measurements outcomes that were involved in the protocol (and that were not discarded). Therefore, if the initial state of the system was $\rho^{AB}$, the final state of the system after outcome $x$ occurred is given by $\mE_x\left(\rho^{AB}\right)/p_x$, where
\be
p_x=\tr\left[\mE_x\left(\rho^{AB}\right)\right]\;.
\ee
If for all $x$, the post measurement state is the same, i.e. $\mE_x\left(\rho^{AB}\right)/p_x=\sigma^{AB}$ for \emph{all} $x$, then $\sigma^{AB}=\sum_{x}\mE_x\left(\rho^{AB}\right)$, and the map $\sum_x\mE_x\eqdef\mE$ is an LOCC CPTP map. In particular, we say that a bipartite state $\rho^{AB}$ is \emph{more entangled} (i.e. more resourceful) than another state $\sigma^{AB}$ if there exists an LOCC CPTP map $\mE$ such that $\sigma^{AB}=\mE(\rho^{AB})$.

In general, given two entangled mixed states $\rho^{AB}$ and $\sigma^{AB}$, it can be very difficult to determine if they are related by an LOCC CPTP map. There are two main reasons for that: as we discussed above, LOCC is a very complex set to characterize, and in addition the states $\rho^{AB}$ and $\sigma^{AB}$ are mixed. It turns out that the entanglement properties of mixed states are very hard to characterized in general. For example, it is known to be NP-hard to determine if a mixed bipartite state $\rho^{AB}$ is entangled or separable. We therefore first treat in this chapter the pure bipartite entanglement manipulations, and postpone the treatment of mixed-state entanglement manipulations to the next chapter.

\subsection{LOCC on Pure Bipartite States}

Here we consider the effect of LOCC operations on a pure bipartite states. The set of all pure bipartite states will be denoted by $\pure(AB)$, and  without loss of generality, we will assume in this section that $|A|=|B|\eqdef d$, since we can always embed any bipartite system\index{bipartite system} in a larger Hilbert space with both local dimensions equal $\max\{|A|,|B|\}$.
Recall that any such state $\psi\in\pure(AB)$ has a Schmidt decomposition of the form (see Exercise~\ref{bipartite})
\be\label{srpsi2}
|\psi^{AB}\ra=\sum_{x\in[d]}\sqrt{p_x}|\psi_{x}^{A}\ra\otimes|\phi_{x}^{B}\ra\eqdef\left(\Lambda^A\otimes I^B\right)|\Omega^{AB}\ra
\ee
where $\{|\psi_x\ra^A\}_{x\in[d]}$ and $\{|\phi_x\ra^B\}_{x\in[d]}$ are orthonormal bases of $A$ and $B$, respectively, $\Lambda^A={\rm Diag}(\sqrt{p_1},\ldots,\sqrt{p_{d}})$ is a diagonal matrix in the basis $\{|\psi_x\ra^A\}_{x\in[d]}$, and $|\Omega^{AB}\ra=\sum_{x\in[d]}|\psi_{x}^{A}\ra\otimes|\phi_{x}^{B}\ra$ is an (unnormalized) maximally entangled state. 

We consider now the effect of a local measurement on Bob side. For this purpose, let
 $N$ be some $d\times d$ complex matrix, and note that
\be\label{npsi}
\left(I^A\otimes N\right)|\psi^{AB}\ra=\left(\Lambda\otimes N\right)|\Omega^{AB}\ra=\left(\Lambda N^T\otimes I^B\right)|\Omega^{AB}\ra\;.
\ee
\begin{exercise}
Show that the matrix $\Lambda N^T$ has same singular values as $N\Lambda $. Hint: Show that for any square matrix $C$, the matrices $C$ and $C^T$ have the same singular values.
\end{exercise}

\noindent Since $\Lambda N^T$ and $N\Lambda $ are two square matrices with the same singular values, it follows from the singular value decomposition that there exists two unitary matrices $U$ and $V$ such that
\be
\Lambda N^T=UN\Lambda V^T
\ee
where the transpose on $V$ is for convenience. Substituting this into~\eqref{npsi} gives
\ba\label{122}
\left(I^A\otimes N\right)|\psi^{AB}\ra&=\left(UN\Lambda V^T\otimes I^B\right)|\Omega^{AB}\ra\\
&=\left(UN\Lambda \otimes V\right)|\Omega^{AB}\ra\\
&=\left(UN\otimes V\right)|\psi^{AB}\ra\;.
\ea
That is, the vectors $\left(I^A\otimes N\right)|\psi^{AB}\ra$ and $\left(N\otimes I^B\right)|\psi^{AB}\ra$ are equivalent up to the local unitary map $U\otimes V$. This observation leads to the following result.

\begin{myt}{\color{yellow} Lo-Popescu's Theorem}\index{Lo-Popescu's theorem}
\begin{theorem}\label{lopop}
The effect of any LOCC map on a pure bipartite state can be simulated by the following protocol:
Alice performs a generalized quantum measurement $\{M_{yx}\}_{x,y}$, sends the result $(x,y)$ to Bob who then performs a local unitary map $V_{yx}$ on his system, and in the final step, Alice and Bob discard the value of $y$. 
\end{theorem}
\end{myt}

\begin{proof}
Consider first a single local instrument, $\{\mE_x\}_{x\in[m]}$, performed by Bob. Any CP map $\mE_x$ can be expressed as $\mE_x(\cdot)=\sum_{y\in[n]}N_{yx}(\cdot)N_{yx}^{*}$, so that
\ba
\mE_x^{B\to B}(\psi^{AB})&=\sum_{y\in[n]}\left(I^A\otimes N_{yx}\right)
\psi^{AB}\left(I^A\otimes N_{yx}^{*}\right)\\
\GG{\eqref{122}}&=\sum_{y\in[n]}\left(U_{yx}N_{yx}\otimes V_{yx}\right)
\psi^{AB}\left(N_{yx}^{*}U_{yx}^*\otimes V_{yx}^* \right)
\ea
where we used~\eqref{122} for each $x\in[m]$ and $y\in[n]$, with $U_{xy}$ and $V_{xy}$ being unitary matrices.
Denoting by $M_{yx}\eqdef U_{yx}N_{yx}$ the above equation becomes
\be
\mE_x^{B\to B}(\psi^{AB})=\sum_{y\in[n]}\left(M_{yx}\otimes V_{yx}\right)
\psi^{AB}\left(M_{yx}^{*}\otimes V_{yx}^* \right)\;.
\ee
Moreover, since $\sum_{x\in[m]}\sum_{y\in[n]}M_{yx}^*M_{yx}=I^A$ we conclude that any quantum instrument\index{quantum instrument} that is performed by Bob, can be simulated with the following protocol: Alice performs a generalized quantum measurement $\{M_{xy}\}_{x\in[m],y\in[n]}$, sends the outcome $(x,y)$ to Bob, who then performs a unitary matrix $V_{xy}$. At the end of the protocol, Alice and Bob discard or forget the value of $y$. 
Therefore, in any LOCC protocol, all the local quantum instruments on Bob's side can be simulated with unitaries and measurements on Alice's side.
Since a sequence of quantum instruments (generalized measurements) on Alice's side can be combined into a single generalized measurement (followed by coarse graining, i.e. discarding of information), we conclude that the most general LOCC protocol on a pure bipartite state can be simulated with a single generalized measurement by Alice's side followed by a unitary on Bob's side that depends on Alice's measurement outcome, and ends with the discarding of partial information of the measurement outcome.
\end{proof}

\begin{exercise}
Show that a sequence of two generalized measurements can be viewed as a single generalized measurement. That is, given two generalized measurement $\{M_x\}_{x\in[m]}$ and $\{N_y\}_{y\in[n]}$ show that the set of matrices $\{L_{xy}\eqdef M_xN_y\}_{x\in[m],y\in[n]}$ is also a generalized measurement.
\end{exercise}

\bex\label{ex:ptom}
Let $\psi\in\pure(AB)$ and $\sigma\in\md(AB)$. Show that if there exists a deterministic LOCC protocol that converts $\psi^{AB}$ to $\sigma^{AB}$, i.e.
$
\psi^{AB}\xrightarrow{\text{\tiny LOCC}} \sigma^{AB}
$,
then there exists a set $\{M_x\}_{x\in[m]}$ of complex matrices in $\ml(A)$, and a set $\{U_x\}_{x\in[m]}$ of unitary matrices in $\ml(B)$ such that
\be
\sigma^{AB}=\sum_{x\in[m]}\left(M_{x}\otimes U_{x}\right)
\psi^{AB}\left(M_{x}\otimes U_{x} \right)^*\;.
\ee
Show further that the above relation can be expressed as 
\be
\sigma^{AB}=\mU^{BX\to B}\circ\mE^{A\to AX}\left(\psi^{AB}\right)\;.
\ee
where $\mE\in\cptp(A\to AX)$ is a quantum instrument and $\mU\in\cptp(BX\to B)$ is a controlled unitary\index{controlled unitary} channel.
Note that in particular this implies that 
$
\psi^{AB}\xrightarrow{\locc_1}\sigma^{AB}\;.
$
\eex

Theorem~\ref{lopop} can be simplified further if we consider only LOCC protocols  that take pure bipartite states to pure bipartite states.
In this case, any LOCC transformation can be simulated by the following simple protocol: Alice performs a generalized measurement $\{M_x\}$, sends the outcome $x$ to Bob, who then performs a local unitary operation $V_x$.
This simplification of LOCC will be crucial for the study of pure-state entanglement theory.

\bex\label{ex:sr0}
The Schmidt rank\index{Schmidt rank} of a pure bipartite state is defined as the number of non-zero Schmidt coefficients; for example, the Schmidt rank of the state given in~\eqref{srpsi2} is the rank of the matrix $\Lambda^A$. We denote the Schmidt rank of a bipartite state $\psi\in\pure(AB)$ by $\sr(\psi)$.
Show that for two bipartite states $\psi,\phi\in\pure(AB)$ with $\sr(\phi)>\sr(\psi)$ it is impossible to convert $\psi$ to $\phi$ by LOCC (not even with probability less than one).
\eex

\subsection{Exact Deterministic Interconversions}

In this section we provide the precise conditions that determine if one quantum state can be converted to another by LOCC. We will use the notation $|\psi^{AB}\ra\xrightarrow{LOCC} |\phi^{AB}\ra$ whenever it is possible to convert a bipartite pure state $\psi\in\pure(AB)$ into the state $\phi\in\pure
(AB)$.
Recall that any bipartite quantum state $\psi\in\pure(AB)$ (with $|A|=|B|\eqdef d$) has a Schmidt decomposition of the form
\be\label{srpsi}
|\psi^{AB}\ra=\sum_{x\in[d]}\sqrt{p_x}|\psi_x\ra^A|\phi_x\ra^B\;,
\ee
where $\{|\psi_x\ra^A\}_{x\in[d]}$ and $\{|\phi_x\ra^B\}_{x\in[d]}$ are orthonormal bases of $A$ and $B$, respectively.
Let $U$ and $V$ be unitary matrices such that $U|\psi_x\ra^A=|x\ra^A$ and $V|\phi_x\ra^B=|x\ra^B$, where $\{|x\ra^A\}$
and $\{|x\ra^B\}$ are the standard bases of $A$ and $B$, respectively. Hence,
\be
U\otimes V|\psi^{AB}\ra=\sum_{x\in[d]}\sqrt{p_x}|xx\ra^{AB}\;.
\ee 
Note also that by applying additional local permutations (which are unitaries) to the state above we can rearrange the order that the Schmidt coefficients. Therefore, 
there exist unitary matrices $U'\in\ml(A)$ and $V'\in\ml(B)$ such that
\be
|\tilde{\psi}^{AB}\ra\eqdef U'\otimes V'|\psi^{AB}\ra=\sum_{x\in[d]}\sqrt{p_x}|xx\ra^{AB}\quad\text{and}\quad p_1\geq p_2\geq\cdots\geq p_d\;.
\ee
The  above form is called the \emph{standard form} of $|\psi^{AB}\ra$. Note that $|\psi^{AB}\ra$ can be converted by LOCC to another state $|\phi^{AB}\ra$ if and only if the standard form of $|\psi^{AB}\ra$ can be converted by LOCC to the standard form\index{standard form} of $|\phi^{AB}\ra$ (see Fig.~\ref{standardform}). Therefore, without loss of generality we will assume here that both $|\psi^{AB}\ra$ and $|\phi^{AB}\ra$ are given in their standard form.

\begin{figure}[h]
\centering
    \includegraphics[width=0.5\textwidth]{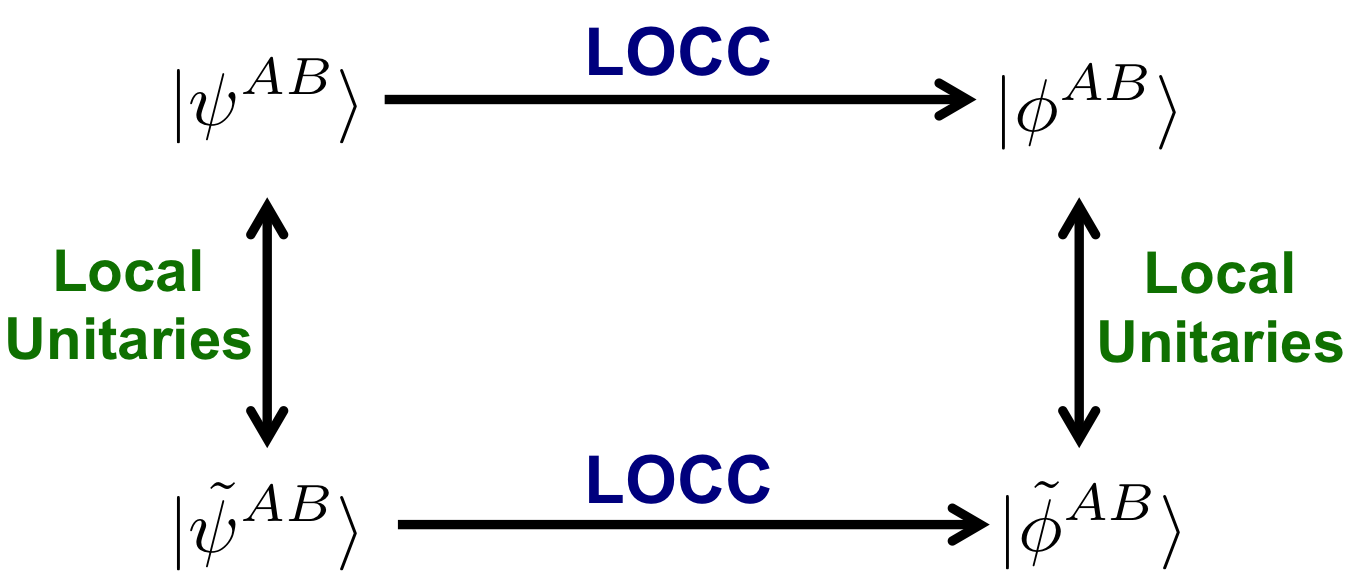}
  \caption{\linespread{1}\selectfont{\small LOCC maps between two pure bipartite states and their standard forms.}}
  \label{standardform}
\end{figure} 

The next theorem provides a connection between LOCC conversions and majorization. For any two density matrices $\rho,\sigma\in\md(A)$, we will say that $\rho$ majorizes $\sigma$, and write $\rho\succ\sigma$, if the probability vectors $\p$ and $\q$, consisting, respectively, of the eigenvalues of $\rho$ and $\sigma$, satisfy $\p\succ\q$.

\begin{myt}{\color{yellow} Nielsen's Majorization Theorem}\index{Nielsen's theorem}
\begin{theorem}\label{nielsen}
Let $\psi,\phi\in\in\pure(AB)$ be two bipartite quantum states, and let $\rho^A\eqdef\tr_B\left[\psi^{AB}\right]$ and $\sigma^A\eqdef\tr_B\left[\phi^{AB}\right]$ be their corresponding reduced density matrices. Then, 
\be
\psi^{AB}\xrightarrow{LOCC} \phi^{AB}\quad\iff\quad \sigma^A\succ\rho^A\;.
\ee 
\end{theorem}
\end{myt}
\begin{proof}
From the argument above we can assume without loss of generality that both $\psi^{AB}$ and $\phi^{AB}$ are given in their standard forms. Furthermore, Exercise~\ref{ex:sr0} implies that we can assume without loss of generality that $\supp(\sigma^A)\subseteq\supp(\rho^A)$. This in turn implies that we can assume without loss of generality  that $\rho^A>0$ since otherwise we embed both $|\psi^{AB}\ra$ and $|\phi^{AB}\ra$ in $\supp(\rho^A)\otimes B$.

Form  Theorem~\ref{lopop} any LOCC map that takes  pure bipartite state $|\psi^{AB}\ra$ to another pure bipartite state $|\phi^{AB}\ra$ can be simulated by 1-way LOCC (i.e. $\locc_1$) of the following form: Alice performs a single generalized measurement $\{M_z\}_{z\in[m]}$ on her system, sends the measurement outcome $z$ to Bob, who then performs the unitary $V_z$. Therefore, after outcome $z$ occurred, the post-measurement state is given by
\be
\frac{1}{\sqrt{t_z}}M_z\otimes V_z|\psi^{AB}\ra\;,
\ee
where $t_z\eqdef\la\psi^{AB}|M_z^*M_z\otimes I^B|\psi^{AB}\ra$ is the probability that Alice's measurement outcome is $z$. Hence, if $|\psi^{AB}\ra$ can be converted by LOCC to $|\phi^{AB}\ra$ with 100\% success rate, then there must exist a generalized measurement $\{M_z\}_{z\in[m]}$ and a collection of unitary matrices $\{V_z\}_{z\in[m]}$ such that 
\be\label{impbasic}
\frac{1}{\sqrt{t_z}}M_z\otimes V_z|\psi^{AB}\ra=|\phi^{AB}\ra\quad\quad\forall\;z\in[m]\;.
\ee
Since we assume without loss of generality that both $|\psi^{AB}\ra$ and  $|\phi^{AB}\ra$ are given in their standard form, we have
\ba
|\psi^{AB}\ra&=\sum_{x\in[d]}\sqrt{p_x}|xx\ra^{AB}=\sqrt{\rho}\otimes I^B|\Omega^{AB}\ra\\
|\phi^{AB}\ra&=\sum_{x\in[d]}\sqrt{q_x}|xx\ra^{AB}=\sqrt{\sigma}\otimes I^B|\Omega^{AB}\ra
\ea
where $\rho$ and $\sigma$ are, respectively, the reduced density matrices of $|\psi^{AB}\ra$ and  $|\phi^{AB}\ra$.
Explicitly, $\rho=\sum_{x\in[d]}p_x|x\lr x|^A$ and $\sigma=\sum_{x\in[d]}q_x|x\lr x|^A$. 
Substituting this into~\eqref{impbasic} gives
\be
\frac{1}{\sqrt{t_z}}\left(M_z\sqrt{\rho}\otimes V_z\right)|\Omega^{AB}\ra=\left(\sqrt{\sigma}\otimes I^B\right)|\Omega^{AB}\ra\;.
\ee
From Exercise~\ref{bipartite} it follows that the above equation hold if and only if 
\be
\frac{1}{\sqrt{t_z}}M_z\sqrt{\rho}V_z^T=\sqrt{\sigma}
\ee
Note that the matrix $U_z\eqdef \left(V_z^{-1}\right)^T$ is unitary. With this notation, the above equation is equivalent to
\be
M_z=\sqrt{t_z}\sqrt{\sigma}U_z\rho^{-1/2}\;.
\ee
The only constraint on $M_z$ is that $\sum_{z\in[m]}M_z^{*}M_z=I^A$. We therefore conclude that $|\psi^{AB}\ra$ can be converted to $|\phi^{AB}\ra$ by LOCC if and only if there exists $m\in\mbb{N}$, unitary matrices $\{U_z\}_{z\in[m]}$, and probabilities $\{t_z\}_{z\in[m]}$ such that
\be
\sum_{z\in[m]}t_z\rho^{-1/2}U_z^*\sigma U_z \rho^{-1/2}=I^A\;,
\ee
or equivalently, 
\be\label{random}
\rho=\sum_{z\in[m]}t_zU_z^*\sigma U_z\;.
\ee
In other words, $|\psi^{AB}\ra$ can be converted to $|\phi^{AB}\ra$ by LOCC if and only if there exists a mixture of unitaries that transforms the reduced density matrix of $|\phi^{AB}\ra$ to the reduced density matrix of $|\psi^{AB}\ra$. Observe that such a random unitary channel is a unital channel. In Section~\ref{unital} we showed that if $\rho=\mE(\sigma)$, with $\mE$ being a unital channel, then there exists a doubly stochastic matrix\index{doubly stochastic matrix} $D$ such that $\p=D\q$ (recall that $\p$ and $\q$ are the probability vectors whose components consist of the eigenvalues of $\rho$ and $\sigma$, respectively). From Theorem~\ref{chmaj} it then follows that $\q\succ\p$ or equivalently, $\sigma\succ\rho$. We therefore conclude that if $\psi^{AB}$ can be converted to $\phi^{AB}$ by LOCC then we must have $\sigma\succ\rho$.

Conversely, if $\sigma\succ\rho$ then from Theorem~\ref{chmaj} we have $\p=D\q$ for some doubly stochastic matrix\index{doubly stochastic matrix} $D$. From Birkhoff/von-Neumann theorem (see Theorem~\ref{birkhoff}) every doubly stochastic matrix\index{doubly stochastic matrix} can we written as a convex combination of permutation matrices. Therefore, there there exists $m\in\mbb{N}$, permutation matrices $\{\Pi_z\}_{z\in[m]}$, and probabilities $\{t_z\}_{z\in[m]}$, such that $\p=\sum_{z}t_z\Pi_z\q$. In Exercise~\ref{shownice} you will show that this relation can be expressed as 
\be
\rho=\sum_{z\in[m]}t_z\Pi_z\sigma \Pi_z^T\;.
\ee
The above equality is equivalent to~\eqref{random} by taking $U_z=\Pi_z^T$. Therefore, $\psi^{AB}$ can be converted to $\phi^{AB}$ by LOCC. This completes the proof.
\end{proof}

\begin{exercise}\label{shownice}
Show that if $\rho$ is a diagonal matrix, and $\p$ is the vector consisting of its diagonal elements, then $\Pi_z\rho\Pi_z^T$ is also a diagonal matrix, with the diagonal elements given by the components of $\Pi_z\p$. Use this to show that $\rho=\sum_{z\in[m]}t_z\Pi_z\sigma \Pi_z^T$ if and only if $\p=\sum_{z\in[m]}t_z\Pi_z\q$.
\end{exercise}

\begin{exercise}
Show that the maximally entangled state $|\Phi^{AB}\ra\eqdef\frac{1}{\sqrt{d}}\sum_{x\in[d]}|xx\ra$ can be converted by LOCC to any other state in $\pure(AB)$. Moreover, show that any state $|\psi\ra\in {AB}$ can be converted by LOCC to any product state of the form $|\phi\ra|\chi\ra\in {AB}$.
\end{exercise}

\begin{exercise}\label{nil}
For any bipartite state $\psi\in \pure(AB)$, and for any $k=1,\ldots,d$, define
\be\label{vidal}
E_{(k)}(\psi^{AB})\eqdef 1-\sum_{x\in[k]}p^{\da}_x=1-\|\p\|_{(k)}\;,
\ee 
where $\p$ is the Schmidt vector of $|\psi^{AB}\ra$, and $\|\p\|_{(k)}$ is the Ky Fan norm of $\p$ (cf.~Definition~\ref{def:kyfan}).
Show that Nielsen Majorization Theorem can be expressed as
\be
\psi^{AB}\xrightarrow{LOCC}\phi^{AB}\quad\iff \quad E_{(k)}(\psi^{AB})\geq E_{(k)}(\phi^{AB})\quad\forall\;k\in[d]\;.
\ee
\end{exercise}

\begin{exercise}\label{faitht}
Let $\psi\in\pure(AB)$, with $m\eqdef|A|=|B|$.
\begin{enumerate}
\item Prove the following theorem, assuming that Alice and Bob share the state $\psi^{AB}$ (but no other entangled systems).
\textbf{Theorem.} Faithful teleportation of a $d$-dimensional qudit is possible if, and only if, 
\begin{equation}
E_{t}(\psi^{AB})\eqdef -\log _{2}p_{\max} 
\geq \log_{2} d\;,
\label{t1}
\end{equation}
where $p_{\max}$ is the largest Schmidt coefficient of $\psi^{AB}$. 
That is, teleportation is possible if, and only if, none of the
Schmidt coefficients are greater than $1/d$.
This also implies that the Schmidt rank
$m$ is greater than or equal to $d$. Hint: Use Nielsen majorization theorem.
\item Find a protocol for faithful teleportation of a \emph{qubit} from Alice's lab to Bob's lab assuming
Alice and Bob share the partially entangled state
\be
|\psi^{AB}\rangle=\frac{1}{\sqrt{2}}|0\rangle_{A}|0\rangle_{B}+\frac{1}{2}|1\rangle_{A}|1\rangle_{B}+
\frac{1}{2}|2\rangle_{A}|2\rangle_{B}\;.
\ee
In particular, determine the projective measurement performed by Alice and the unitary operators
performed by Bob. What is the optimal classical communication cost? That is, how many classical bits Alice
has to send to Bob?
\end{enumerate}
\end{exercise}

\begin{exercise}
Consider $m$ bipartite states $\{\psi^{AB}_{z}\}_{z\in[m]}$ in $\pure(AB)$.
Find an optimal state $\phi^{AB}\in\pure(AB)$ such that:
\begin{enumerate}
\item The state $\phi^{AB}$ can be converted by LOCC to $\psi^{AB}_z$, for all $z\in[m]$.
\item If another state, $\chi\in\pure(AB)$, can be converted by LOCC to $\psi^{AB}_z$, for all $z\in[m]$, then
$\chi^{AB}$ can also be converted to $\phi^{AB}$. That is, $\phi^{AB}$ is optimal.
\end{enumerate}
\end{exercise}

\subsection{Entanglement Catalysis}\label{ecatal}\index{entanglement catalysis}

As we observed earlier, some LOCC conversions of pure bipartite states cannot be realized with a 100\% success rate. This limitation arises because LOCC constitutes a restricted subset of quantum operations. To expand the capabilities of LOCC, one could allow Alice and Bob to temporarily use an entangled state during their LOCC protocols. The condition here is that they must return the entangled systems in their original state at the end of the protocols. At first glance, it might seem that borrowing an entangled system wouldn't provide any advantage for tasks that cannot be accomplished with standard LOCC. However, as we will now demonstrate, Nielsen's majorization theorem\index{Nielsen's theorem} reveals that this entanglement-assisted LOCC (eLOCC) actually represents a significantly broader set of operations compared to LOCC alone.

We start with the following example. Consider the two entangled states
\ba
|\psi^{AB}\ra&=\sqrt{2/5}|00\ra+\sqrt{2/5}|11\ra+\sqrt{1/10}|22\ra+\sqrt{1/10}|33\ra\\
|\phi^{AB}\ra&=\sqrt{1/2}|00\ra+\sqrt{1/4}|11\ra+\sqrt{1/4}|22\ra\;.
\ea
The Schmidt probability vectors associated with the two states above are given by
\be
\p\eqdef\left(
\frac25 , \frac25 , \frac1{10} , \frac1{10}
\right)^T\quad\text{and}\quad\q=\left(
\frac12 , \frac14 , \frac14 , 0
\right)^T\;,
\ee
respectively. In Exercise~\ref{ex551}, you confirmed that neither $\p$ majorizes $\q$ nor $\q$ majorizes $\p$, symbolized as $\p\not\succ\q$ and $\q\not\succ\p$. Consequently, Nielsen's majorization theorem implies that neither $\psi^{AB}$ can be converted to $\phi^{AB}$, nor can $\phi^{AB}$ be converted to $\psi^{AB}$ using LOCC.  Now, consider the state
\be
|\chi^{A'B'}\rangle=\sqrt{3/5}|00\rangle+\sqrt{2/5}|11\rangle\;,
\ee
and let its Schmidt vector be denoted by $\r\eqdef(3/5,2/5)^T$. Interestingly, it is easy to verify that
\be
\q\otimes\r\succ\p\otimes\r\;.
\ee
Therefore, according to Nielsen's theorem, the transformation
\be
|\psi^{AB}\rangle\otimes|\chi^{A'B'}\rangle\xrightarrow{LOCC}|\phi^{AB}\rangle\otimes|\chi^{A'B'}\rangle
\ee
is achievable with a 100\% success rate. In this context, the state $\chi^{AB}$ functions as a catalyst for the conversion of $\psi^{AB}$ into $\phi^{AB}$, and thus is referred to as an \emph{entanglement catalyst}.

\begin{exercise}
Show that there is no entanglement catalyst if both $\psi^{AB}$ and $\phi^{AB}$ have Schmidt rank 3. 
\end{exercise}

\begin{exercise}
Show that the maximally entangled state cannot act as a catalyst in any eLOCC conversions that are not possible by LOCC. 
\end{exercise}

Entanglement catalysis motivates the definition of a new partial order between probability vectors that we studied in Sec.~\ref{sec:tr} and is called the \emph{trumping relation}. Recall that for any $\p,\q\in\prob(n)$ we say that $\q$ trumps $\p$ and write 
\be
\q\succ_{*}\p\;,
\ee
if there exists an integer $m\in\mbb{N}$ and a vector $\r\in\prob(m)$ such that $\q\otimes\r\succ\p\otimes\r$.

\begin{exercise}
Show that any Schur concave function $f:\prob(n)\to\mbb{R}$ that is additive under tensor product, behaves monotonically under the trumping relation. That is, for any $\p,\q\in\prob(n)$ we have
\be
\q\succ_{{}_*}\p\quad\Rightarrow\quad f(\p)\geq f(\q)\;.
\ee
\end{exercise}

A well known family of functions that behaves monotonically under the trumping\index{trumping relation} relation are the R\'enyi entropies. The set of functions
\be\label{falp}
f_{\alpha}(\p)\eqdef\begin{cases}\frac{\text{sign}(\alpha)}{1-\alpha}\log\sum_{x\in[m]}p_x^\alpha &\text{if }0\neq\alpha\in[-\infty,\infty]\\
-\log(p_1\cdots p_m)&\text{if }\alpha=0\;.
\end{cases}
\ee
satisfies the monotonicity under the tramping relation and additivity. For $\alpha\geq 0$ (i.e. $f_\alpha=H_\alpha$ is the R\'enyi entropy) these functions are entropy functions as they also satisfy the normalization condition that they are zero for $\p=(1,0,\ldots,0)$. For $\alpha\leq 0$ they are defined to be $-\infty$ if $\p\not>0$. Such functions are not entropy functions since they do not satisfies the normalization condition, however, they are useful in the characterization of the tramping relation.
Note that any convex combination of the functions above is also additive and monotonic under the trumping relation. 

\bex
Show that the condition~\eqref{stri} of Theorem~\ref{ktt} is equivalent to the condition that
\be
f_\alpha(\p)> f_\alpha(\q)\quad\quad\forall\;\alpha\in[-\infty,\infty]\;,
\ee
where $f_\alpha$ is defined in~\eqref{falp}
\eex

For each $\alpha\in[-\infty,\infty]$ and $\psi\in\pure(AB)$ define
\be\label{ealp}
E_\alpha\left(\psi^{AB}\right)\eqdef f_\alpha(\p)
\ee
where $\p$ is the Schmidt probability vector of $\psi^{AB}$, and $f_\alpha$ is defined in~\eqref{falp}. We will see later on that the functions $\{E_\alpha\}_\alpha$ are measures of entanglement on pure states. From Theorem~\ref{ktt} it follows that these functions can be used to characterize eLOCC transformations.

\begin{myg}{}
\begin{corollary}\label{cor:elocc}
Let $\psi,\phi\in\in\pure(AB)$ be two bipartite quantum states. Then, 
\be
\psi^{AB}\xrightarrow{e\locc} \phi^{AB}\quad\iff\quad E_\alpha\left(\psi^{AB}\right)> E_\alpha\left(\phi^{AB}\right)\quad\quad\forall\;\alpha\in[-\infty,\infty]\;,
\ee
where $\{E_\alpha\}_{\alpha}$ are defined in~\eqref{ealp}.
\end{corollary}
\end{myg}

\bex
Prove the corollary above using Theorem~\ref{ktt}.
\eex

\section{Quantification of Pure Bipartite Entanglement}
How can entanglement be quantified?
According to its definition (see Definition~\ref{ent}), entanglement cannot be created or increased by LOCC. Consequently, entanglement must be quantified using functions that are monotonic under LOCC. In this section, we focus on pure state entanglement and consider a measure of entanglement to be a function
\be
\mathbf{E}:\bigcup_{A,B}\pure(AB)\to\mbb{R}
\ee
that exhibits monotonic behavior under pure-state LOCC transformations.

Any pure bipartite state $\psi\in \pure(AB)$ can be transformed by a local unitary operation (i.e., a reversible LOCC) into the state
\be
|\tilde{\psi}^{AB}\ra\eqdef U\otimes V|\psi^{AB}\ra=\sum_{x\in[d]}\sqrt{p_x}|xx\ra^{AB}\;,
\ee
where $U$ and $V$ are $d\times d$ unitary matrices, $\{p_x\}_{x\in[d]}$ are the Schmidt coefficients of $\psi^{AB}$, and $\{|x\ra^A\}$ and $\{|x\ra^B\}$ are fixed bases of $A$ and $B$. By definition, since the LOCC map $U\otimes V$ is reversible (having an LOCC inverse $U^*\otimes V^*$), we must have for any measure of entanglement on pure states:
\be
\E\left(\psi^{AB}\right)=\E\big(\tilde{\psi}^{AB}\big)=f(\p)\;,
\ee
where $\p\eqdef(p_1,\ldots,p_d)^T$ is the probability Schmidt vector of $\psi^{AB}$ and $f:\prob(d)\to\mbb{R}_{+}$ is some non-negative function of the Schmidt coefficients of $|\psi^{AB}\ra$.  That is, any measure of entanglement on pure states depends only on the Schmidt coefficients of the entangled state.

From Nielsen theorem we have that $|\psi^{AB}\ra\xrightarrow{LOCC} |\phi^{AB}\ra$ if and only if the corresponding Schmidt probability vectors $\p$ and $\q$ of $\psi^{AB}$ and $\phi^{AB}$, respectively, satisfy $\p\prec\q$. On the other hand, by definition, if  $\psi^{AB}\xrightarrow{LOCC} \phi^{AB}$ then any measure of entanglement must satisfy $\E\left(\psi^{AB}\right)\geq \E\left(\phi^{AB}\right)$. Hence, the function $f$ above must be monotonic under majorization; i.e. $f$ must be a Schur concave function. Specifically,
\be
\p\prec\q\;\;\Rightarrow\;\; f(\p)\geq f(\q).
\ee

\subsubsection{Example 1: Entropy of Entanglement}\index{entropy of entanglement}

The entropy of entanglement is arguably the most important measure of pure-state entanglement with several operational interpretations. It is detonated by $E$ and defined for any $\psi\in\pure(AB)$ by
\be\label{eoe}
E\left(\psi^{AB}\right)\eqdef H(\p)\;,
\ee
where $H$ is the Shannon entropy\index{Shannon entropy} and $\p$ is the Schmidt probability vector of $\psi^{AB}$. We will see in the next sections that the entropy of entanglement equals both the entanglement cost and the distillable entanglement in the asymptotic regime. 

\subsubsection{Example 2: $\alpha$-Entropy of Entanglement}

Similar to the entropy of entanglement\index{entropy of entanglement}, for any $\alpha\in[0,\infty]$ the $\alpha$-entropy of entanglement is defined 
for any $\psi\in\pure(AB)$ as
\be
E_\alpha\left(\psi^{AB}\right)\eqdef H_\alpha(\p)\;,
\ee
where $H_\alpha$ is the $\alpha$-R\'enyi entropy\index{R\'enyi entropy} and $\p$ is the Schmidt probability vector of $\psi^{AB}$. In addition to being monotonic under LOCC the $\alpha$-entropy of entanglement is additive under tensor products. We saw in Corollary~\ref{cor:elocc} that this in turn implies that the $E_\alpha$ is also monotonic under eLOCC and therefore can be used to characterize entanglement catalysis\index{entanglement catalysis}.

\subsubsection{Example 3: The concurrence\index{concurrence} monotones}

In~\eqref{531} we saw that the elementary symmetric functions are Schur concave. Therefore, they can be used to define a family of measures of entanglement known as the concurrence monotones. Specifically, they are defined for any $\psi\in\pure(AB)$ and any $k\in[d]$ as
\be\label{1223}
C_k\left(\psi^{AB}\right)\eqdef\left(\frac{f_k(\p)}{f_k(\u)}\right)^{1/k}\quad\text{where}\quad
f_k(\p)\eqdef
\sum_{\substack{x_1<\cdots< x_k\\ x_1,\ldots,x_k\in[n]}}p_{x_1}\cdots p_{x_k}\;,
\ee
$\p$ is the Schmidt probability vector of $\psi^{AB}$, and $\u$ is the $d$-dimensional uniform probability vector. Note that the $k$-concurrence\index{concurrence} is normalized such that it equal one of maximally entangled states. Thus, the concurrence monotones take values between zero and one. The power $1/k$ above is necessary to make these functions entanglement monotones; i.e. it can be shown that these functions not only behave monotonically under LOCC but also non-increasing on average under non-deterministic LOCC (see the following sections below). 

For the two extreme cases $k=2$ and $k=d$, we get for any $\psi\in\pure(AB)$ with reduced density matrix $\rho\eqdef\tr_B\left[\psi^{AB}\right]$, that
\ba\label{congcon}
&C\left(\psi^{AB}\right)\eqdef C_2\left(\psi^{AB}\right)=\sqrt{\frac d{d-1}\left(1-\tr\left[\rho^2\right]\right)}\\
&G\left(\psi^{AB}\right)\eqdef C_d\left(\psi^{AB}\right)=d\left(\det(\rho)\right)^{1/d}\;.
\ea
In literature, $C$ above is called simply the concurrence\index{concurrence}, and $G$ is called the $G$-concurrence since it can be interpreted as the geometric mean\index{geometric mean} of the Schmidt coefficients. Note that for $d=2$ we have $C=G=C_k$.
\bex
Prove the equalities in~\eqref{congcon}.
\eex

\bex\label{ex:preconcurrence}
Let $\{|0\ra,|1\ra\ra\}$ be a basis in which the second Pauli matrix, $\sigma_y$, has the form $\sigma_y=\begin{bmatrix} 0 & i\\
-i & 0\end{bmatrix}$. Show that for any $\psi\in\pure(AB)$ with $|A|=|B|=2$
\be
C\left(\psi^{AB}\right)=\left|\la\bar{\psi}^{AB}|\sigma_y\otimes\sigma_y|\psi^{AB}\ra\right|
\ee
where $\bar{\psi}^{AB}$ is defined such that if $|\psi^{AB}\ra=\sum_{x,y\in\{0,1\}}c_{xy}|x\ra|y\ra$ then
$|\bar{\psi}^{AB}\ra=\sum_{x,y\in\{0,1\}}\bar{c}_{xy}|x\ra|y\ra$.
\eex

\section{Stochastic Interconversions}\index{Stochastic Interconversions}

Quantum mechanics is inherently non-deterministic. This is manifested by quantum measurements that transform a quantum state into several possible post-measurement states. Typically, an LOCC protocol does not convert a quantum state into another quantum state. Instead, it converts it to \emph{ensemble of states} with each state occurring with different probability. Here we consider such probabilistic LOCC transformations among pure bipartite states .

Suppose Alice and Bob share the pure bipartite state $\psi\in \pure(AB)$. After they apply LOCC to this state they end up sharing one out of $n$ states $\{|\phi_z^{AB}\ra\}_{z\in[n]}$ each with corresponding probability 
$t_z$. The resulting ensemble of states  $\{t_z,\;|\phi_z^{AB}\ra\}_{z\in[n]}$ can be described with a cq-state
\be\label{zab}
\sigma^{ZAB}\eqdef\sum_{z\in[n]}t_z|z\lr z|^Z\otimes|\phi^{AB}_z\lr\phi^{AB}_z|\;,
\ee
where $Z$ is a `flag' system registering the value $z$. In this view, the LOCC protocol converted the state $\psi^{AB}$ to the cq-state\index{cq-state} $\sigma^{ZAB}$. The question we study here is to which cq-states the pure state 
$\psi^{AB}$ can be transformed into by LOCC. Since the output state $\sigma^{ZAB}$ is not a pure state, we cannot apply Nielsen majorization theorem. Yet, as we show now, Nielsen theorem is imperative to answer such a question.

\begin{myt}{}
\begin{theorem}\label{thm:dp}
A pure bipartite state $\psi\in \pure(AB)$ can be converted by LOCC to the ensemble $\left\{\phi^{AB}_z,\;t_z\right\}_{z\in[n]}$ of pure states in $\pure(AB)$ if and only if for all $k\in[d]$ ($d\eqdef|A|=|B|$)
\be\label{ekcond}
E_{(k)}\left(\psi^{AB}\right)\geq \sum_{z\in[n]}t_zE_{(k)}\left(\phi^{AB}_z\right)\;,
\ee
where the functions $E_{(k)}$ are defined in~\eqref{vidal}.
\end{theorem}
\end{myt}

\begin{proof}
Let $\p$ be the Schmidt probability vector associated with $\psi^{AB}$. For each $z\in[n]$, define $\q_z\eqdef(q_{1|z},\ldots,q_{d|z})^T$ as the Schmidt probability vector associated with $\phi^{AB}_z$. We can assume without loss of generality that $\p=\p^\da$ and $\q_z=\q_z^{\da}$, since the order of the Schmidt vectors can always be rearranged by applying local unitary (permutation) maps to $|\psi^{AB}\rangle$ and $|\phi_z^{AB}\rangle$.

Also, denote by
\be
\q\eqdef\sum_{z\in[n]}t_z\q_z\quad\text{and by}\quad|\phi^{AB}\ra\eqdef\sum_{x\in[d]}\sqrt{q_x}|xx\ra^{AB}\;,
\ee
the bipartite state whose Schmidt vector is $\q$. Since $\q_z=\q_z^{\da}$ for all $z\in[n]$, it follows that $\q=\q^\da$ as well. The components of $\q$ are thus given by
\be\label{chichi}
q_x=\sum_{z\in[n]}t_zq_{x|z}\quad\quad\forall\;x\in[d]\;.
\ee 
Consequently, for each $k\in[d]$, the entanglement measure $E_{(k)}$ for $\phi^{AB}$ is given by
\ba
E_{(k)}(\phi^{AB})&=\sum_{x=k+1}^{d}q_x\\
\GG{\eqref{chichi}}&= \sum_{z\in[n]}t_z\sum_{x=k+1}^{d}q_{x|z}\\
&=\sum_{z\in[n]}t_zE_{(k)}(\phi^{AB}_z)\;.
\ea
Hence, from Nielsen majorization theorem as expressed in Exercise~\ref{nil}, it follows that~\eqref{ekcond} holds if and only if $\psi^{AB}$ can be converted to $\phi^{AB}$ by LOCC. Therefore, to complete the proof we now show that
$\phi^{AB}$ can be converted by LOCC to the ensemble $\left\{\phi^{AB}_z,\;t_z\right\}_{z\in[n]}$. The conversion is achieved by the following single measurement performed by Alice
\be
M_z\eqdef\sum_{x\in[d]}\sqrt{\frac{t_zq_{x|z}}{q_x}}|x\lr x|^A
\ee
Observe that
\be
\sum_{z\in[n]}M_z^*M_z=\sum_{x\in[d]}\sum_{z\in[n]}\frac{t_zq_{x|z}}{q_x}|x\lr x|=\sum_{x\in[d]}|x\lr x|^A=I^A\;,
\ee
where we used~\eqref{chichi}. Furthermore, note that
\be
\left(M_z\otimes I^B\right)|\phi^{AB}\ra=\sum_{x\in[d]}\sqrt{t_zq_{x|z}}|xx\ra^{AB}=\sqrt{t_z}|\phi^{AB}_z\ra\;.
\ee
That is, Alice's measurement produces the state $|\phi^{AB}_z\ra$ with probability $t_z$. This completes the proof.
\end{proof}

The above theorem generalizes Nielsen majorization theorem to probabilistic transformations, and in addition leads to the following corollary.

\begin{myg}{}
\begin{corollary}\label{cor:vid}
Let $\psi,\phi\in\pure(AB)$ be two pure bipartite entangled states. Then, the maximal probability with which $\psi^{AB}$ can be converted to $\phi^{AB}$ by LOCC is given by
\be\label{1260}
\pr\left(\psi^{AB}\xrightarrow{\text{\tiny LOCC}} \phi^{AB}\right)=\min_{k\in[d]}\frac{E_{(k)}(\psi^{AB})}{E_{(k)}(\phi^{AB})}\;.
\ee
\end{corollary}
\end{myg}

\begin{remark}
Note that for $k=1$, $E_1(\psi^{AB})=1$ for all $\psi\in\pure(AB)$. Therefore, the expression on the right-hand side of the equation above can never exceed one. Note further that the corollary above is a simplification of the formal result given in Corollary~\ref{cor:mono}. That is, for pure bipartite states it is sufficient to check the ratios of only $d$ resource measures in order to compute the maximum probability of conversion.
\end{remark}
\begin{proof}
Consider an optimal LOCC that converts $\psi^{AB}$ to $\phi^{AB}$ with the maximum possible probability. Such LOCC protocol yields $\phi^{AB}$ with probability $p\eqdef \pr\left(\psi^{AB}\xrightarrow{\text{\tiny LOCC}} \phi^{AB}\right)$ and other states with probability $1-p$. All such other states can always be converted deterministically (by LOCC) to the product state $|0\lr 0|^{A}\otimes |0\lr 0|^B$. Therefore, without loss of generality we can assume that $\psi^{AB}$ is converted to $\phi^{AB}$ with probability $p$, and to $|0\lr 0|^{A}\otimes |0\lr 0|^B$ with probability $1-p$. Since $E_{k}(|0\lr 0|^{A}\otimes |0\lr 0|^B)=0$ for all $k=2,\ldots,d$, Theorem~\ref{thm:dp} implies that such an LOCC protocol is possible if and only if
\be
E_{(k)}(\psi^{AB})\geq pE_{(k)}(\phi^{AB})\quad\forall\;k\in\{2,\ldots,d\}\;.
\ee
The proof is concluded by recognizing that the equation above is equivalent to
\be
p\leq\min_{k\in\{2,\ldots,d\}}\frac{E_{(k)}(\psi^{AB})}{E_{(k)}(\phi^{AB})}.
\ee
Therefore, the maximum probability is the one given in~\eqref{1260}.
\end{proof}

\begin{exercise}
Let $|\psi^{AB}\ra=\sqrt{p_1}|11\ra+\ldots+\sqrt{p_{d}}|dd\ra\in\mbb{C}^d\otimes\mbb{C}^d$ be a 2-qudit entangled state.
\begin{enumerate}
\item What is the maximum probability to convert by LOCC $|\psi^{AB}\ra$  to the maximally entangled state $|\Phi^{AB}\ra$.
\item For $d=2$ find the LOCC protocol that achieves the maximum probability you found in Part~1.
\end{enumerate}
\end{exercise}

\begin{exercise}
Show that a maximally entangled state $\Phi^{AB}$ can be converted to any mixed bipartite quantum state $\rho\in\md(AB)$.
\end{exercise}

Theorem~\ref{thm:dp} can also be use to provide necessary and sufficient conditions to determine the conversion of a pure entangled state to a mixed entangled state by LOCC.

\begin{myg}{}
\begin{corollary}\label{cor:1242}
Let $\psi\in\pure(AB)$ and $\sigma\in\md(AB)$ be two  bipartite entangled states with $d\eqdef|A|=|B|$. Then, $\psi^{AB}$ can be converted to $\sigma^{AB}$ by LOCC if and only if
\be\label{1260q}
\max_{\{t_x,\;\phi_x\}_{x\in[m]}}\min_{k\in[d]}\Big\{E_{(k)}(\psi^{AB})-\sum_{x\in[m]}p_xE_{(k)}(\phi^{AB}_x)\Big\}\geq 0\;,
\ee
where the maximum is over all pure-state decompositions of $\sigma$ (i.e.\ over all pure-state ensembles $\{p_x,\;\phi_x\}_{x\in[m]}$ that satisfy $\sigma=\sum_{x\in[m]}p_x\phi_x$).
\end{corollary}
\end{myg}

\begin{proof}
Suppose the condition in~\eqref{1260q} holds. Then, there exists an ensemble of states $\{p_x,\;\phi^{AB}_x\}_{x\in[m]}$ such that $E_{(k)}(\psi^{AB})\geq\sum_{x\in[m]}p_xE_{(k)}(\phi^{AB}_x)$ for all $k\in[d]$. From Theorem~\ref{thm:dp} it follows that $\psi^{AB}$ can be converted by LOCC to the cq-state\index{cq-state} 
\be
\sigma^{XAB}=\sum_{x\in[m]}p_x|x\lr x|^X\otimes\phi_x^{AB}\;.
\ee 
Since tracing out the classical system $X$ is an LOCC operation we conclude that $\psi^{AB}\xrightarrow{\text{\tiny LOCC}} \sigma^{AB}$.

Conversely, suppose $\psi^{AB}\xrightarrow{\text{\tiny LOCC}} \sigma^{AB}$. From Theorem~\ref{lopop} it follows that there exists a generalized measurement on Alice's system $\{M_x\}_{x\in[m]}$ and a set of unitary matrices on Bob's system $\{U_x\}_{x\in[m]}$ such that
\be
\sigma^{AB}=\sum_{x\in[m]}\left(M_x\otimes U_x\right)\psi^{AB}\left(M_x\otimes U_x\right)^*\;.
\ee 
Denote by $|\phi_x^{AB}\ra\eqdef \frac1{\sqrt{p_x}}\left(M_x\otimes U_x\right)|\psi^{AB}\ra$, where $p_x\eqdef \big\la\psi^{AB}\big|M_x^*M_x\otimes I^B\big|\psi^{AB}\big\ra$. Then, from the equation above we get that the ensemble $\{p_x,\;\phi_x^{AB}\}_{x\in[m]}$ form a pure state decomposition of $\sigma^{AB}$. Moreover, by definition, $\psi^{AB}$ can be converted by LOCC to $\phi_x^{AB}$ with probability $p_x$. Therefore, from Theorem~\ref{thm:dp} it follows that 
\be
\min_{k\in[d]}\Big\{E_{(k)}(\psi^{AB})-\sum_{x\in[n]}p_xE_{(k)}(\phi^{AB}_x)\Big\}\geq 0\;.
\ee
Hence, the condition in~\eqref{1260q} holds. This completes the proof.
\end{proof}

\section{Approximate Single-Shot Conversions}\index{single-shot}

In practical scenarios, state transformations are often imperfect. Rather than seeking perfect LOCC conversion of one state into another, it is more realistic to allow the final state to be $\eps$-close to the target state, where $\eps>0$ is a small threshold typically reflective of the inaccuracy inherent in the apparatus used. Allowing for the final state to be any state within $\eps$-proximity to the desired state opens the door to additional state transformations that are not encompassed by Nielsen's majorization condition.

\subsection{The Conversion Distance}\label{sec:cdpures}

Consider two pure bipartite states $\psi,\phi\in\pure(AB)$ (with $d\eqdef|A|=|B|$). The conversion distance of $\psi^{AB}$ into $\phi^{AB}$ is given by (refer to \eqref{cd} with $\mf=\locc$)
\be\label{cd01}
T\left(\psi^{AB}\xrightarrow{\text{\tiny LOCC}}\phi^{AB}\right)\eqdef\min_{\sigma\in\md(AB)}\left\{\frac12\left\|\phi^{AB}-\sigma^{AB}\right\|_1\;:\;\psi^{AB}\xrightarrow{\text{\tiny LOCC}} \sigma^{AB}\right\}\;.
\ee
One might wonder whether this conversion distance changes if we restrict the optimization over $\md(AB)$ to states over $\pure(AB)$. Given that the trace distance between pure states equals the purified distance, constraining the above optimization to pure states yields
\be\label{optih}
P_\star\left(\psi^{AB}\xrightarrow{\text{\tiny LOCC}}\phi^{AB}\right)\eqdef\min_{\varphi\in\pure(AB)}\left\{P\left(\phi^{AB},\varphi^{AB}\right)\;:\;\psi^{AB}\xrightarrow{\text{\tiny LOCC}} \varphi^{AB}\right\}\;.
\ee
It's important to note that the value of $T(\psi^{AB}\xrightarrow{\text{\tiny LOCC}}\phi^{AB})$ is not greater than that of $P_\star(\psi^{AB}\xrightarrow{\text{\tiny LOCC}}\phi^{AB})$. This is because restricting $\sigma^{AB}$ to be a pure state in the calculation of $T(\psi^{AB}\xrightarrow{\text{\tiny LOCC}}\phi^{AB})$ can only increase the minimum value obtained in the optimization process. Furthermore, as we will explore in the next chapter (specifically, in Lemma~\ref{lem1331}), it will be shown that $T(\psi^{AB}\xrightarrow{\text{\tiny LOCC}}\phi^{AB})$ can actually be strictly smaller than $P_\star(\psi^{AB}\xrightarrow{\text{\tiny LOCC}}\phi^{AB})$. However, we will soon discover that the value of $P_{\star}$ remains unchanged even when the optimization is extended from $\pure(AB)$ to the full set of density matrices $\md(AB)$. 

The optimization problem in \eqref{optih} can be simplified as follows. Initially, observe that if $\p,\r\in\prob(d)$ are the Schmidt vectors of $\psi^{AB}$ and $\varphi^{AB}$, respectively, then according to Nielsen's theorem, the condition $\psi^{AB}\xrightarrow{\text{\tiny LOCC}} \varphi^{AB}$ is equivalent to $\r\succ\p$. Thus, for any given Schmidt vector $\r$ of $\varphi^{AB}$, we first perform the optimization over all states $\varphi\in\pure(AB)$ with the same Schmidt vector $\r$. Denoting by $|\tvarphi^{AB}\rangle=\sum_{x\in[d]}\sqrt{r_x}|xx\rangle$, this is equivalent to optimization over the local unitaries $U\otimes V$, such that $|\varphi^{AB}\ra=U\otimes V|\tvarphi^{AB}\ra$. Due to the relationship between trace distance and fidelity, we have
\ba
\min_{U,V\in\muu(d)}P\left(\phi^{AB},\varphi^{AB}\right)=\sqrt{1-\max_{U,V\in\muu(d)}|\la\phi^{AB}|\varphi^{AB}\ra|^2}
\ea
Denoting by $|\phi^{AB}\ra=N\otimes I^B|\Omega^{AB}\ra$ and by $D$ the diagonal matrix with diagonal $\r$, we obtain
\ba\label{cf1273}
\max_{U,V\in\muu(d)}\left|\la\phi^{AB}|\varphi^{AB}\ra\right|&=\max_{U,V\in\muu(d)}\big|\tr[N^*UDV^T]\big|\\
\GG{\substack{\text{von-Neumann trace inequality}\\(see~\eqref{finchi})}}
&= \sum_{x\in[d]}\sqrt{r_x^\da q_{x}^\da}\;,
\ea
where $\q\in\prob(d)$ is the Schmidt vector of $\phi^{AB}$. 
Taking everything into consideration we obtain the following simplification:
\be\label{pdsimple}
P_\star\left(\psi^{AB}\xrightarrow{\text{\tiny LOCC}}\phi^{AB}\right)=\min_{\r\in\prob(d)}\Big\{ P(\q,\r)\;:\;\r\succ\p\Big\}\;,
\ee
where $\p$ is the Schmidt vector of $\psi^{AB}$,  $\q$ the Schmidt vector of $\phi^{AB}$, and $P(\q,\r)\eqdef\sqrt{1-F^2(\q,\r)}$ is the purified distance between probability vectors.

Observe that we added the subscript $\star$ to $P_{\star}$ since the conversion distance between two pure states $\psi,\phi\in\pure(AB)$ as measured by the purified distance is defined as:
\be\label{pcd}
P\left(\psi^{AB}\xrightarrow{\text{\tiny LOCC}}\phi^{AB}\right)\eqdef\min_{\sigma\in\md(AB)}\left\{P\left(\phi^{AB},\sigma^{AB}\right)\;:\;\psi^{AB}\xrightarrow{\text{\tiny LOCC}} \sigma^{AB}\right\}\;.
\ee
At first glance, this conversion distance may seem to be different than $P_\star$, however, the following theorem demonstrate the two are equal.

\begin{myt}{}
\begin{theorem}\label{thmpps}
Let $\psi,\phi\in\pure(AB)$. Then,
\be\label{eq12p70}
P\left(\psi^{AB}\xrightarrow{\text{\tiny LOCC}}\phi^{AB}\right)=P_{\star}\left(\psi^{AB}\xrightarrow{\text{\tiny LOCC}}\phi^{AB}\right)\;.
\ee
\end{theorem}
\end{myt}
\begin{proof}
The inequality
\be\label{12p744}
P\left(\psi^{AB}\xrightarrow{\text{\tiny LOCC}}\phi^{AB}\right)\leq P_{\star}\left(\psi^{AB}\xrightarrow{\text{\tiny LOCC}}\phi^{AB}\right)\;.
\ee
follows trivially since restricting $\sigma^{AB}$ in~\eqref{pcd} to be a pure state $\varphi^{AB}$ can only increase the quantity.
To prove the opposite inequality let $\sigma^{AB}$ be an optimizer of~\eqref{pcd}. Since $\psi^{AB}\xrightarrow{\text{\tiny LOCC}} \sigma^{AB}$ it follows from Corollary~\ref{cor:1242} and its proof that there exists an ensemble $\{t_z,\varphi_z^{AB}\}_{z\in[k]}$ such that $\sigma^{AB}=\sum_{z\in[k]}t_z\varphi^{AB}_z$ and $\psi^{AB}\xrightarrow{\text{\tiny LOCC}} \{t_z,\varphi_z^{AB}\}_{z\in[k]}$. For each $z\in[k]$ let $\r_z$ be the Schmidt vector of $\varphi_z^{AB}$ and define
\be
\r\eqdef\sum_{z\in[k]}t_z\r_z^\da\;.
\ee
Let $\varphi^{AB}$ be a pure state with a Schmidt vector $\r$, so that $\psi^{AB}\xrightarrow{\text{\tiny LOCC}} \varphi^{AB}$. Observe that the square fidelity is given by
\ba\label{12p777}
F\left(\phi^{AB},\sigma^{AB}\right)^2=\la\phi^{AB}|\sigma^{AB}|\phi^{AB}\ra&=\sum_{z\in[k]}t_z\left|\la\phi^{AB}|\varphi^{AB}_z\ra\right|^2\\
\GG{{\it cf.}~\eqref{cf1273}}&\leq \sum_{z\in[k]}t_zF\left(\q^\da,\r_z^\da\right)^2\\
\GG{\eqref{concc}}&\leq F\left(\q^\da,\r^\da\right)^2\;.
\ea
Hence, 
\ba
P\big(\psi^{AB}\xrightarrow{\text{\tiny LOCC}}\phi^{AB}\big)&=\sqrt{1-F\left(\phi^{AB},\sigma^{AB}\right)^2}\\
\GG{\eqref{12p777}}&\geq \sqrt{1-F\left(\q^\da,\r^\da\right)^2}\\
&\geq P_\star\big(\psi^{AB}\xrightarrow{\text{\tiny LOCC}}\phi^{AB}\big)\;.
\ea
Comparing the above inequality with~\eqref{12p744} we get the equality of~\eqref{eq12p70}. 
\end{proof}

\begin{myg}{}
\begin{corollary}\label{corppp}
Let $\psi\in\pure(AB)$ and $\Phi_m\in\pure(A'B')$ be the maximally entangled state with $m\eqdef|A'|=|B'|$. Then,
\be
P\left(\Phi_m\xrightarrow{\text{\tiny LOCC}}\psi^{AB}\right)=\sqrt{E_{(m)}\left(\psi^{AB}\right)}\;,
\ee
where $E_{(m)}$ is the measure of entanglement defined in~\eqref{vidal} for $k=m$.
\end{corollary}
\end{myg}
\begin{remark}
Observe that the corollary above provides an operational meaning to the entanglement monotones $E_{(m)}$. That is, $E_{(m)}(\psi^{AB})$ measures how close (in terms of the square of the purified distance) $\Phi_m$ can reach to $\psi^{AB}$ by LOCC. Furthermore, it's noteworthy that when $m \geq |A|$, the conversion distance $P\left(\Phi_m\xrightarrow{\text{\tiny LOCC}}\psi^{AB}\right)$ equals zero. This outcome arises because, in this scenario, the majorization theorem by Nielsen guarantees that the conversion $\Phi_m\xrightarrow{\text{\tiny LOCC}}\psi^{AB}$ can be accomplished exactly.
\end{remark}
\begin{proof}
Let $\p\in\prob^\da(n)$, with $n\eqdef|A|$, be the Schmidt vector corresponding to $\psi^{AB}$. From Theorem~\ref{thmpps} it follows that
\be
P\left(\Phi_m\xrightarrow{\text{\tiny LOCC}}\psi^{AB}\right)=\min_{\r\in\prob^\da(n)}\Big\{P(\p,\r)\;:\;\r\succ\u^{(m)}\Big\}\;.
\ee
where $\u^{(m)}$ is the uniform density matrix in $\prob(m)$. Now, observe that the condition $\r\succ\u^{(m)}$ holds if and only if $\r$ has at most $m$ non-zero components. Denoting by $|\r|$ the number of non-zero components in $\r$, and using the fact that the square of the purified distance equals one minus the square of the fidelity, we get
\ba
P^2\left(\Phi_m\xrightarrow{\text{\tiny LOCC}}\psi^{AB}\right)&=1-\max_{\r\in\prob(n), |\r|=m}\Big(\sum_{x\in[n]}\sqrt{r_xp_x}\Big)^2\\
&=1-\max_{\r\in\prob(m)}\Big(\sum_{x\in[m]}\sqrt{r_xp_x}\Big)^2\\
\GG{Exercise~\ref{loip}}&=1-\sum_{x\in[m]}p_x\\
&=E_{(m)}\left(\psi^{AB}\right)\;.
\ea
This completes the proof.
\end{proof}

\begin{exercise}\label{loip}
Let $\{s_x\}_{x\in[n]}$ be a set of non-negative real numbers. Show that
\be
\max_{\r\in\prob(n)}\sum_{x\in[n]}\sqrt{r_x}s_x=\|\s\|_2\eqdef\sqrt{\sum_{x\in[n]}s^2_x}
\ee
\end{exercise}

Drawing on the more straightforward formula in \eqref{pdsimple}, a different kind of conversion distance can be defined, one that is based on the trace distance between Schmidt probability vectors. Specifically, let
\be\label{cd02}
T_\star\left(\psi^{AB}\xrightarrow{\text{\tiny LOCC}}\phi^{AB}\right)\eqdef\min_{\r\in\prob(d)}\Big\{ \frac12\|\q-\r\|_1\;:\;\r\succ\p\Big\}\;,
\ee
where $\p$ is the Schmidt vector of $\psi^{AB}$, and $\q$ is the Schmidt vector of $\phi^{AB}$.

This metric can be viewed as the conversion distance of $\psi^{AB}$ into $\phi^{AB}$, in a framework where all bipartite pure states are identified with their corresponding Schmidt vectors. It quantifies the proximity to which the Schmidt vector of $\psi^{AB}$ can approach the Schmidt vector of $\phi^{AB}$ via doubly stochastic matrices. Given that the resource content of pure bipartite entangled states is fully encapsulated in their Schmidt vectors, this conversion distance arguably becomes the most natural way to gauge the effectiveness of interconversion between two pure resources.

In fact, the two conversion distances defined in \eqref{cd01} and \eqref{cd02} are topologically equivalent. This equivalence becomes apparent when considering that the inequalities in \eqref{6263} lead to the following relation:
\begin{align}
&T\left(\psi^{AB}\xrightarrow{\text{\tiny LOCC}}\phi^{AB}\right)\leq P\left(\psi^{AB}\xrightarrow{\text{\tiny LOCC}}\phi^{AB}\right)\leq \sqrt{2T\left(\psi^{AB}\xrightarrow{\text{\tiny LOCC}}\phi^{AB}\right)}\\
&T_\star\left(\psi^{AB}\xrightarrow{\text{\tiny LOCC}}\phi^{AB}\right)\leq P_\star\left(\psi^{AB}\xrightarrow{\text{\tiny LOCC}}\phi^{AB}\right)\leq \sqrt{2T_\star\left(\psi^{AB}\xrightarrow{\text{\tiny LOCC}}\phi^{AB}\right)}\;.
\end{align}
Therefore, combining these inequalities with Theorem~\ref{thmpps}  gives
\begin{align}
& T\left(\psi^{AB}\xrightarrow{\text{\tiny LOCC}}\phi^{AB}\right)\leq \sqrt{2T_\star\left(\psi^{AB}\xrightarrow{\text{\tiny LOCC}}\phi^{AB}\right)}\\
&T_\star\left(\psi^{AB}\xrightarrow{\text{\tiny LOCC}}\phi^{AB}\right)\leq \sqrt{2T\left(\psi^{AB}\xrightarrow{\text{\tiny LOCC}}\phi^{AB}\right)}\;.
\end{align}
Considering the established equivalence between the two conversion distances, it makes sense to primarily use $T_\star$ for further analysis, owing to its following closed-form expression. 

\begin{myt}{\color{yellow} Closed Formula}
\begin{theorem}\label{thm:dlocc}
Let $\psi,\phi\in\pure(AB)$ be two bipartite states with $d\eqdef |A|=|B|$, and let $\p,\q\in\prob(d)$ be the corresponding Schmidt probability vectors of $\psi^{AB}$ and $\phi^{AB}$, respectively. Then,
\be\label{1224t}
T_{\star}\left(\psi^{AB}\xrightarrow{\text{\tiny LOCC}}\phi^{AB}\right)=\max_{k\in[d]}\big\{\|\p\|_{(k)}-\|\q\|_{(k)}\big\}\;.
\ee
\end{theorem}
\end{myt}

\begin{proof}
The proof follows directly from Theorem~\ref{thm:dlocc0}. To see this, observe that
\ba
T_\star\left(\psi^{AB}\xrightarrow{\text{\tiny LOCC}}\phi^{AB}\right)&\eqdef\min_{\r\in{\rm Majo}(\p)}\frac12\|\q-\r\|_1\\
&=T\big(\q,{\rm Majo}(\p)\big)\\
\GG{Theorem~\ref{thm:dlocc0}}&=\max_{k\in[d]}\big\{\|\p\|_{(k)}-\|\q\|_{(k)}\big\}\;.
\ea
This concludes the proof.
\end{proof}

\bex\label{ex1252}
Show that the maximization~\eqref{1224t} over all $k\in[d]$ can be replaced with maximization over all $k\in[r]$, where $r=\sr\left(\psi^{AB}\right)$ is the  Schmidt rank of $\psi^{AB}$.
\eex

\subsection{Distillable Entanglement}\index{single-shot}\index{distillable entanglement}
In this subsection we calculate the single-shot distillable entanglement. 
Consider first the zero-error case in which $\eps=0$, and let $\p=\p^\da\in\prob(d)$ be the Schmidt vector of an entangled pure state $\psi\in\pure(AB)$, with $d\eqdef|A|=|B|$. For the zero-error case, the single-shot distillable entanglement is defined as:
\be
\distill^{\eps=0}\left(\psi^{AB}\right)\eqdef\max_{m\in\mbb{N}}\Big\{\log m\;:\;\psi^{AB}\xrightarrow{\text{\tiny LOCC}}\Phi_m\Big\}\;.
\ee
From Nielsen majorization theorem, $\psi^{AB}\xrightarrow{\text{\tiny LOCC}}\Phi_m$ if and only if $\frac1m\geq p_1$, where $p_1$ is the first component of $\p=\p^\da$. Hence,
\ba
\distill^{\eps=0}\left(\psi^{AB}\right)&=\max_{m\in\mbb{N}}\Big\{\log m\;:\;m\leq\frac1{p_1}\Big\}\\
\GG{{\it m}\text{ is an integer}}&=\log\left\lfloor\frac1{p_1}\right\rfloor\;.
\ea
The quantity $1/p_1$ can be express in terms of the min-entropy\index{min-entropy} of $\p$. Specifically, 
\be\label{forhmin}
\distill^{\eps=0}\left(\psi^{AB}\right)=\log\left\lfloor2^{H_{\min}(A)_\rho}\right\rfloor\;,
\ee
where $H_{\min}(A)_\rho$ is the min-entropy\index{min-entropy} (see~\eqref{qminen}) of the reduced density matrix $\rho^A\eqdef\tr_B\left[\psi^{AB}\right]$.

To extend the formula above for the case that $\eps>0$, we use the computable conversion distance, to calculate the single-shot distillable entanglement\index{distillable entanglement}. 
Specifically, for any $\eps\in(0,1)$ and $\psi\in\pure(AB)$, we define the $\eps$-single-shot distillable entanglement as:
\be\label{combdif}
\distill^\eps\left(\psi^{AB}\right)\eqdef\max_{m\in\mbb{N}}\Big\{\log m\;:\;T_\star\left(\psi^{AB}\xrightarrow{\text{\tiny LOCC}}\Phi_m\right)\leq\eps\Big\}\;.
\ee
From the closed formula for $T_\star$ we get the following result.
\begin{myt}{}
\begin{theorem}\label{thmdias}
Using the same notations as above, for every $\eps\in[0,1)$ and $\psi\in\pure(AB)$ we have
\be\label{123600}
\distill^\eps\left(\psi^{AB}\right)=\log\left\lfloor2^{H_{\min}^\eps(A)_\rho}\right\rfloor\;,
\ee
where $H_{\min}^\eps(A)_\rho$ is the smoothed min-entropy\index{min-entropy} as given in~\eqref{fofhmn}. 
\end{theorem}
\end{myt}

\begin{remark}
In~\eqref{fofhmn} we found a closed form to the smoothed min-entropy. Using this form we can expressed the $\eps$-single-shot distillable entanglement\index{distillable entanglement} of $\psi^{AB}$ as:
\be\label{1236}
\distill^\eps\left(\psi^{AB}\right)=\min_{k\in\{\ell,\ldots,d\}} \log\left\lfloor\frac{k}{\|\p\|_{(k)}-\eps}\right\rfloor\;,
\ee
where $\ell\in[d]$ is the integer satisfying 
$\|\p\|_{(\ell-1)}\leq\eps<\|\p\|_{(\ell)}$. 
\end{remark}

\begin{proof}
From Theorem~\ref{thm:dlocc} and Exercise~\ref{ex1252} we have
\ba\label{12a1}
T_\star\left(\psi^{AB}\xrightarrow{\text{\tiny LOCC}}\Phi_m\right)&=\max_{k\in[d]}\left\{\|\p\|_{(k)}-\|\u^{(m)}\|_{(k)}\right\}\\
&=\max_{k\in[d]}\left\{\|\p\|_{(k)}-\frac km\right\}\;.
\ea
Combining this with the definition in~\eqref{combdif} we obtain
\ba\label{12b12}
\distill^\eps\left(\psi^{AB}\right)&=\max_{m\in\mbb{N}}\left\{\log m\;:\;\|\p\|_{(k)}-\frac km\leq\eps\;\;\quad\forall\;k\in[d]\right\}\\
&=\max_{m\in\mbb{N}}\left\{\log m\;:\;\|\p\|_{(k)}-\frac km\leq\eps\;\;\quad\forall\;k\in\{\ell,\ldots,d\}\right\}\;,
\ea
since from the definition of $\ell$, if $k<\ell$ then the inequality $\|\p\|_{(k)}-\frac km\leq\eps$ holds trivially. 
Finally, observe that for each $k\in\{\ell,\ldots,d\}$, the condition $\|\p\|_{(k)}-\frac km\leq\eps$ can be expressed as $m\leq \frac{k}{\|\p\|_{(k)}-\eps}$, and since $m$ is an integer, this condition is equivalent to $m\leq a_k$, where
\be
a_k\eqdef\left\lfloor\frac{k}{\|\p\|_{(k)}-\eps}\right\rfloor\;.
\ee
Therefore, with this notation Eq.~\eqref{12b12} gives
\ba\label{12b1}
\distill^\eps\left(\psi^{AB}\right)&=\max_{m\in\mbb{N}}\big\{\log m\;:\;m\leq a_k\;\;\forall\;k\in\{\ell,\ldots,d\}\big\}\\
&=\log\min_{k\in\{\ell,\ldots,d\}}\{a_k\}\;.
\ea
This completes the proof.
\end{proof}

\bex
Use the formula in~\eqref{1236} to compute $\distill^\eps(\psi^{AB})$ for the two extreme cases: (1) $\psi^{AB}$ is a maximally entangled state; (2) $\psi^{AB}$ is a product state.
Give a physical interpretation to the results.
\eex

\bex
Let $\psi\in\pure(AB)$. Show that
\be
\lim_{\eps\to 1^-}\distill^\eps\left(\psi^{AB}\right)=\infty\;.
\ee
\eex

\subsection{Entanglement Cost}\index{single-shot}\index{entanglement cost}

In this subsection, we present a closed-form expression for the single-shot entanglement cost. Intriguingly, as we will explore in the next chapter, for the entanglement cost, both conversion distances (referenced in equations \eqref{cd01} and \eqref{cd02}) yield the same entanglement cost. This is a result we could not demonstrate for the distillable entanglement of a pure bipartite state, though we suspect it to be true.

For any $\eps\in(0,1)$ and $\psi\in\pure(AB)$, the $\eps$-single-shot entanglement cost is defined as
\be
\cost^\eps\left(\psi^{AB}\right)\eqdef\min_{m\in\mbb{N}}\Big\{\log m\;:\;T_\star\left(\Phi_m\xrightarrow{\text{\tiny LOCC}}\psi^{AB}\right)\leq\eps\Big\}
\ee
From the closed formula for $T_\star$ we get the following result.
\begin{myt}{}
\begin{theorem}\label{purestatecost}
Let $\eps\in[0,1)$, $\psi\in\pure(AB)$, $d\eqdef\sr(\psi^{AB})$, and $\p\in\prob(d)$ be the Schmidt probability vector of $\psi^{AB}$.
Then, the $\eps$-single-shot entanglement cost of $\psi^{AB}$ is given by
\be\label{1242}
\cost^\eps\left(\psi^{AB}\right)=\log m
\ee
where $m\in[d]$ is the integer satisfying 
$
\|\p\|_{(m-1)}<1-\eps\leq\|\p\|_{(m)}
$.
\end{theorem}
\end{myt}
\begin{proof}
From Theorem~\ref{thm:dlocc} and Exercise~\ref{ex1252} we have for any $m\in\mbb{N}$ 
\be\label{cost1243}
T_\star\left(\Phi_m\xrightarrow{\text{\tiny LOCC}}\psi^{AB}\right)=\max_{k\in[m]}\sum_{x\in[k]}\left(\frac1m-p_x\right)=\max_{k\in[m]}\left\{\frac km-\|\p\|_{(k)}\right\}\;.
\ee
We therefore have
\ba
\cost^\eps\left(\psi^{AB}\right)&=\min_{m\in\mbb{N}}\left\{\log m\;:\;\frac km-\|\p\|_{(k)}\leq\eps\;\;\forall\;k\in[m]\right\}\\
&=\min_{m\in\mbb{N}}\left\{\log m\;:\;m\geq \frac{k}{\|\p\|_{(k)}+\eps}\;\;\forall\;k\in[m]\right\}\\
\GG{Exercise~\ref{ex:4712}}&=\min_{m\in\mbb{N}}\left\{\log m\;:\;m\geq \frac{m}{\|\p\|_{(m)}+\eps}\right\}\\
&=\min_{m\in\mbb{N}}\big\{\log m\;:\;\|\p\|_{(m)}\geq1-\eps\big\}\;.
\ea
%Moreover, if we replace the restriction that $m\geq b_k$ for all $k\in[m]$ with a stronger restriction that $m\geq b_k$ for all $k\in[d]$ we get
%\ba
%E_C^\eps\left(\psi^{AB}\right)&\leq\min_{m\in[d]}\left\{\log m\;:\;m\geq b_k\;\;\forall\;k\in[d]\right\}\\
%&=\log\max_{k\in[d]}b_k\;.
%\ea
This completes the proof.
\end{proof}

\bex\label{ex:4712}
Let \be
b_k\eqdef\frac{k}{\|\p\|_{(k)}+\eps}\quad\quad\forall\;k\in[d]\;.
\ee
Show that
\be
b_{1}\leq b_2\leq\cdots\leq b_d\;.
\ee
\eex

As a simple example of this formula, consider the case that $\eps=0$. In this case, the smallest $m$ for which $\|\p\|_{(m)}\geq 1$ is $m=\sr(\psi^{AB})$. Therefore, as expected, the exact single-shot cost of $\psi^{AB}$ is given by
\be
\cost^{\eps=0}\left(\psi^{AB}\right)=\log\sr\left(\psi^{AB}\right)\;.
\ee

\bex
Use the formula in~\eqref{1242} to compute $\cost^\eps(\psi^{AB})$ for the two extreme cases: (1) $\psi^{AB}$ is a maximally entangled state; (2) $\psi^{AB}$ is a product state.
Give a physical interpretation for your results.
\eex

In the following corollary we provide an operational interpretation for the smoothed max-entropy\index{max-entropy} as the single-shot entanglement cost\index{entanglement cost}. Recall that the max-entropy of $\rho\in\md(A)$ is defined as (cf.~\eqref{qhmax}), 
\be
H_{\max}(A)_\rho\eqdef\log\rank(\rho)\;,
\ee
and its $\eps$-smoothed version as
\be
H_{\max}^\eps(A)_\rho\eqdef\min_{\rho'\in\mb_\eps(\rho)}H_{\max}(A)_{\rho'}\;.
\ee

\begin{myg}{}
\begin{corollary}\label{corhmaxe}
Let $\eps\in[0,1)$, $\psi\in\pure(AB)$, and denote by $\rho^A\eqdef_{B}\left[\psi^{AB}\right]$ its reduced density matrix. Then, the $\eps$-single-shot entanglement cost\index{entanglement cost} of $\psi^{AB}$ is given by
\be
\cost^{\eps}\left(\psi^{AB}\right)=H_{\max}^\eps\left(A\right)_\rho\;.
\ee
\end{corollary}
\end{myg}

\begin{proof}
The proof follows trivially from a combination of the theorem above and the expression for $H_{\max}^\eps$ as given in Lemma~\ref{lemofhmax}.
\end{proof}

\subsection{Embezzlement of Entanglement}\index{embezzlement of entanglement}

In the single-shot regime, we explore the approximate interconversion of resources where the final states are $\eps$-close to the intended states. As previously discussed, this flexibility introduces additional state transformations that go beyond the scope of Nielsen's majorization theorem. Notably, this includes the phenomenon known as \emph{embezzlement of entanglement}.

Embezzlement of entanglement can be regarded as an extreme form of entanglement catalysis\index{entanglement catalysis}. Consider a family of bipartite states $|\chi_n\rangle \in \mathbb{C}^n \otimes \mathbb{C}^n$, defined by
\be\label{chin}
|\chi_n\ra\eqdef\frac{1}{\sqrt{H_n}}\sum_{y\in[n]}\frac1{\sqrt{x}}|x\ra|x\ra\;,
\ee
where the normalization factor $H_n\eqdef\sum_{x\in[n]}\frac1{x}$  is known as the \emph{harmonic number}.
We will demonstrate that $|\chi_n\rangle$ can serve as a catalyst for the \emph{generation} of any arbitrary bipartite state $|\psi\rangle \in \mathbb{C}^m \otimes \mathbb{C}^m$, with $|\chi_n\rangle$ undergoing minimal change. More precisely, for any $\eps>0$, there exists an $n \in \mathbb{N}$ such that
\be
T_\star\left(\chi_n\xrightarrow{\text{\tiny LOCC}}\psi\otimes\chi_n\right)\leq\eps\;.
\ee
This remarkable result implies that it's feasible to `embezzle' a copy of $|\psi\rangle$ from the catalyst $|\chi_n\rangle$, effectively borrowing some of its entanglement while leaving it largely unchanged.

To see how it works, recall from Lemma~\ref{lemmonin} that
\be
T_\star\left(\chi_n\xrightarrow{\text{\tiny LOCC}}\psi\otimes\chi_n\right)\leq T_\star\left(\chi_n\xrightarrow{\text{\tiny LOCC}}\Phi_m\otimes\chi_n\right)\;,
\ee
since the maximally entangled state $|\Phi_m\ra=\frac1{\sqrt{m}}\sum_{x\in[m]}|xx\ra$ can be converted by LOCC to the state $\psi$. It is therefore sufficient to show that the right-hand side of the equation above can be made arbitrarily small as we increase the dimension $n$. Let $\p$ be the Schmidt vector of $\chi_n$ and $\q$ be the Schmidt vector of $\Phi_m\otimes\chi_n$. Observe that $\p\in\prob(n)$ and $\q\in\prob(nm)$. From Theorem~\ref{thm:dlocc} we know that
\be\label{dste}
T_\star\left(\chi_n\xrightarrow{\text{\tiny LOCC}}\Phi_m\otimes\chi_n\right)=\max_{k\in[n]}\left\{\|\p\|_{(k)}-\|\q\|_{(k)}\right\}\;.
\ee
Now, observe that the components of $\q$ have the form $\frac{p_x}m$. Therefore, for any decomposition $k=am+b$, with $a\eqdef\left\lfloor\frac km\right\rfloor$ and some $b\in\{0,1,\ldots,m-1\}$ we have
\be
\|\q\|_{(k)}=\|\p\|_{(a)}+\frac bmp_{a+1}\;.
\ee
Substituting this into~\eqref{dste} gives 
\ba
T_\star\left(\chi_n\xrightarrow{\text{\tiny LOCC}}\Phi_m\otimes\chi_n\right)&= \max_{\substack{k\in[n]\\ b\in\{0,\ldots,m-1\}}}\left\{\|\p\|_{(k)}-\|\p\|_{(\left\lfloor k/m\right\rfloor)}-\frac bmp_{\left\lfloor k/m\right\rfloor+1}\right\}\\
&=\max_{k\in[n]}\left\{\|\p\|_{(k)}-\|\p\|_{(\left\lfloor k/m\right\rfloor)}\right\}\;.
\ea
Now, from the specific form of $\chi_n$ in~\eqref{chin} we have $\|\p\|_{(k)}=H_k/H_n$ so that the above equality is equivalent to
\be
T_\star\left(\chi_n\xrightarrow{\text{\tiny LOCC}}\Phi_m\otimes\chi_n\right)= \max_{k\in[n]}\left\{\frac{H_k-H_{\left\lfloor k/m\right\rfloor}}{H_n}\right\}\;.
\ee
Finally, we use the well known bounds on the harmonic number $H_n$ given by
\be
\ln(n)+\frac{1}{n}\leq H_n\leq\ln(n)+1\;.
\ee
Using these bounds we estimate 
\ba\label{2m2m}
H_k-H_{\left\lfloor k/m\right\rfloor}&\leq \ln(k)+1-\ln\left\lfloor \frac km\right\rfloor-\frac1{\left\lfloor \frac km\right\rfloor}\\
\GG{Exercise~\ref{ex:2m}}&\leq 1+\ln(2m)\;.
\ea
We therefore conclude that
\be
T_\star\left(\chi_n\xrightarrow{\text{\tiny LOCC}}\Phi_m\otimes\chi_n\right)\leq \frac{1+\ln(2m)}{H_n}\xrightarrow{n\to\infty} 0\;,
\ee
since $H_n$ goes to infinity as $n$ goes to infinity. 

\bex\label{ex:2m}
Prove the second inequality of~\eqref{2m2m}.
\eex

\bex
Fix $\alpha\in\mbb{R}$, and consider the bipartite entangled state 
\be
|\varphi_n\ra\eqdef\frac1{\sqrt{N_n}}\sum_{x\in[n]}\sqrt{x^\alpha}|xx\ra\;,
\ee
where $N_n=\sum_{x\in[n]}x^\alpha$ is the normalization factor. Show that only for $\alpha=-1$ the state $|\varphi_n\ra$ can be used to embezzle entanglement.
\eex

In the general case of arbitrary family of pure bipartite states , $\{\chi_n\}_{n\in\mbb{N}}$, observe that for any integer $\ell\leq a\eqdef\left\lfloor n/m\right\rfloor+1$
\ba
\max_{k\in[m(\ell-1),\ldots,m\ell-1]}\left\{\|\p\|_{(k)}-\|\p\|_{(\left\lfloor k/m\right\rfloor)}\right\}&=\max_{k\in[m(\ell-1),\ldots,m\ell-1]}\left\{\|\p\|_{(k)}-\|\p\|_{(\ell-1)}\right\}\\
&=
\|\p\|_{(m\ell-1)}-\|\p\|_{(\ell-1)}
\ea
where we used the convention that $\|\p\|_{(k)}\eqdef 1$ for an integer $k>n$. With this convention,
we conclude that
\be\label{maxiell}
T_\star\left(\chi_n\xrightarrow{\text{\tiny LOCC}}\Phi_m\otimes\chi_n\right)
=\max_{\ell\in[a]}\left\{\|\p\|_{(m\ell-1)}-\|\p\|_{(\ell-1)}\right\}\;.
\ee
Therefore, if $\{\chi_n\}_{n\in\mbb{N}}$ is an embezzling family then in particular it satisfies
\be
\lim_{n\to\infty}\sum_{x=\ell}^{m\ell-1}p_x^{(n)}=0\quad\quad\forall\;\ell\in\mbb{N}\;.
\ee
The above equation holds if and only if
\be
\lim_{n\to\infty}p^{(n)}_x=0\quad\quad\forall\;x\in\mbb{N}\;.
\ee
However, observe that the condition above is in general insufficient to determine if the states $\{\chi_n\}_{n\in\mbb{N}}$ form an embezzling family, since the maximizer $\ell$ in~\eqref{maxiell} can depend on $n$.

\section{Asymptotic Entanglement Theory of Pure States}

In Sec.~\ref{sec:asymptoticregime} we defined the asymptotic resource cost and the asymptotic distillation of a resource. Applying these general definitions to the case of pure bipartite entanglement results in the following definitions of the cost and distillable rates of converting $\psi\in\pure(AB)$ to $\phi\in\pure(AB)$:
\ba
\cost\left(\psi^{AB}\to\phi^{AB}\right)&\eqdef\lim_{\eps\to 0^+}\inf_{n,m\in\mbb{N}}\left\{\frac nm\;:\;T_\star\left(\psi^{\otimes n}\xrightarrow{\text{\tiny LOCC}} \phi^{\otimes m}\right)\leq\eps\right\}\\
\distill\left(\psi^{AB}\to\phi^{AB}\right)&\eqdef\lim_{\eps\to 0^+}\sup_{n,m\in\mbb{N}}\left\{\frac mn\;:\;T_\star\left(\psi^{\otimes n}\xrightarrow{\text{\tiny LOCC}} \phi^{\otimes m}\right)\leq\eps\right\}\;.
\ea
These definitions can then be used to define the entanglement cost\index{entanglement cost} and distillable entanglement\index{distillable entanglement} as:
\be
\cost\left(\psi^{AB}\right)=\cost\left(\Phi_2\to\psi^{AB}\right)\quad\text{and}\quad \distill\left(\psi^{AB}\right)=\distill\left(\psi^{AB}\to\Phi_2\right)\;,
\ee 
where $|\Phi_2\ra\eqdef\frac1{\sqrt{2}}(|00\ra+|11\ra)$ is the $2\times 2$ dimensional maximally entangled state (i.e. a Bell state). In the following subsections we provide closed formulas for these measures of entanglement and discuss their relations to the single-shot quantities $\cost^\eps$ and $\distill^\eps$ that we studied in the previous section.

\subsection{Entanglement Cost}\index{entanglement cost}

From Lemma~\ref{lem:1152}, particularly Exercise~\ref{ex:a11211}, it follows that the entanglement cost as defined above can be expressed for any $\psi\in\pure(AB)$ as
\be\label{125900}
\cost\left(\psi^{AB}\right)=\lim_{\eps\to 0^+}\liminf_{n\to\infty}\frac1n\cost^\eps\left(\psi^{\otimes n}\right)\;.
\ee
In the following theorem we compute the entanglement cost and prove a stronger version of the above relation.

\begin{myt}{}
\begin{theorem}\label{themasycost}
Let $\psi\in\pure(AB)$. Then, for any $\eps\in(0,1)$ 
\be
\cost\left(\psi^{AB}\right)=\lim_{n\to\infty}\frac1n\cost^\eps\left(\psi^{\otimes n}\right)=E\left(\psi^{AB}\right)\;,
\ee
where $E$ is the entropy of entanglement\index{entropy of entanglement} defined in~\eqref{eoe}.
\end{theorem}
\end{myt}

\begin{remark}
Observe that from the theorem above it follows that there is no need to take the limit $\eps\to 0^+$ in~\eqref{125900} ;that is, by taking the limit $n\to\infty$ the dependance on $\eps$ is eliminated (as long as $\eps\in(0,1)$). 
\end{remark}

\begin{proof}
The proof follows directly from a combination of Corollary~\ref{corhmaxe} and the variant of the AEP property given in~\eqref{AEPmaxversion}. Specifically, denoting by $\rho^A\eqdef\tr_B\left[\psi^{AB}\right]$, we get from Corollary~\ref{corhmaxe} that
\ba
\lim_{n\to\infty}\frac1n\cost^\eps\left(\psi^{\otimes n}\right)&=\lim_{n\to\infty}\frac1nH^\eps_{\max}\left(A^n\right)_{\rho^{\otimes n}}\\
\GG{\eqref{AEPmaxversion}}&=H(A)_\rho\;.
\ea
This completes the proof.
\end{proof}

\bex
Provide a more direct proof of the theorem above using the concept of typicality. Hint: The proof is a bit involved and can be found in Appendix~\ref{altprooftyp}.
\eex

\subsection{Distillable Entanglement}\index{distillable entanglement}

From Lemma~\ref{lem:1152}, particularly Exercise~\ref{ex:a11211}, it follows that the distillable entanglement as defined above can be expressed for any $\psi\in\pure(AB)$ as
\be\label{1259}
\distill\left(\psi^{AB}\right)=\lim_{\eps\to 0^+}\limsup_{n\to\infty}\frac1n\distill^\eps\left(\psi^{\otimes n}\right)\;.
\ee
In the following theorem we compute the distillable entanglement and prove a stronger version of the above relation.

\begin{myt}{}
\begin{theorem}\label{thmdistab}
Let $\psi\in\pure(AB)$. Then, for any $\eps\in(0,1)$ 
\be
\distill\left(\psi^{AB}\right)=\lim_{n\to\infty}\frac1n\distill^\eps\left(\psi^{\otimes n}\right)=E\left(\psi^{AB}\right)\;,
\ee
where $E$ is the entropy of entanglement\index{entropy of entanglement} defined in~\eqref{eoe}.
\end{theorem}
\end{myt}

\begin{proof}
The proof follows directly from a combination of Theorem~\ref{thmdias} and the variant of the AEP property given in~\eqref{seeaep}. Specifically, denoting by $\rho^A\eqdef\tr_B\left[\psi^{AB}\right]$, we get from Corollary~\ref{corhmaxe} that
\be
\lim_{n\to\infty}\frac1n\cost^\eps\left(\psi^{\otimes n}\right)=\lim_{n\to\infty}\frac1nH^\eps_{\min}\left(A^n\right)_{\rho^{\otimes n}}\;.
\ee
Moreover, taking $\sigma^A=\u^A$ in~\eqref{seeaep} yields
\be
\lim_{n\to\infty}\frac1nH^\eps_{\min}\left(A^n\right)_{\rho^{\otimes n}}=H(A)_\rho\;.
\ee
This completes the proof.
\end{proof}

For those readers seeking additional insights, an alternative proof utilizing the concept of typicality is provided in Appendix~\ref{altprooftyp}. This proof offers a different perspective and leverages the principles of typicality, which may be of interest to readers who are keen on exploring diverse approaches and methodologies within the field.

\subsection{Reversibility of Pure Bipartite Entanglement Theory}\index{reversibility}

We saw in the last two theorems that both the entanglement cost and the distillable entanglement of  a pure bipartite state are equal to the entropy of entanglement. This equality between the entanglement cost and the distillable entanglement implies that the QRT of pure bipartite entanglement is reversible. Note that we get reversibility under LOCC which is a strict subset of RNG (i.e. ``non-entangling") operations. In the next chapter we will discuss in more details the relationship between non-entangling operations and LOCC.

\bex
Use the theorems above to show that for any $\psi,\phi\in\pure(AB)$
\be
\distill\left(\psi^{AB}\to\phi^{AB}\right)=\frac{E(\psi)}{E(\phi)}\quad\text{and}\quad\cost\left(\psi^{AB}\to\phi^{AB}\right)=\frac{E(\phi)}{E(\psi)}\;.
\ee
\eex

\section{Notes and References}

Comprehensive reviews on entanglement theory can be found in~\cite{HHHH2009} and~\cite{PV2007}. Additional information on LOCC operations can be found in~\cite{Chitambar2014,CDGG2020} and references therein. Theorem~\ref{lopop} is a slight modified version of the one given by~\cite{LP2001}. The relation between entanglement and majorization, specifically Theorem~\ref{nielsen}, was first established by~\cite{Nielsen1999}. The entanglement monotones defined in~\eqref{vidal} were first introduced by~\cite{Vidal2000}. Entanglement catalysis was introduced by~\cite{JP1999b}. The concurrence measure of entanglement was first introduced by~\cite{Wootters1998} for the purpose of computing the entanglement of formation of a mixed bipartite state. Later on it was generalized by~\cite{Gour2005} to the family of entanglement monotones given in~\eqref{1223}. Theorem~\ref{thm:dp} is due to~\cite{JP1999}, while the formula for the maximum probability to convert one state to another by LOCC, i.e.\ Corollary~\ref{cor:vid}, was first proved independently by~\cite{Vidal1999} without the use of Theorem~\ref{thm:dp} as we do here. The second corollary of Theorem~\ref{thm:dp}, i.e.\ Corollary~\ref{cor:1242}, as well as the closed formula for $T_\star$, where first introduced in~\cite{ZTG2023}.
Embezzlement of entanglement was first introduced by~\cite{vanDam2003}.

%%%%%%%%%%%%%%%%%%%%%%%%%%%%%%%%%%%%%%%%%%%%%%%%%%%%%%%%%%%%%%%%%%%%%%%%%%%%%%%%%%%

\chapter{Mixed-State Entanglement}\label{entanglement2}\index{quantum entanglement}

To gain a better understanding of bipartite entanglement theory, we will delve into its most general form in this chapter. The free states of the theory consists of separable states, denoted for any composite system\index{composite system} $AB$ by
\be
\sep(AB)\eqdef\left\{\sum_{x\in[n]}p_x\sigma_x^A\otimes\omega_x^B\;:\;\sigma_x\in\md(A),\;\omega_x\in\md(B),\;\p\in\prob(n),\;n\in\mbb{N}\right\}
\ee
where we used the notation $\p\eqdef(p_1,\ldots,p_n)^T$. Observe that $\sep(AB)$ is a closed convex set. Any quantum state $\rho\in\md(AB)$ that does not belong to $\sep(AB)$ is referred to as an entangled state. This chapter will reveal the intricate structure of entangled states, highlighting the complexity of mixed-state entanglement theory.

\section{Detection of Entanglement}

Unlike pure state entanglement, detecting mixed state entanglement is a challenging task even from a theoretical standpoint. In particular, density matrices in $\md(AB)$ are usually expressed as positive semidefinite matrices with a size of $|AB|\times |AB|$ and a trace of one. Establishing whether these matrices belong to $\sep(AB)$ is a complex undertaking, and in most cases, it falls under the category of NP-hard problems.

\subsection{Entanglement Witnesses}\index{entanglement witness}

One technique for detecting entanglement involves the concept of resource witnesses\index{resource witness}
, which is discussed in Section~\ref{sec:rw}. Spedifically, Definition~\ref{def:wit} of a resource witness can be adapted to apply to entanglement theory.

\begin{myd}{Entanglement Witness}
\begin{definition}\label{def:ewit}
An operator $\Gamma\in\herm(AB)$ is called an entanglement witness if the following two conditions holds:
\begin{enumerate}
\item For any $\sigma\in\sep(AB)$
\be\label{defewit}
\tr\left[\Gamma^{AB}\sigma^{AB}\right]\geq 0\;.
\ee
\item There exists $\rho\in\md(AB)$ such that
\be\label{negae}
\tr\left[\Gamma^{AB}\rho^{AB}\right]<0\;.
\ee
\end{enumerate}
\end{definition}
\end{myd}

From the condition in~\eqref{defewit}, and the fact that $\sep(AB)$ is the convex hull of product states  it follows that if $\Gamma\in\herm(AB)$ is an entanglement witness then for any product state $\psi\otimes\phi\in\pure(AB)$ we must have
\be\label{ew13}
\left\la\psi^A\otimes\phi^B\left|\Gamma^{AB}\right|\psi^A\otimes\phi^B\right\ra\geq 0\;.
\ee
On the other hand, the condition~\ref{negae} also implies that the exists a state $\chi\in\pure(AB)$ such that
\be\label{chigr}
\left\la\chi^{AB}\left|\Gamma^{AB}\right|\chi^{AB}\right\ra< 0\;.
\ee
In other words, the condition~\eqref{negae} implies that $\Gamma^{AB}$ is not positive semidefinite so that we can take $\chi^{AB}$, for example, to be an eigenstate corresponding to a negative eigenvalue of $\Gamma^{AB}$. 

Based on Theorem~\ref{thm:941}, we can conclude that entanglement witnesses are an effective tool for detecting entanglement. Specifically, $\rho\in\md(AB)$ is an entangled state if and only if there exists an entanglement witness $\Gamma\in\herm(AB)$ such that
\be
\tr\left[\Gamma^{AB}\rho^{AB}\right]<0;.
\ee
This characteristic can be employed to demonstrate that the set of separable states occupies a non-zero volume.

\begin{myt}{}
\begin{theorem}
The set $\sep(AB)$ has a non-zero volume in $\md(AB)$. Specifically, there exists $\eps>0$ such that $\mb_{\eps}(\u^{AB})\subset\sep(AB)$, where
$\mb_{\eps}(\u^{AB})$ is the ``ball" of all states in $\md(AB)$ that are $\eps$-close to the maximally mixed state $\u^{AB}=\u^A\otimes\u^B$.
\end{theorem}
\end{myt}

\begin{proof}
Suppose by contradiction that the statement in the theorem is false. Then, there exists a sequence of bipartite entangled states $\{\tau_n^{AB}\}_{n\in\mbb{N}}$ such that
\be
\lim_{n\to\infty}\frac12\left\|\tau_n^{AB}-\u^{AB}\right\|_1=0\;.
\ee
Since we assume that $\tau_n^{AB}$ is entangled, we have $\tau_n\not\in\sep(AB)$. Therefore, there exists an entanglement witness $\Gamma_n^{AB}$ such that
\be
\tr\left[\tau_n^{AB}\Gamma_n^{AB}\right]<0\;.
\ee
Without loss of generality we can assume that for each $n\in\mbb{N}$ the witness $\Gamma_n^{AB}$ is normalized with respect to the Hilbert-Schmidt inner product; i.e.\
\be
\tr\left[\left(\Gamma_n^{AB}\right)^2\right]=1\;.
\ee\index{Hilbert-Schmidt inner product}
Therefore, the sequence $\{\Gamma_n^{AB}\}$ is a sequence of Hermitian operators in the unit sphere of $\herm(AB)$. Since the unit sphere is compact, there exists a subsequence $\{n_k\}_{k\in\mbb{N}}$ of integers such that the limit $\lim_{k\to\infty}\Gamma_{n_k}^{AB}$ exists and equal to some normalized operator $\Gamma^{AB}_\star\in\herm(AB)$.
Since each $\Gamma_{n_k}^{AB}$ is an entanglement witness, the limit $\Gamma^{AB}_\star$ must satisfy
\be\label{witsep2}
\tr\left[\Gamma_{\star}^{AB}\sigma^{AB}\right]\geq 0\quad\quad\forall\;\sigma\in\sep(AB)\;.
\ee 
On the other hand, taking the limit $k\to\infty$ on both sides of the inequality $ \tr\left[\tau_{n_k}^{AB}\Gamma_{n_k}^{AB}\right]<0$ gives
\be
\tr\left[\Gamma_{\star}^{AB}\u^{AB}\right]\leq 0\;,
\ee
so that $\tr\left[\Gamma^{AB}_\star\right]\leq 0$. Now, let $\{|\psi_x\ra^A\}_{x\in[m]}$ be an orthonormal basis of $A$, and $\{|\phi_y\ra^B\}_{y\in[\ell]}$ be an orthonormal basis of $B$. Then,
\be
0\geq\tr\left[\Gamma^{AB}_\star\right]=\sum_{x\in[m]}\sum_{y\in[\ell]}\tr\left[\Gamma_\star^{AB}\left(\psi^A_x\otimes\phi^B_y\right)\right]\;.
\ee
From~\eqref{witsep2} it follows that for each $x\in[m]$ and $y\in[\ell]$ we have $\tr\left[\Gamma_\star^{AB}\left(\psi^A_x\otimes\phi^B_y\right)\right]\geq 0$. We therefore get that for any $x\in[m]$ and $y\in[\ell]$, $\tr\left[\Gamma_\star^{AB}\left(\psi^A_x\otimes\phi^B_y\right)\right]=0$. Finally, since the orthonormal bases $\{|\psi_x\ra^A\}_{x\in[m]}$ and $\{|\phi_y\ra^B\}_{y\in[\ell]}$ where arbitrary, we conclude that
\be
\tr\left[\Gamma_\star^{AB}\left(\psi^A\otimes\phi^B\right)\right]=0\quad\quad\forall\;\psi\in\pure(A)\;\;\;\;\forall\;\phi\in\pure(B)\;.
\ee
However, from Exercise~\ref{ex:povmicb} it follows that the above equation holds if and only if $\Gamma_\star^{AB}=0$ in contradiction with the fact that $\Gamma_\star^{AB}$ is normalized, so in particular, cannot be the zero matrix.
This completes the proof.
\end{proof}

\bex
Let $A_1,\ldots,A_m$ be $m$ physical systems and let $\sep(A_1\cdots A_m)$ be the set of multipartite separable states; i.e. $\sep(A_1\cdots A_m)$ is the convex hull of the set of all $m$-fold product states of the form $\rho_1\otimes\cdots\otimes\rho_m$, with $\rho_x\in\md(A_x)$ for all $x\in[m]$. Show that $\sep(A_1\cdots A_m)$ has a non-zero volume in $\md(A_1\cdots A_m)$.
\eex

The following theorem shows a close connection between entanglement witnesses and positive maps.

\begin{myt}{}
\begin{theorem}\label{thm:1311}
Any entanglement witness is the Choi matrix of a positive map that is not completely positive. Explicitly, $\Gamma\in\herm(AB)$ is an entanglement witness if and only if $\Gamma^{AB}=J_\mE^{AB}$ for some positive map $\mE\in\pos(A\to B)$ that is not completely positive\index{completely positive} (i.e. $\mE\not\in\cp(A\to B)$).
\end{theorem}
\end{myt}

\begin{proof}
Suppose first that $\Gamma^{AB}=J_\mE^{AB}$ for some positive map $\mE\in\pos(A\to B)$ and suppose $\mE\not\in\cp(A\to B)$. Then, for any product state $\rho\otimes\sigma\in\pure(AB)$ we have
\ba
\tr\left[\Gamma^{AB}\left(\rho^A\otimes\sigma^B\right)\right]&=\tr\left[J^{AB}_\mE\left(\rho^A\otimes\sigma^B\right)\right]\\
\Gg{\tr_A\left[J^{AB}_\mE\left(\rho^A\otimes I^B\right)\right]=\mE^{A\to B}\left((\rho^A)^T\right)}&=\tr\left[\sigma^B\mE^{A\to B}\left((\rho^A)^T\right)\right]\\
&\geq 0
\ea
where the last inequality follows from the fact that $\mE\in\pos(A\to B)$  so that $\mE\left(\rho^T\right)\geq 0$. Finally, the existence of a state $\chi\in\pure(AB)$ that satisfies~\eqref{chigr} follows from the fact that $\mE$ is not completely positive\index{completely positive} so its Choi matrix $\Gamma^{AB}$ is not positive semidefinite.

Conversely, suppose $\Gamma^{AB}$ is an entanglement witness and let $\mE\in\ml(A\to B)$ be such that $\Gamma^{AB}=J_\mE^{AB}$ (but we do not assume that $\mE$ is positive). Then, for any $\rho\in\md(A)$ and $\sigma\in\md(B)$ we have
\ba
\tr\left[\sigma^B\mE^{A\to B}\left(\rho^A\right)\right]&=\tr\left[J^{AB}_\mE\left(\left(\rho^A\right)^T\otimes \sigma^B\right)\right]\\
&=\tr\left[\Gamma^{AB}\left(\left(\rho^A\right)^T\otimes \sigma^B\right)\right]\\
&\geq 0
\ea
where the last inequality follows from the fact that $\Gamma^{AB}$ is an entanglement witness and $\left(\rho^A\right)^T\otimes \sigma^B\in\sep(AB)$. Since $\rho^A$ and $\sigma^B$ where arbitrary states, the above inequality implies that $\mE\in\pos(A\to B)$. The map $\mE$ is not completely positive\index{completely positive} since its Choi matrix, $\Gamma^{AB}$, is not positive semidefinite (as $\Gamma^{AB}$ is an entanglement witness).
This completes the proof.
\end{proof}

Observe that the set of all entanglement witnesses consists of all the non-positive semidefinite matrices that are in the dual cone of the set of separable states. Specifically, for the composite system\index{composite system} $AB$, the set of all entanglement witnesses, denoted by $\wit(AB)$, is given by
\be
\wit(AB)=\left\{\Gamma\in\sep(AB)^*\;:\;\Gamma^{AB}\not\geq 0\right\}\;.
\ee
We will now provide two examples of how Theorem~\ref{thm:941}, adapted to entanglement theory with the set $\wit(AB)$ mentioned above, can be used to determine whether a quantum state is entangled.

\subsubsection{Example 1: The Isotropic State}\index{isotropic state}

The isotropic state in $\md(AB)$, with $m\eqdef|A|=|B|$, is defined as
\be\label{iso101}
\rho^{AB}_t=t\Phi_m^{AB}+(1-t)\tau^{AB}
\ee
where $t\in(0,1)$, and 
\be
\tau^{AB}\eqdef\frac{I^{AB}-\Phi_m}{m^2-1}\;.
\ee
Observe that $\Phi_m\tau=\tau\Phi_m=0$, and furthermore, since $\Phi_m^{AB}$ is invariant under the action of the twirling channel $\mG$ defined in~\eqref{gtwi2} also $\rho_t^{AB}$ has this property. In fact, the state $\left\{\rho_t^{AB}\right\}_{t\in[0,1]}$ can be viewed as the set of \emph{all} quantum states that are invariant under $\mG$ (see~\eqref{newform2}). In the following, we will utilize this property to make the argument that the isotropic state\index{isotropic state} $\rho_t^{AB}$ satisfies: 
\be\label{isot}
\rho_t^{AB}\in\sep(AB)\quad\iff \quad t\leq\frac1{m}\;.
\ee

To prove the above statement we follow Theorem~\ref{thm:941}. Specifically, $\rho_t^{AB}$ is separable if and only if $\tr[\Gamma^{AB}\rho^{AB}_t]\geq 0$ for all entanglement witnesses $\Gamma\in\wit(AB)$. The key idea is to use the invariance of $\rho_t^{AB}$ under $\mG$ to get that 
\ba\label{gamoo}
\tr\left[\Gamma^{AB}\rho^{AB}_t\right]&=\tr\left[\Gamma^{AB}\mG\left(\rho^{AB}_t\right)\right]\\
\GG{\mG\text{ is self-adjoint}}&=\tr\left[\mG\left(\Gamma^{AB}\right)\rho^{AB}_t\right]\;.
\ea
Now, observe that if $\sigma\in\sep(AB)$ then also $\mG(\sigma)\in\sep(AB)$ so that 
\be
\tr\left[\mG\left(\Gamma^{AB}\right)\sigma^{AB}\right]=\tr\left[\Gamma^{AB}\mG\left(\sigma^{AB}\right)\right]\geq 0\;.
\ee
Combining this with~\eqref{gamoo} we conclude that $\rho_t^{AB}$ is separable if and only if 
$\tr[\Gamma^{AB}\rho^{AB}_t]\geq 0$ for all entanglement witnesses of the form 
\be
\Gamma^{AB}=\mG\left(\Gamma^{AB}\right)=aI^{AB}+b\Phi_m^{AB}\;,
\ee
where $a,b\in\mbb{R}$. In the final equality, we made use of the fact that $I^{AB}$ and $\Phi_m^{AB}$ spans the subspace of $\mG$-invariant operators in $\herm(AB)$. 

To ensure that the matrix $\Gamma=aI+b\Phi_m$ is an entanglement witness, we need to appropriately specify the coefficients $a$ and $b$. Let's start by noting that $a$ must be non-negative, as evidenced by
\be
a=\tr\left[\Gamma\big(|1\lr 1|\otimes|2\lr 2|\big)\right]\geq 0\;.
\ee
Furthermore, it's important to recognize that $\Gamma$ exhibits two distinct eigenvalues: $a$ with multiplicity $|AB|-1$, and $a+b$ with multiplicity one. Given that an entanglement witness has at least one negative eigenvalue, and considering that $a\geq 0$, it is necessary for $b$ to satisfy $b<-a$. It's also worth mentioning that the scenario where $a=0$ does not yield an entanglement witness (this is an interesting point to ponder – why this is the case?).

Consequently, after rescaling $\Gamma^{AB}$ by a positive factor $a>0$, we can, without loss of generality, assume that $\Gamma^{AB}$ takes the form
\be
W=I^{AB}-r\Phi_m^{AB}\;,
\ee
where $r>1$. From~\eqref{ew13} the matrix $\Gamma^{AB}$ is an entanglement witness if and only if for any product state $\psi\otimes\phi\in\pure(AB)$ 
\be
0\leq\tr\left[\Gamma^{AB}\left(\psi^A\otimes\phi^B\right)\right]=1-r\tr\left[\Phi_m^{AB}\left(\psi^A\otimes\phi^B\right)\right]\;.
\ee
In Exercise~\ref{prove1m} you show that
\be\label{1mproved}
\max_{\substack{\psi\in\pure(A)\\\phi\in\pure(B)}}\tr\left[\Phi_m^{AB}\left(\psi^A\otimes\phi^B\right)\right]=\frac1m\;.
\ee
We therefore conclude that $\Gamma=I-r\Phi_m$ is an entanglement witness if and only if $1<r\leq m$. Hence, $\rho_t^{AB}$ is separable if and only if for all $1<r\leq m$ we have
\be
0\leq\tr\left[\rho^{AB}_t\left(I^{AB}-r\Phi_m^{AB}\right)\right]=1-rt\;.
\ee
The above inequality holds for all $1<r\leq m$ if and only if $t\leq 1/m$. This completes the proof of~\eqref{isot}.

\bex\label{prove1m}
Prove~\eqref{1mproved}.
\eex

\bex
Show that the isotropic state can be expressed as
\be
\rho_t^{AB}=\frac{m^2}{m^2-1}\left((1-t)\u^{AB}+\left(t-\frac1{m^2}\right)\Phi_m^{AB}\right)\;.
\ee
\eex

\subsubsection{Example 2: Werner States}\index{Werner state}

The Werner state is a density matrix in $\md(AB)$ that remains invariant under the action of the $\mG$-twirling map, where $\mG$ in this case is defined as the map given in~\eqref{gtwi}. The Werner state $\rho_{\text{\tiny W}}^{AB}$ is defined for all $p\in[0,1]$ by  (cf.~\eqref{prewerner})
\be\label{werner}
\rho_{\text{\tiny W}}^{AB}\eqdef
p\frac{2}{m(m+1)}\Pi_\sym^{AB}+(1-p)\frac{2}{m(m-1)}\Pi_\asy^{AB}\;.
\ee
Recall that $\Pi_\sym^{AB}=\frac12\left
(I^{AB}+F^{AB}\right)$ and $\Pi_\asy^{AB}=\frac12\left
(I^{AB}-F^{AB}\right)$, where $F^{AB}$ is the flip operator $F^{AB}$ defined in~\eqref{flipoperator}. Therefore, the Werner state\index{Werner state} above can also be expressed more compactly as (see Exercise~\ref{wernercompact} below)
\be\label{wcpact}
\rho_{\text{\tiny W}}^{AB}=\frac1{m(m-\alpha)}\left(I^{AB}-\alpha F^{AB}\right)\;,
\ee
where the new parameter $\alpha\in[-1,1]$ is related to $p$ via
\be
\alpha\eqdef\frac{1+m(1-2p)}{1-2p+m}\;.
\ee
\bex\label{wernercompact}
Prove~\eqref{wcpact} by substituting the expressions $\Pi_\sym^{AB}=\frac12\left(I^{AB}+F^{AB}\right)$ and $\Pi_\asy^{AB}=\frac12\left
(I^{AB}-F^{AB}\right)$ into~\eqref{werner}. 
\eex

Similar to the analysis of the isotropic state, we can determine for which values of $p$ (or $\alpha$) the Werner state\index{Werner state} is entangled. We find that $\rho_{\text{\tiny W}}^{AB}\in\sep(AB)$  if and only if $p\leq\frac12$ or equivalently if and only if $\alpha\leq\frac1m$. We leave the proof as an exercise.
\bex
Prove that $\rho_{\text{\tiny W}}^{AB}\in\sep(AB)$ if and only if $\alpha\leq\frac1m$. Hint: Show first that the Werner state $\rho_{\text{\tiny W}}^{AB}$ is entangled if and only if
$
\tr\left[\Gamma^{AB}\rho_{\text{\tiny W}}^{AB}\right]\geq 0
$
for all $\Gamma\in\wit(AB)$ of the form $\Gamma^{AB}=aI^{AB}+bF^{AB}$. Then find the values of $a$ and $b$ for which $aI^{AB}+bF^{AB}\in\wit(AB)$ and continue from there. 
\eex

\subsubsection{Entanglement Witnesses in Small Dimensions}

For small dimensions, $\wit(AB)$ has the following simple characterization.

\begin{myt}{}
\begin{theorem}\label{thm:atbf}
Let $\Gamma\in\wit(AB)$ with $|AB|\leq 6$. Then, there exists $\eta_1,\eta_2\in\pos(AB)$ such that
\be\label{formatb}
\Gamma^{AB}=\eta^{AB}_1+\mT^{B\to B}\left(\eta^{AB}_2\right)\;,
\ee
where $\mT\in\pos(B\to B)$ is the transpose map.
\end{theorem}
\end{myt}

\begin{proof}
Without loss of generality suppose $|A|=2$ and $|B|\leq 3$. From Theorem~\ref{thm:1311} there exists $\mE\in\pos(A\to B)$ such that
\be
\Gamma^{AB}=\mE^{\tA\to B}\left(\Omega^{A\tA}\right)\;.
\ee
Furthermore, from Theorem~\ref{thm:sw} it follows that $\mE=\mN_1+\mT\circ\mN_2$ for some $\mN_1,\mN_2\in\cp(A\to B)$. Substituting this into the equation above, and denoting by $\eta_j\eqdef\mN_j^{\tA\to B}\left(\Omega^{A\tA}\right)$ for $j=1,2$, we get that $\Gamma^{AB}$ has the form~\eqref{formatb}. Finally, observe that $\eta_1,\eta_2\geq 0$ since $\mN_1$ and $\mN_2$ are completely positive\index{completely positive} maps. This concludes the proof. 
\end{proof}

\subsection{The PPT Criterion}\index{PPT criterion}

In this subsection we consider a simple criterion to detect entanglement. The criterion is known as the Peres-Horodecki criterion or the PPT criterion since it is based on the partial transpose\index{partial transpose}. The criterion states that if $\rho^{AB}=\sum_{x\in[m]}p_x\rho^A_x\otimes\rho^B_x$ is a separable density matrix then its partial transpose 
\be\label{posnpt}
\mT^{B\to B}\left(\rho^{AB}\right)=\sum_{x\in[m]}p_x\rho^A_x\otimes\left(\rho^B_x\right)^T\geq 0\;,
\ee
is a positive semidefinite matrix. We will say that $\rho^{AB}$ has positive partial transpose (PPT) if this property hold, and otherwise, we will say that it has a negative partial transpose\index{partial transpose} (NPT) or simply that the state is an NPT state\footnote{Note that the partial transpose of NPT states can have positive eigenvalues. We use the term NPT only to indicate that the partial transpose of the state has at least one negative eigenvalue.}.

We have seen before that the 2-qubit maximally entangled state is an NPT state, and in Exercise~\ref{petnpt} you showed that all pure entangled states are NPT. Therefore, it is natural to ask if \emph{all} entangled states are NPT. In low dimensions, the following theorem states that this is indeed the case.

\begin{myt}{}
\begin{theorem}\label{thm:2x3}
Let $\rho\in\md(AB)$ be a bipartite density matrix with dimensions of the underlying Hilbert spaces satisfy $|AB|\leq 6$. Then, $\rho^{AB}$ is entangled if and only if it is an NPT state.
\end{theorem} 
\end{myt}

\begin{proof}
If $\rho^{AB}$ is an NPT state then from~\eqref{posnpt} it cannot be separable. Conversely, suppose $\rho^{AB}$ is a PPT state, and recall from Theorem~\ref{thm:941}  (when applied to entanglement theory) that $\rho^{AB}$ is separable if and only if
\be
\tr\left[\rho^{AB}\Gamma^{AB}\right]\geq 0\quad\quad\forall\;\Gamma\in\wit(AB)\;.
\ee
Now, fix $\Gamma\in\wit(AB)$. From Theorem~\ref{thm:atbf} $\Gamma^{AB}$ have the form~\eqref{formatb} for some $\eta_1,\eta_2\in\pos(AB)$. Hence, 
\ba
\tr\left[\rho^{AB}\Gamma^{AB}\right]&=\tr\left[\rho^{AB}\left(\eta^{AB}_1+\mT^{B\to B}\left(\eta^{AB}_2\right)\right)\right]\\
&\geq\tr\left[\rho^{AB}\mT^{B\to B}\left(\eta^{AB}_2\right)\right]\\
\GG{\mT=\mT^*}&=\tr\left[\mT^{B\to B}\left(\rho^{AB}\right)\eta^{AB}_2\right]\\
\GG{\rho^{\it AB}\;is\;PPT}&\geq 0\;.
\ea
Since $\Gamma^{AB}$ was an arbitrary entanglement witness in $\wit(AB)$ we conclude that the above equation holds for all $\Gamma\in\wit(AB)$ so that $\rho^{AB}$ must be a separable state. This concludes the proof.
\end{proof}

The condition $|AB|\leq 6$ in the theorem above is optimal. Indeed, there are examples of PPT entangled states in higher dimensions, including the case $|A|=2$ and $|B|=4$, as well as the case $|A|=|B|=3$.
\subsubsection{Example of a PPT Entangled State}\index{PPT entangled state}
Consider the five product states in $A\otimes B\eqdef\mbb{C}^3\otimes\mbb{C}^3$ given by
\ba\label{5states}
& |\psi_1\ra\eqdef\frac1{\sqrt{2}}|0\ra\big(|0\ra-|1\ra\big)\quad,\quad |\psi_2\ra\eqdef\frac1{\sqrt{2}}\big(|0\ra-|1\ra\big)|2\ra\\
& |\psi_3\ra\eqdef\frac1{\sqrt{2}}|2\ra\big(|1\ra-|2\ra\big)\quad,\quad|\psi_4\ra\eqdef\frac1{\sqrt{2}}\big(|1\ra-|2\ra\big)|0\ra\\
&|\psi_5\ra\eqdef\frac13\big(|0\ra+|1\ra+|2\ra\big)\big(|0\ra+|1\ra+|2\ra\big)\;.
\ea
Observe that the five states above are orthonormal and they are invariant under the partial transpose (i.e. $\mT^{B\to B}(\psi_x^{AB})=\psi_x^{AB}$ for all $x\in[5]$). They also have the property that any pure state in $\mbb{C}^3\otimes\mbb{C}^3$ that is orthogonal to all the five states above must be entangled (see Exercise~\ref{ex:5states}). The set $\{|\psi_x\ra\}_{x\in[5]}$ is consequently called an \emph{unextendible product basis} (UPB) of the subspace $\mH\eqdef\spa\{|\psi_x\ra\}_{x\in[5]}$. It therefore follows that the orthogonal complement of $\mH$ in $\mbb{C}^3\otimes\mbb{C}^3$, denoted by $\mH^\perp$, contains only entangled states. Let $\Pi$ be the projection to $\mH^\perp$; that is,
\be
\Pi^{AB}=I_9-\sum_{x\in[5]}\psi_x^{AB}\;,
\ee
where $I_9$ is the $9\times 9$ identity matrix. It then follows that the bipartite density matrix $\rho\eqdef\frac13\Pi$ is entangled (see Exercise~\ref{ex:5states}). However, the state $\rho$ is also PPT since
\ba
\mT^{B\to B}(\rho^{AB})=\frac13\Big(I^{AB}-\sum_{x\in[5]}\mT^{B\to B}\left(\psi_x^{AB}\right)\Big)=\frac13\Big(I^{AB}-\sum_{x\in[5]}\psi_x^{AB}\Big)=\rho^{AB}\geq0\;.
\ea
Therefore, the two-qutrit state $\rho^{AB}$ is a PPT entangled state.

\bex\label{ex:5states}
Consider the five states defined in~\eqref{5states}.
\ben
\item Show that any non-zero state in $\mbb{C}^3\otimes\mbb{C}^3$ that is orthogonal to all the five states in~\eqref{5states}  must be entangled.
\item Show that the state $\frac13\left(I_9-\sum_{x\in[5]}\psi_x\right)$ is entangled. Hint: Use the fact that $\Pi$ projects into a subspace consisting of only entangled states.
\een
\eex

\bex$\;$
\ben
\item Show that the number elements of every UPB in $\mbb{C}^m\otimes\mbb{C}^n$ cannot be less than $m+n-1$.
\item Let $\mW\subset\mbb{C}^m\otimes\mbb{C}^n$ be a subspace containing no product states (i.e. containing no normalized product vectors).  Show that 
\be
\dim\big(\mW\big)\leq (m-1)(n-1)\;.
\ee
\een
\eex

\bex
Let $\mk$ be the set of PPT operators in $\pos(AB)$. Show that $\mk^*$ consists of all operators $\Gamma\in\herm(AB)$ of the form~\eqref{formatb}.
\eex

\subsection{The Reduction Criterion}\label{sec:reduction}\index{reduction criterion}

In Sec.~\ref{inevitable} we saw that separable states has non-negative conditional entropy. Moreover, in Corollary~\ref{rccor} we saw that if a quantum state $\rho^{AB}$ satisfies $\H(A|B)_\rho\geq 0$ for every measure $\H$ of conditional entropy then $\rho^{AB}$ must satisfy
\be
I^A\otimes\rho^B\geq\rho^{AB}\;.
\ee
In particular, if $\rho^{AB}$ is separable then it must satisfy the condition above.
This criterion for separability is known as \emph{the reduction criterion}. 

The reduction criterion can be expressed in terms of the positive map $\mP\in\pos(A\to A)$
\be\label{pomp}
\mP(\omega^{A})\eqdef \tr\left[\omega^A\right]I^A-\omega^A\quad\quad\forall\;\omega\in\ml(A)\;.
\ee
Recall from Exercise~\ref{exkpos} that the map described above is positive but not 2-positive, and therefore not completely positive. Utilizing this map, the reduction criterion can be expressed as:
\be
\mP^{A\to A}\left(\rho^{AB}\right)\geq 0\;.
\ee
Exercise~\ref{reduction} below provides an alternative expression for the positive map $\mP^{A\to A}$ when $|A|=2$; specifically, for this case
\be\label{2relr}
\mP^{A\to A}\left(\omega^A\right)=\left(\sigma_y\omega^A\sigma_y\right)^T\quad\quad\forall\;\omega\in\ml(A)\;.
\ee
Therefore, in this case the reduction criterion is equivalent to the PPT criterion. 

In general, however, the PPT criterion is a more powerful criterion for detecting entanglement than the reduction criterion, since there exist entangled states that violate the PPT criterion, yet cannot be detected by the reduction criterion, whereas the converse is not true. This means that the PPT criterion can detect a larger class of entangled states than the reduction criterion.
The reason for that is that for $|A|>2$ the map $\mP$ above has the form (see Exercise~\ref{reduction})
\be\label{pp1p2}
\mP=\mP_1+\mT\circ\mP_2\;,
\ee
where $\mP_1,\mP_2\in\cp(A\to A)$ and $\mT\in\pos(A\to A)$ is the transpose map.

Although the reduction criterion may not be as effective as the PPT criterion in detecting entanglement, further investigation reveals that it still holds importance in the field of quantum resource theories. In fact, quantum states that do not satisfy the reduction criterion possess non-zero distillable entanglement, which emphasizes the usefulness of the criterion in other aspects of quantum information. In the upcoming sections, we will explore some of these implications in greater detail.

\bex\label{reduction}
Let $\mP^{A\to A}$ be as in~\eqref{pomp}.
\ben
\item Prove the relation~\eqref{2relr} for the case that $|A|=2$.
\item Prove the relation~\eqref{pp1p2} for the case $|A|\geq 2$. Hint: Show that the partial transpose of the Choi matrix of $\mP^{A\to A}$ is positive semidefinite.
\een
\eex

\subsection{The Realignment Criterion}\index{realignment criterion}

The realignment criterion is another powerful tool used to detect entanglement in quantum systems. 
One of the advantages of the realignment criterion is that it can be used in conjunction with other criteria to strengthen the detection of entanglement. For example, if a state satisfies the PPT criterion but fails the realignment criterion, then the state is entangled. Similarly, if a state fails the PPT criterion but satisfies the realignment criterion, then it is entangled.

The realignment criterion is based on the operator Schmidt decomposition\index{operator Schmidt decomposition} introduced in Exercise~\ref{osd}. Specifically, let $A$ and $B$ be two Hilbert spaces of dimensions $m\eqdef|A|$ and $n\eqdef|B|$ and denote by $k\eqdef\min\{m^2,n^2\}$. Every $\rho\in\md(AB)$ can be expressed in terms of $k$ non-negative real numbers $\{\lambda_x\}_{x\in[k]}$, and two orthonormal sets of Hermitian matrices (w.r.t.\ the Hilbert-Schmidt inner product) $\{\eta_x\}_{x\in[k]}\subset\herm(A)$ and $\{\zeta_y\}_{y\in[k]}\subset\herm(B)$:
\be
\rho^{AB}=\sum_{x\in[k]}\lambda_x\eta_x^A\otimes\zeta_x^B\;.
\ee

\begin{myt}{}
\begin{theorem}
Using the same notations as above, a quantum state $\rho\in\md(AB)$ is entangled if it satisfies
\be\label{lamx}
\sum_{x\in[k]}\lambda_x>1\;.
\ee
\end{theorem}
\end{myt}
\begin{proof}
Since the condition given in~\eqref{lamx} can  be expressed as
\be
\tr\left[\Lambda^{AB}\rho^{AB}\right]<0\quad\text{where}\quad\Lambda\eqdef I^{AB}-\sum_{x\in[k]}\eta_x^A\otimes\zeta_x^B\;,
\ee
it is sufficient to show that $\Lambda^{AB}$ is an entanglement witness. Indeed, let $\psi\in\pure(A)$ and $\phi\in\pure(B)$. Then,
\be
\tr\left[\Lambda^{AB}\left(\psi^A\otimes\phi^B\right)\right]=1-\sum_{x\in[k]}\tr[\psi\eta_x]\tr[\phi\zeta_x]\;.
\ee
Now, let  $\v,\u\in\mbb{R}^k$ be the vectors whose components are $\{\tr[\psi\eta_x]\}_{x\in[k]}$ and $\{\tr[\phi\zeta_x]\}_{x\in[k]}$, respectively. Then, we need to show that $\v\cdot\u\leq 1$. Since $\{\eta_x\}_{x\in[k]}$ is an orthonormal set in $\herm(A)$ (which can be completed to a full orthonormal basis of $\herm(A)$), in terms of the Frobenius norm (i.e. the norm induced by the Hilbert-Schmidt inner product)
\ba
1=\|\psi\|_{2}^2&\geq \Big\|\sum_{x\in[k]}\tr[\psi\eta_x]\eta_x\Big\|_2^2\\
\Gg{\{\eta_x\}_{x\in[k]}\text{ is orthonormal}}&=\sum_{x\in[k]}\left|\tr[\psi\eta_x]\right|^2=\v\cdot\v\;.
\ea
Similarly, we get that $\u\cdot\u\leq 1$. Hence, we must have $\v\cdot\u\leq 1$. This completes the proof.
\end{proof}

The term ``realignment" comes from the fact that the sum of the singular values $\{\lambda_x\}_{x\in[k]}$ can be expressed in terms of the trace norm of a \emph{realigned} version of the state $\rho^{AB}$. For simplicity, suppose $m=|A|=|B|$, let $\rho\in\md(AB)$, and expend $\rho^{AB}$ in the standard basis as
\be
\rho^{AB}=\sum_{x,x',y,y'\in[m]}r_{xx'yy'}|x\lr x'|^A\otimes|y\lr y'|^B\;,
\ee
where $r_{xx'yy'}\in\mbb{C}$. We then define the \emph{realigned} state $\trho^{AB}$ as
\be
\trho^{AB}\eqdef\sum_{x,x',y,y'\in[m]}r_{xx'yy'}|x\lr y|^A\otimes|x'\lr y'|^B\;.
\ee
We then argue (see Exercise~\ref{exm2ex}) that the sum appearing in~\eqref{lamx} can be expressed as
\be\label{xm2lx}
\sum_{x\in[m^2]}\lambda_x=\left\|\trho^{AB}\right\|_1\;.
\ee 
In other words, the realignment criterion can be stated as follows: {\it if the trace-norm of the realigned matrix $\trho^{AB}$ is greater than one (i.e.\ $\|\trho^{AB}\|_1\geq 1$) then the state $\rho^{AB}$ is entangled.}

\bex\label{exm2ex}
Prove the relation~\eqref{xm2lx}.
\eex

\bex
Consider the case $|A|=2$ and $|B|=4$. Let $p\in[0,1]$ and $\rho\in\herm(AB)$ be the matrix
\be
\rho^{AB}=\frac1{7p+1}\bpm pI_4 & p \xi\\
p \xi^T & \eta
\epm
\ee
where $I_4$ is the $4\times 4$ identity matrix, 
\be
\xi\eqdef \bpm0&1& 0&0 \\ 
 0&0&1& 0 \\ 
0&0&0& 1\\ 
 0&0&0& 0
 \epm\;,\quad\text{and}\quad \eta\eqdef\bpm\frac{1+p}2 &0&0&\frac{\sqrt{1-p^2}}2\\ 
 0&p&0&0\\ 
0&0&p&0\\ 
\frac{\sqrt{1-p^2}}2&0&0&\frac{1+p}2
\epm\;.
\ee
\ben
\item Show that $\rho\in\md(AB)$ for all $p\in[0,1]$.
\item Show that $\rho^{AB}$ is PPT for all $p\in[0,1]$.
\item Show that for some $p\in[0,1]$ the state $\rho^{AB}$ is entangled.
\een
Hint: Use the Schur complement (see Sec.~\ref{sec:schurcom}).
\eex

\subsection{The $k$-Extendability Criterion}\index{$k$-extendability criterion}

The concept of symmetric extensions of quantum states provides the most powerful criterion for separability currently known. Consider a state $\sigma^{AB}$ that can be expressed as a convex combination of product states: 
\be\label{rhoissep}\sigma^{AB}=\sum_{x\in[m]}p_x\;\psi_x^A\otimes\phi_x^B\in\sep(AB)\;.\ee We can construct a symmetric extension $\sigma\in\sep(AB\tB)$ of this state by introducing an additional system $\tB$ and defining the extension as \be
\sigma^{AB\tB}=\sum_{x\in[m]}p_x\;\psi_x^A\otimes\phi_x^B\otimes\phi_x^{\tB}\;.
\ee 
This extension is considered symmetric since the original state can be obtained by tracing out either the $B$ or the $\tB$ systems; i.e., the marginals of $\sigma^{AB\tB}$ satisfy
$\sigma^{AB}=\sigma^{A\tB}$. On the other hand, when dealing with entangled states, it is not immediately clear whether a symmetric extension, $\rho^{AB\tB}$, with the property $\rho^{AB}=\rho^{A\tB}$ exists for an entangled state $\rho\in\md(AB)$. While this property holds trivially for separable states, it doesn't hold for all entangled states. 

Note that the extension of the separable state in~\eqref{rhoissep} can also be extended to $k$-copies of $B$ via
\be
\rho^{AB^k}=\sum_{x\in[m]}p_x\;\psi_x^A\otimes\phi_x^{B_1}\otimes\cdots\otimes\phi_x^{B_k}\;,
\ee
where $B\cong B_1\cong\cdots\cong B_k$.
We say that $\rho^{AB^k}$ has a symmetric $k$-extension of $\rho^{AB}$.

\begin{myd}{}
\begin{definition}
We say that $\rho\in\md(AB)$ is $k$-extendible if there exists $\rho\in\md(AB^k)$, with $B^k=B_1\cdots B_k$ and $B\cong B_1\cong\cdots\cong B_k$, such that its marginals  satisfy
\be
\rho^{AB_m}=\rho^{AB_{m'}}=\rho^{AB}\quad\quad\forall\;m,m'\in[k]\;.
\ee
If such a state exists then $\rho^{AB^k}$ is called a symmetric $k$-extension of $\rho^{AB}$.
\end{definition}
\end{myd}

We saw above that every separable quantum state is $k$-extendible for all $k\in\mbb{N}$. Surprisingly, the converse of this statement is also true! This means that if a quantum state $\rho\in\md(AB)$ is $k$-extendible for all $k\in\mbb{N}$, then it must be separable. However, proving this statement requires certain techniques that are beyond the scope of this book. Specifically, it involves the use of the quantum de Finetti theorem. Interested readers can find more information in the `notes and references' section at the end of this chapter.

Given a quantum state $\rho\in\md(AB)$, how can we determine if it is $k$-extendible? Observe that the conditions $\rho^{AB_1}=\rho^{AB_j}$, for all $j\in[k]$, can be expressed as
\be
\tr\left[\rho^{AB^k}\Lambda^{AB^k}_j\right]=0\;,
\ee
for all $\Lambda_j\in\herm(AB^k)$ of the form
\be
\Lambda^{AB^k}_j=\eta^{AB_1}\otimes I^{B_2\cdots B_k}-\xi^{AB_j}\otimes\ I^{B_1\cdots B_{j-1}B_{j+1}\cdots B_k}\;,
\ee
where
$\eta\in\herm(AB_1)$ and $\xi\in\herm(AB_j)$. Note that the linearity of the condition above implies that we can restrict $\eta$ and $\xi$ to belong to orthonormal bases of $\herm(AB_1)$ and $\herm(AB_j)$, respectively. Thus, we conclude that there exists a finite number of operators $\{\Lambda_{j\ell}\}_{j\in[k],\ell\in[n]}$ such that  $\rho^{AB_1}=\rho^{AB_j}$, for all $j\in[k]$, if and only if
\be\label{feaspri}
\tr\left[\rho^{AB^k}\Lambda^{AB^k}_{j\ell}\right]=0\quad\quad\forall\;j\in[k],\;\ell\in[n]\;.
\ee
The conditions specified above indicate that the determination of whether $\rho^{AB}$ is $k$-extendible requires the solution of an SDP feasibility problem. Therefore, the criterion for $k$-extendibility can be computed algorithmically and efficiently.
\bex
Using the same notations as above:
\ben
\item Find an upper bound on $n$.
\item Use Farkas lemma of Exercise~\ref{farkas} to express the dual form of~\eqref{feaspri}.
\een
\eex

\section{Quantification of Entanglement}

Entanglement is quantified by functions that behave monotonically under LOCC. More precisely, adapting Definition~\ref{def:mor} to entanglement theory we get that a \emph{measure of entanglement} is a function 
\be
E:\bigcup_{A,B}\md({AB})\to\mbb{R} 
\ee
that satisfy the following two conditions: 
\begin{enumerate}
\item For any LOCC map $\mE\in\locc(AB\to A'B')$, and any bipartite state $\rho\in \md({AB})$ 
\be
E\big(\mE\left(\rho^{AB}\right)\big)\leq E\left(\rho^{AB}\right)\;.
\ee
\item $E(1)=0$, where $1$ correspond to the only element of $\md(AB)$ when $|A|=|B|=1$.
\end{enumerate}

\begin{exercise}
Show that any measure of entanglement $E$ as defined above satisfies the following two conditions: (1) It is always non-negative, that is, $E(\rho^{AB})\geq 0$ for all $\rho\in\md({AB})$, and (2) it satisfies $E(\sigma^{AB})=0$ for all $\sigma\in\sep({AB})$. 
\end{exercise}

In general, LOCC can be stochastic, in the sense that $\rho^{AB}$ can be converted to $\sigma^{AB}_{x}$ with some probability $p_x$. In this case, the map from $\rho^{AB}$ to $\sigma_{x}^{AB}$ can not be described by a CPTP map. However, by introducing a classical `flag' system $X$, we can view the ensemble $\{\sigma^{AB}_{x},p_x\}_{x\in[m]}$ as a classical quantum state $\sigma^{XAB}\eqdef\sum_{x\in[m]}p_x|x\lr x|^{X}\otimes\sigma_{x}^{AB}$. Hence, if $\rho^{AB}$ can be converted by LOCC to $\sigma^{AB}_{x}$ with probability $p_x$, then there exists map $\mE\in\locc(AB\to XAB)$ such that $\mE(\rho^{AB})=\sigma^{XAB}$. Since the `flag' system $X$ is classical both Alice and Bob have access to it since if Alice holds it she can communicate it to Bob, and vice versa. 
Therefore, the definition above of a measure of entanglement capture also probabilistic transformations. Particularly, $E$ must satisfy $E\left(\sigma^{XAB}\right)\leq E\left(\rho^{AB}\right)$.

Almost all measures of entanglement studied in literature (although not all) satisfy 
\be\label{emon}
E\left(\sigma^{XAB}\right)=\sum_{x\in[m]}p_xE(\sigma_{x}^{AB})\;,
\ee
which is very intuitive since $X$ is just a classical system encoding the value of $x$. We call this relation in~\eqref{emon} the \emph{direct sum property} since, mathematically, $\sigma^{XAB}$ can also be viewed as $\bigoplus_{x\in[m]}p_x\sigma^{AB}_x$. If the direct sum property holds then the condition $E\left(\sigma^{XAB}\right)\leq E\left(\rho^{AB}\right)$ becomes $\sum_{x}p_xE(\sigma_{x}^{AB})\leq E\left(\rho^{AB}\right)$ meaning that LOCC can not increase entanglement on average. Therefore, the direct sum property is in general stronger than the strong monotonicity property~\eqref{1010sm} of a resource measure $\M$. In fact, the condition~\eqref{emon} also implies that $E$ is convex (see Exercise~\ref{ex:dspcon}). We therefore conclude that any measure of entanglement that satisfies the direct sum property is an entanglement monotone.

\begin{exercise}\label{ex:dspcon}
Let $E$ be a measure of entanglement satisfying the direct sum property~\eqref{emon}. Show that $E$ is convex; i.e. for any ensemble of states $\{p_x,\sigma^{AB}_{x}\}_{x\in[m]}$ we have
\be
E\Big(\sum_{x\in[m]}p_x\sigma^{AB}_x\Big)\leq \sum_{x\in[m]}p_xE\left(\sigma^{AB}_x\right)\;.
\ee
\end{exercise}

\subsection{Extension of Entanglement from Pure to Mixed States}\label{cre00}

Quantifying entanglement in mixed states is considerably more complex than quantifying entanglement in pure states. This is partly because there is no equivalent to Nielsen's majorization theorem, and as we discussed earlier, LOCC on mixed bipartite states is much more intricate and difficult to characterize. Nevertheless, several approaches have been developed to help characterize entanglement.

\subsubsection{The Convex Roof Extension}\index{convex roof extension}

Let
\be
E:\bigcup_{A,B}\pure(AB)\to\mbb{R}
\ee
be a measure of pure state entanglement. We can extend the domain of $E$ to mixed states using a method known as the convex roof extension. The method is based on the fact that any bipartite mixed state $\rho^{AB}$ has many pure-state decompositions. Recall that a pure-state decomposition\index{pure-state decomposition} of $\rho^{AB}$ is an ensemble of pure states, $\{p_x,\psi_x^{AB}\}_{x\in[m]}$ (where $\psi_x\in\pure(AB)$ and $\{p_x\}_{x\in[m]}$ is a probability distribution), that satisfies $\rho^{AB}=\sum_{x\in[m]}p_x\psi_x^{AB}$. In Exercise~\ref{ensembles} we saw that every unitary matrix in $\muu(m)$ can be used to define a particular pure state decomposition.

\begin{myd}{Entanglement of Formation}\index{entanglement of formation}
\begin{definition}\label{def:cre}
Let $E$ be a measure of pure state entanglement. The convex roof extension\index{convex roof extension} of $E$, is a function $E_F:\bigcup_{A,B}\md(AB)\to\mbb{R}$ defined on any $\rho\in\md(AB)$ via
\be\label{1331}
E_F\left(\rho^{AB}\right)=\inf\sum_{x\in[m]}p_xE\left(\psi_x^{AB}\right)
\ee
where the infimum is over all pure-state decompositions $\{p_x,\psi_x^{AB}\}_{x\in[m]}$ of $\rho^{AB}$. $E_F$ is called the \emph{entanglement of formation} associated with the pure-state measure $E$.
\end{definition}
\end{myd}
\begin{remark}
The term ``entanglement of formation\index{entanglement of formation}" originated from historical reasons, as it was originally believed that $E_F(\rho^{AB})$, with $E$ taken as the entropy of entanglement\index{entropy of entanglement}, represented the entanglement cost  required to create the state $\rho^{AB}$. However, we will discover later on that it is actually the regularized entanglement of formation that can be interpreted as the entanglement cost of $\rho^{AB}$.
\end{remark}

\bex
Show that if $E$ is the entropy of entanglement then its corresponding entanglement of formation\index{entanglement of formation} satisfies for all $\rho\in\md(AB)$,
\be\label{eofub}
E_F\left(\rho^{AB}\right)\leq\min\left\{H\left(\rho^A\right),H\left(\rho^B\right)\right\}\;.
\ee
\eex

To show that the entanglement of formation is indeed a measure of entanglement, recall that  any measure of pure state entanglement, $E$, can be expressed for all $\psi\in\pure(AB)$ as
\be\label{gcsym}
E\left(\psi^{AB}\right)=g\left(\rho^A\right)\quad\text{with}\quad\rho^A\eqdef\tr_B\left[\psi^{AB}\right]\;,
\ee
for some Schur concave function $g$. A slightly stronger condition than Schur concavity is the condition that $g$ is both symmetric and concave. In this case, the resulting entanglement of formation\index{entanglement of formation} is an entanglement monotone.

\begin{myt}{}
\begin{theorem}\label{thm:em}
Let $E$ be a measure of entanglement on pure states given as in~\eqref{gcsym} with $g$ being a concave symmetric function. Then, its convex roof extension, $E_F$, as defined in Definition~\ref{def:cre}, is an entanglement monotone. 
\end{theorem}
\end{myt}

We first prove the following auxiliary lemma. In this lemma we only consider a quantum instrument on Bob's system and consider a pure initial bipartite state. We show that for this simpler case, the convex roof extension\index{convex roof extension} of $E$ satisfies strong monotonicity.

\begin{myg}{}
\begin{lemma}\label{lem:auab}
Let $\psi\in\pure(AB)$, $\mE^{B\to B'X}\eqdef\sum_{x\in[m]}\mE_x^{B\to B'}\otimes |x\lr x|^X$ be a quantum instrument on Bob's subsystem, and for all $x\in[m]$,
\be
\sigma^{AB'}_x\eqdef \frac1{p_x}\mE_x^{B\to B'}\left(\psi^{AB}\right)\quad\text{where}\quad p_x\eqdef\tr\left[\mE_x^{B\to B'}\left(\psi^{AB}\right)\right]\;.
\ee
Then,
\be
\sum_{x\in[m]}p_xE_F\left(\sigma^{AB'}_x\right)\leq E_F\left(\psi^{AB}\right)\;.
\ee
\end{lemma}
\end{myg}

\begin{remark}
In the proof below, we adopt the notations like $\phi^A \eqdef \tr_B\left[\phi^{AB}\right]$ to denote the reduced density matrix of a pure bipartite state. This notation is instrumental in reducing the number of symbols used, enhancing clarity and conciseness. However, it's crucial to remember that $\phi^A$ in this context represents a mixed state, despite the notation resembling that typically used for pure states. This distinction is important for a correct understanding of the concepts and calculations involved in the proof.
\end{remark}

\begin{proof}
For every $x\in[m]$, let 
\be\label{suy}
\sigma^{AB'}_x=\sum_{y\in[n]}r_{y|x}\phi^{AB'}_{xy}
\ee
where $\{r_{y|x},\phi^{AB'}_{xy}\}_{y\in[n]}$ is the optimal pure-state decomposition\index{pure-state decomposition} of $\sigma^{AB'}_x$. Therefore, for each $x\in[m]$ we have
\ba
E_F\left(\sigma^{AB'}_x\right)&=\sum_{y\in[n]}r_{y|x}E\left(\phi^{AB'}_{xy}\right)\\
\GG{cf.~\eqref{gcsym}}&=\sum_{y\in[n]}r_{y|x}g\left(\phi_{xy}^{A}\right)\\
\GG{{\it g}\;is\;concave}&\leq g\Big(\sum_{y\in[n]}r_{y|x}\phi_{xy}^{A}\Big)\\
\GG{cf.~\eqref{suy}}&=g\left(\sigma_x^A\right)\;.
\ea
Hence,
\ba
\sum_{x\in[m]}p_xE_F\left(\sigma^{AB'}_x\right)&\leq \sum_{x\in[m]}p_xg\left(\sigma_x^A\right)\\
\GG{{\it g}\;is\;concave}&\leq g\Big(\sum_{x\in[m]}p_x\sigma_x^A\Big)\;.
\ea
Finally, observe that the reduced density matrix of $\psi^{AB}$ satisfies 
$\psi^A=\sum_{x\in[m]}p_x\sigma^{A}_x$. Substituting this into the equation above gives
\be
\sum_{x\in[m]}p_xE_F\left(\sigma^{AB'}_x\right)\leq g\left(\psi^A\right)=E\left(\psi^{AB}\right)=E_F\left(\psi^{AB}\right)\;.
\ee
This completes the proof of the lemma.
\end{proof}

We are now ready to prove the theorem.
\begin{proof}[Proof of Theorem~\ref{thm:em}]
Let $\rho\in\md(AB)$ and let $\{p_x,\psi_x^{AB}\}_{x\in[m]}$ be an optimal pure-state decomposition\index{pure-state decomposition} of $\rho^{AB}$ satisfying
\be\label{1328}
E_F\left(\rho^{AB}\right)=\sum_{x\in[m]}p_xE\left(\psi_x^{AB}\right)\;.
\ee

We first prove the strong monotonicity of $E_F$. The proof strategy aims to show that $E_F$ does not increase on average under a general quantum instrument\index{quantum instrument} on Bob's subsystem. We can then apply a similar argument to Alice's side, demonstrating that $E_F$ remains non-increasing under quantum instruments on either subsystem. The significance of this lies in the fact that LOCC consists of such local quantum instruments, coupled with rounds of classical communication, which do not affect the monotonicity property. Therefore, by demonstrating the non-increasing nature of $E_F$ under a quantum instrument on both Bob's side (and, by symmetry arguments, also on Alice's side), it can be concluded that $E_F$ satisfies the strong monotonicity property under LOCC.

Let $\{\mE_z^{B\to B'}\}_{z\in[k]}$ be a quantum instrument on Bob's subsystem, and for each $z\in[k]$, let $\rho_z^{AB'}$ be the post measurement state after outcome $z$ occurred. Moreover, for every $z\in[k]$ and $x\in[m]$ we denote by  $r_{z}\eqdef\tr\left[\mE_z^{B\to B'}\left(\rho^{AB}\right)\right]$,  $t_{z|x}\eqdef\tr\left[\mE_z^{B\to B'}\left(\psi^{AB}_x\right)\right]$, and
\be
\sigma_{xz}^{AB'}\eqdef\frac1{t_{z|x}}\mE_z^{B\to B'}\left(\psi^{AB}_x\right)\;.
\ee 
With these notations, the post-measurement state can be expressed as
\ba
\rho_{z}^{AB'}\eqdef\frac1{r_{z}}\mE_z^{B\to B'}\left(\rho^{AB}\right)&=\frac1{r_{z}}\sum_{x\in[m]}p_x\mE_z^{B\to B'}\left(\psi_x^{AB}\right)\\
&=\sum_{x\in[m]} \frac{p_xt_{z|x}}{r_z}\sigma_{xz}^{AB'}\;.
\ea
We need to show that the average entanglement $\sum_{z\in[k]}r_zE_F\left(\rho_z^{AB'}\right)$ cannot exceed $E_F(\rho^{AB})$.
For this purpose, for each state $\rho^{AB}_z$ we need to find a suitable pure state decomposition that can be related to $\rho^{AB}$. Observe that the equation above involves the mixed states $\{\sigma_{xz}^{AB'}\}_{x\in[m]}$. Therefore, for each $\sigma_{xz}^{AB'}$ we denote by $\{s_{y|xz},\phi^{AB'}_{xyz}\}_{y\in[n]}$ its optimal pure state decomposition, so that
\be\label{1332}
E_F\left(\sigma_{xz}^{AB'}\right)=\sum_{y\in[n]}s_{y|xz}E\left(\phi^{AB'}_{xyz}\right)\;.
\ee
With this final notation, we get our desirable pure-state decomposition\index{pure-state decomposition} of $\rho_z^{AB'}$:
\be
\rho_{z}^{AB'}=\sum_{x\in[m]} \sum_{y\in[n]}\frac{p_xt_{z|x}s_{y|xz}}{r_z}\phi^{AB'}_{xyz}\;.
\ee
Since the above pure-state decomposition of $\rho_z^{AB'}$ is not necessarily optimal, we conclude
\ba
\sum_{z\in[k]}r_zE_F\left(\rho_z^{AB}\right)\leq\sum_{x,y,z}r_z\frac{p_xt_{z|x}s_{y|xz}}{r_z}E\left(\phi^{AB'}_{xyz}\right)
&=\sum_{x,y,z}p_xt_{z|x}s_{y|xz}E\left(\phi^{AB'}_{xyz}\right)\\
\GG{\eqref{1332}}&=\sum_{x,z}p_xt_{z|x}E_F\left(\sigma_{xz}^{AB'}\right)\\
\GG{Lemma~\ref{lem:auab}}&\leq \sum_{x\in[m]}p_xE_F\left(\psi^{AB}_x\right)\\
\GG{\eqref{1328}}&=E_F\left(\rho^{AB}\right)\;.
\ea
We therefore conclude that $E_F$ does not increase on average under a general quantum instrument on Bob's subsystem. This completes the proof of strong monotonicity.

It is therefore left to show that $E_F$ is convex. Indeed, let $\{p_x,\rho^{AB}_x\}_{x\in[m]}$ be an ensemble of bipartite entangled states, and for each $x\in[m]$ let $\{q_{y|x},\psi_{xy}^{AB}\}_{y\in[n]}$ be an optimal pure-state decomposition of $\rho^{AB}_x$ such that
\be\label{1336}
E_F\left(\rho^{AB}_x\right)=\sum_{y\in[n]}q_{y|x}E\left(\psi^{AB}_{xy}\right)\;.
\ee
Now, observe that $\{p_xq_{y|x},\psi_{xy}^{AB}\}_{x,y}$ is a pure-state decomposition of $\sum_xp_x\rho^{AB}_x$. Thus,
\ba
E_F\left(\sum_xp_x\rho^{AB}_x\right)&\leq\sum_{x,y}p_x
q_{y|x}E\left(\psi_{xy}^{AB}\right)\\
\GG{\eqref{1336}}&=\sum_{x\in[m]}p_xE_F\left(\rho^{AB}_x\right)\;.
\ea 
This completes the proof.
\end{proof}

The task of computing convex roof extensions is notably complex, particularly when determining the entanglement of formation\index{entanglement of formation} for a bipartite quantum state. If there were a straightforward, closed formula for the entanglement of formation, identifying whether a bipartite quantum state is entangled would be relatively easy. However, given that this identification task is known to be hard (specifically, NP-hard), it's unrealistic to expect a simple formula for the entanglement of formation. Nevertheless, for two-qubit systems, such a formula does exist, as outlined in the theorem below.

Recall the concurrence\index{concurrence} monotones defined in \eqref{1223} for pure bipartite states . In the case of two-qubit states, all concurrences are equivalent, so we denote them simply by $C$. The concurrence of formation, which is the convex roof extension\index{convex roof extension} of $C$, is then denoted as $C_F$.

\bex
Consider $E$ and $C$ as the entropy of entanglement and concurrence for pure states, respectively. Let $A$ and $B$ be two-qubit systems (i.e., $|A|=|B|=2$), and define the function $g:[0,1]\to [0,1]$ as
\be
g(x)=h_2\left(\frac{1+\sqrt{1-x^2}}{2}\right)
\ee
where $h_2(x)\eqdef-x\log x-(1-x)\log(1-x)$ is the binary Shannon entropy.
\ben
\item Show that for any $\psi\in\pure(AB)$ we have
\be
E\left(\psi^{AB}\right)=g\Big(C\left(\psi^{AB}\right)\Big)\;.
\ee
\item Show that for any $\rho\in\md(AB)$
\be\label{baea3}
E_F\left(\rho^{AB}\right)=g\Big(C_F\left(\rho^{AB}\right)\Big)\;.
\ee
Hint: Show first that the function $g$ is concave.
\een
\eex

The exercise above shows that in order to compute the entanglement of formation of a two-qubit state $\rho^{AB}$ it is sufficient to compute its concurrence\index{concurrence} of formation. In the following theorem we give a closed formula for the concurrence of formation. The closed formula is given in terms of the density matrix
\be\label{rhostart}
\rho^{AB}_\star\eqdef (\sigma_2\otimes \sigma_2)\bar{\rho}^{AB}(\sigma_2\otimes \sigma_2)
\ee
where $\bar{\rho}^{AB}$ is the density matrix whose components
\be
\la xy|\bar{\rho}^{AB}|x'y'\ra\eqdef\overline{\la xy|{\rho}^{AB}|x'y'\ra}\quad\quad\forall\;,x,y,x',y'\in\{0,1\}\;,
\ee
where the orthonormal basis 
$\{|x\ra\}_{x\in\{0,1\}}$ (and similarly $\{|y\ra\}_{y\in\{0,1\}}$) is such that $\sigma_2$ has the form $-i|0\lr 1|+i|1\lr 0|$.

\begin{myt}{\color{yellow} Closed Formula}
\begin{theorem}\label{cfconca}
Let $\rho\in\md(AB)$ be a two qubit mixed state (i.e. $|A|=|B|=2$). 
Then, the concurrence\index{concurrence} of formation of $\rho^{AB}$ is given by
\be\label{formulac}
C_F\left(\rho^{AB}\right)=\max\{0,\lambda_1-\lambda_2-\lambda_3-\lambda_4\}
\ee
where $\{\lambda_1,\ldots,\lambda_4\}$ are the four eigenvalues of the matrix
$\left|\sqrt{\rho}\sqrt{\rho_\star}\right|$,
arranged in a non-increasing order.
\end{theorem}
\end{myt}

In the derivation of the formula~\eqref{formulac}, we will use a bilinear form denoted as $\lp\cdot,\cdot\rp:\mbb{C}^2\otimes\mbb{C}^2\to\mbb{C}$. This bilinear form will not only be instrumental in proving the formula but will also be valuable in the analysis of multipartite entanglement. It is defined for any two vectors $|\psi\ra,|\phi\ra\in AB$ as
\be\label{bilinearf}
\lp|\psi\ra,|\phi\ra\rp\eqdef\la\bar{\psi}^{AB}|\sigma_2\otimes\sigma_2|\phi^{AB}\ra
\ee
where $\bar{\psi}^{AB}$ is defined such that if $|\psi^{AB}\ra=\sum_{x,y\in\{0,1\}}c_{xy}|x\ra|y\ra$ then
$|\bar{\psi}^{AB}\ra=\sum_{x,y\in\{0,1\}}\bar{c}_{xy}|x\ra|y\ra$. Note that $\bar{\psi}^{AB}$ is well define only with respect to some fixed bases $\{|0\ra^A,|1\ra^A\}\subset A$ and $\{|0\ra^B,|1\ra^B\}\subset B$ of $A$ and $B$, respectively. These orthonormal bases are chosen such that $\sigma_2$ has the form $-i|0\lr 1|+i|1\lr 0|$. The relation of the above bilinear form to our study here can be found in Exercise~\ref{ex:preconcurrence} in which you had to show that the concurrence\index{concurrence} of a two-qubit pure state $\psi\in\pure(AB)$ can be expressed as
$
C\left(\psi^{AB}\right)=\big|\lp|\psi\ra,|\psi\ra\rp\big|
$.
In the following exercise you prove several additional properties of this bilinear form.

\bex\label{ex:bilin0}
Consider the bilinear form defined in~\eqref{bilinearf}.
\ben
\item Show the linearity of the bilinear form; that is, show that for any vectors $|\psi\ra$, $|\psi_1\ra$, $|\psi_2\ra$,
$|\phi\ra$, $|\phi_1\ra$, and $|\phi_2\ra$ in $\mbb{C}^2\otimes\mbb{C}^2$
\ba
\lp|\psi_1\ra+|\psi_2\ra,|\phi\ra\rp&=\lp|\psi_1\ra,|\phi\ra\rp+\lp|\psi_2\ra,|\phi\ra\rp\\
\lp|\psi\ra,|\phi_1\ra+|\phi_2\ra\rp&=\lp|\psi\ra,|\phi_1\ra\rp+\lp|\psi\ra,|\phi_2\ra\rp\;.
\ea
\item Show that the bilinear form is symmetric; that is, for any two vectors $|\psi\ra,|\phi\ra\in\mbb{C}^2\otimes\mbb{C}^2$
\be
\lp|\psi\ra,|\phi\ra\rp=\lp|\phi\ra,|\psi\ra\rp\;.
\ee
\item Invariance property. Let $M,N\in SL(2,\mbb{C})$, $\psi,\phi\in\pure(AB)$, and denote $|\tpsi\ra\eqdef M\otimes N|\psi\ra$ and $|\tphi\ra\eqdef M\otimes N|\phi\ra$. Show that
\be
\lp|\tpsi\ra,|\tphi\ra\rp=\lp|\psi\ra,|\phi\ra\rp\;.
\ee
Hint: Use the relation~\eqref{1194}.
\een
\eex

\begin{proof}[Proof of Theorem~\ref{cfconca}]
Consider first the case that $\lambda_1>\lambda_2+\lambda_3+\lambda_4$.
Let $\{p_x,\psi_x\}_{x\in[4]}$ and $\{q_y,\phi_y^{AB}\}_{y\in[n]}$ be two pure-state decompositions of $\rho^{AB}$, and for each $x\in[4]$ and $y\in[n]$ (here $n\geq 4$), let $|\tpsi_x\ra\eqdef\sqrt{p_x}|\psi_x\ra$ and $|\tphi_y^{AB}\ra\eqdef\sqrt{q_y}|\phi_y^{AB}\ra$.
Recall from Exercise~\ref{ensembles} that there exists an $n\times 4$ isometry $V=(v_{yx})$ such that for all $y\in[n]$
\be
|\tphi_y\ra=\sum_{x\in[4]}v_{yx}|\tpsi_x\ra\;.
\ee
We can therefore relate the bilinear forms of the two decompositions as
\be\label{relyyp}
\lp|\tphi_y\ra,|\tphi_{y'}\ra\rp=\sum_{x,x'}v_{yx}v_{y'x'}\lp|\tpsi_x\ra,|\tpsi_{x'}\ra\rp\quad\quad\forall\;y,y'\in[n]\;.
\ee
Denoting by $M_\psi$ and $M_\phi$ the matrices whose components are $\lp\tpsi_x,\tpsi_{x'}\rp$ and $\lp\tphi_y,\tphi_{y'}\rp$, respectively, we get that the above equation can be written as
\be
M_\phi=VM_\psi V^T\;.
\ee
Since $M_\psi$ is symmetric (see second part of Exercise~\ref{ex:bilin0}), there exists a $4\times 4$ unitary matrix $U$ such that $UM_\psi U^T$ is diagonal. Moreover, by appropriate choice of $U$, the diagonal elements of $UM_\psi U^T$ can always be made real and positive (i.e. they are the singular values of $M_\psi$), and arranged on the diagonal of $UM_\psi U^T$ in a non-increasing order. We therefore conclude that there exists a pure-state decomposition\index{pure-state decomposition}  that is diagonal with respect to the bilinear form. For simplicity of the exposition, we take it to be $\{p_x,\psi_x^{AB}\}_{x\in[4]}$ itself; that is, 
\be\label{bildiag}
\lp|\tpsi_x\ra,|\tpsi_{x'}\ra\rp=\lambda_x\delta_{xx'}\quad\quad\forall\;x,x'\in[4]
\ee
with real non-negative numbers $\lambda_1\geq\lambda_2\geq\lambda_3\geq\lambda_4$. From Exercise~\ref{ex:bileig} below it follows that $\lambda_1,\ldots,\lambda_4$ are precisely the eigenvalues of $|\sqrt{\rho}\sqrt{\rho_{\star}}|$. With this specific choice of $\{\tpsi_x\}_{x=1}^4$, let $\{q_y,\phi_y\}_{y\in[n]}$ be another pure-state decomposition of $\rho^{AB}$ (and as before we set $|\tphi_y\ra\eqdef \sqrt{q_y}|\phi_y\ra$). Then, the relation~\eqref{relyyp}
gives 
\be\label{a1361}
\lp|\tphi_y\ra,|\tphi_{y'}\ra\rp=\sum_{x\in[4]}v_{yx}v_{y'x}\lambda_x\;.
\ee
We therefore get that the average concurrence\index{concurrence} of $\{q_y,\phi_y^{AB}\}_{y\in[n]}$ can be expressed as
\ba\label{1361}
\sum_{y\in[n]}q_yC\left(\phi_y^{AB}\right)&=\sum_{y\in[n]}q_y\left|\lp|\phi_y\ra,|\phi_y\ra\rp\right|=\sum_{y\in[n]}\left|\lp|\tphi_y\ra,|\tphi_y\ra\rp\right|\\
\GG{\eqref{a1361}}&=\sum_{y\in[n]}\Big|\sum_{x\in[4]}v_{yx}^2\lambda_x\Big|\\
&\geq \sum_{y\in[n]}\left(|v_{y1}|^2\lambda_1-|v_{y2}|^2\lambda_2-|v_{y3}|^2\lambda_3-|v_{y4}|^2\lambda_4\right)\\
&=\lambda_1-\lambda_2-\lambda_3-\lambda_4\;,
\ea
where we used the inequality $|a+b+c+d|\geq |a|-|b|-|c|-|d|$ for every $a,b,c,d\in\mbb{C}$. Moreover, the inequality above can be saturated by taking $V$ to be the unitary matrix (i.e.\ taking $n=4$) 
\be
V=\frac12\begin{pmatrix}
-1 & i & i &i\\
1 & -i & i & i\\
1 & i & -i & i\\
1 & i & i & -i
\end{pmatrix}\;.
\ee
Indeed, observe that the matrix $V$ above is unitary and has the property that for all $y\in[4]$, $\sum_{x\in[4]}v_{yx}^2\lambda_x=\frac14\big(\lambda_1-\lambda_2-\lambda_3-\lambda_4\big)$.
Therefore, with this $V$ we get from~\eqref{1361} that the average concurrence\index{concurrence} of $\{q_y,\phi_y\}_{y\in[4]}$ is 
\be
\sum_{y\in[4]}\Big|\sum_{x\in[4]}v_{yx}^2\lambda_x\Big|=\lambda_1-\lambda_2-\lambda_3-\lambda_4\;.
\ee

It is therefore left to show that if $\lambda_1\leq\lambda_2+\lambda_3+\lambda_4$ then $C_F(\rho^{AB})=0$. In this case we take
\be
V=\frac12\begin{pmatrix}
-1 & e^{i\theta_2} & e^{i\theta_3} &e^{i\theta_4}\\
1 & -e^{i\theta_2} & e^{i\theta_3} &e^{i\theta_4}\\
1 & e^{i\theta_2} & -e^{i\theta_3} &e^{i\theta_4}\\
1 & e^{i\theta_2} & e^{i\theta_3} &-e^{i\theta_4}
\end{pmatrix}\;,
\ee
where $\theta_2,\theta_3,\theta_4$ are some choices of phases to be determined shortly.
With this $V$, the average concurrence of $\{q_y,\phi_y\}_{y\in[4]}$ is given by (see~\eqref{1361})
\be\label{rhszero}
\sum_{y\in[4]}q_yC\left(\phi_y^{AB}\right)=\big|\lambda_1+\lambda_2e^{2i\theta_2}+\lambda_3e^{2i\theta_3}+\lambda_4e^{2i\theta_4}\big|\;.
\ee
Since we assume that $\lambda_1\leq\lambda_2+\lambda_3+\lambda_4$ (as well as $\lambda_1\geq\lambda_2\geq\lambda_3\geq\lambda_4$) we can always find three angles $\theta_1,\theta_2,\theta_3$ such the right-hand side above is zero (see Exercise~\ref{3angles}). This completes the proof. 
\end{proof}

\bex\label{ex:bileig}
Let $\{\tpsi_x\}_{x=1}^4$ be a set of four 2-qubit sub-normalized states satisfying~\eqref{bildiag} with some non-negative real numbers $\{\lambda_x\}_{x=1}^4$. Show that $\{\lambda_x\}_{x=1}^4$ are the eigenvalues
of $|\sqrt{\rho}\sqrt{\rho_\star}|$, where $\rho\eqdef\sum_{x\in[4]}\tpsi_x$. Hint: Recall from Exercise~\ref{eigfid} that if $\lambda$ is an eigenvalue of $|\sqrt{\rho}\sqrt{\rho_\star}|$ then $\lambda^2$ is an eigenvalue of $\rho\rho_\star$, and compute $\rho\rho_\star|\tpsi_x\ra$.
\eex

\bex\label{3angles}
Show that there for any four non-negative real numbers, $\lambda_1,\ldots,\lambda_4$ that satisfy $\lambda_1\leq\lambda_2+\lambda_3+\lambda_4$ and $\lambda_1\geq\lambda_2\geq\lambda_3\geq\lambda_4$ there exists three angles $\theta_2,\theta_3,\theta_4$ that the right-hand side of~\eqref{rhszero} is zero. Hint: Use a continuity argument.
\eex

\bex
Compute the concurrence of the following two-qubit bipartite mixed states:
\ben
\item The isotropic state\index{isotropic state}
\be
\rho^{AB}_{p}\eqdef p\u^{AB}+(1-p)\Phi^{AB}\;,
\ee
with $p\in[0,1]$.
\item The Werner state\index{Werner state}
\be
\rho^{AB}_t=\frac13t\Pi_\sym^{AB} +(1-t)\Pi_\asy^{AB}\;,
\ee
with $t\in[0,1]$.
\een
\eex

\bex\label{ex:coa} Let $\rho\in\md(AB)$ with $|A|=|B|=2$, and define the quantity
\be\label{coa1}
C_a\left(\rho^{AB}\right)\eqdef\max\sum_{x\in[m]}p_xC(\psi_x^{AB})
\ee
where the maximum is over all pure-state decompositions of $\rho^{AB}$ (i.e. $C_a$ is defined similarly to $C_F$ but with a maximum instead of a minimum). Show that 
\be\label{coa2}
C_a\left(\rho^{AB}\right)=F\left(\rho^{AB},\rho^{AB}_\star\right),
\ee
where $F$ is the fidelity. Hint: Use similar lines as in the proof above. Show also that the square of the fidelity above can be expressed as
\be\label{13p72}
\left\|\sqrt{\rho^{AB}}\sqrt{\rho^{AB}_\star}\right\|_1^2=\tr\left[\rho^{AB}\rho^{AB}_\star\right]\;.
\ee
\eex

\subsubsection{Monotones Based on the Ky Fan Norms}\index{Ky Fan norm}

Let us revisit the entanglement measures introduced in~\eqref{vidal}. According to Nielsen's majorization theorem\index{Nielsen's theorem}, these functions are indicative of whether a pure bipartite state can be transformed into another. In the upcoming sections, we will demonstrate that some of the operational significance of these measures can also be extended to their convex roof extensions. Specifically, we can express the convex roof extension\index{convex roof extension} of the pure-state entanglement measures defined in~\eqref{vidal} as follows:
\be\label{kyfan}
E_{(k)}\left(\rho^{AB}\right)\eqdef\min\sum_{x\in[m]}p_x\left(1-\left\|\rho^A_x\right\|_{(k)}\right)\;,
\ee
where the minimum is over all pure-state decompositions $\rho^{AB}=\sum_xp_x\psi^{AB}_x$, with $\rho_x^A\eqdef\tr_B[\psi_x^{AB}]$, and $\|\cdot\|_{(k)}$ is the Ky Fan norm\index{Ky Fan norm}.

\bex
Show that the functions $E_{(k)}$ as defined above are entanglement monotones.
\eex

Note that for the two-qubit case, the only non-trivial measure $E_{(k)}$ is when $k=1$. In this case
\be
E_{(1)}\left(\rho^{AB}\right)\eqdef\min\sum_{x\in[m]}p_x\lambda_{\min}\left(\rho^A_x\right)\;,
\ee
since each $\rho^{A}_x$ is a qubit so its the minimum eigenvalue $\lambda_{\min}\left(\rho^A_x\right)=1-\left\|\rho^A_x\right\|_{(1)}$. Moreover, observe that 
\ba\label{baea}
\lambda_{\min}\left(\rho^A_x\right)&=\frac12\left(1-\sqrt{1-4\det\left(\rho^A_x\right)}\right)\\
&=\frac12\left(1-\sqrt{1-C^2\left(\psi^{AB}_x\right)}\right)
\ea
where $C^2\left(\psi^{AB}_x\right)$ is the square of the concurrence\index{concurrence} of $\psi^{AB}_x$. In the following exercise you show that the above relation can be used to show that for any two-qubit state $\rho\in\md(AB)$ with $|A|=|B|=2$ we have
\be\label{baea2}
E_{(1)}\left(\rho^{AB}\right)=\frac12\left(1-\sqrt{1-C^2_f\left(\rho^{AB}\right)}\right)\;,
\ee
where $C^2_f\left(\rho^{AB}\right)$ is the square of the concurrence of formation of $\rho^{AB}$. Hence, the closed formula for the concurrence of formation can be used to compute $E_{(1)}$.

\bex
Let $\rho\in\md(AB)$ be a two-qubit state with $|A|=|B|=2$. 
\ben
\item Prove the relation~\eqref{baea}.
\item Prove the relation~\eqref{baea2}. Hint: The proof is similar to the proof of~\eqref{baea3}.
\een
\eex

In Corollary~\ref{cor:1242} we found necessary and sufficient conditions to convert by LOCC a pure bipartite state to a mixed bipartite state. Interestingly, for the case that $d\eqdef|A|=|B|=2$, the minimization in~\eqref{1260q} over $k\in\{1,2\}$ becomes trivial  since for all $\psi\in\pure(AB)$ we have $E_{(2)}\left(\psi^{AB}\right)=0$. We therefore arrive at the following corollary.

\begin{myg}{}
\begin{corollary}\label{cor:1321}
Let $\psi\in\pure(AB)$ and $\sigma\in\md(AB)$ be two bipartite entangled states with $d\eqdef|A|=|B|=2$. Then, $\psi^{AB}$ can be converted to $\sigma^{AB}$ by LOCC if and only if
\be
C\left(\psi^{AB}\right)\geq C_F\left(\sigma^{AB}\right)\;,
\ee
where $C$ is the concurrence\index{concurrence}.
\end{corollary}
\end{myg}

\begin{proof}
As discussed above, taking $k=1$ in~\eqref{1260q} gives that $\psi^{AB}$ can be converted to $\sigma^{AB}$ by LOCC if and only if
\be
E_{(1)}\left(\psi^{AB}\right)\geq E_{(1)}\left(\sigma^{AB}\right)\;.
\ee
The proof is concluded by expressing $E_{(1)}$ on both sides of the equation above in terms of the concurrence (see the relation~\eqref{baea2} between $E_{(1)}$ and the concurrence).
\end{proof}

\bex
Let $\psi\in\pure(AB)$ with $d\eqdef|A|=|B|>2$ and let $\sigma\in\md(A'B')$ with $|A'|=|B'|=2$. Show that $\psi^{AB}$ can be converted to $\sigma^{AB}$ by LOCC if and only if
\be\label{ex:13117}
E_{(1)}\left(\psi^{AB}\right)\geq E_{(1)}\left(\sigma^{A'B'}\right)\;.
\ee
\eex

\subsubsection{Optimal Extensions}

In Chapter~\ref{chadiv} we introduced a method to extend divergences from classical to quantum systems. This method is in fact quite general and can be slightly modified to incorporate extensions of measures of entanglement from pure to mixed states. Specifically, let $E$ be a measure on pure state entanglement. For any $\rho\in\md(AB)$ the maximal extension of $E$ is defined as
\be
\overline{E}\left(\rho^{AB}\right)\eqdef\inf \left\{E\left(\psi^{A'B'}\right)\;:\;\psi^{A'B'}\xrightarrow{\text{\tiny LOCC}} \rho^{AB}\right\}\;,
\ee
where the infimum is over all systems $A'B'$ and all pure states $\psi\in\pure(A'B')$ for which $\psi^{A'B'}$ can be converted by LOCC to $\rho^{AB}$. Similarly, the minimal extension is defined as 
\be
\underline{E}\left(\rho^{AB}\right)\eqdef\sup \left\{E\left(\psi^{A'B'}\right)\;:\;\rho^{AB}\xrightarrow{\text{\tiny LOCC}} \psi^{A'B'}\right\}\;.
\ee
The following exercise demonstrates the optimality of the above definitions.
\bex
Let $E$ be a measure of pure state entanglement, and let $\overline{E}$ and $\underline{E}$ be its maximal and minimal extensions.
\ben
\item Show that $\overline{E}$ and $\underline{E}$ are measures of mixed bipartite entanglement.
\item Show that if $E'$ is a measure of bipartite mixed-state entanglement that reduces to $E$ on pure states, then
\be
\underline{E}\left(\rho^{AB}\right)\leq E'\left(\rho^{AB}\right)\leq\overline{E}\left(\rho^{AB}\right)\;.
\ee
\item Show that if $E$ is additive under tensor products of pure bipartite states, then $\overline{E}$ is sub-additive and $\underline{E}$ super-additive under tensor products of mixed bipartite states.
\een
\eex

The minimal extension $\underline{E}$ is not very useful measure of entanglement since typically a mixed bipartite state  cannot be converted by LOCC to a pure entangled state. Therefore, for such mixed entangled states $\underline{E}$ takes the zero value. On the other hand, the maximal extension is a faithful measure of entanglement (i.e. takes the zero value \emph{only} on separable states).

As an example, recall that the Schmidt rank\index{Schmidt rank} is a measure of entanglement on pure states. Its maximal extension to mixed states is given by
\be\label{srank}
\overline{\sr}\left(\rho^{AB}\right)\eqdef\inf \left\{\sr\left(\psi^{A'B'}\right)\;:\;\psi^{A'B'}\xrightarrow{\text{\tiny LOCC}} \rho^{AB}\right\}\;.
\ee
In general, the condition $\psi^{A'B'}\xrightarrow{\text{\tiny LOCC}} \rho^{AB}$ can be very complicated. However, we can replace $\psi^{A'B'}$ in the equation above with the maximally entangled state $\Phi_k$, where $k\eqdef\sr\left(\psi^{A'B'}\right)=\sr\left(\Phi_k\right)$, since whenever
$\psi^{A'B'}\xrightarrow{\text{\tiny LOCC}} \rho^{AB}$ we also have $\Phi_k\xrightarrow{\text{\tiny LOCC}} \rho^{AB}$. We therefore conclude that
\be\label{srlocc}
{\sr}\left(\rho^{AB}\right)\eqdef\overline{\sr}\left(\rho^{AB}\right)=\min \left\{k\;:\;\Phi_k\xrightarrow{\text{\tiny LOCC}} \rho^{AB}\right\}\;,
\ee
where, for simplicity of the exposition, we removed the over-line symbol from $\overline{\sr}\left(\rho^{AB}\right)$.
\bex\label{ex:sr}
Let $\rho\in\md(AB)$. Show that ${\sr}\left(\rho^{AB}\right)=k$ for some $k\in\mbb{N}$ if and only if the following two conditions hold:
\ben
\item At least one of the states, in any pure-state decomposition of $\rho^{AB}$, has a Schmidt rank\index{Schmidt rank} no smaller than $k$.
\item There exists a pure-state decomposition of $\rho^{AB}$ with all states having Schmidt rank at most $k$.
\een
\eex

\bex
Let $\rho\in\md(AB)$. Show that
\be\label{sreq}
{\sr}\left(\rho^{AB}\right)=\inf\max_{x\in[k]}\sr(\psi_x^{AB})
\ee
where the infimum is over all pure-state decompositions of $\rho^{AB}=\sum_{x\in[k]}p_x\psi^{AB}_x$.
\eex

\subsection{The Relative Entropy of Entanglement}\index{relative entropy of entanglement} 

In Sec.~\ref{sec:dbm} we studied many properties of the relative entropy of a resource, and in Sec.~\ref{sec:crer} we developed a method to compute it. In entanglement theory, the relative entropy of entanglement is defined for any $\rho\in\md(AB)$ as (see Fig.~\ref{dista})
\be
E_R\left(\rho^{AB}\right)\eqdef\min_{\sigma\in\sep(AB)}D\left(\rho^{AB}\big\|\sigma^{AB}\right)\;.
\ee
As discussed in Sec.~\ref{sec:crer}, computing the relative entropy of entanglement can generally be quite challenging. However, in certain special cases, such as with pure states and symmetric states, it is feasible to compute this measure. The complexity in computing the relative entropy of entanglement typically arises from the need to optimize over the large set of separable states, which can be a demanding task for most mixed states. Yet, for pure states and certain states with specific symmetrical properties, this complexity is significantly reduced, making the calculation manageable.

\begin{figure}[h]
\centering
    \includegraphics[width=0.5\textwidth]{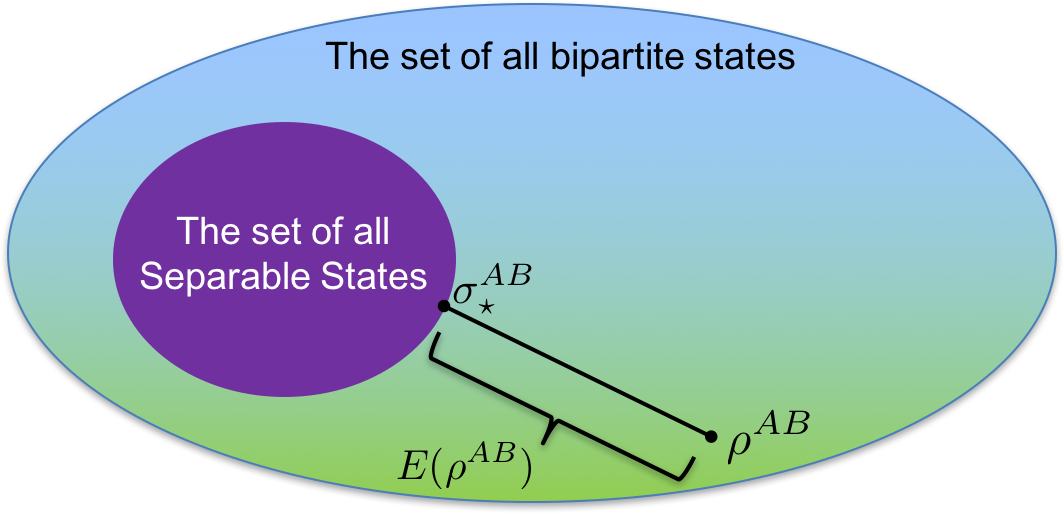}
  \caption{\linespread{1}\selectfont{\small The closest separable state.}}
  \label{dista}
\end{figure}

\subsubsection{Relative Entropy of Entanglement for Pure States}\index{relative entropy of entanglement}

In the next theorem, it's shown that for pure states, the relative entropy of entanglement simplifies to the entropy of entanglement\index{entropy of entanglement}. This finding connects two key entanglement measures, highlighting the elegant underlying structure of quantum entanglement in pure systems.

\begin{myt}{}
\begin{theorem}\label{reoep}
Let $\psi\in\pure(AB)$. Then,
\be
E_R\left(\psi^{AB}\right)=E\left(\psi^{AB}\right)\eqdef H\left(A\right)_\rho\;,
\ee
where $H\left(A\right)_\rho$ is the von-Neumann\index{von-Neumann} entropy of the reduced density matrix $\rho^A\eqdef\tr_B\left[\psi^{AB}\right]$.
\end{theorem}
\end{myt}

\begin{proof}
The proof is based on the closed formula given in Theorem~\ref{2cfre2} for the relative entropy of a resource. Let 
\be
|\psi\ra=\sum_{x\in[n]}\sqrt{p_x}|xx\ra
\ee
be given in its Schmidt form, where $n\eqdef\sr(\psi)$, and let
\be
\sigma_\star\eqdef\sum_{x\in[n]}p_x|xx\lr xx|\;.
\ee
We argue now that $\sigma_\star$ is the closest separable state (see Fig.~\ref{dista}); that is, we argue that
\be\label{ccss}
\min_{\sigma\in\sep(AB)}D\left(\psi^{AB}\big\|\sigma^{AB}\right)=D\left(\psi^{AB}\big\|\sigma_\star^{AB}\right)\;.
\ee
Indeed, from Theorem~\ref{2cfre2} it follows that $\sigma_\star^{AB}$ satisfies the above equality if and only if there exists an entanglement witness $\eta\in\wit(AB)$ such that
\be
\psi^{AB}=\sigma_\star^{AB}-a\mL_{\sigma_\star}^{-1}(\eta)\;.
\ee
The above equality can be expressed as $\mL_{\sigma_\star}^{-1}(a\eta)=\sigma_\star^{AB}-\psi^{AB}$, which is equivalent to
\be
a\eta=\mL_{\sigma_\star}\left(\sigma_\star-\psi\right)=I-\mL_{\sigma_\star}\left(\psi\right)\;.
\ee
That is, $\sigma_\star$ satisfies~\eqref{ccss} if and only if the right-hand side of the equation above is an entanglement witness.
Now, by direct computation with have (see Exercise~\ref{ex1380})
\be\label{1380}
\mL_{\sigma_\star}\left(\psi\right)=\sum_{x,y\in[n]}\sqrt{p_xp_y}\frac{\log p_x-\log p_y}{p_x-p_y}|xx\lr yy|\;.
\ee
Denote by $r_{xy}\eqdef\frac{p_x}{p_y}$ and observe that for any $x,y\in[n]$
\be
c_{xy}\eqdef\sqrt{p_xp_y}\frac{\log p_x-\log p_y}{p_x-p_y}=\frac{\sqrt{r_{xy}}\log(r_{xy})}{r_{xy}-1}\leq 1\;,
\ee 
where the last inequality follows from the fact that logarithm function satisfies $\log(r)\leq (r-1)/\sqrt{r}$ for $r\geq 1$ and opposite inequality for $0<r\leq 1$. We therefore get for any product state $\phi^A\otimes\varphi^B$
\ba
\tr\left[\left(\phi^A\otimes\varphi^B\right)\mL_{\sigma_\star}\left(\psi\right)\right]&=\sum_{x,y\in[n]}c_{xy}\la \phi|x\ra\la \varphi|x\ra\la y|\phi\ra\la y|\varphi\ra\\
&\leq \sum_{x,y\in[n]}c_{xy}\big|\la \phi|x\ra\la \varphi|x\ra\la y|\phi\ra\la y|\varphi\ra\big|\\
\Gg{c_{xy}\leq 1}&\leq \sum_{x,y\in[n]}\big|\la \phi|x\ra\la \varphi|x\ra\la y|\phi\ra\la y|\varphi\ra\big|\\
\Gg{|ab|\leq\frac12\left(|a|^2+|b|^2\right)}&\leq \frac14 \sum_{x,y\in[n]}\left(|\la \phi|x\ra|^2+|\la \varphi|x\ra|^2\right)\left(|\la y|\phi\ra|^2+|\la y|\varphi\ra|^2\right)\\
&=1\;.
\ea
We therefore conclude that for any product state $\phi^A\otimes\varphi^B$
\be
\tr\left[\left(\phi^A\otimes\varphi^B\right)\big(I-\mL_{\sigma_\star}\left(\psi\right)\big)\right]\geq 0\;,
\ee
so that $I-\mL_{\sigma_\star}\left(\psi\right)$ is an entanglement witness. This completes the proof.
\end{proof}

\bex\label{ex1380}
Prove the equality in~\eqref{1380}.
\eex

\subsubsection{Relative Entropy of Entanglement for Symmetric States}\index{relative entropy of entanglement}

Unlike the case of pure bipartite states, there isn't a general, straightforward formula for the relative entropy of entanglement of mixed bipartite states. However, for certain special cases like isotropic states\index{isotropic state} or Werner states\index{Werner state}, such formulas do exist. The underlying reason for this is the symmetry inherent in these states. To understand why a formula exists for these specific states, consider that if there is a bipartite channel $\mG\in\locc(AB\to AB)$ such that $\rho^{AB}=\mG(\rho^{AB})$ (which is true for isotropic and Werner states under various twirling operations), then the relative entropy of entanglement of such a state can be expressed as follows:
\ba
E_R\left(\rho^{AB}\right)&\eqdef\min_{\sigma\in\sep(AB)}D\left(\rho^{AB}\big\|\sigma^{AB}\right)\\
\Gg{\mG\big(\sep(AB)\big)\subseteq\sep(AB)}&\leq\min_{\sigma\in\sep(AB)}D\left(\rho^{AB}\big\|\mG\left(\sigma^{AB}\right)\right)\\
\Gg{\rho^{AB}=\mG(\rho^{AB})}&=\min_{\sigma\in\sep(AB)}D\left(\mG\left(\rho^{AB}\right)\big\|\mG\left(\sigma^{AB}\right)\right)\\
\GG{DPI}&\leq\min_{\sigma\in\sep(AB)}D\left(\rho^{AB}\big\|\sigma^{AB}\right)=E_R\left(\rho^{AB}\right)\;.
\ea
Therefore, all the inequalities must actually be equalities, and particularly:
\be\label{fixedpointgt}
E_R\left(\rho^{AB}\right)=\min_{\sigma\in\sep(AB)}D\left(\rho^{AB}\big\|\mG\left(\sigma^{AB}\right)\right)\;.
\ee
The importance of the above formula lies in the fact that the minimization over separable states of the form $\mG\left(\sigma^{AB}\right)$ might involve significantly fewer parameters compared to the minimization over the entire set of separable states $\sep(AB)$. This simplification is what makes the computation of the relative entropy of entanglement feasible for these symmetric states.

As an example, let's consider the isotropic state\index{isotropic state} defined for all $t\in[0,1]$ (with $m\eqdef|A|=|B|$) as
\be
\rho^{AB}_t=t\Phi_m^{AB}+(1-t)\tau^{AB}\quad\text{where}\quad\tau^{AB}\eqdef\frac{I^{AB}-\Phi_m}{m^2-1}\;.
\ee
Previously, we observed that this state is invariant under the twirling channel described in \eqref{gtwi2} and is separable if and only if $t\leq 1/m$. At one extreme ($t=0$), the isotropic state is the separable state $\tau^{AB}$. At the other extreme ($t=1/m$), the isotropic state is the separable state
\be
\rho_{1/m}^{AB}=\frac m{m+1}\u^{AB}+\frac1{m+1}\Phi_m^{AB}\;.
\ee

For an entangled isotropic state $\rho_t^{AB}$ with $t>1/m$, equation~\eqref{fixedpointgt} yields
\be
E_R\left(\rho^{AB}_t\right)=\min_{t'\in[0,1/m]}D\left(\rho^{AB}_t\big\|\rho_{t'}^{AB}\right)\;,
\ee
since for every separable state $\sigma\in\sep(AB)$, the twirled state $\mG\left(\sigma^{AB}\right)=\rho_{t'}^{AB}$ for some $t'\in[0,1/m]$. This demonstrates how the symmetry of $\rho_t^{AB}$ simplifies the optimization problem. Furthermore, in Exercise~\ref{compys}, you will show that the optimal $t'$ is $t'=1/m$, resulting in
\ba\label{csst}
E_R\left(\rho^{AB}_t\right)&=D\left(\rho^{AB}_t\big\|\rho_{1/m}^{AB}\right)\\
&=\log m+t\log t +(1-t)\log\frac{1-t}{m-1}\;.
\ea
This result is intuitive as the separable state $\rho_{t'}^{AB}$ with $t'=1/m$ is on the boundary of the set of separable states and, roughly speaking, the closest to $\rho_t^{AB}$.
\bex\label{compys}
Prove the equalities in~\eqref{csst}. Hint. Observe that $[\rho_t^{AB},\rho_{t'}^{AB}]=0$ for all $t,t'\in[0,1]$ and use it to show that $D\left(\rho^{AB}_t\big\|\rho_{t'}^{AB}\right)=D(\t\|\t')$, where $\t\eqdef(t,1-t)^T$ and $\t'\eqdef(t',1-t')^T$.
\eex

\bex
Consider the Werner state\index{Werner state} $\rho_W^{AB}$ as given in~\eqref{werner} with $p>1/2$ (i.e., an entangled Werner state). Show that
\be\label{csst2}
E_R\left(\rho^{AB}_W\right)=D\left(\rho^{AB}_W\big\|\omega^{AB}\right)\;,
\ee
where 
\be
\omega^{AB}\eqdef\frac{1}{m(m+1)}\Pi_\sym^{AB}+\frac{1}{m(m-1)}\Pi_\asy^{AB}\;.
\ee
\eex

\subsection{The Robustness of Entanglement}\index{robustness of entanglement}

Section~\ref{robustness} introduces a resource measure called the robustness\index{robustness}. However, this measure is not very useful for affine resource theories, as noted in the Exercise~\ref{robfaith}. On the other hand, since the set of separable states is maximally non-affine (as discussed around~\eqref{txy}), the robustness measure is highly relevant in the context of entanglement theory. Later on, we will explore its operational interpretation.

To apply the definition of the robustness measure to entanglement theory, we define the robustness of entanglement\index{robustness of entanglement} for a bipartite state $\rho\in\md(AB)$ as:
 \be\label{robuste}
 \R\left(\rho^{AB}\right)\eqdef\min\left\{s\geq 0\;:\;\frac{\rho^{AB}+s\omega^{AB}}{1+s}\in\sep(AB)\;\;,\;\;\omega\in\sep(AB)\right\}\;.
 \ee
 The global robustness, $\R_g$, is defined in a manner similar to the definition above, but with the key distinction that the state $\omega$ is chosen from the entire set of density matrices $\md(AB)$, rather than being restricted to separable states $\sep(AB)$.

We have established that the logarithmic global robustness is connected to $\R_g$ through the relationship:
 \be
 D_{\max}\left(\rho^{AB}\big\|\mf\right)=\log\left(1
+\R_g\left(\rho^{AB}\right)\right)\;.
 \ee
In a similar vein, the logarithmic robustness\index{robustness} of a state $\rho^{AB}$ is defined as:
 \be
 \LR\left(\rho^{AB}\right)\eqdef\log\left(1+\R\left(\rho^{AB}\right)\right)\;.
 \ee

 \bex
 Show that in entanglement theory, for all $\rho\in\md(AB)$ there exists a finite $s\geq 0$ and a separable density matrix $\omega^{AB}$ such that the density matrix $\left(\rho^{AB}+s\omega^{AB}\right)/(1+s)$ is separable.
\eex

\bex
Show that the logarithmic robustness is subadditive; that is, for all $\rho\in\md(AB)$ and $\sigma\in\md(A'B')$ we have
\be
\LR\left(\rho^{AB}\otimes\sigma^{A'B'}\right)\leq \LR\left(\rho^{AB}\right)+\LR\left(\sigma^{A'B'}\right)\;.
\ee
\eex

\bex\label{robfaith2}
Let  $\rho\in\md(AB)$. Show that $R\left(\rho^{AB}\right)=0$ if and only if $\rho\in\sep(AB)$.
\eex

\bex
Show that all $\rho\in\md(AB)$ can be written as
\be\label{rhoasie2}
\rho^{AB}=\big(1+\R\left(\rho^{AB}\right)\big)\tau^{AB}-\R\left(\rho^{AB}\right)\omega^{AB}\;,
\ee
for some $\tau,\omega\in\sep(AB)$. The above decomposition of $\rho^{AB}$ is sometimes referred to as a pseudo-mixture of $\rho^{AB}$.
\eex

Note that from the exercise above it follows that $\R\left(\rho^{AB}\right)$ can also be expressed as
\be
 \R\left(\rho^{AB}\right)\eqdef\min\big\{s\geq 0\;:\;\rho^{AB}=(1+s)\tau^{AB}-s\omega^{AB}\;\;,\;\;
\tau,\omega\in\sep(AB)\big\}\;,
 \ee
so that we can think of the pseudo-mixture in~\eqref{rhoasie2} as the optimal one achieved with $s=\R\left(\rho^{AB}\right)$.

\bex
Show that the robustness of entanglement is an entanglement monotone.
\eex

\subsection{Partial Transpose-Based Entanglement Measures}

In the previous sections, we learned that a bipartite quantum state is entangled if the partial transpose\index{partial transpose} of the state has a negative eigenvalue. In this subsection, we will explore how this property can be leveraged to measure the extent of entanglement in a bipartite state. However, entanglement measures constructed in this manner have certain limitations. For instance, they assign a value of zero to all PPT states, including those that are entangled. Despite these shortcomings, these measures possess the advantage of being computationally tractable and can be efficiently computed using SDP algorithms, as we will demonstrate.

\subsubsection{The Negativity\index{negativity} of Entanglement}

 The negativity of a bipartite state $\rho^{AB}$ is a measure of entanglement that is defined as the absolute value of the sum of all the negative eigenvalues of the partial transpose\index{partial transpose} of $\rho^{AB}$. Unlike other measures of entanglement, the negativity of entanglement is relatively easy to compute even for non-qubit systems. Therefore, it is  used quite extensively in literature.

In this section, for any $\rho\in\md(AB)$ we will use the notation $\rho^\Gamma$ to denote the partial transpose of the state $\rho^{AB}$; i.e.,
\be\label{pptg}
\rho^\Gamma\eqdef\mT^{B\to B}\left(\rho^{AB}\right)
\ee
where $\mT\in\pos(B\to B)$ is the transpose map. The intuition behind this notation is that $\rho^\Gamma$ represents half of the full transposed state $\rho^T$.

\begin{myd}{}
\begin{definition}
Let $\rho\in\md(AB)$. The negativity of $\rho^{AB}$ is defined as
\be\label{negativity}
\N\left(\rho^{AB}\right)\eqdef\frac{\left\|\rho^\Gamma\right\|_1-1}{2}\;.
\ee
\end{definition}
\end{myd}
\begin{remark}
We chose as a convention for this book that $\rho^\Gamma$ represents partial transpose\index{partial transpose} on Bob's side. This convention does not effect the definition above since 
\be
\left\|\mT^{B\to B}\left(\rho^{AB}\right)\right\|_1=\left\|\mT^{A\to A}\left(\rho^{AB}\right)\right\|_1\;.
\ee
Therefore, the definition of the negativity above is independent on whether the partial transpose is taken on Bob's side or on Alice's side. 
\end{remark}
Note that $\rho^\Gamma$ has trace one since the partial transpose does not effect the trace. Therefore, if $\lambda_1,\ldots,\lambda_n$ are the eigenvalues of $\rho^\Gamma$ then they sum to one. Suppose, without loss of generality that the first $k$ eigenvalues of $\rho^\Gamma$ are non-negative, and the remaining $n-k$ are negative. We then get 
\ba
\left\|\rho^\Gamma\right\|_1=\sum_{x\in[n]}|\lambda_x|&=\sum_{x\in[k]}\lambda_x-\sum_{x=k+1}^n\lambda_x\\
\Gg{\sum_{x\in[n]}\lambda_x=1}&=1-2\sum_{x=k+1}^n\lambda_x\;.
\ea
Substituting this into~\eqref{negativity} gives
\be
\N\left(\rho^{AB}\right)=-\sum_{x=k+1}^n\lambda_x=\Big|\sum_{x=k+1}^n\lambda_x\Big|\;.
\ee
That is, the negativity of $\rho^{AB}$ is the absolute value of the sum of all the negative eigenvalues of $\rho^\Gamma$. We can therefore express it also as
\be
\N\left(\rho^{AB}\right)=\tr\left[\rho^\Gamma_-\right]\;.
\ee
where $\rho^\Gamma_-\eqdef\left(\rho^\Gamma\right)_-$ is the negative part of $\rho^\Gamma$.
Note that this also demonstrates that the negativity is zero on separable states. 

\bex
Let $\rho\in\md(AB)$. Show that there exists density matrices $\rho_+,\rho_-\in\md(AB)$ such that $\rho_+\rho_-=\rho_-\rho_+=0$ and
\be\label{ntdeco}
\rho^\Gamma=\Big(1+\N\left(\rho^{AB}\right)\Big)\rho_+^{AB}-\N\left(\rho^{AB}\right)\rho_-^{AB}\;.
\ee
\eex

The decomposition~\eqref{ntdeco} of $\rho^\Gamma$ in the exercise above is optimal in the following sense.
Suppose there exists $\sigma,\tau\in\md(AB)$ such that
\be
\rho^\Gamma=(1+a)\sigma^{AB}-a\tau^{AB}\;,
\ee
for some $a\in\mbb{R}_+$. Then, from~\eqref{ntdeco} we have
\be
\Big(1+\N\left(\rho^{AB}\right)\Big)\rho_+^{AB}-\N\left(\rho^{AB}\right)\rho_-^{AB}=(1+a)\sigma^{AB}-a\tau^{AB}\;.
\ee
Let $\Pi_-$ be the projector to the support of $\rho_-^{AB}$.
Multiplying both sides of the equation above by $\Pi_-$
and taking the trace gives
\ba
-\N\left(\rho^{AB}\right)&=\tr\left[\Pi_-\left((1+a)\sigma^{AB}-a\tau^{AB}\right)\right]\\
&\geq-a\tr\left[\Pi_-\tau^{AB}\right]\\
&\geq -a\;.
\ea
Hence, we must have $a\geq \N(\rho^{AB})$. In other words, we can express the negativity of $\rho^{AB}$ as
\be\label{nropt}
\N\left(\rho^{AB}\right)=\inf\Big\{a\in\mbb{R}\;:\;\exists\;\sigma,\tau\in\md(AB)\;\;s.t.\;\; \rho^\Gamma=(1+a)\sigma^{AB}-a\tau^{AB}\Big\}\;.
\ee

\begin{myt}{}
\begin{theorem}
The negativity measure as defined in~\eqref{negativity} is an entanglement monotone.
\end{theorem}
\end{myt}

\begin{proof}
To prove the strong monotonicity property, Let $\{\mE_x\}_{x\in[m]}$ be a quantum instrument on Alice's system, with each $\mE_x\in\cp(A\to A')$ being trace non-increasing and $\sum_{x\in[m]}\mE_x\in\cptp(A\to A')$. For each $x\in[m]$, denote by $\rho_x^{A'B}\eqdef\frac1{p_x}\mE_x^{A\to A'}\left(\rho^{AB}\right)$, where $p_x\eqdef\tr\left[\mE_x^{A\to A'}\left(\rho^{AB}\right)\right]$. Finally, set $\nu\eqdef \N\left(\rho^{AB}\right)$
By definition we have
\ba
\rho_x^{\Gamma}&=\frac1{p_x}\left(\mE_x^{A\to A'}\left(\rho^{AB}\right)\right)^\Gamma\\
\GG{{\small \substack{Partial\; transpose\\ acts\;on\; Bob's\; side}}}&=\frac1{p_x}\mE_x^{A\to A'}\left(\rho^{\Gamma}\right)\\
\GG{\eqref{ntdeco}}&=\frac{1+\nu}{p_x}\mE_x^{A\to A'}\left(\rho_+^{AB}\right)-\frac{\nu}{p_x}\mE_x^{A\to A'}\left(\rho_-^{AB}\right)\;.
\ea
Since the above decomposition of $\rho_x^\Gamma$ is not necessarily optimal (in the sense of~\eqref{nropt})  we must have
\be
\N\left(\rho_x^{A'B}\right)\leq\frac{\nu}{p_x}\tr\left[\mE_x^{A\to A'}\left(\rho_-^{AB}\right)\right]\;.
\ee 
We therefore get that
\ba
\sum_{x\in[m]}p_x\N\left(\rho_x^{A'B}\right)&\leq\nu\sum_{x\in[m]}\tr\left[\mE_x^{A\to A'}\left(\rho_-^{AB}\right)\right]\\
&=\nu=\N\left(\rho^{AB}\right)\;,
\ea
where we used the fact that $\sum_{x\in[m]}\mE_x$ is trace preserving.
That is, the negativity of entanglement cannot increase on average by a quantum instrument\index{quantum instrument} on Alice's side. Since the negativity is not affected if we take the partial transpose\index{partial transpose} on Alice's system (instead of Bob's), using similar arguments as above, we get that the negativity cannot increase on average under any quantum instrument applied on Bob's side. We therefore conclude that the negativity satisfies the strong monotonicity condition of an entanglement monotone.
It is left to show that the negativity is convex.

Let $\{p_x,\rho_{x}^{AB}\}_{x\in[m]}$ be an ensemble of density matrices in $\md(AB)$. Then, by definition,
\ba
\N\Big(\sum_{x\in[m]}p_x\rho_x^{AB}\Big)&=\frac12\Big\|\Big(\sum_{x\in[m]}p_x\rho_x\Big)^\Gamma\Big\|_1-\frac12\\
&=\frac12\Big\|\sum_{x\in[m]}p_x\rho_x^\Gamma\Big\|_1-\frac12\\
&\leq\frac12\sum_{x\in[m]}p_x\big\|\rho_x^\Gamma\big\|_1-\frac12=\sum_{x\in[m]}p_x\N\left(\rho_x^{AB}\right)\;.
\ea
This completes the proof.
\end{proof}

\subsubsection{The Logarithmic Negativity\index{negativity}}

The negativity has many nice properties, but it is not additive. It turns out, that by a small tweak to its definition in~\eqref{negativity}, we can get an additive measure of entanglement.

\bex
Show that the negativity of a pure bipartite state $\psi\in\pure(AB)$ with $m\eqdef|A|=|B|$ is given by
\be
\N(\psi^{AB})=\sum_{\substack{x<y\\ x,y\in[m]}}\sqrt{p_xp_y}
\ee
where $\{p_{x}\}_{x\in[m]}$ are the Schmidt coefficients of $\psi^{AB}$.
\eex

\begin{myd}{The Logarithmic Negativity\index{negativity}}
\begin{definition}
Let $\rho\in\md(AB)$. The logarithmic negativity  of $\rho^{AB}$ is defined as
\be\label{logneg}
\LN\left(\rho^{AB}\right)=\log\left\|\rho^{\Gamma}\right\|_1\;.
\ee
\end{definition}
\end{myd}

Note that the logarithmic negativity can be expressed as a function of the negativity, namely,
\be
\LN\left(\rho^{AB}\right)=\log\left(2\N\left(\rho^{AB}\right)+1\right)\;.
\ee
Therefore, the logarithmic negativity is a measure of entanglement since the negativity is an entanglement monotone. On the other hand, the logarithmic negativity is not an entanglement monotone, in particular, it is in general not convex (Exercise~\ref{notconvex}). 

The logarithmic negativity is additive under tensor products. To see why, let $\rho\in\md(AB)$ and $\sigma\in\md(A'B')$. Then,
\ba
\LN\left(\rho^{AB}\otimes\sigma^{A'B'}\right)&=\log\left\|(\rho\otimes\sigma)^\Gamma\right\|_1\\
&=\log\left\|\rho^\Gamma\otimes\sigma^\Gamma\right\|_1\\
&=\log\left\|\rho^\Gamma\right\|_1\left\|\sigma^\Gamma\right\|_1=\log\left\|\rho^\Gamma\right\|_1+\log\left\|\sigma^\Gamma\right\|_1\\
&=\LN\left(\rho^{AB}\right)+\LN\big(\sigma^{A'B'}\big)\;.
\ea
We will see later on that the logarithmic negativity provides an upper bound to the distillable entanglement.

\bex\label{notconvex}
Show that the logarithmic negativity is not convex.
\eex

\subsubsection{The $\kappa$-Entanglement}\index{$\kappa$-entanglement}

The $\kappa$-Entanglement is another measure of entanglement that is based on the partial transpose\index{partial transpose}. In Sec.~\ref{sec:npt} we will see that the regularized version of this measure has an operational meaning as the zero-error entanglement cost under PPT operations. The $\kappa$-entanglement is defined for all $\rho\in\md(AB)$ as
\be\label{kappae}
E_\kappa\left(\rho^{AB}\right)=\min_{\Lambda\in\pos(AB)}\left\{\log\tr[\Lambda]\;:\;-\Lambda^\Gamma\leq\rho^\Gamma\leq\Lambda^\Gamma\right\}\;.
\ee
In Sec.~\ref{sec:npt} we will see that $E_\kappa$ behaves monotonically under a set of operations that is larger than LOCC. Moreover, if $\rho\in\ppt(AB)$ then we can take in the equation above $\Lambda=\rho$, so that the $\kappa$-Entanglement vanishes on PPT states and in particular on separable states. Therefore, $E_\kappa$ is a measure of entanglement.
\begin{myg}{}
\begin{lemma}
Let $\rho\in\md(AB)$. Then,
\be\label{ineek}
\LN\left(\rho^{AB}\right)\leq E_\kappa\left(\rho^{AB}\right)\leq \min_{\sigma\in\ppt(AB)}D_{\max}\left(|\rho^\Gamma|\big\|\sigma\right)\;.
\ee
\end{lemma}
\end{myg}
\begin{remark}
From the lemma above it follows that the $\kappa$-entanglement equals the logarithmic negativity if 
\be
|\rho^\Gamma|^\Gamma\geq 0\;.
\ee
To see why, note that in this case we have that the state $\rho_\star\eqdef|\rho^\Gamma|/\|\rho^\Gamma\|_1\in\ppt(AB)$, so by taking $\sigma=\rho_\star$ we get that the upper bound
\ba\label{equalityin}
\min_{\sigma\in\ppt(AB)}D_{\max}\left(|\rho^\Gamma|\big\|\sigma\right)&\leq D_{\max}\left(|\rho^\Gamma|\big\|\rho_\star\right)\\
\GG{Exercise~\ref{lnex}}&=\LN(\rho)\;.
\ea
\end{remark}
\begin{proof}
The condition $-\Lambda^\Gamma\leq\rho^\Gamma\leq\Lambda^\Gamma$ in~\eqref{kappae} can also be expressed as
\be
\Lambda^\Gamma\geq\rho^\Gamma\quad\text{and}\quad\Lambda^\Gamma\geq-\rho^\Gamma\;.
\ee
Combining this with the decomposition~\eqref{ntdeco} of $\rho^\Gamma$ gives the following two inequalities:
\be
\Lambda^\Gamma\geq\big(1+\N\left(\rho\right)\big)\rho_+-\N\left(\rho\right)\rho_-\quad\text{and}\quad
\Lambda^\Gamma\geq \N\left(\rho\right)\rho_--\big(1+\N\left(\rho\right)\big)\rho_+\;.
\ee
Let $\Pi_{\pm}$ be the projections to the supports of $\rho_{\pm}$. Then, from the above equations we get
\be\label{pipm}
\Pi_+\Lambda^\Gamma\Pi_+\geq\big(1+\N\left(\rho\right)\big)\rho_+\quad\text{and}\quad
\Pi_-\Lambda^\Gamma\Pi_-\geq \N\left(\rho\right)\rho_-\;.
\ee
Since $\Lambda^\Gamma\geq 0$ we get
\ba
\tr[\Lambda]=\tr\left[\Lambda^\Gamma\right]&\geq\tr\left[\Lambda^\Gamma\left(\Pi_++\Pi_-\right)\right]\\
&=\tr\left[\Pi_+\Lambda^\Gamma\Pi_+\right]+\tr\left[\Pi_-\Lambda^\Gamma\Pi_-\right]\\
\GG{\eqref{pipm}}&\geq 1+2\N(\rho)=\|\rho^\Gamma\|_1\;.
\ea
Since the above inequality holds for all $\Lambda\in\pos(AB)$ that satisfies $-\Lambda^\Gamma\leq\rho^\Gamma\leq\Lambda^\Gamma$, we conclude that the lower bound in~\eqref{ineek} must hold. 

To get an upper bound observe that for all $\rho\in\md(AB)$
\ba
E_\kappa\left(\rho\right)&\leq\min_{\Lambda\in\pos(AB)}\left\{\log\tr[\Lambda]\;:\;\Lambda^\Gamma\geq|\rho^\Gamma|\right\}\\
\GG{\Lambda=t\sigma}&=\min_{\sigma\in\ppt(AB)}\left\{\log(t)\;:\;t\sigma\geq|\rho^\Gamma|\right\}\\
&=\min_{\sigma\in\ppt(AB)}D_{\max}\left(|\rho^\Gamma|\big\|\sigma\right)\;.
\ea
This completes the proof.
\end{proof}

\bex\label{lnex}
Prove the equality in~\eqref{equalityin}.
\eex

\bex
Show that $E_\kappa$ is subadditive. Hint: Show that if $-\Lambda^\Gamma_j\leq\rho^\Gamma_j\leq\Lambda^\Gamma_j$ for $j=1,2$  then
\be
-\Lambda^\Gamma_1\otimes\Lambda_2^\Gamma\leq\rho^\Gamma_1\otimes\rho_2^\Gamma\leq\Lambda^\Gamma_1\otimes\Lambda_2^\Gamma\;.
\ee
\eex

\subsection{The squashed entanglement}\index{squashed entanglement}

In Exercise~\ref{ex:subadditive} we defined the mutual information\index{mutual information} as
\be\label{mi}
I(A:B)_\rho\eqdef D\left(\rho^{AB}\big\|\rho^A\otimes\rho^B\right)=H(A)_\rho+H(B)_\rho-H(AB)_\rho\;.
\ee
This function quantify the total (i.e. both quantum and classical) amount of correlation between Alice and Bob (see Fig.~\ref{CMI}a).
An extension of this quantity, known as the conditional mutual information\index{mutual information} (CMI), is a function on a tripartite density matrix defined by
\be
I(A:B|R)_\rho\eqdef H(A|R)_\rho+H(B|R)_\rho-H(AB|R)_\rho\quad\quad\forall\;\rho\in\md(ABR)\;.
\ee
Since the CMI is defined with respect to the conditional von-Neumann entropy, it can also be expressed for all $\rho\in\md(ABR)$ as
\ba
I(A:B|R)_\rho&=H(A|R)_\rho+H(BR)_\rho-H(ABR)_\rho\\
&=H(A|R)_\rho-H(A|BR)_\rho\\
\GG{\eqref{condileq}}&\geq 0\;.
\ea
Therefore, the CMI cannot be negative. The CMI quantifies the total correlations between Alice and Bob, given that one has access to a reference system $R$ (see Fig.~\ref{CMI}b for an heuristic description).

\begin{figure}[h]
\centering
    \includegraphics[width=0.9\textwidth]{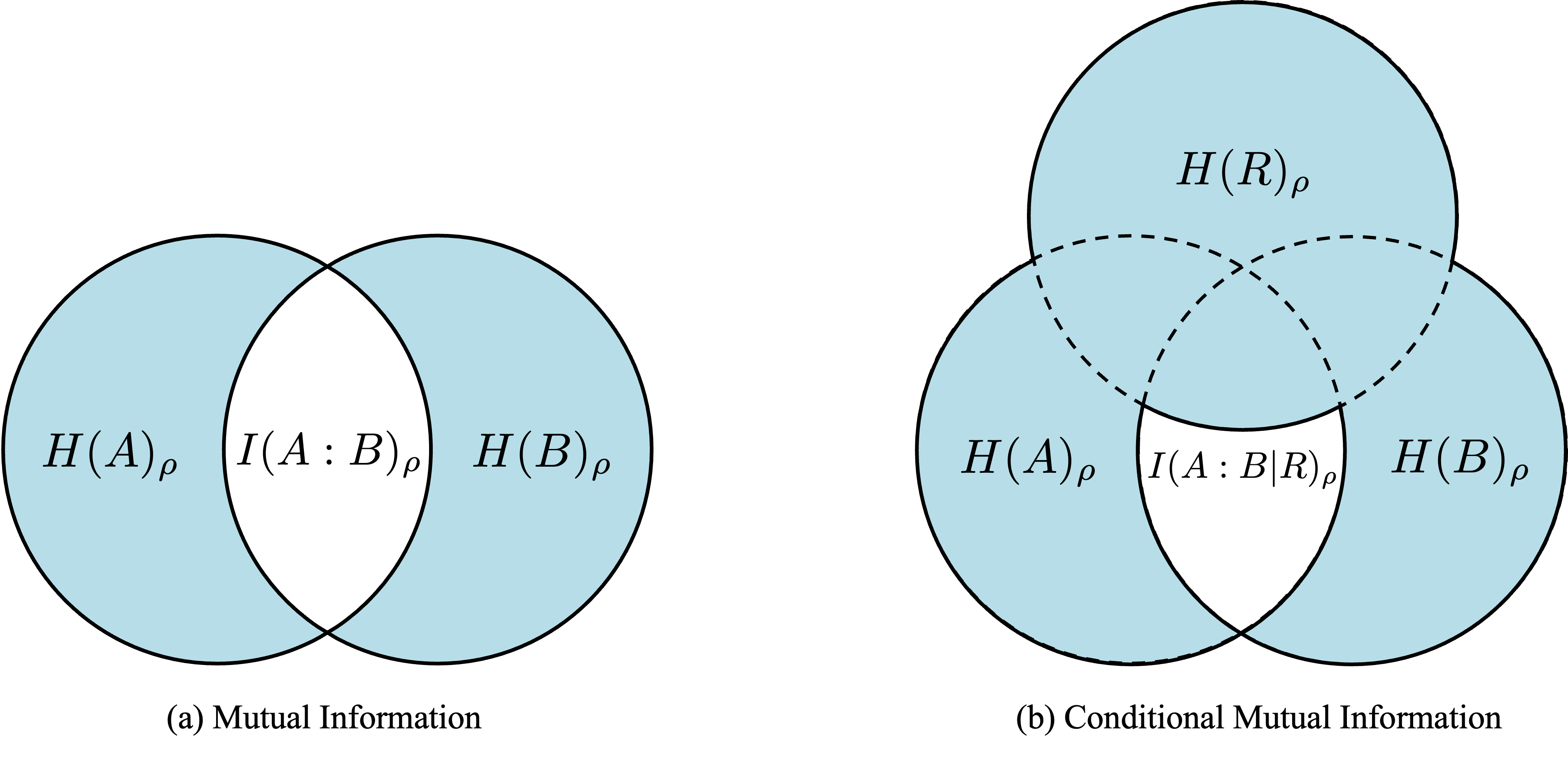}
  \caption{\linespread{1}\selectfont{\small Venn Diagrams. (a) The intersection area (white) illustrates the mutual information\index{mutual information}. (b) The white area illustrates the conditional mutual information.}}
  \label{CMI}
\end{figure} 

\bex[Chain Rule]
Let $\rho\in\md(AA'BB'R)$. 
\ben
\item Show that
\ba\label{chainrule}
&I(AA':B|R)_\rho=I(A':B|R)_\rho+I(A:B|RA')_\rho\\
&I(A:BB'|R)_\rho=I(A:B'|R)_\rho+I(A:B|RB')_\rho\;.
\ea
\item Show that
\be\label{condaaa}
I(AA':B|R)_\rho\geq I(A:B|R)_\rho\;.
\ee
That is, tracing out a local subsystem cannot increase the CMI.
\een
\eex

\bex\label{decocmi}
Show that for any state of the form 
\be
\sigma^{ABRA'}=\sum_{x\in[n]}p_x\sigma^{ABR}_x\otimes |x\lr x|^{A'}\;,
\ee
we have that
\be
I(A:B|RA')_{\sigma}=\sum_{x\in[n]}p_xI(A:B|R)_{\sigma_x}\;.
\ee
\eex

The mutual information\index{mutual information} measures the overall correlations between Alice and Bob, and only takes on the value zero for product states. However, it does not necessarily vanish for separable states that are not products. In contrast, the CMI can be zero even for separable states. For instance, consider the state 
\be\label{rabr}
\rho^{ABR}\eqdef\sum_{x\in[n]}p_x\rho_x^A\otimes\rho_x^B\otimes |x\lr x|^R
\ee
where $p_x$ denotes a probability distribution, $\rho_x^A$ and $\rho_x^B$ are density matrices, and $|x\lr x|^R$ represents a pure state in a third system $R$ that depends on the discrete variable $x$. Although $\rho^{AB}$ is a separable state, its CMI is zero. This is because knowing the value of $x$ through system $R$ allows Alice and Bob to share the product state $\rho_x^A\otimes\rho_x^B$.
The state $\rho^{ABR}$ above belong to a special type of states known as quantum Markov states. 

Quantum Markov states are a type of quantum state that exhibit a special type of correlation structure between different subsystems. In a quantum Markov state, the correlation between two subsystems is entirely mediated by a third subsystem $R$, which serves as a kind of ``bridge" or mediator between $A$ and $B$. This correlation structure is analogous to the Markov property in classical probability theory, where the future state of a system depends only on its present state and not on its past states. 
A quantum Markov state $\rho\in\md(ABR)$ is defined as follows. 

\begin{myd}{}
\begin{definition}\label{markov}
Let $A$, $B$, and $R$ be three quantum systems.
A state $\rho\in\md(ABR)$ is called a quantum Markov state if the following two conditions hold: 
\ben
\item There exists $m\in\mbb{N}$, and two sets of Hilbert spaces $\{R^{(1)}_x\}_{x\in[m]}$ and  $\{R_x^{(2)}\}_{x\in[m]}$ such that $R=\bigoplus_{x\in[m]}R^{(1)}_x\otimes R_x^{(2)}$. 
\item There exists two sets of density matrices $\left\{\rho_x^{AR_x^{(1)}}\right\}_{x\in[m]}$ and $\left\{\rho_x^{BR_x^{(2)}}\right\}_{x\in[m]}$ and a probability vector $\p\in\prob(m)$ such that
\be\label{markovs}
\rho^{ABR}=\bigoplus_{x\in[m]}p_x\rho_x^{AR_x^{(1)}}\otimes \rho_x^{BR_x^{(2)}}\;.
\ee
\een
\end{definition}
\end{myd}

\bex$\;$
\ben
\item Show that the state in~\eqref{rabr} is a quantum Markov state.
\item Show that for any quantum Markov state $\rho\in\md(ABR)$ we have
\be
H(A:B|R)_\rho=0\;.
\ee
Remark: The converse of this statement is also true! That is, any density matrix $\rho\in\md(ABR)$ with zero CMI is necessarily a quantum Markov state.
\een
\eex
The exercise above demonstrate that if $\rho^{AB}$ is a separable state then it has a tripartite extension $\rho^{ABR}$ as in~\eqref{rabr} for which the conditional mutual information\index{mutual information} is zero. This observation motivates the following definition of a measure of entanglement known as the squashed entanglement\index{squashed entanglement}. 

\begin{myd}{The Squashed Entanglement}
\begin{definition}
The squashed entanglement of a bipartite density matrix $\rho\in\md(AB)$ is defined as
\be
E_{\rm sq}\left(\rho^{AB}\right)\eqdef\frac12\inf I(A:B|R)_\omega\;,
\ee
where the infimum is over all finite dimensional systems $R$ and all $\omega\in\md(ABR)$ with marginal $\omega^{AB}=\rho^{AB}$.
\end{definition}
\end{myd}

\begin{remark}
The term ``squashed" entanglement is used because conditioning on a reference system R enables the removal (i.e.\ squash out) of all non-quantum correlations, similar to how conditioning on system R can eliminate classical correlations in separable states. Additionally, since squashed entanglement\index{squashed entanglement} is based on the CMI, it is also known as the CMI entanglement.
\end{remark}

\bex
Show that for $\psi\in\pure(AB)$ we have
\be
E_{\rm sq}\left(\psi^{AB}\right)=H(A)_\psi\;,
\ee
where $H(A)_\psi$ is the von-Neuman entropy of the reduced density matrix on system $A$.
\eex

The squashed entanglement has many desirable properties of measures of entanglement. We start with the fact that it is an entanglement monotone.

\begin{myt}{}
\begin{theorem}
The squashed entanglement is an entanglement monotone.
\end{theorem}
\end{myt}

\begin{proof}
Let $\rho\in\md(AB)$ and let $\omega^{ABR}$ be an extension of $\rho^{AB}$; i.e.\ $\omega^{AB}=\rho^{AB}$.
 Suppose Alice applies on her system $A$ a quantum instrument $\mE\in\cptp(A\to A'A'')$ of the form
\be
\mE^{A\to A'A''}=\sum_{x\in[n]}\mE_x^{A\to A'}\otimes |x\lr x|^{A''}\;.
\ee
Then, the state of the composite system\index{composite system} $ABR$ after Alice's measurement is given by
\be
\sigma^{A'A''BR}\eqdef \mE^{A\to A'A''}\left(\omega^{ABR}\right)\;.
\ee
From Stinespring's dilation theorem there exists an isometry $V:A\to A'A''E$ such that
\be
\sigma^{A'A''BR}=\tr_E\left[V\omega^{ABR}V^*\right]\;.
\ee
Since conditional entropy is invariant under local isometry we get that
\be
\frac12 I(A:B|R)_\omega=\frac12 I(A'A''E:B|R)_{V\omega V^*}\;.
\ee
Combining the equality above with the fact that by tracing out a local subsystem the CMI cannot increase (see~\eqref{condaaa}), we get by tracing out system $E$
\be
\frac12 I(A:B|R)_\omega\geq \frac12 I(A'A'':B|R)_{\sigma}\;.
\ee
Combining this with the chain rule~\eqref{chainrule} gives
\ba
\frac12 I(A:B|R)_\omega&\geq \frac12 I(A'':B|R)_{\sigma}+\frac12 I(A':B|RA'')_{\sigma}\\
\Gg{I(A'':B|R)_{\sigma}\geq 0}&\geq \frac12 I(A':B|RA'')_{\sigma}\\
\GG{Exercise~\ref{decocmi}}&=\frac12 \sum_{x\in[n]}p_xI(A':B|R)_{\sigma_x}\;,
\ea
where $p_x\eqdef\tr\left[\mE^{A\to A'}_x\left(\omega^{ABR}\right)\right]$ and $\sigma^{A'BR}_x\eqdef\frac1{p_x}\mE^{A\to A'}_x\left(\omega^{ABR}\right)$. Finally, by definition, for any $x\in[n]$ we have $\frac12 I(A':B|R)_{\sigma_x}\geq E_{\rm sq}\left(\sigma_x^{A'B}\right)$ so that the inequality above gives
\be
\frac12 I(A:B|R)_\omega\geq \sum_{x\in[n]}p_xE_{\rm sq}\left(\sigma_x^{A'B}\right)\;.
\ee
Since $\omega^{ABR}$ was an arbitrary extension of $\rho^{AB}$ we conclude that
\be
E_{\rm sq}\left(\rho^{AB}\right)\geq \sum_{x\in[n]}p_xE_{\rm sq}\left(\sigma_x^{A'B}\right)\;.
\ee
In other words, the squashed entanglement\index{squashed entanglement} does not increase on average under any local quantum instrument\index{quantum instrument} on system $A$. From symmetry, the same holds for any quantum instrument on Bob's system. Therefore, the squashed entanglement satisfies the strong monotonicity property of an entanglement monotone.

To prove the convexity of $E_{\rm sq}$, let $\rho_1,\rho_2\in\md(AB)$ and $t\in[0,1]$. Let $\omega^{ABR}_1$ and $\omega_2^{ABR}$ be extensions of $\rho_1^{AB}$ and $\rho_2^{AB}$, respectively. Note that in general we can always assume that these extensions has the same reference system $R$ as otherwise we embed the lower dimensional reference system in the higher dimensional one.
Finally, let $R'$ be a qubit system and denote by
\be
\omega^{ABRR'}\eqdef\omega^{ABR}_1\otimes |0\lr 0|^{R'}+(1-t)\omega^{ABR}_2\otimes|1\lr 1|^{R'}\;.
\ee
Since $\omega^{AB}=t\rho^{AB}+(1-t)\sigma^{AB}$ we get that
\ba
E_{\rm sq}\left(t\rho^{AB}+(1-t)\sigma^{AB}\right)&\leq I\left(A:B|RR'\right)_\omega\\
\GG{Exercise~\ref{decocmi}}&=tI\left(A:B|R\right)_{\omega_1}+(1-t)I\left(A:B|R\right)_{\omega_2}\;.
\ea
Since the extensions $\omega_1^{ABR}$ and $\omega_2^{ABR}$ were arbitrary, we conclude that
\be
E_{\rm sq}\left(t\rho^{AB}+(1-t)\sigma^{AB}\right)\leq tE_{\rm sq}\left(\rho^{AB}\right)+(1-t)E_{\rm sq}\left(\sigma^{AB}\right)\;.
\ee
This completes the proof.
\end{proof}

Another interesting property of the squashed entanglement\index{squashed entanglement} is that it is additive. 
\begin{myt}{}
\begin{theorem}
Let $\rho\in\md(AA'BB')$. Then,
\be\label{esqsup}
E_{\rm sq}\left(\rho^{AA'BB'}\right)\geq E_{\rm sq}\left(\rho^{AB}\right)+E_{\rm sq}\left(\rho^{A'B'}\right)\;,
\ee
with equality if $\rho^{AA'BB'}=\rho^{AB}\otimes\rho^{A'B'}$.
\end{theorem}
\end{myt}
\begin{proof}
Let $\omega\in\md(AA'BB'R)$ be a quantum extension of $\rho^{AA'BB'}$. Then, by applying the chain rule in~\eqref{chainrule} one time with respect to Alice's systems and one time with respect to Bob's systems we get
\begin{align}\label{slio}
&I(AA':BB'|R)_\omega=I(A':BB'|R)_\omega+I(A:BB'|RA')_\omega\nonumber\\
\GG{\eqref{chainrule}}&=I(A':B'|R)_\omega+I(A':B|RB')_\omega+I(A:B'|RA')_\omega+I(A:B|RA'B')_\omega\nonumber\\
\GG{CMI\geq 0}&\geq I(A':B'|R)_\omega+I(A:B|RA'B')_\omega\nonumber\\
\GG{By\;definition}&\geq 2E_{\rm sq}\left(\rho^{AB}\right)+2E_{\rm sq}\left(\rho^{A'B'}\right)\;.
\end{align} 
Since $\omega^{AA'BB'R}$ was an arbitrary extension of $\rho^{AA'BB'}$ we conclude that the above inequality implies~\eqref{esqsup}. 

It is left to show that $\rho^{AA'BB'}=\rho^{AB}\otimes\rho^{A'B'}$ we have equality in~\eqref{esqsup}. Let $\omega^{ABR_1}$ and $\omega^{A'B'R_2}$ be extensions of $\rho^{AB}$ and $\rho^{A'B'}$, respectively. Set $R=R_1R_2$ and $\omega^{AA'BB'R}\eqdef\omega^{ABR_1}\otimes\omega^{A'B'R_2}$. In Exercise~\ref{exekal} you show that for such $\omega^{AA'BB'R}$ we have
\be\label{eqkal}
I(A':B|RB')_\omega=I(A:B'|RA')_\omega=0\;.
\ee
Combining this with the second equality in~\eqref{slio} we get
\be
I(AA':BB'|R)_\omega=I(A':B'|R)_\omega+I(A:B|RA'B')_\omega\;.
\ee
Finally, since $\omega^{A'B'R}=\omega^{R_1}\otimes\omega^{A'B'R_2}$  we have $I(A':B'|R)_\omega=I(A':B'|R_2)_\omega$ as conditioning by an additional independent system $R_1$ does not change the CMI. Similarly, 
\be\label{poiu}
I(A:B|RA'B')_\omega=I(A:B|R_1)_\omega\;.
\ee
We therefore get that for any extensions $\omega^{ABR_1}$ and $\omega^{A'B'R_2}$ of $\rho^{AB}$ and $\rho^{A'B'}$ we have
\ba
E_{\rm sq}\left(\rho^{AB}\otimes\rho^{A'B'}\right)&\leq\frac12I(AA':BB'|R)_\omega\\
&=\frac12I(A':B'|R_1)_\omega+\frac12I(A:B|R_2)_\omega\;.
\ea
Since the above inequality holds of any such extensions of $\rho^{AB}$ and $\rho^{A'B'}$, we must have
\be
E_{\rm sq}\left(\rho^{AB}\otimes\rho^{A'B'}\right)\leq E_{\rm sq}\left(\rho^{AB}\right)+E_{\rm sq}\left(\rho^{A'B'}\right)\;.
\ee
Combining this with~\eqref{esqsup} gives an equality.
This completes the proof.
\end{proof}

\bex\label{exekal}
Prove~\eqref{eqkal} and~\eqref{poiu}.
\eex

\subsection{Coherent Information of Entanglement}\label{sec:cioe}\index{coherent information}

In this section, we will explore a measure of entanglement that exhibits monotonic behavior under one-way LOCC. While this measure may not exhibit monotonicity under arbitrary LOCC, it can still be a valuable tool for providing bounds on the distillable entanglement of mixed bipartite states, as we will see in the next section.

The most general one-way LOCC operation that  Alice and Bob can perform is for Alice to apply a quantum instrument $\{\mE_x\}_{x\in[m]}$, with $\mE_x\in\cp(A\to A')$ and $\sum_{x\in[m]}\mE_x\in\cptp(A\to A')$, send the outcome $x$ to Bob, who then applies a quantum channel $\mF_{x}\in\cptp(B\to B')$ that depends on the outcome $x$ received from Alice. The overall operation can be described by the quantum channel
\be\label{13111c}
\mN^{AB\to A'B'}\eqdef\sum_{x\in[m]}\mE_x^{A\to A'}\otimes\mF_x^{B\to B'}\;.
\ee

\begin{myd}{}
\begin{definition}
Let $\rho\in\pure(ABE)$. The coherent information\index{coherent information} of the marginal state $\rho^{AB}$ is defined as
\be
I(A\ra B)_\rho\eqdef -H(A|B)_\rho=H(A|E)_\rho\;,
\ee
where the second equality is due to the duality\index{duality} relation of the conditional von-Neumann\index{von-Neumann} entropy given in~\eqref{sduals}. Moreover, the \emph{coherent information of entanglement} of the state $\rho^{AB}$ is defined as
\be\label{13p149}
E_{\to}\left(\rho^{AB}\right)\eqdef\sup_{\mE\in\cptp(A\to AX)}I\big(A\ra BX\big)_{\mE(\rho)}\;,
\ee
where the supremum is also over all finite dimensions of the classical system $X$.
\end{definition}
\end{myd}

\bex
Let $\rho^{ABX}\eqdef\sum_{x\in[m]}p_x\rho^{AB}_x\otimes|x\lr x|^X$ be a cq-state\index{cq-state} in $\md(ABX)$. 
\ben
\item Show that
\be
I(A\ra BX)_\rho=\sum_{x\in[m]}p_xI(A\ra B)_{\rho_x}\;.
\ee
\item Show that the coherent information is convex. That is, prove that
\be\label{chconvex}
I(A\ra B)_\rho\leq\sum_{x\in[m]}p_xI(A\ra B)_{\rho_x}\;.
\ee
Hint: Use the joint convexity of the relative entropy.
\item Show that for every quantum channel $\mF\in\cptp(B\to B')$ we have
\be\label{chmono}
I(A\ra B')_{\mF(\rho)}\leq I(A\ra B)_\rho\;.
\ee
Hint: Either use the DPI directly, or recall that a single channel on Bob's system is a conditionally mixing operation. 
\een
\eex
\bex\label{sko}
Show that the supremum in~\eqref{13p149} can be restricted quantum channels of the form $\mE^{A\to AX}=\sum_{x\in[n]}\mE_x^{A\to A}\otimes |x\lr x|^X$, where each $\mE_x^{A\to A}$ is a CP map with a single Kraus operator. Hint: Use the joint convexity of the relative entropy.
\eex

\bex\label{spex}
Let $\rho\in\md(AB)$.
\ben
\item Show that $E_\to(\rho^{AB})\geq 0$ with equality if $\rho^{AB}$ is separable.
\item Show that $E_\to(\rho^{AB})\geq I(A\ra B)_{\rho}$.
\een
\eex

Note that while the supremum in the above definition is taken over all dimensions of $X$, we consider only quantum instruments from $A$ to $A$. However, this limitation is not necessary, as the coherent information\index{coherent information} of entanglement for the state $\rho^{AB}$ can be defined as
\be\label{13ppp}
E_{\to}\left(\rho^{AB}\right)\eqdef\sup_{\mE\in\cptp(A\to A'X)}I\big(A'\ra BX\big)_{\mE(\rho)}\;,
\ee
with the supremum extending over all dimensions of system $A'$. To understand this, first consider that if $|A'|\leq |A|$, every channel $\mE\in\cptp(A\to A'B)$ can be embedded in $\cptp(A\to AX)$, as the coherent information is invariant under local isometries (a property shared by all conditional entropies).
Consequently, in this case, the supremum over $\cptp(A\to AX)$ is at least as great as that over $\cptp(A\to A'X)$. 

Conversely, if $|A'|> |A|$, consider a quantum instrument $\mE^{A\to A'X}=\sum_{x\in[n]}\mE_x^{A\to A'}\otimes |x\lr x|^X$, where each $\mE_x^{A\to A'}(\cdot)=M_x(\cdot)M_x^*$ is a CP map with a single Kraus operator $M_x:A\to A'$. Through polar decomposition\index{polar decomposition} , each $M_x$ can be written as $M_x=V_xN_x$, with each $N_x:A\to A$ being part of a generalized measurement, and each $V_x:A\to A'$ an isometry. Due to the invariant property of coherent information\index{coherent information} under isometries, the CP maps $\mE_x^{A\to A'}(\cdot)=M_x(\cdot)M_x^*$ can be substituted with $\mN_x^{A\to A}(\cdot)=N_x(\cdot)N_x^*$, allowing the optimization over all channels in $\cptp(A\to A'X)$ to be replaced with optimization over all quantum instruments in $\cptp(A\to AX)$.

This observation is significant as it can be used to prove that the coherent information\index{coherent information} of entanglement exhibits monotonic behavior under one-way LOCC.

\begin{myt}{}
\begin{theorem}
Let $\rho\in\md(AB)$ and $\mN\in\locc_1(AB\to A'B')$. Then,
\be
E_{\to}\left(\mN^{AB\to A'B'}\left(\rho^{AB}\right)\right)\leq E_{\to}\left(\rho^{AB}\right)\;.
\ee
\end{theorem}
\end{myt}
\begin{proof}
For every quantum instrument $\mE\in\cptp(A'\to A'X)$, $\mE\circ\mN\in\locc_1(AB\to A'B'X)$. Therefore, 
\ba\label{13p153}
E_{\to}\left(\mN^{AB\to A'B'}\left(\rho^{AB}\right)\right)&=\sup_{\mE\in\cptp(A'\to A'X)}I\big(A'\ra B'X\big)_{\mE\circ\mN(\rho)}\\
&\leq \sup_{\mM\in\locc_1(AB\to A'B'X)}I\big(A'\ra B'X\big)_{\mM(\rho)}\;,
\ea
where we replaced $\mE\circ\mN$ with arbitrary $\mM\in\locc_1(AB\to A'B'X)$.
Now, recall that every element of $\locc_1(AB\to A'B'X)$ can be expressed as
\be
\mM^{AB\to A'B'X}\eqdef\sum_{y\in[n]}\mE^{A\to A'X}_y\otimes\mF_y^{B\to B'}\;,
\ee
with each $\mE_y^{A\to A'X}$ being a CP map such that $\sum_{y\in[n]}\mE_y\in\cptp(A\to A'X)$, and each $\mF_y\in\cptp(B\to B')$. 
Thus,
\be
\mM^{AB\to A'B'X}\left(\rho^{AB}\right)=\sum_{y\in[n]}q_y\mF^{B\to B'}_y\left(\sigma_y^{A'BX}\right)\;,
\ee
where $\sigma_y^{A'BX}\eqdef\frac1{q_y}\mE_y^{A\to A'X}\left(\rho^{AB}\right)$ and 
$q_y\eqdef\tr\left[\mE_y^{A\to A'}\left(\rho^{AB}\right)\right]$.
Combining this with the convexity of the coherent information (see~\eqref{chconvex}) we get from~\eqref{13p153} that
\ba
E_{\to}\left(\mN^{AB\to A'B'}\left(\rho^{AB}\right)\right)&\leq \sup_{\mM\in\locc_1}\sum_{y\in[n]}q_yI\big(A'\ra B'X\big)_{\mF_y(\sigma_y)}\\
\GG{cf.~\eqref{chmono}}&\leq \sup_{\mM\in\locc_1}\sum_{y\in[n]}q_yI\big(A'\ra B'X\big)_{\sigma_y}\;.
\ea
Finally, denoting by $Z\eqdef XY$, and by $\mE^{A\to A'Z}\eqdef\sum_{y\in[n]}\mE_y^{A'\to A'X}\otimes |y\lr y|^Y$ we conclude that
\ba
E_{\to}\left(\mN^{AB\to A'B'}\left(\rho^{AB}\right)\right)&\leq \sup_{\mE\in\cptp(A\to A'Z)}I\big(A'\ra BZ\big)_{\mE(\rho)}\\
\GG{\eqref{13ppp}}&=E_{\to}\left(\rho^{AB}\right)\;.
\ea
This completes the proof.
\end{proof}

\bex
Show that $E_{\to}$ is an entanglement monotone under $\locc_1$. That is, prove the strong monotonicity property and convexity.
\eex

\bex\label{esa}
Let $\Phi_m\in\md(AB)$ be the maximally entangled state with $m\eqdef|A|=|B|$. Show that
\be
E_\to\left(\Phi_m^{AB}\right)=\log(m)\;.
\ee 
\eex

The coherent information\index{coherent information} of entanglement is superadditive. That is, for any $\rho\in\md(AB)$ and $\sigma\in\md(A'B')$ we have (see Exercise~\ref{subadco})
\be\label{esubadco}
E_\to\left(\rho^{AB}\otimes\sigma^{A'B'}\right)\geq E_\to\left(\rho^{AB}\right)+E_\to\left(\sigma^{A'B'}\right)\;.
\ee
From Exercise~\ref{ex:reg0} it then follows that the limit in
\be\label{13p162}
E^{\reg}_\to\left(\rho^{AB}\right)\eqdef\lim_{n\to\infty}\frac1nE_\to\left(\rho^{\otimes n}\right)
\ee
exists. We will see in the next section that this regularize coherent information of entanglement has an operational meaning as the one-way distillable entanglement of $\rho^{AB}$.

\bex\label{subadco}
Prove the superadditivity of the coherent information of entanglement as given in~\eqref{esubadco}.
\eex

\section{The Conversion Distance}

The conversion distance in entanglement theory is given by (cf.~\eqref{cd})
\be\label{entcd}
T\left(\rho^{AB}\xrightarrow{\text{\tiny LOCC}} \sigma^{A'B'}\right)\eqdef\frac12\min_{\mN\in\locc}\left\|\mN^{AB\to A'B'}\left(\rho^{AB}\right)-\sigma^{A'B'}\right\|_1\;.
\ee
Computing the above quantity in general is a highly challenging task, so we often rely on establishing lower and upper bounds. In this section, we will narrow our focus to the special cases where either $\rho$ or $\sigma$ is maximally entangled. Recall that these cases are particularly relevant for calculating entanglement distillation and entanglement cost.

\subsection{conversion distance\index{conversion distance} to a Maximally Entangled State}

We start with the following simplification of the expression given in~\eqref{entcd} when $\sigma^{A'B'}$ is maximally entangled.

\begin{myg}{}
\begin{lemma}\label{lem1331}
Let $\rho\in\md(AB)$, and $\Phi_m\in\md(A'B')$ be the maximally entangled state with $m\eqdef|A'|=|B'|$. Then, 
\be\label{13102}
T\left(\rho\xrightarrow{\text{\tiny LOCC}}  \Phi_m\right)=P^2\left(\rho\xrightarrow{\text{\tiny LOCC}}  \Phi_m\right)
=1-\sup_{\mN}\tr\left[\Phi_m\mN\left(\rho\right)\right]\;,
\ee
where the supremum is over all $\mN\in\locc(AB\to A'B')$, and $P$ is the purified distance as given in~\eqref{cf6255}.
\end{lemma}
\end{myg}
\begin{remark}
In Sec.~\ref{sec:cdpures}, we explored various conversion distances among pure bipartite states. It was established that the $P_\star$-conversion distance is equal to the $P$-conversion distance. Additionally, we speculated, albeit without formal proof, that the $T$-conversion distance might be strictly smaller than the $P$-conversion distance.
The lemma above confirms this speculation by demonstrating that, when the target state is $\Phi_m$, the $T$-conversion distance actually aligns with the $P^2$-conversion distance, which is indeed strictly smaller than the $P$-conversion distance. This outcome is based on the understanding that the purified distance is no greater than one; thus, squaring it effectively reduces its magnitude.\end{remark}
\begin{proof}
Let $\mG\in\locc(A'B'\to A'B')$ be the twirling\index{twirling} map given in~\eqref{gtwi2}. That is, for any $\omega\in\md(A'B')$ 
\ba\label{formmg}
\mG\left(\omega\right)&\eqdef\int_{\muu(m)}dU\;(U\otimes\overline{U})\omega(U\otimes\overline{U})^*\\
\GG{\eqref{newform2}}&=\left(1-\tr\left[\Phi_m\omega\right]\right)\tau+\tr\left[\Phi_m\omega\right]\Phi_m\;,
\ea
where $\tau\in\md(A'B')$ is given by $\tau=(I-\Phi_m)/(m^2-1)$.
In particular, observe that for all $\omega\in\md(A'B')$
\be\label{13104}
\frac12\left\|\mG\left(\omega\right)-\Phi_m\right\|_1=\big(1-\tr\left[\Phi_{m}\omega\right]\big)\frac12\left\|\tau-\Phi_{m}\right\|_1=1-\tr\left[\Phi_{m}\omega\right]\;,
\ee
where the last equality follows from the fact that $\tau\Phi_m=\Phi_m\tau=0$.
From the DPI of the trace distance, and the invariance\index{invariance} of $\Phi_{m}$ under the twirling\index{twirling} map $\mG$, it follows that for all $\mN\in\locc(AB\to A'B')$ and all $\rho\in\md(AB)$
\be\label{13105}
\frac12\left\|\mN\left(\rho\right)-\Phi_m\right\|_1\geq \frac12\left\|\mG\circ\mN\left(\rho\right)-\Phi_m\right\|_1\;.
\ee
Since $\mG\circ\mN$ is also an LOCC channel it follows from the inequality above that the conversion distance can be expressed as
\ba 
T\left(\rho\xrightarrow{\text{\tiny LOCC}}  \Phi_m\right)&=\inf_{\mN\in\locc}\frac12\left\|\mG\circ\mN\left(\rho\right)-\Phi_m\right\|_1\\
\GG{\eqref{13104}}&=1-\sup_{\mN\in\locc}\tr\left[\Phi_{m}\mN\left(\rho\right)\right]\;.
\ea
This completes the proof.
\end{proof}

\bex
Let $k\eqdef|A|=|B|$, $m=|A'|=|B'|$, and $\rho\in\md(AB)$. 
Show that 
\be
T\left(\rho^{AB}\xrightarrow{\text{\tiny LOCC}}  \Phi^{A'B'}_m\right)\geq 1-\frac{k}{m}\;.
\ee
Note that this bound is not trivial for $k<m$. Hint: Estimate $T\left(\Phi^{AB}_k\xrightarrow{\text{\tiny LOCC}}  \Phi^{A'B'}_m\right)$.
\eex

\subsubsection{conversion distance\index{conversion distance} Under One-Way LOCC}

The lemma's expression for $T(\rho\xrightarrow{\text{\tiny LOCC}} \Phi_m)$ is complex, making its computation challenging. One contributing factor is the complexity inherent in LOCC. To simplify this, we begin by restricting the channel $\mN$ in \eqref{13102} to one-way LOCC, which yields an upper bound on the conversion distance $T(\rho\xrightarrow{\text{\tiny LOCC}} \Phi_m)$. More explicitly, the one-way conversion distance is defined as:
\be\label{tlocc1}
T\left(\rho\xrightarrow{\text{\tiny LOCC}_1}  \Phi_m\right)=1-\sup_{\mN}\tr\left[\Phi_m\mN\left(\rho\right)\right]\;,
\ee
where the supremum is over all $\mN\in\locc_1(AB\to A'B')$. It is evident that since $\locc_1$ is a subset of $\locc$, we have the following inequality for all $\rho\in\md(AB)$:
\be
T\left(\rho\xrightarrow{\text{\tiny LOCC}} \Phi_m\right)\leq T\left(\rho\xrightarrow{\text{\tiny LOCC}_1}\Phi_m\right)\;.
\ee
Given that one-way LOCC is significantly easier to characterize than LOCC, we can represent  the conversion distance $T(\rho\xrightarrow{\text{\tiny LOCC}_1} \Phi_m)$ as an optimization problem over quantum instruments (see the lemma below). This simplification is advantageous because it reduces the complexity involved in the calculation and allows for a more straightforward analysis of the conversion distance. By focusing on one-way LOCC, we limit the operations to a sequence where one party, say Alice, performs a quantum operation and communicates the outcome classically to the other party (Bob), who then performs a quantum operation based on that information. This constraint narrows down the set of operations to be considered in the optimization problem, making the task of determining the conversion distance\index{conversion distance} more manageable and conceptually clearer.

In the following lemma we relate between the conversion distance under one-way LOCC and the optimized conditional min-entropy\index{conditional min-entropy} $H_{\min}^\ua$ as defined in~\eqref{defcmine}. The relation will be given in terms of the function 
\be\label{defqmin}
Q_{\min}(A'|BX)_\tau\eqdef2^{-H_{\min}^\ua(A'|BX)_\tau}\quad\quad\forall\tau\in\md(A'BX)\;,
\ee
with $X$ being a classical system and $A'$ and $B$ are quantum systems.
\bex\label{excheck}
Let $\rho\in\md(AB)$ and $\mE^{A\to A'X}\eqdef\sum_{x\in[k]}\mE_x^{A\to A'}\otimes |x\lr x|^X$ be a quantum instrument. Show that
\be
Q_{\min}(A'|BX)_{\mE(\rho)}=\sum_{x\in[k]}Q_{\min}(A'|B)_{\mE_x(\rho)}\;,
\ee
where we extended the definition of $Q_{\min}$ to subnormalized states\index{subnormalized states} such that for any  $\sigma\in\md_{\leq}(A'B)$ 
\ba
Q_{\min}(A'|B)_{\sigma}&\eqdef 2^{-H_{\min}^\ua(A'|B)_{\sigma}}\\
&\eqdef\min\Big\{\tr\left[\Lambda^B\right]\;:\;I^{A'}\otimes\Lambda^B\geq\sigma^{A'B},\;\Lambda\in\pos(B)\Big\}\;.
\ea
\eex

\begin{myg}{}
\begin{lemma}\label{lemjjj}
Let $\rho\in\md(AB)$, and $\Phi_m\in\md(A'B')$ be the maximally entangled state ($m\eqdef|A'|=|B'|$). Then, 
\be\label{13p183}
T\left(\rho\xrightarrow{\text{\tiny LOCC}_1} \Phi_m\right)=1-\frac1m\sup_{\mE\in\cptp(A\to A'X)}Q_{\min}(A'|BX)_{\mE(\rho)}\;,
\ee
where $Q_{\min}$ is defined in~\eqref{defqmin}.
\end{lemma}
\end{myg}

\begin{remark}
Note that $Q_{\min}(A'|BX)_{\mE(\rho)}$ depends on $m$ as $m\eqdef|A'|$. Replacing $\cptp(A\to A'X)$ in~\eqref{13p183} with $\cptp(A\to AX)$, we obtain a lower bound (see Exercise~\ref{lbcd}):
\be\label{lbcde}
T\left(\rho\xrightarrow{\text{\tiny LOCC}_1} \Phi_m\right)\geq1-\frac1m\sup_{\mE\in\cptp(A\to AX)}Q_{\min}(A|BX)_{\mE(\rho)}\;.
\ee
\end{remark}

\begin{proof}
Every $\mN\in\locc_1(AB\to A'B')$ can be expressed as:
\be
\mN^{AB\to A'B'}=\mF^{BX\to B'}\circ\mE^{A\to A'X}\;,
\ee
where $X$ is a classical system, and $\mE^{A\to A'X}$ and $\mF^{BX\to B'}$ are quantum channels.  The channel $\mE^{A\to A'X}=\sum_{x\in[k]}\mE_x^{A\to A'}\otimes |x\lr x|^X$ can be viewed as a quantum instrument with $k=|X|$, where we can assume without loss of generality that each $\mE_x^{A\to A'}(\cdot)=M_x(\cdot)M_x^*$ has a single Kraus operator $M_x$ by increasing the dimension of $X$. Thus,
the optimization in~\eqref{tlocc1} over all one-way LOCC channels, $\mN^{AB\to A'B'}$, can be decomposed into optimizations over all $X$, all $\mE^{A\to A'X}$, and all $\mF^{A\to A'X}$. Next, we optimize over $\mF^{A\to A'X}$, while keeping $X$ and $\mE^{A\to A'X}$ fixed. Denoting by $\sigma^{A'BX}\eqdef\mE^{A\to A'X}\left(\rho^{AB}\right)$, we get:
\ba
\max_{\mF\in\cptp}\tr\left[\Phi_m^{A'B'}\mF^{BX\to B'}\circ\mE^{A\to A'X}\left(\rho^{AB}\right)\right]&=\max_{\mF\in\cptp}\tr\left[\Phi_m^{A'B'}\mF^{BX\to B'}\left(\sigma^{A'BX}\right)
\right]\\
\GG{\eqref{r288}}&=\frac1m2^{-H_{\min}^\ua\left(A'|BX\right)_\sigma}\;,
\ea
where we used~\eqref{r288} with $\mF^{BX\to B'}$ replacing $\mE^{B\to \tA}$. Substituting this into~\eqref{tlocc1} we get that the one-way LOCC conversion distance is given as in~\eqref{13p183}.
\end{proof}

\bex\label{lbcd}
Prove~\eqref{lbcde}. Hint: Use the property that any conditional entropy is invariant under local isometries (particularly on Alice's system).
\eex

Building on Lemma~\eqref{lemjjj} and Exercise~\ref{excheck}, the conversion distance can be re-expressed as follows:
\be
T\left(\rho\xrightarrow{\text{\tiny LOCC}_1} \Phi_m\right)=1-\frac1m\sup_{\mE\in\cptp(A\to A'X)}\sum_{x\in[k]}Q_{\min}(A'|B)_{\mE_x(\rho)}\;.
\ee
In Lemma~\ref{lem:hmin1/2}, we established a connection between the optimized conditional min-entropy\index{conditional min-entropy} and the square fidelity. Specifically, let $E$ as be a purifying system and $\rho^{ABE}$ be a purification of $\rho^{AB}$. The marginal of the (subnormalized) \emph{pure} state $\mE_x^{A\to A'}(\rho^{ABE})$ is denoted as $\mE_x^{A\to A'}(\rho^{AE})$. Recall that we can assume each $\mE_x^{A\to A'}$ in the supremum comprises a single Kraus operator. Hence, the relation in Lemma~\ref{lem:hmin1/2} implies:
\be
Q_{\min}(A'|B)_{\mE_x(\rho)}=\max_{\tau\in\md(E)}F^2\left(I^{A'}\otimes\tau^{E},\mE_x^{A\to A'}(\rho^{AE})\right)\;.
\ee
Combining this with the relation $P^2=1-F^2$ between the purified distance and the fidelity, we conclude that~\eqref{13p183} can also be rewritten as:
\be\label{13102b}
T\left(\rho^{AB}\xrightarrow{\text{\tiny LOCC}_1} \Phi_m\right)=\inf_{\{\mE_x\}}\sum_{x\in[k]}\min_{\tau\in\md(E)}P^2\left(\u^{A'}\otimes\tau^{E},\mE_x^{A\to A'}(\rho^{AE})\right)
\;,
\ee
where the infimum is over all $k\in\mbb{N}$ and all quantum instruments $\{\mE_x^{A\to A'}\}_{x\in[k]}$. We will use this form of the conversion distance to get the following upper bound.

\subsubsection{Upper Bound}

The main result of this subsection is the following upper bound on the right-hand side of the equation above. 

\begin{myt}{\color{yellow} Upper Bound}
\begin{theorem}\label{upb1way}
Let $\rho\in\pure(ABE)$ and $m\in\mbb{N}$. Then, 
\be\label{13p171}
T\left(\rho^{AB}\xrightarrow{\text{\tiny LOCC}_1} \Phi_m\right)\leq \sqrt{m}2^{-\frac{1}{2}\tH_2^\ua(A|E)_\rho}\;,
\ee
where $\tH_2^\ua(A|E)_\rho\eqdef-\min_{\omega\in\md(E)}\tD_2\left(\rho^{AE}\|I^{A}\otimes\rho^E\right)$ is the optimized conditional entropy as defined with respect to the quantum Sandwich divergence of order $\alpha=2$.
\end{theorem}
\end{myt}
\begin{remark}
When $m\geq |A|$, the upper bound above is trivial, since in this case
\be
\sqrt{m}2^{-\frac{1}{2}\tH_2^\ua(A|E)\rho}=2^{\frac{1}{2}\left(\log (m)-\tH_2^\ua(A|E)\rho\right)}\geq 1\;,
\ee
where we used the fact that $\tH_2^\ua(A|E)_\rho\leq\log|A|\leq\log(m)$. However, as we will soon see, this upper bound is very useful when $|A|> m$. Specifically, we will use it to derive a tight lower bound on the distillable entanglement.
\end{remark}

\begin{proof}
We get the upper bound on the conversion distance\index{conversion distance} in two stages. First, by taking $\tau^E=\rho^E$ in~\eqref{13102b} we get 
\ba\label{wqs}
T\left(\rho^{AB}\xrightarrow{\text{\tiny LOCC}_1} \Phi^{A'B'}_m\right)
\leq\inf_{\{\mE_x\}}\sum_{x\in[k]}P^2\left(\u^{A'}\otimes\rho^{E},\mE_x^{A\to A'}(\rho^{AE})\right)\;.
\ea
Second, we replace the maximization above over all quantum instruments $\{\mE_x\}_{x\in[k]}$ with a specific choice of a quantum instrument\index{quantum instrument} to get a simpler upper bound. We will denote by $n\eqdef|A|$ and assume that $m\leq n$ (see the remark above).

Observe that the expression in~\eqref{wqs} has a form that is somewhat similar to the decoupling theorem\index{decoupling theorem}  studied in Sec.~\ref{sec:decoupling}.  Therefore, our strategy is to choose $\{\mE_x\}_{x\in[k]}$ in such a way that we will be able to use the upper bound given in the decoupling theorem. For this purpose, recall that the twirling\index{twirling} operation $\mG\in\cptp(A\tA\to A\tA)$ as defined in~\eqref{twirlingw} can be express as a finite convex combination of product unitary channels as given in~\eqref{7222}. With these $k\in\mbb{N}$, $\p\in\prob(k)$, and $\{U_x\}_{x\in[k]}\subset\muu(A)$, we define
\be
\mE_x^{A\to A'}\eqdef p_x\; \mN^{A\to A'}\circ\mU_x^{A\to A}\;,
\ee
where $\mU_x(\cdot)=U_x(\cdot)U_x^*$, and $\mN^{A\to A'}(\cdot)\eqdef\frac{n}{m}V^*(\cdot)V$, where $V:A'\to A$ is some isometry. 

We now discuss the properties of the set $\{\mE_x^{A\to A'}\}_{x\in[k]}$. First, observe that by definition, the channel
\be
\mR^{A\to A}\eqdef\sum_{x\in[k]}p_x\;\mU_x^{A\to A}\;,
\ee 
corresponds to the completely randomizing channel that outputs the maximally mixed state irrespective on the input state. This follows from the fact that both~\eqref{twirlingw} and~\eqref{7222} corresponds to the same twirling channel, so their marginal channels are also the same (see~\eqref{8249}). With this at hand, we get that
\be
\mE^{A\to A'}\eqdef \sum_{x\in[k]}\mE_x^{A\to A'}=\mN^{A\to A'}\circ\mR^{A\to A}\;.
\ee
Next, we argue that $\mE^{A\to A'}$ is trace preserving so that $\{\mE_x^{A\to A'}\}_{x\in[k]}$ as defined above is indeed a quantum instrument. To see it, observe that for all $\omega\in\ml(A)$ we have
\ba
\tr\left[\mE^{A\to A'}(\omega^A)\right]&=\tr\left[\mN^{A\to A'}\circ\mR^{A\to A}\left(\omega^A\right)\right]\\
\Gg{\mN(\u^A)=\frac1n\mN(I^A)=\frac1mVV^*}&=\tr\left[\mN^{A\to A'}\left(\tr[\omega^A]\u^A\right)\right]\\
&=\tr[\omega^A]\frac1{m}\tr[VV^*]\\
&=\tr[\omega^A]\;,
\ea
where in the last line we used the fact that $VV^*$ is a projection of rank $m=|A'|$ since $V:A'\to A$ is an isometry.

Therefore, with this choice of quantum instrument, Eq.~\eqref{wqs} becomes
\ba\label{d1way}
T\left(\rho^{AB}\xrightarrow{\text{\tiny LOCC}_1}\Phi^{A'B'}_m\right)&\leq \sum_{x\in[k]}p_xP^2\left(\u^{A'}\otimes\rho^{E},\mN^{A\to A'}\circ\mU^{A\to A}_x(\rho^{AE})\right)\\
\Gg{P^2(\rho,\sigma)\leq \|\rho-\sigma\|_1}&\leq\sum_{x\in[k]}p_x\left\|\mN^{A\to A'}\Big(U^A_x\rho^{AE}\left(U^A_x\right)^*\Big)-\u^{A'}\otimes\rho^{E}\right\|_1\;,
\ea
where the last line follows from~\eqref{fitr}. Finally, to apply the decoupling theorem\index{decoupling theorem}  as given in~\eqref{corimpi}, we define
\be
\tau^{AA'}\eqdef\frac1{n}J_\mN^{AA'}=\frac1{n}\mN^{\tA\to A'}\left(\Omega^{A\tA}\right)=\frac1m\left(I^{A}\otimes V\right)^*\Omega^{A\tA}\left(I^{A}\otimes V\right)\;.
\ee
Note in particular that the marginal $\tau^{A'}=\u^{A'}$ so that the right-hand side of Eq.~\eqref{d1way} has the exact same form as given on the left-hand side of~\eqref{corimpi} (in the decoupling theorem). Hence, we can apply the decoupling theorem to get
\be
T\left(\rho^{AB}\xrightarrow{\text{\tiny LOCC}_1} \Phi^{A'B'}_m\right)\leq 2^{-\frac{1}{2}\big(\tH_2^\ua(A|E)_\rho+\tH_2^\ua(A|A')_\tau\big)}\;.
\ee
Moreover, since $\tau^{AA'}$ is maximally entangled we get that
\be
\tH_2^\ua(A|A')_\tau=-\log(m)\;.
\ee
Substituting this into the previous equation gives~\eqref{13p171}. This completes the proof.
\end{proof}

\subsection{conversion distance\index{conversion distance} from a Maximally Entangled State}

The conversion distance in~\eqref{entcd} is defined with respect to the trace distance. Since all metrics in finite dimensional Hilbert spaces are topologically equivalent, we can always select a different metric to simplify computations. In this section, we find that the square of the purified distance is a more convenient measure to work with. Recall from Lemma~\ref{lem1331} that the $T$-conversion distance (measured with respect to the trace distance $T$) is equivalent to the $P^2$-conversion distance when computing the conversion distances to a maximally entangled state.  However, it is important to note that this equivalence does not hold when the source/input state is maximally entangled.

 The $P^2$-conversion distance from $\sigma\in\md(A'B')$ to $\rho\in\md(AB)$ is defined as
\be
P^2\left(\sigma^{A'B'}\xrightarrow{\text{\tiny LOCC}} \rho^{AB} \right)\eqdef\max_{\mE\in\locc}P^2\left(\mE^{A'B'\to AB}\big(\sigma^{A'B'}\big),\rho^{AB}\right)\;.
\ee
In what follows we take the source state $\sigma^{A'B'}$ to be the maximally entangled state $\Phi_m^{A'B'}$.

\bex
Show that for any $\rho\in\md(AB)$ and any $\sigma\in\md(A'B')$ we have
\be
T^2\left(\sigma\xrightarrow{\text{\tiny LOCC}} \rho \right)\leq P^2\left(\sigma\xrightarrow{\text{\tiny LOCC}} \rho \right)\leq 2T\left(\sigma\xrightarrow{\text{\tiny LOCC}} \rho \right)\;.
\ee
Hint: See Theorem~\ref{pdthm}.
\eex

\begin{myt}{}
\begin{theorem}\label{lem1291}
Let $\rho\in\md(AB)$ and $m\in\mbb{N}$. Then,
\be
P^2\left(\Phi_m\xrightarrow{\text{\tiny LOCC}} \rho^{AB} \right)=E_{(m)}\left(\rho^{AB}\right)
\ee
where $E_{(m)}\left(\rho^{AB}\right)$ is the entanglement monotone defined in~\eqref{kyfan} with $k=m$.
\end{theorem}
\end{myt}

\begin{proof}
Recall first that for the case that $\rho^{AB}$ is a pure pure state, the theorem follows from Corollary~\ref{corppp}.
We therefore need to generalize this result to the case  that $\rho^{AB}$ is a mixed state. We start by showing that
\be\label{twosets}
\Big\{\mE\left(\Phi_m\right)\;:\;\mE\in\locc(A'B'\to AB)\Big\}=\Big\{\omega\in\md(AB)\;:\;\sr\left(\omega^{AB}\right)\leq m\Big\}\;,
\ee
where $\sr\left(\omega^{AB}\right)$ is the Schmidt rank\index{Schmidt rank} as defined in~\eqref{srank}. Indeed,
since the Schmidt rank $\sr$ as defined in~\eqref{srank} is a measure of entanglement  it follows that for $\omega^{AB}=\mE\left(\Phi_m\right)$
\be
\sr\left(\omega^{AB}\right)\leq\sr\left(\Phi_m\right)=m\;.
\ee
Therefore, the left-hand side of~\eqref{twosets} is contained in the right-hand side. On the other hand, from Exercise~\ref{ex:sr} it follows that every state $\omega^{AB}$ with Schmidt rank\index{Schmidt rank} no greater than $m$ has a pure state decomposition with all states having Schmidt rank no greater than $m$. 
As a consequence of Nielsen's theorem, such a state $\omega^{AB}$ can be generated by LOCC from $\Phi_{m}$. That is, the right-hand side of~\eqref{twosets} is contained in the left-hand side. This completes the proof of the equality in~\eqref{twosets}.

Let $E$ be a purifying system of dimension $n\eqdef|E|\leq |AB|$. From the equivalency of the two sets in~\eqref{twosets} we get
\ba\label{13pp13}
\max_{\mE\in\locc(A'B'\to AB)}F^2\left(\mE\left(\Phi_m\right),\rho^{AB} \right)&=\max_{\omega\in\md(AB),\;\sr(\omega)\leq m}F^2\left(\omega^{AB},\rho^{AB} \right)\\
\GG{Uhlmann's\; theorem}&=\max_{\substack{\psi,\phi\in\pure(ABE)\\
\sr(\phi^{AB})\leq m,\;\psi^{AB}=\rho^{AB}}}\left|\la\phi^{ABE}|\psi^{ABE}\ra\right|^2\;.
\ea
Let $\{|x\ra^E\}_{x\in[n]}$ be a fixed orthonormal basis of the purifying system $E$, and observe that every purification $\psi^{ABE}$ of $\rho^{AB}$ can be expressed as 
\be\label{pabe}
|\psi^{ABE}\ra\eqdef\sum_{x\in[n]}\sqrt{p_x}|\psi_x^{AB}\ra|x\ra^E\;,
\ee
where  $\{p_x,\psi_x^{AB}\}_{x\in[n]}$ is a pure state decomposition of $\rho^{AB}$.
Specifically, there is a one-to-one correspondence between all purifications $\psi^{ABE}$ of $\rho^{AB}$ that have the form~\eqref{pabe}, and all pure states decompositions $\{p_x,\psi_x^{AB}\}_{x\in[n]}$ of  $\rho^{AB}=\sum_{x\in[n]}p_x\psi_x^{AB}$.

Similarly, for every $\omega^{AB}$ with Schmidt rank\index{Schmidt rank} $\sr(\omega^{AB})\leq m$ let 
\be
|\phi^{ABE}\ra\eqdef\sum_{x\in[n]}\sqrt{q_x}|\phi_x\ra^{AB}|x\ra^E\;,
\ee 
be a purification of $\omega^{AB}$ with the property that each $|\phi_x\ra^{AB}$ has a Schmidt rank no greater than $m$ (see Exercise~\ref{ex:sr}). 
With this at hand, we get from~\eqref{13pp13} that
\be
\max_{\mE\in\locc(A'B'\to AB)}F^2\left(\mE\left(\Phi_m\right),\rho^{AB} \right)=\max\Big|\sum_{x\in[n]}\sqrt{q_xp_x}\la\phi_x^{AB}|\psi_x^{AB}\ra\Big|^2\;,
\ee
where the maximum on the right-hand side is over all pure states decompositions of $\rho^{AB}=\sum_xp_x\psi_x^{AB}$, all probability vectors $\q\in\prob(n)$, and all pure states $\{\phi_x^{AB}\}_{x\in[n]}$ with Schmidt rank\index{Schmidt rank} no greater than $m$. Now, from Corollary~\ref{corppp} it follows that (see Exercise~\ref{ex:pofe}) for every $\psi\in\pure(AB)$
\be\label{purestateex}
\max_{\substack{\phi\in\pure(AB)\\\sr(\phi)\leq m}}\left|\la\phi^{AB}|\psi^{AB}\ra\right|^2=\|\psi^A\|_{(m)}\;.
\ee
Therefore, there exists $\{\phi_x^{AB}\}_{x\in[n]}$ such that
\be
\la\phi_x^{AB}|\psi_x^{AB}\ra=\left|\la\phi_x^{AB}|\psi_x^{AB}\ra\right|=\sqrt{\|\psi_x^{A}\|_{(m)}}\;,
\ee
and that $\left|\la\phi_x^{AB}|\psi_x^{AB}\ra\right|$ cannot exceed this value. We therefore conclude that
\ba
\max_{\mE\in\locc(A'B'\to AB)}F^2\left(\mE\left(\Phi_m\right),\rho^{AB} \right)
&=\max_{\{p_x,\psi_x\}}\max_{\q\in\prob(n)}\Big(\sum_{x\in[n]}\sqrt{q_xp_x\left\|\psi_x^{A}\right\|_{(m)}}\Big)^2\\
\GG{Exercise~\ref{loip}}&= \max_{\{p_x,\psi_x\}}\sum_{x\in[n]}p_x\left\|\psi_x^{A}\right\|_{(m)}
\ea
where the maximum on the right-hand sides stands for a maximum over all pure state decompositions $\{p_x,\psi_x^{AB}\}_{x\in[n]}$ of 
$\rho^{AB}=\sum_{x\in[n]}p_x\psi_x^{AB}$.
In terms of the purified distance we have
\ba
\max_{\mE\in\locc(A'B'\to AB)}P^2\left(\mE\left(\Phi_m\right),\rho^{AB} \right)&=1-\max\sum_{x\in[n]}p_x\left\|\psi_x^{A}\right\|_{(m)}\\
&=\min\sum_{x\in[n]}p_x\left(1-\left\|\psi_x^{A}\right\|_{(m)}\right)\\
&=E_{(m)}\left(\rho^{AB}\right)\;,
\ea
where both the min and max above are over all pure-state decompositions $\{p_x,\psi_x^{AB}\}_{x\in[n]}$ of $\rho^{AB}$. This completes the proof.
\end{proof}

\bex\label{ex:pofe}
Use Corollary~\ref{corppp} to prove the equality in~\eqref{purestateex}.
\eex

\section{Single-Shot Distillable Entanglement}\index{single-shot}\index{distillable entanglement}

The $\eps$-single-shot distillable entanglement is then defined as (cf.~\eqref{epssins})
\be\label{edone}
\distill^{\eps}\left(\rho^{AB}\right)\eqdef\max\Big\{\log m\;:\;T\left(\rho^{AB}\xrightarrow{\text{\tiny LOCC}}  \Phi_m\right)\leq\eps\Big\}\;.
\ee
Since the computation of the distillable entanglement is hard, we start with the single-shot one-way distillable entanglement. As $\locc_1$ is a subset of LOCC, any lower bound on the one-way distillable entanglement will automatically provide a lower bound on the distillable entanglement defined above.

\bex
Let $k\eqdef|A|=|B|$, $m=|A'|=|B'|$, and $\rho\in\md(AB)$. 
Show that
\be
\distill^{\eps}\left(\rho^{AB}\right)\leq \log(k)-\log(1-\eps)\;.
\ee
\eex

\subsection{One-Way Single-Shot Distillable Entanglement}\index{single-shot}\index{distillable entanglement}

Similar to~\eqref{edone}, we define the $\eps$-single-shot distillable entanglement under 1-way LOCC as
\be\label{f1131}
\distill_{\to}^{\eps}\left(\rho^{AB}\right)\eqdef\max\Big\{\log m\;:\;T\left(\rho^{AB}\xrightarrow{\text{\tiny LOCC}_1}  \Phi_m^{A'B'}\right)\leq\eps\Big\}\;.
\ee
A simple formula for the above expression is not presently available. However, we can provide some useful lower and upper bounds.

In the following theorem, we present an upper bound on the single-shot one-way distillable entanglement\index{distillable entanglement}. It is worth noting that the upper bound given in~\eqref{ub2way} will not be helpful in this case, as we are considering a subset of LOCC. Therefore, we can expect to obtain a tighter upper bound, particularly since the upper bound given in~\eqref{ub2way} remains valid even if we replace LOCC with non-entangling operations.

The upper bound presented in the following theorem is expressed in terms of the coherent information\index{coherent information} of entanglement, denoted as $E_\to$. This particular measure of entanglement has been defined and extensively examined in Sec.~\ref{sec:cioe}. 

\begin{myt}{}
\begin{theorem}\label{epiu}
Let $\rho\in\md(AB)$ and $\eps\in(0,1/2)$. Then, the one-way $\eps$-single-shot distillable entanglement is bounded by
\be\label{singleed}
\distill^\eps_{\to}\left(\rho^{AB}\right)\leq\frac1{1-2\eps}E_\to\left(\rho^{AB}\right)+\frac{1+\eps}{1-2\eps}h\left(\frac\eps{1+\eps}\right)\;,
\ee
where $h(x)\eqdef-x\log x-(1-x)\log(1-x)$ is the binary Shannon entropy.
\end{theorem}
\end{myt}

\begin{proof}
Let $m\in\mbb{N}$ be such that $\distill^\eps_{\to}\left(\rho^{AB}\right)=\log m$ so that $T\left(\rho^{AB}\xrightarrow{\text{\tiny LOCC}_1}  \Phi^{A'B'}_m\right)\leq\eps$. This means that $\rho^{AB}\xrightarrow{\locc_1}\sigma^{A'B'}$ for some state $\sigma\in\md(A'B')$ that is $\eps$-close to $\Phi_m^{A'B'}$. Therefore, from the monotonicity of $E_\to$ under one-way LOCC we get that
\be\label{rsapbp}
E_\to\left(\rho^{AB}\right)\geq E_\to\left(\sigma^{A'B'}\right)\;.
\ee
Next, we use the fact that $\sigma^{A'B'}$ is $\eps$-close the $\Phi_m^{A'B'}$ to show that the right-hand side of the equation above cannot be much smaller than $\log(m)$. Indeed, combining
the continuity property of the function $I(A'\ra B')_\rho\eqdef-H(A'|B')_\rho$ (see~\eqref{condentcon}), with the second part of Exercise~\ref{spex}, gives
\ba
E_\to\left(\sigma^{A'B'}\right)&\geq I(A'\ra B')_\sigma\\
\GG{\eqref{condentcon}}&\geq I(A'\ra B')_{\Phi_m}-2\eps\log m-(1+\eps)h\left(\frac\eps{1+\eps}\right)\\
&=(1-2\eps)\log(m)-(1+\eps)h\left(\frac\eps{1+\eps}\right)\;.
\ea
The proof is concluded by noting that the inequality above in conjunction with the inequality~\eqref{rsapbp} yields the desired inequality~\eqref{singleed}.
\end{proof}

\bex
Use similar lines as in the proof above to prove the following bound on the $\eps$-shot distillable entanglement
\be
\distill^\eps\left(\rho^{AB}\right)\leq\frac1{1-2\eps}\sup_{\mE\in\locc(AB\to A'B')}I\big(A'\ra B'\big)_{\mE(\rho)}+\frac{1+\eps}{1-2\eps}h\left(\frac\eps{1+\eps}\right)\;.
\ee
where the supremum is also over all systems $A'$ and $B'$.
\eex

In the context where the single-shot distillable entanglement (and the entanglement cost) is expressed as $\log(m)$ for some integer $m \in \mathbb{N}$, it is useful to introduce a specific notation, $\gtrapprox$, to denote a particular type of inequality between two real numbers $a, b \in \mathbb{R}$. This notation is defined as follows:
\be
a \gtrapprox b \quad \iff \quad a \geq \log\left\lfloor 2^b \right\rfloor\;.
\ee
This definition provides a convenient way to express inequalities that are relevant in the quantification of entanglement, especially in scenarios involving logarithmic expressions and integer values.  With this in mind,
our next goal is a lower bound on the $\distill^\eps_{\to}\left(\rho^{AB}\right)$. The lower bound given below is known as the single-shot \emph{hashing bound}. \index{hashing bound}

\begin{myt}{}
\begin{theorem}\label{sinhash}
Let $\rho\in\pure(ABE)$ and $\eps\in(0,1)$. Then, for every $0<\delta<\eps$ we have
\be\label{lbths}
\distill_{\to}^{\eps}\left(\rho^{AB}\right)\gtrapprox H_{\min}^{\delta}(A|E)_{\rho}+\log(\eps-\delta)^2\;.
\ee
\end{theorem}
\end{myt}
\begin{remark}
Observe that unlike the upper bound which is given in terms of the coherent information of entanglement, the lower bound above does not involve an optimization over channels in $\cptp(A\to AX)$. 
\end{remark}

\begin{proof}
The main strategy of the proof is to use the upper bound given in Theorem~\ref{upb1way} for the one-way convcersion distance. Specifically, let $\delta\in(0,1)$ and let $\trho\in\mb_\delta(\rho^{AE})$ be such that 
$H_{\min}^{\ua\delta}(A|E)_\rho=H_{\min}^\ua(A|E)_{\trho}$. With these definitions we get
\ba
\sqrt{m}2^{-\frac{1}{2}H_{\min}^{\ua\delta}(A|E)_\rho}&=\sqrt{m}2^{-\frac{1}{2}H_{\min}^\ua(A|E)_{\trho}}\\
\Gg{H_{\min}^\ua\leq H_2^\ua}&\geq\sqrt{m}2^{-\frac{1}{2}\tH_2^\ua(A|E)_{\trho}}\\
\GG{\eqref{13p171}}&\geq T\left(\trho^{AB}\xrightarrow{\text{\tiny LOCC}_1}  \Phi^{A'B'}_m\right)\\
\GG{Lemma~\ref{cdub}}&\geq T\left(\rho^{AB}\xrightarrow{\text{\tiny LOCC}_1}  \Phi^{A'B'}_m\right)-\delta\;.
\ea
That is,
\be\label{13p173}
T\left(\rho^{AB}\xrightarrow{\text{\tiny LOCC}_1}  \Phi^{A'B'}_m\right)\leq \sqrt{m}2^{-\frac{1}{2}H_{\min}^{\ua\delta}(A|E)_\rho}+\delta\;.
\ee
Therefore, for any $\eps\in(0,1)$ and $0<\delta<\eps$ we get that the one-way $\eps$-distillable entanglement satisfies
\ba
\distill_{\to}^{\eps}\left(\rho^{AB}\right)&\eqdef\max\Big\{\log m\;:\;T\left(\rho^{AB}\xrightarrow{\text{\tiny LOCC}_1}  \Phi^{A'B'}_m\right)\leq\eps\Big\}\\
\GG{\eqref{13p173}}&\geq \max\Big\{\log m\;:\;\sqrt{m}2^{-\frac{1}{2}H_{\min}^\delta(A|E)_\rho}+\delta\leq\eps\Big\}\\
&=\max\Big\{\log m\;:\;\log m\leq H_{\min}^{\delta}(A|E)_{\rho}+\log(\eps-\delta)^2\Big\}\\
&= \log\left\lfloor (\eps-\delta)^22^{H_{\min}^{\delta}(A|E)_{\rho}}\right\rfloor\;.
\ea
This completes the proof.
\end{proof}

\bex
Show that for any $\rho\in\pure(ABE)$ and $\eps\in(0,1)$ we have
\be
\distill_{\to}^{\eps}\left(\rho^{AB}\right)\gtrapprox \tH_{2}^\ua(A|E)_{\rho}+2\log\eps\;.
\ee
\eex

\section{Asymptotic Distillable Entanglement}\index{distillable entanglement}

The asymptotic distillable entanglement is defined for any $\rho\in\md(AB)$ as (cf.~\eqref{dist5} and~\eqref{dist2})
\be
\distill\left(\rho^{AB}\right)\eqdef\lim_{\eps\to 0^+}\sup_{n,m\in\mbb{N}}\left\{\frac mn\;:\;T\left(\rho^{\otimes n}\xrightarrow{\text{\tiny LOCC}}  \Phi_2^{\otimes m}\right)\leq\eps\right\}\;.
\ee
From ~\eqref{disteq} it follows that the asymptotic distillable entanglement can be expressed as
\be
\distill\left(\rho^{AB}\right)=\lim_{\eps\to 0^+}\limsup_{n\to\infty}\frac1n\distill^\eps\left(\rho^{\otimes n}\right)\;.
\ee
Similarly, the one-way $\eps$-distillable entanglement is given by
\be\label{13135}
\distill_\to\left(\rho^{AB}\right)=\lim_{\eps\to 0^+}\limsup_{n\to\infty}\frac1n\distill_\to^\eps\left(\rho^{\otimes n}\right)\;.
\ee
\bex
Show that for any $n\in\mbb{N}$ and any $\rho\in\md(AB)$ we have
\be\label{13136a}
\distill\left(\rho\right)\geq \frac1n\distill\left(\rho^{\otimes n}\right)\quad\text{and}\quad \distill_\to\left(\rho\right)\geq \frac1n\distill_\to\left(\rho^{\otimes n}\right)\;.
\ee
\eex

\subsection{Simple Lower Bound on the Distillable Entanglement}\index{distillable entanglement}

Our analysis begins with a simple lower bound.

\begin{myt}{\color{yellow} The Hashing Bound}\index{hashing bound}
\begin{theorem}
Let $\rho\in\md(AB)$. Then,
\be\label{hashing}
\distill_\to\left(\rho^{AB}\right)\geq I(A\ra B)_\rho\;.
\ee
\end{theorem}
\end{myt}
\begin{remark}
Since the distillable entanglement is always no smaller than the one-way distillable entanglement, we also have \be
\distill\left(\rho^{AB}\right)\geq I(A\ra B)_\rho\;.
\ee
\end{remark}

\begin{proof}
Let $\rho^{ABE}\in\pure(ABE)$ be a purification of $\rho^{AB}$,
and let $\eps,\delta\in(0,1)$ be such that $\delta<\eps$. From the lower bound in~\eqref{lbths} we get 
\ba\label{13202}
\liminf_{n\to\infty}\frac1n\distill_\to^\eps\left(\rho^{\otimes n}\right)&\geq\liminf_{n\to\infty}\frac1n \Big(H_{\min}^{\delta}(A^n|E^n)_{\rho^{\otimes n}}+2\log(\eps-\delta)\Big)\\
&=\liminf_{n\to\infty}\frac1n H_{\min}^{\delta}(A^n|E^n)_{\rho^{\otimes n}}\\
\GG{Theorem~\ref{eandr}}&=H(A|E)_\rho\\
\GG{Duality\;relation~\eqref{sduals}}&=-H(A|B)_\rho=I(A\ra B)_\rho\;.
\ea
Since the equation above holds for all $\eps\in(0,1)$, it also holds if we take the limit $\eps\to 0^+$. This completes the proof.
\end{proof}

It is worth noting that the hashing bound\index{hashing bound} reveals that the distillable entanglement\index{distillable entanglement} is non-zero whenever the conditional entropy of $\rho^{AB}$ is negative. As we discussed earlier, the conditional entropy can only be negative for entangled states, which aligns with the fact that only entangled states can possess non-zero distillable entanglement. However, in the upcoming sections, we will discover that the converse statement is not true. Specifically, there exist entangled states with zero distillable entanglement.

\subsection{One-Way Distillable Entanglement}\index{distillable entanglement}

Although the hashing bound\index{hashing bound} is a useful and straightforward lower bound on the distillable entanglement, its tightness in general is not very clear. Interestingly, on pure states, the hashing bound is \emph{equal} to the distillable entanglement. This is because, for any pure state $\psi\in\pure(AB)$, the coherent information\index{coherent information} is given by
\be
I(A\ra B)_\psi=H(A)_\psi\;,
\ee
which is equal to the entropy of entanglement\index{entropy of entanglement} of the pure state $\psi^{AB}$.  For mixed states, we have the following expression for the one-way distillable entanglement.

\begin{myt}{}
\begin{theorem}\label{oproneway}
Let $\rho\in\md(AB)$. Then, the one-way distillable entanglement\index{distillable entanglement} of $\rho^{AB}$ is given by
\be\label{eq143}
\distill_\to\left(\rho^{AB}\right)= E_\to^{\reg}\left(\rho^{AB}\right)\;,
\ee
where $E_\to^\reg$ is the regularized coherent information\index{coherent information} of entanglement as defined in~\eqref{13p162}.
\end{theorem}
\end{myt}
\begin{remark}
It is worth noting that the theorem  provides an operational interpretation for the coherent information of entanglement as the one-way distillable entanglement. 
\end{remark}

\begin{proof}
For the direct part (i.e. achievability), observe that any quantum instrument $\mE\in\cptp(A\to AX)$ can be considered as a special type of $\locc_1$ (i.e. Alice implies the instrument $\{\mE_x\}_{x\in[m]}$ on her system and sends the outcome $x$ to Bob). Since the single-shot one-way distillable entanglement behaves monotonically under such $\locc_1$ we get that for all $\eps\in(0,1)$
\be
\distill_\to^\eps\left(\rho^{AB}\right)\geq \distill_\to^\eps\left(\sigma^{ABX}\right)\;,
\ee
where $\sigma^{ABX}\eqdef\mE^{A\to AX}\left(\rho^{AB}\right)$. Since for every $n\in\mbb{N}$ the equation above also holds with $n$ copies of $\rho$ and $\sigma$ we get
\ba\label{13144}
\liminf_{n\to\infty}\frac1n\distill_\to^\eps\left(\rho^{\otimes n}\right)
&\geq\liminf_{n\to\infty}
\frac1n\distill_\to^\eps\left(\sigma^{\otimes n}\right)\\
\GG{\eqref{13202}}&\geq I(A\ra BX)_\sigma
\ea
Since the inequality above holds for all $\mE\in\cptp(A\to AX)$ we conclude that
\be\label{13p207}
\liminf_{n\to\infty}\frac1n\distill_\to^\eps\left(\rho^{\otimes n}\right)\geq E_\to\left(\rho^{AB}\right)\;.
\ee
We would like to replace the right-hand side above with the regularized version of $E_\to$. For this purpose, fix $k\in\mbb{N}$ and observe that by applying the above inequality for $\rho^{\otimes k}\in\md\left(A^kB^k\right)$ we get
\be\label{13p208}
\liminf_{n\to\infty}\frac1{kn}\distill_\to^\eps\left(\rho^{\otimes kn}\right)\geq \frac1kE_\to\left(\rho^{\otimes k}\right)\;.
\ee
We next show that the left-hand side of the two equations above coincide. Indeed, by definition of the $\liminf$, the left-hand side of~\eqref{13p207} is no greater than the left-hand side of~\eqref{13p208}. For the converse, let $\{n_j\}_{j\in\mbb{N}}$ be a subsequence of integers such that
\be\label{bdn2}
\liminf_{n\to\infty}\frac1n\distill_\to^\eps\left(\rho^{\otimes n}\right)=\lim_{j\to\infty}\frac1{n_j}\distill_\to^\eps\left(\rho^{\otimes n_j}\right)\;.
\ee
Now, for any $j\in\mbb{N}$, set $m_j\eqdef k\left\lfloor \frac{n_j}{k}\right\rfloor$; i.e.\ $m_j$ is the largest multiple of $k$ that is no greater than $n_j$. In particular, note that $n_j-k<m_j\leq n_j$. Then,
\ba
\liminf_{n\to\infty}\frac1{kn}\distill_\to^\eps\left(\rho^{\otimes kn}\right)&\leq \liminf_{j\to\infty}\frac1{m_j}\distill_\to^\eps\left(\rho^{\otimes m_j}\right)\\
\Gg{m_j\leq n_j}&\leq \liminf_{j\to\infty}\frac1{m_j}\distill_\to^\eps\left(\rho^{\otimes n_j}\right)\\
\Gg{n_j-k<m_j\leq n_j}&=\lim_{j\to\infty}\frac1{n_j}\distill_\to^\eps\left(\rho^{\otimes n_j}\right)\\
\GG{\eqref{bdn2}}&=\liminf_{n\to\infty}\frac1n\distill_\to^\eps\left(\rho^{\otimes n}\right)\;,
\ea
where the first inequality follows from the fact that $\{m_j\}_{j\in\mbb{N}}$ is a subset of $\{kn\}_{n\in\mbb{N}}$, the second inequality from the fact that $\distill_\to^\eps\left(\rho^{\otimes m_j}\right)\geq \distill_\to^\eps\left(\rho^{\otimes n_j}\right)$ as $m_j\geq n_j$, and the third in equality from the fact that $\frac1{m_j}=\frac{n_j}{m_j}\frac1 {n_j}$ and
$
\lim_{j\to\infty}{n_j}/{m_j}=1
$
(since $n_j-k< m_j\leq n_j$). This completes the proof that the left-hand side of~\eqref{13p207} equals the left-hand side of~\eqref{13p208} so that 
\be
\liminf_{n\to\infty}\frac1{n}\distill_\to^\eps\left(\rho^{\otimes n}\right)\geq \frac1kE_\to\left(\rho^{\otimes k}\right)\;.
\ee
Since the equation above holds for all $k\in\mbb{N}$ it must also hold for the limit $k\to\infty$; that is, we conclude that
\be\label{ineq146}
\liminf_{n\to\infty}\frac1{n}\distill_\to^\eps\left(\rho^{\otimes n}\right)\geq E^\reg_{\to}\left(\rho^{AB}\right)\;.
\ee

For the converse inequality we apply the upper bound~\eqref{singleed} with $\rho^{\otimes n}$ instead of $\rho$. Explicitly, observe that for any $\eps\in(0,1)$
\ba
\limsup_{n\to\infty}\frac1n\distill_\to^\eps\left(\rho^{\otimes n}\right)
&\leq \limsup_{n\to\infty}\frac1n\left\{\frac1{1-2\eps}E_\to\left(\rho^{\otimes n}\right)+\frac{1+\eps}{1-2\eps}h\left(\frac\eps{1+\eps}\right)\right\}\\
&=\frac1{1-2\eps}E_\to^\reg\left(\rho^{AB}\right)\;.
\ea
Taking the limit $\eps\to 0^+$ and combining the resulting inequality with~\eqref{ineq146} we conclude that the equality in~\eqref{eq143} holds. 
\end{proof}

In the final step of the proof just discussed, we had to consider the limit as $\eps$ approaches $0$ (from the positive side). It remains unclear to the author whether this step is essential, and whether the following equality holds true:
\be
\lim_{n\to\infty}\frac1n\distill_\to^\eps\left(\rho^{\otimes n}\right)=E_\to^\reg\left(\rho^{AB}\right)
\ee
for all $\eps$ within the interval $(0,1)$.
This point raises an interesting question in the study of quantum information theory, particularly regarding the behavior of the distillable entanglement under asymptotic conditions. The uncertainty here revolves around whether the regularized entanglement measure, $E_\to^\reg$, aligns with the distillable entanglement\index{distillable entanglement} rate, $\frac1n\distill_\to^\eps\left(\rho^{\otimes n}\right)$, for any non-zero $\eps$. Resolving this would contribute to a deeper understanding of entanglement properties in quantum systems.

\subsection{Bound Entanglement}\index{bound entanglement}

The following theorem presents a fascinating discovery in quantum mechanics, demonstrating that certain entangled states have a distillable entanglement\index{distillable entanglement} of zero. This means that their entanglement is bounded, and they cannot be transformed into a pure bipartite form, making them an example of \emph{bound} entanglement.
Despite its inability to be distilled, bound entanglement still has practical applications in quantum information theory.

One important application of bound entanglement is in quantum cryptography, where it can be used as a resource for various cryptographic protocols. For example, bound entangled states can be used to implement quantum key distribution (QKD) protocols, which allow two parties to establish a secure shared key even in the presence of an eavesdropper.
Another potential application of bound entanglement is in quantum communication networks, where it can be used as a resource for distributing entanglement between multiple parties. This could be useful for tasks such as quantum teleportation, where entanglement is used to transmit quantum information between distant locations.

Overall, while bound entanglement cannot be used for some tasks that pure entanglement can, it still has practical uses in quantum information processing and communication. From a fundamental point of view,
this remarkable result challenges our understanding of entanglement and highlights the complexity of quantum systems. It underscores the importance of exploring and understanding the different types of entanglement that can exist in quantum systems. It also presents a new avenue for research, as scientists continue to investigate the properties and applications of bound entanglement.

\begin{myt}{}
\begin{theorem}\label{thm:pptdist}
Let $\rho\in\md(AB)$ be an entangled state with a positive partial transpose. Then,
\be
\distill\left(\rho^{AB}\right)=0\;.
\ee
\end{theorem}
\end{myt}

\begin{proof}
Suppose by contradiction that $\distill\left(\rho^{AB}\right)>0$. This in particular means that there exists $n\in\mbb{N}$ such that
\be\label{xlocc}
\left(\rho^{AB}\right)^{\otimes n}\xrightarrow{\text{\tiny LOCC}} \sigma^{A'B'}\;,
\ee
for some two-qubit entangled state $\sigma\in\md(A'B')$ with $|A'|=|B'|=2$. However, since the logarithmic negativity is additive we get
\ba
\LN\left(\rho^{\otimes n}\right)&=n\LN\left(\rho^{AB}\right)\\
\GG{\rho^{\it AB}\;is\;PPT}&=0\;,
\ea
whereas
\be
\LN\left(\sigma^{A'B'}\right)>0
\ee
since $\sigma^{A'B'}$ is a two-qubit entangled state and from Theorem~\ref{thm:2x3} it follows that it is NPT (recall that the logarithmic robustness\index{robustness} is strictly positive on NPT states).
We therefore get that 
\be
\LN\left(\rho^{\otimes n}\right)<\LN\left(\sigma^{A'B'}\right)
\ee
in contradiction with~\eqref{xlocc} and the fact that the logarithmic robustness is a measure of entanglement and therefore cannot increase by LOCC. This completes the proof.
\end{proof}

By applying the hashing bound\index{hashing bound}, we can deduce that if $H(A|B)\rho<0$, then the distillable entanglement of $\rho^{AB}$ is strictly positive. Combining this with the theorem mentioned above, we can conclude that if $\rho\in\md(AB)$ is PPT, then its conditional entropy $H(A|B)\rho\geq 0$. This observation is consistent with the reduction criterion discussed in Section~\ref{sec:reduction}, as Corollary~\ref{rccor} and Theorem~\ref{thm:neg-cond-ent} show that states satisfying the reduction criterion (particularly PPT states) have non-negative conditional entropy.

\bex
Let $\rho\in\pure(ABC)$ be a tripartite pure state. Show that if its marginals satisfy $I^A\otimes\rho^C>\rho^{AC}$ then $\distill\left(\rho^{AB}\right)>0$. 
\eex

Theorem~\ref{thm:pptdist} states that PPT entangled states have zero distillable entanglement. This result is an important insight into the relationship between entanglement and the partial transpose\index{partial transpose} operation. However, it raises the question of whether the converse of this property also holds. That is, are all entangled states with zero distillable entanglement (i.e., bound entangled states) necessarily PPT? This is one of the most challenging and long-standing open problems in quantum information theory, and despite significant efforts over the past two decades, the answer is still unknown.
Despite the current lack of a definitive answer to this question, research in this area continues to progress, with new insights and techniques being developed to study the properties of entangled states and their relation to the partial transpose operation.

\section{Single-Shot  Entanglement Cost}\index{single-shot}\index{entanglement cost}

In this section we compute the single-shot entanglement cost of a bipartite mixed entangled state $\rho\in\md(AB)$.
For any $\eps\in(0,1)$, we define the $\eps$-single-shot entanglement cost as
\be\label{econe}
\cost^{\eps}\left(\rho^{AB}\right)\eqdef\min\Big\{\log m\;:\;P^2\left(\Phi_m\xrightarrow{\text{\tiny LOCC}}\rho^{AB}\right)\leq \eps\Big\}\;,
\ee
where $\Phi_m$ is the maximally entangled state in $\md(A'B')$ with $m\eqdef|A'|=|B'|$. Since all metrics in finite dimensional Hilbert spaces are topologically equivalent, we chose the square of the purified distance as it is easier to work with, and in particular, has the form given in Theorem~\ref{lem1291}. 

\subsubsection{The Zero-Error Cost}\index{entanglement cost}

Before providing a formula for this $\eps$-single-shot entanglement cost, we first consider the case $\eps=0$. In this case, observe that 
\ba\label{zeroeps}
\cost^{\eps=0}\left(\rho^{AB}\right)&=\min\Big\{\log m\;:\;\Phi_m\xrightarrow{\text{\tiny LOCC}}\rho^{AB}\Big\}\\
\GG{\eqref{srlocc}}&=\log\sr\left(\rho^{AB}\right)\;.
\ea
That is, the logarithm of the Schmidt rank\index{Schmidt rank} of $\rho^{AB}$
  offers an operational interpretation as the zero-error entanglement cost of $\rho^{AB}$. Following this, we demonstrate that $\sr(\rho^{AB})$ bears a close relationship to the conditional max-entropy. To elaborate further, let's first present an alternative method for describing the convex roof extension\index{convex roof extension}.

\begin{myd}{}
\begin{definition}
Let $\rho\in\md(AB)$ be a bipartite density matrix, and let $X$ be a classical system of dimension $k$. We say that $\rho^{XAB}$ is a \emph{regular extension} of $\rho^{AB}$ if there exists a pure state decomposition of $\rho^{AB}=\sum_{x\in[k]}p_x\psi_x^{AB}$ such that
\be\label{givenx}
\rho^{XAB}\eqdef\sum_{x\in[k]}p_x|x\lr x|^X\otimes\psi_{x}^{AB}\;.
\ee
\end{definition}
\end{myd}

Note that if $\{\psi_x^{AB}\}_{x\in[k]}$ in the definition above were composed of mixed states instead of pure states, then $\rho^{XAB}$ would not necessarily qualify as a regular extension, even though it would still be an extension of $\rho^{AB}$. Moreover, we show now that the Schmidt rank\index{Schmidt rank} of $\rho^{AB}$ can be expressed as an optimization problem over all marginal cq-states $\rho^{XA}$ that results from regular extensions.  Explicitly, if $\rho^{XAB}$ is a regular extension of $\rho^{AB}$, as given in~\eqref{givenx}, then the marginal cq-state, $\rho^{XA}$, has the form
\be\label{samenot}
\rho^{XA}=\sum_{x\in[k]}p_x|x\lr x|^X\otimes\rho_{x}^{A}\;,
\ee
where $\rho_x^A\eqdef\tr_B\left[\psi_x^{AB}\right]$. By definition, $\sr(\psi_x^{AB})=\tr\left[\Pi_{\rho_x}^A\right]$, where $\Pi_{\rho_x}\in\pos(A)$ is the projection in $A$ to the support of $\rho_x^A$. Combining this with the relation~\eqref{sreq} we can express the Schmidt rank\index{Schmidt rank} of $\rho^{AB}$ as
\be\label{obser4}
{\sr}\left(\rho^{AB}\right)=\inf_{\rho^{ABX}}\max_{x\in[k]}\tr\left[\Pi_{\rho_x}^A\right]\;,
\ee
where the maximum is over all regular extensions $\rho^{ABX}$ of $\rho^{AB}$. The expression above can be rewritten in terms of the conditional max-entropy\index{conditional max-entropy} of the state $\rho^{AX}$. To see this, let $\Pi_\rho^{XA}$ be the projection to the support of $\rho^{XA}$ defined as
\be
\Pi_\rho^{XA}=\sum_{x\in[k]}|x\lr x|^X\otimes\Pi_{\rho_x}^A\;,
\ee
and observe that from its definition
\ba\label{exhmaxs}
H_{\max}(A|X)_\rho&=\max_{\tau\in\md(X)}-D_{\min}\left(\rho^{XA}\big\|\tau^X\otimes I^A\right)\\
&=\max_{\tau\in\md(X)}\log\tr\left[\Pi_{\rho}^{XA}\left(\tau^X\otimes I^A\right)\right]\\
&=\log\max_{x\in[k]}\tr\left[\Pi_{\rho_x}^A\right]
\ea
where in the last line we replaced the maximum over all $\tau\in\md(X)$ with a maximum over all $x\in[k]$.
Using this observation in conjunction with~\eqref{obser4}, we can express the logarithm of the Schmidt rank\index{Schmidt rank} of $\rho^{AB}$ as
\be\label{xox3}
\log{\sr}\left(\rho^{AB}\right)=\inf_{\rho^{ABX}}H_{\max}(A|X)_\rho\;.
\ee
We therefore conclude that the zero-error entanglement cost of $\rho^{AB}$ is given by
\be\label{cost0re}
\cost^{\eps=0}\left(\rho^{AB}\right)=\inf_{\rho^{ABX}}H_{\max}(A|X)_\rho\;.
\ee
where the infimum is over all classical systems $X$ and all regular extensions $\rho^{ABX}$ of $\rho^{AB}$. In the following exercise you show that we can remove the restriction to regular extensions.

\bex\label{regularer}
Show that~\eqref{cost0re} still holds even if we take the infimum over all classical systems $X$ and all extensions $\rho^{ABX}$ of $\rho^{AB}$.  
\eex

\bex\label{eofc}
Let $\rho\in\md(AB)$. Show that the entanglement of formation of $\rho^{AB}$ can be expressed as
\be
E_F\left(\rho^{AB}\right)\eqdef\inf_{\rho^{ABX}} H(A|X)_\rho
\ee
where $H(A|X)_\rho$ is the von-Neumann conditional entropy, and the infimum is over all classical systems $X$ and all extensions $\rho^{ABX}$ of $\rho^{AB}$.
\eex

\bex\label{ex:snsn}
Let $\rho\in\md(AB)$ and $n\in\mbb{N}$. 
\ben
\item Show that if $\rho^{ABX}$ is a regular extension of $\rho^{AB}$ then $\left(\rho^{ABX}\right)^{\otimes n}$ is a regular extension of $\left(\rho^{AB}\right)^{\otimes n}$. 
\item Show that there exists a density matrix $\omega^{A^nB^nX^n}$ that is a regular extension of $\left(\rho^{AB}\right)^{\otimes n}$ but it does not have the form $\left(\sigma^{ABX}\right)^{\otimes n}$ (for some $\sigma\in\md(XAB)$).
\een
\eex

\subsubsection{The General Case}

From the above discussion, for every $\eps\in(0,1)$ the $\eps$-single-shot entanglement cost\index{entanglement cost} can be expressed as
\ba\label{xox1}
\cost^{\eps}\left(\rho^{AB}\right)&=\min\left\{\log\sr\left(\omega^{AB}\right)\;:\;P^2\left(\omega^{AB},\rho^{AB}\right)\leq\eps\;,\;\omega\in\md(AB)\right\}\\
&=\inf_{\omega^{ABX}}\left\{H_{\max}(A|X)_\omega\;:\;P^2\left(\omega^{AB},\rho^{AB}\right)\leq\eps\right\}
\ea
where the second infimum is over all density matrices $\omega^{AB}$, all classical systems $X$, \emph{and} over all regular extensions $\omega^{ABX}$ of $\omega^{AB}$.
Given the complexity of the above expression, we'll transition to an alternative approach, specifically employing the formula presented in Theorem~\ref{lem1291} for the conversion distance.

\begin{myt}{}
\begin{theorem}\label{cmaint}
Let $\rho\in\md(AB)$. Then, the $\eps$-single-shot entanglement cost is given by
\be\label{mainhmax}
\cost^\eps\left(\rho^{AB}\right)= \inf_{\rho^{ABX}} H_{\max}^{\eps}(A|X)_\rho\;.
\ee
where $H_{\max}^{\eps}(A|X)_\rho$ is the smoothed conditional max-entropy (see remark below), and the infimum is over all classical systems $X$ and all extensions $\rho^{ABX}$ of $\rho^{AB}$. 
\end{theorem}
\end{myt}
\begin{remark}
Exercise~\ref{regularer} allows us to limit the infimum in~\eqref{mainhmax} to \emph{regular} extensions $\rho^{ABX}$ of $\rho^{AB}$. Additionally, from \eqref{exhmaxs}, the smoothed version of $H_{\max}(A|X)_\rho$ is expressed as:
\be\label{omex}
H_{\max}^{\eps}(A|X)_\rho=\min_{\omega\in\mb_{\eps}\left(\rho^{XA}\right)}\max_{x\in[k]}\log\tr\left[\Pi_{\omega_x}^A\right]\;,
\ee
where $\omega_x^A\eqdef\frac1{q_x}\tr_X\left[\left(|x\lr x|^X\otimes I^A\right)\omega^{XA}\right]$ with $q_x\eqdef\tr\left[\left(|x\lr x|^X\otimes I^A\right)\omega^{XA}\right]$.
Interestingly, the optimization's minimizer $\omega^{XA}$ can be an $m$-pruned version of $\rho^{XA}$ (as referenced in \eqref{cfrm}). Therefore, before delving into the proof of Theorem~\ref{cmaint}, we first establish that the minimizer $\omega^{XA}$ in the above optimization is the $m$-pruned state of the cq-state\index{cq-state} $\rho^{XA}=\sum_{x\in[k]}p_x|x\lr x|\otimes\rho_x^A$. This pruned state is given by:
\be\label{cqmpruned}
\rho^{(m)}=\sum_{x\in[k]}p_x|x\lr x|\otimes\rho_x^{(m)}\;,
\ee
where each $\rho^{(m)}_x$ is the $m$-pruned version of $\rho_x$ as defined in~\eqref{cfrm}.
\end{remark}

\bex\label{logmex}
Use~\eqref{1260} to show that if $\rho^{(m)}\neq\rho^{XA}$ then $H(A|X)_{\rho^{(m)}}=\log m$.
\eex

\begin{myg}{}
\begin{lemma}
Let $\eps\in(0,1)$, $d\eqdef|A|$, $\rho\in\md(XA)$, and for all $m\in[d]$ let $\rho^{(m)}\in\md(XA)$ be the $m$-pruned version of $\rho^{XA}$ as defined in~\eqref{cqmpruned}. Then, 
\be\label{rhscan}
H_{\max}^{\eps}(A|X)_\rho=\min_{m\in[d]}\left\{\log m\;:\;\rho^{(m)}\approx_\eps\rho^{XA}\right\}
\ee
\end{lemma}
\end{myg}
\begin{remark}
Observe that the trace distance between $\rho^{XA}$ and its $m$-pruned version is given by
\ba
\frac12\left\|\rho^{XA}-\rho^{(m)}\right\|_1&=\sum_{x\in[k]}p_x\frac12\left\|\rho_x-\rho_x^{(m)}\right\|_1\\
\GG{\eqref{tdprun}}&=\sum_{x\in[k]}p_x\left(1-\|\rho_x\|_{(m)}\right)\\
&=1-\sum_{x\in[k]}p_x\|\rho_x\|_{(m)}\;.
\ea
Therefore, \eqref{rhscan} can also be written as
\be\label{rhsxk}
H_{\max}^{\eps}(A|X)_\rho=\min_{m\in[d]}\Big\{\log m\;:\;\sum_{x\in[k]}p_x\|\rho_x\|_{(m)}\geq 1-\eps\Big\}\;.
\ee
\end{remark}
\begin{proof}
By definition,
\ba
H_{\max}^{\eps}(A|X)_\rho&=\min\left\{H_{\max}(A|X)_\omega\;:\;\omega^{XA}\approx_\eps\rho^{XA}\right\}\\
\Gg{\text{Restricting}\;\omega=\rho^{(m)}}&\leq 
\min_{m\in[d]}\left\{H_{\max}(A|X)_{\rho^{(m)}}\;:\;\rho^{(m)}\approx_\eps\rho^{XA}\right\}\\
\GG{Exercise~\ref{logmex}}&=\min_{m\in[d]}\left\{\log m\;:\;\rho^{(m)}\approx_\eps\rho^{XA}\right\}\;.
\ea
For the converse direction, suppose by contradiction that (see~\eqref{rhsxk}) 
\be\label{minirhs}
H_{\max}^{\eps}(A|X)_\rho<\min_{m\in[d]}\Big\{\log m\;:\;\sum_{x\in[k]}p_x\|\rho_x\|_{(m)}\geq 1-\eps\Big\}\;.
\ee
Let $m$ be the minimizer of the right-hand side above, so that $\sum_{x\in[k]}p_x\|\rho_x\|_{(m)}\geq 1-\eps$. Further, let $\omega\in\md(XA)$ be such that $H_{\max}^\eps(A|X)_\rho=H_{\max}(A|X)_\omega$. Then, by the assumption
\ba\label{cgf}
\log m&>H_{\max}(A|X)_\omega\\
\GG{cf.~\eqref{exhmaxs}}&=\log\max_{x\in[k]}\tr\left[\Pi_{\omega_x}^A\right]\;,
\ea
so that $\tr\left[\Pi_{\omega_x}^A\right]<m$ for all $x\in[k]$. Thus, since the rank of each $\omega_x$ is no greater than $m-1$ we get that for any $x\in[k]$, $\left\|\rho^A_x\right\|_{(m-1)}\geq \tr\left[\rho^A_x\Pi_{\omega_x}^A\right]$. Therefore, denoting $\Pi^{XA}_\omega\eqdef\sum_{x\in[k]}|x\lr x|^X\otimes\Pi_{\omega_x}^A$ we obtain (cf.~\eqref{5p164})
\ba
\sum_{x\in[k]}p_x\left\|\rho^A_x\right\|_{(m-1)}&\geq  \tr\left[\rho^{XA}\Pi^{XA}_\omega\right]\\
\Gg{\tr\left[\omega^{XA}\Pi^{XA}_\omega\right]=1}&=1+\tr\left[\left(\rho^{XA}-\omega^{XA}\right)\Pi^{XA}_\omega\right]\\
\Gg{\eta\geq-(\eta)_-\quad\forall\eta\in\herm(XA)}&\geq 1-\tr\left[\left(\rho^{XA}-\omega^{XA}\right)_-\Pi^{XA}_\omega\right]\\
\Gg{\Pi^{XA}_\omega\leq I^{XA}}&\geq 1-\tr\left(\rho^{XA}-\omega^{XA}\right)_-\\
&\geq 1-\eps\;,
\ea
where we used the fact that $\tr\left(\rho^{XA}-\omega^{XA}\right)_-=\frac12\|\omega^{XA}-\rho^{XA}\|_1\leq\eps$. Therefore we get a contradiction with the assumption that $m$ was the minimizer of the right-hand side of~\eqref{minirhs}. This completes the proof.
\end{proof}

We are now ready to prove Theorem~\ref{cmaint}.

\begin{proof}[Proof of Theorem~\ref{cmaint}]
From Theorem~\ref{lem1291} it follows that the conversion distance\index{conversion distance} that appears in~\eqref{econe} can be expressed as
\be
P^2\left(\Phi_m\xrightarrow{\text{\tiny LOCC}} \rho^{AB} \right)=\min_{\rho^{XAB}}\sum_{x\in[k]}p_x\left(1-\left\|\rho^A_x\right\|_{(m)}\right)
\ee
where the minimum is over all regular extensions $\rho^{XAB}$ of $\rho^{AB}$, with the same notations as in~\eqref{samenot}.
We therefore get that
the $\eps$-single-shot entanglement cost\index{entanglement cost} as defined in~\eqref{econe} can be expressed as 
\ba\label{13p253}
\cost^\eps\left(\rho^{AB}\right)&=\min_{m\in[d]}\Big\{\log m\;:\;\max_{\rho^{XAB}}\sum_{x\in[k]}p_x\left\|\rho^A_x\right\|_{(m)}\geq 1-\eps\Big\}\\
\GG{Exercise~\ref{ex1206}}&=\min_{\rho^{XAB}}\min_{m\in[d]}\Big\{\log m\;:\;\sum_{x\in[k]}p_x\left\|\rho^A_x\right\|_{(m)}\geq 1-\eps\Big\}\\
\GG{\eqref{rhsxk}}&=\min_{\rho^{XAB}}H_{\max}^{\eps}(A|X)_\rho\;,
\ea
where the maximum is overl all regular extensions $\rho^{XAB}$ of $\rho^{AB}$.
 This completes the proof.
\end{proof}

\begin{exercise}\label{ex1206}
Let $\lambda\in\mbb{R}_+$, $\mk_1,\mk_2\subseteq\md(A)$ be two subsets of density matrices, and $f:\md(A)\to\mbb{R}_+$ and $g:\md(A)\times\md(A)\to\mbb{R}_+$ be two functions. 
\ben
\item Show that
\be
\inf_{\rho\in\mk_1}\Big\{f(\rho)\;:\;\sup_{\sigma\in\mk_2}g(\rho,\sigma)\geq\lambda\Big\}=\inf_{\sigma\in\mk_2}\inf_{\rho\in\mk_1}\Big\{f(\rho)\;:\;g(\rho,\sigma)\geq\lambda\Big\}
\ee
\item Use similar lines to prove the second equality in~\eqref{13p253}
\een
\end{exercise}

\bex
Consider the two formulas given in~\eqref{1242} and~\eqref{mainhmax} for the $\eps$-single-shot
entanglement cost of pure and mixed states, respectively. 
\ben
\item Show that the two formulas coincide when $\rho^{AB}=\psi^{AB}$ is a pure state. 
\item Without assuming~\eqref{mainhmax}, use Theorem~\ref{0distdm0} to show (by direct calculation) that for the pure state case, the formula in~\eqref{1242} can be expressed as 
\be
\cost^\eps\left(\psi^{AB}\right)=H_{\max}^\eps(\rho^A)\eqdef\inf_{\sigma\in\mb_\eps(\rho)}H_{\max}(\sigma^A)\;,
\ee
where $\rho^A\eqdef\tr_B\left[\psi^{AB}\right]$.
\een
\eex

\section{Asymptotic Entanglement Cost}\index{entanglement cost}

The asymptotic entanglement cost is defined for any $\rho\in\md(AB)$ as (cf.~\eqref{cost5} and~\eqref{cost2})
\be
\cost\left(\rho^{AB}\right)\eqdef\lim_{\eps\to 0^+}\inf_{n,m\in\mbb{N}}\left\{\frac nm\;:\;T\left( \Phi_2^{\otimes n}\xrightarrow{\rm {\tiny LOCC}}\rho^{\otimes m}\right)\leq\eps\right\}
\ee
Recall from Exercise~\ref{equivdistcost} that the above cost does not change if we replace the trace distance with the square of the purified distance, as the two metrics are topologically equivalent. Thus,
from~\eqref{deseq} it follows that the asymptotic entanglement cost can be expressed as
\be
\cost\left(\rho^{AB}\right)\eqdef\lim_{\eps\to 0^+}\liminf_{n\to\infty}\frac1n \cost^{\eps}\left(\rho^{\otimes n}\right)\;.
\ee

\begin{myt}{}
\begin{theorem}\label{ecost}
Let $\rho\in\md(AB)$. Then, the entanglement cost\index{entanglement cost} of $\rho^{AB}$ can be expressed as
\be
\cost\left(\rho^{AB}\right)=E_F^\reg\left(\rho^{AB}\right)\eqdef\lim_{n\to\infty}\frac1n E_F\left(\rho^{\otimes n}\right)
\ee
where $E_F$ is the entanglement of formation\index{entanglement of formation} (see Definition~\ref{def:cre}).
\end{theorem}
\end{myt}

\begin{proof}
We first prove that $\cost\left(\rho^{AB}\right)\leq E_F^\reg\left(\rho^{AB}\right)$.
From Theorem~\ref{cmaint} we have 
\ba
\cost^{\eps}\left(\left(\rho^{AB}\right)^{\otimes n}\right)&= \inf_{\rho_n}H_{\max}^\eps(A^n|Y_n)_{\rho_n}\\
\Gg{\text{taking}\;\rho_n=\left(\rho^{XAB}\right)^{\otimes n}}&\leq \inf_{\rho^{XAB}} H_{\max}^\eps(A^n|X^n)_{\rho^{\otimes n}}
\ea
where the first infimum is over all classical systems $Y_n$ and all regular extensions $\rho_n\in\md(Y_nA^nB^n)$ of $\left(\rho^{AB}\right)^{\otimes n}$, and the second infimum is over all regular extensions $\rho\in\md(XAB)$ of $\rho^{AB}$. For the inequality above we used the fact that if $\rho^{XAB}$ is a regular extension of $\rho^{AB}$ then $\left(\rho^{XAB}\right)^{\otimes n}$ is a regular extension of $\left(\rho^{AB}\right)^{\otimes n}$ (see Exercise~\ref{ex:snsn}).
Therefore, the entanglement cost satisfies
\ba
\cost\left(\rho^{AB}\right)&\leq \lim_{\eps\to 0^+}\inf_{\rho^{XAB}} \liminf_{n\to\infty}\frac1n H_{\max}^\eps(A^n|X^n)_{\rho^{\otimes n}}\\
\Gg{\text{AEP of the form}~\eqref{hmaxaep}}&=\inf_{\rho^{XAB}} H(A|X)_{\rho}\\
\GG{Exercise~\ref{eofc}}&=E_F\left(\rho^{AB}\right)\;.
\ea
Thus, $E_F\left(\rho^{AB}\right)\geq \cost\left(\rho^{AB}\right)$. Repeating the same argument with $m\in\mbb{N}$ copies of $\rho^{AB}$ gives $E_F\left(\rho^{\otimes m}\right)\geq \cost\left(\rho^{\otimes m}\right)$.
Combining this with~\eqref{11120} we get
\be
\cost\left(\rho^{AB}\right)\leq \frac1m \cost\left(\rho^{\otimes m}\right)\leq \frac1m E_F\left(\rho^{\otimes m}\right)\;.
\ee
Since the inequality above holds for all integers $m$, it also holds in the limit $m\to\infty$. Hence, $\cost\left(\rho^{AB}\right)\leq E_F^\reg\left(\rho^{AB}\right)$.

For the converse inequality, observe that
\ba\label{sintog}
\cost^{\eps}\left(\left(\rho^{AB}\right)^{\otimes n}\right)&= \inf_{\rho_n}H_{\max}^\eps(A^n|Y_n)_{\rho_n}\\
\GG{By \;definition}&= \inf_{\rho_n} \min_{\omega\in\mb_\eps\left(\rho_n^{Y_nA^n}\right)}H_{\max}(A^{n}|Y_n)_\omega\\
\Gg{H_{\max}(A|X)\geq H(A|X)}&\geq \inf_{\rho_n}\min_{\omega\in\mb_\eps\left(\rho_n^{Y_nA^n}\right)}H(A^{n}|Y_n)_\omega\;.
\ea
Now, the asymptotic continuity\index{asymptotic continuity} property~\eqref{condentcon2} of the conditional entropy gives for any $\omega\in\mb_{\eps}\left(\rho_n^{Y_nA^n}\right)$
\be
H(A^{n}|Y_n)_\omega\geq H(A^{n}|Y_n)_{\rho_n}-n\log|A|f(\eps)\;.
\ee
Substituting this into~\eqref{sintog} gives
\ba
\cost^\eps\left(\rho^{\otimes n}\right)&\geq \inf_{\rho_n} H(A^{n}|Y_n)_{\rho_n}-n\log|A|f(\eps)\\
\GG{Exercise~\ref{eofc}}&=E_F\left(\rho^{\otimes n}\right)-n\log|A|f(\eps)\;.
\ea
Dividing both sides by $n$ and taking the limit $n\to\infty$ followed by $\eps\to 0^+$ gives $\cost\left(\rho^{AB}\right)\geq E_F^\reg\left(\rho^{AB}\right)$. This completes the proof.
\end{proof}

If the entanglement of formation\index{entanglement of formation} ($E_F$) were additive under tensor products, determining $E_F^\reg\left(\rho^{AB}\right)$ would be a more straightforward task. For quite some time, a prevalent belief among researchers in the field was that $E_F$ is indeed additive, implying its equivalence to the entanglement cost\index{entanglement cost}. However, in a pivotal development in 2008, Hastings refuted this additivity\index{additivity} conjecture, demonstrating that the entanglement of formation is generally not additive. From its definition, for any states $\rho\in\md(AB)$ and $\sigma\in\md(A'B')$, the following inequality holds:
\be\label{subadeof}
E_F\left(\rho^{AB}\otimes\sigma^{A'B'}\right)\leq E_F\left(\rho^{AB}\right)+E_F\left(\sigma^{A'B'}\right)\;.
\ee
Hastings' result indicates that this inequality can be strict, even when $\rho$ is equal to $\sigma$. Notably, Hastings' proof is existential, meaning it establishes the existence of such non-additivity without providing an explicit counterexample. To date, an explicit example where $E_F$ is not additive has not been identified, but Hastings' contribution significantly altered our understanding on this problem (further details can be found in the "Notes and References" section at the end of this chapter).

While the entanglement of formation\index{entanglement of formation} (EoF) is not generally additive, it can be additive for certain specific states, allowing for efficient computation of their entanglement cost. An interesting concept relevant in this context is that of an ``entanglement breaking subspace".

\begin{myd}{}
\begin{definition}
A bipartite subspace $\mK\subset AB$ is said to be entanglement breaking\index{entanglement breaking} subspace (EBS) if for every other bipartite system\index{bipartite system} system $A'B'$ and any pure state in $\mK\otimes A'B'$, its reduced density matrix that obtained after tracing out system $B$ can be written as a separable state between system $A$ and the joint system $A'B'$.
\end{definition}
\end{myd}

To be more precise, a subspace $\mathcal{K}$ is an EBS if, for every $\psi\in\pure\left(\mathcal{K}\otimes A'B'\right)$, we have:
\be
\tr_B\left[\psi^{ABA'B'}\right]=\sum_{x\in[m]}p_x\phi_x^A\otimes\varphi_x^{A'B'}\;,
\ee
where $m\in\mathbb{N}$, $\p=(p_1,\ldots,p_m)^T\in\prob(m)$, and $\phi_x\in\pure(A)$ and $\varphi_x\in\pure(A'B')$ for each $x\in[m]$. Note that requiring $\mathcal{K}$ to be a subspace is a strong condition. Even if two states in $\pure\left(\mathcal{K}\otimes A'B'\right)$ satisfy the above condition, it is not clear why all their linear combinations would satisfy it as well. This raises the question of whether EBSs exist. In the following discussion, we present a couple of examples of EBSs.

We start with an example in two qubits. Let $\mK$ be the subspace spanned by the two product states $|00\ra$ and $|11\ra$. We argue now that this subspace of $\mbb{C}^2\otimes\mbb{C}^2$ is indeed an EBS. Indeed, any state in $\psi\in\pure\left(\mK\otimes A'B'\right)$ can be expressed as
\be
|\psi^{ABA'B'}\ra=|00\ra^{AB}\otimes|\phi_0^{A'B'}\ra+|11\ra^{AB}\otimes|\phi_1^{A'B'}\ra
\ee
where $\phi_0^{AB}$ and $\phi_1^{AB}$ are some subnormalized pure states. Clearly, the reduced density matrix $\psi^{AA'B'}$ has the form
\be
\psi^{AA'B'}=|0\lr 0|^{A}\otimes\phi_0^{A'B'}+|1\lr 1|^{A}\otimes\phi_1^{A'B'}\;,
\ee
which is a separable state between $A$ and $A'B'$.

In the second example, we consider the 2-qutrit space $\mbb{C}^3\otimes\mbb{C}^3$. In Exercise~\ref{exebs} you show that its antisymmetric  subspace that is spanned by the three states
\ba
|\chi_1^{AB}\ra&\eqdef\frac1{\sqrt{2}}(|01\ra-|10\ra)\\
|\chi_2^{AB}\ra&\eqdef\frac1{\sqrt{2}}(|12\ra-|21\ra)\\
|\chi_3^{AB}\ra&\eqdef\frac1{\sqrt{2}}(|20\ra-|02\ra)\\
\ea
is an EBS.
\bex\label{exebs}
Let $\mK$ be the subspace spanned by the three vectors above. Show that:
\ben
\item For all $x,y\in[3]$ we have $\tr_B\left[|\chi_x^{AB}\lr\chi_y^{AB}|\right]=\left(\delta_{xy}I^A-|y\lr x|^A\right)$.
\item The reduced density matrix of every $\phi^{AB}\in\pure\left(\mK\right)$ of the form $|\phi^{AB}\ra=\sum_{x\in[3]}\lambda_x|\chi^{AB}_x\ra$ (with $\lambda_x\in\mbb{C}$)  can be expressed as
\be
\tr_B[\phi^{AB}]=\frac12 I^A-\frac12\varphi^T
\ee
where $|\varphi^{A}\ra\eqdef\sum_{x\in[3]}\lambda_x|x\ra^A$.
\item The subspace $\mK$ is EBS.
\een
\eex

\begin{myt}{}
\begin{theorem}\label{addif}
Let $\rho\in\pure(AB)$ be a bipartite quantum state whose support is an EBS. Then, for every bipartite system\index{bipartite system} $A'B'$ and every $\sigma\in\md(A'B')$ we have
\be
E_F\left(\rho^{AB}\otimes\sigma^{A'B'}\right)=E_F\left(\rho^{AB}\right)+E_F\left(\sigma^{A'B'}\right)\;.
\ee
In particular, the entanglement cost\index{entanglement cost} of $\rho^{AB}$ equals its entanglement of formation\index{entanglement of formation}.
\end{theorem}
\end{myt}
\begin{proof}
Due to~\eqref{subadeof} it is sufficient to prove that
\be
E_F\left(\rho^{AB}\otimes\sigma^{A'B'}\right)\geq E_F\left(\rho^{AB}\right)+E_F\left(\sigma^{A'B'}\right)\;.
\ee
Denote by $\mK\eqdef\supp\left(\rho^{AB}\right)$. The key idea is to first show that for every $\psi\in\pure\left(\mK\otimes A'B'\right)$ we have
\be
E\left(\psi^{ABA'B'}\right)\geq E_F\left(\psi^{AB}\right)+E_F\left(\psi^{A'B'}\right)\;,
\ee
where $E$ on the left-hand side is the entropy of entanglement\index{entropy of entanglement} between systems $AA'$ and $BB'$, and for simplicity of notations we use $\psi^{AB}$ and $\psi^{A'B'}$ to denote the \emph{mixed} marginal states of $\psi^{ABA'B'}$. To see why the above inequality holds, recall that $\psi^{ABA'B'}$ belongs to an EBS so that we can express it as
\be
|\psi^{ABA'B'}\ra\eqdef\sum_{x\in[m]}\sqrt{p_x}|x\ra^B\otimes|\phi_x^A\ra\otimes|\varphi_x^{A'B'}\ra
\ee
where $m\in\mathbb{N}$, $\p=(p_1,\ldots,p_m)^T\in\prob(m)$, and $\phi_x\in\pure(A)$ and $\varphi_x\in\pure(A'B')$ for each $x\in[m]$. We therefore get from the definition of the entropy of entanglement that
\ba\label{haap}
E\left(\psi^{ABA'B'}\right)=H\left(AA'\right)_\psi
\ea
where $H(AA')_\psi$ is the von-Neumann\index{von-Neumann} entropy of the marginal state
\be
\psi^{AA'}\eqdef\sum_{x\in[m]}p_x\phi_x^A\otimes\varphi_x^{A'}\quad\text{where}\quad\varphi_x^{A'}\eqdef\tr_{B'}\left[\varphi^{A'B'}_x\right]\;.
\ee
Now, let $C$ be an $m$-dimensional system and let $\sigma\in\md(AA'C)$ be the state
\be
\sigma^{AA'C}\eqdef\sum_{x\in[m]}p_x\phi_x^A\otimes\varphi_x^{A'}\otimes|x\lr x|^C\;.
\ee
Then, from the strong subadditivity\index{subadditivity} as given in~\eqref{strsa} and the fact that the marginal $\sigma^{AA'}=\psi^{AA'}$ we get 
\ba\label{somest1}
H(AA')_\psi&=H(AA')_\sigma\\ 
\GG{Strong\;subadditivity~\eqref{strsa}}&\geq H(A)_\sigma+H(AA'C)_\sigma-H(AC)_\sigma\\
\GG{Exercise~\ref{somest}}&=H\left(\psi^A\right)+\sum_{x\in[m]}p_xH\left(\varphi_x^{A'}\right)\;.
\ea
 Now, from~\eqref{eofub} we have $H\left(\psi^A\right)\geq E_F\left(\psi^{AB}\right)$. Moreover, by definition, for every $x\in[m]$ we have $H\left(\varphi_x^{A'}\right)=E\left(\varphi_x^{A'B'}\right)$. Combining this with the equation above and with~\eqref{haap} gives
 \ba
 E\left(\psi^{ABA'B'}\right)&\geq E_F\left(\psi^{AB}\right)+\sum_{x\in[m]}p_xE\left(\varphi_x^{A'B'}\right)\\
 &\geq E_F\left(\psi^{AB}\right)+E_F\left(\psi^{A'B'}\right)\;.
 \ea
  This completes the proof.
  \end{proof}

\bex\label{somest}
Prove the last equality in~\eqref{somest1}.
\eex

\bex
Compute the entanglement cost of the state
\be
\rho^{AB}\eqdef p\Phi_+^{AB}+(1-p)\Phi_-^{AB}\;,
\ee
for all $p\in[0,1]$, where
\be
|\Phi^{AB}_{\pm}\ra\eqdef\frac1{\sqrt{2}}\left(|00\ra\pm|11\ra\right)\;.
\ee
\eex

\section{Beyond LOCC: Non-Entangling Operations}\index{non-entangling operations}\label{beyond1}

As we have discussed earlier, the study of mixed-state entanglement is significantly more complex than that of pure bipartite entanglement, leading to numerous unresolved issues. The intricate nature of LOCC is the main contributing factor to this difficulty. To address this challenge, this section focuses on examining entanglement theory under  different sets of operations that are comparatively simpler to handle. Specifically, we explore briefly in this section the theory of entanglement under RNG operations, also referred to in entanglement theory as \emph{non-entangling operations}.

A quantum channel $\mE\in\cptp(AB\to A'B')$ is considered non-entangling if, for any $\sigma\in\sep(AB)$, $\mE(\sigma)\in\sep(A'B')$. It should be emphasized that the relative entropy of entanglement is a measure of entanglement that behaves monotonically under non-entangling operations. This subsection is dedicated to the calculation of the single-shot entanglement cost, as well as the distillable entanglement, under non-entangling operations. The calculation reveals that (1) the operational meaning of the relative entropy of a resource, as measured by hypothesis testing divergence, is the single-shot distillable entanglement and (2) the smoothed logarithmic robustness\index{robustness} can be interpreted as the single-shot entanglement cost.

\subsection{The Single-Shot Entanglement Cost}\index{single-shot}\index{entanglement cost}

In order to compute the single-shot entanglement cost under non-entangling operations, we first need simplify the expression for the conversion distance:
\be\label{mesa}
T\left(\Phi^{A'B'}_m\xrightarrow{\text{\tiny RNG}}\rho^{AB} \right)
=\frac12\min_{\mE\in\rng}\left\|\rho^{AB}-\mE^{A'B'\to AB}\left(\Phi_m^{A'B'}\right)\right\|_1\;.
\ee
Since $\Phi_m^{A'B'}$ is invariant under the action of the (self-adjoint) twirling map
\be
\mG\left(\omega^{A'B'}\right)=\int dU\;\left(U\otimes\overline{U}\right)\omega^{A'B'}\left(U\otimes\overline{U}\right)^*\quad\quad\forall\;\omega\in\ml(A'B')\;,
\ee
we can replace $\mE$ in~\eqref{mesa} with $\mE\circ\mG$, or in other words, we can assume without loss of generality that $\mE=\mE\circ\mG$. Any such non-entangling (RNG) operation has the form (see~\eqref{om1om2})
\be\label{asgiven2}
\mE(\omega)=\tr\left[(I-\Phi_m)\omega\right]\sigma^{AB}+\tr\left[\Phi_m\omega\right]\eta^{AB}\quad\quad\forall\;\omega\in\md(A'B')\;,
\ee
where $\sigma,\eta\in\md(AB)$. Note that the channel $\mE$ is RNG (i.e., non-entangling) if and only if 
\be\label{condaes}
\tr\left[(I-\Phi_m)\omega\right]\sigma+\tr\left[\Phi_m\omega\right]\eta\in\sep(AB)\quad\quad\forall\;\omega\in\sep(A'B')\;.
\ee
Now, recall that the density state $\tau\eqdef(I-\Phi_m)/(m^2-1)$ is a separable isotropic state\index{isotropic state} (see~\eqref{isot}). Taking $\omega=\tau$ above we get
$\mE(\tau)=\sigma$. Therefore, since $\tau$ is a separable state we get that $\sigma$ must be separable as well. More generally, from~\eqref{1mproved} we have that $\tr\left[\Phi_m^{AB}\omega^{AB}\right]\leq\frac1m$
for all separable states $\omega\in\sep(AB)$. Therefore, the condition in~\eqref{condaes} holds if and only if 
\be\label{sepsig}
\frac1{m}\eta^{AB}+\frac {m-1}m\sigma^{AB}\in\sep(AB)\;.
\ee
To summarize, the channel $\mE$ as given in~\eqref{asgiven2} is RNG if and only if $\sigma\in\sep(AB)$ and the equation above holds. Hence,
\be
T\left(\Phi^{A'B'}_m\xrightarrow{\text{\tiny RNG}}\rho^{AB} \right)=\min\frac12\left\|\rho^{AB}-\eta^{AB}\right\|_1
\ee
where the minimum is over all $\eta\in\md(AB)$ that satisfies~\eqref{sepsig} with some $\sigma\in\sep(AB)$. 

The robustness\index{robustness} measure of entanglement of the state $\eta^{AB}$ is defined as (see~\eqref{robustness})
\be
\R(\eta^{AB})\eqdef\min\left\{s\geq 0\;:\;\frac{\eta+s\sigma}{1+s}\in\sep(AB)\;\;,\;\;\sigma\in\sep(AB)\right\}\;.
\ee
Comparing this with the expression for the conversion distance above we get 
that conversion distance can be expressed compactly as
\be
T\left(\Phi^{A'B'}_m\xrightarrow{\text{\tiny RNG}}\rho^{AB} \right)=\frac12\min_{\substack{\eta\in\md(AB)\\ \R(\eta)\leq m-1}}\left\|\rho^{AB}-\eta^{AB}\right\|_1\;.
\ee
In other words, the conversion distance\index{conversion distance} above can be interpreted as the distance of $\rho^{AB}$ to the set of states with robustness no greater than $m-1$.

Using the compact expression for the conversion distance above, we get that for any $\eps\in(0,1)$, the $\eps$-single-shot entanglement cost\index{entanglement cost} under non-entangling operations is given by
\ba
\cost^\eps(\rho^{AB})&=\min_{m\in\mbb{N}}\Big\{\log m \;:\;\frac12\left\|\rho^{AB}-\eta^{AB}\right\|_1\leq\eps \;,\;\R\left(\eta^{AB}\right)\leq m-1\Big\}.
\ea
That is,
\be
\cost^\eps\left(\rho^{AB}\right)=\log \left(1+\R^\eps\left(\rho^{AB}\right)\right)=\LR^{\eps}\left(\rho^{AB}\right)\;.
\ee
The formula above provides an operational interpretation for the smoothed logarithmic robustness\index{robustness} as the single-shot entanglement cost\index{entanglement cost} under non-entangling operations.

\subsection{Single-Shot Distillable Entanglement}\index{single-shot}\index{distillable entanglement}

The conversion distance is given by
\be
T\left(\rho^{AB}\xrightarrow{\text{\tiny RNG}} \Phi^{A'B'}_m\right)
=1-\sup_{\mE\in\rng}\tr\left[\Phi_m^{A'B'}\mE^{AB\to A'B'}\left(\rho^{AB}\right)\right]\;.
\ee
Due to the symmetry of $\Phi_m$ we can assume without loss of generality that $\mE=\mG\circ\mE$ so that the non-entangling operation $\mE$ has the form (see~\eqref{8248})
\be
\mE(\omega)=\big(1-\tr\left[\Lambda\omega\right]\big)\tau^{A'B'}+\tr\left[\Lambda\omega\right]\Phi_m^{A'B'}\quad\quad\forall\;\omega\in\ml(AB)\;,
\ee
where $\Lambda\in\eff(AB)$, and 
\be
\tau^{A'B'}\eqdef\frac{I^{A'B'}-\Phi_m}{m^2-1}\;.
\ee
Observe that for $\omega\in\md(AB)$ the state $\mE(\omega)$ is an isotropic state\index{isotropic state}, and therefore is entangled if and only if $\tr[\Lambda\omega]\leq\frac1m$ (we used again~\eqref{isot}). Since $\Lambda=\mE^*(\Phi_m)$ we conclude that
\be
T\left(\rho\xrightarrow{\text{\tiny RNG}} \Phi_m\right)
=1-\sup_{\Lambda\in\eff(AB)}\left\{\tr\left[\Lambda\rho\right]\;:\;\tr\left[\Lambda\sigma\right]\leq\frac1m,\;\forall\;\sigma\in\sep(AB)\right\}\;.
\ee
Therefore, the distillation is given by
\ba\label{optimm}
\distill^\eps\left(\rho\right)&\eqdef\sup_{m\in\mbb{N}}\left\{\log m\;:\;T\left(\rho\xrightarrow{\text{\tiny RNG}} \Phi_m\right)\leq\eps\right\}\\
&=\sup_{m\in\mbb{N}}\left\{\log m\;:\;\tr\left[\Lambda\rho\right]\geq 1-\eps,\;\max_{\sigma\in\sep(AB)}\tr\left[\Lambda\sigma\right]\leq\frac1m,\;,\; \Lambda\in\eff(AB)\right\}\;.
\ea
The condition that $m$ is no greater than the reciprocal of $\max_{\sigma\in\sep(AB)}\tr\left[\Lambda\sigma\right]$ implies the following:
\ba
\distill^\eps\left(\rho\right)&
\leq-\log\min_{\substack{\Lambda\in\eff(AB)\\ \tr\left[\Lambda\rho\right]\geq 1-\eps}}\max_{\sigma\in\sep(AB)}\tr\left[\Lambda\sigma\right]\\
&=-\log\max_{\sigma\in\sep(AB)}\min_{\substack{\Lambda\in\eff(AB)\\ \tr\left[\Lambda\rho\right]\geq 1-\eps}}\tr\left[\Lambda\sigma\right]\\
&=\min_{\sigma\in\sep(AB)}D_{\min}^{\eps}\left(\rho\|\sigma\right)\\
&=D_{\min}^\eps\left(\rho\|\sep\right)\;.
\ea

\bex
Using the same notations as above, show that
\be
\distill^\eps\left(\rho\right)=\log\left\lfloor2^{D_{\min}^\eps\left(\rho\|\sep\right)}\right\rfloor
\ee
Hint: Observe that the optimal $m$ in~\eqref{optimm} is the floor of the reciprocal of $\max_{\sigma\in\sep(AB)}\tr\left[\Lambda\sigma\right]$.
\eex

\subsection{The Asymptotic Regime}

The results obtained in the single-shot regime lead directly to the following expressions for the cost and distillation of a bipartite state $\rho\in\md(AB)$ under non-entangling operations:
\ba
&\cost\left(\rho^{AB}\right)=\lim_{\eps\to 0}\liminf_{n\to\infty}\frac1n\LR^{\eps}\left(\rho^{\otimes n}\right)\\
&\distill\left(\rho^{AB}\right)=\lim_{\eps\to 0}\limsup_{n\to\infty}\frac1nD_{\min}^\eps\left(\rho^{\otimes n}\|\sep\right)\;.
\ea
From these expressions, it becomes evident that if the generalized quantum Stein's lemma\index{Stein's lemma} (as proposed in Conjecture~\ref{gsl}) is valid, then the distillable entanglement under non-entangling operations would be equal to the regularized relative entropy of entanglement.

In contrast, regarding the entanglement cost, it is known (as detailed in the section on `Notes and References' at the end of this chapter) that there are states for which the entanglement cost is strictly greater than the distillable entanglement. This implies that even under a broad range of non-entangling operations, the reversibility of mixed state entanglement is not guaranteed. In essence, this reflects a fundamental asymmetry in the processes of creating and extracting entanglement from quantum systems.

\section{Beyond LOCC: NPT-Entanglement Theory}\label{sec:npt}\index{NPT-entanglement}

In this subsection, we delve into the quantum resource theory where $\mf(AB)$ is defined as the set of PPT states within $\md(AB)$, and $\mf(AB\to A'B')$ as the set of completely PPT preserving quantum channels. The exploration of this resource theory is not only intriguing from a theoretical standpoint but is also driven by the fact that the set of completely PPT preserving quantum channels encompasses LOCC. Consequently, the entanglement cost and distillation rates determined under these operations offer lower and upper bounds, respectively, on the corresponding rates under LOCC.

In the framework of completely PPT-preserving operations, entangled states that exhibit a positive partial transpose are regarded as free resources. Accordingly, in this resource theory, the focus is on \emph{NPT-entanglement}, emphasizing interest in entangled states with a negative partial transpose (NPT), which are states whose partial transpose is not positive semidefinite. In the rest of this section, we will use the notation $\ppt(AB)$ to refer to the set of all density matrices in $\md(AB)$ that have a positive partial transpose. Thus, we obtain the following relation:
\be
\sep(AB)\subset\ppt(AB)\subset\md(AB)\;.
\ee

\subsection{The Set of Completely PPT Preserving Operations}
\index{PPT operations}

\begin{myd}{}
\begin{definition}
Let $\mE\in\ml(AB\to A'B')$ be a linear map. We say that $\mE$ is PPT preserving if 
\be
\mE(\rho)\in\ppt\left(A'B'\right)\quad\quad\forall\;\rho\in\ppt(AB)\;.
\ee
Moreover, we say the $\mE^{AB\to A'B'}$ is completely PPT preserving if $\id^{A''B''}\otimes\mE^{AB\to A'B'}$ is PPT preserving for every bipartite system\index{bipartite system} $A''B''$.
\end{definition}
\end{myd}

To characterize the set of completely PPT preserving operations, we begin by defining the partial transpose\index{partial transpose} of a linear map. Let $\mE\in\ml(AB\to A'B')$ be a linear map. The partial transpose of $\mE$ is defined as follows:
\be
\mE^\Gamma(\omega)\eqdef\Big(\mE\left(\omega^\Gamma\right)\Big)^\Gamma\quad\quad\forall\;\omega\in\ml(AB)\;,
\ee
where the superscript $\Gamma$ indicates partial transpose w.r.t.\ Bob's systems (see~\eqref{pptg}). In the following exercise you prove some of the key properties of this extension of partial transpose to linear maps.

\bex
Let $\mE\in\ml(AB\to A'B')$ be a bipartite linear map, $\mE^\Gamma$ be its partial transpose, and $J_\mE\eqdef J_{\mE}^{ABA'B'}$ be its Choi matrix. Prove the following statements:
\ben
\item $\mE$ is PPT preserving if and only if $\mE^\Gamma$ is PPT preserving.
\item $\mE$ is completely PPT preserving if and only if $\mE^\Gamma$ is completely PPT preserving.
\item $\mE$ satisfies:
\be\label{ex1343}
J_{\mE^\Gamma}=J_{\mE}^\Gamma\;,
\ee
where on the right-hand side the superscript $\Gamma$ denotes the partial transpose of $J_\mE^{ABA'B'}$ with respect to both $B$ and $B'$.
\item $\mE$ satisfies:
\be\label{megamma}
\left(\mE^*\right)^\Gamma=\left(\mE^\Gamma\right)^*
\;.
\ee
\een
\eex

\begin{myt}{}
\begin{theorem}\label{alsocptp}
Let $\mE\in\cptp(AB\to A'B')$ be a bipartite quantum channel. Then, the following are equivalent:
\ben
\item $\mE$ is completely PPT preserving.
\item $\mE^\Gamma$ is a quantum channel.
\een
\end{theorem}
\end{myt}

\begin{proof}

We can use~\ref{ex1343} to observe that $\mE^\Gamma$ is a quantum channel if and only if $J_\mE^\Gamma\geq 0$. Thus, it suffices to show that $\mE$ is completely PPT preserving if and only if its Choi matrix is PPT.
To see this, note that the Choi matrix of $\mE^{AB\to A'B'}$ can be expressed as
\begin{align}
J_\mE^{ABA'B'}=\mE^{\tA\tB\to A'B'}\left(\Omega^{(AB)(\tA\tB)}\right),
\end{align}
where $\Omega^{(AB)(\tA\tB)}$ is an unnormalized maximally entangled state between system $\tA\tB$ and system $AB$. Furthermore, we can write
\begin{align}
\Omega^{(AB)(\tA\tB)}=\Omega^{A\tA}\otimes \Omega^{B\tB},
\end{align}
where $\Omega^{A\tA}$ and $\Omega^{B\tB}$ are unnormalized maximally entangled states between the respective systems. Since we take the partial transpose with respect to system $B\tB$, we get from the equation above that $\Omega^{(AB)(\tA\tB)}$ is PPT with respect to $B\tB$.
Therefore, if $\mE$ is completely PPT preserving, then its Choi matrix $J_\mE^{ABA'B'}$ must  be PPT, since $\Omega^{ABA'B'}$ is PPT. 

Conversely, suppose $J_\mE^\Gamma\geq 0$ (i.e. $\mE^\Gamma$ is a quantum channel) and let $\rho^{A''B''AB}$ be a PPT state with respect to system $B''B$. For simplicity of the exposition here, we use the superscript $\Gamma$ to indicate partial transpose on all systems on Bob's side (i.e. for $\rho^{AB}$, the superscript $\Gamma$ in $\rho^{\Gamma}$ stands for partial transpose w.r.t.\ $B$, and for $\rho^{A''B''AB}$, the superscript in $\rho^\Gamma$ stands for partial transpose\index{partial transpose} w.r.t.\ to system $B''B$).
Then,
\ba
\left(\mE^{AB\to A'B'}\big(\rho^{A''B''AB}\big)\right)^\Gamma=\mE^{\Gamma}\left(\rho^{\Gamma}\right)\geq 0\;,
\ea
since $\rho^\Gamma\geq 0$ and $\mE^\Gamma$ is completely positive. Therefore, the state $\mE^{AB\to A'B'}\left(\rho^{A''B''AB}\right)$ is PPT, and since $\rho^{A''B''AB}$ was an arbitrary PPT state in $\md(A''B''AB)$ we conclude that $\mE$ is completely PPT preserving. This completes the proof.
\end{proof}

We denote by $\ppt(AB\to A'B')$ the set of all completely PPT preserving channels in $\cptp(AB\to A'B')$.

\bex
Recall that $LOCC(AB\to A'B')$ is a subset of $\sep(AB\to A'B')$; see~\eqref{ex12p6}. Show that
\be
\sep(AB\to A'B')\subset\ppt(AB\to A'B')\;.
\ee
\eex

\bex
Show that if $\mE\in\ppt(AB\to A'B')$ then there exists a channel $\mN\in\ppt(AB\to A'B')$ such that $\mE=\mN^\Gamma$.
\eex

\subsubsection{Monotonicity of Entanglement Measures}

The above theorem states that a quantum channel $\mE\in\cptp(AB\to A'B')$ is completely PPT preserving if and only if its partial transpose $\mE^\Gamma$ is also completely positive. Using this property, we can show that measures of entanglement based on the partial transpose\index{partial transpose} behave monotonically under completely PPT preserving operations. Indeed,
for $\rho\in\md(AB)$ and $\mE\in\ppt(AB\to A'B')$, we have $(\mE(\rho))^\Gamma=\mE^\Gamma(\rho^\Gamma)$, which gives $\|\big(\mE(\rho)\big)^\Gamma\|_1=\|\mE^\Gamma(\rho^\Gamma)\|_1\leq\|\rho^\Gamma\|_1$ by DPI and the fact that $\mE^\Gamma$ is a quantum channel. Therefore, both the negativity and the logarithmic negativity behave monotonically under completely PPT preserving operations.

To see that $E_\kappa$ also behaves monotonically, consider an optimal operator $\Lambda\in\pos(AB)$ that satisfies $-\Lambda^\Gamma\leq\rho^\Gamma\leq\Lambda^\Gamma$ and $E_\kappa(\rho)=\log\tr[\Lambda]$. Then, for all $\mE\in\ppt(AB\to A'B')$, we have:
\be
-\mE^\Gamma\left(\Lambda^\Gamma\right)\leq\mE^\Gamma\left(\rho^\Gamma\right)\leq\mE^\Gamma\left(\Lambda^\Gamma\right)
\ee
since $\mE^\Gamma$ is CP. Since $(\mE(\rho))^\Gamma=\mE^\Gamma(\rho^\Gamma)$ we get that the above equation is equivalent to
\be
-\Lambda'^\Gamma\leq\big(\mE(\rho)\big)^\Gamma\leq \Lambda'^\Gamma\quad\text{where}\quad\Lambda'\eqdef\mE(\Lambda)\;.
\ee
Therefore, $\Lambda'$ is a feasible solution in the optimization problem of $E_\kappa\big(\mE(\rho)\big)$ so that
\be
E_\kappa\big(\mE(\rho)\big)\geq\log\tr\left[\Lambda'\right]=\log\tr[\Lambda]=E_\kappa(\rho)\;.
\ee
That is, $E_\kappa$ behaves monotonically under completely PPT preserving operations.

\bex
Consider the linear map $\mE\in\cptp(AB\to AB)$ with $m\eqdef|A|=|B|$ defined for all $\omega\in\ml(AB)$ as
\be
\mE(\omega)\eqdef\tr[\Lambda\omega]\u^{AB}+\tr[(I-\Lambda)\omega]\Phi_m^{AB}\;,
\ee
where
\be
\Lambda\eqdef\frac1{m+1}\left(I^{AB}-\Phi_m^{AB}\right)^\Gamma\;.
\ee
\ben
\item Show that $\Lambda\in\eff(AB)$.
\item Show that $\mE$ is PPT preserving (but not necessarily completely PPT preserving).
\item Show that for $m>3$ we have $N\left(\mE\left(\rho_{\text{\tiny W}}^{AB}\right)\right)>N\left(\rho_{\text{\tiny W}}^{AB}\right)$, where $\rho_{\text{\tiny W}}^{AB}$ is the maximally entangled Werner state (see~\eqref{wcpact} with $\alpha=1$)
\be
\rho_{\text{\tiny W}}^{AB}=\frac{1}{d(d-1)}\left(I^{AB}-F^{AB}\right)\;.
\ee
\een
\eex
The exercise above demonstrates that the negativity measure, in general, does not exhibit monotonic behavior under PPT-preserving operations, but as we saw earlier, it does exhibit monotonic behavior under \emph{completely} PPT-preserving operations.

\subsection{The PPT Conversion Distance}

In this section, we aim to simplify the calculation of the conversion distance\index{conversion distance} under PPT operations. Since our focus is on the cost and distillation of entanglement under PPT operations, we will only consider the conversion distance $T(\rho\xrightarrow{\text{\tiny PPT}}\sigma)$ when either $\rho$ or $\sigma$ is maximally entangled. We begin with the case where $\sigma$ is maximally entangled.

\subsubsection{The PPT Conversion Distance to a Maximally Entangled State}

Very similar to Lemma~\ref{lem1331}, the conversion distance under completely PPT preserving operations is given by
\be\label{convdlem}
T\left(\rho^{AB}\xrightarrow{\text{\tiny PPT}} \Phi^{A'B'}_m\right)
=1-\sup_{\mE\in\ppt}\tr\left[\Phi_m^{A'B'}\mE^{AB\to A'B'}\left(\rho^{AB}\right)\right]\;.
\ee
In order to simplify the expression above for the conversion distance, we will need the following lemma.

\begin{myg}{}
\begin{lemma}
Let $A,B,A',B'$ be four quantum systems $m\eqdef|A'|=|B'|$, and let $\Lambda\in\herm(AB)$. The following are equivalent:
\ben
\item There exists $\mE\in\ppt(AB\to A'B')$ such that
$
\Lambda^{AB}=\mE^*\left(\Phi_m^{A'B'}\right)
$.
\item $\Lambda\in\eff(AB)$ and 
$
\left\|\Lambda^{\Gamma}\right\|_\infty\leq\frac1m
$.
\een
\end{lemma}
\end{myg}

\begin{proof}
Suppose first that $\Lambda=\mE^*(\Phi_m)$ for some $\mE\in\ppt(AB\to A'B')$. Since $\mE$ is a CPTP map it follows that $\mE^*$ is a unital CP map. Therefore, since $0\leq\Phi_m\leq I^{A'B'}$ we have $0\leq \Lambda\leq I^{AB}$. Moreover, observe that
\ba
\Lambda^\Gamma=\Big(\mE^*(\Phi_m)\Big)^\Gamma
&=\left(\mE^*\right)^\Gamma\left(\Phi_m^\Gamma\right)\\
\GG{\eqref{megamma}}&=\left(\mE^\Gamma\right)^*\left(\Phi_m^\Gamma\right)\\
&=\frac1m\left(\mE^\Gamma\right)^*\left(F^{AB}\right)\;,
\ea
where $F^{AB}$ is the flip operator. Since $\mE^\Gamma$ is also a CPTP map (see Theorem~\ref{alsocptp}) it follows that $\left(\mE^\Gamma\right)^*$ is a unital CP map. Combining this with the fact that $-I^{AB}\leq F^{AB}\leq I^{AB}$ (recall $F^2=I^{AB}$) we conclude that 
\be
-\frac{1}{m}I^{AB}\leq\Lambda^\Gamma\leq\frac{1}{m}I^{AB}\;,
\ee
which is equivalent to $
\left\|\Lambda^{\Gamma}\right\|_\infty\leq\frac1m
$.

Conversely, suppose $\Lambda\in\eff(AB)$ and 
$
\left\|\Lambda^{\Gamma}\right\|_\infty\leq\frac1m
$. Define the measurement-prepare channel\index{measurement-prepare channel} $\mE\in\cptp(AB\to A'B')$ as
\be
\mE\left(\omega\right)\eqdef\tr\left[\Lambda\omega\right]\Phi_m+\tr\left[(I-\Lambda)\omega\right]\tau\quad\quad\forall\;\omega\in\ml(AB)\;,
\ee
where $\tau\eqdef(I-\Phi_m)/(m^2-1)\in\md(A'B')$.  Observe that $\Lambda=\mE^{*}(\Phi_m)$ (can you see why?). The intuition behind the definition above comes from the observation that the optimization over the PPT channels in~\eqref{convdlem} can be further restricted to channels that satisfy $\mE=\mG\circ\mE$ which according to~\eqref{8248} have the form of the channel above. It is therefore left to show that $\mE$ as defined above is a PPT quantum channel. 

Indeed, observe that by definition for all $\omega\in\ml(AB)$ we have
\ba\label{megs}
\mE^\Gamma(\omega)&=\Big(\mE\left(\omega^\Gamma\right)\Big)^\Gamma\\
&=\tr\left[\Lambda\omega^\Gamma\right]\Phi_m^\Gamma+\tr\left[(I-\Lambda)\omega^\Gamma\right]\tau^\Gamma\\
\GG{\substack{\text{The partial transpose}\\\text{is self-adjoint}}}&=\tr\left[\Lambda^\Gamma\omega\right]\Phi_m^\Gamma+\tr\left[\left(I-\Lambda^\Gamma\right)\omega\right]\tau^\Gamma\;.
\ea
Now, from~\eqref{sw1} and \eqref{sw2} we get that
\be\label{pmg}
\Phi_m^\Gamma=\frac1mF^{A'B'}=\frac1m\left(\Pi_\sym-\Pi_\asy\right)\;,
\ee
and (recall $I=\Pi_\sym+\Pi_\asy$)
\be
\tau^\Gamma=\frac{I-\Phi_m^\Gamma}{m^2-1}
=\frac1{m(m+1)}\Pi_\sym+\frac1{m(m-1)}\Pi_\asy\;.
\ee
Substituting these expressions for $\Phi_m^\Gamma$ and $\tau^\Gamma$ into~\eqref{megs} (and rearranging terms) gives (see Exercise~\ref{furtherdetails})
\be\label{fdtls}
\mE^\Gamma(\omega)=\tr\left[\Lambda'\omega\right]\sigma_{\sym}+\tr\left[\left(I-\Lambda'\right)\omega\right]\sigma_\asy\;,
\ee
where
\be
\Lambda'\eqdef\frac12\big(I+m\Lambda\big)\;\;,\;\;\sigma_\sym\eqdef\frac2{m(m+1)}\Pi_\sym\;\;,\;\;
\sigma_\asy\eqdef\frac2{m(m-1)}\Pi_{\asy}\;.
\ee
Hence, since $\|\Lambda\|_{\infty}\leq1/m$ we get that $\Lambda'\in\eff(A'B')$ so that $\mE^\Gamma$ is itself a measurement-prepare quantum channel. That is, $\mE$ is indeed a PPT channel. This completes the proof.
\end{proof}

\bex\label{furtherdetails}
Give more details for the derivation of~\eqref{fdtls}.
\eex

We therefore get the following corollary.
\begin{myg}{}
\begin{corollary}
Using the same notations as above, the PPT conversion distance to a maximally entangled state is given by
\be\label{cdppt}
T\left(\rho^{AB}\xrightarrow{\text{\tiny PPT}} \Phi^{A'B'}_m\right)
=1-\max_{\substack{\Lambda\in\eff(AB)\\\left\|\Lambda^\Gamma\right\|_\infty\leq\frac1m}}\tr\left[\Lambda^{AB}\rho^{AB}\right]\;.
\ee
\end{corollary}
\end{myg}

\bex
Use the lemma above to prove the corollary.
\eex
 
\subsubsection{The PPT Conversion Distance from a Maximally Entangled State}

The conversion distance\index{conversion distance} from a maximally entangled state is given by
\be
T\left(\Phi^{A'B'}_m\xrightarrow{\text{\tiny PPT}} \rho^{AB} \right)=\frac12\min_{\mE\in\ppt}\left\|\rho^{AB}-\mE^{A'B'\to AB}\left(\Phi_m^{A'B'}\right)\right\|_1\;.
\ee
In the next lemma we provide a characterization of the density matrix $\mE^{A'B'\to AB}\left(\Phi_m^{A'B'}\right)$.
\begin{myg}{}
\begin{lemma}
Let $A,B,A',B'$ be four quantum systems $m\eqdef|A'|=|B'|$, and let $\sigma\in\md(AB)$. The following are equivalent:
\ben
\item There exists $\mE\in\ppt(AB\to A'B')$ such that
$
\sigma^{AB}=\mE\left(\Phi_m^{A'B'}\right)
$.
\item There exists $\omega\in\md(AB)$ such that 
\be\label{m+1}
(1-m)\omega^\Gamma\leq\sigma^\Gamma\leq(1+m)\omega^\Gamma\;.
\ee
\een
\end{lemma}
\end{myg}
\begin{remark}
The condition presented in equation~\eqref{m+1} leads to the implication that $(m+1)\omega^\Gamma \geq (1-m)\omega^\Gamma$. This inequality can be simplified and equivalently restated as $m\omega^\Gamma \geq 0$. Thus, the partial transpose of $\omega$ is positive semidefinite, so that $\omega\in\ppt(AB)$.
\end{remark}
\begin{proof}
Let $\mG\in\cptp(A'B'\to A'B')$ be the twirling channel as defined in~\eqref{gtwi2}. Recall that $\mG$ is an LOCC channel, and thus it is completely PPT preserving. If $\sigma=\mE(\Phi_m)$ for some PPT channel $\mE$, we can assume without loss of generality that $\mE=\mE\circ\mG$. Otherwise, we can replace $\mE$ with $\mE'=\mE\circ\mG$, which has the desired property. From~\eqref{om1om2} it then follows that for all $\eta\in\ml(AB)$
\be
\mE(\eta)=\tr\left[\eta\Phi_m\right]\sigma+\tr\left[\left(I-\Phi_m\right)\eta\right]\omega\;,
\ee 
where we replaced $\omega_1$ in~\eqref{om1om2} with $\mE(\Phi_m)=\sigma$ and renamed the density matrix $\omega_2$ as $\omega\in\md(AB)$. The partial transpose of $\mE$ is given for all $\eta\in\ml(AB)$ as
\ba
\mE^\Gamma(\eta)&=\tr\left[\eta^\Gamma\Phi_m\right]\sigma^\Gamma+\tr\left[\left(I-\Phi_m\right)\eta^\Gamma\right]\omega^\Gamma\\
\GG{\substack{\text{Partial transpose}\\ \text{is self-adjoint}}}&=\tr\left[\eta\Phi_m^\Gamma\right]\sigma^\Gamma+\tr\left[\left(I-\Phi_m^\Gamma\right)\eta\right]\omega^\Gamma\;.
\ea
We next use~\eqref{pmg} to express $\Phi_m^\Gamma$ in terms of the symmetric and antisymmetric projectors. Hence,
\ba
\mE^\Gamma(\eta)&=\frac1m\tr\left[\eta\left(\Pi_\sym-\Pi_\asy\right)\right]\sigma^\Gamma+\frac1m\tr\left[\big((m-1)\Pi_\sym+(m+1)\Pi_\asy\big)\eta\right]\omega^\Gamma\\
&=\frac1m\tr\left[\eta\Pi_\sym\right]\left(\sigma^\Gamma+\left(m-1\right)\omega^\Gamma\right)+\frac1m\tr\left[\eta\Pi_\asy\right]\left(\left(1+m\right)\omega^\Gamma-\sigma^\Gamma\right)\;.
\ea
Hence, $\mE^\Gamma$ is completely positive\index{completely positive} if and only if the matrices $\sigma$ and $\omega$ satisfy~\eqref{m+1}. This completes the proof.
\end{proof}

From the lemma above it follows that the conversion distance can be expressed as
\be\label{cdcos}
T\left(\Phi^{A'B'}_m\xrightarrow{\text{\tiny PPT}} \rho^{AB} \right)=\min_{\omega,\sigma\in\md(AB)}\left\{\frac12\|\rho-\sigma\|_1\;:\;(1-m)\omega^\Gamma\leq\sigma^\Gamma\leq(1+m)\omega^\Gamma\right\}\;.
\ee

Now that we have obtained formulas for the two types of PPT-conversion distances in this subsection, we can use them for the operational tasks of entanglement distillation and entanglement cost.

\subsection{Distillation of NPT Entanglement}\index{hypothesis testing}\index{single-shot}

In this subsection, we express the single-shot distillable NPT-entanglement in terms of the hypothesis testing divergence and show that it can be computed using an SDP program. We then use it to derive a tight upper bound on the asymptotic distillable NPT-entanglement, known as Rains' bound. Since the distillable NPT-entanglement is not smaller than the LOCC distillable entanglement\index{distillable entanglement}, we can conclude that this upper bound also applies to the distillable entanglement under LOCC.

\subsubsection{Single-Shot Distillable NPT-Entanglement}\index{single-shot}

For every $\eps\in(0,1)$, the $\eps$-single-shot distillable NPT-entanglement of a bipartite state $\rho\in\md(AB)$ is defined as
\be
\distill^\eps\left(\rho^{AB}\right)=\sup_{m\in\mbb{N}}\Big\{\log m\;:\;T\left(\rho^{AB}\xrightarrow{\text{\tiny PPT}}  \Phi^{A'B'}_m\right)\leq\eps\Big\}\;.
\ee

\begin{myt}{}
\begin{theorem}\index{hypothesis testing}
Let $\rho\in\md(AB)$ and $\eps\in(0,1)$. Then, the $\eps$-single-shot distillable NPT-entanglement is given by
\be\label{nptdist}
\distill^\eps\left(\rho^{AB}\right)=\min_{\substack{\|\eta^\Gamma\|_1\leq 1\\\eta\in\herm(AB) }}\log\left\lfloor 2^{D_{\min}^\eps\left(\rho^{AB}\|\eta^{AB}\right)}\right\rfloor\;,
\ee
where the definition of the hypothesis testing divergence above has been extended to operators that are not necessarily density matrices.

\end{theorem}
\end{myt}
\begin{proof}
Using the expression for the conversion distance given in~\eqref{cdppt} we get
\ba\label{minimalva}
\distill^\eps\left(\rho\right)&=\sup_{m\in\mbb{N}}\Big\{\log m\;:\;\tr\left[\Lambda\rho\right]\geq 1-\eps,\;\left\|\Lambda^\Gamma\right\|_\infty\leq\frac1m,\;\Lambda\in\eff(AB)\Big\}\\
\Gg{m=\left\lfloor\frac1{\left\|\Lambda^\Gamma\right\|_\infty}\right\rfloor}&=\max\Big\{\log\left\lfloor\frac1{\left\|\Lambda^\Gamma\right\|_\infty}\right\rfloor\;:\;\tr\left[\Lambda\rho\right]\geq 1-\eps,\;\Lambda\in\eff(AB)\Big\}
\ea
Now, from Exercise~\ref{optrrel} we have that 
\ba
\left\|\Lambda^\Gamma\right\|_\infty=\max_{\substack{ \|\eta\|_1\leq 1\\\eta\in\herm(AB)}}\tr\left[\Lambda^\Gamma\eta\right]\\
\GG{\Gamma\text{ is self-adjoint}}=\max_{\substack{ \|\eta\|_1\leq 1\\\eta\in\herm(AB)}}\tr\left[\Lambda\eta^\Gamma\right]
\ea
We therefore get that the minimal value that $\left\|\Lambda^\Gamma\right\|_\infty$ can take under the constraints given in~\eqref{minimalva} is given by:
\ba
\min_{\substack{\tr\left[\Lambda\rho\right]\geq 1-\eps\\ \Lambda\in\eff(AB)}}\left\|\Lambda^\Gamma\right\|_\infty
&=\min_{\substack{\tr\left[\Lambda\rho\right]\geq 1-\eps\\ \Lambda\in\eff(AB)}}\;\max_{\substack{\|\eta\|_1\leq 1\\\eta\in\herm(AB) }}\tr\left[\Lambda\eta^\Gamma\right]\\
\GG{minmax\; theorem}&=\max_{\substack{\|\eta\|_1\leq 1\\\eta\in\herm(AB) }}\;\min_{\substack{\tr\left[\Lambda\rho\right]\geq 1-\eps\\ \Lambda\in\eff(AB)}}\tr\left[\Lambda\eta^\Gamma\right]\\
&=\max_{\substack{\|\eta\|_1\leq 1\\\eta\in\herm(AB) }}2^{-D_{\min}^\eps\left(\rho\|\eta^\Gamma\right)}\\
&=\max_{\substack{\|\eta^\Gamma\|_1\leq 1\\\eta\in\herm(AB) }}2^{-D_{\min}^\eps\left(\rho\|\eta\right)}\;.
\ea
Substituting this into~\eqref{minimalva} we obtain~\eqref{nptdist}. This completes the proof.
\end{proof}

\bex
Show that the $\eps$-single shot distillable NPT-entanglement can be be computed with an SDP program and find the dual problem\index{dual problem} of~\eqref{nptdist}.
\eex

\bex
Let $\eps\in(0,1)$, $\rho\in\md(AB)$, and $\eta\in\herm(AB)$. Show that for all $\mE\in\cptp(AB\to A'B')$ we have
\be
D_{\min}^\eps\left(\mE(\rho)\big\|\mE(\eta)\right)\leq D_{\min}^\eps(\rho\|\eta)\;.
\ee
That is, the DPI with $D_{\min}^\eps$ still holds even if $\eta$ is not positive semidefinite.
\eex

For states that have a certain symmetry, the optimization problem given in~\eqref{nptdist} can be performed analytically. For example, consider the Werner state\index{Werner state}, $\rho^{AB}_{\text{\tiny W}}$, as defined in~\eqref{werner}. This state is invariant under the twirling map $\mG\in\cptp(AB\to AB)$ as defined in~\eqref{twirlingw}. Let $\eta\in\herm(AB)$ be an optimal matrix such that
\be
\distill^\eps\left(\rho^{AB}_{\text{\tiny W}}\right)=D_{\min}^\eps\left(\rho^{AB}_{\text{\tiny W}}\big\|\eta^{AB}\right)\;.
\ee
Due to the invariance\index{invariance} property of $\rho_{\text{\tiny W}}^{AB}$ we have
\ba
D_{\min}^\eps(\rho_{\text{\tiny W}}\|\eta)&\geq D_{\min}^\eps\left(\mG(\rho_{\text{\tiny W}})\big\|\mG(\eta)\right)\\
&=D_{\min}^\eps\left(\rho_{\text{\tiny W}}\big\|\mG(\eta)\right)\;.
\ea
Moreover, since $\mG\in\ppt(AB\to AB)$ and $\|\eta^\Gamma\|_1\leq 1$ we get that also $\zeta\eqdef\mG(\eta)$ satisfies
\ba
\left\|\zeta^\Gamma\right\|_1=\left\|\mG^\Gamma\left(\eta^\Gamma\right)\right\|_1\\
\GG{DPI\to}\leq \left\|\eta^\Gamma\right\|_1\leq 1\;.
\ea
Therefore, since $\eta$ was optimal, we must have $D_{\min}^\eps(\rho_{\text{\tiny W}}\|\eta)=D_{\min}^\eps(\rho_{\text{\tiny W}}\|\zeta)$, which means that $\zeta$ is also optimal. To summarize, without loss of generality, we can restrict the optimization in~\eqref{nptdist} to Hermitian matrices $\eta\in\herm(AB)$ that satisfy both $\|\eta^\Gamma\|_1\leq 1$ and $\mG(\eta)=\eta$.
This additional condition implies that $\eta$ can be written as a linear combination of $I^{AB}$ and $F^{AB}$, or equivalently, $\eta^\Gamma$ can be expressed as
\be\label{ptbs}
\eta^{\Gamma}=a\Phi_m^{AB}+b\tau^{AB}_m\;,
\ee
for some $a,b\in\mbb{R}$, and $\tau_m^{AB}\eqdef(I^{AB}-\Phi_m^{AB})/(m^2-1)$. 
\bex
Let $\rho_{\text{\tiny W}}^{AB}$ be the Werner state\index{Werner state} with $\alpha>\frac1m$ (i.e., $\rho_{\text{\tiny W}}^{AB}$ is entangled). Show that for $m>2$
\be
\distill^\eps\left(\rho^{AB}_{\text{\tiny W}}\right)=\log\left\lfloor\frac{m(m+\alpha)-2}{m(m-\alpha)(1-\eps)}\right\rfloor\;.
\ee
Hint: Optimize over $a,b\in\mbb{R}$ that appear in~\eqref{ptbs} and note that since $\Phi_m^{AB}\tau^{AB}_m=0$, we get
$
\left\|\eta^\Gamma\right\|_1=|a|+|b|\;,
$
so the condition $\|\eta^\Gamma\|_1\leq 1$ becomes equivalent to $|a|+|b|\leq 1$.
\eex

In the optimization problem in~\eqref{nptdist}, the operator $\eta\in\herm(AB)$ does not have to be positive semidefinite. However, if we restrict it to be in $\pos(AB)$ (instead of $\herm(AB)$) we get the following upper bound.

\begin{myg}{}
\begin{corollary}\label{sscor}
Let $\rho\in\md(AB)$ and $\eps\in(0,1)$. Then, the $\eps$-single-shot distillable NPT-entanglement is bounded from above by
\be\label{nptdist2}
\distill^\eps\left(\rho^{AB}\right)\leq\min_{\sigma\in\md(AB)}\Big\{D_{\min}^\eps\left(\rho\|\sigma\right)+\LN(\sigma)\Big\}\;,
\ee
where $\LN$ stands for the logarithmic negativity as defined in~\eqref{logneg}.
\end{corollary}
\end{myg}
\begin{proof}
By removing the floor function, and replacing $\herm(AB)$ in the right-hand side of~\eqref{nptdist} with the smaller set $\pos(AB)$, we get 
\ba
\distill^\eps(\rho)&\leq \min_{\substack{\|\eta^\Gamma\|_1\leq 1\\\eta\in\pos(AB) }}D_{\min}^\eps\left(\rho\|\eta\right)\\
\Gg{\eta=t\sigma}&=\min_{\substack{t\|\sigma^\Gamma\|_1\leq 1\\\sigma\in\md(AB),\;t\geq 0}}\Big\{D_{\min}^\eps\left(\rho\|\sigma\right)-\log t\Big\}\\
\GG{\substack{\text{Taking the largest}\\ \text{possible}\;t=1/\|\sigma^\Gamma\|_1}}&=\min_{\sigma\in\md(AB)}\Big\{D_{\min}^\eps\left(\rho\|\sigma\right)+\log \|\sigma^\Gamma\|_1\Big\}\;.
\ea
Finally, observe that the second term on the right-hand side of the equation above is the logarithmic negativity of $\sigma^{AB}$ as defined in~\eqref{logneg}. This completes the proof.
\end{proof}

\subsubsection{Asymptotic Distillable NPT-Entanglement}

Using Corollary~\ref{sscor}, we can get the following upper bound on the distillable NPT-entanglement.

\begin{myg}{}
\begin{corollary}
Let $\rho\in\md(AB)$. Then, the distillable NPT-entanglement is bounded from above by
\be\label{nptdist3}
\distill\left(\rho^{AB}\right)\leq \min_{\sigma\in\md(AB)}\Big\{D\left(\rho\|\sigma\right)+\LN(\sigma)\Big\}\;.
\ee
The right-hand side is known as the Rains' bound.
\end{corollary}
\end{myg}

\begin{proof}
The proof follows directly from a combination of Corollary~\ref{sscor} and the quantum Steins' lemma given in~\eqref{7171}. Explicitly, let $\eps\in(0,1)$ and $\sigma\in\md(AB)$. Then, from Corollary~\ref{sscor} we get
\ba
\limsup_{n\to\infty}\frac1n\distill^\eps\left(\rho^{\otimes n}\right)&\leq\limsup_{n\to\infty}\left\{\frac1nD_{\min}^\eps\left(\rho^{\otimes n}\big\|\sigma^{\otimes n}\right)+\frac1n\LN\left(\sigma^{\otimes n}\right)\right\}\\
\GG{\eqref{7171}+\text{Additivity of }\LN}&=D(\rho\|\sigma)+\LN(\sigma)\;.
\ea
Since the inequality above  holds for all $\sigma\in\md(AB)$ we can take the minimum over all density matrices so that
\be
\limsup_{n\to\infty}\frac1n\distill^\eps\left(\rho^{\otimes n}\right)\leq \min_{\sigma\in\md(AB)}\Big\{D\left(\rho\|\sigma\right)+\LN(\sigma)\Big\}\;.
\ee
Note that the inequality above is in fact stronger than~\eqref{nptdist3} in the sense that it holds for all $\eps\in(0,1)$. This completes the proof.
\end{proof}

\bex
Show that the Rains bound is a measure of entanglement and in particular does not increase under completely PPT preserving operations.
\eex

\bex
Prove that the Rains' bound for a pure bipartite state $\psi\in\pure(AB)$ is equal to the entropy of entanglement $E(\psi^{AB})$, which means that on pure bipartite states, the Rains' bound is equal to the distillable entanglement. Hint: Take a look at the proof of Theorem~\ref{reoep}.
\eex

\subsection{Cost of NPT Entanglement}\index{single-shot}
Using the conversion distance\index{conversion distance} provided in~\eqref{cdcos}, we can express the $\eps$-single-shot NPT-entanglement cost\index{entanglement cost} as
\be\label{nptcost}
\cost^\eps\left(\rho^{AB}\right)=\min_{m\in\mbb{N}}\left\{\log(m)\;:\;(1-m)\omega^\Gamma\leq\sigma^\Gamma\leq(1+m)\omega^\Gamma,\;\omega\in\md(AB),\;\sigma\in\mb_\eps(\rho)\right\}
\ee
The expression above can be simplified by replacing $1\pm m$ with $\pm m$. This change does not change much the entanglement cost.

\begin{myt}{}
\begin{theorem}
For all $\eps\in(0,1)$ and all $\rho\in\md(AB)$ we have
\be\label{boundsek}
\log_2\left(2^{E_\kappa^\eps\left(\rho^{AB}\right)}-1\right)\leq\cost^\eps\left(\rho^{AB}\right)\leq \log_2\left(2^{E_\kappa^\eps\left(\rho^{AB}\right)}+1\right)\;,
\ee
where 
\be
E_\kappa^\eps\left(\rho^{AB}\right)=\min_{\rho'\in\mb_\eps(\rho)}E_\kappa^\eps\left(\rho'^{AB}\right)
\ee
is the smoothed version of the $\kappa$-entanglement\index{$\kappa$-entanglement} defined in~\eqref{kappae}.
\end{theorem}
\end{myt}

\begin{proof}
Recall from Exercise~\ref{smoothedvers} that
\be
\cost^\eps\left(\rho\right)=\min_{\rho'\in\mb_\eps(\rho)}\cost^{\eps=0}_\mf\left(\rho'\right)
\ee
(this can also be verified directly from the expression in~\eqref{nptcost}). Therefore, it is sufficient to prove the lemma for the case $\eps=0$.
For $\eps=0$ the entanglement cost\index{entanglement cost} given in~\eqref{nptcost} takes the form
\be
\cost^{\eps=0}\left(\rho^{AB}\right)=\min_{m\in\mbb{N}}\left\{\log(m)\;:\;(1-m)\omega^\Gamma\leq\rho^\Gamma\leq(1+m)\omega^\Gamma,\;\omega\in\md(AB)\right\}\;.
\ee
Now, observe that every $m\in\mbb{N}$ and $\omega\in\ppt(AB)$ that satisfy $(1-m)\omega^\Gamma\leq\rho^\Gamma$ also satisfy
$-(1+m)\omega^\Gamma\leq\rho^\Gamma$. Therefore, we get that
\ba
\cost^{\eps=0}\left(\rho^{AB}\right)&\geq\min_{m\in\mbb{N}}\left\{\log(m)\;:\;-(1+m)\omega^\Gamma\leq\rho^\Gamma\leq(1+m)\omega^\Gamma,\;\omega\in\md(AB)\right\}\\
\Gg{\Lambda\eqdef(1+m)\omega}&=\min_{\Lambda\in\pos(AB)}\left\{\log\big(\tr\left[\Lambda\right]-1\big)\;:\;-\Lambda^\Gamma\leq\rho^\Gamma\leq \Lambda^\Gamma\right\}\\
\GG{\eqref{kappae}}&=\log_2\left(2^{E_\kappa\left(\rho^{AB}\right)}-1\right)\;.
\ea
Similarly, for the other inequality observe that every $m\in\mbb{N}$ and $\omega\in\ppt(AB)$ that satisfy $\rho^\Gamma\leq (m-1)\omega^\Gamma$ also satisfy
$\rho^\Gamma\leq (m+1)\omega^\Gamma$. Therefore, we get that
\ba
\cost^{\eps=0}\left(\rho^{AB}\right)&\leq\min_{m\in\mbb{N}}\left\{\log(m)\;:\;-(1-m)\omega^\Gamma\leq\rho^\Gamma\leq(1-m)\omega^\Gamma,\;\omega\in\md(AB)\right\}\\
\Gg{\Lambda\eqdef(m-1)\omega}&=\min_{\Lambda\in\pos(AB)}\left\{\log\big(\tr\left[\Lambda\right]+1\big)\;:\;-\Lambda^\Gamma\leq\rho^\Gamma\leq \Lambda^\Gamma\right\}\\
\GG{\eqref{kappae}}&=\log_2\left(2^{E_\kappa\left(\rho^{AB}\right)}+1\right)\;.
\ea
This completes the proof.
\end{proof}

The theorem above demonstrates that the single-shot NPT-entanglement cost\index{entanglement cost} is essentially given by the smoothed version of the $\kappa$-entanglement\index{$\kappa$-entanglement}. From the bounds in~\eqref{ineek} it follows that
\be\label{ineek2}
\LN^\eps\left(\rho^{AB}\right)\leq E_\kappa^\eps\left(\rho^{AB}\right)\leq \min_{\substack{\sigma\in\ppt(AB)\\\omega\in\mb_\eps(\rho)}}D_{\max}\left(|\omega^\Gamma|\big\|\sigma\right)\;.
\ee 
While it is true that the lower and upper bounds above may appear to be simpler than computing the $E_\kappa^\eps$ directly using an SDP program, it is important to note that this may not always be the case. In fact, in many instances, computing the bounds may require solving non-trivial optimization problems themselves, and as such may not necessarily be any easier to compute than the original quantity $E_\kappa^\epsilon$. Therefore, while the bounds can be a useful tool for gaining insight into the behavior of $E_\kappa^\epsilon$, they should not be relied upon exclusively as a substitute for computing the quantity directly using an SDP program. Moreover, it is not clear to the author how these bounds can be used in deriving computable bounds for the asymptotic NPT-entanglement cost. 

\subsubsection{The Asymptotic NPT-Entanglement Cost}\index{entanglement cost}

From the lower and upper bounds given in~\eqref{boundsek} we get the following expression for the NPT-entanglement cost.
\begin{myt}{}
\begin{theorem}
Let $\rho\in\md(AB)$. The NPT-entanglement cost of $\rho^{AB}$ is given by
\be
\cost\left(\rho^{AB}\right)=\lim_{\eps\to 0^+}\lim_{n\to\infty}\frac1nE_\kappa^\epsilon\left(\rho^{\otimes n}\right)\;.
\ee
\end{theorem}
\end{myt}
\bex
Prove the theorem above, and in particular show that the limit
\be
\lim_{n\to\infty}\frac1nE_\kappa^\epsilon\left(\rho^{\otimes n}\right)
\ee
exists for all $\rho\in\md(AB)$ and all $\eps\in(0,1)$.
\eex
From the theorem above it follows that
\be\label{compek}
\cost\left(\rho^{AB}\right)\leq E_\kappa\left(\rho^{AB}\right)\;,
\ee
since for all $n\in\mbb{N}$ and all $\eps\in(0,1)$ we have
\ba
E_\kappa^\epsilon\left(\rho^{\otimes n}\right)&\leq E_\kappa\left(\rho^{\otimes n}\right)\\
\GG{E_\kappa\;\text{is subadditive}}&\leq nE_\kappa(\rho)\;.
\ea
Therefore, Eq.~\eqref{compek} provides a computable upper bound on the NPT-entanglement cost.

\section{Notes and References}

For those interested to learn more about entanglement theory, several comprehensive reviews are available. Two recommended sources are~\cite{HHHH2009} and~\cite{PV2007}. If you're specifically interested in entanglement detection and entanglement witnesses, we suggest checking out~\cite{GT2009} and~\cite{CS2014}. For a study of the entanglement properties of isotropic states and reduction criteria, we recommend~\cite{HH1999}. The Werner state, a key example in entanglement theory, was first introduced in~\cite{Werner1989}. 

In 1996, the positive partial transpose (PPT) criterion, was proposed in~\cite{Peres1996}. In the same year, the proof that a state is separable if and only if it is PPT (as stated in Theorem~\ref{thm:2x3}) was given in~\cite{HHH1996}. Another useful technique for constructing PPT entangled states is to use unextendible product bases (UPB), as demonstrated in the example given in~\eqref{5states}, which was first proposed by~\cite{BDM+1999}. 

The realignment criterion, first introduced in~\cite{CW2003}, and was developed further in~\cite{Rudolph2003}. If you're interested in the $k$-extendability criterion, take a look at the original paper~\cite{DPS2004}.

To learn more about entanglement monotones and convex roof extensions, please refer to~\cite{Vidal2000}. The closed formulas for the entanglement of formation~\eqref{baea3} and the concurrence of formation~\eqref{formulac} were discovered by~\cite{Wootters1998}. The characterization of pure-to-mixed state conversions, as given in Corollary~\ref{cor:1321} and~\eqref{ex:13117}, can be attributed to~\cite{ZTG2023}. The Schmidt rank measure of mixed-state entanglement was initially introduced in~\cite{TH2000}, and its interpretation as the maximal extension from the Schmidt rank on pure states was introduced in~\cite{GT2020}.

The concept of relative entropy of entanglement was initially introduced in~\cite{VP1998} (also see the review article~\cite{Vedral2002}). The robustness of entanglement was first introduced in~\cite{VT1999}. The negativity was first introduced in~\cite{VW2002}, its logarithmic version in~\cite{APE2003,Plenio2005}, and the $\kappa$-entanglement in~\cite{WW2020}. The squashed entanglement, also referred to as the CMI entanglement, was initially mentioned in Eq.(42) of~\cite{Tucci2001}. However, it was later proven to be an additive measure of entanglement in~\cite{CW2004}. 
More recently, in~\cite{Brandao2011}, it was shown that the squashed entanglement is a faithful measure of entanglement. 

The study of single-shot distillable entanglement was first conducted in~\cite{BD2010} using the quantum spectrum method, which we did not employ in this book. Instead, we utilized other techniques such as the decoupling theorem to obtain the bounds presented in Theorem~\ref{sinhash} and Theorem~\ref{epiu}.

The first protocols for asymptotic distillation of mixed states were introduced in the seminal paper~\cite{BBP+1996}. The (asymptotic) hashing bound~\eqref{hashing} and the formula for one-way distillable entanglement, as presented in Theorem~\ref{oproneway}, were discovered in~\cite{DW2005}.

Bound entanglement was discovered in~\cite{HHH1998}. Since then, extensive research has been conducted on bound entanglement, with various applications in quantum cryptography, such as those presented in~\cite{HHHO2005}, and in distributing entanglement in quantum networks, such as those outlined in~\cite{Miyake2003}.

The study of single-shot entanglement cost was conducted in~\cite{BD2011}, where tight lower and upper bounds for single-shot entanglement cost were identified. Here, in Theorem~\ref{cmaint}, we have slightly improved upon those findings by providing the precise expression for the single-shot entanglement cost. The proof for the asymptotic entanglement cost, specifically in Theorem~\ref{ecost}, was presented in~\cite{HHT2001}.

The additivity\index{additivity} of entanglement of formation under tensor products had been an unresolved problem since the mid-90s, and it was discovered to be related to three other conjectures, namely the classical capacity of a quantum channel, and the minimum entropy output of a quantum channel, with the latter being the easiest to approach. In a remarkable development~\cite{Hastings2009}, it was established that the minimum entropy output of a quantum channel is not additive, disproving all conjectures, including the additivity of entanglement of formation. Nevertheless, an explicit example that demonstrates the non-additivity of entanglement of formation is still missing.

Theorem~\ref{addif} and the concept of an `entanglement breaking subspace' were first introduced in~\cite{VDC2002}. More recent developments on this topic can be found in~\cite{ZC2019}.

The concept of non-entangling operations (i.e., operations beyond LOCC) was introduced in~\cite{BP2008} for the asymptotic regime, and further developed for the single-shot regime in~\cite{BD2011}. More recently, in~\cite{LR2023}, it was demonstrated that the entanglement cost can be strictly greater than the distillable entanglement under non-entangling operations, thereby illustrating that entanglement theory is not reversible even under the broad set of non-entangling operations.

Completely PPT preserving operations (sometimes referred to as PPT operations) were introduced in~\cite{Rains1999}. In the same work, among many other findings, the Rains' bound~\eqref{nptdist3} on distillable NPT-entanglement was discovered. The monotonicity of negativity and logarithmic negativity under PPT operations was proven in~\cite{Plenio2005}. The expression presented in~\eqref{nptdist} for the one-shot distillable NPT-entanglement was initially discovered in~\cite{FWTD2019} (see also~\cite{RFWG2019} for additional results on distillation beyond LOCC). Lastly, the NPT-entanglement cost was first studied in~\cite{APE2003} and developed further in~\cite{WW2020}.

%%%%%%%%%%%%%%%%%%%%%%%%%

%%%%%%%%%%%%%%%%%%%%%%%%%%%

\chapter{Multipartite Entanglement}\index{quantum entanglement}

Thus far, our focus has been on entanglement that is shared solely between two parties. However, entanglement is not restricted to bipartite systems\index{bipartite system} and can exist among any number of parties. In this section, we will examine the properties of multipartite entanglement, comparing and contrasting it with bipartite entanglement. It's important to note that the theory of multipartite entanglement can be quite complex. Therefore, in this chapter, we will restrict ourselves to pure multipartite states, and concentrate on simpler cases involving three and four qubits in greater detail.

\section{Stochastic LOCC (SLOCC)}\index{SLOCC}

Consider a Hilbert space $A^n$ that is composed of $n$ subsystems, denoted as $A_1, A_2, \cdots, A_n$. In the previous chapter, we discussed the bipartite case where $n=2$, and showed that the Nielsen majorization theorem can be utilized to determine state-conversion under LOCC.
However, as we will see in the upcoming sections, this theorem cannot be applied for $n>2$, and even for two pure states $\psi$ and $\phi$ that belong to $\pure(A^n)$, deterministic conversion from $\psi$ to $\phi$ under LOCC is almost always impossible, unless the states are related by local unitaries. Hence, the majority of the research on multipartite entanglement has been dedicated to studying stochastic interconversions\index{Stochastic Interconversions}.

For two pure states $\psi$ and $\phi$ belonging to $\pure(A^n)$, we say that $\psi$ can be converted to $\phi$ via stochastic LOCC, or SLOCC, denoted as $\psi\xrightarrow{\slocc}\phi$, if there exists a matrix $M_x\in\ml(A_x)$ for each $x\in[n]$ such that $M_x^*M_x\leq I^{A_x}$ and
\be\label{mpesl}
|\phi\ra=M_1\otimes\cdots\otimes M_n|\psi\ra;.
\ee
The relation above implies that $\psi$ can be converted into $\phi$ through local measurements with some non-zero probability, since each $M_x$ can be considered as one Kraus element of a local generalized measurement on system $A_x$.

The set of all states $|\phi\ra$ in $A^n$ that can be obtained from $|\psi\ra$ as in~\eqref{mpesl} is called the SLOCC class of $\psi$. The SLOCC class of $\psi$ comprises two types of states: those that can be written in the form~\eqref{mpesl} with \emph{invertible} matrices $M_1,\ldots,M_n$, and those in which at least one of the matrices $M_x$ is non-invertible. In the former case, $\psi$ can be converted to $\phi$ and vice versa using SLOCC, whereas in the latter, the resulting state $|\phi\ra$ cannot be converted back to $\psi$ via SLOCC.

If $|A_x|=2$ for some $x\in[n]$ (i.e. the $x$-th subsystem is a qubit) and $M_x$ is non-invertible, then the resulting state $|\phi\ra$ is a product state between the qubit system $A_x$ and the remaining $n-1$ subsystems. To demonstrate this, assume $x=1$ for simplicity, so that $M_1$ is a $2\times 2$ non-invertible matrix. Since $M_1$ is a rank-one matrix, it can be written as $M_1=|u\lr v|$ where $|u\ra$ is an unnormalized vector in $A_1$, and $|v\ra$ is a normalized vector in $A_1$. The state $|\psi\ra$ can be expressed as
\be
|\psi\ra^{A^n}=a|v\ra^{A_1}|\psi_1\ra^{A_2\cdots A_n}+b|v^\perp\ra^{A_1}|\psi_2\ra^{A_2\cdots A_n};,
\ee
where $a,b\in\mbb{C}$, $|v^\perp\ra\in A_1$ is an orthogonal vector to $|v\ra$, and $\psi_1$ and $\psi_2$ are some pure states in $A_2\cdots A_n$. Thus,
\be
M_1\otimes M_2\otimes\cdots\otimes M_n|\psi\ra^{A^n}=a|u\ra^{A_1}\otimes \left(M_2\otimes\cdots\otimes M_n|\psi_1\ra^{A_2\cdots A_n}\right);,
\ee
which is a product state between system $A_1$ and system $A_2\cdots A_n$.

\bex
Let $\psi,\phi\in\pure(A^n)$ and suppose there exists matrices $M_1,\ldots,M_n$ such that~\eqref{mpesl} holds. Let $B\eqdef A_2\cdots A_n$, $d\eqdef|A_1|$, and suppose $\det(M_1)=0$. Show that 
\be
\sr\left(\phi^{A_1B}\right)\leq d-1\;,
\ee 
where we view $\phi^{A_1B}$ as a bipartite state between $A_1$ and $B$.
\eex

As we observed in the previous exercise, if any of the matrices $M_1,\ldots,M_n$ is non-invertible, then the resulting state $|\phi\ra$ belongs to a class of states with lower Schmidt rank. Hence, we can categorize $n$-partite entanglement in two stages:
\ben
\item Determine the $n$ Schmidt ranks between each subsystem and the other $n-1$ subsystems.
\item Classify the $n$-partite entanglement based on a fixed set of $n$ Schmidt ranks obtained in the first step.
\een

It is worth noting that for the second step mentioned above, we only need to consider reversible SLOCC conversions where all matrices $M_1,\ldots,M_n$ in~\eqref{mpesl} are invertible. Hence, states $\psi,\phi\in\pure(A^n)$ belong to the same reversible SLOCC class if and only if there exists a matrix \be\label{14p5}
M\in GL^n\eqdef GL(A_1)\times\cdots\times GL(A_n)\;,
\ee
such that $|\phi\rangle=M|\psi\rangle$. Here, for each $x\in[n]$, the set $GL(A_x)$ represents the group of invertible matrices in $\ml(A_x)$. It is also noteworthy that $M$ takes the form of $M_1\otimes\cdots\otimes M_n$, where $M_x\in GL(A_x)$ for each $x\in[n]$. Please note that our notation $GL^n$ does not explicitly specify the system $A^n=A_1\cdots A_n$. However, in the rest of this chapter we will assume that the context makes it clear which system we are referring to.

With these observations, we can use certain tools from representation theory to characterize the reversible SLOCC class of $|\psi\ra$; i.e., the set of states ${M|\psi\ra}$. To achieve this, we begin by relaxing the normalization condition that $\big\|M|\psi\ra\big\|=1$ and allowing each $M_x$ to vary over any element of $SL(A_x)$. The group $SL(A_x)$ is a subgroup of $GL(A_x)$ with the property that the determinant of its elements is one. The limitation to this group can only affect the normalization of the states in ${M|\psi\ra}$. Therefore, we will consider the ``orbit" of $\psi$ with respect to the group
\be\label{slgroup}
SL^n\eqdef SL(A_1)\times\cdots\times SL(A_n)
\ee
Mathematically, the SL-orbit of a pure state $\psi\in\pure(A^n)$ is defined as
\be
SL^n|\psi\ra\eqdef\Big\{M|\psi\ra\;:\;M\in SL^n\Big\};.
\ee
By working with SL-orbits rather than GL-orbits, we can classify multipartite entanglement using SL-invariant polynomials.

\section{SL-Invariant Polynomials}\index{SLIP}

Let us recall the bilinear form defined in~\eqref{bilinearf}. One of its properties (as shown in Exercise~\ref{ex:bilin}) is that for any two-qubit pure states $|\psi\ra$ and $|\phi\ra$
\be
\lp M|\psi\ra,M|\phi\ra\rp=\lp|\psi\ra,|\phi\ra\rp\quad\quad\forall\;M\in SL^2\;.
\ee
The function $f:\mbb{C}^2\otimes\mbb{C}^2\to\mbb{C}$ defined by
\be\label{14p9}
f(|\psi\ra)\eqdef\lp|\psi\ra,|\psi\ra\rp\quad\quad\forall\;\psi\in\mbb{C}^2\otimes\mbb{C}^2\;,
\ee
is a polynomial in the coefficients of $|\psi\ra\eqdef\sum_{x,y\in\{0,1\}}a_{xy}|xy\ra$. Specifically, it is the polynomial
\be\label{slipd22}
f(|\psi\ra)=a_{00}a_{11}-a_{01}a_{10}\;.
\ee
This polynomial is invariant under the group $SL^2$; that is, $f(M|\psi\ra)=f(|\psi\ra)$ for any $M\in SL^2$. Hence, we refer to it as an SL-invariant polynomial. Note that if $f(|\psi\ra)\neq 0$, then $\psi$ is not a product state, and furthermore, all states in the orbit $SL^2|\psi\ra$ are not product states. On the other hand, if $|\phi\ra\not\in SL^2|\psi\ra$, then $|\phi\ra$ is necessarily a product state. Therefore, in the case of two qubits, there exist precisely two classes of states: $SL^2|00\ra$ and $SL^2|\psi\ra$ for some two-qubit non-product state $|\psi\ra$ (i.e., $f(|\psi\ra)\neq 0$). Similarly, as we will see shortly, for more general multipartite systems, SL-invariant polynomials can be used to classify all reversible SLOCC classes.

\begin{myd}{}
\begin{definition}
A polynomial $f:A^n\to\mbb{C}$ is called SL-invariant polynomial, or in short SLIP, if for any $M\in{SL^n}$ and any $\psi\in\pure(A^n)$ we have $f(M|\psi\ra)=f(|\psi\ra)$. The set of all SLIPs on $A^n$ with $f(\0)=0$ is denoted by $\slip(A^n)$.
\end{definition}
\end{myd}

\begin{remark}
The condition that $f(\0)=0$ is a convention that we will adopt to eliminate trivial SLIPs that are constant for all vectors in $A^n$. Furthermore, we will see shortly that this convention implies that SLIPs vanish on product states.
\end{remark}

The set of all SLIPs forms a vector space over $\mbb{C}$. Additionally, the following exercise reveals that this vector space has a basis consisting of \emph{homogeneous} SLIPs. Therefore, we will concentrate on homogeneous SLIPs of some fixed degree $k\in\mbb{N}$. For instance, the SLIP in~\eqref{slipd22} is homogeneous of degree $2$ since it satisfies $f(c|\psi\ra)=c^2f(|\psi\ra)$ for any $c\in\mbb{C}$. The dimension of the space of all homogeneous SLIPs of a fixed degree $k$ is finite, but, as we will see, it grows exponentially with $n$.

\bex
Show that the vector space of SLIPs has a basis consisting of homogeneous SLIPs.
\eex

\bex
Let $AB$ be a bipartite system\index{bipartite system} with $d\eqdef |A|=|B|$. Show that the function 
\be
f(|\psi\ra)=\det(M)\quad\quad\forall\;|\psi\ra=M\otimes I^B|\Omega^{AB}\ra\;,
\ee
is an homogeneous SLIP of degree $d$.
\eex

The degrees of homogeneous SLIPs have a close relationship with the local dimensions of the subsystems. To understand this connection, consider a multipartite system $A^n=A_1\cdots A_n$, where $m_x\eqdef|A_x|$ for each $x\in[n]$. Suppose $f_k:A^n\to\mbb{C}$ is a homogeneous SLIP of degree $k\in\mbb{N}$. Note that if $c\in\mbb{C}$ satisfies $c^{m_x}=1$ for some $x\in[n]$, then the matrix $cI^{A_x}$ has a determinant of one, implying that $cI^{A^n}\in{SL^n}$. Hence, we obtain the following relationship: for every $|\psi\ra\in A^n$
\be
f_k\left(|\psi\ra\right)=f_k\left(cI^{A^n}|\psi\ra\right)=c^kf_k\left(|\psi\ra\right)\;,
\ee
where the first equality is due to the SL-invariance property of $f_k$, and the second equality is due to the homogeneity of $f_k$.
Therefore, as long as $f_k$ is not the zero polynomial it follows that $c^k=1$ for any complex number $c$ that satisfies $c^{m_x}=1$. This means that $m_x$ must divide $k$. Since this holds for all $x$ we can conclude that $k$ is divisible by the least common multiple $r\eqdef{\rm lcm}(m_1,\ldots,m_n)$.

\bex\label{vanpro}
Let $|\psi\ra\in A^n$ be a product state; i.e.\ $|\psi\ra^{A^n}=|\psi_1\ra^{A_1}\otimes\cdots\otimes|\psi_n\ra^{A_n}$. Show that for any $f\in\slip(A^n)$ we have $f(|\psi\ra)=0$.
\eex

Remember that for $\psi,\phi\in\pure(A^n)$, the relation $\psi\xrightarrow{\slocc}\phi$ holds if there exists a matrix $M_x\in\ml(A_x)$ for every $x\in[n]$ such that both $M_x^*M_x\leq I^{A_x}$ and
$
|\phi\ra = M|\psi\ra
$
are satisfied, where $M\eqdef M_1\otimes\cdots\otimes M_n$. The subsequent lemma establishes that if any of the matrices in the set $\{M_x\}_{x\in[n]}$ exhibits rank deficiency, then any SLIP will be nullified for $|\phi\ra$.

\begin{myg}{}
\begin{lemma}\label{rankd}
Let $f:A^n\to\mbb{C}$ be a SLIP and $|\psi\ra$ and $|\phi\ra$ be as above. If there exists $x\in[n]$ such that $\det(M_x)=0$ then $f(|\phi\ra)=0$.
\end{lemma}
\end{myg}

\begin{proof}
Since every SLIP can be expressed as a linear combination of homogeneous SLIPs, we will assume without loss of generality that $f$ is a homogeneous SLIP of degree $k\in\mbb{N}$. Denote by $d\eqdef|A^n|$, and for every $\eps\in[0,1)$, define $M_\eps\eqdef M_1(\eps)\otimes\cdots\otimes M_n(\eps)$, where each $M_x(\eps)$ is a slight perturbation of $M_x$ ensuring that $\det(M_x(\eps))\neq0$ for all $\eps\in(0,1)$. As a consequence, $\mu_\eps\eqdef\det(M_\eps)\neq 0$ for all $\eps\in(0,1)$, which implies
$
N_\eps\eqdef M_\eps/\mu_\eps^{1/d}
$
is an element of $SL^n$. By definition, we can express:
\ba
f\left(M_\eps|\psi\ra\right) &= f\left(\mu_\eps^{1/d}N_\eps|\psi\ra\right)\\
\Gg{f\;\text{is homogeneous of degree }k}&= \mu_\eps^{k/d}f\left(N_\eps|\psi\ra\right)\\
\Gg{f\in\slip(A^n)\;,\;N_\eps\in SL^n}&= \mu_\eps^{k/d}f\left(|\psi\ra\right).
\ea
Upon taking the limit as $\eps\to 0^+$ on both sides and noting that $M=\lim_{\eps\to 0^+}M_\eps$ and $\lim_{\eps\to 0^+}\mu_\eps=\det(M)=0$, we infer that $f(M|\psi\ra)=0$, or equivalently, $f(|\phi\ra)=0$.
This concludes the proof.
\end{proof}

\subsection{SLIPs of $n$-Qubits}

SLIPs, in general, involve cumbersome expressions. However, for systems of $n$-qubits, there exists a broad class of slips that can be expressed elegantly due to the property~\eqref{1194} of the symplectic matrix $J_2\eqdef-i\sigma_2=|0\lr 1|-|1\lr 0|$.
Recall the bilinear form $\lp\cdot,\cdot\rp$ as defined in~\eqref{bilinearf}. This form can be extended to any number of qubits. Specifically, for any $n\in\mbb{N}$, we define the bilinear form $\lp\cdot,\cdot\rp_n:A^n\times A^n\to\mbb{C}$ as follows:\be\label{bilinearfn}
\lp\psi^{A^n},\phi^{A^n}\rp_n\eqdef\big\la\bar{\psi}^{A^n}\big|\underbrace{J_2 \otimes \cdots\otimes J_2}_{n-\text{times}}\big|\phi^{A^n}\big\ra\quad\quad\forall\;|\psi\ra,|\phi\ra\in A^n\;,
\ee
where for convenience we replaced the second Pauli matrix $\sigma_2$ that appear in~\eqref{bilinearf} with $J_2\eqdef-i\sigma_2=|0\lr 1|-|1\lr 0|$.

\bex\label{ex:bilin}
Consider the bilinear form in~\eqref{bilinearfn}.
\ben
\item Use the relation~\eqref{1194} to show that for any $M\in {SL^n}$, and any vectors $|\psi\ra,|\phi\ra\in A^n$
\be
\lp M|\psi\ra,M|\phi\ra\rp_n=\lp|\psi\ra,|\phi\ra\rp_n\;.
\ee
\item Show that for any odd integer $n\in\mbb{N}$ we have
\be
\lp|\psi\ra,|\psi\ra\rp_n=0\;.
\ee
\een
\eex

The bilinear form defined above can be used to define SLIPs on systems of $n$ qubits. For an even number of qubits, it follows from the exercise above that 
\be\label{14p16}
f(|\psi\ra)\eqdef \lp|\psi\ra,|\psi\ra\rp_n\;,
\ee
is an homogeneous SLIP of degree two. 

For an odd number of qubits, one can define a SLIP of degree four as follows: For any $|\psi\ra\in A^n$ we write 
\be
|\psi^{A^n}\ra=|0\ra^{A_1}|\psi_0\ra^{A_2\cdots A_n}+|1\ra^{A_1}|\psi_1\ra^{A_2\cdots A_n}
\ee 
where $|\psi_0\ra,|\psi_1\ra\in A_2\cdots A_n$. The basis $|0\ra^{A_1}$ and $|1\ra^{A_1}$ is chosen such that $J_2$ has the standard form\index{standard form}. With these choices the determinant
\be\label{g4slip}
g_4(|\psi\ra)\eqdef\det\begin{pmatrix}
\lp|\psi_0\ra,|\psi_0\ra\rp_{n-1} &\lp|\psi_0\ra,|\psi_1\ra\rp_{n-1}\\
\lp|\psi_1\ra,|\psi_0\ra\rp_{n-1} &\lp|\psi_1\ra,|\psi_1\ra\rp_{n-1}
\end{pmatrix}
\ee
is an homogeneous SLIP of degree 4. To see that the above function is a SLIP, observe that for any $M^{(n)}=M_1\otimes M^{(n-1)}\in{SL^n}$, we have
$g_4\left(M^{(n)}|\psi\ra\right)=g_4\left(M_1\otimes I^{A_2\cdots A_n}|\psi\ra\right)$ since $\lp\cdot,\cdot\rp_{n-1}$ is invariant under the action of $M^{(n-1)}$. Now, let $M_1=\begin{pmatrix} a & b\\
c & d\end{pmatrix}$ and observe that:
\ba
M_1\otimes I^{A_2\cdots A_n}|\psi\ra&=\big(a|0\ra+c|1\ra\big)|\psi_0\ra+\big(b|0\ra+d|1\ra\big)|\psi_1\ra\\
&=|0\ra\big(a|\psi_0\ra+b|\psi_1\ra\big)+|1\ra\big(c|\psi_0\ra+d|\psi_1\ra\big)\;.
\ea
For each $x,y\in\{0,1\}$ we denote by $\mu_{xy}\eqdef\lp|\psi_x\ra,|\psi_y\ra\rp_{n-1}$. With these notations
\begin{align}
&g_4\left(M_1\otimes I^{A_2\cdots A_n}|\psi\ra\right)\nonumber\\&=
\det\begin{pmatrix}
\lp a|\psi_0\ra+b|\psi_1\ra,a|\psi_0\ra+b|\psi_1\rp &\lp a\psi_0\ra+b|\psi_1\ra,c|\psi_0\ra+d|\psi_1\ra\rp\nonumber\\
\lp c|\psi_0\ra+d|\psi_1\ra,a|\psi_0\ra+b|\psi_1\ra\rp &\lp c|\psi_0\ra+d|\psi_1\ra,c|\psi_0\ra+d|\psi_1\ra\rp
\end{pmatrix}\nonumber\\
&=(a^2\mu_{00}+b^2\mu_{11}+2ab\mu_{01})(c^2\mu_{00}+d^2\mu_{11}+2cd\mu_{10})-\big(ac\mu_{00}+bd\mu_{11}+(ad+cb)\mu_{01}\big)^2\nonumber\\
&=(ad-bc)^2\left(\mu_{00}\mu_{11}-\mu_{01}^2\right)
\end{align}
where the last line follows from direct algebraic simplification of all the terms involved.
Since $M_1\in SL(2,\mbb{C})$ we have $ad-bc=1$ so that
\be
g_4\left(M_1\otimes I^{A_2\cdots A_n}|\psi\ra\right)=\mu_{00}\mu_{11}-\mu_{01}^2=g_4(|\psi\ra)\;.
\ee

To get other SLIPs, let $A^n$ be a system of $n$ qubits, and for any choice of $m<n$ of its qubits, we associate a \emph{bipartite cut} denoted as $\mbb{A}_m\otimes\mbb{B}_{n-m}$, where $\mbb{A}_m$ is a system of $m$ qubits of $A^n$, and $\mbb{B}_{n-m}$ is the system comprising of the remaining $n-m$ qubits of $A^n$. With respect to this bipartite cut, any vector $|\psi\ra\in A^n$ can be expressed as
\ba\label{elxy}
|\psi^{A^n}\ra&=\sum_{x\in[2^m]}\sum_{y\in[2^{n-m}]}\lambda_{xy}|u_x\ra^{\mbb{A}_m}|v_y\ra^{\mbb{B}_{n-m}}\\
&=\Lambda\otimes I^{\tilde{\mbb{B}}_{n-m}}\big|\Omega^{\mbb{B}_{n-m}\tilde{\mbb{B}}_{n-m}}\big\ra
\ea
where $\{|u_x\ra^{\mbb{A}_m}\}$ and $\{|v_y\ra^{\mbb{B}_{n-m}}\}$ are orthonormal bases of $\mbb{A}_m$ and $\mbb{B}_{n-m}$, respectively, and each coefficient $\lambda_{xy}\in\mbb{C}$. Therefore, and vector $|\psi\ra\in A^n$, and every bipartite cut $\mbb{A}_m\otimes\mbb{B}_{n-m}$ of $A^n$ defines a matrix $\Lambda\eqdef(\lambda_{xy})$ with $x\in[2^m]$ and $y\in[2^{n-m}]$. 

\begin{myt}{}
\begin{theorem}
Let $A^n$ be a system of $n$ qubits, $m\in[n-1]$, and $\mbb{A}_m\otimes\mbb{B}_{n-m}$ be a bipartite cut of $A^n$. For any $|\psi\ra\in A^n$ let $\Lambda=(\lambda_{xy})$ be the $2^m\times 2^{n-m}$ matrix defined in~\eqref{elxy}, and set $J_2\eqdef|0\lr 1|-|1\lr 0|$. Then, for any $\ell\in\mbb{N}$ the function
\be
f_\ell(|\psi\ra)\eqdef\tr\left[\left(J_2^{\otimes m}\Lambda J_2^{\otimes (n-m)}\Lambda^T\right)^\ell\right]
\ee
is a homogeneous SLIP of degree $2\ell$.
\end{theorem}
\end{myt}

\begin{proof}
Let $M\in{SL^n}$. We need to show that $f(M|\psi\ra)=f(|\psi\ra)$. Let $N\in SL^m$ and $L\in SL^{n-m}$ be such that $M=N\otimes L$. Then,
\ba
M|\psi^{A^n}\ra&=N\Lambda\otimes L\big|\Omega^{\mbb{B}_{n-m}\tilde{\mbb{B}}_{n-m}}\big\ra\\
&=N\Lambda L^T\otimes I^{\tilde{\mbb{B}}_{n-m}}\big|\Omega^{\mbb{B}_{n-m}\tilde{\mbb{B}}_{n-m}}\big\ra
\ea
We therefore get that
\ba
f_\ell\left(M|\psi^{A^n}\ra\right)&=\tr\left[\left(J_2^{\otimes m}N\Lambda L^TJ_2^{\otimes (n-m)}L\Lambda^T N^T\right)^\ell\right]\\
\GG{Cyclic\; permutation}&=\tr\left[\left(N^TJ_2^{\otimes m}N\Lambda L^TJ_2^{\otimes (n-m)}L\Lambda^T\right)^\ell\right]\;,
\ea
where we have used the invariance of the trace under cyclic permutation (note that the power over $\ell$ does not effect this property). To complete the proof we now argue that $N^TJ_2^{\otimes m}N=J_2^{\otimes m}$ and $L^TJ_2^{\otimes (n-m)}L=J_2^{\otimes (n-m)}$ so that the right-hand side above equals $f\left(|\psi^{A^n}\ra\right)$. Indeed, observe that $N=N_1\otimes\cdots\otimes N_m$, where $N_x\in SL(2,\mbb{C})$ so that
\ba
N^TJ_2^{\otimes m}N&=N_1^TJ_2N_1\otimes\cdots\otimes N_mJ_2N_m\\
\GG{\eqref{1194}}&=J_2\otimes\cdots\otimes J_2=J_2^{\otimes m}\;.
\ea
 In the same way, one can prove that $L^TJ_2^{\otimes (n-m)}L=J_2^{\otimes (n-m)}$. This completes the proof.
\end{proof}

\subsubsection{Examples:}
\ben
\item The case $n=2$. In this case the only non-trivial $m$ is $m=1$. In this case for any two-qubit state $|\psi\ra=\sum_{x,y\in\{0,1\}}\lambda_{xy}|xy\ra$ we get
\ba
f_\ell(|\psi\ra)&=\tr\left[\left(J_2\Lambda J_2\Lambda^T\right)^\ell\right]\\
\GG{\eqref{1194}}&=\tr\left[\left(J_2\det(\Lambda) J_2\right)^\ell\right]\\
\Gg{J_2^2=I}&=2\big(\det(\Lambda)\big)^\ell
\ea
Note that in this case $\det(\Lambda)=\lp|\psi\ra,|\psi\ra\rp_2$ which is the same SLIP we already discussed before.
\item The case $n=3$. Consider the bipartite cut $(A_1A_2)\otimes A_3$ of the three-qubit system $A^3$. With respect to this bipartite cut, any state $|\psi\ra\in A^3$ can be expressed as
\be
\big|\psi^{A^3}\big\ra=\Lambda\otimes I^{\tA_3}\big|\Omega^{A_3\tA_3}\big\ra\;,
\ee 
where $\Lambda:A_3\to A_1A_2$ is a $4\times 2$ matrix. It will be convinient to view the matrix $\Lambda$ in a block form 
\be
\Lambda=\begin{bmatrix} \Lambda_1 \\ \Lambda_2\end{bmatrix}
\ee
where $\Lambda_1$ and $\Lambda_2$ are both $2\times 2$ matrices. For this choice of bipartite cut we have
\be
f_\ell(|\psi\ra)=\tr\left[\left(J_2^{\otimes 2}\Lambda J_2\Lambda^T\right)^\ell\right]\;.
\ee
The term
\ba
\Lambda J_2\Lambda^T&=\begin{bmatrix} \Lambda_1 \\ \Lambda_2\end{bmatrix} J_2[\Lambda_1^T\;\Lambda_2^T]=\begin{bmatrix} \Lambda_1J_2\Lambda_1^T & \Lambda_1J_2\Lambda_2^T\\ \Lambda_2J_2\Lambda_1^T & \Lambda_2J_2\Lambda_2^T
\end{bmatrix}\;.
\ea
Hence, combining this with $J_2^{\otimes 2}=\begin{bmatrix} 0 & J_2\\
-J_2 & \end{bmatrix}$
gives
\be
J_2^{\otimes 2}\Lambda J_2\Lambda^T=
\begin{bmatrix} 
J_2\Lambda_2J_2\Lambda_1^T &
J_2\Lambda_2J_2\Lambda_2^T\\
-J_2\Lambda_1J_2\Lambda_1^T &
-J_2 \Lambda_1J_2\Lambda_2^T 
\end{bmatrix}\;.
\ee
Since the trace of the matrix above is zero the case $\ell=1$ is trivial. For the case $\ell=2$ we have
\ba
\tr\left[\left(J_2^{\otimes 2}\Lambda J_2\Lambda^T\right)^2\right]&=\tr\left[J_2\Lambda_2J_2\Lambda_1^TJ_2\Lambda_2J_2\Lambda_1^T\right] \\
&-\tr\left[J_2\Lambda_2J_2\Lambda_2^TJ_2\Lambda_1J_2\Lambda_1^T\right]\\
&+\tr\left[J_2 \Lambda_1J_2\Lambda_2^TJ_2 \Lambda_1J_2\Lambda_2^T\right]\\
&-\tr\left[J_2\Lambda_1J_2\Lambda_1^TJ_2\Lambda_2J_2\Lambda_2^T\right]\;.
\ea
and due to the cyclic permutation of the trace we have
\be
f_{\ell=2}(|\psi\ra)=2\tr\Big[J_2\Lambda_2J_2\Lambda_1^TJ_2\Lambda_2J_2\Lambda_1^T-J_2\Lambda_2J_2\Lambda_2^TJ_2\Lambda_1J_2\Lambda_1^T\Big]\;.
\ee
The above expression is an homogeneous SLIP of degree 4. Since for three qubits there are no homogeneous  SLIPs of degree two, any other SLIP must be proportional to some power of $f_{\ell=2}$.
Hence, for three qubits, the above SLIP is essentially the only one.
\item The case $n=4$. Let $\mbb{A}_2\otimes\mbb{B}_2$ be a bipartite cut with exactly two qubits on each side and let $|\psi\ra\in A^4$ be given as $|\psi\ra=\Lambda\otimes I^{\mbb{B}_2}|\Omega^{\mbb{A}_2\mbb{B}_2}\ra$, where $\Lambda$ is a $4\times 4$ matrix. Then, the function
\be
f_\ell(|\psi\ra)=\tr\left[\left(J_2^{\otimes 2}\Lambda J_2^{\otimes 2}\Lambda^T\right)^\ell\right]\;,
\ee
is an homogeneous SLIP of degree $2\ell$.
Specifically, consider the four-qubit state
\be
|\psi\ra=\lambda_1|\Psi_+\ra|\Psi_+\ra+\lambda_2|\Psi_{-}\ra|\Psi_{-}\ra+\lambda_3|\Phi_+\ra|\Phi_+\ra+\lambda_4|\Phi_-\ra|\Phi_-\ra\;,
\ee
where $\lambda_x\in\mbb{C}$ for all $x\in[4]$ and the two-qubit states $|\Psi_{\pm}\ra$ and $|\Phi_{\pm}\ra$ form the Bell basis\index{Bell basis} of $\mbb{C}^2\otimes\mbb{C}^2$.
For this states it follows that
\be\label{rellam}
f_\ell(|\psi\ra)=\lambda_1^{2\ell}+\lambda_2^{2\ell}+\lambda_3^{2\ell}+\lambda_4^{2\ell}\;.
\ee
\bex
Prove the relation~\eqref{rellam}.
\eex
\een

\subsection{The Set of All Homogeneous SLIPs}

Homogeneous polynomials of degree one on $A^n$ can be defined using an inner product of the form $f_\chi(\psi)\eqdef\la\chi|\psi\ra$ for all $|\psi\ra\in A^n$, where $\chi$ is a fixed coefficient vector in $A^n$. Similarly, homogeneous polynomials of degree $k\in\mbb{N}$ can be expressed as 
\be\label{dkp} 
f_\chi(\psi)=\left\la\chi\big|\psi^{\otimes k}\right\ra\quad\quad\forall\;\psi\in A^n\;,
\ee 
where the coefficient vector $|\chi\ra\in \left(A^n\right)^{\otimes k}$. However, this relation is not one-to-one; $f_\chi$ is equal to $f_{\chi'}$ if the coefficient vectors $|\chi\ra$ and $|\chi'\ra$ are related by a permutation matrix. This permutation is with respect to the $k$ copies of $A^n$. As a result, there exists an isomorphism between the space of homogeneous polynomials of degree $k$ and the subspace $\sym_k(A^n)$ of $\left(A^n\right)^{\otimes k}$(see definition in~\eqref{defsym}).

The polynomial $f_\chi$ above is SLIP if and only if for any $M\in{SL^n}$ for all $|\psi\ra\in A^n$ we have $f_\chi(M|\psi\ra)=f_\chi(|\psi\ra)$, which is equivalent to
\be
\left\la\chi|M^{\otimes k} \big|\psi^{\otimes k}\right\ra=\left\la\chi\big|\psi^{\otimes k}\right\ra\;.
\ee
Since the above equation has to hold for all $|\psi\ra$, and since if $M\in{SL^n}$ then $M^*\in{SL^n}$ we conclude that $f_\chi$  is SLIP if and only if (see Exercise~\ref{sympolyhom})
\be\label{1440b}
M^{\otimes k}|\chi\ra=|\chi\ra\quad\quad\forall\;M\in{SL^n}\;.
\ee
In other words, $|\chi\ra$ is a ${SL^n}$-fixed vector under the representation $\pi_k:{SL^n}\mapsto\ml\left(\left(A^n\right)^{\otimes k}\right)$ given by $\pi_k(M)=M^{\otimes k}$. Denoting by $\slip_k(A^n)$ the set of all homogeneous SLIPs of degree $k$, and by $V\eqdef (A^n)^{\otimes k}$ we conclude that
\be
\slip_k(A^n)=\Big\{f_\chi\;:\;|\chi\ra\in V^{{SL^n}}\Big\}\;,
\ee
where we used the same notations as in~\eqref{8213}.
Our task is therefore to characterize $V^{{SL^n}}$. For this purpose, observe that the vector space $V$ is isomorphic to
\be
V=(A^n)^{\otimes k}\cong A_1^{\otimes k}\otimes\cdots\otimes A_n^{\otimes k}\;.
\ee
Let $P:V\to A_1^{\otimes k}\otimes\cdots\otimes A_n^{\otimes k}$ be this isomorphism (permutation) map. Under this isomorphism any matrix $M^{\otimes k}$, with $M\eqdef M_1\otimes\cdots\otimes M_n\in{SL^n}$, goes to
\be
PM^{\otimes k}P^{-1}=P\big(M_1\otimes\cdots\otimes M_n\big)^{\otimes k}P^{-1}=M_1^{\otimes k}\otimes\cdots\otimes M_n^{\otimes k}\;.
\ee
Therefore, for any $|\chi\ra\in V$ that satisfies~\eqref{1440b} the vector $|\phi\ra\eqdef P|\chi\ra$ satisfies
\be\label{mphik}
M_1^{\otimes k}\otimes\cdots\otimes M_n^{\otimes k}|\phi\ra=|\phi\ra\;.
\ee

\bex\label{sympolyhom}
Let $|\chi\ra\in\left(A^n\right)^{\otimes k}$. Show that $f_\chi$ as defined in~\eqref{dkp} is an homogeneous SLIP of degree $k$ if and only if~\eqref{1440b} holds. 
\eex

\begin{myg}{}
\begin{lemma}
Let $|\phi\ra\in A_1^{\otimes k}\otimes\cdots\otimes A_n^{\otimes k}$. Then, $|\phi\ra$ satisfies~\eqref{mphik} if and only if 
\be
|\phi\ra\in W\eqdef \left(A_1^{\otimes k}\right)^{SL(A_1)}\otimes\cdots\otimes \left(A_n^{\otimes k}\right)^{SL(A_n)}\;,
\ee
where we used the notations given in~\eqref{8213}.
\end{lemma}
\end{myg}
\begin{proof}
Clearly, by definition, if $|\phi\ra\in W$ then $|\phi\ra$ satisfies~\eqref{mphik}. Conversely, suppose that $|\phi\ra$ satisfies~\eqref{mphik}, and let
$\{|v_{x}\ra\}$ be an orthonormal basis of $\left(A_1^{\otimes k}\right)^{SL(A_1)}$, and $\{|u_{y}\ra\}$ be an orthonormal basis of the orthogonal complement of $\left(A_1^{\otimes k}\right)^{SL(A_1)}$ in $A_1^{\otimes k}$. Finally, let $\{|\varphi_z\ra\}$ be an orthonormal basis of $A_2^{\otimes k}\otimes\cdots\otimes A_n^{\otimes k}$. With these notations, since $|\phi\ra\in A_1^{\otimes k}\otimes\cdots\otimes A_n^{\otimes k}$ it can be expressed as
\be
|\phi\ra=\sum_{x,z}\lambda_{xz}|v_{x}\ra|\varphi_{z}\ra+\sum_{y,z}\mu_{yz}|u_{y}\ra|\varphi_z\ra\;,
\ee
where $\lambda_{xz},\mu_{yz}\in\mbb{C}$. Using this expression in~\eqref{mphik}, and taking a special case in which $M_2=I^{A_2}$,\ldots,$M_n=I^{A_n}$, gives
\be
\sum_{x,z}\lambda_{xz}M_1^{\otimes k}|v_{x}\ra|\varphi_{z}\ra+\sum_{y,z}\mu_{yz}M_1^{\otimes k}|u_{y}\ra|\varphi_z\ra=\sum_{x,z}\lambda_{xz}|v_{x}\ra|\varphi_{z}\ra+\sum_{y,z}\mu_{yz}|u_{y}\ra|\varphi_z\ra\;.
\ee
Since $M_1\in SL(A_1)$ we have for all $x$,  $M_1^{\otimes k}|v_{x}\ra=|v_x\ra$ (by definition of $|v_x\ra$). Hence, the above equation can be simplified to
\be
\sum_{y,z}\mu_{yz}M_1^{\otimes k}|u_{y}\ra|\varphi_z\ra=\sum_{y,z}\mu_{yz}|u_{y}\ra|\varphi_z\ra\;.
\ee
Since the vectors $\{|\varphi_z\ra\}$ are orthonormal, it follows that for all $z$ and all $M_1\in SL(A_1)$ we have
\be
\sum_{y}\mu_{yz}M_1^{\otimes k}|u_{y}\ra=\sum_{y}\mu_{yz}|u_{y}\ra\;.
\ee
The above equation implies that for each $z$ the vector
$\sum_{y}\mu_{yz}|u_{y}\ra$ belongs to the subspace $\left(A_1^{\otimes k}\right)^{SL(A_1)}$. However, by definition, the vectors $\{|u_y\ra\}$ belong to the orthogonal complement of $\left(A_1^{\otimes k}\right)^{SL(A_1)}$. Therefore, the coefficients $\{\mu_{yz}\}$ must be zero, so that
\be
|\phi\ra=\sum_{x,z}\lambda_{xz}|v_{x}\ra|\varphi_{z}\ra\;.
\ee
Denoting by $C\eqdef A_2^{\otimes k}\otimes\cdots\otimes A_n^{\otimes k}$, the above equation can be expressed as $\Pi_1\otimes I^{C}|\phi\ra=|\phi\ra$, where
\be
\Pi_1\eqdef\sum_x|v_x\lr v_x|
\ee
is the orthogonal projection to the subspace $\left(A_1^{\otimes k}\right)^{SL(A_1)}$. Denoting by the $\Pi_x$ the orthogonal projection to the subspace $\left(A_x^{\otimes k}\right)^{SL(A_x)}$, and repeating the same argument for any $x\in[n]$ we conclude that
\be
\Pi_1\otimes\cdots\otimes\Pi_n|\phi\ra=|\phi\ra\;.
\ee
That is, $|\phi\ra\in W$. This completes the proof.
\end{proof}

The lemma above shows that characterizing $V^{{SL^n}}$ can be done by characterizing $\left(A_x^{\otimes k}\right)^{SL(A_x)}$ or the orthogonal projection $\Pi_x$. Therefore, the problem of characterizing all SLIPs of degree $k$ can be reduced to characterizing $\left(A^{\otimes k}\right)^{SL(A)}$, which is a classic representation theory problem that uses the Schur-Weyl duality. This duality connects the irreducible representations (irreps) of $SL(A)$ to the symmetric group on $k$ elements, with a natural action. Further information can be found in the `Notes and References' section at the end of this chapter.

\subsection{SLIPs and Multipartite Entanglement Monotones}

The following theorem illustrates the usefulness of SLIPs in quantifying entanglement.

\begin{myt}{}
\begin{theorem}
Let $f_k:A^n\to\mbb{R}_+$ be an homogenous SLIP of degree $k\in\mbb{N}$, and define for any $\psi\in\pure(A^n)$
\be\label{efk}
E\left(\psi^{A^n}\right)\eqdef\left|f_k\left(\big|\psi^{A^n}\big\ra\right)\right|^{2/k}\;.
\ee
Then, $E$ is an entanglement monotone on pure multipartite states; i.e. it is zero on product states and it does not increase on average under LOCC.
\end{theorem}
\end{myt}

\begin{proof}
From Exercise~\ref{vanpro}, we deduce that $E$ vanishes on product states. Consider an arbitrary $m\in\mbb{N}$ and $\psi\in\pure(A^n)$. If there exists an LOCC protocol that transforms $\psi^{A^n}$ to $\phi_x^{A^n}\in\pure(A^n)$ with a probability $p_x$, where $x\in[m]$, then each $\phi_x^{A^n}$ can be represented as:
\be
\big|\phi_x^{A^n}\big\ra=\frac1{\sqrt{p_x}}M_x\big|\psi^{A^n}\big\ra\;,
\ee
where each matrix $M_x$ is a tensor product of the form $M_x=\Lambda_{x1}\otimes\cdots\otimes\Lambda_{xn}$, and for every $y\in[n]$, $\Lambda_{xy}\in\ml(A_y)$. Additionally, we have the relation $\sum_{x\in[m]}M_x^*M_x=I^{A^n}$.

Leveraging Lemma~\ref{rankd}, we observe that $f(M_x|\psi\ra)=0$ when $\det(M_x)=0$. We can then categorize the set $\{M_x\}_{x\in[m]}$ into two subsets: the matrices that are rank deficient and those that possess full rank. Without loss of generality, let's assume the first $r\in[m]$ matrices $\{M_x\}_{x\in[r]}$ are all of full rank, while the subsequent matrices, for all $x=r+1,\ldots,m$, satisfy the condition $\det(M_x)=0$.

Thus, for each $x\in[r]$, aside from a scalar coefficient, $M_x$ can be interpreted as a member of ${SL^n}$. More precisely, we can express $M_x$ as $M_x=\mu_xN_x$, where $\mu_x\eqdef\left(\det(M_x)\right)^{1/d}$, $d\eqdef|A^n|$, and the normalized matrix $N_x\eqdef\frac1{\mu_x}M_x$ belongs to $SL^n$. With these notations, we can proceed as follows:\ba\label{14552}
\sum_{x\in[m]}p_xE\left(\phi^{A^n}_x\right)=\sum_{x\in[r]}p_x\left|f_k\left(\frac1{\sqrt{p_x}}M_x\big|\psi^{A^n}\big\ra\right)\right|^{2/k}
&=\sum_{x\in[r]}p_x\left|f_k\left(\frac{\mu_x}{\sqrt{p_x}}N_x\big|\psi^{A^n}\big\ra\right)\right|^{2/k}\\
\GG{{\it f_k}\;is\;homogenous\;of\;degree\;{\it k}}&=\sum_{x\in[r]}|\mu_x|^2\left|f_k\left(N_x\big|\psi^{A^n}\big\ra\right)\right|^{2/k}\\
\GG{{\it {SL^n}}\;invariance}&=\sum_{x\in[r]}|\mu_x|^2\left|f_k\left(\big|\psi^{A^n}\big\ra\right)\right|^{2/k}\\
&=E\left(\psi^{A^n}\right)\sum_{x\in[r]}|\mu_x|^2\;.
\ea
Now, observe that
\be
\sum_{x\in[r]}|\mu_x|^2\leq \sum_{x\in[m]}|\mu_x|^2=\sum_{x\in[m]}\big|\det(M_x)\big|^{2/d}=\sum_{x\in[m]}\Big(\det\left(M_x^{*}M_x\right)\Big)^{1/d}\;,
\ee
where we removed the absolute value since $M_x^*M_x\geq 0$. From the geometric-arithmetic inequality we have that $\Big(\det\left(M_x^{*}M_x\right)\Big)^{1/d}\leq\frac1d\tr\left[M_x^*M_x\right]$. Hence, substituting this into the equation above gives
\ba
\sum_{x\in[r]}|\mu_x|^2\leq \sum_{x\in[m]}\frac1d\tr\left[M_x^*M_x\right]=\frac1d\tr\left[I^{A_n}\right]=1\;.
\ea
Combining this with~\eqref{14552} we conclude that
\be
\sum_{x\in[m]}p_xE\left(\phi^{A^n}_x\right)\leq E\left(\psi^{A^n}\right)\;.
\ee
This completes the proof.
\end{proof}

To extend the definition of $E$ as defined in~\eqref{efk} to mixed multipartite states, we can employ the convex roof extension\index{convex roof extension}. In particular, for a homogeneous SLIP of degree $k$, we can use the following approach:
\be\label{meman}
E\left(\rho^{A^n}\right)\eqdef\min\sum_{x\in[m]}p_x\left|f_k\left(\big|\psi^{A^n}_x\big\ra\right)\right|^{2/k}\;,
\ee
where the minimum is over all pure state decompositions of $\rho^{A^n}=\sum_{x\in[m]}p_x\psi^{A^n}$. The above theorem implies that $E$ is an entanglement monotone on mixed states.
\bex
Show that $E$ as defined in~\eqref{meman} is an entanglement monotone on multipartite mixed states.
\eex

If we set $n=k=2$ and choose $f_k$ to be the specific SLIP~\eqref{14p9} in~\eqref{meman}, we can see that $E$ corresponds to the concurrence\index{concurrence}. Although there is a simple closed formula for the concurrence of mixed bipartite states, it may not be immediately clear whether similar formulas exist for multipartite entanglement. Interestingly, there is one known example of such a formula for an even number of qubits.

\bex
Let $n\in\mbb{N}$ be an even integer,  $\rho\in\md(A^n)$, $\lp\cdot,\cdot\rp_n:A^n\times A^n\to\mbb{C}$ be the bilinear form as defined in~\eqref{bilinearfn}, and $f$ be the SLIP of degree 2 as defined in~\eqref{14p16}. Finally, let $E$ be the entanglement monotone as defined in~\eqref{meman} but with $f$ replacing $f_k$. Show that
\be
E\left(\rho^{A^n}\right)=\max\Big\{0,\lambda_1-\sum_{x=2}^\ell\lambda_x\Big\}
\ee
where $\ell\eqdef 2^n$, and $\{\lambda_x\}_{x\in[\ell]}$ are the eigenvalues of the matrix $\left|\sqrt{\rho}\sqrt{\rho_\star}\right|$ arranged in non-decreasing order. The matrix $\rho_\star$ is defined similarly to~\eqref{rhostart} as
\be
\rho^{A^n}_\star\eqdef \sigma_2^{\otimes n} \bar{\rho}^{AB}\sigma_2^{\otimes n}\;.
\ee
Hint: Follow the exact same lines as in the proof of Theorem~\ref{cfconca} by with $\lp\cdot,\cdot\rp_n$ replacing the bilinear form given in~\eqref{bilinearf}.
\eex

\section{Characteristics of Multipartite Entanglement}

In this section, we will present several key results from representation theory and algebraic geometry that help characterize the structure of multipartite entangled states. While some of these results are presented without their full proofs, as they go beyond the scope of this book, interested readers can find more information in the book by~\cite{Wallach2017}.
By leveraging these powerful mathematical tools, we can gain deeper insight into the properties of multipartite entangled states.

\subsection{Critical States}

Let ${\rm Lie}({SL^n})$ be the Lie algebra of the group ${SL^n}$ defined in~\eqref{slgroup}. Define the set of \emph{critical} states in $A^n$ to be
\be
{\rm Crit}(A^n)\eqdef\Big\{|\psi\ra\in A^n\;:\;\la\psi|X|\psi\ra=0\quad\forall\;X\in{\rm Lie}\left({SL^n}\right)\Big\}\;.
\ee
The reason for this terminology, is that any state $|\psi\ra\in {\rm Crit}(A^n)$ is a critical point of the function
$f:{SL^n}|\psi\ra\to\mbb{R}_+$ defined by $f(|\phi\ra)\eqdef\||\phi\ra\|$. In fact, we have something that is a bit stronger.

\begin{myt}{\color{yellow} Kempf-Ness Theorem (Part I)}\index{Kempf-Ness theorem}
\begin{theorem}\label{0knt}
Let $|\psi\ra\in A^n$. The following statements are equivalent:
\ben
\item The state $|\psi\ra\in {\rm Crit}(A^n)$. 
\item For any $M\in {SL^n}$ we have $\big\|M|\psi\ra\big\|\geq \big\||\psi\ra\big\|$.
\item For any $x\in[n]$, the reduced density matrix of $\psi^{A^n}$ on the $x^{\text{th}}$-sub-system is proportional to the identity matrix $I^{A_x}$.
\een
\end{theorem}
\end{myt}
\begin{proof}
We start by proving that $1\Rightarrow 2$. Suppose $|\psi\ra\in {\rm Crit}(A^n)$ and observe that for any $M\in {SL^n}$ also $M^*M\in {SL^n}$. We can therefore write $M^*M=e^X$ for some $X\in{\rm Lie}\left({SL^n}\right)$. Hence,
\ba
\|M|\psi\ra\|^2&=\la\psi|e^X|\psi\ra\\
\Gg{e^X\geq I+X}&\geq\la\psi|\left(I+X\right)|\psi\ra\\
\GG{|\psi\ra\;is\;critical}&=\la\psi|\psi\ra=\||\psi\ra\|^2\;.
\ea
To prove that $2\Rightarrow 1$, suppose that for any $M\in {SL^n}$ we have $\big\|M|\psi\ra\big\|\geq \big\||\psi\ra\big\|$. Then, for any $X\in{\rm Lie}\left({SL^n}\right)$ and $t\in\mbb{R}$ we have
\be
f(t)\eqdef\left\|e^{\frac12tX}|\psi\ra\right\|^2=\la\psi|e^{tX}|\psi\ra\geq \la\psi|\psi\ra=f(0)\;.
\ee
Hence, $t=0$ must be a critical point of the function $f(t)$ so that $f'(0)=\la\psi|X|\psi\ra=0$. Since this holds for any $X\in{\rm Lie}\left({SL^n}\right)$ we conclude that $|\psi\ra\in {\rm Crit}(A^n)$.

We next prove the equivalence of 1 and 3. Recall that any $X\in{\rm Lie}\left({SL^n}\right)$ can be written as a linear combination of matrices, that up to a permutation of the subsystems of $A^n$, have the form $X_1\otimes I^{A_2}\otimes\cdots\otimes I^{A_n}$. Now, if $X=X_1\otimes I^{A_2}\otimes\cdots\otimes I^{A_n}$ then the condition $\la\psi|X|\psi\ra=0$ is equivalent to \be\tr\left[\rho^{A_1}X_1\right]=0,
\ee 
where $\rho^{A_1}\eqdef\tr_{A_2\cdots A_n}\left[\psi^{A^n}\right]$. Since the above condition has to hold for all $X_1\in {\rm Lie}\big({SL}(A_1)\big)$, we conclude that $\rho^{A_1}$ is proportional to the identity matrix. In other words, $|\psi\ra\in{\rm Crit}(A^n)$ if and only if for any $x\in[n]$, the reduced density matrix of $\psi^{A^n}$ on the $x$th-subsystem is proportional to the identity matrix $I^{A_x}$. This completes the proof.
\end{proof}

\bex\label{addknt}
Let $|\psi\ra\in{\rm Crit}(A^n)$ and let $M\in SL^n$ be such that $\big\|M|\psi\ra\big\|= \big\||\psi\ra\big\|$. 
\ben 
\item Show that there exists a local unitary matrix; i.e., $U\in SU(d_1)\times\cdots\times SU(d_n)$ such that $M|\psi\ra=U |\psi\ra$.
\item Show that if in addition, $M>0$, then $M|\psi\ra=|\psi\ra$.
\een
\eex

\subsection{The Null Cone}\index{Null Cone}

\begin{myd}{}
\begin{definition}
Let $A^n$ be a multipartite system. 
The \emph{null cone} of $A^n$, denoted by ${\rm Null}(A^n)$, is the set of all vectors in $A^n$ on which all SLIPs vanish. That is, $|\psi\ra\in{\rm Null}(A^n)$ if and only if
\be
f(|\psi\ra)=0\quad\quad\forall\;f\in\slip(A^n)\;.
\ee
\end{definition}
\end{myd}
For two qubit states the null cone consists only of product states. This can be easily verified by noting that the SLIP given by $f(|\psi\ra)\eqdef\lp|\psi\ra,|\psi\ra\rp_2$ is zero if and only if $|\psi\ra\in\mbb{C}^2\otimes\mbb{C}^2$ is a product state. For higher number of qubits the null cone is not trivial. As an example, consider the three-qubit state, known as the W-state,
\be\label{wstate}
|W\ra\eqdef\frac1{\sqrt{3}}\big(|100\ra+|010\ra+|001\ra\big)\;.
\ee
This state has the property that for any $0\neq t\in\mbb{C}$ the matrix $M_t\eqdef \begin{pmatrix}
t & 0\\
0 & t^{-1}
\end{pmatrix}^{\otimes 3}$ satisfies
\be\label{mtlim}
M_t|W\ra=t|W\ra\;.
\ee
Since $M_t\in SL^3$, for any homogeneous SLIP $f_k$ of degree $k$ we have
\be
f_k(|W\ra)=f_k(M_t|W\ra)
=f_k(t|W\ra)
=t^kf_k(|W\ra)\;.
\ee
Since $t\neq 0$ this means $f_k(|W\ra)=0$. Since $f_k$ is an arbitrary homogenous SLIP, this implies that for any $f\in\slip(A^3)$ (where $A$ is a qubit; i.e. $|A|=2$) we have $f(|W\ra)=0$. Therefore, the W-state belong to the null cone of three qubits.

From~\eqref{mtlim} of the example above it follows that
\be
\lim_{t\to 0}M_t|W\ra=\lim_{t\to 0}t|W\ra=\0\;.
\ee
That is, the orbit $SL^3|W\ra$ contains a sequence of vectors approaching the zero vector. This is precisely the key property of states in the null cone.

\begin{myt}{\color{yellow} The Hilbert-Mumford Theorem}\index{Hilbert-Mumford theorem}
\begin{theorem}
Let $|\psi\ra\in A^n$. The following statements are equivalent:
\ben
\item The vector $|\psi\ra\in{\rm Null}(A^n)$.
\item There exists a sequence of vectors $\{|\psi_k\ra\}_{k\in\mbb{N}}\subset {SL^n}|\psi\ra$ such that
\be
\lim_{k\to\infty}|\psi_k\ra=\0\;.
\ee
\een
\end{theorem}
\end{myt}

The direction that $2\Rightarrow 1$ is relatively simple to show. Indeed, suppose there is a sequence of vectors ${|\psi_k\ra}_{k\in\mbb{N}}\subset{SL^n}|\psi\ra$ that approaches the zero vector in the limit $k\to\infty$. Let $f\in\slip(A^n)$. Then, for any $k\in\mbb{N}$ we have $f(|\psi_k\ra)=f(|\psi\ra)$. Since this holds for any integer $k$ it must hold also for the limit $k\to\infty$. Combining this with the continuity of polynomial functions we get
\be
f(|\psi\ra)=\lim_{k\to\infty}f(|\psi_k\ra)=f(\0)=0\;.
\ee
As $f$ was an arbitrary SLIP we conclude that $|\psi\ra\in{\rm Null}(A^n)$. The other direction can be found in Theorem~43 of~\cite{Wallach2017}. 

\bex
Let $\lambda_1,\ldots,\lambda_n\in\mbb{C}$, and let 
\be
|\psi\ra\eqdef \lambda_1|10\ldots0\ra+\lambda_2|01\ldots0\ra+\cdots+\lambda_n|00\ldots1\ra\in A^n\;.
\ee
Show that $|\psi\ra\in{\rm Null}(A^n)$.
\eex

\subsection{Stable States}\index{stable states}

\begin{myd}{}
\begin{definition}
Let $A^n$ be a multipartite system. 
A state $\psi\in\pure(A^n)$ is said to be \emph{stable} if its orbit ${SL^n}|\psi\ra$ is closed. The set of all stable states is denoted $\stable(A^n)$.
\end{definition}
\end{myd}

Note that the orbit ${SL^n}|\psi\ra$ is closed if for any sequence of states $\{|\phi_k\ra\}_{n\in\mbb{N}}\subset {SL^n}|\psi\ra$ with a limit $\lim_{k\to\infty}|\phi_k\ra=|\phi\ra$ we have that the limit $|\phi\ra$ is also in ${SL^n}|\psi\ra$. Therefore, states in the null cone are not stable since if $|\psi\ra\in{\rm Null}(A^n)$ is a non-zero vector then ${SL^n}|\psi\ra$ does not contain the zero vector. Still, ${SL^n}|\psi\ra$ contains a sequence of vectors with zero limit since $|\psi\ra$ is in the null cone. Hence, the null cone and the set of stable states forms two disjoint set of states in $A^n$. The following theorem shows that any state in $A^n$ can be written as a linear combination of these two set of states.

\begin{myt}{}
\begin{theorem}
Let $A^n\eqdef A_1\cdots A_n$ be a multipartite system and $\psi\in A^n$. Then, there exists $|\phi\ra\in\stable(A^n)$ and $|\chi\ra\in{\rm Null}(A^n)$ such that
\be
|\psi\ra=|\phi\ra+|\chi\ra\;.
\ee
\end{theorem}
\end{myt}

The above result follows from a variant of the Hilbert-Mumford theorem given in Theorem~45 of~\cite{Wallach2017}.
The theorem above states that the vector space $A^n$ can be decomposed into the direct sum
\be
A^n={\rm Stable}(A^n)\oplus{\rm Null}(A^n)\;.
\ee
In addition, it can be shown that almost all vectors in $A^n$ are stable in the sense that the closure of ${\rm Stable}(A^n)$ is the whole space; i.e.\
\be\label{almosteverything}
A^n=\overline{{\rm Stable}(A^n)}\;.
\ee
Therefore, much of the characterization in literature of multipartite entanglement is focused on stable states.

\begin{myt}{\color{yellow} The Kempf-Ness Theorem (Part II)}\index{Kempf-Ness theorem}
\begin{theorem}\label{knt}
Let $|\psi\ra\in A^n$. Then, $|\psi\ra\in{\rm Stable}(A^n)$ if and only if ${SL^n}|\psi\ra$ contains a critical state.
\end{theorem}
\end{myt}
\begin{proof}
Suppose $|\psi\ra$ is stable so that ${SL^n}|\psi\ra$ is closed. Then, there exists a state $|\phi\ra\in {SL^n}|\psi\ra$ with minimal norm; that is, for any $M\in {SL^n}$
\be
\big\|M|\psi\ra\big\|\geq \big\||\phi\ra\big\|\;.
\ee
But since $|\phi\ra=N|\psi\ra$ for some $N\in {SL^n}$ we can express the above equation as
\be
\big\|MN^{-1}|\phi\ra\big\|\geq \big\||\phi\ra\big\|\quad\quad\forall\;M\in {SL^n}\;.
\ee
Since any $M'\in {SL^n}$ can be expressed as $M'=MN^{-1}$ for some $M\in {SL^n}$ we conclude that
$\big\|M'|\phi\ra\big\|\geq \big\||\phi\ra\big\|$ for all $M'\in {SL^n}$. From Theorem~\ref{0knt} it then follows that $|\phi\ra$ is a critical state. That is, the orbit ${SL^n}|\psi\ra$ contains a critical state. The proof of the converse part can be found in Theorem 47 of~\cite{Wallach2017}.
\end{proof}

\subsection{Characterization of SLOCC Classes}

In this subsection we show that SLOCC classes of states can be characterized with SLIPs. Specifically, let $\psi,\phi\in\pure(A^n)$. How can we determine if these two states belong to the same SLOCC class? According to the discussion above~\eqref{14p5} it is sufficient to consider reversible SLOCC classes. Hence, $\psi^{A^n}$ and $\phi^{A^n}$ belong to the same (reversible) SLOCC class if and only if there exists $\theta\in[0,2\pi)$ and $M\in {SL^n}$ such that
\be\label{14p75}
\big|\phi^{A^n}\big\ra=e^{i\theta}\frac{M\big|\psi^{A^n}\big\ra}{\left\|M\big|\psi^{A^n}\big\ra\right\|}\;.
\ee 
Observe that if $f\in\slip_k(A^n)$ for some $k\in\mbb{N}$ then the above equation gives
\be
f\left(\big|\phi^{A^n}\big\ra\right)=\frac{e^{i\theta k}}{\left\|M\big|\psi^{A^n}\big\ra\right\|^k}f\left(\big|\psi^{A^n}\big\ra\right)\;.
\ee
Therefore, if $h\in\slip_k(A^n)$ is another homogenous SLIP of degree $k$ such that $h\left(\big|\psi^{A^n}\big\ra\right)\neq 0$ then the above equation is equivalent to
\be\label{ratiofh}
\frac{f\left(\big|\phi^{A^n}\big\ra\right)}{h\left(\big|\phi^{A^n}\big\ra\right)}=\frac{f\left(\big|\psi^{A^n}\big\ra\right)}{h\left(\big|\psi^{A^n}\big\ra\right)}\;.
\ee
That is, if $\psi^{A^n}$ and $\phi^{A^n}$ belong to the same reversible SLOCC then the above equation must hold. The following theorem demonstrates that the converse is also true for almost all states in $A^n$.

\begin{myt}{}
\begin{theorem}
Let $|\psi\ra,|\phi\ra\in{\rm Stable}(A^n)$. Then, there exists $\theta\in[0,2\pi)$ and $M\in {SL^n}$ such that~\eqref{14p75} holds if and only if~\eqref{ratiofh} holds for all $k\in \mbb{N}$ and all $f,h\in\slip_k(A^n)$ with $h\left(\big|\psi^{A^n}\big\ra\right)\neq 0$.
\end{theorem}
\end{myt}

\begin{remark}
Note that in the theorem above, $k$ is unbounded. However, since it is known that the space of SLIPs has a finite dimension, it is possible to restrict $k$, although the best upper bound is unknown.
\end{remark}

\begin{proof}
We showed above that~\eqref{14p75} implies~\eqref{ratiofh}. It is therefore left to show the converse. If there exists $h\in\slip_k(A^n)$ such that $h\left(\big|\psi^{A^n}\big\ra\right)\neq 0$ but $h\left(\big|\phi^{A^n}\big\ra\right)= 0$ then clearly $\psi^{A^n}$ and $\phi^{A^n}$ are not in the same invertible SLOCC class. We therefore assume without loss of generality that there exists $k\in\mbb{N}$ and $h\in\slip_k(A^n)$ such that both $h\left(\big|\psi^{A^n}\big\ra\right)\neq 0$ and $h\left(\big|\phi^{A^n}\big\ra\right)\neq 0$, and denote by
\be\label{lamlam}
\lambda\eqdef\frac{h\left(\big|\phi^{A^n}\big\ra\right)}{h\left(\big|\psi^{A^n}\big\ra\right)}\neq 0\;.
\ee
With this notation, our assumption in~\eqref{ratiofh} implies that for all $f\in\slip_k(A^n)$, 
\be
f\left(\big|\phi^{A^n}\big\ra\right)=\lambda f\left(\big|\psi^{A^n}\big\ra\right)=f\left(\lambda^{1/k}\big|\psi^{A^n}\big\ra\right)\;.
\ee
Our first goal is to show that up to some phase factors, $f$ above can be replaced with \emph{any} SLIP (even not homogeneous). For this purpose,
consider the subgroup $\G_{n,k}\subset GL(A^n)$ defined by
\be
\G_{n,k}\eqdef\left\{\mu M\;:\;\mu^k=1\;\;,\;\;M\in {SL^n}\;,\;\;\mu\in\mbb{C}\right\}\;,
\ee
and observe that in addition for being ${SL^n}$-invariant polynomial (i.e., SLIP), $h$ is also $\G_{n,k}$-invariant polynomial. Moreover, the degree of any homogeneous $\G_{n,k}$-invariant polynomial must be divisible by $k$. To see this, let $g$ be a $\G_{n,k}$-invariant polynomial of degree $m$. Then, since for any $\mu\in\mbb{C}$ such that $\mu^k=1$ we have $\mu I\in\G_{n,k}$, it follows that 
$g(|\psi\ra)=g(\mu I|\psi\ra)=\mu^mg(|\psi\ra)$, so that $\mu^m=1$. Since $m$ satisfies this property for any such $\mu$ (i.e.\ any $k$-th root of unity), we conclude that $m=kr$ for some $r\in\mbb{N}$.

Now, fix $k$, and let $g$ be a homogenous $G_{n,k}$-invariant polynomial
of degree $kr$ for some $r\in\mbb{N}$. Since $g$ is a SLIP, from the assumption of the theorem
\be\label{ratiokell2}
\frac{g\left(\big|\phi^{A^n}\big\ra\right)}{h^r\left(\big|\phi^{A^n}\big\ra\right)}=\frac{g\left(\big|\psi^{A^n}\big\ra\right)}{h^r\left(\big|\psi^{A^n}\big\ra\right)}\;.
\ee
Combining this with~\eqref{lamlam} yields
\be
g\left(\big|\phi^{A^n}\big\ra\right)=\lambda^{r} g\left(\big|\psi^{A^n}\big\ra\right)=g\left(\lambda^{1/k}\big|\psi^{A^n}\big\ra\right)\;,
\ee
where in the last equality we used the fact that $g$ is homogeneous of degree $kr$. Since the above equation holds for any \emph{homogenous} ${SL^n}$-invariant polynomial $g$ (recall that $r$ was arbitrary), it must also hold for all (possibly non-homogeneous) ${SL^n}$-invariant polynomials. Hence, using a result from invariant theory that closed  orbits of a reductive
algebraic subgroup of $GL(A^n)$ are separated by their invariant polynomials, we conclude that there exists $\mu\in\mbb{C}$ with $\mu^k=1$ and $M\in {SL^n}$ such that $\big|\phi^{A^n}\big\ra=\lambda^{1/k}\mu M\big|\psi^{A^n}\big\ra$. The upshot is $\big|\phi^{A^n}\big\ra=c M\big|\psi^{A^n}\big\ra$ for some $c\in\mbb{C}$, and the normalization $\left\|\big|\phi^{A^n}\big\ra\right\|=1$ gives $c=e^{i\theta}/ \left\|M\big|\psi^{A^n}\big\ra\right\|$. This completes the proof.

\end{proof}

The theorem above demonstrates that SLIPs can be used to classify multipartite entanglement. We give two examples of such classifications in three and four qubits systems.

\section{Multipartite Entanglement of Three and Four Qubits}

\subsection{Classification of Three Qubit Entanglement}\index{three-qubit entanglement}

\subsubsection{Canonical Form\index{canonical form} }

In the previous chapters, we learned that pure bipartite states can always be represented in their Schmidt form. Specifically, for a two-qubit system $AB$, any state $\psi\in\pure(AB)$ can be expressed, up to local unitaries, as
\begin{equation}
|\psi^{AB}\rangle = \sqrt{p}|00\rangle + \sqrt{1-p}|11\rangle,
\end{equation}
where $p\in[0,1]$. We refer to this representation as the canonical form\index{canonical form}  of the state $\psi^{AB}$.

Now, our goal is to find a canonical form for any three-qubit state in $ABC$ where $|A|=|B|=|C|=2$. To achieve this, we will utilize the following property presented in the following exercise.

\bex
Let $AB$ be a two-qubit system and let $|\psi_0\ra,|\psi_1\ra\in AB$ be two pure bipartite vectors. Show that if the vectors $|\psi_0^{AB}\ra$ and $|\psi_1^{AB}\ra$ are linearly independent then there exists numbers $a,b\in\mbb{C}$ such that $a|\psi_0^{AB}\ra+b|\psi_1^{AB}\ra$ is a product (i.e. non-entangled) state.
Hint: Denote by $c\eqdef\frac ab$ and view the determinant of the reduced density matrix of the (non-normalized) state $c|\psi_0^{AB}\ra+|\psi_1^{AB}\ra$ as a quadratic polynomial in $c$. Recall that over the complex field, all quadratic polynomials have roots.
\eex

\begin{myt}{}
\begin{theorem}\label{canonform}
Let $ABC$ be a three-qubit system, and let $\psi\in\pure(ABC)$. Then, up to a local unitary matrix in $\ml(ABC)$, the state $\big|\psi^{ABC}\big\ra$ can be expressed as
\be\label{14p85}
\big|\psi^{ABC}\big\ra=\lambda_0|000\ra+\lambda_1e^{i\theta}|100\ra+\lambda_2|101\ra+\lambda_3|110\ra+\lambda_4|111\ra\;,
\ee
where $\lambda_0,\ldots,\lambda_4\in\mbb{R}_+$ and $\theta\in[0,\pi]$.
\end{theorem}
\end{myt}

\begin{remark}
The normalization of $\psi^{ABC}$ implies that 
$\lambda_0^2+\cdots+\lambda_4^2=1$. In the proof below, the fact that we can restrict $\theta$ to the domain $[0,\pi]$ will be left as an exercise.
\end{remark}

\begin{proof}
Every three-qubit state $|\psi\ra\in ABC$ can be expressed as
\be\label{fcitp0}
\big|\psi^{ABC}\big\ra=|0\ra^A|\psi_0^{BC}\ra+|1\ra^A|\psi_1^{BC}\ra\;,
\ee
where $|\psi_0\ra,|\psi_1\ra\in BC$ are two orthogonal (possibly unnormalized) vectors. From the exercise above it follows that there exists two complex numbers $a,b\in\mbb{C}$ such that $a|\psi_0^{BC}\ra+b|\psi_1^{BC}\ra$ is a product state. Note that without loss of generality we can assume that $|a|^2+|b|^2=1$. Therefore, the matrix $U=\bpm
 a & b\\
 -\bar{b} & \bar{a}
\epm
$
is a unitary matrix, so that by applying $U$ to the first qubit of $|\psi^{ABC}\ra$ we get
\ba
U^A\otimes I^{BC}\big|\psi^{ABC}\big\ra&=\left(a|0\ra^A-\bar{b}|1\ra^{A}\right)\big|\psi_0^{BC}\big\ra+\left(b|0\ra^A+\bar{a}|1\ra^A\right)|\psi_1^{BC}\ra\\
&=|0\ra^A\left(a\big|\psi_0^{BC}\big\ra+b\big|\psi_1^{BC}\big\ra\right)+|1\ra^A\left(\bar{a}\big|\psi_1^{BC}\big\ra-\bar{b}\big|\psi_0^{BC}\big\ra\right)\;.
\ea
Since  $a\big|\psi_0^{BC}\big\ra+b\big|\psi_1^{BC}\big\ra$ is a (possibly unnormalized) product state, there exists a local unitary on $BC$ that transform it to the state $\lambda_0|00\ra^{BC}$ where $\lambda_0\in\mbb{C}$ is some normalization factor. We therefore conclude that, up to local unitaries, the state $|\psi^{ABC}\ra$ can be expressed as
\be\label{lam01d}
|\psi^{ABC}\big\ra=\lambda_0|000\ra+|1\ra|\phi^{BC}\ra
\ee
where $|\phi^{BC}\ra$ is some vector in $BC$. Let $\lambda_1,\ldots,\lambda_4\in\mbb{C}$ be such that
\be
|\phi^{BC}\ra=\lambda_1|00\ra+\lambda_2|01\ra+\lambda_3|10\ra+\lambda_4|11\ra\;.
\ee
Note that by applying to the state above, the local unitary 
\be
U^{BC}\eqdef\begin{pmatrix} e^{i\theta_1} & 0\\
0 & e^{i\theta_2}
\end{pmatrix}\otimes
\begin{pmatrix} e^{i\theta_3} & 0\\
0 & e^{i\theta_4}
\end{pmatrix}\;,
\ee
we get
\be
U^{BC}|\phi^{BC}\ra=\lambda_1e^{i(\theta_1+\theta_3)}|00\ra+\lambda_2e^{i(\theta_1+\theta_4)}|01\ra+\lambda_3e^{i(\theta_2+\theta_3)}|10\ra+\lambda_4e^{i(\theta_2+\theta_4)}|11\ra\;.
\ee
Therefore, by choosing appropriately the four phases $\theta_1,\theta_2,\theta_3,\theta_4$ we can make three of the $\lambda$s non-negative real numbers. We choose them to be $\lambda_2,\lambda_3,\lambda_4\in\mbb{R}_+$. Moreover, observe that by applying
$e^{i\theta}|0\lr 0| + |1\lr 1|$ to system $A$ in~\eqref{lam01d} we can add a phase to $\lambda_0$. Therefore, we can assume without loss of generality that $\lambda_0$ is a real non-negative number.
\end{proof}

\bex
Complete the proof above by showing that $\theta$ in~\eqref{14p85} can be restricted to $[0,\pi]$.
\eex

\bex\label{ex:3marginals}
Let $\big|\psi^{ABC}\big\ra$ be the three-qubit state given in~\eqref{14p85}. Show that its three local marginals (i.e. reduced density matrices) are given by
\be
\psi^A=
\bpm 
\lambda_0^2 & \lambda_0\lambda_1e^{-i\theta}\\
\lambda_0\lambda_1e^{-i\theta} & 1-\lambda_0^2
\epm 
\quad,\quad
\psi^B=
\bpm 
\lambda_0^2+\lambda_1^2+\lambda_2^2 & \lambda_1\lambda_3e^{i\theta}+\lambda_2\lambda_4\\
\lambda_1\lambda_3e^{-i\theta}+\lambda_2\lambda_4 & \lambda_3^2+\lambda_4^2
\epm
\ee 
and
\be
\psi^C=
\bpm 
\lambda_0^2+\lambda_1^2+\lambda_3^2 & \lambda_1\lambda_2e^{i\theta}+\lambda_3\lambda_4\\
\lambda_1\lambda_2e^{-i\theta}+\lambda_3\lambda_4 & \lambda_2^2+\lambda_4^2
\epm\;.
\ee
\eex

From Theorem~\ref{canonform} and the exercise above we get that up to local unitaries, there is  only one normalized critical state given by the GHZ state
\be
|GHZ\ra\eqdef\frac1{\sqrt{2}}\big(|000\ra+|111\ra\big)\;.
\ee

\begin{myg}{}
\begin{corollary}
Let $ABC$ be a composite system\index{composite system} of three qubits. Then, if $|\psi\ra\in{\rm Crit}(ABC)$ is normalized then there exists a local unitary matrix $U_1\otimes U_2\otimes U_3\in\ml(ABC)$ such that
\be
\big|\psi^{ABC}\big\ra=U_1\otimes U_2\otimes U_3|GHZ\ra\;.
\ee
\end{corollary}
\end{myg}
\begin{proof}
From the properties of critical states (see Theorem~\ref{0knt}) we know that if $|\psi\ra\in{\rm Crit}(ABC)$ is normalized then all three local marginals $\psi^A$, $\psi^B$, and $\psi^C$ must be maximally mixed. Now, from Theorem~\ref{canonform} we know that up to local unitaries the state $\psi^{ABC}$ can be expressed as in~\eqref{14p85}.
Hence, using this form, we get from Exercise~\ref{ex:3marginals} that the condition $\psi^A=\frac12 I^A$ holds if and only if $\lambda_0^2=\frac12$ and $\lambda_1=0$. The condition $\psi^B=\frac12 I^B$ gives in particular $\lambda_0^2+\lambda_1^2+\lambda_2^2=\frac12$. Therefore, also $\lambda_2=0$. Finally, the condition $\psi^C=\frac12 I^C$ gives $\lambda_3=0$ and $\lambda_4^2=\frac12$. Hence, the state $\psi^{ABC}$ as given in~\eqref{14p85} is critical if and only in it is the GHZ state. This concludes the proof.
\end{proof}

Recall the homogeneous SLIP of degree four as defined in~\eqref{g4slip} for odd number of qubits. For three qubit system $ABC$ (with $|A|=|B|=|C|=2$), its absolute value is called \emph{the 3-tangle}, and it is given for any vector\index{Tangle}
\be\label{14p83}
|\psi^{ABC}\ra\eqdef |0\ra^A|\psi_0^{BC}\ra+|1\ra^A|\psi_1^{BC}\ra\in ABC
\ee
by
\be\label{tan4slip}
{\rm Tangle}\left(\big|\psi^{ABC}\big\ra\right)\eqdef\left|\det\begin{pmatrix}
\lp|\psi_0\ra,|\psi_0\ra\rp &\lp|\psi_0\ra,|\psi_1\ra\rp\\
\lp|\psi_1\ra,|\psi_0\ra\rp &\lp|\psi_1\ra,|\psi_1\ra\rp
\end{pmatrix}\right|
\ee
where  $\lp|\psi_x\ra,|\psi_y\ra\rp\eqdef\la\bar{\psi}^{BC}_x| J_2\otimes J_2|\psi^{BC}_y\ra$ for each $x,y\in\{0,1\}$; recall that $J_2\eqdef|0\lr 1|-|1\lr 0|$.

From the corollary discussed earlier, it follows that all stable vectors in $ABC$ are, up to normalization, contained in the $\G_3$ orbit of the GHZ state $|GHZ\rangle$. In other words, almost all three-qubit normalized states are in the SLOCC class of the GHZ state. This, in turn, implies that almost all three-qubit states have a non-zero 3-tangle, which is consistent with the formula for the 3-tangle given below.

\bex
Show that the 3-tangle of the state $\psi^{ABC}$ in~\eqref{14p85} is given by
\be
{\rm Tangle}\left(\big|\psi^{ABC}\big\ra\right)=\lambda_0\lambda_4\;.
\ee
\eex

The formula presented in the exercise above shows that the 3-tangle is zero when $\lambda_0 = 0$, which makes sense because in this case, the state $\psi^{ABC}$ is a product state between system $A$ and system $BC$. This implies that the state is in the null cone, i.e., it has no genuine tripartite entanglement.
On the other hand, if $\lambda_4 = 0$, then the 3-tangle is also zero. In this case, the state $\psi^{ABC}$ can be expressed as
\be
\big|\psi^{ABC}\big\ra=\lambda_0|000\ra+\lambda_1e^{i\theta}|100\ra+\lambda_2|101\ra+\lambda_3|110\ra\;.
\ee
If we apply the flip operator $|1\0|+|0\lr 1|$ to the first qubit of the state $\psi^{ABC}$, the resulting state takes the form:
\be
\big|\psi^{ABC}\big\ra=\lambda_1e^{i\theta}|000\ra+\lambda_0|100\ra+\lambda_2|001\ra+\lambda_3|010\ra\;.
\ee 
Moreover, observe that by applying the local unitary matrix $\left(e^{-i\theta/3}|0\lr 0|+ e^{2i\theta/3}|1\lr 1|\right)^{\otimes 3}$ we can eliminate the phase attached to the $|000\ra$ term. Therefore, after renaming the coefficients $\lambda_0,\ldots,\lambda_3$ we conclude that unless the state $\psi^{ABC}$ is a product state between $A$ and $BC$, its 3-tangle is zero if and only if, up to local unitaries,
it can be expressed as\index{$W$-state}
\be\label{4lamd}
|\psi^{ABC}\ra=\lambda_0|000\ra+\lambda_1|100\ra+\lambda_2|010\ra+\lambda_3|001\ra\;,
\ee
with $\lambda_0,\ldots,\lambda_3\in\mbb{R}_+$.

\bex
Show that for any three-qubit pure state $\psi^{ABC}$ of the form~\eqref{4lamd}, there exists three matrices $M,N,L\in GL(2,\mbb{C})$ such that
\be
\big|\psi^{ABC}\big\ra=M\otimes N\otimes L\big|W\big\ra
\ee
where $|W\ra$ is the W-state as defined in~\eqref{wstate}.
\eex

The preceding discussion and exercise demonstrate that the SLOCC class of the $W$-state consists of all states whose 3-tangle vanishes. Furthermore, since the $W$-state lies in the null cone (as shown in the discussion below equation~\eqref{wstate}), we can conclude that the null cone precisely consists of the SLOCC class of the $W$-state. This, in turn, implies that a three-qubit vector lies in the null cone if and only if its 3-tangle vanishes. This also implies that any other SLIP must be proportional to a power of the 3-tangle. Therefore, the 3-tangle is essentially the absolute value of the only SLIP in three qubits.

In summary, we can divide the space of three qubits into six invertible SLOCC classes:

\begin{itemize}
\item The ``genuine" tripartite entanglement classes: the GHZ class and the W-class.
\item Three bipartite entanglement classes: the three $SL^3$-orbits generated by $|0\rangle^A|\Phi^{BC}\rangle$, $|0\rangle^B|\Phi^{AC}\rangle$, and $|\Phi^{AB}\rangle|0\rangle^C$.
\item The unentangled class generated by $|000\rangle$.
\end{itemize}

\subsection{Four-Qubit Entanglement}\index{four-qubit entanglement}

In this section, our goal is to analyze the classification of SLOCC classes of four-qubit states by seeking their canonical forms under both local unitaries and SLOCC. We will find that characterizing these forms is considerably more complex than in the three-qubit case. Therefore, we will focus our attention on the set of critical states, which is somewhat simpler to characterize. For interested readers, we refer to the references listed in the section `Notes and References' at the end of this chapter for more details on this topic. 

\subsubsection{The canonical form\index{canonical form}  Under Local Unitaries}

In this section, we consider the composite system\index{composite system} ABCD consisting of four qubits, where $|A|=|B|=|C|=|D|=2$. The primary technique employed in the study of four-qubit entanglement is the ``accident" in Lie-group theory, which states an isomorphism between the special orthogonal group $SO(4)$ (consisting of $4\times4$ real orthogonal matrices with determinant one) and the group $SU(2)\otimes SU(2)$. We denote this isomorphism as
\be
SU(2)\otimes SU(2)\cong SO(4)\;.
\ee
In the following exercise you prove this isomorphism.

\bex\label{ex:amazingiso}
Consider the $4\times 4$ complex matrix
\be\label{unit}
T\eqdef\frac1{\sqrt{2}}
\bpm
1 & 0 & 0 &1\\
0 & i & i & 0\\
0 & -1 & 1 & 0\\
i & 0 & 0 & -i
\epm\;.
\ee
\ben
\item Show that $T$ is a unitary matrix.
\item Show that for all $U_1,U_2\in SU(2)$ we have
\be
T\big(U_1\otimes U_2\big)T^*\in SO(4)\;.
\ee
\een
Hint: Show that $T^TT=J\otimes J$, where $J=|0\lr 1|-|1\lr 0|$ is the matrix that satisfies~\eqref{1194}. 
\eex

We can use the above isomorphism to get the canonical
form of a four-qubit state. Let $\big|\psi^{ABCD}\big\ra\in ABCD$ be a four qubit state, and let $M:AB\to AB$ be the $4\times 4$ complex matrix defined via
\be\label{mpsin}
\big|\psi^{ABCD}\big\ra=M\otimes I^{CD}\big|\Omega^{(AB)(CD)}\big\ra
\ee 
where
\be
\big|\Omega^{(AB)(CD)}\big\ra=\sum_{x,y\in\{0,1\}}|xy\ra^{AB}|xy\ra^{CD}\;.
\ee
In other words, we view four-qubit states as $4\times 4$ complex matrices. Consider now a state 
\be\big|\phi^{ABCD}\big\ra \eqdef U_1\otimes U_2\otimes U_3\otimes U_4\big|\psi^{ABCD}\big\ra
\ee
with each $U_x\in SU(2)$. That is, $|\psi\ra$ and $|\phi\ra$ are related by local unitaries. Let $N$ be the $4\times 4$ matrix representing $\big|\phi^{ABCD}\big\ra$ similarly to~\eqref{mpsin}. Then, from the second part of Exercise~\ref{bipartite} we get that $M$ and $N$ are related by 
\ba
N&=(U_1\otimes U_2)M(U_3\otimes U_4)^T\\
&=T^*O_1TMT^*O_2T\;,
\ea
where $T$ is the unitary matrix~\eqref{unit} and
\be
O_1\eqdef T\big(U_1\otimes U_2\big)T^*\quad\text{and}\quad O_2\eqdef T\big(U_3\otimes U_4\big)^TT^*\;.
\ee
Observe that $O_1,O_2\in SO(4)$. 

Continuing, let $TMT^*=M_1+iM_2$, where $M_1$ and $M_2$ are matrices with real coefficients. Then, $TNT^*=N_1+iN_2$, where
$N_1\eqdef O_1M_1O_2$ and $N_2\eqdef O_1M_2O_2$ are real matrices. Finally, observe that by appropriate choice of $O_1$ and $O_2$, the matrix $N_1$ (or $N_2$) can be made diagonal using the (real) singular value decomposition. The resulting $N=N_1+iN_2$ and consequently $\big|\phi^{ABCD}\big\ra$ can be viewed as the canonical form\index{canonical form}  of $\big|\psi^{ABCD}\big\ra$. However,
this canonical form of four-qubit states is not very useful as it involves too many parameters. Specifically, even if $N_1$ is diagonal, the matrix $N_2$ is not. 

\bex\label{promms}
Prove that any 4-qubit state can be expressed, up to local unitary transformations, in the form given by equation~\eqref{mpsin}, where the matrix $M$ satisfies the condition
\be\label{msme}
M^*M=D+i\Lambda;,
\ee
and $D,\Lambda\in\mbb{R}^{4\times 4}$, with $D$ being a diagonal matrix with non-negative diagonal elements and $\Lambda$ being a skew-symmetric matrix.
\eex

\bex
Prove that if $\Lambda_1,\Lambda_2\in SL(2,\mbb{C})$ then
\be\label{sl2cso4}
T(\Lambda_1\otimes\Lambda_2)T^*\in SO(4,\mbb{C})\;,
\ee
where $SO(4,\mbb{C})$ is the (non-compact) special orthogonal group over $\mbb{C}$; i.e. $O\in SO(4,\mbb{C})$ if and only if $O\in\mbb{C}^{4\times 4}$, $O^TO=I_4$, and $\det(O)=1$. Here $T$ is the same matrix that was used in Exercise~\ref{ex:amazingiso}.
\eex

\subsubsection{Critical States}

In this subsection, we will characterize the set ${\rm Crit}(ABCD)$ of critical states in the four-qubit system by leveraging the isomorphism described in~\eqref{sl2cso4}. Specifically, we begin by considering $\Lambda=\Lambda_1\otimes\Lambda_2\otimes\Lambda_3\otimes\Lambda_4\in \G_4$, $\big|\psi\big\ra\in ABCD$, and the matrix $M$ defined in~\eqref{mpsin}. We observe that
\be
\Lambda\big|\psi^{ABCD}\big\ra=N\otimes I^{CD}\big|\Omega^{(AB)(CD)}\big\ra\quad\text{where}\quad N=\big(\Lambda_1\otimes \Lambda_2\big)M\big(\Lambda_3\otimes \Lambda_4\big)^T\;.
\ee
Next, under the isomorphism in~\eqref{sl2cso4}, the matrix $M$ is transformed into $\tilde{M}=TMT^*$ and $N$ into $\tilde{N}=TNT^*$. We can then express the relation between $\tilde{M}$ and $\tilde{N}$ as $\tilde{N}=O_1\tilde{M}O_2$, where\be
O_1\eqdef T\big(\Lambda_1\otimes \Lambda_2\big)T^*\quad\text{and}\quad O_2\eqdef T\big(\Lambda_3\otimes \Lambda_4\big)^TT^*\;.
\ee 
Note that if $O_1$ and $O_2$ were unitaries, we could have diagonalized $\tilde{M}$ using the singular value decomposition. However, since they are orthogonal, this is not always possible. Nevertheless, a somewhat cumbersome canonical form\index{canonical form}  does exist (see, e.g.,~\cite{VDDV2002}).

We now focus on four-qubit states in $ABCD$ whose corresponding $\tilde{M}$ matrix has the form $O_1'DO_2'$, where $D$ is a $4\times 4$ complex diagonal matrix, and $O_1'$ and $O_2'$ are $4\times 4$ orthogonal complex matrices. We will show that all critical states in four qubits belong to this class. Therefore, by the Kempf-Ness theorem (Theorem~\ref{knt}) in conjunction with~\eqref{almosteverything}, this class of states is dense in $ABCD$. In other words, almost all four-qubit pure states have this property.

We begin by noting that the diagonalizable property of $\tilde{M}$ remains invariant under the action of $\G_4$. This is because we have already shown that for every $|\psi\ra\in ABCD$, the transformation $|\psi\ra \to\Lambda|\psi\ra$ translates, under the isomorphism, to the transformation of $\tilde{M}$ to $O_1\tilde{M}O_2=O_1O_1'DO_2'O_2$, which is of the form $Q_1DQ_2$, where $Q_1=O_1O_1'$ and $Q_2=O_2O_2'$ are two orthogonal matrices.

Next, for a fixed diagonal matrix $D={\rm Diag}(\lambda_1,\lambda_2,\lambda_3,\lambda_4)$, where each $\lambda_x\in\mbb{C}$ ($x\in[4]$), we take the state corresponding to $\tilde{M}=D$ to represent this $\G_4$ orbit. Note that $M=T^*\tilde{M}T=T^*DT$, so the representative state has the form
\ba
\big|\psi^{ABCD}\big\ra&=T^*DT\otimes I^{CD}\big|\Omega^{(AB)(CD)}\big\ra\\
&=(T^*D\otimes T^T)\big|\Omega^{(AB)(CD)}\big\ra.
\ea
For any $j\in[4]$ with binary representation $(x,y)$ (with $x,y\in{0,1}$), we define $|u_j^{AB}\ra\eqdef T^{*}|xy\ra^{AB}$ and $|v_j^{CD}\ra\eqdef T^{T}|xy\ra^{CD}$. With these notations
\be
\big|\psi^{ABCD}\big\ra=\sum_{j\in[4]}\lambda_j|v_j^{AB}\ra|u_j^{CD}\ra\;.
\ee
It is simple to check that for each $j\in[4]$ the states  $|u_j^{AB}\ra$ and $|v_j^{CD}\ra$ are maximally entangled. Hence, up to local unitary matrices, the state above can be expressed as
\be\label{psilambda}
\big|\psi^{ABCD}_{\boldsymbol{\lambda}}\big\ra=\lambda_1\big|\Phi^{AB}_+\big\ra|\Phi^{CD}_+\big\ra+
\lambda_2\big|\Phi_-^{AB}\big\ra|\Phi_-^{CD}\big\ra+\lambda_3\big|\Psi_+^{AB}\big\ra|\Psi_+^{CD}\big\ra+\lambda_4\big|\Psi_{-}^{AB}\big\ra|\Psi_{-}^{CD}\big\ra\;,
\ee
where $\{|\Phi_{\pm}\ra,|\Psi_{\pm}\ra\}$ denotes the Bell basis\index{Bell basis} of maximally entangled states in four qubits.

Note that if there exists another diagonal matrix $D'={\rm Diag}(\lambda_1',\lambda_2',\lambda_3',\lambda_4')$ such that $D'=O_1DO_2$, we then must have
\be
{D'}^2=(D')^TD'=O_2^TD^2O_2\;.
\ee
Therefore, the coefficients $\lambda_1,\ldots,\lambda_4$ must be equal to the coefficients $\lambda_1',\ldots,\lambda_4'$ up to a plus/minus sign. This means that the states $|\psi_{\boldsymbol{\lambda}}\ra$ and $|\psi_{\boldsymbol{\lambda}'}\ra$ belong to the same $\G_4$ orbit if and only if there exists a permutation $\pi$ on four elements such that for all $j\in[4]$, $\lambda_j'=\lambda_{\pi(j)}$ or $\lambda_j'=-\lambda_{\pi(j)}$.

Since local unitaries do not form a subgroup of $\G_4$, we cannot directly translate the above conclusion to the language of SLOCC classes. However, we can order the coefficients ${\lambda_x}$ such that $\lambda_1$ has the maximal absolute value and apply a global phase to the state $|\psi_{\boldsymbol{\lambda}}\ra$ to remove the phase of $\lambda_1$, so that $\lambda_1$ is a positive real number. With this convention, we arrive at the following result.

\begin{myt}{}
\begin{theorem}
Let $|\psi_{\boldsymbol{\lambda}}\ra$ and $|\psi_{\boldsymbol{\lambda}'}\ra$ be two four qubit states as given in~\eqref{psilambda}, with the coefficients $\lambda_1$ and $\lambda_1'$ being real positive such that for all $x=2,3,4$ we have $\lambda_1\geq |\lambda_x|$ and $\lambda_1'\geq |\lambda_x'|$. Then,
the two states $|\psi_{\boldsymbol{\lambda}}\ra$ and $|\psi_{\boldsymbol{\lambda}'}\ra$ belong to the same SLOCC class if and only if $\lambda_1=\lambda_1'$ and there exists a permutation, $\pi$, on three elements such that for each $x\in\{2,3,4\}$ we have $\lambda_x'=\lambda_{\pi(x)}$ or $\lambda_x'=-\lambda_{\pi(x)}$.
\end{theorem}
\end{myt}

The theorem above highlights a stark contrast between three-qubit systems and four-qubit systems. While three-qubit systems have a finite number of SLOCC classes, the same cannot be said for four-qubit systems. In fact, the theorem demonstrates that four-qubit systems have an uncountable number of SLOCC classes.

This has significant implications, as it means that converting $|\psi_{\boldsymbol{\lambda}}\ra$ to $|\psi_{\boldsymbol{\lambda}'}\ra$ by LOCC is impossible, even with a probability less than one, unless ${\boldsymbol{\lambda}'}={\boldsymbol{\lambda}}$ up to a permutation and a sign change of the components of ${\boldsymbol{\lambda}'}$ and ${\boldsymbol{\lambda}}$. In simpler terms, the components of ${\boldsymbol{\lambda}'}$ and ${\boldsymbol{\lambda}}$ must be identical except for a rearrangement and possibly a change in sign.

\bex
Show that the state $|\psi_{\boldsymbol{\lambda}}\ra$ in~\eqref{psilambda} is a critical state. Specifically, show that if $\big|\psi^{ABCD}_{\boldsymbol{\lambda}}\big\ra$ is normalized then its four local marginals are maximally mixed; i.e. show that
\be
\psi^{A}_{\boldsymbol{\lambda}}=\psi^{B}_{\boldsymbol{\lambda}}=\psi^{C}_{\boldsymbol{\lambda}}=\psi^{D}_{\boldsymbol{\lambda}}=\frac12 I_2\;.
\ee
\eex
It is worth noting that in~\cite{VDD2003,VDDV2002,Wallach2004} it has been shown that up to local unitaries, the set
\be
\mathfrak{C}\eqdef\left\{\big|\psi^{ABCD}_{\boldsymbol{\lambda}}\big\ra\;:\;\lambda_1,\lambda_2,\lambda_3,\lambda_4\in\mbb{C}\quad,\quad\sum_{x\in[4]}|\lambda_x|^2=1\right\}
\ee
is the set of \emph{all} critical states in four qubits.

\section{Deterministic Interconversions of Multipartite Entanglement}

As discussed in the previous section, four-qubit systems exhibit an uncountable number of SLOCC classes. This means that deterministic and non-deterministic LOCC conversions between two randomly selected states is typically not possible in multipartite systems. However, for two states that belong to the same SLOCC class, it may be possible to convert one state to another deterministically via LOCC. Recall, however, that LOCC operations can be very complex in multipartite systems. As such, it will be more convenient to consider the larger set of separable operations instead. These operations can be used to transform a state into any other state in the same SLOCC class, and are generally easier to handle than LOCC operations.

A quantum channel $\mE\in\cptp(A^n\to A^n)$ is said to be separable if it has an operator sum\index{operator sum} representation of the form
\be
\mE\left(\cdot\right)=\sum_{k\in[m]}M_k(\cdot)M_k^*\quad\text{where}\quad M_k=N_1^{(k)}\otimes N_2^{(k)}\otimes\cdots\otimes N_n^{(k)}\;,
\ee
and for each $j\in[n]$, $N_j^{(k)}\in\ml(A_j)$. We denote the set of all such channels by $\sep(A^n\to A^n)$. While the matrices $N_j^{(k)}$ (and, by extension, $M_k$) might not always be invertible, in this section, our attention is specifically on the conversion between one pure state to another under the assumption that all $\{M_k\}{k\in[m]}$ are non-singular. We use the notation $\sep_{1}(A^n\to A^n)\subset\sep(A^n\to A^n)$ to represent all separable channels of this kind. In essence, our focus is restricted to separable operations as defined earlier, with each $M_k$ being an element of $GL^n$. For further insights and references into the relations between LOCC, $\sep_1$, and SEP, readers interested are referred to the concluding section of this chapter, titled ``Notes and references."

\subsection{The Stabilizer Group}\index{stabilizer group}

In order to fully characterize conversions among pure multipartite states, it is necessary to introduce the concept of the stabilizer group. This group plays a crucial role in determining the properties and symmetries of a given state, and can help us identify which states can be transformed into one another via separable operations. 

It is worth noting that, unlike the stabilizer formalism commonly used in quantum error correction codes, the stabilizer group discussed here is not necessarily a subgroup of the Pauli\index{Pauli} group. Instead, it is a subgroup of $GL^n$, a much larger group that includes the Pauli group as a special case. This difference is important because the stabilizer group in quantum error correction codes is designed to protect against certain types of errors, whereas the stabilizer group in multipartite quantum systems reflects the underlying symmetries and properties of the system itself. By understanding the structure and properties of this group, we can develop new insights into the behavior of multipartite quantum systems and discover new ways to manipulate and control them.

\begin{myd}{}
\begin{definition}
Let $\psi\in\pure(A^n)$. The stabilizer group of $\psi^{A^n}$ is a subgroup of $GL^n$ defined by
\be
\stab(\psi)\eqdef\big\{\Lambda\in GL^n\;:\;\Lambda|\psi\ra=|\psi\ra\big\}
\ee
\end{definition}
\end{myd}

Note that the set $\stab(\psi)$ is not empty since the identity matrix belongs to it.

\bex
Let $\psi\in\pure(A^n)$ and consider the stabilizer group $\stab(|\psi\ra)$.
\ben
\item Show that $\stab(|\psi\ra)$ is indeed a group.
\item Show that for any $\Lambda\in GL^n$ and $|\phi\ra\eqdef\Lambda|\psi\ra$ we have
\be
\stab(\phi)=\Lambda\;\stab(\psi)\;\Lambda^{-1}\;.
\ee
\een
\eex

\bex
Let $AB$ be a bipartite system\index{bipartite system} with $|A|=|B|$. Find the stabilizer group of the maximally entangled state $|\Phi^{AB}\ra$.
\eex

The stabilizer group for $\psi$ is a subgroup of $GL^n$. One may naturally wonder how this group is related to the same group, but with $GL^n$ replaced by $SL^n$. The following theorem demonstrates that, unless $\psi$ is in the null cone of $A^n$, every element in the stabilizer group of $\psi$ lies in $SL^n$ up to a factor given by a root of unity.

\begin{myt}{}
\begin{theorem}\label{thm:0052}
Let $\psi\in\pure(A^n)$ and suppose that $|\psi\ra\not\in{\rm Null}(A^n)$. Then, there exists $m\in\mbb{N}$ such that
\be\label{g2g4}
\stab(\psi)\subset\G_m\eqdef\left\{e^{i\frac{2\pi k}{m}}\Lambda\;:\;k\in[m]\;,\;\Lambda\in SL^n\right\}\;.
\ee
\end{theorem}
\end{myt}

\begin{proof}
By definition, since $|\psi\ra\not\in{\rm Null}(A^n)$ there exists a homogeneous SLIP, $f$, with the property that $f(|\psi\ra)\neq 0$. Let $m$ be the degree of $f$. Now, let $\Lambda'\in\stab(\psi)$ and observe that since $\stab(\psi)\subset GL^n$ there exists $a\in\mbb{C}$ such that $\Lambda'=a\Lambda$, where $\Lambda\in SL^n$. Thus, the property $\Lambda'|\psi\ra=|\psi\ra$ gives
\be
f(|\psi\ra)=f\left(\Lambda'|\psi\ra\right)=f\left(a\Lambda|\psi\ra\right)=a^mf\left(\Lambda|\psi\ra\right)=a^mf(|\psi\ra)\;.
\ee
Since $f(|\psi\ra)\neq 0$ we must have $a^m=1$ so that $\Lambda'=a\Lambda\in\G_m$. This completes the proof.
\end{proof}

\begin{myg}{}
\begin{corollary}\label{cor:cyclic}
Let $\psi\in\pure(A^n)$ and suppose there exists an homogeneous SLIP of degree $m\in\mbb{N}$ that is not vanishing on $\psi$. Let $\G\eqdef\stab(\psi)\cap SL^n$. Then, the quotient group $\stab(\psi)/\G$ is a  group of order at most $m$.
\end{corollary}
\end{myg}

\begin{proof}
We first need to show that $\G$ is a normal subgroup of $\stab(\psi)$. To see why, recall that for any $M\in\stab(\psi)$ we have $M=e^{i\frac{2\pi k}{m}}\Lambda$ where $\Lambda\in SL^n$. Therefore, for any $\Gamma\in\G$ we get
\be
M\Gamma=e^{i\frac{2\pi k}{m}}\Lambda\Gamma\Lambda^{-1}\Lambda=\Gamma'M
\ee
where $\Gamma'\eqdef\Lambda\Gamma\Lambda^{-1}$. We therefore need to show that $\Gamma'\in\G$. Since $\Lambda'$ is a product of three matrices in $SL^n$ it is itself in $SL^n$. To show that $\Gamma'\in\stab(\psi)$ observe that by definition $\Gamma'$ can also be expressed as $\Gamma'=M\Lambda M^{-1}$ which is a product of three matrices in $\stab(\psi)$ and therefore also $\Lambda'$ is in $\stab(\psi)$. Hence, $\G$ is a normal subgroup of $\stab(\psi)$. Finally, note that any $M=e^{i\frac{2\pi k}{m}}\Lambda$ as above satisfies $M^m\in SL^n$ so that $M^m\in \G$. This completes the proof.
\end{proof}

In the second corollary of Theorem~\ref{thm:0052}, we use the notation $d_x\eqdef|A_x|$ to represent the local dimension of subsystem $A_x$ for each $x\in[n]$. Additionally, we denote by 
\be
SU^n\eqdef SU(d_1)\times\cdots\times SU(d_n)
\ee
and for any $m\in\mbb{N}$
\be
\mathbf{K}_m\eqdef\left\{e^{i\frac{2\pi k}{m}}U\;:\;k\in[m]\;,\;U\in SU^n\right\}\;.
\ee 

\begin{myg}{}
\begin{corollary}\label{cor:0051}
Let $\psi\in{\rm Crit}(A^n)$ be such that $\stab(\psi)$ is a finite group. Then, there exists $m\in\mbb{N}$ such that
\be
\stab(\psi)\subset \mathbf{K}_m\;.
\ee
\end{corollary}
\end{myg}

\begin{proof}
Let $M\in\stab(\psi)$. Since $\psi$ is a critical state it is not in the null cone of $A^n$, so that from Theorem~\ref{thm:0052} there exists $N\in SL^n$, and $a\in\mbb{C}$ with $a^m=1$, such that $M=aN$. Moreover, using the polar decomposition\index{polar decomposition}  we can further express $N$ as $N=U\Lambda$, where $U\in SU^n$ and $\Lambda>0$ is a positive matrix in $SL^n$.
Hence,
\be\label{14p131}
\big\||\psi\ra\big\|=\big\|M|\psi\ra\big\|=\big\|aU\Lambda|\psi\ra\big\|=\big\|\Lambda|\psi\ra\big\|\;.
\ee
As $\Lambda \in SL^n$ is positive definite, the Kempf-Ness theorem (as described in Exercise~\ref{addknt}) implies that $\Lambda|\psi\ra = |\psi\ra$. In other words, $\Lambda$ belongs to the stabilizer group $\stab(\psi)$. Since $\stab(\psi)$ is a finite group, the sequence $\{\Lambda^k\}_{k\in\mbb{N}}$ must contain elements that are equal to each other, and therefore, there exists $k\in\mbb{N}$ such that $\Lambda^k=I^{A^n}$. Since $\Lambda>0$ we must have $\Lambda=I^{A^n}$.
Hence, $M=aU\in\mathbf{K}_m$. Since $M$ was an arbitrary element of $\stab(\psi)$ we conclude that all the elements of $\stab(\psi)$ belong to $\mathbf{K}_m$. This completes the proof. 
\end{proof}

\subsubsection{Example 1: The Stabilizer Group of the 3-Qubit GHZ State}

Let $\phi\eqdef\frac1{\sqrt{2}}(|000\ra+|111\ra)$ be the GHZ state of three qubits, and let
$
\G\eqdef\stab(\phi)\cap SL^3
$.
A straightforward calculation (see Exercise~\ref{rtsl}) shows that $\G$ is given by
\be\label{rtsl0}
\G=\left\{\bpm s_1 & 0\\ 0 & s_1^{-1}\epm\otimes \bpm s_2 & 0\\ 0 & s_2^{-1}\epm\otimes \bpm s_3 & 0\\ 0 & s_3^{-1}\epm\;:\;s_1s_2s_3=1\;,\;\;s_1,s_2,s_3\in\mbb{C}\right\}
\ee
Now, recall that in three qubits, the 3-tangle is defined in terms of an homogeneous SLIP of degree 4. Therefore,
from Corollary~\ref{cor:cyclic} we get that the quotient group $\stab(\phi)/\G$ is a group of order at most four. However, note that if $\Lambda\in SL^3$ then also $-\Lambda\in SL^3$ so that $\G_4=\G_2$, where the groups $\G_2$ and $\G_4$ are defined in~\eqref{g2g4}. We therefore conclude that 
$\stab(\phi)/\G$ is a group of order at most two. Since $X\otimes X\otimes X\in\stab(\phi)/\G$ we conclude that $\stab(\phi)/\G$ contains only the identity matrix and the flip matrix $X\otimes X\otimes X$ so that $\stab(\phi)$ is the union of $\G$ and the coset $(X\otimes X\otimes X)\G$.

\bex\label{rtsl}
Prove~\eqref{rtsl0} by direct calculation.
\eex

\bex
Find the stabilizer group of the W-state of three qubits. Is it compact?
\eex

\subsubsection{Example 2: The Stabilizer Group of a Generic State in  Four Qubits}

As a second example, let $\psi$ by the four-qubit state
\be\label{lpsi}
|\psi\ra=\lambda_1|\Phi_+\ra|\Phi_+\ra+
\lambda_2|\Phi_-\ra|\Phi_-\ra+\lambda_3|\Psi_+\ra|\Psi_+\ra+\lambda_4|\Psi_{-}\ra|\Psi_{-}\ra\;,
\ee
where $\{|\Phi_{\pm}\ra,|\Psi_{\pm}\ra\}$ is the Bell basis\index{Bell basis} of two qubits, $\lambda_1,\lambda_2,\lambda_3,\lambda_4\in\mbb{C}$, and $\lambda_x^2\neq\lambda_{x'}^2$ for all $x\neq x'\in[4]$.\index{Bell basis} For this four-qubit state, it can be shown (see~\cite{GW2010} and~\cite{Wallach2004}) that the group $\G=\stab(\psi)\cap SL^4$ is the Klein group consisting of only four elements:
\be\label{klein}
\G=\{I\;,\;X\otimes X\otimes X\otimes X\;,\;Y\otimes Y\otimes Y\otimes Y\;,\;Z\otimes Z\otimes Z\otimes Z\}
\ee
where $X,Y,Z$ are the three Pauli\index{Pauli} matrices.

Now, recall the homogeneous SLIP of degree 2 given in~\eqref{rellam}. This polynomial is non-zero if
\be\label{lambdas}
\lambda_1^2+\lambda_2^2+\lambda_3^2+\lambda_4^2\neq 0\;.
\ee
Therefore, if the coefficients of $\psi$ in~\eqref{lpsi} satisfy the relation~\eqref{lambdas} then from Corollary~\ref{cor:cyclic} there is no difference between $\G$ and $\stab(\psi)$ so we get the stabilizer of $\psi$ is the Klein group given in~\eqref{klein}.

\bex
Consider the 4-qubit state:
\be
|\psi\ra\eqdef\frac1{\sqrt{3}}\Big(|\Phi_+\ra|\Phi_+\ra+
\omega|\Phi_-\ra|\Phi_-\ra+\overline{\omega}|\Psi_+\ra|\Psi_+\ra\Big)\;,
\ee
where $\omega=e^{i\frac{2\pi}{3}}$.
\ben
\item Show that $\G\neq\stab(\psi)$.
\item Find all the elements of $\stab(\psi)$.
\een
\eex

\subsection{Generalization of Nielsen Majorization Theorem}

In this subsection, we extend Theorem~\ref{nielsen} to the multipartite scenario by defining a positive semidefinite operator for each stable state in $A^n$. Recall that Theorem~\ref{knt} establishes that any stable state $\psi\in\pure(A^n)$ is related to some critical state $\chi\in\pure(A^n)$ by an $SL^n$-orbit. Therefore, there exist $\theta\in[0,2\pi)$ and $M\in SL^n$ such that
\be
|\psi\ra=e^{i\theta}\frac{M|\chi\ra}{\|M|\chi\ra\|}\;.
\ee
We say that $\Lambda\in\pos(A^n)$ is an associated positive (semidefinite) operator (APO) of $\psi$ if $\Lambda$ can be expressed as
\be
\Lambda\eqdef\frac{M^*M}{\la\chi |M^*M|\chi\ra}\;.
\ee
As demonstrated by the following exercise, the APO of the state $\psi$ above is not unique.
\bex
Show that if $S\in\stab(\chi)$ then both $\Lambda$ and $S^*\Lambda S$ are APOs of $\psi$.
\eex

\bex\label{ex0rdm}
Consider a bipartite state $\psi\in\pure(AB)$ with $|A|=|B|$. Show that one of its APOs is given by $\rho^A\otimes I^B$, where $\rho^A\eqdef\tr_B\left[\psi^{AB}\right]$.
\eex

\begin{myt}{}
\begin{theorem}\label{intercm}
Let $\psi_1,\psi_2\in\pure(A^n)$ be two multipartite states in the same reversible SLOCC class of $\chi\in\pure(A^n)$. Let $\Lambda_1$ and $\Lambda_2$ be any APOs corresponding to $\psi_1$ and $\psi_2$, respectively. Then, $\psi_1\xrightarrow{\text{\tiny SEP}_1} \psi_2$ if and only if there exists a probability distribution $\{p_x\}_{x\in[m]}$ along with operators $\{S_x\}_{x\in[m]}\subset\stab(\chi)$ such that
\be
\Lambda_1=\sum_{x\in[m]}p_xS_x^*\Lambda_2 S_x\;.
\ee
\end{theorem}
\end{myt}
\begin{proof}
Let  $N_1,N_2\in {SL^n}$ be such that for $j=1,2$
\be\label{psiji}
|\psi_j\ra=e^{i\theta_j}\frac{N_j|\chi\ra}{\|N_j|\chi\ra\|}\;,
\ee
and denote by
\be\label{l1l2}
\Lambda_1=\frac{N_1^*N_1}{\la\chi |N_1^*N_1|\chi\ra}\quad\text{and}\quad\Lambda_2=\frac{N_2^*N_2}{\la\chi |N_2^*N_2|\chi\ra}\;.
\ee
Now, $\psi_1\xrightarrow{\text{\tiny SEP}_1} \psi_2$ if and only if there exists  $M_x\in GL^n$ satisfying $\sum_xM_x^*M_x=I^{A^n}$ such that 
\be
M_x|\psi_1\ra=c_x|\psi_2\ra
\ee
for some $c_x\in\mbb{C}$. After combining the relation above with~\eqref{psiji}, and performing some algebra, we obtain:
\be
\frac1{c_x}\frac{\|N_2|\chi\ra\|}{\|N_1|\chi\ra\|}N^{-1}_2M_xN_1|\chi\ra=|\chi\ra\;.
\ee
Thus, for each $x$ we have
\be
\frac1{c_x}\frac{\|N_2|\chi\ra\|}{\|N_1|\chi\ra\|}N^{-1}_2M_xN_1=S_x\;,
\ee
where $S_x\in\stab(\chi)$. Isolating $M_x$ gives
\be
M_x=c_x\frac{\|N_1|\chi\ra\|}{\|N_2|\chi\ra\|}N_2S_xN_1^{-1}\;.
\ee
Using the notations~\eqref{l1l2}, it can be verified easily that $\sum_x M_x^*M_x=I^{A^n}$ if and only if
\be
\Lambda_1=\sum_xp_xS_x^*\Lambda_2 S_x\;,
\ee
with $p_x\eqdef|c_x|^2$. Hence, $\psi_1\xrightarrow{\text{\tiny SEP}_1} \psi_2$ if and only if $\Lambda_1$ and $\Lambda_2$ satisfy the relation above. This completes the proof.
\end{proof}

Let $\chi\in{\rm Crit}(A^n)$ have a finite stabilizer, i.e., $\stab(\chi)={U_{x}}_{x\in[m]}$ is a finite set of unitaries, as established by Corollary~\ref{cor:0051}. In this situation, we can define the $\stab(\chi)$-twirling operation as:
\be
\mG\left(\omega^{A^n}\right)=\frac1m\sum_{x\in[m]}U_x\omega^{A^n}U_x^*\quad\quad\forall\;\omega\in\ml(A^n)\;.
\ee 
Now, according to Theorem~\ref{intercm} $\psi_1\xrightarrow{\text{\tiny SEP}_1} \psi_2$ if and only if there exists a probability distribution $\{p_x\}_{x\in[m]}$ such that
\be\label{pxua}
\Lambda_1=\sum_{x\in[m]}p_xU_x^*\Lambda_2 U_x\;.
\ee
By taking the twirling map $\mG$ on both sides of the equation above we get that if $\psi_1\xrightarrow{\text{\tiny SEP}_1} \psi_2$ then
\be
\mG\left(\Lambda_1\right)=\mG\left(\Lambda_2\right)\;.
\ee
In other words, the condition above is a necessary condition for the conversion $\psi_1\xrightarrow{\text{\tiny SEP}_1} \psi_2$ (but not always sufficient). Moreover, if $\Lambda_1$ is symmetric, meaning $\mG(\Lambda_1)=\Lambda_1$, then $\psi_1\xrightarrow{\text{\tiny SEP}_1} \psi_2$ if and only if 
$
\Lambda_1=\mG\left(\Lambda_2\right)
$.

One special case that $\Lambda_1$ is symmetric is the case that $\psi_1=\chi$. In this case, $\Lambda_1=I^{A^n}$ and $\chi\xrightarrow{\text{\tiny SEP}_1} \psi_2$ if and only if $\mG\left(\Lambda_2\right)=I^{A^n}$. Conversely, if $\psi_2=\chi$ then $\Lambda_2=I^{A^n}$, so the condition~\eqref{pxua} becomes $\Lambda_1=I^{A^n}$. In other words, $\psi_1\xrightarrow{\text{\tiny SEP}_1} \chi$ if and only if up to local unitaries $\psi_1=\chi$. This is consistent with the intuition that the critical state is the maximally entangled state of the SLOCC orbit.

To demonstrate how the theorem mentioned above generalizes Nielsen's majorization theorem\index{Nielsen's theorem}, we now apply it to the bipartite case. Let us consider a bipartite system\index{bipartite system} $AB$ with $m\eqdef|A|=|B|$. The only critical state of this system is the maximally entangled state $\Phi_m$, and its stabilizer is given by:
\be
\stab(\Phi_m)\eqdef\left\{S^{-1}\otimes S^T\;:\;S\in GL\left(m\right)\right\}\;.
\ee
The stabilizer group mentioned above is clearly not compact. However, in the derivation of Nielsen's majorization theorem, we employed Lo-Popescu's Theorem\index{Lo-Popescu's theorem} (Theorem~\ref{lopop}) which limits Bob's operations to be unitary operations. Thus, we can limit $S$ to be a unitary matrix without loss of generality.

Consider two bipartite states $\psi_1,\psi_2\in\pure(AB)$. As per Exercise~\ref{ex0rdm}, an APO of $\psi_1$ has the form $\Lambda_1=\rho_1^A\otimes I^B$, where $\rho_1^A$ is the reduced density matrix of $\psi^{AB}_1$. Similarly, an APO of $\psi_2$ has the form $\Lambda_2=\rho_2^A\otimes I^B$, where $\rho_2^A$ is the reduced density matrix of $\psi^{AB}_2$. Therefore, from Theorem~\ref{intercm}, we can infer that $\psi_1^{AB}\xrightarrow{\text{\tiny SEP}_1} \psi_2^{AB}$ if and only if there exists a probability distribution $\{p_x\}_{x\in[k]}$
along with $k$ unitary matrices $\{U_x\}_{x\in[k]}\subset U(m)$ such that:
\be
\rho_1^A\otimes I^B=\sum_{x\in[k]}p_xU_x^*\rho_2^AU_x\otimes I^B\;.
\ee
This condition is precisely the same as the condition we obtained in~\eqref{random} for Nielsen's majorization criterion.

\bex
Let $\psi$ be the 4-qubit state~\eqref{lpsi} and suppose it satisfies~\eqref{lambdas}. Classify all the 4-qubit states $\phi$ for which $\psi\xrightarrow{\text{\tiny SEP}_1}\phi$ holds.
\eex

\section{Entanglement of Assistance}\index{entanglement of assistance}

In the previous sections, we have seen that local operations and classical communication (LOCC) alone are limited in their ability to manipulate multipartite entanglement. For instance, the classification of four-qubit entanglement involves infinitely many SLOCC classes, highlighting the complexity of entanglement manipulation under LOCC for generic multipartite entangled pure states in $\pure(A^n)$, where $n\geq 4$. More generally, it has been shown that if two generic multipartite entangled pure states $\psi$ and $\phi$ cannot be locally converted into each other by a unitary operation, then the conversion cannot be achieved by LOCC as well. However, this situation can change significantly if the target state $\phi$ is not generic.

While bipartite entanglement is the most useful form of entanglement in quantum information, the concentration of multipartite systems into two parties has also been extensively studied. In this section, we focus on the conversion of a tripartite pure state into an ensemble of bipartite states. We begin by considering the following conversion of a tripartite state $\psi\in\pure(ABR)$ into an ensemble of pure states, or equivalently, a cq-state\index{cq-state} of the form
\be
\sigma^{ABX}=\sum_{x\in[m]}p_x\phi_x^{AB}\otimes |x\lr x|^X\;,
\ee
where $\p\eqdef(p_1,...,p_m)^T\in\prob(m)$, and each $\phi_x\in\pure(AB)$.

\begin{myg}{}
\begin{lemma}\label{lem:steer}
Let $\psi\in\pure(AB)$ and let $\rho^{B}\eqdef\tr_
A\left[\psi^{AB}\right]$ be its reduced density matrix on system $B$. Then, for every pure state decomposition $\rho^B=\sum_{x\in[m]}p_x\phi_x^B$, there exists a POVM on Alices system, $\{\Lambda_x\}_{x\in[m]}\subset\eff(A)$, such that for all $x\in[m]$
\be\label{14130}
\phi^{B}_x=\frac1{p_x}\tr_A\left[\left(\Lambda_x^A\otimes I^B\right)\psi^{AB}\right]\quad\text{and}\quad p_x=\tr\left[\left(\Lambda_x^A\otimes I^B\right)\psi^{AB}\right]\;.
\ee
That is, every pure state decomposition of $\rho^B$ can be realized by a generalized measurement on Alice's system.
\end{lemma}
\end{myg}

\begin{proof}
Consider the pure state 
\be
|\tpsi^{RB}\ra\eqdef\sum_{x\in[n]}\sqrt{p_x}|x\ra^R|\phi_x^B\ra\;.
\ee
Since both $\psi^{AB}$ and $\tpsi^{RB}$ are purifications of $\rho^B$ there exists an isometry $V:A\to R$ such that $|\tpsi^{RB}\ra=V\otimes I^B|\psi^{AB}\ra$. Therefore, taking
\be
\Lambda_x^A\eqdef V^*|x\lr x|^RV\quad\quad\forall\;x\in[m]\;,
\ee
gives
\be
\tr_A\left[\left(\Lambda_x^A\otimes I^B\right)\psi^{AB}\right]=\tr_R\left[\left(|x\lr x|^R\otimes I^B\right)\tpsi^{AB}\right]=\sqrt{p_x}\phi_x^B\;.
\ee
Therefore, with this choice of the POVM $\{\Lambda_x^A\}_{x\in[m]}$ we get~\eqref{14130}. This completes the proof.
\end{proof}

Let $\psi\in\pure(ABR)$ be a tripartite pure state with marginal $\rho^{AB}\eqdef\tr_R\left[\psi^{ABR}\right]$. From Lemma~\ref{lem:steer}, we know that every pure state decomposition of $\rho^{AB}$ can be realized by a generalized measurement on the reference system $R$. Thus, for a given measure of bipartite entanglement $E$, we define the \emph{entanglement of assistance} to be
\be
E_a\left(\rho^{AB_1}\right)\eqdef \sup\sum_{x\in[m]}p_xE\left(\phi_x^{AB_1}\right)\;,
\ee
where the supremum is taken over all pure state decompositions of $\rho^{AB}=\sum_{x\in[m]}p_x\phi_x^{AB}$. Note that this definition is similar to the definition of the entanglement of formation given in \eqref{1331}, except that we take the supremum instead of the infimum as taken in \eqref{1331}.

\bex\label{exeoan}
Compute the entanglement of assistance of the maximally mixed state $\u^{AB}$ and conclude that the entanglement of assistance is not a measure of entanglement.
\eex

\bex
Let $E$ be a measure of pure bipartite entanglement, and let $E_F$ and $E_a$ be its corresponding entanglement of formation and assistance respectively. Let $\psi\in\pure(AB_1B_2)$ be a tripartite pure state with marginal $\rho^{AB_1}\eqdef\tr_{B_2}\left[\psi^{AB_1B_2}\right]$. Show that if $\psi^{AB_1B_2}$ satisfies the disentangling condition~\eqref{dec} then
\be
E_F\left(\rho^{AB}\right)=E_a\left(\rho^{AB}\right)\;.
\ee
\eex

Exercise~\ref{exeoan} demonstrates that entanglement of assistance is not a measure of entanglement as a function of the bipartite state on system $AB$. However, one may wonder if the entanglement of assistance is a measure of tripartite pure entanglement. Specifically, for any $\psi\in\pure(ABR)$, we define
\be
\tilde{E}_a\left(\psi^{ABR}\right)\eqdef E_a\left(\rho^{AB}\right)\;,
\ee
where $\rho^{AB}\eqdef\tr_R\left[\psi^{ABR}\right]$. One can then ask whether it is impossible to increase $\tilde{E}_a$ by LOCC. In other words, can an LOCC prior to the measurement on the reference system increase the entanglement of assistance? Surprisingly, we now show that such a pre-LOCC can increase the entanglement of assistance, and therefore $\tilde{E}_a$ is also not an entanglement measure.

Let $A=A_1A_2$ with $|A_1|=2$, $|A_2|=|B|=4$, and $|R|=2$, and let $\psi\in\pure(ABR)$ be the state
\be\label{notem}
|\psi^{ABR}\ra\eqdef\frac12\left(|0\ra^{A_1}|\Phi^{A_2B}_0\ra|0\ra^R+|0\ra^{A_1}|\Phi^{A_2B}_1\ra|1\ra^R+|1\ra^{A_1}|\Phi^{A_2B}_2\ra|+\ra^R+|1\ra^{A_1}|\Phi^{A_2B}_3\ra|-\ra^R\right)\;,
\ee 
where $\{\Phi_x^{A_2B}\}_{x\in\{0,1,2,3\}}$ are four maximally entangled states to be determined shortly.
Consider a protocol where Alice measures system $A_1$ in the ${|0\ra,|1\ra}$ basis and sends the classical outcome to the referee. Based on the outcome, the referee performs either the measurement in the ${|0\ra^R,|1\ra^R}$ basis (if Alice's outcome is zero) or the measurement in the ${|+\ra^R,|-\ra^R}$ basis (if Alice's outcome is one). Regardless of the measurement outcomes of both Alice and the referee, Alice and Bob end up with one of the four maximally entangled states $\{\Phi_x^{A_2B}\}_{x\in{0,1,2,3}}$. In particular, by local unitary operation, Alice and Bob can transform any of these four states into the maximally entangled state $\Phi^{A_2B}$. Thus, we conclude that
\be
\psi^{ABR}\xrightarrow{\text{\tiny LOCC}} \Phi^{A_2B}\;.
\ee
However, we now show that for certain choices of maximally entangled states $\{\Phi_x^{A_2B}\}_{x\in{0,1,2,3}}$, the transformation above cannot be achieved (even with probability less than one) if we only allow system $R$ to perform a measurement.

The reduced density matrix $\rho^{AB}\eqdef\tr_R\left[\psi^{ABR}\right]$ of the state above can be expressed as
\be
\rho^{AB}=\varphi_0^{AB}+\varphi^{AB}_1
\ee
where
\be
|\varphi_0^{AB}\ra\eqdef\frac12|0\ra^{A_1}|\Phi^{A_2B}_0\ra+\frac1{2\sqrt{2}}|1\ra^{A_1}\left(|\Phi^{A_2B}_2\ra+|\Phi^{A_2B}_3\ra\right)
\ee
and
\be
|\varphi_1^{AB}\ra\eqdef\frac12|0\ra^{A_1}|\Phi^{A_2B}_1\ra+\frac1{2\sqrt{2}}|1\ra^{A_1}\left(|\Phi^{A_2B}_2\ra-|\Phi^{A_2B}_3\ra\right)
\ee
We argue that there exists four maximally entangled states $\{\Phi_x^{A_2B}\}_{x\in\{0,1,2,3\}}$ such that any linear combination of $|\varphi_0^{AB}\ra$ and $|\varphi_1^{AB}\ra$ is not maximally entangled.
Indeed, take 
\be
|\Phi^{A_2B}_0\ra=|\Phi^{A_2B}_2\ra=\frac12(|00\ra+|11\ra+|22\ra+|33\ra)
\ee
and
\ba
&|\Phi^{A_2B}_1\ra=\frac12(|00\ra-i|11\ra-|22\ra+i|33\ra)\\
&|\Phi^{A_2B}_3\ra=\frac12(|00\ra-i|11\ra+|22\ra-i|33\ra)\;.
\ea
With these choices we get by direct calculation that for any $a,b\in\mbb{C}$ the linear combination
\be\label{avarphi}
a|\varphi_0^{AB}\ra+b|\varphi^{AB}_1\ra
\ee
is not proportional to the maximally entangled state (see Exercise~\ref{checkassi}). Therefore, the state $\psi^{ABR}$ cannot be converted to $\Phi^{AB}$ (even with probability less than one) by a local measurement on system $R$.
Alternatively, none of the pure-state decompositions of $\rho^{AB}$ contains a maximally entangled state (i.e. 2-ebits).

\bex\label{checkassi}
Show that for any choice of $a,b\in\mbb{C}$ with $|a|^2+|b|^2=1$ the state in~\eqref{avarphi} is not maximally entangled. Hint: Write the state in~\eqref{avarphi} as a linear combination $\sum_{x=0}^3|\phi_x^{A}\ra|x\ra^B$ and show that the vectors $\{|\phi_x^A\ra\}_{x}$ cannot  all have the same norm and also orthogonal to each other.
\eex

\subsection{Entanglement of Collaboration}

Since pre-LOCC operations performed by Alice and Bob can increase the entanglement of assistance, we can modify the definition of entanglement of assistance by including arbitrary LOCC operations performed by all parties. This modification results in a new measure called the ``entanglement of collaboration", which more comprehensively quantifies the amount of bipartite entanglement available for collaborative tasks among all parties.

\begin{myd}{Entanglement of Collaboration}\index{entanglement of collaboration}
\begin{definition}
Let $E$ be a measure of bipartite entanglement for mixed states. Its corresponding measure of tripartite entanglement, known as the "entanglement of collaboration" and denoted by $E_c$, is defined as:
\be
E_c\left(\rho^{ABR}\right)\eqdef\sup_{\mN\in\locc}E\left(\mN^{ABR\to A'B'}\left(\rho^{ABR}\right)\right)\quad\quad\forall\;\rho\in\md(ABR)\;,
\ee
where the supremum is over all quantum systems $A'$ and $B'$, and all LOCC channels $\mN^{ABR\to A'B'}$.
\end{definition}
\end{myd}

In other words, the entanglement of collaboration is the maximal bipartite entanglement (as measured by $E$) that can be shared between Alice and Bob after all parties collaborate via LOCC. Observe that the entanglement of collaboration is never smaller than the entanglement of assistance. Indeed, suppose $\rho^{ABR}=\psi^{ABR}$ is a pure state and take $A'=A$, $B'=BX$ where $X$ is a classical system corresponding to the outcome of a POVM performed on system $R$. Then, taking the LOCC channel $\mN^{ABR\to ABX}=\id^{AB}\otimes\mE^{R\to X}$, where $\mE\in\cptp(R\to X)$ is a POVM Channel\index{POVM channel} , we get that
  \be
  \mN^{ABR\to ABX}\left(\psi^{ABR}\right)=\sigma^{ABX}\eqdef\sum_{x\in[m]}p_x\phi^{AB}_x\otimes |x\lr x|^X
  \ee
  is a cq-state, where $\{p_x,\phi_x^{AB}\}_{x\in[m]}$ is one of the pure-state decompositions of $\rho^{AB}$. Therefore, for this choice of $\mN^{ABR\to A'B'}$ we get
  \ba
  E\left(\mN^{ABR\to A'B'}\left(\rho^{ABR}\right)\right)&=E\left(\sigma^{ABX}\right) \\ 
  \GG{Assuming\;{\it E}\;is\;convex\;linear~\eqref{0107}}&=\sum_{x\in[m]}p_xE(\phi^{AB}_x)\;.
  \ea
Therefore, since entanglement of collaboration is defined as a supremum over all such LOCC channels $\mN^{ABR\to A'B'}$ it must be no smaller than the entanglement of assistance. Furthermore, unlike entanglement of assistance, entanglement of collaboration is a measure of tripartite entanglement.

\bex
Show that the entanglement of collaboration is a measure of tripartite entanglement.
\eex

\begin{myg}{}
\begin{lemma}
Let $E$ be a measure of bipartite entanglement. Then, for all $\rho\in\md(ABR)$
\be
E_c\left(\rho^{ABR}\right)\leq\min\Big\{E\left(\rho^{A(BR)}\right),E\left(\rho^{B(AR)}\right)\Big\}
\ee
where the parenthesis in $\rho^{A(BR)}$ indicates that the entanglement is computed between system $A$ and the composite system\index{composite system} $BR$.
\end{lemma}
\end{myg}

\bex
Prove the lemma above.
\eex

One of the most fundamental questions in entanglement theory is the distillation of Bell states from multiple copies of a bipartite entangled state. As we saw earlier, for a given pure state $\psi^{AB}$, the distillable entanglement is determined by the von-Neumann entropy of the reduced density matrix $\psi^A$. A similar question arises in the multipartite regime: given many copies of a tripartite pure entangled state $\psi^{ABR}$, how many Bell states can be distilled between Alice and Bob by LOCC of all three parties sharing the state?

The above lemma asserts that the optimal distillation rate cannot exceed the minimum between the entropies of system $A$ and system $B$. Remarkably, it has been shown that this upper bound can be attained. However, we will present the proof of this statement in volume 2 of this book, after introducing the quantum state merging protocol from quantum Shannon theory. 

\section{Monogamy of Entanglement}\index{monogamy of entanglement}

Monogamy of entanglement is a fundamental property of multipartite quantum systems, whereby the amount of entanglement between two subsystems limits the amount of entanglement that each subsystem can have with the rest of the system. Unlike classical correlation, entanglement cannot be freely shared among multiple parties, making this principle a fundamental concept in quantum information theory. For instance, if Alice's system is maximally entangled with Bob's system, it cannot be simultaneously entangled with another third system. That is, sharing entanglement is limited. 

This principle has far-reaching implications in various areas, including quantum communication and cryptography.  Overall, monogamy of entanglement is a fundamental property of quantum systems with significant implications in both theoretical and practical aspects of quantum information theory.

\subsection{Quantification of Monogamy of Entanglement}

We can quantify the phenomenon of monogamy of entanglement as follows. Let $B=B_1B_2$ be a composite system, and let $E$ be a measure of entanglement. We say that $E$ is monogamous if for all $\rho\in\md(AB)$
\be\label{monogamy}
E\left(\rho^{AB}\right)\geq E\left(\rho^{AB_1}\right)+E\left(\rho^{AB_2}\right)\;.
\ee 
If $E$ is monogamous and $A$ and system $B_1$ are maximally entangled, then we must have $E\left(\rho^{AB}\right)=E\left(\rho^{AB_1}\right)$. From the inequality above, we can conclude that $E\left(\rho^{AB_2}\right)=0$, meaning that $B_2$ has no entanglement with $A$.

It is worth noting that not all measures of entanglement satisfy the inequality above, and it is not immediately clear that monogamous measures of entanglement exist. Therefore, to demonstrate the monogamy of entanglement, we must first show that there are measures of entanglement that satisfy \eqref{monogamy}.

\begin{myt}{}
\begin{theorem}
The squashed entanglement\index{squashed entanglement} is a monogamous measure of entanglement satisfying~\eqref{monogamy}.
\end{theorem}
\end{myt}

\begin{proof}
Let $\rho^{ABR}$ be an extension of the state $\rho^{AB}$. Using the chain rule of the conditional mutual information\index{mutual information} (see the second equality in~\eqref{chainrule}) we get
\ba
\frac12I(A:B|R)_\rho&=\frac12I(A:B_1|R)_\rho+\frac12I(A:B_2|RB_1)_\rho\\
\GG{By\;definition}&\geq E_{\rm sq}\left(\rho^{AB_1}\right)+E_{\rm sq}\left(\rho^{AB_2}\right)\;.
\ea
Since the above inequality holds for all extensions $\rho^{ABR}$ of $\rho^{AB}$ we conclude that 
\be
E_{\rm sq}\left(\rho^{AB}\right)\geq E_{\rm sq}\left(\rho^{AB_1}\right)+E_{\rm sq}\left(\rho^{AB_2}\right)\;.
\ee
This completes the proof.
\end{proof}

It is important to recognize that not all measures of entanglement adhere to the monogamy condition specified by \eqref{monogamy}. Nevertheless, the squashed entanglement\index{squashed entanglement} is currently the only known measure of entanglement that satisfies \eqref{monogamy} in all finite dimensions, which is a remarkable property that highlights the unique nature of this measure. Other measures satisfy~\eqref{monogamy} on fixed dimensions. For example, on qubit systems, the square of the concurrence\index{concurrence} is also a monogamous measure of entanglement that satisfies~\eqref{monogamy} when $|A|=|B_1|=|B_2|=2$.

\subsubsection{Qubit Monogamy Relations}

For qubit systems one can use the concurrence\index{concurrence} to quantify entanglement. Let $\psi\in\pure(ABC)$ be a 3-qubit pure state; that is, $|A|=|B|=|C|=2$. In this case, from~\eqref{13p72} and the fact that $\rho^{AB}$ is at most rank 2, the concurrence of assistance (as defined in Exercise~\ref{ex:coa}; see~\eqref{coa1} and~\eqref{coa2}) can be expressed 
\ba\label{directc}
C_a\left(\rho^{AB}\right)^2&=\tr\left[\rho^{AB}\rho^{AB}_\star\right]\\
\GG{Exercise~\ref{ex:ckw}}&=2\left(\det\left(\rho^A\right)+\det\left(\rho^B\right)-\det\left(\rho^C\right)\right)\;.
\ea
\bex\label{ex:ckw}
Use a direct calculation to prove the equality in~\eqref{directc}.
\eex

The equality above implies that for the pure state $\psi^{ABC}$ with marinals $\rho^{AB}$ and $\rho^{AC}$ we have
\be
C_a\left(\rho^{AB}\right)^2+C_a\left(\rho^{AC}\right)^2=4\det\left(\rho^A\right)=C\left(\psi^{A(BC)}\right)^2\;.
\ee
Denoting by $\tau\left(\rho^{AB}\right)\eqdef C\left(\rho^{AB}\right)^2$ (the square of the concurrence of formation, also know as the $2$-tangle) and using the fact that $C\left(\rho^{AB}\right)\leq C_a\left(\rho^{AB}\right)$ we arrive at the following monogamy inequality
\be\label{ckw}
\tau\left(\psi^{A(BC)}\right)\geq\tau\left(\rho^{AB}\right)+\tau\left(\rho^{AC}\right)\;,
\ee
where
\be
\tau\left(\psi^{A(BC)}\right)\eqdef 2\left(1-\tr\left[\left(\rho^A\right)^2\right]\right)=4\det\left(\rho^A\right)\;.
\ee
The relation~\eqref{ckw} is widely known in the literature as the Coffman-Kundu-Wootter (CKW) monogamy relation. This seminal result introduced the concept of monogamy of entanglement for the first time, demonstrating that the amount of entanglement between two subsystems is limited by the amount of entanglement between each subsystem and a third subsystem. 
\bex
Show that 
\be
\tau\left(\psi^{A(BC)}\right)-\tau\left(\rho^{AB}\right)-\tau\left(\rho^{AC}\right)={\rm Tangle}\left(\psi^{ABC}\right)
\ee
where the right-hand side is the 3-tangle defined in~\eqref{tan4slip}.
\eex

\subsection{Monogamy Without Inequalities}

The definition of a monogamous measure of entanglement, as provided in equation ~\eqref{monogamy}, only captures a partial aspect of monogamy of entanglement. This is due to the fact that many important measures of entanglement do not satisfy this relation, and some of them are not even additive under tensor product. Therefore, the summation in the right-hand side of \eqref{monogamy} is only a convenient choice and not a necessity.
For instance, it is well-known that if $E$ does not satisfy this relation, it is still possible to find a positive exponent $\alpha>0$ such that the function $E^\alpha$ satisfies the relation. This is already evident in the CKW relation, where $E$ is taken to be the square of the concurrence\index{concurrence} instead of the concurrence itself.

Moreover, there exist measures of entanglement that are multiplicative under tensor product, which implies that they fail to satisfy \eqref{monogamy} or any power of it. This highlights the necessity for a more nuanced and refined definition of monogamy of entanglement that encompasses these peculiarities. This becomes especially important when dealing with the range of measures of entanglement available, which exhibit varying properties.

One approach to address the issues with the current definition of a monogamous measure of entanglement is to replace the relation given in~\eqref{monogamy} with a family of monogamy relations of the form:
\be\label{monogamy2a}
E\left(\rho^{AB}\right)\geq f\Big(E\left(\rho^{AB_1}\right),E\left(\rho^{AB_2}\right)\Big)\;,
\ee 
where $f$ is a function of two variables that satisfies certain conditions. While this family of monogamy relations may be more flexible than the original definition, it still lacks a clear theoretical foundation. Thus, a more desirable solution would be to derive the monogamy relations from more basic principles, which would provide a deeper understanding of the nature of this phenomenon.

Recently, such approach to monogamy of entanglement has been proposed, which is more ``fine-grained" in nature and avoids the need for introducing a function $f$. This approach does not involve monogamy relations such as \eqref{monogamy} or \eqref{monogamy2a}. Instead, it defines a measure of entanglement $E$ to be monogamous if it satisfies a certain condition that does not involve inequalities.
In particular,
this approach takes into account the fact that different measures of entanglement have varying properties and limitations, rather than attempting to impose a one-size-fits-all definition.
By adopting this more nuanced approach, we can gain a deeper understanding of the monogamy of entanglement and how it manifests itself across different measures.

\begin{myd}{}
\begin{definition}\label{main00}
Let $B=B_1B_2$ denote a composite system, and let $E$ be a measure of bipartite entanglement.  $E$ is said to be \emph{monogamous} if for any bipartite state $\rho\in\md(AB)$ that satisfies
\be\label{dec}
E\left(\rho^{AB}\right)=E\left(\rho^{AB_1}\right)
\ee
we have that $E\left(\rho^{AB_2}\right)=0$.
\end{definition}
\end{myd}

The condition expressed in equation~\eqref{dec} is considered to be a strong one, and is usually not met by most states within $\md(AB)$. It is often referred to as the ``disentangling condition". As we will explore further below, quantum Markov states (defined in Definition~\ref{markov}) are always guaranteed to satisfy this equality for any entanglement monotone $E$. Additionally, the condition presented in equation~\eqref{monogamy} is even stronger than the one defined in Definition~\ref{main00}. Specifically, if $E$ satisfies~\eqref{monogamy}, then any $\rho^{AB}$ that satisfies~\eqref{dec} must have $E(\rho^{AB_2})=0$. Nonetheless, Definition~\ref{main00} still captures the essence of monogamy, by stipulating that if system $A$ shares the maximum amount of entanglement with subsystem $B_1$, it is left with no entanglement to share with $B_2$.

In Definition~\ref{main00} we do not invoke a particular monogamy relation such as~\eqref{monogamy}.
Instead, we propose a minimalist approach which is not quantitative, in which we only require what 
is essential from a measure of entanglement to be monogamous. Yet, this requirement is sufficient 
to generate a more quantitative monogamy relation as demonstrated in the following theorem.

\begin{myt}{}
\begin{theorem}
Let $E$ be a continuous measure of entanglement. Then, $E$ is monogamous 
according to Definition~\ref{main00} if and only if for every $d\in\mbb{N}$, and every systems $A$ and $B=B_1B_2$ with $|AB|=d$,  there exists
$0<\alpha<\infty$ such that 
	\be\label{power}
	E^\alpha(\rho^{AB})\geq E^\alpha(\rho^{AB_1})+E^\alpha(\rho^{AB_2})\quad\quad\forall \;\rho\in\md(AB)\;.	
	\ee
\end{theorem}
\end{myt}

\begin{proof}
We leave it as an exercise to show that if $E$ satisfies~\eqref{power} then $E$ is a monogamous measure of entanglement (according to Definition~\ref{main00}). We therefore assume now that $E$ is monogamous and prove the relation~\eqref{power}.
	Since $E$ is a measure of entanglement,
	it is non-increasing under partial traces, and therefore for all $\rho\in\md(AB)$
	\be\label{eqws}
	E(\rho^{AB})\geq \max\{E(\rho^{AB_1}),E(\rho^{AB_2})\}\;.
	\ee 
	Without loss of generality we assume that $E(\rho^{AB})>0$ and set 
\be
	x_1\eqdef \frac{E(\rho^{AB_1})}{E(\rho^{AB})}
	\quad\text{and}\quad x_2\eqdef \frac{E(\rho^{AB_2})}{E(\rho^{AB})}\;.
\ee	
From~\eqref{eqws} we get that $x_1,x_2\in[0,1]$. Moreover, since $E$ is monogamous we get that if $x_1=1$ then we must have $x_2=0$ and vice versa. Therefore,
	there exists $\mu>0$ such that
	\be\label{equality}
	x_1^\mu+x_2^\mu\leq 1\;,
	\ee
	since either $x_j^\mu\rightarrow 0$ when $\mu$ increases, or if 
	$x_1=1$ then by assumption $x_2=0$ and similarly if $x_2=1$ then $x_1=0$. Let $f:\md(AB)\to\mbb{R}_+$ denote a function defined such that $f(\rho^{AB})$ represents the smallest value of $\mu$ that satisfies equality in~\eqref{equality}.
	Since $E$ is continuous, so is $f$, and the compactness of $\md(AB)$ gives:
	\be\label{optimal}
	\alpha\eqdef\max_{\rho\in\md(AB)}f(\rho^{AB})<\infty\;.
	\ee
	By definition, $\alpha$ satisfies the condition in~(\ref{power}).
\end{proof}

It is important to note that the relation given in~\eqref{power} is \emph{not} of the form
given in~\eqref{monogamy2a}, since the monogamy exponent $\alpha$ in~\eqref{power} 
depends on the dimension $d$, whereas $f$ is considered universal in the 
sense that it does not depend on the dimension. Therefore, if a measure of entanglement such 
as the entanglement of formation\index{entanglement of formation} is not monogamous according to the class of relations given 
in~\eqref{monogamy2a}, it does not necessarily mean that it is not monogamous according to Definition~\ref{main00}. 

In the next theorem we show that all quantum Markov states satisfy the disentangling condition. For this purpose, we will rename $B_1$ as $B$ and $B_2$ as $B'$, since the theorem involves further decomposition of system $B$ into subsystems. Specifically, an entangled Markov quantum state $\rho\in\md(ABB')$ is a state of the form (cf.~\eqref{markovs})
\be\label{qms}
\rho^{ABB'}=\bigoplus_{x\in[m]}p_x\rho_x^{AB_x^{(1)}}\otimes \rho_x^{B_x^{(2)}B'}
\ee
where 
\be
B=\bigoplus_{x\in[m]}B^{(1)}_x\otimes B_x^{(2)}\;,
\ee 
and for each $x\in[m]$, $\rho_x^{AB_x^{(1)}}$ and $\rho_x^{B_x^{(2)}B'}$ are density matrices in $\md(AB_x^{(1)})$ and $\md(AB_x^{(2)})$, respectively.

\begin{myt}{}
\begin{theorem}
Let $E$ be an entanglement monotone. Then, $E$ satisfies the disentangling 
condition $E\left(\rho^{ABB'}\right)=E\left(\rho^{AB}\right)$ (cf.~\eqref{dec}) for all quantum Markov states of the form~\eqref{qms}.
\end{theorem}
\end{myt}

\begin{proof}	
	Since local ancillary systems are free in entanglement theory, one can append 
	a classical ancillary system $X$ that encodes the orthogonality of the 
	subspaces $B_{x}^{(1)}\otimes B_{x}^{(2)}$. This can be done with an 
	isometry that maps states in $B_{x}^{(1)}\otimes B_{x}^{(2)}$ to states 
	in $B^{(1)}\otimes B^{(2)}\otimes|x\lr x|^{X}$, where systems $B^{(1)}$ and 
	$B^{(2)}$ have dimensions $\max_x\big|B_{x}^{(1)}\big|$ and  $\max_x\big|B_{x}^{(2)}\big|$, respectively.
Therefore, without loss of generality we can write the above Markov state as
\be\label{mainform00}
\sigma^{ABB'X}=\sum_{x\in[m]}p_x\;\rho^{AB^{(1)}}_{x}\otimes\rho^{B^{(2)}B'}_{x}\otimes |x\lr x|^X\;.
\ee
Now, note that with any entanglement monotone $E$, the entanglement between $A$ and $BB'$ is measured by
\ba
E\left(\rho^{ABB'}\right)&=E\left(\sigma^{ABB'X}\right)\\
\GG{\eqref{emon}}&=\sum_{x\in[m]}p_x E\left(\rho^{AB^{(1)}}_{x}\otimes\rho^{B^{(2)}B'}_{x}\right)\\
&=\sum_{x\in[m]}p_x E\left(\rho^{AB^{(1)}}_{x}\right)\;.
\ea
Similarly, the entanglement between $A$ and $B$ is measured by
\ba
E\left(\rho^{AB}\right)&=E\left(\sigma^{ABX}\right)\\
&=\sum_{x\in[m]}p_x E\left(\rho^{AB^{(1)}}_{x}\otimes\rho^{B^{(2)}}_{x}\right)\\
&=\sum_{x\in[m]}p_x E\left(\rho^{AB^{(1)}}_{x}\right)\;.
\ea
We therefore obtain $E\left(\rho^{ABB'}\right)=E\left(\rho^{AB}\right)$.
This completes the proof.
\end{proof}

%\bex
%Let $\rho^{ABB'}$ be a quantum Markov state as in~\eqref{qms} and let $\mE\in\cptp(B\to B'')$ be a local channel on Bob side. Consider the state $\sigma^{AB''B'}\eqdef\mE^{B\to B''}\left(\rho^{ABB'}\right)$. Show that for any entanglement monotone, $E$, we have $E\left(\sigma^{AB''B'}\right)=E\left(\sigma^{AB''}\right)$.
%\eex

The Markov state mentioned in the theorem has an important property: the marginal state $\rho^{AB'}$ is separable, which implies $E(\rho^{AB'})=0$. Therefore, Markov states always satisfy the condition given in Definition~\ref{main00}. However, one might question whether the converse of the statement in the theorem holds true. In other words, if a state $\rho^{ABB'}$ satisfies $E(\rho^{ABB'})=E(\rho^{AB})$, is it necessarily a Markov state? For mixed tripartite states, the answer is obviously ``no" because all separable states between system $A$ and $BB'$ satisfy $E(\rho^{ABB'})=E(\rho^{AB})$, but not all separable states are Markov states. Nonetheless, in the following theorem, we will see that under mild assumptions, the converse of the above theorem holds for pure tripartite states.

In Section~\ref{cre00}, we observed that every entanglement monotone takes the form~\eqref{gcsym} when evaluated on pure states. Specifically, the entanglement monotone $E$ can be expressed as follows:
\be\label{gcsym00}
E\left(\psi^{AB}\right)=g\left(\rho^A\right)\quad\text{with}\quad\rho^A\eqdef\tr_B\left[\psi^{AB}\right]\;,
\ee
where the function $g:\md(A)\to\mbb{R}_+$ is Schur concave. Furthermore, we noted that if $g$ is symmetric (i.e., invariant under unitary channels) and concave, then the convex roof extension of $E$ corresponds to an entanglement monotone. As a reminder, given any measure of entanglement $E$ on mixed states, we can construct its convex roof extension\index{convex roof extension} as 
\be\label{ooo1}
E_F\left(\rho^{AB}\right)\eqdef\min\sum_{x\in[m]}p_xE\left(\psi_x^{AB}\right)\quad\quad\forall\;\rho\in\md(AB)\;,
\ee
where the minimum is over all pure state decompositions of $\rho^{AB}=\sum_{x\in[m]}p_x\psi_x^{AB}$. Moreover, if $E$ is convex (e.g., entanglement monotone) then $E\left(\rho^{AB}\right)\leq E_F\left(\rho^{AB}\right)$ for all $\rho\in\md(AB)$.

\begin{myt}{}
\begin{theorem}\label{puretrimon}
Let $E$ be an entanglement monotone and $\psi\in\pure(ABB')$. Suppose $g$ as defined in~\eqref{gcsym00} is strictly concave. The following statements are equivalent:
\ben
\item $E\left(\psi^{ABB'}\right)=E\left(\rho^{AB}\right)$, where $\rho^{AB}\eqdef\tr_{B'}\left[\psi^{ABB'}\right]$.
\item There exists subsystems $B_1$ and $B_2$, $\chi\in\pure(AB_1)$, $\phi\in\pure(B_2B')$, and an isometry $U:B_1B_2\to B$ such that 
$
\psi^{ABB'}=U\left(\chi^{AB_1}\otimes\phi^{B_2B'}\right)U^*
$.
\een
\end{theorem}
\end{myt}
\begin{remark}
The above theorem states that if the pure state $\psi^{ABB'}$ satisfies the disentangling condition, then it is a Markov state (up to local unitary on system $B$). Additionally, keep in mind that since $E$ measures entanglement, the function $g$ defined in~\eqref{gcsym00} is invariant under unitary channels. As $g$ is also strictly concave, the convex roof extension of $E$ yields an entanglement monotone, as stated in Theorem~\ref{thm:em}. However, we don't assume $E$ to be equal to its convex roof extension. Instead, we observe that since $E$ is convex, it is always lower than or equal to its convex roof extension\index{convex roof extension}.
\end{remark}

\begin{proof}
We only prove the implication from the first statement to the second, as the converse is straightforward and left as an exercise for the reader. The first statement implies that
\be
E\left(\psi^{ABB'}\right)=E\left(\rho^{AB}\right)\leq E_F\left(\rho^{AB}\right)\;,
\ee
where we used the fact that $E$ is no greater than its convex roof extension\index{convex roof extension}.
On the other hand, from Lemma~\ref{lem:steer} we get that every pure-state decomposition\index{pure-state decomposition} of $\rho^{AB}=\sum_{x\in[m]}p_x\psi_x^{AB}$ has a corresponding measurement on system $B'$ of $\psi^{ABB'}$, where the outcome $x$ occurs with probability $p_x$ and the post-measurement state on system $AB$ is $\psi^{AB}_x$. When combined with the fact that $E$ is an entanglement monotone, this implies that
\be
E\left(\psi^{ABB'}\right)\geq\sum_{x\in[m]}p_xE\left(\psi_x^{AB}\right)\;.
\ee

The two equations above lead to the very strong conclusion that \emph{all} pure-state decompositions of $\rho^{AB}$ have the \emph{same} average entanglement, which equals $E\left(\psi^{ABB'}\right)$. In other words, the inequality in the above equation is actually an equality, and it holds for every pure-state decomposition\index{pure-state decomposition} $\{p_x,\psi_x^{AB}\}_{x\in[m]}$ of $\rho^{AB}$. This equality can be expressed in terms of the function $g$ as follows:
\be
g(\rho^A)=\sum_{x\in[m]}p_xg(\rho_x^A)\;,
\ee
where $\rho_x^A\eqdef\tr_{B}\left[\psi_x^{AB}\right]$.
Since $g$ is strictly concave, the equation above holds if and only if $\rho^A=\rho_x^A$ for all $x\in[m]$. Let $B_1$ be a system of dimension $r\eqdef|B_1|=\rank(\rho^A)$, and let $\chi\in\pure(AB_1)$ be a purification of $\rho^A$. Since each $|\psi_x^{AB}\ra$ is also a purification of $\rho^A=\rho^A_x$, we can infer that there exists an isometry $V_x:B_1\to B$ such that
\be\label{pp11}
|\psi_x^{AB}\ra=I^A\otimes V_x^{B_1\to B}|\chi^{AB_1}\ra\;.
\ee
Our first goal is to show that if $\{|\psi_x^{AB}\ra\}_{x\in[m]}$ are the eigenvectors of $\rho^{AB}$, then $V_{x'}^*V_x=\delta_{xx'}I^{B_1}$.

To prove it, let $\{q_y,\phi_y^{AB}\}_{x\in[m]}$ be another pure-state decomposition\index{pure-state decomposition} of $\rho^{AB}$, also with $m$ elements). Then, for the exact same reasons as stated above, for each $y\in[m]$ there exists an isometry $W_y:B_1\to B$ such that
\be\label{pp22}
|\phi_y^{AB}\ra=I^A\otimes W_y^{B_1\to B}|\chi^{AB_1}\ra\;.
\ee
Recall from Exercise~\ref{ensembles} that the ensemble $\{q_y,\;\phi_y^{AB}\}_{x\in[m]}$ can be related to the spectral decomposition of $\rho^{AB}$ by a unitary matrix $U=(u_{yx})$ as follows:
\be\label{specdc}
\sqrt{q_y}|\phi_y^{AB}\ra=\sum_{x\in[m]}u_{yx}\sqrt{p_x}|\psi_x^{AB}\ra\quad\quad\forall\;y\in[m]\;.
\ee
Combining this with~\eqref{pp11} and~\eqref{pp22} gives
\be
\left(I^A\otimes \sqrt{q_y}W_y\right)|\chi^{AB_1}\ra=\Big(I^A\otimes\sum_{y\in[m]}u_{yx}\sqrt{p_x}V_x\Big)|\chi^{AB_1}\ra\;.
\ee
By multiplying both sides by $\left(\rho^A\right)^{-1/2}$ we can replace $|\chi^{AB_1}\ra$ on both sides of the equation above with the (unnormalized) maximally entangled state $|\Omega^{\tB_1B_1}\ra$. Therefore, the equation above gives
\be
\sqrt{q_y}W_y=\sum_{x\in[m]}u_{yx}\sqrt{p_x}V_x\;.
\ee
Since $W_y$ is an isometry we get
\ba\label{gfio}
q_yI^{B_1}&=\sum_{x,x'\in[m]}\bar{u}_{yx'}u_{yx}\sqrt{p_{x'}p_x}V_{x'}^*V_x\\
&=\sum_{x\in[m]}p_x|u_{yx}|^2I^{B_1}+\sum_{\substack{x\neq x'\\ x,x'\in[m]}}\bar{u}_{yx'}u_{yx}\sqrt{p_{x'}p_x}V_{x'}^*V_x\;.
\ea
Now, using the fact that $\{|\psi_x^{AB}\ra\}_{x\in[m]}$ forms an orthonormal set of vectors we get from~\eqref{specdc} that 
\be
q_y=\left\|\sqrt{q_y}|\phi_y^{AB}\ra\right\|^2_2=\sum_{x\in[m]}p_x|u_{yx}|^2\;.
\ee
Combining this with~\eqref{gfio} gives
\be
\sum_{\substack{x\neq x'\\ x,x'\in[m]}}\bar{u}_{yx'}u_{yx}\sqrt{p_{x'}p_x}V_{x'}^*V_x=\0\;.
\ee
The equation above holds for all unitary matrices $U=(u_{yx})$ and all $y\in[m]$. Setting $y=1$, and choosing $U$ to be a unitary matrix with its first row as $\frac{1}{\sqrt{2}}(1,1,0,\ldots,0)$ gives $V_{2}^*V_2+V_{1}^*V_2=\0$. Similarly, choosing the first row of $U$ to be $\frac{1}{\sqrt{2}}(1,i,0,\ldots,0)$ gives $V_{2}^*V_2-V_{1}^*V_2=\0$. Thus, we obtain $V_1^*V_2=V_2^*V_1=\0$. By repeating the same argument with permuted versions of $\frac{1}{\sqrt{2}}(1,1,0,\ldots,0)$ and $\frac{1}{\sqrt{2}}(1,i,0,\ldots,0)$, we conclude that for all $x,x'\in[m]$ such that $x\neq x'$, we have $V_{x'}^*V_x=\0$.

Let $\{|z\ra\}_{z\in[r]}$ be an orthonormal basis of $B_1$, and define $|\varphi_{xz}^B\ra\eqdef V_x|z\ra$ for all $x\in[m]$ and $z\in[r]$. Using the fact that $V_{x'}^*V_x=\delta_{xx'}I^{B_1}$, we can derive that $\la\varphi_{x'z'}^B|\varphi_{xz}^B\ra=\delta_{xx'}\delta_{zz'}$ for all $x\in[m]$ and $z\in[r]$. Let $\mk$ be the subspace spanned by the orthonormal vectors $\{|\varphi_{xz}^B\ra\}$, with $x\in[m]$ and $z\in[r]$, and note that the dimension of $\mk$ is $mr$. Thus, there exists a subspace $B_2$ of $B$ with $|B_2|=m$ such that $\mk$ is isomorphic to $B_1\otimes B_2$. This isomorphism implies that there exists an isometry $U:B_1B_2\to B$ such that $|\varphi_{xz}^B\ra=U^{B_1B_2\to B}|z\ra^{B_1}|x\ra^{B_2}$. Combining this with the definition $|\varphi_{xz}^B\ra\eqdef V_x|z\ra^{B_1}$ gives $V_x|z\ra^{B_1}=U|z\ra^{B_1}|x\ra^{B_2}$ for all $x\in[m]$ and all $z\in[r]$. Hence, $V_x^{B_1\to B}=U^{B_1B_2\to B}\left(I^{B_1}\otimes|x\ra^{B_2}\right)$ so that
$
|\psi_x^{AB}\ra=U^{B_1B_2\to B}|\chi^{AB_1}\ra|x\ra^{B_2}
$
and
\be\label{purioft}
\rho^{AB}=U\left(\chi^{AB_1}\otimes\sigma^{B_2}\right)U^*\quad
\text{where}\quad
\sigma^{B_2}\eqdef\sum_{x\in[m]}p_x|x\lr x|^{B_2}\;.
\ee
Observe that the state in~\eqref{purioft} has a purification of the form
$
U^{B_1B_2\to B}|\chi^{AB_1}\ra|\phi^{B_2B'}\ra
$,
where 
\be
|\phi^{B_2B'}\ra=\sum_{x\in[m]}\sqrt{p_x}|x\ra^{B_2}|x\ra^{B'}\;.
\ee
Therefore, since $|\psi^{ABB'}\ra$ is also a purification of $\rho^{AB}$ we conclude that up to a local unitary on system $B$ and on system $B'$, the state $\psi^{ABB'}$ has the form $|\chi^{AB_1}\ra|\phi^{B_2B'}\ra$.
\end{proof}

\bex
Consider a monotonically increasing convex function  $h:\mbb{R}_+\to\mbb{R}_+$ with the property that $h(t)=0$ if and only if $t=0$. For any state $\rho\in\md(AB)$, let
\be
E_h\left(\rho^{AB}\right)\eqdef\min\sum_{x\in[m]}p_xh\left(E\left(\psi^{AB}_x\right)\right)\;,
\ee
where the minimum is over all pure-state decompositions of $\rho^{AB}=\sum_{x\in[m]}p_x\psi_x^{AB}$. Show that if $E_h$ is monogamous on pure tripartite states, then $E$ is also monogamous on pure tripartite states.
\eex

Many operational measures of entanglement, such as the relative entropy of entanglement, entanglement cost, and distillable entanglement, reduce to the entropy of entanglement\index{entropy of entanglement} for bipartite pure states. Recall that the entropy of entanglement is given in terms of the von Neumann entropy of the reduced state, $H(\rho)=-\tr[\rho\log\rho]$. Since the von Neumann entropy is known to be strictly concave, these measures are monogamous on pure tripartite states. In the following theorem, we show that their convex roof extensions are also monogamous for mixed tripartite states.

\begin{myt}{}
\begin{theorem}
Let $\rho\in\md(ABB')$, and $E$ and $E_F$ be as in~\eqref{gcsym00} and~\eqref{ooo1}, respectively, where $g$ is both symmetric and strictly concave. Then, the convex roof extension\index{convex roof extension} $E_F$ is monogamous.
\end{theorem}
\end{myt}

\begin{proof}
Let $\{p_x,|\psi^{ABB'}\ra\}_{x\in[m]}$ be the optimal pure state decomposition of $\rho^{ABB'}$ satisfying
\be
E_F\left(\rho^{ABB'}\right)=\sum_{x\in[m]}p_xE\left(\psi_x^{ABB'}\right)\;.
\ee
Denoting by $\rho_x^{AB}\eqdef\tr_{B'}\left[\psi^{ABB'}_x\right]$ we have that
\be
\rho^{AB}=\sum_{x\in[m]}p_x\rho_x^{AB}\;.
\ee
Suppose $E_F\left(\rho^{ABB'}\right)=E_F\left(\rho^{AB}\right)$. Combining this with the two equations above gives
\ba
\sum_{x\in[m]}p_xE\left(\psi_x^{ABB'}\right)&=E_F\left(\rho^{AB}\right)\\
\Gg{\text{Convexity of }E_F}&\leq \sum_{x\in[m]}p_xE_F\left(\rho_x^{AB}\right)\;.
\ea
On the other hand, for every $x\in[m]$ we have 
\be
E\left(\psi_x^{ABB'}\right)\geq E_F\left(\rho_x^{AB}\right)\;,
\ee
since $E_F$ is a measure of entanglement (in fact, an entanglement monotone) and does not increase under the tracing out of the local system $B'$. Hence, from the two inequalities above we get that for all $x\in[m]$ we have
\be
E\left(\psi_x^{ABB'}\right)= E_F\left(\rho_x^{AB}\right)\;.
\ee
Now, from Theorem~\ref{puretrimon} we get that for each $x\in[m]$ there exists systems $B^{(1)}_x$ and $B^{(2)}_x$, and an isometry $V_x:B^{(1)}_xB^{(2)}_x\to B$ such that
\be
|\psi_x^{ABB'}\ra=V_x^{B^{(1)}_xB^{(2)}_x\to B}\big|\chi^{AB_x^{(1)}}_x\big\ra\big|\phi^{B_x^{(2)}B'}_x\big\ra\;,
\ee
for some $\chi_x\in\pure(AB_x^{(1)})$ and $\phi_x\in\pure\left(B_x^{(2)}B'\right)$. Tracing out system $B$ on both sides of the equation above gives
\be
\psi_x^{AB'}=\chi_x^A\otimes\phi_x^{B'}\;.
\ee
Therefore, the marginal state $\rho^{AB'}=\sum_{x\in[m]}p_x\psi_x^{AB'}$ is separable so that $E\left(\rho^{AB'}\right)=0$.
This completes the proof.
\end{proof}

\section{Notes and References}

A comprehensive review of multipartite entanglement can be found in Chapter 17 of the book~\cite{BZ2006}. The classification of all homogeneous SL-invariant polynomials using the Schur-Weyl duality is presented in~\cite{GW2013}.

Critical states, also known as normal forms, are discussed in detail in~\cite{VDD2003} and~\cite{Wallach2004} for mathematically inclined readers. The canonical form\index{canonical form}  of 3-qubit states as given in Theorem~\ref{canonform} is due to~\cite{AAC+2000}. The classification of SLOCC classes in three qubits was done by~\cite{DVC2000}, while that for four qubits was done by~\cite{VDDV2002}. A detailed analysis of all 4-qubit maximally entangled states can be found in~\cite{GW2010, SDK2016}. The classification of maximally entangled sets of multipartite systems is presented in~\cite{DSK2013}.

The generalization of Nielsen's majorization theorem to the multipartite case, as presented in Theorem~\ref{intercm}, is due to~\cite{GW2011}. 
The generalization of this theorem to the full set $\sep$ can be found in~\cite{HES+2021}.
Several other generalizations of this result, particularly deterministic interconversions of multipartite entanglement under various operations including LOCC can be found in~\cite{SDSK2017} and~\cite{DSSK2017}.
It is worth mentioning that in~\cite{GKW2017} and~\cite{SWGK2018}, it was shown that the stabilizer group of almost all multipartite entangled states is trivial, and consequently, LOCC conversion between two states in the \emph{same} SLOCC class is almost never possible. Nevertheless, certain multipartite states that have symmetry (e.g., GHZ states, graph states, stabilizer states, etc.) have a non-trivial stabilizer group and consequently rich entanglement properties~\cite{LSH+2023}.

In~\cite{HES+2021}, it was demonstrated that any pure-state transformation attainable by $LOCC$ using a finite number of communication rounds can also be accomplished using $\sep_1$. However, not every pure-state transformation possible with SEP is achievable with $\sep_1$. These findings underscore that $\sep_1$ serves as a robust outer approximation of LOCC, particularly given that infinite rounds of classical communication are less feasible in practice.

The concept of entanglement of assistance was first introduced in~\cite{DFM+1998}, and the example given in~\eqref{notem} that demonstrates that it is not a tripartite entanglement monotone was taken from~\cite{GS2006}. The result that asymptotic entanglement of assistance is equal to the smaller of its two local entropies was discovered in~\cite{SVW2005}.
The concept of localizable entanglement was first introduced in the context of spin chains in~\cite{VPC2004}, and its comparison with entanglement of collaboration can be found in~\cite{Gour2006}.

Monogamy of entanglement was first introduced in~\cite{CKW2000}, in which the CKW monogamy relation was discovered. The monogamy of the squashed entanglement was discovered in~\cite{CW2004}. The concept of "monogamy of entanglement without inequalities" was first introduced in~\cite{GG2018} and developed further in~\cite{GG2019}. Additional references on monogamy of entanglement can be found in those papers.

\part{Additional Examples of Static Resource Theories}

\chapter{The Resource Theory of Asymmetry}\index{asymmetry}\label{Ch:Asymmetry}

In Shannon's theory, information is considered to be ``fungible," meaning it can be encoded into any physical system's degree of freedom, and the information's content is independent of the encoding method. For instance, a simple yes/no message can be transmitted equally well by a 5/0-volt potential difference across a circuit element or by flipping a coin to heads/tails. This type of information is known as ``speakable information," as it can be conveyed through speech or symbols.

However, there are also non-fungible types of information, such as a direction in space, the time of an event, or the relative phase between two quantum states in a superposition. Such information is referred to as ``unspeakable information" because it cannot be conveyed verbally without a shared coordinate system, a synchronized clock, or a common phase reference. For example, directional information can only be transmitted between two parties through the exchange of a physical system whose state represents the direction itself, such as a classical gyroscope, in the absence of a common gravitational field or stellar background.

Unspeakable information can be communicated verbally when a reference frame is present, which is true for both classical and quantum information. However, despite speakable information being fungible, multiple parties must first agree on how to encode/decode this information in a physical system, which implicitly necessitates a common reference frame. As a result, quantum information processing tasks assume the existence of a shared reference frame, and the lack of this shared frame significantly restricts what can be accomplished.

The absence or deterioration of a common reference frame is a natural constraint that frequently occurs in the study of multiple physical systems. Consequently, this constraint gives rise to a resource theory of reference frames, which can be more broadly classified as a resource theory of asymmetry.

\section{Free States and Free Operations}

Consider two parties (Alice and Bob) who do not share a reference frame. Mathematically, we represent the information about the frame by an element $g$ of a compact group $\G$.  For instance, $g\in \G$ could correspond to a particular orientation in space, clock synchronization, phase information, etc.  

In this chapter, it is assumed without explicit statement that $\G$ is either a finite group or a compact Lie group. Additionally, readers who are not well-versed in representation theory are advised to first read Appendix~\ref{sec:rep} before proceeding with this chapter. The same notations as those in Appendix~\ref{sec:rep} will be utilized, and frequent references to this appendix will be made throughout the chapter.

\subsubsection{$\G$-Invariant States}\index{$\G$-invariant states}

In the resource theory of asymmetry, each element $g\in \G$ is denoted by a unitary matrix $U_g$. If $\rho\in\md(A)$ represents the density matrix of a quantum system with respect to Alice's reference frame, then the state of the same physical system with respect to Bob's reference frame is given by
\begin{equation}
\label{Eq:Unitary_conjugation}
\mU_g(\rho)\eqdef U_g\rho U_{g}^{*}
\end{equation}
If Alice and Bob are unaware of the element $g\in \G$ that establishes the relation between their reference frames, then the states that Alice can prepare relative to Bob's reference frame are those satisfying $\rho=\mU_g(\rho)$ for all $g\in \G$. Such states are referred to as $\G$-invariant and satisfy $[\rho, U_g]=0$, as indicated by Definition~\ref{ginva}.

The absence of a shared reference frame places a limitation on the types of states that Alice can generate relative to Bob's reference frame. She is only capable of creating $\G$-invariant states, which comprise the free states in the QRT of reference frames, denoted as
\begin{equation}
\mf(A)=\inv_\G(A)\eqdef\left\{\rho\in\md(A)\;:\;\mU_g(\rho)=\rho\;\;\;\forall g\in \G\right\}\;.
\end{equation}

For instance, suppose the group $\G=U(1)$ corresponds to an optical phase reference or to dynamics with rotational symmetry around a fixed axis (in which case the group is SO$(2)$, which is known to be isomorphic to the group $U(1)$). In this scenario, a unitary representation of $\G$ is provided by $U_{\theta}=e^{i\hat{N}\theta}$, where $\theta\in U(1)$ and $\hat{N}$ is the total number operator (or in the case of rotational symmetry, $\hat{N}$ can be replaced with $L_{\n}$, the angular momentum operator in the $\n$ direction).
In this instance, the free states are given by states of the form $\sum_{n}p_n|n\lr n|$, where $|n\ra$ corresponds to the eigenvectors of $\hat{N}$.

More generally, it will be observed that the absence of a shared reference frame enforces a \emph{superselection rule} regarding the types of states that Alice can generate. This superselection rule is characterized by the fact that coherent superpositions between states in specific subspaces are not feasible. For instance, coherent superpositions of $U(1)$ states among the eigenstates of the number operator are not free and cannot be prepared by Alice.

\subsubsection{$\G$-Covariant Channels}\index{$\G$-covariant channel}

The set of free operations in the QRT of reference frames can be defined similarly to the free states. Let $\sigma\in\md(B)$ be an arbitrary density matrix of system $B$ described in Bob's reference frame. Suppose Alice performs a quantum operation on this system described by the channel $\mE\in\cptp(A\to A)$ in her reference frame. How would this operation be described in Bob's reference frame? If Bob knows that their reference frames are linked by an element $g\in \G$, then $\mU_g^*(\sigma)$ is Alice's description of the initial state, and $\mE(\mU_g^*(\sigma))$ is her description of the final state. Therefore, the final state in Bob's reference frame is given by $\mU_g\circ\mE\circ\mU_g^*(\sigma)$, and his description of Alice's operation is $\mU_g\circ\mE\circ\mU_g^*$.
 
Hence, if Alice and Bob are unaware of the value of $g\in \G$, they will have a similar description of the CPTP map $\mE$ only if $\mE$ satisfies
\be\label{15p3i}
\mU_g\circ\mE\circ\mU_{g}^{*}=\mE \quad\quad\forall\;g\in \G\;.
\ee 
Quantum channels of this kind are referred to as \emph{$\G$-covariant}, and they represent the free operations in the QRT of asymmetry. Similar to $\G$-invariant states, a quantum channel is $\G$-covariant if and only if it commutes with $\mU_g$ for all $g\in \G$.

Therefore, the set of free operations in the QRT of reference frames can be expressed as
\begin{equation}
\mf(A\to A)=\cov_\G(A\to A)\eqdef\left\{\mE\in\cptp(A\to A)\;:\;[\mE,\mU_g]=0\quad\forall\;g\in \G\right\}.
\end{equation}
where $[\mE,\mU_g]\eqdef\mE\circ\mU_g-\mU_g\circ\mE$ (see Fig. \ref{gcova} below).   For instance, for $\G=U(1)$, a $\G$-covariant quantum channel, $\mE\in\cptp(A\to A)$, satisfies for all $\theta\in[0,2\pi)$ and all $\rho\in\md(A)$
\be
\mE\left(e^{i\theta\hat{N}}\rho e^{-i\theta\hat{N}}\right)=e^{i\theta\hat{N}}\mE(\rho) e^{-i\theta\hat{N}}\;.
\ee

In subsequent sections, we'll delve into characterizations of $\G$-covariant channels, employing covariant adaptations of the operator sum\index{operator sum} representation and the Stinespring representation of a quantum channel. Before we proceed to these broader characterizations, let's focus initially on the specific case of unitary channels. Consider a covariant unitary channel $\mE(\cdot)=V(\cdot)V^*$, where $V:A\to A$ is a unitary matrix. According to condition~\eqref{15p3i}, for every element $g\in\G$ and for any state $\rho\in\md(A)$, the channel satisfies:
\be
\left(U_gVU_g^*\right)\rho^A\left(U_gVU_g^*\right)^*=V\rho^A V^*\;.
\ee
This implies that for every $g\in\G$, there exists a phase $\omega_g\in\mbb{C}$ with $|\omega_g|=1$ such that 
\be\label{15p7}
U_gVU_g^*=\omega_gV\;.
\ee

Since this equation holds for all $g\in\G$, it follows that the map $g\mapsto\omega_g$ is a 1-dimensional representation of $\G$. Specifically, when $g=e$ is the identity element, we have $U_g=I$ which gives $\omega_g=1$. Furthermore, for $g,h\in\G$, we have
\ba
\omega_{gh}V&=U_{gh}VU_{gh}^*\\
&=U_{g}U_hVU_{h}^*U_g^*\\
&=\omega_hU_{g}VU_g^*\\
&=\omega_h\omega_{g}V\;,
\ea
which implies that $\omega_{gh}=\omega_g\omega_h$. In other words, the set of all $\G$-covariant unitary channels can be characterized by unitary matrices that are ``almost" $\G$-invariant, meaning they commute with the elements of the group up to a phase, where this phase itself forms a 1-dimensional representation of $\G$.

In conclusion, we have observed that in the QRT of reference frames, the set of free states is the set of \emph{symmetric} states (i.e., those states that commute with $U_g$ for all $g\in \G$), and the set of free operations is the set of \emph{symmetric} operations (i.e., those operations that commute with $\mU_g$ for all $g\in \G$). Symmetric evolutions are prevalent in physics and may arise in various contexts, not just from the absence of a shared reference frame. Therefore, the set of $\G$-covariant operations defines a resource theory with applications extending beyond quantum reference frames. It may be referred to as a QRT of \emph{asymmetry} because in any QRT in which $\mf$ specifies a set of $\G$-covariant operations, asymmetric states and asymmetric operations are the resources of the theory.

Thus far, we have only examined $\G$-covariant channels with the same input and output dimensions. More generally, a quantum channel $\mE:\cptp(A\to B)$ is $\G$-covariant with respect to two (unitary) representations of $\G$, $\{U_g^A\}_{g\in \G}$ and $\{U_g^B\}_{g\in \G}$, if
\be\label{gcov}
\mE^{A\to B}\circ\mU_{g}^{A}=\mU_g^B\circ\mE^{A\to B}\quad\forall g\in \G.
\ee
Refer to Fig.~\ref{gcova} for an illustrative depiction of $\G$-covariant operations. 
The set of all $\G$-covariant quantum channels in $\cptp(A\to B)$ will be denoted by $\cov_\G(A\to B)$. It is worth noting that this notation does not explicitly specify the two unitary representations of $\G$, $\{U_g^A\}_{g\in \G}$ and $\{U_g^B\}_{g\in \G}$. The representations used will be clear from the context.

\begin{figure}[h]
\centering
    \includegraphics[width=0.3\textwidth]{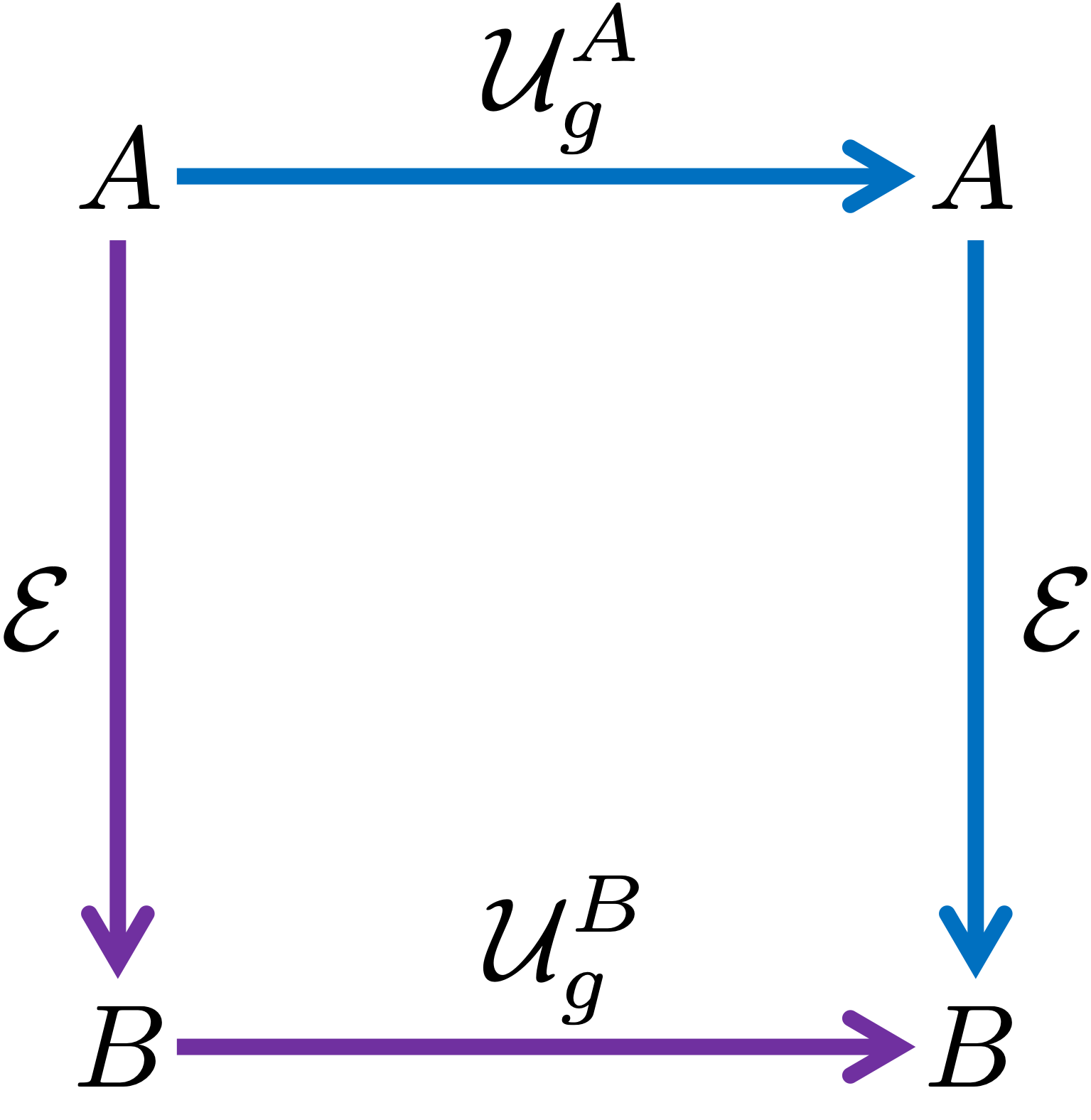}
  \caption{\linespread{1}\selectfont{\small Heuristic description of $\G$-covariant operations. The channel $\mE$ is $\G$-covariant if for every choice of group element $g\in \G$, the two different pathways yield the same outcome.}}
  \label{gcova}
\end{figure} 

\subsubsection{$G$-Covariant Measurements}

The result obtained from a quantum measurement, often referred to as the classical outcome, provides a form of information known as ``speakable information." This type of information can be effectively communicated between parties who do not share a common reference frame. Let's consider an example where Alice and Bob do not have a shared Cartesian reference frame. Suppose Alice performs a measurement on the spin of an electron in the $z$-direction relative to her reference frame and obtains an outcome of ``up" (indicating that the electron's spin is pointing in the positive $z$-direction). Alice can then transmit this outcome to Bob, allowing him to determine that the electron's spin is aligned with the positive $z$-direction in relation to Alice's frame. Therefore, even though the specific information about the $z$-direction itself cannot be conveyed, the measurement outcome, i.e., the ``up"/``down" information, can be effectively communicated between the parties involved. 

 Consequently, we make the assumption that the group $\G$ associated with the resource theory of asymmetry has a trivial action on classical systems that represent measurement outcomes. Moving forward, in Section \ref{qinst}, we observed that a general quantum measurement can be characterized by a quantum instrument denoted as $\mE\in\cptp(A\to BX)$, where $X$ represents the classical outcome of the measurement. We refer to $\mE$ as a $\G$-covariant quantum instrument if it satisfies the condition:
\be\label{covqi}
\mE^{A\to BX}\circ\mU_g^{A\to A}=\mU_g^{B\to B}\circ\mE^{A\to BX}\quad\quad\forall\;g\in\G\;.
\ee
The collection of all such $\G$-covariant quantum instruments is denoted by $\cov_\G(A\to BX)$.

Every quantum instrument\index{quantum instrument} $\mE^{A\to BX}$ as discussed above can be expressed as
\be
\mE^{A\to BX}=\sum_{x\in[m]}\mE_x^{A\to B}\otimes |x\lr x|^X
\ee
where $m\in\mbb{N}$, and each $\mE_x\in\cp(A\to B)$. If $\mE^{A\to BX}$ is $\G$-covariant the relation above in conjunction with~\eqref{covqi} implies that for all $x\in[m]$ we have
\be\label{covqi2}
\mE^{A\to B}_x\circ\mU_g^{A\to A}=\mU_g^{B\to B}\circ\mE^{A\to B}_x\quad\quad\forall\;g\in\G\;.
\ee
In other words, the quantum instrument\index{quantum instrument} $\mE^{A\to BX}$ is $\G$-covariant if and only if each CP map $\mE_x^{A\to B}$ is $\G$-covariant.

A special type of $\G$-covariant quantum instrument is a $\G$-covariant POVM. We get $\G$-covariant POVM by taking above $B$ to be the trivial system (i.e., $|B|=1$) so that for each $x\in[m]$ and every $\rho\in\ml(A)$, $\mE_x^{A\to B}(\rho^A)=\tr[\Lambda_x^A\rho^A]$ for some $\Lambda_x\in\eff(A)$ and the set $\{\Lambda_x^A\}_{x\in[m]}$ is a POVM. Now, for a trivial system $B$, the condition given in~\eqref{covqi2} becomes equivalent to
\be
\tr[\Lambda_x\rho]=\tr[\Lambda_x\mU_g\left(\rho\right)]=\tr[\mU_g^*\left(\Lambda_x\right)\rho]\quad\quad\forall\;g\in\G.
\ee
Since the condition above holds for all $\rho\in\ml(A)$ we must have $\Lambda_x=\mU_g^*(\Lambda_x)$ for all $g\in\G$ and $x\in[m]$. In other words, a POVM $\{\Lambda_x\}_{x\in[m]}$ is $\G$-covariant if and only if each element $\Lambda_x$ is $\G$-invariant; i.e., each $\Lambda_x$ satisfies $[\Lambda_x, U_g]=0$ for all $g\in\G$.

\section{Distinctive Concepts in the QRT of Asymmetry}

In this subsection, we introduce several mathematical tools and concepts that are distinctive to the resource theory of asymmetry and do not appear in other resource theories. 

\subsection{The $\G$-Twirling\index{twirling} Operation}\label{secgt}

If Alice lacks the information about $g$, her description of Bob's density matrix is obtained by averaging over all possible values of $g$. The uniform Haar measure over the group $\G$ is denoted by $dg$, and this average can be expressed as follows:
\be
\mG(\rho)\eqdef\int_{\G} dg\;\mU_g(\rho)\;.
\ee 
The averaging CPTP map is known as the $\G$-twirling map (see Sec.~\ref{sec:inv}). If the group $\G$ is finite, the integral is replaced by a discrete sum over the $|\G|$ elements of the group, that is, $\mG(\rho)=\frac{1}{|\G|}\sum_{g\in \G}\mU_g(\rho)$.

The free states in this QRT have a very particular structure. First, note that $\rho\in\mf(A)$ if and only if it is $\G$-invariant, meaning that $\mU_g(\rho)=\rho$ for all $g$. In particular, $\mG(\rho)=\rho$ for all $\rho\in\mf(A)$. Combining this with the definition of $\mf(A)$ implies that $\G$-twirling\index{twirling} is a resource-destroying map (see Definition~\ref{def:RDM}). Additionally, one can characterize the free states using techniques from representation theory. In particular, Theorem~\ref{deop} states that $\rho\in\md(A)$ is free if and only if $\rho^A$ has the following form:
\be\label{invcon}
\rho^A=\bigoplus_{\lambda\in\irr(U)} \u^{B_\lambda}\otimes \rho^{C_\lambda}_{\lambda}
\quad\text{where}\quad
 \rho^{C_\lambda}_\lambda\eqdef\tr_{B_\lambda}\left[\Pi^{A_\lambda}\rho^A\Pi^{A_\lambda}\right]\;.
\ee 
Moreover, note that the above expression implies the following corollary.

\begin{myg}{}
\begin{corollary}\label{gcovcor}
The $\G$-twirling map can be decomposed as
\be
\mG=\bigoplus_{\lambda\in\irr(U)}\left(\mR^{B_\lambda}\otimes\id^{C_\lambda}\right)\circ\mP^{A_\lambda}
\ee
where $\mP^{A_\lambda}(\cdot)=\Pi^{\lambda}(\cdot)\Pi^{\lambda}$, $\Pi^{A_\lambda}:A\to A$ is a projection to subspace $A_\lambda$, $\mR^{B_\lambda}\in\cptp(B_\lambda\to B_\lambda)$ is the completely depolarizing (randomizing) channel, and $\id^{C_\lambda}\in\cptp(C_\lambda\to C_\lambda)$ is the identity channel.
\end{corollary}
\end{myg}

The above corollary demonstrates that the $\G$-twirling operation eliminates any correlations among distinct irreducible representations. For example, let us consider the case where $\G=\muu(1)$. As this group is Abelian, it has only one-dimensional irreps (i.e., $|B_\lambda|=1$). The irreps of $\muu(1)$ are labeled by integers $\lambda=k\in\mbb{Z}$, and the $k$-th irrep $u_k:\muu(1)\mapsto\mbb{C}$ is of the form:
\be
u_k(\theta)=e^{ik\theta}\quad\quad\forall\theta\in \muu(1)\;.
\ee
In this context, we will consider an infinite dimensional (separable) Hilbert space denoted by $A$ with basis vectors $|n\ra$ where $n$ belongs to the set of integers $\mbb{Z}$. The ``number" operator which generates the $\muu(1)$ symmetry can be defined as follows:
\be
\hat{N}\eqdef\sum_{n\in\mbb{Z}}n|n\lr n|\;,
\ee
Note that we allow negative values of $n$ and work with the representation $\theta\mapsto e^{i\hat{N}\theta}$.

For each irrep on a single copy of $A$, the multiplicity space is trivial (i.e., $|C_\lambda|=1$), and the $\G$-twirling operation can be easily represented as
\be
\mG(\cdot)=\sum_{k\in\mbb{Z}}|k\lr k|(\cdot)|k\lr k|\;,
\ee
which means that $\mG$ is the completely dephasing channel with respect to the basis ${|k\ra}_{k\in\mbb{Z}}$ in this case.

However, when considering $\ell$ copies of $A$, the multiplicity space of a given irrep is usually not trivial, and as a result, the $\G$-twirling operation is not equivalent to the dephasing channel.  Specifically, let $\hat{N}_x$ be the number operators associated with system $A_x$ for each $x\in[\ell]$. Consider the unitary representation on $A^n=(A_1,\ldots,A_\ell)$ defined by
\be
\theta\mapsto \bigotimes_{x\in[\ell]}e^{i\hat{N}_x\theta}=e^{i \hat{N}_\tot\theta},\quad\text{where}\quad \hat{N}_\tot\eqdef\sum_{x\in[\ell]}\hat{N}_x\;.
\ee
In this case, the irreps are denoted by the eigenvalues $n\in\mbb{Z}$ of the total number operator $\hat{N}\tot$. While the representation space $B_{n}$ (i.e., $B_\lambda$ with $\lambda=n$) is trivial (i.e., one dimensional) for every irrep $\lambda=n$, the multiplicity space $C_{n}$ is not. Let
\be
\Pi_n^{(\ell)}\eqdef\sum_{\substack{k_1+\cdots+k_\ell=n\\k_1,\ldots,k_\ell\in\mbb{Z}}}|k_1\lr k_1|\otimes\cdots\otimes|k_\ell\lr k_\ell|
\ee
be the projection onto the eigenspace of $\hat{N}_{\tot}$ corresponding to the eigenvalue $n$. Using this notation, the $\G$-twirling operation can be expressed as
\be\label{n15p18}
\mG_\ell(\cdot)=\sum_{n\in\mbb{Z}}\Pi_n^{(\ell)}(\cdot)\Pi_n^{(\ell)}\;.
\ee
Note that $\Pi_n^{(\ell)}$ is the projection onto the multiplicity space $C_n$. This space is often referred to as the \emph{decoherence-free subspace}, as any pure state $\psi\in\pure(C_n)$ is $U(1)$-invariant, i.e., $\mG(\psi)=\psi$. For example, if $\ell=3$, any linear combination of $|011\ra$, $|101\ra$, and $|110\ra$ is an eigenvector of $\hat{N}_\tot$ corresponding to an eigenvalue of $2$. Therefore, the coherence of any state in the span of these three vectors remains unaffected by the $\G$-twirling operation.

The $\G$-twirling operation can be applied to quantum channels as well. Suppose Alice applies a quantum operation $\mE\in\ml(A\to A)$ to her system. If Bob knows the relation between their reference frames, then he can describe the operation relative to his own system as $\mU_g\circ\mE\circ\mU_g^*$, where $\mU_g$ is a unitary operator that relates Alice's and Bob's frames. However, in the absence of a shared reference frame, Bob cannot use this description. Instead, the channel $\mE$ appears to him as a mixture of the form $\int_\G dg \;\mU_g\circ\mE\circ\mU_{g}^{\dag}$. In order for Alice and Bob to have the same description of the channel, the condition $\mE=\int_\G dg \;\mU_g\circ\mE\circ\mU_{g}^{\dag}$ must be satisfied. This integral is a type of twirling operation applied to the channel $\mE$.

\begin{exercise}
Show that the $\G$-twirling map is unital and idempotent; i.e. $\mG\circ\mG=\mG$.
\end{exercise}

For compact Lie groups the $\G$-twirling is defined in terms of an integral over the group. From Carath\'eodory theorem (see Theorem~\ref{carath}) it follows that the $\G$-twirling can be expressed as a finite convex combination of unitary channels $\mU_g$. To see why, for every $g\in\G$ let $|\psi_g^{A\tA}\ra\eqdef U_g^A\otimes I^{\tA}\big|\Omega^{A\tA}\big\ra$, and let $\mc$  be the convex hull of the set $\big\{\psi_g^{A\tA}\big\}_{g\in\G}$. Note that $\mc\subset\mr$, where $\mr$ is a subspace of $\herm(A\tA)$ given by
\be
\mr\eqdef\left\{\Lambda^{A\tA}\in\herm(A\tA)\;:\; \Lambda^A\propto I^A,\quad \Lambda^{\tA}\propto I^{\tA}\right\}\;,
\ee
where the symbol $\propto$ stands for `proportional to'.

Since we assume that $\G$ is a compact Lie group, the set $\big\{\psi_g^{A\tA}\big\}_{g\in\G}$ is also compact in $\mr$. Consequently, also its convex hull is compact (this follows from Carath\'eodory's theorem, see Exercise~\ref{compacthull}).
Therefore, the Choi matrix of the $\mG$, $J_\mG^{A\tA}$, also belongs to $\mc$, and from Carath\'eodory's theorem it can be expressed as a convex combination of at most $d$ elements of $\big\{\psi_g^{A\tA}\big\}_{g\in\G}$, where $d\eqdef\dim(\mr)+1$. We therefore conclude that there exists $\p\in\prob(d)$ and $d$ group elements $\{g_x\}_{x\in[d]}\subset\G$ such that
\be
J_\mG^{A\tA}=\sum_{x\in[d]}p_x\psi^{A\tA}_{g_x}\;.
\ee
In other words, $\mG$ can be expressed as
\be\label{15p15n}
\mG(\cdot)=\sum_{x\in[d]}p_x U_{g_x}(\cdot)U_{g_x}^*\;.
\ee

\bex\label{ubdim}
Let $m\eqdef|A|$.  Show that 
\be\label{15p22}
d\eqdef \dim(\mr)+1=(m^2-1)^2+2\leq m^4\;.
\ee
\eex

\bex\label{etr}
Let $\mG_k\in\cptp(A^k\to A^k)$ be the $\G$-twirling map as defined on $k$-copies of $A$. That is, for any $\rho\in\md(A^k)$ we have
\be
\mG_k(\rho^{A^k})\eqdef\int_\G dg\; U_g^{\otimes k}\rho^{A^k}\left(U_g^{\otimes k}\right)^*\;.
\ee
Show that
\be
\mG_1^{\otimes k}\circ\mG_k=\mG_1^{\otimes k}\;.
\ee
\eex

\bex\label{affcon}
Let $\rho\in\md(A)$ and $\alpha\in\mbb{R}_+$. Show that if $\rho$ is $\G$-invariant then also $\rho^\alpha/\tr[\rho^\alpha]$. Hint: Use~\eqref{invcon}. 
\eex

\bex\label{randomig}
Let $g\mapsto U_g$ be a projective unitary representation of a finite or compact Lie group $\G$. For each $\lambda\in\irr(U)$, let $U_g^{(\lambda)}$ be the reduction of $U_g$ to the space $B_\lambda$ as given in~\eqref{lamir}. Show that for every $\rho\in\ml(B_\lambda)$ we have
\be
\int dg U_g^{(\lambda)}\rho^{B_\lambda}U_g^{*(\lambda)}=\tr\left[\rho^{B_\lambda}\right] I^{B_\lambda}\;.
\ee 
\eex

\subsubsection{The Weighted $\G$-Twirling\index{twirling}}

The weighted $\G$-twirling is a variant of the $\G$-twirling that also plays an important role in the resource theory of asymmetry. It is defined as follows:
Let $\p:\G\to\mbb{R}_+:g\mapsto p(g)$ a probability distribution that is normalized such that $\int_\G dg\;p(g)=1$. With respect to this distribution, the weighted $\G$-twirling is  defined as
\be\label{15p30i}
\mG_\p(\cdot)\eqdef\int_\G dg\;p(g)U_g(\cdot)U_g^*\;.
\ee
When $p$ is uniform, i.e., $p(g)=1$, we get that $\mG_\p=\mG$. 

It is worth noting that by choosing different probability densities $\p$, we can obtain any convex combination of the unitary channels $\mU_g(\cdot)\eqdef U_g(\cdot)U_g^*$. Additionally, if $\rho\in\inv_\G(A)$, then $\mG_\p(\rho)=\rho$. However, the converse is not true; that is, in general, there exists a density matrix $\rho\not\in\inv_\G(A)$ such that $\mG_\p(\rho)=\rho$. One such example is obtained by taking $p(g)=\delta(g)$  to be the Dirac delta function (i.e., $p(g)=0$ for all $g\neq e$) so that $\mG_\p=\id^A$ is the identity channel. In this extreme example, every resource state $\rho\not\in\inv_\G(A)$ satisfies $\mG_\p(\rho)=\rho$.

Unlike like the $\G$-twirling, the weighted $\G$-twirling is not necessarily $\G$-covariant. This depends on the group $\G$ and on the distribution $p(g)$. For example, for abelian group the weighted $\G$-twirling is covariant since $\mG_\p\circ\mU_g=\mU_g\circ\mG_\p$ for all $g\in\G$. For non-abelian groups, the weighted $\G$-twirling is $\G$-covariant if $p(g)$ is a class function as introduced in Definition~\ref{funclass}.

\begin{myt}{}
\begin{theorem}
Let $\p:\G\to\mbb{R}_+:g\mapsto p(g)$ be a normalized probability distribution. If $p(g)$ is a class function then $\mG_\p$ is $\G$-covariant.
\end{theorem}
\end{myt}

\begin{proof}
Let $h\in\G$ and $\rho\in\ml(A)$. Then,
\ba
\mU_h^*\circ\mG_\p\circ\mU_h(\rho)&=\int_\G dg\;p(g)U_{h^{-1}}U_gU_h\rho U_h^*U_g^*U_{h^{-1}}^*\\
&=\int_\G dg\;p(g)U_{h^{-1}gh}\rho U_{hgh^{-1}}^*\\
\Gg{g'\eqdef hgh^{-1}}&=\int_\G dg'\;p(hg'h^{-1})U_{g'}\rho U_{g'}^*\\
\GG{{\it p}\;\text{is a class function}}&=\int_\G dg'\;p(g')U_{g'}\rho U_{g'}^*=\mG_\p(\rho)\;.
\ea
Since $\rho\in\ml(A)$ was arbitrary we conclude that $\mU_h^*\circ\mG_\p\circ\mU_h=\mG_\p$ for all $h\in\G$. This completes the proof. 
\end{proof}

In the definition of the weighted $\G$-twirling we assumed that $p(g)$ is an arbitrary probability density over $\G$. However, as we saw earlier, thanks to Carath\'eodory's theorem, the $\G$-twirling can be expressed as a finite convex combination of unitary channels of the form  $U_g(\cdot)U_g^*$. The same arguments can be applied to $\mG_\p$, so we can assume, without loss of generality, that $\p$ is a discrete probability distribution, i.e., $\p\in\prob(d)$ for some $d\leq m^4$, where $m\eqdef|A|$. Hence, the weighted $\G$-twirling takes the form:
\be
\mG_\p(\cdot)=\sum_{x\in[d]}p_x U_{g_x}(\cdot)U_{g_x}^*\;.
\ee

\bex
Give an example of a group $\G$, a probability distribution $p(g)\neq \delta(g)$, and a state $\rho\in\md(A)$ such that $\mG_\p(\rho)=\rho$ but $\rho\not\in\inv_\G(A)$.
\eex

\subsection{Three Representations of $\G$-Covariant Maps}

\subsubsection{Covariant Operator Sum Representation}

In this subsection, we introduce the concept of an irreducible tensor operator and use it to characterize the operator sum\index{operator sum} representation of a covariant channel. The irreducible tensor operator is defined with respect to a projective representation, $g\mapsto U_g^E$, of the group $\G$. This projective representation induces a particular structure on the Hilbert space $E$ (see Theorem~\ref{thdco}). Specifically, $E$ can be decomposed as 
\be
E=\bigoplus_{\lambda\in\irr(U^E)} B_\lambda\otimes C_\lambda
\ee
where $B_\lambda$ is an irreducible $G$-invariant subspace of $A$,
and $C_\lambda$ is the multiplicity subspace. According to Theorem~\ref{thdco} for all $g\in\G$ we can also decompose $U_g^E$ as
\be\label{de1513}
U_g^E\cong\bigoplus_{\lambda\in\irr(U^E)} U_g^{(\lambda)}\otimes I^{C_\lambda}\;,
\ee
where each $U_g^{(\lambda)}$ acts irreducibly on $B_\lambda$.
We will denote by $u^{(\lambda)}_{m'm}(g)$ the $m'm$-component of $U_g^{(\lambda)}$, and use the indices $\lambda$, $m$, and $x$ to label the basis of $E$ as given in~\eqref{ginvbasis}. The index $x$ corresponds to the multiplicity index.

\begin{myd}{}
\begin{definition}
Let $g\mapsto U_g^E$ be as above, and let $g\mapsto U^A_g$ and $g\mapsto U_g^B$ be two additional projective unitary representations of $\G$. We say that a set of operators $\{K_{\lambda,m,x}\}_{\lambda,m,x}\subset\ml(A,B)$ is \emph{an irreducible tensor operator}\index{irreducible tensor operator} with respect to the three projective unitary representations on systems $A$, $B$, and $E$, if its elements are orthonormal and satisfy for all $\lambda$, $m$, and $x$,
\be\label{fortitude}
U_g^{B}K_{\lambda,m,x}U_g^{*A}=\sum_{m'}u^{(\lambda)}_{m'm}(g)K_{\lambda,m',x}\quad\quad\forall\;g\in\G\;,
\ee
where $u^{(\lambda)}_{m'm}(g)$ is $m'm$ component of the matrix $U^{(\lambda)}_g$ that appear in~\eqref{de1513}.
\end{definition}
\end{myd}
\begin{remark}
The orthonormality condition for an irreducible tensor operator is defined in terms of the Hilbert-Schmidt inner product. Specifically, we make the assumption that:\index{Hilbert-Schmidt inner product}
\be
\tr\left[K^{*}_{\lambda',m',x'}K_{\lambda,m,x}\right]=\delta_{\lambda\lambda'}\delta_{mm'}\delta_{xx'}\;.
\ee
\end{remark}

The condition stated in Equation~\eqref{fortitude} imposes constraints not only on the elements of the irreducible tensor operator, but also on the three representations $U_g^A$, $U_g^B$, and $U_g^E$. This can be illustrated by the following exercise, where it is shown that the cocycle of the map $g\mapsto U_g^E$ is entirely determined by the cocycles of $g\mapsto U_g^A$ and $g\mapsto U_g^B$.

\bex
Suppose the representations $g\mapsto U^A_g$ and $g\mapsto U_g^B$ have cocycles 
\be\label{cocycles}
\left\{e^{i\theta^A(g,h)}\right\}_{g,h\in G}\quad\text{and}\quad\left\{e^{i\theta^B(g,h)}\right\}_{g,h\in G}\;,
\ee 
respectively. Show that if~\eqref{fortitude} holds then $g\mapsto U_g^E$ has a cocycle \be\left\{e^{i\left(\theta^B(g,h)-\theta^A(g,h)\right)}\right\}_{g,h\in G}\;.\ee 
\eex

It is important to note that if $A=B$ in the definition given above, then the exercise shows that the representation $g\mapsto U_g^E$ is non-projective. Additionally, we emphasize that the irreps $\lambda\in\irr(U^E)$ used in the definition of the irreducible tensor operator may not necessarily be the same irreps that appear in the decompositions of $U_g^A$ or $U_g^B$. Therefore, the dimension of the system $E=\mathrm{span}\{|\lambda,m,x\ra^E\}$ depends on the irreps $\lambda\in\irr(U^E)$ that appear in the decomposition of $U_g^E$. Specifically, the components $u^{(\lambda)}_{mm'}(g)$ of $U^{(\lambda)}_g$ appear in the decomposition of $U^E_g$ and not in the decompositions of $U_g^A$ or $U_g^B$.

\begin{myt}{}
\begin{theorem}\label{covkra}
Let $\mE\in\ml(A\to B)$ be trace preserving. The map $\mE\in\cov_G(A\to B)$ if and only if there exists a projective unitary representation $g\mapsto U_g^E$ of $G$, and a corresponding orthonormal set of irreducible tensor operators $\{K_{\lambda,m,x}\}_{\lambda,m,x}$ in $\ml(A,B)$ such that
\be\label{lftex}
\mE(\rho)=\sum_{\lambda,m,x}K_{\lambda,m,x}\rho K_{\lambda,m,x}^*\quad\quad\forall\;\rho\in\ml(A)\;.
\ee
\end{theorem}
\end{myt}

\begin{proof}
We first prove that the channel given in~\eqref{lftex} is G-covariant. Indeed, for any $\rho\in\ml(A)$ we have
\ba
\mU_g^B\circ\mE\circ\mU^{*A}_g(\rho)&=\sum_{\lambda,m,x}U_g^BK_{\lambda,m,x}U_g^{*A}\rho \left(U_g^BK_{\lambda,m,x}U_g^{*A}\right)^*\\
\GG{\eqref{fortitude}}&=\sum_{\lambda,m,x,k,k'}u^{(\lambda)}_{km}(g)\bar{u}^{(\lambda)}_{k'm}(g)K_{\lambda,k,x}\rho K_{\lambda,k',x}^*\\
\GG{{\it U_g^{(\lambda)}}\;is\;a\;unitary\;matrix}&=\sum_{\lambda,x,k,k'}\delta_{kk'}K_{\lambda,k,x}\rho K_{\lambda,k',x}^*=\mE(\rho)\;.
\ea
Hence, $\mE$ is $\G$-covariant.

Conversely, suppose $\mE$ is a $G$-covariant quantum channel. Let $\{K_x\}_{x\in[n]}\in\ml(A,B)$ be a canonical\index{canonical representation}  Kraus decomposition of $\mE$ (see Corollary~\ref{canonical}). Since $\mE$ is $G$-covariant it follows that for all $g\in G$ and $\rho\in\ml(A)$
\be
\mE(\rho)=\mU_g^B\circ\mE\circ\mU_g^{A*}(\rho)=\sum_{x\in[n]}\left(U_g^B K_x U_{g}^{*A}\right)\rho \left(U_g^B K_x U_{g}^{*A}\right)^*\;.
\ee
Therefore, the set $\{U_g^B K_x U_{g}^{*A}\}_{x\in[n]}$ also form a canonical\index{canonical representation} Kraus decomposition of $\mE$. Now, recall from Sec.~\ref{osr} that every two operator sum representations of $\mE$ that have the same number of elements are related by a unitary matrix. Therefore, for any $g\in G$ there exists an $n\times n$  unitary matrix $U_g^E=\big(u_{xz}(g)\big)\in\ml(E)$ with $n\eqdef|E|$ such that for all $x\in[n]$, we have
\be\label{1518}
U_g^B K_x U_{g}^{*A}=\sum_{z\in[n]}u_{zx}(g)K_z\;.
\ee
Furthermore, since $\{K_z\}_{z\in[n]}$ are linearly independent (as they are orthonormal in the Hilbert-Schmidt inner product), it follows that for every $g\in\G$, there is a unique $U_g^E$ that satisfies the equation~\eqref{1518}. Additionally, using the notation given in~\eqref{cocycles} for the cocycles, we get that for all $g,h\in\G$
\ba
U_{gh}^B K_x U_{gh}^{*A}&=e^{i\left(\theta^B(g,h)-\theta^A(g,h)\right)}U_g^BU_h^B K_x U_{h}^{*A}U_{g}^{*A}\\
\GG{\eqref{1518}}&= e^{i\left(\theta^B(g,h)-\theta^A(g,h)\right)}\sum_{z\in[n]}u_{zx}(h)U_g^BK_z U_{g}^{*A}\\
\GG{Using~\eqref{1518}~once~more}&=e^{i\left(\theta^B(g,h)-\theta^A(g,h)\right)}\sum_{z,z'\in[n]}u_{zx}(h)u_{z'z}(g)K_{z'}\\
&=e^{i\left(\theta^B(g,h)-\theta^A(g,h)\right)}\sum_{z'\in[n]}(U^{E}_gU^{E}_h)_{z'x}K_{z'}
\ea
We therefore conclude that
\be
U^{E}_{gh}=e^{i\left(\theta^B(g,h)-\theta^A(g,h)\right)}U^{E}_gU^{E}_h\;.
\ee
That is, the mapping $g\mapsto U^{E}_g$ is a projective unitary representation of the group $G$. Finally, using the unitary freedom in the choice of the canonical\index{canonical representation} Kraus decomposition $\{K_z\}_{z\in[n]}$ of $\mE$ (see Exercise~\ref{canfree}), we choose it in such a way that $U^{E}_g$ is block-diagonal with respect to the irreps of $G$. In this basis, $U_g^E=\bigoplus_\lambda U_g^{(\lambda)}\otimes I^{C_\lambda}$ so we can denote the Kraus operators by $\{K_{\lambda,m,x}\}$ (with $\lambda$ the irrep label and $x$ the multiplicity index). This completes the proof. 
\end{proof}

\bex
Extend the theorem above to CP maps that are not necessarily trace preserving. That is, show that $\mE\in\cp(A\to B)$ is $\G$-covariant if and only if it can be expressed as in~\eqref{lftex}.
\eex

To illustrate the theorem above, we will provide a few examples. Let's begin with the case of a covariant unitary channel. As we mentioned earlier, a unitary channel $\mE(\cdot)=V(\cdot)V^*$, where $V:A\to A$ is a unitary matrix, is covariant if and only if~\eqref{15p7} holds for all $g\in\G$. Here, $g\mapsto\omega_g$ is a 1-dimensional unitary representation of $\G$. As we will illustrate now, the theorem mentioned above can be used to derive the same conclusion.

Indeed, since a unitary channel has only one Kraus operator, the unitary representation $g\mapsto U_g^E$ must be an irreducible representation (therefore, a single $\lambda$), one-dimensional (thus, a single $m$), and with no multiplicity (a single $x$). This implies that $|E|=1$, so we have $U_g^E=\omega_g$ for some $\omega_g\in\mbb{C}$ where $|\omega_g|=1$. Let us denote the single Kraus operator of $\mE$ by $V=K_{\lambda,m,x}$. In this case, the relation~\eqref{fortitude} can be expressed as follows:
\be
U_g^AVU_g^{*A}=\omega_gV\;.
\ee

It's worth noting that if there exists $g\in\G$ such that $\omega_g\neq 1$ (i.e., $V$ is not $\G$-invariant), then $\mG(V)=0$. To see why, consider taking the integral over $\G$ (with respect to the Haar measure) on both sides of the equation above:
\be
\mG(V)=cV\quad\text{where}\quad c\eqdef\int_\G dg\; \omega_g\;.
\ee
If $c\neq 0$, then we have $V=\frac1c\mG(V)=\mG\left(\frac1cV\right)$, which is $\G$-invariant. This implies that $\omega_g=1$ for all $g\in\G$, and $c=1$ in this case. Therefore, if $V$ is not $\G$-invariant, we must have $c=0$, and consequently, $\mG(V)=\0$.

As another example, let's consider the group $U(1)$. 
The Kraus operators of a $U(1)$ covariant channel can be labeled as $K_{k,\alpha}\in\ml(A)$, where $\alpha$ is the multiplicity index. Then, from~\eqref{fortitude} we get that
\be\label{formkop}
e^{i\theta \hat{N}}K_{k,\alpha}e^{-i\theta \hat{N}}=e^{ik\theta}K_{k,\alpha}\quad\quad\forall\;\theta\in U(1)\;.
\ee
Note that the irreducible representations of $U(1)$ are one-dimensional. As a result, the Kraus operators are not mixed with one another under the action of $U(1)$. This provides a significant simplification compared to the non-Abelian case. 

Any Kraus operator $K_{k,\alpha}$ that satisfies~\eqref{formkop} must have the form (see Exercise~\ref{shiftop})
\be\label{shifty}
K_{k,\alpha}=S_kD_{k,\alpha}\;,
\ee
where
\be
S_k\eqdef\sum_{n\in\mbb{Z}}|n+k\lr n|\;,
\ee
is the ``shift" operator, and $D_{k,\alpha}$ are diagonal operators in $\ml(A)$; i.e., $\la n|D_{k,\alpha}|n'\ra=0$ for $n\neq n'$. Note that in the infinite-dimensional Hilbert space $A$, the shift operator $S_k$ is unitary. Therefore, in the QRT of $U(1)$-asymmetry, the set of free unitary operations consists of diagonal unitaries, shift operators, and combinations of the two.

\bex\label{shiftop}
Prove~\eqref{shifty}. Hint: Substitute $K_{k,\alpha}=\sum_{n,n'}c_{nn'}^{(k,\alpha)}|n\lr n'|$ in~\eqref{formkop}.
\eex

\bex
Let $\mG$ be the U(1)-twirling map, and $S_k$ the shift operator. Show that $S_k$ is not U(1)-invariant by showing that $\mG(S_k)=0$. Still, we emphasize that $\mE(\cdot)=S_k(\cdot)S_k^*$ is U(1)-covariant.
\eex

\bex
Let $A$ be an $n$-dimensional Hilbert space, $\mbb{Z}_n$ be the cyclic group of $n$-elements, and let for any $k\in\mbb{Z}_n$ let 
\be
G_k\eqdef\sum_{x\in[n]}|x+k\;({\rm mod}\; n)\lr x|\;.
\ee
Find the general form of the Kraus operators constituting the operator sum representation of a $\mbb{Z}_n$-covariant channel with respect to the representation $k\mapsto G_k$.
\eex

Theorem~\ref{covkra} demonstrates that every $G$-covariant channel $\mE\in\cov_\G(A\to B)$ induces an auxiliary system $E$ and a projective unitary representation $g\mapsto U_g\in\ml(E)$ that corresponds to its Kraus decomposition with irreducible tensor operators. The terms ``induced space" and ``induced representation" can be used to refer to $E$ and $g\mapsto U_g^E$, respectively. These concepts are important in the study of symmetry in physics, as they enable us to understand the structure of $G$-covariant channels in terms of irreducible tensor operators. Furthermore, the induced space $E$ appears in the covariant version of Stinespring's representation.

\subsubsection{Covariant Stinespring Representation}

In Theorem~\ref{thm:stine}, we learned that it is possible to represent every quantum channel as the action of an isometry followed by a partial trace. In this context, we now demonstrate that the Stinespring representation of a covariant channel involves an isometry that is itself $\G$-invariant.

\begin{myt}{}
\begin{theorem}\label{covstine}
Let $\mE\in\ml(A\to B)$ be a linear map. The map $\mE\in\cov_G(A\to B)$ if and only if there exists a system $E$, a projective unitary representation $g\mapsto U_g^E$, and an (intertwiner) isometry $V:A\to BE$ such that $\mE(\rho)=\tr_{E}\left[V\rho V^*\right]$ and for all $g\in G$
 \be\label{ginvst}
 \left(U_g^B\otimes \overline{U}_g^{E}\right)V=VU_g^A\;.
\ee
\end{theorem}
\end{myt}
\begin{remark}
If $\mE$ is $G$-covariant then in the proof below we will see that the representation $g\mapsto U_g^E$ is given by the induced representation of $\mE$. The components of the matrix $\overline{U}_g^E=\left(U^{*E}_g\right)^T$ equals the complex conjugate of the corresponding components of $U_g^E$. Recall that $g\mapsto \overline{U}_g^E$ is also a projective unitary representation (see Exercise~\ref{pur}).
\end{remark}

\begin{proof}
Suppose $\mE$ has the form $\mE(\rho)=\tr_{E}\left[V\rho V^*\right]$, where the isometry $V$ satisfies~\eqref{ginvst}. Using the standard Stinesprings dilation theorem, we know that $\mE\in\cptp(A\to B)$. Moreover, for all $g\in\G$ and all $\rho\in\ml(A)$ we have
\ba
\mE^{A\to B}\circ\mU^A_g(\rho)&=\tr_{E}\left[V\mU^A_g(\rho) V^*\right]\\
\GG{\eqref{ginvst}}&=\tr_{E}\left[\mU^B_g\otimes\overline{\mU}^E_g\left(V\rho V^*\right)\right]\\
&=\mU^B_g\left(\tr_{E}\left[\left(V\rho V^*\right)\right]\right)\\
&=\mU^B_g\circ\mE^{A\to B}(\rho)\;,
\ea
where $\overline{\mU}^E_g(\cdot)\eqdef \overline{U}^E_g(\cdot)(U^E_g)^T$.

Conversely, suppose $\mE\in\cov_G(A\to B)$. Let $\{K_{\lambda,m,x}\}$ be its canonical\index{canonical representation} covariant Kraus decomposition as given in Theorem~\ref{covkra}, and define the isometry $V:A\to BE$ as
\be
V\eqdef\sum_{\lambda,m,x}K_{\lambda,m,x}\otimes|\lambda,m,x\ra^E
\ee
where $E\eqdef\spa\{|\lambda,m,x\ra^E\}$ is the induced space of $\mE$ decomposed according to the irreps of $G$ that appear in the induced representation of $\mE$. By definition, for all $\rho\in\ml(A)$ we have
\ba
\tr_E\left[V\rho V^*\right]&=\sum_{\lambda,m,x}K_{\lambda,m,x}\rho K_{\lambda,m,x}^*\\
\GG{\eqref{lftex}}&=\mE(\rho)\;.
\ea
Therefore, it is left to show that $V$ satisfies~\eqref{ginvst}. Indeed, taking $g\mapsto U_g^E$ to be the induced representation of $\mE$
we get
\ba
\left(U^B_g\otimes \overline{U}_{g}^{E}\right)VU_{g}^{*A}&=\sum_{\lambda,m,x}U_g^BK_{\lambda,m,x}U_g^{*A}\otimes \overline{U}_{g}^{E}|\lambda,m,x\ra^E\\
\GG{\eqref{fortitude}}&=\sum_{\lambda,x}\sum_{m,m'}u^{(\lambda)}_{m'm}(g)K_{\lambda,m',x}\otimes \overline{U}_{g}^{E}|\lambda,m,x\ra^E\\
\Gg{\substack{\sum_{m}u^{(\lambda)}_{m'm}(g)|\lambda,m,x\ra\\=\left(U_{g}^{E}\right)^T|\lambda,m',x\ra}}&=\sum_{\lambda,x}\sum_{m'}K_{\lambda,m',x}\otimes \overline{U}_{g}^{E}\left(U_{g}^{E}\right)^T|\lambda,m',x\ra^E\\
\Gg{\overline{U}_{g}^{E}\left(U_{g}^{E}\right)^T=I^E}&=V\;.
\ea 
Therefore, $V$ is an intertwiner isometry.
\end{proof}

When considering a channel $\mE\in\cov_\G(A\to A)$, a slightly different version of the covariant Stinespring dilation theorem is obtained. Note that in this case, not only is the output system $B$ replaced with the input system $A$, but the same projective unitary representation is also considered on both the input and output systems of $\mE$. This enables us to obtain a covariant Stinespring dilation theorem that involves a $G$-invariant unitary matrix.

\begin{myt}{}
\begin{theorem}\label{1321}
Let $\mE\in\ml(A\to A)$ be a linear map. The map $\mE$ is a $G$-covariant quantum channel, i.e. $\mE\in\cov_\G(A\to A)$, if and only if there exists a system $E$, $\G$-invariat state $|0\lr 0|\in\pure(E)$, and $\G$-invariant unitary $W:AE\to AE$ such that
\be\label{ginvst2}
\mE^{A\to A}(\rho^A)=\tr_{E}\left[W^{AE}\left(\rho^A\otimes |0\lr 0|^E\right) W^{*AE}\right]\;.
\ee
\end{theorem}
\end{myt}
\begin{remark}
The matrix $W^{AE}$ in the theorem above is $\G$-invariant with respect to the projective unitary representation $g\mapsto U_g^A\otimes \overline{U}_g^E$, where the representation $g\mapsto U_g^E$ is the induced representation of $\mE$. Moreover, from Theorem~\ref{vgv} it follows that a $\G$-invariant pure state always exists. To see it, using the same notations as in Theorem~\ref{vgv} we get for all $\psi\in\pure(E)$ the vector $\Pi|\psi\ra$ is proportional to a $\G$-invariant state.
\end{remark}
\begin{proof}
The proof that~\eqref{ginvst2} implies that $\mE$ is $\G$-covariant follows similar lines as the ones appear in the proof of Theorem~\ref{covstine} (we leave the details to Exercise~\ref{ex:1c1}). For the converse, if $\mE\in\cov_\G(A\to A)$, Theorem~\ref{covstine} states that there exists an intertwiner isometry $V:A\to AE$ such that
\be
 \left(U_g^A\otimes \overline{U}_g^{E}\right)V=VU_g^A
\quad
\text{and}  
\quad
\mE(\rho)=\tr_{E}\left[V\rho V^*\right]\;.
\ee
Now, let $|0\ra\in E$ be a $\G$-invariant state and define $\tA\eqdef\{|\psi^A\ra\otimes|0\ra^E\;:\;|\psi\ra\in A\}$. Clearly $\tA$ is a subspace of $AE$ and we define the isometry $\tV:\tA\to AE$ via
\be\label{vtv}
\tV|\psi\ra^A|0\ra^E\eqdef V|\psi\ra^A\;.
\ee
With this definition we have
\be\label{oiuy}
\mE(\rho)=\tr_{E}\left[\tV\left(\rho^A\otimes|0\lr 0|^E\right) \tV^*\right]
\ee
and from Exercise~\ref{ex2c2} it follows that $\tV$ is a $\G$-invariant isometry.
Denote by $\Pi\eqdef I^A\otimes |0\lr 0|^E$ the projection (in $AE$) to $\tA$ . Then, Theorem~\ref{compuni} states that there exists a $\G$-invariant unitary $W\in\muu(AE)$ such that $\tV\Pi=W\Pi$. We therefore get from~\eqref{oiuy} that~\eqref{ginvst2} must hold with this choice of $W$.
\end{proof}

\begin{exercise}\label{ex:1c1}
Show that if $W:AE\to AE$ is $\G$-invariant unitary matrix and $|0\lr 0|\in\pure(E)$ is $\G$ invariant state then the map $\mE$ as defined in~\eqref{ginvst2} is $\G$-covariant.
\end{exercise}

\begin{exercise}\label{ex2c2}
Show that the operator $\tV$ as defined in~\eqref{vtv} is a $\G$-invariant isometry. {\it Hint:\ Show that the operator $\tV\Pi:AE\to AE$ with $\Pi\eqdef I^A\otimes |0\lr 0|^E$ is $\G$-invariant.} 
\end{exercise}

\subsection{Alignment of References Frames}\label{rfa}\index{reference frame}

As we discussed earlier, when parties are separated, they often need to establish a common reference frame. For example, they may need to synchronize their clocks or align their Cartesian frames. While lacking a shared reference frame doesn't completely hinder tasks such as communication and computation, it does impose limitations, reducing their practical efficiency. This often requires more advanced encodings. Hence, parties may prioritize allocating communication resources to establish a shared reference frame initially. Later, they can utilize a standard encoding instead of continuously circumventing its absence with a relational encoding.

In tasks aimed at establishing a shared reference frame, parties can employ quantum particles to encode information regarding the relative orientation of their frames. For instance, spin-1/2 particles, like electrons, can encode the orientation of Cartesian frames, while exchanging quantum states of an optical mode can align phase references. Hence, in the realm of quantum reference frames, which involve quantum particles holding information about a shared reference frame, the usefulness of a quantum state is determined by the amount of information that can be extracted from it to establish such a reference frame.

The above discussion illustrates that the resource theory of quantum reference frames introduces certain aspects that differ from what we have encountered thus far. Specifically, when Alice and Bob do not share a reference frame, Alice can gain at least partial information about Bob's reference frame by receiving a resource in the form of a quantum state that encodes it. As a result, the set of free operations (i.e., $\G$-covariant operations) needs to be updated to incorporate this partial information. For instance, instead of using regular $\G$-twirling operations, weighted $\G$-twirling operations can be employed, taking into account that the parties have partial knowledge about the element $g\in\G$ that relates their reference frames.

We now discuss the general approach to align reference frames making the notions discussed above rigorous.  Consider two parties, Alice and Bob, who doesn't share a reference frame with $\G$ being the corresponding group describing the reference frame. The goal is for Bob to learn the element $g\in\G$ that relates between his reference frame and Alice's reference frame. To accomplish that, Alice sends Bob a quantum reference frame (e.g., spin-1/2 particles pointing in the $z$-direction of her Cartesian reference frame)  in a form of a quantum state $\rho\in\md(A)$.  From Bob's perspective, he received one of the states $\{U_g\rho U_g^*\}_{g\in\G}$, all occurring with uniform prior.

To determine the specific state he possesses, i.e., to identify the group element $g\in\G$, Bob conducts a POVM, $\{\Lambda_g\}_{g\in\G}$, on his system. Consequently, the probability that Bob guesses the group element as $g'$, given the actual element is $g$, is denoted by:
\be
q(g'|g)\eqdef\tr\left[\Lambda_{g'}U_g\rho U_g^*\right]\;.
\ee
In order to quantify how much information Bob gained after the measurement, consider first the case that $\G$ is a finite group. In this senario, we can use the probability that Bob guess $g$ correctly as our \emph{figure of merit} and maximize this function over all states and all POVMs. For a given state $\rho\in\md(A)$ and a POVM $\{\Lambda_g\}_{g\in\G}$, this probability is given by:
\be\label{gpfiducial}
\pr_{\text{guess}}\left(\rho,\{\Lambda_g\}_{g\in\G}\right)\eqdef\frac1{|G|}\sum_{g\in\G}p(g|g)=\frac1{|G|}\sum_{g\in\G}\tr\left[\Lambda_{g}U_g\rho U_g^*\right]\;.
\ee
Thus, Alice and Bob's objective is to maximize this guessing probability\index{guessing probability} across all possible $\rho$ (referred to as the \emph{fiducial} state) and all POVMs $\{\Lambda_g\}_{g\in\G}$.

Conversely, if $\G$ is a compact Lie group, the chance of Bob correctly inferring Alice's reference frame becomes infinitesimally small. In such instances, the direct likelihood or guessing probability cannot serve as an effective figure of merit. Instead, the \emph{maximum likelihood} of a correct guess is adopted as the figure of merit. This maximum likelihood, akin to the formula above but integrated over the group, is defined as:
\be\label{gpg}
\mu_{\max}\eqdef\max\int_\G dg\;p(g|g)=\max\int_\G dg\; \tr\left[\Lambda_{g}U_g\rho U_g^*\right]\;,
\ee
with the maximization conducted over all fiducial states $\rho\in\md(A)$ and all POVMs $\{\Lambda_g\}_{g\in\G}$. Given that the guessing probability\index{guessing probability} in \eqref{gpg} is linear in $\rho$, the maximal value can always be achieved with a pure state, allowing us to assume, for simplification, that the fiducial state $\rho=\psi$ is pure.

 From Theorem~\ref{thdco} it follows that the Hilbert space $A$ can be decomposed as $A=\bigoplus_{\lambda\in\irr(U)} B_\lambda\otimes C_\lambda$, where for each irrep $\lambda$, $B_\lambda$ denotes the representation space, and $C_\lambda$ denotes the multiplicity space. We will denote by $d_\lambda\eqdef|B_\lambda|$ and $m_\lambda\eqdef|C_\lambda|$. With these notations we can write any pure state $\psi\in\pure(A)$ as
\be\label{0wq}
|\psi\ra=\bigoplus_{\lambda\in\irr(U)}c_\lambda|\psi_\lambda\ra
\ee
where each $\psi_\lambda\in\pure(B_\lambda C_\lambda)$, and $c_\lambda\in\mbb{C}$ with $\sum_{\lambda\in\irr(U)}|c_\lambda|^2=1$. In the following theorem we consider a projective unitary representation, $g\mapsto U_g^A$, and denote by $d_\lambda\eqdef|B_\lambda|$ and $m_\lambda\eqdef|C_\lambda|$  the dimensions of the representation and multiplicity spaces (respectively) associated with the irrep $\lambda\in\irr(U)$. We also use the notation $n_\lambda\eqdef\min\{m_\lambda,d_\lambda\}$ for all $\lambda\in\irr(U)$.

 \begin{myt}{}
 \begin{theorem}\label{maxlike}
Using the notations above, the maximum likelihood $\mu_{\max}$ as defined in~\eqref{gpg} is given by 
\be 
\mu_{\max}=\sum_{\lambda\in\irr(U)}d_\lambda n_\lambda\;. 
\ee
 \end{theorem}
 \end{myt}
 
 \begin{proof}
 Let $\psi$ and $\{\Lambda_g\}_{g\in\G}$ be the optimal state and POVM that maximizes the maximum likelihood. Expressing $\psi$ as in~\eqref{0wq}, observe that due to the Schmidt decomposition of $|\psi_\lambda\ra$ there exists an orthogonal projector $\Pi_\lambda^{C_\lambda}$ such that $\tr\left[\Pi_\lambda^{C_\lambda}\right]=n_\lambda$ and 
 $I^{B_\lambda}\otimes \Pi^{C_\lambda}_\lambda|\psi_\lambda\ra=|\psi_\lambda\ra$. Therefore, the operator 
 \be
 \Pi\eqdef\bigoplus_{\lambda\in\irr(U)}I^{B_\lambda}\otimes \Pi_\lambda^{C_\lambda}
 \ee 
 satisfies $\Pi|\psi\ra=|\psi\ra$ and $\tr[\Pi]=\sum_{\lambda\in\irr(U)}d_\lambda n_\lambda$. From~\eqref{lamir} we also get that $[\Pi,U_g]=0$ for all $g\in\G$. Therefore, 
 \ba
 U_g\psi U_g^*&=U_g\Pi\psi \Pi U_g^*\\
 \Gg{[\Pi,U_g]=0}&=\Pi U_g\psi U_g^*\Pi\\
 \Gg{U_g\psi U_g^*\leq I^A}&\leq \Pi\;.
 \ea
Substituting the above inequality into~\eqref{gpg} we get that the maximum likelihood is bounded from above by:
 \ba
 \mu_{\max}&\leq \max\int_\G dg\; \tr\left[\Lambda_{g}\Pi\right]\\
 \Gg{\int_\G dg\; \Lambda_{g}=I}&=\tr[\Pi]=\sum_{\lambda\in\irr(U)}d_\lambda n_\lambda\;.
 \ea
 
For the converse, take $\rho=\psi$ to be the pure state
\be\label{optstate}
|\psi\ra=\frac1{\sqrt{\nu}}\bigoplus_{\lambda\in\irr(U)}\sqrt{d_\lambda n_\lambda}|\Phi_\lambda\ra\quad\text{where}\quad|\Phi_\lambda\ra\eqdef\frac{1}{\sqrt{n_\lambda}}\sum_{x\in[n_\lambda]}|x\ra^{B_\lambda}|x\ra^{C_\lambda}\;,
\ee
and $\nu\eqdef\sum_{\lambda\in\irr(U)}d_\lambda n_\lambda$. For the POVM we define $\Lambda_g\eqdef\nu U_g\psi U_g^*$, and observe that
\ba
\int_\G dg\;\Lambda_g&=\nu\mG(\psi)\\
\GG{Corollary~\ref{gcovcor}}&=\bigoplus_{\lambda\in\irr(U)}d_\lambda n_\lambda(\mR^{B_\lambda}\otimes\id^{C_\lambda})(\Phi_\lambda)\\
&=\bigoplus_{\lambda\in\irr(U)}I^{B_\lambda}\otimes\Pi^{C_\lambda}_\lambda,
\ea
where $\Pi^{C_\lambda}_\lambda\eqdef\sum_{x\in[n_\lambda]}|x\lr x|^{C_\lambda}$. Let $\Pi$ be the projector appearing on the right hand side of the equation above and observe that it satisfies $\Pi|\psi\ra=|\psi\ra$ and $[U_g,\Pi]=0$. Therefore, the set $\{I^A-\Pi\}\cup\{\Lambda_g\}_{g\in\G}$ is a POVM. Moreover, the measurement outcome corresponding to the element $I^A-\Pi$ occur with probability
\be
\tr\left[\left(I^A-\Pi\right)U_g\psi U_g^*\right]=\tr\left[U_g\left(I^A-\Pi\right)\psi U_g^*\right]=0\;\quad\quad\forall\;g\in\G.
\ee
 In other words, the outcome corresponding to $I^A-\Pi$ never occur. We therefore conclude that for this choice of POVM and fiducial state the maximum likelihood is given by 
 \be
\int_\G dg\tr\left[\Lambda_{g}U_g\psi U_g^*\right]=\nu\int_\G dg\tr\left[\psi^2\right]=\nu\;.
\ee
Hence $\mu_{\max}\geq \nu$ and since we already saw that $\mu_{\max}\leq \nu$ we conclude that $\mu_{\max}=\nu$. This completes the proof.
 \end{proof}

\subsubsection{Example: Alignment of Cartesian Frame}

As an example, we explore the alignment of an entire three-dimensional coordinate system using the communication of 1/2-spin particles. Suppose the Cartesian frames of Alice and Bob  are related by a group element $g\in SO(3)$. Every group element of $SO(3)$ represents a rotation by some angle $\theta$ and along a unit vector $\n\in\mbb{R}^3$.  In Sec.~\ref{0qubit0} we studied the representation of $SO(3)$ in $\mbb{C}^2$ and showed that $g\eqdef(\theta,\n)\mapsto U_g\eqdef e^{\frac12\theta\n\cdot\boldsymbol{\sigma}}$ is a unitary representation of $SO(3)$ on $A\eqdef\mbb{C}^2$.

Here we are interested in the representation of $SO(3)$ on the space $A^n\eqdef\left(\mbb{C}^{2}\right)^{\otimes n}$ for some integer $n$. We extend here the group of rotations $SO(3)$ to the group $SU(2)$ to allow for spinor representations. For simplicity, we will assume that $n$ is even,  and use some well-known results from representation theory. Specifically, the representation $g\mapsto U_g^{\otimes n}$ can be decomposed into a direct sum of $SU(2)$ irreps, labeled by the total angular momentum $j$ ranging from $0$ to $n/2$. The decomposition~\eqref{decoal},  for the representation of $SU(2)$ on $A^n$, has been extensively studied in representation theory, and is given by
\be
A^n=\bigoplus_{j=0}^{n/2}B_j\otimes C_j
\ee
where $|B_j|=2j+1$ and 
\be
|C_j|={n\choose n/2-j}\frac{2j+1}{n/2+j+1}\;.
\ee
From the formula above we see that $|C_j|\geq |B_j|$ for all $j$ except the case $j=n/2$ for which $|B_j|\geq |C_j|$. Therefore, according to Theorem~\ref{maxlike}, the maximum likelihood of a correct guess is given by
\be
\mu_{\max}=\sum_{j=0}^{n/2-1}(2j+1)^2+(n+1)=\frac16n^3+\frac56n+1\;.
\ee
Moreover, in this case the optimal state~\eqref{optstate} that achieves the maximum likelihood above is given by
\be
|\psi\ra=\frac1{\sqrt{\mu_{\max}}}\left(\sum_{j=0}^{n/2-1}(2j+1)|\Phi_j\ra+\sqrt{n+1}|n/2,n/2\ra\right)\;.
\ee
where 
\be
|\Phi_j\ra\eqdef\frac{1}{\sqrt{2j+1}}\sum_{m=-j}^j|j,m\ra^{B_j}|\phi_{m}^{C_j}\ra
\ee
and $\{|\phi_m^{C_j}\ra\}_{m\in\{-j,\ldots,j\}}$ is an orthonormal set of vectors in $C_j$. 

As a specific example, suppose $n=2$. In this case we have two irreps, corresponding to total angular momentum $j=0$ and $j=1$. In this case $|C_j|=1$ for both $j=0$ and $j=1$, and the representation space $B_0=\spa\{|0,0\ra\}$ is one dimensional spanned by the singlet state
\be
|0,0\ra^{B_0}\eqdef|\Psi_-^{A\tA}\ra\eqdef\frac1{\sqrt{2}}\left(|01\ra-|10\ra\right)
\ee
whereas $B_1$ is three dimensional spanned by the triplet states $|1,1\ra^{B_1}\eqdef|0\ra^A|0\ra^A$, $|1,0\ra^{B_1}=|\Psi_+^{A\tA}\ra$, and $|1,-1\ra\eqdef |1\ra^A|1\ra^A$. Therefore, the formula above implies to the two 1/2-spin particle state
\be
|\psi\ra=\frac12\left(|0,0\ra^{B_0}+\sqrt{3}|1,1\ra^{B_1}\right)=\frac12\left(|\Psi_-^{A\tA}\ra+\sqrt{3}|11\ra^{A\tA}\right)
\ee
achieves the largest maximum likelihood $\mu_{\max}=4$ as defined in~\eqref{gpg}. It is worth pointing out that the state above is not unique, and replacing $|1,1\ra^{B_1}$ with any other normalized state in $B_1$ would still give the maximum likelihood $\mu_{\max}=4$.

\bex
Let $\varphi\in\pure(B_1)$ be a pure state in the triplet space of two spin-1/2 particles. Show that there exists a POVM $\{\Lambda_g\}_{g\in\G}$ such that the state
\be
|\psi\ra=\frac12\left(|0,0\ra^{B_0}+\sqrt{3}|\varphi^{B_1}\ra\right)
\quad
\text{satisfies} 
\quad
\int_\G dg\; \tr\left[\Lambda_{g}U_g\rho U_g^*\right]=4\;.
\ee
\eex

\section{Quantification of Asymmetry}

The reformulation of symmetric dynamics in the context of a resource theory has significant implications. Often, the dynamics of a system can be so complex that a complete characterization of its evolution becomes impractical. Instead, by understanding the symmetries of the Hamiltonian, partial information about its dynamics can be gained. Noether's Theorem is one way to do this, as it states that a differentiable symmetry of the action of a physical system has a corresponding conservation law. However, Noether's theorem is not applicable to open systems\index{open systems}, and as a result, it does not capture all the consequences of symmetric evolution of mixed states.

Recently, it was demonstrated that the QRT of asymmetry provides a systematic approach to capturing all the outcomes of symmetric evolution. The fundamental idea is that the conserved quantities of closed systems can be substituted with resource monotones in open systems. These resource monotones\index{resource monotone} measure the degree of asymmetry in a quantum state, and they cannot increase under symmetric evolution.

In this section, we will explore different measures of asymmetry and their properties. We will focus on three different types of measures: 
\ben
\item \textbf{Measures of Quantum Frameness:} Measures that quantify how well a resource state can be utilized for the alignment of quantum reference frames.
\item \textbf{Relative Entropies of Asymmetry:} Measures that are derived from the general framework of resource theories using different choices of relative entropies.
\item \textbf{Derivatives of Asymmetry:} Measures that are unique to the theory of asymmetry and involve taking derivatives of quantum divergences.
\een

While some measures, like the relative entropy of asymmetry, are derived from the general framework of quantum resource theories, they have certain drawbacks, such as not being additive under tensor products and having zero regularized versions. To overcome these limitations, we introduce a new technique to construct measures of asymmetry that involve taking derivatives of quantum divergences. We refer to these measures as \emph{derivatives of asymmetry}.
The concept of derivatives of asymmetry encompasses significant measures like the quantum Fisher information and the Wigner-Yanase-Dyson skew information. These measures play pivotal roles in fields like quantum metrology, where precision and sensitivity are paramount. By exploring the derivatives of asymmetry, we can gain a deeper understanding of asymmetry in quantum systems and their applications in diverse fields beyond quantum information.

\subsection{Measures of Quantum Frameness}\index{frameness}

In this section we introduce a subset of measures of asymmetry that we call ``measures of quantum frameness". These are measures that appear as ``figure of merits" in the  context of reference frame alignment, and they quantify the uncertainty (or more correctly, the certainty) 
that Bob has about Alice's reference frame. In other words, they quantify the
distinguishability of the elements in the set $\ms(\rho)\eqdef\{U_g \rho U_g^*\}_{g\in\G}$. Note that since $\ms(\rho)=\ms(U_g\rho U_g^*)$, measures of quantum frameness must be invariant under the action $\rho\mapsto U_g\rho U_g^*$.

In any reference frame alignment scheme, Alice sends Bob a quantum state $\rho\in\md(A)$ that is described relative to her reference frame.  From Bob's perspective, he received one of the states $\{U_g\rho U_g^*\}_{g\in\G}$, all occurring with uniform prior.
To learn $g\in\G$, Bob performs a POVM, $\{\Lambda_g\}_{g\in\G}$ on his system so that the probability that he guesses $g'\in\G$, given that the actual element relating the frames is $g$, is given by
\be
q(g'|g)\eqdef\tr\left[\Lambda_{g'}U_g\rho U_g^*\right]\;.
\ee
In order to quantify how much information Bob gained after the measurement, let $X$ be the random variable corresponding to the element $g\in\G$ that relates between Alice and Bob's reference, and let $Y$ be the random variable associated with Bob's measurement outcome $g'\in\G$. With these notations, any measure of conditional uncertainty can be used to quantify the uncertainty of $X$ given that Bob has access to $Y$. Let $S(X|Y)_\q$, with $\q\eqdef\{q(g'|g)\}_{g,g'\in\G}$ denotes the probability distribution, be some measure of conditional certainty such as the negative of the conditional entropy $H(X|Y)_\q$. Then, a measure of quantum frameness associated with $S(X|Y)_\q$ is defined for all $\rho\in\md(A)$ as
\be\label{optframe}
\F(\rho)\eqdef\max_{\{\Lambda_g\}}S(X|Y)_\q
\ee
where the maximum is over all POVMs that Bob can perform on his system. In other words, Bob chooses a POVM that maximizes his certainty about $X$. We say that $F$ is a measure of quantum reference frame only if it can be written in this way.

\begin{myt}{}
\begin{theorem}
Every measure of quantum frameness is a measure of asymmetry.
\end{theorem}
\end{myt}

\begin{proof}
Let $\rho\in\md(A)$ and $\{\Gamma_{g}\}_{g\in\G}\subset\eff(A)$. Let $\mN\in\cov_\G(A\to B)$ and observe that since $\mN$ is $\G$-covariant we get that
\ba
\tr\left[\Gamma_{g'}U_g^B\mN(\rho) U^{*B}_g\right]
&=\tr\left[\Gamma_{g'}\mN\left(U_g^A\rho U^{*A}_g\right)\right]\\
&=\tr\left[\mN^*\left(\Gamma_{g'}\right)U_g\rho U_g^*\right]\;.
\ea
Therefore, 
\be
\F\big(\mN(\rho)\big)=\max_{\{\mN^*(\Gamma_g)\}}S(X|Y)_\q\leq \max_{\{\Lambda_g\}}S(X|Y)_\q=\F(\rho)\;,
\ee
where the first maximum is over all POVMs of the form $\{\mN^*(\Gamma_g)\}_{g\in\G}$ which is a subset of all possible POVMs $\{\Lambda_g\}_{g\in\G}$. This completes the proof.
\end{proof}

Note that in the proof above we did not need to use any of properties of the function $S(X|Y)_\q$. However, since $S(X|Y)_\q$ measures the conditional certainty of $X$ given that Bob has access to $Y$, measures of quantum frameness has additional properties.  
In Chapter~\ref{sec:ce} we saw that all measures of conditional uncertainty has to behaves monotonically under conditional majorization. However, in Chapter~\ref{sec:ce} we only considered finite dimensional, discrete probability distributions. Therefore, for finite groups in which $X$ and $Y$ are discrete random variables with $|X|=|Y|=|\G|$,
the function $S(X|Y)_\q$ must behaves monotonically  under conditional majorization. We called such functions in Sec.~\ref{sec:csc} conditionally Schur convex functions. 

The extension of conditional majorization to continuous probability distributions is a complex and currently unresolved issue in the field. However, various functions, such as the family of conditional R'enyi entropies (including the conditional von-Neumann\index{von-Neumann} entropy), can be utilized to measure the conditional entropy of continuous distributions. For the purpose of our discussion here, we only need to focus on one common property shared by all such functions that quantify conditional uncertainty: their invariance under the action of the group, both from the left and from the right.

Let's recall that $S(X|Y)_\q$ represents a function of the conditional probability distribution $q(g'|g)$. Now, suppose Bob rotates his reference frame by an element $h\in\G$, causing the corresponding element $g\in\G$ relating his frame to Alice's frame to change to $h^{-1}g$. Consequently, the outcome of the measurement $g'$ transforms to $h^{-1}g'$ under this change in Bob's reference frame. Since such a transformation should not affect Bob's uncertainty about $g$, we deduce that both distributions $q(g'|g)$ and $r(g'|g)\eqdef q(h^{-1}g'|h^{-1}g)$ represent the same conditional uncertainty. Hence, any function $S(X|Y)_\p$ that quantifies Bob's certainty about $X$ must be \emph{left-invariant}, meaning that $S(X|Y)_\q=S(X|Y)_\r$ holds for all conditional distributions $\q$ and all $h\in\G$.

Similarly, let's consider the scenario where Bob changes his reference frame such that $g\mapsto gh$ and $g'\mapsto g'h$. As before, such a transformation should not affect Bob's uncertainty about $g$. Consequently, both distributions $q(g'|g)$ and $r(g'|g)\eqdef q(g'h|gh)$ represent the same conditional uncertainty. Thus, any function $S(X|Y)_\q$ that quantifies Bob's certainty about $X$ must also be \emph{right-invariant}, meaning that $S(X|Y)_\q=S(X|Y)_\r$ holds for all conditional distributions $\q$ and all $h\in\G$.

Many of the functions $S(X|Y)_\q$ are not linear in $\q$ which makes the optimization in~\eqref{optframe} very difficult for such choices. We therefore focus here on measure of conditional certainty that are linear in $\q$. We start with the maximum likelihood that we already encountered in Sec.~\ref{rfa}

\subsubsection{The Maximum Likelihood}\index{maximum likelihood}

In the previous section we introduced a figure of merit, called maximum likelihood, that characterize how well a quantum state can be used to establish a shared reference frame. In particular, for a quantum state $\rho\in\md(A)$ we define
\be
\mu(\rho)\eqdef\max\int_\G dg\; \tr\left[\Lambda_{g}U_g\rho U_g^*\right]
\ee
where the maximum is over all POVMs $\{\Lambda_g\}_{g\in\G}\subset\eff(A)$. Observe that $\mu_{\max}\eqdef\max_{\rho\in\md(A)}\mu(\rho)$ is the maximum likelihood for Bob's correct guess of Alice's reference frame. 

\bex
Consider the maximum likelihood measure as defined above.
\ben
\item Show that $\mu(\rho)$ can be expressed as in~\eqref{optframe} with
\be\label{fxyllr}
S(X|Y)_\q\eqdef\int_\G dg\;q(g|g)\;.
\ee
\item Show that the function in~\eqref{fxyllr} is both left and right-invariant.
\een
\eex

In the theorem below we will see that the maximum likelihood $\mu(\rho)$ can be expresses in terms of the max relative entropy\index{max relative entropy}. Among other things, this result demonstrates that the function $\mu(\rho)$ behaves monotonically under $\G$-covariant operations and therefore can be used to define a measure for asymmetry. 
To be more precise, since for $\rho\in\inv_\G(A)$ we have $\mu(\rho)=1$, the function $\rho\mapsto\mu(\rho)-1$ is a measure of asymmetry (in fact, it can be shown to be an asymmetry monotone, see the exercise below).

\begin{myt}{}
\begin{theorem}\label{thm:mur}
Using the same notations as above, for any $\rho\in\md(A)$
\be
\log\mu(\rho)=\min_{\sigma\in\inv_\G(A)}D_{\max}\left(\rho\|\sigma\right)\;.
\ee
\end{theorem}
\end{myt}

\begin{proof}
For any POVM $\{\Lambda_g\}_{g\in\G}$ denote by
\be
\Lambda\eqdef\int_\G dg\;U_g^*\Lambda_{g}U_g\;,
\ee
so that
\be\label{mlik}
\int_\G dg\; \tr\left[\Lambda_{g}U_g\rho U_g^*\right]=\tr\left[\Lambda\rho\right]\;.
\ee
Since $\{\Lambda_g\}_{g\in\G}$ is a POVM we get that 
\be
\mG(\Lambda)=\int_\G dg\;\mG\left(U_g^*\Lambda_{g}U_g\right)
=\int_\G dg\;\mG(\Lambda_{g})
=\mG\left(\int_\G dg\;\Lambda_{g}\right)=\mG(I^A)=I^A\;.
\ee
Conversely, let $\Lambda\in\pos(A)$ be such that $\mG(\Lambda)=I^A$, and define , $\Lambda_g\eqdef U_g\Lambda U_g^*$ for every $g\in\G$, so that $\int_\G dg\;\Lambda_g=\mG(\Lambda)=I^A$.
By definition, this POVM $\{\Lambda_g\}_{g\in\G}$ satisfies~\eqref{mlik}.

We can therefore express $\mu(\rho)$ as the following SDP:
\be\label{opty}
\mu(\rho)=\max_{\substack{\Lambda\in\pos(A)\\\mG(\Lambda)=I}}\tr\left[\Lambda\rho\right]
\;.
\ee 
The above optimization problem is an SDP. As such, it has a dual given by (see Sec.~\ref{app:sdp})
\be\label{optyd}
\mu(\rho)=\min\Big\{t\geq0\;:\;t\sigma\geq\rho\;\;,\;\;\sigma\in\inv_\G(A)\Big\}\;.
\ee
That is,
\be
\log\mu(\rho)=\min_{\sigma\in\inv_\G(A)}D_{\max}\left(\rho\|\sigma\right)\;.
\ee
This completes the proof. 
\end{proof}

\bex
Use the duality relations discussed in Sec.~\ref{app:sdp} to show that the dual of~\eqref{opty} is given by the expression in~\eqref{optyd}.
\eex

\bex
Show that the maximum likelihood $\mu(\rho)$ is an asymmetry monotone.
\eex

We now use the expression in the theorem above to compute the maximum likelihood of a pure state $\psi\in\md(A)$. For this purpose, we will use the fact that
any quantum state $\sigma\in\inv_\G(A)$ has the form
\be\label{fqginv}
\sigma^A=\bigoplus_{\lambda\in\irr(U)}I^{B_\lambda}\otimes\sigma_\lambda^{C_\lambda}\;,
\ee
for some $\sigma_\lambda\in \pos(C_\lambda)$. Let $s_\lambda\eqdef \tr\left[\sigma_\lambda\right]$ and observe that since $\sigma$ is normalized we must have
\be\label{traces}
\sum_{\lambda\in\irr(U)}d_\lambda s_\lambda=1\;,
\ee
where $d_\lambda\eqdef|B_\lambda|$.
We will also use the fact that any pure state $\psi\in\pure(A)$ can be expressed as
\be\label{0wq00}
|\psi\ra=\bigoplus_{\lambda\in\irr(U)}\sqrt{p_\lambda}|\psi_\lambda\ra
\ee
where $\{p_\lambda\}_{\lambda\in\irr(U)}$ is a probability distribution, and each $\psi_\lambda\in\pure(B_\lambda C_\lambda)$ is a pure state in the tensor product of the representation and multiplicity spaces.

\begin{myt}{}
\begin{theorem}\label{thm:dutoi}
The maximum likelihood of a pure state $\psi\in\pure(A)$  is given by
\be\label{formulamu}
\log\mu(\psi)=H_{1/2}\big(\mG(\psi)\big)\eqdef2\log\tr\left[\sqrt{\mG(\psi)}\right]\;,
\ee
where $H_{1/2}$ is the R\'enyi entropy\index{R\'enyi entropy} of order $\alpha=1/2$.
\end{theorem}
\end{myt}

\begin{proof}
Let $t\in\mbb{R}_+$ and $\sigma\in\md(A)$ be such that $t\sigma\geq\psi$. This condition holds if and only if 
\be
tI^A\geq \sigma^{-1/2}|\psi\lr\psi|\sigma^{-1/2}\;,
\ee
(where all inverses are understood as generalized inverses).
The above condition holds if and only if $t\geq\la\psi|\sigma^{-1}|\psi\ra$. Therefore,
\be
\mu(\psi)=\min_{\sigma\in\inv_\G(A)}\la\psi|\sigma^{-1}|\psi\ra\;.
\ee
Now, a density matrix $\sigma\in\inv_\G(A)$ if and only if it has the form~\eqref{fqginv}. Therefore, the maximum likelihood of $\psi$ is given by
\be
\mu(\psi)=\min\sum_{\lambda\in\irr(U)}p_\lambda \big\la\psi_\lambda\big|I^{B_\lambda}\otimes \sigma^{-1}_\lambda\big|\psi_\lambda\big\ra
\ee
where the minimum is over all $\sigma_\lambda\in\pos(C_\lambda)$ whose traces satisfy~\eqref{traces}. Using the notations $s_\lambda\eqdef\tr\left[\sigma_\lambda\right]$ and $\eta_\lambda\eqdef\frac1{s_\lambda}\sigma_\lambda$, we split the minimization into two parts: first, we fix the numbers $\{s_\lambda\}_{\lambda\in\irr(U)}$ and minimize the expression over all $\eta_\lambda\in\md(C_\lambda)$, and then we minimize the resulting expression over all $\{s_\lambda\}_{\lambda\in\irr(U)}$ that satisfy~\eqref{traces}. That is,
\be
\mu(\psi)=\min_{\{s_\lambda\}}\sum_{\lambda\in\irr(U)}p_\lambda s_\lambda^{-1}\min_{\eta_\lambda\in\md(C_\lambda)}\big\la\psi_\lambda\big|I^{B_\lambda}\otimes \eta^{-1}_\lambda\big|\psi_\lambda\big\ra\;.
\ee
Denote the reduced density matrix of $\psi_\lambda$ by $\rho_\lambda^{C_\lambda}\eqdef\tr_{B_\lambda}\left[\psi^{B_\lambda C_\lambda}_\lambda\right]$, and
observe that
\ba
\big\la\psi_\lambda\big|I^{B_\lambda}\otimes \eta^{-1}_\lambda\big|\psi_\lambda\big\ra&=\tr\left[\rho_\lambda\eta^{-1}_\lambda\right]\\
\Gg{\tau_\lambda\eqdef\frac{\sqrt{\rho_\lambda}}{\tr\left[\sqrt{\rho_\lambda}\right]}}&=\left(\tr\left[\sqrt{\rho_\lambda}\right]\right)^2\tr\left[\tau_\lambda^2\eta^{-1}_\lambda\right]\\
\GG{Definition~\ref{def:petz}\;with\;\alpha=2}&=\left(\tr\left[\sqrt{\rho_\lambda}\right]\right)^2 2^{D_2(\tau_\lambda\|\eta_\lambda)}\;.
\ea
Therefore, the minimum of the expression above over all $\eta\in\md(C_\lambda)$ is obtained when $\eta=\tau_\lambda$. Hence, 
\be
\mu(\psi)=\min_{\{s_\lambda\}}\sum_{\lambda\in\irr(U)}p_\lambda s_\lambda^{-1} \left(\tr\left[\sqrt{\rho_\lambda}\right]\right)^2\;.
\ee
For the remaining of the optimization problem,  for each $\lambda\in\irr(U)$ we denote by $r_\lambda \eqdef p_\lambda \left(\tr\left[\sqrt{\rho_\lambda}\right]\right)^2$ and $q_\lambda\eqdef\frac1s s_\lambda$, where $s\eqdef \sum_{\lambda\in\irr(U)}s_\lambda$.
Observe that from~\eqref{traces} we get that $s^{-1}=\sum_{\lambda\in\irr(U)}d_\lambda q_\lambda$. Therefore, with these notations we get that
\be
\mu(\psi)=\min\sum_{\lambda\in\irr(U)}d_\lambda q_\lambda\sum_{\lambda\in\irr(U)}r_\lambda q_\lambda^{-1}\;.
\ee
where the minimum is over all probability distributions $\{q_\lambda\}_{\lambda\in\irr(U)}$.
From the Cauchy-Schwarz inequality we get that the minimum is given by
\be
\mu(\psi)=\Big(\sum_{\lambda\in\irr(U)}\sqrt{d_\lambda r_\lambda}\Big)^2\;.
\ee
Taking the log on both sides and substituting the expression $p_\lambda \left(\tr\left[\sqrt{\rho_\lambda}\right]\right)^2$ for $r_\lambda$ gives
\be
\log\mu(\psi)=2\log\sum_{\lambda\in\irr(U)}\sqrt{d_\lambda p_\lambda}\tr\left[\sqrt{\rho_\lambda}\right]\;.
\ee
Finally, observe the expression inside the log on the right-hand side of the equation above is given by $\tr\left[\sqrt{\mG(\psi)}\right]$. Hence, $\log\mu(\psi)$ is given by~\eqref{formulamu}.
This completes the proof.
\end{proof}

\bex
Use the formula in~\eqref{formulamu} to prove Theorem~\ref{maxlike}.
\eex

\subsubsection{The Weighted Maximum Likelihood}

In Sec.~\ref{rfa} we highlighted the importance of maximizing the likelihood of making the correct guess. However, when it comes to practical considerations, relying solely on the maximum likelihood density as a measure of success is not particularly advantageous, as it only rewards a completely accurate guess. A
 more general approach is to employ a payoff function $f:\G\times\G\mapsto\mbb{R}_+$, denoted as $f(g', g)$, which determines the reward or payoff associated with guessing group element $g'$ when the true group element that relates between the parties' reference frames is $g$. By assuming a uniform prior for the signal states, the figure of merit for the alignment scheme can be defined as the average payoff:
\be\label{mufr}
\mu_f(\rho)\eqdef\max\int_\G dg\int_\G dg'\; f(g,g')q(g'|g)\quad\quad\forall\;\rho\in\md(A)\;,
\ee
where the maximum is over all POVMs $\{\Lambda_{g'}\}_{g'\in\G}\subset\eff(A)$, and $q(g'|g)\eqdef \tr\left[\Lambda_{g'}U_g\rho U_g^*\right]$ is the probability of guessing $g'$ given that the actual element that relates between the parties' reference frames is $g$. Note that by taking $f(g',g)=\delta(g'g^{-1})$ to be the Dirac delta function we can get back the maximum likelihood function. We therefore call the function above the weighted maximum likelihood.

Note that we can write $\mu_f(\rho)=\max S(X|Y)_\q$ as given in~\eqref{optframe}, where $\q\eqdef\{q(g'|g)\}_{g,g'\in\G}$, the maximum is over all POVM as above, and $S(X|Y)_\q=L_f(\q)$ is taken to be the linear functional
\be
L_f(\q)\eqdef\int_\G dg\int_\G dg'\; f(g,g')q(g'|g)\;.
\ee
Since the function $L_f(\q)$ represents the certainty that Bob has about $g$, it has to be (see the discussion above) both left and right invariant. Fix $h\in\G$ and denote by $r(g'|g)\eqdef q(hg'|hg)$. Since $L_f$ is left invariant we have $L_f(\r)=L_f(\q)$ for all conditional distributions $\p$. Since
\be
L_f(\r)\eqdef\int_\G dg\int_\G dg'\; f(g,g')q(h^{-1}g'|h^{-1}g)=\int_\G dg\int_\G dg'\; f(hg,hg')q(g'|g)\;,
\ee
the condition  $L_f(\r)=L_f(\q)$ is equivalent to
\be
\int_\G dg\int_\G dg'\; \Big(f(hg,hg')-f(g,g')\Big)q(g'|g)=0\;.
\ee
As the above condition holds for all conditional probability distributions $q(g'|g)$ , we conclude that $f$ itself is \emph{left-invariant}. That is,
\be\label{li}
f(hg,hg')=f(g,g')\quad\quad\forall\;g,g',h\in\G\;.
\ee
The left-invariance\index{invariance} property of $f$ is consistent with the intuition that the payoff function should exclusively depend on the \emph{relative} transformation between the transmitted state, characterized by the group element $g$, and the measurement outcome, represented by the group element $g'$.

Following similar arguments as above, the right-invariance property of $L_f$ implies that the function $f$ itself is also \emph{right-invariant}, that is, 
\be
f(kh^{-1},k'h^{-1})=f(k,k')\quad\quad\forall\;k,k',h\in\G\;.
\ee
The fact that $f$ is both right and left invariant has the following consequences. 

First, by taking $h=g^{-1}$ in~\eqref{li} we get that
\be
f(g,g')=f(e,g^{-1}g')\;.
\ee
That is, $f(g,g')$ can be viewed as a function of $g^{-1}g'$. We will denote this function by $p$ so that $f(g,g')=p(g^{-1}g')$. Now, since $f(g,g')$ is also right invariant we get that for all $h,g\in\G$ we have $p(hgh^{-1})=p(g)$. That is, $p$ is a class function as introduced in Definition~\ref{funclass}.

Since $f$ is non-negative so is $p$, and consequently, it is natural to normalize $p$ such that $\int_\G dg\; p(g)=1$. That is, $\{p(g)\}_{g\in\G}$ is a probability distribution over the group. 
Moreover, for a function $f$ that is both left and right invariant we have
\ba
\mu_f(\rho)&=\max\int_\G dg\int_\G dg'\; p(g^{-1}g')\tr\left[\Lambda_{g'}U_g\rho U_g^*\right]\\
\Gg{h\eqdef g^{-1}g'}&=\max\int_\G dg'\int_\G dh\; p(h)\tr\left[\Lambda_{g'}U_{g'}U_{h}^*\rho U_hU_{g'}^*\right]\\
\GG{renaming\;{\it g'}\;as\;{\it g}}&=\max\int_\G dg\;\tr\left[U_g^*\Lambda_{g}U_g\;\mG_\p(\rho)\right]\;,
\ea
where $\mG_\p$ is the weighted $\G$-twirling as defined in~\eqref{15p30i}.
We therefore conclude that 
\be
\mu_f(\rho)=\mu\big(\mG_\p(\rho)\big)=\min_{\sigma\in\inv_\G(A)}D_{\max}\left(\mG_\p(\rho)\big\|\sigma\right)\;,
\ee
where the last equality follows from Theorem~\ref{thm:mur}

\bex
Explain why for $f$ that is not left and right invariant, the function $\mu_f$ is not necessarily a measure of asymmetry.
\eex

\subsection{The Relative Entropy of Asymmetry}\index{relative entropy of asymmetry}

For every normalized quantum divergence $\D$ the ``distance" (as measured by $\D$) of a state $\rho\in\md(A)$ to the set $\inv_\G(A)$ is a measure of asymmetry. That is, the function
\be
\asy_\D(\rho)=\min_{\sigma\in\inv_\G(A)}\D(\rho\|\sigma)
\ee
is a measure of asymmetry. For certain choices of the divergence $\D$, the function above can be hard to compute. However, for the relative entropy it has a very simple form.
 
For any $\alpha\in[0,2]$, the $\alpha$-R\'enyi relative entropy of asymmetry is defined as
\be\label{asydal}
\asy_{\alpha}\left(\rho\right)\eqdef\min_{\sigma\in\inv_\G(A)}D_{\alpha}(\rho\|\sigma)\quad\quad\forall\;\rho\in\md(A)\;.
\ee
In Exercise~\ref{affcon}  you showed  that if $\rho$ is $\G$-invariant then for all $\alpha\in[0,2]$ the state $\rho_\alpha\eqdef\rho^\alpha/\tr[\rho^\alpha]$ is also $\G$ invariant. Therefore, from Theorem~\ref{cfrelative} it follows that the $\alpha$-R\'enyi relative entropy of asymmetry is given by
\ba\label{cfae}
\asy_{\alpha}\left(\rho\right)&=\frac{1}{\alpha-1}\log\left\|\mG\left(\rho^\alpha\right)\right\|_{1/\alpha}\\
\GG{\eqref{cutefor}}&=H_{1/\alpha}\big(\mG(\rho_\alpha)\big)-H_\alpha(\rho)\;,
\ea
where $H_\alpha$ is the $\alpha$-R\'enyi entropy, and $\mG$ is the $\G$-twirling map that is also the resource destroying map\index{resource destroying map} of the QRT of asymmetry. The special case of $\alpha=1$ is also known as the $\G$-asymmetry of the state $\rho\in\md(A)$ and is given by
\be\label{defasy}
\asy(\rho)=H\big(\mG(\rho)\big)-H(\rho)\;.
\ee

From Theorem~\ref{thm:mur} it follows that the log of the maximum likelihood, $\log\mu(\rho)$, can be viewed as the max relative entropy\index{max relative entropy} of asymmetry, in which $D_\alpha$ in~\eqref{asydal} is replace by $D_{\max}$. Since $D_{\max}$ is the largest relative entropy, the formula in~\eqref{cfae} with $\alpha=2$ can be used to provide a lower bound for $\log\mu(\rho)$. Specifically, we have
\be
\log\mu(\rho)\geq H_{1/2}\left(\mG\left(\frac{\rho^2}{\tr[\rho^2]}\right)\right)-H_2(\rho)\;.
\ee
Remarkably, due to Theorem~\ref{thm:dutoi}, the inequality above becomes an equality on \emph{all} pure states.

Despite the elegant expression above for the $\G$-asymmetry, in general, the $\G$-asymmetry is not additive under tensor products. In fact, in the following theorem we show that its regularization\index{regularization} is zero!
\begin{myt}{}
\begin{theorem}\label{asygasy}
Let $\G$ be a finite or compact Lie group and let $\rho\in\md(A)$. Then,
\be
\lim_{n\to\infty}\frac1n\asy\left(\rho^{\otimes n}\right)=0\;.
\ee
\end{theorem}
\end{myt}

\begin{remark}
The theorem above underscores a notable constraint associated with using the $\G$-asymmetry as a measure of asymmetry in quantum systems. It signals the necessity to investigate other measures capable of surmounting this limitation, particularly in the asymptotic regime where numerous copies of asymmetric states are considered. We will see that venturing into alternative measures will pave the way for a broader and more nuanced comprehension of asymmetry's nature and characteristics when approached from the perspective of the asymptotic domain.
\end{remark}

\begin{proof}
According to~\eqref{15p15n} the action of the $\G$-twirling on the state $\rho$ is given by
\be
\mG(\rho)=\sum_{x\in[d]}p_x U_{g_x}(\rho)U_{g_x}^*\;,
\ee
where $d$ is an integer satisfying $d\leq m^4$ (see Exercise~\ref{ubdim}). Therefore, from the von-Neumman property~\eqref{convone} we get that
\be
H\big(\mG(\rho)\big)\leq H(\rho)+H(\p)\leq H(\rho)+\log(d)\;.
\ee
Thus, combining this with the definition in~\eqref{defasy} gives $\asy(\rho)\leq\log(d)$. 

Now, fix $n\in\mbb{N}$ and consider the action of the $\G$-twirling on $\rho^{\otimes n}$:
\be
\mG_n\left(\rho^{\otimes n}\right)\eqdef \int_\G dg\;U_g^{\otimes n}\rho^{\otimes n} U_g^{\otimes n}\;.
\ee
Observe that the support of $\rho^{\otimes n}$ is a subspace of the symmetric subspace $\sym_n(A)$. Thus, we can view $\rho^{\otimes n}$ as a positive semidefinite operator acting on $\sym_n(A)$. Moreover, the map $g\mapsto U_g^{\otimes n}$ can also be seen as a projective unitary representation of $\G$ on the space $\sym_n(A)$. Therefore, if we repeat the same steps that led to the inequality $\asy(\rho)\leq\log(d)$ but with $\rho^{\otimes n}$ instead of $\rho$, we obtain:
\be
\asy\left(\rho^{\otimes n}\right)\leq\log(d_n)\;,
\ee
where $d_n$ is an integer no greater than the dimension of $\sym_n(A)$ to the power four (see~\eqref{15p22}). Combining this with the formula~\eqref{dsim} for the dimension of the symmetric subspace we arrive 
at
\ba\label{growlog}
\asy\left(\rho^{\otimes n}\right)&\leq4\log{n+m-1\choose n}\\
\GG{\eqref{typeb}}&\leq 4m\log(n+1)\;.
\ea
Hence,
\be
\lim_{n\to\infty}\frac1n\asy\left(\rho^{\otimes n}\right)\leq 4m\lim_{n\to\infty}\frac1n\log(n+1)=0\;.
\ee
This completes the proof.
\end{proof}

To illustrate the theorem mentioned above, let's consider the group U(1) and the pure state $\psi$ belonging to $\pure(A)$, given by
\be
|\psi^A\ra=\sum_{x\in[m]}c_x|x\ra\;,
\ee
where $\{|x\ra\}_{x\in[m]}$ are the eigenvectors of the number operator $\hat{N}$, and each $c_x\in\mbb{C}$. We can express $n$ copies of $\psi$ as
\be
|\psi^{\otimes n}\ra=\sum_{j\in[mn]}a_j|\phi_j^{A^n}\ra
\ee
where $a_j\in\mbb{C}$ and $|\phi_j^{A^n}\ra$ is the eigenvector of the total number operator $\hat{N}_\tot$ corresponding to the eigenvalue $j$, for each $j\in[mn]$. By applying the $\G$-twirling to $\psi^{\otimes n}$, we obtain (see~\eqref{n15p18})
\be
\mG_n\left(\psi^{\otimes n}\right)=\sum_{j\in[mn]}|a_j|^2\phi_j^{A^n}\;.
\ee
Denoting by $\p\in\prob(mn)$ with components $p_j\eqdef|a_j|^2$ for each $j\in[mn]$, we conclude that
\be
H\left(\mG_n\left(\psi^{\otimes n}\right)\right)=H(\p)\leq\log(mn)\;.
\ee
Therefore,
\be
\lim_{n\to\infty}\frac1nH\left(\mG_n\left(\psi^{\otimes n}\right)\right)\leq\lim_{n\to\infty}\frac1n\log(nm)=0\;.
\ee
The key observation in this example is that the rank of $\mG\left(\psi^{\otimes n}\right)$ grows linearly with $n$.

\subsubsection{The Weighted $\G$-Asymmetry}

Using the weighted $\G$-twirling, we define the wighted $\G$-asymmetry as:
\be\label{wga}
\asy_\p(\rho)\eqdef H\big(\mG_\p(\rho)\big)-H(\rho)\;.
\ee
\begin{myt}{}
\begin{theorem}
The weighted $\G$-asymmetry as defined in~\eqref{wga} is a measure of asymmetry.
\end{theorem}
\end{myt}
\begin{proof}
We begin by expressing $\asy_\p(\rho)$ as the mutual information of the state $\sigma^{XA}$, defined as
\be\label{sigxaeq}
\sigma^{XA}\eqdef\sum_{x\in[d]}p_x|x\lr x|^X\otimes U_{g_x}\rho^A U_{g_x}^*\;,
\ee
where $X$ is a classical system of dimension $d$.
By definition, the mutual information is given by (see~\eqref{mi}) 
\be
D\left(\sigma^{XA}\big\|\sigma^A\otimes \sigma^X\right)=H\left(\sigma^X\right)+H\left(\sigma^A\right)-H\left(\sigma^{XA}\right)\;,
\ee
where $\sigma^A=\mG_\p\left(\rho^A\right)$ and $H\left(\sigma^X\right)=H(\p)$. From Exercise~\ref{ex:ubent}, we have 
\ba
H\left(\sigma^{XA}\right)&=H(\p)+\sum_xp_xH\left(U_{g_x}\rho^A U_{g_x}^*\right)\\
\Gg{H\left(U_{g_x}\rho^A U_{g_x}^*\right)=H\left(\rho^A\right)}&=H(\p)+H\left(\rho^A\right)\;.
\ea
Combining these results, we get
\be\label{15p74}
\asy_\p\left(\rho^A\right)=D\left(\sigma^{XA}\big\|\sigma^X\otimes \sigma^A\right)\;.
\ee
Next, let $\mN\in\cov_\G(A\to B)$ and observe that since $\mN^{A\to B}\circ\mU_g^A=\mU_g^B\circ\mN^{A\to B}$ for all $g\in\G$, we have
\ba
\mN^{A\to B}\left(\sigma^{XA}\right)&=\mN^{A\to B}\circ\mE^{A\to XA}\left(\rho^{A}\right)\\
&=\sum_{x\in[d]}p_x|x\lr x|^X\otimes \mU_{g_x}^B\circ\mN^{A\to B}\left(\rho^{A}\right)\;.
\ea
Therefore,
\ba\label{15p76i}
\asy_\p\left(\mN^{A\to B}\left(\rho^A\right)\right)&=D\left(\mN^{A\to B}\left(\sigma^{XA}\right)\big\| \sigma^X\otimes\mN^{A\to B}\left(\sigma^A\right)\right)\\
\GG{DPI}&\leq D\left(\sigma^{XA}\big\|\sigma^X\otimes \sigma^A\right)\\
&=\asy_\p\left(\rho^A\right)\;.
\ea
This completes the proof.
\end{proof}

\subsubsection{The Mutual Information of Asymmetry}\index{mutual information}

The relation~\eqref{15p74} can be generalized to an arbitrary divergence $\D$ in the following way. For any $\rho \in \md(A)$ and $\p \in \prob(d)$, we define the \emph{mutual information of asymmetry} of $\rho^A$ with respect to $\p$ as:
\be\label{15p77i}
\mbb{I}_{\p}\left(\rho^A\right)\eqdef\D\left(\sigma^{XA}\big\|\sigma^X\otimes\sigma^A\right)\;,
\ee
where  $\sigma^{XA}$ is defined as in~\eqref{sigxaeq}.
Note that the same argument used in~\eqref{15p76i} can be repeated with $\D$ replacing $D$. Therefore, $\mbb{I}_{\p}$ is also a measure of asymmetry.

By tracing out system $X$ in~\eqref{15p77i}, we obtain the function (recall that the marginal $\sigma^A=\mG_\p\left(\rho^A\right)$)
\be
{\A}_{\p}(\rho)\eqdef\D\left(\rho\big\|\mG_\p(\rho)\right)\quad\quad\forall\;\rho\in\md(A)\;.
\ee
This function is also a measure of asymmetry, since if $\rho\in\inv_\G(A)$ then $\mbb{A}_\p(\rho)=0$, and for any $\mN\in\cov_\G(A\to B)$ we have
\ba
{\A}_{\p}\big(\mN(\rho)\big)&=\D\left(\mN(\rho)\big\|\mG_\p\circ\mN(\rho)\right)\\
\GG{\mN\;is\;\G-covariant}&=\D\left(\mN(\rho)\big\|\mN\circ\mG_\p(\rho)\right)\\
\GG{DPI}&\leq \D\left(\rho\|\mG_\p(\rho)\right)={\A}_{\p}(\rho)\;.
\ea
Therefore, $\A_\p$ is a valid measure of asymmetry.

\bex
Show that for every divergence $\D$, all $\rho\in\md(A)$, and all $\p\in\prob(d)$ we have
\be
{\A}_{\p}(\rho)\leq \mbb{I}_{\p}\left(\rho\right)\;.
\ee
\eex

\subsection{Derivatives of Asymmetry}\index{derivatives of asymmetry}

Consider the last measure of asymmetry $\mbb{A}_\p$ that we discussed in the previous subsection, and take $\p$ to be the vector $\p=(1,0,\ldots,0)^T$. Further, denote by $g\eqdef g_1$ so that $\A_\p$ can be denoted as $\A_g$ and is given by 
\be
{\A}_{g}(\rho)\eqdef\D\left(\rho\big\|U_g\rho U_g^*\right)\quad\quad\forall\;\rho\in\md(A)\;.
\ee

Since a Lie group is a differentiable manifold, we can take $g$ to be infinitesimally close to the identity element. Specifically, we can always choose $U_g=e^{i t \Lambda}$, where $\Lambda$ is some generator of the representation $g\mapsto U_g$, and $t\geq 0$ is some phase. The derivative of asymmetry with respect to the divergence $\D$ and generator $\Lambda$ is defined for all $\rho\in\md(A)$ as 
\be\label{15p83i}
\D_\Lambda(\rho)\eqdef\frac d{dt}\D\left(\rho\big\|e^{it \Lambda}\rho e^{-it \Lambda}\right)\Big|_{t=0}\;.
\ee
As we will see shortly that quite often the derivative on the right-hand side yields the constant zero function. In such cases, $\D_\Lambda(\rho)$ is defined in terms of the second derivative as
\be
\D_\Lambda(\rho)\eqdef\frac12\frac {d^2}{dt^2}\D\left(\rho\big\|e^{it \Lambda}\rho e^{-it \Lambda}\right)\Big|_{t=0}\;.
\ee
\bex
Show that the derivative of asymmetry as defined above is a measure of asymmetry. Hint: Use the fact that for each $g\in\G$ the function $\A_g$ is a measure of asymmetry.
\eex

In order to compute the derivatives above, we will use the expension 
\be
e^{it \Lambda}\rho e^{-it \Lambda}=\rho+it[\Lambda,\rho]-\frac12t^2\big[\Lambda,[\Lambda,\rho]\big]+O(t^3)\;.
\ee
It will be convenient to use the notations $\sigma\eqdef i[\Lambda,\rho]$ and $\eta\eqdef-\frac12\big[\Lambda,[\Lambda,\rho]\big]$ so that
\be
e^{it \Lambda}\rho e^{-it \Lambda}=\rho+t\sigma+t^2\eta+O(t^3)\;.
\ee
We now use this expansion to compute the derivatives of asymmetry for several examples.

\subsubsection{Differential Trace Distance of Asymmetry}

As our first example, let $\D=T$ be the trace distance.
In this case, $\D_\Lambda=T_\Lambda$ is called the \emph{differential trace distance of asymmetry} and is given by
\ba
T_\Lambda(\rho)&=\lim_{t\to 0^+}\frac1t T\left(\rho,\rho+t\sigma\right)\\
&=\frac12\lim_{t\to 0^+}\frac1t\big\|t\sigma\big\|_1\\
&=\frac12\big\|[\rho,\Lambda]\big\|_1
\ea
The differential trace distance  measures the asymmetry in a state $\rho$ relative to a subgroup of $\G$ associated with a generator $\Lambda$. This measure depends on the coherence of $\rho$ over the eigenspaces of $\Lambda$, which is indicated by the non-zero commutator $[\rho,\Lambda]$. The question then arises as to which norm should be used to measure the commutator $[\rho,\Lambda]$ and thus, the asymmetry of $\rho$. While the answer to this question may not be immediately apparent, the above discussion indicated that the trace norm is the most appropriate measure for this purpose.

\bex
Let $\psi\in\pure(A)$ and $\Lambda\in\herm(A)$. Show that 
\be
T_\Lambda(\psi)=\sqrt{\la\psi|\Lambda^2|\psi\ra-\la\psi|\Lambda|\psi\ra^2}\;.
\ee
That is, on pure states, the differential trace distance of asymmetry reduces to the variance of the observable $\Lambda$.
\eex

\subsubsection{The Wigner-Yanase-Dyson Skew Information}\index{Wigner-Yanase-Dyson}

For our second example, we take $\D=D_\alpha$ to be the Petz\index{Petz} quantum R\'enyi divergence\index{R\'enyi divergence} of over $\alpha\in[0,2]$ (see Definition~\ref{def:petz}). In this case, we will see that the first derivative in~\eqref{15p83i} is zero, so we will have to expend the function $D_\alpha\left(\rho\big\|e^{it \Lambda}\rho e^{-it \Lambda}\right)$  up to second order in $t$. Up to second order in $t$ we have
\ba
D_\alpha\left(\rho\big\|\rho+t\sigma+t^2\eta\right)=\frac{1}{\alpha-1} \log\tr\left[\rho^\alpha\left(\rho+t\sigma+t^2\eta\right)^{1-\alpha}\right]\;.
\ea
In what follows, we will consider the spectral decomposition of $\rho^A=\sum_{x\in[m]}p_x|x\lr x|^A$ (where $\{|x\ra\}_{x\in[m]}$ is the basis of $A$ consisting of the eigenvectors of $\rho^A$), and make use of the divided difference approach discussed in Appendix~\ref{A:DD}. Particularly, the trace above has the form given in Corollary~\ref{cordd} with $g(t)\eqdef t^\alpha$ and $f(t)\eqdef t^{1-\alpha}$. Therefore, the function $h(t)$ as defined in Corollary~\ref{cordd} is given by
\be
h(t)\eqdef g(t)f'(t)=1-\alpha\;,
\ee
so that $h(\rho)=(1-\alpha)I^A$ is a constant function. As such, $\tr\left[h(\rho)\sigma\right]=(1-\alpha)\tr[\sigma]=0$ and similarly $\tr\left[h(\rho)\eta\right]=(1-\alpha)\tr[\eta]=0$. Observe further that since $h$ is a constant function, $\mL_h(\sigma)=0$. Substituting all this into Corollary~\ref{cordd} we conclude that 
\ba
D_\alpha\left(\rho\big\|\rho+t\sigma+t^2\eta\right)&=\frac{1}{\alpha-1} \log\left(1-\frac12t^2\tr\left[\mL_f(\sigma)\mL_g(\sigma)\right]\right)+O(t^3)\\
&=\frac12\frac{t^2}{1-\alpha}\tr\left[\mL_f(\sigma)\mL_g(\sigma)\right]+O(t^3)\;,
\ea
where the self-adjoint\index{self-adjoint} linear maps $\mL_f,\mL_g\in\herm(A\to A)$ are defined by (see Appendix~\ref{A:DD} for more details)
\ba
\la x|\mL_f(\sigma)|y\ra&=\frac{p_x^{1-\alpha}-p_y^{1-\alpha}}{p_x-p_y}\la x|\sigma|y\ra\\
&=-i\left(p_x^{1-\alpha}-p_y^{1-\alpha}\right)\la x|\Lambda|y\ra\;,
\ea
and similarly
\be
\la y|\mL_g(\sigma)|x\ra=-i\left(p_y^{\alpha}-p_x^{\alpha}\right)\la y|\Lambda|x\ra\;.
\ee
Therefore,
\ba
\tr\left[\mL_f(\sigma)\mL_g(\sigma)\right]&=\sum_{x,y\in[m]}\left(p_y^{1-\alpha}-p_x^{1-\alpha}\right)\left(p_y^{\alpha}-p_x^{\alpha}\right)|\la x|\Lambda|y\ra|^2\\
&=2\sum_{x,y\in[m]}p_x|\la x|\Lambda|y\ra|^2-2\sum_{x,y\in[m]}p_x^{1-\alpha}p_y^{\alpha}|\la x|\Lambda|y\ra|^2\\
&=2\tr\left[\rho\Lambda^2\right]-2\tr\left[\rho^{1-\alpha}\Lambda\rho^\alpha\Lambda\right]\;.
\ea
Hence, for all $\alpha\in[0,2]$, $\rho\in\md(A)$ and $\Lambda\in\herm(A)$, the differential $\alpha$-R\'enyi divergence\index{R\'enyi divergence} of asymmetry is given by
\be
D_{\Lambda,\alpha}(\rho)=\frac1{1-\alpha}\left(\tr\left[\rho\Lambda^2\right]-\tr\left[\rho^{1-\alpha}\Lambda\rho^\alpha\Lambda\right]\right)\;.
\ee
The expression in the parenthesis above (i.e. without the factor $1/(1-\alpha)$) is known as the Wigner-Yanase-Dyson skew information. Note that as the previous example, on pure states the Wigner-Yanase-Dyson skew information reduces to the variance of $\Lambda$.

 \bex
 Let $\rho=\psi\in\pure(A)$ and $\Lambda\in\herm(A)$. Show that
\be
D_{\Lambda,\alpha}(\psi)=\frac1{1-\alpha}\left(\la\psi|\Lambda^2|\psi\ra-\la\psi|\Lambda|\psi\ra^2\right)\;.
\ee
\eex

\bex
Show that for $\alpha\in(0,1)$ the function $D_{\Lambda,\alpha}(\rho)$ is concave in $\rho$. Hint: Use Lieb's Concavity Theorem (see Theorem~\ref{lieb}).
\eex

\bex
Show that for $\alpha=1$ and $\rho\in\md(A)$ we have
\be
D_{\Lambda}(\rho)\eqdef\lim_{\alpha\to 1}D_{\Lambda,\alpha}(\rho)=\tr\left[\Lambda^2\rho\log\rho\right]-\tr\left[\Lambda\rho\Lambda\log\rho\right]\;.
\ee
\eex

\subsubsection{The Quantum Fisher Information}\index{Fisher information}

As our third example, we take $\D=\tD_{\alpha}$ to be the minimal quantum divergence. In this case, we use the invariance of every relative entropy under unitary operations to get that
\be
\tD_\alpha\left(\rho\|U_g\rho U_g^*\right)=\tD_\alpha\left(U_g^*\rho U_g\|\rho\right)\;.
\ee
Since  $U_g^*\rho U_g=\rho-t\sigma+t^2\eta+O(t^3)$ we get that
\be
\tD_\alpha\left(\rho\|U_g\rho U_g^*\right)=\D_\alpha(\rho-t\sigma+t^2\eta\|\rho)+O(t^3)\;.
\ee
By definition,
\ba
\D_\alpha(\rho-t\sigma+t^2\eta\|\rho)&=\frac1{\alpha-1}\log\tr\left[\left(\rho^{\frac{1-\alpha}{2\alpha}}\left(\rho-t\sigma+t^2\eta\right)\rho^{\frac{1-\alpha}{2\alpha}}\right)^\alpha\right]\\
&=\frac1{\alpha-1}\log\tr\left[\left(\trho-t\tsigma+t^2\teta\right)^\alpha\right]
\ea
where
\be
\trho\eqdef\rho^{\frac{1-\alpha}{2\alpha}}\rho\rho^{\frac{1-\alpha}{2\alpha}}=\rho^{1/\alpha},\quad \tsigma\eqdef \rho^{\frac{1-\alpha}{2\alpha}}\sigma\rho^{\frac{1-\alpha}{2\alpha}},\quad\text{and}\quad\teta\eqdef\rho^{\frac{1-\alpha}{2\alpha}}\eta\rho^{\frac{1-\alpha}{2\alpha}}\;.
\ee
The trace $\tr\left[\left(\trho-t\tsigma+t^2\teta\right)^\alpha\right]$ has the form given in Corollary~\ref{cordd} with $g(t)\eqdef 1$ and $f(t)\eqdef t^{\alpha}$. Therefore, 
$
h(t)\eqdef g(t)f'(t)=\alpha t^{\alpha-1}
$,
so that 
\be
\tr\left[h(\trho)\tsigma\right]=\alpha\tr[\trho^{\alpha-1}\tsigma]=\alpha\tr[\sigma]=0
\ee 
and similarly $\tr\left[h(\trho)\teta\right]=\alpha\tr[\eta]=0$. Observe further that since $g$ is a constant function, $\mL_g(\sigma)=0$. Substituting all this into Corollary~\ref{cordd} we conclude that 
\be
\tr\left[\left(\trho-t\tsigma+t^2\teta\right)^\alpha\right]=1+\frac12t^2\tr\left[\tsigma\mL_h(\tsigma)\right]+O(t^3)\;.
\ee
It will be convenient to denote $s\eqdef\frac1\alpha$. Since we assume that $\alpha\in[1/2,\infty]$ we have that $s\in[0,2]$. Working with the eigenbasis of $\rho$ we get for all $x,y\in[m]$
\ba\label{15a1}
\la x|\tsigma|y\ra&=p_x^{\frac{1-\alpha}{2\alpha}}p_{y}^{\frac{1-\alpha}{2\alpha}}\la x|\sigma|y\ra\\
\Gg{s\eqdef1/\alpha,\;\sigma\eqdef i[\Lambda,\rho]}&=ip_x^{(s-1)/2}p_{y}^{(s-1)/2}(p_y-p_x)\la x|\Lambda|y\ra\;.
\ea
Furthermore, since the eigenvalues of $\trho$ are $\{p_x^{s}\}_{x\in[m]}$ we get by definition of $\mL_h$ that
\be\label{15a2}
\la y|\mL_h(\tsigma)|x\ra=\frac{h\left(p_y^{s}\right)-h\left(p_x^{s}\right)}{p_y^{s}-p_x^{s}}\la y|\tsigma|x\ra=\frac1s\frac{p_y^{1-s}-p_x^{1-s}}{p_y^{s}-p_x^{s}}\la y|\tsigma|x\ra\;,
\ee
where the case $p_x=p_y$ is understood in terms of the limit
\be
\lim_{p_y\to p_x}\frac{p_y^{1-s}-p_x^{1-s}}{p_y^{s}-p_x^{s}}=\frac{1-s}{s}p_x^{1-2s}\;.
\ee
With these expressions we get
\ba
\tr\left[\tsigma\mL_h(\tsigma)\right]&=\sum_{x,y\in[m]}\la x|\tsigma|y\ra\la y|\mL_h(\tsigma)|x\ra\\
\GG{\eqref{15a1},\eqref{15a2}}&=\frac1s\sum_{x,y\in[m]}\frac{p_x^{s-1}-p_y^{s-1}}{p_y^{s}-p_x^{s}}(p_y-p_x)^2|\la x|\Lambda|y\ra|^2\;.
\ea
Note that since the limit $p_y\to p_x$ of the components in the sum above is zero, we can restrict the sum above to all $x,y\in[m]$ that satisfies $p_x\neq p_y$.
Hence, we conclude that
\be\label{familytd}
\tD_{\Lambda,s}(\rho)\eqdef\tD_{\Lambda,\alpha}(\rho)=\frac1{2(s-1)}\sum_{\substack{x,y\in[m]\\p_x\neq p_y}}\frac{p_x^{s-1}-p_y^{s-1}}{p_x^{s}-p_y^{s}}(p_x-p_y)^2|\la x|\Lambda|y\ra|^2\;.
\ee

If $\rho$ is given by the pure state $\psi=|1\lr 1|\in\pure(A)$ then $p_x=\delta_{1x}$ for all $x\in[m]$. In this case, for all $s\in[0,2]$
\ba
\tD_{\Lambda,s}(\psi)&=\frac1{s-1}\sum_{y=2}^{m}|\la \psi|\Lambda|y\ra|^2+\frac1{s-1}\sum_{x=2}^{m}|\la x|\Lambda|\psi\ra|^2\\
&=\frac2{s-1}\sum_{x=2}^{m}|\la x|\Lambda|\psi\ra|^2\\
\Gg{\sum_{x=2}^m|x\lr x|^A=I^A-\psi^A}&=\frac2{s-1}\left\la\psi^A\left|\Lambda\left(I^A-\psi^A\right)\Lambda\right|\psi^A\right\ra\\
&=\frac2{s-1}\left(\la\psi|\Lambda^2|\psi\ra-\la\psi|\Lambda|\psi\ra^2\right)\;.
\ea

The Fisher information is a measure of asymmetry that is obtained by setting $s=2$ in the family of asymmetry monotones given in equation~\eqref{familytd}, which yields:
\be
F_{\Lambda}(\rho)\eqdef4\tD_{\Lambda,2}(\rho)=2\sum_{x,y\in[m]}\frac{(p_x-p_y)^2}{p_x+p_y}|\la x|\Lambda|y\ra|^2\;.
\ee
The Fisher information is a fundamental concept in statistics and information theory with numerous applications in quantum metrology and quantum information. It plays a crucial role in studying the ultimate limits of precision in quantum measurements, commonly referred to as the quantum Cram\'er-Rao bound. Moreover, the Fisher information is employed to measure the distinguishability of quantum states, to characterize the entanglement properties of multipartite systems, and to devise optimal quantum measurement strategies. In the field of quantum thermodynamics, it has an operational interpretation as the coherence cost of preparing a system in a particular state without any restrictions on work consumption.

\bex
Show that for $s=\alpha=1$
\be 
\tD_{\Lambda}(\rho)\eqdef\lim_{\alpha\to 1}\tD_{\Lambda,\alpha}(\rho)=\sum_{x,y\in[m]}(\log p_x-\log p_y)(p_x-p_y)|\la x|\Lambda|y\ra|^2\;.
\ee
\eex

\bex
Show that $\tD_{\Lambda,s}$ can be expressed as
\be
\tD_{\Lambda,s}(\rho)=\tr\left[\Lambda\mL_\rho(\Lambda)\right]\quad\quad\forall\;\Lambda\in\herm(A),\;\;\forall\;\rho\in\md(A)\;,
\ee
where $\mL_\rho\in\herm(A\to A)$ is a self-adjoint linear map.
\eex

\bex
Show that the Fisher information upper bound the Wigner-Yanase skew information; that is, show that for all $\rho\in\md(A)$ and $\Lambda\in\herm(A)$ we have
\be
F_\Lambda(\rho)\geq 4\left(\tr\left[\rho\Lambda^2\right]-\tr\left[\sqrt{\rho}\Lambda\sqrt{\rho}\Lambda\right]\right)\;.
\ee
Hint: Recall that $\tD_\alpha(\rho\|\sigma)\leq D_{\alpha}(\rho\|\sigma)$ for all $\rho,\sigma\in\md(A)$.
\eex

\section{Manipulation of Pure Asymmetry}

In this section, we aim to investigate the circumstances in which it is possible to transform an asymmetric pure state $\psi$ into another pure state $\phi$, using $\G$-covariant maps. To achieve this, we must first identify all the pure states that are equivalent under symmetric (i.e., $\G$-covariant) operations. This characterization is crucial for understanding the limitations and possibilities of pure-state transformations in this resource theory.

\subsection{Characterization of Asymmetry Equivalence Classes}  

\begin{myd}{}
\begin{definition}Consider a projective unitary representation $g\mapsto U_g\in\ml(A)$, where $\G$ is a finite or compact Lie group.
\begin{enumerate}
\item We say that two states $\rho,\sigma\in\md(A)$ are $\G$-equivalent if there exists $\mE,\mF\in\cov_\G(A\to A)$ such that $\rho=\mE(\sigma)$ and $\sigma=\mF(\rho)$.
\item Two pure states $\psi,\phi\in\pure(A)$ are called unitarily $\G$-equivalent if there exists a $\G$-invariant unitary matrix $V:A\to A$ such that $V|\psi\ra=|\phi\ra$.
\end{enumerate}
\end{definition}
\end{myd}

We will also refer to the set of all states $\sigma\in\md(A)$ that are $\G$-equivalent to $\rho$ as the \emph{$\G$-equivalence class of $\rho$}. In this subsection, our focus is on characterizing the $\G$ equivalence class of a pure state. To achieve this goal, we begin by characterizing unitarily $\G$-equivalent states. Note that if $[V,U_g]=0$ holds for all $g$, then $[V^*,U_g]=0$ holds for all $g$ as well. Therefore, we can replace the condition $V|\psi\ra=|\phi\ra$ in the definition of unitarily $\G$-equivalent states with $|\psi\ra=V|\phi\ra$.

\bex
Let $g\mapsto\G$ be a projective unitary representation of $\G$ and for every $\rho\in\md(A)$ let 
\be
\sym_\G(\rho)\eqdef\{g\in\G\;:\;U_g\rho U_g^*=\rho\}\;.
\ee 
\ben
\item Show that $\sym_\G(\rho)$ is a subgroup of $\G$.
\item Show that if $\rho\xrightarrow{\G-\cov}\sigma$ for some $\rho,\sigma\in\md(A)$ then $\sym_\G(\rho)$ is a subgroup of  $\sym_\G(\sigma)$.
\item Show that if $\rho$ and $\sigma$ are $\G$-equivalent then $\sym_\G(\rho)=\sym_\G(\sigma)$.
\een 
\eex

Every projective unitary representation, $g\mapsto U_g^A$, corresponds to a decomposition of the Hilbert space $A$ as given in~\eqref{decoal}. Specifically, $A=\bigoplus_{\lambda\in\irr(U)}A_\lambda$, where $A_\lambda=B_\lambda\otimes C_\lambda$. Accordingly, every two pure states $\psi,\phi\in\pure(A)$ can be expressed as
\be\label{plpl}
|\psi^A\ra=\sum_{\lambda\in\irr(U)}|\psi_\lambda^{B_\lambda C_\lambda}\ra\quad\text{and}\quad|\phi^A\ra=\sum_{\lambda\in\irr(U)}|\phi_\lambda^{B_\lambda C_\lambda}\ra\;,
\ee
where $|\psi_\lambda\ra,|\phi_\lambda\ra\in B_\lambda  C_\lambda$ are subnormalized states\index{subnormalized states} in $A_\lambda$. In the following theorem we show that $\psi^A$ and $\phi^A$ are unitarily $\G$ equivalent if the marginals of $\phi_\lambda^{B_\lambda C_\lambda}$ and $\phi_\lambda^{B_\lambda C_\lambda}$ on $B_\lambda$ are the same.
In addition, the theorem characterizes states that are unitarily $\G$-equivalent in terms of their characteristic functions. In Sec.~\ref{cf}, we discuss various properties of characteristic functions\index{characteristic function}, and we encourage readers who are unfamiliar with this material to read Sec.\ref{cf} before proceeding to the theorem below.

\begin{myt}{}
\begin{theorem}\label{thm1541}
Let $\psi,\phi\in\pure(A)$. The following statements are equivalent:
\ben
\item The states $\psi$ and $\phi$ are unitarily $\G$-equivalent.
\item There exists a unitary matrix $V\in\muu(A)$ such that $VU_g|\psi\ra=U_g|\phi\ra$, $\forall\;g\in \G$.
\item The characteristic functions of $\psi$ and $\phi$ are the same: $\chi_\psi(g)=\chi_\phi(g)$, $\forall\;g\in\G$.
\item Using~\eqref{plpl}, for all $\lambda\in\irr(U)$,
$
\tr_{C_\lambda}\left[\psi_\lambda^{B_\lambda C_\lambda}\right]=\tr_{C_\lambda}\left[\phi_\lambda^{B_\lambda C_\lambda}\right]
$.
\een
\end{theorem}
\end{myt}
\begin{proof}
{\it The implication $1\Rightarrow 2$:}
Suppose that $\psi$ and $\phi$ are unitarily $\G$-equivalent. Then there exists a $\G$-invariant unitary matrix $V:A\to A$ such that $|\phi\ra=V|\psi\ra$. Since $V$ is $\G$-invariant, after multiplying both sides by $U_g$ from the left we get
$U_g|\phi\ra=U_gV|\psi\ra=VU_g|\psi\ra$.

{\it The implication $2\Rightarrow 3$:} For all $g\in \G$ we have
\ba
\la\psi|U_g|\psi\ra&=\la\psi|V^*VU_g|\psi\ra\\
\Gg{V|\psi\ra=|\phi\ra}&=\la\phi|VU_g|\psi\ra\\
\Gg{VU_g|\psi\ra=U_g|\phi\ra}&=\la\phi|U_g|\phi\ra\;.
\ea

{\it The implication $3\Rightarrow 4$:} Since we assume that $\chi_\psi(g)=\chi_\phi(g)$ for all $g\in\G$, we get  from Theorem~\ref{th158}
\ba
\tr_{C_\lambda}\left[\psi_\lambda^{B_\lambda C_\lambda}\right]&=|B_\lambda|\int_\G dg\;\chi_{\psi}(g^{-1})U^{(\lambda)}_g\\
\Gg{\chi_\psi(g^{-1})=\chi_\phi(g^{-1})}&=|B_\lambda|\int_\G dg\;\chi_{\phi}(g^{-1})U^{(\lambda)}_g\\
\GG{\eqref{nli}}&=\tr_{C_\lambda}\left[\phi_\lambda^{B_\lambda C_\lambda}\right]\;.
\ea

{\it The implication $4\Rightarrow 1$:} Since for each $\lambda\in\irr(U)$ the states $\psi_\lambda^{B_\lambda C_\lambda}$ and $\phi_\lambda^{B_\lambda C_\lambda}$ have the same marginal on representation space $B_\lambda$, there exists a unitary matrix $V_\lambda:C_\lambda\to C_\lambda$ such that
\be
I^{B_\lambda}\otimes V_\lambda^{C_\lambda}\big|\psi_\lambda^{B_\lambda C_\lambda}\big\ra=\big|\phi_\lambda^{B_\lambda C_\lambda}\big\ra\;.
\ee
Let $V:A\to A$ be the unitary matrix
\be
V\eqdef\bigoplus_{\lambda\in\irr(U)}I^{B_\lambda}\otimes V_\lambda^{C_\lambda}\;.
\ee
Then, by definition, $|\phi^A\ra=V|\psi^A\ra$, and since for each $g\in\G$ the unitary matrix $U_g$ has the form~\eqref{lamir} we get that $[V,U_g]=0$. Hence, $\psi^A$ and $\phi^A$ are unitarily $\G$-equivalent. This completes the proof. 
\end{proof}

\bex
Let $\psi\in\pure(A)$ and $\phi\in\pure(B)$, where $|B|\neq|A|$, and consider two projective unitary representations $g\mapsto U_g^A$ and $g\mapsto U_g^B$ in $A$ and $B$, respectively. Show that if $\chi_\psi(g)=\chi_\phi(g)$ for all $g\in\G$ then $\psi$ and $\phi$ are $\G$-equivalent. 
\eex

The above theorem characterizes unitarily $\G$-equivalent states. However, from a resource theory perspective, two states belong to the same resource equivalence class if they are $\G$-equivalent (not necessarily unitarily). Our next theorem characterize this $\G$-equivalence class, assuming that the states involved are $\G$-regular.
\begin{myd}{}
\begin{definition}\label{gregular}
Let $\psi,\phi\in\pure(A)$ and $\G$ be a group. We say that $\psi$ and $\phi$ are $\G$-regular\index{$\G$-regular} with respect to a representation $g\mapsto U_g$ if one of the following two conditions holds:
\ben
\item The group $\G$ is finite and there is no $g\in\G$ such that $\chi_\psi(g)=\chi_\phi(g)= 0$; that is, the functions $\chi_\psi,\chi_\phi:\G\to\mbb{C}$ cannot take the zero value simultaneously.
\item The group $\G$ is a compact Lie group and there is no open set (other than the trivial one) $\mc$ of $\G$ for which $\chi_\psi(g)=\chi_\phi(g)= 0$ for all $g\in\mc$.
\een
\end{definition}
\end{myd}

It is worth pointing out that every \emph{connected} compact Lie group $\G$ satisfies the second condition above. In fact, if $\G$ is connected, for any state $\psi\in\pure(A)$, there cannot be an open neighbourhood $\mc$ of $\G$ for which $\chi_\psi(g)= 0$ for all $g\in\mc$. To see why, by contradiction, suppose that $\chi_\psi(g)= 0$ for all $g\in\mc$. Since the function $\chi_\psi:\G\to\mbb{C}$ is analytic, the identity theorem in complex analysis implies that $\chi_\psi$ is the zero function, which contradicts the fact that $\chi_\psi(e)=1$ for the identity element $e\in\G$.

\bex
Let $\G$ be a compact Lie group and let $\psi,\phi\in\pure(A)$ be such that for all $g\in \G$ there exists elements $h,h'\in\mathbf{H}$ of a connected subgroup $\mathbf{H}$ of $\G$ for which $|\chi_\psi(hgh')|+|\chi_\phi(hgh')|\neq 0$. Show that $\psi$ and $\phi$ are $\G$-regular.
\eex

To clarify the notion of $\G$-regular states, let's consider the group $O(2)$ of $2\times 2$ real orthogonal matrices. This group is a compact Lie group, but it is not connected because matrices with determinant one are not continuously connected to matrices with determinant minus one. Let $\mathbf{H}\eqdef SO(2)$ be the subgroup of $O(2)$ consisting of all the elements of $O(2)$ with determinant one. The question we want to answer is: Is there a state $\psi\in\pure(\mbb{C}^2)$ such that $\chi_\psi(g)=0$ for all $g\not\in\mathbf{H}$?

To answer this question, we first observe that all the matrices $g\in O(2)$ with $\det(g)=-1$ have the form $\bpm \cos\theta & \sin\theta\\ \sin\theta & -\cos\theta\epm$ for some $\theta\in[0,2\pi]$. Therefore, $\chi_\psi(g)=0$ for all $g\not\in\mathbf{H}$ if and only if
\be
\big\la\psi\big|\bpm \cos\theta & \sin\theta\\ \sin\theta & -\cos\theta\epm\big|\psi\big\ra=0\quad\quad\forall\;\theta\in[0,2\pi]\;.
\ee
The only pure state that satisfies the above equation is $|\psi\ra=\frac1{\sqrt{2}}(|0\ra+i|1\ra)$. Therefore, in this example, the second condition in Definition~\ref{gregular} is satisfied except in the case where $|\psi\ra=|\phi\ra=\frac1{\sqrt{2}}(|0\ra+i|1\ra)$.

\begin{myt}{}
\begin{theorem}\label{connected}
Let $g\mapsto U_g$ be a projective unitary representation of $\G$, and let $\psi,\phi\in\pure(A)$ be $\G$-regular states. The states $\psi$ and $\phi$ are $\G$-equivalent if and only if there exists a 1-dimensional unitary representation of $\G$, $\{e^{i\theta_g}\}_{g\in \G}$, such that for all $g\in \G$
\be\label{mconm}
\la\psi|U_g|\psi\ra=e^{i\theta_g}\la\phi|U_g|\phi\ra\;.
\ee
\end{theorem} 
\end{myt}
\begin{remark}
We will see in the proof below that for any finite or compact (not necessarily connected) Lie group $\G$, if~\eqref{mconm} holds, then $\phi$ and $\psi$ are $\G$-equivalent. Therefore, we only need the assumption that $\psi$ and $\phi$ are $\G$-regular for the converse part. In Sec.~\ref{app:regular} of the appendix we provide additional observations for the case that $\psi$ and $\phi$ are not $\G$-regular. Moreover, it is worth noting that semi-simple compact Lie groups, such as $SU(2)$, do not have any non-trivial 1-dimensional representation. Therefore, it follows from the theorem above and the preceding theorem that for such groups, the following statements are all equivalent: 
\begin{enumerate}
\item $\psi$ and $\phi$ are $\G$-equivalent.
\item $\psi$ and $\phi$ are unitarily $\G$-equivalent.
\item $\psi$ and $\phi$ have the same characteristic function.
\end{enumerate}
\end{remark}

\begin{proof}
We first prove that the condition~\eqref{mconm} implies that $\psi$ and $\phi$ are $\G$-equivalent. Let $E$ be a qubit system (i.e., $|E|=2$) with an orthonormal basis $\{|\varphi_1\ra,|\varphi_2\ra\}$, and let $g\mapsto U_g^E$ be the non-projective unitary representation of $\G$ with
\be
U_g^E\eqdef\varphi_1^E+e^{i\theta_g}\varphi_2^E\;,
\ee
where $g\mapsto e^{i\theta_g}$ is the 1-dimensional representation of $\G$ that appear in~\eqref{mconm}. 
Clearly, with respect to the representation $g\mapsto U_g^E$, the state $\varphi_1^E$ and $\varphi_2^E$ are $\G$-invariant, and for all $g\in \G$, $\chi_{\varphi_1}(g)=1$ and $\chi_{\varphi_2}(g)=e^{i\theta_g}$. 

Combining this with~\eqref{mconm} we conclude that $|\psi\ra^A|\varphi_1\ra^E$ has the same characteristic function as $|\phi\ra^A|\varphi_2\ra^E$. From Theorem~\ref{thm1541} it follows that there exists a $\G$-invariant unitary $V$ (with respect to the representation $g\mapsto U_g^A\otimes U_g^E$) such that
\be
\psi^A\otimes\varphi_1^E=V\left(\phi^A\otimes\varphi_2^E\right)V^*\;.
\ee
Define a quantum channel $\mE(\rho^A)\eqdef\tr_E\left[V\left(\rho^A\otimes\varphi^E_2\right)V^*\right]$ for all $\rho\in\ml(A)$. By definition $\mE(\phi^A)=\psi^A$, and the $\G$-invariance\index{invariance} of $V$ and $\varphi^E_2$ implies that $\mE\in\cov_\G(A\to A)$. Therefore, $\phi$ can be converted to $\psi$ by symmetric (i.e. covariant) operations. Following similar lines we also get that $\psi$ can be converted to $\phi$ by $\G$-covariant operations. Hence, $\psi$ and $\phi$ are $\G$-equivalent.

For the converse part of the proof, suppose there exists a $\G$-covariant channel mapping $\psi$ to $\phi$ and another $\G$-covariant channel that maps $\phi$ to $\psi$. From the covariant version of Stinespring delation theorem (see Theorem~\ref{covstine}) there exists two isometries $V_1:A\to AE$ and $V_2:A\to A\tE$, each satisfying~\eqref{ginvst} for all $g\in\G$, and with the property that
\begin{align}
&V_1|\psi\ra^A=|\phi\ra^A|\varphi_{1}\ra^E\label{fg1}\\
&V_2|\phi\ra^A=|\psi\ra^A|\varphi_2\ra^{\tE}\;,
\end{align}
for some $\varphi_1,\varphi_2\in\pure(E)$.
Since, $V_1$ and $V_2$ satisfy~\eqref{ginvst} for all $g\in\G$,
the two equations above imply that for all $g\in \G$
\ba\label{cpc}
&\chi_\psi(g)=\chi_\phi(g)\chi_{\varphi_1}(g)\\
&\chi_\phi(g)=\chi_\psi(g)\chi_{\varphi_2}(g)\;.
\ea
Combining these two equations yields in particular that
\ba\label{abso}
&\chi_\psi(g)=\chi_\psi(g)\chi_{\varphi_1}(g)\chi_{\varphi_2}(g)\\
&\chi_\phi(g)=\chi_\phi(g)\chi_{\varphi_1}(g)\chi_{\varphi_2}(g)
\ea
First, if for all $g\in \G$
 $\chi_\psi(g)\neq 0$ and/or $\chi_\phi(g)\neq 0$ then $\chi_{\varphi_1}(g)\chi_{\varphi_2}(g)=1$. Since the absolute value of characteristic functions cannot exceed one, it follows that $|\chi_{\varphi_1}(g)|=|\chi_{\varphi_2}(g)|=1$ for all $g\in\G$. Therefore, from Lemma~\ref{lem:c71} we get that the states $\varphi_1$ and $\varphi_2$ are $\G$-invariant in this case. 
Second, suppose $\G$ is a compact Lie group and suppose by contradiction that there exists $g\in\G$ such that $\chi_{\varphi_1}(g)\chi_{\varphi_2}(g)\neq 1$. Then, from the continuity of the characteristic function, there exists a neighbourhood $\mc\subset\G$ of $g$ such that for all $g'\in\mc$  we have $\chi_{\varphi_1}(g')\chi_{\varphi_2}(g')\neq 1$. From~\eqref{abso} it then follows that $\chi_\psi(g')=\chi_\phi(g')=0$ for all $g'\in\mc$ in contradiction with the assumption that $\psi$ and $\phi$ are $\G$-regular. Therefore, also in this case $\chi_{\varphi_1}(g)\chi_{\varphi_2}(g)=1$ for all $g\in\G$, so that $\varphi_1$ and $\varphi_2$ are $\G$-invariant.
 
To summarize, in both cases we can express the characteristic functions of $\varphi_1$ and $\varphi_2$ as
$\chi_{\varphi_1}(g)=e^{i\theta_g}$ and $\chi_{\varphi_2}(g)=e^{-i\theta_g}$, where $g\mapsto e^{i\theta_g}$ is a 1-dimensional unitary representations of $\G$ (see Exercise~\ref{ex:c71}). Substituting this into~\eqref{cpc} completes the proof.
\end{proof}

As an example, consider the group $U(1)$ and an arbitrary state 
\be
|\tpsi\ra=\sum_{n\in[m]}\lambda_n|n\ra\;,
\ee
where $\{|n\ra\}_{n\in\mbb{Z}}$ is the eigenbasis of the number operator. The characteristic function of $\tpsi$ is given by
\be
\chi_\tpsi(\theta)=\big\la\tpsi\big|e^{i\theta\hat{N}}\big|\tpsi\big\ra=\sum_{n\in[m]}|\lambda_n|^2e^{i\theta n}\;.
\ee
Since the characteristic function of $\tpsi$ depends only on the absolute values of the coefficients $\{\lambda_n\}_{n\in[m]}$, we get from Theorem~\ref{thm1541} (particularly, the equivalence of the first and third statements of this theorem) that $\tpsi$ is unitarily $\G$-equivalent to the state
\be
|\psi\ra=\sum_{n\in[m]}\sqrt{p_\psi(n)}|n\ra
\ee
where $p_\psi(n)\eqdef|\lambda_n|^2$. Therefore, similar to the Schmidt decomposition in entanglement theory, in the QRT of $U(1)$ asymmetry, all pure states are unitarily $U(1)$-equivalent to a state of the form above. Therefore, in this resource theory, the resource is characterized by the probability distribution $p_\psi:\mbb{Z}\to[0,1]$.

Every 1-dimensional unitary representation of $U(1)$ is determined by some integer $k\in\mbb{Z}$ and a mapping $\theta\mapsto e^{i\theta k}$. Therefore, according to Theorem~\ref{connected} two pure state $\psi$ and $\phi$ are $\G$-equivalent if and only if there exists $k\in\mbb{Z}$ such that $\chi_\psi(\theta)=e^{i\theta k}\chi_\phi(\theta)$ for all $\theta\in U(1)$. This condition can be written as
\be\label{ft00}
\sum_{n\in\mbb{Z}}p_\psi(n)e^{i\theta n}=e^{i\theta k}\sum_{n\in\mbb{Z}}p_\phi(n)e^{i\theta n}\quad\quad\forall\;\theta\in U(1)\;,
\ee
where $p_\psi,p_\phi:\mbb{Z}\to [0,1]$ are the probability distributions associated with $\psi$ and $\phi$, respectively. Using the Fourier transform (see Exercise~\ref{ex:ft}) we get that the above condition can be expressed as
\be\label{ft11}
p_\psi(n)=p_\phi(n+k)\;.
\ee
As a specific example, observe that the states $|\psi\ra=\frac1{\sqrt{2}}(|0\ra+|1\ra)$ and $|\phi\ra=\frac1{\sqrt{2}}(|1\ra+|2\ra)$ are $\G$-equivalent since in this case $p_\psi(n)=p_\phi(n-1)$.

\bex\label{ex:ft}
Show that the condition in~\eqref{ft00} is equivalent to one in~\eqref{ft11}. Hint: Apply a Fourier transform on both sides of~\eqref{ft00}.
\eex

\subsection{Deterministic Transformations}
In this subsection, we will explore the conditions under which a pure asymmetric state can be deterministically converted into another state by $\G$-covariant operations. In entanglement theory, we learned that such conversions under LOCC are determined by Nielsen's majorization theorem. However, we will show that in the QRT of asymmetry convertibility is actually determined by the concept of a positive-definite function on a group. For readers who are not familiar with this topic, we provide a review in Sec.~\ref{pdfg} of the appendix.

\begin{myt}{}
\begin{theorem}\label{purestateconv}
Let $\psi,\phi\in\pure(A)$ be pure states. There exists a $\G$-covariant map $\mE\in\cov_\G(A\to A)$ such that $\phi=\mE(\psi)$ if and only if there exists a positive definite function $f:\G\to\mbb{C}$  such that $\chi_\psi(g)=\chi_\phi(g)f(g)$ for all $g\in\G$.
\end{theorem}
\end{myt}
\begin{remark}
If $\chi_\phi(g)\neq 0$ for all $g\in\G$ then the theorem above states in this case that $\psi$ can be converted to $\phi$ by symmetric operations if and only if $\chi_\psi(g)/\chi_\phi(g)$ is a positive definite function over $\G$.
\end{remark}

\begin{proof}
Suppose first that $\psi\xrightarrow{\G-\cov}\phi$. In the derivation of the relation in~\eqref{cpc}, using the covariant Stinespring dilation theorem we showed that the condition $\psi\xrightarrow{\G-\cov}\phi$ implies that there exists a pure state $\varphi\in\pure(A)$ such that
\be
\chi_\psi(g)=\chi_\phi(g)\chi_\varphi(g)\;.
\ee
Hence, taking $f(g)\eqdef\chi_{\varphi}(g)$ we get $\chi_\psi(g)=\chi_\phi(g)f(g)$. Finally, observe that the characteristic function $\chi_{\varphi}:\G\mathbf{C}$ is a normalized positive definite function over $\G$ (see Theorem~\ref{pdd}). 

Conversely, suppose $\chi_\psi(g)=\chi_\phi(g)f(g)$ for some positive definite function $f$. Since for $g=e$ we get $f(e)=\chi_\psi(e)/\chi_\phi(e)=1$ the function $f$ is normalized so that according to Theorem~\ref{pdd} it corresponds to some characteristic function $f(g)=\la\varphi|U_g^E|\varphi\ra$, where $g\mapsto U_g^E$ is some unitary representation of $\G$ on some Hilbert space $E$. Moreover, there exists a $\G$-invariant state $|0\ra\in E$ whose characteristic function is  constant and equal to one for all group elements. Therefore, from the relation $\chi_\psi(g)=\chi_\phi(g)f(g)$ we get that the states $|\psi\ra^A|0\ra^E$ and $|\phi\ra^A|\varphi\ra^E$ have the same characteristic function.
Therefore, there exists a $\G$-invariant unitary $V:AE\to AE$ such that $V\left(|\psi\ra^A|0\ra^E\right)=|\phi\ra^A|\varphi\ra^E$. Taking the trace over $E$ on both sides demonstrates that $\psi$ can be converted to $\phi$ by a $\G$-covariant channel.
\end{proof}

\bex
Consider two states $\psi,\phi\in\pure(A)$ and suppose $\psi$ has the property that $\chi_\psi(g)=0$ for all $g\in\G$ such that $g\neq e$. Show that $\psi\xrightarrow{\G-\cov}\phi$. In other words, $\psi$ with such a property is a maximal resource state.
\eex

\subsubsection{Example: The Cyclic Group $\mbb{Z}_n$}

Let $n\in\mbb{N}$ be a fixed integer, and consider the group $\mbb{Z}_n$, which represents a cyclic group of order $n$. The group $\mbb{Z}_n$ is the group of integers $\{0,\ldots,n-1\}$, where the group operation is addition modulo $n$, and $0$ is the identity element of the group. It is well known that every cyclic group or order $n$ is isomorphic to $\mbb{Z}_n$. Since $\mbb{Z}_n$ is an Abelian group,  all of its irreps are one-dimensional. Each irrep can be uniquely identified by an integer $y\in\mbb{Z}_n$, under the action $x\mapsto e^{i\frac{2\pi yx}{n}}$ for all $x\in\mbb{Z}_n$.

Consider a (non-projective) unitary representation of $\mbb{Z}_n$ in the space $A=\mbb{C}^n$, given for all $x\in\mbb{Z}_n$ by $x\mapsto U_x$, where 
\be
U_x\eqdef\sum_{y\in\mbb{Z}_n}e^{i\frac{2\pi yx}{n}}|y\lr y|\;.
\ee 
Note that the above unitary representation of $\mbb{Z}_n$ composed of a direct sum of its irreps, each occurring with multiplicity one.

We would like to find the conditions under which the quantum pure state $|\psi\ra=\sum_{x=0}^{n-1}\sqrt{p_x}|x\ra$ can be converted to another pure state $|\phi\ra\eqdef\sum_{x=0}^{n-1}\sqrt{q_x}|x\ra$ by $\mbb{Z}_n$-covariant operations. Observe that
the characteristic function of $|\psi\ra$ is given for any $x\in\mbb{Z}_n$ by
\be
\chi_\psi(x)=\la\psi|U_x|\psi\ra=\sum_{y\in\mbb{Z}_n}p_ye^{i\frac{2\pi yx}{n}}\;.
\ee
Similarly, $\chi_\phi(x)$ can be expressed as above with $q_y$ replacing $p_y$. The above equation demonstrates that the characteristic function is nothing but the discrete Fourier transform of the sequence $\{p_0,\ldots,p_{n-1}\}$.

The theorem above implies that $\psi$ can be converted to $\phi$ by $\mbb{Z}_n$-covariant operations if and only if the function $x\mapsto \chi_\psi(x)/\chi_\phi(x)$ is a positive definite function over $\mbb{Z}_n$. From Exercise~\ref{ftex} we have that a function $f:\mbb{Z}_n\to\mbb{C}$ is positive definite if and only if its (discrete) Fourier transform is positive. We therefore conclude that $\psi\xrightarrow{\mbb{Z}_n-\cov}\phi$ if and only if
\be\label{0alt}
\sum_{x\in\mbb{Z}_n}\frac{\chi_\psi(x)}{\chi_\phi(x)}e^{i\frac{2\pi xy}{n}}\geq 0\quad\quad\forall\;y\in\mbb{Z}_n\;.
\ee

To illustrate the condition above, we consider now the case $n=2$. For $n=2$ the condition above gives for $y\in\mbb{Z}_2=\{0,1\}$
\be
0\leq\frac{\chi_\psi(0)}{\chi_\phi(0)}+(-1)^y\frac{\chi_\psi(1)}{\chi_\phi(1)}=1+(-1)^y\frac{p_0-p_1}{q_0-q_1}\;.
\ee
The condition above can be expressed as
\be
\frac{|p_0-p_1|}{|q_0-q_1|}\leq 1
\ee
which is equivalent to 
\be
\max\{p_0,p_1\}\leq\max\{q_0,q_1\}\;.
\ee

The condition we obtained  for the case $n=2$ can be expressed also as $\psi\xrightarrow{\mbb{Z}_2-\cov}\phi$ if and only if $\q\succ\p$ where $\p\eqdef(p_0,p_1)^T$ and $\q\eqdef(q_0,q_1)^T$. More generally, for arbitrary integer $n\in\mbb{N}$ we have that if $\psi\xrightarrow{\mbb{Z}_2-\cov}\phi$ then necessarily $\q\succ\p$. To see why,
observe that the relation $\chi_\psi(x)=\chi_\phi(x)f(x)$ implies that $f(0)=1$ and $f(x)$ itself can be expressed as a Fourier series
\be
f(x)=\sum_{z\in\mbb{Z}_n}r_ze^{\frac{i2\pi zx}{n}}
\ee
where $r_z\in\mbb{R}$. Since
$f$ is positive definition over $\mbb{Z}_n$, we must have that $r_z\geq 0$ for all $z\in\mbb{Z}_n$. Since $f(0)=1$ we conclude that $\{r_z\}_{z\in\mbb{Z}_n}$ is a probability distribution. Substituting the above expression for $f(x)$ into the relation $\chi_\psi(x)=\chi_\phi(x)f(x)$ gives
\be
\sum_{y\in\mbb{Z}_n}p_ye^{i\frac{2\pi yx}{n}}=\sum_{w,z\in\mbb{Z}_n}q_wr_ze^{i\frac{2\pi (z+w)x}{n}}\;.
\ee
Considering all summations and subtractions to be modulus $n$, we change variables on the right hand side of the equation above by denoting $y=z+w$ so that
\be
\sum_{y\in\mbb{Z}_n}p_ye^{i\frac{2\pi yx}{n}}=\sum_{y,w\in\mbb{Z}_n}q_{w}r_{y-w}e^{i\frac{2\pi yx}{n}}\;.
\ee
Hence, the equation above implies that for all $y\in\mbb{Z}_n$
\be\label{alternative}
p_y=\sum_{w\in\mbb{Z}_n}q_{w}r_{y-w}\;.
\ee
Denoting by $\p\eqdef(p_0,\ldots,p_{n-1})^T$, $\q\eqdef(q_0,\ldots,q_{n-1})^T$ and by $R$ the $n\times n$ matrix whose $(y,w)$ component is $r_{y-w}$ we can express the equation above as $\p=R\q$.
Observe that by definition $R$ is doubly stochastic so that $\q\succ\p$.

The relation given in~\eqref{alternative} can be used to provide alternative necessary and sufficient conditions for $\psi\xrightarrow{\mbb{Z}_n-\cov}\phi$. First, note that by rewriting~\eqref{altenative} with $z\eqdef y-w$ we get that $\psi\xrightarrow{\mbb{Z}_n-\cov}\phi$ if and only if there exists $\r=(r_0,\ldots,r_{n-1})^T\in\prob(n)$ such that
\be\label{alternative}
p_y=\sum_{z\in\mbb{Z}_n}q_{y-z}r_{z}\;.
\ee
Next, observe that the equation above can be expressed simply as $\p=Q\r$, where $Q$ is an $n\times n$ matrix whose $(y,z)$ component is $q_{y-z}$. Hence, assuming $Q$ is invertible we conclude that $\psi\xrightarrow{\mbb{Z}_n-\cov}\phi$ if and only if $Q^{-1}\p\geq 0$, where the inequality is entry-wise.
In order to avoid the computation of $Q^{-1}$ we can also use the Cramer's rule as we discuss now.

The matrix $Q$ as defined above is known as a \emph{circulant} matrix. The eigenvalues of such matrices are given by the discrete Fourier transforms. Specifically, the $x$-th eigenvalue of $Q$ is given by  \be
\lambda_x(Q)=\chi_\phi(x)=\sum_{y\in\mbb{Z}_n}p_ye^{i\frac{2\pi yx}{n}}\quad\quad\forall\;x\in\mbb{Z}_n\;.
\ee
The matrix $Q$ is also doubly stochastic so its determinant is in the interval $[0,1]$. Therefore, as long as $\chi_\phi(x)\neq 0$ for all $x\in\mbb{Z}_n$ we have $\det(Q)>0$. 
Next, for any $x\in\mbb{Z}_n$ let $Q_x$ be the matrix obtained from $Q$ by replacing the $x$-th column with the column $(p_0,p_1,\ldots,p_{n-1})^T$. Then, assuming $\det(Q)>0$ we get from Cramer's rule  that 
$\psi\xrightarrow{\mbb{Z}_n-\cov}\phi$ if and only if $\det(Q_x)\geq 0$ for all $x\in\mbb{Z}_n$.

 As a specific example, consider the case $n=3$. For this case the matrix $Q$ has the form
\be
Q=\bpm 
q_0 & q_2 & q_1\\
q_1 & q_0 & q_2\\
q_2 & q_1 & q_0
\epm\;.
\ee
Observe that $\det(Q)\geq 0$ with equality if and only if $q_0=q_1=q_2$. That is, if $|\phi\ra\neq\frac1{\sqrt{3}}(|0\ra+|1\ra+|2\ra)$ then $\det(Q)>0$. Hence, $\psi\xrightarrow{\mbb{Z}_3-\cov}\phi$ if and only if the following three conditions hold:
\ba
&\det(Q_0)=p_0(q_0^2-q_1q_2)+p_1(q_1^2-q_0q_2)+p_2(q_2^2-q_0q_1)\geq 0\\
&\det(Q_1)=p_0(q_2^2-q_0q_1)+p_1(q_0^2-q_1q_2)+p_2(q_1^2-q_0q_2)\geq 0\\
&\det(Q_2)=p_0(q_1^2-q_0q_2)+p_1(q_2^2-q_0q_1)+p_2(q_0^2-q_1q_2)\geq 0\;.
\ea 

\bex
Show that for every $\psi\in\pure(\mbb{C}^n)$ we have
$\Phi\xrightarrow{\mbb{Z}_n-\cov}\psi$, where $|\Phi\ra\eqdef\frac1{\sqrt{n}}\sum_{x=0}^{n-1}|x\ra$. In other words, $\Phi$ is a state with maximal $\mbb{Z}_n$-asymmetry.
\eex

\bex
Consider the case $n=3$, and let $|\psi\ra=\sum_{x=0}^3\sqrt{p_x}|x\ra$ and $|\phi\ra=\sum_{x=0}^3\sqrt{q_x}|x\ra$.
\ben
\item Show that if $q_1=q_2$ then $\psi\xrightarrow{\mbb{Z}_3-\cov}\phi$ if and only if $\q\succ\p$.
\item Show that for $\p=(5/12,7/24,7/24)^T$ and $\q=(5/12,1/3,1/4)^T$ it is not possible to convert $\psi$ to $\phi$ by $\mbb{Z}_3$-covariant operations even though $\q\succ\p$. 
\een
\eex

\subsection{Catalysis}\index{catalysis of asymmetry}

In every resource theory, if the state $\psi$ cannot be deterministically transformed into the state $\phi$ using the limited set of operations, the use of a catalyst provides a potential solution. As we explored in earlier chapters, a catalyst refers to an additional system that is initially prepared in a state not compatible with the constraints of the resource theory but must be restored to its original state at the conclusion of the process. An illustrative example can be found in the resource theory of entanglement, as discussed in Sec.~\ref{ecatal}, where we observed that certain conversions between states are prohibited under LOCC. However, by employing LOCC alongside a suitable catalyst, such conversions become achievable.

This notion of catalysis vividly demonstrates the significant variations encountered within the resource theory of asymmetry, contingent upon the choice of groups involved.  Specifically, we will demonstrate that a catalyst holds no utility for a connected compact Lie group, whereas for a finite group, a catalyst always exists.

\begin{myt}{}
\begin{theorem}
Let $\psi,\phi\in\pure(A)$, and $\G$ be a group with a projective unitary representation $g\mapsto U_g^A$. Suppose further $\psi$ cannot be converted to $\phi$ by $\G$-covariant operations. Then,
\ben
\item If $\G$ is a finite group, then there exists an ancillary system $C$ along with a projective unitary representation $g\mapsto U_g^C$, and a state $\varphi\in\pure(C)$, such that
\be\label{catalysisg}
\psi^A\otimes\varphi^C\xrightarrow{\G-\cov}\phi^A\otimes\varphi^C\;.
\ee
\item If $\G$ is a connected compact Lie group then~\eqref{catalysisg} can never hold.
\een
\end{theorem}
\end{myt} 

\begin{proof}
Suppose first that $\G$ is a finite group, and let $g\mapsto U_g^C$ be the regular representation of $\G$ on the space $C\eqdef\mbb{C}^{|\G|}=\spa\{|g\ra\;:\;g\in\G\}$ (see Sec.~\ref{sec:rr}). Fix an element $h\in\G$ and let $|\varphi^C\ra\eqdef|h\ra^C$. By the definition of the regular representation, we have that $\chi_\varphi(g)=\delta_{e,g}$, so that~\eqref{holdstri} holds trivially. Therefore, for any $h\in\G$, the state $|\varphi^C\ra =|h\ra$ satisfies~\eqref{catalysisg}. 

We next prove that if $\G$ is a connected compact Lie group then the relation~\eqref{catalysisg} never holds. Suppose by contradiction that~\eqref{catalysisg} does hold. Then, from Theorem~\ref{purestateconv} there exists a positive-definite function $f:\G\to\mbb{C}$ such that $\chi_{\psi\otimes\varphi}(g)=\chi_{\phi\otimes\varphi}(g)f(g)$ for all $g\in\G$. Since the representation on system $AC$ is given by $g\mapsto U_g^A\otimes U_g^C$, we have $\chi_{\psi\otimes\varphi}(g)=\chi_{\psi}(g)\chi_{\varphi}(g)$ and similarly $\chi_{\phi\otimes\varphi}(g)=\chi_{\phi}(g)\chi_{\varphi}(g)$ so that
\be\label{holdstri}
\chi_{\psi}(g)\chi_{\varphi}(g)=\chi_{\phi}(g)\chi_{\varphi}(g)f(g)\quad\quad\forall\;g\in\G\;.
\ee
As discussed  below Definition~\ref{gregular}, since $\G$ is a connected compact Lie group, there exists a neighbourhood, $\mc$, around the identity element of the group such that $\chi_{\varphi}(g)\neq 0$ for all elements $g\in\mc$. Combining this with the equation above gives
\be
\chi_{\psi}(g)=\chi_{\phi}(g)f(g)\quad\quad\forall\;g\in\mc\;.
\ee
However, since the functions $\chi_\psi$, $\chi_\phi$, and $f$, are all analytic, the identity theorem in complex analysis implies that the equality above holds for all $g\in\G$. Hence, from Theorem~\ref{purestateconv} we get that $\psi\xrightarrow{\G-\cov}\phi$ in contradiction with the asumption of the theorem that $\psi$ cannot be converted to $\phi$ by $\G$-covariant operations. Hence, the relation~\eqref{catalysisg} cannot hold if $\G$ is a connected compact Lie group.
\end{proof}

The existence of a catalyst for finite groups is a consequence of the fact that for finite groups, it is possible to completely overcome the lack of a shared reference frame by sending a single resource from Alice to Bob. In the proof presented above, the state $|\varphi^C\ra\eqdef|h\ra^C$ serves as an ``ultimate" resource that removes the restriction to $\G$-covariant operations. To understand why, let's revisit the guessing probability\index{guessing probability} given in~\eqref{gpfiducial}.

Taking $\rho=\varphi^C$ with the regular representation $U_g|h\lr h|U_g^*=|gh\lr gh|$ yields
\be
\pr_{\text{guess}}\left(\rho,\{\Lambda_g\}_{g\in\G}\right)=\frac1{|G|}\sum_{g\in\G}\tr\left[\Lambda_{g}|gh\lr gh|\right]\;.
\ee
Therefore, by choosing $\Lambda_g\eqdef |gh\lr gh|$, we obtain $\pr_{\text{guess}}\left(\rho,\{\Lambda_g\}_{g\in\G}\right)=1$. Consequently, upon receiving the state $\varphi^C$, Bob can determine the group element $g$ that relates his reference frame to Alice's reference frame.

The above discussion demonstrates that for finite groups, the notion of catalysis is not very useful as it merely reflects the existence of an ultimate resource for such groups, specifically in relation to the regular representation. On the other hand, for connected Lie groups, catalysis does not exist, and it remains an open problem whether catalysis can exist for arbitrary compact Lie groups.

\subsection{Probabilistic Transformations} 

In this subsection we study the conversion of a pure state  to an ensemble of pure states by $\G$-covariant operations. Recall that every ensemble of pure states $\{p_x,\phi_x^A\}_{x\in[m]}$ can characterized with a cq-state\index{cq-state} in $\md(AX)$ of the form
\be
\sigma^{AX}\eqdef\sum_{x\in[m]}p_x\phi_x^A\otimes |x\lr x|^X\;.
\ee
Given a pure state $\psi\in\pure(A)$ we want to find the conditions under which the conversion
$\psi^A\xrightarrow{\G-\cov}\sigma^{AX}$ is posible.

\begin{myt}{}
\begin{theorem}
Using the same notations as above, $\psi^A\xrightarrow{\G-\cov}\sigma^{AX}$ if and only if there exists normalized positive-definite and continuous (in the case of Lie group) functions $f_x:\G\to\mbb{C}$ such that 
\be\label{ptr0}
\chi_\psi(g)=\sum_{x\in[n]}p_xf_x(g)\chi_{\phi_x}(g)\;.
\ee
\end{theorem}
\end{myt}

\begin{proof}
From the covariant version of Stinespring dilation theorem, $\mE\in\cov_\G(A\to AX)$ if and only if there exists a system $E$, a projective unitary representation $g\mapsto U_g^E$, and an intertwiner isometry $V:A\to AXE$ such that for all $\eta\in\ml(A)$ we have $\mE(\eta)=\tr_E(V\eta V^*)$. Therefore, $\psi^A\xrightarrow{\G-\cov}\sigma^{AX}$ if and only if there exists an intertwiner isometry $V:A\to AXE$ 
such that
\be
\sigma^{AX}=\mE^{A\to AX}\left(\psi^A\right)=\tr_E\left[V\psi^A V^*\right].
\ee

We first assume that such a covariant channel $\mE^{A\to AX}$ exists, and prove the relation~\eqref{ptr0}. Indeed, the equation above implies that $V|\psi^A\ra$ is a purification of $\sigma^{AX}$ and therefore have the form
\be\label{ts0}
V|\psi^A\ra=\sum_{x\in[m]}\sqrt{p_x}|\phi_x^{A}\ra|x\ra^X|\varphi_x^E\ra\;,
\ee
for some orthonormal set $\{|\varphi_x^E\ra\}_{x\in[m]}$ in $E$. Since $\G$ acts trivially on system $X$, and since $V$ is an intertwiner we get that
\ba\label{ts1}
VU_g^A|\psi^A\ra&=\left(U_g^B\otimes I^X\otimes U_g^E\right)V|\psi^A\ra\\
&=\sum_{x\in[m]}\sqrt{p_x}U_g^A|\phi_x^{A}\ra\otimes |x\ra^X\otimes U_g^E|\varphi_x^E\ra
\ea
Finally, taking the inner product between the two states in~\eqref{ts0} and~\eqref{ts1} gives
\be
\chi_\psi(g)=\sum_{x\in[m]}p_x \chi_{\phi_x}(g)\chi_{\varphi_x}(g)\;.
\ee
Since $f_x(g)\eqdef\chi_{\varphi_x}(g)$ is a positive-definitive function (see Theorem~\ref{pdd}) we get that~\eqref{ptr0} holds.

Conversely, suppose~\eqref{ptr0} holds. From Theorem~\ref{pdd} $f_x$ can be expressed as the characteristic function of some state $\varphi_x^E$. Without loss of generality we can assume that the states $\{|\varphi_x^E\ra\}_{x\in[m]}$ are orthonormal since otherwise we can replace each $|\varphi_x^E\ra$ with $|\varphi_x^E\ra|x\ra^{E'}$, where $E'$ is another ancillary system upon which the group $\G$ acts trivially (so that $|\varphi_x^E\ra$ and $|\varphi_x^E\ra|x\ra^{E'}$ have the same characteristic function). With this in mind, let 
\be
|\phi^{AXE}\ra\eqdef \sum_{x\in[m]}\sqrt{p_x}|\phi_x^{A}\ra|x\ra^X|\varphi_x^E\ra\;.
\ee
Then, from~\eqref{ptr0} we get that $\chi_\psi(g)=\chi_{\phi}(g)$ for all $g\in\G$.
Moreover, there exists a $\G$-invariant state $|0\ra\in XE$ whose characteristic function is  constant and equal to one for all group elements. Therefore, from the relation $\chi_\psi(g)=\chi_\phi(g)$ we get that the states $|\psi^A\ra|0\ra^{XE}$ and $|\phi^{AXE}\ra$ have the same characteristic function.
Therefore, there exists a $\G$-invariant unitary $V:AXE\to AXE$ such that $V\left(|\psi^A\ra|0\ra^{XE}\right)=|\phi^{AXE}\ra$. Taking the trace over $E$ on both sides demonstrates that $\psi^A$ can be converted to $\sigma^{AX}$ by a $\G$-covariant channel. This completes the proof.
\end{proof}

\bex
Prove the following corollary to the theorem above: The conversion\\ $\psi^A\xrightarrow{\G-\cov}\phi^{A}$ can be achieved with probability $q$ if and only if there exists a normalized positive definition function $f:\G\to \mbb{C}$ such that $\chi_\psi(g)-qf(g)\chi_\phi(g)$ is positive definite.
\eex

\section{Manipulation of Mixed Asymmetry}\index{single-shot}

In this section we study interconversions among asymmetric mixed states in the single-shot regime. Unlike like the pure-state case, for mixed state there is no simple criterion to determine in one state can be converted to another by $\G$-covariant operations. However, we will see that the problem can be casted as an SDP optimization problem.

We start with the following observation about the
Choi matrix $J^{AB}_\mE$ of a $\G$-covariant channel $\mE\in\cov_\G(A\to B)$. The $G$-covariance property implies that
\be
\mE^{A\to B}=\mU^{B}_g\circ\mE^{A\to B}\circ\mU^{*A}_g\quad\quad\forall\;g\in\G\;.
\ee
Applying both sides of this equation to the maximally entangled state $|\Omega^{A\tA}\ra$ gives 
\ba
J^{AB}_\mE&=\mU^{B}_g\circ\mE^{\tA\to B}\circ\mU^{*\tA}_g\left(\Omega^{A\tA}\right)\\
\GG{\eqref{ptrans}}&=\bar{\mU}^{A}_g\otimes\mU^{B}_g\circ\mE^{\tA\to B}\left(\Omega^{A\tA}\right)\\
&=\bar{\mU}^{A}_g\otimes\mU^{B}_g\left(J^{AB}_\mE\right)
\ea
Therefore, the matrix $J^{AB}_\mE$ is a Choi matrix of a $\G$-covariant channel if and only if
\be \label{chogasy}
\left(\bar{U}_g^{A}\otimes U_g^B\right) J^{AB}_\mE\left(\bar{U}_g^{A}\otimes U_g^{B}\right)^*=J^{AB}_\mE\quad\quad\forall\; g\in\G.
\ee
That is, the Choi matrix $J^{AB}_\mE$ is symmetric with respect to the projective unitary representation $g\mapsto \bar{U}_g^{A}\otimes U_g^B$. In this section, we will denote by $\mG\in\cptp(AB\to AB)$ the $\G$-twirling operation with respect to this representation, so that $\mE$ is $\G$-covariant if and only if $\mG\left(J_\mE^{AB}\right)=J^{AB}_\mE$.

With this property we can use Theorem~\ref{thm:1111} to get necessary and sufficient conditions for a conversion of one mixed state to another by $\G$-covariant operations. 
To apply Theorem~\ref{thm:1111} for the case that $\mf(A\to B)=\cov_\G(A\to B)$, observe that 
\ba\label{extras}
\sup_{\mE\in\cov_\G(A\to B)}\tr\left[\eta^B\mE^{A\to B}\left(\rho^A\right)\right]&=\sup_{\mE\in\cov_\G(A\to B)}\tr\left[J_\mE^{AB}\left(\rho^T\otimes \eta^B\right)\right]\\
&=\sup_{\substack{J\in\pos(AB)\\ J^A=I^A}}\tr\left[J^{AB}\mG^{AB\to AB}\left(\rho^T\otimes \eta^B\right)\right]\\
\GG{\eqref{bb2b}}&=2^{-H_{\min}^\ua(B|A)_{\mG\left(\rho^T\otimes \eta\right)}}\;.
\ea
Therefore, Theorem~\ref{thm:1111} implies the following characterization of $\rho^A\xrightarrow{\G-\cov}\sigma^{B}$.

\begin{myg}{}
\begin{corollary}
Let $\rho\in\md(A)$ and $\sigma\in\md(B)$. The following are equivalent:
\ben
\item $\rho^A\xrightarrow{\G-\cov}\sigma^{B}$.
\item For all $\eta\in\md(B)$ we have
$
H_{\min}^\ua(B|A)_{\mG\left(\rho^T\otimes \eta\right)}\leq H_{\min}^\ua(B|\tB)_{\mG\left(\sigma^T\otimes \eta\right)}
$.
\een
\end{corollary}
\end{myg}

While the condition outlined in the corollary holds theoretical significance, it falls short of offering a practical methodology for assessing whether a quantum state $\rho^A$ can be transformed into another state $\sigma^B$ through $\G$-covariant operations. To address this gap, a more applicable criterion is derived from the condition presented in~\eqref{abcxz}. For the context at hand, this criterion is articulated in a specific format, which we encapsulate as a theorem for clarity and ease of application.

\begin{myt}{}
\begin{theorem}\label{thcgp}
Let $\rho\in\md(A)$, $\sigma\in\md(B)$, and define
\be
f\left(\rho,\sigma\right)\eqdef\min_{\tau\in\pos(A),\;\eta\in\md(B)}\Big\{\tr[\tau-\sigma\eta]\;:\;\tau^A\otimes I^{B}\geq\mG\left(\rho^T\otimes \eta^B\right)\Big\}\;.
\ee
Then, $\rho^A\xrightarrow{\G-\cov}\sigma^{B}$ if and only if $f\left(\rho,\sigma\right)\geq 0$.
\end{theorem}
\end{myt}

\begin{remark}
The optimization of the function $f(\rho,\sigma)$ can be solved efficiently and algorithmically with an SDP program.
\end{remark}

The proof of the theorem above is based on the fact that $\sigma=\mE(\rho)$ if and only if for all $\Lambda\in\herm(B)$ we have $\tr[\Lambda\sigma]=\tr[\Lambda\mE(\rho)]$. This relation can be expressed as
\be
\tr[\Lambda\sigma]=\tr\left[J_\mE^{AB}\left(\rho^T\otimes \Lambda^B\right)\right]\;.
\ee
In the following exercise you use this to complete the proof.

\bex
Use~\eqref{abcxz} and~\eqref{extras} to prove the theorem above.
\eex

\bex
Let $\rho\in\md(A)$ and $\sigma\in\md(B)$. Show that $\rho^A\xrightarrow{\G-\cov}\sigma^{B}$ if and only if there exists $\mF\in\cptp(A\to B)$ such that for all $g\in \G$
\be\label{fug}
\mF\big(\mU_g(\rho)\big)=\mU_g(\sigma)\;.
\ee 
\eex

\bex
The conversion distance from $\rho\in\md(A)$ to $\sigma\in\md(B)$ is defined as
\be
T\left(\rho\xrightarrow{\mf}\sigma\right)\eqdef \min_{\mE\in\cov_G(A\to B)}\frac12\left\|\sigma^B-\mE^{A\to B}(\rho^A)\right\|_1\;.
\ee
Use the trace distance property
\be
\frac12\left\|\sigma^B-\mE^{A\to B}(\rho^A)\right\|_1=\min_{\substack{\Lambda\in\pos(B)\\\Lambda\geq \sigma^B-\mE^{A\to B}(\rho^A)}}\tr\left[\Lambda\right]\;,
\ee
to show that the conversion distance can be expressed as the following SDP: 
\be\label{ascd}
T\left(\rho\xrightarrow{\mf}\sigma\right)=\min \tr\left[\Lambda\right]
\ee
subject to:
\begin{enumerate}
\item $\Lambda^B\geq \sigma^B-\tr_A\left[J^{A\tA}\left(\rho^T\otimes I^{\tA}\right)\right]$.
\item $J^A=I^A$.
\item $\mG(J^{A\tA})=J^{A\tA}$.
\item $\Lambda\in\pos(B),J\in\pos(A\tA)$.
\end{enumerate}
\eex

\section{Time Translation Symmetry}\label{sectts}\index{time translation}

Time-translation symmetry, also known as time-translational invariance\index{invariance}, is a fundamental concept in physics that relates to the behavior of physical systems under shifts or translations in time. It is a principle that states that the laws of physics remain unchanged or invariant over time.
In simpler terms, time-translation symmetry implies that the fundamental laws of physics do not depend on the specific moment in time at which they are applied. This means that if a physical experiment or process is performed today or tomorrow, under the same conditions, the outcome should be the same.

The concept of time-translation symmetry is closely related (via Noether's theorem) to the conservation of energy, where the total energy of a closed system remains constant over time. It provides a foundation for many important principles and theories in physics, including the laws of motion, quantum mechanics, and relativity. Therefore, time-translation symmetry is a fundamental symmetry in the universe, and it plays a crucial role in our understanding of the laws governing the behavior of matter and energy over time. Given its importance, we devote this section to study the resource theory of asymmetry with respect to this time-translation symmetry. 

\subsection{Time-Translation Covariant Operations}

We say that a quantum state $\rho\in\md(A)$ is time-translation invariant with respect to a Hamiltonian $H^A\in\pos(A)$ if, for all $t\in\mbb{R}$, the following equation holds:
\be
e^{-iH^At}\rho^A e^{iH^At}=\rho^A\;.
\ee
Note that $\rho^A$ is time-translation invariant if and only if it commutes with the Hamiltonian. This is why it is sometimes referred to in the literature as ``quasi-classical," as both the state and the Hamiltonian are diagonal with respect to the same basis. Throughout this section, we will alway work with the eigenbases of the Hamiltonians. 

The Hamiltonian $H^A$ can be decomposed as 
\be\label{ahamil}
H^A=\sum_{x\in[m]}a_x\Pi_x^A\;,
\ee 
where $\{a_x\}_{x\in[m]}$ is the set of distinct eigenvalues of $H^A$ and each $\Pi_x^A$ is a projection to the eigenspace of $a_x$. Without loss of generality, we will assume that $a_1<a_2<\cdots<a_m$ (noting that they are all distinct, allowing us to arrange $\{a_x\}_{x\in[m]}$ in increasing order). With the above form of $H^A$, the state $\rho^{A}$ is time-translation invariant, if and only if it takes the form:
\be\label{formpxqw}
\rho=\sum_{x\in[m]}p_x\rho_x\;,
\ee 
where $\p\in\prob(m)$, $\rho_x\in\md(A)$, and $\rho_x\rho_y=0$ for every $x\neq y$. We will use the notation $\inv(A)$ to denote the set of states in $\md(A)$ that are time-translation invariant.

\bex
Prove the above form of $\rho$. Hint: $\supp(\rho_x)\subseteq\supp(\Pi_x)$.
\eex

Consider a quantum channel $\mN\in\cptp(A\to B)$, where systems $A$ and $B$ have corresponding Hamiltonians $H^A\in\pos(A)$ and $H^B\in\pos(B)$. The channel $\mN$ is said to be \emph{time-translation covariant} if for all $t\in\mbb{R}$
\be\label{ttcfor}
\mN^{A\to B}\left(e^{-iH^At}\rho^A e^{iH^At}\right)=e^{-iH^Bt}\mN^{A\to B}(\rho^B) e^{iH^Bt}\quad\quad\forall\;t\in\mbb{R}\;.
\ee
We will use the notation $\cov(A\to B)$ to denote the set of all time-translation covariant channels in $\cptp(A\to B)$. 

In the Choi representation, the property given in~\eqref{ttcfor} can be expressed as (see the relation~\eqref{chogasy})
	\be
	\left(e^{-i\bar{H}^At}\otimes e^{iH^Bt}\right)J^{AB}_{\mN}\left(e^{-i\bar{H}^At}\otimes e^{iH^Bt}\right)^*=J^{AB}_{\mN}\quad\quad\forall\;t\in\mbb{R}\;.
	\ee
Note that $\bar{H}^A$ has the same eigenvalues as $H^A$. For our purposes, we can replace $\bar{H}^A$ (in the equation above)  with ${H}^A$, since it will not make any difference in our analysis. Therefore, $\mN\in\cov(A\to B)$ if and only if $J_{\mN}^{AB}$ commutes with the operator 
\ba\label{totalhamil}
\xi^{AB}\eqdef H^A\otimes I^B-I^A\otimes H^B\;.
\ea
Therefore, the degeneracy of the energy levels of the operator $\xi^{AB}$ will play a key role in the resource theory of time-translation asymmetry.

\subsubsection{The Pinching Channel}

The set of channels $\cov(A\to A)$ contains the pinching channel\index{pinching channel} $\mP_H$ as defined in Sec.~\ref{sec:pinching}. For $H^A$ as in~\eqref{ahamil}, the pinching channel on system $A$ is given by
\be\label{gtwirl}
\mP^{A\to A}_H\left(\rho^A\right)\eqdef\sum_{x\in[m]}\Pi_x^A\rho^A\Pi_x^A\;.
\ee
This pinching channel, also known as the ``twirling channel" (as it is the $\G$-twirling map with respect to the group $\G=\{e^{iH^At}\}_{t\in\mbb{R}}$), has the property that a state $\rho\in\md(A)$ is quasi-classical if and only if $\mP_H(\rho)=\rho$. 

\bex
Show that the condition $\mP_H(\rho)=\rho$ is equivalent to the condition that $\rho$ has the form given in~\eqref{formpxqw}.
\eex

\bex
Let $\mP\in\cptp(A\to A)$ and $\mP'\in\cptp(A'\to A')$ be the pinching channel\index{pinching channel} associated with the Hamiltonians $H^A$ and $H^{A'}$, respectively. Further, let $\mN\in\cptp(A\to A')$.
\ben\label{expart2}
\item  Show that $\mP'\circ\mN\circ\mP\in\cov(A\to A')$.
\item Show that if $\mN\in\cov(A\to A')$ then $\mP'\circ\mN=\mN\circ\mP$. 
\item Show that if the Hamiltonian $H^A$ is non-degenerate then
$
\mP^{A\to A}_H=\Delta^{A\to A}\;,
$
where $\Delta^{A\to A}$ is the completely dephasing channel as defined in Sec.~\ref{cdcha}.
\een
\eex

Covariant channels can also be characterized in terms of the pinching channel. Consider $\mN\in\cptp(A\to B)$ and let $\mP_\xi\in\cov(AB\to AB)$ be the pinching channel\index{pinching channel} associated with the operator $\xi^{AB}$ given in~\eqref{totalhamil}. Then, the quantum channel $\mN^{A\to B}$ is time-translation covariant if and only if its Choi matrix satisfies
\be
\mP^{AB\to AB}_\xi\left(J_\mN^{AB}\right)=J_\mN^{AB}\;.
\ee
This follows from Exercise~\ref{compinch} and our earlier observation that $\mN\in\cov(A\to B)$ if and only if its Choi matrix commutes with $\xi^{AB}$.

The twirling channel can also be used to quantify time-translation asymmetry. For example, the relative entropy distance of a quantum state $\rho\in\md(A)$ to its twirled state $\mP(\rho)$ is a time-translation asymmetry (sometimes referred to as coherence) measure given by
\be\label{coherence}
C(\rho)\eqdef D\left(\rho\big\|\mP_H(\rho)\right)=H\big(\mP_H(\rho)\big)-H(\rho)\;,
\ee
where $D(\rho\|\sigma)\eqdef\tr[\rho\log\rho]-\tr[\rho\log\sigma]$ is the Umegaki\index{Umegaki} relative entropy and $H(\rho)\eqdef-\tr[\rho\log\rho]$ is the von-Neumann\index{von-Neumann} entropy. The above function is non-increasing under time-translation covariant operations, and achieves its maximal value of $\log d$ (where $d\eqdef|A|$) on the maximally coherent state $|+\ra\eqdef\frac1{\sqrt{d}}\sum_{x\in[d]}|x\ra$, where $\{|x\ra\}_{x\in[d]}$ is the energy eigenbasis.
We next move to characterize the set $\cov(A\to B)$ in three different cases that depends on the level of degeneracy of the Hamiltonians involved.

\subsubsection{The Case of Relatively Non-Degenerate Hamiltonians}

Let $H^{A}$ and $H^{B}$ be the Hamiltonians of two systems $A$ and $B$, of dimensions  $m\eqdef |A|$ and $n\eqdef |B|$. The Hamiltonians can be expressed in their spectral decomposition as
	\be\label{hamilab}
	H^A=\sum_{x\in[m]}a_x|x\lr x|^A\quad\text{and}\quad H^B=\sum_{y\in[n]}b_y|y\lr y|^B\;,
	\ee
	where $\{a_x\}$ and $\{b_y\}$ are the energy eigenvalues of $H^A$ and $H^B$, respectively. 

\begin{myd}{}
\begin{definition}
We say that the Hamiltonians $H^A$ and $H^B$, as defined in~\eqref{hamilab}, are \emph{relatively non-degenerate} if for all $x,x'\in[m]$ and $y,y'\in[n]$ we have
\be
a_x-a_{x'}=b_y-b_{y'}\quad\Rightarrow\quad x=x'\;\text{ and }\;y=y'\;.
\ee
If the condition above does not hold we say that the Hamiltonians are relatively degenerate.
\end{definition}
\end{myd}

Note that if $H^A$ and $H^B$ are relatively non-degenerate, then each of them is also non-degenerate. For example, suppose $H^A$ is degenerate with $a_x=a_{x'}$ for some $x\neq x'\in[m]$. Then, for $y=y'$ we get $a_x-a_{x'}=0=b_y-b_{y'}$ even though $x\neq x'$. Therefore, relative non-degeneracy is a stronger notion than non-degeneracy. In fact, relative non-degeneracy of $H^A$ and $H^B$ is equivalent to the non-degeneracy of the operator $\xi^{AB}$ as defined in~\eqref{totalhamil}.
Moreover, in the generic case in which $H^A$ and $H^B$ are arbitrary (chosen at random) the Hamiltonians are relatively non-degenerate. 	
For this case, time-translation covariant channels have a very simple characterization.

\begin{myt}{}
\begin{theorem}\label{thm1}
{\rm Let $A$ and $B$ be two physical systems with relatively non-degenerate Hamiltonians. Then, $\mN\in\cptp(A\to B)$ is a time-translation covariant channel if and only if
\be\label{ccgasy}
\mN^{A\to B}=\Delta^{B\to B}\circ\mN^{A\to B}\circ\Delta^{A\to A}\;,
\ee
where $\Delta^{A\to A}$ and $\Delta^{B\to B}$ are the completely dephasing channels of systems $A$ and $B$, respectively. In other words, for physical systems with relatively non-degenerate Hamiltonians only classical channels are time-translation covariant.}
\end{theorem}
\end{myt}
\begin{proof}
Since we assume that the Hamiltonians $H^A$ and $H^B$ are relatively non-degenerate we get that the joint operator, $\xi^{AB}$, is non-degenerate. Hence, $J_\mN^{AB}$ is diagonal in the same eigenbasis $\{|x\ra^A|y\ra^B\}_{x\in[m],y\in[n]}$ of $\xi^{AB}$, so that
\be
\Delta^{A\to A}\otimes\Delta^{B\to B}\left(J_\mN^{AB}\right)=J_\mN^{AB}\;.
\ee
The above equation describes the same relation as the one given in~\eqref{ccgasy}. Hence, $\mN^{A\to B}$ is a classical channel. This completes the proof. 
\end{proof}

\subsubsection{The Case of Bohr Spectrum}\index{Bohr spectrum}

We consider now the case in which $A=B$. Therefore, since $H^A=H^B$  we cannot apply the characterization theorem above to this case. Instead, we will assume that $H^A$ has a non-degenerate Bohr spectrum. In its spectral decomposition $H^A$ has the form 
\be\label{someha}
H^A=\sum_{x\in[m]}a_x|x\lr x|^A\;,
\ee 
where $\{a_x\}_{x\in[m]}$ is the set of distinct eigenvalues of $H^A$.

\begin{myd}{}
\begin{definition}\label{curres}	
We say that $H^A$ as given in~\eqref{someha} has a non-degenerate Bohr spectrum if it has the property  that for any $x,y,x',y'\in[m]$ 
\be
a_x-a_y=a_{x'}-a_{y'}\quad\iff\quad x=x'\text{ and }y=y'\quad
\text{or}\quad x=y\text{ and }x'=y'\;;\nonumber
\ee
that is, there are no degeneracies in the nonzero
differences of the energy levels of $H^A$.
\end{definition}
\end{myd}

\bex
Show that $H^A$ has a non-degenerate Bohr spectrum if and only if all the non-zero eigenvalues of the operator
\be
\xi^{A\tA}\eqdef H^A\otimes I^{\tA}-I^A\otimes H^{\tA}
\ee
are distinct. In other words, $H^A$ has a non-degenerate Bohr spectrum if and only if the zero eigenvalue of $\xi^{A\tA}$ is the sole eigenvalue with a multiplicity greater than one.
\eex

It is noteworthy that the vast majority of Hamiltonians exhibit a non-degenerate Bohr spectrum, indicating that Hamiltonians lacking this feature are exceptionally rare, constituting a set of measure zero. This observation segues into a focused interest in time-translation covariant channels that are compatible with non-degenerate Bohr spectra, offering a unique area for characterization.

\begin{myt}{}
\begin{theorem}\label{iilem1}
Consider a Hamiltonian, $H^A$, as outlined in \eqref{someha}, which is characterized by having a non-degenerate Bohr spectrum, and let $\mN\in\cptp(A\to A)$. The channel $\mN\in\cov(A\to A)$ if and only if for all $x,x',y,y'\in[m]$, $\la xx'|J_\mN^{A\tA}|yy'\ra=0$ unless $x=x'$ and $y=y'$, 
\text{or} $x=y$ and $x'=y'$.
\end{theorem}
\end{myt}
\begin{remark}
We will see below that even if the spectrum of the Hamiltonian $H^A$ has degeneracies, any quantum channel $\mN\in\cptp(A\to A)$ whose Choi matrix has the form~\eqref{bohr} is necessarily time-translation covariant. 
\end{remark}

\begin{proof}
Following the same lines as in Theorem~\ref{thm1}, by replacing $H^B$ with $H^A$ everywhere, we get that a quantum channel $\mN\in\cptp(A\to A)$ is time-translation covariant if and only if its Choi matrix $J^{A\tA}_\mN$ commutes with the operator
\be
\xi^{A\tA}\eqdef H^A\otimes I^{\tA}-I^{A}\otimes H^{\tA}=\sum_{x,y\in[m]}(a_x-a_y)|x\lr x|^A\otimes|y\lr y|^{\tA}\;.
\ee
Since $H^A$ has a non-degenerate Bohr spectrum, the set $\{a_x-a_y\}$ that appear in the sum above consists of distinct eigenvalues, as we only consider indices $x,y\in[m]$ that satisfy $y\neq x$. We therefore conclude that the pinching channel $\mP_\xi\in\cptp(A\tA\to A\tA)$ associated with the operator $\xi^{A\tA}$ is given by 
\be
\mP_\xi(\cdot)=\Pi(\cdot)\Pi+\sum_{\substack{x,y\in[m]\\ x\neq y}}P_{xy}(\cdot)P_{xy}\;,
\ee
where
\be\label{15p247}
P_{xy}\eqdef|xy\lr xy|\quad\text{and}\quad\Pi\eqdef\sum_{x\in[m]}|xx\lr xx|\;.
\ee
Observe that $\Pi$ is the projection to the zero eigenspace of $\xi^{A\tA}$.
With these notations the condition $J_\mN^{A\tA}=\mP_\xi\left(J_\mN^{A\tA}\right)$ is equivalent to
\be\label{jchoiinto2}
J_\mN^{A\tA}=\Pi J_\mN^{A\tA}\Pi+\sum_{\substack{x,y\in[m]\\ x\neq y}}P_{xy}J_\mN^{A\tA}P_{xy}\;\;.
\ee
Observe that Choi matrix $J_\mN^{A\tA}$ satisfies the condition above if and only if $\la xx'|J_\mN^{A\tA}|yy'\ra=0$ unless $x=x'$ and $y=y'$, 
\text{or} $x=y$ and $x'=y'$.
This completes the proof.
\end{proof}

The condition in the theorem is equivalent to the statement that the Choi matrix has the form
\be\label{bohr}
	J^{A\tA}_\mN=\sum_{x,y\in[m]}\left(p_{y|x}|xy\lr xy|^{A\tA}+(1-\delta_{xy})q_{xy}|xx\lr yy|^{A\tA}\right)\;,
	\ee
where $q_{xy}\eqdef \big\la xx\big|J_\mN^{A\tA}\big|yy\big\ra$ and $p_{y|x}\eqdef\la xy|J_{\mN}^{A\tA}|xy\ra$. Observe that by definition $p_{x|x}=q_{xx}$ for all $x\in[m]$.  Given that $J_{\mN}^{AB}$ is the Choi matrix of a quantum channel, it implies certain properties for the coefficients $\{p_{y|x}\}_{x,y\in[m]}$ and the matrix $Q_\mN$, which consists of the components $q_{xy}$.
Specifically, the first term in the right hand side of the equation above corresponds to $\Pi J_{\mN}^{A\tA}\Pi$, and the second term is a sum over all  $P_{xy}J_{\mN}^{A\tA}P_{xy}$. Therefore, from the condition $\Pi J_{\mN}^{A\tA}\Pi\geq 0$ we get that $Q_\mN\geq 0$, where $Q_\mN$ is the matrix whose components are $q_{xy}$. Similarly, the condition $P_{xy}J_{\mN}^{A\tA}P_{xy}\geq 0$ implies that $p_{y|x}\geq 0$. Thus, we conclude that $J_{\mN}^{A\tA}\geq 0$ if and only if $Q_{\mN}\geq 0$ and each $p_{y|x}\geq 0$.
Finally, the remaining condition $J_\mN^{A}=I^A$ implies that for all $x\in[m]$ we have $\sum_{y\in[m]}p_{y|x}=1$. To summarize, the theorem above implies that $\mN\in\cov(A\to A)$ if its Choi matrix has the form~\eqref{bohr}, with $Q_\mN\geq 0$ and $\{p_{y|x}\}_{x,y\in[m]}$ being a conditional probability distribution. 

\subsection{Exact State Conversion}\label{sec:exactasy}

In this section, we examine the precise state conversions for each of the two distinct degeneracies we discussed above of the Hamiltonians involved. Similar to previous sections, we denote by $\rho\xrightarrow{\cov}\sigma$ the conversion of a quantum state $\rho$ to another quantum state $\sigma$ through covariant operations. It is worth mentioning that there are numerous other significant examples of Hamiltonians whose spectra are either degenerate or do not satisfy the condition stated in Definition~\ref{curres}. The QRT of time-translation asymmetry with such Hamiltonians is still in the process of being fully developed and remains an active area of research.

\subsubsection{The Case of Relatively Non-Degenerate Hamiltonians}	

First, we consider the conversion of	
a state $\rho\in\md(A)$ to a state $\sigma\in\md(B)$ by covariant operation, with Hamiltonians $H^A$ and $H^B$ that are relatively non-degenerate. As we saw in Theorem~\ref{thm1} the set $\cov(A\to B)$ consists of all classical channels in $\cptp(A\to B)$, with respect to the eigen-bases of the Hamiltonians $H^A$ and $H^B$. Therefore, in this case,  $\rho^A\xrightarrow{\cov}\sigma$ if and only if $\sigma^B$ is classical.

\bex
Prove the statement above. That is, show that $\rho^A\xrightarrow{\cov}\sigma^B$ if and only if $\sigma^B=\Delta^B(\sigma^B)$.
\eex

\subsubsection{The Case of Bohr Spectrum}

Next, we explore the exact single-shot interconversions of systems whose Hamiltonians possess a non-degenerate Bohr spectrum. This problem is more challenging compared to the similar problem involving relatively non-degenerate Hamiltonians. We will use the notation $\{|x\ra^A\}_{x\in[m]}$ to represent the energy eigenbasis of a Hamiltonian $H^A$, and consider two density matrices in $\md(A)$:
\be\label{12}
\rho^A=\sum_{x,x'\in[m]}r_{xx'}|x\lr x'|^A\;\text{ and }\;\sigma^A=\sum_{x,x'\in[m]}s_{xx'}|x\lr x'|^A
\ee
with components $\{r_{xx'}\}$ and $\{s_{xx'}\}$, respectively. In the theorem below we assume that $r_{xx'}\neq 0$ for all $x,x'\in[m]$, and define the $m\times m$ matrix $Q$, with components
\be\label{1369}
q_{xy}\eqdef\begin{cases} 
\min\left\{1,\frac{s_{xx}}{r_{xx}}\right\} & \text{ if }x=y\\
\frac{s_{xy}}{r_{xy}}& \text{ otherwise.}
\end{cases}
\ee

\begin{myt}{}	
\begin{theorem}\label{ttcs}
Let $\rho,\sigma\in\md(A)$ be as in~\eqref{12} with $r_{xx'}\neq0$ for all $x,x'\in[m]$, and suppose the Hamiltonian $H^A$ has a non-degenerate Bohr spectrum.
Then, the following statements are equivalent:
\ben
\item There exists $\mE\in\cov(A\to A)$ such that $\sigma=\mE(\rho)$. 
\item The matrix $Q$ as defined in~\eqref{1369} is positive semidefinite.
\een
\end{theorem}
\end{myt}
\begin{remark}
We will see in the proof below that the second statement implies the first statement even if the Hamiltonian $H^A$ has a degenerate Bohr spectrum.
Moreover, we will see that if $r_{xy}=0$ for some off diagonal terms (i.e. $x\neq y$) then $s_{xy}$ must also be zero. However, in this case, for any $x\neq y\in[m]$ with $r_{xy}=0$, the components of $q_{xy}$ can be arbitrary. This means that in this case the condition becomes cumbersome, as we will need to require that there \emph{exists} $Q$ as defined above but with no restriction on the components $q_{xy}$ for which $r_{xy}=0$.
\end{remark}
\begin{proof}	
	 From Theorem~\ref{iilem1} and the preceeding discussion below~\eqref{bohr}, it follows that there exists $\mN\in\cov(A\to A)$ such that $\sigma=\mN(\rho)$ if and only if there exists a conditional probability distribution $\{p_{y|x}\}_{x,y\in[m]}$, and an $m\times m$ positive semidefinite matrix $Q$, such that
	\ba
	\sigma=\mN(\rho)&=\tr_A\left[J^{A\tA}_\mN(\rho^T\otimes I^{\tA})\right]\\
	&=\sum_{x, y\in[m]}p_{y|x}r_{xx}|y\lr y|+\sum_{\substack{x\neq y\\ x,y\in[m]}}q_{xy}r_{xy}|x\lr y|
	\ea
	That is, $\sigma=\mN(\rho)$ if and only if 
	\ba
	&s_{yy}=\sum_{x\in[m]}p_{y|x}r_{xx}\quad\quad\forall\;y\in[m]\quad\text{and}\\
	&s_{xy}=q_{xy}r_{xy}\quad\quad\forall\;x\neq y\in[m]\;.
	\ea
	Hence, for the off diagonal terms, $s_{xy}=0$ whenever $r_{xy}=0$. Since
	we assume that  that all the off-diagonal terms of $\rho$ are non-zero, i.e.\ $r_{xy}\neq 0$ for $x\neq y$, there is no freedom left in the choice of the off diagonal terms of $Q_\mN$ and we must have $q_{xy}=\frac{s_{xy}}{r_{xy}}$.
Since $Q_\mN$ must be positive semidefinite we will maximize its diagonal terms $\{p_{x|x}\}_{x\in[m]}$ given the constraint that $s_{yy}=\sum_{x\in[m]}p_{y|x}r_{xx}$. This constraint immediately gives $s_{yy}\geq p_{y|y}r_{yy}$ so that we must have $p_{y|y}\leq\frac{s_{yy}}{r_{yy}}$. Clearly, we also have $p_{y|y}\leq 1$ so we conclude that
	\be
	p_{y|y}\leq\min\left\{1,\frac{s_{yy}}{r_{yy}}\right\}\;.
	\ee
	Remarkably, this condition is sufficient since there exists conditional probabilities $\{p_{y|x}\}$, with both $p_{y|y}=\min\left\{1,\frac{s_{yy}}{r_{yy}}\right\}$ and 	$s_{yy}=\sum_{x\in[m]}p_{y|x}r_{xx}$.	Indeed,  for simplicity set $r_x\eqdef r_{xx}$ and $s_x\eqdef s_{xx}$, and define
	\be\label{coedf}
	p_{y|x}\eqdef
	\begin{cases}
	\min\left\{1,\frac{s_{x}}{r_{x}}\right\} &\text{if }x=y\\
	\frac{1}{\mu r_x}(s_{y}-r_{y})_+(r_x-s_x)_+ & \text{otherwise}
	\end{cases}
	\ee
	where
	\be
	\mu\eqdef\sum_{y\in[m]}(s_{y}-r_y)_+=\frac12\|\s-\r\|_1\;,
	\ee
	and we used the notation $(s_{y}-r_y)_+\eqdef s_y-r_y$ if $s_y\geq r_y$ and $(s_{y}-r_y)_+\eqdef 0$ if $s_y< r_y$. Clearly, $p_{y|x}\geq 0$, and it is straightforward to check that $\sum_{y\in[m]}p_{y|x}=1$ and $s_{y}=\sum_{x\in[m]}p_{y|x}r_{x}$; that is, the above conditional probability distribution satisfies all the required conditions. This completes the proof. 
\end{proof}

Observe that if $H^A$ has degenerate Bohr spectrum and $Q\geq 0$ then	we still get that the Choi matrix of the form~\eqref{bohr} (with $p_{y|x}$ as in~\eqref{coedf} and $q_{xy}$ as in~\eqref{1369}) corresponds to a quantum channel $\mN\in\cptp(A\to A)$ with the property that $\sigma=\mN(\rho)$.
As discussed below the proof of Theorem~\ref{iilem1}, all channels with a Choi matrix of the form~\eqref{bohr} are time-translation covariant. Hence, $\mN\in\cov(A\to A)$.
		
\bex
In the proof above we saw that if $r_{xy}=0$ for some $x\neq y$ then $\sigma=\mE(\rho)$ for some $\mE\in\cov(A\to A)$ only if $s_{xy}=0$. Use this to show  that if $\rho$ has a block diagonal form $\rho=\begin{pmatrix}\trho & \0\\
\0 & \0
\end{pmatrix}$, and if it can be converted by a time-translation covariant channel to $\sigma$, then $\sigma$ must have the form $\sigma=\begin{pmatrix}\tsigma & \0\\
\0 & D
\end{pmatrix}$ where $D$ is some diagonal matrix.
\eex

\begin{exercise}
Show that $J^{AB}$ as given in~\eqref{bohr} is positive semidefinite if and only if both $p_{y|x}\geq 0$ for all $x$ and $y$, and $Q\geq 0$.
\end{exercise}

\begin{exercise}
Show that the coefficients $\{p_{y|x}\}$ as defined in~\eqref{coedf} satisfy 
\be
\sum_{y\in[m]}p_{y|x}=1\quad\text{and}\quad s_{y}=\sum_{x\in[m]}p_{y|x}r_{x}\quad\quad\forall\;y\in[m]\;.
\ee
\end{exercise}

\subsubsection{Example: The Qubit Case}

For the case that $|A|=2$, all non-degenerate Hamiltonians (i.e., Hamiltonians with two distinct eigenvalues) have a Bohr spectrum.
Let 
\be\label{rsoff}
\rho=\begin{pmatrix} a & z\\\bar{z} &1-a  &\end{pmatrix}\quad\text{and}\quad\sigma=\begin{pmatrix} b & w\\\bar{w} &1-b  &\end{pmatrix} 
\ee
be two qubit states.
Without loss of generality suppose that $a\geq b$.
In this case the matrix $Q$ can be expressed as
\be
Q=\begin{pmatrix}
\frac ba\; &\; \frac wz\\
\; &\; \\
\frac{\bar{w}}{\bar{z}}\; &\; 1
\end{pmatrix}\;,
\ee
and $Q\geq 0$ if and only if
\be\label{coddd}
\frac ba\geq\left|\frac wz\right|^2\;.
\ee
Therefore, $\rho\xrightarrow{\cov}\sigma$ if and only if $\nu(\rho)\geq\nu(\sigma)$, where $\nu:\md(A)\to \mbb{R}_+$ is a measure of qubit time-translation-asymmetry defined on every density matrix of the form~\eqref{rsoff} as
\be
\nu(\rho)\eqdef\frac{|z|^2}{a}\;.
\ee

If $\rho$ is a pure state, so that $|z|=\sqrt{a(1-a)}$, then $\nu(\rho)\geq\nu(\sigma)$ holds if and only if
$|w|^2\leq b(1-a)$. Note that $|w|^2\leq b(1-b)$ since $\sigma\geq 0$. Therefore, by taking 
\be
a\in\left[b,1-\frac{|w|^2}{b}\right]
\ee
we get $|w|^2\leq b(1-a)$ and also $a\geq b$. Hence, for any mixed state $\sigma$ there exists a pure state $\psi$ that can be converted to $\sigma$.

On the other hand, if $\sigma$ is pure (i.e. $|w|^2=b(1-b)$) and $\rho$ arbitrary qubit, then the condition in~\eqref{coddd} becomes
\be
\left|z\right|^2\geq a(1-b)\;.
\ee
Since $\rho\geq 0$ we also have $|z|^2\leq b(1-b)$. Combining both equation we find that the only way $\rho$ can be converted to a pure qubit state $\sigma$ is if
$b=a$ (since $a\geq b$ was the initial assumption) and $|z|^2=a(1-a)$. That is, $\rho$ is a pure state itself, and up to a diagonal unitary equals to $\sigma$. Hence, pure coherence cannot be obtained from mixed coherence, and deterministic interconversion among inequivalent pure resources is not possible.

The example above shows that there is no unique ``golden unit\index{golden unit}" that can be used as the ultimate resource in two dimensional systems. Instead, any pure resource (i.e. pure state that is not an energy eigenstate) is maximal in the sense that there is no other resource that can be converted into it. However, the set of all pure qubit resources is maximal (i.e. any mixed state can be reached from some pure state by translation covariant operations). We now show that this latter property holds in general.

\begin{myg}{}
\begin{corollary}\label{puretomix}
Let $\sigma\in\md(A)$ be an arbitrary state, and denote by $p_x\eqdef\la x|\sigma|x\ra$ the diagonal elements of $\sigma$ in the energy eigenbasis $\{|x\ra\}_{x\in[m]}$ of system $A$. Then, the pure quantum state
\be
|\psi\ra\eqdef\sum_{x\in[m]}\sqrt{p_x}|x\ra
\ee
can be converted to $\sigma$ by a time-translation covariant channel.
\end{corollary}
\end{myg}
\begin{proof}
Observe that the diagonal elements of $Q$ are all 1, and the off-diagonal terms are given by
\be
q_{xy}=\frac{\sigma_{xy}}{\sqrt{p_xp_y}}\quad\quad\forall\;x,y\in[m]\;,\;x\neq y.
\ee
Therefore, we can express $Q=D_\p^{-1}\sigma D_{\p}^{-1}$, where $D_{\p}$ is the diagonal matrix whose diagonal is $(\sqrt{p_1},...,\sqrt{p_m})$. Since $D_\p>0$ and $\sigma\geq 0$ it follows that $Q\geq 0$. This completes the proof.
\end{proof}

\begin{exercise}
Show that if $\rho$ and $\sigma$ are two distinct pure states and both have non-zero off-diagonal terms (with respect to the energy eigenbasis) then the matrix $Q$ is not positive semidefinite. \end{exercise}

\section{Notes and References}

An outstanding review article on reference frames and superselection rules in quantum information can be found in~\cite{BRS2007}. The theory of quantum reference frames as a resource theory was initially introduced in~\cite{GS2008} and further developed as the resource theory of asymmetry in~\cite{MS2013,MS2014}. The Kraus representation of a $\G$-covariant map was presented in~\cite{GS2008}, which was later utilized in~\cite{Marvian2012} to derive the covariant version of Stinespring's dilation theorem.

The review article~\cite{BRS2007} provides numerous references on the advancement of techniques for aligning reference frames. In particular, we adopted the group theoretical approach to frame alignment developed in~\cite{CDS2005}.

The concept of relative entropy of asymmetry, initially referred to as $\G$-asymmetry, was introduced in~\cite{VAWJ2008} and further developed in~\cite{GMS2009}. Other measures of asymmetry, including several derivatives of asymmetry, were investigated in~\cite{MS2014b}.

The study of pure-state asymmetry originated in~\cite{GS2008} for specific groups and was subsequently extended to all finite or compact Lie groups in~\cite{MS2013,MS2014}. However, the manipulation of mixed-state asymmetry, particularly in the context of approximate state conversions, remains poorly understood. Nonetheless, the recognition that the exact state conversion problem can be solved using semidefinite programming (refer to Theorem~\ref{thcgp}) was first introduced in~\cite{GJB+2018}.

Exact conversions under time-translation covariant transformations were investigated in~\cite{Gour2022}. The positivity of the matrix $Q$ in Theorem~\ref{ttcs} was discovered earlier in~\cite{NG2015} through a slightly different approach. Lastly, the asymptotic regime under periodic Hamiltonians was recently studied in~\cite{Marvian2022}, where it was demonstrated that the quantum Fisher information can be interpreted as the cost of time-translation asymmetry.

%%%%%%%%%%%%%%%%%%%%%%%%%%%%%%%%%%%%%%%%%%%%%%%%%%%%%%%%%%%%%%%%%%%%%%%%%%%%%%%%%%%%%%%%%%%%%%%%%%%%

\chapter{The Resource Theory of Nonuniformity}\index{nonuniformity}\label{ch:nonuniformity}

This chapter introduces a specialized sub-theory of quantum thermodynamics, which will receive further attention in the subsequent chapter. This theory operates under the premise that the environment exhibits a high level of ``noise," leading physical systems to inherently evolve towards a maximally mixed state. Within this framework, systems that have reached maximally mixed states are deemed readily accessible and free, whereas pure states are considered valuable resources. Consequently, this theory is often termed the resource theory of purity or nonuniformity\index{nonuniformity}, emphasizing that a state's value increases with its deviation from the uniform (maximally) mixed state.

In comparison to the broader field of quantum thermodynamics, the QRT of nonuniformity presents a more streamlined approach, facilitating simpler calculations of conversion rates and resource monotones. Notably, the profound link between majorization and relative majorization, explored in Sec.~\ref{secrm}, allows numerous findings in the QRT of athermality to be directly inferred as corollaries from theorems within the QRT of nonuniformity. This interconnection suggests a logical progression: by initially laying down the principles and methodologies inherent to the QRT of nonuniformity, we can seamlessly apply these insights to thermodynamic systems. This approach not only streamlines our exploration but will also enriche our overall comprehension of quantum thermodynamics.

\section{The Free Operations}

As discussed above, for any system $A$, we consider the maximally mixed state $\u^A\eqdef\frac1{|A|}I^A$ to be the free state of the theory. That is, the set of free states $\mf(A)\eqdef\{\u^A\}$ consists of a single state. The set of free operations of the QRT of nonuniformity consists of physically implementable operations (see Sec.~\ref{Sect:PhysicallyImplementable}) relative to this set of states. They are called completely factorizable channels.

\subsection{Completely Factorizable Channels}\index{completely factorizable}

\begin{myd}{}
\begin{definition}
A quantum channel $\mN\in\cptp(A\to A')$ is said to be completely factorizable if there exist systems $B$ and $B'$ and a unitary channel $\mU\in\cptp(AB\to A'B')$ such that $|AB|=|A'B'|$ and
\be
\mN^{A\to A'}\left(\rho^A\right)=\tr_{B'}\left[\mU^{AB\to A'B'}\left(\rho^A\otimes\u^B\right)\right]\quad\quad\forall\;\rho\in\ml(A)\;.
\ee
\end{definition}
\end{myd}

Note that a completely factorizable channel $\mN^{A\to A'}$ has the property that $\mN^{A\to A'}\left(\u^A\right)=\u^{A'}$. That is, factorizable channels take maximally mixed states to maximally mixed states. In particular, if $|A|=|A'|$ then a completely factorizable channel is unital. However, as we will see shortly, not all unital channels are completely factorizable. 

\begin{myt}{}
\begin{theorem}\label{mixnoi}
Let $\mN\in\cptp(A\to A)$ be a mixture of unitaries of the form
\be
\mN^{A\to A}=\sum_{x\in[\ell]} p_x\;\mU^{A\to A}_x
\ee
where each $\mU^{A\to A}_x$ is a unitary channel, $\ell\in\mbb{N}$, and 
$\p\eqdef(p_1,\ldots,p_\ell)^T$ is a probability vector in $\mbb{Q}^n$ (i.e. each $p_x$ is a non-negative rational number). Then, $\mN^{A\to A}$ is completely factorizable.
\end{theorem}
\end{myt}
\begin{proof}
Since all the $\{p_x\}_{x\in[\ell]}$ are rational, there exists a common denominator $m\in\mbb{N}$ and $\ell$ integers $\{m_x\}_{x\in[\ell]}$ such that $p_x=\frac{m_x}m$, and in particular $\sum_{x\in[\ell]} m_x=m$ since $\sum_{x\in[\ell]} p_x=1$. Set $n\eqdef |A|$, and let $B$ be a system with dimension $|B|=m$. Define a unitary matrix $U:AB\to AB$ via its action on a basis element $|xy\ra\in AB$ with $x\in[n]$ and $y\in[m]$ as
\be
U^{AB}|x\ra^A|y\ra^B=U_{k_y}^A|x\ra^A|y\ra^B
\ee
where $k_y$ is the integer in $[\ell]$ 
satisfying
\be
\sum_{j\in[k_y-1]}m_j\leq y<\sum_{j\in[k_y]}m_{j}
\ee
(note that $k_y$ depends on $y$). That is, $U^{AB}$ is a controlled unitary\index{controlled unitary} that its action on $A$ depends on the input of system $B$.
Using the notation $\mU^{AB\to AB}\eqdef U^{AB}(\cdot)U^{*AB}$ get that for all $\omega\in\ml(A)$
\ba
\tr_{B}\left[\mU^{AB\to AB}\left(\omega^A\otimes\u^B\right)\right]
&=\frac1m\sum_{y\in[m]}\tr_{B}\left[\mU^{AB\to AB}\left(\omega^A\otimes|y\lr y|^B\right)\right]\\
&=\frac1m\sum_{y\in[m]}\mU^{A\to A}_{k_y}\left(\omega^A\right)\;,
\ea
where we used the definition of $U^{AB}$ above. Now, observe that from the definition of $k_y$, for any $x\in[\ell]$ there exists $m_x$ values of $y\in[m]$ for which $k_y=x$. Therefore, continuing from the last line above we get
\ba
\tr_{B}\left[\mU^{AB\to AB}\left(\omega^A\otimes\u^B\right)\right]
&=\sum_{x\in[\ell]}\frac{m_{x}}{m}\;\mU_x^{A\to A}\left(\omega^A\right)\\
&=\sum_{x\in[\ell]} p_x\;\mU_{x}^{A\to A}(\omega^A)\;.
\ea
Hence, $\sum_{x\in[\ell]} p_x\mU_{x}^{A\to A}$ is a noisy operation.
This completes the proof.
\end{proof}

\subsection{Noisy Operations}\index{noisy operations}

In the proof of Theorem~\eqref{mixnoi}, we made the assumption that the coefficients $\{p_x\}_{x\in[\ell]}$ are rational numbers. Additionally, it should be noted that the dimension $m$ of system $B$ is determined by the common denominator of these rational coefficients. Consequently, we cannot employ a continuity\index{continuity} argument to establish that any mixture of unitaries, potentially with irrational coefficients, qualifies as a completely factorizable channel. This limitation arises because the dimension of system $B$ tends to infinity when the rational coefficients $\{p_x\}_{x\in[\ell]}$ approach irrational numbers.

The aforementioned issue stems from the fact that the system $B$ appearing in the definition of completely factorizable channels, while finite, is yet unbounded. Consequently, it is possible, in principle, to define a sequence of completely factorizable channels $\mN_j\in\cptp(A\to A')$, which can only be realized using systems $B_j$ of increasing dimensions as $j$ grows. Specifically, we observe that $\lim_{j\to\infty}|B_j|=\infty$.
In other words, the set encompassing all completely factorizable channels in $\cptp(A\to A')$ is not closed. However, this unphysical attribute can be addressed by considering the closure of the set of completely factorizable channels.
\begin{myd}{}
\begin{definition}
A quantum channel $\mN\in\cptp(A\to A')$ is called a \emph{noisy operation} if there exists a sequence of completely factorizable channels $\{\mN_k\}_{k\in\mbb{N}}\subset\cptp(A\to A')$ such that
\be\label{limit16p7}
\lim_{k\to\infty}\mN^{A\to A'}_k=\mN^{A\to A'}\;.
\ee
The set of all noisy operations in $\cptp(A\to A')$ is denoted by $\noisy(A\to A')$.
\end{definition}
\end{myd}

\begin{remark}
The limit~\eqref{limit16p7} is understood in terms of the Choi matrices. That is, the relation~\eqref{limit16p7} means that
\be
\lim_{k\to\infty}\left\|J_{\mN_k}^{AA'}-J_{\mN}^{AA'}\right\|_1=0\;.
\ee
\end{remark}

Note that by definition the set of noisy operations is closed. Moreover, the set of noisy operations in $\cptp(A\to A)$ forms a subset of unital channels (see Exercise~\ref{noisyisunital}). However, it can be shown that not every unital channel is a noisy operation, so that noisy operations forms a strict subset of unital channels.

\bex\label{noisyisunital}
Show that if $\mN\in\noisy(A\to A)$ then $\mN^{A\to A}$ is a unital channel.
\eex

\begin{myt}{}
\begin{theorem}\label{anymixture}
Any mixture of unitary channels in $\cptp(A\to A)$ is a noisy operation.
\end{theorem}
\end{myt}

\bex
Use Theorem~\ref{mixnoi} and the definition of noisy operations to prove the theorem above.
\eex

\subsection{The Structure of the QRT of Nonuniformity}

We define the resource theory of nonuniformity as a framework where the set of free operations is noisy operations; i.e., for any pair of systems $A$ and $A'$, $\mf(A\to A')=\noisy(A\to A')$. In this resource theory, we can treat all states and operations as classical without any loss of generality. To see this, observe first that any quantum state $\rho^A$ can be converted into a diagonal state in the same basis by applying a unitary channel. Such a unitary channel constitutes a reversible noisy operation within the set of free operations. Hence, all resource states in $\md(A)$ can be represented by diagonal density matrices in the same basis.
We therefore fix a basis and denote the completely dephasing channel in this basis as $\Delta^A\in\cptp(A\to A)$. To summarize, without loss of generality we can assume that all resources are characterized by states satisfying $\Delta^A(\rho^A)=\rho^A$.

Now, suppose we can convert $\rho\in\md(A)$ to $\sigma\in\md(B)$ using a free noisy operation $\mN\in\cptp(A\to B)$. Since we assume that $\rho^A=\Delta^A(\rho^A)$ and $\sigma^B=\Delta^B(\sigma^B)$, it follows that
\ba
\sigma^B&=\mN^{A\to B}\left(\rho^A\right)\\
\GG{\sigma\text{ is diagonal}}&=\Delta^B\circ\mN^{A\to B}\left(\rho^A\right)\\
\GG{\rho\text{ is diagonal}}&=\Delta^B\circ\mN^{A\to B}\circ\Delta^A\left(\rho^A\right)\;.
\ea
Hence, if $\rho^A$ can be converted to $\sigma^B$ using a free quantum channel $\mN\in\noisy(A\to B)$, then $\rho^A$ can also be mapped to $\sigma^B$ using the \emph{classical} channel $\Delta^B\circ\mN^{A\to B}\circ\Delta^A$. The latter channel is also a noisy operation since $\Delta^A$ and $\Delta^B$ are themselves noisy operations; they can be expressed as random unitary channels (see Exercise~\ref{pinchex}). Therefore, all resources and free channels can be characterized using classical systems and classical channels.

Finally, considering that all pure states in a fixed dimension are equivalent, we select the qubit pure state $|0\lr 0|$, where $|0\ra\in\mbb{C}^2$, as the chosen reference state for this resource theory, serving as the golden unit.

\section{Measures of Nonuniformity}

A measure of nonuniformity was defined earlier in~Def.~\ref{mesnon}. On the other hand, according to the definition of a resource measure, a measure of nonuniformity is a function 
\be
g: \bigcup_{A} \md(A) \to \mbb{R} \cup \{\infty\}
\ee 
that is non-increasing under noisy operations and take the value zero on free states. To see that both definitions are equivalent, observe first that since we consider only diagonal states (in the same basis) we can replace $\md(A)$ above with the classical set $\prob(d)$, where $d\eqdef|A|$.

Due to Corollary~\ref{noimaj}, the monotonicity of $g$ under noisy operation is equivalent to the Schur concavity of $g$ and to the third condition  in Def.~\ref{mesnon}. The only additional assumption that we added in Def.~\ref{mesnon} is that $g$ is continuous. This assumption is crucial for the bijection\index{bijection} between divergences and measures of non-uniformity (see Theorem~\ref{biji}), and we will assume it also here. 

The bijection\index{bijection} given in Theorem~\ref{biji} demonstrates that all measures of nonuniformity can be expressed as
\be
g(\p)=\D\left(\p\big\|\u^{(d)}\right)\quad\quad\forall\;d\in\mbb{N}\quad\forall\;\p\in\prob(d)\;.
\ee
where $\D$ is a classical divergence. Therefore, all the divergences and relative entropies that introduced in Chapters~\ref{chadiv} and~\ref{ch:relent} can be used to quantify nonuniformity. A particular useful one is the nonuniformity measure obtained by taking $\D$ to be the KL-divergence. In this case, for all $\p\in\prob(n)$ we have
\be
g(\p)=D\left(\p\big\|\u^{(d)}\right)=\log(d)-H(\p)\;,
\ee
where $H$ is the Shannon entropy\index{Shannon entropy}.  Similarly, for the R\'enyi divergences we have for all $\alpha\in[0,\infty]$
\be
g_\alpha(\p)=D_\alpha\left(\p\big\|\u^{(d)}\right)=\log(d)-H_\alpha(\p)\;.
\ee 
It is worth mentioning that for pure states, specifically when taking $\p=(1,0,\ldots,0)^T$, we get that $g_\alpha(\p)=\log(d)$. This implies that the nonuniformity of pure states increases with the dimension $d$.

\section{Interconversions in the Single-Shot Regime}\index{single-shot}

In this section, we delve into the interconversions between nonuniformity states in the single-shot regime, considering both exact and approximate scenarios. We will demonstrate that, similar to pure bipartite entanglement, majorization plays a crucial role in determining these interconversions. We will use the notations $\rho\xrightarrow{\noisy}\sigma$ whenever $\sigma=\mN(\rho)$ for some noisy operation $\mN\in\noisy(A\to A)$.

\subsection{Exact Deterministic Conversions}

\begin{myt}{}
\begin{theorem}\label{noimaj}
Let $\rho,\sigma\in\md(A)$. Then, $\rho\xrightarrow{\noisy}\sigma$ if and only if $\rho\succ\sigma$.
\end{theorem}
\end{myt}
\begin{proof}
 Suppose $\sigma=\mN(\rho)$ for some noisy operation $\mN\in\noisy(A\to A)$. Since a noisy operation $\mN\in\noisy(A\to A)$ is also a unital channel, from Section~\ref{unital} it follows that $\rho\succ\sigma$. In the same subsection we also proved that $\rho\succ\sigma$ if and only if there exist a random unitary channel that take $\rho$ to $\sigma$. From the Theorem~\ref{anymixture}, this random unitary is also a noisy operation, so the proof is concluded.
\end{proof}

The theorem above can be slightly modified to accommodate systems of different dimensions. In particular, if $\rho\in\md(A)$ and $\sigma\in\md(B)$ then $\sigma^B=\mN^{A\to B}(\rho^A)$ for some noisy operation $\mN\in\noisy(A\to B)$ if and only if
$\rho^A\otimes\u^B\succ\u^A\otimes\sigma^B$. This is because appending a maximally mixed state is a reversible free operation.

From here onward we consider the `states' of the QRT of nonuniformity to be probability vectors in $\prob(d)$. Therefore, from the theorem and the discussion above it follows that for two given states $\p\in\prob(d)$ and $\q\in\prob(d')$ we have
\be
\p\xrightarrow{\noisy}\q\quad\iff\quad \left(\p,\u^{(d)}\right)\succ\left(\q,\u^{(d')}\right)\;.
\ee
That is, conversion under noisy operations induce a pre-order that can be characterized with relative majorization. Note that if $d=d'$ this pre-order reduces to the standard definition of majorization, however, for $d\neq d'$ it is not equivalent to majorization between $\p$ and $\q$. In particular, embedding a state, say $\p\in\prob(d)$, in a higher dimensional space $\prob(d')$ with $d'>d$ (by adding zero components) can increase the resourcefulness of $\p$. Therefore, such embeddings are not free.

\bex
Let $\q=(1/2,1/2,0,0)^T$ be the vector obtained from the uniform state $\u^{(2)}$ by adding two zeros. Show that $\q$ can be converted by noisy operations to any state in $\md(2)$.
\eex

\subsection{The Conversion Distance}

Following the general definition given in~\eqref{cd}, we define the conversion distance of nonuniformity, from a state $\p\in\prob(d)$ into a state $\q\in\prob(d')$ as
\be\label{offk}
T\left(\p\xrightarrow{\noisy} \q\right)\eqdef\min_{\r\in\prob(d')}\left\{\frac12\left\|\q-\r\right\|_1\;:\;(\p,\u^{(d)})\succ(\r,\u^{(d')})\right\}\;.
\ee
From the properties of the conversion distance\index{conversion distance} (see for example Lemma~\ref{lemmonin}), it follows that $T(\p\xrightarrow{\noisy} \q)$ remains invariant under any permutation of the components of $\p$ or $\q$. Therefore, in the rest of this chapter we will always assume without loss of generality that $\p=\p^\da$ and $\q=\q^\da$.

\begin{myt}{}
\begin{theorem}\label{thm:1632}
Let $\p,\q\in\prob(d)$ be two probability vectors. Then,
\be\label{fordn}
T\left(\p\xrightarrow{\noisy} \q\right)=\max_{\ell\in[d]}\big\{\|\q\|_{(\ell)}-\|\p\|_{(\ell)}\big\}\;.
\ee
\end{theorem}
\end{myt}

\begin{remark}
The case that $\p\in\prob(d)$ and $\q\in\prob(d')$ with $d\neq d'$ can be solved by applying the theorem above to the vectors $\p\otimes\u^{(d')}$ and $\u^{(d)}\otimes\q$. Specifically,
 \be
T\left(\p\xrightarrow{\noisy} \q\right)=\max_{\ell\in[dd']}\Big\{\left\|\u^{(d)}\otimes\q\right\|_{(\ell)}-\left\|\p\otimes\u^{(d')}\right\|_{(\ell)}\Big\}\;.
\ee
\end{remark}

\begin{proof}
Since we consider the case that both $\p$ and $\q$ are $d$-dimensional, the conversion distance can be expressed as
\be
T\left(\p\xrightarrow{\noisy} \q\right)=\min_{\r\in\prob(d)}\left\{\frac12\left\|\q-\r\right\|_1\;:\;\p\succ\r\right\}\;.
\ee
The above expression for the conversion distance represents the distance of $\q$ to the set ${\rm majo}(\p)$ as defined in~\eqref{defmaj6} (with $\p$ replacing $\q$). Hence,
\ba
T\left(\p\xrightarrow{\noisy} \q\right)&=T\big(\q,{\rm majo}(\p)\big)\\
\GG{Theorem~\ref{thm:majog}}&=\max_{\ell\in[n]}\left\{\|\q\|_{(\ell)}-\|\p\|_{(\ell)}\right\}\;.
\ea
This completes the proof.
\end{proof}

\subsection{The Single-Shot Nonuniformity Cost}

In order to define the single-shot nonuniformity cost of a resource state $\p\in\prob^\da(d)$, we first recall that the vector $\e^{(m)}_1$, defined as $(1, 0, \ldots, 0)^T \in \prob^\da(m)$, represents the maximal resource in dimension $m$. Moreover, the resourcefulness of $\e^{(m)}_1$ increases with $m$, and the set $\{\e^{(m)}_1\}_{m\in\mbb{N}}$ forms a golden unit\index{golden unit} according to Definition~\ref{def:gu}. Therefore, we define the $\eps$-nonuniformity cost of $\p$ as:
\be
\cost^{\eps}(\p)\eqdef\min\left\{\log m\;:\;T\left(\e_1^{(m)}\xrightarrow{\noisy}\p\right)\leq\eps\right\}\;.
\ee
Clearly, $\cost^{\eps}(\p)\leq\log d$ since $T\left(\e_1^{(d)}\xrightarrow{\noisy}\p\right)=0$. We will therefore assume (implicitly) in the rest of this section that $m\leq d$. In the following theorem we use the notation $H_{\min}^\eps(\p)$ for the smoothed min-entropy\index{min-entropy} of $\p$. In~\eqref{fofhmn} we found a closed form for this smoothed entropy given by:
\be
H_{\min}^\eps(\p)=-\log\max_{\ell\in[d]}\left\{\frac{\|\p\|_{(\ell)}-\eps}{\ell}\right\}\;.
\ee

\begin{myt}{}
\begin{theorem}\label{thm:costnoisy}
Let $\eps\in(0,1)$ and $\p\in\prob^\da(d)$.
The $\eps$-nonuniformity cost of $\p$ is given by
\be\label{16p28}
\cost^{\eps}(\p)=\log\left\lceil d2^{-H_{\min}^\eps(\p)}\right\rceil\;.
\ee
\end{theorem}
\end{myt}
\begin{proof}
We first prove the theorem for the case $\eps=0$. In this case,
\be
\cost^{\eps=0}(\p)\eqdef\min\left\{\log m\;:\;\e_1^{(m)}\xrightarrow{\noisy}\p\right\}\;.
\ee
The condition $\e_1^{(m)}\xrightarrow{\noisy}\p$ is equivalent to $(\e_1^{(m)},\u^{(m)})\succ(\p,\u^{(d)})$. Moreover, in Exercise~\ref{exshop1e} you show that the condition $(\e_1^{(m)},\u^{(m)})\succ(\p,\u^{(d)})$ is equivalent to $p_1\leq\frac md$. Since the smallest integer that satisfies this condition is
$m=\left\lceil dp_1\right\rceil$ we conclude that
\ba\label{cuteen}
\cost^{\eps=0}(\p)&=\log\left\lceil dp_1\right\rceil\\
\GG{cf.~\eqref{qminen}}&=\log\left\lceil d2^{-H_{\min}(\p)}\right\rceil\;.
\ea
This completes the proof for the case $\eps=0$. For $\eps>0$ we use~\eqref{smooths} to get
\ba
\cost^{\eps}(\p)&=\min_{\p'\in\mb_\eps(\p)}\cost^{\eps=0}(\p')\\
\GG{\eqref{cuteen}}&=\min_{\p'\in\mb_\eps(\p)}\log\left\lceil d2^{-H_{\min}(\p')}\right\rceil\\
&=\log\left\lceil d2^{-H_{\min}^\eps(\p)}\right\rceil\;,
\ea
where the last line follows from the definition of $H_{\min}^\eps(\p)$. This completes the proof.
\end{proof}

\bex\label{exshop1e}
Show that the condition $(\e_1^{(m)},\u^{(m)})\succ(\p,\u^{(d)})$ is equivalent to $p_1\leq\frac md$.
\eex

\bex
Let $\p\in\prob^\da(d)$ and $m\in[d]$.
\ben
\item Show that
\be
T\left(\e_1^{(m)}\xrightarrow{\noisy}\p\right)=f_\p\left(\frac md\right)
\ee
where $f_{\p}(t)\eqdef\sum_{x\in[d]}(p_x-t)_+$ is the function studied at the end of Sec.~\ref{sec:flattest}.
\item Provide a direct proof of the theorem above using the above conversion distance and the explicit expression given in~\eqref{1440} for $f_\p^{-1}$.
\item Show that the conversion distance above can also be expressed as
\be
T\left(\e_1^{(m)}\xrightarrow{\noisy}\p\right)=\frac12\left\|\p-m\u^{(d)}\right\|_1-\frac{m-1}{2}\;.
\ee
\een
\eex

\bex\label{cor1441}
Show that the single-shot $\eps$-nonuniformity cost of $\p$ is bounded by
\be
\log \left(\|\p\|_{(k)}-\eps\right)\leq \cost^{\eps}(\p)-\log(d/k)\leq
\log \left(\|\p\|_{(k)}-\eps+\frac kd\right)
\ee
where $k\in[d]$ is the integer satisfying $\eps\in(r_k,r_{k+1}]$, where $r_k$ is defined in~\eqref{defrz}.
\eex

\subsection{The Single-Shot Distillable Nonuniformity}

For any $\eps\in(0,1)$ and $\p\in\prob^\da(d)$, we define the $\eps$-single-shot distillable nonuniformity of $\p$ as
\be
\distill^{\eps}(\p)\eqdef\max\left\{\log m\;:\;T\left(\p\xrightarrow{\noisy}\e_1^{(m)}\right)\leq\eps\right\}\;.
\ee
Unlike the case for resource cost, an analogous formula to \eqref{smooths} does not exist for resource distillation. Therefore, the calculation of single-shot distillable nonuniformity necessitates a direct computation of the conversion distance $T\left(\p\xrightarrow{\noisy}\e_1^{(m)}\right)$. In the following lemma we provide a closed formula of this conversion distance in terms of the coefficient $\mu_m$ which is defined for all $m\in\mbb{N}$ as
\be\label{16p41}
\mu_m\eqdef \|\p\|_{\left(\left\lfloor d/m\right\rfloor\right)}+\left(d/m-\left\lfloor d/m\right\rfloor\right)p_{\left\lfloor d/m\right\rfloor+1}\;,
\ee 
with the convention that $\|\p\|_{(0)}=0$ so that $\mu_m=\frac dmp_1$ if $m>d$.
\begin{myg}{}
\begin{lemma}
Let $d,m\in\mbb{N}$, $\p\in\prob^\da(d)$, and $\mu_m$ as defined in~\eqref{16p41}. Then,
\be\label{bay}
T\left(\p\xrightarrow{\noisy}\e_1^{(m)}\right)=1-\mu_m\;.
\ee
\end{lemma}
\end{myg}
\begin{proof}
The case $m>d$ is left as an exercise, and we assume here that $m\leq d$.
From the previous section, the conversion distance can be expressed as
\be\label{0optik0}
T\left(\p\xrightarrow{\noisy}\e_1^{(m)}\right)=\max_{k\in[dm]}\sum_{j\in[k]}\left((\e_1^{(m)}\otimes\u^{(d)})^\da_j-(\u^{(m)}\otimes\p)^\da_j\right)\;.
\ee
Since the vector $\e_1^{(m)}\otimes\u^{(d)}$ has exactly $n$ non-zero components (all equal to $1/d$), we get that the optimizer $k$ above must satisfy $k\leq d$. Moreover, the $j$th term in the sum above have the form
\be
(\e_1^{(m)}\otimes\u^{(d)})^\da_j-(\u^{(m)}\otimes\p)^\da_j=\frac1d-\frac{p_x}m
\ee
where $x= \left\lceil\frac jm\right\rceil$. Since $\p=\p^\da$ the terms in the equation above are non-decreasing with $j$. We therefore conclude that the optimal $k$ in~\eqref{0optik0} must be $k=d$.
Denoting $a\eqdef\left\lfloor\frac dm\right\rfloor$ and $b\eqdef d-am$ (hence $d=am+b$) we get
\ba
T\left(\p\xrightarrow{\noisy}\e_1^{(m)}\right)&=1-\Big(\sum_{x\in[a]}p_x+b\frac{p_{a+1}}m\Big)\\
\Gg{b= d-am}&=1-\|\p\|_{(a)}-\left(\frac dm-a\right)p_{a+1}\\
&=1-\mu_m\;.
\ea
This completes the proof.
\end{proof}

Combining the definition of the $\eps$-single-shot distillable nonuniformity with the lemma above we obtain the following closed form for $\distill^{\eps}(\p)$.

\begin{myt}{}
\begin{theorem}\label{thm:16p3p4}
Let $\eps\in(0,1)$, $\p\in\prob^\da(d)$, $m\in[d]$, and $\mu_m$ as defined in~\eqref{16p41}. 
If $p_1>1-\eps$ then 
$
\distill^{\eps}(\p)\eqdef\left\lfloor dp_1/(1-\eps)\right\rfloor
$.
Otherwise, the $\eps$-single-shot  distillable nonuniformity is given by
\be
\distill^{\eps}(\p)\eqdef\max_{m\in[d]}\left\{\log m\;:\;\mu_m\geq1-\eps\right\}\;.
\ee
\end{theorem}
\end{myt}

\bex
Use the closed form in~\eqref{bay} to prove Theorem~\ref{thm:16p3p4}.
\eex

\bex
Show that for $\eps=0$ the single-shot distillable nonuniformity of $\p\in\prob(d)$ is given by
\be
\distill^{\eps=0}(\p)=\log(d)-H_{\max}(\p)\;,
\ee
where $H_{\max}$ is the max-entropy given by $H_{\max}(\p)\eqdef\log(k)$, where $k$ is the number of non-zero components of $\p$.
\eex

The formula in Theorem~\ref{thm:16p3p4} is somewhat cumbersome. One can get somewhat simpler bounds on the single-shot distillable entanglement\index{distillable entanglement} by removing the floor functions that appear in the definition of $\mu_m$. These simpler bounds can be expressed in terms of the formula for the smoothed max-entropy\index{max-entropy} given in Lemma~\ref{lemofhmax}. Specifically, from Lemma~\ref{lemofhmax} it follows that the smoothed max-entropy can be expressed as the logarithm of an integer $k$ satisfying
\be\label{pkkk}
\|\p\|_{(k-1)}< 1-\eps\leq \|\p\|_{(k)}
\ee
with the convention $\|\p\|_{(0)}\eqdef 0$. 

\begin{myg}{}
\begin{corollary}\label{cor:101}
Let $\eps\in(0,1)$, $\p\in\prob(d)$, and set $k\eqdef 2^{H_{\max}^\eps(\p)}$.
Then,
\be\label{1450}
\log(d-1-k)-\log\left(1+k\right)\leq \distill^{\eps}(\p)\leq\log(d)-\log\left(k-1\right)
\ee
\end{corollary}
\end{myg}
\begin{proof}
First, observe that
\be\label{pmup}
\|\p\|_{\left(\left\lfloor d/m\right\rfloor\right)}\leq\mu_m\leq\|\p\|_{\left(\left\lfloor d/m\right\rfloor+1\right)}
\ee
Therefore,
\be
\distill^{\eps}(\p)\leq \max_{m\in[d]}\left\{\log m\;:\;\|\p\|_{(\floor{d/m}+1)}\geq1-\eps\right\}\;.
\ee
Now, observe that $\left\lfloor\frac dm\right\rfloor\leq\frac dm$ so that $m\leq \frac d{\floor{d/m}}$. Hence,
\ba
\distill^{\eps}(\p)&\leq \max_{m\in[d]}\left\{\log \frac d{\floor{d/m}}\;:\;\|\p\|_{(\floor{d/m}+1)}\geq1-\eps\right\}\\
\GG{\substack{Replacing\;{\it \floor{d/m}}\\with\;arbitrary\;\ell\in{\it [d]}}}&\leq\max_{\ell \in[d]}\left\{\log \frac d{\ell}\;:\;\|\p\|_{(\ell+1)}\geq1-\eps\right\}\\
\GG{\eqref{pkkk}}&=\log\frac d{k-1}\;.
\ea
For the lower bound, observe that~\eqref{pmup} gives
\be
\distill^{\eps}(\p)\geq \max_{m\in[d]}\left\{\log m\;:\;\|\p\|_{(\floor{d/m})}\geq1-\eps\right\}\;.
\ee
Now, for the lower bound we cannot replace $\floor{d/m}$ with arbitrary integer $\ell\in[d]$ since this will increase the right-hand side above. Instead, we use the fact that for any $s\in[\frac1d,1]$ there exists a unique $m\in[d]$ such that
\be
s-\frac1d<\frac md\leq s\;.
\ee
Observe further that  for any such $s\in[\frac 1d,1]$ and $m\in[d]$,  if in addition $\|\p\|_{(\floor{s^{-1}})}\geq 1-\eps$ then also $\|\p\|_{(\floor{d/m})}\geq1-\eps$ since $s^{-1}\leq \frac dm$. Moreover, since such $m$ and $s$ also satisfy $\log m\geq \log (ds-1)$ we get that
\be
\distill^{\eps}(\p)\geq \max_{s\in[\frac1d,1]}\left\{\log (ds-1)\;:\;\|\p\|_{(\floor{s^{-1}})}\geq1-\eps\right\}\;.
\ee
In the last step, denote by $\ell\eqdef\floor{s^{-1}}\in[d]$ and use the fact that $s^{-1}\leq\ell+1$ to get $s\geq\frac1{\ell+1}$. Substituting this to the right-hand side of the equation above gives
\ba
\distill^{\eps}(\p)&\geq \max_{\ell\in[d]}\left\{\log\left( \frac d{1+\ell}-1\right)\;:\;\|\p\|_{(\ell)}\geq1-\eps\right\}\\
&=\log\left( \frac d{1+k}-1\right)\;.
\ea
This completes the proof.
\end{proof}

\section{Asymptotic Conversions}

The distillation rate of a nonuniformity state $\q$ from another nonuniformity state $\p$ is defined as
\be\label{drnoisy}
\distill(\p\to\q)\eqdef\lim_{\eps\to 0^+}\sup_{n,m\in\mbb{N}}\left\{\frac mn\;:\;T\left(\p^{\otimes n}\xrightarrow{\noisy} \q^{\otimes m}\right)\leq\eps\right\}
\ee
We will show in this section is that the above conversion rate has the following simple formula. 
\begin{myt}{}
\begin{theorem}\label{pryo}
Let $\p\in\prob(d)$ and $\q\in\prob(d')$ be two probability vectors.
The conversion rate in~\eqref{drnoisy} is given by
\be
\distill(\p\to\q)=\frac{D\left(\p\|\u^{(d)}\right)}{D\left(\q\|\u^{(d')}\right)}=\frac{\log(d)-H(\p)}{\log(d')-H(\q)}\;.
\ee
\end{theorem}
\end{myt}
\begin{remark}
Note that the formula for the asymptotic conversion rate demonstrates that the resource theory of nonuniformity is reversible. Specifically, note that for any $\p$ and $\q$ as above, $\distill(\p\to\q)\distill(\q\to\p)=1$.
\end{remark}

We prove the theorem above by computing separately the nonuniformity cost and the distillable nonuniformity.
Recall from the discussion in Sec.~\ref{sec:asymptoticregime}, specifically~\eqref{deseq}, that the asymptotic cost of a nonuniformity state $\p\in\prob(k)$ is given by
\be
\cost(\p)\eqdef\lim_{\eps\to 0^+}\liminf_{n\to\infty}\frac1n\cost^{\eps}\left(\p^{\otimes n}\right)\;.
\ee
Therefore, we can use the results from the single-shot case to compute this asymptotic rate.
 
\begin{myg}{}
\begin{lemma}
Let $\p\in\prob(d)$ and $\eps\in(0,1)$. Then, the asymptotic nonuniformity cost of $\p$ is given by
\be\label{costkhp}
\cost(\p)=\lim_{n\to\infty}\frac1n\cost^{\eps}\left(\p^{\otimes n}\right)=\log(d)-H(\p)\;.
\ee
\end{lemma}
\end{myg}
\begin{proof}
From the result in the single-shot case, specifically~\eqref{16p28}, we obtain
\ba
\lim_{n\to\infty}\frac1n\cost^{\eps}\left(\p^{\otimes n}\right)&=\lim_{n\to\infty}\frac1n\log\left\lceil d^n2^{-H_{\min}^\eps\left(\p^{\otimes n}\right)}\right\rceil\\
\GG{AEP~\eqref{vergaep}}&=\log(d)-H(\p)\;.
\ea
This completes the proof. 
\end{proof}

Similarly, from the discussion in Sec.~\ref{sec:asymptoticregime}, specifically~\eqref{disteq}, the asymptotic distillation rate of a nonuniformity state $\p\in\prob(d)$ is given by
\be
\distill(\p)=\lim_{\eps\to 0^+}\limsup_{n\to\infty}\frac1n\distill^\eps\left(\p^{\otimes n}\right)\;.
\ee
As before, we can use the results from the single-shot regime to compute this expression.

\begin{myg}{}
\begin{lemma}
Let $\p\in\prob(d)$ and $\eps\in(0,1)$. Then, the asymptotic distillable nonuniformity  is given by
\be
\distill(\p)=\lim_{n\to\infty}\frac1n\distill^\eps(\p^{\otimes n})=\log(d)-H(\p)\;.
\ee
\end{lemma}
\end{myg}

\begin{proof}
From the upper bound in~\eqref{1450} we get
\ba
\limsup_{n\to\infty}\frac1n\distill^\eps(\p^{\otimes n})&\leq \limsup_{n\to\infty}\frac1n\log\left(\frac{d^n}{2^{H_{\max}^\eps\left(\p^{\otimes n}\right)}-1}\right)\\
\GG{AEP~\eqref{AEPmaxversion}}&=\log(d)-H(\p)\;.
\ea
Similarly, from the lower bound in~\eqref{1450} we get
\ba
\liminf_{n\to\infty}\frac1n\distill^\eps(\p^{\otimes n})&\geq\liminf_{n\to\infty}\frac1n\log\left( \frac {d^n}{1+2^{H_{\max}^\eps\left(\p^{\otimes n}\right)}}-1\right)\\
\GG{AEP~\eqref{AEPmaxversion}}&=\log(d)-H(\p)\;.
\ea
Comparing the two inequalities in the two equations above we conclude that 
\be
\lim_{n\to\infty}\frac1n\distill^\eps(\p^{\otimes n})=\log(d)-H(\p)\;.
\ee
This completes the proof.
\end{proof}

\bex
Use the Lemmas above to prove Theorem~\ref{pryo}.
\eex

\section{Notes and References}

To the best of the author's knowledge, the inaugural paper on the resource theory of non-uniformity was presented by the work of \cite{HHO2003}. This paper introduced the term ``Noisy Operations." The study of factorizable channels, in the context of von Neumann algebra, was carried out independently, as can be seen in, for instance, \cite{HM2011}. It was only much later that this resource theory was identified as a subset of the QRT of quantum athermality. For a comprehensive review and additional references, one can refer to the review article \cite{GMN+2015}.

\chapter{Quantum Thermodynamics}\label{ch:thermal}

Thermodynamics stands as one of the most influential theories in physics, finding applications across a wide range of disciplines. Initially focused on steam engines, its relevance has expanded to encompass fields such as biochemistry, nanotechnology, and black hole physics, among others~\cite{GMM2004,BCG+2018,DC2019}. Despite its immense success, the foundational aspects of thermodynamics continue to be a subject of controversy. There persists a pervasive confusion regarding the relationship between macroscopic and microscopic laws, particularly concerning reversibility and time-symmetry. Furthermore, there is a lack of consensus on the optimal formulation of the second law. As early as 1941, Nobel laureate Percy Bridgman noted, ``there are almost as many formulations of the Second Law as there have been discussions of it," and unfortunately, little progress has been made in resolving this situation since then. In recent years, researchers have taken a fresh perspective on these fundamental issues by approaching thermodynamics as a resource theory. This viewpoint considers a system that is not in equilibrium with its environment as a valuable resource known as ``athermality." Athermality serves as the fuel utilized in work extraction, computational erasure operations, and other thermodynamic tasks.

The resource-theoretic approach to thermodynamics delves into the quantification of a state's deviation from equilibrium and explores its utility in quantum thermodynamics. It also investigates the necessary and sufficient conditions for transforming one state into another. Within this framework, different notions of state conversion can be examined, including exact and approximate conversions, single-copy and multiple-copy scenarios, and conversions with or without the aid of a catalyst.

These quantum-information techniques have brought forth numerous novel insights, particularly considering the historical importance of information in foundational topics such as Maxwell's demon~\cite{MNV2009}, the thermodynamic reversibility of computation~\cite{Bennett1973,Bennett1982}, Landauer's principle regarding the work cost of erasure~\cite{Landauer1961,JW2000}, and Jaynes's utilization of maximum entropy principles in deriving statistical mechanics~\cite{Jaynes1957a,Jaynes1957b}.

Furthermore, the resource-theoretic approach to thermodynamics reveals that the conventional formulation of the second law of thermodynamics, which focuses on entropy non-decrease, is insufficient as a criterion for determining the feasibility of a given state conversion. However, we will discover that it is possible to identify a set of measures quantifying the degree of nonequilibrium (including entropy) such that a state conversion is feasible if and only if all of these measures do not increase.

\section{Thermal States and Athermality States}\index{Gibbs state}

We use the term `thermal bath' or  `thermal reservoir' to indicate a thermodynamic system with the property that any amount of reasonable heat that is added or extracted from it does not change its temperature. In other words, its heat capacity is extremely large. For a heat bath that is held at a fixed inverse temperature $\beta\eqdef\frac1{k_BT}$, the   state,
\be
\gamma^B\eqdef\frac{e^{-\beta H^B}}{\tr\left[e^{-\beta H^B}\right]}\;,
\ee
is the thermal equilibrium state known as the Gibbs state. The Gibbs state, $\gamma^B$, is also referred to as the thermal state of the system $B$, and the normalization factor
\be
\mZ^B\eqdef\tr\left[e^{-\beta H^B}\right]
\ee 
is called the \emph{partition function}.

The free states in the resource theory of athermality corresponds to physical systems that are in thermal equilibrium with their surrounding. We will therefore consider Gibbs states to be free. Note however that the Gibbs state of a system $B$  is defined with respect to the Hamiltonian $H^B$ of the system (or heat bath). As it follows from the exercise below, any density matrix is a Gibbs state with respect to some Hamiltonian.

\begin{exercise}
Show that for any density matrix $\rho\in\md_{>0}(A)$ there exists a Hamiltonian $H\in\pos(A)$ such that $\rho^A$ is the Gibbs state with respect to this Hamiltonian.
\end{exercise}

\subsection{Optimality of the Gibbs state}\index{Gibbs state}

Energy and entropy are the two cornerstones of thermodynamics. Any quantum (or classical) system tends to evolve into equilibrium state with its environment. In such a spontaneous evolution the second law of thermodynamics states that the entropy associated with the system cannot decrease, or alternatively, heat can never pass from a colder system to a warmer body without some external work that put into the system. One can therefore think of the equilibrium state of a system with a given (fixed) entropy as the state that has the lowest amount of energy, so that no further heat-exchanged can occur  with the environment. 

Explicitly, let $A$ be  a quantum system with Hamiltonian $H^A$ and a given entropy $S$. Among all states $\rho\in\md(A)$ with entropy $H(\rho)=S$, we want to find the one that has the smallest amount of energy. That is, we want to minimize $\tr[\rho^AH^A]$ given the constraint that the entropy of $\rho^A$ equals $S$. This problem can be solved by minimizing the Lagrangian
\be
\mL(\rho,\lambda)\eqdef\tr[H\rho]+\lambda(\tr[\rho\log\rho]+S)
\ee
over all $\rho\in\md(A)$, where $\lambda$ is a Lagrange multiplier. Let $\rho$ be the optimal density matrix that minimizes the Lagrangian above. Then, any other state in $\md(A)$ can be written as $\rho+tY$ for some $t\in\mbb{R}$ and $Y\in\herm(A)$ is a traceless matrix (we can also assume without loss of generality that $\|Y\|_\infty\leq 1$ although we will not need it).
Since $\rho$ is optimal we must have for any such $Y$
\ba\label{dird2}
0&=\frac d{dt}\mL(\rho+tY,\lambda)\Big|_{t=0}\\
\GG{Exercise~\ref{dird}}&=\tr[HY]+\lambda\tr[Y\log\rho]\;.
\ea
That is, an optimal $\rho$ must satisfy
\be
\tr[Y(H+\lambda\log\rho)]=0
\ee
for all traceless matrices $Y\in\herm(A)$, so that $H+\lambda\log\rho$ is orthogonal (in the Hilbert-Schmidt inner product) to the subspace of all traceless matrices in $\herm(A)$. Consequently, $H+\lambda\log\rho$ must be proportional to the identity matrix; i.e., there exists $c\in\mbb{R}$ such that $H+\lambda\log\rho=cI$. Hence, the optimal $\rho$ has the form
\be
\rho=e^{\frac c\lambda}e^{-\frac{H}{\lambda}}=\frac{e^{-\beta H}}{\mZ}\;,
\ee
where in the last equality we denoted by $\beta\eqdef\frac1\lambda$, and used the fact that $\tr[\rho]=1$ so that $e^{\frac c\lambda}=1/\tr[e^{-\beta H}]$.

\bex\label{dird}
Use Corollary~\ref{cordd} to prove the expression for the directional derivative given in~\eqref{dird2}.
\eex
  
\bex
Let $\alpha\in[0,\infty]$. Find the state $\rho_\alpha\in\md(A)$ that minimizes $\tr[H^A\rho^A]$ while keeping the $\alpha$-R\'enyi entropy fixed.
\eex

\subsection{Passive States}\label{passive}\index{passive state}

In deriving the Gibbs state mentioned above, we sought a state that minimizes the average energy while maintaining a constant von Neumann entropy. Although this method is mathematically sounds, it does not offer a convincing rationale for designating the Gibbs state as the only free state within the model. This raises the question of whether there exists a more operational method to derive the Gibbs state, one that does not rely on the concept of von Neumann entropy. As we will explore here, such an approach does indeed exist.

Let $A$ be a physical system with a Hamiltonian $H^A$.
Suppose that the system $A$ is described by the density matrix $\rho\in\md(A)$. Therefore, the energy of the system is given by $\tr[H^A\rho^A]$. Under a closed (i.e. unitary) evolution/process, the system $A$ can be evolved into another state of the form $U\rho^AU^*$, where $U\in\muu(A)$ is some unitary matrix. The maximal amount of work that can be extracted from such a system cannot exceed the energy difference between the initial and final states. Therefore, we define the maximal extractable work from a system in a state $\rho^A$ as
\be
W_{\max}\left(\rho^A\right)=\max_{U}\tr\left[H^A\left(\rho^A-U\rho^AU^*\right)\right]
\ee
where the maximum is over all unitary matrices $U\in\muu(A)$.
Interestingly, the above optimization problem can be solved analytically.

\begin{myg}{}
\begin{lemma}
Let $H^A=\sum_{x\in[m]}a_x|x\lr x|^A$ be the Hamiltonian of system $A$, with the energy eigenvalues arranged in non-decreasing order; i.e. $a_1\leq a_2\leq\cdots\leq a_m$. Then,
\be\label{cothe}
W_{\max}\left(\rho^A\right)=\tr\left[H^A\rho^A\right]-\tr\left[H^A\sigma^A_\rho\right]\quad\quad\forall\;\rho\in\md(A)\;,
\ee
where $\sigma^A_\rho$ is defined with respect to the eigenvalues $\{p_x\}_{x\in[m]}$ of $\rho^A$ as
\be
\sigma_\rho^A\eqdef\sum_{x\in[m]}p_x^\da|x\lr x|^A\;.
\ee
\end{lemma}
\end{myg}

\begin{proof}
From a variant of the von-Neumann trace inequality, known as the Ruhe’s Trace Inequality as given in Theorem~\ref{rti},  it follows that
\ba
\tr\left[H^AU\rho^AU^*\right]&\geq\sum_{x\in[m]}a_xp_x^\da\\
&=\tr\left[H^A\sigma_\rho^A\right]\;,
\ea
where we used the lower bound in~\eqref{p168} with $N\eqdef H^A$ and $M\eqdef U\rho^AU^*$.
The proof is then concluded with the observation that there exists a unitary matrix $U$ satisfying $U\rho^AU^*=\sigma_\rho^A$.
\end{proof}

\begin{exercise}\label{ex1713}
Let $n,m\in\mbb{N}$ and $\rho,\sigma\in\md(A)$. 
\ben
\item Show that
\be
W_{\max}\left(\rho\otimes\sigma\right)\geq W_{\max}\left(\rho\right)+W_{\max}\left(\sigma\right)\;.
\ee
\item Show that if $n\geq m$ then
\be
\frac 1nW_{\max}\left(\rho^{\otimes n}\right)\geq \frac1mW_{\max}\left(\rho^{\otimes m}\right)\;.
\ee
\een
\end{exercise}

In the following lemma, we demonstrate that the maximum extractable work can never exceed the difference between the energy of the system and the energy of the system at equilibrium.

\begin{myg}{}
\begin{lemma}\label{lem1712}
Let $\gamma^A$ be a Gibbs state of system $A$ with a temperature $\beta$ for which $H(\rho^A)=H(\gamma^A)$.
Then,
\be\label{17p13}
W_{\max}\left(\rho^A\right)\leq\tr\left[H^A\rho^A\right]-\tr\left[H^A\gamma^A\right]\;.
\ee
\end{lemma}
\end{myg}
\begin{proof}
The Gibbs state $\gamma^A$ is the state with the smallest energy that has an entropy $H(\rho^A)=H(\sigma_\rho^A)$. Therefore, the state $\sigma_\rho^A$ has higher energy than $\gamma^A$ so that
\be
\tr\left[H^A\gamma^A\right]\leq \tr\left[H^A\sigma_\rho^A\right]\;.
\ee
Combining this with~\eqref{cothe} completes the proof.
\end{proof}

The lemma mentioned above establishes an additive upper bound. Consequently, it can be inferred from the lemma that:
\ba\label{17p14}
W_{\max}^\reg\left(\rho^{A}\right)&\eqdef\lim_{n\to\infty}\frac1nW_{\max}\left(\rho^{\otimes n}\right)\\
\GG{\eqref{17p13}\text{ applied to system }{\it A^n}}&\leq\lim_{n\to\infty}\frac1n\left(\tr\left[H^{A^n}\rho^{\otimes n}\right]-\tr\left[H^{A^n}\gamma^{A^n}\right]\right)\\
&=\tr\left[H^A\rho^A\right]-\tr\left[H^A\gamma^A\right]\;,
\ea
where we used the relations $\tr\left[H^{A^n}\rho^{\otimes n}\right]=n\tr\left[H^A\rho^A\right]$ and $\tr\left[H^{A^n}\gamma^{A^n}\right]=n \tr\left[H^A\gamma^A\right]$.
We now demonstrate that the aforementioned inequality actually holds as an equality.
\begin{myt}{}
\begin{theorem}\label{thm:passive}
Let $\gamma^A$ be a Gibbs state of system $A$ with a temperature $\beta$ for which $H(\rho^A)=H(\gamma^A)$.
Then,
\be
W_{\max}^\reg\left(\rho^A\right)=\tr\left[H^A\rho^A\right]-\tr\left[H^A\gamma^A\right]\;.
\ee
\end{theorem}
\end{myt}
\begin{proof}
From~\eqref{17p14} it is sufficient to prove that
\be\label{17p16}
W_{\max}^\reg\left(\rho^A\right)\geq\tr\left[H^A\rho^A\right]-\tr\left[H^A\gamma^A\right]\;.
\ee
Let
\be
f(\rho)\eqdef\min_{U}\tr[H^AU\rho^A U^*]\;,
\ee
and observe that the inequality in~\eqref{17p16} is equivalent to
\be
f^\reg(\rho)\eqdef\lim_{n\to\infty}\frac1nf(\rho^{\otimes n})\leq\tr[H^A\gamma^A]\;.
\ee
Our goal is therefore to prove this inequality.

Since $f$ is invariant under unitaries we can assume without loss of generality that $\rho$ is diagonal and we denote its diagonal by $\p=(p_1,\ldots,p_m)^T$. We also denote by $X$ the random variable that takes the value $X=x\in[m]$ with probability $p_x$.
Further, let $\gamma'\in\md(A)$ be another Gibbs state of system $A$, with a diagonal $\g'=(g_1',\ldots,g_m')^T$ and with inverse temperature $\beta'<\beta$ (i.e. $\gamma'$ corresponds to a Gibbs state with the same Hamiltonian $H^A$ and with a higher temperature). Further, let $Y$ be the random variable that takes the value $Y=y\in[m]$ with probability $g'_y$. With these notations we have
\be
H(Y)=H(\gamma')>H(\gamma)=H(\rho)=H(X)\;.
\ee
Now, let $\eps>0$ and recall that the number of strongly $\eps$-typical sequences drawn from an i.i.d.$\sim\p$ source scales predominantly as $2^{nH(X)}$. Therefore, since $H(X)<H(Y)$ we get that for sufficiently small $\eps>0$ and sufficiently large $n$ we have $|\mt_{\eps}^{\st}(X^n)|<|\mt_{\eps}^{\st}(Y^n)|$. In particular, there exists a one-to-one function $\pi_n: [m]^n\to [m]^n$, with the property that for any $x^n\in\mt_{\eps}^{\st}(X^n)$ we have $\pi_n(x^n)\in\mt_{\eps}^{\st}(Y^n)$. Define the unitary $U_n\in\ml(A^n)$ by its action on basis elements of $A^n$ as
$
U_n|x^n\ra\eqdef|\pi_n(x^n)\ra
$
for all $x^n\in[m]^n$. Since $U_n$ is not necessarily optimal we get
\ba
f^\reg(\rho)&\leq\lim_{n\to\infty}\frac1n \tr\left[H^{A^n}U_n\rho^{\otimes n}U_n^*\right]\\
&= \lim_{n\to\infty} \frac1n \sum_{x^n\in[m]^n}p_{x^n}\tr\left[H^{A^n}|\pi(x^n)\lr \pi(x^n)|\right]\;.
\ea
Next, observe that
\be
H^{A^n}=\sum_{x^n\in[m]^n}\left(a_{x_1}+\cdots+a_{x_n}\right)|x^n\lr x^n|=n\sum_{x^n\in[m]^n}\t(x^n)\cdot\a|x^n\lr x^n|\;,
\ee
where $\t(x^n)\in\type(n,m)$ is the type of the sequence $x^n$ and $\a\eqdef(a_1,\ldots,a_m)^T$.
Substituting this into the previous equation gives
\ba\label{sety2}
f^\reg(\rho)&\leq\lim_{n\to\infty}\sum_{x^n\in[m]^n}p_{x^n}\t\left(\pi_n(x^n)\right)\cdot\a\\
\GG{Exercise~\ref{sety}}&=\lim_{n\to\infty}\sum_{x^n\in\mt_{\eps}^{\st}(X^n)}p_{x^n}\t\left(\pi_n(x^n)\right)\cdot\a\;,
\ea
where in the second line we restricted $x^n$ to the set of strongly $\eps$-typical sequences. The theorem of strongly typical sequences ensures that the contribution of non-typical sequences vanishes in the limit $n\to\infty$ (see Exercise~\ref{sety}). Since the above inequality holds for all $\eps\in(0,1)$, taking the limit $\eps\to 0^+$ gives
\ba
f^\reg(\rho)&\leq\lim_{\eps\to 0^+}\lim_{n\to\infty}\sum_{x^n\in\mt_{\eps}^{\st}(X^n)}p_{x^n}\t\left(\pi_n(x^n)\right)\cdot\a\\
&=\g'\cdot\a=\tr\left[H^A\gamma'^{A}\right]\;,
\ea
where we used the fact that $\pi_n(x^n)\in\mt_{n}^{\eps,\st}(\g')$ so that $\t\big(\pi_n(x^n)\big)\to\g'$ as $n\to\infty$. 
Finally, since we proved that $f^\reg(\rho)\leq \tr\left[H^A\gamma'^{A}\right]$ for any Gibbs state with inverse temperature $\beta'>\beta$ it follows that the inequality also hold for $\beta'=\beta$. This completes the proof.
\end{proof}

\bex\label{sety}
Prove the relation~\eqref{sety2}. Hint: Use Theorem~\ref{stypi} in conjunction with the fact that $\t\left(\pi_n(x^n)\right)\cdot\a$ is bounded from above; e.g., $\t\left(\pi_n(x^n)\right)\cdot\a\leq a_m$ since $\t\left(\pi_n(x^n)\right)$ is a probability vector.
\eex

A state $\rho\in\md(A)$ characterized by $W^{\reg}{\max}(\rho)=0$ is identified as a \emph{completely passive state}. Such states are inherently unable to facilitate work extraction, irrespective of their quantity. As inferred from Exercise~\ref{ex1713}, if $W^{\reg}{\max}(\rho)=0$, then it follows that:
\be
W_{\max}\left(\rho^{\otimes n}\right)=0\quad\quad\forall\;n\in\mbb{N}\;.
\ee
This insight, derived from Theorem~\ref{thm:passive}, establishes the Gibbs state as the unique completely passive state. Consequently, this finding compellingly supports the designation of the Gibbs state, or thermal state, as the exclusive free state in the domain of quantum thermodynamics.

\bex
Give full details why the only state that is completely passive is the Gibbs state.
\eex

\subsection{Athermality States}\index{athermality state}

An athermality state of system $A$ cannot be solely characterized by $\rho^A$ since the resourcefulness of the state also depends on the Hamiltonian. Therefore, in quantum thermodynamics, every thermodynamic state comprises a quantum state $\rho \in \md(A)$ that acts on the Hilbert space $A$, and a time-independent Hamiltonian $H^A \in \pos(A)$ that governs the dynamics of system $A$. In other words, a state of athermality can be characterized by a pair $(\rho^A, H^A)$. This characterization is widely used in the literature.

However, from a resource-theoretic perspective, this characterization has several drawbacks. Firstly, it is not invariant under an energy shift of the form $H^A \mapsto H^A + cI^A$, where $c \in \mathbb{R}$ is a constant. Indeed, the choice of setting the minimal energy of a system to be zero is somewhat arbitrary. Secondly, we will observe that the resourcefulness of the state $\rho^A$ is determined in relation to its deviation from the Gibbs state $\gamma^A$ of system $A$.

Therefore, it appears more natural to characterize athermality states (i.e., the ``objects" of this theory) as pairs of the form $(\rho^A, \gamma^A)$. It is worth noting that all the relevant information about the Hamiltonian $H^A$ is contained in the Gibbs state $\gamma^A$, which is invariant under energy shifts. Using this notation, the Gibbs state can be represented as $(\gamma^A, \gamma^A)$. In the following lemma, we demonstrate that the Gibbs state can also be utilized to determine if a given unitary evolution exhibits time-translation symmetry.

\begin{myg}{}
\begin{lemma}\label{comgib}
Let $U\in\muu(A)$ be a unitary matrix. Then $U^A$ commutes with a Hamiltonian $H\in\pos(A)$ if and only if $U^A$ commutes with the Gibbs state $\gamma^A$.
\end{lemma}
\end{myg}
\begin{proof}
By definition, the Gibbs state $\gamma^A$ commutes with $U^A$ if and only if $e^{-\beta H^A}$ commutes with $U^A$. Therefore, if $[U^A,H^A]=0$ then clearly $U^A$ commutes with $\gamma^A$. Conversely, suppose $U^A$ commutes with $e^{-\beta H^A}$ and express
$H^A=\sum_{x}\lambda_x P_x$, where $\{P_x\}$ are orthogonal projections satisfying $P_xP_y=\delta_{xy}P_x$ and $\{\lambda_x\}$ are \emph{distinct} eigenvalues of $H^A$. 
Then,
\be
\sum_{x}e^{-\beta\lambda_x} UP_x=Ue^{-\beta H^A}=e^{-\beta H^A}U=\sum_{y}e^{-\beta\lambda_y} P_yU\;.
\ee
Multiplying by $P_x$ from the right and $P_y$ from the left we get
\be
e^{-\beta\lambda_x} P_yUP_x=e^{-\beta\lambda_y} P_yUP_x\;.
\ee
since $\lambda_x\neq\lambda_y$ we conclude that $P_yUP_x=0$ for all $x\neq y$. Hence, $U$ commutes with $H^A$.
\end{proof}

Note that in the lemma above, the condition that $U^A$ commutes with $\gamma^A$ can be expressed as
\be
\mU^{A\to A}\left(\gamma^A\right)\eqdef U^A\gamma^A U^{*A}=\gamma^A\;.
\ee
That is, the unitary matrix $U^A$ commutes with the Hamiltonian if and only if the unitary channel $\mU^{A\to A}$ preserves the Gibbs state.

Suppose now that system $B$ is comprised of two subsystems $B_1$ and $B_2$ and that the total Hamiltonian of system $B$ can be expressed as
\be
H^B=H^{B_1}\otimes I^{B_2}+I^{B_1}\otimes H^{B_2}\;.
\ee 
In this case, the Gibbs state $\gamma^B$ of the composite system\index{composite system} can be expressed as a tensor product of the two Gibbs states of the subsystems. Indeed, we have
\be
\gamma^B=\frac{e^{-\beta \left(H^{B_1}\otimes I^{B_2}+I^{B_1}\otimes H^{B_2}\right)}}{\tr\left[e^{-\beta \left(H^{B_1}\otimes I^{B_2}+I^{B_1}\otimes H^{B_2}\right)}\right]}\;,
\ee
and since
\be
e^{-\beta \left(H^{B_1}\otimes I^{B_2}+I^{B_1}\otimes H^{B_2}\right)}
=e^{-\beta H^{B_1}}\otimes e^{-\beta  H^{B_2}}
\ee
we conclude that $\gamma^B=\gamma^{B_1}\otimes\gamma^{B_2}$, where for each $j=1,2$, $\gamma^{B_j}\eqdef e^{-\beta H^{B_j}}/\tr\left[e^{-\beta H^{B_j}}\right]$ is the Gibbs state of subsystem ${B_j}$.

\section{The Free Operations}

There have been various formulations in the literature regarding the free operations of the QRT of thermodynamics, all of which involve operations conserving certain extensive quantities like energy, particle number, charge, etc. In this chapter, we focus solely on free operations corresponding to energy conservation, giving rise to the resource theory of athermality. However, these operations can be analogously extended to conserve other extensive quantities. For further details, we direct the reader to the section titled 'History and Further Reading.'

Let us denote the Hilbert space associated with the thermal bath with a fixed temperature $T$ as $B$. Additionally, we will consider a quantum system $A$ that interacts with the heat bath (see Fig.~\ref{heatbath}).

\begin{figure}[h]\centering    \includegraphics[width=0.6\textwidth]{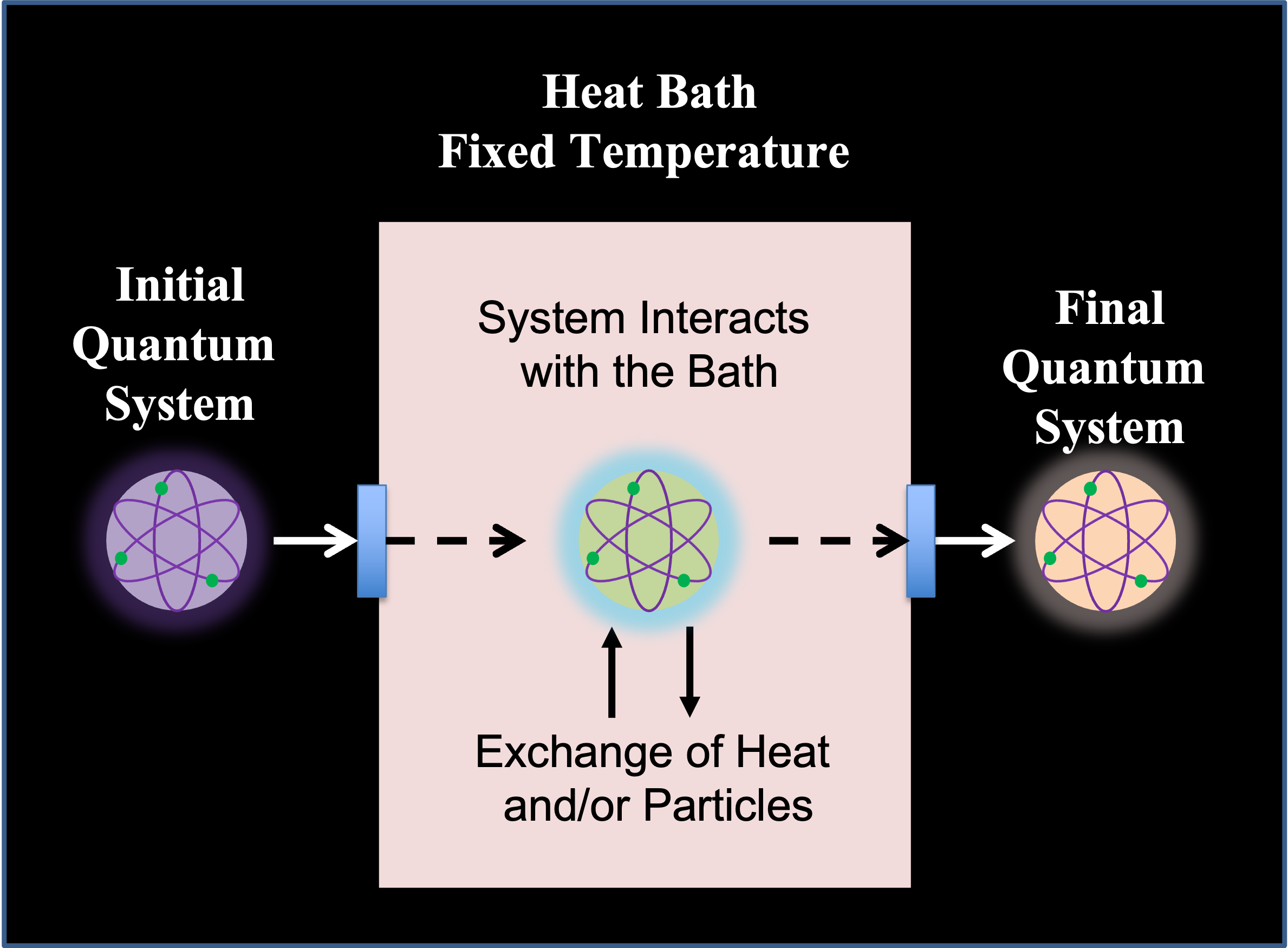}
  \caption{\linespread{1}\selectfont{\small A quantum system interacting with a heat bath.}}
  \label{heatbath}
\end{figure} 

\subsection{Thermal Operations}\index{thermal operations}

The set of free operations relative to a background heat bath at temperature $T$ comprise of three basic steps:
\begin{enumerate}
\item Thermal equilibrium. Any subsystem $B$, with Hamiltonian $H^{B}\in\pos(B)$, can be prepared in its thermal Gibbs state $\gamma^B$.
\item Conservation of energy.  Unitary operation on a composite physical system that commutes with the total Hamiltonian can be implemented. 
\item Discarding subsystems. It is possible to trace over any subsystem of a composite system.
\end{enumerate} 
\begin{remark}
For the second step of conservation of energy, it is assumed that the couplings among the subsystems of the composite system\index{composite system} is controlled entirely by the experimenter. Therefore, 
in the absence of such intervention, it will be assumed that the total Hamiltonian is decoupled, and can be expressed as a sum of the free Hamiltonians of the subsystems.
\end{remark}

Any CPTP map comprising of the above three steps is called \emph{thermal operation} (TO). TO forms the set of free operations in the QRT of athermality.
To investigate more closely these operations, let $(\rho^A,\gamma^A)$ be an athermality state with $\rho\in\md(A)$ and corresponding Gibbs state $\gamma\in\md(A)$ (or equivalently a Hamiltonian $H\in\pos(A)$). From the first step above it follows that the transformation
\be
\rho^A\to\rho^A\otimes\gamma^B
\ee
is a free operation, where $B$ is some ancillary system in the Gibbs state $\gamma^B$ and Hamiltonian $H^B$. The total Hamiltonian of system $AB$ is given by $H^{AB}\eqdef H^A\otimes I^B+I^A\otimes H^B$. Its corresponding Gibbs state is given by $\gamma^{AB}\eqdef\gamma^{A}\otimes\gamma^B$. According to the second step above, any unitary matrix $U: AB\to AB$ that commutes with the total Hamiltonian $H^{AB}$ yields a permissible evolution of the system $AB$. Combining this with Lemma~\ref{comgib} we conclude that a unitary evolution $\mU\in\cptp(AB\to AB)$ is free if and only if it preserves the Gibbs state $\gamma^{AB}$. For such a Gibbs preserving unitary channel $\mU^{AB\to AB}$ we get that the transformation
\be
\rho^A\otimes\gamma^B\to \mU^{AB\to AB}\left(\rho^A\otimes\gamma^B\right)
\ee
can be implemented by TO. Finally, if systems $A=A_1\cdots A_n$ and $B=B_1\cdots B_m$ are themselves comprised of several subsystems $A_1,\ldots,A_n$ and $B_1,\ldots,B_m$ then tracing out several of these subsystems is a free operation. We therefore conclude that any thermal operation $\mE\in\cptp(A\to A')$ can be expressed as
\be\label{reato}
\mE^{A\to A'}(\rho^A)=\tr_{B'}\left[\mU^{AB\to AB}\left(\rho^A\otimes\gamma^B\right)\right]
\ee
where $A'$ and $B'$ are (possibly composite) subsystems of $AB$ such that $AB= A'B'$.

In the following lemma we show that the requirement $AB=A'B'$ is not necessary, only that $AB\cong A'B'$. This is not completely trivial since systems $A'$ and $B'$ can correspond to completely different physical systems.

\begin{myg}{}
\begin{lemma}\label{lem1721}
Let $AB$, $A'B'$ be two composite physical systems with corresponding Gibbs states $\gamma^{AB}\eqdef\gamma^A\otimes\gamma^B$ and $\gamma^{A'B'}\eqdef\gamma^{A'}\otimes\gamma^{B'}$. Suppose $|AB|=|A'B'|$ and let $\mU^{AB\to A'B'}\in\cptp(AB\to A'B')$ be a unitary channel satisfying
\be\label{17p11}
\mU^{AB\to A'B'}\left(\gamma^{AB}\right)=\gamma^{A'B'}\;.
\ee 
Then, the map
\be\label{therm0}
\mN^{A\to A'}\left(\omega^{A}\right)\eqdef\tr_{B'}\left[\mU^{AB\to A'B'}\left(\omega^A\otimes\gamma^B\right)\right]\quad\quad\forall\;\omega\in\ml(A)
\ee  
is a thermal operation. 
\end{lemma}
\end{myg}
\begin{proof}
Let $\gamma^{ABA'B'}\eqdef \gamma^{AB}\otimes \gamma^{A'B'}$ and let $\mV\in\cptp(ABA'B'\to ABA'B')$ be the unitary matrix given by
\be
\mV\eqdef \mU^{AB\to A'B'}\otimes \mU^{*A'B'\to AB}\;.
\ee 
In the exercise below you show that $\mV$ preserves the joint Gibbs state $\gamma^{ABA'B'}$.
Hence, the channel
\ba
\tr_{ABB'}\left[\mV\left(\omega^A\otimes\gamma^{BA'B'}\right)\right]
&=\tr_{ABB'}\left[\mU^{AB\to A'B'}\left(\omega^A\otimes\gamma^B\right)\otimes\mU^{*A'B'\to AB}\big(\gamma^{A'B'}\big)\right]\\
&=\tr_{B'}\left[\mU^{AB\to A'B'}\left(\omega^A\otimes\gamma^B\right)\right]
\ea 
is a thermal operation.
\end{proof}

\begin{exercise}
Show that the matrix $\mV$ as defined in the proof above is indeed Gibbs preserving. Hint: Apply $\mU^*$ to both sides of~\eqref{17p11} to show that $\mU^*$ is Gibbs preserving.
\end{exercise}

\bex
Consider the unitary matrix $U:AB\to A'B'$ associated with the unitary channel $\mU^{AB\to A'B'}$ mentioned in the lemma above (i.e. $\mU^{AB\to A'B'}(\cdot)\eqdef U(\cdot)U^*$. Demonstrate that the condition~\eqref{17p11} is satisfied if and only if
\be\label{17p38n}
UH^{AB}=H^{A'B'}U\;.
\ee
\eex

Recall that a density matrix $\rho\in\md(A)$ can be viewed as an athermality state only when the Hamiltonian or Gibbs state of system $A$ is specified.
Similarly, a quantum channel $\mN\in\cptp(A\to A')$ on its own cannot be considered a thermal operation without specifying  the Gibbs state associated with systems $A$ and $A'$. We will therefore view a thermal operation as a triple $(\mN^{A\to A'},\gamma^A,\gamma^{A'})$, where $\mN\in\cptp(A\to A')$, $\gamma^A$ is the input Gibbs state, and $\gamma^{A'}$ is the output Gibbs state.  We use this perspective in the  following formal definition of thermal operations.

\begin{myd}{}
\begin{definition}
Let $\mN\in\cptp(A\to A')$ and $\gamma^A$ and $\gamma^{A'}$ be two density matrices. The triple $(\mN^{A\to A'},\gamma^A,\gamma^{A'})$ is called a \emph{thermal operation} if there exists a unitary channel $\mU\in\cptp(AB\to A'B')$ (with $|AB|=|A'B'|$), and density matrices $\gamma^B$ and $\gamma^{B'}$, such that both~\eqref{17p11} and~\eqref{therm0} hold.
\end{definition}
\end{myd}

In the following lemma we show that the triple $(\mN^{A\to A'},\gamma^A,\gamma^{A'})$ in the definition above is not independent.

\begin{myg}{}
\begin{lemma}
Let  $(\mN^{A\to A'},\gamma^A,\gamma^{A'})$ be a thermal operation. Then, 
$\mN^{A\to A'}$ is Gibbs preserving; i.e.,
\be\label{gp}
\mN^{A\to A'}(\gamma^A)=\gamma^{A'}\;.
\ee
\end{lemma}
\end{myg}

\begin{proof}
Observe that since $\mU^{AB\to A'B'}$ in~\eqref{therm0} is Gibbs preserving it follows that
\be
\mN^{A\to A'}(\gamma^A)=\tr_{B'}\left[\mU^{AB\to A'B'}\left(\gamma^{AB}\right)\right]=\tr_{B'}\left[\gamma^{A'B'}\right]=\gamma^{A'}\;.
\ee
This completes the proof.
\end{proof}
 
We denote by $\tho(A\to A')$ the set of all quantum channels $\mN\in\cptp(A\to A')$ such that $(\mN^{A\to A'},\gamma^{A},\gamma^{A'})$ is a thermal operation. For simplicity of the notation $\tho(A\to A')$, we do not explicitly specify the input and output Gibbs states. However, throughout this chapter, the physical systems $A$ and $A'$ will always possess well-defined Hamiltonians (and consequently Gibbs states).
We also denote by $\gp(A\to A')$ the set of all Gibbs preserving CPTP maps; that is, 
\be
\gp(A\to A')\eqdef\left\{\mN\in\cptp(A\to A')\;:\;\mN^{A\to A'}(\gamma^A)=\gamma^{A'}\right\}\;.
\ee
Clearly, from their definitions and the lemma above it follows that
\be
\tho(A\to A')\subseteq\gp(A\to A')\;.
\ee

\begin{myt}{}
\begin{theorem}
The set $\tho(A\to A')$ is convex.
\end{theorem}
\end{myt}
\begin{proof}
Let $\{\mN_x\}_{x\in[m]}$ be a set of $m$ channels in $\tho(A\to A')$, and consider a convex combination of these $m$ channels:
\be
\mN^{A\to A'}\eqdef\sum_{x\in[m]} p_x\mN_x^{A\to A'}
\ee
where $\p\eqdef(p_1,\ldots,p_m)^T\in\prob(m)$. Since each $\mN_x$ is a thermal operation, it can be expressed as
\be\label{17p43}
\mN_x^{A\to A'}\left(\omega^A\right)\eqdef\tr_{B_x'}\left[\mU_x^{AB_x\to A'B_x'}\left(\omega^{A}\otimes\gamma^{B_x}\right)\right]\quad\quad\forall\;\omega\in\ml(A)\;,
\ee
where for each $x\in[m]$, $B_x$ and $B_x'$ are auxiliary thermal bathes, and $\mU_x$ is a Gibbs preserving unitary channel.
Let
\be
B\eqdef\bigoplus_{x\in[m]}B_{x}\;,\;\;B'\eqdef\bigoplus_{x\in[m]}B_{x}'\quad\text{and}\quad\gamma^B\eqdef\bigoplus_{x\in[m]}p_x\gamma^{B_{x}}\;.
\ee
Finally, we define the unitary channel $\mU\in\cptp(AB\to A'B')$ as
\be
\mU^{AB\to A'B'}(\eta^{AB})\eqdef\bigoplus_{x\in[m]}\mU_{x}^{AB_{x}\to A'B_{x}'}(\eta^{AB_{x}})\quad\quad\forall\;\eta^{AB}\eqdef\bigoplus_{x\in[m]}\eta^{AB_{x}}\in\md(AB)\;.
\ee
With these definitions we get for all $\omega\in\ml(A)$
\ba
\tr_{B'}\left[\mU^{AB\to A'B'}\left(\omega^{A}\otimes\gamma^{B}\right)\right]&=\sum_{x\in[m]}p_x\tr_{B'_{x}}\left[\mU^{AB_{x}\to A'B'_{x}}\left(\omega^{A}\otimes\gamma^{B_{x}}\right)\right]\\
\GG{\eqref{17p43}}&=\sum_{x\in[m]}p_x\mN^{A\to A'}_x(\omega^A)=\mN^{A\to A'}(\omega^A)\;.
\ea
Therefore, $\mN^{A\to A'}$ is a thermal operation.
This completes the proof.
\end{proof}

\subsection{Closed Thermal Operations}\index{closed thermal operations}

Thermal operations comprise of all quantum channels of the form given in~\eqref{therm0}.
In general, these operations do not form a topologically closed set, as the dimension of system $B$ in~\eqref{therm0} is unbounded (but finite since we will only consider finite dimensional systems). It is therefore possible that some given state cannot be converted to another by thermal operations, and yet, for \emph{any} $\eps>0$ the conversion is possible up to an $\eps$-error. This unphysical property can be tailored back to noisy operations which were defined as the closure of factorizable channels. We will therefore define \emph{closed} thermal operations (CTO) to be the closure of thermal operations. 

\begin{myd}{}
\begin{definition}
Let $A$ and $A'$ be two physical systems. The set of \emph{closed thermal operations}, denoted as $\cto(A\to A')$, is defined as 
\be
\cto(A\to A')\eqdef\overline{\tho(A\to A')}\;.
\ee
\end{definition} 
\end{myd}
 \begin{remark}
By definition, $\mN\in\cto(A\to A')$ if and only if there exists a sequence of thermal operations $\left\{\mN^{A\to A'}_n\right\}_{n\in\mbb{N}}\subset\tho(A\to A')$ such that
\be\label{limch0}
\lim_{n\to\infty}\mN^{A\to A'}_n=\mN^{A\to A'}\;.
\ee
The limit in~\eqref{limch0} is equivalent to
\be
\lim_{n\to\infty}\left\|J_{\mN_n}^{AA'}-J_\mN^{AA'}\right\|_1=0\;,
\ee
where $J_{\mN_n}^{AA'}$ and $J_{\mN}^{AA'}$ are the Choi matrices of $\mN_n$ and $\mN$, respectively.
\end{remark}
The physical justification for CTO is obvious: for $\eps\eqdef10^{-100}$ being one over googol, it is impossible to discriminate between states or channels that are $\eps$-close to each other and for all practical purposes the states or channels can be considered identical. We will see however, that this assumption has significant consequences particularly in the quasi-classical regime.

\bex
Show that {\rm CTO} is closed under concatenation. That is, show that if $(\mN,\gamma^A,\gamma^B)$ and $\big(\mM,\gamma^B,\gamma^C)$ are closed thermal operations then $(\mM\circ\mN,\gamma^A,\gamma^C)$ is also a closed thermal operation.
\eex 

%We will use the symbol $\xleftrightarrow{\tho\;}_\eps$ to indicate equivalence between two athermality states that tolerates a small $\eps>0$ error. That is, we write
%\be
%(\rho^A,\gamma^A)\xleftrightarrow{\tho\;}_\eps(\sigma^{A'},\gamma^{A'})
%\ee
%if $(\rho^A,\gamma^A)$ can be converted by thermal operations to a state that is $\eps$-close to
%$(\sigma^{A'},\gamma^{A'})$, and vice versa. 

\begin{myg}{}
\begin{lemma}\label{lem1715}
Let $(\rho^A,\gamma^A)$ and $(\sigma^{A'},\gamma^{A'})$ be two athermality states. Then the following statements are equivalent (see Fig.~\ref{figcto}):
\ben
\item The state $\left(\rho^A,\gamma^A\right)$ can be converted to $\left(\sigma^{A'},\gamma^{A'}\right)$ by CTO.
\item For any $\eps>0$ there exists states $\trho^{A}$ and $\tsigma^{A'}$ that are $\eps$-close to $\rho^A$ and $\sigma^{A'}$, respectively, such that 
\be\label{convtho}
\tsigma^{A'}=\mN^{A\to A'}\left(\trho^A\right)\text{ for some }\mN\in\tho(A\to A')\;.
\ee
\een
\end{lemma}
\end{myg}

\begin{proof}
The proof that $1\Rightarrow 2$ is left as an exercise, and we prove that $2\Rightarrow 1$.
Let $\{\eps_k\}_{k\in\mbb{N}}$ be a sequence of positive numbers with zero limit, and for each $k\in\mbb{N}$, let $(\rho_{k}^A,\gamma^A)$ be an athermality state that can be converted to a state that is $\eps_k$-close to $(\sigma^{A'},\gamma^{A'})$. That is, for each $k$ there exists a thermal operation $\mN_k\in\tho(A\to A')$ with the property that
\be\label{17p47}
\mN_k^{A\to A'}\left(\rho_{k}^A\right)\approx_{\eps_k}\sigma^{A'}\;.
\ee
Since the set $\cptp(A\to A')$ is compact, there exists a converging subsequence of $\{\mN_k\}_{k\in\mbb{N}}$. For simplicity of the exposition here, we assume without loss of generality that the sequence $\{\mN_k\}_{k\in\mbb{N}}$ itself is converging (otherwise, we have to replace $k$ with a subsequence $\{n_k\}_{k\in\mbb{N}}$) and set $\mN\eqdef\lim_{k\to\infty}\mN_k$. By definition, $\mN\in\cto(A\to A')$ since each $\left(\mN_k^{A\to A'},\gamma^{A},\gamma^{A'}\right)$ is a thermal operation. Moreover, observe that
\be
\mN^{A\to A'}\left(\rho^A\right)=\lim_{k\to\infty}\mN_k^{A\to A'}\left(\rho_{k}\right)=\sigma^{A'}\;,
\ee
where we used~\eqref{17p47}. Hence, $\left(\rho^A,\gamma^A\right)$ can be converted to $\left(\sigma^{A'},\gamma^{A'}\right)$ by CTO. This completes the proof.
\end{proof}

\begin{figure}[h]\centering    \includegraphics[width=0.5\textwidth]{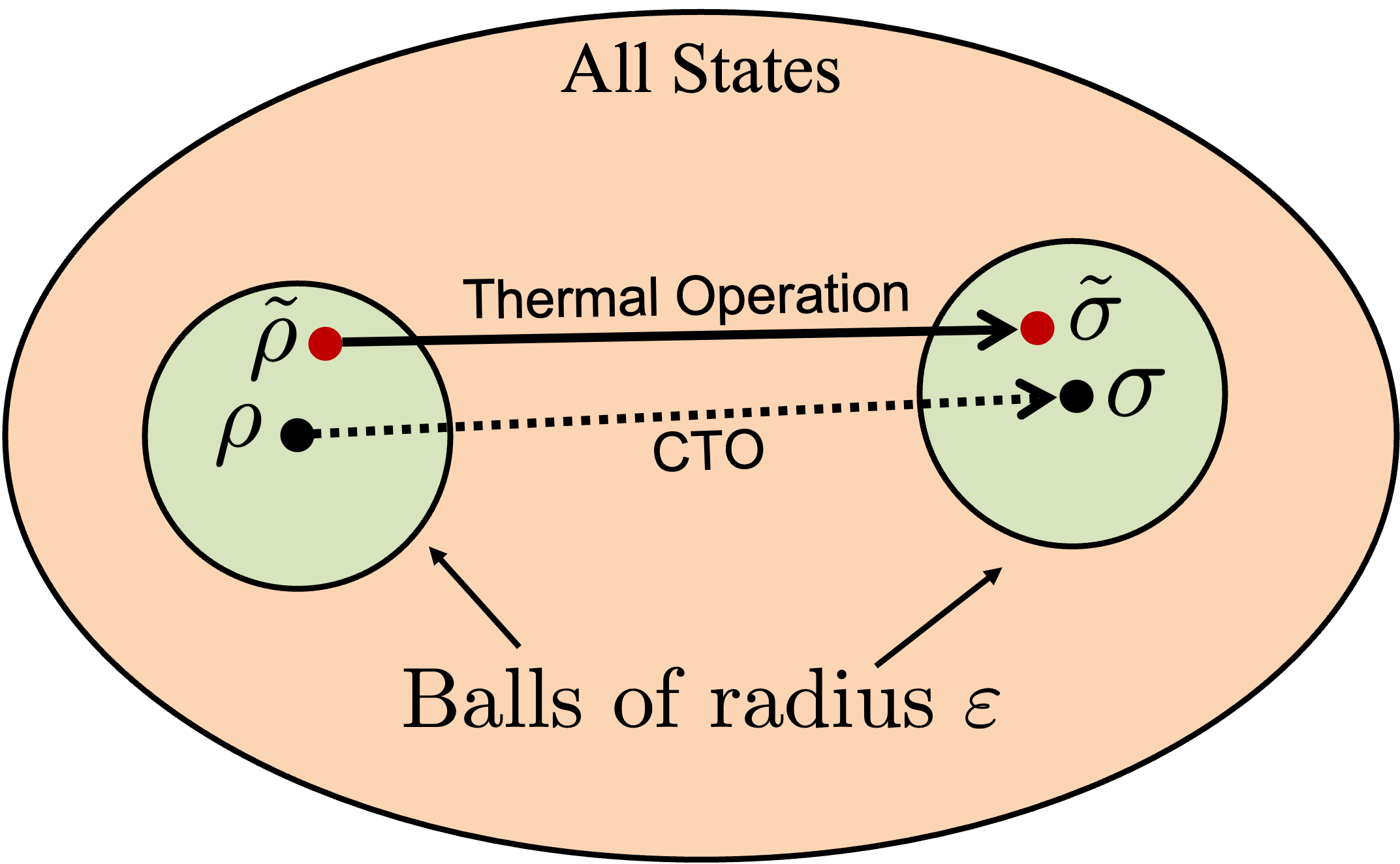}
  \caption{\linespread{1}\selectfont{\small The conversion of $\rho$ to $\sigma$ by CTO. For any $\eps>0$ there exists states $\tilde{\rho}$ and $\tsigma$ that are $\eps$-close to $\rho$ and $\sigma$, respectively, such that $\trho$ can be converted to $\tsigma$ by thermal operation.}}
  \label{figcto}
\end{figure}

\bex
Prove that $1\Rightarrow 2$ in the lemma above.
\eex

\bex
Show that the lemma above still holds even if we replace in~\eqref{convtho} $\mN\in\tho(A\to A')$ with $\mN\in\cto(A\to A')$.
\eex

\subsection{Gibbs-Preserving Covariant Operations}\index{GPC operations}

In Sec.~\ref{sectts} we studied time-translation covariant channels. Such channels are defined with respect to the input and output Hamiltonians associated with the channel. Specifically, let $H^A\in\pos(A)$ and $H^{A'}\in\pos(A')$ be two Hamiltonians, and let $\mE\in\cptp(A\to A')$. Then, we say that $\mE^{A\to A'}$ is time-translation covariant with respect to the Hamiltonians $H^A$ and $H^{A'}$ if for all $t\in\mbb{R}$
\be\label{gpc}
\mU^{A'\to A'}_t\circ\mE^{A\to A'}=\mE^{A\to A'}\circ\mU_t^{A\to A}\quad\quad\forall t\in\mbb{R}\;.
\ee
where $\mU_t^{A\to A}(\cdot)\eqdef U^A_t(\cdot)U_t^{*A}$ is the unitary channel defined with $U^{A}_t\eqdef e^{-itH^A}$ and similarly $\mU_t^{A'\to A'}(\cdot)\eqdef U^{A'}_t(\cdot)U_t^{*A'}$, where $U^{A'}_t\eqdef e^{-itH^{A'}}$. We show here that thermal operations are time-translation covariant.

\begin{myt}{}
\begin{theorem}
Let  $\gamma\in\md(A)$ be a Gibbs state, and let $\mE\in\cto(A\to A')$. Then, $\mE^{A\to A'}$ is time-translation covariant with respect to the Hamiltonians of systems $A$ and $A'$. 
\end{theorem}
\end{myt}

\begin{proof}
Suppose first that $\mE\in\tho(A\to A')$, and it has the form~\eqref{therm0} with $AB\cong A'B'$.
To see that $\mE^{A\to A'}$ is time-translation covariant, observe that
\ba\label{17p55n}
\mE^{A\to A'}\left(e^{-itH^{A}}\rho^Ae^{itH^{A}}\right)&=
\tr_{B'}\left[\mU^{AB\to A'B'}\left(e^{-itH^{A}}\rho^Ae^{itH^{A}}\otimes\gamma^B\right)\right]\\
\Gg{[\gamma^B,H^B]=0}&=\tr_{B'}\left[\mU^{AB\to A'B'}\left(e^{-itH^{A}}\rho^Ae^{itH^{A}}\otimes e^{-itH^{B}}\gamma^Be^{itH^{B}}\right)\right]\\
&=\tr_{B'}\left[\mU^{AB\to A'B'}\circ\mV^{AB\to AB}_t\left(\rho^A\otimes\gamma^B\right)\right]
\;,
\ea
where
$
V^{AB}_t\eqdef e^{-itH^A}\otimes e^{-itH^B}=e^{-itH^{AB}}
$,
$H^{AB}\eqdef H^{A}\otimes I^B+I^A\otimes H^B$ is the total Hamiltonian, and  $\mV^{AB\to AB}_t\eqdef V^{AB}_t(\cdot)V^{*AB}_t$. From~\eqref{17p38n} we get that the unitary channel $\mU^{AB\to AB}\eqdef U^{AB}(\cdot)U^{*AB}$ satisfies 
\be
\mU^{AB\to A'B'}\circ\mV^{AB\to AB}_t=\mV^{A'B'\to A'B'}_t\circ \mU^{AB\to A'B'}\;,
\ee
where $\mV_t^{A'B'\to A'B'}(\cdot)\eqdef e^{-itH^{A'B'}}(\cdot)e^{itH^{A'B'}}$. Combining this with~\eqref{17p55n} we get
for any
$\rho\in\md(A)$ 
\be
\mE^{A\to A'}\left(e^{-itH^{A}}\rho^Ae^{itH^{A}}\right)=\tr_{B'}\left[\mV^{A'B'\to A'B'}_t\circ\mU^{AB\to A'B'}\left(\rho^A\otimes\gamma^B\right)\right]\;.
\ee
Finally, observe that $\mV_t^{A'B'\to A'B'}=\mU_t^{A'\to A'}\otimes\mU_t^{B'\to B'}$, where $\mU_t^{A'\to A'}(\cdot)\eqdef e^{-itH^{A'}}(\cdot)e^{itH^{A'}}$ and $\mU_t^{B'\to B'}(\cdot)\eqdef e^{-itH^{B'}}(\cdot)e^{itH^{B'}}$. Substituting this into the equation above gives
\ba
\mE^{A\to A'}\left(e^{-itH^{A}}\rho^Ae^{itH^{A}}\right)&=
\mU^{A'\to A'}_t\left(\tr_{B'}\left[\mU^{AB\to A'B'}\left(\rho^A\otimes\gamma^B\right)\right]\right)\\
&=e^{-itH^{A'}}\mE^{A\to A'}\left(\rho^A\right)e^{itH^{A'}}\;.
\ea
This completes the proof for $\mE\in\tho(A\to A')$. The case $\mE\in\cto(A\to A')$ follows from the fact that the limit of time-translation covariant channels is itself time-translation covariant (see Exercise~\ref{ttcex}).
\end{proof}

\bex\label{ttcex}
Let $G$ be a group, and let $\{\mE_n\}_{n\in\mbb{N}}$ be a sequence of channels in $\cov_{\G}(A\to A')$ (with respect to some unitary representations of $G$ on $A$ and $A'$). Show that if the limit $\mE\eqdef\lim_{n\to\infty}\mE_n$ exists then also $\mE\in\cov_G(A\to A')$.
\eex

\begin{myd}{}
\begin{definition}
Let $\gamma^A$ and $\gamma^{A'}$ be two Gibbs states. A channel $\mN\in\gp(A\to A')$ is called a Gibbs-preserving covariant operation (in short, GPC operation) if in addition of being Gibbs preserving it is also time-translation covariant satisfying~\eqref{gpc}.
We denote by $\gpc(A\to A')$ the set of all such GPC channels in $\gp(A\to A')$. 
\end{definition}
\end{myd}

From its definition $\gpc(A\to A')$ form a subset of $\gp(A\to A')$. Furthermore, from the theorem above we have
\be
\cto(A\to A')\subseteq \gpc(A\to A')\;.
\ee
\bex
Show that $\gpc(A\to A')$ is convex.
\eex

The following exercise shows another (possibly strictly) subclass of $\gpc$ operations. We say that isometry channel $\mV\in\cptp(A\to A')$ is \emph{time-translation covariant} if it satisfies
\be
VH^A=H^{A'}V
\ee
where $H^A$ and $H^{A'}$ are the Hamiltonians of systems $A$ and $A'$.
\bex\label{isogpc}
Let $A$, $B$, $A'$, and $B'$ be four physical systems with corresponding Hamiltonians $H^{A}$, $H^{B}$, $H^{A'}$, and $H^{B'}$, and let $\mV^{AB\to A'B'}(\cdot)=V(\cdot)V^*$ be a time-translation covariant isometry channel. Denote by 
\be\label{therm00}
\mE^{A\to A'}\left(\omega^{A}\right)\eqdef\tr_{B'}\left[\mV^{AB\to A'B'}\left(\omega^A\otimes\gamma^B\right)\right]\quad\quad\forall\;\omega\in\ml(B)\;,
\ee
and set $t\eqdef\frac{\mZ^{AB}}{\mZ^{A'B'}}$. Show that the map 
\be
\mN^{A\to A'}\left(\omega^{A}\right)=t\mE^{A\to A'}\left(\omega^{A}\right)+\left(\gamma^{A'}-t\mE^{A\to A'}\left(\gamma^{A}\right)\right)\tr\left[\omega^A\right]
\ee 
is a thermal operation (and in particular a quantum channel). 
Hint: Start with the covariance property  
$
e^{-\beta H^{AB}}=V^*e^{-\beta H^{A'B'}}V\;.
$
to get
\be
\mZ^{AB}=\tr\left[VV^*e^{-\beta H^{A'B'}}\right]=\mZ^{A'B'}\tr\left[VV^*\gamma^{A'B'}\right]\leq\mZ^{A'B'}\;,
\ee
with equality if and only if $|AB|=|A'B'|$ (in which case $V$ is a unitary matrix), and conclude that
\be
\tau^{A'}\eqdef \frac{\gamma^{A'}-t\mE^{A\to A'}\left(\gamma^{A}\right)}{1-t}
\ee
is a density matrix. 
\eex

\section{Quasi-Classical Athermality}\index{quasi-classical}

In this section we examine a scenario in which every resource $(\rho^A, \gamma^A)$ is diagonal in the eigenbasis of $H^A$; i.e. $[\rho^A,\gamma^A]=0$. This implies that we are considering physical systems that lack quantum coherence between different energy levels. In particular, the Gibbs state, being commutative with the Hamiltonian, also lacks coherence between energy levels. Therefore, we refer to this scenario as the quasi-classical case.   We start by showing that in the quasi-classical case thermal operations has the same capability for interconversions as Gibbs-preserving operations.

\subsection{CTO vs GPO}

In the semi-classical regime, it is convenient to denote an athermality state $(\rho^A, \gamma^A)$ (where we assume that $[\rho, \gamma] = 0$) by $(\p, \g)$, where $\p, \g \in \prob(m)$ are probability vectors of dimension $m \eqdef |A|$, and their components comprise the eigenvalues of $\rho$ and $\gamma$, respectively. It then follows that in the quasi-classical regime, a state $(\p, \g)$ can be converted to $(\p', \g')$ by GPO if and only if there exists an $m' \times m$ column stochastic matrix, $E$, where $m \eqdef |A|$ and $m' \eqdef |A'|$, such that
\be
\p'=E\p\quad\text{and}\quad\g'=E\g\;.
\ee
Note that $E$ corresponds to a Gibbs preserving channel. The relation above corresponds precisely to the definition of relative majorization\index{relative majorization} (see Section~\ref{secrm}).
Therefore, we conclude that
\be\label{gp17}
(\p,\g)\xrightarrow{\gp}(\p',\g')\quad\iff\quad \left(\p,\g\right)\succ \left(\p',\g'\right)\;.
\ee
Remarkably, the relation above remains unchanged even if we replace the set GPO with CTO.

\begin{myt}{}
\begin{theorem}\label{thm:quasi}
Let $(\rho,\gamma)$ and $(\rho',\gamma')$ be two quasi-classical states of systems $A$ and $A'$, respectively. The following statements are equivalent:
\begin{enumerate}
\item $(\rho,\gamma)$ can be converted to $(\rho',\gamma')$ by CTO.
\item $(\rho,\gamma)$ can be converted to $(\rho',\gamma')$ by GPO.
\end{enumerate}
\end{theorem}
\end{myt}
\begin{remark}
Note that the theorem above does not state that CTO=GPO, only that they have the same conversion power. In general, we have CTO$\subseteq$GPO since GPO is a closed set of operations containing thermal operations. Therefore, the implication $1\Rightarrow 2$ is trivial, and we only need to prove the direction $2\Rightarrow 1$.
\end{remark}

The proof of the theorem above is technically involved and extensive; it has been deferred to Appendix~\ref{app:proof}. It remains a compelling open challenge to discover a more concise and straightforward proof for this theorem.

\subsubsection{The Church of the Trivialized Hamiltonian} 

The theorem above, in conjunction with~\eqref{gp17},  implies that interconversions under CTO can be characterized with relative majorization.
\begin{myg}{}
\begin{corollary}
Let $(\p,\g)$ and $(\p',\g')$ be two athermality states (in the quasi-classical regime) of systems $A$ and $A'$, respectively. Then,
\be\label{gp18}
(\p,\g)\xrightarrow{\cto}(\p',\g')\quad\iff\quad \left(\p,\g\right)\succ \left(\p',\g'\right)\;.
\ee
\end{corollary}
\end{myg}

We can therefore apply all the machinery of the theory of (relative) majorization to the theory of athermality. 
In particular, one of the immediate consequences of the corollary above is that in the quasi-classical regime, there exists a bijection\index{bijection} between the resource theory of athermality and the resource theory of nonuniformity. This remarkable connection between the two theories essentially states that in the quasi-classical regime athermality \emph{is} nonuniformity. This equivalence follows from Theorem~\ref{onlyr}. 

Specifically, suppose $(\p,\g)$ is an athermality state in the quasi-classical regime, and suppose that $\g$ has only rational components. 
Then, we can write the components of $\g$ as $g_x=\frac{n_x}n$ with $x\in[m]$, $n_x\in\mbb{N}$, and $n\eqdef\sum_{x\in[m]}n_x$, and we have 
\be\label{pag1}
\left(\p,\g\right)\sim\left(\r,\u^{(n)}\right)
\quad
\text{where}
\quad
\r\eqdef\bigoplus_{x=1}^mp_x\u^{(n_x)}\;.
\ee
That is, there exists an $n$-dimensional system $R$ in some state $\r\in\prob(n)$, with trivial Hamiltonian (i.e.\ uniform Gibbs states), such that $(\p,\g)\sim(\r,\u^{(n)})$. Combining this with Theorem~\ref{thm:quasi} we conclude that
\be
(\p,\g)\xrightarrow{\cto}(\r,\u^{(n)})\quad\text{and}\quad (\r,\u^{(n)})\xrightarrow{\cto}(\p,\g)\;.
\ee
In other words, $(\p,\g)$ and $(\r,\u^{(n)})$ corresponds to the same resource, so that the athermality of $(\p,\g)$ can be interpreted as the nonuniformity of $(\r,\u^{(n)})$.

\bex
Let $\eps>0$ and $(\p,\g)$ be an athermality state (in the quasi-classical regime). We do not assume that $\g$ has rational components. Show that there exists an $n$-dimensional system $R$ with trivial Hamiltonian, and two states $\r_1,\r_2\in\prob(n)$ that satisfies $\frac12\|\r_1-\r_2\|_1\leq\eps$ and
\be
(\r_1,\u^{(n)})\succ(\p,\g)\succ(\r_2,\u^{(n)})\;.
\ee
Hint. Use Sec.~\ref{TGC}.
\eex

\begin{exercise}
Prove that the relation~\eqref{pag1} implies that
for any $k\in\mbb{N}$ we also have
\be\label{pag2}
\left(\p^{\otimes k},\g^{\otimes k}\right)\sim\left(\r^{\otimes k},\left(\u^{(n)}\right)^{\otimes k}\right)\;.
\ee
\end{exercise}

The equivalence between athermality and non-uniformity give rise to the following property.

\begin{myt}{\color{yellow} The Many Second Laws of Thermodynamics}\index{second laws}
\begin{theorem}
Let $(\p,\g)$ and $(\p',\g')$ be two thermal states. The following statements are equivalent.
\begin{enumerate}
\item For every $\eps>0$ there exists a thermal catalyst $\kappa\eqdef(\r,\tg)$ such that
\be\label{starcat}
(\p_\eps,\g)\otimes\kappa\xrightarrow{\cto}(\p'_\eps,\g')\otimes\kappa\;,
\ee
for some $\p_\eps\in\mb_\eps(\p)$ and $\p'_\eps\in\mb_\eps(\p')$.
\item For all $\alpha\geq\frac12$
		\be
		D_{\alpha}(\p\|\g) \geq D_{\alpha}(\p'\|\g')\quad\text{and}\quad D_{\alpha}(\g\|\p) \geq D_{\alpha}(\g'\|\p')\;.
		\ee
\end{enumerate}
\end{theorem}
\end{myt}

\begin{remark}
Very recently (see the notes and references at the end of this section) it was shown that the theorem above can be strengthened by replacing $\p_\eps^A$ with $\p^A$ so that~\eqref{starcat} becomes
\be
(\p,\g)\otimes\kappa\xrightarrow{\cto}(\p'_\eps,\g')\otimes\kappa\;.
\ee 
This improvement makes the result somewhat more physical, and furthermore, provides simple characterization for catalytic majorization\index{catalytic majorization} (cf.~Lemma~\ref{oneside}):  $(\p,\g)\succ_c(\p',\g')$ if and only if for every $\eps>0$ there exists $\p_\eps'\in\mb_\eps(\p')$ such that $(\p,\g)\succ_*(\p'_\eps,\g')$. The proof of this improvement involves techniques not covered in this book and the interested reader can find the relevant references at the last section of this chapter.
\end{remark}

\begin{proof}
From Theorem~\eqref{thm:quasi} the condition in~\eqref{starcat} is equivalent to:
\be
(\p_\eps,\g)\succ_*(\p'_\eps,\g')
\ee
Therefore, from Lemma~\ref{oneside} the condition above is equivalent to:
\be
(\p,\g)\succ_c(\p',\g')\;.
\ee
Hence, the equivalence of the two conditions in the theorem follows from Theorem~\ref{thm:char2}. This completes the proof.
\end{proof}

\section{Quantification of Athermality}

Measures of athermality are functions that take athermality states to the real numbers and behave monotonically under CTO. Recall that in the theory of athermality, any physical system is described by pair of states of the form $(\rho^A,\gamma^A)$, and consequently measures of athermality are functions of such pair of states. Since we consider Hamiltonians with bounded energy, all Gibbs states are positive-definite (i.e., we assume $\gamma^A>0$).

\begin{myd}{}
\begin{definition}
A measure of athermality is a function
\be
\mbD:\bigcup_{A}\md(A)\times\md_{>0}(A)\to\mbb{R}
\ee
that satisfies the following two conditions:
\begin{enumerate}
\item  Monotonicity\index{monotonicity}: For any two athermality states $(\rho^A,\gamma^A)$ and $(\sigma^{A'},\gamma^{A'})$ 
\be
(\rho^A,\gamma^A)\xrightarrow{\cto}(\sigma^{A'},\gamma^{A'})\quad\Rightarrow\quad \mbD(\rho^A,\gamma^A)\geq \mbD(\sigma^{A'},\gamma^{A'})\;.
\ee
\item Normalization: On the trivial system $|A|=1$, $\mbD(1,1)=1$.
\end{enumerate}
\end{definition}
\end{myd}

We have chosen the symbol $\mbD$ to denote a measure of athermality, given that every normalized quantum divergence $\mbD$ also serves as a measure of athermality. Such measures of athermality behave monotonically under the larger set of Gibbs-preserving operations. However, it's worth noting that not all measures of athermality are quantum divergences, as they only need to exhibit monotonic behavior under CTO.

\bex
Show that every normalized quantum divergence is a measure of athermality.
\eex

In the quasi-classical regime, athermality measures are applied to pairs of probability vectors, with the stipulation that the second vector remains strictly positive due to the Gibbs states' inability to contain zero components (assuming finite energies). Furthermore, as previously discussed, two athermality states $(\p, \g)$ and $(\p', \g')$ satisfy $(\p, \g) \xrightarrow{\cto} (\p', \g')$ if and only if $\left(\p, \g\right) \succ \left(\p', \g'\right)$. Thus, within the quasi-classical framework, the earlier definition of an athermality measure essentially transforms into the definition of a divergence. This implies that, in the quasi-classical domain, athermality measures are indeed divergences. Additionally, the direct correlation between classical divergences and non-uniformity measures extends to form a bijection\index{bijection} between nonuniformity measures and athermality measures, further intertwining these concepts.

\subsection{A Complete Family of Monotones}

In section Sec.~\ref{sec:geta} we introduced a complete family of resource monotones.
Taking $\mf=\gpc$ we compute these monotones for the theory of athermality. In order to apply Theorem~\ref{thm:1111} into our case here, we will view each state as a pair $(\rho^A,\gamma^A)$  and the free operations as channels of the form $\mE\oplus\mE$, with $\mE\in\cov(A\to A')$. With these identifications, the state $\eta$ in~\eqref{tri111} is replaced by $\eta\eqdef(\eta_0,\eta_1)$ with $\eta_0,\eta_1\in\pos(A)$ so that~\eqref{9146} becomes 
\be\label{getapa}
G_{\eta}(\rho,\gamma)\eqdef\max\tr\left[J^{AA'}\left(\rho^T\otimes \eta_0^{A'}+\gamma^{A}\otimes \eta_1^{A'}\right)\right]-\left\|\eta_0^{A'}+\eta_1^{A'}\right\|_\infty
\ee
where the maximum is over all $J\in\pos(AA')$ subject to:
\begin{enumerate}
\item $\mP_\xi(J^{AA'})=J^{AA'}$, where $\mP_\xi$ is the pinching channel associated with the operator (cf.~\eqref{totalhamil})
\be
\xi^{AA'}\eqdef H^A\otimes I^{A'}-I^A\otimes H^{A'}\;.
\ee
\item $J^A=I^A$.
\end{enumerate}

The above optimization problem is an SDP, and consequently has a dual given by (see Exercise~\ref{exghout})
\be\label{pirmal}
G_{\eta}(\rho,\gamma)=\min\tr\left[\sigma^A\right]-\left\|\eta_0^{A'}+\eta_1^{A'}\right\|_\infty
\ee
where the minimum is over all $\sigma\in\pos(A)$ subject to
\be 
\sigma^A\otimes I^{A'}\geq\omega^{AA'}\eqdef\mP_\xi\left(\rho^T\otimes \eta_0^{A'}+\gamma^A\otimes\eta_1^{A'}\right)\;.
\ee
Hence, $G_{\eta}(\rho,\gamma)$ can be expressed as 
\be\label{athemono9}
G_{\eta}(\rho,\gamma)=2^{-H_{\min}^\ua(A'|A)_{\omega}}-\left\|\eta_0^{A'}+\eta_1^{A'}\right\|_\infty\;.
\ee

If the Hamiltonians $H^A$ and $H^{A'}$ are non-degenerate then the operator $\xi^{AA'}$ is non-degenerate so that $\mP_\xi$ is the completely dephasing channel in the energy eigenbasis. In this case, $\omega^{AA'}$ is diagonal and therefore we can assume without loss of generality that also $\eta_0$ and $\eta_1$ are diagonal. For this case, for every choice of $\eta$ we have
\be
G_{\eta}(\rho,\gamma)=G_{\eta}(\Delta(\rho),\gamma)\;,
\ee
where $\Delta\in\cptp(A\to A)$ is the energy dephasing channel. Therefore, for such a choice of system $A'$, $G_\eta$ depends only on the diagonal elements of $\rho$.

\bex\label{exghout}
Express the optimization problem in~\eqref{getapa} as a conic linear programming of the form~\eqref{dual123} (i.e., as a dual problem) and then use the primal problem~\ref{primal} to obtain~\eqref{pirmal}.
\eex

\bex
Show that if $A= A'$ and $H^A$ has a non-degenerate Bohr spectrum, then without loss of generality we can assume that $\eta_1$ is diagonal in the energy eigenbasis (i.e., $G_\eta$ depends only on the diagonal elements of $\eta_1$).
\eex

\bex
Let $\eta_0,\eta_1\in\pos(A')$, $\rho,\gamma\in\md(A)$, and $\tilde{\omega}^{AA'}\eqdef \gamma^A\otimes\left(\eta_0^{A'}+\eta_1^{A'}\right)$. 
\ben
\item Show that
\be
H_{\min}^\ua(A'|A)_{\tilde{\omega}}=H_{\min}(A')_{\omega}=-\log\left\|\eta_0^{A'}+\eta_1^{A'}\right\|_{\infty}\;.
\ee
\item Show that the function
\be\label{cfwithit}
f_\eta(\rho,\gamma)\eqdef H_{\min}(A')_{\omega}-H_{\min}^\ua(A'|A)_{\omega}
\ee
is a measure of athermality.
\een
\eex

\bex\label{exom}
Show that for the case $\mf=\gp$, the athermality monotones $G_\eta(\rho,\gamma)$ are given as in~\eqref{athemono9}, but with
\be
\omega^{AA'}\eqdef\rho^T\otimes \eta_0^{A'}+\gamma^A\otimes\eta_1^{A'}\;.
\ee
\eex

\subsection{The Free Energy}\index{free energy}

We saw in this book that the Umegaki\index{Umegaki} relative entropy has several operational interpretations and play a key role in quantum resource theories.
In particular, we will see in the following sections that under Gibbs preserving operations the relative entropy is the unique measure of athermality in the asymptotic setting, as in this setting both the distillable rate of athermality and the athermality cost are given in terms of the relative entropy.
Therefore, it is not a surprise that the relative entropy to the Gibbs state is related to an important quantity in thermodynamics known as the \emph{free energy}. 

In thermodynamics, the free energy is a fundamental concept that represents the potential energy available in a system to do useful work. It is a state function, meaning its value depends only on the current state of the system and not on how the system reached that state. Free energy is denoted by the symbol ``F" and for an athermality state $(\rho,\gamma)$ of system $A$ the free energy is defined as the energy available to do useful work and is given by:
\be
F(\rho)\eqdef\tr\left[\rho\hat{H}\right]-TH(\rho)
\ee
where $T$ is the temperature, and we added here the `hat' symbol to the Hamiltonian $\hat{H}$ of the system, in order to distinguish it from the entropy symbol $H(\rho)$, which stands for the von-Neumann\index{von-Neumann} entropy of $\rho$.

\bex
Show that the free energy of the Gibbs state $\gamma\eqdef\frac1\mZ e^{-\beta \hat{H}}$ is given by
\be\label{zpartition}
F(\gamma)=-T\log\mZ\;.
\ee
\eex

To see the relation of the free energy to the relative entropy, observe that the relative entropy of athermality is given by:
\ba
D(\rho\|\gamma)&=-H(\rho)-\tr\left[\rho\log\left(\frac1\mZ e^{-\beta \hat{H}}\right)\right]\\
&=\log\mZ-H(\rho)+\beta\tr\left[\rho\hat{H}\right]\\
&=\beta F(\rho)+\log\mZ\\
\GG{\eqref{zpartition}}&=\beta\big(F(\rho)-F(\gamma)\big)\;.
\ea
Hence, the free energy is the key factor that directly governs the optimal rate of  interconversions of athermality.

The Umegaki relative entropy of  athermality has another interesting representation. For a quantum athermality state $(\rho,\gamma)$, with Hamiltonian $\hat{H}$, we can express $D(\rho\|\gamma)$ as:
\ba\label{18p26}
D\left(\rho\|\gamma\right)&=-H(\rho)-\tr\left[\rho\log\gamma\right]\\
&=-H(\rho)-\tr\left[\mP_{\hat{H}}(\rho)\log\gamma\right]\\
&=D\left(\mP_{\hat{H}}(\rho)\big\|\gamma\right)+H\big(\mP_{\hat{H}}(\rho)\big)-H(\rho)\\
&=D\left(\mP_{\hat{H}}(\rho)\big\|\gamma\right)+C(\rho)\;,
\ea
where $\mP_{\hat{H}}$ is the pinching channel\index{pinching channel} associated with the Hamiltonian $\hat{H}$, and $C(\rho)$ is the coherence measure defined in~\eqref{coherence} ($C(\rho)$ is also known as the $\G$-asymmetry of the state $\rho$ as defined in~\ref{defasy}, where $\G$ stands for the group of time-translation symmetry).
That is, the athermality of the state $(\rho,\gamma)$ can be decomposed into two components:
\begin{enumerate}
\item Its nonuniformity that is quantified by $D\left(\mP_{\hat{H}}(\rho)\big\|\gamma\right)$.
\item Its asymmetry (or coherence between energy eigenspaces) that is quantified by the coherence measure $C(\rho)$.
\end{enumerate}
We will see later on that this decomposition has an operational meaning, in which (roughly speaking) $D\left(\mP_{\hat{H}}(\rho)\big\|\gamma\right)$ is the cost to prepare the athermality state $(\mP_{\hat{H}}(\rho),\gamma)$ and $C(\rho)$ is the cost to `rotate' $\mP_{\hat{H}}(\rho)$ to $\rho$. Moreover, since the regularization\index{regularization} of the $C(\rho)$ vanishes (see Theorem~\ref{asygasy}), we conclude that
\be\label{18p27}
\lim_{n\to\infty}\frac1nD\left(\mP_n\left(\rho^{\otimes n}\right)\big\|\gamma^{\otimes n}\right)=D\left(\rho\|\gamma\right)\;,
\ee
where $\mP_n$ is the pinching channel\index{pinching channel} associated with the total Hamiltonian of system $A^n$.

\section{Single-Shot Exact Interconversions}\index{single-shot}

In this section we study the condition under which an athermality state $(\rho^A,\gamma^A)$ can be converted to another state $(\sigma^A,\gamma^A)$ by either CTO, GPC, or GPO. We will work in the basis $\{|x\ra\}_{x\in[m]}$ in which the Hamiltonian is diagonalized, and denote by $\hat{H}=\sum_{x\in[m]}a_x|x\lr x|$ the Hamiltonian of system $A$, where $\{a_x\}_{x\in[m]}$ are the energy eigenvalues of $\hat{H}$.

\subsection{Exact Conversions under GPO}

The state $(\rho,\gamma)$ can be converted to $(\rho',\gamma')$ by GPO if and only if there exists  a  channel $\mN\in\cptp(A\to {A'})$ such that $\mN(\rho)=\rho'$ and $\mN(\gamma)=\gamma'$. In the previous chapter we saw that in the quasi-classical case these conditions are equivalent to relative majorization. Here we study its quantum version.

\begin{myd}{}
\begin{definition}
Let $\rho,\gamma\in\md(A)$ and $\rho',\gamma'\in\md(A')$. We say that the pair $(\rho,\gamma)$ relatively majorizes the pair $(\rho',\gamma')$, and write
\be
(\rho,\gamma)\succ(\rho',\gamma')
\ee
if there exists a channel $\mN\in\cptp(A\to A')$ such that
\be\label{twoconrel}
\mN(\rho)=\rho'\quad\text{and}\quad\mN(\gamma)=\gamma'\;.
\ee
\end{definition}
\end{myd}

The two conditions in~\eqref{twoconrel} are equivalent to the existence of a Choi matrix $J\in\pos(A{A'})$ that satisfies
\be
\tr_{A'}\left[J^{A{A'}}\left(\rho^T\otimes I^{A'}\right)\right]=\rho'\quad\text{and}\quad
\tr_{A'}\left[J^{A{A'}}\left(\gamma^T\otimes I^{A'}\right)\right]=\gamma'\;.
\ee
This problem, of determining whether or not such a Choi matrix $J^{AA'}$ exists is an SDP feasibility problem that can be solved efficiently and algorithmically using techniques from semi-definite programming. However, unlike the classical case, where relative majorization\index{relative majorization} can be characterized with Lorenz curves, it is not known in the fully quantum case whether a similar geometrical characterization exists.

Observe that any quantum divergence behaves monotonically under quantum relative majorization. Specifically, if $\D$ is a quantum divergence then
\be
(\rho,\gamma)\succ(\rho',\gamma')\quad\Rightarrow\quad\D(\rho\|\gamma)\geq \D\left(\rho'\big\|\gamma'\right)\;.
\ee
The converse to the above property also holds. That is, if for any choice of a quantum divergence $\D$ we have 
$\D(\rho\|\gamma)\geq \D\left(\rho'\big\|\gamma'\right)$ then we must have $(\rho,\gamma)\succ(\rho',\gamma')$. In fact,  we show now that this assertion still holds even if we restrict $\D$ to have a very specific form.

Recall the complete family of monotones given in~\eqref{athemono9} with $\omega^{AA'}\eqdef\rho^T\otimes\eta_0^{A'}+\gamma^A\otimes\eta_1^{A'}$ given as in Exercise~\ref{exom}. From the completeness of the family of monotones, it follows that $(\rho,\gamma)\succ(\rho',\gamma')$ if and only if $G_{\eta}(\rho,\gamma)\geq G_{\eta}(\rho',\gamma')$ for all $\eta_0,\eta_1\in\pos(A')$. Similar to~\eqref{cfwithit}, for every $\eta_0,\eta_1\in\pos(A')$ we define
\be
D_\eta(\rho\|\gamma)\eqdef H_{\min}(A')_{\omega}-H_{\min}(A'|A)_\omega\;.
\ee
The above functions forms a family of normalized quantum divergences that can be used to characterize quantum relative majorization.

\bex
Show that for every $\eta_0,\eta_1\in\pos(A')$, the function $D_\eta$ as defined above is a quantum divergence.
\eex

\begin{myt}{}
\begin{theorem}\label{chadeta}
Let $\rho,\gamma\in\md(A)$ and $\rho',\gamma'\in\md(A')$. Then, the following are equivalent:
\ben
\item $(\rho,\gamma)\succ(\rho',\gamma')$.
\item For any $\eta_0,\eta_1\in\pos(A')$ we have $D_\eta(\rho\|\gamma)\geq D_\eta(\rho'\|\gamma')$.
\een
\end{theorem}
\end{myt}

\bex
Consider Theorem~\ref{chadeta}.
\ben
\item Prove the theorem.
\item Show that the theorem holds even if we restrict $\eta_0$ and $\eta_1$ to satisfy $\tr[\eta_0+\eta_1]=1$ (hence, we can assume without loss of generality that $\omega^{AA'}$ is a density matrix).
\item Show that the theorem holds even if we restrict $\eta_0$ and $\eta_1$ to satisfy $\tr[\eta_0]=\tr[\eta_1]=1/2$.
\een
\eex

While the aforementioned theorem offers a characterization of quantum relative majorization, it doesn't provide a straightforward method for determining whether one pair of states relatively majorizes another due to the necessity of verifying an infinite number of conditions. Instead, as previously mentioned, one can employ Semidefinite Programming (SDP) feasibility algorithms from convex analysis to address this challenge.

Nevertheless, when dealing with qubit states $\rho$ and $\gamma$, a far simpler method exists to characterize quantum relative majorization. In the upcoming theorem, we leverage the fidelity function $F(\eta,\zeta)\eqdef|\eta^{1/2}\zeta^{1/2}|_1$ for all $\eta,\zeta\in\pos(A)$, including unnormalized states, to provide a more accessible approach to understanding quantum relative majorization.

\begin{myt}{}
\begin{theorem}\label{qubitchar}
Let $\rho,\gamma\in\md(A)$ and $\rho',\gamma'\in\md(A')$, and suppose $|A|=2$. Furthermore, let $a\eqdef2^{D_{\max}(\rho\|\gamma)}$ and $b\eqdef2^{D_{\max}(\gamma\|\rho)}$.
Then, $(\rho,\gamma)\succ(\rho',\gamma')$ if and only if the following conditions hold:
\ben
\item $D_{\max}(\rho\|\gamma)\geq D_{\max}(\rho'\|\gamma')$.
\item $D_{\max}(\gamma\|\rho)\geq D_{\max}(\gamma'\|\rho')$.
\item $
F\left(a\gamma'-\rho',b\rho'-\gamma'\right)\geq F\big(a\gamma-\rho,b\rho-\gamma\big)
$.
\een
\end{theorem}
\end{myt}

\begin{remark}
By definition of $a$ and $b$ we have  $a\gamma-\rho\geq 0$ and $b\rho-\gamma\geq 0$. Moreover, since $\rho$ and $\gamma$ are qubits, $a\gamma-\rho$ and $b\rho-\gamma$ are rank one. Finally, note that from the first two conditions above we also get $a\gamma'-\rho'\geq 0$ and $b\rho'-\gamma'\geq 0$.
\end{remark}

\begin{proof}
The case $\rho=\gamma$ is left as an exercise, and we assume now that $\rho\neq\gamma$ so that both $a$ and $b$ are strictly greater than one.
The necessity of the first two conditions in the theorem follows from the fact that $D_{\max}$ satisfies the DPI. Similarly, the necessity of the third condition follows from the DPI of the fidelity. 
It is therefore left to show the sufficiency of the conditions.
 Define
\be\label{18p32}
\psi\eqdef\frac{a\gamma-\rho}{a-1}\quad\text{and}\quad\phi\eqdef\frac{b\rho-\gamma}{b-1}\;.
\ee
As discussed in the remark above, $\psi$ and $\phi$ are \emph{pure} states. Denoting by
\be\label{18p33}
\eta\eqdef\frac{a\gamma'-\rho'}{a-1}\quad\text{and}\quad\zeta\eqdef\frac{b\rho'-\gamma'}{b-1}\;,
\ee
we can express the third condition as $F(\eta,\zeta)\geq F(\psi,\phi)$. From Uhlmann's theorem there exists purifications of $\eta$ and $\zeta$, denoted by $\psi',\phi'\in\pure(A'\tA')$ such that $F(\eta,\zeta)=|\la\psi'|\phi'\ra|$, so that we get
\be
|\la\psi'|\phi'\ra|\geq |\la\psi|\phi\ra|\;.
\ee
The main trick of the proof is to introduce two states $\varphi_1,\varphi_2\in \pure(R)$, where $R$ is some Hilbert space (which we can choose to be two dimensional), that satisfies
\be
\la\psi|\phi\ra=\la\psi'|\phi'\ra\la\varphi_1|\varphi_2\ra\;.
\ee
The above invariant overlap implies that the matrix $V:A\to A'\tA' R$ defined by
\be
V|\psi\ra\eqdef|\psi'\ra|\varphi_1\ra\quad\text{and}\quad V|\phi\ra\eqdef |\phi'\ra|\varphi_2\ra\;,
\ee
is an isometry. Finally, let $\mN\in\cptp(A\to A')$ be the channel defined via
\be
\mN^{A\to A'}\left(\omega^A\right)\eqdef\tr_{\tA'R}\left[V\omega^AV^*\right]\quad\quad\forall\;\omega\in\ml(A)\;.
\ee
To see that the channel above satisfies the desired properties, first observe that by isolating $\rho$ and $\gamma$ from~\eqref{18p32} we get 
\be
\rho=\frac{(a-1)\psi+a(b-1)\phi}{ab-1}\quad\text{and}\quad\gamma=\frac{b(a-1)\psi+(b-1)\phi}{ab-1}\;.
\ee
Therefore,
\ba
\mN(\rho)&=\frac{(a-1)\mN(\psi)+a(b-1)\mN(\phi)}{ab-1}\\
\GG{By\;definition}&=\frac{(a-1)\eta+a(b-1)\zeta}{ab-1}\\
\GG{\eqref{18p33}}&=\rho'
\ea
Using similar lines it can be shown that $\mN(\gamma)=\gamma'$ (see Exercise~\ref{complep}). This completes the proof.
\end{proof}

\bex\label{complep}
Prove the assertion in the proof above that 
$\mN(\gamma)=\gamma'$.
\eex

\bex
Let $\rho,\sigma,\gamma\in\md(A)$ be three qubit states (i.e. $|A|=2$). Show that 
\be
\left(\rho,\gamma\right)\xrightarrow{\gp}\left(\sigma,\gamma\right)
\ee
if and only if
\be
D_{\max}\left(\rho\big\|\gamma\right)\geq D_{\max}\left(\sigma\big\|\gamma\right)\quad\text{and}\quad
D_{\max}\left(\gamma\big\|\rho\right)\geq D_{\max}\left(\gamma\big\|\sigma\right)\;.
\ee
That is, the third fidelity condition of the theorem above is unnecessary in this case.
\eex

\subsection{Exact Conversions under CTO and GPC}

Given two athermality states $(\rho,\gamma)$ of system $A$, and $(\rho',\gamma')$ of system $A'$, the condition $\left(\rho,\gamma\right)\xrightarrow{\gpc}\left(\rho',\gamma'\right)$ is equivalent to the existence of a Choi matrix $J\in\pos(A{A'})$ that satisfies the following four conditions:
\ben
\item $\tr_{A'}\left[J^{A{A'}}\left(\rho^T\otimes I^{A'}\right)\right]=\rho'$
\item $\tr_{A'}\left[J^{A{A'}}\left(\gamma^T\otimes I^{A'}\right)\right]=\gamma'$
\item $\mP_\xi\left(J^{AA'}\right)=0$, where $\mP_\xi$ is the pinching channel\index{pinching channel} of $\xi^{AA'}\eqdef H^A\otimes I^{A'}-I^{A}\otimes H^{A'}$.
\item $J^{A}=I^A$.
\een
Similar to the GPO case, this problem, of determining whether or not such a Choi matrix $J^{AA'}$ exists, is an SDP feasibility problem that can be solved efficiently and algorithmically using techniques from semi-definite programming. However, for certain choices of Hamiltonians, there exists a much simpler way to characterize the conversion $\left(\rho,\gamma\right)\xrightarrow{\gpc}\left(\rho',\gamma'\right)$.

\subsubsection{The Case of Relatively Non-Degenerate Hamiltonians}

Theorem~\ref{thm1} of Sec.~\ref{sectts}  has the following implication in thermodynamics.
 \begin{myg}{}
 \begin{corollary}\label{cor:rgsg}
 Let $A$ and $A'$ be two physical systems with relatively non-degenerate Hamiltonians. Let $(\rho,\gamma)$ and $(\rho',\gamma')$ be two athermality states on system $A$ and $A'$, respectively. Further, ler $\r$, $\r'$, $\g$, and $\g'$, be the probability vectors whose components are the elements on the diagonals of $\rho$, $\sigma'$, $\gamma$ and $\gamma'$, respectively. Then, for $\mf$ being CTO or GPC, the following are equivalent:
 \ben
 \item $\left(\rho,\gamma\right)\xrightarrow{\mf}\left(\rho',\gamma'\right)$.
 \item $\rho'$ is diagonal and $\left(\r,\g\right)\succ\left(\r',\g'\right)$.
 \een
 \end{corollary}
 \end{myg}
 
 \bex
 Use Theorem~\ref{thm1} to prove the corollary above.
 \eex
 
 Note that in the general case of relatively non-degenerate Hamiltonians, CTO and GPC operations can only disrupt the coherence between the energy levels of the input state $\rho^A$. In such scenarios, coherence cannot be manipulated, but only destroyed. Therefore, for the remainder of this chapter, we will focus on Hamiltonians that exhibit relative degeneracy.

When considering the conversion of one athermality state $(\rho,\gamma)$ to another athermality state
$(\rho',\gamma')$ we will use the properties
\be
\left(\rho,\gamma\right)\xleftrightarrow{\mf}\left(\rho\otimes\gamma',\gamma\otimes\gamma'\right)\quad\text{and}\quad\left(\rho',\gamma'\right)\xleftrightarrow{\mf}\left(\gamma\otimes\rho',\gamma\otimes\gamma'\right)\;.
\ee
The equivalence relations above follow from the fact that appending or removing a Gibbs state is a free operation in the theory of athermality.
Therefore, the conversion $\left(\rho,\gamma\right)\xrightarrow{\mf}\left(\rho',\gamma'\right)$ between a state of system $A$ and a state of system $A'$ is equivalent to the conversion $\left(\rho\otimes\gamma',\gamma\otimes\gamma'\right)\xrightarrow{\mf}\left(\gamma\otimes\rho',\gamma\otimes\gamma'\right)$ between two states of system $AA'$; see Fig.~\ref{addrem}. In other words,
interconversions among states with the same dimensions (i.e. states with $|A|=|A'|$) is general enough to capture also interconversions with $|A'|\neq |A|$ (as long as we do not impose non-degeneracy constraints). We will therefore focus here on interconversions among states that are all in $\md(A)$. 

\begin{figure}[h]
\centering
    \includegraphics[width=0.5\textwidth]{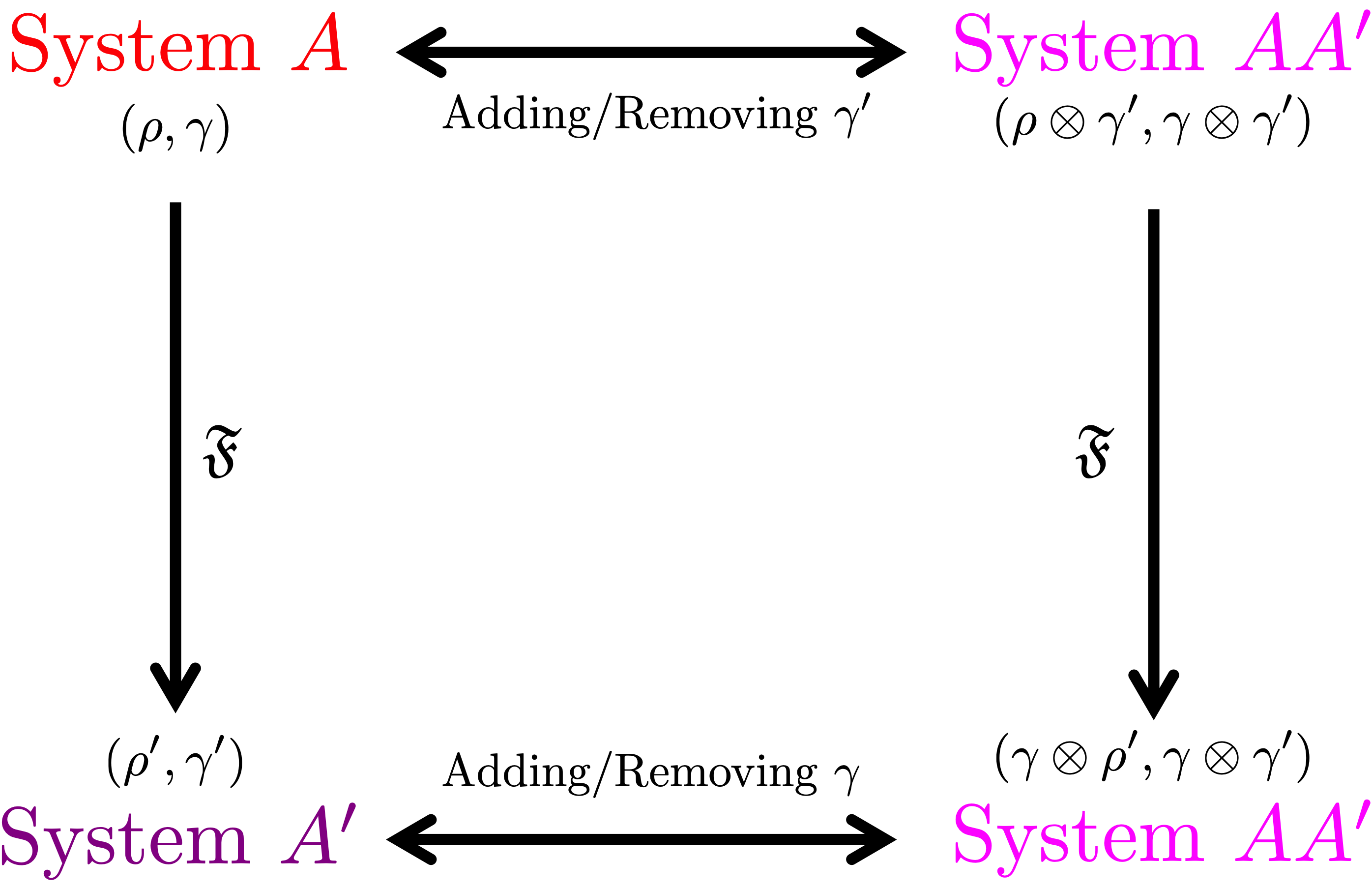}
  \caption{\linespread{1}\selectfont{\small Equivalence of a conversion from $A$ to $A'$ and a conversion from $AA'$ to itself.}}
  \label{addrem}
\end{figure}

\subsubsection{The Case of Bohr Spectrum}\index{Bohr spectrum}

Consider a physical system $A$ whose Hamiltonian, $\hat{H}=\sum_{x\in[m]}a_x|x\lr x|$, has a non-degenerate Bohr spectrum; i.e. 
$a_x-a_{x'}=a_y-a_{y'}$  if and only if $x=x'$ and $y=y'$, or $x=y$ and $x'=y'$.
Further, consider a conversion of the form $\left(\rho,\gamma\right)\xrightarrow{\gpc}\left(\sigma,\gamma\right)$, where \emph{all} the off-diagonal terms of $\rho$ are non-zero.
In this case, Theorem~\ref{ttcs} states that $\rho$ can be converted to $\sigma$ by a time-translation covariant channel if and only if the matrix $Q$ as defined in~\eqref{1369} is positive semidefinite.  Since GPC channels are in particular covariant under the time-translation group, the condition $Q\geq 0$ is a necessary (but not sufficient) condition for
$\left(\rho,\gamma\right)\xrightarrow{\gpc}\left(\sigma,\gamma\right)$. 

To establish the full set of necessary and sufficient conditions, let $J^{AB}$ be the Choi matrix of a time-translation covariant channel $\mE\in\cov(A\to A)$ that satisfies $\mE(\rho)=\sigma$ and $\mE(\gamma)=\gamma$. Denoting by $\{r_{xy}\}_{x,y\in[m]}$ and $\{s_{xy}\}_{x,y\in[m]}$ the components of $\rho$ and $\sigma$, respectively, we get that the Choi matrix of $\mE$ has the form (cf.~\eqref{bohr})
\be
	J^{A\tA}=\sum_{x, y}p_{y|x}|x\lr x|^A\otimes |y\lr y|^{\tA}+\sum_{x\neq y}\frac{s_{xy}}{r_{xy}}|x\lr y|^A\otimes |x\lr y|^{\tA}\;,
	\ee
where $P=(p_{y|x})$ is some column stochastic matrix, and we assumed that the off diagonal terms of $\rho^A$ are non-zero. Let $\r$ and $\s$ be the probability vectors consisting of the diagonals of $\rho$ and $\sigma$, and identify the diagonal matrix $\gamma$ with the Gibbs vector $\g$ consisting of its diagonal. Then, the Choi matrix above corresponds to such a GPC channel $\mE$ if and only if it is positive semidefinite \emph{and} 
\be\label{1621}
P\r=\s\quad\text{and}\quad P\g=\g\;.
\ee
The above condition  implies that $(\r,\g)\succ(\s,\g)$, however, it is not sufficient since we also require that $J^{A\tA}\geq 0$. This latter condition is equivalent to the requirement that the matrix obtained by replacing the diagonal elements of $Q$ (as defined in~\eqref{1369}) with $\{p_{x|x}\}_{x\in[m]}$ is positive semidefinite. 
We summarize these considerations in the following exercise.

\bex\label{lem1841}
Let $(\rho,\gamma)$ and $(\sigma,\gamma)$ be two athermality states of a system $A$, whose Hamiltonian $\hat{H}$ has a non-degenerate Bohr spectrum. Suppose also that the off diagonal terms of $\rho$ are non-zero. Show that
 \be
 (\rho,\gamma)\xrightarrow{\rm GPC}(\sigma,\gamma)
 \ee
 if and only if there exists a column stochastic matrix $P$ that satisfies both~\eqref{1621} and the matrix
 \be\label{matrix18}
\sum_{x\in[m]}p_{x|x}|x\lr x|+\sum_{x\neq y\in[m]}\frac{s_{xy}}{r_{xy}}|x\lr y|\geq 0\;.
 \ee
\eex

The exercise above does not offer significant computational simplification compared to the SDP feasibility problem discussed at the beginning of this section. This is because determining the existence of a column stochastic matrix $P$ itself constitutes an SDP problem. However, the exercise's significance lies in its ability to highlight the role of quantum coherence in converting athermality, as demonstrated by the following theorem.
Furthermore, we will observe later that in the qubit case, the exercise above provides a straightforward criterion for exact inter-conversions under GPC.

\begin{myt}{}
\begin{theorem}\label{thm:gpcco}
Let $(\rho,\gamma)$ and $(\sigma,\gamma)$ be two quantum athermality states of dimension $m\eqdef|A|$. For any $x,y\in[m]$ let $r_{xy}\eqdef\la x|\rho|y\ra$ and $s_{xy}\eqdef\la x|\sigma|y\ra$ be the $xy$-component of $\rho$ and $\sigma$, respectively.
Suppose that $r_{xy}\neq 0$ for all $x,y\in[m]$ and that $r_{xx}=s_{xx}$ for all $x\in[m]$. Then,
\be
 (\rho,\gamma)\xrightarrow{\rm GPC}(\sigma,\gamma)
 \quad\iff\quad
Q\eqdef I+\sum_{x\neq y\in[m]}\frac{s_{xy}}{r_{xy}}|x\lr y|\geq 0\;.
 \ee
\end{theorem}
\end{myt}

\begin{proof}
Since the diagonals of $\rho$ and $\sigma$ are the same, we get that if $Q\geq 0$ then by taking the stochastic matrix $P$ to be the identity matrix, all the conditions in Exercise~\ref{lem1841} are satisfied so that $(\rho,\gamma)\xrightarrow{\rm GPC}(\sigma,\gamma)$. Conversely, if $(\rho,\gamma)\xrightarrow{\rm GPC}(\sigma,\gamma)$ then by Exercise~\ref{lem1841} there exists a stochastic matrix $P$ with a diagonal $\{p_{x|x}\}$ that satisfies~\eqref{matrix18}. By adding the positive semidefinite matrix $\sum_{x\in[m]}(1-p_{x|x})|x\lr x|$ to the matrix in~\eqref{matrix18} we get that also $Q\geq 0$. This completes the proof.
\end{proof}

In simple terms, the condition stated in the theorem above, that $\rho$ and $\sigma$ share the same diagonals, implies that they have the same non-uniformity and only differ in their coherence (asymmetry) properties. Interestingly, the condition $Q\geq 0$ turns out to be identical to the condition given in Theorem~\ref{ttcs} when $\rho$ and $\sigma$ have the same diagonal elements. Thus, in this case, we can state that $(\rho,\gamma)\xrightarrow{\rm GPC}(\sigma,\gamma)$ if and only if $\rho$ can be transformed into $\sigma$ through time-translation covariant operations.
It is noteworthy that the Gibbs state, $\gamma$, does not play a role in such conversions because $\rho$ and $\sigma$ share the same non-uniformity (i.e., same diagonal elements).

\begin{myg}{}
\begin{corollary}\label{puretomixgpc}
Let $\sigma\in\md(A)$ be an arbitrary state, and denote by $p_x\eqdef\la x|\sigma|x\ra$ the diagonal elements of $\sigma$ in the energy eigenbasis $\{|x\ra\}_{x\in[m]}$ of system $A$. Then, the pure quantum state
\be
|\psi\ra\eqdef\sum_{x\in[m]}\sqrt{p_x}|x\ra
\ee
can be converted to $\sigma$ by GPC. That is, $(\psi,\gamma)\xrightarrow{\gpc}(\sigma,\gamma)$.
\end{corollary}
\end{myg}

\bex
Prove the corollary above. Hint: See the proof of Corollary~\ref{puretomix}.
\eex

\bex\label{pureenough}
Show that the corollary above still holds even if we replace $\gpc$ with $\cto$. Hint: Use the fact that $\psi$ can be converted to $\sigma$ by time-translation covariant channel, and then use Theorem~\ref{1321}.
\eex

\subsubsection{The Qubit Case}

In this subsection we use the considerations above to provide the analytical conditions for inter-conversions of athermality when the systems invloved are qubits.
Specifically, let $\rho,\sigma,\gamma\in\md(A)$ with $|A|=2$. Denote
\be\label{18p79}
\rho=\begin{pmatrix}
r & a \\
\bar{a} & 1-r
\end{pmatrix}
\quad\text{,}\quad
\sigma=\begin{pmatrix}
s & b \\
\bar{b} & 1-s
\end{pmatrix}
\quad\text{and}\quad
\gamma=\begin{pmatrix}
g & 0 \\
0 & 1-g
\end{pmatrix}\;.
\ee 
We also denote the diagonals of the matrices above by $\r\eqdef(r,1-r)^T$, $\s\eqdef(s,1-s)^T$ and $\g=(g,1-g)^T$, respectively. We would like to find the conditions under which $(\rho,\gamma)\xrightarrow{\gpc}(\sigma,\gamma)$. Recall that if $a=0$ then we must have $b=0$ since GPC cannot generate coherence between energy levels. Therefore, the case $a=0$ has already been covered by the quasi-classical regime. We will therefore assume in the rest of this subsection that $a\neq 0$.

\begin{myt}{}
\begin{theorem}\label{qubitthm}
Let $\rho,\sigma,\gamma\in\md(A)$ be three qubit states as above and suppose $a\neq 0$ and $\gamma\neq\u$. Then, for $\r\neq \g$, $(\rho,\gamma)\xrightarrow{\gpc}(\sigma,\gamma)$ if and only if $(\r,\g)\succ (\s,\g)$ and
\be\label{18p78}
\frac{|b|^2}{|a|^2}\leq\frac{s-g}{r-g}+g(1-g)\left(\frac{r-s}{r-g}\right)^2\;.
\ee
For $\r=\g$, $(\rho,\gamma)\xrightarrow{\gpc}(\sigma,\gamma)$ if and only if $\s=\g$ and $|a|\geq |b|$.
\end{theorem}
\end{myt}

\begin{proof}
From Exercise~\ref{lem1841} it follows that $(\rho,\gamma)$ can be converted to $(\sigma,\gamma)$ by GPC if and only if there exists a $2\times 2$ column stochastic matrix $P=\{p_{y|x}\}_{x,y\in\{0,1\}}$ that satisfies
$P\r=\s$, $P\g=\g$, and 
\be
\begin{pmatrix}
p_{0|0} & b/a\\
\bar{b}/\bar{a} & p_{1|1}
\end{pmatrix}\geq 0\;.
\ee 
Note that this last condition is equivalent to
\be\label{18p80}
\frac{|b|^2}{|a|^2}\leq p_{0|0}p_{1|1}\;.
\ee
The conditions $P\r=\s$ and $P\g=\g$ can be expressed as the following linear systems of equations
\be\label{linsys}
\begin{bmatrix}
r & 1-r \\
g & 1-g
\end{bmatrix}\begin{bmatrix}
p_{0|0}  \\
p_{0|1} 
\end{bmatrix}=\begin{bmatrix}
s  \\
g 
\end{bmatrix}
\quad\text{and}\quad
\begin{bmatrix}
r & 1-r \\
g & 1-g
\end{bmatrix}\begin{bmatrix}
p_{1|0}  \\
p_{1|1} 
\end{bmatrix}=\begin{bmatrix}
1-s  \\
1-g 
\end{bmatrix}\;.
\ee
Note that the equations involving $p_{1|0}$ and $p_{1|1}$ follows trivially from the ones involving $p_{0|0}$ and $p_{0|1}$ since $P$ is column stochastic.
From Cramer's rule it then follows that for the case that $r\neq g$
\be
p_{0|0}=\frac{\det\begin{pmatrix}
s & 1-r \\
g & 1-g
\end{pmatrix}}{\det\begin{pmatrix}
r & 1-r \\
g & 1-g
\end{pmatrix}}
\quad\text{and}\quad
p_{1|1}=\frac{\det\begin{pmatrix}
r & 1-s \\
g & 1-g
\end{pmatrix}}{\det\begin{pmatrix}
r & 1-r \\
g & 1-g
\end{pmatrix}}\;.
\ee
Finally, substituting the above expression in~\eqref{18p80} gives (after some simple algebra) the inequality~\eqref{18p78}. 

For the case that $r=g$ we also have $s=g$ (otherwise, $(\r,\g)\not\succ(\s,\g)$) and the linear system of equations in~\eqref{linsys} has a unique solution given by $p_{0|0}=p_{1|1}=1$. Therefore, in this case, \eqref{18p80} gives $|b|\leq |a|$. This completes the proof.
\end{proof}

\bex
Show that if $s=g$ in~\eqref{18p79} then $(\rho,\gamma)$ can be converted to $(\sigma,\gamma)$ by GPC if and only if
\be
\frac{|b|^2}{|a|^2}\leq\det(\gamma)\;.
\ee
\eex

From the exercise above it follows that already in the qubit case, conversions under GPC have a certain type of discontinuity. To see this, consider the case $s=g$, and observe that the condition $|a|^2\det(\gamma)\geq|b|^2$ is stronger than the condition $|a|\geq |b|$ that one obtains if also $r=g$. In particular, observe that $\det(\gamma)\leq\frac14$. Hence, there exists an $\eps>0$ and $\rho,\sigma,\gamma\in\md(A)$ such that for any $\rho\in\mb_{\eps}(\sigma)$ the state $(\rho,\gamma)$ cannot be converted by GPC to $(\sigma,\gamma)$ unless $\rho=\sigma$.

\bex
Find explicit example of three qubit states $\rho,\sigma,\gamma$, and $\eps>0$ such that for any $\rho\in\mb_{\eps}(\sigma)$, $(\rho,\gamma)\not\xrightarrow{\gpc}(\sigma,\gamma)$ unless $\rho=\sigma$.
\eex

\section{The Conversion Distance of Athermality}

We consider first the case that $\mf=\gp$, and define the conversion distance\index{conversion distance} as
\be\label{1624}
T\left((\rho,\gamma)\xrightarrow{\gp}(\rho',\gamma')\right)\eqdef \min_{\mE\in\cptp(A\to {A'})}\left\{\frac12\left\|\rho'-\mE(\rho)\right\|_1\;:\;\gamma'=\mE\left(\gamma\right)\right\}
\ee
Using the fact that the trace distance between two density matrices can be expressed as
\be
\frac12\left\|\rho'-\mE(\rho)\right\|_1=\min_{\substack{\Lambda\in\pos({A'})\\\Lambda\geq \rho'-\mE(\rho)}}\tr\left[\Lambda\right]\;,
\ee
we can express the conversion distance\index{conversion distance} as the following SDP:
\be
T\left((\rho,\gamma)\xrightarrow{\gp}(\rho',\gamma')\right)=\min \tr\left[\Lambda\right]
\ee
where the minimum is over all $\Lambda\in\pos(A')$ that satisfy the following conditions:
\begin{enumerate}
\item $\Lambda\geq \rho'-\tr_A\left[J^{A{A'}}\left(\rho^T\otimes I^{A'}\right)\right]$.
\item $\gamma'=\tr_A\left[J^{A{A'}}\left(\gamma^A\otimes I^{A'}\right)\right]$.
\item $J\in\pos(A{A'})$ and $J^A=I^A$.
\end{enumerate}
For the case that $\mf=\gpc$ the conversion distance is evaluated exactly as above with the additional constraint on $J^{AA'}$ that
\be
\mP_\xi\left(J^{AA'}\right)=J^{AA'}\;,\quad\text{where }\xi^{AA'}\eqdef H^{A}\otimes I^{A'}-I^{A}\otimes H^{A'}\;.
\ee
Note that this additional condition is still in a form suitable for SDP.

The preceding discussion demonstrates that the conversion distance of athermality can be computed numerically. However, the formulation presented above for the conversion distance lacks insight and does not offer any practical means to calculate the distillable athermality or the athermality cost of a state $(\rho,\gamma)$.
Therefore, we now turn our attention to the case where the target state is quasi-classical, and show that for this case there exists an analytical formula for the conversion distance.

\begin{myt}{}
\begin{theorem}
Let $(\rho,\gamma)$ be an arbitrary state of system $A$ and $\left(\rho',\gamma'\right)$ be a quasi-classical state of system $A'$. Then,
\be\label{17p122}
T\left((\rho,\gamma)\xrightarrow{\gpc}(\rho',\gamma')\right)=T\left((\mP(\rho),\gamma)\xrightarrow{\gpc}(\rho',\gamma')\right)\;,
\ee
where $\mP$ denotes the pinching channel\index{pinching channel} associated with the Hamiltonian of system $A$.
\end{theorem}
\end{myt}

\begin{remark}
Note that on the right-hand side, we have a conversion distance between two quasi-classical states. In the next subsection, we will demonstrate that for such cases, an analytical formula exists.
\end{remark}

\begin{proof}
Let $\mP$ and $\mP'$ be the pinching channels associated with the Hamiltonians of systems $A$ and $A'$, respectively, and observe that for any $\mE\in\cov(A\to A')$ that satisfies $\gamma'=\mE(\gamma)$ we have 
\ba\label{gppe}
\gamma'=\mP'(\gamma')&=\mP'\circ\mE(\gamma)\\\GG{Part~2 \;of\; Exercise~\ref{expart2}}&=\mE\circ\mP(\gamma)\;.
\ea
Therefore,
\ba\label{18p93b}
T\left((\mP(\rho),\gamma)\xrightarrow{\gpc}(\rho',\gamma')\right)&= \min_{\mE\in\cov(A\to A')}\left\{\frac12\left\|\rho'-\mE\circ\mP(\rho)\right\|_1\;:\;\gamma'=\mE\left(\gamma\right)\right\}\\
\GG{\eqref{gppe}}&=\min_{\mE\in\cov(A\to A')}\left\{\frac12\left\|\rho'-\mE\circ\mP(\rho)\right\|_1\;:\;\gamma'=\mE\circ\mP\left(\gamma\right)\right\}\\
\GG{\mN\eqdef\mE\circ\mP}&\geq\min_{\mN\in\cov(A\to A')}\left\{\frac12\left\|\rho'-\mN(\rho)\right\|_1\;:\;\gamma'=\mN\left(\gamma\right)\right\}\\
&=T\left((\rho,\gamma)\xrightarrow{\gpc}(\rho',\gamma')\right)\;.
\ea
For the converse inequality, observe that by using Part 2 of Exercise~\ref{expart2} we get that for every $\mE\in\cov(A\to A')$
\ba\label{dpiou}
\left\|\rho'-\mE\circ\mP(\rho)\right\|_1&=\left\|\rho'-\mP'\circ\mE(\rho)\right\|_1\\
\Gg{\mP'(\rho')=\rho'}&=\left\|\mP'\big(\rho'-\mE(\rho)\big)\right\|_1\\
\GG{DPI}&\leq \left\|\rho'-\mE(\rho)\right\|_1\;.
\ea
Combining this inequality with the definition of the conversion distance, specifically with the first equality in~\eqref{18p93b}, gives
\ba\label{18p93a}
T\left((\mP(\rho),\gamma)\xrightarrow{\gpc}(\rho',\gamma')\right)&\leq\min_{\mE\in\cov(A\to A')}\left\{\frac12\left\|\rho'-\mE(\rho)\right\|_1\;:\;\gamma'=\mE\left(\gamma\right)\right\}\\
&=T\left((\rho,\gamma)\xrightarrow{\gpc}(\rho',\gamma')\right)\;.
\ea
Combining the two inequalities in~\eqref{18p93b} and~\eqref{18p93b} gives the equality in~\eqref{17p122}. This completes the proof.
\end{proof}

\bex
Use Lemma~\ref{lemmonin} to provide a shorter proof of the inequality in~\eqref{18p93b}.
\eex

\bex\label{hamilpinch}
Let $\mP$ and $\mP'$ be the pinching channel associated with the Hamiltonians of systems $A$ and $A'$, respectively, and let $\mN\eqdef\mP'\circ\mE$, where
$\mE\in\cptp(A\to A')$. Show that $\mN\in\cov(A\to A')$ if and only if
\be
\mN=\mN\circ\mP\;.
\ee 
\eex

\subsection{The Conversion Distance Between Quasi-Classical States}\label{sec:cdqc}\index{quasi-classical}

Consider two quasi-classical athermality states  $(\p,\g)$ of system $A$ and $(\p',\g')$ of system $A'$. Since applying a permutation on both components of $\p$ and $\g$ is a reversible thermal operation, without loss of generality we assume that the components of the probability vectors are ordered as
\be\label{dorderi}
\frac{p_1}{g_1}\geq\frac{p_2}{g_2}\geq\cdots\geq\frac{p_m}{g_m}\quad\text{and}\quad
\frac{p_1'}{g_1'}\geq\frac{p_2'}{g_2'}\geq\cdots\geq\frac{p_n'}{g_n'}\;.
\ee
where $m\eqdef|A|$ and $n\eqdef|A'|$. Suppose first that the 
Gibbs states $\g$ and $\g'$ have rational coefficients, and denote by
$k\in\mbb{N}$ the common denominator of all the components of $\g$ and $\g'$. That is, for each $x\in[m]$ and $y\in[n]$, we have $g_x=\frac{a_x}{k}$ and $g'_y=\frac{b_y}{k}$, where $a_x,b_y\in\mbb{N}$ are integers satisfying 
\be
\sum_{x\in[m]}a_x=\sum_{y\in[n]}b_y=k.
\ee
Recall that from Theorem~\ref{onlyr}, there exists $\r,\s\in\prob(k)$ such that $(\p,\g)\sim(\r,\u^{(k)})$ and $(\p',\g')\sim(\r',\u^{(k)})$. Specifically, 
\be\label{rsk}
\r\eqdef\bigoplus_{x\in[m]}p_x\u^{(a_x)}\;\;\text{and}\;\;\r'\eqdef\bigoplus_{y\in[n]}p'_y\u^{(b_y)}\;.
\ee
From~\eqref{dorderi} we have $\r=\r^\da$ and $\r'=\r'^\da$, and moreover,
\be
(\p,\g)\succ(\p',\g')\iff \r\succ\r'\;.
\ee
With these notations and the assumption that the Gibbs vectors have rational components, we have the following closed formula for the conversion distance.

\begin{myt}{}
\begin{theorem}\label{qclaq}
 Let $(\p,\g)$, $(\p',\g')$, and $\r,\r'\in\prob(k)$ be as above. Then,
\be\label{exconv}
T\left((\p,\g)\xrightarrow{\cto}(\p',\g')\right)=\max_{\ell\in[k]}\big\{\|\r'\|_{(\ell)}-\|\r\|_{(\ell)}\big\}\;.
\ee
\end{theorem}
\end{myt}
\begin{proof}
Let $E$ be a $k\times n$ column stochastic matrix defined on every $\q\in\prob(n)$ 
as (cf.~\eqref{cf149})
\be\label{rssp}
E\q\eqdef\bigoplus_{y\in[n]}q_y\u^{(b_y)}\;.
\ee
Observe that $\r'=E\p'$ and that $\|\p'-\q\|_1=\|\r'-E\q\|_1$ (see Exercise~\ref{showsq}). Thus, 
\ba\label{139rp17}
T\left((\p,\g)\xrightarrow{\cto}(\p',\g')\right)&=\min_{\q\in\prob(n)}\left\{\frac12\left\|\r'-E\q\right\|_1\;:\;\r\succ E\q\right\}\\
\Gg{\s\eqdef E\q}&\geq \min_{\s\in\prob(k)}\left\{\frac12\left\|\r'-\s\right\|_1\;:\;\r\succ \s\right\}\\
\Gg{cf.~\eqref{offk}}&\eqdef T\left(\r\xrightarrow{\noisy}\r'\right)\;.
\ea
We next show that the above inequality is in fact an equality.

Let 
\be
D\eqdef\bigoplus_{y\in[n]}D^{(b_y)}\quad\text{with}\quad D^{(b_y)}\eqdef\frac1{b_y}\begin{bmatrix} 1 & \cdots & 1\\
\vdots & \ddots & \vdots\\
1 & \cdots & 1
\end{bmatrix}\;,
\ee
and observe that $D$ is a $k\times k$ doubly stochastic matrix\index{doubly stochastic matrix} satisfying $D\r'=\r'$. Thus, using the data processing inequality with the matrix $D$ in the definition of $T\left(\r\xrightarrow{\noisy}\r'\right)$ above gives
\ba
T\left(\r\xrightarrow{\noisy}\r'\right)&\geq \min_{\s\in\prob(k)}\left\{\frac12\left\|D\r'-D\s\right\|_1\;:\;\r\succ \s\right\}\\
\Gg{D\r'=\r'\text{ and }\s\succ D\s}&\geq\min_{\s\in\prob(k)}\left\{\frac12\left\|\r'-D\s\right\|_1\;:\;\r\succ D\s\right\}
\ea
since $\r\succ D\s$ is a weaker constraint than $\r\succ \s$. By definition $D\s$ has the form $E\q$ for some $\q\in\prob(n)$ (see Exercise~\ref{showsq}). Therefore,
\be 
T\left(\r\xrightarrow{\noisy}\r'\right)\geq \min_{\q\in\prob(n)}\left\{\frac12\left\|\r'-E\q\right\|_1\;:\;\r\succ E\q\right\}=
T\left((\p,\g)\xrightarrow{\cto}(\p',\g')\right) \;.
\ee
Combining this with~\eqref{139rp17} we conclude
\ba\label{1560}
T\left((\p,\g)\xrightarrow{\cto}(\p',\g')\right)&= T\left(\r\xrightarrow{\noisy}\r'\right)\\
\GG{Theorem~\ref{thm:1632}}&=\max_{\ell\in[k]}\big\{\|\r'\|_{(\ell)}-\|\r\|_{(\ell)}\big\}\;.
\ea
This completes the proof.
\end{proof}

\bex\label{showsq}
Using the same symbols as in the proof above, show that $\|\p'-\q\|_1=\|\r'-E\q\|_1$, and that for every $\s\in\prob(k)$ there exists $\q\in\prob(n)$ such that $D\s=E\q$.
\eex

In the theorem above we assumed that the Gibbs states $\g$ and $\g'$ have rational components. In Appendix~\ref{continuity} we show that the conversion distance\index{conversion distance} is continuous in $\g$ and $\g'$. This in turns implies that one can use the theorem above to estimate the conversion distance up to an arbitrary precision even for the case that $\g$ and $\g'$ have irrational components.

\subsection{The Golden Unit of Athermality}

We start by discussing the golden unit\index{golden unit} of athermality in the quasi-classical regime, and use the notation $\{\e_1,\ldots,\e_m\}$ for the standard basis of $\mbb{R}^m$.
From Exercise~\ref{themaj}, the maximal resource of a system $A$ of dimension $|A|=m$, with a fixed Hamiltonian $H^A\eqdef\sum_{x\in[m]}a_x^\ua|x\lr x|^A$ (or equivalently with a fixed Gibbs state $\g$) is given by $(\e_m,\g)$, where $\e_m$ corresponds to the maximal eigenvalue $a_m$ of $H^A$. In general, we cannot call $(\e_m,\g)$ the ``golden unit" of system $A$ since it depends on the Hamiltonian $H^A$. That is, without specifying the Hamiltonian of system $A$, and as long as the maximal energy $a_m<\infty$, we cannot specify a resource that is maximal on all systems with the same dimension $|A|=m$.    

On the other hand, if we do take $a_m=\infty$, so that the Gibbs state has a zero $m$-component, then the resource $(\e_m,\g)$ is an infinite resource in the sense that all other systems $A'$ in \emph{any} dimension and any quasi-classical athermality state $(\p',\g')$ will satisfy $(\e_m,\g)\succ(\p',\g')$. Hence, such a system cannot serve as the golden unit since it is an infinite resource. In other words, even in dimension $m=2$, by maximizing over all two dimensional Hamiltonians, we get an infinite resource, by taking $a_2=\infty$ so that $\g^A=\e_1$ and the thermal state is the infinite resource $(\e_2,\e_1)$.

We therefore choose the golden unit for a fixed dimension $|A|=m$, to be of the form $(\e_m,\g)$, but instead of choosing the Hamiltonian that maximizes  the resource in that dimension we chooses the one that minimizes it. That is, we are looking for a Gibbs state $\g$ that satisfies
\be\label{17p103}
(|m\lr m|,\tg)\succ(|m\lr m|,\g)\quad\quad\forall\;\tg\in\prob(m)\;.
\ee
The only vector $\g$ that satisfies the above relation is the uniform vector $\u^{(m)}$. Hence, the golden unit\index{golden unit} for any system $A$ will be chosen as
\be
(\e_m,\u^{(m)})\sim(|0\lr 0|^A,\u^A)\;,
\ee
where on the right-hand side we used the quantum notations, with $|0\ra$ denoting any pure state of system $A$. Recall that when the Gibbs state is uniform all pure states are equivalent under thermal operations (which are equivalent to noisy operations in this case).

\bex
Show that a vector $\g\in\prob(m)$ satisfies~\eqref{17p103} for all $\tg\in\prob(m)$ if and only if $\g=\u^{(m)}$.
\eex

In the fully quantum case, under GPC and CTO, coherence among energy level is a resource that cannot be measured by the golden unit $(|0\lr 0|^A,\u^A)$. The reason is that this golden unit is quasi-classical, and it cannot be converted by GPC (or CTO) to any athermality state that is not quasi-classical (even if we take $m\eqdef |A|=\infty$). This means that in the QRT of quantum athermality, there exists another type of resource, namely, time-translation asymmetry, that can not be quantified by the golden unit\index{golden unit} $(|0\lr 0|^A,\u^A)$. We conclude that quantum athermality can be viewed as a resource comprising of two types:
\ben
\item Nonuniformity (since in the quasi-classical regime athermality can be viewed as nonuniformity)
\item Time-translation asymmetry (also referred as quantum coherence).
\een

In contrast to GPC and CTO, GPO has the capability to induce coherence between energy levels. Consequently, as demonstrated in the subsequent exercise, we can retain the state $(|0\lr 0|^A,\u^A)$ as the golden unit\index{golden unit} of the resource theory.
\bex
Let $m\eqdef|A|$ and $(\rho',\gamma')$ be an athermality state of system $A'$. Show that for sufficiently large $m$
\be
(|0\lr 0|^A,\u^{A})\xrightarrow{\gp}(\rho',\gamma')\;.
\ee
\eex

\begin{exercise}
Show that under $\gp$ operations, the resource $(|0\lr 0|^A,\u^A)$ is equivalent to
the resource $\left(|0\lr 0|^X,\u^X_m\right)$, where $X$ is a two-dimensional classical system, $m\eqdef|A|$, and
\be
\u_m^X\eqdef \frac1m|0\lr 0|^X+\frac{m-1}{m}|1\lr 1|^X\;.
\ee
\end{exercise}

The exercise above demonstrates that we can always consider the golden unit to be a qubit. Moreover, note that $\u_m^X$ is well defined even if $m$ is not an integer. 
This can help simplifying certain expressions, and we will therefore consider also the states $\left(|0\lr 0|^X,\u^X_m\right)$ with $m\in\mbb{R}_+$. We will use the notation
\be
\Upsilon_m\eqdef\left(|0\lr 0|^X,\u^X_m\right)
\ee
to denote this golden unit\index{golden unit}.

\bex
Show that $\{\Upsilon_m\}_{m\in\mbb{N}}$ satisfies the conditions of a golden unit outlined in~Definition~\ref{def:gu}.
\eex

\subsection{The Conversion Distance to and from the Golden Unit}

The conversion distance\index{conversion distance} (under GPO) from an arbitrary state $(\rho,\gamma)$ of system $A$ to the golden unit $\Upsilon_m$ is given by
\be
T\left((\rho,\gamma)\xrightarrow{\gp}\Upsilon_m\right)= \min_{\mE\in\cptp(A\to X)}\left\{\frac12\left\||0\lr 0|^X-\mE(\rho^A)\right\|_1\;:\;\u^X_m=\mE\left(\gamma^A\right)\right\}
\ee
Observe that since any $\mE\in\cptp(A\to X)$ is a binary POVM Channel\index{POVM channel} , it can be expressed as
\be
\mE(\omega)=\tr\big[\omega\Lambda\big]|0\lr 0|^X+\tr\big[\omega\left(I-\Lambda\right)\big]|1\lr 1|^X\quad\quad\forall\;\omega\in\ml(A)\;,
\ee
for some effect $0\leq \Lambda\leq I^A$. We therefore get the simplification (see Exercise~\ref{00x})
\be\label{00xmx}
\frac12\left\||0\lr 0|^X-\mE(\rho^A)\right\|_1=1-\tr[\rho\Lambda]
\ee
so that
\be\label{18p55}
T\left((\rho,\gamma)\xrightarrow{\gp}\Upsilon_m\right)= \min_{\Lambda\in\eff(A)}\left\{1-\tr[\rho\Lambda]\;:\;\tr[\Lambda\gamma]=\frac1m\right\}
\ee
Note that the expression above is somewhat similar to the hypothesis testing divergence (see Exercise~\ref{simihtd}). Moreover, since the golden unit $\Upsilon_m$ is quasi-classical we get from~\eqref{17p122} that under $\gpc$ we have
\be\label{18p55i}
T\left((\rho,\gamma)\xrightarrow{\gpc}\Upsilon_m\right)= \min_{\Lambda\in\eff(A)}\left\{1-\tr[\mP(\rho)\Lambda]\;:\;\tr[\Lambda\gamma]=\frac1m\right\}
\ee
where $\mP$ is the pinching channel\index{pinching channel} associated with the Hamiltonian of system $A$.

\bex\label{00x}
Prove the equality~\eqref{00xmx}.
\eex

\bex\label{simihtd}
Set $\eps\eqdef1-\frac1m$. Show that
\be
T\left((\rho,\gamma)\xrightarrow{\gp}\Upsilon_m\right)=2^{D_{\min}^\eps(\gamma\|\rho)}\;.
\ee
\eex

We next consider the conversion distance from the golden unit $\Upsilon_m$ to an arbitrary state $(\rho,\gamma)$ of system $A$. Here we only consider GPO since GPC cannot generate coherence. By definition, 
\be
T\left(\Upsilon_m\xrightarrow{\gp}(\rho,\gamma)\right)= \min_{\mE\in\cptp(X\to A)}\left\{\frac12\left\|\rho^A-\mE(|0\lr 0|^X)\right\|_1\;:\;\gamma^A=\mE\left(\u^X_m\right)\right\}
\ee
Denoting by $\omega=\mE(|0\lr 0|)$ and $\tau=\mE(|1\lr 1|)$, the conversion distance can be simplified as
\ba\label{cd1859}
T\left(\Upsilon_m\xrightarrow{\gp}(\rho,\gamma)\right)&= \min_{\omega,\tau\in\md(A)}\left\{\frac12\left\|\rho-\omega\right\|_1\;:\;\gamma=\frac1m\omega+\frac{m-1}m\tau\right\}\\
&=\min_{\omega\in\md(A)}\left\{\frac12\left\|\rho-\omega\right\|_1\;:\;m\gamma\geq\omega\right\}\\
&=\min_{\omega\in\md(A)}\left\{\frac12\left\|\rho-\omega\right\|_1\;:\;D_{\max}(\omega\|\gamma)\leq\log m\right\}\;.
\ea
This expression will be instrumental in our calculations regarding the cost of athermality under  GPO.

\bex
Let $(\rho,\gamma)$ be an athermality state of system $A$. Show that under GPC for any $m\in\mbb{N}$
\be
T\left(\Upsilon_m\xrightarrow{\gpc}(\rho,\gamma)\right)\geq\min_{\sigma\in\md(A)}\frac12\left\|\rho-\Delta(\sigma)\right\|_1
\ee
where $\Delta\in\cptp(A\to A)$ is the completely dephasing channel (with respect to the basis  of the Hamiltonian of system $A$).
\eex

\section{Distillation and Cost in the Single-Shot Regime}\index{single-shot}

\subsection{Distillation of Athermality}

As discussed earlier, for any $\eps\in[0,1]$ the $\eps$-approximate single-shot distillation of an athermality state $(\rho,\gamma)$ of system $A$ is defined by
\be\label{18p54}
\distill^{\eps}\left(\rho,\gamma\right)\eqdef\log\sup_{0<m\in\mbb{R}}\left\{m\;:\;T\left((\rho,\gamma)\xrightarrow{\gp}\Upsilon_m\right)\leq\eps\right\}\;.
\ee
Integrating this with the formulas from the preceding subsection that pertain to the conversion distance, we arrive at the subsequent outcome. We denote by $\mP\in\cptp(A\to A)$ the pinching channel associated with the Hamiltonian of system $A$, and by $D_{\min}^\eps$  the quantum hypothesis testing divergence as defined in~\eqref{dehl}.

\begin{myt}{}
\begin{theorem}\label{dmindistill}
Let $\eps\in[0,1]$. For any athermality state $(\rho,\gamma)$ of a quantum system $A$, the $\eps$-approximate single-shot distillation of athermality is given by:
\ben
\item Under GPO:
$
\distill^{\eps}\left(\rho,\gamma\right)=D_{\min}^\eps\left(\rho\|\gamma\right)
$.
\item Under GPC and CTO: $
\distill^{\eps}\left(\rho,\gamma\right)=D_{\min}^\eps\left(\mP(\rho)\|\gamma\right)\;.
$
\een
\end{theorem}
\end{myt}
\begin{proof}
From~\eqref{18p54} we get 
\ba
\distill^{\eps}\left(\rho,\gamma\right)&
=-\log\inf_{0<m\in\mbb{R}}\left\{\frac1m\;:\;T\left((\rho,\gamma)\xrightarrow{\gp}\Upsilon_m\right)\leq\eps\right\}\\
\GG{\eqref{18p55}}&=-\log\inf_{0<m\in\mbb{R}}\left\{\frac1m\;:\;1-\tr[\rho\Lambda]\leq\eps\;,\;\;\tr[\Lambda\gamma]=\frac1m\;,\;\;\Lambda\in\eff(A)\right\}\\
&=-\log\inf\Big\{\tr[\Lambda\gamma]\;:\;1-\tr[\rho\Lambda]\leq\eps\;,\;\;\Lambda\in\eff(A)\Big\}\\
&=D_{\min}^\eps(\rho\|\gamma)\;.
\ea
This completes the proof of the first part. The second part of the proof follows from the first part in conjunction with~\eqref{17p122}. This concludes the proof.
\end{proof}

Observe that when we take $\eps=0$ we get that the exact single-shot distillation is given by
\be
\distill^{0}\left(\rho^A,\gamma^A\right)=D_{\min}\left(\rho^A\big\|\gamma^A\right)\;,
\ee
This result give a physical meaning to the min relative entropy\index{min relative entropy} as the exact single-shot distillation rate under GPO.

\subsection{Athermality Cost Under $\gp$}

For any $\eps\in[0,1]$, 
the $\eps$-single-shot cost of an athermality state $(\rho,\gamma)$ is defined as
\be\label{cost1857}
\cost^{\eps}\left(\rho,\gamma\right)\eqdef\log\inf_{0<m\in\mbb{R}}\Big\{m\;:\;T\left(\Upsilon_m\xrightarrow{\gp}(\rho,\gamma)\right)\leq\eps\Big\}\;.
\ee

\begin{myt}{}
\begin{theorem}\label{costatq}
Let $\eps\in[0,1]$. For any athermality state $(\rho,\gamma)$ of system $A$, the $\eps$-single-shot distillation (under $\gp$) is given by
\be
\cost^{\eps}\left(\rho,\gamma\right)=D_{\max}^\eps(\rho\|\gamma)\;.
\ee
\end{theorem}
\end{myt}

\begin{proof}
Combining the expression~\eqref{cd1859} for the conversion distance together with the definition~\eqref{cost1857} gives
\ba
\cost^{\eps}\left(\rho,\gamma\right)&=\inf_{0<m\in\mbb{R}}\Big\{\log m\;:\;\frac12\|\rho-\omega\|_1\leq\eps\;,\;\;D_{\max}(\omega\|\gamma)\leq\log m\;,\;\;\omega\in\md(A)\Big\}\\
&=\inf\Big\{D_{\max}(\omega\|\gamma)\;:\;\frac12\|\rho-\omega\|_1\leq\eps\;,\;\;\omega\in\md(A)\Big\}\\
&=D_{\max}^\eps\left(\rho\|\gamma\right)\;.
\ea
This completes the proof.
\end{proof}

Observe that for $\eps=0$ we get the exact single-shot athermality cost 
\be
\cost^{0}\left(\rho,\gamma\right)=D_{\max}\left(\rho\|\gamma\right)\;.
\ee
This result provides a physical meaning to the max relative entropy\index{max relative entropy} as the exact single-shot cost under GPO.

\bex
Let $\gamma\in\md_{>0}(A)$ be the Gibbs state of system $A$ with eigenvalues $g_1,\ldots,g_m$. Let $\psi_\gamma\in\pure(A)$ be the pure state
\be
|\psi_\gamma\ra\eqdef\sum_{x\in[m]}\sqrt{g_x}|x\ra\;.
\ee
Show that the exact single-shot athermality cost of $(\psi_\gamma,\gamma)$ is equal to $\log(m)$.
\eex

\section{The Asymptotic Regime}

A primary goal of resource theories is to attain reversibility in the asymptotic inter-conversions of resources. This entails that the cost-rate, in an asymptotic context, for generating a specific  resource should align with the rate at which golden units can be extracted from it. Reversibility characteristics hold significant importance in the realm of quantum information, given the value of quantum resources. They guarantee that resources are not wasted  during quantum information processing tasks. Nonetheless, the pursuit of reversibility frequently necessitates the contemplation of a broader array of permissible operations.\index{reversibility}

Unlike GPO, both thermal operations and GPC are incapable of generating coherence between energy levels. This means that even for very large $m$, the golden unit $\Upsilon_m$ cannot be converted into a single copy of an athermality state $(\rho, \gamma)$ that exhibits coherence across energy levels. Nevertheless, as we will soon discover, this irreversibility --- highlighted by the significant cost of preparing the $(\rho, \gamma)$ state in contrast to the finite rate at which it can be utilized to distill golden units of athermality --- can be mitigated by introducing a modest degree of coherence into the system.

In certain cases, reversibility can be attained by allowing the use of a sublinear amount of resources. For instance, in the resource theory of pure bipartite entanglement, we observed that distillation requires no communication, whereas formation necessitates a sublinear amount of classical communication. Thus, reversibility is achieved in this theory through local operations and a sublinear amount of classical communication. The concept of adding a sublinear amount of a specific resource to achieve reversibility is highly appealing because the rate at which such resources are consumed diminishes in the asymptotic limit. We will employ this idea when studying the asymptotic cost of athermality under thermal operations. However, we begin by examining the asymptotic distillation of athermality.

\subsection{Distillation of Athermality}

In this section we compute the asymptotic distillable athermality under either $\gp$, $\gpc$, or $\cto$. The asymptotic distillable rate of an athermality state $(\rho,\gamma)$ is related to the single-shot quantity via (cf.~\eqref{disteq})
\be
\distill\left(\rho,\gamma\right)=\lim_{\eps\to 0+}\limsup_{n\to\infty}\frac1n\distill^{\eps}\left(\rho^{\otimes n},\gamma^{\otimes n}\right)\;.
\ee 
Recall from Theorem~\ref{dmindistill} that in the single-shot regime, for any $\eps\in(0,1)$, the distillable athermality under GPO is given by
\be
\distill^{\eps}\left(\rho,\gamma\right)=D_{\min}^\eps\left(\rho\big\|\gamma\right)\;.
\ee
The regularization\index{regularization} of the formula above is given by
\ba
\lim_{n\to\infty}\frac1n\distill^{\eps}\left(\rho^{\otimes n},\gamma^{\otimes n}\right)&=\lim_{n\to\infty}\frac1nD_{\min}^\eps\left(\rho^{\otimes n}\big\|\gamma^{\otimes n}\right)\\
\GG{\small The\;Quantum\;Stein's\;Lemma}&=D(\rho\|\gamma)\;.
\ea
Note that in this case we did not need to take the limsup over $n$ since the limit exists.
Therefore, under GPO, the asymptotic distillable athermality is given by the relative entropy $D(\rho\|\gamma)$. Remarkably, this is also the distillable rate under GPC and CTO.

\begin{myt}{}
\begin{theorem}\label{droat}
Let $(\rho,\gamma)$ be an athermality state of a quantum system $A$, and let $\eps\in(0,1)$. Then, the distillable athermality under either CTO or GPC is given by
\be
\distill\left(\rho,\gamma\right)=\limsup_{n\to\infty}\frac1n\distill^{\eps}\left(\rho^{\otimes n},\gamma^{\otimes n}\right)=D\left(\rho\|\gamma\right)\;.
\ee
\end{theorem}
\end{myt}

\begin{proof}
Let $\eps\in(0,1)$ and recall from Theorem~\ref{dmindistill} that the $\eps$-single-shot distillable athermality under GPC or CTO is given by
\be
\distill^{\eps}\left(\rho,\gamma\right)=D_{\min}^\eps\left(\mP(\rho)\big\|\gamma\right)\;,
\ee
where $\mP$ is the pinching channel corresponding to the Hamiltonian of system $A$.
Since $\mP(\gamma)=\gamma$ we have
\ba
\distill^{\eps}\left(\rho,\gamma\right)&=D_{\min}^\eps\left(\mP(\rho)\big\|\mP(\gamma)\right)\\
\GG{DPI}&\leq D_{\min}^\eps\left(\rho\|\gamma\right)\;.
\ea
Thus,
\ba
\limsup_{n\to\infty}\frac1n\distill^{\eps}\left(\rho^{\otimes n},\gamma^{\otimes n}\right)&\leq\limsup_{n\to\infty}\frac1nD_{\min}^\eps\left(\rho^{\otimes n}\big\|\gamma^{\otimes n}\right)\\
\GG{\small The\;Quantum\;Stein's\;Lemma}&=D(\rho\|\gamma)\;.
\ea
To get the opposite inequality, for every $n\in\mbb{N}$ let $\mP_n\in\cto(A^n\to A^n)$ denotes the pinching channel associated with the Hamiltonian of system $A^n$. Now, fix $k\in\mbb{N}$ and observe that for every $\eps\in(0,1)$
\ba
\limsup_{n\to\infty}\frac1n\distill^{\eps}\left(\rho^{\otimes n},\gamma^{\otimes n}\right)&=\limsup_{n\to\infty}\frac1nD_{\min}^\eps\left(\mP_n(\rho^{\otimes n})\big\|\gamma^{\otimes n}\right)\\
&\geq
\limsup_{n\to\infty}\frac1{nk}D_{\min}^\eps\left(\mP_{nk}(\rho^{\otimes nk})\big\|\gamma^{\otimes nk}\right)\\
\GG{DPI}&\geq \limsup_{n\to\infty}\frac1{nk}D_{\min}^\eps\left(\mP_k^{\otimes n}\circ\mP_{nk}(\rho^{\otimes nk})\big\|\mP_k^{\otimes n}\left(\gamma^{\otimes nk}\right)\right)\;,
\ea
where in the last line we used the data processing inequality with the channel $\mP_k^{\otimes n}$. Now, the Gibbs state is invariant under the pinching channel\index{pinching channel} and in particular $\mP_k^{\otimes n}\left(\gamma^{\otimes nk}\right)=\gamma^{\otimes nk}$. Moreover,
from Exercise~\ref{etr} it follows that $\mP_k^{\otimes n}\circ\mP_{nk}=\mP_k^{\otimes n}$. We therefore get that
\ba
\limsup_{n\to\infty}\frac1n\distill^{\eps}\left(\rho^{\otimes n},\gamma^{\otimes n}\right)&\geq\limsup_{n\to\infty}\frac1{nk}D_{\min}^\eps\left(\mP_k^{\otimes n}(\rho^{\otimes nk})\big\|\gamma^{\otimes nk}\right)\\
&=\frac1k\limsup_{n\to\infty}\frac1{n}D_{\min}^\eps\left(\left(\mP_k(\rho^{\otimes k})\right)^{\otimes n}\big\|\left(\gamma^{\otimes k}\right)^{\otimes n}\right)\\
&=\frac1kD\left(\mP_k\left(\rho^{\otimes k}\right)\big\|\gamma^{\otimes k}\right)\;,
\ea
where in the last line we used the quantum Stein's lemma\index{Stein's lemma}.
The above inequality can also be understood physically by observing that the state
$\sigma_k\eqdef\mP_k\left(\rho^{\otimes k}\right)$ is quasi-classical, and consequently, it has a distillable athermality rate given by $D(\sigma_k\|\gamma^{\otimes k})$. Now, since the above inequality holds for all $k\in\mbb{N}$ we conclude that
\ba
\limsup_{n\to\infty}\frac1n\distill^{\eps}\left(\rho^{\otimes n},\gamma^{\otimes n}\right)&\geq\limsup_{k\to\infty}\frac1kD\left(\mP_k\left(\rho^{\otimes k}\right)\big\|\gamma^{\otimes k}\right)\\
\GG{\eqref{18p27}}&=D(\rho\|\gamma)\;.
\ea
This completes the proof.
\end{proof}

\subsection{Athermality Cost}

We begin our discussion by examining the cost of athermality under GPO. In this context, the asymptotic cost rate of an athermality state $(\rho,\gamma)$ connects to the single-shot quantity as follows:
\be
\cost\left(\rho,\gamma\right) = \lim_{\eps\to 0+}\liminf_{n\to\infty}\frac{1}{n}\cost^{\eps}\left(\rho^{\otimes n},\gamma^{\otimes n}\right)\;.
\ee
Integrating this with Theorem~\ref{costatq}, we arrive at the equation:
\ba
\lim_{n\to\infty}\frac{1}{n}\cost^{\eps}\left(\rho^{\otimes n},\gamma^{\otimes n}\right) &= \lim_{n\to\infty}\frac{1}{n}D_{\max}^\eps\left(\rho^{\otimes n}\big\|\gamma^{\otimes n}\right)\\
\GG{AEP} &= D(\rho\|\gamma)\;.
\ea
Hence, under GPO, both the asymptotic cost and the distillable athermality are given by the relative entropy $D(\rho\|\gamma)$. This signifies that, within the framework of GPO, the QRT of athermality exhibits reversibility. We will now proceed to explore the athermality cost under CTO.\index{reversibility}

As discussed above, the golden unit $\Upsilon_m$ cannot be used to generate states with coherence among energy levels. Therefore, the QRT of athermality under CTO is irreversible. In this section we show how reversibility can be restored by appending the free operations with resources that are asymptotically negligible. To see how it is done, we first need to introduce a few concepts.

\subsubsection{Scaling of Time-Translation Asymmetry}

Let $A$ be a physical system with Hamiltonian $H^A=\sum_{x\in[m]}a_x|x\lr x|$, where $m=|A|$, and
let $|\psi\ra=\sum_{x\in[m]}\sqrt{p_x}|x\ra$  be given in its standard form\index{standard form}. 
For any $n\in\mbb{N}$, the state $\psi^{\otimes n}$ has the form 
\ba
|\psi^{\otimes n}\ra&=\sum_{x^n\in[m]^n}\sqrt{p_{x^n}}|x^n\ra=\sum_{x^n\in[m]^n}2^{-\frac n2\big(H(\t(x^n))+D\left(\t(x^n)\|\p\right)\big)}|x^n\ra
\ea
where we used~\eqref{tpxn}. For any type $\t\in\type(n,m)$ define
\be
|\t\ra^{A^n}\eqdef \frac{1}{{n\choose nt_1,\ldots,nt_m}^{1/2}}\sum_{x^n\in X^n(\t)}|x^n\ra\;,
\ee
where the sum runs over all sequences $x^n\in[m]^n$ of the same type $\t$.
With the above notations
\be\label{psi15}
|\psi\ra^{\otimes n}=\sum_{\t\in\type(n,m)}\sqrt{q_{\t,n}}|\t\ra^{A^n}
\ee
where
\be
q_{\t,n}\eqdef {n\choose nt_1,\ldots,nt_m}2^{-n\big(H(\t)+D\left(\t\|\p\right)\big)}\;.
\ee
Note that the vectors $|\t\ra^{A^n}$ are eigenvectors of the Hamiltonian of system $A^n$. Specifically,
\be\label{15p106}
H^{A^n}|\t\ra^{A^n}=n\sum_{x\in[m]}t_xa_x|\t\ra^{A^n}\;,
\ee
so that the energy in the state $|\t\ra^{A^n}$ is $n$ times the average energy with respect to the type $\t$. 

\bex
Consider the generic case, in which the energy eigenvalues $\{a_1,\ldots,a_m\}$ are rationally independent; i.e. for any set of $m$ integers $\ell_1,\ldots,\ell_m\in\mbb{Z}$ we have 
\be
\ell_1a_1+\cdots+\ell_ma_m=0\quad\iff \quad\ell_1=\ell_2=\cdots=\ell_m=0\;.
\ee
Show that under this mild assumption (which we will \emph{not} assume in the text), for every $n\in\mbb{N}$, the number of distinct eigenvalues of $H^{A^n}$ equals $|\type(n,m)|$. That is, each energy eigenvalue of $H^{A^n}$ corresponds to exactly one type.
\eex 

Given that each $|\t\ra^{A^n}$ is an energy eigenstate, it naturally follows from~\eqref{psi15} that we can express $|\psi^{\otimes n}\ra$ as a linear combination of up to $|\type(n,m)|\leq (n+1)^m$ energy eigenstates. In simpler terms, the coherence inherent in $|\psi^{\otimes n}\ra$ can be compactly represented within an $(n+1)^m$ dimensional vector (dimension polynomial in $n$).

This observation leads to a notable implication. As established in Corollary~\ref{puretomixgpc}, for any mixed state in $\md(A)$ there exists a  pure state in $\pure(A)$ that can be converted into it via GPC. When we couple this insight with the aforementioned observation, a significant deduction emerges: the pure state coherence cost for preparing $\rho^{\otimes n}\in\md(A^n)$ must not surpass $m\log(n+1)$. To put it differently, the rate of asymmetry cost – the coherence expense per instance of $\rho$ – cannot outpace $m\frac{\log(n+1)}{n}$, a ratio that approaches zero in the limit as $n\to\infty$.
In contrast, the non-uniformity cost does \emph{not} go to zero in the asymptotic limit since the energy of $\rho^{\otimes n}$ grows linearly with $n$.

In summary, athermality is made up of two main resources: nonuniformity and time-translation asymmetry, the latter of which is often referred to as coherence. Because of this, the costs related to athermality states can be categorized into two parts: the cost of nonuniformity and the cost of coherence. However, the coherence cost decreases and approaches zero in the asymptotic limit, necessitating a unique form of rescaling. This complexity lends a subtle character to the resource theory of quantum athermality, leaving several critical questions within the theory still unresolved.

\subsubsection{The Energy Spread}\index{energy spread}

The energy spread of a given pure state $\psi\in\pure(A)$ is defined as the difference between the maximal and minimal energies that appear when writing $\psi$ as a superposition of energy eigenvectors.
In the discussion above we saw that $n$ copies of a state $\psi\in\pure(A)$ can be expressed as a linear combination of no more that $(n+1)^m$ energy eigenvectors. Among these energy eigenvectors are 
the zero energy eigenvector (corresponding to the type $\t=(1,0,\ldots,0)^T$) and the maximal energy eigenvector (corresponding to the type $\t=(0,\ldots,0,1)^T$). Therefore, since the energy in the decomposition~\eqref{psi15} spreads from zero to $na_m$ (where $a_m$ is the maximal energy of a single copy of system $A$), we conclude that the energy spread of $\psi^{\otimes n}$ is $na_m$.

The energy spread can be reduced significantly if one allows for a small deviation from the state $\psi^{\otimes n}$. Specifically, let $\eps\in(0,1)$ and denote by $\ms_{n,\eps}$ the set of all types\index{types (method)} in $\type(n,m)$ for which $\frac12\|\t-\p\|_1\leq\eps$. We also denote by $\ms_{n,\eps}^c$ the complement of the set $\ms_{n,\eps}$ in $\type(n,m)$. With these notations, for any $\eps\in(0,1)$ we can split $|\psi^{\otimes n}\ra$ into two parts
\be\label{splitp}
|\psi\ra^{\otimes n}=\sum_{\t\in\ms_{n,\eps}}\sqrt{q_{\t,n}}|\t\ra^{A^n}+\sum_{\t\in\ms_{n,\eps}^c}\sqrt{q_{\t,n}}|\t\ra^{A^n}\;.
\ee
From Lemma~\ref{lem721} it follows that the coefficients $q_{\t,n}$
satisfies
\be
\frac1{(n+1)^m}2^{-nD(\t\|\p)}\leq q_{\t,n}\leq 2^{-nD(\t\|\p)}\;.
\ee
Therefore, the fidelity of $|\psi^{\otimes n}\ra$ with the second term on the right-hand side of~\eqref{splitp} is given by
\ba\label{pinsker}
\sum_{\t\in\ms_{n,\eps}^c}q_{\t,n}&\leq\sum_{\t\in\ms_{n,\eps}^c}2^{-nD(\t\|\p)}\\
\GG{Pinsker's\; inequality}&\leq \sum_{\t\in\ms_{n,\eps}^c}2^{-2n\eps^2}\\
&\leq 2^{-2n\eps^2}\big|\type(n,m)\big|\leq 2^{-2n\eps^2}(n+1)^m\xrightarrow{n\to\infty}0\;.
\ea
Therefore, for any $\eps>0$ and sufficiently large $n$, the state $|\psi\ra^{\otimes n}$ can be made arbitrarily close to the state
\be\label{18p112}
|\psi_\eps^n\ra\eqdef\frac1{\sqrt{\nu_\eps}}\sum_{\t\in\ms_{n,\eps}}\sqrt{q_{\t,n}}|\t\ra^{A^n}\quad\text{where}\quad\nu_\eps\eqdef\sum_{\t\in\ms_{n,\eps}}q_{\t,n}\;.
\ee

From~\eqref{15p106}  the energy of any type $\t\in\type(n,m)$ is given by $\mu_\t\eqdef n\sum_{x\in[m]}t_xa_x$. In the sum above, the type $\t$ belong to $\ms_{n,\eps}$ so that $\frac12\|\t-\p\|_1\leq\eps$. Consequently, each component $x\in[m]$ of the vector $\t-\p$ satisfies $|t_x-p_x|\leq 2\eps$. Using this property, we get that
\be
|\mu_\t-\mu_\p|\leq n\sum_{x\in[m]}a_x|t_x-p_x|\leq 2n\eps\sum_{x\in[m]}a_x.
\ee
Therefore, for any two types\index{types (method)} $\t,\t'\in\type(n,m)$ that are $\eps$-close to $\p$ we have
\be\label{18p107}
|\mu_\t-\mu_{\t'}|\leq 4n\eps\sum_{x\in[m]}a_x
\ee
In other words, the energy spread of $|\psi^n_\eps\ra$ is no greater than $4n\eps\sum_{x\in[m]}a_x$.

Note that by taking $\eps>0$ sufficiently small we can make the energy spread $4n\eps\sum_{x\in[m]}a_x$ much smaller that $na_m$. However, we still get that the energy spread of $\psi_\eps^n$ is linear in $n$. We show now that by taking $\eps$ to depend on $n$, we can find states in $\pure(A^n)$ that are very close to $\psi^{\otimes n}$ but with energy spread that is sublinear in $n$. 

\begin{myg}{}
\begin{lemma}\label{lem:nchin}
Let $\psi\in\pure(A)$ and $\alpha\in\left(1/2,1\right)$. There exists a sequence of pure state $\{\chi_n\}_{n\in\mbb{N}}$ in $\pure(A^n)$ with the following properties:
\begin{enumerate}
\item The limit
\be\label{18p108}
\lim_{n\to\infty}\left\|\psi^{\otimes n}-\chi_n\right\|_1=0\;.
\ee
\item The state $\chi_n$ can be expressed as a linear combination of no more than $(n+1)^m$ energy eigenstates.
\item The energy spread of $\chi_n$ is no more than $4n^{\alpha}\sum_{x\in[m]}a_x$.
\end{enumerate}  
\end{lemma}
\end{myg}

\begin{proof}
Let $\eps_n=n^{\alpha-1}$. Since $\alpha\in\left(\frac12,1\right)$ we have $\lim_{n\to\infty}\eps_n=0$ and $\lim_{n\to\infty}n\eps_n^2=\infty$. The latter implies that if we replace $\eps$ in~\eqref{pinsker} with $\eps_n$ we still get the zero limit of~\eqref{pinsker}. Hence, the pure state $\chi_n\eqdef\psi_{\eps_n}^n$ satisfies~\eqref{18p108}.
Since for all $\eps>0$ we have that $\psi_\eps^n$ can be expressed as a linear combination of no more than $(n+1)^m$ energy eigenvectors, it follows that also $\chi_n$ have this property. Finally, from~\eqref{18p107} we get that the energy spread of $\chi_n$ cannot exceed 
\be
4n\eps_n\sum_{x\in[m]}a_x=4n^{\alpha}\sum_{x\in[m]}a_x\;.
\ee
This completes the proof.
\end{proof}

\bex\label{ex:glini}
Show that the average energy
\be
\left\la\chi_n\left|H^{\otimes n}\right|\chi_n\right\ra
\ee
grows linearly with $n$.
\eex

\subsubsection{Sublinear Athermality Resources}

We saw in the Lemma above that the state $\psi^{\otimes n}$ is very close to a state $\chi_n$, whose energy spread is sublinear in $n$. On the other hand,  the average energy $\la\chi_n|H^{\otimes n}|\chi_n\ra$ grows linearly in $n$ (see Exercise~\ref{ex:glini}). 
We will therefore consider systems whose energy grows sub-linearly in $n$ as asymptotically negligible resources. Note however that such resources can contain significant coherence among energy levels since the coherence grows logarithmically with $n$. Indeed, as we will see shortly, such resources makes the QRT of athermality reversible.

\begin{myd}{}
\begin{definition}
A sublinear athermality resource (SLAR) is a sequence of quantum athermality systems $\{R_n\}_{n\in\mbb{N}}$, such that $|R_n|$ grows polynomially with $n$, and there exists two constants independent of $n$, $0\leq \alpha<1$ and $c>0$, such that
\be
\left\|H^{R_n}\right\|_{\infty}\leq cn^\alpha\quad\quad\forall\;n\in\mbb{N}\;.
\ee
\end{definition}
\end{myd}
The key assumption in the given definition is that the energy of systems $R_n$ grows sublinearly with $n$. Consequently, as $n$ approaches infinity in the asymptotic limit, the resourcefulness of any states $\big\{(\omega^{R_{n}},\gamma^{R_n})\big\}_{n\in\mathbb{N}}$ becomes insignificant compared to the resourcefulness of $n$ copies of the golden unit $
\Upsilon_2\eqdef\left(|0\lr 0|^X,\u^X_2\right)
$. We will soon discover that this small amount of athermality resource is sufficient to restore reversibility.

\bex\label{exdisslar}
Show that the distillation rate of athermality as given in Theorem~\ref{droat} does not change if we replace CTO by CTO+SLAR. In other words, show that SLAR cannot increase the distillation rate of athermality.
\eex

\subsubsection{The Cost of Athermality}

The type of free operations that we consider here are CTO assisted with SLAR.  In order to define the cost under such operations, for any system $R$ and $\eps\in(0,1)$, we define the $R$-assisted $\eps$-single-shot cost as
\begin{align}\label{def17p}
&\cost^\eps_R\left(\rho^A,\gamma^A\right)\\&\eqdef\inf_{0<m\in\mbb{R}}\left\{\log m\;:\;\min_{\phi\in\md(R)}T\left(\Upsilon_m\otimes\left(\phi^R,\gamma^R\right)\xrightarrow{\cto}\left(\rho^A,\gamma^A\right)\right)\leq\eps\right\}\;.\nonumber
\end{align}
From Corollary~\ref{puretomixgpc} and Exercise~\ref{pureenough} it follows that 
any mixed state in $\md(A)$ can be obtained by thermal operations from a pure state in $\pure(A)$. Thus, we can restrict the minimum above over all density matrices $\phi\in\md(A)$ to a minimum over all pure states $\phi\in\pure(A)$.

\bex\label{1twoe}
Let $\eps\in(0,1/2)$, $\rho,\sigma,\gamma\in\md(A)$, and suppose that $\rho\approx_\eps\sigma$. Show that for any system $R$
\be
\cost^{2\eps}_R\left(\rho,\gamma\right)\leq\cost^\eps_R\left(\sigma,\gamma\right)\;.
\ee
\eex

With the above definition of the $R$-assisted single-shot athermality cost,
we define the asymptotic SLAR-assisted athermality cost as
\be\label{slar2}
\cost\left(\rho,\gamma\right)\eqdef\inf_{\{R_n\}}\lim_{\eps\to 0^+}\liminf_{n\to\infty}\frac1n\cost^\eps_{R_n}\left(\rho^{\otimes n},\gamma^{\otimes n}\right)\;,
\ee
where the infimum is over all SLARs, $\{R_n\}_{n\in\mbb{N}}$. 
We show now that for pure states the above cost can be expressed in terms of the relative entropy. The proof of the mixed-state case is far more complicated (see the discussion in the `Notes and References' section at the end of this chapter).

\begin{myt}{}
\begin{theorem}\label{purecostf}
Let $(\psi,\gamma)$ be an athermality state with $\psi\in\pure(A)$. Then, the SLAR-assisted athermality cost of $(\psi,\gamma)$ is given by
\be
\cost\left(\psi,\gamma\right)=D\left(\psi\|\gamma\right)\;,
\ee
where $D$ is the Umegaki\index{Umegaki} relative entropy.
\end{theorem}
\end{myt}

\begin{proof}
Since the cost of athermality under CTO assisted with SLAR can not be smaller that the distillation rate under the same operations, we get from Exercise~\ref{exdisslar} that 
\be
\cost\left(\psi,\gamma\right)\geq D\left(\psi\big\|\gamma\right)\;.
\ee
Our goal is therefore to prove the opposite inequality.

Let $\eps\in(0,1/2)$ and $\{\chi_n\}_{n\in\mbb{N}}$ be the sequence of pure states that satisfies all the properties outlined in Lemma~\ref{lem:nchin}. In particular, each $\chi_n$ is very close to $\psi^{\otimes n}$ (for $n$ sufficiently large) so that for for sufficiently large $n$ we have (see Exercise~\ref{1twoe})
\be\label{17p197}
\cost_{R_n}^{2\eps}\left(\psi^{\otimes n},\gamma^{\otimes n}\right)\leq \cost_{R_n}^{\eps}\left(\chi_n,\gamma^{\otimes n}\right)\;.
\ee
Therefore, we focus now on finding upper bound on $\cost_{R_n}^{\eps}\left(\chi_n,\gamma^{\otimes n}\right)$.

By definition, the energy spread of $\chi_n$ is given by $4n^\alpha\sum_{x\in[m]}a_x$ for some $\alpha\in(\frac12,1)$, and each $\chi_n$ has the form (cf.~\eqref{18p112})
\be\label{chisuper}
|\chi_n\ra=\sum_{\t\in\ms_n}\sqrt{q_{\t}}|\t\ra^{A^n}\;,
\ee
where $\ms_n$ is the set of all types\index{types (method)} $\t\in\type(n,m)$ that satisfies $\frac12\|\t-\p\|_1\leq n^{\alpha-1}$ (i.e., using the same notations discussed above~\eqref{splitp} we have $\ms_n\eqdef\ms_{n,\eps_n}$ with $\eps_n\eqdef n^{\alpha-1}$), and $\{q_\t\}_{\t\in\ms_n}$ form a probability distribution over the set of types in $\sigma_n$.
Let $k_n$ be the number of terms in the superposition above (hence $k_n\leq (n+1)^m$). Furthermore,  let the set $\{\mu_j\}_{j\in[k_n]}$ denote the energy eigenvalues of the Hamiltonian $H^{A^n}$. These eigenvalues correspond to the energy eigenvectors $|\t\ra^{A^n}$ that appear in the superposition~\eqref{chisuper}. That is, each $j\in[\ell]$ corresponds exactly to one type $\t$ that appears in the superposition~\eqref{chisuper}. Although the energies eigenvalues $\{\mu_j\}$ depend also on $n$, we did not add a subscript $n$ to ease on the notations. Without loss of generality we also assume that $\mu_1\leq\cdots\leq\mu_{k_n}$, so that the energy spread of $\chi_n$ is $\mu_{k_n}-\mu_1\leq 4n^\alpha\sum_{x\in[m]}a_x$ (see Lemma~\ref{lem:nchin}). We will also denote by $\s^{(n)}\in\ms_n$ the type that corresponds to the smallest energy $\mu_1$, and by $z^n\in[m]^n$ the sequence of type $\s^{(n)}$ so that $H^{A^ n}|z^n\ra^{A^n}=\mu_1|z^n\ra^{A^n}$.

With these notations, we are ready to define the SLAR system $R_n$ to be a $k_n$-dimensional quantum system whose Hamiltonian is given by
\be
H^{R_n}=\sum_{j\in[k_n]}(\mu_j-\mu_1)|j\lr j|^{R_n}\;.
\ee
Note that the Hamiltonian $H^{R_n}$ has the same eigenvalues as the energies that appears in $\chi_n$ shifted by $\mu_1$.  Observe that $|1\lr 1|^{R^n}$ is a zero-energy state of system $R_n$, and the maximal energy of $H^{R_n}$ is given by $\mu_{k_n}-\mu_1\leq 4n^\alpha\sum_{x\in[m]}a_x$
so that $\{R_n\}_{n\in\mbb{N}}$ is indeed a SLAR.
We take the SLAR of system $R_n$ to be
\be
|\phi^{R_n}\ra\eqdef \sum_{j\in[k]}\sqrt{q_j}|j\ra^{R_n}
\ee
where $q_{j}\eqdef q_{\t}$ with $\t$ being the type that corresponds to the energy $\mu_j$. By construction,  the state
\be
\phi^{R_n}\otimes|z^n\lr z^n|^{A^n}
\ee
has the exact same energy distribution as the state
\be
|1\lr 1|^{R_n}\otimes \chi_n^{A^n}
\ee
(recall that $|1\ra^{R_n}$ corresponds to the zero energy of system ${R_n}$). Hence, the above two states are equivalent resources and can be converted from one to the other by reversible thermal operations (i.e.\ an energy preserving unitary). We now use this resource equivalency to compute the cost of $\chi_n$ in terms of the cost of the quasi-classical state $|z^n\lr z^n|$. We do it in three steps:
\ben
\item Replacing $\chi_n^{A^n}$ with $|1\lr 1|^{R_n}\otimes\chi_n^{A^n}$: By adding the resource $(|1\lr 1|^{R_n},\gamma^{R_n})$ we can only increase the cost. Therefore,
\be
\cost_{R_n}^{\eps}\left(\chi_n^{A^n},\gamma^{A^n}\right)
\leq\cost_{R_n}^{\eps}\left(|1\lr 1|^{R_n}\otimes\chi_n^{A^n},\gamma^{R_nA^n}\right)\;.
\ee
\item Replacing $|1\lr 1|^{R_n}\otimes\chi_n^{A^n}$ with $\phi^{R_n}\otimes|z^n\lr z^n|^{A^n}$: As discussed above, these two states are equivalent resources so that
\be
\cost_{R_n}^{\eps}\left(|1\lr 1|^{R_n}\otimes\chi_n^{A^n},\gamma^{R_nA^n}\right)=\cost_{R_n}^{\eps}\left(\phi^{R_n}\otimes|z^n\lr z^n|^{A^n},\gamma^{R_nA^n}\right)\;.
\ee
\item Replacing $\phi^{R_n}\otimes|z^n\lr z^n|^{A^n}$ with $|z^n\lr z^n|^{A^n}$: The cost of $|z^n\lr z^n|$ without the assistance of $R_n$ cannot be smaller than the cost of $\phi^{R_n}\otimes|z^n\lr z^n|$ with the assistance of $R_n$, since the latter is defined in terms of a minimum over all states in $\md(R_n)$ (see the minimization in~\eqref{def17p}). Therefore,
\be
\cost_{R_n}^{\eps}\left(\phi^{R_n}\otimes|z^n\lr z^n|^{A^n},\gamma^{R_nA^n}\right)
\leq \cost^{\eps}\left(|z^n\lr z^n|^{A^n},\gamma^{A^n}\right)\;.
\ee
\een
Combining all the three steps above with~\eqref{17p197}, and using the fact in the quasi-classical regime GPO has the same conversion power as CTO (see~Theorem~\ref{thm:quasi}), we get that
\ba
\cost_{R_n}^{2\eps}\left(\psi^{\otimes n},\gamma^{\otimes n}\right)&\leq\cost^{\eps}\left(|z^n\lr z^n|,\gamma^{\otimes n}\right)\\
\GG{Theorem~\ref{dmindistill}}&=D_{\max}^\eps\left(|z^n\lr z^n|\big\|\gamma^{\otimes n}\right)\\
&\leq D_{\max}\left(|z^n\lr z^n|\big\|\gamma^{\otimes n}\right)\;,
\ea
where in the last inequality we used the fact that $D_{\max}$ is always no smaller than its smoothed version. Now, observe that
\ba\label{18p126}
D_{\max}\left(|z^n\lr z^n|\big\|\gamma^{\otimes n}\right)&=-\log\left\la z^n\left|\gamma^{\otimes n}\right|z^n\right\ra\\
&=-\sum_{x\in[m]}ns_x^{(n)}\log\la x|\gamma|x\ra\;,
\ea
where in the last equality we used the fact that the sequence $z^n$ has a type $\s^{(n)}$.
Hence, the cost per each copy of $\psi$ can not exceed
\ba\label{18p127}
\limsup_{n\to\infty}\frac1n\cost_{R_n}^{2\eps}\left(\psi^{\otimes n},\gamma^{\otimes n}\right)&\leq\limsup_{n\to\infty}\frac1nD\left(|z^n\lr z^n|\big\|\gamma^{\otimes n}\right)\\
&=-\lim_{n\to\infty}\sum_{x\in[m]}s_x^{(n)}\log\la x|\gamma|x\ra\\
\Gg{\frac12\left\|\p-\s^{(n)}\right\|_1\leq n^{\alpha-1}}&=-\sum_{x\in[m]}p_x\log\la x|\gamma|x\ra\\
&=D(\psi\|\gamma)\;.
\ea
This completes the proof.
\end{proof}

\bex
Prove explicitly the second line in~\eqref{18p126}.
\eex

\section{Notes and References}

There exist several strong operational justifications that the Gibbs state the free state of the theory of athermality. First, in~\cite{RGE2012} it was shown that the Gibbs state is the unique
equilibrium state that a quantum system will evolve to under weak coupling with the thermal bath.
Second, in~\cite{BHN+2015,HR2016} it was shown that if, in the implementation of a thermal operation,
one could freely introduce any other density operator
$\sigma$ inequivalent to the Gibbs state of the ancillary system,
then the QRT would become trivial. More precisely, it
would be possible to freely generate any density matrix $\rho$ to arbitrary precision by consuming many copies of $\sigma$.
The final, and perhaps most compelling reason for considering the Gibbs state to be free involves work extraction and the notion of passivity introduced in Sec.~\ref{passive}. Theorem~\ref{thm:passive} and its corollary that  a
state is completely passive if and only if it is the Gibbs
state is due to~\cite{Lenard1978,PW1978}. It is worth noting that one can also consider a resource theory of thermodynamics where \emph{all} states, including the Gibbs state, are considered resources. Such a resource theory was investigated in~\cite{SOF2017}.

In this chapter, we have expounded on the resource theory of athermality, which revolves around the principle of energy conservation. However, its extension to other conserved observables follows in a similar manner~\cite{HR2016, Halpern2018}, encompassing non-commuting observables as well~\cite{HFOW2016, GPS+2016, LJR2017}.

Thermal operations and closed thermal operations were first introduced in~\cite{JW2000} although the terminology used here was given much later in~\cite{BHO+2013,HO2013}. The refinement of thermal operations as given in Lemma \ref{lem1721} is due to~\cite{Gour2022}.
The statement that in the quasi-classical regime, CTO and GPO have the same conversion power (see Theorem~\ref{thm:quasi}) was first proved in~\cite{JW2000}.
 However, for the convertibility among general states (i.e., those not commuting with the
Hamiltonian), in~\cite{FOR2015} an example was given, demonstrating that GPO are strictly
more powerful than CTO.
The set of Gibbs-Preserving Covariant (GPC) operations were introduced in~\cite{LKJT2015}.

The characterization of quantum relative majorization in terms of semi-definite programming can be found in~\cite{GJB+2018}. Moreover, in~\cite{BG2017} partial characterization of quantum relative majorization was given in terms of an extension of Lorenz curves to the quantum domain. The elegant characterization of quantum relative majorization in the (partially) qubit case (i.e., Theorem~\ref{qubitchar}) is due to~\cite{HJRW2012}. Another characterization in which all states are qubits was given in~\cite{AU1980}. 

Corollaries~\ref{cor:rgsg} and~\ref{puretomixgpc}, and Theorems~\ref{thm:gpcco} and~\ref{qubitthm}, can be found in~\cite{Gour2022}. More information on coherences in the theory of athermality, along with another set of constraints similar to the one given in Theorem~\ref{qubitthm} can be found in~\cite{CSHO2015}. More details on the SDP formulation of exact interconversions in the theory of athermality can be found in~\cite{GJB+2018}.

In our proof of Theorem~\ref{purecostf}, we primarily drew from the work presented in~\cite{Gour2022}. Although the proof for the mixed state variant of the theorem was initially introduced in~\cite{BHO+2013}, a more comprehensive and rigorous proof was later provided in the broader context of~\cite{SOF2017}. It's important to highlight that the proof outlined in~\cite{SOF2017} (specifically, Theorem~1) stipulates that the ancillary system, referred to (in this book) as the SLAR, should possess a dimension of $2^{\sqrt{n\log n}}$. Consequently, a lingering question remains regarding the possibility of reducing this dimension to Poly$(n)$, as is feasible in the pure-state scenario.

\part{Appendices}

\begin{appendix}

\chapter{Elements of Convex Analysis}\label{secconvex}

%{Convex sets in $\mbb{R}^n$}\label{A:1}

We describe here a few properties of convex sets in a finite dimensional (real) Hilbert space (e.g. $\mbb{R}^n$) that are used quite often in quantum information. 
A set $\mc\subset\mbb{R}^n$ is said to be \emph{convex} if for any two elements $\v,\u\in\mc$ and any
$t\in[0,1]$ the vector $t\v+(1-t)\u\in\mc$. Consequently, if $\v_1,\ldots,\v_m\in\mc$ and $p_1,\ldots,p_m$ are non-negative with $\sum_{x\in[m]}p_x=1$ then
\be
\sum_{x\in[m]}p_x\v_x\in\mc.
\ee

\section{The Hyperplane Separation Theorem}

A hyperplane in $\mbb{R}^n$ is the set of all vectors $\v\in\mbb{R}^n$ that satisfy $\n\cdot\v=c$ for some fixed (constant) $c\in\mbb{R}$ and a fixed (normal) vector $\n\in\mbb{R}^n$. It generalizes the equation of a plane in $\mbb{R}^3$ which has the form $n_1x+n_2y+n_3y=c$ for all points $(x,y,z)$ in a plane whose normal vector is $\n=(n_1,n_2,n_3)^T$.
The hyperplane separation theorem basically states that any two convex sets with empty intersection can always be separated by a hyperplane (see Fig.~\ref{figa1}). In its most generality (i.e. including the infinite dimensional case) it is also known as the Hahn-Banach separation theorem which is itself a variant of the Hahn-Banach theorem. 
\begin{figure}[h]\centering    \includegraphics[width=0.6\textwidth]{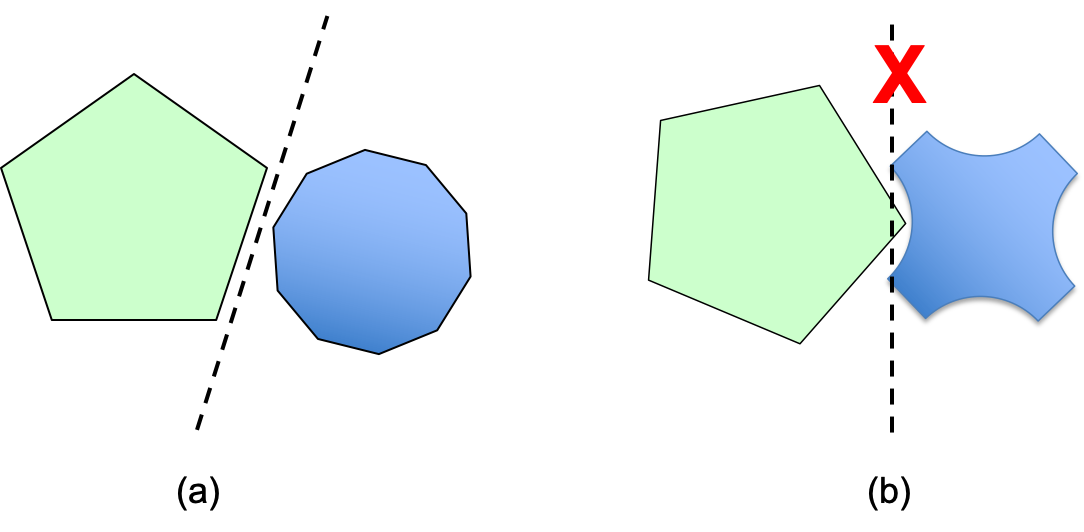}
  \caption{\linespread{1}\selectfont{\small (a) A separating hyperplane between two polytopes. (b) A separating hyperplane does not exists since one of the sets is not convex.}}
  \label{figa1}
\end{figure}

\begin{myt}{\color{yellow} The Hyperplane Separation Theorem}
\begin{theorem}\label{hypert}
Let $\mc_1$ and $\mc_2$ be two disjoint convex subsets of $\mbb{R}^n$. Then there exist a nonzero vector $\n\in\mbb{R}^n$ and a real number $c\in\mbb{R}$ such that
\be\label{hyper}
\n\cdot\r_2\leq c\leq\n\cdot\r_1\quad\forall \r_1\in\mc_1 \quad\text{and}\quad\forall \r_2\in\mc_2.
\ee
That is, $\n$ is the normal vector of the  hyperplane $\{\v\in\mbb{R}^n\;:\;\n\cdot\v = c\}$ that separates $\mc_1$ and $\mc_2$. Moreover, if the sets $\mc_1$ and $\mc_2$ are also closed and at least one of them is compact then one can replace the inequalities above with strict inequalities.
\end{theorem}
\end{myt}

\begin{remark}
The hyperplane separation theorem has numerous applications in convex analysis and beyond. Consequently, it has many variants and also has several proofs. Since this theorem has been used many times in this book, we provide below its proof for the purpose of self-containment. This is by no means aims to replace a more thorough study of this subject. A reader interested in more details can follow standard textbooks on convex analysis. 
\end{remark} 

\begin{proof}
We will define the vector $\n$ and then show that it has all the desired properties. The key idea is to use the fact (see the proof below) that if $\mc\subseteq\mbb{R}^n$ is closed and convex then there exists a unique vector in $\mc$ with a minimum (Euclidean) norm. Then, the vector $\n$ will be taken to be the vector with minimal norm in the closer of $\mc_1-\mc_2$. We now discuss the details.

Let $\mc\eqdef\overline{\mc_1-\mc_2}$ be the closure of the set $\{\r_1-\r_2\;:\;\r_1\in\mc_1\;,\;\r_2\in\mc_2\}$. Since the later is convex , also its closure, $\mc$, is convex (see Exercise~\ref{exmc1}). 
Let $d\eqdef\inf\{\|\n\|^2\;:\;\n\in\mc\}$. Geometrically, $d$ is the distance between the two sets. Note that since $\mc_1$ and $\mc_2$ are disjoint, the set $\mc_1-\mc_2$ does not contain the zero vector. However, its closer may contain it. We will first consider the case that $d>0$, and later treat the case $d=0$. 

By definition of $d$, there exists a sequence $\n_j\in\mc$ such that $\|\n_j\|\to d$. 
This sequence is a Cauchy sequence since
\be
\|\n_j-\n_k\|^2=2\|\n_j\|+2\|\n_k\|^2-\|\n_j+\n_k\|^2
\ee
and $\|\n_j+\n_k\|^2=4\|(\n_j+\n_k)/2\|^2\geq 4d$ since the convex combination $(\n_j+\n_k)/2\in\mc$. Hence,
\be
\|\n_j-\n_k\|^2\leq 2\|\n_j\|+2\|\n_k\|^2-4d
\ee
which goes to zero as $j,k\to\infty$. We define $\n\in\mc$ to be the limit of $\{\n_j\}_{j\in\mbb{N}}$. Next, let $\r_1\in\mc_1$ and $\r_2\in\mc_2$, and observe that since both $\r_1-\r_2$ and $\n$ are elements of $\mc$, any convex combination $t(\r_1-\r_2)+(1-t)\n$ with $t\in(0,1)$ is also in $\mc$. Therefore, its square norm cannot be smaller than $d$. Hence,
\ba
d&\leq\|t(\r_1-\r_2)+(1-t)\n\|^2\\
&=t^2\|\r_1-\r_2\|^2+2t(1-t)(\r_1-\r_2)\cdot\n+(1-t)^2d
\ea 
Subtracting both sides by $(1-t)^2d$ and dividing by $t$ gives
\be
(2-t)d\leq t\|\r_1-\r_2\|^2+2(1-t)(\r_1-\r_2)\cdot\n\;.
\ee
Finally, since the above inequality holds for all $t\in(0,1)$ it must also hold for $t=0$. That is,
\be\label{a:d}
d\leq (\r_1-\r_2)\cdot\n\quad \forall \r_1\in\mc_1 \quad\text{and}\quad\forall \r_2\in\mc_2\;.
\ee
Note that if $d>0$ this implies~\eqref{hyper} (see Exercise~\ref{exd0}). It is therefore left to check the case $d=0$. 

Suppose first that the interior of $\mc_1-\mc_2$ is not empty. Therefore, there exists a sequence 
$
\mk_1\subset\mk_2\subset\cdots
$
of non-empty closed subsets of the interior of $\mc_1-\mc_2$ such that their union is the interior of $\mc_1-\mc_2$.
Since  $\mc_1-\mc_2$ does not contains the zero vector (recall that $\mc_1$ and $\mc_2$ are disjoint sets) each $\mk_j\subseteq\mc_1-\mc_2$ does not contains the zero vector. Moreover, since $\mk_j$ is closed  it contains a non-zero vector $\n_j\in\mk_j$ with minimal norm. 

We now apply the same argument leading to~\eqref{a:d} with $\mc_1$ replaced with $\mk_j$ and $\mc_2$ replaced with the zero set $\{\0\}$ (which is disjoint from $\mk_j$). For such choices, $d$ in~\eqref{a:d} equals $\|\n_j\|^2$ so that~\eqref{a:d} becomes $0\leq \|\n_j\|^2\leq \v\cdot\n_j$ for all $\v\in\mk_j$. We can therefore normalize all $\{\n_j\}$ and argue that they satisfies  $\v\cdot\n_j\geq 0$ for all $\v\in\mk_j$. Finally, the sequence of normalized vectors $\{\n_j\}$ contains a convergence subsequence (since the sphere in $\mbb{R}^n$ is compact), and therefore its limit $\n$ also satisfies $\v\cdot\n\geq 0$ for all $\v$ in the interior of $\mc_1-\mc_2$. Hence, by continuity, the inequality $\v\cdot\n\geq 0$ must also hold for all $\v$ in $\mc_1-\mc_2$ itself. This completes the proof for the case that the interior of $\mc_1-\mc_2$ is not empty.

If the interior of $\mc_1-\mc_2$ is empty then its span has a dimension strictly smaller than the dimension of the whole space. Therefore, it is contained in some hyperplane $\{\v\in\mbb{R}^n\;:\;\v\cdot\n=c\}$ so that $\v\cdot\n\geq c$ for all $\v$ in $\mc_1-\mc_2$. As we argued before, this implies~\eqref{hyper}. The remaining part of the proof for the case that $\mc_1,\mc_2$ are closed and compact is left as an exercise.
\end{proof}

\begin{exercise}\label{exmc1}
Show that if $\mc_1$ and $\mc_2$ are two convex subsets of $\mbb{R}^n$ then $\overline{\mc_1-\mc_2}$ is also convex.
\end{exercise}

\begin{exercise}\label{exd0}
Show that if $d>0$ then~\eqref{a:d} implies~\eqref{hyper}.
\end{exercise}

\begin{exercise}
Complete the proof above. That is, show that~\eqref{hyper} holds with strict inequalities if $\mc_1$ and $\mc_2$ are closed and at least one of them is compact.
\end{exercise}

\begin{exercise}
Show that if $\mc_1$ and $\mc_2$ are two disjoint convex subsets of $\mbb{R}^n$, and if $\mc_1$ is open in $\mbb{R}^n$, then there exist a nonzero vector $\n\in\mbb{R}^n$ and a real number $c\in\mbb{R}$ such that
\be\label{hyperg}
\n\cdot\r_2\leq c<\n\cdot\r_1\quad\forall \r_1\in\mc_1 \quad\text{and}\quad\forall \r_2\in\mc_2.
\ee
Hint: Use the theorem above and the fact that separating hyperplanes cannot intersect the interiors of convex sets.
\end{exercise}

\section{Convex Hulls, Faces, and Polytopes}\label{secpoly}

The \emph{convex hull} of a set $\mk\in\mbb{R}^n$, denoted by $\conv(\mk)$, is the smallest convex set in $\mbb{R}^n$ that contains $\mk$. Equivalently, it is the intersection of all convex sets containing $\mk$. If $\mk$ contains a finite number of vectors, i.e. its cardinality is $|\mk|<\infty$, then $\conv(\mk)$ is called a \emph{polytop}, and it is the set containing all convex combinations of the vectors in $\mk$. That is, if $\mk=\{\v_1,\ldots,\v_m\}$ then
\be
\conv(\mk)\eqdef\left\{\sum_{x\in[m]}p_x\v_x\;:\;0\leq p_x\in\mbb{R}\;\;,\;\;\sum_{x\in[m]}p_x=1\right\}\;.
\ee
Note that by definition, the convex hull of a single vector $\v\in\mbb{R}^n$ consists of just the vector $\v$. Hence, the set of all $m$-dimensional probability vectors is a polytope in $\mbb{R}^m$.

As a simple example, consider the set $\prob(m)$ consisting of all probability vectors in $\mbb{R}^m$. That is, $\prob(m)$ denotes the set of all $m$-dimensional vectors with non-negative components that sum to one. It is simple to check that
\be
\prob(m)=\conv\{\e_1,\ldots,\e_m\}\;,
\ee
where $\{\e_x\}_{x\in[m]}$ is the standard basis of $\mbb{R}^m$.

The closed and open intervals between two vectors $\v_1,\v_2\in\mbb{R}^n$ is defined respectively as
\ba
& [\v_1,\v_2]\eqdef\left\{t\v_1+(1-t)\v_2\;:\;0\leq t\leq 1\right\}\\
& (\v_1,\v_2)\eqdef\left\{t\v_1+(1-t)\v_2\;:\;0< t< 1\right\}\;.
\ea

\begin{myd}{Face}
\begin{definition}
Consider a convex set $\mc\subseteq\mbb{R}^n$. A subset $\mf\subseteq\mc$ is called a \emph{face} of $\mc$ if for any $\v\in\mf$ and any  $\v_1,\v_2\in\mc$ such that $\v\in(\v_1,\v_2)$ we have $\v_1,\v_2\in\mf$. 
\end{definition}
\end{myd}
In other words, $\mf$ is a face of $\mc$ if for any $\v_1,\v_2\in\mc$ we have that
\be
\mf\cap(\v_1,\v_2)\neq\O\quad\Rightarrow\quad \v_1,\v_2\in\mf.
\ee
To have a better understanding of this definition, let
$\mc\eqdef\conv\{\v_1,\ldots,\v_m\}$ be the convex hull of $m$ vectors in $\mbb{R}^n$ (i.e. $\mc$ is a polytope), and let $\sum_{x\in[m]}p_x\v_x$ be a vector that belongs to a face $\mf$ of the polytope $\mc$. Then, for any $x\in[m]$ with $p_x\in(0,1)$ we must have $\v_x\in\mf$. Hence, any face of $\mc$ must be a convex hull of a subset of $\{\v_1,\ldots,\v_m\}$. 
Note, however, that the converse is not necessarily true. That is, a convex hull of a subset of $\{\v_1,\ldots,\v_m\}$ is not necessarily a face.

For any $x\in[m]$ the set $\{\v_x\}$ (consisting of a single vector) is a face of the convex polytope $\mc\subset\mbb{R}^n$. It is also called a \emph{vertex} of the polytope. Any face $\mf$ of $\mc$ that can be expressed as $\mf=\conv\{\v_x,\v_y\}$, where $x,y\in[m]$ and $x\neq y$ is called an \emph{edge} of the polytope $\mc$. Note that we do \emph{not} claim that $\conv\{\v_x,\v_y\}$ is necessarily a face, only that if it is a face, then it is called an edge. Finally, a \emph{facet} of $\mc$ is a face that can be expressed as a convex hull of $n-1$ distinct vectors in $\{\v_1,\ldots,\v_m\}$. Therefore, faces of convex sets generalize the notion of vertices, edges and facets of polytopes (see Fig.~\ref{faces}).

 \begin{figure}[h]\centering    \includegraphics[width=0.4\textwidth]{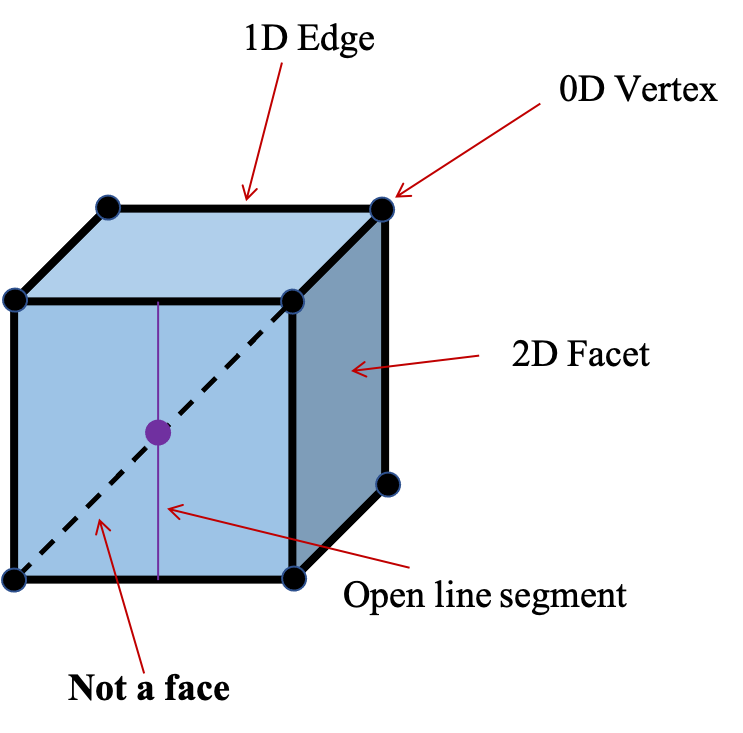}
  \caption{\linespread{1}\selectfont{\small Faces of a 3D cube. The dashed line is not a face since it contains points in open intervals with end points that are outside of the dashed line.}}
  \label{faces}
\end{figure}

Every vector $\w\in\mbb{R}^n$ can be used to define a face of a compact convex set $\mc\subset\mbb{R}^n$ given by
\be\label{face}
\mf_\w\eqdef\big\{\v\in\mc\;:\;\w\cdot\v=\max_{\u\in\mc}\w\cdot\u\big\}\;.
\ee
To show that this set is indeed a face of $\mc$, observe first that $\mf_\w$ is non-empty since $\mc$ is a compact set. Now, let $\v=t\v_1+(1-t)\v_2$ where $\v\in\mf_w$, $t\in(0,1)$, and $\v_1,\v_2\in\mc$. Then, by definition
\ba
\max_{\u\in\mc}\w\cdot\u=\w\cdot\v&=t\w\cdot\v_1+(1-t)\w\cdot\v_2\\
&\leq t\max_{\u\in\mc}\w\cdot\u+(1-t)\max_{\u\in\mc}\w\cdot\u\\
&=\max_{\u\in\mc}\w\cdot\u\;.
\ea
Hence, the inequality above must be an equality which can only hold if both $\w\cdot\v_1=\max_{\u\in\mc}\w\cdot\u$ and $\w\cdot\v_2=\max_{\u\in\mc}\w\cdot\u$. That is, $\v_1,\v_2\in\mf_\w$.

\begin{exercise}
Show that if $\v\in\mf_\w$ then any vector $\v'\in\mc$ with the property that 
\be
(\v-\v')\cdot\w=0
\ee 
is also in $\mf_\w$.
\end{exercise}

\section{Extreme Points}\label{sec:extreme}

\begin{myd}{Extreme Point}
\begin{definition}
An extreme point of a convex set $\mc\subseteq\mbb{R}^n$ is a point that does not belong to any open interval of $\mc$.
\end{definition}
\end{myd}

In other words, an extreme point is a point that cannot be expressed as $t\v+(1-t)\w$, for some $t\in(0,1)$ and two \emph{distinct} vectors $\v,\w\in\mc$ (i.e. $\v\neq\w$). Observe that by definition if a convex set $\mc=\{\v\}$ consists of a single vector $\v\in\mbb{R}^n$ then $\v$ is an extreme point of $\mc$.

\begin{exercise}\label{ex135}
Let $\w\in\mbb{R}^n$, $\mc\subseteq\mbb{R}^n$ be a compact convex set, and let $\mf$ be a face of $\mc$. Show that if $\v\in\mbb{R}^n$ is an extreme point of $\mf$ then it is also an extreme point of $\mc$. Conversely, show that if $\v\in\mf$ is an extreme point of $\mc$ then it is also an extreme point of $\mf$.
\end{exercise}

\begin{myt}{\color{yellow} Krein–Milman theorem}
\begin{theorem}\label{kmt}
Every compact convex set of $\mbb{R}^n$ equals to the closed convex hull of its extreme points.
\end{theorem}
\end{myt}
\begin{remark}
The theorem above indicates the significance and importance of extreme points in convex analysis. The theorem implies in particular that the set of extreme points  of a compact convex set in $\mbb{R}^n$ is non-empty. In its proof below we make use of the Zorn's lemma from set theory. 
\end{remark}

\begin{proof}
Let $\mc\subseteq\mbb{R}^n$ be a non-empty compact convex set. We first prove that the set of extreme points of $\mc$ is non-empty. If $\mc$ consists of a single vector then we are done. Otherwise, let $\v_1,\v_2\in\mc$ be two distinct vectors (i.e. $\v_1\neq\v_2$). From the hyperplane separation theorem (see Theorem~\ref{hypert}) there exists a vector $\w_1\in\mbb{R}^n$ such that
$\w_1\cdot\v_1>\w_1\cdot\v_2$. This implies that the face $\mf_{\w_1}$ of $\mc$ does not contain the point $\v_2$ (see the definition of $\mf_\w$ in~\eqref{face}).
We next apply the same procedure to $\mf_{\w_1}$. Specifically, if this set contains a single point then that point is an extreme point, and from Exercise~\ref{ex135} it is also an extreme point of $\mc$ so that we are done. Otherwise, the face $\mf_{\w_1}$ contains two vectors (that are not the same) that can be separated by 
a hyperplane with a normal vector $\w_2$. Hence, the face $
\mf_{\w_2}\eqdef\{\v\in\mf_{\w_1}\;:\;\w_2\cdot\v=\max_{\u\in\mf_{\w_1}}\w_2\cdot\u\}$ of $\mf_{\w_1}$ does not contain one of the two vectors. Continuing in this way, if the process does not stop at some step $j$ for which $\mf_{\w_j}$ contains a single point (and therefore it must be an extreme point), then we get an infinite sequence of faces $\{\mf_{\w_j}\}_{\j=1}^\infty$ that are ordered by strict inclusion
\be
\mf_{\w_1}\supset\mf_{\w_2}\supset\cdots\supset\mf_{\w_j}\supset\cdots
\ee
Such a sequence of compact closed convex sets has a minimal element  (Zorn's lemma) which we denote by $\mf$.
From the Exercise~\ref{exface} below, it follows that $\mf$ is itself a face of $\mc$. Therefore, if it contains more than one point then we can continue with the same procedure to get a strict subset $\mf'\subset\mf$ in contradiction with the minimality of $\mf$. Hence, $\mf$ must contain exactly one element. This element must be an extreme point. 
We therefore proved that the set of extreme point of $\mc$ is non-empty.

To complete the proof, denote by $\mk$ the closed convex hull of the extreme points of $\mc$. Suppose by contradiction that there exists a vector $\u\in\mc$ that is not in $\mk$. Since the one element set $\{\u\}$ and the set $\mk$ are two closed disjoint sets, it follows from the hyperplane separation theorem (Theorem~\ref{hypert}) that there exists a vector $\w\in\mbb{R}^n$ such that
\be
\w\cdot\v<\w\cdot\u\quad\quad\forall\;\v\in\mk\;.
\ee
The above inequality implies that the face $\mf_\w$ as defined in~\eqref{face} does not intersect the set $\mk$. But from the first part of the proof, it follows that $\mf_\w$ contains an extreme point. This point is not in $\mk$ which is a contradiction to the definition of $\mk$. This completes the proof.
\end{proof}

\begin{exercise}\label{exface}
Show that for any $j\in\mbb{N}$, the minimal set $\mf$ is a face of $\mf_{\w_j}$.
\end{exercise}

As an example,
consider the set $\stoc(m,n)$ of all $m\times n$ column stochastic matrices. This set is clearly convex. Its extreme points are the column stochastic matrices that has in each column exactly one component equals to one and the rest are zero (can you prove it?). Denote by $\{F
_{j}\}_{j\in[m^n]}$ the $m^n$ extreme points of $\stoc(m,n)$. Then, any matrix $M\in\stoc(m,n)$ can be expressed as 
\be
M=\sum_{j\in[m^n]}t_jF_j
\ee
where $\t=(t_1,\ldots,t_{m^n})^T$ is a probability vector.

The following theorem shows that it is always possible to bound the number of elements in a convex combinations of vectors in $\mbb{R}^n$.

\begin{myt}{\color{yellow} Carath\'eodory's Theorem}
\begin{theorem}\label{carath}  Let $\mk$ be a subset of $\mbb{R}^{n}$. If $\v\in\conv(\mk)$ then $\v$ can be written as a convex combination of at most $n+1$ elements of $\mk$.  
\end{theorem}
\end{myt}
\begin{proof}
Let $\v\in\conv(\mk)$. Then, there exists $m\in\mbb{N}$, an $m$-dimensional probability vector $\p\eqdef (p_1,\ldots,p_m)^T$, and $m$ vectors $\w_1,\ldots,\w_m\in\mk$ such that
\be
\v=\sum_{x\in[m]}p_x\w_x\;.
\ee
If $m\leq n+1$ then we are done. Otherwise, $m> n+1$ so that the vectors $\w_2-\w_1,\ldots.,\w_m-\w_1$ must be linearly dependent (since there are $m-1>n$ of them).
Let $\lambda_2,\ldots,\lambda_m\in\mbb{R}$ be $m-1$ numbers, not all zero, such that
\be
\sum_{x=2}^m\lambda_x(\w_x-\w_1)=0\;.
\ee
Now, denote $\lambda_1\eqdef-\sum_{x=2}^m\lambda_x$ so that the equation above becomes 
\be\label{sumxshave}
\sum_{x\in[m]}\lambda_x\w_x=0\;.
\ee 
Observe that since $\sum_{x\in[m]}\lambda_x=0$ the set $\{\lambda_x\}_{x\in[m]}$ contains at least one strictly positive number (as we assume that not all of them are zero). We can therefore define
\be\label{0mu00}
\mu\eqdef\min\left\{\frac{p_x}{\lambda_x}\;:\;\lambda_x>0\;,\;x\in[m]\right\}\;.
\ee
By definition, $\mu$ has the property that $q_x\eqdef p_x-\mu\lambda_x\geq 0$ for all $x\in[m]$. Observe also that $\sum_{x\in[m]}q_x=1$ so that $\q=(q_1,\ldots,q_m)^T$ is a probability vector. In addition, from the definition of $\mu$, there exists at least one $y\in[m]$ (the minimizer of~\eqref{0mu00}) such that 
$q_y=p_y-\mu\lambda_y=0$. Without loss of generality suppose that $y=m$. We then get that the convex combination
\ba
\sum_{x\in[m-1]}q_x\w_x=\sum_{x\in[m]}q_x\w_x&=\sum_{x\in[m]}(p_x-\mu\lambda_x)\w_x\\
\GG{\eqref{sumxshave}}&=\sum_{x\in[m]}p_x\w_x=\v\;.
\ea
Hence, $\v$ can be expressed as a convex combination of $m-1$ vectors in $\mk$. We then repeat the process until we express $\v$ as a convex combination of $n+1$ vectors in $\mk$.
\end{proof}

\begin{exercise}\label{ex:2by2}
Let
\be
\mc\eqdef\Big\{N\in\mbb{R}^{n\times n}\;:\;\|N\r\|_2\leq 1 \quad\forall\;\r\in\mbb{R}^n\;\text{ s.t. }\;\|\r\|_2= 1\Big\}\;.
\ee
\begin{enumerate}
\item Show that $\mc$ is a compact convex set in $\mbb{R}^{n\times n}$.
\item Show that a matrix $O\in\mbb{R}^{n\times n}$ is an extreme point of $\mc$ if and only if $O$ is an orthogonal matrix (i.e. $O^TO=I_n$). 
Hint: Show first that $N\in\mc$ if and only if $N^TN\leq I_n$.
\item Show that every $N\in\mc$ can be expressed as a convex combination of a finite number of orthogonal matrices.
\end{enumerate}
\end{exercise}

\bex\label{compacthull}
Let $\mc\in\mbb{R}^n$ be a compact set (i.e., closed and bounded). Show that it's convex hull, $\conv(\mc)$, is also compact. Hint: Use Carath\'eodory's theorem.
\eex

\section{Polyhedrons}

\begin{myd}{Polyhedron}
\begin{definition}
Let $\r_1,\ldots,\r_m\in\mbb{R}^n$ be $m$ vectors, and $c_1,\ldots,c_m\in\mbb{R}^m$. The set
\be\label{polyhd}
\mc\eqdef\Big\{\v\in\mbb{R}^n\;:\;\v\cdot\r_x\leq c_x\quad\quad\forall\;x\in[m]\big\}\;.
\ee
is called a convex \emph{polyhedron}. 
\end{definition}
\end{myd}

\begin{exercise}\label{ccpoly}
Show that $\mc$ in~\eqref{polyhd} is a closed convex set.
\end{exercise}

The extreme points of polyhedrons are called vertices and our next goal is to characterize them. Intuitively, one would expect that an extreme point $\e$ of the polyhedron $\mc$ as defined above should saturate some of inequalities given in~\eqref{polyhd}. That is, we would expect that $\e\cdot\r_x=c_x$ at least for some $x\in[m]$. The following theorem makes this intuition rigorous.

\begin{myt}{}
\begin{theorem}\label{polyhdr}
Let $\mc\subseteq\mbb{R}^n$ be the polyhedron as defined in~\eqref{polyhd}. Then, a point $\e\in\mc$ is an extreme point of $\mc$ if and only if the set
\be
\mk\eqdef\Big\{\r_x\;:\;\e\cdot\r_x=c_x\;\;,\;\;x\in[m]\Big\}
\ee
span $\mbb{R}^n$; i.e. $\spa\{\mk\}=\mbb{R}^n$.
In particular, if $\e$ is a vertex then $|\mk|\geq n$.
\end{theorem} 
\end{myt}
\begin{proof}
Let $\e\in\mc$ and suppose first that $\spa\{\mk\}\neq\mbb{R}^n$. Then, there exists a vector $\v\in\mbb{R}^n$ such that $\v\cdot\r_x=0$ for all $\r_x\in\mk$. Since $\e\in\mc$, for all $\r_x\not\in\mk$ we must have $\r_x\cdot\e<c_x$ . Thus, for sufficiently small $\eps>0$ we have for all $x\in[m]$
\be
(\e+\eps\v)\cdot\r_x\leq c_x \quad\text{and}\quad (\e-\eps\v)\cdot\r_x\leq c_x\;.
\ee
The equations above implies that $\w_{\pm}\eqdef\e\pm\eps\v\in\mc$. On the other hand, $\e=\frac12(\w_++\w_-)$, so that $\e$ is not an extreme point.

Suppose now that $\spa\{\mk\}=\mbb{R}^n$, and suppose that $\e=t\v+(1-t)\w$ for some two vectors $\v,\w\in\mc$ and $t\in(0,1)$. By definition of $\mc$, we have $\r_x\cdot\v\leq c_x$ and $\r_x\cdot\w\leq c_x$ for all $x\in[m]$.
On the other hand, $\r_x\cdot\e=c_x$ for all $\r_x\in\mk$. That is, for any $\r_x\in\mk$ we have
\be
c_x=\r_x\cdot(\underbrace{t\v+(1-t)\w}_{\e})=t\r_x\cdot\v+(1-t)\r_x\cdot\w\leq tc_x+(1-t)c_x=c_x\;.
\ee
The above equation implies that 
$\r_x\cdot\v=\r_x\cdot\w=c_x$ for all $\r_x\in\mk$. Since $\spa\{\mk\}=\mbb{R}^n$ we must have that the linear system of equations $\r_x\cdot \v=c_x$, over all $x\in[m]$ with $\r_x\in\mk$ (here the components of $\v$ are viewed as the variables of the linear system of equations) has a unique solution. This means that $\v=\u=\e$, so that $\e$ is an extreme point. 
\end{proof}

It is natural to ask what is the relationship between polytopes and polyhedrons. Remarkably, if a polyhedron is bounded then it is a polytope.

\begin{myg}{}
\begin{corollary}\label{corpoly}
Convex polyhedrons have a finite number of extreme points and if they are bounded then they are polytopes (i.e. they are convex hulls of finitely many vertices).
\end{corollary}
\end{myg}

\begin{proof}
Let $\mc$ be a polyhedron as in~\eqref{polyhd}. From the Theorem~\ref{polyhdr} we know that $\e$ is an extreme point of $\mc$ if and only if $\e$ is a solution to the linear system of equations, $\r_x\cdot\e=c_x$, where $x$ is running over all $x\in[m]$ such that $\r_x\in\mk$. Since $\spa\{\mk\}=\mbb{R}^n$, the solution to each such linear system of equations is unique, and moreover, since $|\mk|\geq n$ there can be no more than ${m\choose n}$ extreme points (${m\choose n}$ is the number of $n$ distinct vectors that can be chosen from the set $\{\r_1,\ldots,\r_m\}$). Hence, polyhedrons have a finite number of extreme points. Now, if $\mc$ is also bounded it must be compact since convex polyhedrons are closed (see Exercise~\ref{ccpoly}). Hence, in this case, from Krein–Milman theorem (i.e. Theorem~\ref{kmt}) $\mc$ is the convex hull of its extreme points. Since we proved that $\mc$ has a finite number of vertices, $\mc$ must be a polytope.
\end{proof}

\section{Affine Subspaces and the Birkhoff Polytope}\label{sec:bir}

\begin{myd}{Affine Subspace}
\begin{definition}
Let $A$ be a subspace of a vector space $V$, and let $\v\in V$. The translation
\be
\mA\eqdef A+\v\eqdef\Big\{\w+\v\;:\;\w\in A\Big\}
\ee
is called an \emph{affine} subspace of $V$. The dimension of $\mA$ is defined as the dimension of $A$.
\end{definition}
\end{myd}
From its definition, it is clear that if $\v\in A$ then $\mA=A$.
The relevance of affine subspaces to our study here is that shifting a subspace by a fixed vector does not change any of the key properties of convex sets. Therefore, many of the theorems already covered in this chapter, can be generalized in a straightforward manner to incorporate affine subspaces. For example, in Theorem~\ref{polyhdr} we assume that the polyhedron $\mc$ is in $\mbb{R}^n$. Clearly, since all $n$-dimensional  vectors spaces over $\mbb{R}$ isomorphic to $\mbb{R}^n$ we can replace $\mbb{R}^n$ with any $n$ dimensional subspace $A$ of some vector space $V$, and moreover the theorem still holds if we replace $A$ with an affine subspace $\mA$ since by shifting a polyhedron by a fixed vector we do not change any of its properties. 

An affine subspace $\mA$ has the property that for any $\v_1,\ldots,\v_m\in\mA$ and any $m$ real numbers $t_1,\ldots,t_m\in\mbb{R}$ that satisfies $\sum_{x\in[m]}t_x=1$ we have
\be
\sum_{x\in[m]}t_x\v_x\in\mA\;.
\ee
Note that the coefficients $\{t_x\}$ can be negative (hence, in general, they do not form a probability vector). 

As an example of an affine subspace, consider the subspace $A\subset\mbb{R}^{n\times n}$ consisting of all the $n\times n$ real matrices whose rows and columns sum to zero. That is, $N=(\nu_{xy})\in A$ if and only if
\be
\sum_{x'\in[n]}\nu_{x'y}=\sum_{y'\in[n]}\nu_{xy'}=0\quad\quad\forall\;x,y\in[n]\;.
\ee
Then, the set 
\be\label{exaffine}
\mA\eqdef A+I_n\eqdef\Big\{N+I_n\;:\;N\in A\Big\}\;,
\ee
is an affine subspace of $\mbb{R}^{n\times n}$.
\begin{exercise}\label{exaf}
Let $A$ and $\mA$ be as above.
\begin{enumerate}
\item Show that $A$ above is indeed a subspace, and show that $|A|=(n-1)^2$.
\item Show that $M\eqdef(\mu_{xy})\in\mA$ if and only if its components satisfy
\be
\sum_{x'=1}^n\mu_{x'y}=\sum_{y\in[n]}\mu_{xy'}=1\quad\quad\forall\;x,y\in[n]\;.
\ee
\end{enumerate}
\end{exercise}

The affine subspace $\mA$ as defined in~\eqref{exaffine} contains the set of all doubly stochastic matrices. A doubly stochastic matrix\index{doubly stochastic matrix} is an $n\times n$ matrix whose components are non-negative and has the property that the entries of each row and column sums to one. The set of all $n\times n$ doubly stochastic matrices is a polytope in the real vector space $\mbb{R}^{n\times n}$, and we will denote it by $\mB_n$ (after Birkhoff). Doubly stochastic matrices appear quite often in several resource theories. 

\begin{exercise}
Let $\mB_n$ be the Birkhoff polytope.
\begin{enumerate}
\item Show that $\mB_n$ is indeed a polytope. Hint: Show first that it is a bounded polyhedron in $\mA$ as defined in~\eqref{polyhd} (with the dot product replaced by the Hilbert Schmidt inner product) and then use Corollary~\ref{corpoly}.
\item Show that any permutation matrix is an extreme point of $\mB_n$. Recall that the entries in each row or column of a permutation matrix consists of zeros except for one entry being equal to 1.
\end{enumerate}
\end{exercise}

The exercise above states that any permutation matrix is a vertex of $\mB_n$. It turns out that there are no other vertices for $\mB_n$.

\begin{myt}{\color{yellow} The Birkhoff-von Neumann Theorem}
\begin{theorem}\label{birkhoff}
The vertices (extreme points) of $\mB_n$ are exactly the permutation matrices.
\end{theorem}
\end{myt}

\begin{proof}
We will prove the theorem by induction. The case $n=1$ is trivial, so we assume now that $n>1$ and that the theorem holds for $(n-1)\times(n-1)$ doubly stochastic matrices. We will denote by $\mA$ the affine subspace~\eqref{exaffine}. Therefore, $M\eqdef(\mu_{xy})\in\mA$ is in $\mB_n$ if and only if its entries satisfies $\mu_{xy}\geq 0$. These $n^2$ inequalities defines the polyhedron $\mB_n$.
Now, according to Theorem~\ref{polyhdr}, if $M$ is an extreme point then the total number of equalities $\mu_{xy}=0$ must be at least $|\mA|=(n-1)^2$ (see first part of Exercise~\ref{exaf}). Now, since $\sum_{y\in[n]}\mu_{xy}=1$ for all $x\in[n]$, $M$ cannot contain a row (or column) with all zeros. On the other hand, suppose each row of $M$ has at least two non-zero components. In this case, the number of zero components of $M$ would not exceed $n(n-2)$ which is strictly smaller than $(n-1)^2$ (so that this case is not possible). We therefore conclude that at least one of the rows, say the $y$-row, has exactly one non-zero component, say the $x$-component of the $y$-row. This $(x,y)$-component must be equal to $1$ since the row sums to $1$. This in turn implies that in the $y$-column, except for the $x$-component, all the other components are zero. Therefore, crossing out the $x$-row and $y$-column results with an $(n-1)\times (n-1)$ doubly stochastic matrix\index{doubly stochastic matrix} that is also an extreme point. The proof is then concluded by the induction assumption.
\end{proof}

\begin{exercise}\label{ex:mun}
Show that any $n\times n$ doubly stochastic matrix can be expressed as a convex combination of $m\leq (n-1)^2+1$ permutation matrices. Hint: Use the above arguments in conjunction with Carath\'eodory's Theorem. 
\end{exercise}

\section{Polarity and Half Spaces}

\begin{myd}{}
\begin{definition}
Let $\mc\in\mbb{R}^n$ be a non-empty set. The set
\be
\mc^{\circ}\eqdef\Big\{\w\in\mbb{R}^n\;:\;\w\cdot\v\leq 1\;\text{ for all }\;\v\in\mc\big\}
\ee
is called the \emph{polar} of $\mc$.
\end{definition}
\end{myd}

Note that irrespective of the set $\mc$, the polar $\mc^{\circ}$ is always a closed convex set that contains the zero vector. It is also straightforward to check that the polar of $\mbb{R}^n$ is the zero vector and the polar of the set consisting only of the zero vector is the whole space $\mbb{R}^n$.

\begin{exercise}\label{expolar}
Let $\mc,\mk\in\mbb{R}^n$ be two non-empty sets. 
\begin{enumerate}
\item Show that if $\mc\subseteq\mk$ then $\mk^{\circ}\subseteq\mc^{\circ}$.
\item Show that $\mc\subseteq\left(\mc^{\circ}\right)^{\circ}$.
\item Suppose $\mc=\conv\{\v_1,\ldots,\v_m\}$ is a polytope with $m$ vertices $\v_1,\ldots,\v_m\in\mbb{R}^n$. Show that
\be\label{polyh}
\mc^{\circ}\eqdef\Big\{\w\in\mbb{R}^n\;:\;\w\cdot\v_x\leq 1\;\text{ for all }\;x\in[m]\big\}
\ee
\end{enumerate}
\end{exercise}
Note that from the third part of the exercise above we see that the polar of a polytope is a polyhedron.

\begin{myt}{\color{yellow}The Bipolar Theorem}\label{bipolar}
\begin{theorem}\label{bipolarthm}
Let $\mc\subseteq\mbb{R}^n$ be a closed convex set that contains the zero vector. Then, $\mc=\left(\mc^{\circ}\right)^{\circ}$.
\end{theorem}
\end{myt}
\begin{proof}
The part $\mc\subseteq\left(\mc^{\circ}\right)^{\circ}$ was given in the exercise above. We therefore prove here that $\left(\mc^{\circ}\right)^{\circ}\subseteq\mc$. Suppose by contradiction that there exists a vector $\w\in\left(\mc^{\circ}\right)^{\circ}$ that is not in $\mc$. Then, since $\mc$ is closed convex set, from the hyperplane separation theorem (see Theorem~\ref{hyper}) there exists a vector $\r\in\mbb{R}^n$ and a constant $c\in\mbb{R}$ such that $\w\cdot\r>c>\v\cdot\r$ for all $\v\in\mc$. Since the zero vector belongs to $\mc$, by taking $\v=\0$ we get $c>0$. Therefore, defining $\s\eqdef\frac1c\r$ we conclude that
both $\w\cdot\s>1$ and $1>\v\cdot\s$ for all $\v\in\mc$. The latter implies that $\s\in\mc^{\circ}$, but then the former implies that $\w\not\in(\mc^{\circ})^{\circ}$ in contradiction with our assumption. This completes the proof.
\end{proof}

\begin{exercise}\label{expolar2}
Let $\eps>0$ and $\mb_{\eps}(\0)\eqdef\{\v\in\mbb{R}^n\;:\;\|\v\|\leq\eps\}$ be a ball of radius $\eps$ (in the Euclidean norm). Show that
\be
\mb_{\eps}(\0)^{\circ}=\mb^{1/\eps}(\0)\;.
\ee
\end{exercise}

\begin{myt}{\color{yellow} The Polar of a Polytope}
\begin{theorem}\label{popoly}
Let $\mc$ be a polytope in $\mbb{R}^n$. Then, $\mc^{\circ}$ is also a polytope in $\mbb{R}^n$.
\end{theorem}
\end{myt}

\begin{proof}
Without loss of generality we assume that the interior of $\mc$ is not empty, and furthermore we assume that the zero vector is in the interior of $\mc$ (otherwise,  we can shift $\mc$ so that the origin of the coordinate system is in its interior). Therefore, there exists $\eps>0$ such that $\mb_{\eps}(\0)\subset\mc$. From Exercise~\ref{expolar} 
this implies that 
\be
\mc^{\circ}\subset\mb_{\eps}(\0)^{\circ}=\mb^{1/\eps}(\0)\;,
\ee
where the last equality follows from Exercise~\ref{expolar2}.
That is, $\mc^{\circ}$ is a bounded polyhedron.  From Corrolary~\ref{corpoly} it must be a polytope. 
\end{proof}

Every hyperplane separates $\mbb{R}^n$ into two half spaces.
A closed half-space of $\mbb{R}^n$ is therefore the set of all vector $\v\in\mbb{R}^n$ that satisfies $\n\cdot\v\leq c$ for some fixed $c\in\mbb{R}$ and a fixed (normal) vector $\n\in\mbb{R}^n$. Therefore, a convex polyhedron as defined in~\eqref{polyhd} can be viewed as the intersection of finitely many half-spaces. Similarly, in the following theorem we show that a convex polytope can also be expressed as the intersection of finitely many half-spaces (the half-spaces that are defined by its facets); see Fig.~\ref{figa2}. This means in particular that every polytope is a polyhedron (recall that the converse of this assertion is also true if the polyhedron is bounded). 

\begin{figure}[h]\centering    \includegraphics[width=0.2\textwidth]{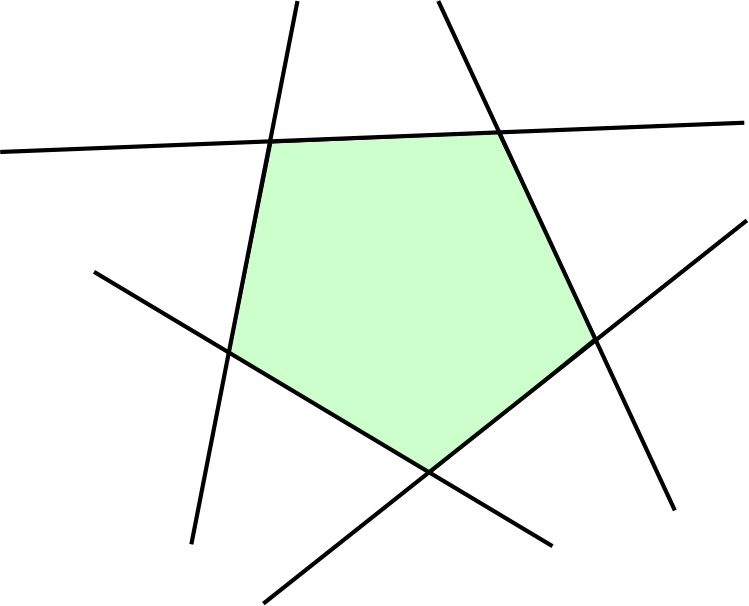}
  \caption{\linespread{1}\selectfont{\small Intersection of half-spaces}}
  \label{figa2}
\end{figure} 

\begin{myt}{}
\begin{theorem}\label{mink}
Let $\mc\eqdef\conv\{\v_1,\ldots,\v_m\}$ be a convex polytope in $\mbb{R}^n$ that contains the zero vector. Then there exist $k\in\mbb{N}$ vectors $\s_1,\ldots,\s_k\in\mbb{R}^n$ such that $\v\in\mc$ if and only if
\be
\s_x\cdot\v\leq 1\quad\forall x=1,\ldots,k\;.
\ee
In other words, $\mc$ is the intersection of $k$ half-spaces, and the facets of $\mc$ are given by
\be
\mf_x\eqdef\Big\{\v\in\mc\;:\;\v\cdot\s_x=1\quad\quad\forall\;x\in[k]\Big\}\;.
\ee
\end{theorem}
\end{myt}
\begin{remark}
The condition that $\mc$ contains the zero vector is just for convenience and in fact unnecessary. Specifically, if $\0\not\in\mc$ then the theorem still hold if we replace the equations $\s_x\cdot\v\leq 1$  with $\s_x\cdot\v\leq r_x$, where $\r_x$ are some real numbers (see Exercise~\ref{exsxrx}).
\end{remark}

\begin{proof}
From Theorem~\ref{popoly}, the polar of $\mc$ is itself a polytope. Therefore, there exists $k\in\mbb{N}$, and $\s_1,\ldots,\s_k\in\mbb{R}^n$, such that $\mc^{\circ}=\conv\{\s_1,\ldots,\s_k\}$. From the bipolar theorem (Theorem~\ref{bipolarthm}) we get
\ba
\mc&=\left(\mc^{\circ}\right)^{\circ}\\
\GG{\eqref{polyh}}&=\Big\{\v\in\mbb{R}^n\;:\;\v\cdot\s_x\leq 1\quad\quad\forall\;x\in[k]\Big\}\;.
\ea
This completes the proof.
\end{proof}

\bex\label{exsxrx}
Show that the theorem above still holds even if $\0\not\in\mc$ as long as  the equations $\s_x\cdot\v\leq 1$ are replaced with $\s_x\cdot\v\leq r_x$, where $\r_x$ are some real numbers
\eex

\begin{myt}{\color{yellow} The Supporting Hyperplane Theorem}
\begin{theorem}\label{supporting}
Let $\mc\subset\mbb{R}^n$ be a convex set and let $\v$ be a point on its boundary (i.e.\ $\v$ belongs to the closure of $\mc$ and not belong to the interior of $\mc$). Then, there exists $\s\in\mbb{R}^n$ with the property that
\be
\v'\cdot\s\geq\v\cdot\s\quad\quad\forall\;\v'\in\mc\;.
\ee
\end{theorem}
\end{myt}
\bex
Prove the supporting hyperplane theorem above. Hint: Use Theorem~\ref{hyper}. 
\eex

\section{Support Functions}\label{ss:support}\index{support function}

\begin{myd}{}
\begin{definition}
Let $\mc$ be a compact subset of $\mbb{R}^n$. Then, the \emph{support function} of $\mc$, $f_\mc:\mbb{R}^n\to\mbb{R}$ is defined for all $\v\in\mbb{R}^n$ as
\be
f_\mc(\v)=\max\big\{\v\cdot\r\;:\;\r\in\mc\big\}\;.
\ee
\end{definition}
\end{myd}

Note that if $\mc=\conv(\r_1,\ldots,\r_k)$ is the convex hull of $k$ vectors then
\be
f_\mc(\v)=\max_{j\in[k]}\v\cdot\r_j\;,
\ee
is a sublinear functional. One of the most useful facts about support functions is the following theorem.

\begin{myt}{}
\begin{theorem}\label{inclusion}
Let $\mc_1$ and $\mc_2$ be two subsets of $\mbb{R}^n$. Suppose also that $\mc_1$ is closed, compact, and convex.
Then,
\be
\mc_1\supseteq\mc_2\iff f_{\mc_1}(\v)\geq f_{\mc_2}(\v)\quad\forall\;\v\in\mbb{R}^n\;.
\ee
\end{theorem}
\end{myt}

\begin{proof}
The direction that $\mc_1\supseteq\mc_2$ implies $f_{\mc_1}\geq f_{\mc_2}$ follows trivially from the definition.
On the other hand, if $\mc_1\not\supseteq\mc_2$ then there exists a vector $\r\in\mc_2$ such that $\r\not\in\mc_1$. Hence, the sets $\{\r\}$ and $\mc_1$ are two disjoint closed compact convex sets of $\mbb{R}^n$. From the hyperplane separation theorem (see Theorem~\ref{hyper}) there exists $c\in\mbb{R}$ and a vector $\n\in\mbb{R}^n$ such that 
\be
\n\cdot\r'<c<\n\cdot\r\quad\quad\forall\;\r'\in\mc_1
\ee
Taking the maximum over $\r'\in\mc_1$ gives
\be
f_{\mc_1}(\n)< c<\n\cdot\r\leq f_{\mc_2}(\n)\;.
\ee
Hence, $f_{\mc_1}\not\geq f_{\mc_2}$. This completes the proof.

\end{proof}
\begin{myg}{}
\begin{lemma}\label{sumation}
Let $\mc_1$ and $\mc_2$ be two compact convex sets of $\mbb{R}^n$. Then, their support functions satisfy
\be
f_{\mc_1+\mc_2}(\v)=f_{\mc_1}(\v)+ f_{\mc_2}(\v)\quad\forall\;\v\in\mbb{R}^n\;.
\ee
\end{lemma}
\end{myg}

\begin{proof}
By direct calculation we have
\ba
f_{\mc_1+\mc_2}(\v)&=\max_{\r_1\in\mc_1\;,\; \r_2\in\mc_2}\v\cdot(\r_1+\r_2)\\
&=\max_{\r_1\in\mc_1}\v\cdot\r_1+\max_{\r_2\in\mc_2}\v\cdot\r_2\\
&=f_{\mc_1}(\v)+ f_{\mc_2}(\v)\;.
\ea
\end{proof}

\section{Convex Cones}\label{sec:dualcone}

\begin{myd}{Cone and Dual Cone}
\begin{definition}
A subset $\mk\subseteq A$ of a real Hilbert space $A$ is called a cone if for any non-negative real number $t\geq 0$, and any $\v\in\mk$ we have $t\v\in\mk$. The dual of a cone $\mk$ in $A$ is the set
\be\label{conedual}
\mk^*\eqdef\left\{\v\in A\;:\;\v\cdot\w\geq 0\;\text{ for all }\;\w\in\mk\right\}\;.
\ee
\end{definition}
\end{myd}
It is simple to check (see Exercise~\ref{excone}) that $\mk^*$ is both closed and convex.

\begin{exercise}\label{excone}
Let $A$ be a Hilbert space.
\begin{enumerate}
\item Show that if $\mk\subseteq A$ is a cone then $\mk^*$ is closed and convex. 
\item Show that if $\mk_1,\mk_2\subseteq A$ are two cones such that $\mk_1\subseteq\mk_2$ then $\mk_2^*\subseteq\mk_1^*$.
\end{enumerate}
\end{exercise}

\begin{example}
Let $A$ be a Hilbert space and consider the space $\herm(A)$. Recall that $\herm(A)$ represents the (real) vector space of all Hermitian matrices
acting on a Hilbert space $A$. Since $\herm(A)\cong\mbb{R}^{n}$, with $n\eqdef|A|^2$, the definition of a cone and dual cone can be applied to the vector space $\herm(A)$. An important example of a cone in this space is the cone of positive semidefinite matrices, $\mk\eqdef\pos(A)$. This is a cone since if $\Lambda\in\herm(A)$ is positive semidefinite, i.e. $\Lambda\geq 0$, then also $t\Lambda\geq 0$ for all $t\geq 0$. Intrestingly, this cone is a \emph{self dual} cone in the sense that $\mk^*=\mk$ (see the exercise below).
\end{example}

\begin{exercise}
Let $A$ be a Hilbert space and consider the space $\herm(A)$.
\begin{enumerate}
\item Show that the cone of positive semidefinite matrices is a self dual cone.
\item Show that the dual cone of the whole space $\mk\eqdef\herm(A)$ is $\mk^*=\{\0\}$ where $\0$ is the zero matrix in $\herm(A)$.
\end{enumerate}
\end{exercise}

\begin{myt}{}
\begin{theorem}\label{thm:closedconv}
Let $\mk\subseteq A$ be a cone in a Hilbert space $A$. Then, $\mk^{**}$ is the closer of the smallest convex cone containing $\mk$. In particular, if $\mk$ is a closed convex cone then $\mk^{**}=\mk$.
\end{theorem}
\end{myt}

\begin{proof}
Let $\mc$ be the closer of the smallest convex set containing $\mk$.
By the definition of a dual cone in~\eqref{conedual}, if $\w\in\mk$ then for all $\v\in\mk^*$ we must have $\w\cdot\v\geq 0$. On the other hand,
\be
\mk^{**}\eqdef\left\{\u\in A\;:\;\u\cdot\v\geq 0\;\text{ for all }\;\v\in\mk^*\right\}\;.
\ee
Therefore, if $\w\in\mk$ we must have $\w\in\mk^{**}$ so that $\mk\subseteq\mk^{**}$. Since $\mk^{**}$ is a closed convex set it must contain $\mc$.  Now, suppose by contradiction that  the inclusion $\mc\subseteq\mk^{**}$ is strict. That is, there exists $\v\in\mk^{**}$ that is not in $\mc$. Then, from the hyperplane separation theorem (see Theorem~\ref{hyper}) there exists a vector $\w\in\mbb{R}^n$ such that
\be\label{beee}
\w\cdot\v<\mu\eqdef\min_{\u\in\mc}\w\cdot\u\;.
\ee  
Since the zero vector belongs of $\mc$ we have in particular that $\mu\leq 0$ and $\w\cdot\v<0$. We argue next that $\mu$ must be zero. Otherwise, $\mu<0$ so that there exists $\r\in\mc$ with $\w\cdot\r<0$. But since $\mc$ is a cone also $t\r\in\mc$ for any $t>0$ so we get from the definition of $\mu$ that 
$
\mu\leq\w\cdot(t\r)
$
which goes to $-\infty$ as $t\to\infty$. This is not possible since according to~\eqref{beee} $\mu$ is bounded from below. We therefore conclude that $\mu=0$. This in turn implies that $\w\in\mc^*\subseteq\mk^{*}$ (where we used the second part of Exercise~\ref{excone} in conjunction with the fact that $\mk\subseteq\mc$). However, recall that $\v\in\mk^{**}$ which implies in particular that $\v\cdot\w\geq 0$, in contradiction with $\w\cdot\v<0$. Therefore, our initial assumption that $\v\not\in\mc$ was incorrect. This completes the proof.
\end{proof}

\section{Conic Linear Programming and Semidefinite Programming}\label{app:sdp}\index{linear programming}

Conic linear programming (CLP) and particularly semidefinite programming (SDP) have been used quite often in the field of quantum information science, as many of the optimization problems involve linear functions. In this short subsection, we will only mention a few useful properties of the vast field of CLP and SDP that will be useful to us later on. A reader who is interested in more details on the subject will find some useful references in the section on `History and further reading' at the end of this chapter.

In this book we will encounter many optimization problems that can be expressed in terms of two cones.
Specifically, let $A_1$ and $A_2$ be two Hilbert spaces, let  $\mk_1\subseteq {V}_1\subseteq \herm(A_1)$ and $\mk_2\subseteq V_2\subseteq \herm(A_2)$ be two convex cones in two subspaces of Hermitian matrices $V_1$ and $V_2$, and let $\mN:V_1\to V_2$ be a linear map between the two vector spaces.   Let also $H_1\in V_1$ and $H_2\in V_2$ be two (fixed) Hermitian matrices. With these notations, many of the optimization problems that we will encounter in this book can be expressed as the following problem which we will call here the \emph{primal} problem. 

\begin{myd}{The Primal Problem}
\ba
\text{Find}\quad & \alpha\eqdef\inf\tr\left[\eta H_1\right]\label{primal}\\
\text{Subject to}\quad &\mN(\eta)-H_2\in\mk_2\quad\text{and}
\quad \eta\in\mk_1
\ea
\end{myd}
\begin{remark}
The primal problem above has been expressed with respect to two vector spaces of Hermitian matrices $V_1$ and $V_2$ since these are what we typically encounter in quantum physics. However, everything that we will discuss in this section is also applicable for \emph{any} finite dimensional abstract Hilbert spaces
$V_1$ and $V_2$ by replacing the Hilbert-Schmidt inner product above with the inner product of the vector space $V_1$. 
\end{remark}
Any $\eta\in\mk_1$ that satisfies $\mN(\eta)-H_2\in\mk_2$ is called a \emph{feasible plane} or a \emph{primal feasible plane}. If there are no feasible planes then by convention $\alpha\eqdef+\infty$. Moreover, a primal feasible plane $\eta$ is said to be \emph{optimal} if $\tr[\eta H_1]=\alpha$.

The optimization problem above is  a conic linear program (CLP), and
if the cones $\mk_1$ and $\mk_2$ are $\pos(A_1)$ and $\pos(A_2)$, respectively, then~\eqref{primal} is known as an SDP optimization problem. The latter can be solved efficiently with the help of one of the many SDP algorithms studied in literature. While we will not study in this book the SDP algorithms themselves, whenever an optimization problem can be formulated in an SDP we will simply say that the problem can be solved efficiently (and algorithmically) on a computer.

There are many types of CLP and SDP optimization problems. Still, almost all of them can be formulated as above. For example, consider the SDP optimization problem:
\ba
\text{Find}\quad & \alpha\eqdef\inf\tr\left[\eta H\right]\\
\text{Subject to}\quad &\tr[\eta\omega_j]=c_j\quad\forall\;j\in[m]\nonumber\\
&\eta\in\herm(A)\;,\nonumber
\ea
where $A$ is some Hilbert space, $\{c_j\}$ are some constants in $\mbb{R}$, and $H$ and $\{\omega_j\}$ are fixed Hermitian matrices in $\herm(A)$. We would like to show that this problem can be formulated as in~\eqref{primal}. For this purpose, define $A_1\eqdef A$, and $\mk_1\eqdef\herm(A)$. Let $V_2$ be the vector space of $m\times m$ diagonal (Hermitian) matrices, and define a linear map $\mN:\herm(A)\to V_2$ via
\be
\mN(\eta)\eqdef\diag\Big(\tr[\eta\omega_1] ,\tr[\eta\omega_2],\ldots, \tr[\eta\omega_m]\Big)
\quad\quad\forall\;\eta\in\herm(A)\;.
\ee
Note that this map is indeed linear.
Denoting by $C$ the diagonal matrix with diagonal $(c_1,\ldots,c_m)$, and by $\mk_2$ the cone in $V_2$ that contains only the zero matrix, we can express the problem above as
\ba
\text{Find}\quad & \alpha\eqdef\inf\tr\left[\eta H\right]\\
\text{Subject to}\quad &\mN(\eta)-C\in\mk_2\quad\text{and}\quad\eta\in\mk_1
\ea
which has the same form as~\eqref{primal} after identifying $C$ with $H_2$ and $H$ with $H_1$. 

The primal problem~\eqref{primal} can also be expressed with a \emph{single} cone. This can be done as follows. Consider the vector space $V\eqdef V_1\oplus V_2$ and the cone $\mk=\mk_1\times\mk_2\subseteq V$. Denote also by $\tilde{H}_1\eqdef H_1\oplus\0\in V$, and by $\tilde{\mN}:V\to V_2$ the linear map
\be
\tilde{\mN}(\xi)\eqdef \mN(\eta)-\zeta\quad\quad\forall\;\xi\eqdef\eta\oplus\zeta\in V\;.
\ee  
Note that with these notations we have that $\tr[\eta H_1]=\tr[\xi\tilde{H}_1]$, and the condition $\mN(\eta)-H_2\in\mk_2$ is equivalent to $\tilde{\mN}(\xi)=H_2$ since the latter is equivalent to $\mN(\eta)-H_2=\zeta\in\mk_2$ (we assume that $\xi$ is arbitrary element of $\mk$). Hence, we get that the primal problem above can be expressed as
\ba
\text{Find}\quad & \alpha\eqdef\inf\tr\left[\xi \tilde{H}_1\right]\label{primal2}\\
\text{Subject to}\quad &\tilde{\mN}(\xi)=H_2\quad\text{and}
\quad \xi\in\mk%\nonumber
\ea
That is, we were able to express the primal problem as an optimization problem that involves a single cone. In fact, note that the optimization problem above has the exact same form as the primal problem if we take $\mk_2$ in the primal problem to be the cone consisting only of the zero matrix.

\subsection{Duality}
 
Every primal CLP optimization problem has a dual problem. The dual problem\index{dual problem} of the primal CLP problem given in~\eqref{primal} is given as follows.

\begin{myd}{The Dual Problem}
\ba
\text{Find}\quad & \beta\eqdef\sup\tr\left[\zeta H_2\right]\\
\text{Subject to}\quad &H_1-\mN^*(\zeta)\in\mk_1^*\quad\text{and}\quad\zeta\in\mk_2^*\label{dual123}
\ea
\end{myd}

Any $\zeta\in\mk_2^*$ that satisfies $H_1-\mN^*(\zeta)\in\mk_1^*$ is called a \emph{dual feasible plane}. If there are no dual feasible planes than by convention $\beta\eqdef-\infty$.

\begin{exercise}\label{exdual2}
Show that with the notations of~\eqref{primal2}, the dual problem can be expressed as
\ba
\text{Find}\quad & \beta\eqdef\sup\tr\left[\zeta H_2\right]\\
\text{Subject to}\quad &\tilde{H}_1-\tilde{\mN}^*(\zeta)\in\mK^*\quad\text{and}\quad\zeta\in V_2\nonumber
\ea
\end{exercise}

The significance of the dual problem\index{dual problem} is that quite frequently $\alpha=\beta$. We start first by showing that $\alpha\geq\beta$.

\begin{myg}{Weak Duality} 
\begin{lemma}
For any primal feasible plane $\eta$, and dual feasible plane $\zeta$,  we have
\be\label{0h3}
\tr\left[\eta H_1\right]\geq  \tr\left[\zeta H_2\right]
\ee
That is, $\alpha\geq\beta$.
\end{lemma}
\end{myg}

\begin{proof}
Let $\eta$ and $\zeta$ be as in the lemma. Since $H_1-\mN^*(\zeta)\in\mk_1^*$ and $\eta\in\mk_1$ we have from the definition of a dual cone that the inner product
\be
\tr\left[\big(H_1-\mN^*(\zeta)\big)\eta\right]\geq 0\;.
\ee
This inequality can be expressed as
\be\label{0h1}
\tr[H_1\eta]\geq\tr[\zeta\mN(\eta)]\;.
\ee
On the other hand, since $\mN(\eta)-H_2\in\mk_2$, and $\zeta\in\mk_2^*$ we have that the inner product
\be
\tr\left[\big(\mN(\eta)-H_2\big)\zeta\right]\geq 0\;.
\ee
The above inequality can be expressed as
\be\label{0h2}
\tr[\zeta\mN(\eta)]\geq \tr\left[\zeta H_2\right]\;.
\ee
Combining~\eqref{0h1} and~\eqref{0h2} produce~\eqref{0h3}. This completes the proof.
\end{proof}

\begin{exercise}
Let $\eta\in\mk_1$ be such that $\mN(\eta)-H_2\in\mk_2$, and let $\zeta\in\mk_2^*$ be such that $H_1-\mN^*(\zeta)\in\mk_1^*$. Show that if in addition
\be
\tr\left[\big(H_1-\mN^*(\zeta)\big)\eta\right]=\tr\left[\big(\mN(\eta)-H_2\big)\zeta\right]= 0
\ee
then $\alpha=\beta$.
\end{exercise}

\begin{exercise}\label{exdualinf}
Show that if $\alpha=-\infty$ there are no dual feasible planes, and if $\beta=+\infty$ there are no primal feasible planes. 
\end{exercise}

The following theorem, known also as the strong duality theorem, is the key result of this section that we will use quite often in the book. It provides a sufficient condition for $\alpha=\beta$ to hold. We will use the notation ${\rm int}(\mk)$ to denote the \emph{interior} of a cone $\mk$.

\begin{myt}{\color{yellow} Strong Duality} 
\begin{theorem}
We have $\alpha=\beta$ if one of the following two conditions hold:
\begin{enumerate}
\item $\mk_1$ and $\mk_2$ are closed convex cones and there exists a primal feasible plane.
\item There exists $\eta\in{\rm int}(\mk_1)$ that satisfies $\mN(\eta)-H_2\in{\rm int}(\mk_2)$ (this is known as Slater's condition), and here exists a primal optimal plane. 
\end{enumerate}
\end{theorem}
\end{myt}

It turns out that in all the problems that we will consider in this book these mild conditions (also known as Slater's conditions) will hold so that $\alpha=\beta$.

\begin{proof}
If $\alpha=-\infty$ then from Exercise~\ref{exdualinf} there are no dual feasible planes so that by convention $\beta=-\infty$. Hence, in this case $\alpha=\beta$. We therefore consider now the case $\alpha>-\infty$ (i.e. $\alpha$ is bounded from below) and prove the sufficiency of the first condition.

From~\eqref{primal2} and Exercise~\ref{exdual2} it is sufficient to prove the theorem for the case $\mk_2=\{\0\}$. We will therefore denote by $\mk\eqdef\mk_1$ and assume that it is closed. Consider the convex cone
\be
\mc\eqdef\Big\{\big(\mN(\eta),\tr[\eta H_1]\big)\;:\;\eta\in\mk\Big\}\subset V_2\oplus\mbb{R}\;.
\ee
Since the set $\mk$ is closed, also the set $\mc$ is closed in $V_2\oplus\mbb{R}$ (recall that we are working in finite dimensions).
Note that any $\eta\in\mk$ that satisfies $\mN(\eta)=H_2$ results with a point $(H_2,\tr[\eta H_1])\in\mc$. We therefore interested in the intersection of the cone $\mc$ with the line
\be
\ml\eqdef\Big\{(H_2,t)\;:\;t\in\mbb{R}\Big\}\;.
\ee
The intersection $\mc\cap\ml$ consists of points of the form $\{(H_2,\tr[\eta H_1])\}$ over all primal feasible planes $\eta$. This intersection is closed (since both $\ml$ and $\mc$ are closed), and is not empty since there is a primal feasible plane. Moreover, since the set  of numbers $\{\tr[\eta H_1]\}$ over all primal feasible planes $\eta$ is bounded from below (recall $\alpha>-\infty$), there exists a feasible optimal plane $\eta_0$ such that $\alpha=\tr[\eta_0 H_1]$. In the rest of the proof, $\eta_0$ will denote this feasible \emph{optimal} plane.

From the Weak Duality Lemma we know that $\alpha\geq\beta$. To show the converse, we will show that for any $\eps>0$ we have $\beta\geq\alpha-\eps$ so that we must have $\alpha=\beta$. Set $\eps>0$ and
observe that from its definition, the point $(H_2,\alpha-\eps)\not\in\mc$. Therefore, from the hyperplane separation theorem (Theorem~\ref{hyper}) there exists a hyperplane $(\zeta,s)\in V_2\oplus\mbb{R}$ and a constant $c\in\mbb{R}$ such that
\be\label{shyps}
\tr[\mN(\eta) \zeta]+s\tr[\eta H_1]<c<\tr[\zeta H_2]+s(\alpha-\eps)\quad\quad\forall\;\eta\in\mk\;.
\ee
Note that on the left-hand side we have the inner product between $(\zeta,s)$ and $(\mN(\eta),\tr[\eta H_1])\in\mc$, and on the right-hand side the inner product between $(\zeta,s)$ and $(H_2,\alpha-\eps)$. 
Since we can take $\eta=\0$ we must have $c>0$. On the other hand, if we take $\eta=\eta_0$ the left-hand side becomes $\tr[H_2\zeta]+s\alpha$ and when comparing it with the right-hand side we conclude that $s<0$. Moreover, since the rescaling $\zeta\mapsto\frac1{|s|}\zeta$ and $s\mapsto\frac{s}{|s|}$ does not change the inequalities above, we can assume without loss of generality that $s=-1$. Therefore, since $c>0$ the right-hand side of the equation above gives
\be
\tr[\zeta H_2]>\alpha-\eps\;.
\ee
It is therefore left to show that $\zeta$ is a dual feasible plane so that $\beta$, which is defined as the supremum of $\tr[\zeta H_2]$ over all dual feasible planes, is also greater than $\alpha-\eps$. Indeed, since $\mk$ is a cone, we must have 
\be
\tr[\mN(\eta) \zeta]-\tr[\eta H_1]\leq 0\quad\quad\forall\;\eta\in\mk\;.
\ee
Otherwise, if for some $\eta\in\mk$ the left-hand side above is positive, then the inequality on the left-hand side of~\eqref{shyps} (with $s=-1$) will be violated for $t\eta$ with $t$ a positive real number that is sufficiently large. 
The equation above can be expressed as
\be
\tr\Big[\eta\big(H_1-\mN^*(\zeta)\big)\Big]\geq 0\quad\quad\forall\;\eta\in\mk\;,
\ee
which is equivalent to $H_1-\mN^*(\zeta)\in\mk^*$. Hence, $\zeta$ is a dual feasible plane. This completes the proof of the sufficiency of the first condition. For the second condition see Exercise~\ref{slater}.
\end{proof}

\begin{exercise}\label{slater}
Prove the sufficiency of the second condition (slater's condition) in the theorem above. Hint: Define $\mc$ as in the proof above but with ${\rm int}(\mk)$ replacing $\mk$, and use the version in~\eqref{hyperg} of the hyperplane separation theorem.
\end{exercise}

\section{Fixed-Point Theorem}

We end this short review on convex analysis with an important result in analysis known as  \emph{Brouwer's fixed-point theorem}. We will only state the theorem without proving it as the proof involves material that is not covered in this book. 

\begin{myt}{\color{yellow} Brouer's Fixed-Point Theorem} 
\begin{theorem}\label{fixedpoint}
Let $\n\in\mbb{N}$, $\mc\subset\mbb{R}^n$ a compact convex set, and $f:\mc\to\mc$ be a continuous function. Then, there exists $\v\in\mc$ such that
\be
f(\v)=\v\;.
\ee
\end{theorem}
\end{myt}

In quantum information this theorem is typically used for functions from density matrices to density matrices. One example of such \emph{linear} functions are quantum channels. However, observe that the theorem above holds for all continuous functions (not only linear ones).

\section{Notes and References}

There are many excellent books on convex analysis. We followed~\cite{Barvinok2002} in the presentation of conic linear programming. Also~\cite{Watrous2018} gives an overview on semidefinite programming with a focus on problems in quantum information.

%%%%%%%%%%%%%%%%%%%%%
%%%%%%%%%%%%%%%%%%%%%%

\chapter{Operator Monotonicity and Operator Convexity}\label{sec:oc}

Let $\mI$ be an interval in $\mbb{R}$, and consider a real function $f:\mI\to\mbb{R}$.
Any such function can be extended to Hermitian matrices whose eigenvalues are in $\mI$. This can be done as follows. First, if $D\eqdef {\rm Diag}(\lambda_1,\ldots,\lambda_n)$ is an $n\times n$ diagonal matrix with the diagonal entries $\lambda_1,\ldots,\lambda_n\in\mI$ then $f(D)\eqdef{\rm Diag}\big(f(\lambda_1),\ldots,f(\lambda_n)\big)$. If $\eta$ is an Hermitian matrix with eigenvalues $\lambda_1,\ldots,\lambda_n\in\mI$, it can be expressed as $\eta=UDU^*$, where $D\eqdef {\rm Diag}(\lambda_1,\ldots,\lambda_n)$ and $U$ is a unitary matrix.  We then define $f(\eta)\eqdef Uf(D)U^*$. In the rest of this section, we will always assume this extension of a real function $f:\mI\to\mbb{R}$ to Hermitian matrices.

\begin{exercise}\label{relfmmst}
Let $M\in\mbb{C}^{n\times n}$ be a square complex matrix, and let $\mI$ be an interval in $\mbb{R}$ containing the eigenvalues of $MM^*$. Show that for any function $f:\mI\to\mbb{R}$ we have
\be\label{b1ex}
Mf(M^*M)=f(MM^*)M\;.
\ee
Hint: Use the singular value decomposition  $M=UDV$ (and in particular $M^*=V^*DU^*$), and express separately both sides of the equation above in terms of $U$, $D$, and $V$.
\end{exercise}

\section{Definitions and Basic Properties}

\begin{myd}{}
\begin{definition}
Let $\mI$ be an interval in $\mbb{R}$, and $f:\mI\to\mbb{R}$ a real function. 
\begin{enumerate}
\item We say that $f$ is \emph{operator monotone} if for every Hilbert space $A$ and any $\eta,\zeta\in\herm(A)$ that satisfies $\eta\geq\zeta$ we have
$
f(\eta)\geq f(\zeta)
$.
\item We say that $f$ is \emph{operator convex} if for every Hilbert space $A$, $\eta,\zeta\in\herm(A)$, and  $t\in[0,1]$ 
\be\label{ocon}
f(t\eta+(1-t)\zeta)\leq tf(\eta)+(1-t)f(\zeta)\;.
\ee
\end{enumerate}
\end{definition}
\end{myd}
\begin{remark}
In addition, we say that $f$ is \emph{operator anti-monotone} if $-f$ is operator monotone, and \emph{operator concave} if $-f$ is operator convex.
\end{remark}

Not every monotonic function is operator monotone, and not every convex function is operator convex. For example, consider the function $f(r)=r^2$. This function is monotonically increasing in the domain $[0,\infty)$. However, it is \emph{not} operator monotone. To see why, consider the matrices 
\be
\eta=\begin{pmatrix}2 & 1\\
1 & 1\end{pmatrix}\quad\text{and}\quad \zeta\eqdef\begin{pmatrix}1 & 1\\
1 & 1\end{pmatrix}\;.
\ee
Observe that $\eta-\zeta=\begin{pmatrix}1 & 0\\
0 & 0\end{pmatrix}\geq 0$ so that $\eta\geq\zeta$. On the other hand, a simple calculation reveals that
\be
f(\eta)-f(\zeta)=\eta^2-\zeta^2=\begin{pmatrix}3 & 1\\
1 & 0\end{pmatrix}\not\geq 0\;.
\ee
Hence, $f$ is not operator monotone.

As another example, consider the function $f(r)=r^3$. This function is convex on the interval $[0,\infty)$. However, it is not operator convex on that interval.
\begin{exercise}
Show that the function $f(r)=r^3$ is not operator convex on the interval $[0,\infty)$.
Hint: Take $\eta=\begin{pmatrix}3 & 1\\
1 & 1\end{pmatrix}$, $\zeta=\begin{pmatrix}1 & 1\\
1 & 1\end{pmatrix}$, and $t=\frac12$.
\end{exercise}

\begin{exercise}
Show that the function $f(r)=a+br$ (defined on any interval) is operator monotone for any $a\in\mbb{R}$ and $b\geq 0$. Show that it is operator convex on any $a,b\in\mbb{R}$. 
\end{exercise}
\begin{exercise}
Let $f_1,f_2:\mI\to\mbb{R}$ be two real functions and define for any $r\in\mI$, $f(r)\eqdef af_1(r)+bf_2(r)$ for some fixed non-negative real numbers $a,b\in\mbb{R}_+$.
\begin{enumerate}
\item Show that if $f_1$ and $f_2$ are operator convex then $f$ is operator convex.
\item Show that if $f_1$ and $f_2$ are operator monotones then $f$ is operator monotone.
\end{enumerate}
\end{exercise}

In this book we will only work with continuous functions. In this case, the condition~\eqref{ocon} for operator convexity can be replaced with a more special condition in which we take $t=\frac12$.

\begin{myg}{}
\begin{lemma}
Let $f:\mI\to\mbb{R}$ be a continuous function. Then, $f$ is operator convex if and only if for any Hilbert space $A$, any $\eta,\zeta\in\herm(A)$ 
\be\label{hocon}
f\left(\frac{\eta+\zeta}2\right)\leq \frac{f(\eta)+f(\zeta)}2\;.
\ee
\end{lemma} 
\end{myg}
\begin{proof}
Clearly, if $f$ satisfies~\eqref{ocon} then it satisfies~\eqref{hocon}. We therefore show that~\eqref{hocon} implies~\eqref{ocon}.
Let $\eta,\zeta\in\herm(A)$ and suppose~\eqref{hocon} holds. Observe that for $t=1/4$ we get
\ba
f\left(\frac14\eta+\frac 34\zeta\right)&=f\left(\frac12\left(\frac12\eta+\frac 12\zeta\right)+\frac12\zeta\right)\\
\GG{\eqref{hocon}}&\leq \frac12f\left(\frac12\eta+\frac 12\zeta\right)+\frac12f(\zeta)\\
\GG{\eqref{hocon}}&\leq \frac14\big(f(\eta)+f(\zeta)\big)+\frac12f(\zeta)\\
&=\frac14f(\eta)+\frac 34f(\zeta).
\ea
Hence, the condition~\eqref{ocon} holds for $t=\frac14$ and $t=\frac34$. Similarly, by repetition (e.g. taking convex combinations $\frac12\eta+\frac12\left(\frac14\eta+\frac 34\zeta\right)$, etc) it follows that~\eqref{ocon} must hold for all dyadic rationals, i.e. numbers of the form $t=\frac{m}{2^n}$ where $n\in\mbb{N}$ is arbitrary and $m$ is any integer in $[2^n]$. Since the set of such dyadic rationals is dense in $[0,1]$, it follows from the continuity of $f$ that~\eqref{ocon} holds for all $t\in[0,1]$. This completes the proof.
\end{proof}

\begin{exercise}\label{alpha2}
Use the lemma above to prove that the function $f(t)=t^2$ is operator convex on any interval. Hint: Show that the difference between $\frac{f(\eta)+f(\zeta)}2$ and $f\left(\frac{\eta+\zeta}2\right)$ can be expressed as a square of an Hermitian matrix. 
\end{exercise}

\section{Key Examples}

Consider the function $f(r)=r^\alpha$ for some $\alpha\in\mbb{R}$.  We already saw that for $\alpha=2$ the function is operator convex on any interval (see Exercise~\ref{alpha2}). Another relatively simple case, is the case $\alpha=-1$. In this case,
the function $f(r)=\frac1r$ is operator anti-monotone in the interval $\mI=(0,\infty)$. To see this, suppose $\eta\geq\zeta>0$. Then, conjugating both sides of $\eta\geq\zeta$ by $\eta^{-\frac12}$ gives $I\geq\eta^{-\frac12}\zeta\eta^{-\frac12}$. This means that all the eigenvalues of $\eta^{-\frac12}\zeta\eta^{-\frac12}$ are no greater than one. Therefore, all the eigenvalues of its inverse are at least one. Hence, 
\be
\left(\eta^{-\frac12}\zeta\eta^{-\frac12}\right)^{-1}\geq I\;.
\ee
The above inequality is equivalent to $\eta^{\frac12}\zeta^{-1}\eta^{\frac12}\geq I$ and after conjugating both sides by $\eta^{-\frac12}$ we get $\zeta^{-1}\geq\eta^{-1}$.

As another example, consider the case $\alpha=\frac12$. In this case we need to show that if $\eta\geq\zeta\geq 0$ then $\eta^{\frac12}\geq\zeta^{\frac12}$. Suppose first that $\eta>0$. In this case, by conjugating both sides of $\eta\geq\zeta$ by $\eta^{-\frac12}$ gives
\be
I\geq \eta^{-\frac12}\zeta\eta^{-\frac12}=N^*N\quad\text{where}\quad N\eqdef \zeta^{\frac12}\eta^{-\frac12}\;.
\ee
The above inequality implies in particular that the maximal eigenvalue of the complex matrix $N$ cannot exceed one.
Observe that the matrix $N$ is similar to the matrix $\eta^{-\frac14}N\eta^{\frac14}=\eta^{-\frac14}\zeta^{\frac12}\eta^{-\frac14}$ which is Hermitian. Since similar matrices has the same eigenvalues we conclude that
$I\geq \eta^{-\frac14}\zeta^{\frac12}\eta^{-\frac14}$. Conjugating both sides by $\eta^{\frac14}$ we conclude that $\eta^{\frac12}\geq\zeta^{\frac12}$.
The case that $\eta$ is not strictly positive (but still positive semidefinite) follows from the fact that $\eta\geq\zeta$ implies that $\eta+\eps I\geq\zeta$ for any $\eps>0$. Hence, since $\eta+\eps I>0$ we conclude from the argument above that 
$(\eta+\eps I)^{\frac12}\geq\zeta^{\frac12}$. Since this inequality holds for all $\eps>0$ it must also hold for $\eps=0$. This completes the proof that the function $f(r)=\sqrt{r}$ is operator monotone in the domain $[0,1]$.

It is possible to show that for any $\alpha\in[0,1]$ this function is operator monotone on the domain $[0,\infty)$. In Table~\ref{table:1} we summarized everything that is known in literature about the operator monotonicity and convexity of the function $f(r)=r^\alpha$. In the section `History and further readings' we give more information about where the proofs can be found.

\begin{table}[ht]
\caption{The Function $f:\mI\to\mbb{R}:r\mapsto r^\alpha$} % title of Table
\centering % used for centering table
\begin{tabular}{|c| c| c| c| c| c|} % centered columns (4 columns)
\hline %inserts double horizontal lines
$\substack{\text{Interval}\\ \mI}$ & $\substack{\text{Range of}\\ \alpha}$ & $\substack{\text{Operator}\\ \text{Monotone}}$ & $\substack{\text{Operator}\\ \text{Convex}}$ 
& $\substack{\text{Operator Anti-}\\ \text{Monotone}}$ & $\substack{\text{Operator}\\ \text{Concave}}$\\ [0.5ex] % inserts table
%heading
\hline % inserts single horizontal line
$[0,\infty)$ & $[0,1]$ & Yes & No & No & Yes \\ % inserting body of the table
$(0,\infty)$ & $[-1,0)$ & No & Yes & Yes & No \\
$[0,\infty)$ & $[1,2]$ & No & Yes & No & No\\
$(0,\infty)$ & $(-\infty,-1)\cup(2,\infty)$ & No & No & No & No\\ [1ex] % [1ex] adds vertical space
\hline %inserts single line
\end{tabular}
\label{table:1} % is used to refer this table in the text
\end{table}

Other important examples of functions that appears a lot in applications are the log function $f(r)=\log(r)$ defined on the interval $(0,\infty)$ as well as the function $f(r)=-r\log r$. The former is known to be both operator concave and operator monotone, while the latter is known to be operator concave. 

\section{Trace Functions}

A trace function is a function from $\herm(A)$ to the real line, of the form 
\be
\eta\mapsto\tr[f(\eta)]\quad\quad\forall\;\eta\in\herm(A)\;,
\ee
where $f:\mbb{R}\to\mbb{R}$. Such functions appear in many applications, and we will see later on that certain key quantities in quantum information, such as entropies and relative entropies, are defined in terms of trace functions. For our purposes, we will always assume that $f$ is continuous.

\bex\label{exfirstder}
Let $f:\mbb{R}\to\mbb{R}$ be continuously differentiable, fix $\eta,\zeta\in\herm(A)$, and define
\be
g(t)\eqdef\tr[f(\eta+t\zeta)]\quad\quad\forall\;t\in\mbb{R}\;.
\ee
Use the divided difference approach discussed in Appendix~\ref{A:DD} to show that the function $g(t)$ is continuously differentiable and
\be
g'(t)=\tr\left[\zeta f'\left(\eta+t\zeta\right)\right]\;.
\ee
\eex

\begin{myt}{}
\begin{theorem}\label{tracefunction}
Let $f:\mbb{R}\to\mbb{R}$ be continuous, and $A$ a finite dimensional Hilbert space.
\ben
\item If the function $f(t)$ is monotonically non-decreasing in $\mbb{R}$ then the function $\eta\mapsto\tr[f(\eta)]$ is monotonically non-decreasing in $\eta\in\herm(A)$.
\item If the function $f(t)$ is convex in $\mbb{R}$ then the function $\eta\mapsto \tr[f(\eta)]$ is convex in $\eta\in\herm(A)$.
\een 
\end{theorem}
\end{myt}

\begin{proof}
{\it Part 1.} Suppose first that $f$ is differentiable so that $f'(t)\geq 0$ for all $t\in\mbb{R}$. Under this assumption we have that $f'(\xi)\geq 0$ for any $\xi\in\herm(A)$ (note that $\xi$ does not have to be positive semidefinite). Let $\eta,\zeta\in\herm(A)$ be such that $\eta\geq\zeta$. We need to show that $\tr[f(\eta)]\geq\tr [f(\zeta)]$. Set $\rho\eqdef\eta-\zeta$ and observe that $\rho\in\pos(A)$. For any $t\in[0,1]$ define the function 
\be
g(t)\eqdef\tr \left[f(\zeta+t\rho)\right]\;,
\ee
so that $g(0)=\tr[f(\zeta)]$ and $g(1)=\tr [f(\eta)]$.
Therefore,
\ba
g(1)-g(0)&=\int_{0}^1g'(t)dt\\
\GG{Exercise~\ref{exfirstder}}&=\int_{0}^1\tr\left[\rho f'\left(\zeta+t\rho\right)\right]dt\\
\Gg{\rho\geq 0}&=\int_{0}^1\tr\left[\rho^{1/2}f'\left(\zeta+t\rho\right)\rho^{1/2}\right]dt
\ea
Finally, since $f'(\zeta+t\rho)\geq 0$, also $\rho^{1/2}f'\left(\zeta+t\rho\right)\rho^{1/2}\geq 0$ so that the integrand on the right-hand side of the equation above is non-negative. Hence, $g(1)\geq g(0)$. This completes the proof for the case that $f$ is differentiable. The proof of the case that $f$ is only continuous (but not necessarily differentiable) follows from continuity by taking a sequence of continuously differentiable functions whose limit is $f$ (such a sequence always exists). 

{\it Part 2.} Consider the spectral decomposition of $\eta=\sum_{x\in[m]}\lambda_x\Pi_x$, where $\Pi_x\eqdef|x\lr x|$, $\{|x\ra\}_{x\in[m]}$ form an orthonormal eigenbasis of $A$, and each $\lambda_x\in\mbb{R}$. Let $\{|\psi_y\ra\}_{y\in[m]}$ be another orthonormal basis of $A$. Then,
\ba
\tr[f(\eta)]&=\sum_{y\in[m]}\Big\la \psi_y\Big|\sum_{x\in[m]}f(\lambda_x)\Pi_x\Big|\psi_y\Big\ra=\sum_{y\in[m]}\sum_{x\in[m]}f(\lambda_x)\la \psi_y|\Pi_x|\psi_y\ra\\
\GG{{\it f} \;is\; convex}&\geq\sum_{y\in[m]}f\Big(\sum_{x\in[m]}\lambda_x\la \psi_y|\Pi_x|\psi_y\ra\Big)\;.
\ea
Since $\eta=\sum_{x\in[m]}\lambda_x\Pi_x$ we conclude that
\be\label{bex1m}
\tr[f(\eta)]\geq \sum_{y\in[m]}f\big(\la \psi_y|\eta|\psi_y\ra\big)\quad\quad\forall\;\eta\in\herm(A)\;.
\ee
Now, let $t\in[0,1]$, $\eta,\zeta\in\herm(A)$, and $\{|\psi_y\ra\}_{y\in[n]}$ be an orthonormal basis of $A$ consisting of the eigenvalues of $t\eta+(1-t)\zeta$. For these choices we get
\ba
\tr\big[f(t\eta+(1-t)\zeta)\big]&=\sum_{y\in[m]}\la \psi_y|f(t\eta+(1-t)\zeta)|\psi_y\ra\\
\Gg{\substack{|\psi_y\ra\text{ is an eigenvector}\\\text{of }t\eta+(1-t)\zeta}}&=\sum_{y\in[m]}f\Big(\big\la \psi_y\big|\big(t\eta+(1-t)\zeta\big)\big|\psi_y\big\ra\Big)\\
&=\sum_{y\in[m]}f\Big(t\la \psi_y\big|\eta|\psi_y\ra+(1-t)\la\psi_y|\zeta|\psi_y\ra\Big)\\
\GG{{\it f}\;is\;convex}&\leq t\sum_{y\in[m]}f\big(\la \psi_y|\eta|\psi_y\ra\big)+(1-t)\sum_{y\in[m]}f\big(\la \psi_y|\zeta|\psi_y\ra\big)\\
\GG{\eqref{bex1m}}&\leq t\tr[f(\eta)]+(1-t)\tr[f(\zeta)]\;.
\ea
This completes the proof.
\end{proof}

\begin{exercise}\label{tracefunc}
Let $K\in\ml(A)$, $\alpha\in(0,\infty)$, and define the function $f:\pos(A)\to\mbb{R}$ via
\be
f(\rho)\eqdef\tr\left[\left(K^*\rho K\right)^\alpha\right]\quad\quad\forall\;\rho\in\pos(A)\;.
\ee
Show that if $\rho,\sigma\in\pos(A)$ satisfy $\rho\geq\sigma$ then $f(\rho)\geq f(\sigma)$. 
\end{exercise}

\begin{myt}{\color{yellow} von Neumann's Trace Inequality}
\begin{theorem}
Let  $M,N\in\mbb{C}^{n\times n}$ be two complex matrices with singular values $\mu_1\geq\cdots\geq\mu_n$ and $\nu_1\geq\cdots\geq\nu_n$, respectively. Then,
\be\label{1p180}
\big|\tr[MN]\big|\leq\sum_{x\in[n]}\mu_x\nu_x\;.
\ee
\end{theorem}
\end{myt}

\begin{proof}
Using the singular value decomposition, we have $M=U_1D_1V_1$ and $N=U_2D_2V_2$, where 
$D_1\eqdef\diag\{\mu_1,\ldots,\mu_n\}$, $D_2\eqdef\diag\{\nu_1,\ldots,\nu_n\}$, and $U_1,U_2,V_1,V_2$ are four unitary matrices. We therefore need to show that
\be
\big|\tr[D_1UD_2V]\big|\leq\tr[D_1D_2]\;,
\ee
where $U\eqdef V_1U_2$ and $V\eqdef V_2U_1$ are two unitary matrices.
For any $k\in[n]$ let
\be
\Pi_k\eqdef\sum_{x\in[k]}|x\lr x|\;,
\ee
and observe that the diagonal matrices $D_1$ and $D_2$ can be expressed as
\be
D_1=\sum_{k\in[n]}(\mu_k-\mu_{k+1})\Pi_k\quad\text{and}\quad D_2=\sum_{k\in[n]}(\nu_k-\nu_{k+1})\Pi_k\;,
\ee
with the convention that $\mu_{n+1}=\nu_{n+1}=0$. Denoting by $a_k\eqdef\mu_k-\mu_{k+1}$ and $b_k\eqdef\nu_k-\nu_{k+1}$, and using the triangle inequality we get that
\be
\big|\tr[D_1UD_2V]\big|\leq \sum_{k,\ell\in[n]}a_kb_\ell\big|\tr[\Pi_kU\Pi_\ell V]\big|\quad\text{and}\quad\tr[D_1D_2]=\sum_{k,\ell\in[n]}a_kb_\ell\tr[\Pi_k\Pi_\ell]\;.
\ee
Therefore, the proof will be concluded by showing that for each $k,\ell\in[n]$
\be\label{166}
\big|\tr[\Pi_kU\Pi_\ell V]\big|\leq \tr[\Pi_k\Pi_\ell]\;.
\ee
Without loss of generality suppose $\ell\geq k$. Then,
\be
\big|\tr[\Pi_kU\Pi_\ell V]\big|=\left|\sum_{x\in[k]}\la x|U\Pi_\ell V|x\ra\right|\leq\sum_{x\in[k]}\big|\la x|U\Pi_\ell V|x\ra\big|\leq k\;,
\ee
where the last inequality follows from the fact that $\big|\la x|U\Pi_\ell V|x\ra\big|\leq 1$ (see Exercise~\ref{exupv}). Since $\tr[\Pi_k\Pi_\ell]=k$ the equation above implies~\eqref{166}. This completes the proof.
\end{proof}

\bex\label{exupv}
Use Cauchy-Schwarz inequality to show that $\big|\la x|U\Pi_\ell V|x\ra\big|\leq 1$.
\eex

\bex
Using the same notations as above, show that
\be\label{finchi}
\max_{U,V\in\muu(n)}\big|\tr[MUNV]\big|=\sum_{x\in[n]}\mu_x\nu_x\;.
\ee
\eex

\begin{myt}{\color{yellow} Ruhe’s Trace Inequality}
\begin{theorem}\label{rti}
Let  $M,N\in\herm(A)$ be two Hermitian matrices with eigenvalues $\mu_1\geq\cdots\geq\mu_n$ and $\nu_1\geq\cdots\geq\nu_n$, respectively. Then,
\be\label{p168}
\sum_{x\in[n]}\mu_x\nu_{n+1-x}\leq\tr[MN]\leq\sum_{x\in[n]}\mu_x\nu_x\;.
\ee
\end{theorem}
\end{myt}

\begin{proof}
Since $M$ is Hermitian we can work in its eigenbasis so that without loss of generality we will assume that $M=D_1\eqdef\diag(\mu_1,\ldots,\mu_n)$ is a diagonal matrix. We will also decompose $N=UD_2U^*$, where $D_2\eqdef\diag(\nu_1,\ldots,\nu_n)$, and $U$ is unitary. Thus,
\be
\tr[MN]=\tr\left[D_1UD_2U^*\right]=\sum_{k,\ell\in[n]}a_kb_\ell\tr\left[\Pi_kU\Pi_\ell U^*\right]\;,
\ee
where $a_k$, $b_\ell$, and $\Pi_k$, are the same as in the proof of the von-Neumann\index{von-Neumann} trace inequality above. Note that while the eigenvalues $\{\mu_k\}$ and $\{\nu_k\}$ can be negative, the differences $a_k\eqdef\mu_k-\mu_{k+1}$ and $b_k\eqdef\nu_k-\nu_{k+1}$ are non-negative for all $k\in[n]$. Combining this with~\eqref{166} and the equation above we conclude that
\be
\tr\left[D_1UD_2U^*\right]\leq\sum_{k,\ell\in[n]}a_kb_\ell\tr\left[\Pi_k\Pi_\ell\right]=\tr[D_1D_2]=\sum_{x\in[n]}\mu_x\nu_x\;.
\ee
This completes the proof of the upper bound in~\eqref{p168}.

For the proof of the lower bound in~\eqref{p168}, we repeat the exact same lines as above with $D_1$ remain unchanged, but with $D_2\eqdef\diag(\nu_n,\ldots,\nu_1)$. This change of order in the diagonal of $D_2$ implies that for each $k\in[n]$ the coefficient $b_k\leq 0$. We therefore get in this case that
\ba
\tr\left[D_1UD_2U^*\right]&=\sum_{k,\ell\in[n]}a_kb_\ell\tr\left[\Pi_kU\Pi_\ell U^*\right]
=\sum_{k,\ell\in[n]}a_k|b_\ell|\left(-\tr\left[\Pi_kU\Pi_\ell U^*\right]\right)\\
\GG{\eqref{166}}&\geq \sum_{k,\ell\in[n]}a_k|b_\ell|\left(-\tr\left[\Pi_k\Pi_\ell\right]\right)=\sum_{k,\ell\in[n]}a_kb_\ell\tr\left[\Pi_k\Pi_\ell\right]=\tr[D_1D_2]\;.
\ea
Finally, observe that $\tr[D_1D_2]$ equals to the left-hand side of~\eqref{p168}. This completes the proof.
\end{proof}

\section{Characterization of Operator Convexity}\label{opjens}

In this subsection we characterize operator convex functions in terms of Jensen's inequality. These characterizations will be used numerous times in this book, particularly in the study of quantum divergences and R\'enyi relative entropies.

\begin{myt}{\color{yellow} The Operator Jensen's Inequality (Isometry Form)}
\begin{theorem}
Let  $f:\mI\to\mbb{R}$ be a real function. Then, $f$ is operator convex on $\mI$ if and only if  
for any isometry $V:A\to B$ (with $|A|\leq |B|<\infty$), and any $\rho\in\herm(B)$, 
\be\label{jensen}
f\big(V^*\rho V\big)\leq V^*f(\rho)V\;.
\ee
\end{theorem}
\end{myt}
\begin{remark}
Note that for $|A|=1$ and isometry $V=|\psi\ra\in B$ one obtains the more familiar Jensen's inequality
\be\label{jensen2}
f\big(\la\psi |\rho |\psi\ra\big)\leq \la\psi|f(\rho)|\psi\ra\;.
\ee
In this case, it is sufficient to require that $f$ is convex.
\end{remark}

\begin{proof}
Suppose first that $f$ is operator convex, and let $V:A\to B$ be an isometry. For simplicity denote by $m=|A|$ and $n=|B|$, and let $U$ be a unitary matrix obtained from the $n\times m$ isometry $V$ by adding $n-m$ columns to $V$. That is, $U$ can be expressed as
\be
U=\big[V\;N\big]
\ee
where $N$ is an $n\times (n-m)$ matrix. 

Every matrix $M\in\ml(B)$ can be expressed in a block matrix form as $M=\begin{bmatrix}M_{11} & M_{12}\\
M_{21} & M_{22}\end{bmatrix}$, where the block matrix $M_{11}$ is $m\times m$ and the rest of the block matrices are such that $M$ is $n\times n$.
With this in mind, note that for any $\rho\in\herm(B)$
\be
U^*\rho U=\begin{bmatrix} V^*\\ N^*
\end{bmatrix}\rho\big[V\;N\big]=\begin{bmatrix}V^*\rho V & V^*\rho N\\
N^*\rho V & N^*\rho N\end{bmatrix}\;.
\ee
Finally, we define a linear map $\mE:\herm(B)\to\herm(B)$ via
\be
\mE(\sigma)=\frac{1}{2}\sigma+\frac{1}{2}Z\sigma Z\quad\forall\;\sigma\in\herm(B)\;,
\ee
where $Z=\begin{bmatrix} I_m & \0\\
\0 & -I_{n-m}\end{bmatrix}$. 
A key property of this map is that it acts as a type of a dephasing map (in fact, it belongs to a type of quantum channels known as the pinching channels). Particularly, note that
\be\label{zyx}
\mE(U^*{\rho}U)=\begin{bmatrix}V^*\rho V & \0\\
\0 & N^*\rho N\end{bmatrix}\quad\quad\forall\;\rho\in\herm(B)\;.
\ee 
This also implies that for any $\rho\in\herm(B)$
\be\label{zyx1}
f\big(\mE(U^*{\rho}U)\big)=\begin{bmatrix}f(V^*\rho V) & \0\\
\0 & f(N^*\rho N)\end{bmatrix}\;.
\ee
With these notations we get from~\eqref{zyx1}
\ba
f\big(V^*\rho V\big)&=\Big(f\big(\mE(U^*{\rho} U)\big)\Big)_{11}\\
\GG{{\it f}\text{ is operator convex}}&\leq \left(\frac{1}{2}f\left(U^*{\rho} U\right)+\frac{1}{2}f\left(ZU^*{\rho} UZ\right)\right)_{11}\\
\GG{{\it U}\text{ and }{\it UZ}\text{ are unitaries}}&=\left(\frac{1}{2}U^*f\left({\rho} \right)U+\frac{1}{2}ZU^*f\left({\rho} \right)UZ\right)_{11}=\Big(\mE\big(U^*f(\rho)U\big)\Big)_{11}\\
\GG{\eqref{zyx}}&=V^*f(\rho)V\;.
\ea
Therefore, $f$ satisfies the condition given in~\eqref{jensen}.

We next assume that $f$ satisfies~\eqref{jensen} and use it to show that $f$ is operator convex. Let $A$ be a Hilbert space, $t\in[0,1]$, and define
$V:A\to A\oplus A$ to be the matrix
\be\label{iso18}
V=\begin{bmatrix} t^{\frac12}I^A \\ (1-t)^{\frac12}I^A\end{bmatrix}\;,
\ee
where $I^A$ is the identity matrix on $A$. It is simple to check that $V^*V=I^A$ so that $V$ is an isometry. Moreover, a direct calculation gives for any $\rho,\sigma\in\herm(A)$
\be\label{118}
V^*\begin{bmatrix}
\rho & \0\\
\0 & \sigma
\end{bmatrix}V=t\rho+(1-t)\sigma\;.
\ee
We therefore get that
\ba
f\big(t\rho+(1-t)\sigma\big)&=f\left(V^*\begin{bmatrix}
\rho & \0\\
\0 & \sigma
\end{bmatrix}V\right)\\
\GG{\eqref{jensen}}&\leq V^*f\left(\begin{bmatrix}
\rho & \0\\
\0 & \sigma
\end{bmatrix}\right)V
=V^*\begin{bmatrix}
f(\rho) & \0\\
\0 & f(\sigma)
\end{bmatrix}V\\
\GG{\eqref{118}}&=tf(\rho)+(1-t)f(\sigma)\;.
\ea
Hence, $f$ is operator convex. This completes the proof.
\end{proof}

\begin{myt}{\color{yellow} The Operator Jensen's Inequality (Algebraic Form)}
\begin{theorem}
Let  $f:\mI\to\mbb{R}$ be a real function, and suppose that $0\in\mI$ and $f(0)\leq 0$. Then, $f$ is operator convex on $\mI$ if and only if for any Hilbert space $A$, any $\rho,\sigma\in\herm(A)$, and any $M_1,M_2\in\ml(A)$ such that $M_1^*M_1+M^*_2M_2\leq I$ we have
\be\label{109}
f(M_1^*\rho M_1+M_2^*\sigma M_2)\leq M_1^*f(\rho) M_1+M_2^*f(\sigma) M_2\;.
\ee
\end{theorem}
\end{myt}
\begin{remark}
The condition that $f(0)\leq 0$ cannot be removed from the theorem above. This condition is necessary for this version of Jensen's inequality, since by taking $|A|=1$ and setting $M_1=M_2=0$ in~\eqref{109} we get that $f(0)\leq 0$.
\end{remark}

\begin{proof}

Suppose first that $f$ is operator convex, and let $\rho,\sigma\in\herm(A)$, and $M_1,M_2\in\ml(A)$ be such that $M_1^*M_1+M^*_2M_2\leq I$. Define $M_3\eqdef\sqrt{I-M_1^*M_1-M_2^*M_2}$ so that $\sum_{x\in[3]}M_x^*M_x=I$.
Finally, denote by
\be\label{v9v}
V\eqdef\bpm
M_1 \\ M_2 \\ M_3
\epm\quad\text{and}\quad\omega\eqdef\bpm \rho & \0 & \0\\
\0 & \sigma & \0\\
\0\ & \0 & \0
\epm\;.
\ee
Observe that the matrix $V:A\to A\oplus A\oplus A$
is an isometry since $V^*V=\sum_{x\in[3]}M_x^*M_x=I$.
We therefore get
\ba
f(M_1^*\rho M_1+M_2^*\sigma M_2)&=f(V^*\omega V)\\
\GG{\eqref{jensen}}&\leq V^*f(\omega)V\\
\GG{(\ref{v9v})}&=M_1^*f(\rho) M_1+M_2^*f(\sigma) M_2+M_3^*f(\0)M_3\\
\Gg{f(0)\leq 0}&\leq M_1^*f(\rho) M_1+M_2^*f(\sigma) M_2\;.
\ea
This complete the first direction of the theorem.

For the converse, suppose that~\eqref{109} holds for all $M_1$ and $M_2$ as in the theorem above. Then, by taking $M_2$ to be the zero matrix, and $M_1$ to be a projection in $\herm(A)$, we get that the inequality~\eqref{109} implies, in particular, that for any projection $\Pi\in\herm(A)$ and any $\rho\in\herm(A)$ we have 
\be\label{125}
f(\Pi\rho\Pi)\leq\Pi f(\rho)\Pi\;. 
\ee
Now, define the unitary matrix $U:A\oplus A\to A\oplus A$ via
\be
U\eqdef\begin{pmatrix} t^{1/2}I^A & -(1-t)^{1/2}I^A\\ (1-t)^{1/2}I^A & t^{1/2}I^A\end{pmatrix}
\ee
This matrix is the unitary extension of the isometry $V$ as defined in~\eqref{iso18}.
Particularly, note that $U$ and the isometry $V$ in~\eqref{iso18} satisfy the relation
\be
U\Pi =\begin{pmatrix} t^{1/2}I^A & \0 \\ (1-t)^{1/2}I^A & \0\end{pmatrix}=\bpm V & \0^A\epm\quad\text{where}\quad \Pi\eqdef\begin{pmatrix}
I^A & \0^A\\
\0^A & \0^A
\end{pmatrix}\;.
\ee
Combining this with~\eqref{118} yields
\ba
\begin{pmatrix}
f\big(t\rho+(1-t)\sigma\big) & \0\\
\0 & f(\0)
\end{pmatrix}
&=f\left(\Pi U^*\begin{bmatrix}
\rho & \0\\
\0 & \sigma
\end{bmatrix}U\Pi\right)\\
\GG{\eqref{125}}&\leq \Pi f\left(U^*\begin{bmatrix}
\rho & \0\\
\0 & \sigma
\end{bmatrix}U\right)\Pi\\
&=\Pi U^*\begin{bmatrix}
f(\rho) & \0\\
\0 & f(\sigma)
\end{bmatrix}U\Pi \\
\GG{\eqref{118}}&=\begin{pmatrix} tf(\rho)+(1-t)f(\sigma) & \0\\
\0 & \0
\end{pmatrix}\;.
\ea
Hence, $f$ is operator convex. This completes the proof.
\end{proof}

\begin{exercise}
Show that a function $f:\mI\to\mbb{R}$ with $0\in\mI$ and $f(0)\leq 0$ is operator convex if and only if for all $n\in\mbb{N}$, all $\rho_1,\ldots,\rho_n\in\herm(A)$, and all $M_1,\ldots,M_n\in\ml(A)$ such that 
$
\sum_{x\in[n]}M^*_xM_x\leq I^A
$
we have
\be
f\left(\sum_{x\in[n]}M^*_x\rho_x M_x\right)\leq \sum_{x\in[n]}M^*_xf\left(\rho_x\right) M_x \;.
\ee
\end{exercise}

\section{The Kubo-Ando Operator Mean}

Operator convexity can also be characterized in terms of the following operator mean.

\begin{myd}{}
\begin{definition}\label{kaom}
Let $f:[0,\infty)\to[0,\infty)$ be a continuous function.
The Kubo-Ando operator mean (also known as the perspective function associated with $f$) is a map
\be
\#_f:\pos(A)\times\pos_{>0}(A)\to\pos_{>0}(A)
\ee
defined for any $\rho\in\pos(A)$ and $\sigma\in\pos_{>0}(A)$ by
\be\label{kaom2}
\rho\#_f\sigma\eqdef \sigma^{\frac12}f\left(\sigma^{-\frac12}\rho\sigma^{-\frac12}\right)\sigma^{\frac12}
\ee
Moreover, if $f(t)=\sqrt{t}$ then $\#_f$ is denoted simply by $\#$ and is called the operator geometric mean.
\end{definition}
\end{myd}
\begin{remark}
We exchanged the roll of $\sigma$ and $\rho$ from the original definition as it will be more convenient in the context of quantum information to work with this definition.
\end{remark}

\begin{myt}{}
\begin{theorem}\label{equivjoint}
Let $f:[0,\infty)\to[0,\infty)$ be a continuous function. The following are equivalent:
\begin{enumerate}
\item $f$ is operator convex.
\item $\#_f$ is jointly convex.
\end{enumerate}
\end{theorem}
\end{myt}

\begin{proof}
We first prove the direction $1\Rightarrow2$. Let $\rho=t\rho_1+(1-t)\rho_2$ and $\sigma=t\sigma_1+(1-t)\sigma_2$ with $t\in(0,1)$, $\rho_1,\rho_2\in\pos(A)$ and $\sigma_1,\sigma_2\in\pos_{>0}(A)$. Define the matrices $M_1\eqdef(t\sigma_1)^{\frac12}\sigma^{-\frac12}$ and $M_2\eqdef\big((1-t)\sigma_2\big)^{\frac12}\sigma^{-\frac12}$. Observe that these matrices form a generalized measurement; i.e. $M_1^*M_1+M_2^*M_2=I^A$. Moreover, in terms of these matrices we can express the term $\sigma^{-\frac12}\rho\sigma^{-\frac12}$ in~\eqref{kaom2} as
\be
\sigma^{-\frac12}\rho\sigma^{-\frac12}=
M_1^*\sigma^{-\frac12}_1\rho_1\sigma^{-\frac12}_1M_1+ 
M_2^*\sigma^{-\frac12}_2\rho_2\sigma^{-\frac12}_2M_2
\ee
Now, from Jensen's operator inequality~\eqref{109} it follows that
\be
f\left(\sigma^{-\frac12}\rho\sigma^{-\frac12}\right)\leq M_1^*f\left(\sigma^{-\frac12}_1\rho_1\sigma^{-\frac12}_1\right)M_1+ 
M_2^*f\left(\sigma^{-\frac12}_2\rho_2\sigma^{-\frac12}_2\right)M_2
\ee
Conjugating both sides by $\sigma^{\frac12}(\cdot)\sigma^{\frac12}$ and recalling that $M_1\sigma^{\frac12}=(t\sigma_1)^{\frac12}$ and $M_2\sigma^{\frac12}=\big((1-t)\sigma_2\big)^{\frac12}$ gives
\be
\rho\#_{f}\sigma\leq t\rho_1\#_f\sigma_1+(1-t)\rho_2\#_f\sigma_2\;.
\ee
That is, $\#_f$ is jointly convex.
For the direction $2\Rightarrow 1$ observe that for $\sigma=I^A$ we get $\rho\#_f\sigma=f(\rho)$ so that the convexity of $f$ follows from the joint convexity\index{joint convexity} of $\#_f$. 
\end{proof}

For the function $f(t)=t^\alpha$ for some $\alpha\in\mbb{R}$ we use the notation
\be
\rho\#_\alpha\sigma\eqdef \sigma^{\frac12}\left(\sigma^{-\frac12}\rho\sigma^{-\frac12}\right)^\alpha \sigma^{\frac12}
\ee
Observe that if $\rho$ and $\sigma$ commutes then 
\be
\rho\#_\alpha\sigma=\rho^\alpha\sigma^{1-\alpha}
\ee

The Kubo-Ando operator mean can also be applied to operators on the vector space of super operators consisting of all linear transformations from $\ml(A)$ to itself. Particularly, in the proof of the theorem below, for any $\rho\in\ml(A)$ we will consider the linear operators
\be
\mL_{\rho}(\omega)\eqdef\rho\omega\;\text{ and }\;\mR_{\rho}(\omega)\eqdef\omega\rho\quad\quad\forall\;\omega\in\ml(A)\;.
\ee
Observe that $\mL_\rho,\mR_\rho:\ml(A)\to\ml(A)$ are linear operators belonging to the Hilbert space $\ml(A\to A)$.
\begin{exercise}
Let $\rho,\sigma\in\ml(A)$, $\alpha\in[0,\infty)$, and consider the left and right operators, $\mL_\rho$ and $\mR_\sigma$ as define above. Show that:
\begin{enumerate}
\item Commutativity; $\mL_\rho\circ \mR_\sigma=\mR_\sigma \circ\mL_\rho$.
\item If $\rho,\sigma\in\herm(A)$ then $\mL_\rho$ and $\mR_\sigma$ are self-adjoint with respect to the Hilbert-Schmidt inner product. 
\item If $\rho\in\pos_{>0}(A)$ then $\mL_\rho$ and $\mR_\rho$ are invertible with inverses
\be
\mL_\rho^{-1}=\mL_{\rho^{-1}}\quad\text{and}\quad \mR_\rho^{-1}=\mR_{\rho^{-1}}\;.
\ee
\item If $\rho\geq 0$ then
\be
\mL_\rho^\alpha=\mL_{\rho^\alpha}\quad\text{and}\quad \mR_\rho^\alpha=\mR_{\rho^\alpha}\;.
\ee
\item If $\rho\geq 0$ and $\sigma>0$ then
\be\label{xgx}
\mR_\rho\#_\alpha \mL_\sigma\eqdef \mR_{\rho^\alpha}\mL_{\sigma^{1-\alpha}}\;.
\ee
\end{enumerate}
\end{exercise}

\section{Lieb's Concavity Theorem}

\begin{myt}{\color{yellow} Lieb's Concavity Theorem}\label{lct}
\begin{theorem}\label{lieb}
Let $K\in\mbb{C}^{m\times n}$ and $\alpha\in(0,1)$. Then, the function
\be
f(\rho,\sigma)\eqdef\tr\left[K^*\rho^\alpha K\sigma^{1-\alpha}\right]
\ee
is jointly concave.
\end{theorem}
\end{myt}

\begin{proof}
Observe that 
\ba
f(\rho,\sigma)&=\la K\;,\;\rho^\alpha K\sigma^{1-\alpha}\ra_{HS}\\
&=\left\la K\;,\;\mL_{\rho^\alpha}\mR_{\sigma^{1-\alpha}}(K)\right\ra_{HS}\\
{\color{red} \text{\eqref{xgx}}\rightarrow}&=\left\la K\;,\;\mR_\rho\#_\alpha \mL_\sigma(K)\right\ra_{HS}\;.
\ea
Therefore, $f$ is jointly concave since from Theorem~\ref{equivjoint} it follows that $\#_\alpha$ is jointly concave for $\alpha\in(0,1)$.
\end{proof}

When combining Lieb's theorem above with the Young's inequality~\eqref{young} we get the following result.
 
\begin{myg}{}
\begin{corollary}
Let $\eta,\rho\in\pos(A)$ and $\alpha\in(0,1)$. Then, the function
\be
\rho\mapsto\tr\left[\big(\eta\rho^\alpha\eta\big)^{\frac1\alpha}\right]
\ee
is concave. 
\end{corollary}\label{yl01}
\end{myg}
\begin{proof}
Set $M\eqdef\eta\rho^\alpha\eta$, $p\eqdef\frac1\alpha$ and $q\eqdef\frac1{1-\alpha}$. Finally, let $\sigma\in\pos(A)$ and denote by $N\eqdef\sigma^{1-\alpha}$. Then,
by definition
\ba
\tr\left[\eta\rho^\alpha \eta\sigma^{1-\alpha}\right]&=\tr[MN]\\
\GG{Young's\; inequality~\eqref{young}}&\leq\frac1p\tr[M^p]+\frac1q\tr[N^q]\\
&=\alpha\tr\left[\big(\eta\rho^\alpha\eta\big)^{\frac1\alpha}\right]+(1-\alpha)\tr[\sigma]\;.
\ea
 Therefore, isolating the term $\tr\left[\big(\eta\rho^\alpha\eta\big)^{\frac1\alpha}\right]$  gives
\be
\tr\left[\big(\eta\rho^\alpha\eta\big)^{\frac1\alpha}\right]\geq \frac1\alpha \tr\left[\eta\rho^\alpha \eta\sigma^{1-\alpha}\right]-\frac{1-\alpha}\alpha\tr[\sigma]
\ee
Now, recall that the Young's inequality achieves equality for $N^q=M^p$ which is equivalent to $\sigma= \big(\eta\rho^\alpha\eta\big)^{\frac1\alpha}$. Combining this with the inequality above we conclude that
\be
\tr\left[\big(\eta\rho^\alpha\eta\big)^{\frac1\alpha}\right]=\max_{\sigma\in\pos(A)}\Big\{ \frac1\alpha \tr\left[\eta\rho^\alpha \eta\sigma^{1-\alpha}\right]-\frac{1-\alpha}\alpha\tr[\sigma]\Big\}\;.
\ee
Now, from Lieb's theorem, the first term on the right-hand side is jointly concave in $\rho$ and $\sigma$, whereas the second term is linear in $\sigma$ and in particular concave. Hence, this immediately implies that the term on the left-hand side is concave in $\rho$ (see Exercise~\ref{younglieb} below for more details on this last assertion).
\end{proof}

\begin{exercise}\label{younglieb}
Let $f:\pos(A)\times\pos(A)\to\mbb{R}$ be a jointly concave function. Show that the function
\be
g(\rho)\eqdef\max_{\sigma\in\pos(A)}f(\rho,\sigma)
\ee
is concave in $\rho$, assuming the maximum above is achievable (for all $\rho$).
\end{exercise}

\section{The Quantum Weighted Geometric Mean Inequality}\index{geometric mean}\label{secacm}

For any two numbers $a$ and $b$, the weighted-geometric mean $a^{1-s}b^s$, with $s\in[0,1]$, can never be smaller than  the minimum between $a$ and $b$. This inequality is sometimes used in proofs related to hypothesis testing (particularly, in the proof of the classical Chernoff bound). Here we discuss a generalization of this inequality when $a$ and $b$ are replaced with matrices $M$ and $N$. Note that a minimum between two positive operators do not exists. We therefore express the minimum as 
\be
\min\{a,b\}=\frac{1}{2}\left(a+b-|a-b|\right)
\ee
so that the right-hand side can be generalized to operators. With this identification in mind, we can generalize the inequality $\min\{a,b\}\leq a^{1-s}b^s$ as follows.
\begin{myt}{}
\begin{theorem}
For any two positive semidefinite matrices $M,N\geq 0$ (of the same finite dimension) and any $0\leq \alpha\leq 1$ the following inequality holds
\be\label{acm}
\frac{1}{2}\tr\Big[M+N-\big|M-N\big|\Big]\leq \tr\big[M^{1-s}N^s\big]\;.
\ee
\end{theorem}
\end{myt}
\begin{exercise}
Show that if~\eqref{acm} holds for all $s\in[1/2,1]$ then it must also hold  for all $s\in[0,1/2]$.
\end{exercise}

\begin{proof}
Since the term $|M-N|$ can be expressed as $|M-N|=2(M-N)_+-(M-N)$, the inequality~\eqref{acm} is equivalent to
\be
\tr(M-N)_+\geq \tr[M]-\tr\big[M^{1-s}N^{s}\big]\;.
\ee
The identity $M-N=(M-N)_+-(M-N)_-$ gives
\be
M\leq M+(M-N)_-=N+(M-N)_+\;.
\ee
Combining the above inequality with the operator monotonicity of the function $f(t)=t^s$ for $s\in[0,1]$ gives
\be\label{mlmm}
M^s\leq \big(N+(M-N)_+\big)^s\quad\quad\forall\;s\in[0,1]\;.
\ee
With this inequality at hand, we get
\ba
\tr[M]-\tr\big[M^{1-s}N^{s}\big]&=\tr\left[M^{1-s}\left(M^s-N^{s}\right)\right]\\
\GG{\eqref{mlmm}}&\leq \tr\left[M^{1-s}\Big(\big(N+(M-N)_+\big)^s-N^{s}\Big)\right]\\
\GG{\eqref{mlmm}\;with\;\text{$1-s$}}&\leq \tr\left[\big(N+(M-N)_+\big)^{1-s}\Big(\big(N+(M-N)_+\big)^s-N^{s}\Big)\right]\\
&=\tr[N]+\tr(M-N)_+-\tr\left[\big(N+(M-N)_+\big)^{1-s}N^{s}\right]\\
\GG{see~\eqref{193}~below}&\leq \tr(M-N)_+
\ea 
where on the last inequality we used the fact that $N+(M-N)_+\geq N$ so that 
\be\label{193}
\big(N+(M-N)_+\big)^{1-s}\geq N^{1-s}\;.
\ee 
This completes the proof.
\end{proof}

\section{The Schur Complement}\label{sec:schurcom}

We conclude this section with a very useful tool that among other things determines the positive definiteness of a block matrix in terms of its block matrices.
We will also see that it can be used to prove operator convexity of certain functions.

Consider an Hermitian matrix
\be\label{146}
M=\begin{pmatrix}
\rho & \eta\\
\eta^* & \sigma
\end{pmatrix}\;,
\ee
where $\rho\in\herm(A)$, $\sigma\in\herm(B)$ and $\eta\in\mbb{C}^{|A|\times |B|}$. Then, the Schur complement of the block $\sigma$ of $M$ is defined as the matrix
\be\label{schurcom2}
M/\sigma\eqdef \rho-\eta\sigma^{-1}\eta^*\;,
\ee
where $\sigma^{-1}$ is taken to be the generalized inverse if the inverse of $\sigma$ does not exists~\footnote{The generalize inverse of a complex matrix $\sigma$ is the matrix $\sigma^{-1}$ that satisfies $\sigma\sigma^{-1}\sigma=\sigma$.} Similarly, 
the Schur complement of the block $\rho$ of $M$ is defined as the matrix
\be
M/\rho\eqdef \sigma-\eta^*\rho^{-1}\eta\;.
\ee
The significance of this definition is given in the following theorem.

\begin{myt}{}
\begin{theorem}\label{schurc}
Let $M$ be the Hermitian block matrix given in~\eqref{146}. Then, $M\geq 0$ if and only if at least one of the following two conditions holds:
\begin{enumerate}
\item $\rho\geq 0$ and $M/\rho\geq 0$.
\item $\sigma\geq 0$ and $M/\sigma\geq 0$.
\end{enumerate}
\end{theorem}
\end{myt}

\begin{proof}
We will show the equivalence of the second condition with $M\geq 0$.
The main idea of the proof is to define the matrix
\be
L\eqdef\begin{bmatrix}
I^A & \0\\
\sigma^{-1}\eta^*& I^B
\end{bmatrix}\;,
\ee
and to observe that $L$ is invertible with inverse
\be\label{1150}
L^{-1}=\begin{bmatrix}
I^A & \0\\
-\sigma^{-1}\eta^*& I^B
\end{bmatrix}\;.
\ee
Then, by direct calculation,
we have (see Exercise~\ref{ex:schur})
\be\label{1151}
M=L^*\begin{bmatrix}
\rho-\eta\sigma^{-1}\eta^* & \0\\
\0 & \sigma
\end{bmatrix}L
\ee
and since $L$ is invertible we must have $M\geq 0$ if and only if $\begin{bmatrix}
\rho-\eta\sigma^{-1}\eta^* & \0\\
\0 & \sigma
\end{bmatrix}\geq 0$. This completes the proof.
\end{proof}

\begin{exercise}\label{ex:schur}$\;$
\begin{enumerate}
\item Verify that $L^{-1}$ as given in~\eqref{1150} is indded the inverse of $L$. 
\item Verify the equality in~\eqref{1151}.
\item Show the equivalence of $M\geq 0$ with the first condition of the theorem above (in the proof above we only showed the equivalence with the second condition).
\end{enumerate}
\end{exercise}

\begin{mye}{}
\begin{corollary}\label{corschur}
Let $f:\pos(A)\times\ml(A)\to\pos(A)$ be a function defined by
\be
f(\rho,\eta)\eqdef \eta^*\rho\eta\quad\quad\forall\rho\in\pos(A)\;,\;\forall\eta\in\ml(A)\;.
\ee
Then, $f$ is jointly convex.
\end{corollary}
\end{mye}

\begin{proof}
Let $\rho_0,\rho_1\in\pos(A)$ and $\eta_0,\eta_1\in\ml(A)$, and define
\be
M_0\eqdef\begin{pmatrix}
\rho_0 & \eta_0\\
\eta^*_0 & \eta^*_0\rho_0^{-1}\eta_0
\end{pmatrix}
\quad\text{and}\quad M_1\eqdef\begin{pmatrix}
\rho_1 & \eta_1\\
\eta^*_1 & \eta^*_1\rho_1^{-1}\eta_1
\end{pmatrix}\;.
\ee
From Theorem~\ref{schurc} above $M_1,M_2\geq 0$. Therefore, for any $t\in[0,1]$ we have
\be
0\leq tM_0+(1-t)M_1=\begin{pmatrix}
t\rho_0+(1-t)\rho_1 & t\eta_0+(1-t)\eta_1\\
\eta^*_0+(1-t)\eta_1^* & t\eta^*_0\rho_0^{-1}\eta_0+(1-t)\eta^*_1\rho_0^{-1}\eta_1
\end{pmatrix}\;.
\ee
Since the matrix above is positive semidefinite,  its Schur complements is also positive semidefinite (i.e.\ we are using Theorem~\ref{schurc} once again). Hence, in particular, the Schur complement
\be
t\eta^*_0\rho_0^{-1}\eta_0+(1-t)\eta^*_1\rho_0^{-1}\eta_1-\big(t\eta_0+(1-t)\eta_1\big)^*\big(t\rho_0+(1-t)\rho_1\big)^{-1}\big(t\eta_0+(1-t)\eta_1\big)\geq 0\;.
\ee
In terms of the function $f$, the above equation can be expressed as
\be
tf(\rho_0,\eta_0)+(1-t)f(\rho_1,\eta_1)\geq f\big(t_0\rho_0+(1-t)\rho_1,t\eta_0+(1-t)\eta_1\big)\;.
\ee
Hence, $f$ is jointly convex.
\end{proof}

\begin{exercise}$\;$
\begin{enumerate}
\item Show that the function $f:\pos(A)\to\pos(A)$ given by $f(\rho)\eqdef\rho^{-1}$ for all $\rho\in\pos(A)$ is convex.
\item Show that the function $f:\ml(A)\to\pos(A)$ given by $f(\eta)\eqdef\eta^*\eta$ for all $\eta\in\ml(A)$ is convex.
\end{enumerate}
\end{exercise}

\begin{exercise}
Use Theorem~\ref{equivjoint} to prove Corollary~\ref{corschur}.
\end{exercise}

\section{Notes and References}

A review on operator monotonicity and convexity is given in~\cite{Bhatia1997}, and the more recent short course by~\cite{Carlen2010} can be helpful. The proofs of the assertions given in Table~\ref{table:1} can be found in these references. In the proof of Theorem~\ref{equivjoint} we followed~\cite{ENE2011}, and the extremely short proof of Lieb's concavity theorem (Theorem~\ref{lct}) as presented here is due to~\cite{NEE2013}.

%%%%%%%%%%%%%%%%%%%%%%%
%%%%%%%%%%%%%%%%%%%%%%%%

\chapter{Elements of Representation Theory}\label{sec:rep}

In this chapter we provide a relatively short review of group theory and representation theory. We only review concepts from representation theory that are particularly useful for applications in quantum information theory, and that we are using in this book. Therefore,  this section does not attempt to provide an extensive review of the exceedingly vast subject of representation theory. Further, much of the material discussed here can be found in standard textbooks on representation theory. Yet, a reader not familiar with groups and their representations will find this section self-contained and sufficient for the understanding of the material discussed in this book. Particularly, most of the material in this section is used in the study of the resource theory of asymmetry (see Chapter~\ref{Ch:Asymmetry}).

\section{Groups}

\begin{myd}{}
\begin{definition}
A group $\G$ is a set of objects equipped with internal composition operation from $\G\times\G$ to $\G$ satisfying the following three axioms:
\begin{enumerate}
\item Existence of identity. There exists an element $e\in\G$ satisfying for all $g\in\G$,
$
ge=eg=g
$.
\item Existence of an inverse. For every $g\in\G$ there exists an element $g^{-1}\in\G$  satisfying
$
g^{-1}g=gg^{-1}=e
$.
\item Associativity. For any $g_1,g_2,g_3\in\G$, 
$
(g_1g_2)g_3=g_1(g_2g_3)
$.
\end{enumerate}
A group $\G$ is called abelian, or commutative, if in addition to the above properties for any $g,h\in\G$, $gh=hg$.
\end{definition}
\end{myd}

As a very simple example of a group, consider the set of all integers in $\mbb{Z}$. This set together with the `addition' operation forms a group. That is, for any $a,b\in\mbb{Z}$ we have $a+b\in\mbb{Z}$ and the $+$ operation satisfies all the axioms in the definition above. In particular, $0$ is considered as the identity element.

Another example of a group is the Cyclic group, denoted by $\mbb{Z}_n$ for some $n\in\mbb{N}$. This is a finite group consisting of the $n$ integers $\{0,1,\ldots,n-1\}$. The group composition operation is addition modulo $n$, and the identity element is $0$. We will discuss additional examples shortly.

\begin{myd}{}
\begin{definition}
Let $\G_1$ and $\G_2$ be two groups. A group homomorphism is a map $f:\G_1\to \G_2$ that preserves the group operation. That is, for any $g,h\in \G_1$, the map $f$ satisfies
\be
f(g_1g_2)=f(g_1)f(g_2) 
\ee
where the product on the left-hand side is in $G_1$, and the product on the right-hand side is in $G_2$.
\end{definition}
\end{myd}
Note that a group homomorphism maps the identity element $e_1\in\G_1$ to the identity element $e_2\in G_2$; i.e. $f(e_1)=e_2$. This in turn implies that a homomorphism satisfies $\big(f(g)\big)^{-1}=f(g^{-1})$ for all $g\in\G_1$ since 
\be
f(g)f(g^{-1})=f(gg^{-1})=f(e_1)=e_2\;. 
\ee

A subset $\G'\subseteq \G$ is called a subgroup of $\G$ if $\G'$ is itself a group. 
In particular, given a homomorphism $f:\G_1\to \G_2$ between two groups, the image of $\G_1$
\be
\im(f)\eqdef\{f(g)\;:\;g\in\G_1\}
\ee
 is a subgroup of $\G_2$, and the group kernel 
 \be
 \ker(f)\eqdef f^{-1}(e_2)\eqdef\{g\in\G_1\;:\;f(g)=e_2\}
 \ee 
 is a subgroup of $\G_1$. 
 
 \begin{exercise}
 Prove that $\im(f)$ and $\ker(f)$ are indeed subgroups of $\G_2$ and $\G_1$, respectively.
 \end{exercise}

In this book we will consider two types of groups, finite groups and Lie groups.
Finite groups are groups with a finite number of elements. For example, the set of all bijections from a given finite set to itself form a group known as the permutation group (or symmetric group) denoted by $S_n$.  It is known (Cayley's theorem) that every finite group $\G$ is isomorphic to a subgroup of the symmetric group acting on the elements of $\G$. Consequently, the symmetric group plays an important role in various areas of theoretical physics.

Lie groups, on the other hand, are groups that are also smooth differentiable manifolds. That is, a Lie group can be parametrized with a chart of local coordinates, and the smoothness of the manifold means that for any $g,h\in G$ the inversion map $g\mapsto g^{-1}$ and the multiplication map $(g,h)\mapsto gh$ are smooth maps. Here are several examples of Lie groups that are most popular in physics:
\begin{enumerate}
\item The rotation group in $\mbb{R}^n$, denoted by $SO(n)$, is a collection of all $n\times n$ matrices that corresponds to rotations in $\mbb{R}^n$. That is,
\be
SO(n)\eqdef\left\{O\in\mbb{R}^{n\times n}\;:\;O^TO=I_n\;,\;\det(O)=1\right\}\;.
\ee
The group composition operation is the multiplication between matrices, and the identity element of the group is the $n\times n$ identity matrix.
Note that in the trivial case that $n=1$ the group $SO(1)$ consist of a single element (i.e. the identity element), the number 1. In the case that $n=2$ the group can be parametrized with a single parameter, specifically
\be
SO(2)\eqdef\left\{\begin{pmatrix} \cos\theta & -\sin\theta\\ \sin\theta &\cos\theta
\end{pmatrix}\;:\;\theta\in[0,2\pi)\right\}
\ee
As a manifold, this group is isomorphic to the circle. Note also that the inversion of a group element corresponds to $\theta\mapsto2\pi-\theta$ which is clearly a smooth (differentiable) map. Similarly, the composition of two group elements corresponds to the mapping $(\theta_1,\theta_2)\mapsto\theta_1+\theta_2 \mod2\pi$ which is a differentiable map. 

The case $n=3$ corresponds to the group $SO(3)$ which corresponds to rotations in $\mbb{R}^3$. Each rotation in $\mbb{R}^3$ can be described as a rotation by an angle $\theta\in[0,2\pi)$ along some axis of rotation. Let $\n\in\mbb{R}^3$ be the unit vector pointing in the direction of the axis of rotation, and denote by $w\eqdef\cos(\theta/2)$, and
$(x,y,z)^T\eqdef\sin(\theta/2)\n$. Then, $SO(3)$ is a collection of all matrices $R^{(\n)}_{\theta}$ that can be expressed as:
\be\label{orthogonalmatrix}
R^{(\n)}_{\theta}=\begin{pmatrix}
1-2y^2-2z^2 & 2xy-2zw & 2xz+2yw\\
2xy+2zw & 1-2x^2-2z^2 & 2yz-2xw\\
2xz-2yw & 2yz+2xw & 1-2x^2-2y^2
\end{pmatrix}
\ee
It can be shown that if $\v\in\mbb{R}^3$ then $R^{(\n)}_{\theta}\v$ is a vector obtained from $\v$ after rotating it by an angle $\theta$ along the axis of the direction $\n$. The group $SO(3)$ can also be parametrized with the three Euler's angles, as opposed to the axis parametrization above.
\item The unitary group of degree $n$, denoted $U(n)$, is the group of all $n\times n$ unitary matrices. Note that the determinant can be viewed as a group homomorphism
$
\det:U(n)\to U(1)
$
since any unitary matrix has a determinant equals to $e^{i\theta}$ for some $\theta\in[0,2\pi)$. Observe that the kernel of this group consists of all $n\times n$ unitary matrices with determinant one. This subgroup of $U(n)$ is denoted by $SU(n)$ and is called the \emph{special unitary group}. In quantum mechanics, the case $n=2$ corresponds to rotations of $\frac12$-spin particles and therefore plays an important role in physics. This group be expressed as
\be\label{su2}
SU(2)\eqdef\left\{\begin{pmatrix} a & b\\ -\bar{b} & \bar{a}
\end{pmatrix}\;:\;|a|^2+|b^2|=1\;,\;\;\;a,b\in\mbb{C}\right\}
\ee
Denoting by $a=s_0+is_1$ and $b=s_2+is_3$, and condition $|a|^2+|b|^2=1$
becomes $\s\cdot\s=s_0^2+s_1^2+s_2^2+s_3^2=1$. Therefore, the expression above reveals that the underlying manifold of $SU(2)$ is the three-sphere $S^3$, i.e. the sphere or radius one in $\mbb{R}^4$. Such a sphere can be characterized with hyperspherical coordinates $\alpha,\beta\in[0,\pi]$ and $\gamma\in[0,2\pi)$ satisfying
\begin{align}
&s_0=\cos\alpha\nonumber\\
&s_1=\sin\alpha\cos\beta\nonumber\\
&s_2=\sin\alpha\sin\beta\cos\gamma\nonumber\\
&s_3=\sin\alpha\sin\beta\sin\gamma\;.\label{sphsu2}
\end{align}

Alternatively, in quantum physics it is more convinient to parameterize the group elements via $a=r_0+ir_3$ and $b=r_2+ir_1$ where $r_0,r_1,r_2,r_3\in\mbb{R}$. With this parametrization, any matrix $U\in SU(2)$ of the form~\eqref{su2} can be expressed as
\be\label{unitar2}
U=r_0I+i\left(r_1\sigma_1+r_2\sigma_2+r_3\sigma_3\right)
\ee
where $\sigma_1$, $\sigma_2$, and $\sigma_3$, are the three Pauli\index{Pauli} matrices defined in Exercise~\ref{pauli}. 
Given that $r_{0}^2+r_{1}^2+r_{2}^2+r_{3}^2=1$, it is convenient to denote by $\cos(\theta)\eqdef r_0$ and by
$\n\eqdef \frac{1}{\sqrt{1-r_{0}^2}}(r_1,r_2,r_3)^T$ so that
\ba
U&=\cos(\theta)I+i\sin(\theta)\left(n_1\sigma_1+n_2\sigma_2+n_3\sigma_3\right)\\&= \cos(\theta)I+i\sin(\theta)\n\cdot\bs{\sigma}\;.
\ea
where we introduced the notation $\n\cdot\bs{\sigma}$ to mean $n_1\sigma_1+n_2\sigma_2+n_3\sigma_3$. Therefore, any element of SU$(2)$ has the above form.
\item The \emph{general linear group}, denoted $GL(n,\mbb{F})$ (in short $GL(n)$ or $GL(A)$, where $A$ is a Hilbert space of dimension $|A|=n$), is defined as the set of all $n\times n$ invertible matrices. This set is a group under matrix multiplication. An important subgroup of $GL(n)$ that appears for example in multipartite entanglement theory, is the \emph{special linear group} $SL(n)$. It consists of all $n\times n$ matrices with determinant one.
\item The symplectic group is defined for any $n\in\mbb{N}$ as
\be\label{sp2n}
Sp(2n,\mbb{F})\eqdef\left\{M\in\mbb{F}^{2n\times 2n}\;:\;M^TJ_{2n}M=J_{2n}\right\}
\ee
where $J_{2n}\eqdef\begin{pmatrix} 0 & I_n\\ -I_n & 0
\end{pmatrix}$ and the field $\mbb{F}$ can be either $\mbb{R}$ or $\mbb{C}$. It is straightforward to check that the above set is indeed a group under matrix multiplication. The symplectic group comes up in several contexts in physics including both classical and quantum mechanics. For example, it represents symmetries of canonical coordinates that preserve the Poisson bracket of classical mechanics. One of its remarkable properties is that for $n=1$ we have the equality $Sp(2,\mbb{F})=SL(2,\mbb{F})$. This equality follows from a simple observation that any $2\times 2$ complex matrix $M=\begin{pmatrix}
a & b\\
c & d
\end{pmatrix}$ with $a,b,c,d\in\mbb{F}$ satisfies
\be\label{1194}
M^TJ_2M=\begin{pmatrix} 0 & ad-bc\\ bc-ad & 0
\end{pmatrix}=\det(M)J_2\;,
\ee
where $J_2\eqdef\begin{pmatrix} 0 & 1\\ -1 & 0
\end{pmatrix}$.
\item The real line $\mbb{R}$ under addition is a (non-compact) Lie group.
\end{enumerate}

In the examples above, the groups $SO(n)$, $U(n)$, $SU(n)$, are compact, whereas $GL(n)$ or the real line $\mbb{R}$ for example are not compact. Compact Lie groups are the simplest examples of continuous groups, and as such, plays an important role in numerous applications in physics.

\begin{exercise}
Show that 
\be\label{nsig}
e^{i\theta(\n\cdot\bs{\sigma})}=\cos(\theta)I+i\sin(\theta)\n\cdot\bs{\sigma}\;.
\ee
\end{exercise}

From the exercise above it follows that all the elements of SU$(2)$ can be expressed as $e^{i\theta (\n\cdot\bs{\sigma})}$. It will be convenient (see the next exercise) to parametrize the elements of SU$(2)$ with the matrices
\be\label{c15}
T_\theta^{(\n)}\eqdef e^{-i\frac\theta2 (\n\cdot\bs{\sigma})}
\ee
where $\theta\in[0,4\pi)$ and $\n\in\mbb{R}^3$ is a unit vector. Note that we divided $\theta$ by $-2$ to obtain the following relation between SU$(2)$ and SO$(3)$.

\begin{exercise}\label{2to1}
Consider the map $f:{\rm SU}(2)\to{\rm SO}(3)$ defined by 
\be
f\left[e^{-i\frac{\theta}{2}(\n\cdot\bs{\sigma})}\right]=R^{(\n)}_{\theta}\;.
\ee 
\begin{enumerate}
\item Show that $f$ is a group homomorphism. That is,
given two unit vectors $\n_1$ and $\n_2$, and two rotation angles $\theta_1$ and $\theta_2$, 
\be\label{corres}
f\left[e^{-i\frac{\theta_1}{2}(\n_1\cdot\bs{\sigma})}e^{-i\frac{\theta_2}{2}(\n_2\cdot\bs{\sigma})}\right]=
f\left[e^{-i\frac{\theta_1}{2}(\n_1\cdot\bs{\sigma})}\right]f\left[e^{-i\frac{\theta_2}{2}(\n_2\cdot\bs{\sigma})}\right]\;.
\ee
\item Show that $f$ is $2:1$ (two-to-one) and onto. That is,  every element in {\rm SO}$(3)$ corresponds exactly to two elements in {\rm SU}$(2)$. Hint: Denote $w\eqdef\cos(\theta/2)$ and 
$(x,y,z)^T=\sin(\theta/2)\n$, and use the fact that $R^{(\n)}_{\theta}$ can be expressed as in~\eqref{orthogonalmatrix}.
\end{enumerate}
\end{exercise}

\section{Group Representations}
\begin{myd}{}
\begin{definition}
Let $A$ be a Hilbert space and $\G$ a group. A homomorphism $\pi:\G\to\ml(A)$ mapping every element $g\in \G$ to a matrix $\pi(g)\in\ml(A)$ is called a \emph{representation} of $\G$.
\end{definition}
\end{myd}
\begin{remark}
The image of $\pi$ in the definition above is a subset of $\ml(A)$ consisting of $|A|\times |A|$ invertible matrices. Therefore, one can replace $\ml(A)$ in the definition above with the general linear group $GL(A)$. Note that since $\pi$ is a homomorphism it follows that $\pi(e)=I^A$. Moreover, note that for any group $\G$ and any Hilbert space $A$, there exists a group representation $\pi(g)\eqdef I^A$ for all $g\in\G$. This representation is called the \emph{trivial} representation. 
\end{remark}

For a given representation $\pi:\G\to\ml(A)$, a subspace $B\subseteq A$ is said to be a $G$-invariant subspace of $A$ if for all $|\psi\ra\in B$ and all $g\in\G$ the vector
$\pi(g)|\psi\ra$ is also in $B$. Observe that if $B$ is a $G$-invariant subspace of $A$ then the restriction of $\pi(g)$ to the subspace $B$, denoted by $\pi(g)\big|_B$, can itself be viewed as a representation of $\G$, taking elements in $\G$ to elements in $\ml(B)$. This representation is called a \emph{subrepresentation} of $\pi$. Note that the mapping from $\G$ to $\ml(B)$ given by $g\mapsto\pi(g)\big|_B$ is a subrepresentation of $\pi$ if and only if $B$ is a $G$-invariant subspace of $A$. Therefore, quite often we refer to a $G$-invariant subspace of $A$ as a subrepresentation of $A$, where we think of $A$ (somewhat implicitly) as the representation $\pi:\G\to\ml(A)$.

According to this definition, every representation $A$ has at least two subrepresentations. These are the trivial representations in which $B=\{\0\}$ or $B=A$ (in the latter, meaning that $\pi$ is a subrepresentation of itself). 
An \emph{irreducible representation}, in short called \emph{irrep}, is a representation that has no proper (i.e. non-trivial) subrepresentation. Therefore, if $B\subseteq A$ is a $G$-invariant subspace w.r.t.\ an irrep then $B=\{\0\}$ or $B=A$.

For example, consider a representation of $SO(2)$ in $\mbb{R}^4$ defined by
\be\label{simex}
\theta\mapsto\begin{pmatrix}
\cos\theta & -\sin\theta & 0 & 0\\ 
\sin\theta &\cos\theta & 0 & 0\\
0 & 0 & \cos\theta & \sin\theta\\
0 & 0 &-\sin\theta &\cos\theta
\end{pmatrix}
\ee
Clearly, the above representation has two proper subrepresentations of $\mbb{R}^4$. Each of these two subrepresentations cannot be reduced further, so they are irreps. However, these two irreps are equivalent.

\begin{myd}{Equivalent Representations}
\begin{definition}
Two representations or subrepresentations, $\pi_1:\G\to\ml(A)$ and $\pi_2:\G\to\ml(B)$, are said to be \emph{equivalent} if there exists an \emph{isomorphism}  $\eta:A\to B$ (in particular, $|A|=|B|$ and $\eta$ is invertible) such that
\be\label{ce}
\pi_1(g)\eta=\eta\pi_2(g)\;.
\ee
If there is no such an \emph{intertwiner} map $\eta$ we say that the two representations are inequivalent.
\end{definition}
\end{myd}

Note that in particular, since $\eta$ is invertible, for each $g\in\G$ the matrix $\pi_1(g)$ in~\eqref{ce} is similar to the matrix $\pi_2(g)$.
In Fig.~\ref{fintertwiner} we drew a commutativity diagram describing the equivalence of two representations that are related via~\eqref{ce}. Note that the action of the representation $\pi_1$ on the Hilbert space $A$ is mirrored by $\eta$ to the Hilbert space $B$ in which it takes the form of $\pi_2$. Note also that the directions of all the arrows in the figure are reversible.

\begin{figure}[h]\centering    \includegraphics[width=0.2\textwidth]{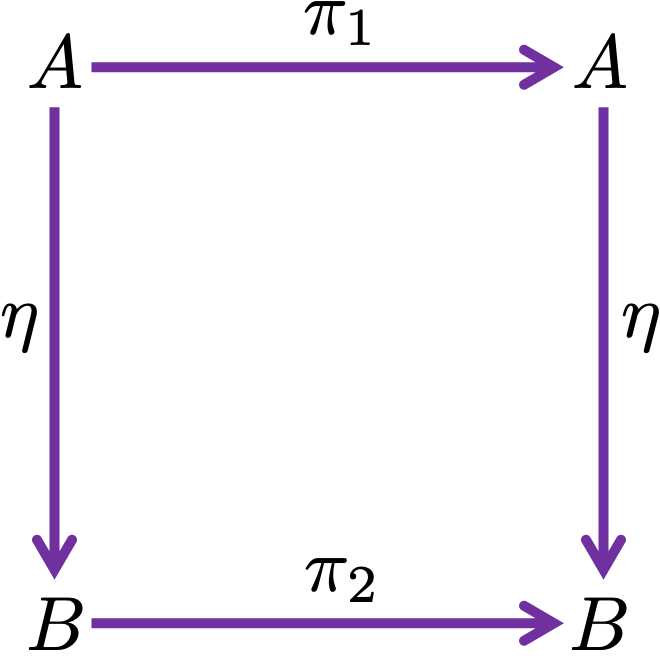}
  \caption{\linespread{1}\selectfont{\small A commutativity diagram for two equivalent representations. Each arrow is reversible.}}
  \label{fintertwiner}
\end{figure} 

\begin{myt}{\color{yellow} Schur's Lemma}
\begin{theorem}\label{schur}
Let $\G$ be a group, and $A_1$ and $A_2$ be two Hilbert spaces. Also let 
$\pi_1:\G\to\ml(A_1)$ and $\pi_2:\G\to\ml(A_2)$ be two irreducible representations of $\G$, and suppose there exists a complex matrix (linear transformation) $T:A_1\to A_2$ that is equivalent under the action of $\G$; that is, $T\pi_1(g)=\pi_2(g)T$ for all $g\in\G$. Then,
\begin{enumerate}
\item If $\pi_1$ and $\pi_2$ are inequivalent representations then $T=0$.
\item If $\pi_1=\pi_2\eqdef\pi$ (in particular $A_1=A_2\eqdef A$) then $T=\lambda I^A$ for some $\lambda\in\mbb{C}$.
\end{enumerate}
\end{theorem}
\end{myt}
\begin{proof}
{\it Part 1.} The idea of the proof is to look at the kernel and image of $T$. Let $|\psi\ra\in\ker(T)$. Then, from the commutativity property of $T$, for all $g\in\G$
\be
T\pi_1(g)|\psi\ra=\pi_2(g)T|\psi\ra=\pi_2(g)\0=\0\;.
\ee
That is, if $|\psi\ra\in\ker(T)$ then also $\pi_1(g)|\psi\ra\in\ker(T)$ for all $g\in\G$. In other words, $\ker(T)$ is a $G$-invariant subspace of $A_1$. Now, recall that $\pi_1$ is an irrep, and therefore since $\ker(T)$ is a $G$-invariant subspace of $A_1$ we must have $\ker(T)=\{\0\}$ or $\ker(T)=A_1$. 

Next, let $|\phi\ra\in\im(T)$. Then, there exists $|\psi\ra\in A_1$ such that $T|\psi\ra=|\phi\ra$. Therefore, using the commutativity property of $T$ we get that for all $g\in\G$ 
\be
\pi_2(g)|\phi\ra=\pi_2(g)T|\psi\ra=T\pi_1(g)|\psi\ra\;.
\ee
That is, if $|\phi\ra\in\im(T)$ then also $\pi_2(g)|\phi\ra\in\im(T)$ for all $g\in\G$.
Hence, $\im(T)$ is a $G$-invariant subspace of $A_2$, and since $\pi_2$ is an irrep we must have either $\im(T)=\{\0\}$ or $\im(T)=A_2$.

Combining everything, we conclude that there are two options: (1) $\ker(T)=\{\0\}$ and $\im(T)=A_2$, or (2)
$\ker(T)=A_1$ and $\im(T)=\{\0\}$. From Exercise~\ref{exschur}  $(1)$ can only hold if $A_1=A_2$ and $T$ is invertible. This option is not possible since we assume in Part 1 that $\pi_1$ and $\pi_2$ are inequivalent. Option (2) on the other hand implies that $T=\0$ (see Exercise~\ref{exschur}). This completes the first part of the proof.

{\it Proof of Part 2.} The proof is based on the fundamental theorem of algebra that states that every non-constant single-variable polynomial with complex coefficients has at least one complex root. Therefore, there exists $\lambda\in\mbb{C}$ that is a root for the characteristic polynomial of $T$; i.e. $\det\left(T-\lambda I^A\right)=0$. This means that there exists a non-zero vector $|\psi\ra\in\ker\left(T-\lambda I^A\right)$. We then get for all $g\in\G$
\be
\left(T-\lambda I^A\right)\pi(g)|\psi\ra=\pi(g)\left(T-\lambda I^A\right)|\psi\ra=0\;,
\ee
where we used the commutativity of $T$ with $\pi(g)$. The above equation implies that $\pi(g)|\psi\ra\in\ker\left(T-\lambda I^A\right)$ for all $g\in\G$ and all $|\psi\ra\in\ker\left(T-\lambda I^A\right)$. Therefore, $\ker\left(T-\lambda I^A\right)$ is a $G$-invariant subspace of $A$, and since $\pi$ is an irrep we must have $\ker\left(T-\lambda I^A\right)=A$ (recall that we already ruled out the case $\ker\left(T-\lambda I^A\right)=\{\0\}$). We therefore conclude that $T=\lambda I^A$.
\end{proof}

\begin{exercise}\label{abelian}
Show that all the irreps (over a complex field) of an abelian group $\G$ are 1-dimensional.
\end{exercise}

The following theorem demonstrates that all representations of a finite group can be decomposed into a direct sum of irreps.

\begin{myt}{}
\begin{theorem}\label{findecom}
Let $G$ be a \emph{finite} group.
Then, every representation $\pi:\G\mapsto\ml(A)$ can be decomposed into a direct sum of irreps of $A$.
\end{theorem}
\end{myt}
\begin{proof}
If there are no proper (i.e. non-trivial) subrepresentation of $A$ then $\pi$ is itself an irrep and the proof is done. Therefore, suppose $A_1$ is a proper $G$-invariant subspace corresponding to the proper subrepresentation $\pi_1:\G\to\ml(A_1)$ (i.e. $\pi_1$ is subrepresentation of $\pi$). Let $P:A\to A$ be the projection to the subspace $A_1$, and define the operator $T:A\to A$ as
\be\label{findecom2}
T\eqdef\frac1{|G|}\sum_{g\in G}\pi(g)P\pi(g^{-1})\;.
\ee
Observe that the operator $T$ is an intertwiner, that is, for any $h\in\G$
\ba
T\pi(h)&=\frac1{|G|}\sum_{g\in G}\pi(g)P\pi(g^{-1})\pi(h)\\
&=\frac1{|G|}\sum_{g\in G}\pi(g)P\pi(g^{-1}h)\\
\Gg{a\eqdef h^{-1}g}&=\frac1{|G|}\sum_{a\in G}\pi\left(ha\right)P\pi(a^{-1})\\
&=\pi(h)\frac1{|G|}\sum_{a\in G}\pi\left(a\right)P\pi(a^{-1})\\
&=\pi(h)T\;.
\ea
Moreover, we argue now that $T$ is a projection.
First, observe that if $|\psi\ra\in A_1$ also $\pi(g)|\psi\ra\in A_1$ since $A_1$ is a $G$-invariant subspace. This in particular implies that $P\pi(g)|\psi\ra=\pi(g)|\psi\ra$ so we conclude that for all $|\psi\ra\in A_1$
\ba
T|\psi\ra&=\frac1{|G|}\sum_{g\in G}\pi(g^{-1})P\pi(g)|\psi\ra\\
\Gg{P\pi(g)|\psi\ra=\pi(g)|\psi\ra\to}
&=\frac1{|G|}\sum_{g\in G}\pi(g^{-1})\pi(g)|\psi\ra=\frac1{|G|}\sum_{g\in G}|\psi\ra=|\psi\ra\;.
\ea
Second, observe  that for any $|\psi\ra\in A$ (not necessarily in $A_1$) we have 
$T|\psi\ra\in A_1$. Combining this with the above equation gives $T^2|\psi\ra=T|\psi\ra$ for all $|\psi\ra\in A$. Hence, $T^2=T$; i.e. $T:A\to A$ is a projection and an intertwiner. Therefore, both $\im(T)=A_1$ and  $A_0\eqdef\ker(T)$ are $G$-invariant subspaces and we have $A=A_1\oplus A_0$ (as representations). Repeating the process we can continue in this way to decompose $A_0$ and $A_1$ into direct sum of $G$-invariant subspaces until we decompose $A$ into a direct sum of irreducible $G$-invariant subspaces.
\end{proof}

\section{Unitary Projective Representations}

A unitary representation of a group $\G$, is a group representation $\pi:\G\to\ml(A)$ in which all the elements $\pi(g)$ are unitary operators in $\ml(A)$. In this case we denote the representation $\pi$ by $U$ and the elements $\pi(g)$ by $U_g$. The set of unitaries $\{U_g\}$ is itself a group in $\ml(A)$. Therefore, if the mapping $g\mapsto U_g$ is one-to-one then the group $\{U_g\}$ is isomorphic to $\G$. For example, the elements of the group $SU(2)$ as defined in~\ref{su2} are themselves unitary matrices and therefore the group $SU(2)$ equals its unitary representation on $\ml(\mbb{C}^2)$.

In quantum mechanics, quantum states are represented with a density matrix $\rho\in\md(A)$. Therefore, we will mostly be interested in the action of a group $\G$ on density matrices. Such an action (i.e. representation of $\G$) can still be unitary but with a slight modification given the fact that we are not intrested in maps of the form $|\psi\ra\to U_g|\psi\ra$ but instead in maps of the form $|\psi\lr\psi|\to U_g|\psi\lr\psi|U_g^*$ or more generally
\be
\rho\mapsto\mU_g(\rho)\eqdef U_g\rho U_g^*\;.
\ee
The above mapping is the typical way a symmetry is represented in the space of quantum states (we will discuss it in much more details in Chapter~\ref{Ch:Asymmetry}). We will therefore be mostly interested in group representations of the form $g\mapsto\mU_g$, where $\mU_g$ is defined as above such that for each $g\in\G$ the matrix $U_g$ is unitary. Note that if the mapping $g\mapsto U_g$ is a unitary representation of $\G$ then also $g\mapsto \mU_g$ is a group representation. However, the latter does not implies the former since $\mU_g$ is insensitive to phases, i.e.\ it is invariant under $U_g\mapsto e^{i\theta}U_g$. Therefore, in order to describe the representation of a symmetry in quantum mechanics, we need to relax a bit the requirement that a representation is unitary, and instead require it to be ``only" a \emph{projective} unitary representation.

\begin{myd}{Projective Unitary Representation}
\begin{definition}
Let $\G$ be a group and $A$ a Hilbert space.
A representation $U:\G\to\ml(A)$, defined by the mapping $g\mapsto U_g$, is called a projective unitary representation of $\G$ if for each $g\in \G$ the matrix $U_g$ is unitary, and 
\ba\label{ugh}
U_{g}U_h&=\omega(g,h)U_{gh}\quad\forall g,h\in\G\;,\\
U_e&=I^A
\ea
where $\omega(g,h)\in\mbb{C}$ with $|\omega(g,h)|=1$. The phase factor $\omega(g,h)$ is also called a \emph{cocycle}.
\end{definition}
\end{myd}
\begin{remark}
The cocycle must satisfy
\be\label{co1}
\omega(g,e)=\omega(e,g)=1\quad\quad\forall\;g\in\G\;,
\ee
since 
\be
U_g=U_gU_e=\omega(g,e)U_{ge}=\omega(g,e)U_{g}\;.
\ee 
Similarly, the cocycles must satisfy the cocycle equation
\be\label{co2}
\omega(a,bc)\omega(b,c)=\omega(a,b)\omega(ab,c)\quad\quad\forall\;a,b,c\in\G\;,
\ee
since on one hand
\be
U_aU_bU_c=\omega(b,c)U_{a}U_{bc}=\omega(b,c)\omega(a,bc)U_{abc}
\ee
and on the other hand
\be
U_aU_bU_c=\omega(a,b)U_{ab}U_{c}=\omega(a,b)\omega(ab,c)U_{abc}\;.
\ee
\end{remark}

It is important to note that two projective unitary representations of the \emph{same} group can have \emph{different} cocycles. If two projective unitary representations $g\mapsto U_g$ and $g\mapsto V_g$ have the same cocycle, i.e. $\omega_U(g,h)=\omega_V(g,h)$ for all $g,h\in\G$, then we say that the two representations are in the same factor system. Note also that two subrepresentations of a given projective unitary representation $U$ always have the same cocycle as $U$.

\bex\label{pur}
Let $g\mapsto U_g$ be a projective unitary representation of a group $\G$ with a cocycle $\omega(g,h)$ for all $g,h\in\G$. Show that $g\mapsto \bar{U}_g$ is also a projective unitary representation with cocycle $\overline{\omega(g,h)}$. Why (in general) $g\mapsto U^*_g$ is not a representation of $\G$?
\eex

As an example, consider the group $G=\mbb{Z}_n\times\mbb{Z}_n$ consisting of the cartesian products of two copies of the cyclic group. The elements of the group are pair of integers $(p,q)$ with both $p,q\in\{0,1,\ldots,n-1\}$ and the group operation, denoted by the addition symbol +, of two pairs is given by
\be
(p,q)+(p',q')\eqdef\big(p+p'\;({\rm mod}\;n)\;,\;q+q'\;({\rm mod}\;n)\big)\;.
\ee
Define the phase and shift operators $T,S:\mbb{C}^n\to\mbb{C}^n$ by
\be
T|x\ra=e^{i\frac{2\pi x}{n}}|x\ra\quad,\quad S|x\ra\eqdef\big|x+1\;({\rm mod}\;n)\big\ra\quad\forall\;x\in\{0,1,\ldots,n-1\}
\ee
Note that both $S$ and $T$ are unitary matrices. We define a projective unitary representation $W:G\to\ml(\mbb{C}^n)$ via 
\be\label{wpq}
(p,q)\mapsto W_{p,q}\eqdef S^pT^q\quad\quad\forall\;(p,q)\in G\;.
\ee 
Since $p$ and $q$ are integers we have that $W_{p,q}$ is a unitary matrix. Observe that 
\be
ST|x\ra=e^{i\frac{2\pi x}{n}}S|x\ra=e^{i\frac{2\pi x}{n}}|x+1\;({\rm mod}\;n)\big\ra
\ee
whereas
\be
TS|x\ra=T|x+1\;({\rm mod}\;n)\big\ra=e^{i\frac{2\pi (x+1)}{n}}|x+1\;({\rm mod}\;n)\big\ra\;.
\ee
Therefore, we conclude that
\be\label{st=}
ST=e^{i\frac{2\pi}{n}}TS\;.
\ee
In the exercise below you show that $\{W_{p,q}\}$ is a projective unitary representation of $G$. The operators $W_{p,q}$ are known as the Hiesenberg-Weyl operators.
\begin{exercise}
Use the relation~\eqref{st=} to show that the mapping $(p,q)\mapsto W_{p,q}$ forms a projective unitary representation of $\mbb{Z}_n\times\mbb{Z}_n$. Find its cocycle.
\end{exercise}

\subsection{Direct Sum Decompositions}

For finite groups, we saw in Theorem~\ref{findecom} that group representations can be decomposed into a direct sum of irreps. We show now that this also holds for all projective unitary representations.
\begin{myt}{}
\begin{theorem}\label{findecom3}
Let $\G$ be a group, and let $U:\G\to\ml(A):g\mapsto U_g$ be a projective unitary representation of $G$.
Then, $U$ can be decomposed into a direct sum of irreps of $A$.
\end{theorem}
\end{myt}
\begin{proof}
If there are no proper (i.e. non-trivial) subrepresentation of $A$ then $U$ is itself an irrep and the proof is done. Therefore, suppose $B$ is a proper $\G$-invariant subspace of $A$. Let $B^\perp$ be the orthogonal complement of $B$ in $A$.
Let $|\phi\ra\in B^\perp$, and observe that since $B$ is $G$-invariant we have that for all $|\psi\ra\in B$ and all $g\in\G$, $U_{g}|\psi\ra\in B$. This means that for any $|\psi\ra\in B$ we have
\be
\la\psi|U_g|\phi\ra=\overline{\la\phi|U_{g^{-1}}|\psi\ra}=0\;.
\ee
That is, $U_g|\phi\ra\in B^\perp$ for all $g\in\G$, so that $B^\perp$ is $\G$-invariant.
Repeating the process we can continue in this way to decompose $B$ and $B^\perp$ into direct sum of proper $G$-invariant subspaces until we decomposed $A$ into a direct sum of \emph{irreducible} $G$-invariant subspaces.
\end{proof}

The theorem above states that a projective unitary representation $g\mapsto U_g$ can be decomposed as
\be
U_g=\bigoplus_{j=1}^kU_g^{(j)}\quad\quad\forall\;g\in\G
\ee
where each $U^{(j)}:g\mapsto U_g^{(j)}$ is an irrep of $U$. Note that some of these irreps may be equivalent. It will be convenient  to group such equivalent irreps, and denote by $\irr(U)$ the set of all equivalent classes of irreps that appear in the decomposition above. That is, any $\lambda\in\irr(U)$ represents a class of equivalent irreps. The number $m_\lambda$ of equivalent irreps in the same $\lambda$-equivalence class is called \emph{multiplicity}.
With these notations, the above decomposition can be expressed as
\be\label{decoug}
U_g=\bigoplus_{\lambda\in\irr(U)}\bigoplus_{x\in[m_\lambda]}U_g^{(\lambda,x)}\;,
\ee
where for each $x\in[m_\lambda]$ the map $U^{(\lambda,x)}:g\mapsto U_g^{(\lambda,x)}$ is an irrep belonging to the $\lambda$-equivalence class.

For example, consider the unitary representation $\theta\mapsto U_\theta$ of $SU(2)$ in $\mbb{R}^4$, where $U_\theta$ is the $4\times 4$ matrix given in~\eqref{simex}. Clearly, we can express $U_\theta=U_\theta^{(1)}\oplus U_\theta^{(2)}$, where
\be
U_\theta^{(1)}\eqdef\begin{pmatrix}
\cos\theta & -\sin\theta \\ 
\sin\theta &\cos\theta 
\end{pmatrix}
\quad\text{and}\quad U_\theta^{(2)}\eqdef
\begin{pmatrix}
\cos\theta & \sin\theta\\
-\sin\theta &\cos\theta
\end{pmatrix}
\ee
This is the direct sum decomposition into irreps of $\theta\mapsto U_\theta$. Note that in this case we have only one equivalence class, without loss of generality we can name it $\lambda=1$, and this equivalence class contains two irreps given by $U_\theta^{(1)}$ and $U_\theta^{(2)}$, so that the multiplicity of this irrep is $m_1=2$ (i.e. $m_{\lambda=1}=2$).

As another example, consider the group $U(1)$ and its representation $\theta\mapsto U_\theta$, where
\be\label{utk}
U_\theta=\sum_{k\in[n]}e^{i\theta k}|k\lr k|\;.
\ee
Clearly, this representation already written as direct sum of irreps. Observe that 
for each $k$, the map $\theta\mapsto e^{i\theta k}|k\lr k|$ defines a 1-dimensional irrep of $U_\theta$ (recall that for abelian groups all irreps are 1-dimensional; see Exercise~\ref{abelian}). In this case the equivalence class of irreps is labeled by $\lambda=k$ and the multiplicity $m_k=1$ for all $k\in[n]$. 

The following theorem slightly simplify the decomposition~\eqref{decoug}.
\begin{myt}{}
\begin{theorem}\label{thdco}
Let $A$ be a Hilbert space and $g\mapsto U_g$ be a projective unitary representation of a group $\G$ in $\ml(A)$. The representation $U_g$ induces the following structure on $A$
\be\label{decoal}
A=\bigoplus_{\lambda\in\irr(U)} A_{\lambda}\eqdef\bigoplus_{\lambda\in\irr(U)} B_\lambda\otimes C_\lambda\;,
\ee
where $B_\lambda$ is an irreducible $G$-invariant subspace of $A$,
and $C_\lambda=\mbb{C}^{m_\lambda}$. Moreover, under this decomposition, each $U_g$ takes the form
\be\label{lamir}
U_g\cong\bigoplus_{\lambda\in\irr(U)} U_g^{(\lambda)}\otimes I^{C_\lambda}\;,
\ee
where $U_g^{(\lambda)}$ acts irreducibly on $B_\lambda$, and $I^{C_\lambda}$ is the identity matrix on $C_\lambda$.
\end{theorem}
\end{myt}
\begin{remark}
The subspace $B_\lambda$ is called \emph{the representation space}, and the subspace $C_\lambda$ is called \emph{the multiplicity space}. They are mathematical objects and we will think about them later on as virtual subsystems. 
Moreover, the above decomposition of $A$ means that there exists an orthonormal basis $\{|\lambda,m,x\ra^{A_\lambda}\}_{\lambda,m,x}$ whose elements are 
\be\label{ginvbasis}
|\lambda,m,x\ra^{A_\lambda}\eqdef |\lambda,m\ra^{B_\lambda}|x\ra^{C_\lambda}
\ee
where $\{|x\ra^{C_\lambda}\}_{x\in[m_\lambda]}$ is an orthonormal basis of the multiplicity space $C_\lambda$, and $\{|\lambda,m\ra^{B_\lambda}\}_{m=1}^{d_\lambda}$ is an orthonormal basis of representation space $B_{\lambda}$, where $d_\lambda\eqdef|B_{\lambda}|$.
\end{remark}
\begin{proof}
We first argue that without loss of generality the intertwiner map between two irreps $U_g^{(\lambda,x)}$ and $U_g^{(\lambda,x')}$ in the decomposition~\eqref{decoug}can be taken to be unitary. Indeed, by definition of equivalent representations, if $T$ is the intertwiner between $U_g^{(\lambda,x)}$ and $U_g^{(\lambda,x')}$ then $U_g^{(\lambda,x)}T=TU_g^{(\lambda,x')}$. Since both $U_g^{(\lambda,x)}$ and $U_g^{(\lambda,x')}$ are unitary matrices we must have
\ba
T^*T&=T^*U_g^{*(\lambda,x)}U_g^{(\lambda,x)}T\\
&=U_g^{^*(\lambda,x')}T^*TU_g^{(\lambda,x')}\quad\quad\forall\;g\in\G\;.
\ea
Therefore, since $U_g^{(\lambda,x')}$ is an irrep of $G$, from the second part of Schur's Lemma (see Theorem~\ref{schur}) it follows that $T^*T=\lambda I$ for some $\lambda\in\mbb{C}$. Since $T^*T> 0$ (recall that $T$ is invertible) we can redefine $T\mapsto\frac1{\sqrt{\lambda}}T$ so that the new $T$ is unitary.

Now, denote by $U^{(\lambda)}_g\eqdef U_g^{(\lambda,1)}$ and by $T_x^{(\lambda)}$ the unitary intertwiner satisfying
\be
U_g^{(\lambda,x)}=T_x^{(\lambda)}U_g^{\lambda}T_x^{*(\lambda)}\;.
\ee
Taking the direct sum over $x\in[m_\lambda]$ on both sides of the equation above gives
\be\label{tlamdir}
\bigoplus_{x\in[m_\lambda]}U_g^{(\lambda,x)}= T^{\lambda}\left(U_{g}^{\lambda}\otimes I_{m_{\lambda}}\right)T^{*\lambda}
\ee
where $I_{m_{\lambda}}$ is the $m_{\lambda}\times m_{\lambda}$ identity matrix, $T^\lambda\eqdef\oplus_{x\in[m_\lambda]}T_x^{(\lambda)}$ is a unitary matrix, and $U_{g}^{\lambda}\otimes I_{m_{\lambda}}$ is viewed as 
\be
U_{g}^{\lambda}\otimes I_{m_{\lambda}}=\underbrace{U_{g}^{\lambda}\oplus\cdots\oplus U_{g}^{\lambda}}_{m_\lambda\text{-times}}\;.
\ee
Finally, taking the direct sum over all $\lambda\in\irr(U)$ on both sides of~\eqref{tlamdir}, we get that the unitary matrix $T\eqdef \oplus_{\lambda}T^{\lambda}$ satisfies
\be
\bigoplus_{\lambda\in\irr(U)}\bigoplus_{x\in[m_\lambda]}U_g^{(\lambda,x)}=T\Big(\bigoplus_{\lambda\in\irr(U)}\left(U_{g}^{\lambda}\otimes I_{m_{\lambda}}\right)\Big)T^*\;.
\ee
The proof is concluded with the identification of the subspaces $B_\lambda$ and $C_\lambda$ as the subspaces on which $U_{g}^{\lambda}$ and $ I_{m_{\lambda}}$ act upon (with $I_{m_\lambda}=I^{C_\lambda}$).
\end{proof}

\subsection{$\G$-Invariant Operators}

\begin{myd}{}\begin{definition}\label{ginva}
Let $g\mapsto U_g$ be a projective unitary representation of a group $\G$ in $\ml(A)$. An operator $\rho\in\ml(A)$ is called $\G$-invariant with respect to this representation if $[\rho,U_g]=0$ for all $g\in\G$.
\end{definition}
\end{myd}

Invariant states, often called symmetric states, plays an important role in physics, particularly in the resource theory of asymmetry. The following theorem provide a simple characterization of such states with respect to the decomposition~\eqref{decoal} of the underlying Hilbert space.

\begin{myt}{\color{yellow} Characterization of $\G$-Invariant Operators}
\begin{theorem}\label{deop}
Let $\rho\in\ml(A)$ and $U:\G\to\ml(A)$ be a projective unitary representation. Then, $\rho$ is $\G$-invariant if and only if 
$\rho$ can be decomposed as
\be\label{formra}
\rho^A=\bigoplus_{\lambda\in\irr(U)} \u^{B_\lambda}\otimes \rho^{C_\lambda}_{\lambda}
\ee
where $\u^{B_\lambda}=\frac1{|B_\lambda|}I^{B_\lambda}$ is the maximally mixed (uniform) state on system $B_\lambda$, and
\be\label{ouni}
 \rho^{C_\lambda}_\lambda\eqdef\tr_{B_\lambda}\left[\Pi^{A_\lambda}\rho^A\Pi^{A_\lambda}\right]\;,
\ee
where $\Pi^{A_\lambda}$ is the projection to the subspace $A_\lambda$ as defined in~\eqref{decoal}.
\end{theorem}
\end{myt}

\begin{proof}
We are working in a basis in which $U_g$ has the form~\eqref{lamir}. Therefore, if $\rho^A$ has the form~\eqref{formra} then it clearly commutes with $U_g$ for all $g\in\G$ so that $\rho^A$ is $\G$-invariant. Conversely, suppose $\rho^A$ is $\G$-invariant, and denote 
\be
\rho^A=\sum_{\lambda,\lambda'\in\irr(U)}\rho_{{}_{\lambda\lambda'}}\quad\text{where}\quad \rho_{{}_{\lambda\lambda'}}\eqdef\Pi^{A_{\lambda'}}\rho^A\Pi^{A_{\lambda}}\;.
\ee
Note that $\rho_{{}_{\lambda\lambda'}}$ is a linear map from $A_{\lambda}\to A_{\lambda'}$.
Since $\rho$ commutes with $U_g$ for all $g\in\G$ it follows immediately from the form~\eqref{lamir} of $U_g$ that  
\ba
0=[\rho^A, U_g]&=\sum_{\lambda,\lambda'}\left[\rho_{{}_{\lambda\lambda'}},U_g\right]\\
&=\sum_{\lambda,\lambda'}\left(\rho_{{}_{\lambda\lambda'}}\left(U_g^{(\lambda)}\otimes I^{C_{\lambda}}\right)-\left(U_g^{(\lambda')}\otimes I^{C_{\lambda'}}\right)\rho_{{}_{\lambda\lambda'}}\right)
\ea
Multiplying both sides of the equation above by $\Pi^{A_{\lambda'}}$ from the right, and  $\Pi^{A_{\lambda}}$ from the left, we get for all $\lambda$ and $\lambda'$ 
\be
\rho_{{}_{\lambda\lambda'}}\left(U_g^{(\lambda)}\otimes I^{C_{\lambda}}\right)=\left(U_g^{(\lambda')}\otimes I^{C_{\lambda'}}\right)\rho_{{}_{\lambda\lambda'}}\;.
\ee
Now, by multiplying from the left both sides with $I^{B_{\lambda}}\otimes T^{C_{\lambda'}\to C_{\lambda}}$, for some $m_{\lambda}\times m_{\lambda'}$ matrix $T\in\ml(C_{\lambda'}, C_{\lambda})$ and taking the partial trace over $C_{\lambda}$ gives
\be
\omega_{{}_{\lambda\lambda'}}U_g^{(\lambda)}=U_g^{(\lambda')}\omega_{{}_{\lambda\lambda'}}\quad\text{where}\quad \omega_{{}_{\lambda\lambda'}}\eqdef\tr_{C_{\lambda}}\left[\left(I^{B_{\lambda}}\otimes T\right)\rho_{{}_{\lambda\lambda'}}\right]\;.
\ee
Finally, from the first part of Schur's lemma it follows that unless $\lambda=\lambda'$ we get $\omega_{{}_{\lambda\lambda'}}=0$. Since this holds for all $T\in\ml(C_{\lambda'}, C_{\lambda})$ we conclude from Exercise~\ref{abapbp} that also $\rho_{{}_{\lambda\lambda'}}=0$ for $\lambda\neq\lambda'$. Moreover, from the second part of Schur's lemma we get that $\omega_{{}_{\lambda\lambda}}=\tr_{C_{\lambda}}\left[\left(I^{B_{\lambda}}\otimes T\right)\rho_{{}_{\lambda\lambda'}}\right]$ is proportional to the identity matrix for all $T\in\ml(C_\lambda)$. Hence, from Exercise~\ref{propounit} we conclude that
$\rho_{\lambda\lambda}=\u^{B_\lambda}\otimes\rho_{\lambda}^{C_\lambda}$. This completes the proof.
\end{proof}

The theorem above apply to any operator $\rho\in\ml(A)$. In this book we will only consider $\G$-invariant quantum states; i.e.\ $\G$-invariant operators in $\md(A)$. For the case that $\rho$ is a pure quantum state we have the following corollary.

\begin{myg}{}
\begin{corollary}
Let $g\mapsto U_g\in\ml(A)$ be a projective unitary representation of $\G$ and let $\psi\in\pure(A)$ be a pure state. The following are equivalent:
\begin{enumerate}
\item $\psi$ is $\G$-invariant; i.e. $U_g\psi U_g^*=\psi$ for all $g\in\G$. 
\item There exists an irrep $\lambda$ with $|B_\lambda|=1$ such that $\psi\in\pure(C_\lambda)$.
\end{enumerate}
\end{corollary}
\end{myg}

\begin{proof}
Taking $\rho=\psi$ in~\eqref{formra}, it follows that the right-hand side of~\eqref{formra} is a rank one matrix only if and only if the direct sum consists of a single non-zero term, denoted by $\lambda$, for which $|B_\lambda|=1$. This completes the proof.
\end{proof}

\section{Invariant Measures Over a Lie Group}\label{sec:inv}

In the proof of Theorem~\ref{findecom}, particularly Eq.~\eqref{findecom2} we used a type of group average over all elements of the group. This type of averaging turns out to be extremely useful in applications, and in particular, its extension to continuous groups can be used to prove a variant of Theorem~\ref{findecom} that is applicable to compact Lie groups.

A measure on a Lie group, denoted here by $\mu$, is a map that assign a volume or size to any subset of $\G$. Typically, subsets of $\G$ have volume since $\G$ is a manifold. More rigorously, let $\Sigma_\G$ denotes all subsets of $\G$ including $\G$ itself (in mathematics $\Sigma_\G$ is called $\sigma$-algebra). We then define a measure on $\G$ as follows.

\begin{myd}{}
\begin{definition}
A function 
$\mu:\Sigma_\G\to\mbb{R}$ is called a \emph{measure} on $\G$ if it satisfies the following three conditions: 
\begin{enumerate}
\item $\mu$ non-negative, i.e. $\mu(S)\geq 0$ for all $S\in\Sigma_\G$
\item On the empty set  $\mu(\emptyset)=0$. 
\item $\mu$ is countable additive, meaning that for any countable collections $\{S_x\}_{x=1}^\infty$ of pairwise disjoint sets of $\G$
\be
\mu\left(\bigcup_{x=1}^{\infty}S_x\right)=\sum_{x=1}^{\infty}\mu(S_x)\;.
\ee
\end{enumerate}
\end{definition}
\end{myd}
The definition above is consistent with what one would expect from a function that quantify the volume or size of a region on a manifold. However, since we are interested here in measures on Lie groups, we would like the measure also to be invariant under the action of the group.

By definition of Lie groups, for a fixed group element $h\in\G$, the map $g\mapsto hg$ is an isomorphism between smooth manifolds (also known dif and only ifeomorphism). Such a map transform any region $S\subseteq\G$ to the region $hS\eqdef\{hg\;:\;g\in \G\}$.
We then say that $\mu$ is left-invariant if $\mu(hS)=\mu(S)$ for all $S\subseteq\G$ and all $h\in \G$. Similarly, we say that $\mu$ is right-invariant if $\mu(Sh)=\mu(S)$ for all $S\subseteq\G$ and all $h\in \G$.

All groups have a left-invariant and right-invariant measures. This result is known as Haar's Theorem (the proof of Haar's theorem goes beyond the scope of this book). For compact groups these Haar measures are finite and unique up to a multiplicative constant. If the two invariant-measures of a Lie group equals each other up to a multiplicative constant then the group is said to be {\it unimodular}. All compact Lie groups are unimodular, and also many non-compact groups that appear in applications in physics are unimodular. In this book we will only consider unimodular Lie groups. Moreover, when the group is compact, so that $\mu(\G)<\infty$, we will always implicitly assume that the Haar measure is normalized; i.e. $\mu(\G)=1$.

\begin{exercise}
Let $\mu_L$ be a left-invariant measure on a group $\G$. Prove the existence of right-invariant measure. {\it Hint: Define $\mu_R(S)\eqdef\mu_L(S^{-1})$ for any subset $S\subseteq\G$, where $S^{-1}\eqdef\{g^{-1}\;:\;g\in S\}$.} 
\end{exercise}

\subsubsection{Examples:}
\begin{enumerate}
\item Consider the group $U(1)\eqdef\{e^{i\theta}\;:\;\theta\in[0,2\pi)\}$. This group is clearly homomorphic to the group with elements in $[0,2\pi)$ under group operation of addition modulo $2\pi$. For any set $S\subseteq[0,2\pi)$ the Haar measure of $U(1)$ is given by
\be
\mu(S)=\frac1{2\pi}\int_Sd\theta\;,
\ee
or equivalently, $d\mu(g)=\frac1{2\pi}d\theta$.
Since $U(1)\cong SO(2)$ this is also the Haar measure of $SO(2)$ 
\item The Haar measure of $SU(2)$. Recall from~\eqref{sphsu2} that the group elements of $SU(2)$ can be characterize in terms of the Hyperspherical coordinates $(\alpha,\beta,\gamma)$. It turns out that the Haar measure of a region $R\subseteq SU(2)$ is given by
\be
\mu(R)=\int_R\sin(2\alpha)d\alpha d\beta d\gamma
\ee 
\end{enumerate}

The Haar measure can be used to define various averages over a group. For example, consider a function $f:\G\to\mbb{C}$.  One can define the average of the function $f$ over the compact group $G$ as 
\be
\int_\G dg\;f(g)\;,
\ee
where we use the short notation $dg$ for the Haar measure $d\mu(g)$.
Given a projective unitary representation $g\mapsto U_g$ one can also define averages over elements of $\ml(A)$ as
\be\label{gtwirl}
\mG(\rho)\eqdef\int_\G dg\; U_g\rho U_g^*\quad\quad\;\forall\;\rho\in\ml(A)\;.
\ee 
The map $\mG:\ml(A)\to\ml(A)$ is linear and is known as the $\G$-twirling map. 

\begin{remark}
If the group $\G$ is finite we can always replace that averages above with summations. In particular, for finite group the integral $\int_\G dg$ can be simply replaced with a sum $\frac1{|\G|}\sum_{g\in\G}$, and under this replacement, all the theorems and statements below that apply for compact Lie group, also apply for finite groups.
\end{remark}

\begin{exercise}\label{ginvg}
Consider the $\G$-twirling map $\mG$. 
\begin{enumerate}
\item Use the invariance property of the Haar measure to show that for any $\rho\in\ml(A)$
\be
[\mG(\rho),U_g]=0\quad\quad\forall\;g\in\G\;.
\ee
\item Show that $\mG\circ\mG=\mG$. That is, for all $\rho\in\ml(A)$ we have $\mG\big(\mG(\rho)\big)=\mG(\rho)$.
\end{enumerate}
\end{exercise}

\begin{exercise}
Let $\G$ be a compact group and let $\pi:\G\to\ml(A)$ be a group representation (not necessarily unitary).
Define a map $(\cdot,\cdot)_\G: A\times A\to\mbb{C}$ as
\be
(\phi,\psi)_\G\eqdef\int_Gdg\;\la\pi(g)\phi|\pi(g)\psi\ra\quad\quad\forall\;\psi,\phi\in A\;,
\ee
where $dg$ is the (unique) right-invariant Haar measure.
\begin{enumerate}
\item Show that $(\cdot,\cdot)_\G$ defines an inner product.
\item Show that 
\be
(\pi(h)\phi,\pi(h)\psi)_\G=(\phi,\psi)_\G\quad\quad\;\forall\;h\in\G\;;
\ee
i.e. all the matrices $\pi(h)$ are unitaries with respect to the inner product $(\cdot,\cdot)_\G$.
\item Use the previous parts of this question in conjunction with Theorem~\ref{findecom3} to show that $\pi$ can be decomposed into a direct sum of irreps.
\end{enumerate}
\end{exercise}
\begin{myt}{}
\begin{theorem}\label{ginva2}
Let $\rho\in\ml(A)$ and $U:\G\to\ml(A)$ be a projective unitary representation. An operator $\rho\in\ml(A)$ is $G$-invariant if and only if $\mG(\rho)=\rho$.
\end{theorem}
\end{myt}
\begin{proof}
Suppose $\mG(\rho)=\rho$. Then, for all $h\in\G$
\ba
U_h\rho=U_h\mG(\rho)&=\int_\G dg\; U_hU_g\rho U_g^*=\int_\G dg\; U_{hg}\rho U_g^*\\
\GG{g'\eqdef hg}&=\int_\G dg'\; U_{g'}\rho U_{h^{-1}g'}^*=\int_\G dg'\; U_{g'}\rho U_{g'}^*U_{h^{-1}}^*\\
&=\mG(\rho)U_h=\rho U_h\;.
\ea
Conversely, if $\rho$ is $\G$-invariant then
\be
\mG(\rho)=\int_\G dg\; U_g\rho U_g^*=\int_\G dg\rho=\rho\;.
\ee
\end{proof}

In the next theorem we show that the average of $\{U_g\}$ over the group (w.r.t. the Haar measure) is an orthogonal projection.
\begin{myt}{}
\begin{theorem}\label{vgv}
Let $\G$ be a finite or compact Lie group and let $g\mapsto U_g$ be a unitary representation acting on $A$. Define
\be
A^\G\eqdef\Big\{|\psi\ra\in A\;:\;U_g|\psi\ra=|\psi\ra\quad\forall\;g\in\G\Big\}
\quad
and
\quad
\Pi\eqdef\int_\G dg\;U_g\;.
\ee
Then, $\Pi$ is an orthogonal projector onto $A^\G$.
\end{theorem}
\end{myt}
\begin{proof}
For any $g\in\G$ we have
\be
U_g\Pi=\int_\G dh\;U_gU_h=\int_\G dh\;U_{gh}=\int_\G dh\;U_{h}
\ee
where in the last equality we used the fact that the Haar measure $dh$ is invariant under the group action. Hence $U_g\Pi=\Pi$ for all $g\in\G$ and we get that
\be
\Pi^*\Pi=\int_\G dg\;U_{g^{-1}}\Pi=\int_\G dg\;\Pi =\Pi\;.
\ee
Therefore, from the first part of Exercise~\ref{ppp} it follows that $\Pi$ is an orthogonal projection. Moreover, since $U_g\Pi=\Pi$ we get that $U_g\Pi|\psi\ra=\Pi|\psi\ra$ for all $g\in\G$. Hence, $\Pi|\psi\ra\in A^\G$ for all $|\psi\ra\in A$. Finally, to show that $\Pi$ is not a projection to a proper subspace of $A^{\G}$, observe that for every $|\psi\ra\in A^\G$ we have
\be
\Pi|\psi\ra=\int_\G dg\;U_g|\psi\ra=\int_\G dg\;|\psi\ra=|\psi\ra\;.
\ee
This completes the proof.
\end{proof}

\section{Orthogonality Between Irreps}

A remarkable result known as the Wyle theorem, follows directly from Theorems~\ref{deop} and~\ref{ginva2}. It states that the components of the matrices $\{U_g\}$ satisfy an orthogonality relation. Note that from~\eqref{lamir} we know that in the basis of~\eqref{ginvbasis}
\be
\la\lambda,m,x|U_g|\lambda',m',x'\ra=\delta_{\lambda\lambda'}\delta_{xx'}\la\lambda,m,x|U_g|\lambda,m',x\ra\;.
\ee
The following theorem states additional orthogonality condition satisfied by the matrix elements
\be\label{compo}
u_{mm'}^{\lambda}(g)\eqdef\la\lambda,m,x|U_g|\lambda,m',x\ra=\la\lambda,m|U_g^{(\lambda)}|\lambda,m'\ra\;.
\ee
In the equation above, the set $\{|\lambda,m,x\ra\}_{m,x}$ form the basis of $A_\lambda$, whereas $\{|\lambda,m\ra\}_{m}$ forms a basis of $B_\lambda$.
In particular, on the left-hand side of the equation above there is no index $x$, since from~\eqref{lamir} the components $u_{mm'}^{\lambda}(g)$ do not depend on $x$.

\begin{myt}{}
\begin{theorem}
Let $g\mapsto U_g$ be a projective unitary representation of a group $\G$ in $\ml(A)$, and let $u_{mm'}^{\lambda}(g)$ be the components of $U_g$ as defined in~\eqref{compo}. Then, 
 \be\label{ortrr}
\int_Gdg\;{u}_{mm'}^{\lambda}(g)\bar{u}_{kk'}^{\lambda'}(g)=\frac{\delta_{\lambda\lambda'}\delta_{mk}\delta_{m'k'}}{|B_\lambda|}\;.
\ee
\end{theorem}
\end{myt}
\begin{proof}
Take
\be
\rho=|\lambda,m',x\lr\lambda',k',x|=|\lambda,m'\lr\lambda',k'|\otimes |x\lr x|
\ee
and observe that
\ba\label{xxc1}
\la\lambda,m,x|\mG(\rho)|\lambda',k,x\ra&=\int_Gdg\;\la\lambda,m,x|U_g|\lambda,m',x\lr\lambda',k',x|U_g^*|\lambda',k,x\ra\\
&=\int_Gdg\;{u}_{mm'}^{\lambda}(g)\bar{u}_{kk'}^{\lambda'}(g)\;.
\ea
Now, denote by $\sigma\eqdef\mG(\rho)$ and for any irrep $\mu$ denote by $B_\mu$ the representation space, and by $C_\mu$ the multiplicity space. Then,
\ba\label{xxc2}
\sigma_\mu^{C_\mu}\eqdef\tr_{B_\mu}\left[\Pi^{A_\mu}\sigma\right]&=\int_\G dg\;\tr_{B_\mu}\left[\Pi^{A_\mu}U_g(|\lambda,m',x\lr\lambda',k',x|)U_g^*\right]\\
\GG{\eqref{lamir}}&=\int_\G dg\;\tr_{B_\mu}\left[\Pi^{A_\mu}\left(U_g^{(\lambda)}|\lambda,m'\lr\lambda',k'|U_g^{*(\lambda')}\otimes|x\lr x|^{C_\mu}\right)\right]\\
&=\delta_{\mu\lambda}\delta_{\mu\lambda'}\int_\G dg\;\tr\left[U_g^{(\lambda)}|\mu,m'\lr\mu,k'|U_g^{*(\lambda')}\right]|x\lr x|^{C_\mu}\\
&=\delta_{\mu\lambda}\delta_{\mu\lambda'}\delta_{m'k'}|x\lr x|^{C_\mu}\;.
\ea
Since $\sigma=\mG(\rho)$ is $\G$-invariant (see Theorem~\ref{ginva2})  we get form~\eqref{formra} (when applied to $\sigma$) 
\ba
\mG(\rho)&=\bigoplus_{\mu\in\irr(U)} \u^{B_\mu}\otimes \sigma^{C_\mu}_{\mu}\\
\GG{\eqref{xxc2}}&=\delta_{\lambda\lambda'}\delta_{m'k'}\u^{B_\lambda}\otimes|x\lr x|^{C_\lambda}
\ea
so that
\be
\la\lambda,m,x|\mG(\rho)|\lambda',k,x\ra=\frac{\delta_{\lambda\lambda'}\delta_{mk}\delta_{m'k'}}{|B_\lambda|}\;.
\ee
Comparing this with~\eqref{xxc1} concludes the proof.
\end{proof}

Note that the orthogonality relations in the theorem above can be used to obtain other types of relations. For example, the relations~\eqref{ortrr} implies that (see Exercise~\ref{ortt})
\be\label{glm}
\int_\G dg\;\bar{u}_{mm'}^{\lambda}(g)U_{g}^{(\lambda')}=\frac{\delta_{\lambda\lambda'}}{|B_\lambda|}|\lambda,m\lr\lambda,m'|^{B_{\lambda}}
\ee
Moreover, this relation can be extended to $U_g=\bigoplus_{\lambda\in\irr(U)} (U_g^{(\lambda)}\otimes I^{C_\lambda})$ (see~\eqref{lamir}) via
\be\label{1126}
\int_\G dg\;\bar{u}_{mm'}^{\lambda}(g)U_{g}=\frac{\delta_{\lambda,\irr(U)}}{|B_\lambda|}|\lambda,m\lr\lambda,m'|^{B_{\lambda}}\otimes I^{C_\lambda}\;.
\ee
where $\delta_{\lambda,\irr(U)}\eqdef\begin{cases}1\text{ if }\lambda\in\irr(U)\\ 0 \text{ otherwise}\end{cases}$. Taking $m'=m$ and summing over $m$ results in the relation
\be
\int_\G dg\;\bar{\chi}_{\lambda}(g)U_{g}=\frac{\delta_{\lambda,\irr(U)}}{|B_\lambda|}I^{B_{\lambda}}\otimes I^{C_\lambda}\;,
\ee
where $\chi_{\lambda}(g)\eqdef\tr\left[U_{g}^{(\lambda)}\right]$ is called the $\lambda$-character. The orthogonality relations discussed in this section have several other interesting consequences. One of them, which we will discuss later on in this chapter, 
is a one-to-one correspondence between a reduction of a matrix $\rho\in\ml(A)$ onto its $\lambda$-irrep and its characteristic function.

\begin{exercise}\label{ortt}
Prove the relation~\eqref{glm} and the equality $\bar{u}_{mm'}^{\lambda}(g)={u}_{mm'}^{\lambda}(g^{-1})$. {\it Hint:\ For the former, express $U^{(\lambda)}_g=\sum_{k,k'}{u}_{kk'}^{\lambda}(g)|\lambda,k\lr\lambda,k'|$ and use the orthogonality relations~\eqref{ortrr}.}
\end{exercise}

\begin{exercise}[Orthogonality of Characters]
Use the orthogonality relations above to show that the characters are orthogonal; i.e. show that for any $\lambda,\mu\in\irr(U)$
\be
\int_\G dg\;\bar{\chi}_\lambda(g)\chi_\mu(g)=\delta_{\lambda\mu}
\ee
Use the above orthogonality relation to show that the character $\chi(g)\eqdef\tr[U_g]$ satisfies
\be\label{exreg}
\int_\G dg\;\bar{\chi}_\lambda(g)\chi(g)=\begin{cases}m_{\lambda} &\text{ if }\lambda\in\irr(U)\\
0 & \text{ otherwise}
\end{cases}\;.
\ee
\end{exercise}

\section{The Regular Representation}\label{sec:rr}

\subsection{Finite Groups}

\begin{myd}{}
\begin{definition}
Let $\G$ be a finite group, and let  $\{\omega(g,h)\}_{g,h\in\G}$ be a cocycle of $\G$ satisfying~\eqref{co1} and ~\eqref{co2}. The regular representation $g\mapsto U_g^\reg$ is a unitary projective representation of $\G$ on the Hilbert space $\mbb{C}^{|\G|}=\spa\{|g\ra\;:\;g\in\G\}$ defined by the relation
\be
U_g^\reg|h\ra\eqdef\omega(g,h)|gh\ra\quad\quad\forall\;g,h\in\G\;.
\ee
\end{definition}
\end{myd}

Note that for any fixed $g\in\G$, $U_g^\reg$ maps the basis $\{|h\ra\}_{h\in\G}$ to itself (up to phases) and therefore $U_g^\reg$ must be a unitary matrix.
Furthermore, for any $g_1,g_2,h\in\G$ we have by definition
\ba
U_{g_1}^\reg U_{g_2}^\reg|h\ra&=\omega(g_2,h) U_{g_1}^\reg|g_2h\ra\\
&=\omega(g_2,h)\omega(g_1,g_2h)|g_1g_2h\ra\\
\GG{\eqref{co2}}&=\omega(g_1,g_2)\omega(g_1g_2,h)|g_1g_2h\ra\\
&=\omega(g_1,g_2)U_{g_1g_2}^\reg|h\ra\;,
\ea
and since the equation above holds for all $h$ we get that
\be
U_{g_1}^\reg U_{g_2}^\reg=\omega(g_1,g_2)U_{g_1g_2}^\reg\;.
\ee
That is, $g\mapsto U_g^\reg$ is indeed a unitary projective representation of $\G$ with cocycle $\{\omega(g,h)\}_{g,h\in\G}$.
Moreover, note that $U_g^\reg$ can be expressed as
\be
U_g^\reg=\sum_{h\in\G}\omega(g,h)|gh\lr h|\;,
\ee
so that its character is given by
\be\label{dge}
\chi^\reg(g)\eqdef\tr[U^\reg_g]=|\G|\delta_{g,e}\;.
\ee

The regular representation of $U^\reg$ depends only on the group $\G$ and the cocycle $\omega$. Therefore, we will denote the set of equivalence classes of irreps of $U^\reg$ by $\irr(\G,\omega)$. From~\eqref{exreg} it follows that the dimension of the multiplicity space of any irrep $\lambda\in\irr(\G,\omega)$ is given by
\ba
m_\lambda&=\frac1{|\G|}\sum_{g\in\G}\bar{\chi}_\lambda(g)\chi^\reg(g)\\
\GG{\eqref{dge}}&=\bar{\chi}_\lambda(e)\\
&=\tr\left[I^{B_\lambda}\right]=|B_\lambda|\;.
\ea
That is, the multiplicity space has the same dimension as the representation space.
This equality has the following remarkable application.  
Recall that according to~\eqref{decoal}, the Hilbert space $\mbb{C}^{|\G|}$ can be decomposed with respect to the irreps of $U^\reg$ such that
\be
\mbb{C}^{|\G|}=\bigoplus_{\lambda\in\irr(\G,\omega)} B_\lambda\otimes C_\lambda\;.
\ee
Since $d_\lambda\eqdef|B_\lambda|=m_\lambda$ we conclude that
\be
|\G|=\sum_{\lambda\in\irr(\G,\omega)}d_\lambda^2\;.
\ee
The above relation implies that the vectors $\{\v_g\}_{g\in\G}$ defined by
$\v_g\eqdef\left\{\frac1{\sqrt{d_\lambda}}{u}_{kk'}^{\lambda}(g)\right\}_{k,k'\in[d_\lambda]}^{\lambda\in\irr(\G,\omega)}$ belong to $\mbb{C}^{|\G|}$ (since they have exactly $|\G|$ components). Moreover, using this in conjunction with the orthogonality relations~\eqref{ortrr} we conclude that $\{\v_g\}_{g\in\G}$ is an orthonormal basis of $\mbb{C}^{|G|}$.

\subsection{Compact Lie Groups}

In order to define the regular representation on compact Lie groups it is necessary to introduce first the Hilbert space $L^2(\G)$ of square integrable functions over a compact Lie group $\G$. The space $L^2(\G)$ consists of all integrable functions $f:\G\to\mbb{C}$ such that
\be
\int_\G dg\;|f(g)|^2<\infty\;.
\ee
It forms a Hilbert space under the inner product
\be
\la f_1|f_2\ra\eqdef\int_\G dg\;\overline{f_1(g)}f_2(g)\quad\quad\forall\;f_1,f_2\in L^2(\G)\;.
\ee

As the notation for the inner product above suggests, we will use the Dirac notation to denotes the elements of $L^2(\G)$.\index{$L^2(\G)$} This will make the analogy with the case of finite groups much more apparent. Hence, the vector $|f\ra$ for example corresponds to the function $f(g)\in L^2(\G)$. We also denote by $\delta(g)$ the Dirac-delta on the group, defined by the relation
\be
\la \delta|f\ra=\int_\G dg\;\overline{\delta(g)}f(g)=f(e)\quad\forall\;f\in L^2(\G)\;.
\ee
We will therefore denote $|e\ra\eqdef|\delta\ra$, so that $f(e)=\la e|f\ra$. We point out that while $\delta(g)\not\in L^2(\G)$ there is a way to make the concepts we discuss below mathematically rigorous via the introduction of a \emph{rigged Hilbert space}. However, this topic goes beyond the scope of this book, and since we only use the Dirac delta function in this subsection we will not elaborate on it here. For more information on this subject, we refer the reader to the section ``Notes and References"  at the end of this chapter.

Continuing, for any $h\in\G$ we denote by $|h\ra$ the function $\delta(h^{-1}g)$ so that
\be
\la h|f\ra=\int_\G dg\;\overline{\delta(h^{-1}g)}f(g)=f(h)\quad\forall\;f\in L^2(\G)\;.
\ee
Observe also that 
\be
\la h|g\ra=\int_\G dg'\;\overline{\delta(h^{-1}g')}\delta(g^{-1}g')=\delta(h^{-1}g)=\delta(g^{-1}h)\;.
\ee
With these notations, given a cocycle $\omega$, we define the regular representation $g\mapsto U_g^\reg$ of a compact Lie group (in analogy with its definition on finite groups) as
\be
U_g^\reg|h\ra\eqdef\omega(h,g)|gh\ra\;.
\ee
The above definition implies that $U_g^\reg$ can be expressed as
\be\label{ugreg}
U_g^\reg\eqdef\int_\G dh\;\omega(g,h)|gh\lr h|\;,
\ee
and therefore we get for example that
\be\label{cregd}
\chi^\reg(g)\eqdef\tr\left[U_g^\reg\right]=\int_\G dh\;\omega(g,h)\delta(g)=\delta(g)
\ee

To illustrate the above definitions, consider the group $U(1)$ and for simplicity consider the trivial cocycle $\omega(\theta,\theta')=1$ for all $\theta,\theta'\in U(1)\cong[0,2\pi)$. The Hilbert space $L^2\big(U(1)\big)$ is the set of all square integrable functions on $[0,2\pi)$. Every function $f\in L^2\big(U(1)\big)$ can be expressed as $f(\theta)=\la\theta|f\ra$, and in particular, the functions
\be\label{fnn}
f_n(\theta)\eqdef\la\theta|n\ra\eqdef e^{in\theta}\quad\quad n\in\mbb{Z}\;,
\ee
forms an orthonormal basis of $L^2\big(U(1)\big)$ (due to the Fourier expansion).
Indeed,
\be
\la n|n'\ra\eqdef \la f_n|f_{n'}\ra=\frac1{2\pi}\int_0^{2\pi}d\theta\; e^{-in\theta}e^{in'\theta}=\delta_{nn'}\;.
\ee
Note that the inner product has the factor of $\frac1{2\pi}$ since the Haar measure in this case is $\frac{d\theta}{2\pi}$. Hence, the functions in~\eqref{fnn} are normalized with respect to this inner product.  
In this example, the regular representation~\eqref{ugreg} takes the form
\be
U_\theta^{\reg}=\frac1{2\pi}\int_0^{2\pi} d\theta'\;|\theta+\theta'\lr \theta'|\;,
\ee
where the summation $\theta+\theta'$ is mod $2\pi$. The matrix components of $U_\theta^{\reg}$ in the $f_n$-basis $\{|n\ra\}$ is given by
\ba
\la n|U_\theta^{\reg}|n'\ra&=\frac1{2\pi}\int_0^{2\pi} d\theta'\;\la n|\theta+\theta'\lr \theta'|n'\ra\\
&=\frac1{2\pi}\int_0^{2\pi} d\theta'\; e^{in(\theta+\theta')}e^{-i n'\theta'}\\
&=e^{i\theta n}\delta_{nn'}\;.
\ea
Hence, with respect to the basis $\{|n\ra\}$ we can express the regular representation as
\be\label{untt}
U_\theta^{\reg}=\sum_{n\in\mbb{Z}}e^{in\theta}|n\lr n|\;.
\ee
That is, the regular representation is a direct sum of \emph{all} the irreps of $U(1)$ (cf.~\eqref{utk}).

\subsection{Fourier Expansion in $L^2(\G)$}

In Theorem~\ref{thdco} we proved that any projective unitary representation on a finite dimensional space can be decomposed into a direct some of irreps. For compact groups this theorem can be extended also to projective unitary representations on infinite dimensional space (see subsection `Notes and References'). Therefore, the regular representation $U_g^\reg$ can be decomposed into a direct sum of finite dimensional irreps. Denoting as before by $\irr(\G,\omega)$ the set of all equivalence classes of irreps, we can decompose $U^\reg_g$
as
\be\label{uregdeco}
U^\reg_g\cong\bigoplus_{\lambda\in\irr(\G,\omega)} U_g^{(\lambda)}\otimes I^{C_\lambda}\;,
\ee
where for each $\lambda$ the dimension $d_\lambda\eqdef|B_\lambda|<\infty$.
From~\eqref{exreg} it follows that for any $\lambda\in\irr(\G,\omega)$ the dimension of the multiplicity space $C_\lambda$ is given by
\ba
m_\lambda&=\int_\G dg\;\bar{\chi}_\lambda^\reg(g)\chi^\reg(g)\\
\GG{\eqref{cregd}}&=\int_\G dg\;\bar{\chi}_\lambda^\reg(g)\delta(g)\\
&=\bar{\chi}_\lambda^\reg(e)=\tr[I^{B_\lambda}]=d_\lambda\;.
\ea
This remarkable result also implies that the Hilbert space $L^2(\G)$ can be decomposed as
\be\label{lgdeco}
L^2(\G)\cong\bigoplus_{\lambda\in\irr(\G,\omega)}B_\lambda\otimes C_\lambda\quad\text{with}\quad B_\lambda\cong C_\lambda\cong\mbb{C}^{d_\lambda}\;.
\ee
Now, define for any $\lambda\in\irr(\G,\omega)$ and any $k,k'\in[d_\lambda]$ the functions
\be
f_{kk'}^\lambda(g)\eqdef\frac1{\sqrt{d_\lambda}}u_{kk'}^{\lambda}(g)\;,
\ee
where $u_{kk'}^\lambda(g)$ are the matrix elements of $U_g^{(\lambda)}$ as appear in~\eqref{uregdeco}. From the orthogonality relations~\eqref{ortrr} we have that $\{f_{kk'}^\lambda(g)\}$ is an orthonormal set of functions in $L^2(\G)$, and from~\eqref{lgdeco} we conclude that $\{f_{kk'}^\lambda(g)\}$ is an orthonormal basis of $L^2(\G)$. We therefore arrive at the following theorem.
\begin{myt}{\color{yellow} Fourier Expansion}
\begin{theorem}\label{clig}
Let $\G$ be a compact Lie group and let $\omega$ be a cocycle.  
Any function $f(g)\in L^2(\G)$ can be expanded as
\be\label{c113}
f(g)=\sum_{\lambda\in\irr(\G,\omega)}\sum_{k,k'=1}^{d_\lambda}c_{kk'}^\lambda \bar{u}_{kk'}^\lambda(g)\;, 
\ee 
where the coefficients $c_{kk'}^\lambda\in\mbb{C}$ can be expressed as
\be\label{ckkk}
c_{kk'}^\lambda=\frac1{d_\lambda}\int_\G dg\;u_{kk'}^\lambda(g) f(g)\;.
\ee
\end{theorem}
\end{myt}
\begin{remark}
The relation above is the generalization of Fourier series. To see this, consider the group $U(1)$ whose elements are parametrized by $\theta\in[0,2\pi)$. In this case we denote the irreps by integers $\lambda=n$, and we know that they are all one dimensional. Therefore, $u_{kk'}^\lambda(g)$ becomes $u^\lambda(g)$ (since $d_\lambda=1$ so that $k=k'=1$) and recall that $\lambda=n$. In other words, $u_{kk'}^\lambda(g)$ can be replaced with $f_n(\theta)\eqdef e^{i\theta n}$ (see \eqref{untt}), and $c_{kk'}^\lambda$ are replaced with $c_n$. Hence, for $\G=U(1)$ the two equations in the theorem above simplify to
\be
f(\theta)=\sum_{n\in\mbb{Z}}c_n e^{in\theta}\quad\text{and}\quad c_n=\frac1{2\pi}\int_0^{2\pi} d\theta e^{in\theta}f(\theta)\;
\ee
where we replaced $g$ by $\theta$ and the Haar measure $dg$ by $\frac{d\theta}{2\pi}$. This is precisely the Fourier expansion of periodic function (with $2\pi$ period). The theorem above demonstrate that the Fourier expansion is not a special feature of the group $U(1)$ but it exists for any compact Lie group.
\end{remark}

\begin{exercise}
Prove Theorem~\ref{clig} in full details. Hint: Use the arguments above it.
\end{exercise}

\subsubsection{Class Functions}

\begin{myd}{}
\begin{definition}\label{funclass}
A function $f\in L^2(\G)$ is called a \emph{class function} if for all $h,g\in\G$ we have
\be
f(hgh^{-1})=f(g)\;.
\ee
\end{definition}
\end{myd}

In other words, a function $f:\G\to\mbb{C}$ is a class function if it is constant on conjugacy classes. The character $\chi_{\lambda}(g)\eqdef\tr\left[U_{g}^{(\lambda)}\right]$ discussed above is an example of a class function since 
\ba
\chi_{\lambda}(hgh^{-1})=\tr\left[U_{hgh^{-1}}^{(\lambda)}\right]&=\tr\left[U_{h}^{(\lambda)}U_{g}^{(\lambda)}U_{h^{-1}}^{(\lambda)}\right]\\
&=\tr\left[U_{g}^{(\lambda)}\right]=\chi_{\lambda}(g)\;.
\ea
In the next theorem we show that the orthogonality between irreps implies that all class functions are linear combinations of the characters.

\begin{myt}{}
\begin{theorem}\label{classfun}
Let $f:\G\to\mbb{C}$ be a class function. Then, $f$ can be written as
\be
f(g)=\sum_{\lambda\in\irr(U)}a^\lambda\chi^\lambda(g)\;,
\ee
for some coefficients $a_\lambda\in\mbb{C}$.
\end{theorem}
\end{myt}

\begin{proof}
For every $\lambda\in\irr(U)$ let
\be
M_\lambda\eqdef\int_\G dg\;f(g)U_g^{(\lambda)}\;.
\ee
Since $f$ is a class function, the matrix $M_\lambda$ satisfies for all $h\in\G$
\ba
U_h^{(\lambda)}M_\lambda U_h^{*(\lambda)}=\int_\G dg\;f(g)U_h^{(\lambda)}U_g^{(\lambda)}U_h^{*(\lambda)}
&=\int_\G dg\;f(g)U_{hgh^{-1}}^{(\lambda)}\\
\Gg{{g'}\eqdef hgh^{-1}}&=\int_\G d{g'}\;f(h^{-1}{g'}h)U_{{g'}}^{(\lambda)}\\
\Gg{f(h^{-1}{g'}h)=f({g'})}&=\int_\G d{g'}\;f({g'})U_{{g'}}^{(\lambda)}=M_\lambda\;.
\ea
That is, $[M_\lambda,U_h^{(\lambda)}]=0$ for all $h\in\G$. Since $h\mapsto U_h^{(\lambda)}$ is an irrep, we get from Schur's lemma that $M_\lambda=b^\lambda I^{B_\lambda}$ for some $a^\lambda\in\mbb{C}$. In terms the components, this relation can be expressed as
\be
\int_\G dg\;f(g)u_{kk'}^\lambda=b^\lambda\delta_{kk'}\quad\quad\forall\;k,k'\in[d_\lambda]\;.
\ee
In other words, the coefficients $c_{kk'}^\lambda$ given in~\eqref{ckkk} satisfies $c_{kk'}^\lambda=\frac{b^\lambda}{d_\lambda}\delta_{kk'}$. Denoting by $a^\lambda\eqdef \frac{b^\lambda}{d_\lambda}$ we get from~\eqref{c113} that
\be
f(g)=\sum_{\lambda\in\irr(\G,\omega)}\sum_{k,k'=1}^{d_\lambda}c_{kk'}^\lambda \bar{u}_{kk'}^\lambda(g)=\sum_{\lambda\in\irr(\G,\omega)}a_\lambda\sum_{k,=1}^{d_\lambda} \bar{u}_{kk}^\lambda(g)=\sum_{\lambda\in\irr(\G,\omega)}a_\lambda\chi^\lambda(g)\;.
\ee
This completes the proof.
\end{proof}

\section{The Characteristic Function}\label{cf}

In Chapter~\ref{Ch:Asymmetry} we will encounter certain functions known as characteristic functions that play a major role in the QRT of asymmetry. We therefore discuss some of their properties here.

\begin{myd}{}
\begin{definition}\label{chara}
Let $\rho\in\ml(A)$ and $g\mapsto U_g$ be a projective unitary representation of a group $\G$ in $\ml(A)$. 
The characteristic function of $\rho$ is the function
\be
\chi_{\rho}(g)\eqdef \tr\left[\rho U_{g}\right]\;.
\ee
\end{definition}
\end{myd}

%We discuss here some of the key properties of characteristic functions.

\noindent\textbf{Three basic properties of characteristic functions:}
\begin{enumerate}
\item Invariance. Suppose $V\in\muu(A)$ is a $\G$-invariant unitary matrix, i.e. $[V,U_g]=0$ for all $g\in \G$. Then,
\be
\chi_{V\rho V^*}(g)=\tr\left[V\rho V^*U_g\right]=\tr\left[\rho V^*VU_g\right]=\tr\left[\rho U_g\right]=\chi_\rho(g)\;.
\ee 
\item Multiplicativity. Let $\rho\in\md(A)$ and $\sigma\in\md(B)$. Then,
\be
\chi_{\rho\otimes\sigma}(g)=\tr\left[\left(\rho^A\otimes\sigma^B\right)\left(U_g^A\otimes U_g^B\right)\right]=\chi_\rho(g)\chi_\sigma(g)\;.
\ee 
Here we assumed that the characteristic function on $AB$ is defined with respect to the representation $g\mapsto U_g^A\otimes U_g^B$.
\item Boundedness. For all $g\in \G$ and all $\rho\in\md(A)$ we have 
\be
|\chi_\rho(g)|=\big|\tr\left[\rho U_g \right]\big|\leq 1\;,
\ee
with equality if $g=e$ is the identity element of $\G$.
\end{enumerate}

In the next lemma we show that characteristic functions can be used to characterize $\G$-invariant states.

\begin{myg}{}
\begin{lemma}\label{lem:c71}
Let $\psi\in\pure(A)$. Then, the following statements are equivalent.
\begin{enumerate}
\item $|\chi_\psi(g)|=1$ for all $g\in \G$.
\item $\psi$ is $\G$-invariant.
\end{enumerate} 
\end{lemma}
\end{myg}
\begin{proof}
If $\psi$ is $\G$-invariant then by definition $U_g\psi U_g^*=\psi$ so that $U_g|\psi\ra=e^{i\theta_g}|\psi\ra$ for some $\theta_g\in[0,2\pi)$, so that $|\chi_\psi(g)|=1$. Conversely, suppose that $|\chi_\psi(g)|=1$ for all $g\in \G$. This means that there exist phases $\theta_g\in[0,2\pi)$ such that
\be
\la\psi|U_g|\psi\ra=e^{i\theta_g}\;.
\ee
Since $\psi$ is a pure normalized state, the above equation implies that
\be
U_g|\psi\ra=e^{i\theta_g}|\psi\ra\;.
\ee
That is, $\psi$ is $\G$-invariant.
\end{proof}

\begin{exercise}\label{ex:c71}
Show that in both directions of the proof above $\{e^{i\theta_g}\}$ must be a 1-dimensional representation of $\G$. Use this to conclude that $\psi$ is $\G$-invariant if and only if $\chi_\psi(g)$ is a 1-dimensional unitary representation of $\G$.
\end{exercise}

\begin{exercise}
Let $L$ be the representation of a generator of a Lie group $\G$ and $\rho\in\md(A)$. 
\begin{enumerate}
\item Show that for any $n\in\mbb{N}$
\be
\tr\left[\rho L^n\right]=i^{-n}\frac{\partial^n}{\partial\theta^n}\chi_\rho\left(e^{i\theta L}\right)\Big|_{\theta=0}
\ee
\item Let $\kappa_L^{(n)}$ denotes the $n$-th order cumulant defined as
\be
\kappa_L^{(n)}\eqdef i^{-n}\frac{\partial^n}{\partial\theta^n}\log\Big(\chi_\rho\left(e^{i\theta L}\right)\Big)\Big|_{\theta=0}
\ee
Show that the first and second order cumulants are the mean and the variance of the observable\index{observable} (i.e. Hermitian matrix) $L$.
\end{enumerate}
\end{exercise}

Let $\rho\in\ml(A)$ and $g\mapsto U_g$ be a projective unitary representation of a group $\G$ in $\ml(A)$. 
The reduction of $\rho$ onto the $\lambda$-irrep is the matrix
\be
 \rho^{B_\lambda}_\lambda\eqdef\tr_{C_\lambda}\left[\Pi^{A_\lambda}\rho^A\Pi^{A_\lambda}\right]\;.
\ee
Note that $\rho_\lambda^{B_\lambda}$ above is the marginal of $\Pi^{A_\lambda}\rho^A\Pi^{A_\lambda}$ in the representation space $B_\lambda$, whereas $\rho_\lambda^{C_\lambda}$ as define in~\eqref{ouni} for $\G$-invariant matrices is the marginal of $\Pi^{A_\lambda}\rho^A\Pi^{A_\lambda}$ in the multiplicity space $C_\lambda$.

\begin{myt}{}
\begin{theorem}\label{th158}
Let $\rho\in\ml(A)$, and let $g\mapsto U_g$ be a projective unitary representation of a compact group $\G$ in $\ml(A)$. Then, there is a one-to-one correspondence between $\rho^{B_\lambda}_\lambda$ and $\chi_{\rho}(g)$ given by
\ba\label{nli}
&\chi_{\rho}(g)=\sum_{\lambda}\tr\left[\rho^{B_\lambda}_\lambda U^{(\lambda)}_g\right]\\
&\rho^{B_\lambda}_\lambda=|B_\lambda|\int_\G dg\;\chi_{\rho}(g^{-1})U^{(\lambda)}_g
\ea
\end{theorem}
\end{myt}
\begin{remark}
The relationship between the characteristic function of $\rho$ and its reduction onto the $\lambda$-irrep is known as the Fourier transform over the group.
\end{remark}

\begin{proof}
The first equality follows directly from~\eqref{lamir} since
\be\label{chara0}
\chi_{\rho}(g)=\tr\left[\rho^A U_{g}\right]=\tr\left[\rho^A \bigoplus_\lambda (U_g^{(\lambda)}\otimes I^{C_\lambda})\right]=\sum_{\lambda}\tr\left[\rho^{B_\lambda}_\lambda U^{(\lambda)}_g\right]\;.
\ee
For the second equality we will use the relation~\eqref{1126}.  
Multiplying both sides of~\eqref{1126}  by $\rho\in\ml(A)$ and taking the trace gives
\be
\int_\G dg\;\bar{u}_{mm'}^{\lambda}(g)\chi_\rho(g)=\frac{1}{|B_\lambda|}\big\la\lambda,m'\big|\rho^{B_\lambda}_\lambda\big|\lambda,m\big\ra
\ee
Since the above equation holds for all $m,m'\in[|B_\lambda|]$ we conclude that
\be
\rho^{B_\lambda}_\lambda=|B_\lambda|\int_\G dg\;\chi_{\rho}(g)U^{(\lambda)}_{g^{-1}}
\ee
where we used the fact that $\bar{u}_{mm'}^{\lambda}(g)={u}_{mm'}^{\lambda}(g^{-1})$ (see Exercise~\ref{ortt}).
\end{proof}

\begin{exercise}
Show that if $\rho\in\ml(A)$ is $\G$-invariant then its reduction the $\lambda$-irrep is given by
\be
\rho^{B_\lambda}_\lambda=\tr\left[\rho^A\Pi^{A_\lambda}\right]\u^{B_\lambda}\;.
\ee
\end{exercise}

\begin{exercise}
Let $\rho\in\ml(A)$, and $g\mapsto U_g$ be a projective unitary representation over a compact group $\G$. Define 
\be
\trho^A\eqdef\bigoplus_\lambda\rho_\lambda^{B_\lambda}\otimes \frac1{|B_\lambda|}I^{C_\lambda}\;.
\ee
Show that $\trho$  satisfies
\be
\trho^{A}=\int_\G dg\;\chi_{\rho}(g^{-1})U_{g}
\ee
\end{exercise}

\section{Positive Definite Functions on a Group}\label{pdfg}
The characteristic function discussed above is closely related to the concept of positive definite function over a group.

\begin{myd}{}
\begin{definition}
A complex function $f:\G\to \mbb{C}$ is said to be \emph{positive definite} if for all choices of $n\in\mbb{N}$, $g_1,\ldots,g_n\in\G$, and $c_1,\ldots,c_n\in\mbb{C}$
\be\label{defcx}
\sum_{x\in[n]}\sum_{y\in[n]}\bar{c}_xc_yf(g_x^{-1}g_y)\geq 0\;.
\ee
Moreover, $f$ is said to be \emph{normalized} if $f(e)=1$.
\end{definition}
\end{myd}
\begin{remark}
For the case that $\G$ is a compact Lie group and $f$ is continuous, the definition above is equivalent to the statement that 
\be
\int_\G dg\int_\G dh\;\bar{c}(g)f(g^{-1}h)c(h)\geq 0
\ee
where $c\in L^2(\G)$. 
\end{remark}

\begin{exercise}
Show that a complex function $f:\G\to \mbb{C}$ is positive definite if for all choices of $n\in\mbb{N}$, $g_1,\ldots,g_n\in\G$, and $c_1,\ldots,c_n\in\mbb{C}$
\be
\sum_{x\in[n]}\sum_{y\in[n]}\bar{c}_xc_yf(g_yg_x^{-1})\geq 0\;.
\ee
Note that we replace $f(g_x^{-1}g_y)$ in~\eqref{defcx} with $f(g_yg_x^{-1})$.
\end{exercise}

\bex\label{ftex}
Let $\G=\mbb{Z}_n=\{0,1,\ldots,n-1\}$ be the cyclic group with the group operation being addition modulo $n$. Show that a function $f:\mbb{Z}_n\to\mbb{C}$ is positive definite if and only if
\be
\sum_{x\in\mbb{Z}_n}f(x)e^{i\frac{2\pi xy}{n}}\geq 0\quad\quad\forall\;y\in\mbb{Z}_n\;.
\ee
Observe that the condition above also implies that the left hand side is real.
\eex

If a complex function $f:\G\to \mbb{C}$ is a characteristic function, i.e. $f(g)=\tr[\rho U_g]$ for some $\rho\in\md(A)$ and some (non-projective) unitary representation $g\mapsto U_g$ acting on $A$, then for any $c_1,\ldots,c_n\in\mbb{C}$ and $g_1,\ldots,g_n\in\G$
\ba
\sum_{x\in[n]}\sum_{y\in[n]}\bar{c}_xc_yf(g_x^{-1}g_y)&=\sum_{x\in[n]}\sum_{y\in[n]}\bar{c}_xc_y\tr\left[\rho U_{g_x}^*U_{g_y}\right]\\
\GG{{\it C\eqdef\sum_{x\in[n]}c_xU_{g_x}}}&=\tr\left[C\rho C^*\right]
\geq 0\;.
\ea
In other words, all characteristic functions are positive definite functions over the group. Conversely, we will see below that every normalized positive definite function $f$ over a group is a characteristic function. 

\begin{exercise} Let  $\G$ be a compact Lie group and $f:\G\to\mbb{C}$ be a positive definite function on $\G$. For any two functions $f_1,f_2\in L^2(\G)$ define 
\be
\la f_1|f_2\ra_{f}\eqdef\int_\G dg\int_\G dh\; \overline{f_1(g)}f_2(h)f(g^{-1}h)\;.
\ee
Show that $\la f_1|f_2\ra_f$ is an inner product.
\end{exercise}

Recall that any function $f(g)\in L^2(\G)$ on a compact Lie group can be expressed as (see Theorem~\eqref{clig})
\be\label{ro2}
f(g)=\sum_{\lambda\in\irr(\G)}\sum_{k,k'=1}^{d_\lambda}c_{kk'}^\lambda \bar{u}_{kk'}^\lambda(g)\;, 
\ee 
where we chose the trivial cocycle $\omega(g,h)=1$ and denoted $\irr(\G)\eqdef\irr(\G,\omega=1)$.  In the following theorem we use it to show that normalized positive definite functions are characteristic functions.

\begin{myt}{}
\begin{theorem}\label{pdd}
Let $\G$ be a finite or compact Lie group and $f(g)\in L^2(\G)$. The following are equivalent:
\begin{enumerate}
\item There exists a Hilbert space $A$, a state $\psi\in\pure(A)$, and a (non-projective) unitary representation $g\mapsto U_g\in\ml(A)$ such that 
\be
f(g)=\chi_\psi(g)=\la\psi|U_g|\psi\ra\;.
\ee
\item $f$ is a normalized positive definite function on $\G$.
\end{enumerate}
\end{theorem}
\end{myt}

\begin{proof}[Proof of Theorem~\ref{pdd}] Since we already saw that all characteristic functions are positive definite. It is therefore left to show that $2\Rightarrow 1$. Suppose $f$ is a normalized positive definite function on $\G$. Recall that if $f$ is also a characteristic function of some state $\rho$ then $f$ and $\rho$ satisfies~\eqref{nli} with $f$ replacing $\chi_\rho$. However, since we need to prove that $f$ is a characteristic function we use this relationship as a definition. That is, for any $\lambda\in\irr(\G)$ we define the operator
\be\label{ro1}
\rho^{B_\lambda}_\lambda\eqdef d_\lambda\int_\G dg\;f(g^{-1})U^{(\lambda)}_{g}\;,
\ee
where $g\mapsto U^{(\lambda)}_g$ is the $\lambda$-irrep of the regular representation of $\G$. We first show that the operator above is positive semidefinite.
Let $\eta\in\ml(B_\lambda)$, multiply both sides of the equation above by $\eta\eta^*$, and take the trace to get
\be\label{1319}
\tr\left[\eta\eta^*\rho^{B_\lambda}_\lambda\right]=d_\lambda\int_\G dg\;f(g^{-1})\chi_{{}_{\eta\eta^*}}(g)\;.
\ee
We next decompose $\chi_{{}_{\eta\eta^*}}$ into two characteristic functions.
To do that, first observe that since $\eta,\eta^*\in\ml(B_\lambda)$ we have
\be
\chi_\eta(g)=\tr\left[\eta U_g\right]=\tr\left[\eta U^{(\lambda)}_g\right]\quad\text{and}\quad\chi_{\eta^*}(g)=\tr\left[\eta^* U_g\right]=\tr\left[\eta^* U^{(\lambda)}_g\right]\;.
\ee
Next, consider the second relation in~\eqref{nli} with $\eta$ replacing $\rho$ and $h$ replacing $g$; that is,
\be
\eta=d_\lambda\int_\G dh\;\chi_{\eta}(h^{-1})U^{(\lambda)}_h\;.
\ee
Multiplying both of its sides by $ \eta^*U_g^{(\lambda)}$, with some $g\in\G$, and taking the trace on both sides gives
\ba
\chi_{{}_{\eta\eta^*}}(g)&=d_\lambda\int_\G dh\;\chi_{{}_\eta}(h^{-1})\tr\left[U^{(\lambda)}_gU^{(\lambda)}_h\eta^*\right]\\
&=d_\lambda\int_\G dh\;\chi_{\eta}(h^{-1})\chi_{{}_{\eta^*}}(gh)\;.
\ea
Substituting this into~\eqref{1319} and using the fact that $\chi_{{}_{\eta^*}}(gh)=\bar{\chi}_{{}_{\eta}}(h^{-1}g^{-1})$ gives
\be
\tr\left[\eta\eta^*\rho^{B_\lambda}_\lambda\right]=d_\lambda^2\int_\G dh\int_\G dg\;f(g^{-1})\chi_{{}_{\eta}}(h^{-1})\bar{\chi}_{{}_{\eta}}(h^{-1}g^{-1})
\ee
Finally, by changing variables $k_1\eqdef h^{-1}$ and $k_2\eqdef h^{-1}g^{-1}$ the above equation becomes
\ba
\tr\left[\eta\eta^*\rho^{B_\lambda}_\lambda\right]
&=d_\lambda^2\int_\G dk_1\int_\G dk_2\;{\chi}_{{}_{\eta}}(k_1)f(k^{-1}_1k_2)\bar{\chi}_{{}_{\eta}}(k_2)\\
\GG{Since\; {\it f}\; is\; positive\; definite}&\geq 0\;.
\ea
Since $\eta$ was arbitrary we conclude that $\rho_\lambda^{B_\lambda}\geq 0$. 

We next construct the operator
\be
\rho^A\eqdef\bigoplus_{\lambda\in\irr(\G)} \rho_\lambda^{B_\lambda}\otimes\u^{C_\lambda}\;.
\ee
From the analysis above this operator is positive semidefinite. We show next that its trace is one (i.e.\ it is a density matrix) and that $f$ can be expressed as the characteristic function of $\rho^A$. To see this, recall that since $f(g)\in L^2(\G)$, it can be expressed as a linear combination of the basis elements $\{{u}_{k'k}^\mu(g)\}$ as
\be\label{ro2q}
f(g)=\sum_{\mu\in\irr(\G)}\sum_{k,k'=1}^{d_\mu}a_{kk'}^\mu {u}_{k'k}^\mu(g)\;,
\ee
where each $a_{kk'}^\mu\in\mbb{C}$ (for convenience we used ${u}_{k'k}^\mu(g)$ instead of $\bar{u}_{kk'}^\mu(g)$, so the coefficients $a_{kk'}^\mu$ are different than the coefficients $c_{kk'}^\mu$ of~\eqref{ro2}).
Substituting this into~\eqref{ro1} gives
\be
\rho^{B_\lambda}_\lambda= d_\lambda\sum_{\mu\in\irr(\G)}\sum_{k,k'=1}^{d_\mu}a_{kk'}^\mu \int_\G dg\;\bar{u}_{kk'}^\mu(g)U^{(\lambda)}_{g}
\ee
Combining this with~\eqref{glm} we conclude that
\be\label{rhobl}
\rho^{B_\lambda}_\lambda=\sum_{k,k'=1}^{d_\lambda}a_{kk'}^\lambda|\lambda,k\lr\lambda,k'|\;.
\ee
Finally, combining this with the expression~\eqref{chara0} for the characteristic function, we get
\ba
\chi_\rho(g)&=\sum_{\lambda}\tr\left[\rho^{B_\lambda}_\lambda U^{(\lambda)}_g\right]\\
\GG{\eqref{rhobl}}&=\sum_{k,k'=1}^{d_\lambda}a_{kk'}^\lambda u_{k'k}^\lambda(g)\\
\GG{\eqref{ro2q}}&=f(g)\;.
\ea
Moreover, since $f$ is normalized, i.e. $f(e)=1$, we have $\sum_\lambda\tr\big[\rho_\lambda^{B_\lambda}\big]=1$, so that $\rho^A$ 
is a quantum state. Finally, let $|\psi^{A\tA}\ra\eqdef\sqrt{\rho^A}\otimes I^{\tA}|\Omega^{A\tA}\ra$ be a purification of $\rho^A$, and define the unitary representation $g\mapsto U_{g}^\reg\otimes I^{\tA}$. Then, with respect to this representation $\chi_\psi(g)=f(g)$.
\end{proof}

\section{$\G$-Invariant Isometries}

Let $V:A'\to A$ be an isometry; i.e. $V^*V=I^{A'}$ and $|{A'}|\leq |A|$.
One of the key properties of isometries is that they can be completed to a unitary matrix by adding additional columns to make it a square unitary matrix. This property is very useful in applications. Here we ask the question if it is possible to complete a $\G$-invariant isometry $V:{A'}\to A$ into a $\G$-invariant unitary matrix $W:A\to A$. However, we first need to define what we mean by $\G$-invariant isometry.

Let $g\mapsto U_g^{A'}\in\ml({A'})$ and $g\mapsto U_g^A\in\ml(A)$ be two projective unitary representations of a group $\G$ on spaces ${A'}$ and $A$.
Recall that an intertwiner is a linear transformation $T:{A'}\to A$ that satisfies
\be
TU_g^{A'}=U_g^AT\quad\quad\forall\;g\in\G\;.
\ee
One can define a $\G$-invariant isometry as the above intertwiner that is also an isometry. In this definition, the $\G$-invariance property of the isometry is given with respect to two representations, namely $U_g^A$ and $U_g^{A'}$. Here we discuss another way to define a $\G$-invariant isometry that depends only on the representation $g\mapsto U_g^A$.

As we are interested in extending an isometry $V:{A'}\to A$ into a unitary $W:A\to A$ we will assume that ${A'}$ is a subspace of $A$ and denote by $\Pi:A\to A'$ the projection onto ${A'}$. We say that $V:{A'}\to A$ is $\G$-invariant with respect to the projective unitary representation $g\mapsto U_g\in\ml(A)$ if the operator $\tilde{V}\eqdef V\Pi:A\to A$ is $\G$-invariant. 
Since the domain of $\tilde{V}$ is the whole space $A$, the operator/matrix $\tilde{V}$ belongs to $\ml(A)$ so that the statement that it is $\G$-invariant is well defined and is equivalent to the condition
\be
\tilde{V}U_g=U_g\tilde{V}\quad\quad\forall\;g\in\G\;.
\ee
Note that in this definition, the $\G$-invariance property is defined with respect to a single representation $g\mapsto U_g^A$, and there is no need to consider another representation on system ${A'}$ (i.e. $g\mapsto U_g^{A'}$).
 
\begin{myt}{}
\begin{theorem}\label{compuni}
Let $g\mapsto U_g\in\ml(A)$ be a projective unitary representation acting on the Hilbert space $A$, and let $V:A'\to A$ be a $\G$-invariant isometry with respect to this representation of $\G$. Then, there exists a $\G$-invariant unitary matrix $W:A\to A$ such that $VP=WP$, where $P:A\to A'$ is the projection onto the subspace $A'$ of $A$.
\end{theorem}
\end{myt}

\begin{proof}
Since $\tilde{V}\eqdef VP$ commutes with $U_g$ it follows that also $\tilde{V}^*$ commutes with $U_g$. Therefore, $P=\tilde{V}^*\tilde{V}$ also commutes with $U_g$. Now, consider the irrep decomposition of the Hilbert space $A=\bigoplus_\lambda B_\lambda\otimes C_\lambda$.
From Theorem~\ref{deop} it follows that
\be
\tilde{V}=\bigoplus_\lambda I^{B_\lambda}\otimes \tilde{V}_\lambda^{C_\lambda}\quad\text{and}\quad P=\bigoplus_\lambda I^{B_\lambda}\otimes \Pi_\lambda^{C_\lambda}\;,
\ee
where $\tilde{V}_\lambda^{C_\lambda}\eqdef \frac1{|B_\lambda|}\tr_{B_\lambda}\left[\Pi^{B_\lambda}\tilde{V}\right]$ and $\Pi_\lambda^{C_\lambda}\eqdef \frac1{|B_\lambda|}\tr_{B_\lambda}\left[\Pi^{B_\lambda}P\right]$, with $\Pi^{B_\lambda}$ being the projection onto the space $B_\lambda$.
Now, observe that since $P$ is a projection the condition $PP=P$ gives $\Pi_\lambda^{C_\lambda}\Pi_\lambda^{C_\lambda}=\Pi_\lambda^{C_\lambda}$ so that each $\Pi_\lambda^{C_\lambda}$ is itself a projection in the space $C_\lambda$. Moreover,
since $P=\tilde{V}^*\tilde{V}$ we conclude that $\Pi_\lambda^{C_\lambda}=\tilde{V}_\lambda^{*C_\lambda}\tilde{V}_\lambda^{C_\lambda}$. Therefore, from Exercise~\ref{partialiso} it follows that for each $\lambda$, $\tilde{V}_\lambda^{C_\lambda}$ can be completed to a unitary  $W_\lambda:C_\lambda\to C_\lambda$. That is, there exists a unitary matrix $W_\lambda\in\ml(C_\lambda)$ satisfying $W_\lambda\Pi_\lambda^{C_\lambda}=\tilde{V}_\lambda^{C_\lambda}\Pi_\lambda^{C_\lambda}$. 
Define the matrix $W\in\ml(A)$ by
\be
W\eqdef\bigoplus_\lambda I^{B_\lambda}\otimes W_\lambda^{C_\lambda}\;.
\ee
Then, clearly $W^*W=I^A$ so that $W$ is a unitary matrix and from its definition $WP=\tilde{V}P=VP$. This completes the proof.
\end{proof}

\section{The Symmetric Subspace}\label{sec:symsub}

In this section, we embark on an exploration of the symmetric subspace, an integral concept in quantum information theory. We initiate our discourse with an essential grounding in the symmetric subspace, tailored from a quantum information viewpoint, to lay a solid foundation for further exploration. Due to space constraints, this discussion will not encompass all aspects of the symmetric subspace, particularly its significant role in quantum phenomena like state estimation, optimal cloning, and the de Finetti theorem. For those readers keen on a deeper dive into these areas, we recommend consulting the ``Notes and References" section at the end of this chapter for additional resources. Our primary objective here is to succinctly review some of the fundamental properties of the symmetric subspace that are crucial for the concepts developed in this book.

Let $\mathbf{S}_n$ be the finite symmetric group consisting of the $n!$ permutations on $n$ symbols. Recall that for any Hilbert space $A$, we use the notation $A^n\eqdef A^{\otimes n}$.
For any permutation $\pi\in\mathbf{S}_n$ we define the unitary representation $\pi\mapsto P^{A^n}_\pi$, where $A$ is a $m$-dimensional Hilbert space, and $P^{A^n}_\pi$ is a unitary matrix in $\ml(A^n)$ given by
\be\label{ppi}
P_\pi^{A^n}\eqdef\sum_{x^n\in[m]^n}\big|x_{\pi^{-1}(1)}\cdots x_{\pi^{-1}(n)}\big\ra\big\la x_1\cdots x_n\big|\;.
\ee

\begin{exercise}
Show that $\{P_\pi^{A^n}\}_{\pi\in\mathbf{S}_n}$ is indeed a unitary representation of $\mathbf{S}_n$.
\end{exercise}

\begin{myd}{}
The symmetric subspace of $A^n$ is denoted by $\sym_n(A)$ and is defined as
\be\label{defsym}
\sym_n(A)\eqdef\Big\{|\psi\ra\in A^n\;:\;P_\pi^{A^n}|\psi\ra=|\psi\ra\quad\quad\forall\;\pi\in\mathbf{S}_n\Big\}
\ee
\end{myd}
According to Theorem~\ref{vgv} in the appendix, the orthogonal projection to the symmetric subspace $\sym_n(A)$, denote by $\Pi_\sym^{A^n}$, is given by 
\be
\Pi_\sym^{A^n}=\frac1{n!}\sum_{\pi\in\mathbf{S}_n}P_\pi^{A^n}\;.
\ee
In order to calculate the dimension of $\sym_n(A^n)$, observe that the action of the projection above on any element $|x^n\ra\eqdef|x_1\cdots x_n\ra\in A^n$ of the standard basis of $A^n$ gives the symmetric vector
\be\label{symxny}
\Pi_\sym^{A^n}|x^n\ra=\frac1{n!}\sum_{\pi\in\mathbf{S}_n}|x_{\pi(1)}\cdots x_{\pi(n)}\ra\;.
\ee
Since the type of each sequence $(x_{\pi(1)}\cdots x_{\pi(n)})$ equals to the type of $x^n$,
the state above is uniquely determined by the type of $x^n$. Recall that $\t(x^n)$ denotes the type of $x^n$, and $X^n(\t)$ denotes the set of all sequences $x^n\in[m]^n$ whose type is $\t$. Keeping this in mind, we define for any type $\t\in\type(n,m)$ the unit vector
\be\label{phit}
|\varphi_{{}_\t}\ra\eqdef\frac1{\sqrt{k_\t}}\sum_{x^n\in X^n(\t)}|x^n\ra\;,
\ee
where 
\be
k_\t\eqdef|X^n(\t)|={n\choose nt_1,\ldots,nt_m}\;.
\ee
Observe that the state in~\eqref{symxny} is proportional to $|\varphi_{{}_\t}\ra$.
Therefore, since the image of $\Pi_\sym^{A^n}$ is $\sym_n(A)$, the set of vectors
$\{|\varphi_{{}_\t}\ra\}_{\t\in\type(n,m)}$ is an orthonormal basis of $\sym_n(A)$. This implies that the dimension of the symmetric subspace is given by
\ba\label{dsim}
\left|\sym_n(A)\right|&=\left|\type(n,m)\right|\\
\GG{\eqref{typeb}}&={n+m-1\choose n}\leq (n+1)^m\;.
\ea

\subsection{Characterization of the Symmetric Subspace}

The symmetric subspace has another interesting characterization. We now show that every state in $\sym_n(A)$ can be written a linear combination of states of the form $|\psi\ra^{\otimes n}$. 
\begin{myt}{}
\begin{theorem}\label{ntenn}
Let $A$ be a Hilbert space and $n\in\mbb{N}$. Then,
\be\label{rhdv}
\sym_n(A)=\spa\Big\{|\psi\ra^{\otimes n}\;:\;|\psi\ra\in A\Big\}\;.
\ee
\end{theorem}
\end{myt}
\begin{proof}
Let $B\eqdef\spa\{|\psi\ra^{\otimes n}\;:\;|\psi\ra\in A\}$ be the vector space on the right-hand side of~\eqref{rhdv},
and observe that  $B\subseteq \sym_n(A)$. We therefore need to show that $\sym_n(A)\subseteq B$. Let $|\psi\ra=\sum_{x\in[m]}v_x|x\ra$, and for every $x^n\in[m]^n$ let $v_{x^n}\eqdef v_{x_1}\cdots v_{x_n}$. Then,
\be\label{form31}
|\psi\ra^{\otimes n}=\sum_{x^n\in[m]^n}v_{x^n}|x^n\ra
=\sum_{\t\in\type(n,m)}v_{1}^{nt_1}\cdots v_m^{nt_m}|\chi_{{}_\t}\ra
\ee
where $|\chi_{{}_t}\ra= \sum_{x^n\in X^n(\t)}|x^n\ra$ is an unnormalized version of the normalized state defined in~\eqref{phit}. 

Next, we define the polynomial $f:\mbb{C}^m\to B$ as
\be
f(\v)\eqdef \sum_{\t\in\type(n,m)}v_{1}^{k_1}\cdots v_m^{k_m}|\chi_{{}_\t}\ra\quad\quad\forall\;\v\in\mbb{C}^m\;,
\ee
where $k_j\eqdef nt_j\in\mbb{N}$ for all $j\in[m]$. Observe that the integers $\{k_j\}_{j\in[m]}$ depends on $\t$ (for simplicity of the exposition we did not add a subscript to indicate that).
From~\eqref{form31} we have ${f}(\v)\in B$ for all $\v\in\mbb{C}^n$. We now argue that this implies that $|\chi_{{}_\t}\ra\in B$ for all $\t\in\type(n,m)$ so that $\sym_n(A)\subseteq B$. Indeed, observe that for any $\t$ and corresponding integers $k_1,\ldots,k_m$ we have
\be
|\chi_{{}_\t}\ra\propto\frac{\partial^n{f}(v_1,\ldots,v_m)}{\partial v_1^{k_1}\cdots\partial v_m^{k_m}}\Big|_{\v=0}\;.
\ee
Now, since $f(\v)\in V$ for all $\v\in\mbb{C}^m$, and since all partial derivatives are limits of linear combinations of $f(\v)$ at different points $(v_1,\ldots,v_m)$, we conclude that $|\chi_{{}_\t}\ra\in B$. This completes the proof.
\end{proof}

\begin{myt}{}
\begin{theorem}
Let $U\mapsto U^{\otimes n}$ be the natural unitary representation of $\muu(A)$ acting on $A^n$. With respect to this representation  $\sym_n(A)$ is an irreducible subspace of $A^n$.
\end{theorem}
\end{myt}
\begin{proof}
Let $|\psi_1\ra,|\psi_2\ra\in \sym_n(A)$ be two non-zero vectors in the symmetric subspace.
To show that $\sym_n(A)$ does not have an irreducible subspace, it will be enough to show that there exists $U\in\muu(A)$ such that
\be\label{unisat}
\left\la\psi_1\left|U^{\otimes n}\right|\psi_2\right\ra\neq 0\;.
\ee
From Lemma~\ref{ntenn} it follows that both $|\psi_1\ra$ and $|\psi_2\ra$ can be expressed as  linear combinations of states of the form $|\varphi\ra^{\otimes n}$. Hence, there exists $|\varphi_1\ra,|\varphi_2\ra\in A$ such that $\la\psi_1|\varphi_1^{\otimes n}\ra\neq 0$ and $\la\psi_2|\varphi_2^{\otimes n}\ra\neq 0$. For $j=1,2$ denote 
\be
\G_j\eqdef\Big\{U\in \muu(A)\;:\;U|\varphi_j\ra=|\varphi_j\ra\Big\}
\ee
and observe that both $\G_1$ and $\G_2$ are subgroups of $\muu(A)$. By definition, $|\varphi_j\ra$ is an eigenvector corresponding to the eigenvalue one for any $U\in\G_j$. Therefore, denoting $m\eqdef|A|$, from the spectral decomposition of such $U$, we conclude that  every $U\in\G_j$ has the form $U=\tilde{U}\oplus |\varphi_j\lr\varphi_j|$, where $\tilde{U}$ is a unitary matrix acting on the $(m-1)$-dimensional subspace orthogonal to $|\varphi_j\ra$.  In other words, for every $j\in\{1,2\}$
\be
\G_j=\Big\{\tilde{U}\oplus |\varphi_j\lr\varphi_j|\;:\;\tilde{U}\in \muu(m-1)\Big\}\;.
\ee
Now, from Theorem~\ref{vgv} we get that for $j=1,2$ (with $dV_j$ being the Haar measure on $\G_j$)
\be
\Pi_j\eqdef\int_{\G_j}dV_j\;V^{\otimes n}_j\;,
\ee
are, respectively, the orthogonal projections to the subspaces
\be\label{8213}
(A^{n})^{\G_j}\eqdef\Big\{|\psi\ra\in A^n\;:\;V^{\otimes n}_j|\psi\ra=|\psi\ra\quad\forall\;V_j\in\G_j\Big\}\;.
\ee 
We next show that the dimension $\left|(A^{n})^{\G_j}\right|=1$. First, observe that $|\varphi_j\ra^{\otimes n}\in(A^{n})^{\G_j}$ so the dimension of $(A^{n})^{\G_j}$ is at least one. Let $|\psi\ra\in(A^{n})^{\G_j}$ so that $V^{\otimes n}_j|\psi\ra=|\psi\ra$ for all $V_j\in\G_j$. Since each such $V_j$ has the form $V_j=\tilde{V_j}\oplus |\varphi_j\lr\varphi_j|$ we can take in particular $\tilde{V}_j=e^{i\theta}P_j$ where where $P_j$ is the projection to the subspace orthogonal to $|\varphi_j\ra$, and $\theta$ is any phase in $[0,2\pi]$. For this choice we get $(e^{i\theta}P_j\oplus |\varphi_j\lr\varphi_j|)^{\otimes n}|\psi\ra=|\psi\ra$ for all $\theta\in[0,2\pi]$.  But this is only possible for a state $|\psi\ra$ that is proportional to $|\varphi_j\ra^{\otimes n}$.
That is, up to a proportionality coefficient, the \emph{only} element of 
$(A^{n})^{\G_j}$ is $|\varphi_j\ra^{\otimes n}$. Hence, the projection to $(A^{n})^{\G_j}$ is
\be\label{pij}
\Pi_j=|\varphi_j\lr\varphi_j|^{\otimes n}\;.
\ee
Finally, let $W$ be any unitary matrix in $U(m)$ that satisfies $W|\varphi_2\ra=|\varphi_1\ra$. Then, we get that
\ba
&\int_{\G_1} dV_1\int_{\G_2}dV_2\;\left\la\psi_1\left|(V_1WV_2)^{\otimes n}\right|\psi_2\right\ra\\
&=\Big\la\psi_1\Big|\left(\int_{\G_1} dV_1\;V_1^{\otimes n}\right)W^{\otimes n}\left(\int_{\G_2} dV_2\;V_2^{\otimes n}\right)\Big|\psi_2\Big\ra\\
\GG{\eqref{pij}}&=\left\la\psi_1\big|\varphi_1^{\otimes n}\right\ra\left(\la\varphi_1|W|\varphi_2\ra\right)^n\left\la\varphi_2^{\otimes n}\big|\psi_2\right\ra\\
\GG{{\it W|\varphi_2\ra=|\varphi_1\ra}}&=\left\la\psi_1\big|\varphi_1^{\otimes n}\right\ra\left\la\varphi_2^{\otimes n}\big|\psi_2\right\ra\neq 0\;.
\ea
Therefore, there must exists at least one unitary matrix $V_1$ and one unitary matrix $V_2$ such that $U=V_1WV_2$ is satisfying~\eqref{unisat}. This completes the proof that $\sym_n(A)$ is an irreducible subspace.
\end{proof}

The space $A^n$ has another useful subspace called the \emph{antisymmetric subspace} which we denoted by $\asy(A^n)$. It is defined by
\be
\asy(A^n)\eqdef\Big\{|\psi\ra\in A^n\;:\;(-1)^{\sign(\pi)}P_\pi^{A^n}|\psi\ra=|\psi\ra\quad\forall\;\pi\in\mathbf{S}_n\Big\}
\ee
\begin{exercise}
Show that $\{(-1)^{\sign(\pi)}P_\pi^{A^n}\}_{\pi\in\mathbf{S}_n}$ is a unitary representation of $\mathbf{S}_n$.
\end{exercise}
From Theorem~\ref{vgv} it follows that the projection to the antisymmetric subspace is given by
\be
\Pi_\asy^{A^n}=\frac1{n!}\sum_{\pi\in\mathbf{S}_n}(-1)^{\sign(\pi)}P_\pi^{A^n}\;.
\ee
Now, observe that
\be\label{permu}
\Pi_\asy^{A^n}|x^n\ra=\sum_{\pi\in\mathbf{S}_n}(-1)^{\sign(\pi)}|x_{\pi(1)}\cdots x_{\pi(n)}\ra\;.
\ee
This vector is zero unless $n\leq m$. Otherwise, if $n>m$ any sequence $x^n$ must contain at least two components that are equal to each other. Without loss of generality suppose $x_1=x_2$. Then, for any permutation $\pi\in\mathbf{S}_n$ the permutation $\pi'$ defined by 
\be
\pi'(j)\eqdef\begin{cases}
\pi(2) &\text{if }j=1\\
\pi(1) &\text{if }j=2\\
\pi(j) &\text{if }j>2
\end{cases}
\ee
has the opposite sign than $\pi$, but yet have
\be
|x_{\pi'(1)}\cdots x_{\pi'(n)}\ra=|x_{\pi(1)}\cdots x_{\pi(n)}\ra\;.
\ee
Therefore, $\pi$ and $\pi'$ cancel their contributions in~\eqref{permu}.

The analysis above also tells us that $\Pi_\asy^{A^n}|x^n\ra=0$ whenever $x^n$ contains two components that are equal to each other. Therefore, the dimension of $\asy(A^n)$ is given by the number of ways to choose $n$-distinct entries from $\{1,\ldots,m\}$. That is, for $m\geq n$ the dimension of the antisymmetric subspace is given by
\be
\left|\asy(A^n)\right|={m\choose n}\;.
\ee

\subsection{The Bipartite Symmetric Subspace}\label{subsec:bip}

The case $n=2$ is very unique since in this case we get that
\be
\left|\sym(A^2)\right|+\left|\asy(A^2)\right|={d+1\choose 2}+{d\choose 2}=d^2=|A|^2\;.
\ee
This means that the space $A^2\eqdef A\otimes A$ can be decomposed as
\be\label{irdeco}
A^2=\sym(A^2)\oplus\asy(A^2)\;.
\ee
We already saw that $\sym(A^2)$ is an irreducible subspace of $A^2$, and we will see shortly  that the above decomposition is a decomposition into the two irreps of the ``natural" representation of the group $U(m)$ (or $SU(m)$) on the space $A^2$.

In the case that $n=2$ the set of all permutations on two elements, $\mS_2$, consists of only two permutations. Therefore, from~\eqref{ppi} we get that the projection to the symmetric subspace takes the simple form
\be\label{sw1}
\Pi_\sym^{A^2}=\frac12\left(I^{A^2}+F\right)
\ee
where $F:A^2\to A^2$ is known as the \emph{swap} operator given by
\be\label{flipoperator}
F\eqdef\sum_{x,y\in[m]}|x\lr y|\otimes |y\lr x|\;.
\ee
Similarly, the projection to the antisymmetric subspace is given by
\be\label{sw2}
\Pi_\asy^{A^2}=\frac12\left(I^{A^2}-F\right)\;.
\ee
\begin{exercise}
Show that for any unitary matrix $U\in \ml(A)$ the swap operator commutes with $U\otimes U$.
\end{exercise}
The exercise above implies that 
\be
F=\left(U\otimes U\right) F \left(U\otimes U\right)^*=\sum_{x,y\in[m]}U|x\lr y|U^*\otimes U|y\lr x|U^*
\ee
so that $F$ is independent on the choice of basis of $A$. 

The antisymmetric subspace has an orthonormal basis $\{|\psi_{xy}^{-}\ra\}_{x<y\in[m]}$ given by
\be
|\psi_{xy}^-\ra\eqdef\frac1{\sqrt{2}}\left(|xy\ra-|yx\ra\right)\quad\quad\forall\;1\leq x<y\leq m\;.
\ee
Similarly, the symmetric subspace has an orthonormal basis $\{|\psi_{xy}^{+}\ra\}_{x\leq y\in[m]}$ given by
\be
|\psi_{xy}^+\ra\eqdef\begin{cases}\frac1{\sqrt{2}}\left(|xy\ra+|yx\ra\right)
&\text{if }1\leq x<y\leq m\\
|xx\ra&\text{if }x=y\in[m]
\end{cases}\;.
\ee
\begin{myt}{}
\begin{theorem}
The subspace $\asy(A^2)$ is an irreducible subspace of $A^2$ under the natural representation of $U(m)$; i.e.\ under $U\mapsto U\otimes U$.
\end{theorem}
\end{myt}

\begin{proof}
Suppose by contradiction that $\asy(A^2)$ is not an irreducible subspace. Therefore, there must exists non-zero vectors $|\psi_1\ra,|\psi_2\ra\in\asy(A^2)$ such that
\be\label{conper}
\la\psi_1|U\otimes U|\psi_2\ra=0\quad\quad\forall\;U\in U(m)\;.
\ee
Denote
\be
|\psi_1\ra=\sum_{x<y}a_{xy}|\psi_{xy}^-\ra\quad\text{and}\quad|\psi_2\ra=\sum_{x<y}b_{xy}|\psi_{xy}^-\ra\;.
\ee
Now, take $U=DP_\pi $, where $D=\sum_{x\in[m]}e^{i\theta_x}|x\lr x|$ is a diagonal unitary with $\theta_x\in[0,2\pi]$, and $P_\pi\eqdef\sum_{x\in[m]}|\pi^{-1}(x)\lr x|$ is a permutation matrix with $\pi\in\mS_m$. Then,
\be
(U\otimes U)|\psi_2\ra=\sum_{x<y}b_{xy}(U\otimes U)|\psi_{xy}^-\ra=\sum_{x<y}b_{xy}e^{i(\theta_{\pi^{-1}(x)}+\theta_{\pi^{-1}(y)})}|\psi_{\pi^{-1}(x)\pi^{-1}(y)}^-\ra
\ee
Therefore, from~\eqref{conper} we get that for all permutations $\pi\in\mS_m$ and all phases $\{\theta_x\}_{x\in[m]}$ we have
\be
0=\la\psi_1|U\otimes U|\psi_2\ra=\sum_{x\in[m]}e^{i\theta_x}\sum_{y=x+1}^me^{i\theta_y}a_{xy}^*b_{\pi(x)\pi(y)}\;.
\ee
Taking the derivative of both sides above with respect to $\theta_1$ gives
\be
\sum_{y=2}^me^{i\theta_y}a_{1y}^*b_{\pi(1)\pi(y)}=0\;.
\ee
Since the equation above holds for all $\theta_y$ we must have $a_{1y}^*b_{\pi(1)\pi(y)}=0$ for all $y=2,\ldots,m$ and all permutations $\pi\in\mS_m$. Since $|\psi_2\ra\neq 0$, for each $y\in\{2,\ldots,m\}$ there exists $\pi\in\mS_m$ such that $b_{\pi(1)\pi(y)}\neq 0$. Hence, we must have $a_{1y}=0$ for all $y=2,\ldots,d$. Next, observe that the relation~\eqref{conper} becomes
\be
\sum_{x=2}^me^{i\theta_x}\sum_{y=x+1}^me^{i\theta_y}a_{xy}^*b_{\pi(x)\pi(y)}=0\;.
\ee
Therefore, taking the derivative with respect to $\theta_2$ and repeating similar lines as above we conclude that $a_{2y}=0$ for all $y=3,\ldots,m$. Continuing in this way we get that $a_{xy}=0$ for all $1\leq x<y\leq m$ in contradiction with the assumption that $|\psi_1\ra\neq 0$. This concludes the proof.
\end{proof}

\begin{exercise}
Extend the proof above for the case that $n>2$. That is, prove that $\asy(A^n)$ is an irreducible subspace of $A^n$ under the natural representation of $U(m)$ in $A^n$.
\end{exercise}

\section{Notes and References}

Representation theory is a vast area. For a more standard mathematical treatment we refer e.g.~\cite{HF1991}. We adapted the notations to ones that are more commonly used in physics. A review of some of the topics covered here including applications to quantum estimation theory can also be found in~\cite{Chiribella2006}.  A review on the properties of the symmetric subspace and their applications in quantum information are given in~\cite{Harrow2013}. In the Appendix~D we also introduce the symmetric version of Uhlmann's theorem (Theorem~\ref{symuhl}) which is due to~\cite{BP2010}.

%%%%%%%%%%%%%%%%%%%%%%%
%%%%%%%%%%%%%%%%%%%%%%%%

\chapter{Miscellany}

\section{The Divided difference}\label{A:DD}

We will make use of the notion of the \emph{divided difference} for $f$, for which we refer the reader to~\cite[(6.1.17)]{HJ1999}
 for more details. The divided difference for a function $f:\mbb{C}\to\mbb{C}$,
 given a sequence of distinct complex points, 
 $\alpha_i\in\mbb{C}, i=1,\ldots,n$, is defined for $i=0,1$ by
 \begin{align}\label{defdivdif1}
 & f[\alpha_1]:=f(\alpha_1)\\
 & f[\alpha_1,\alpha_2]\eqdef \frac{f(\alpha_1)-f(\alpha_2)}{\alpha_1-\alpha_2},
 \end{align}
 and defined inductively by
 \be
f[\alpha_1,\ldots,\alpha_i,\alpha_{i+1}]=
\frac{f[\alpha_1,\ldots,\alpha_{i-1},\alpha_{i}] - f[\alpha_1,\ldots,\alpha_{i-1},\alpha_{i+1}]}
 {\alpha_i-\alpha_{i+1}},
 \ee
 for $i=2,3,\ldots,n$.  It is well known that $f[\alpha_1,\ldots,\alpha_i,\alpha_{i+1}]$ is a symmetric function in $\alpha_1,\ldots,\alpha_{i+1}$,
 e.g. \cite[p.~392]{HJ91}.  For points that are not distinct it is defined by an appropriate limit. For example, for $x\ne y$ we have
 \begin{align}\label{g1g}
 &  f[x,x]=f'(x)\nonumber\\ 
 & f[x,x,y]=\frac{f'(x)}{(x-y)} -\frac{f(x)-f(y)}{(x-y)^2}\\ 
 & f[x,x,x]=\frac{1}{2} f''(x).\label{ququ}
 \end{align}
 Note that~\eqref{ququ} can be obtained from~\eqref{g1g} by setting $h\eqdef y-x\to 0$ and expanding
 $f(y)=f(x+h)=f(x)+hf'(x)+\frac{1}{2}h^2f''(x)+O(h^3)$.

 \begin{theorem}\label{taylor}  Let $A={\rm Diag}(\alpha_1,\ldots,\alpha_n)\in\mbb{C}^{n\times n}$ be a diagonal square matrix, and
 $B=[b_{ij}]\in\mbb{C}^{n\times n}$ be a complex square matrix. Assume that $f(x):\mbb{C}\to\mbb{C}$ satisfy one of the following conditions:
 \begin{enumerate}
 \item\label{smoothcase1}  $f(x)$ is an analytic function in some domain $\md\subset \mbb{C}$ which contains $\alpha_1,\ldots,\alpha_n$,
 and can be approximated uniformly in $\md$ by polynomials.
 \item\label{smoothcase2}  $\alpha_1,\ldots,\alpha_n$ are in a real open interval $(a,b)$ and $f$ has $n$ continuous derivatives in $(a,b)$.
 \end{enumerate}
 Then
 \begin{equation}\label{polcase1}
 f(A+tB)=f(A)+tL_{A}^{(1)}(B)+t^2L_{A}^{(2)}(B)+\cdots+t^nL_{A}^{(n)}(B)+O(t^{n+1})
 \end{equation}
 Here $L_{A}^{(1)}:\mbb{C}^{n\times n}\to\mbb{C}^{n\times n}$ is a linear operator, and $L_{A}^{(2)}:\mbb{C}^{n\times n}\to \mbb{C}^{n\times n}$ is a quadratic homogeneous noncommutative
 polynomial in $B$, etc. For $i,j=1,\ldots,n$ we have
  \begin{align}\label{LABform}
& [L_{A}^{(1)}(B)]_{ij}=f[\alpha_i,\alpha_j]b_{ij}=\frac{f(\alpha_i)-f(\alpha_j)}{\alpha_i-\alpha_j} b_{ij}\\
 \label{QABform}
& [L_{A}^{(2)}(B)]_{ij}=\sum_{k\in[n]} f[\alpha_i,\alpha_k,\alpha_j] b_{ik}b_{kj}.
 \end{align}
 And more generally, for $1\leq m\leq n-1$
 \be
 [L_{A}^{(m+1)}(B)]_{ij}=\sum_{k_1=1}^n\cdots\sum_{k_m=1}^n f[\alpha_i,\alpha_{k_1},\alpha_{k_2},\ldots,\alpha_{k_{m}},\alpha_j] 
 b_{ik_{1}}b_{k_1k_2}\cdots b_{k_{m-1}k_{m}}b_{k_{m}j}
 \ee
 In particular
 \begin{align}
&  \tr(L_{A}^{(1)}(B))=\sum_{j\in[n]}f'(\alpha_j)b_{jj}\\
 \label{tracQAB}
& \tr(L_{A}^{(2)}(B))=\frac{1}{2}\sum_{i,j=1}^n f'[\alpha_i,\alpha_j]b_{ij}b_{ji}=
\sum_{i,j=1}^n \frac{f'(\alpha_i)-f'(\alpha_j)}{2(\alpha_i-\alpha_j)}b_{ij}b_{ji}.
 \end{align}
 and more generally for $1\leq m\leq n$
 \be
 \tr(L_{A}^{(m)}(B))=\frac{1}{m}\sum_{j_1,j_2,\ldots,j_m=1}^n f'[\alpha_{j_1},\ldots,\alpha_{j_m}]b_{j_1j_2}b_{j_2j_3}\cdots b_{j_{m-1}j_m}b_{j_{m}j_1}\;.
 \ee
\end{theorem}
\begin{remark}
The expansion above can be naturally generalized to higher than the second order, but for the purpose of this article, we will
only need to expand $f(A+tB)$ up to the second order in $t$. Moreover, for our purposes we will only need to assume that the $\alpha_i$ are real and the condition 2 on $f$ holds. We kept condition 1 on $f$ in the theorem just to be a bit more general. 
\end{remark}
Note that in all the expressions above, one must identify $\alpha_i=\alpha_j$ with the limit $\alpha_j\to\alpha_i$.
For example, the term
\be
\frac{f'(\alpha_i)-f'(\alpha_j)}{2(\alpha_i-\alpha_j)}=\frac{1}{2}f''(\alpha_i)\;\;\;\text{for}\;\;\alpha_i=\alpha_j\;.
\ee
In particular, note that if $B$ is diagonal, Eq.~\eqref{tracQAB} gives the known second order term of the Taylor expansion.
 \proof  
 From the conditions on $f$, it is enough to prove the theorem assuming $f$ is a polynomial.
 By linearity, it is enough to prove all the claims for $f(x)=x^m$.
 Clearly, in the expansion
 \be
 (A+tB)^{m}=A^{m}+tL_A(B)+t^2Q_A(B)+O(t^3)
 \ee
 we must have 
 \begin{align}
 & L_A(B)=\sum_{0\le p,q,\; p+q=m-1} A^pBA^q,\label{121}\\ 
 & Q_A(B)=\sum_{0\le p,q,r,\; p+q+r=m-2} A^pBA^qBA^r,
 \label{LABQABfor}
 \end{align}
 where we expanded $(A+tB)^m$ up to first and second order in $t$. All that is left to show is that 
 these matrices coincide with the ones defined in~Eqs.~(\ref{LABform},\ref{QABform}).
 
 Indeed, since $A$ is diagonal, the matrix elements of the $L_{A}(B)$ in Eq.(\ref{121}) are given by
 \be
 [L_A(B)]_{ij}=\sum_{0\le p,q,\; p+q=m-1}\alpha_{i}^{p}\alpha_{j}^{q}b_{ij}=\frac{\alpha_{i}^{m}-\alpha_{j}^{m}}{\alpha_{i}-\alpha_{j}}b_{ij}\;,
 \ee
 which is equal to the exact same matrix elements given in Eq.(\ref{LABform}).
 
In the same way, since $A$ is diagonal, observe that the matrix elements of the $Q_{A}(B)$ in Eq.(\ref{LABQABfor}) are given by
 \be
 [Q_A(B)]_{ij}=\sum_{k\in[n]}\sum_{0\le p,q,r,\; p+q+r=m-2}\alpha_{i}^{p}\alpha_{k}^{q}\alpha_{j}^{r}b_{ik}b_{kj}\;.
 \ee
 On the other hand, a straightforward calculation gives for $f(x)=x^m$
 \be
 x^m[\alpha_i,\alpha_k,\alpha_j]=\sum_{0\le p,q,r,\; p+q+r=m-2} \alpha_{i}^{p} \alpha_{k}^{q} \alpha_{j}^{r}.
 \ee
 Thus, the expressions in Eq.~(\ref{QABform}) and Eq.~(\ref{LABQABfor}) for $Q_{A}(B)$ are the same.

 We now prove Eq.~\eqref{tracQAB}.  Observe first that Eq.~\eqref{QABform} yields
 \begin{equation}\label{tracQAB1}
 \tr(Q_A(B))=\sum_{i,j=1}^n  f[\alpha_i,\alpha_i, \alpha_j]b_{ij}b_{ji}\;,
 \end{equation}
 where we have used the symmetry $f[\alpha_i,\alpha_j, \alpha_i]=f[\alpha_i,\alpha_i, \alpha_j]$.
 Now, since $b_{ij}b_{ji}$ is symmetric under an exchange between $i$ and $j$, we can replace 
 $f[\alpha_i,\alpha_i, \alpha_j]$ in Eq.~(\ref{tracQAB1}) with 
 \be
 \frac{1}{2}\left(f[\alpha_i,\alpha_i, \alpha_j]+ f[\alpha_j,\alpha_j, \alpha_i]\right)
 =\frac{1}{2} f'[\alpha_i,\alpha_j]\;,
 \ee
 where for the last equality we used Eq.~\eqref{g1g}. \qed
 
 \begin{myg}{}
 \begin{corollary}\label{cordd}
 Let $f:\mbb{R}\to\mbb{R}$ be as above, $g:\mbb{R}\to\mbb{R}$ be another function, and $h:\mbb{R}\to\mbb{R}$ be the function $h(t)=g(t)f'(t)$. Then, for any $\rho,\sigma\in\herm(A)$ and $t\in(0,1)$
 \ba
 \tr[g(\rho)f(\rho+t\sigma&+t^2\eta)]=\tr\left[g(\rho)f(\rho)\right]+t\tr\left[h(\rho)\sigma\right]\\
 &+\frac12t^2\Big(\tr\left[h(\rho)\eta\right]+\tr\left[\sigma\mL_h(\sigma)\right]-\tr\left[\mL_f(\sigma)\mL_g(\sigma)\right]\Big)
 \ea
 \end{corollary}
 \end{myg}
 
 \begin{proof}
 Let $\xi\eqdef\sigma+t\eta$ and observe that
 \ba
f(\rho+\theta\xi)&=f(\rho)+t\mL_\rho^{(1)}(\xi)+t^2\mL_\rho^{(2)}(\xi)+O(t^3)\\
\Gg{\xi\eqdef \sigma+t\eta}&=f(\rho)+t\mL_\rho^{(1)}(\sigma)+t^2\left(\mL_\rho^{(1)}(\eta)+\mL_\rho^{(2)}(\sigma)\right)+O(t^3)
\ea
Since the map $\mL^{(1)}_\rho$ is self-adjoint we get
\ba
\tr\left[g(\rho)\mL_\rho^{(1)}(\sigma)\right]&=\tr\left[\mL_\rho^{(1)}\big(g(\rho)\big)\sigma\right]\\
\Gg{\mL_\rho^{(1)}\big(g(\rho)\big)=g(\rho)f'(\rho)}&=\tr\left[g(\rho)f'(\rho)\sigma\right]=\tr[h(\rho)\sigma]\;,
\ea
where we use the fact that $\mL_\rho^{(1)}\left(g(\rho)\right)=g(\rho)f'(\rho)$ (we leave it as an exercise). Similarly,
\be
\tr\left[g(\rho)\mL_\rho^{(1)}(\eta)\right]=\tr\left[h(\rho)\eta\right]\;.
\ee
Finally,
\ba
\tr\left[g(\rho)\mL^{(2)}_\rho(\sigma)\right]&=\sum_{x\in[m]}g(\alpha_x)\la x|\mL^{(2)}_\rho(\sigma)|x\ra\\
&=\sum_{x,y\in[m]}g(\alpha_x)f[\alpha_x,\alpha_y,\alpha_x]|\la x|\sigma|y\ra|^2\\
\GG{\eqref{g1g}}&=\sum_{x,y\in[m]}g(\alpha_x)\left(\frac{f'(\alpha_x)}{\alpha_x-\alpha_y}-\frac{f(\alpha_x)-f(\alpha_y)}{(\alpha_x-\alpha_y)^2}\right)|\la x|\sigma|y\ra|^2\\
\Gg{\sigma=\sigma^*}&=\frac12\sum_{x,y\in[m]}\left(\frac{h(\alpha_x)-h(\alpha_y)}{\alpha_x-\alpha_y}-\frac{(g(\alpha_x)-g(\alpha_y))(f(\alpha_x)-f(\alpha_y))}{(\alpha_x-\alpha_y)^2}\right)|\la x|\sigma|y\ra|^2\\
&=\frac12\sum_{x,y\in[m]}\Big(\la x|\sigma|y\ra\left[\mL_h(\sigma)\right]_{yx}-\left[\mL_g(\sigma)\right]_{xy}\left[\mL_f(\sigma)\right]_{yx}\Big)\\
&=\frac12\tr\left[\sigma\mL_h(\sigma)\right]-\frac12\tr\left[\mL_f(\sigma)\mL_g(\sigma)\right]\;.
\ea
This completes the proof.
 \end{proof}

\section{The Maximal $f$-Divergence\index{$f$-divergence}: Singular Case}\label{singularfd}

We prove here the formula given in~\eqref{7124} for the maximal $f$-Divergence\index{$f$-divergence} for the case that $\sigma$ is singular. For this purpose, we consider $\rho,\sigma\in\md(A)$, and we express $\rho$ and $\sigma$  in a block form 
\be\label{decomrs}
\rho=\begin{pmatrix}\rho_{11} & \zeta\\
\zeta^* & \rho_{22}\end{pmatrix}\quad\text{and}\quad\sigma=\begin{pmatrix}\sigma_{11} & \0_{12}\\
\0_{21} & \0_{22}\end{pmatrix}
\ee
where $\sigma_{11}>0$ and $\0$ denotes a zero matrix. Note that we can always find a basis in which $\rho$ and $\sigma$ has the above form. Moreover, unless specified otherwise, all inverses of matrices will be understood as generalized inverses. For example, the inverse of $\sigma$ is understood as
$
\sigma^{-1}\eqdef\begin{pmatrix}\sigma_{11}^{-1} & \0_{12}\\
\0_{21} & \0_{22}\end{pmatrix}
$.
Recall that from~\eqref{schurcom2} that the Schur complement of the block $\rho_{22}$ of $\rho$ is $\rho/\rho_{22}=\rho_{11}-\zeta\rho_{22}^{-1}\zeta^*$.
\begin{myg}{}
\begin{lemma}\label{lemuse}
Let $\D$ be a classical divergence, $\rho,\sigma\in\md(A)$ be as in~\eqref{decomrs}, and denote by $\trho\eqdef\rho_{11}-\zeta\rho_{22}^{-1}\zeta^*$ and by $\tsigma\eqdef\sigma_{11}$. 
Then, the maximal quantum extension of $\D$ can be expressed as
\be\label{v}
\oD(\rho\|\sigma)=\inf\D\left(\p\Big\|\begin{bmatrix}\q \\ 0\end{bmatrix}\right)
\ee
where the infimum is over all $1<n\in\mbb{N}$, all $\p\in\prob(n)$, and all POVMs $\{\Lambda_x\}_{x\in[n-1]}$ acting on the support of $\sigma$ that satisfy the following constraints:
For all $x\in[n-1]$, $q_x\eqdef\tr[\tsigma \Lambda_x]>0$ and 
\item
\be\label{ineq21}\tsigma^{-\frac12}\trho\tsigma^{-\frac12}\geq\sum_{x\in[n-1]}\frac{p_x}{q_x}\Lambda_x\;.\ee
\end{lemma}
\end{myg}

\begin{proof}
From Lemma~\ref{zeros} and the optimization in~\eqref{maxdpq} and~\eqref{rsrs}, it follows that
$\oD(\rho\|\sigma)$ can be expressed as in~\eqref{v}, where the infimum is over all $1<n\in\prob(n)$, $\p\in\prob(n)$ and $0<\q\in\prob(n-1)$, 
such that there exists $n-1$ density matrices $\{\omega_x\}_{x\in[n-1]}\subset\md(A)$ satisfying
\be\label{grgr}
\rho\geq\sum_{x\in[n-1]}p_x\omega_x\quad\text{and}\quad\sigma=\sum_{x\in[n-1]}q_x\omega_x\;,
\ee
where we used the fact that the first relation above holds if and only if $\rho=\sum_{x\in[n-1]}p_x\omega_x+p_n\omega_n$, for some density matrix $\omega_n$. Note that since $\sigma$ has the form given in~\eqref{decomrs}, it follows from the second relation above and the fact that $\q>0$ that also the density matrices $\{\omega_x\}$ have the  form $\omega_x=\begin{pmatrix}(\omega_x)_{11} & \0_{12}\\
\0_{21} & \0_{22}\end{pmatrix}$ for all $x\in[n-1]$.

%  We will denote by $\Pi_{\sigma}$ the projection to the support of $\sigma$, so that $\sigma=\Pi_{\sigma}\sigma\Pi_\sigma$ and $\Pi_{\sigma}\rho\Pi_\sigma=\begin{pmatrix}\rho_{11} & 0\\
%0 & 0\end{pmatrix}$. 
Define the matrix
\be
L\eqdef\begin{pmatrix}I_{11} & -\zeta\rho^{-1}_{22}\\
\0_{21} & I_{22}\end{pmatrix}\;.
\ee
This type of matrix has been used in the first chapter when we studied the Schur complement of positive semidefinite matrices (see Sec.~\ref{sec:schurcom}).
It has an inverse given by $L^{-1}=\begin{pmatrix}I_{11} & \zeta\rho^{-1}_{22}\\
\0_{21} & I_{22}\end{pmatrix}$ and it satisfies
\be\label{gaq}
L\rho L^{*}=\begin{pmatrix}\rho_{11}-\zeta\rho_{22}^{-1}\zeta^* & \0_{12}\\
\0_{21} & \rho_{22}\end{pmatrix},\quad L\sigma L^*=\sigma,\;\text{ and}\quad L \omega_x L^*=\omega_x\quad\forall\;x\in[n-1]\;.
\ee
Therefore, by applying the conjugation $L(\cdot)L^*$ on both sides of~\eqref{grgr} we get
\be\label{needpinching}
\begin{pmatrix}\trho & \0_{12}\\
\0_{21} & \rho_{22}\end{pmatrix}\geq\sum_{x\in[n-1]}p_x\omega_x\quad\text{and}\quad\sigma=\sum_{x\in[n-1]}q_x\omega_x\;.
\ee
Since $L$ is invertible, we can conjugate the above relations by $L^{-1}(\cdot)(L^*)^{-1}$ to get back~\eqref{grgr}. Therefore the above relations are equivalent to~\eqref{grgr}.
Moreover, since $\rho_{22}\geq 0$, it follows that the relation above holds if and only if
\be\label{gaq2}
\trho \geq\sum_{x\in[n-1]}p_x\tomega_x\quad\text{and}\quad\tsigma=\sum_{x\in[n-1]}q_x\tomega_x\;.
\ee
where $\tomega_x\eqdef(\omega_x)_{11}$ and $\tsigma\eqdef\sigma_{11}$.
Finally, denoting by
$\Lambda_x\eqdef q_x\tsigma^{-\frac12}\tomega_x\tsigma^{-\frac12}$, and applying the conjugation $\tsigma^{-\frac12}(\cdot)\tsigma^{-\frac12}$ to both sides of~\eqref{gaq2}  gives the relations
\be
\tsigma^{-\frac12}\trho\tsigma^{-\frac12}\geq\sum_{x\in[n-1]}\frac{p_x}{q_x}\Lambda_x\quad\text{and}\quad\sum_{x\in[n-1]}\Lambda_x=I^A\;.
\ee
This completes the proof.
\end{proof}

\begin{exercise}
Prove that $\trho$ in the lemma above has trace no greater than one.
\end{exercise}

When $\D$ equals the $f$-Divergence\index{$f$-divergence}, Eq.~\eqref{ffdiv} implies (together with the lemma above) that the infimum in~\eqref{maxdpq} can be expressed as 
\be\label{odeq}
\bD_f(\rho\|\sigma)\eqdef\inf_{1<n\in\mbb{N}}\left\{ \sum_{x\in[n-1]}q_xf\left(\frac{p_x}{q_x}\right)+\tilde{f}(0)p_{n}\right\}\;,
\ee
where $\tilde{f}$ is defined in~\eqref{tildef}, and the infimum above is subject to the same conditions given in Lemma~\ref{lemuse}.
Observe that if $\tilde{f}(0)=\infty$ then $p_n$ in~\eqref{odeq} can be taken to be zero (otherwise, $\bD_f(\rho\|\sigma)=\infty$). This means that the first relation in~\eqref{grgr} must also hold with equality (recall the original condition~\eqref{rsrs}). But from the second relation in~\eqref{grgr}, and the fact that $q_x>0$ for all $x\in[n-1]$, this is possible only if $\supp(\rho)\subseteq\supp(\sigma)$. That is, in the case $\tilde{f}(0)=\infty$ we have $\bD_f(\rho\|\sigma)=\infty$ for $\supp(\rho)\not\subseteq\supp(\sigma)$ and for $\supp(\rho)\subseteq\supp(\sigma)$ we can take $p_n=0$ in the optimization above and also replace the inequality sign of~\eqref{ineq21} with an equality. 
 
Going back to the general case,
one natural choice/guess for the optimal $n$, $\p$ and $\{\Lambda_x\}_{x\in[n-1]}$, is to choose them such that we have equality in~\eqref{ineq21}. This is possible for example by taking $n=r+1$, where $r$ is the dimension of the support of $\sigma$, and for any $x\in[r]$ to take $\Lambda_x=|\psi_x\lr\psi_x|$ with $|\psi_x\ra$ being the 
$x$-eigenvector of $\tsigma^{-\frac12}\trho\tsigma^{-\frac12}$ corresponding to the eigenvalue $p_x/q_x$ (i.e. $\p$ is chosen such that $p_x/\tr[\tsigma \Lambda_x]$ is the $x$-eigenvalue of $\tsigma^{-\frac12}\trho\tsigma^{-\frac12}$). For this choice we have 
\be\label{6116}
\tsigma^{-\frac12}\trho\tsigma^{-\frac12}=\sum_{x\in[r]}\frac{p_x}{q_x}|\psi_x\lr\psi_x|
\ee
which forces $p_n$ to be
\ba
p_n&=1-\sum_{x\in[r]}p_x=1-\sum_{x\in[r]}\frac{p_x}{q_x}\la\psi_x|\tsigma|\psi_x\ra=1-\tr[\trho]\;,
\ea
where the last equality follows by multiplying both sides of~\eqref{6116} by $\tsigma$ and taking the trace.
Moreover, for these choices of $n$, $\p$ and $\{\Lambda_x\}$, we have
\ba
\sum_{x\in[n]}q_xf\left(\frac{p_x}{q_x}\right)&=\sum_{x\in[n]}\tr[\tsigma|\psi_x\lr\psi_x|] f\left(\frac{p_x}{q_x}\right)\\
\Gg{\forall t\geq 0\;\;f(t|\psi_x\lr\psi_x|)=f(t)|\psi_x\lr\psi_x|\rightarrow}&=\sum_{x\in[n]}\tr\left[\tsigma f\left(\frac{p_x}{q_x}|\psi_x\lr\psi_x|\right)\right]\\
\Gg{\{|\psi_x\ra\}\text{ is orthonormal }\rightarrow}&=\tr\left[\tsigma f\left(\sum_{x\in[n]}\frac{p_x}{q_x}|\psi_x\lr\psi_x|\right)\right]\\
&=\tr\left[\tsigma f\left(\tsigma^{-\frac12}\trho\tsigma^{-\frac12}\right)\right]\;.
\ea 
Note that we obtained the formula above for a particular choice of $n$, $\p$ and $\{E_x\}$. Therefore, since this is not necessarily the optimal choice (recall $\bD_f$ is defined in terms of an infimum), we must have 
\be\label{showed}
\bD_f(\rho\|\sigma)\leq \tr\left[\tsigma f\left(\tsigma^{-\frac12}\trho\tsigma^{-\frac12}\right)\right]+(1-\tr[\trho])\tilde{f}(0)\;.
\ee
Interestingly, to get this upper bound we did not even assume that $f$ is convex, but if $f$ is operator convex we get an equality. 

Before we state the theorem below, we point out a remarkable result from matrix analysis that we will use below. Suppose  $f:[0,\infty)\to \mbb{R}$ is operator convex. In Sec.~\ref{sec:oc} we saw that operator convexity is a strong condition, as it provides a lot of information about $f$. In particular,  in~\cite{HMPB2011} it was shown that if in addition $f$ satisfies $\tilde{f}(0)\eqdef\lim_{\eps\to 0^+}\eps f(\frac1\eps)<\infty$, then $f$ necessarily have the form
\be\label{formoff}
f(r)=f(0)+\tilde{f}(0)r+g(r)\quad\quad\forall\;r\in[0,\infty)\;,
\ee
where $g:[0,\infty)\to\mbb{R}$ is an operator monotone decreasing function.

\begin{myt}{\color{yellow} Closed Formula of The Maximal $f$-Divergence\index{$f$-divergence}} 
\begin{theorem}\label{t633}
Let $\rho,\sigma\in\md(A)$ given as in~\eqref{decomrs} with $\trho\eqdef\rho_{11}-\zeta\rho_{22}^{-1}\zeta^*$ and $\tsigma\eqdef\sigma_{11}$, and let $f\eqdef(0,\infty)\to\mbb{R}$ be operator convex, with $f(0)\eqdef\lim_{\eps\to 0^+}f(\eps)$ and $f(1)=0$.
Then,
\be\label{mainfd}
\bD_f(\rho\|\sigma)= \tr\left[\tsigma f\left(\tsigma^{-\frac12}\trho\tsigma^{-\frac12}\right)\right]+(1-\tr[\trho])\tilde{f}(0)\;,
\ee
where $\tilde{f}(0)\eqdef\lim_{\eps\to 0^+}\eps f(\frac1\eps)$.
\end{theorem}
\end{myt}

\begin{remark}
We use the convention $0\cdot\infty=0$ in the case that both $\tr[\trho]=1$ and $\tilde{f}(0)=\infty$. The inverse of $\rho_{22}$ in the theorem above is a generalized inverse (in case $\rho_{22}$ is not invertible). This in particular implies that if $\supp(\rho)\subseteq\supp(\sigma)$ then $\rho_{22}^{-1}=0$ and $\trho=\rho=\rho_{11}$ so that $\tr[\trho]=1$ and the second term on the right-hand side of~\eqref{mainfd} vanishes. Moreover,  the requirement that $f(1)=0$ is not necessary, but if $f(1)\neq 0$ then the resulting divergence $\bD_f$ will not be normalized (i.e. we will get $\bD_f(1\|1)\neq0$). Finally, observe that~\eqref{mainfd} can be expressed in terms of the Kubo-Ando operator mean $\#_f$ (see Definition~\ref{kaom}) as
\be
\bD_f(\rho\|\sigma)= \tr\left[\trho\#_f\tsigma\right]+(1-\tr[\trho])\tilde{f}(0)\;.
\ee
\end{remark}

\begin{proof}
We have already shown in~\eqref{showed} that $\bD_f(\rho\|\sigma)$ cannot be great than the right-hand side of~\eqref{mainfd}. Therefore, it is left to show the opposite inequality. Let $1<n\in\mbb{N}$, $\{\Lambda_x\}_{x\in[n-1]}$ be a POVM acting on the support of $\sigma$, and  $\p\in\prob(n)$. Suppose the conditions in Lemma~\eqref{lemuse} hold with these elements. From Naimark's\index{Naimark} theorem (see Theorem~\ref{naimark}) there exists a tuple of mutual orthonormal projectors $\{P_x\}_{x\in[n-1]}\subset\pos(B)$, where $B$ is the extended Hilbert space, and an isometry $V:A\to B$  such that $\Lambda_x=V^*P_xV$ for all $x\in[n-1]$. We use this to compute
\ba
\sum_{x\in[n-1]}q_xf\left(\frac{p_x}{q_x}\right)&=\sum_{x\in[n-1]}\tr[\Lambda_x\tsigma]f\left(\frac{p_x}{q_x}\right)\\
&=\tr\Big[\sum_{x\in[n-1]}f\left(\frac{p_x}{q_x}\right)E_x\tsigma\Big] \\
&=\tr\Big[\sum_{x\in[n-1]}f\left(\frac{p_x}{q_x}\right)P_xV\tsigma V^{*}\Big]\;,\nonumber
\ea
where we put the sum inside the trace. Now, since $\{P_x\}$ are orthogonal projectors we have $\sum_{x\in[n-1]}f\left(\frac{p_x}{q_x}\right)P_x=f\left(\sum_{x\in[n-1]}\frac{p_x}{q_x}P_x\right)$. Combining this with the cyclic property of the trace we get
\ba\label{seque}
 \sum_{x\in[n-1]}q_xf\Big(\frac{p_x}{q_x}\Big)&=\tr\left[V^*f\Big(\sum_{x\in[n-1]}\frac{p_x}{q_x}P_x\Big)V\tsigma \right]\\
\GG{\text{Jensen's Inequality~\eqref{jensen}}\rightarrow}&\geq \tr\left[f\Big(\sum_{x\in[n-1]}\frac{p_x}{q_x}V^{*}P_xV\Big)\tsigma \right]\\
&=\tr\left[f\Big(\sum_{x\in[n-1]}\frac{p_x}{q_x}\Lambda_x\Big)\tsigma \right]\;.
\ea 
To continue, we first consider the case that $\tilde{f}(0)=\infty$. From the remark below~\eqref{odeq} we know that $\bD_f(\rho\|\sigma)=\infty$ unless $\supp(\rho)\subseteq\supp(\sigma)$. 
Furthermore, if $\supp(\rho)\subseteq\supp(\sigma)$ then we 
can replace the inequality sign of~\eqref{ineq21} with an equality, so that the above equation gives the desired inequality (recall that $\tsigma=\sigma$ and $\trho=\rho$ if $\sigma>0$ which in the context here $\sigma>0$ is effectively the same statement as $\supp(\rho)\subseteq\supp(\sigma)$ since we can restrict all computations to the support of $\sigma$)
\be
\sum_{x\in[n-1]}q_xf\left(\frac{p_x}{q_x}\right)\geq  \tr\left[\sigma f\left(\sigma^{-\frac12}\rho\sigma^{-\frac12}\right)\right]\;.
\ee
Note that we did not include on the left-hand side of the equation above the term $p_n\tilde{f}(0)$ since in this case (i.e. the case $\tilde{f}(0)=\infty$) we must have $p_n=0$ so the term $p_n\tilde{f}(0)=0\cdot\infty=0$ by convention.

Next, we consider the case $\tilde{f}(0)<\infty$. In this case we know that $f$ has the form~\eqref{formoff}. Hence, continuing from~\eqref{seque} we get
\ba
\sum_{x\in[n-1]}q_xf\left(\frac{p_x}{q_x}\right)&\geq f(0)+\tilde{f}(0)\tr\Big[\sum_{x\in[n-1]}\frac{p_x}{q_x}\Lambda_x\tsigma \Big]+\tr\Big[g\Big(\sum_{x\in[n-1]}\frac{p_x}{q_x}\Lambda_x\Big)\tsigma \Big]\\
\Gg{\substack{\text{from~\eqref{ineq21} and operator}\\ \text{monotonicity decreasing of } g}\rightarrow}&\geq f(0)+\tilde{f}(0)\sum_{x\in[n-1]}p_x+\tr\left[g\left(\tsigma^{-\frac12}\trho\tsigma^{-\frac12}\right)\tsigma \right]\\
\Gg{\substack{\text{using again the relation}\\ g(r)=f(r)-f(0)-\tilde{f}(0)r}\rightarrow}&=\tilde{f}(0)\Big(\sum_{x\in[n-1]}p_x-\tr[\trho]\Big)+\tr\left[f\left(\tsigma^{-\frac12}\trho\tsigma^{-\frac12}\right)\tsigma \right]
\ea
Hence,
\be
\sum_{x\in[n-1]}q_xf\left(\frac{p_x}{q_x}\right)+\tilde{f}(0)p_n\geq \tilde{f}(0)\left(1-\tr[\trho]\right)+\tr\left[f\left(\tsigma^{-\frac12}\trho\tsigma^{-\frac12}\right)\tsigma \right]\;.
\ee
Since the above inequality holds for any choice of $1<n\in\mbb{N}$, POVM $\{\Lambda_x\}_{x\in[n-1]}$, and $\p\in\prob(n)$ that satisfy the conditions in Lemma~\eqref{lemuse}, we conclude that $\bD_f(\rho\|\sigma)$ is no smaller than the right-hand side of the equation above. This concludes the proof.
\end{proof}

As an example, consider the function $f_\alpha(r)=\frac{r^\alpha-r}{\alpha(\alpha-1)}$ which is known to be operator convex for $\alpha\in(0,2]$. For this function we have
\be
\tilde{f}_\alpha(0)=\lim_{\eps\to 0^+}\frac{\eps\left(\frac1{\eps^\alpha}-\frac{1}{\eps}\right)}{\alpha(\alpha-1)}=\lim_{\eps\to 0^+}\frac{\eps^{1-\alpha}-1}{\alpha(\alpha-1)}=\begin{cases} \frac1{\alpha(1-\alpha)} &\text{if }\; 0<\alpha<1\\ \infty &\text{if }\;1\leq \alpha\leq 2 \end{cases}
\ee
For $\alpha\in[1,2]$, unless $\supp(\rho)\subseteq\supp(\sigma)$ we have $\bD_f(\rho\|\sigma)=\infty$. For the case $\supp(\rho)\subseteq\supp(\sigma)$ we have 
\ba\label{6130}
\bD_{f_\alpha}(\rho\|\sigma)&=\frac{1}{\alpha(\alpha-1)}\tr\left[\sigma\left( \left(\sigma^{-\frac12}\rho\sigma^{-\frac12}\right)^\alpha-\sigma^{-\frac12}\rho\sigma^{-\frac12}\right)\right]\\
&=\frac{1}{\alpha(\alpha-1)}\left(\tr\left[\sigma\left(\sigma^{-\frac12}\rho\sigma^{-\frac12}\right)^\alpha\right]-1\right)\;.
\ea
On the other hand, for the case $\alpha\in(0,1)$, Eq.~\eqref{mainfd} gives
\ba\label{6131}
\bD_{f_\alpha}(\rho\|\sigma)&=\frac{1}{\alpha(\alpha-1)}\left(\tr\left[\tsigma\left(\tsigma^{-\frac12}\trho\tsigma^{-\frac12}\right)^\alpha\right]-\tr[\trho]\right)+\frac{1-\tr[\trho]}{\alpha(1-\alpha)}\\
&=\frac{1}{\alpha(\alpha-1)}\left(\tr\left[\tsigma\left(\tsigma^{-\frac12}\trho\tsigma^{-\frac12}\right)^\alpha\right]-1\right)\;.
\ea
Combining everything we conclude that for any $\alpha\in(0,2]$ we have
\be
\bD_{f_\alpha}(\rho\|\sigma)=\begin{cases}\frac{\tr\left[\tsigma\left(\tsigma^{-\frac12}\trho\tsigma^{-\frac12}\right)^\alpha\right]-1}{\alpha(\alpha-1)} &\text{if }\;\alpha\in(0,1)\;\text{ or }\;\supp(\rho)\subseteq\supp(\sigma)\\
\infty & \text{otherwise}
\end{cases}
\ee

\begin{exercise}\label{a01}
Show that for $\alpha\in(0,1)$ we have for any $\rho,\sigma\in\md(A)$
\be
\bD_{f_\alpha}(\rho\|\sigma)=\lim_{\eps\to 0^+}\bD_{f_\alpha}(\rho\|\sigma+\eps I)
\ee
\end{exercise}

\section{Smoothing with the Second Variable of $D_{\max}$}\label{SmoothSec}

\begin{myg}{}
\begin{lemma}\label{tzetze}
Let $\rho,\omega\in\md(A)$, $\eta\in\pos(A)$, and $\delta\in(0,1)$. Set $\eps\eqdef\sqrt{2\delta}$ and suppose that $\rho\leq \eta+\delta\omega$.  Then, there exists $\rho'\in\mb_{\eps}(\rho)$ such that 
\be\label{10p}
\rho'\leq\frac1{1-\delta}\eta\;.
\ee
\end{lemma}
\end{myg}
\begin{remark}
The condition in~\eqref{10p} can be written as 
\be\label{10121}
D_{\max}\left(\rho'\big\|p^{-1}\eta\right)\leq\log(p)-\log(1-\delta)
\ee
where $p\eqdef\tr[\eta]$. Since $\rho'$ is $\eps$-close to $\rho$ we can conclude that
\be
D_{\max}^\eps\left(\rho\big\|p^{-1}\eta\right)\leq\log(p)-\log(1-\delta)\;.
\ee
\end{remark}

\begin{proof}
Let $G\eqdef \eta^{\frac12}(\eta+\delta\omega)^{-\frac12}$ and observe that 
\ba
\tr[G^*G\rho]&=\tr\left[(\eta+\delta\omega)^{-\frac12}\eta(\eta+\delta\omega)^{-\frac12}\rho\right]\\
\GG{\eta=\eta+\delta\omega-\delta\omega}&=1-\delta\tr\left[(\eta+\delta\omega)^{-\frac12}\omega(\eta+\delta\omega)^{-\frac12}\rho\right]\\
\GG{\rho\leq\eta+\delta\omega}&\geq 1-\delta\;.
\ea
Moreover, define $\rho'\eqdef G\rho G^*/\tr[G^*G\rho]$ and note that the above inequality implies that
\ba
\rho'&\leq\frac{G\rho G^*}{1-\delta}\\
&=\frac{\eta^{\frac12}(\eta+\delta\omega)^{-\frac12}\rho (\eta+\delta\omega)^{-\frac12}\eta^{\frac12}}{1-\delta}\\
\GG{\rho\leq\eta+\delta\omega}&\leq\frac{\eta}{1-\delta}\;.
\ea
It is left to show that $\rho'\in\mb_{\eps}(\rho)$. Using similar arguments as in~\eqref{10114r} and~\eqref{10115} we get that $G^*G\leq I^A$ and $P\eqdef\frac12(G+G^*)\leq I^A$. Moreover, following similar arguments as in~\eqref{10116} we get that $F(\rho,\rho')\geq\tr[\rho P]$. Hence,
\ba
F(\rho,\rho')&\geq1-\tr[\rho (I-P)]\\
\GG{\rho\leq\eta+\delta\omega\;\text{ and }\;I-P\geq 0}&\geq 1-\tr[(\eta+\delta\omega) (I-P)]\\
&=1-\delta-\tr[\eta]+\tr[(\eta+\delta\omega) P]\\
\GG{\text{by definition of }P}&=1-\delta-\tr[\eta]+\tr[\eta^{\frac12}(\eta+\delta\omega)^{\frac12} ]\\
\GG{(\eta+\delta\omega)^{\frac12}\geq\eta^{\frac12}}&\geq 1-\delta\;.
\ea
Using the upper bound in~\eqref{fitr} we conclude that 
\ba
\frac12\|\rho-\rho'\|_1&\leq \sqrt{1-F(\rho,\rho')^2}\\
&\leq \sqrt{1-(1-\delta)^2}\\
&=\sqrt{2\delta-\delta^2}\\
&\leq\sqrt{2\delta}=\eps\;.
\ea
This completes the proof.
\end{proof}

The above lemma can be used to bound the smoothed max relative entropy\index{max relative entropy} in terms of its following variant, defined via
\be
D_{\max}^{(\eps)}(\rho\|\sigma)\eqdef\min_{\sigma'\in\mb_{\eps}(\sigma)}D_{\max}(\rho\|\sigma')\quad\quad\forall\;\rho,\sigma\in\md(A)\;.
\ee 
That is, we use the brackets on $\eps$ to indicate that the smoothing is done with respect to the second argument of $D_{\max}$.
\begin{myg}{}
\begin{lemma}
Let $\rho,\sigma\in\md(A)$ be such that $r\eqdef 2^{D_{\max}(\rho\|\sigma)}<\infty$.
Then, for any $0<\eps<\frac1r$
\be 
D_{\max}^{\sqrt{2\eps}}(\rho\|\sigma)\leq D_{\max}^{(\eps)}(\rho\|\sigma)-\log\left(1-\eps r\right)\;.
\ee
\end{lemma}
\end{myg}
\begin{proof}
Let $\sigma'$ be such that $D_{\max}^{(\eps')}(\rho\|\sigma)=D_{\max}(\rho\|\sigma')$. Since $\sigma'\in\mb_{\eps}(\sigma)$ there exists $0\leq\delta'\leq\eps'$ and $\omega',\omega\in\md(A)$ such that (cf.~\eqref{decomom})
\ba
\sigma'+\delta'\omega'=\sigma+\delta'\omega\;.
\ea
Denote by $t\eqdef 2^{D_{\max}(\rho\|\sigma')}=2^{D_{\max}^{(\eps')}(\rho\|\sigma)}$.
Then from its definition we have
\ba
\rho&\leq t\sigma'\\
&\leq t(\sigma'+\delta'\omega')\\
&= t\sigma+t\delta'\omega\;.
\ea
Hence, from Lemma~\ref{tzetze} there exists $\rho'\in\mb_{\eps}(\rho)$, with $\eps\eqdef \sqrt{2\delta'}\leq\sqrt{2\eps'}$ such that 
\be
\rho'\leq \frac t{1-t\delta'}\sigma\leq \frac t{1-t\eps'}\sigma\;.
\ee
We therefore conclude that
\ba
D_{\max}^\eps(\rho\|\sigma)&\leq D_{\max}(\rho'\|\sigma)\\
&\leq\log t-\log(1-t\eps')\\
&=D_{\max}^{(\eps')}(\rho\|\sigma)-\log\left(1-\eps'2^{D_{\max}^{(\eps')}(\rho\|\sigma)}\right)\\
&\leq D_{\max}^{(\eps')}(\rho\|\sigma)-\log\left(1-\eps'2^{D_{\max}(\rho\|\sigma)}\right)\;.
\ea
This completes the proof.
\end{proof}

\section{Two Proofs of the Classical Stein's Lemma}\label{acht}

Below we give a proof, which to the author's knowledge, appears for the first time in this book, and is based on the bounds~\eqref{blbounds}. A more standard proof follows it.

\begin{proof}[Proof of Theorem~\ref{th:sl}]
Suppose first that all the components of $\q$ consists of rational numbers. That is, there exists $k_1,\ldots,k_m\in\mbb{N}$ such that $\q=\left(\frac{k_1}k,\ldots,\frac{k_m}k\right)^T$, where $k\eqdef k_1+\cdots+k_m$. From Theorem~\ref{onlyr} the vector \be\r\eqdef\bigoplus_{x\in[m]}p_x\u^{(k_x)}=\Big(\underbrace{\frac{p_1}{k_1},\ldots,\frac{p_1}{k_1}}_{k_1\text{-times}},\underbrace{\frac{p_2}{k_2},\ldots,\frac{p_2}{k_2}}_{k_2\text{-times}},\ldots,
\underbrace{\frac{p_m}{k_m},\ldots,\frac{p_m}{k_m}}_{k_m\text{-times}}\Big)\ee satisfies $(\p,\q)\sim(\r,\u^{(k)})$. Without loss of generality we assume that the components of $\p$ and $\q$ are ordered as in~\eqref{order}. Note that this is equivalent to $\r=\r^\da$.
Moreover, observe that the relation $(\p,\q)\sim(\r,\u^{(k)})$ also implies that for any $n\in\mbb{N}$ we have $(\p^{\otimes n},\q^{\otimes n})\sim(\r^{\otimes n},\u^{(k^n)})$. We therefore get that for all $n\in\mbb{N}$
\be
\frac1nD_{\min}^{\eps}\left(\p^{\otimes n}\big\|\q^{\otimes n}\right)
=\frac1nD_{\min}^{\eps}\left(\r^{\otimes n}\big\|\u^{(k^n)}\right)\;.
\ee
Combining this with the upper bound in~\eqref{blbounds} we get that
\be
\frac1nD_{\min}^{\eps}\left(\p^{\otimes n}\big\|\q^{\otimes n}\right)\leq-\log b_{\ell_n}=-\log\frac{\ell_n}{k^n}=\log(k)-\frac1n\log(\ell_n)\;,
\ee
where $b_{\ell_n}\eqdef\left\|\u^{(k^n)}\right\|_{(\ell_n)}=\frac{\ell_n}{k^n}$, and $\ell_n\in\{0,1,\ldots,k^n-1\}$ is the integer satisfying 
\be\label{rnsums}
\left\|\r^{\otimes n}\right\|_{(\ell_n)}< 1-\eps\leq \left\|\r^{\otimes n}\right\|_{(\ell_n+1)}\;.
\ee
Similarly, from the lower bound in~\eqref{blbounds} we get
\be
\frac1nD_{\min}^{\eps}\left(\p^{\otimes n}\big\|\q^{\otimes n}\right)\geq\log(k)-\frac1n\log(\ell_n+1)\;.
\ee
Hence, from the two bounds above we conclude that
\be\label{limitde}
\lim_{n\to\infty}\frac1nD^\eps_{\min}\left(\p^{\otimes n}\big\|\q^{\otimes n}\right)=\log(k)-\lim_{n\to\infty}\frac1n\log(\ell_n)\;,
\ee
where we will see shortly that the limits above exists.
It is therefore left to estimate $\ell_n$. For this purpose, we will estimate the sums in~\eqref{rnsums} using the notion of (weak) typicality. Observe first that $j$ in~\eqref{rnsums} can be expressed as a sequence $x^n=(x_1,\ldots,x_n)\in[m]^n$ (so that the components of $\r^{\otimes n}$ can be expressed as $r_{x^n}\eqdef r_{x_1}\cdots r_{x_n}$). Denote by $\ms_{\ell_n}$ the set of the $\ell_n$ sequences that correspond to the largest probabilities $r_{x^n}$. With these notations we have
\be
\sum_{j\in[\ell_n]}\left(\r^{\otimes n}\right)^\da_j=\sum_{x^n\in\ms_{\ell_n}}r_{x^n}\;.
\ee
Let $\delta>0$ be arbitrary small number. From the definition of $\ms_{\ell_n}$ it follows that if there exists $x^n\in\ms_{\ell_n}$ such that $r_{x^n}< 2^{-n(H(\r)+\delta)}$ then the set $\ms_{\ell_n}$ contains the set of $\delta$-typical sequences, $\mt_{n,\delta}(X)$. Therefore, in this case the sum above is greater than $\pr\left(\mt_{n,\delta}(X)\right)$. However, for sufficiently large $n$ this probability exceed $1-\eps$ in contradiction with~\eqref{rnsums}. Therefore, without loss of generality we can assume that for sufficiently large $n$ all the sequences $x^n\in\ms_{\ell_n}$ satisfies $r_{x^n}\geq 2^{-n(H(\r)+\delta)}$. Combining this with the first bound in~\eqref{rnsums} 
we have
\ba
1-\eps&> \sum_{x^n\in\ms_{\ell_n}}r_{x^n}\\
&\geq \ell_n2^{-n(H(\r)+\delta)}\;.
\ea
Hence,
$
\ell_n\leq (1-\eps)2^{n(H(\r)+\delta)}
$
which gives
\be\label{bbb1}
\limsup_{n\to\infty}\frac1n\log(\ell_n)\leq H(\r)+\delta\;.
\ee
Next, from the second bound in~\eqref{rnsums} we have
\ba
1-\eps&\leq \sum_{x^n\in\ms_{\ell_n+1}}r_{x^n}\\
&\leq\sum_{x^n\in\ms_{\ell_n+1}\cap\mt_{n,\delta}(X)}r_{x^n}+\sum_{x^n\not\in\mt_{n,\delta}(X)}r_{x^n}\\
&\leq (\ell_n+1)2^{-n(H(\r)-\delta)}+\pr\left(\mt_{n,\delta}^c(X)\right)
\ea
where $\mu_n\eqdef\pr\left(\mt_{n,\delta}^c(X)\right)$ is the probability that a sequence is not $\delta$-typical which is goes to zero as $n$ goes to infinity. Hence,
$
\ell_n\geq(1-\eps-\delta_n)2^{n(H(\r)-\delta)}
$
so that
\be\label{bbb2}
\liminf_{n\to\infty}\frac1n\log(\ell_n)\geq H(\r)-\delta\;.
\ee
Since the two bounds in~\eqref{bbb1} and~\eqref{bbb2} holds for all $\delta>0$ they must hold also for $\delta=0$. Hence,
\be
\lim_{n\to\infty}\frac1n\log(\ell_n)=H(\r)\;,
\ee
so that~\eqref{limitde} gives
\be
\lim_{n\to\infty}\frac1nD^\eps_{\min}\left(\p^{\otimes n}\big\|\q^{\otimes n}\right)=\log(k)-H(\r)=D\left(\r\|\u^{(k)}\right)=D(\p\|\q)\;.
\ee
This completes the proof for the case that $\q$ has rational components. For the general case, let $\{\s_k\},\{\r_k\}\in\prob(m)\cap\mbb{Q}^m$ be two sequences of probability vectors with rational components\index{rational components} such that both $\s_k\to\q$
and $\r_k\to\q$, and in addition
\be
(\p,\s_k)\succ(\p,\q)\succ(\p,\r_k)\;.
\ee
The existence of such sequences follows from Exercise~\ref{1lem1}. Therefore, since $D_{\min}^\eps$ is a divergence we have
\be\label{limitsinfsup}
\frac1nD^\eps_{\min}\left(\p^{\otimes n}\big\|\s^{\otimes n}_k\right)\geq\frac1nD^\eps_{\min}\left(\p^{\otimes n}\big\|\q^{\otimes n}\right)\geq\frac1nD^\eps_{\min}\left(\p^{\otimes n}\big\|\r^{\otimes n}_k\right)\;.
\ee
Taking on all sides  the limits $n\to\infty$ followed by $k\to\infty$ completes the proof.
\end{proof}

\begin{exercise}
Give more details for the last argument involving~\eqref{limitsinfsup} and the limits $n\to\infty$ and $k\to\infty$. For the limit $n\to\infty$, consider two cases involving $\liminf$ and $\limsup$ and then conclude at the end that the limit exists.
\end{exercise}

Next, we provide an alternative more traditional proof. This proof is based on the AEP property.

\begin{proof}[Alternative proof of Theorem~\ref{th:sl}]
Let $\eps>0$ and $\t\in[m]^n$ be a probabilistic hypothesis test satisfying $\alpha_n(\t)\leq\eps$. Observe first that any probabilistic hypothesis test satisfies
\ba
\beta_n(\t)&=\sum_{x^n\in[m]^n}t_{x^n}q_{x^n}\\
&\geq \sum_{x^n\in\mathfrak{R}_{n}^\eps }t_{x^n}q_{x^n}\\
\GG{\eqref{ineqrne}}&\geq 2^{-n(D(\p\|\q)+\eps)}\sum_{x^n\in \mathfrak{R}_n^\eps }t_{x^n}p_{x^n}\;.
\ea
On the other hand, since $\alpha_n(\t)\leq\eps$ we have
\ba
\eps&\geq 1-\sum_{x^n\in[m]^n}t_{x^n}p_{x^n}\\
&=1-\sum_{x^n\in\mathfrak{R}_n^\eps}t_{x^n}p_{x^n}-\sum_{x^n\not\in\mathfrak{R}_n^\eps}t_{x^n}p_{x^n}\\
\Gg{t_{x^n}\leq 1}&\geq 1-\sum_{x^n\in\mathfrak{R}_n^\eps}t_{x^n}p_{x^n}-\sum_{x^n\not\in\mathfrak{R}_n^\eps}p_{x^n}\\
&=1-\sum_{x^n\in\mathfrak{R}_n^\eps}t_{x^n}p_{x^n}-\left(1-\pr(\mr_n^\eps)_\p\right)\;.
\ea
Combining the two equations above gives
\be
\beta_n(\t)\geq 2^{-n(D(\p\|\q)+\eps)}\left(\eps+\pr(\mr_n^\eps)_\p\right)
\ee
This proves that $\limsup_{n\to\infty}D^\eps_{\min}\left(\p^{\otimes n}\big\|\q^{\otimes n}\right)\leq D(\p\|\q)$ since $\pr(\mr_n^\eps)_\p\to 1$ as $n\to\infty$.

For the opposite inequality (i.e. the achievability part of the proof), let $0<\eps'<\eps$, and take $\t\in[m]^n$ to be the vector whose $x^n$-component is one if $x^n\in\mr_n^{\eps'}$ and zero otherwise. For this choice we have $\alpha_n(\t)=1-\pr(\mathfrak{R}_n^{\eps'})_\p\leq\eps'\leq\eps$ for sufficiently large $n$ (see Exercise~\ref{ex:tt}). Hence, with this choice of $\t$ we get
\ba
\liminf_{n\to\infty}D^\eps_{\min}\left(\p^{\otimes n}\big\|\q^{\otimes n}\right)&\geq \liminf_{n\to\infty}-\frac{1}{n}\log\beta_n(\t)\\
&=\liminf_{n\to\infty}-\frac{1}{n}\log\pr(\mathfrak{R}_n^{\eps'})_\q\\
\GG{\eqref{7196b}}&\geq \lim_{n\to\infty}-\frac{1}{n}\log 2^{-n(D(\p\|\q)-\eps')}\\
&=D(\p\|\q)-\eps'\;.
\ea
Since the above inequality holds for all $0<\eps'<\eps$, the proof is completed. 
\end{proof}

\section{Alternative (direct) proofs of Theorem~\ref{themasycost} and Theorem~\ref{thmdistab}}\label{altprooftyp}

\noindent\textbf{Theorem.}
{\it Let $\psi\in\pure(AB)$. Then, for any $\eps\in(0,1)$ 
\be
\cost\left(\psi^{AB}\right)=\lim_{n\to\infty}\frac1n\cost^\eps\left(\psi^{\otimes n}\right)=E\left(\psi^{AB}\right)\;,
\ee
where $E$ is the entropy of entanglement defined in~\eqref{eoe}.}

\begin{proof}
For any $\eps\in(0,1)$ we get from~\eqref{1242} that
\be
\cost^\eps\left(\psi^{\otimes n}\right)=\min_{m\in[d^n]}\left\{\log m\;:\;\|\p^{\otimes n}\|_{(m)}\geq 1-\eps\right\}
\ee
where $d\eqdef\sr(\psi^{AB})$ and $\p\in\prob(d)$ is the Schmidt probability vector of $\psi^{AB}$. The components of $\p^{\otimes n}$ are the probabilities $p_{x^n}\eqdef p_{x_1}\cdots p_{x^n}$, where $x_1,\ldots,x_n\in[d]$. We can therefore think of $x^n$ as a sequence drawn from an i.i.d.$\sim\p$-source. From the second inequality of Theorem~\ref{tts} we know that for any $\delta\in(0,1)$, the number of such sequences that are $\delta$-typical cannot exceed $2^{n(E(\psi^{AB})+\delta)}$. Taking $m\eqdef\left\lceil2^{n(E(\psi^{AB})+\delta)}\right\rceil$ we get that the sum of the $m$ largest components of $\p^{\otimes n}$, i.e.
$
\|\p^{\otimes n}\|_{(m)}
$, is greater than the probability of the set of $\delta$-typical sequences. For sufficiently large $n$ this probability exceed $1-\eps$. We therefore conclude that for sufficiently large $n$
\be
\cost^\eps\left(\psi^{\otimes n}\right)\leq \log\left\lceil2^{n(E(\psi^{AB})+\delta)}\right\rceil\;.
\ee
Dividing by $n$ and taking the limit $n\to\infty$ we get that for all $\delta>0$
\be
\limsup_{n\to\infty}\frac1n\cost^\eps\left(\psi^{\otimes n}\right)\leq E(\psi^{AB})+\delta\;.
\ee
Since the above inequality holds for all $\delta>0$ we must have for all $\eps\in(0,1)$
\be\label{nin1}
\limsup_{n\to\infty}\frac1n\cost^\eps\left(\psi^{\otimes n}\right)\leq E\left(\psi^{AB}\right)\;.
\ee

Conversely, fix $\eps\in(0,1)$ and for each $n\in\mbb{N}$, let $m_n\in[d^n]$ be the smallest integer satisfying $\|\p^{\otimes n}\|_{(m_n)}\geq 1-\eps$. Denote by $\mathbf{S}_n\subset[d]^n$ the set of $m_n$ sequences $x^n$ with the highest probabilities $p_{x^n}$. By definition, we have $\cost^\eps\left(\psi^{\otimes n}\right)=\log(m_n)$, $\|\p^{\otimes n}\|_{(m_n)}=\pr(\mathbf{S}_n)$, and $|\mathbf{S}_n|=m_n$.
Suppose now, by contradiction, that there exists $r<E\left(\psi^{AB}\right)$ such that
\be
\liminf_{n\to\infty}\frac1n\cost^\eps\left(\psi^{\otimes n}\right)\leq r\;.
\ee
This means that for any $k\in\mbb{N}$ there exists $n\geq k$ such that $|\mathbf{S}_n|=m_n\leq 2^{r n}$. However, from the third part of Theorem~\ref{tts} (particularly Exercise~\ref{vtts3}) it follows that for any $\delta>0$ there exists $n$ sufficiently large such that $\pr(\mathbf{S}_n)\leq \delta$. Taking $0<\delta<1-\eps$ we get a contradiction since by definition $\pr(\mathbf{S}_n)=\|\p^{\otimes n}\|_{(m_n)}\geq 1-\eps$. We therefore conclude that
\be\label{nin2}
\liminf_{n\to\infty}\frac1n\cost^\eps\left(\psi^{\otimes n}\right)\geq E\left(\psi^{AB}\right)\;.
\ee
The two inequalities in~\eqref{nin1} and~\eqref{nin2} gives
\be
\lim_{n\to\infty}\frac1n\cost^\eps\left(\psi^{\otimes n}\right)=E\left(\psi^{AB}\right)\;.
\ee
This concludes the proof.
\end{proof}

\noindent\textbf{Theorem.} {\it 
Let $\psi\in\pure(AB)$. Then, for any $\eps\in(0,1)$ 
\be
\distill\left(\psi^{AB}\right)=\lim_{n\to\infty}\frac1n\distill^\eps\left(\psi^{\otimes n}\right)=E\left(\psi^{AB}\right)\;,
\ee
where $E$ is the entropy of entanglement defined in~\eqref{eoe}.
}

\begin{proof}
Recall that for any $\eps\in(0,1)$ we get from~\eqref{1236} that
\be\label{firfor}
\distill^\eps\left(\psi^{\otimes n}\right)=\log\min_{k\in\{\ell_n,\ldots,d^n\}}\left\lfloor\frac{k}{\left\|\p^{\otimes n}\right\|_{(k)}-\eps}\right\rfloor
\ee
where $d\eqdef\sr(\psi^{AB})$, $\p\in\prob(d)$ is the Schmidt probability vector of $\psi^{AB}$, and $\ell_n$ is the integer satisfying $\left\|\p^{\otimes n}\right\|_{(\ell_n)}>\eps$ and $\left\|\p^{\otimes n}\right\|_{(\ell_n-1)}\leq\eps$. Let $\delta\in(0,1)$ and recall from the proof of the previous theorem that for  $k=k_n\eqdef\left\lceil2^{n(E(\psi^{AB})+\delta)}\right\rceil$ we get that the sum of the $k_n$ largest components of $\p^{\otimes n}$, i.e.
$
\|\p^{\otimes n}\|_{(k_n)}
$, is greater than the probability of the set of $\delta$-typical sequences. In the limit $n\to\infty$ this probability goes to one and in particular exceeds $\eps$ so that $\ell_n\leq k_n$ for sufficiently large $n$. We therefore get from~\eqref{firfor} that
\be
\distill^\eps\left(\psi^{\otimes n}\right)\leq\log\left\lfloor\frac{k_n}{\left\|\p^{\otimes n}\right\|_{(k_n)}-\eps}\right\rfloor
\ee
Hence,
\ba
\limsup_{n\to\infty}\frac1n\distill^\eps\left(\psi^{\otimes n}\right)&\leq\limsup_{n\to\infty}\frac1n\log\left\lfloor\frac{k_n}{\left\|\p^{\otimes n}\right\|_{(k_n)}-\eps}\right\rfloor\\
\Gg{\left\|\p^{\otimes n}\right\|_{(k_n)}\xrightarrow{n\to\infty}1\;}&=\limsup_{n\to\infty}\frac1n\log\left\lfloor\frac{k_n}{1-\eps}\right\rfloor\\
&=E\left(\psi^{AB}\right)+\delta\;.
\ea
Since the above inequality holds for all $\delta\in(0,1)$ we must have
\be\label{liin}
\limsup_{n\to\infty}\frac1n\distill^\eps\left(\psi^{\otimes n}\right)\leq E\left(\psi^{AB}\right)\;.
\ee

Conversely, suppose by contradiction that there exists $r<\distill\left(\psi^{AB}\right)$ such that
\be
\liminf_{n\to\infty}\frac1n\distill^\eps\left(\psi^{\otimes n}\right)\leq r\;.
\ee
For each $n\in\mbb{N}$ let $k_n$ be such that 
\be
\distill^\eps\left(\psi^{\otimes n}\right)=\log\left\lfloor\frac{k_n}{\left\|\p^{\otimes n}\right\|_{(k_n)}-\eps}\right\rfloor
\ee
Then, from the two equations above we get that for any $a\in\mbb{N}$ there exists  $n\geq a$ such that
\be
\left\lfloor\frac{k_n}{\left\|\p^{\otimes n}\right\|_{(k_n)}-\eps}\right\rfloor\leq 2^{rn}
\ee
Since $\left\|\p^{\otimes n}\right\|_{(k_n)}\leq 1$ it follows that also
\be
\frac{k_n}{1-\eps}-1\leq\left\lfloor\frac{k_n}{1-\eps}\right\rfloor\leq 2^{rn}
\ee
so that
\be
k_n\leq 2^{n\left(r+\frac1n\log(1-\eps)\right)}+1-\eps\;.
\ee
Hence, for sufficiently large $n$ we have $k_n\leq 2^{nr'}$ for some $r'<\distill\left(\psi^{AB}\right)$. 
For each $n\in\mbb{N}$ denote by $\mathbf{S}_n\subset[d]^n$ the set of $k_n$ sequences $x^n$ with the highest probabilities $p_{x^n}$. To summarize, we got that for any $a\in\mbb{N}$ there exists $n\geq a$ such that  $\|\p^{\otimes n}\|_{(k_n)}=\pr(\mathbf{S}_n)$, and $|\mathbf{S}_n|=k_n\leq 2^{nr'}$. However, from the third part of Theorem~\ref{tts} (particularly Exercise~\ref{vtts3}) it follows that there exists $n$ sufficiently large such that $\pr(\mathbf{S}_n)\leq \eps$ in contradiction with the assumption that $\|\p^{\otimes n}\|_{(k_n)}>\eps$. Hence, we must have
\be
\liminf_{n\to\infty}\frac1n\distill^\eps\left(\psi^{\otimes n}\right)\geq \distill\left(\psi^{AB}\right)\;.
\ee
The proof is concluded by comparing the above inequality with~\eqref{liin}.
\end{proof}

\section{Beyond States that are $\G$-Regular}\label{app:regular}

In Theorem~\ref{connected}, to demonstrate that $\G$-equivalent states meet the criteria of \eqref{mconm}, it was essential to posit that both $\psi$ and $\phi$ are $\G$-regular. Even in the absence of this assumption, we can still discern a relationship between the characteristic functions $\chi_\psi(g)$ and $\chi_\phi(g)$. To explore this, we introduce a subset, denoted as $\bS\subset\G$, and given by: 
\be\label{subs}
\bS\eqdef\big\{g\in\G\;:\;\chi_\psi(g)\neq 0\;\text{ or }\;\chi_\phi(g)\neq 0\big\}\;.
\ee
Note that the identity element belongs to $\bS$ and since $\chi_\psi(g^{-1})=\overline{\chi_\psi(g)}$ we get that if $g\in\bS$ also $g^{-1}\in\bS$. Let $\bH$ be the group generated by $\bS$; that is,
\be\label{subh}
\bH\eqdef\la\bS\ra\eqdef\big\{g_1\cdots g_n\;:\;g_1,\ldots,g_n\in\bS,\;\;n\in\mbb{N}\big\}\;.
\ee
With these definitions we get that if $\psi$ and $\phi$ are $\G$-equivalent, then~\eqref{mconm} still holds if we replace $\G$ by $\bH$.

\begin{myt}{}
\begin{theorem}
Let $\G$ be a finite or compact Lie group, $\psi,\phi\in\pure(A)$, and $\bH$ be the subgroup of $\G$ as defined in~\eqref{subh} and~\eqref{subs}. If $\psi$ and $\phi$ are $\G$-equivalent then there exists a 1-dimensional unitary representation of $\bH$, $\{e^{i\theta_h}\}_{h\in \bH}$, such that 
\be\label{mconm2}
\la\psi|U_h|\psi\ra=e^{i\theta_h}\la\phi|U_h|\phi\ra\quad\quad\forall\;h\in\bH\;.
\ee
\end{theorem}
\end{myt}
\begin{remark}
Note that if it is possible to extend the one dimensional representation $\{e^{i\theta_h}\}_{h\in \bH}$ from $\bH$ to $\G$ then in such cases~\eqref{mconm2} holds for all $h\in\G$ (recall that $\la\psi|U_g|\psi\ra=\la\phi|U_g|\phi\ra=0$ for $g\not\in\bH$). Therefore, in such cases we get again the equivalence of Theorem~\ref{connected}. However, such extensions of $\{e^{i\theta_h}\}_{h\in \bH}$ are not always exist (see Exercise~\ref{notalways}).
\end{remark}
\begin{proof}
Following the same steps in the proof of Theorem~\ref{connected} that led to~\eqref{abso}, it follows that $|\chi_{\varphi_1}(h)|=|\chi_{\varphi_2}(h)|=1$ for all $h\in\bS$ and in particular
\be
\la\varphi_1|U_h|\varphi_1\ra=e^{i\theta_h}\;,
\ee
for some phases $\theta_h\in[0,2\pi)$. Note that the equation above implies that for all $h\in\bS$
\be
U_g|\varphi_1\ra=e^{i\theta_h}|\varphi_1\ra\;.
\ee
Furthermore, if $g,h\in \bS$ are such that $gh\in \bS$ then 
\be
e^{i\theta_{gh}}|\varphi_1\ra=U_{gh}|\varphi_1\ra=U_gU_h|\varphi_1\ra=e^{i\left(\theta_g+\theta_h\right)}|\varphi_1\ra
\ee
and we get
\be
\theta_{gh}=\theta_g+\theta_h\mod 2\pi\;.
\ee
Therefore, the set $\{e^{i\theta_h}\;:\;h\in \bS\}$ can be completed to a 1-dimensional unitary representation of $\bH\eqdef\la\bS\ra$. Indeed, for any element $h\in\bH$ that is not in $\bS$, there exists $n\in\mbb{N}$ and $g_1,\ldots,g_n\in \bS$ such that $h=g_1\cdots g_n$. For such $h$ we define
\be
\theta_{h}\eqdef\theta_{g_1}+\cdots+\theta_{g_n}\mod 2\pi\;.
\ee
Note that $\theta_h$ above is well defined since if we also have $h=k_1\cdots k_m$ for some $k_1,\ldots,k_m\in\bS$ then 
\begin{align}
& U_h|\varphi_1\ra=U_{g_1}\cdots U_{g_n}|\varphi_1\ra=e^{i\sum_{x\in[n]}\theta_{g_x}}|\varphi_1\ra\quad\text{and}\\
&U_h|\varphi_1\ra=U_{k_1}\cdots U_{k_m}|\varphi_1\ra=e^{i\sum_{y\in[m]}\theta_{k_y}}|\varphi_1\ra\;,
\end{align}
so that we must have $\sum_{y\in[m]}\theta_{k_y}=\sum_{x\in[n]}\theta_{g_x}$ mod $2\pi$.
We therefore conclude that $h\mapsto e^{i\theta_h}$ is a 1-dimensional representation of the subgroup $\bH$ of $\G$. The proof is concluded with the observation that for any $g\in\bH$ that is not in $\bS$ we have  by the definition of $\bS$ that $\chi_\psi(g)=\chi_\phi(g)=0$. Therefore, in this case~\eqref{mconm2} holds trivially.
\end{proof}

\bex\label{notalways}
Consider the group $SU(2)$ and consider its 2-element subgroup $\bH=\{I_2,-I_2\}$. Show that the 1-dimensional unitary representation that takes $I_2$ to $1$ and $-I_2$ to $-1$ is incompatible with a 1-dimensional representation on $SU(2)$. In other words, show that there is no unitary representation of $SU(2)$ that maps $I_2$ to $1$ and $-I_2$ to $-1$.  
\eex

\section{Proof of Theorem~\ref{thm:quasi}}\label{app:proof}

\begin{figure}[h]\centering    \includegraphics[width=0.8\textwidth]{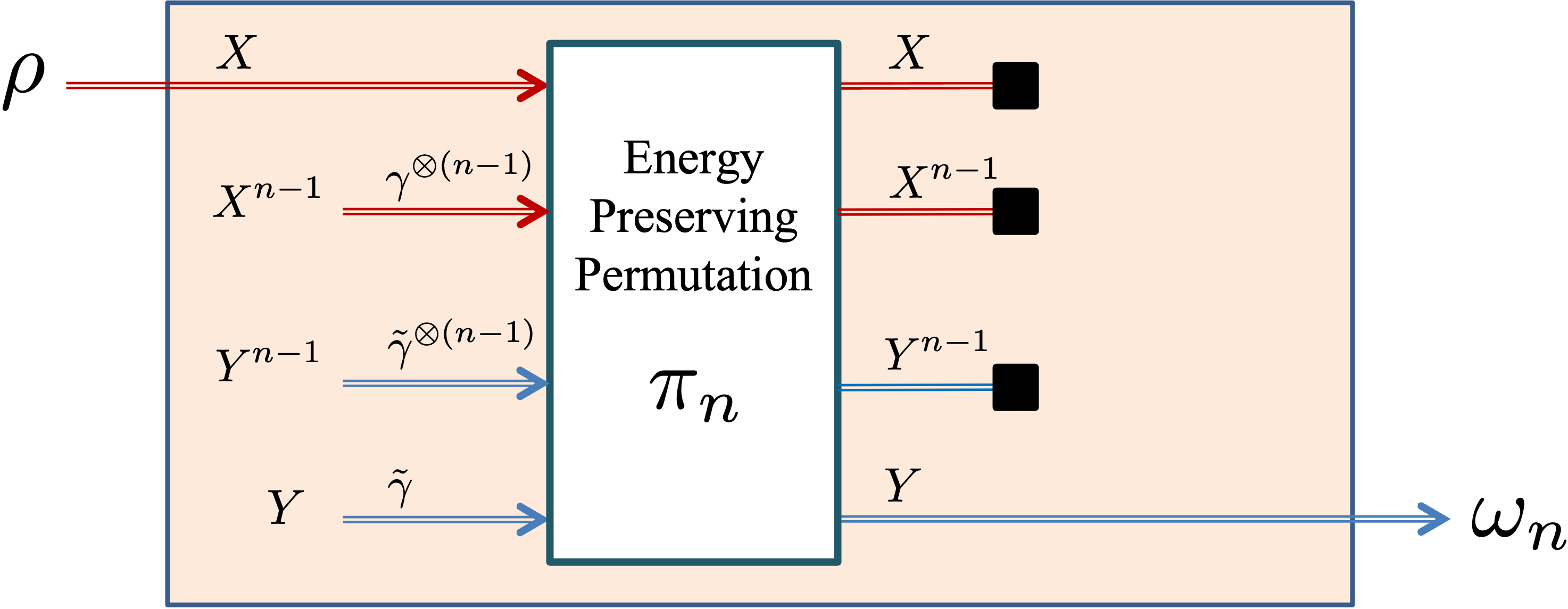}
  \caption{\linespread{1}\selectfont{\small Thermal operations that generate the state $\omega_n$.}}
  \label{thermalpermutation}
\end{figure} 

\begin{myt}{}
\textbf{Theorem.}
Let $(\rho,\gamma)$ and $(\sigma,\tgamma)$ be two quasi-classical states of systems $A$ and $A'$, respectively. The following statements are equivalent:
\begin{enumerate}
\item $(\rho,\gamma)$ can be converted to $(\sigma,\tgamma)$ by CTO.
\item $(\rho,\gamma)$ can be converted to $(\sigma,\tgamma)$ by GPO.
\end{enumerate}
\end{myt}

\begin{proof}
Let $\p,\q,\g,\tg$ be the probability vectors whose components are the diagonals of $\rho,\sigma,\gamma,\tgamma$, respectively. From the second statement of the theorem and~\eqref{gp17} we have that $(\p,\g)\succ(\q,\tg)$. To show that $(\p,\g)\xrightarrow{\cto}(\q,\tg)$ we will construct a sequence of thermal operations such that their limit maps $(\p,\g)$ to $(\q,\tg)$. For convenience, we will think of $X\eqdef A$ and $Y\eqdef A'$ as two classical systems, and consider the Gibbs state of system $X^nY^n$.
This Gibbs state can be written as
\be
\gamma^{\otimes n}\otimes\tgamma^{\otimes n}=\sum_{\substack{x^{n}\in[m]^{n}\\y^{n}\in[k]^{n}}}g_{x^{n}}\tilde{g}_{y^{n}}|x^{n}y^n\lr x^{n}y^n|\;.
\ee
We also consider the initial state 
\be\label{17p76}
\rho\otimes\gamma^{\otimes (n-1)}\otimes\tgamma^{\otimes n}=\sum_{\substack{x^{n}\in[m]^{n}\\y^{n}\in[k]^{n}}}\frac{p_{x_1}}{g_{x_1}}g_{x^n}\tilde{g}_{y^{n}}|x^{n}y^n\lr x^{n}y^n|\;.
\ee
Our goal is to construct an energy preserving unitary (in fact a permutation) $\mU$ such that the state
\be
\omega_n\eqdef\tr_{X^nY^{n-1}}\left[\mU\left(\rho\otimes\gamma^{\otimes (n-1)}\otimes\tgamma^{\otimes (n-1)}\otimes\tgamma\right)\right]
\ee
goes to $\sigma$ as $n$ goes to infinity; see Fig.~\ref{thermalpermutation}. We take three steps towards that goal:\\

\noindent\textbf{Step 1: Projection to a Typical Subspace}\\

We project the state in~\eqref{17p76} to the strongly typical subspace. Specifically, let $\mt_n(X)$ and $\mt_n(Y)$ be the sets of all $\eps$-strongly-typical sequences with $\eps=\frac1{n^{1/3}}$; i.e.,
\ba
&\mt_{n}(X)\eqdef\left\{x^{n}\in[m]^n\;:\;\left\|\t(x^{n})-\g\right\|_\infty\leq n^{-1/3}\right\}\\
&\mt_{n}(Y)\eqdef\left\{y^{n}\in[k]^n\;:\;\left\|\t(y^{n})-\tg\right\|_\infty\leq n^{-1/3}\right\}\;.
\ea
Then, the projection of the initial state in~\ref{17p76} to the corresponding typical subspace is given by the sub-normalized state
\be\label{1779}
\eta^{X^nY^n}\eqdef\sum_{\substack{x^{n}\in\mt_n(X)\\ y^n\in\mt_n(Y)}}\frac{p_{x_1}}{g_{x_1}}g_{x^n}\tilde{g}_{y^{n}}|x^{n}y^n\lr x^{n}y^n|\;.
\ee
From the theorem of strongly typical sequences, the state above has the following property.
\begin{myg}{}
\begin{lemma}
\be\label{lemeta}
\frac12\left\|\eta^{X^nY^n}-\rho\otimes\gamma^{\otimes (n-1)}\otimes\tgamma^{\otimes n}\right\|_1\xrightarrow{n\to\infty}0\;.
\ee
\end{lemma}
\end{myg}
\begin{proof}
First, observe that $\frac1{g_{x_1}}g_{x^n}=g_{x^{n-1}}$ with $x^{n-1}\eqdef(x_2,...,x_{n})$ so that
\ba
\tr\left[\eta^{X^nY^n}\right]&\eqdef\sum_{x^{n}\in\mt_n(X)} p_{x_1}g_{x^{n-1}}\sum_{y^n\in\mt_n(Y)}\tilde{g}_{y^{n}}\\
\GG{\text{Hoeffding's}\; inequality~\eqref{2m2ne}}&\geq(1-e^{-2n^{1/3}})
\sum_{x^{n}\in\mt_n(X)} p_{x_1}g_{x^{n-1}}
\ea
Moreover, denoting by $\eps_n\eqdef n^{-1/3}$ and by $\eps_n'\eqdef \eps_n-\frac1{n}$, we get from~\eqref{gili9} that
\be
\sum_{x^{n}\in\mt_n(X)} p_xg_{x^n}\geq1-e^{-2(n-1)\eps_n^{\prime 2}}\;\;
\xrightarrow{n\to\infty}1\;,
\ee
where the limit follows from the fact that for very large $n$ we have $(n-1)\eps_n'^2\approx n^{1/3}$. 

The two equations above implies that $\lim_{n\to\infty}\tr\left[\eta^{X^nY^n}\right]=1$, and therefore, from the gentle operator lemma (see Lemma~\ref{gentle2}) it follows that~\eqref{lemeta} holds. This completes the proof of the Lemma.
\end{proof}

Setting 
\be
\omega_n^{\prime Y}\eqdef\tr_{X^nY^{n-1}}\left[\mU\left(\eta^{X^nY^n}\right)\right]
\ee
we get that
\be
\lim_{n\to\infty}\frac12\left\|\omega_n^{\prime Y}-\omega_n^Y\right\|_1=0\;.
\ee
It is therefore sufficient to show that there exists a permutation channel $\mU\in\cptp(X^nY^n\to X^nY^n)$ such that $\lim_{n\to\infty}\omega_n'=\sigma$.\\

\noindent\textbf{Step 2: Construction of the Unitary $\mU^{X^nY^n\to X^nY^n}$}\\
 
For any $\s\in\type(n,m)$ and $\t\in\type(n,k)$ we denote by $\lp x,\s,\t,\cdot\rp$ the set of all sequences $(x,x^{n-1},y^n)$ with the same $x$, same type $\s$ of $x^{n}$, and the same type $\t$ of $y^{n}$. Similarly, $\lp\cdot,\s,\t,y\rp$ is used to denote the set of all components $(x^n,y^{n-1},y)$ with the same $y$, same type $\s$ of $x^{n}$, and the same type $\t$ of $y^{n}$. Fix $\s\in\type(n,m)$ and $\t\in\type(n,k)$, and denote the cardinality of these sets  by $a_x\eqdef\left|\lp x,\s,\t,\cdot\rp\right|$ and $b_y\eqdef\left|\lp\cdot,\s,\t,y\rp\right|$. Observe that (see Exercise~\ref{ex1769})
\be\label{1769}
a_x=s_x|x^n(\s)||y^n(\t)|\quad\text{and}\quad b_y=t_y|x^n(\s)||y^n(\t)|\;.
\ee

Now, let $R=(r_{y|x})$ be the column stochastic matrix satisfying $R\p=\q$ and $R\g=\tg$. For any $x\in[m]$ and $y\in[k]$ we define by induction on $x$ 
\be\label{1770}
\ell_{yx}\eqdef\min\left\{b_y-\sum_{x'<x}\ell_{yx'}\;,\;\left\lfloor r_{y|x}a_x\right\rfloor\right\}
\ee
where for $x=1$ we use the convention that $\sum_{x'<1}\ell_{yx'}\eqdef 0$. Observe that 
each integer $\ell_{yx}\geq 0$ (see Exercise~\ref{ex1769}) and the sum
\be
\sum_{y\in[k]}\ell_{yx}\leq \sum_{y\in[k]}\left\lfloor r_{y|x}a_x\right\rfloor\leq \sum_{y\in[k]}r_{y|x}a_x=a_x\;.
\ee
Therefore, for each $x\in[m]$ there exists $k$ disjoint sets $\{\ml_{yx}\}_{y\in[k]}$ with each $\ml_{yx}\subset\lp x,\s,\t,\cdot\rp$ and with $|\ml_{yx}|=\ell_{yx}$. Now, observe also that from their definition, the integers $\{\ell_{yx}\}$  satisfy
\be
\sum_{x\in[m]}\ell_{yx}\leq b_y\;.
\ee
Therefore, there exists an injective map 
\be
f:\bigcup_{x\in[m],y\in[k]}\ml_{yx}\to\lp \cdot,\s,\t,\cdot\rp
\ee
with the property that
\be
f\left(\ml_{yx}\right)\subset\lp \cdot,\s,\t,y\rp\;.
\ee
We can extend $f$ to a bijection\index{bijection} 
\be
f:\lp \cdot,\s,\t,\cdot\rp\to\lp \cdot,\s,\t,\cdot\rp
\ee
Observe that the bijection\index{bijection} $f=f_{\s,\t}$ as defined above can be defined for any $\s\in\type(n,m)$ and $\t\in\type(n,k)$. Therefore, the set of bijections $\{f_{\s,\t}\}_{\s,\t}$ can be used to define the bijection\index{bijection} map
\be
\pi_n\left(x^{n}y^{n}\right)\eqdef f_{\s,\t}\left(x^{n}y^{n}\right)
\ee
where $\t$ is the type of $x^{n}$ and $\s$ is the type of $y^{n}$. Observe that $\pi_n$ is a thermal operation since it does not change $\s$ and $\t$ (hence preserves the energy). We define the unitary (permutation) channel $\mU\in\cptp(X^nY^n\to X^nY^n)$ as
\be
\mU\left(|x^ny^n\lr x^ny^n|\right)\eqdef|\pi_n(x^{n}y^{n})\lr\pi_n(x^{n}y^{n})|\quad\quad\forall\;x^n\in[m]^n,\;\;\forall\;y^n\in[k]^n\;.
\ee
We are now ready to estimate $\omega_n'$ with the above choice of $\mU$.\\

\noindent\textbf{Step 3: Estimation of $\omega_n'$}\\

Observe that the $(x^{n},y^{n})$ eigenvalue of $\rho\otimes\gamma^{\otimes (n-1)}\otimes\tgamma^{\otimes n}$ is given by
\be
\frac{p_x}{g_x}g_{x^{n}}\tilde{g}_{y^{n}}=\frac{p_x}{g_x}2^{-n\big(H(\s)+H(\t)+D\left(\s\|\g\right)+D\left(\t\|\tg\right)\big)}
\ee 
where $x_1\eqdef x$ and we used~\eqref{tpxn} twice with $\t$ and $\s$ being the types of $x^{n}$ and $y^{n}$, respectively. Denoting by $c_{\s,\t}\eqdef 2^{-n\left(H(\s)+H(\t)+D\left(\s\|\g\right)+D\left(\t\|\tg\right)\right)}$ we get from the definition of $\eta^{X^nY^n}$ in~\eqref{1779} that $\omega_n'$ can be expressed as 
\be\label{17p88}
\omega_n'=\sum_{(\s,\t)\in\mc_n}c_{\s,\t}\sum_{x\in[m]}\frac{p_x}{g_x}\sum_{(x^{n},y^{n})\in\langle x,\s,\t,\cdot\rangle}\tr_{X^nY^{n-1}}\left[|\pi_n(x^{n}y^{n})\lr\pi_n(x^{n}y^{n})|\right]\;.
\ee
where (recall $\eps_n\eqdef n^{-1/3}$)
\be
\mc_n\eqdef\big\{(\s,\t)\;:\;\s\in\type(n,m),\;\t\in\type(n,k),\;\s\approx_{\eps_n}\g,\;\t\approx_{\eps_n}\tg\big\}\;.
\ee
Next, instead of summing over all the elements of $\lp x,\s,\t,\cdot\rp$ (of the third sum above), we will restrict the summation only to sequences $(x^n,y^n)$ that belong to the subset $\bigcup_{y\in[k]}\ml_{yx} \subset\lp x,\s,\t,\cdot\rp$ and we will show that the remaining terms are negligible (i.e. goes to zero as $n$ goes to infinity). That is, we define
\be\label{17p88}
\omega_n''=\sum_{(\s,\t)\in\mc_n}\sum_{x\in[m]}\sum_{y\in[k]}\sum_{(x^{n},y^{n})\in\ml_{yx}^{\s,\t}}c_{\s,\t}\frac{p_x}{g_x}\tr_{X^nY^{n-1}}\left[|\pi_n(x^{n}y^{n})\lr\pi_n(x^{n}y^{n})|\right]\;,
\ee
where we added $\s,\t$ superscript to $\ml_{yx}$ since it depends on the types $\s$ and $\t$.
We show now that $\omega_n''\to\sigma$ as $n\to\infty$, and since $\tr[\sigma]=1$  we must have $\|\omega_n'-\omega_n''\|_1\to 0$ as $n\to\infty$.
Now, observe that for $(x^{n},y^{n})\in\ml_{yx}^{\s,\t}$ we have that $\pi_n(x^{n}y^{n})\in\lp \cdot,\s,\t,y\rp$, so that the last component of the sequence $(x^n,y^n)$ is $y_n=y$. Therefore, we get that
\ba
\omega_n''&=\sum_{(\s,\t)\in\mc_n}\sum_{x\in[m]}\sum_{y\in[k]}\sum_{(x^{n},y^{n})\in\ml_{yx}^{\s,\t}}c_{\s,\t}\frac{p_x}{g_x}|y\lr y|^Y\\
&=\sum_{(\s,\t)\in\mc_n}\sum_{x\in[m]}\sum_{y\in[k]}c_{\s,\t}\frac{p_x}{g_x}\ell_{yx}^{\s,\t}|y\lr y|^Y\;.
\ea
Observe that we added explicitly the dependance of $\ell_{yx}$ on $\s$ and $\t$.
\begin{myg}{}
\begin{lemma}
Let $\{\t_n,\s_n\}_{n\in\mbb{N}}$ be a sequence  of pair of types such that $(\t_n,\s_n)\in\mc_n$. Denote by $\ell_{xy}^{(n)}$ the coefficients~\eqref{1770} that corresponds to the pair of types $(\s_n,\t_n)$. Then,
\be\label{ryx}
r_{y|x}=\lim_{n\to\infty}\frac{\ell_{yx}^{(n)}}{a_x^{(n)}}\;.
\ee
\end{lemma}
\end{myg}
\begin{proof}
By definition, $\lim_{n\to\infty}\s_n=\g$ and $\lim_{n\to\infty}\t_n=\tg$ since $(\s_n,\t_n)\in\mc_n$. Denote by $\{s_x^{(n)}\}_{x\in[m]}$ and $\{t_y^{(n)}\}_{y\in[k]}$ the components of $\s_n$ and $\t_n$, respectively. Then, for $x=1$ we get
\be
\lim_{n\to\infty}\frac{\ell_{y1}^{(n)}}{a_1^{(n)}}=\lim_{n\to\infty}\min\left\{\frac{t_y^{(n)}}{s_1^{(n)}},\frac{\left\lfloor r_{y|1}a_1^{(n)}\right\rfloor}{a_1^{(n)}}\right\}=\min\left\{\frac{\tilde{g}_y}{g_1},r_{y|1}\right\}=r_{y|1}\;,
\ee
where the last equality follows from the fact that $\tg=R\g$ so that
\be
\tilde{g}_y=\sum_{x\in[m]}r_{y|x}g_x\geq r_{y|1}g_1\;.
\ee
Now, fix $x\in[m]$ and suppose that the limit~\eqref{ryx} (with $x$ being replaced by $x'$) holds for all $x'<x$. We need to show that it also holds for $x$. Indeed, 
\ba\label{lxya}
\lim_{n\to\infty}\frac{\ell_{yx}^{(n)}}{a_{x}^{(n)}}&=\lim_{n\to\infty}\min\left\{\frac{t_y^{(n)}}{s_{x}^{(n)}}-\frac{\sum_{x'<x}\ell_{yx'}^{{(n)}}}{a_{x}^{(n)}}\;\;,\;\;\frac{\left\lfloor r_{y|x}a_{x}^{(n)}\right\rfloor}{a_{x}^{(n)}}\right\}\\
&=\min\left\{\frac{\tilde{g}_y}{g_{x}}-\sum_{x'<x}\lim_{n\to\infty}\frac{\ell_{yx'}^{{(n)}}}{a_{x}^{(n)}}\;\;,\;\;r_{y|x}\right\}\;.
\ea
Since we assume by induction that~\eqref{ryx} holds if we replace $x$ with $x'<x$ we get
\ba
\lim_{n\to\infty}\frac{\ell_{yx'}^{{(n)}}}{a_{x}^{(n)}}&=\lim_{n\to\infty}\frac{a_{x'}^{(n)}}{a_{x}^{(n)}}\frac{\ell_{yx'}^{{(n)}}}{a_{x'}^{(n)}}\\
&=\lim_{n\to\infty}\frac{s_{x'}^{(n)}}{s_{x}^{(n)}}\frac{\ell_{yx'}^{\s,\t}}{a_{x'}^{(n)}}\\
&=\frac{g_{x'}}{g_x}\lim_{n\to\infty}\frac{\ell_{yx'}^{{(n)}}}{a_{x'}^{(n)}}\\
\GG{By\;induction}&=\frac{g_{x'}}{g_x}r_{y|x'}
\ea
Substituting this into~\eqref{lxya} we get that
\be
\lim_{n\to\infty}\frac{\ell_{yx}^{(n)}}{a_{x}^{(n)}}=\min\left\{\frac{\tilde{g}_y}{g_{x}}-\frac1{g_x}\sum_{x'<x}r_{y|x'}g_{x'}\;\;,\;\;r_{y|x}\right\}=r_{y|x}
\ee
where the last equality follows from
\be
\tilde{g}_y=\sum_{w\in[m]}r_{y|w}g_w\geq \sum_{x'<x}r_{y|x'}g_{x'}+r_{y|x}g_x\;.
\ee
This completes the proof of the limit in~\eqref{ryx}.
\end{proof}

Finally, we compute the limit of $\omega_n''$ as $n$ goes to infinity. For this purpose, we first observe that $h_{\s,\t}\eqdef c_{\s,\t}|x^n(\s)||y^n(\t)|$ forms a probability distribution over all pair of types $(\s,\t)\in\type(n,m)\times\type(n,k)$ with the property that
$
\sum_{(\s,\t)\in\mc_n} h_{\s,\t}
$ is the probability that a given pair of sequences $(x^n,y^n)$ is $\eps_n$-typical. Therefore, from the theorem of strong-typicality this probability goes to one in the limit $n\to\infty$. With this in mind, observe that
\ba
\lim_{n\to\infty}\omega_n''&=\lim_{n\to\infty}\sum_{(\s,\t)\in\mc_n}h_{\s,\t}\sum_{y\in[k]}\sum_{x\in[m]}p_x\frac{\ell_{yx}^{\s,\t}}{a_x^{\s,\t}}|y\lr y|^Y\\
\GG{\eqref{ryx}}&=\lim_{n\to\infty}\sum_{(\s,\t)\in\mc_n}h_{\s,\t}\sum_{y\in[k]}\sum_{x\in[m]}p_xr_{y|x}|y\lr y|^Y\\
\Gg{\lim_{n\to\infty}\sum_{(\s,\t)\in\mc_n}h_{\s,\t}=1}&=\sum_{y\in[k]}\Big(\sum_{x\in[m]}r_{y|x}p_x\Big)|y\lr y|^Y\\
\Gg{\q=R\p}&=\sum_{y\in[k]}q_y|y\lr y|^Y=\sigma^Y\;.
\ea
This completes the proof.
\end{proof}

\bex\label{ex1769}$\;$
\ben
\item Prove the relations in~\eqref{1769}.
\item Show that the coefficients $\ell_{xy}$ as defined in~\eqref{1770} are non-negative.
\een
\eex

\section{Continuity}\label{continuity}

In Section~\ref{sec:cdqc}, we calculated the conversion distance between two athermality states within the quasi-classical regime. Notably, in Theorem~\ref{qclaq}, we postulated that the Gibbs states $\g$ and $\q'$ possess positive rational components. In this section, we demonstrate that the conversion distance is continuous. Consequently, one can utilize Theorem~\ref{qclaq} to approximate the conversion distance with arbitrary precision, even when $\g$ and $\g'$ have irrational components.

For this purpose, we fix two probability vectors $\p^A\in\prob(m)$ and $\q^B\in\prob(n)$ and define for all $\g^A\in\prob(m)$ and $\g^B\in\prob(m)$ the function:
\be
f\left(\g^A,\g^B\right)\eqdef T\left((\p^A,\g^A)\xrightarrow{\mf}(\q^B,\g^B)\right)\;.
\ee
Moreover, fix $\g^A,\g'^A\in\prob(m)$ and $\g^B,\g'^B\in\prob(n)$, and denote by $\delta\eqdef\frac12\left\|\g^A-\g'^A\right\|_1$ and $\eps\eqdef\frac12\left\|\g^B-\g'^B\right\|_1$. Furthermore, let $g_{\min}^B$ and $g_{\min}^{\prime B}$ be the smallest components of $\g^B$ and $\g'^B$, respectively. With these notations we prove the following continuity\index{continuity} lemma.

\begin{myg}{}
\begin{lemma}\label{lem:gbmin}
Using the same notations as above,
\be\label{gbmin1}
\Big|f\left(\g^A,\g'^B\right)-f\left(\g^A,\g^B\right)\Big|\leq\frac{\eps}{\min\{g_{\min}^B,g_{\min}^{\prime B}\}}\;,
\ee
and if $0<\delta<\frac12g_{\min}^B$ then we also have
\be\label{gbmin2}
\Big|f\left(\g'^A,\g^B\right)-f\left(\g^A,\g^B\right)\Big|\leq\frac{2\delta}{g_{\min}^B}\;.
\ee
\end{lemma}
\end{myg}

\begin{proof}
Let $\q'^B$ be optimal such that
\be\label{pgqg2}
f\left(\g^A,\g^B\right)=\frac12\left\|\q^B-\q'^B\right\|_1\quad\text{and}\quad(\p^A,\g^A)\succ(\q'^B,\g^B)\;.
\ee 
From Lemma~\ref{4lem40} there exists $\q''^B$ such that 
\be
\left(\q'^B,\g^B\right)\succ\left(\q''^B,\g'^B\right)\quad\text{and}\quad\frac12\left\|\q'^B-\q''^B\right\|_1\leq\frac\eps{g_{\min}^B}\;.
\ee
Since $\left(\p^A,\g^A\right)\succ\left(\q'^B,\g^B\right)$ we also have
$\left(\p^A,\g^A\right)\succ\left(\q''^B,\g'^B\right)$. Hence,
\ba
f\left(\g^A,\g'^B\right)&\leq\frac12\left\|\q^B-\q''^B\right\|_1\\
\GG{Triangle\;inequality}&\leq\frac12\left\|\q^B-\q'^B\right\|_1+\frac12\left\|\q'^B-\q''^B\right\|_1\\
&\leq f\left(\g^A,\g^B\right)+\frac\eps{g_{\min}^B}\;.
\ea
For the converse inequality, observe that by repeating the exact same lines as above, exchanging everywhere $\g^B$ with $\g'^B$, we get
\be
f\left(\g^A,\g^B\right)\leq f\left(\g^A,\g'^B\right)+\frac\eps{g_{\min}^{\prime B}}
\ee
This completes the proof of the inequality~\eqref{gbmin1}.

For the proof of~\eqref{gbmin2}, as before, let $\q'^B$ be optimal such that~\eqref{pgqg2} holds. We would like to find a vector $\q''^B$ that is close to $\q'^B$ and that satisfies $(\p^A,\g'^A)\succ(\q''^B,\g^B)$. Since $(\p^A,\g^A)\succ(\q'^B,\g^B)$ there exists a column stochastic matrix such that $E\q^A=\q'^B$ and $E\g^A=\g^B$. Denote by $\g'^B=E\g'^A$, and observe that since
 $\g^A$ is $\delta$-close to $\g'^A$, also $\g^B$ is $\delta$-close to $\g'^B$ (DPI under $E$).  Moreover, by definition we have
 $(\p^A,\g'^A)\succ(\q'^B,\g'^B)$. Now,
from Lemma~\ref{4lem40} there exists $\q''^B$ such that 
\be
\left(\q'^B,\g'^B\right)\succ\left(\q''^B,\g^B\right)\quad\text{and}\quad\frac12\left\|\q'^B-\q''^B\right\|_1\leq\frac\delta{g_{\min}^{\prime B}}\;.
\ee
Since $\left(\p^A,\g'^A\right)\succ\left(\q'^B,\g'^B\right)$ we also have
$\left(\p^A,\g'^A\right)\succ\left(\q''^B,\g^B\right)$. Hence,
\ba
f\left(\g'^A,\g^B\right)&\leq\frac12\left\|\q^B-\q''^B\right\|_1\\
\GG{Triangle\;inequality}&\leq\frac12\left\|\q^B-\q'^B\right\|_1+\frac12\left\|\q'^B-\q''^B\right\|_1\\
&\leq f\left(\g^A,\g^B\right)+\frac\delta{g_{\min}^{\prime B}}\\
\Gg{\g'^B\approx_\delta\g^B}&\leq f\left(\g^A,\g^B\right)+\frac\delta{g_{\min}^{B}-\delta}\\
\Gg{\delta\leq\frac12g_{\min}^B}&\leq f\left(\g^A,\g^B\right)+\frac{2\delta}{g_{\min}^{B}}\;.
\ea
The opposite inequality can be obtained using the exact same lines as above by exchanging the roles of $\g^A$ and $\g'^A$. This completes the proof of~\eqref{gbmin2}. 
\end{proof}

\section{Alternative Proof of Blackwell Theorem}\label{AlternativeBP}

The main text presents a proof of the Blackwell theorem using the relation specified in \eqref{onlyr}. Historically, this relation wasn't identified, and alternative proofs for the theorem were formulated using convex analysis techniques. To offer a comprehensive perspective, this appendix section furnishes a more "traditional" (albeit lengthier) proof of the Blackwell theorem. This approach employs convex analysis, echoing the methodologies found in~\cite{Dahl1999}.

First, we aim to demonstrate that the relative majorization relation can be supplanted by an inclusion relationship among convex sets. Specifically, for each $k\in\mbb{N}$, we denote the following set of two-column matrices:
\be\label{inc1}
\mathfrak{M}(\p,\q,k)\eqdef \Big\{[E\p\;E\q]\;:\;E\in\stoc(k,n)\Big\}\subset\mbb{R}^{k\times 2}\;.
\ee
Observe that the set $\mathfrak{M}(\p,\q,k)$ is convex and consists of all two-column matrices $[\r\;\s]$ with probability vectors $\r,\s\in\prob(k)$ for which $(\p,\q)\succ(\s,\r)$.

\begin{myg}{}
\begin{lemma}
Let $n,m\in\mbb{N}$, $\p,\q\in\prob(n)$, and $\p',\q'\in\prob(m)$.
The following statements are equivalent:
\begin{enumerate}
\item $(\p,\q)\succ(\p',\q')$
\item For all $k\in\mbb{N}$ 
\be\label{5p111}
\mathfrak{M}(\p,\q,k)\supseteq\mathfrak{M}(\p',\q',k)\;.
\ee
\end{enumerate}
\end{lemma}
\end{myg}

\begin{proof}
Suppose first that $(\p,\q)\succ(\p',\q')$ and fix $k\in\mbb{N}$. Then, if the two column matrix $[\r'\;\s']\in\mathfrak{M}(\p',\q',k)$ we get by definition that $(\p',\q')\succ(\r',\s')$ so that from the transitivity of relative majorization we get that also $(\p,\q)\succ(\r',\s')$. That is, the two column matrix $[\r'\;\s']\in\mathfrak{M}(\p,\q,k)$. Hence, the inclusion in~\eqref{5p111} must hold.

Conversely, suppose~\eqref{5p111} holds for all $k\in\mbb{N}$. In particular, it holds for $k=m$. In this case we have 
\be
[\p'\;\q']\in\mathfrak{M}(\p',\q',m)\subseteq\mathfrak{M}(\p,\q,m)\;,
\ee
so that there exists $E\in\stoc(m,n)$ such that $[\p\;\q]=[E\p\;E\q]$; i.e. $(\p,\q)\succ(\p',\q')$. 
\end{proof}

We therefore proved the equivalence between relative majorization and the inclusion relations between the sets in~\eqref{5p111}. The significance of this observation is that now we can make use of the fact that inclusion relation between compact sets is related to inequalities between their support functions. Specifically, from Theorem~\ref{inclusion} we know that two compact sets $\mc_1$ and $\mc_2$ satisfy $\mc_1\subseteq\mc_2$ if and only if their support functions $f_{\mc_1}$ and $f_{\mc_2}$ satisfy $f_{\mc_1}\leq f_{\mc_2}$ everywhere in their domain. 
The support function of $\mathfrak{M}(\p,\q,k)$, denoted by $f_{\mathfrak{M}(\p,\q,k)}:\mbb{R}^{k\times 2}\to\mbb{R}$, is given for any two-column matrix $S=[\s_1\; \s_2]\in\mbb{R}^{k\times 2}$ by
\ba\label{supportfu}
f_{\mathfrak{M}(\p,\q,k)}(S)&\eqdef\max\big\{\tr\big(S^TT\big)\;:\;T\in\mathfrak{M}(\p,\q,k)\big\}\\
&=\max\big\{\tr\big(S^T[E\p\;E\q]\big)\;:\;E\in\stoc(k,n)\big\}\\
&=\max_{E\in\stoc(k,n)}\big\{\s_1^TE\p+\s_2^TE\q\big\}\;.
\ea
Therefore, the lemma above in conjunction with Theorem~\ref{inclusion} implies that $(\p,\q)\succ(\p',\q')$ if and only if for all $k\in\mbb{N}$ and all $S\in\mbb{R}^{k\times 2}$
\be\label{fcsup}
f_{\mathfrak{M}(\p,\q,k)}(S)\geq f_{\mathfrak{M}(\p',\q',k)}(S)\;.
\ee
Our next goal is therefore to compute the optimization problem given in~\eqref{supportfu} for the support function.

\begin{myg}{}
\begin{lemma}\label{chararm2}
Let $n,m\in\mbb{N}$, $\p,\q\in\prob(n)$, and $\p',\q'\in\prob(m)$.
For every $x\in[n]$ and $y\in[m]$ denote by $\r_x,\r_y\in\mbb{R}^2$ the vectors $\r_x\eqdef(p_x,q_x)^T$ and $\r_y'\eqdef(p_y',q_y')^T$. The following statements are equivalent:
\begin{enumerate}
\item $(\p,\q)\succ(\p',\q')$
\item For any $k\in\mbb{N}$ and any set of vectors $\v_1,\ldots,\v_k\in\mbb{R}^2_{+}$
\be\label{ineq518}
\sum_{x\in[n]}\max_{z\in[k]}\{\r_x\cdot\v_z\}\geq \sum_{y\in[m]}\max_{z\in[k]}\{\r_y'\cdot\v_z\}\;.
\ee
\end{enumerate}
\end{lemma}
\end{myg}
\begin{remark}
The expressions appearing on the right-hand side and left-hand side of~\eqref{ineq518} are precisely the support functions given in~\eqref{fcsup}. They are known as \emph{sublinear functionals}.
This lemma provides a characterization for relative majorization in terms of sublinear functionals (see subsection~\ref{ss:support}).
\end{remark}

\begin{proof}
Denoting by $\{e_{z|x}\}_{x\in[n],\;z\in[k]}$ the components (conditional probabilities) of $E$ and by $\{s_{z1}\}_{z\in[k]}$ and $\{s_{z2}\}_{z\in[k]}$ the components of $\s_1$ and $\s_2$, we continue from~\eqref{supportfu}
\ba\label{deri1}
f_{\mathfrak{M}(\p,\q,k)}(S)&=\max_{E\in\stoc(k,n)}\sum_{z\in[k]}\sum_{x\in[n]}\big(s_{z1}e_{z|x}p_x+s_{z2}e_{z|x}q_x\big)\\
&=\max_{E\in\stoc(k,n)}\sum_{x\in[n]}\sum_{z\in[k]}e_{z|x}\big(s_{z1}p_x+s_{z2}q_x\big)\\
\GG{Exercise~\ref{lastline}}&=\sum_{x\in[n]}\max_{z}\big(s_{z1}p_x+s_{z2}q_x\big)\;.
\ea
Finally, denoting the rows of $S$ by $\v_{z}\eqdef (s_{z1},s_{z2})^T$ we conclude that 
\be\label{mpqk}
f_{\mathfrak{M}(\p,\q,k)}(S)=\sum_{x\in[n]}\max_{z\in[k]}\{\r_x\cdot\v_z\}\;.
\ee
Combining this with~\eqref{fcsup} completes the proof of the equivalence between $(\p,\q)\succ(\p',\q')$ and~\eqref{ineq518} with arbitrary vectors $\v_1,\ldots,\v_k\in\mbb{R}^2$. It is left to show that we can assume that $\v_1,\ldots,\v_k\in\mbb{R}^2_+$. Indeed, for each $j\in[k]$ let $\v_j'\eqdef\v_j+(r,r)^T\geq 0$ for some sufficiently large $r>0$. In the Exercise~\ref{ex5p3p11} below you show that the inequality in~\eqref{ineq518} holds with $\v_1,\ldots,\v_k$ if and only if it holds with $\v_1',\ldots,\v_k'$. Hence, without loss of generality we can assume that all the vectors $\v_1,\ldots,\v_k$ have non-negative components.
\end{proof}

\begin{exercise}\label{lastline}
Explain in more details the derivation in the last line of~\eqref{deri1}.
\end{exercise}

\begin{exercise}\label{ex5p3p11}
Show that for sufficiently large $r>0$, the inequality in~\eqref{ineq518} holds with $\v_1,\ldots,\v_k$ if and only if it holds with $\v_1',\ldots,\v_k'$, where for each $j\in[k]$, $\v_j'\eqdef\v_j+(r,r)^T\geq 0$. 
\end{exercise}

The next Lemma is a crucial simplification of the previous lemma. Specifically, we show that it is sufficient to take $k=2$ in Lemma~\ref{chararm2}.

\begin{myg}{}
\begin{lemma}\label{chararm3}
Using the same notations as in Lemma~\ref{chararm2}, the following statements are equivalent:
\begin{enumerate}
\item $(\p,\q)\succ(\p',\q')$.
\item For any two vectors $\v_1,\v_2\in\mbb{R}^2_{+}$
\be\label{ineq518b}
\sum_{x\in[n]}\max\{\r_x\cdot\v_1,\r_x\cdot\v_2\}\geq \sum_{y\in[m]}\max\{\r_y'\cdot\v_1,\r_y'\cdot\v_2\}\;.
\ee
\item $\mathfrak{M}(\p,\q,2)\supseteq\mathfrak{M}(\p',\q',2)$.
\end{enumerate}
\end{lemma}
\end{myg}

\begin{proof}
We start by proving the equivalency of 1 and 2. From Lemma~\ref{chararm2} it is sufficient to show that if~\eqref{ineq518b} holds then~\eqref{ineq518} holds for all $k\geq 3$ (the case $k=1$ is trivial). Let $\v_1,\ldots,\v_k\in\mbb{R}^2$ with $k\geq 3$. 
In order to prove the inequality in~\eqref{ineq518}, we first observe that the term
$\max_{z\in[k]}\{\r_x\cdot\v_z\}$ is the support function of the polytope $\mathfrak{C}=\conv(\v_1,\ldots,\v_k)$. We order the set of vertices $\{\v_1,\ldots,\v_k\}$ such that for any $x\in\{2,\ldots,k\}$ the vector $\v_x-\v_{x-1}$ is on the boundary of $\mc$ (see Fig.~\ref{zonotope}a).
Specifically, recall that the support function\index{support function} of $\mc$ is given by $f_{\mathfrak{C}}(\s)=\max_{x\in[k]}\s\cdot\v_x$ for all $\s\in\mbb{R}^2$. Therefore, the left-hand side of~\eqref{ineq518} can be expressed as
\be
\sum_{x\in[n]}\max_{z\in[k]}\{\r_x\cdot\v_z\}=\sum_{x\in[n]}f_{\mc}(\r_x)\;.
\ee
The key idea of the proof is to use the property of support functions under addition of sets (see Theorem~\ref{sumation}).
For this purpose, it would have been useful if it was possible to write $\mc$ as a sum of convex sets with each set in the sum being the convex hull of only two vectors (so that~\eqref{ineq518b} can be applied). While it is not possible to decompose $\mc$ in this way, we define now a set with the same support function\index{support function} as $\mc$, but for which such a decomposition is possible.

\begin{figure}[h]\centering
    \includegraphics[width=1\textwidth]{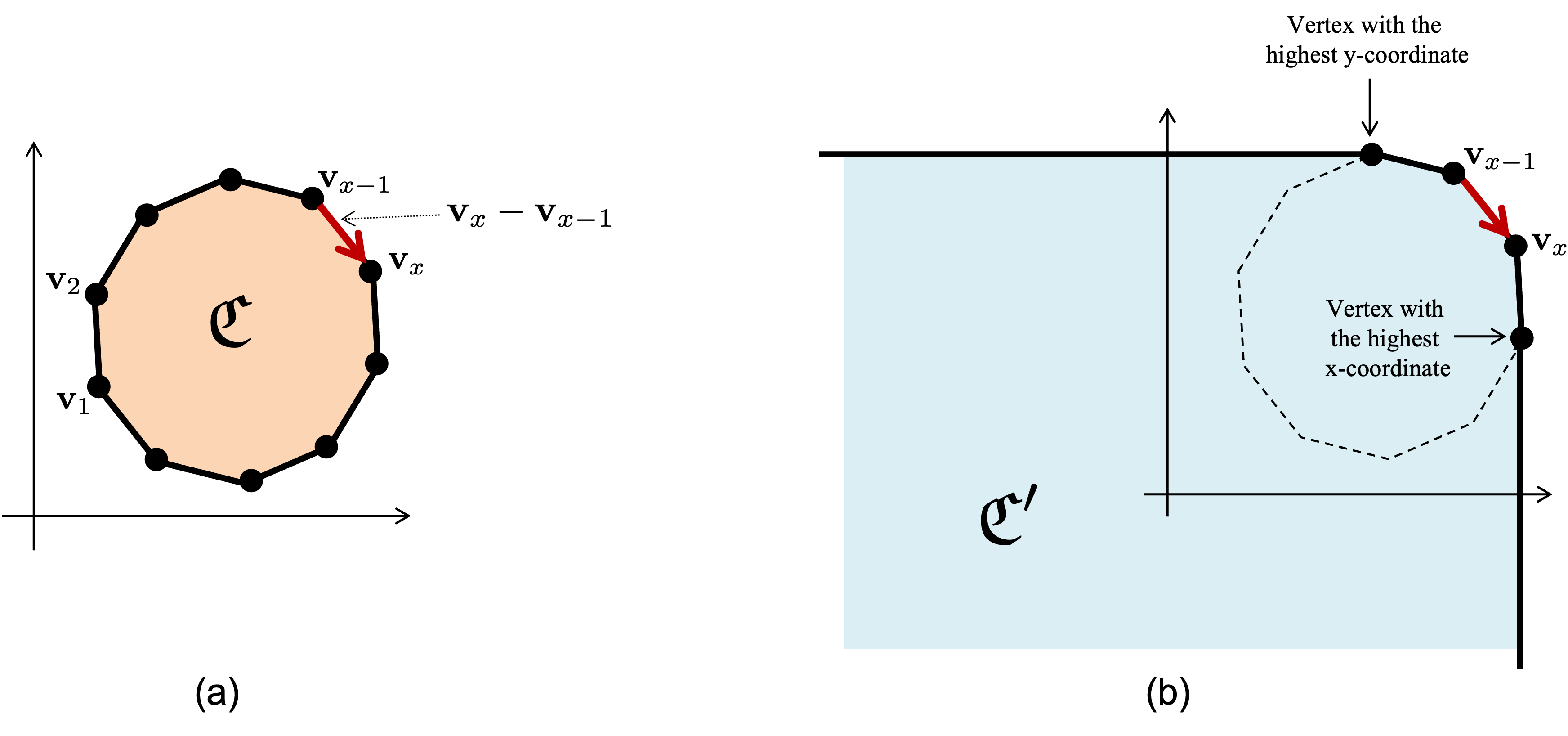}
  \caption{(a) The polytope $\mc$. The red arrow represents the vector $\v_x-\v_{x-1}$. (b) The polyhedron $\mc'\eqdef\mc-\mbb{R}^2_+$ is described with the blue area.}
  \label{zonotope}
\end{figure}

We define the desired set in two steps. First, we define the set \be
\mc'\eqdef\mc-\mbb{R}_+^2\eqdef\big\{\v-\p\;:\;\v\in\mc\;,\;\;\p\in\mbb{R}^2_+\big\}\;.
\ee 
That is, the set $\mc'$ is an unbounded polyhedron consisting of all the vectors $\r\in\mbb{R}^2$ for which there exists $\v\in\mc$ with the property that $\r\leq\v$. By definition, the support function\index{support function} of $\mc'$ equals that of $\mc$ (see Exercise~\ref{supfccp}). We will see now that the set $\mc'$ is a bit simpler to work with as it contains only a few of the vertices of $\mc$ (the ones that will be relevant for the computation of the support of $\mc$). 

In Fig.~\ref{zonotope}b we a depicted the set $\mc'$ and its relation to $\mc$. Observe that the set $\mc'$ is bounded by (1) the vertical line that passes through the vertex with the highest $x$-coordinate, (2) the horizontal line that  passes though the vertex with the highest $y$-coordinated, and (3) the portion of the boundary of $\mc$ that connects between the vertex with the highest $y$-coordinate and the one with the highest $x$-coordinate. In particular, observe that the set of all vertices of $\mc'$ is a subset of $\{\v_1,\ldots,\v_k\}$. For simplicity of notations, we take $\{\v_1,\ldots,\v_{k'}\}$ with $k'\leq k$, to be the set of vertices of $\mc'$. Moreover, observe that we can always arrange the set $\{\v_1,\ldots,\v_{k'}\}$ such that for each $x\in\{2,3,\ldots,k'-1\}$ the vector $\v_x$ is a ``neighbour" of $\v_{x-1}$ and $\v_{x+1}$. That is, each vector
$\v_x-\v_{x-1}$ is on the boundary of $\mc'$ and its angle with the $x$-axis is in the interval $[-\pi/2,0]$ (see Fig.~\ref{zonotope}b). Further, the angle of  $\v_x-\v_{x-1}$ with the $x$-axis is non-increasing in $x\in\{2,\ldots,k'\}$ (i.e., the angle becomes closer to $-\pi/2$ as $x$-increases).

With the above ordering of  the vertices $\{\v_1,\ldots,\v_{k'}\}$ we are now ready to construct the second convex set. Denote by $\v_0=\0$ the zero vector in $\mbb{R}^2$ and for each $y\in[k']$ let $\mk_y$ be the convex hull of the zero vector $\0\in\mbb{R}^2$ and the vector $\v_y-\v_{y-1}$. That is, for each $y\in[k']$ we can express $\mk_y=\{t(\v_y-\v_{y-1})\;:\;t\in[0,1]\}$. We then define
\ba
\mk&\eqdef\mk_1+\cdots+\mk_{k'}\\
&=\Big\{\sum_{x\in[k']}t_x(\v_x-\v_{x-1})\;:\;\t=(t_1,\ldots,t_{k'})^T\in[0,1]^{k'}\Big\}\;.
\ea
Clearly, the set $\mk$ is convex and its support function is given for any $\s\in\mbb{R}^2$ by
\be\label{maxins}
f_{\mk}(\s)=\max_{\t\in[0,1]^{k'}}\sum_{x\in[k']}t_x(\v_x-\v_{x-1})\cdot\s\;.
\ee
Now, recall that we are only interested in $\s\geq 0$, so that the angle of $\s$ with the $x$-axis is in the interval $[0,\pi/2]$. Therefore, the angle between $\s$ and the vector  $\v_x-\v_{x-1}$ is non-decreasing in $x$, and this angle is in the interval $[0,\pi]$. Therefore, there exists $\ell\in[k']$ such that the dot product $(\v_x-\v_{x-1})\cdot\s$ is non-negative for all $x\in[\ell]$ and it is negative for all $x\in\{\ell+1,\ldots,k'\}$. Therefore, the maximum in~\eqref{maxins} is obtained by taking $t_x=1$ for $x\in[\ell]$ and $t_x=0$ for $x\not\in[\ell]$. That is, we get that for $\s\geq 0$
\be
f_{\mk}(\s)=\sum_{x\in[\ell]}(\v_x-\v_{x-1})\cdot\s=\v_\ell\cdot\s\;.
\ee
On the other hand, observe that for \emph{all} $\ell'\in[k']$, by taking $t_x=1$ for $x\in[\ell']$ and $t_x=0$ for $x\not\in[\ell']$, Eq.~\eqref{maxins} gives
\be
f_{\mk}(\s)\geq\sum_{x\in[\ell']}(\v_x-\v_{x-1})\cdot\s=\v_{\ell'}\cdot\s\;.
\ee
From the two equations above we therefore conclude that 
that for all $\s\geq 0$ we have
\be
f_{\mk}(\s)=\max_{x\in[k']}\v_x\cdot\s=f_{\mc'}(\s)=f_{\mc}(\s)\;.
\ee
We therefore found a convex set $\mk$ that has the same support function as $\mc$ on vectors in $\mbb{R}^2_+$, and that can be expressed as $\mk=\mk_1+\cdots+\mk_{k'}$. From the property of support functions under addition of sets (see Theorem~\ref{sumation}), we therefore get that for all $\s\in\mbb{R}^2_+$
\be
f_{\mc}(\s)=f_{\mk}(\s)=\sum_{y\in[k']}f_{\mk_y}(\s)\;.
\ee
Combining everything, for each $k\geq 3$ we get
\be
\sum_{x\in[n]}\max_{z\in[k]}\{\r_x\cdot\v_z\}=\sum_{x\in[n]}f_{\mc}(\r_x)=\sum_{x\in[n]}f_{\mk}(\r_x)=\sum_{x\in[n]}\sum_{y\in[k']}f_{\mk_y}(\r_x)\;.
\ee
Now, from the definition of each $\mk_y$ we get
\ba
\sum_{x\in[n]}f_{\mk_y}(\r_x)&=\sum_{x\in[n]}\max\big\{0,\r_x\cdot(\v_y-\v_{y-1})\big\}\\
\GG{\eqref{ineq518b}}&\geq \sum_{x\in[n]}\max\big\{0,\r_x'\cdot(\v_y-\v_{y-1})\big\}=\sum_{x\in[n]}f_{\mk_y}(\r_x')
\ea
since we assume that~\eqref{ineq518} holds for $k=2$ (and for each fixed $y$ we can write the term $\max\big\{0,\r_x\cdot(\v_y-\v_{y-1})\big\}$ in the form $\max_{j\in[2]}\r_x\cdot \u_j$, where $\u_1\eqdef 0$ and $\u_2\eqdef \v_y-\v_{y-1}$). Combining this with the previous equation we conclude that
\ba
\sum_{x\in[n]}\max_{z\in[k]}\{\r_x\cdot\v_z\}&\geq \sum_{x\in[n]}\sum_{y\in[k']}f_{\mk_y}(\r_x')=\sum_{x\in[n]}f_{\mk}(\r_x')\\
&=\sum_{x\in[n]}f_{\mc}(\r_x')=\sum_{x\in[n]}\max_{z\in[k]}\{\r_x'\cdot\v_z\}\;.
\ea
Hence, \eqref{ineq518} holds for all $k\in\mbb{N}$, so that from Lemma~\ref{chararm2} we get that $(\p,\q)\succ(\p',\q')$.

To prove the equivalency of the second and third statements of the lemma,
recall that due to~\eqref{mpqk} the condition~\eqref{ineq518b} is equivalent to
\be\label{noy}
f_{\mathfrak{M}(\p,\q,2)}(S)\geq f_{\mathfrak{M}(\p',\q',2)}(S)\quad\quad\forall\; S\in\mbb{R}_+^{2\times 2}\;.
\ee
As argued below~\eqref{mpqk}, the condition~\eqref{noy}  holds if and only if it holds for all $S$ in $\mbb{R}^{2\times 2}$ (not necessarily $\mbb{R}_+^{2\times 2}$). We can therefore conclude from Theorem~\ref{inclusion} that the condition above is equivalent to $\mathfrak{M}(\p,\q,2)\supseteq\mathfrak{M}(\p',\q',2)$ so that the second and third statements of the lemma are equivalent.
 This completes the proof.
\end{proof}

\bex\label{supfccp}
Show that the support function of $\mc$ equals the support function of $\mc'$.
\eex

In the following exercise you simplify even further the expression given in~\eqref{ineq518b}.

\begin{exercise}\label{extwo}
Using the same notations as in Lemma~\ref{chararm2}, show that
\be\label{jkl}
(\p,\q)\succ (\p',\q')\quad\iff\quad\sum_{x\in[n]}\max\{0,\r_x\cdot\v\}\geq \sum_{y\in[m]}\max\{0,\r_y'\cdot\v\}\quad\forall\;\v\in\mbb{R}^2\;.
\ee 
Hint: Use the formula $\max\{a,b\}=\frac{a+b}{2}+\frac{|a-b|}2$ in~\eqref{ineq518b}, and recall that $\sum_x\r_x=(1,1)^T$.
\end{exercise}

\begin{proof}[\textbf{Alternative Proof of Theorem~\ref{chararm}}]
To see that 1 and 2 are equivalent, observe that
\ba
\mathfrak{M}(\p,\q,2)&\eqdef \Big\{[E\p\;E\q]\;:\;E\in\stoc(2,n)\Big\}\\
&=\left\{\begin{bmatrix}
\t\cdot \p\; & \;\t\cdot\q\\
1-\t\cdot\p\; & \;1-\t\cdot\q
\end{bmatrix}\;:\;\t\in[0,1]^n\right\}
\ea 
where $\t\in[0,1]^n$ is the first row of $E$, and $\textbf{1}_n-\t$ is its second row (recall that $E$ is a $2\times n$ column stochastic matrix).
Note that $(\t\cdot\p,\t\cdot\q)$ are precisely the elements of $\mt(\p,\q)$. Therefore, the inclusion $\mt(\p,\q)\supseteq\mt(\p',\q')$ 
is equivalent to the inclusion $\mathfrak{M}(\p,\q,2)\supseteq\mathfrak{M}(\p',\q',2)$. We already saw in Lemma~\ref{chararm3} that this latter inclusion is equivalent to $(\p,\q)\succ(\p',\q')$. Hence, we proved the equivalence of the first and second statement of the theorem.

Finally, it is left to show the equivalence between the first and third statement of the theorem. From Exercise~\eqref{extwo} it follows that $(\p,\q)\succ(\p',\q')$ is equivalent to the condition that for any $a,b\in\mbb{R}$ we have
\be
\sum_{x\in[n]}\max\{0,ap_x+bq_x\}\geq \sum_{y\in[m]}\max\{0,ap_y'+bq_y'\}
\ee
where we took $\v=(a,b)^T$ in~\eqref{jkl}. Using on both sides of the equation above the fact that for any $r\in\mbb{R}$, $\max\{0,r\}=\frac12\big(r+|r|\big)$ we conclude that the equation above is equivalent to the statement that
\be\label{132}
\sum_{x\in[n]}|ap_x+bq_x|\geq \sum_{y\in[m]}|ap_y'+bq_y'|\;,
\ee
where we removed from both sides the constant term \be\frac12\sum_{x\in[n]}(ap_x+bq_x)=\frac12\sum_{y\in[m]}(ap_y'+bq_y')=\frac12(a+b)\;.\ee
Finally, dividing both sides of~\eqref{132} by $a$ (we can assume without loss of generality that $a\neq 0$), and denoting $t\eqdef-b/a$ gives
\be
\sum_{x\in[n]}|p_x-tq_x|\geq \sum_{y\in[m]}|p_y'-tq_y'|\;.
\ee
This completes the proof of the equivalence between first and third statements of the theorem.
\end{proof}

\section{Symmetric Purification}

Let $\rho\in\md(A^n)$ be a density matrix that is symmetric under permutations of the $n$-subsystems. If $\rho^{A^n}$ is not a pure state then it can be purified. In Exercise~\ref{purification} we saw that all purifications of
$\rho^{A^n}$ are related by isometries. In this section we show that there exists a purification of $\rho^{A^n}$ that is itself symmetric under permutations, and satisfies in addition some useful properties.
We will denote by $\mG:\ml(A^n)\to\ml(A^n)$ the twirling map
\be
\mG_n(\omega^{A^n})\eqdef\frac1{n!}\sum_{\pi\in\mathbf{S}_n}P^{A^n}_{\pi^{-1}}\omega^{A^n}P^{A^n}_\pi\quad\quad\forall\;\omega\in\ml(A^n)\;.
\ee
In this section we say that a density matrix $\sigma_n\in\ml(A^n)$ is symmetric if $\mG_n(\sigma_n)=\sigma_n$ or equivalently if $\sigma_nP^{A^n}_\pi=P^{A^n}_\pi\sigma_n$ for all permutations $\pi\in\mathbf{S}_n$.

\begin{myt}{\color{yellow} Symmetric Uhlmann's Theorem}
\begin{theorem}\label{symuhl}
Let $\rho\in\md(A)$, $|\psi^{A\tA}\ra\eqdef I\otimes\sqrt{\rho}\big|\Omega^{A\tA}\ra$ be a purification of $\rho$, and $\sigma_n\in\md(A^n)$ be a symmetric density matrix. Then, there exists a symmetric purification  of $\sigma_n$, denoted by $\phi_n\in\md(A^n\tA^n)$, that satisfies
\be\label{flron}
F\left(\rho^{\otimes n},\sigma_n\right)=\left|\big\la\psi^{\otimes n}\big|\phi_n\big\ra\right|\;,
\ee
where $F$ is the fidelity.
\end{theorem}
\end{myt}

\begin{remark}
The symmetry of the state $\phi_n\in\md\big((A\tA)^n\big)$ is with respect to permutations among the $n$ copies of $A\tA$. That is, the state $\phi_n$ satisfies
\be
[\phi_n,P^{A^n}_\pi\otimes P^{\tA^n}_\pi]=0\quad\quad\forall\;\pi\in\mathbf{S}_n\;.
\ee
\end{remark}

\begin{proof}
From the polar decomposition\index{polar decomposition}  there exists a unitary matrix $U\in\ml(A^n)$ such that
\be\label{stdfg}
\sqrt{\sigma_n}\sqrt{\rho^{\otimes n}}=U\left|\sqrt{\sigma_n}\sqrt{\rho^{\otimes n}}\right|\;.
\ee
We then define
\be\label{defpni}
|\phi_n\ra\eqdef I^{A^n}\otimes\sqrt{\sigma_n}U|\Omega^{A^n\tA^n}\ra\;.
\ee
It is straightforward to check that this $\phi_n$ satisfies~\eqref{flron} (see Exercise~\ref{ex:lsdi}). 
Moreover, note that $\phi_n$ is indeed a purification of $\sigma_n$. It is therefore left to show that $\phi_n$ is symmetric.

For this purpose, we first show that $U$ can be taken to be symmetric. Since both matrices $\sqrt{\sigma_n}\sqrt{\rho^{\otimes n}}$ and $\left|\sqrt{\sigma_n}\sqrt{\rho^{\otimes n}}\right|$ are symmetric, from Theorem~\ref{deop} it follows that they can be expressed as
\be
\sqrt{\sigma_n}\sqrt{\rho^{\otimes n}}=\bigoplus_{\lambda} I^{B_\lambda}\otimes \eta^{C_\lambda}_{\lambda}\quad\text{and}\quad
\left|\sqrt{\sigma_n}\sqrt{\rho^{\otimes n}}\right|=\bigoplus_{\lambda} I^{B_\lambda}\otimes \zeta^{C_\lambda}_{\lambda}\;,
\ee
where $\eta^{C_\lambda}_{\lambda}$ and $\zeta^{C_\lambda}_{\lambda}$ are operators on the multiplicity space of the irrep $\lambda$. Define
\be
V\eqdef \sqrt{\sigma_n}\sqrt{\rho^{\otimes n}}\left|\sqrt{\sigma_n}\sqrt{\rho^{\otimes n}}\right|^{-1}=\bigoplus_{\lambda} I^{B_\lambda}\otimes \eta^{C_\lambda}_{\lambda}\left(\zeta^{C_\lambda}_{\lambda}\right)^{-1}
\ee
where all inverses are generalized inverses. Since $V$ is a partial isometry\index{partial isometry} (see Exercise~\ref{ex:piso}) it follows that each $\eta^{C_\lambda}_{\lambda}\left(\zeta^{C_\lambda}_{\lambda}\right)^{-1}$ is a partial isometry. Therefore, we can complete each $\eta^{C_\lambda}_{\lambda}\left(\zeta^{C_\lambda}_{\lambda}\right)^{-1}$ to a unitary matrix $U_\lambda\in\ml(C_\lambda)$. Defining
\be
U\eqdef \bigoplus_{\lambda} I^{B_\lambda}\otimes U^{C_\lambda}_{\lambda}
\ee
we get that $U$ is symmetric and satisfies~\eqref{stdfg}. Finally, since $U$ is symmetric we get that for all $\pi\in\mathbf{S}_n$
\ba
P^{A^n}_\pi\otimes P^{\tA^n}_\pi|\phi_n^{A^n\tA^n}\ra&=P^{A^n}_\pi\otimes P^{\tA^n}_\pi\sqrt{\sigma_n}U|\Omega^{A^n\tA^n}\ra\\
\GG{\sigma_{\it n}\;is\;symmetric}&=P^{A^n}_\pi\otimes \sqrt{\sigma_n}P^{\tA^n}_\pi U|\Omega^{A^n\tA^n}\ra\\
\GG{{\it U}\;is\;symmetric\to}\;&=P^{A^n}_\pi\otimes \sqrt{\sigma_n}UP^{\tA^n}_\pi|\Omega^{A^n\tA^n}\ra\\
\GG{\Omega^{\it A^n\tA^n}\;is\;symmetric}&=I^{A^n}_\pi\otimes \sqrt{\sigma_n}U|\Omega^{A^n\tA^n}\ra\\
&=|\phi_n^{A^n\tA^n}\ra\;.
\ea
Hence, $\phi_n$ is symmetric.
\end{proof}

\bex\label{ex:lsdi}
Show explicitly that $\phi_n$, as defined in~\eqref{defpni} satisfies~\eqref{flron}
\eex

\bex\label{ex:piso}
Let $\Lambda\in\ml(A)$ and define $V=\Lambda|\Lambda|^{-1}$ where the inverse is a generalized inverse. Show that $V$ is a partial isometry.
\eex

\end{appendix}

{\pagestyle{plain}
\bibliography{QRT}
\cleardoublepage}

\printindex
	
\end{document}